\let\mypdfximage\pdfximage
\def\pdfximage{\immediate\mypdfximage}

\documentclass[
 twocolumn,
 superscriptaddress,
 showpacs,
 amsmath,
 amssymb,
 aps,
]{revtex4-1}

\usepackage[dvipdfmx]{graphicx}
\usepackage{dcolumn}
\usepackage{bm}

\usepackage{newfloat}
\usepackage{etoolbox}

\DeclareMathOperator*{\argmin}{arg\,min}

\usepackage{mathtools}

\usepackage{amsthm}
\newtheorem{theorem}{Theorem}

\usepackage{subcaption}

\usepackage{xcolor}

\usepackage{diagbox}
\usepackage{multirow}

\usepackage{algcompatible}
\usepackage[size=small]{caption}
\DeclareCaptionFormat{algorithm}{\hrulefill\vskip-8pt\hrulefill\par#1#2#3\vskip-4pt\hrulefill\vskip-8pt}
\usepackage[size=small]{caption}
\AtEndEnvironment{algorithm}{\vskip-8pt\noindent\hrulefill\par\nobreak\vskip-8pt\hrulefill}
\DeclareFloatingEnvironment[
  fileext=loa,
  listname=List of Algorithms,
  name=ALGORITHM,
  placement=tbhp,
]{algorithm}

\usepackage{braket}

\newcommand{\vast}{\bBigg@{3.0}}
\newcommand{\Vast}{\bBigg@{3.5}}

\allowdisplaybreaks

\usepackage{hyperref}

\usepackage{MnSymbol}

\usepackage{accents}

\usepackage{slashed}

\usepackage{tikz}
\usetikzlibrary{quantikz}

\usepackage{soul}

\extrafloats{200}
\begin{document}


\title{Ansatz-Independent Variational Quantum Classifier}

\author{Hideyuki Miyahara}

\email{miyahara@g.ucla.edu, hmiyahara512@gmail.com}

\affiliation{
Department of Electrical and Computer Engineering,
Samueli School of Engineering,
University of California, Los Angeles, California 90095
}

\author{Vwani Roychowdhury}

\email{vwani@g.ucla.edu}

\affiliation{
Department of Electrical and Computer Engineering,
Samueli School of Engineering,
University of California, Los Angeles, California 90095
}

\date{\today}

\begin{abstract}
The paradigm of variational quantum classifiers (VQCs) encodes \textit{classical information} as quantum states, followed by quantum processing and then measurements to generate classical predictions.
VQCs are promising candidates for efficient utilization of a near-term quantum device: classifiers involving $M$-dimensional datasets can be implemented with only $\lceil \log_2 M \rceil$ qubits by using an amplitude encoding.
A general framework for designing and training VQCs, however, has not been proposed, and a fundamental understanding of its power and analytical relationships with classical classifiers are not well understood.
An encouraging specific embodiment of VQCs, quantum circuit learning (QCL), utilizes an ansatz: it expresses the quantum evolution operator as a circuit with a predetermined topology and parametrized gates; training involves learning the gate parameters through optimization.
In this letter, we first address the open questions about VQCs and then show that they, including QCL, fit inside the well-known kernel method.
Based on such correspondence, we devise a design framework of efficient ansatz-independent VQCs, which we call the unitary kernel method (UKM): it directly optimizes the unitary evolution operator in a VQC.
Thus, we show that the performance of QCL is bounded from above by the UKM.
Next, we propose a variational circuit realization (VCR) for designing efficient quantum circuits for a given unitary operator. By combining the UKM with the VCR, we establish an efficient framework for constructing high-performing circuits.
We finally benchmark the relatively superior performance of the UKM and the VCR via extensive numerical simulations on multiple datasets.
\end{abstract}

\maketitle

\begin{figure*}[tp]
\centering
\includegraphics[scale=0.50]{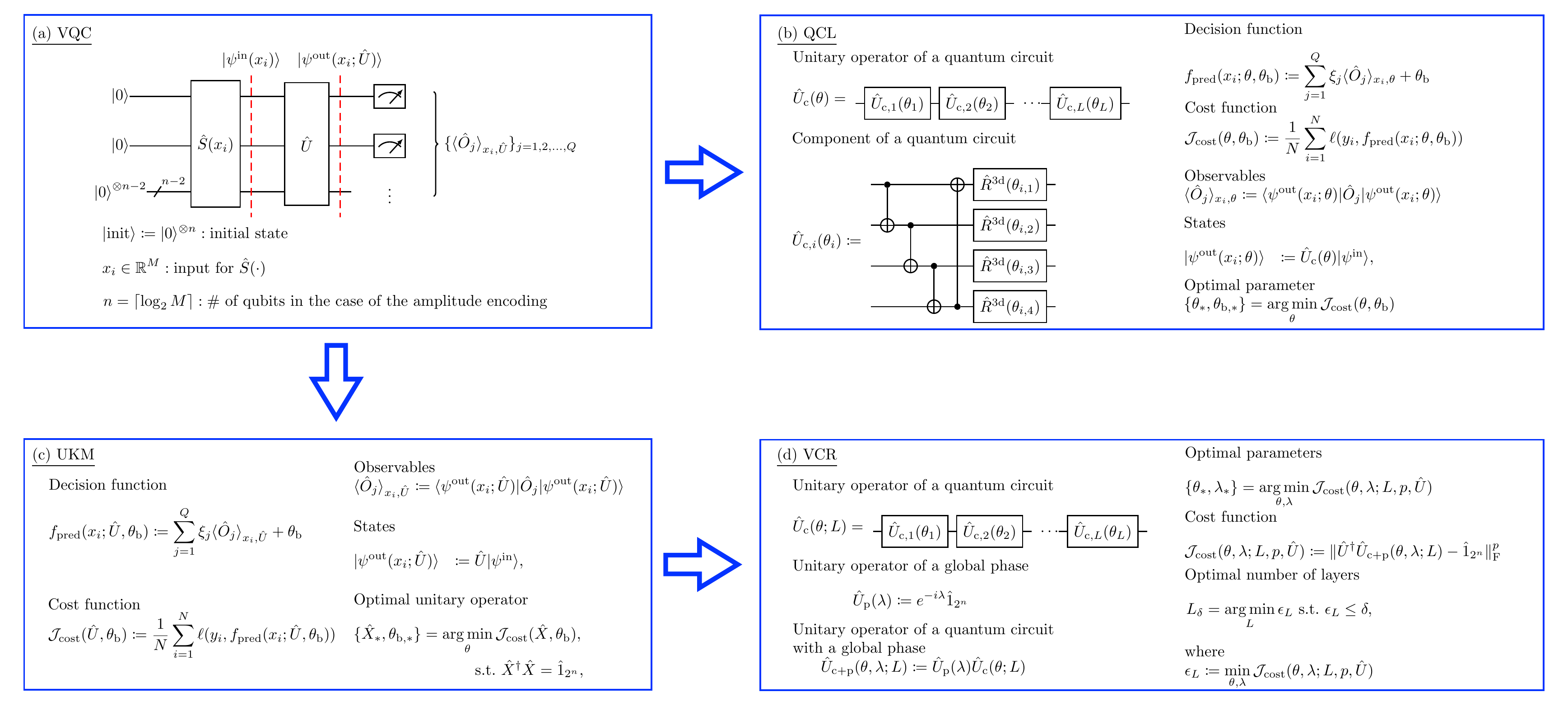}
\caption{Schematic of the algorithms discussed in this letter: (a) A general form of a hybrid quantum-classical classifier, which we refer to as a VQC, (b) QCL, (c) UKM, and (d) VCR. (a) In the architecture of a VQC, the initial state is $| \mathrm{init} \rangle \coloneqq | 0 \rangle^{\otimes n}$. We first encode a given classical vector $x_i$: $| \psi^\mathrm{in} (x_i) \rangle \coloneqq \hat{S} (x_i) | \mathrm{init} \rangle$. As mentioned in the discussions, one can embed $x_i$ into a higher dimensional vector $\phi(x_i) \in \mathbb{R}^L$ with $L= \mathcal{O} (M^c)$ and then use the rest of the framework; the number of qubits $n$ will still be $\mathcal{O} (\log M)$, thus retaining any potential quantum advantage. Second, we apply $\hat{U}$: $| \psi^\mathrm{out} (x_i; \hat{U}) \rangle \coloneqq \hat{U} | \psi^\mathrm{in} (x_i) \rangle$. Third, we perform measurements with respect to $\{ \hat{O}_j \}_{j=1, 2, \dots, Q}$. Finally, we make a prediction on the label of $x_i$ by using the outputs of the measurements. (b) In QCL, we assume a circuit geometry parameterized by $\theta$ for $\hat{U}$: $\hat{U}_\mathrm{c} (\theta)$. In most cases, a circuit used for QCL is composed of single- and two-qubit operators and has a layered structure. A typical example is shown. (c) In the UKM, we directly optimize $\hat{U}$. (d) In the VCR, we decompose a unitary operator into a quantum circuit by assuming a layered structure for a quantum circuit. For a circuit realization, a simpler circuit is preferable; so, we explicitly denote the number of layers $L$.}
\label{main-arXiv-schematic-quantum-classifer}
\end{figure*}

\textit{Introdction.}---
Since the discovery of Shor's algorithm~\cite{Shor001}, much effort has been devoted to the development of quantum algorithms and quantum computers~\cite{Arute001}.
To exploit a near-term quantum device, several variational quantum algorithms~\cite{Cerezo01} have been proposed, including the quantum approximate optimization algorithm~\cite{Farhi001} and the variational quantum eigensolver~\cite{McClean001}.
Then, quantum circuit learning (QCL) was proposed in Refs.~\cite{Mitarai001, Schuld001} and is now considered to be a promising candidate to utilize a near-term quantum device efficiently for application.
QCL itself, however, is a special case of a larger set of hybrid quantum-classical classifiers -- a class that we refer to as variational quantum classifiers (VQCs) -- since it assumes an ansatz, where the circuit topology is fixed and only the gates are parameterized; See Fig.~\ref{main-arXiv-schematic-quantum-classifer}.
Thus, several questions remain unanswered, including (i) can one get better performance than QCL by systematically designing an ansatz-independent VQC, and (ii) given that a VQC and QCL perform end-to-end classical machine learning (ML) tasks, are they related to any well-known classical ML algorithms and if the classical counterparts can perform better.

In this letter, we first discuss the correspondence between a VQC and the well-known kernel method~\cite{Bishop001, Murphy001}.
Then we propose an ansatz-independent VQC, which we call the unitary kernel method (UKM).
By using the UKM, we show that the performance of QCL is bounded from above by the UKM.

To effectively use the unitary operator obtained by the UKM, we propose a unitary decomposition method to create a circuit geometry, which we call the variational circuit realization (VCR).
By combining the UKM and the VCR, we can efficiently construct a circuit geometry that works well for classification problems.

Fig.~\ref{main-arXiv-schematic-quantum-classifer} presents a schematic of a general VQC, introduces and compares QCL and the UKM and explains the role of the VCR.

\textit{Variational quantum classifier.}---
We first introduce an analytical formalism for a VQC.
Suppose that we are given a $n$-qubit system and a classical dataset $\mathcal{D} \coloneqq \{ x_i, y_i \}_{i=1}^N$, where $x_i \in \mathbb{R}^M$ is a feature vector and $y_i \in \{1, -1\}$ is the corresponding label for $i = 1, 2, \dots, N$.
In this letter, we consider the amplitude encoding~\cite{Schuld001}; thus we fix $n = \lceil \log_2 M \rceil$.
As mentioned in the discussions, one can embed $x_i$ into a higher dimensional vector $\phi(x_i) \in \mathbb{R}^L$ with $L= \mathcal{O} (M^c)$ and then use the rest of the framework; the number of qubits $n$ will still be $\mathcal{O} (\log M)$, thus retaining any potential quantum advantage.
Here, $\mathcal{O} (\cdot)$ is the big-O notation and $c$ is a certain constant.
In this letter, however, we stick to the amplitude encoding case.
A schematic of a VQC with amplitude encoding of $x_i$'s is shown in Fig.~\ref{main-arXiv-schematic-quantum-classifer}(a).
Let us consider making a prediction on $y_i$ by the following function:
\begin{align}
  f_\mathrm{pred} (x_i; \hat{U}, \theta_\mathrm{b}) &\coloneqq \sum_{j=1}^Q \xi_j \langle \hat{O}_j \rangle_{x_i, \hat{U}} + \theta_\mathrm{b}, \label{main-arXiv-prediction-UKM-001-001}
\end{align}
where
$| \psi^\mathrm{in} (x_i) \rangle \coloneqq \hat{S} (x_i) | \mathrm{init} \rangle$, $| \psi^\mathrm{out} (x_i; \hat{U}) \rangle \coloneqq \hat{U} | \psi^\mathrm{in} (x_i) \rangle$, and $\langle \hat{O}_j \rangle_{x_i, \hat{U}} \coloneqq \langle \psi^\mathrm{out} (x_i; \hat{U}) | \hat{O}_j | \psi^\mathrm{out} (x_i; \hat{U}) \rangle$.
While  $\{ \xi_j \}_{j=1}^Q$ can also be learned and optimized, the convention in \cite{Mitarai001, Schuld001} is to treat them a fixed parameter and $\theta_\mathrm{b}$ is a bias term to be estimated.

In a VQC, we estimate $\hat{U}$ and $\theta_\mathrm{b}$ in Eq.~\eqref{main-arXiv-prediction-UKM-001-001} imposing the unitarity constraint on $\hat{U}$ as follows:
\begin{align}
  \{ \hat{U}_*, \theta_{\mathrm{b}, *} \} &= \argmin_{\hat{U}, \theta_\mathrm{b}} \mathcal{J}_\mathrm{cost} (\hat{U}, \theta_\mathrm{b}), \nonumber \\
  & \quad \mathrm{subject \ to} \ \hat{U}^\dagger \hat{U} = \hat{1}_{2^n}, \label{main-arXiv-VQC-optimization-001}
\end{align}
where
\begin{align}
  \mathcal{J}_\mathrm{cost} (\hat{U}, \theta_\mathrm{b}) &\coloneqq \frac{1}{N} \sum_{i=1}^N \ell (y_i, f_\mathrm{pred} (x_i; \hat{U}, \theta_\mathrm{b})). \label{main-arXiv-cost-function-UKM-001-001}
\end{align}
Here, $\ell (\cdot, \cdot)$ is a loss function, such as the mean-squared error function or the hinge function~\cite{Bishop001, Murphy001}, and $\hat{1}_n$ is the $n$-dimensional identity operator.
As explained later, we consider a parameterized unitary operator and optimize it in QCL and we directly optimize the unitary operator in the UKM.

\textit{Correspondence between a VQC and the kernel method.}---
Next, we discuss the relationship between a VQC and the kernel method~\cite{Bishop001, Murphy001}.
Introducing $\psi_l^\mathrm{in} (x_i) \coloneqq \langle l | \psi^\mathrm{in} (x_i) \rangle$, $O_j^{k, l} \coloneqq \langle k | \hat{O}_j | l \rangle$, and $u_{k, l} \coloneqq \langle k | \hat{U} | l \rangle$ for $k, l = 1, 2, \dots, 2^n$, $\langle \hat{O}_j \rangle_{x_i, \hat{U}}$ is rewritten as
\begin{align}
  \langle \hat{O}_j \rangle_{x_i, \hat{U}} &= \sum_{\substack{k_1, k_2, \\ k_3, k_4}} [ \psi_{k_1}^\mathrm{in} (x_i) ]^* u_{k_1, k_2}^* O_j^{k_2, k_3} u_{k_3, k_4} \psi_{k_4}^\mathrm{in} (x_i). \label{main-arXiv-observable-rewriten-001-001}
\end{align}
Then, by using the penalty method~\cite{Boyd001, Fletcher001}, we can rewrite a VQC as
\begin{align}
  \lim_{\lambda \to \infty} \min_{\{ u_{k, l} \}_{k, l = 1}^{2^n}, \theta_\mathrm{b}} \mathcal{J}_\mathrm{pm} (\{ u_{k, l} \}_{k, l = 1}^{2^n}, \theta_\mathrm{b}; \lambda),
\end{align}
where
\begin{align}
  & \mathcal{J}_\mathrm{pm} (\{ u_{k, l} \}_{k, l = 1}^{2^n}, \theta_\mathrm{b}; \lambda) \nonumber \\
  & \quad \coloneqq \frac{1}{N} \sum_{i=1}^N \ell \bigg( y_i, \sum_{j=1}^Q \xi_j \tilde{O}_j (\{ u_{k, l} \}_{k, l = 1}^{2^n}) + \theta_\mathrm{b} \bigg) \nonumber \\
  & \qquad + \frac{\lambda}{2} \sum_{k, l = 1, 2, \dots, 2^n} \big[ u_k^\mathrm{H} u_l - \delta_{k, l} \big]^2, \label{main-arXiv-cost-QCL-002-001}
\end{align}
and we have denoted the right-hand side of Eq.~\eqref{main-arXiv-observable-rewriten-001-001} by $\tilde{O}_j (\{ u_{k, l} \}_{k, l = 1}^{2^n})$.
Here, $u_k \coloneqq [u_{1, k}, u_{2, k}, \dots, u_{2^n, k}]^\mathrm{H}$ for $k = 1, 2, \dots, 2^n$ and $(\cdot)^\mathrm{H}$ is the Hermitian conjugate.

Then, we turn our attention to the kernel method~\cite{Bishop001, Murphy001}.
In the conventional kernel method, we make a prediction on $y_i$ by
\begin{align}
  f_\mathrm{pred} (x; v) &\coloneqq v^\intercal \phi (x),
\end{align}
where $\phi (\cdot) \coloneqq [\phi_1 (\cdot), \phi_2 (\cdot), \dots]^\intercal$ and $v \coloneqq [v_1, v_2, \dots]^\intercal$ are the vector of feature maps and a real vector, respectively.
We minimize the following function to determine $v$:
\begin{align}
  \mathcal{J}_\mathrm{cost} (v) &\coloneqq \frac{1}{N} \sum_{i=1}^N \ell (y_i,f_\mathrm{pred} (x_i; v)) + \frac{\lambda}{2} \| v \|_\mathrm{F}^2. \label{main-arXiv-cost-kernel-001-001}
\end{align}
Here, the second term in the right-hand side of Eq.~\eqref{main-arXiv-cost-kernel-001-001} is the regularization term and $\| \cdot \|_\mathrm{F}$ is the Frobenius norm.
To clarify the correspondence between a VQC and the kernel method, we consider a modified version of the kernel method; that is, let us consider the following functions as the classifier and the cost function, respectively:
\begin{align}
  f_\mathrm{pred} (x; w) &\coloneqq \sum_{k, l} w_{k, l} \tilde{\phi}_k (x) \tilde{\phi}_l (x), \\
  \mathcal{J}_\mathrm{cost} (w) &\coloneqq \frac{1}{N} \sum_{i=1}^N \ell (y_i,f_\mathrm{pred} (x_i; w)) + \frac{\lambda}{2} | w_{k, l} |^2, \label{main-arXiv-cost-quad-001-001}
\end{align}
where $\{ w_{k, l} \}_{k, l}$ are complex variables.
Comparing Eqs.~\eqref{main-arXiv-cost-QCL-002-001} with $\theta_\mathrm{b} = 0$ and \eqref{main-arXiv-cost-quad-001-001}, a VQC and the kernel method have the following relationships:
\begin{subequations}
\begin{align}
  \psi_l^\mathrm{in} (x_i) &\Leftrightarrow \tilde{\phi}_l (x_i), \\
  \sum_j \sum_{k_1, k_2} \xi_j u_{k, k_1}^* O_j^{k_1, k_2} u_{k_2, l} &\Leftrightarrow w_{k, l}.
\end{align}
\label{main-arXiv-correspondence-QCL-kernel-method-001-001}%
\end{subequations}
Eq.~\eqref{main-arXiv-correspondence-QCL-kernel-method-001-001} provides us a clear relationship between a VQC and the kernel method~\footnote{The kernel method is discussed in Sec.~\ref{supp-arXiv-sec-kernel-001} of the SI and the relationship between a VQC and the kernel method is discussed in Sec.~\ref{supp-arXiv-sec-correpondence-QCL-kernel-001} of the SI in detail.} and implicitly states that a VQC is a subset of the kernel method.
Furthermore, Eq.~\eqref{main-arXiv-correspondence-QCL-kernel-method-001-001} motivates us to devise the UKM, which lies between a VQC and the kernel method.

\textit{Quantum circuit learning.}---
Here, we review QCL proposed in Refs.~\cite{Schuld001, Mitarai001} from the viewpoint of a VQC.
The schematic of QCL is demonstrated in Fig.~\ref{main-arXiv-schematic-quantum-classifer}(b).
In QCL, we assume a parameterized unitary operator $\hat{U}_\mathrm{c} (\theta)$ as $\hat{U}$ and optimize $\theta$~\footnote{Refer to Sec.~\ref{supp-arXiv-sec-quantum-circuit-001-001} of the SI for the details of quantum circuits.}.
We then compute $| \psi^\mathrm{out} (x_i; \theta) \rangle \coloneqq \hat{U}_\mathrm{c} (\theta) | \psi^\mathrm{in} (x_i) \rangle$.
Then, we make a prediction on $x_i$ by
\begin{align}
  f_\mathrm{pred} (x_i; \theta, \theta_\mathrm{b}) &\coloneqq \sum_{j=1}^Q \xi_j \langle \hat{O}_j \rangle_{x_i, \theta} + \theta_\mathrm{b}, \label{main-arXiv-prediction-QCL-001-001}
\end{align}
where
$\langle \hat{O}_j \rangle_{x_i, \theta} \coloneqq \langle \psi^\mathrm{out} (x_i; \theta) | \hat{O}_j | \psi^\mathrm{out} (x_i; \theta) \rangle$.
Similarly to Eq.~\eqref{main-arXiv-prediction-UKM-001-001}, $\{ \xi_j \}_{j=1}^Q$ are fixed parameters and $\theta_\mathrm{b}$ is a bias term to be estimated.
The second step of QCL is to update $\theta$ and $\theta_\mathrm{b}$ by
\begin{align}
   \{ \theta_*, \theta_{\mathrm{b}, *} \} &= \argmin_{\theta, \theta_\mathrm{b}} \mathcal{J}_\mathrm{cost} (\theta, \theta_\mathrm{b}),
\end{align}
where
\begin{align}
  \mathcal{J}_\mathrm{cost} (\theta, \theta_\mathrm{b}) &\coloneqq \frac{1}{N} \sum_{i=1}^N \ell (y_i, f_\mathrm{pred} (x_i; \theta, \theta_\mathrm{b})), \label{main-arXiv-cost-QCL-001-001}
\end{align}
and $\ell (\cdot, \cdot)$ is a loss function~\footnote{For details, refer to Sec.~\ref{supp-arXiv-sec-QCL-001-001} of the SI.}.
For this purpose, we often use the Nelder-Mead method~\cite{Nelder001} and other sophositicated numerical methods~\cite{Boyd001, Fletcher001}.

As mentioned above, QCL assumes a parameterized unitary operator $\hat{U}_\mathrm{c} (\theta)$; thus, its performance heavily depends on the topology of $\hat{U}_\mathrm{c} (\theta)$.
An assumed circuit topology is also called an ansatz; thus we can say that QCL is an ansatz-dependent VQC.
This fact strongly motivates us to devise an ansatz-independent VQC, that is, the UKM.
Furthermore, Ref.~\cite{McClean002} pointed out the difficulty of learning parameters of quantum circuits, which they call the barren plateaus problem.
Then, a VQC that is free of the barren plateaus problem is of interest.

\textit{Method of splitting orthogonal constraints.}---
The UKM heavily relies on the method of splitting orthogonal constraints (SOC); then, we review the method of SOC~\cite{Lai001}.
The method of SOC is a method for solving an optimization problem under an orthogonal constraints:
\begin{subequations}
\begin{align}
  \min_X \ & \mathcal{J}_\mathrm{cost} (X), \\
  \mathrm{subject \ to} \ & X^\intercal A X = I,
\end{align}
\label{main-arXiv-problem-SOC-matrix-001-001}%
\end{subequations}
where $A \succ O$.
Here, $I$ and $O$ are the identity matrix and the zero matrix, respectively.
By splitting the constraints, we first rewrite Eq.~\eqref{main-arXiv-problem-SOC-matrix-001-001} as
\begin{subequations}
\begin{align}
  \min_{X, P} \ & \mathcal{J}_\mathrm{cost} (X), \\
  \mathrm{subject \ to} \ & L X = P, \\
  & P^\intercal P = I,
\end{align}
\label{main-arXiv-problem-SOC-matrix-002-001}%
\end{subequations}
where $L$ is a matrix that satisfies
\begin{align}
  A &= L^\intercal L.
\end{align}
Then we solve Eq.~\ref{main-arXiv-problem-SOC-matrix-002-001} by
\begin{subequations}
\begin{align}
  \{ X_k, P_k \} &= \argmin_{X, P} \mathcal{J}_\mathrm{SOC} (X; P, D_{k-1}), \nonumber \\
  & \quad \mathrm{subject \ to} \ P^\intercal P = I, \label{main-arXiv-algorithm-SOC-001-011} \\
  D_k &= D_{k-1} + L X_k - P_k,
\end{align}
\label{main-arXiv-algorithm-SOC-001-001}%
\end{subequations}
where
\begin{align}
 \mathcal{J}_\mathrm{SOC} (X; P, D) &\coloneqq \mathcal{J}_\mathrm{cost} (X) + \frac{r}{2} \| L X - P + D \|_\mathrm{F}^2. \label{main-arXiv-cost-SOC-001-001}
\end{align}
Here, $r$ is a positive constant.
The performance of the method of SOC depends on $r$; thus, we need to find an appropriate value of $r$.
In the  derivation of Eq.~\eqref{main-arXiv-cost-SOC-001-001}, the idea of Bregrman iterative regularization~\cite{Osher001, Yin001} is used~\footnote{See Sec.~\ref{supp-arXiv-sec-BIR-001} in the SI for details.}.

It is still difficult to perform the computation of Eq.~\eqref{main-arXiv-algorithm-SOC-001-011}.
By splitting it and using operator unitarization (OU), which is explained later, we then have the following formula:
\begin{subequations}
\begin{align}
  X_k &= \argmin_X \mathcal{J}_\mathrm{SOC} (X; P_{k-1}, D_{k-1}), \label{main-arXiv-algorithm-SOC-002-011} \\
  P_k &= \argmin_P \frac{r}{2} \| P - (L X_k + D_{k-1}) \|_\mathrm{F}^2, \nonumber \\
  & \quad \mathrm{subject \ to} \ P^\intercal P = I, \label{main-arXiv-algorithm-SOC-002-012} \\
  D_k &= D_{k-1} + L X_k - P_k. \label{main-arXiv-algorithm-SOC-002-013}
\end{align}
\label{main-arXiv-algorithm-SOC-002-001}%
\end{subequations}
Eq.~\eqref{main-arXiv-algorithm-SOC-002-001} is the final form of the method of SOC~\footnote{See Sec.~\ref{supp-arXiv-sec-SOC-001-001} for details.}.
Note that in the original paper of the method of SOC~\cite{Lai001}, only real matrices are considered; however, in the UKM, we utilize its complex version.
The complex version of the method of SOC follows directly by noting that the SVD is also applicable to a complex matrix and the product of unitary matrices leads to a unitary matrix.
Furthermore, in the original version of the method of SOC, we optimize a real-valued function of a real matrix.
To optimize a real-valued function of a complex matrix, we optimize the real and complex parts of the complex matrix independently~\footnote{See Sec.~\ref{supp-arXiv-sec-UKM-001-002} for details.}.

\textit{Operator unitarization.}---
OU is used in the method of SOC and will be used to obtain a unitary operator from a non-unitary operator.
We briefly explain OU here.
Let us consider the following optimization problem:
\begin{align}
  \hat{P}_* &= \argmin_{\hat{P}} \frac{1}{2} \| \hat{P} - \hat{Y} \|, \nonumber \\
  & \quad \mathrm{subject \ to} \ \hat{P}^\dagger \hat{P} = \hat{1}. \label{main-arXiv-minimization-on-P-Y-complex-001-001}
\end{align}
The solution of Eq.~\eqref{main-arXiv-minimization-on-P-Y-complex-001-001} is given by~\cite{Lai001}
\begin{align}
\hat{P}_* &= \hat{K}_1 \hat{K}_2^\dagger, \label{main-arXiv-OU-main-001-001}
\end{align}
where $\hat{K}_1$ and $\hat{K}_2^\dagger$ are unitary operators that satisfy
\begin{align}
  \hat{Y} = \hat{K}_1 \hat{\Sigma} \hat{K}_2^\dagger, \label{main-arXiv-OU-sub-001-001}
\end{align}
and $\hat{\Sigma}$ is a diagonal operator in the sense that $\{ \langle i | \hat{\Sigma} | i \rangle \}_i$ are the singular values of $\hat{Y}$ and $\langle i | \hat{\Sigma} | j \rangle = 0$ for $i \ne j$.
In this letter, we call Eq.~\eqref{main-arXiv-OU-main-001-001} with Eq.~\eqref{main-arXiv-OU-sub-001-001} OU~\footnote{See Secs.~\ref{supp-arXiv-sec-SOC-001-001} and \ref{supp-arXiv-sec-UKM-001-001} of the SI for details.}.

\textit{Unitary kernel method.}---
We here describe the UKM, which is one of the main algorithms in this letter.
The schematic of the UKM is demonstrated in Fig.~\ref{main-arXiv-schematic-quantum-classifer}(c).
In the UKM, we directly minimize Eq.~\eqref{main-arXiv-cost-function-UKM-001-001}.
To this end, we employ the unitary version of the method of SOC~\cite{Lai001}.
Hereafter, we denote, by $\hat{X}$, an operator obtained via the method of SOC.
We introduce $\hat{P}$ and $\hat{D}$ and iterate update equations for $\hat{X}$, $\hat{P}$, and $\hat{D}$ until convergence.
Hereafter, we denote, by $\hat{X}_k$, $\hat{P}_k$, $\hat{D}_k$, and $\theta_{\mathrm{b}, k}$, $\hat{X}$, $\hat{P}$, $\hat{D}$, and $\theta_\mathrm{b}$ at the $k$-th iteration, respectively.
At the first step of the $k$-th iteration, we compute $\hat{X}_k$ and $\theta_{\mathrm{b}, k}$ by
\begin{align}
& \{ \hat{X}_k, \theta_{\mathrm{b}, k} \} \nonumber \\
& \quad = \argmin_{\hat{X}, \theta_\mathrm{b}} \bigg[ \mathcal{J}_\mathrm{cost} (\hat{X}, \theta_\mathrm{b}) + \frac{r}{2} \| \hat{X} - \hat{P}_{k-1} + \hat{D}_{k-1} \|_\mathrm{F}^2 \bigg]. \label{main-arXiv-quantum-kernel-method-001-011}
\end{align}
Next, we compute $\hat{P}_k$ by
\begin{align}
  \hat{P}_k &= \hat{K}_{1, k} \hat{K}_{2, k}^\dagger, \label{main-arXiv-quantum-kernel-method-001-013}
\end{align}
where $\hat{K}_{1, k}$ and $\hat{K}_{2, k}^\dagger$ are unitary operators that satisfy $\hat{K}_{1, k} \hat{\Sigma}_k \hat{K}_{2, k}^\dagger = \hat{X}_k + \hat{D}_{k-1}$ and $\hat{\Sigma}_k$ is a diagonal operator.
At the end of the $k$-th iteration, we compute
\begin{align}
  \hat{D}_k &= \hat{D}_{k-1} + \hat{X}_k - \hat{P}_k. \label{main-arXiv-quantum-kernel-method-001-014}
\end{align}
We repeat the above equations, Eq.~\eqref{main-arXiv-quantum-kernel-method-001-011}, \eqref{main-arXiv-quantum-kernel-method-001-013}, and \eqref{main-arXiv-quantum-kernel-method-001-014}, until convergence.
We call this method the UKM.
In Algo.~\ref{main-arXiv-quantum-kernel-method-001-001}, the UKM is summarized~\footnote{For the details of the UKM, refer to Sec.~\ref{supp-arXiv-sec-UKM-001-001} of the SI.}.
\begin{algorithm}[t]
\caption{Unitary kernel method (UKM)} \label{main-arXiv-quantum-kernel-method-001-001}
\begin{algorithmic}[1]
\STATE set $\hat{P}_0$ and $\hat{D}_0$
\FOR{$k = 1, 2, \dots, K$}
\STATE compute $\hat{X}_k$ and $\theta_{\mathrm{b}, k}$ by Eq.~\eqref{main-arXiv-quantum-kernel-method-001-011}
\STATE compute $\hat{P}_k$ by Eq.~\eqref{main-arXiv-quantum-kernel-method-001-013}
\STATE compute $\hat{D}_k$ by Eq.~\eqref{main-arXiv-quantum-kernel-method-001-014}
\ENDFOR
\end{algorithmic}
\end{algorithm}

It is clear from the formulation of SOC that $\hat{X}$ does not strictly satisfy the unitarity constraint; instead, $\hat{P}$ and OU of $\hat{X}$ does. Thus, using the optimal value of $\hat{X}$ obtained from UKM leads to a classical classifier (it cannot be implemented using a quantum circuit), and will in general have higher performance than the unitary transformations given by $\hat{P}$ and OU of $\hat{X}$ that approximate $\hat{X}$.
Thus, we compute the success rates for the training and test datasets by using all the versions: $\hat{X}$, $\hat{P}$, and OU of $\hat{X}$~\footnote{OU is explained in Sec.~\ref{supp-arXiv-sec-UKM-001-001} of the SI.}, of which only $\hat{P}$ and OU of $\hat{X}$ correspond to VQC.

\textit{Variational circuit realization.}---
There are a few studies on decomposing unitary operators into quantum circuits~\cite{Shende001, Nielsen001, Plesch001}, including Knill's decomposition and the quantum Shannon decomposition (QSD).
In these methods, however, the number of CNOT gates scales exponentially in the number of qubits $n$.

Here we propose an alternate method: the target circuit is comprised of $L$ layers of a parameterized sub-circuit with parameterized gates and a fixed topology; similar to the ansatz used in QCL.
We then solve for the minimum number of layers $L$, such that the optimized circuit approximates the given unitary operator with a specified precision of $\delta$.
We refer to this circuit methodology as the Variational Circuit Realization technique.
The schematic of the VCR is demonstrated in Fig.~\ref{main-arXiv-schematic-quantum-classifer}(d).
Let $\hat{U}$ and $\hat{U}_\mathrm{c} (\theta; L)$ be a target unitary operator and a unitary operator realized by a quantum circuit that has $L$ layers, respectively.
Typically, the target unitary operator is obtained by the UKM discussed above.
Furthermore, we define the global phase unitary operator $\hat{\Phi}_{2^n} (\lambda) \coloneqq e^{- i \lambda} \hat{1}_{2^n}$.
When $\hat{U}$ and $\hat{U}_\mathrm{c+p} (\theta, \lambda; L) \coloneqq \hat{\Phi}_{2^n} (\lambda) \hat{U}_\mathrm{c} (\theta; L)$ are identical, we have
\begin{align}
  \hat{U}^\dagger \hat{U}_\mathrm{c+p} (\theta, \lambda; L) &= \hat{1}_{2^n}.
\end{align}
Then, we can estimate $\theta$, for any $p > 0$, by
\begin{align}
  \{ \theta_*, \lambda_* \} &= \argmin_{\theta, \lambda} \mathcal{J}_\mathrm{cost} (\theta, \lambda; L, p, \hat{U}), \label{main-arXiv-optimization-VCR-001}
\end{align}
where
\begin{align}
  \mathcal{J}_\mathrm{cost} (\theta, \lambda; L, p, \hat{U}) &\coloneqq \| \hat{U}^\dagger \hat{U}_\mathrm{c+p} (\theta, \lambda; L) - \hat{1}_{2^n} \|_\mathrm{F}^p. \label{main-arXiv-theta-lambda-VCR-001-001}
\end{align}
We call this method the VCR.

In a circuit realization, the complexity of a circuit is of great interest.
In this letter, we assume a layered structure for a quantum circuit.
Thus, given an error threshold $\delta$, it is convenient to define $L_\delta$:
\begin{align}
  L_\delta &\coloneqq \argmin_L \epsilon_L, \nonumber \\
  & \quad \mathrm{subject \ to} \ \epsilon_L \le \delta,
\end{align}
where
\begin{align}
  \epsilon_L &\coloneqq \min_{\theta, \lambda} \mathcal{J}_\mathrm{cost} (\theta, \lambda; L, p, \hat{U}).
\end{align}

\textit{Numerical simulation of the UKM.}---
We first show the numerical results of QCL and the UKM for the cancer dataset ($0$ or $1$)~\footnote{The iris dataset in the UCI repository~\cite{Dua001} has two labels: (0) `B' and (1) `M.' In the cancer dataset ($0$ or $1$), we consider the classification problem between the 0 label and the 1 label. Furthermore, we relabel $0$ with $-1$ to adjust labels with the eigenvalues of $\hat{\sigma}_z$. For the numerical results for other datasets, refer to Sec.~\ref{supp-arXiv-sec-datasets-UKM-001-001} of the SI.} in the UCI repository~\cite{Dua001}.
The results for multiple datasets with different dimensions, $M$, are presented in Table~\ref{main-arXiv-table-all-001}
\begin{table*}[tb]
  \begin{tabular}{cccc|ccc|cc}
    \hline \hline
    & & & & & -- Quantum -- & & \multicolumn{2}{c}{-- Classical --} \\
    Dataset & $N$ & $M$ & $n$ & UKM ($\hat{P}$) & UKM (OU of $\hat{X}$) & QCL & UKM ($\hat{X}$) & Kernel method \\
    \hline
    Iris ($0$ or $1$)         &  100 &   4 & 2 & 1.0000/\textbf{1.0000} & 1.0000/\textbf{1.0000} & 1.0000/\textbf{1.0000} & 1.0000/1.0000 & 1.0000/1.0000 \\
    Iris ($0$ or non-$0$)     &  150 &   4 & 2 & 1.0000/0.9987 & 1.0000/\textbf{1.0000} & 1.0000/\textbf{1.0000} & 1.0000/1.0000 & 1.0000/1.0000 \\
    Iris ($1$ or non-$1$)     &  150 &   4 & 2 & 0.7880/0.7789 & 0.7953/\textbf{0.7994} & 0.6801/0.5872 & 0.9781/0.9618 & 0.9751/0.9666 \\
    Cancer ($0$ or $1$)       &  569 &  30 & 5 & 0.9194/\textbf{0.9131} & 0.9184/0.9115 & 0.8797/0.8768 & 0.9218/0.9160 & 0.9618/0.9568 \\
    Sonar ($0$ or $1$)        &  208 &  60 & 6 & 0.9159/\textbf{0.7985} & 0.9175/0.7909 & 0.7455/0.6924 & 0.8903/0.7774 & 1.0000/0.8198 \\
    Wine ($0$ or non-$0$)     &  178 &  14 & 4 & 0.9200/\textbf{0.9185} & 0.9212/0.9171 & 0.9155/0.9126 & 0.9364/0.9313 & 0.9987/0.9955 \\
    Semeion ($0$ or $1$)      &  323 & 256 & 8 & 1.0000/0.9943 & 1.0000/\textbf{0.9945} & 0.9210/0.9099 & 1.0000/0.9957 & 1.0000/1.0000 \\
    Semeion ($0$ or non-$0$)  & 1593 & 256 & 8 & 0.9988/0.9949 & 0.9990/\textbf{0.9953} & 0.8989/0.8982 & 0.9969/0.9925 & 1.0000/0.9955 \\
    MNIST256 ($0$ or $1$)     &  569 & 256 & 8 & 0.9991/\textbf{0.9969} & 1.0000/0.9951 & 0.9511/0.9459 & 0.9985/0.9966 & 1.0000/1.0000 \\
    MNIST256 ($0$ or non-$0$) & 2766 & 256 & 8 & 0.9922/0.9871 & 0.9927/\textbf{0.9889} & 0.9053/0.9050 & 0.9894/0.9859 & 0.9992/0.9953 \\
    \hline \hline
  \end{tabular}
\caption{Results of 5-fold CV with 5 different random seeds of the UKM ($\hat{X}$, $\hat{P}$, and OU of $\hat{X}$), QCL, and the kernel method for all the datasets. The cells are of the format ``training performance/test performance." We choose the model that shows the best test performance for each algorithm. For the UKM, we consider the complex and real cases with and without the bias term. We set $r = 0.010$. For QCL, we consider the CNOT-based, CRot-based, 1d-Heisenberg, and FC-Heisenberg circuits with and without the bias term for the iris, cancer, sonar, and wine datasets, and the CNOT-based and CRot-based circuits with and without bias term for the semeion and MNIST256 datasets. We set the number of layers $L$ to 5. For the kernel method, we consider the linear functions and the second-order polynomial functions with and without the bias term for $\lambda = 10^{-2}, 10^{-1}, 1$. The values of the best VQC for each dataset are printed in bold. The numbers of data points $N$ and dimensions $M$ of the datasets are shown. The number of qubits $n$ required for the amplitude encoding is also shown. Note that $n = \lceil \log_2 M \rceil$.}
\label{main-arXiv-table-all-001}
\end{table*}

Before getting into the numerical results, we state the numerical setup~\footnote{For the details of numerical settings, refer to Sec.~\ref{supp-arXiv-sec-numerical-settings-UKM-001} of the SI.}.
For the UKM, we put $r = 0.010$ and set $K = 30$ in Algo.~\ref{main-arXiv-quantum-kernel-method-001-001}.
Furthermore, we use the conjugate gradient (CG) method to find the solution of Eq.~\eqref{main-arXiv-quantum-kernel-method-001-011} and run the CG iteration $10$ times~\footnote{Refer to Sec.~\ref{supp-arXiv-sec-CG-001-001} of the SI for the details of the CG method and Sec.~\ref{supp-arXiv-sec-UKM-CG-001-001} of the SI for the details of the UKM with the CG method.}.
The UKM is closed in real matrices; then we consider the real and complex cases.
For QCL, we consider four types of quantum circuits: the CNOT-based circuit, the CRot-based circuit, the 1-dimensional (1d) Heisenberg circuit, and the fully-connected (FC) Heisenberg circuit~\footnote{The definitions of the CNOT-based circuit, the CRot-based circuit, the 1d Heisenberg circuit, and the FC Heisenberg circuit are given in Sec.~\ref{supp-arXiv-sec-quantum-circuit-001-001} of the SI.}, and run iterations $300$ times.
To accelerate QCL, we utilize the stochastic gradient descent method~\cite{Murphy001}.
In both cases, we use the squared error function $\ell_\mathrm{SE} (a, b) \coloneqq \frac{1}{2} | a - b |^2$ for $\ell (\cdot, \cdot)$ in Eqs.~\eqref{main-arXiv-cost-function-UKM-001-001} and \eqref{main-arXiv-cost-QCL-001-001}, and set $Q = 1$ and $\xi_1 = 1$ in Eqs.~\eqref{main-arXiv-prediction-UKM-001-001} and \eqref{main-arXiv-prediction-QCL-001-001}.
Furthermore, we consider two cases with the bias term and without the bias term in Eqs.~\eqref{main-arXiv-prediction-UKM-001-001} and \eqref{main-arXiv-prediction-QCL-001-001}.

\begin{figure}[t]
\centering
\includegraphics[scale=0.45]{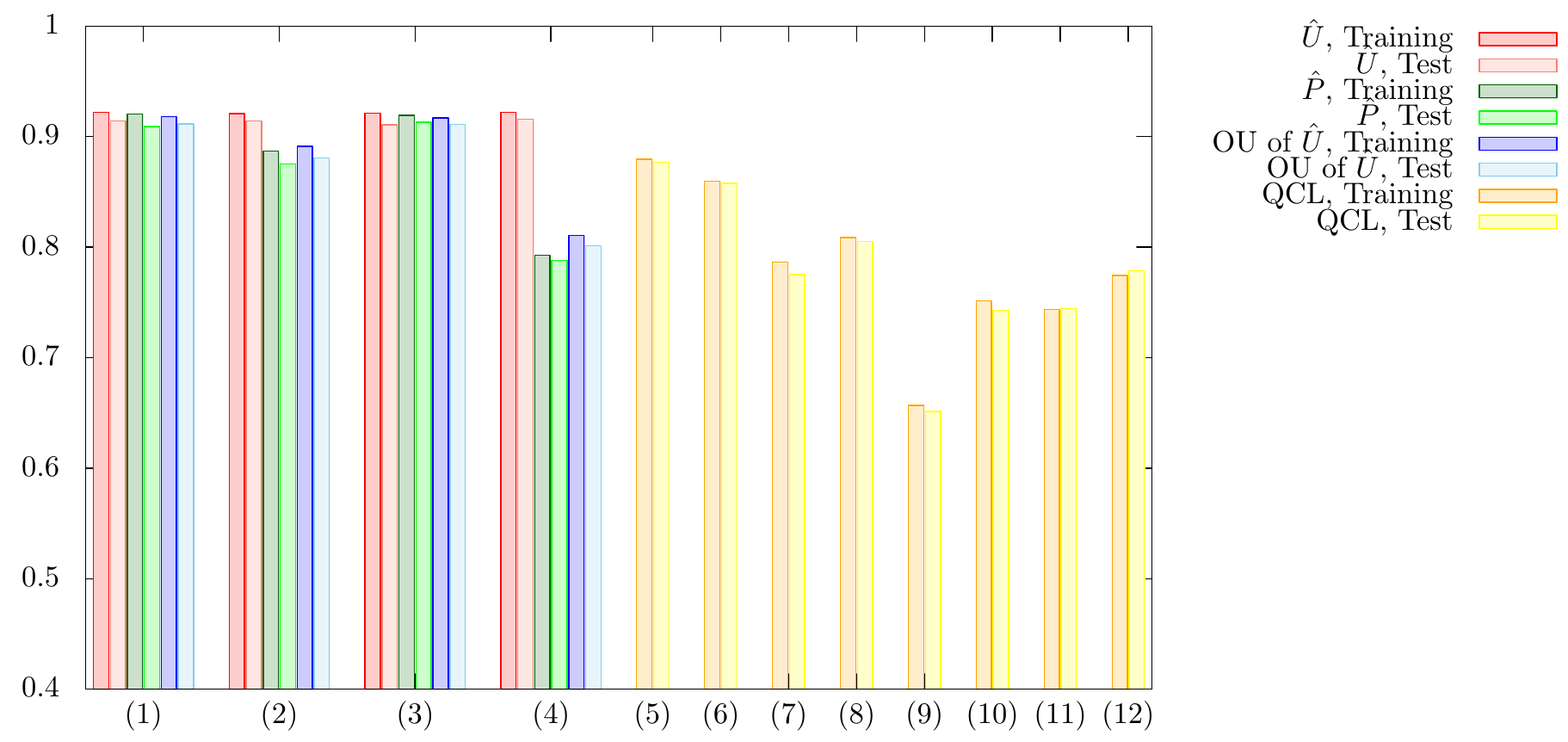}
\caption{Results of 5-fold CV with 5 different random seeds for the cancer dataset ($0$ or $1$). For the UKM, we put $r = 0.010$ and $K = 30$. We repeat the CG iterations for Eq.~\eqref{main-arXiv-quantum-kernel-method-001-011} $10$ times in each step of the method of SOC. For QCL, the number of layers $L$ is 5 and the number of iterations is $300$. The numerical settings are as follows: (1) complex matrix without bias, (2) complex matrix with bias, (3) real matrices without the bias term, (4) real matrices with the bias term, (5) CNOT-based circuit without the bias term, (6) CNOT-based circuit with the bias term, (7) CRot-based circuit without the bias term, (8) CRot-based circuit with the bias term, (9) 1d Heisenberg circuit without the bias term, (10) 1d Heisenberg circuit with the bias term, (11) FC Heisenberg circuit without the bias term, and (12) FC Heisenberg circuit with the bias term.}
\label{main-arXiv-numerical-result-performance-UKM-QCL-UCI-cancer-0-1}
\end{figure}

We summarize the results of 5-fold cross-validation (CV) with 5 different random seeds of QCL and the UKM in Tables~\ref{main-arXiv-table-UCI-cancer-0-1-002} and \ref{main-arXiv-table-UCI-cancer-0-1-001}, respectively.
For each method, we select the best model for the training dataset over iterations to compute the performance.
\begin{table}[tb]
  \begin{tabular}{cc|cc}
    \hline \hline
    Algo. & Condition & Training & Test \\
    \hline
    QCL & CNOT-based, w/o bias    & 0.8797 & 0.8768 \\
    QCL & CNOT-based, w/ bias     & 0.8597 & 0.8577 \\
    \hline
    QCL & CRot-based, w/o bias    & 0.7866 & 0.7752 \\
    QCL & CRot-based, w/ bias     & 0.8085 & 0.8052 \\
    \hline
    QCL & 1d Heisenberg, w/o bias & 0.6568 & 0.6512 \\
    QCL & 1d Heisenberg, w/ bias  & 0.7515 & 0.7427 \\
    \hline
    QCL & FC Heisenberg, w/o bias & 0.7435 & 0.7444 \\
    QCL & FC Heisenberg, w/ bias  & 0.7744 & 0.7789 \\
    \hline \hline
  \end{tabular}
\caption{Results of 5-fold CV with 5 different random seeds for the cancer dataset ($0$ or $1$). The number of layers $L$ is 5 and the number of iterations is $300$. As shown in Fig.~\ref{main-arXiv-numerical-result-layers-dependence-QCL-UCI-cancer-0-1}, increasing the number of layers $L$ does not lead to better performance, and can in fact decrease performance of QCL.}
\label{main-arXiv-table-UCI-cancer-0-1-002}
\end{table}
\begin{table}[tb]
  \begin{tabular}{cc|cc}
    \hline \hline
    Algo. & Condition & Training & Test \\
    \hline
    UKM & $\hat{X}$, complex, w/o bias       & 0.9219 & 0.9143 \\
    UKM & $\hat{P}$, complex, w/o bias       & 0.9204 & 0.9093 \\
    UKM & OU of $\hat{X}$, complex, w/o bias & 0.9184 &	0.9115 \\
    \hline
    UKM & $\hat{X}$, complex, w/ bias        & 0.9207 & 0.9143 \\
    UKM & $\hat{P}$, complex, w/ bias        & 0.8870 & 0.8753 \\
    UKM & OU of $\hat{X}$, complex, w/ bias  & 0.8912 & 0.8805 \\
    \hline
    UKM & $\hat{X}$, real, w/o bias          & 0.9213 & 0.9107 \\
    UKM & $\hat{P}$, real, w/o bias          & 0.9194 & 0.9131 \\
    UKM & OU of $\hat{X}$, real, w/o bias    & 0.9170 & 0.9112 \\
    \hline
    UKM & $\hat{X}$, real, w/ bias           & 0.9218 & 0.9160 \\
    UKM & $\hat{P}$, real, w/ bias           & 0.7929 & 0.7879 \\
    UKM & OU of $\hat{X}$, real, w/ bias     & 0.8107 & 0.8014 \\
    \hline \hline
  \end{tabular}
\caption{Results of 5-fold cross-validation (CV) with 5 different random seeds for the cancer dataset ($0$ or $1$). We set $r = 0.010$ and $K = 30$. We repeat the CG iterations for Eq.~\eqref{main-arXiv-quantum-kernel-method-001-011} $10$ times in each step of the method of SOC. Due to the inherent formulation of SOC, $\hat{X}$ does not strictly satisfy the unitarity condition; $\hat{P}$ and OU of $\hat{X}$ strictly satisfy the unitarity condition, yielding VQC. The overall higher performance of $\hat{X}$ can be attributed to it being a classical classifier; a special case of a kernel method. Note, however, that the UKM models without bias yield better performance than the best QCL results, as shown in Table~\ref{main-arXiv-table-UCI-cancer-0-1-002}}
\label{main-arXiv-table-UCI-cancer-0-1-001}
\end{table}
In Fig.~\ref{main-arXiv-numerical-result-performance-UKM-QCL-UCI-cancer-0-1}, we plot the data shown in Tables~\ref{main-arXiv-table-UCI-cancer-0-1-002} and \ref{main-arXiv-table-UCI-cancer-0-1-001}.
As shown in Fig.~\ref{main-arXiv-numerical-result-performance-UKM-QCL-UCI-cancer-0-1}, the performance of the UKM is better that that of QCL in several numerical setups.

Given our analytical results showing that a kernel method is a super-set of VQC, we next present the performance of the kernel method~\footnote{Particularly, we use Ridge classification as the kernel method. In Ridge classification, we make prediction on the label of $x_i$ by $f_\mathrm{pred} (x_i; v) \coloneqq \frac{1}{N} \sum_j v_j \phi_j (x_i)$ where $v \coloneqq [v_1, v_2, \dots v_G]^\intercal$ is a parameter and use the squared error function. For the details of Ridge classification, see Sec.~\ref{supp-arXiv-sec-Ridge-001} of the SI. Furthermore, the kernel method is described in Sec.~\ref{supp-arXiv-sec-kernel-001} of the SI. Ref.~\cite{Bishop001, Murphy001} are also helpful.}.
We summarize the results of 5-fold CV with 5 different random seeds of the kernel method in Table~\ref{main-arXiv-table-UCI-cancer-0-1-003}.
We set $\lambda = 10^{-1}$, which is the coefficient of the regularization term, and consider the linear functions and the second-order polynomial functions for a feature map with and without normalization.
\begin{table}[tb]
  \begin{tabular}{cc|cc}
    \hline \hline
    Algo. & Condition & Training & Test \\
    \hline
    Kernel method & Linear, w/o normalization & 0.9623 & 0.9549 \\
    Kernel method & Linear, w/ normalization  & 0.9205 & 0.9176 \\
    \hline
    Kernel method & Poly-2, w/o normalization & 0.9936 & 0.9361 \\
    Kernel method & Poly-2, w/ normalization  & 0.9210 & 0.9195 \\
    \hline \hline
  \end{tabular}
\caption{Results of 5-fold CV with 5 different random seeds of the kernel method for the cancer dataset ($0$ or $1$). We set $\lambda = 10^{-1}$.}
\label{main-arXiv-table-UCI-cancer-0-1-003}
\end{table}

\begin{figure}[t]
\centering
\includegraphics[scale=0.45]{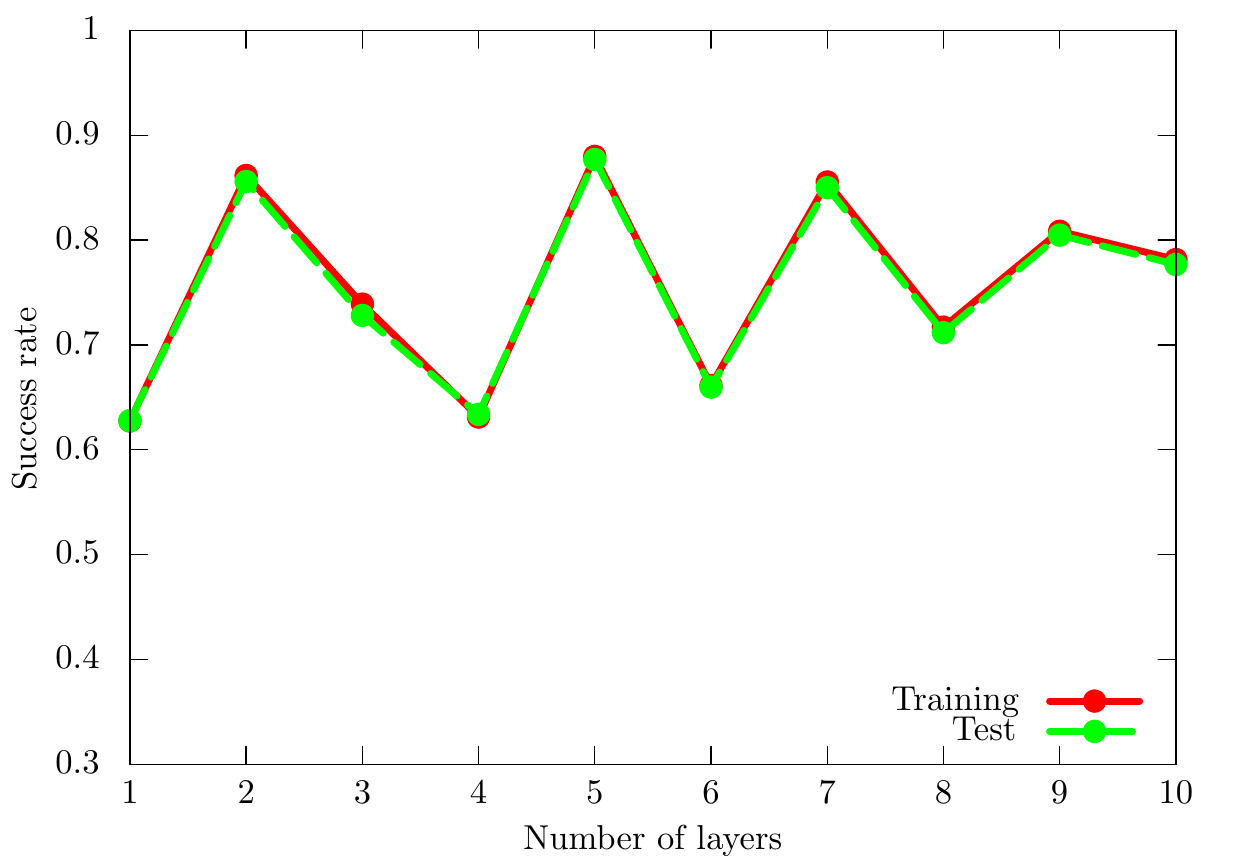}
\caption{Performance dependence of QCL on the number of layers $L$ for the cancer dataset ($0$ or $1$). We use the CNOT-based circuit geometry and set $\theta_\mathrm{bias} = 0$. We iterate the computation $300$ times.}
\label{main-arXiv-numerical-result-layers-dependence-QCL-UCI-cancer-0-1}
\end{figure}

We then explore the performance dependence of QCL on the number of layers $L$.
The result is shown in Fig.~\ref{main-arXiv-numerical-result-layers-dependence-QCL-UCI-cancer-0-1}.
One would naturally expect that increasing the number of layers $L$ leads to better performance. In general, a circuit with $L+1$ layers can clearly do at least as well as the circuit with $L$ layers: pick the same parameters for the first $L$ layers, and choose parameters to create an identity operator with the last layer. But Fig.~\ref{main-arXiv-numerical-result-layers-dependence-QCL-UCI-cancer-0-1} shows it is not the case. Rather, the test performance gets worse when we increase the number of layers $L$. This variability is potentially related to the structure of the cost function landscape: as the number of parameters is increased by adding an extra layer, there are potentially more local minima or the landscape develops what has been referred to as a ``barren plateau'' in Ref.~\cite{McClean002}.

\begin{figure}[t]
\centering
\includegraphics[scale=0.45]{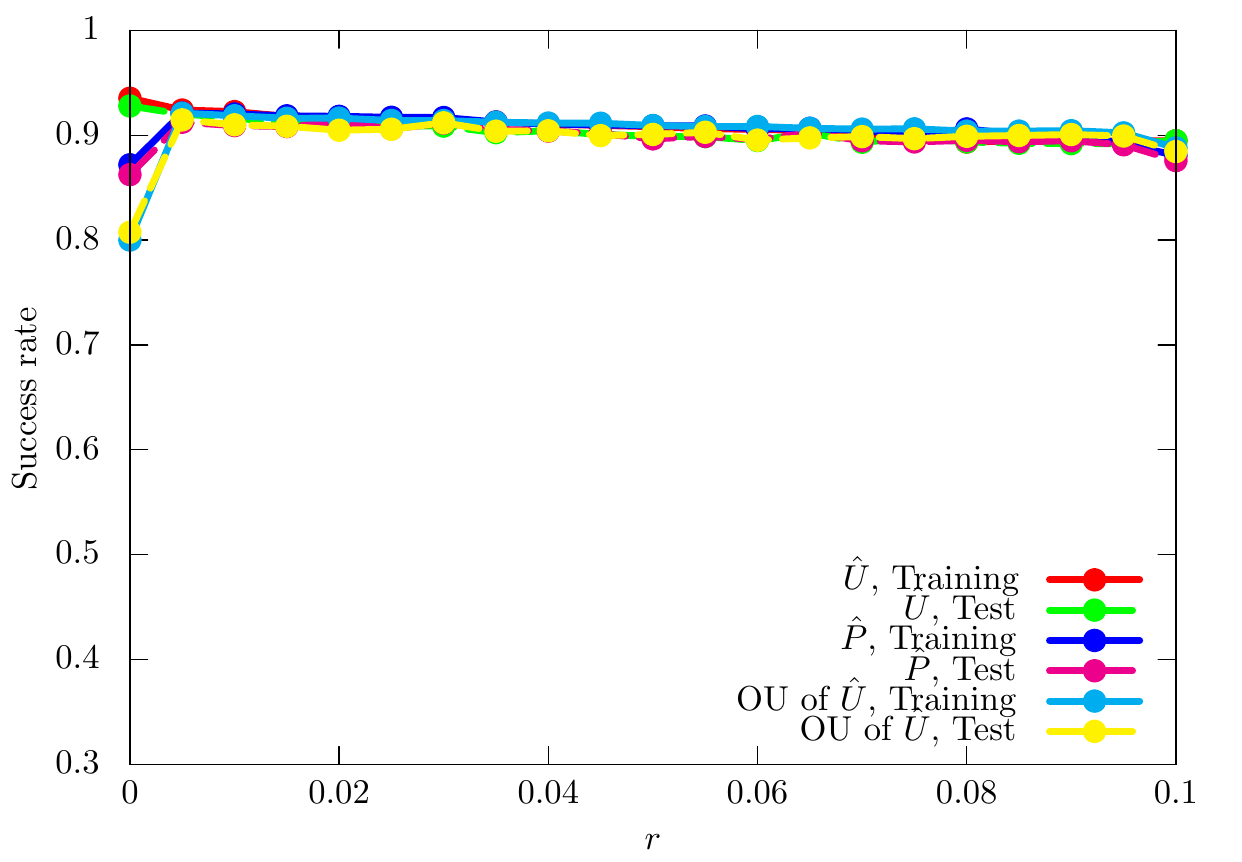}
\includegraphics[scale=0.45]{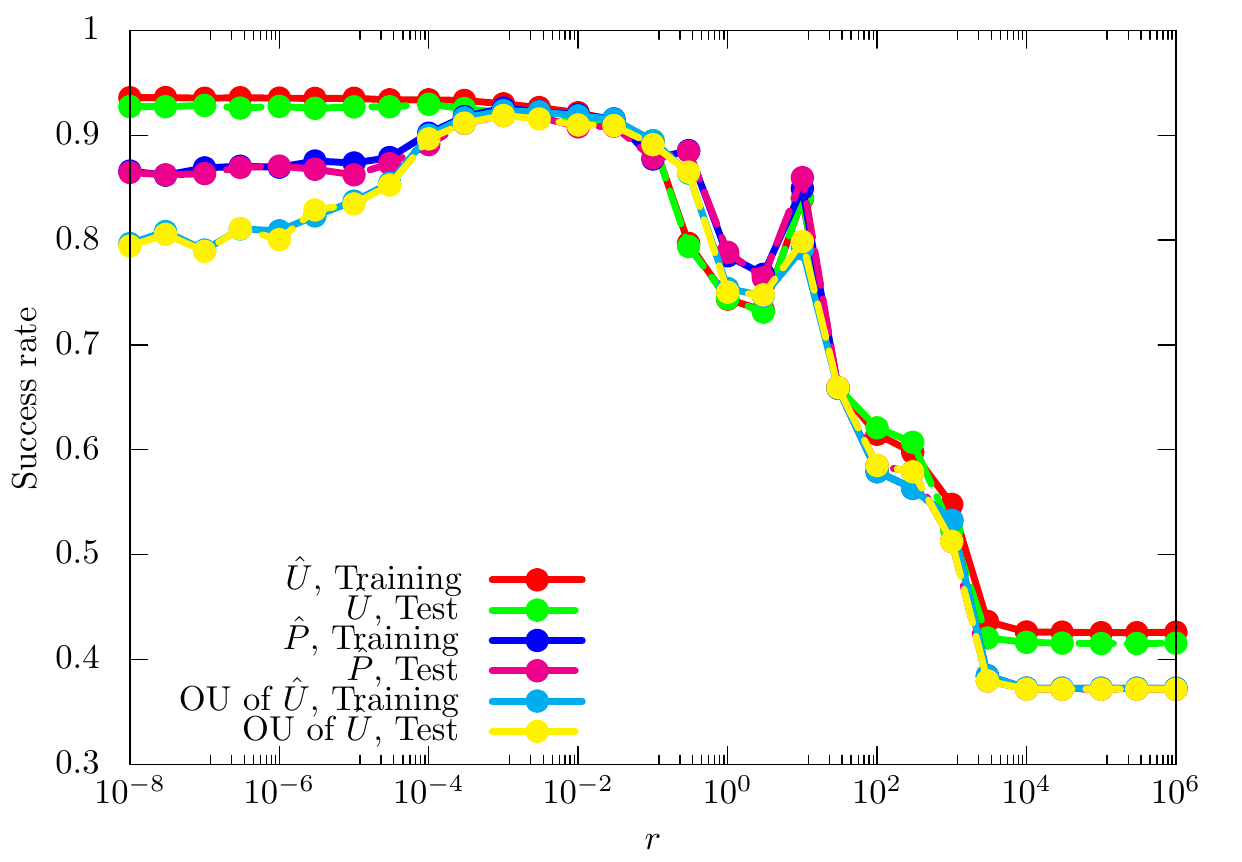}
\caption{Performance dependence of the UKM on $r$, which is the coefficient of the second term in the right-hand side of Eq.~\eqref{main-arXiv-quantum-kernel-method-001-011} for the cancer dataset ($0$ or $1$). We use complex matrices and set $\theta_\mathrm{bias} = 0$. We put $r = 0.010$ and $K = 30$. We repeat the CG iterations for Eq.~\eqref{main-arXiv-quantum-kernel-method-001-011} $10$ times in each step of the method of SOC.}
\label{main-arXiv-numerical-result-r-dependence-UKM-UCI-cancer-0-1}
\end{figure}

We also see the performance dependence of the UKM on $r$, which is the coefficient of the second term in the right-hand side of Eq.~\eqref{main-arXiv-quantum-kernel-method-001-011}.
The result is shown in Fig.~\ref{main-arXiv-numerical-result-r-dependence-UKM-UCI-cancer-0-1}.
For small $r$, $\hat{X}$ in the UKM deviates from unitary matrices and the performance gets better.
On the other hand, for large $r$, $\hat{X}$ in the UKM becomes closer to unitary matrices but the performance gets worse.
Thus, we should choose an appropriate value of $r$.

\begin{figure}[tp]
\centering
\includegraphics[scale=0.45]{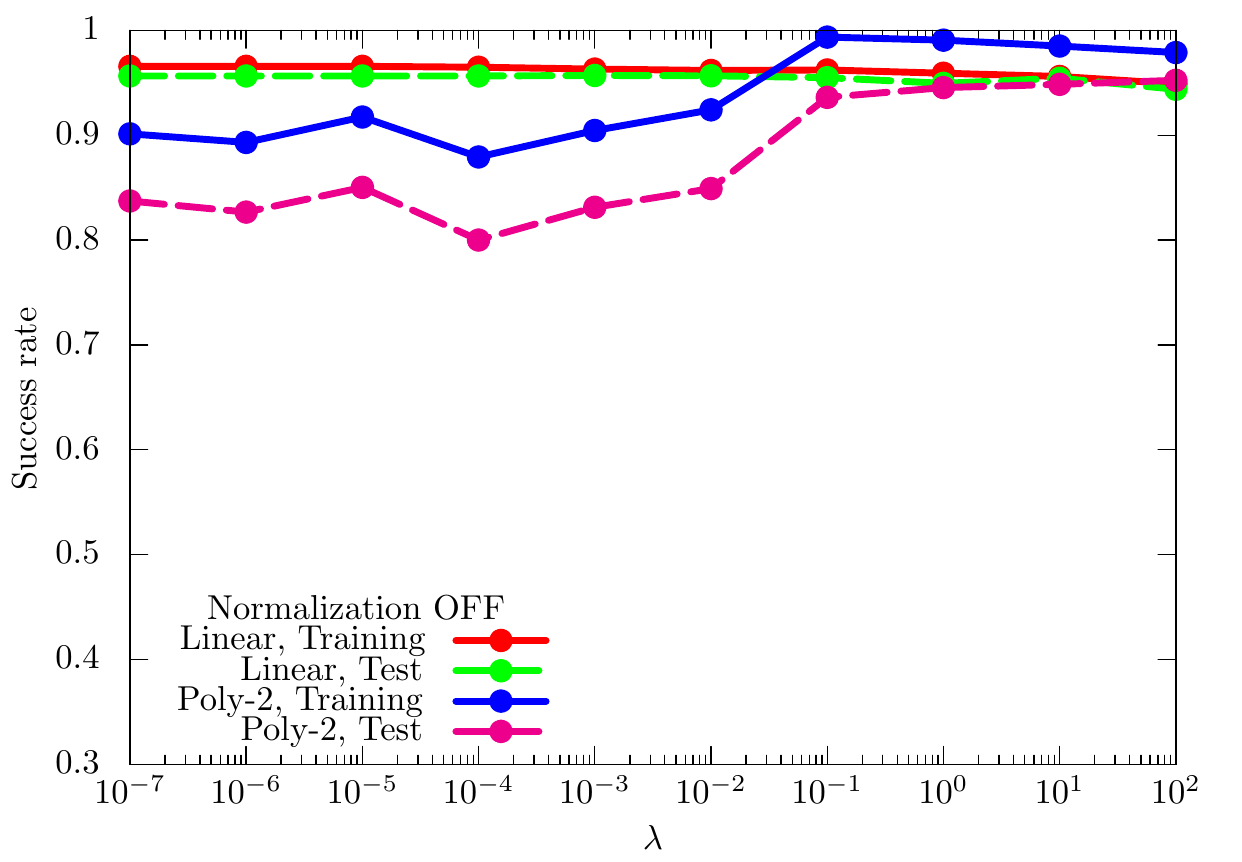}
\includegraphics[scale=0.45]{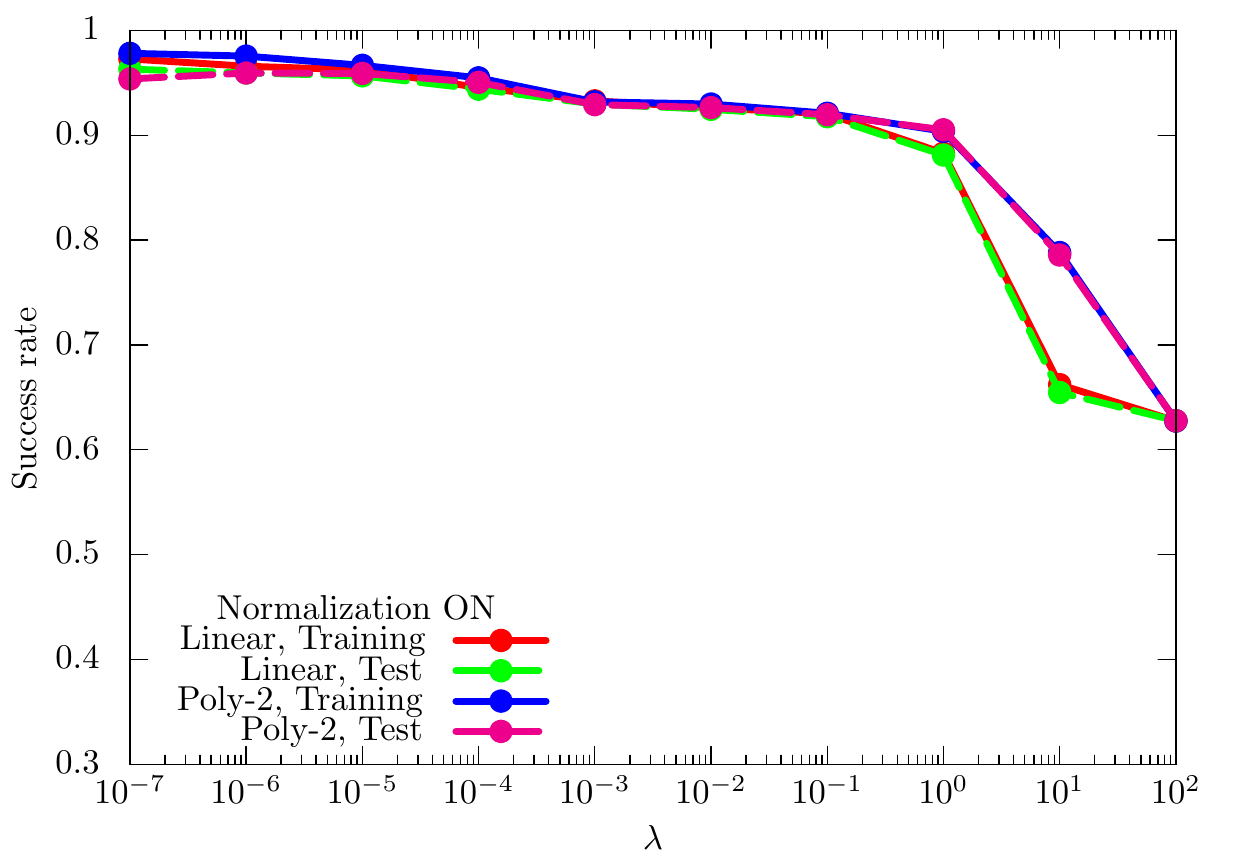}
\caption{Performance dependence of the kernel method on $\lambda$, which is the coefficient of the regularization term for the cancer dataset ($0$ or $1$). For $\phi (\cdot)$, we use the linear functions and the second-degree polynomial functions with and without normalization. We use the squared error function.}
\label{main-arXiv-numerical-result-lambda-dependence-kernel-method-cancer-0-1}
\end{figure}

In Fig.~\ref{main-arXiv-numerical-result-lambda-dependence-kernel-method-cancer-0-1}, we show the performance dependence of the kernel method on $\lambda$, which is the coefficient of the regularization term.
Like $r$ in the UKM, we also need to choose an appropriate $\lambda$ to realize good performance.

In Table.~\ref{main-arXiv-table-all-001}, we summarize the performance of the UKM, QCL, and the kernel method for all the datasets investigated in this study.
We choose the model that shows the best test performance for each algorithm.
For the UKM, we consider the complex and real cases with and without the bias term.
We set $r = 0.010$.
For QCL, we consider the CNOT-based, CRot-based, 1d-Heisenberg, and FC-Heisenberg circuits with and without the bias term for the iris, cancer, sonar, and wine datasets, and the CNOT-based and CRot-based circuits with and without bias term for the semeion and MNIST256 datasets.
We set the number of layers $L$ to 5.
For the kernel method, we consider the linear functions and the second-order polynomial functions with and without the bias term for $\lambda = 10^{-2}, 10^{-1}, 1$.
The numerical results support the claim that the UKM lies between the kernel method and QCL.
We also show the detailed numerical results for all the datasets in the supplemental information (SI)~\footnote{In Sec.~\ref{supp-arXiv-sec-numerical-result-UKM-001} of the SI, the numerical results for other datasets are shown.}.
The results shown in the SI are consistent with this letter.
Finally, we note the difference between the squared error and hinge functions.
\textit{In this letter, we have used the squared error function; we show the results of the hinge function in the SI.
The results are qualitatively same as obtained with the squared error function, and the statements about the relative performances of UKM, QCL, and Kernel methods do not change.}

\textit{Discussions on the UKM.}---
As shown in this letter, the performance of QCL is bounded from above by the UKM.
One of the biggest reasons is a difference in the degrees of freedom in QCL and the UKM.
In the UKM, we have $\mathcal{O} (M^2)$ parameters to estimate; on the other hand, the number of parameters in the QCL is $\mathcal{O} (L \ln M)$.
This difference implies that a circuit ansatz introduces a strong bias in QCL, and may restrict the performance of QCL considerably.
Thus, by designing the UKM, we can explore the ultimate power of QCL and at least, for the case of a small number of qubits $n$, the numerical results in this letter show that the ultimate power of QCL is limited.
However, by using a limited number of layers $L$, QCL can often get good performance, while requiring a fewer number of gate operations.
Thus, if we can really develop a quantum computer with a sufficiently large number of qubits $n$, QCL may be useful for applications.

As noted earlier, we can also address the potential limitations of QCL  from the viewpoint of optimization.
Fig.~\ref{main-arXiv-numerical-result-layers-dependence-QCL-UCI-cancer-0-1} implies the difficulty of optimizing parameters in QCL.
Otherwise, the success rates in Fig.~\ref{main-arXiv-numerical-result-layers-dependence-QCL-UCI-cancer-0-1} should be more smooth and monotonically increasing.
This phenomenon may come from the barren plateaus problem~\cite{McClean002}.
On the other hand, the performance of the UKM is very high and close to that of the kernel method in \ref{main-arXiv-numerical-result-r-dependence-UKM-UCI-cancer-0-1}; thus, we can say that the UKM does not suffer from a similar optimization problem.

We also note a practical difficulty in implementing QCL.
If we can properly choose circuit geometry, the performance of QCL and the UKM should be the same or very close.
But, in reality, there is a gap in the performance of QCL and the UKM; thus, it implies that choosing a good circuit geometry is difficult.

\textit{Numerical simulation of the VCR.}---
We then show numerical simulations on the VCR.
For $\hat{U}_\mathrm{c} (\theta)$, we use the CNOT-based circuit geometry.
Furthermore, we use the BFGS method~\cite{Fletcher001} to solve Eq.~\eqref{main-arXiv-optimization-VCR-001}.

\begin{figure}[t]
\centering
\includegraphics[scale=0.45]{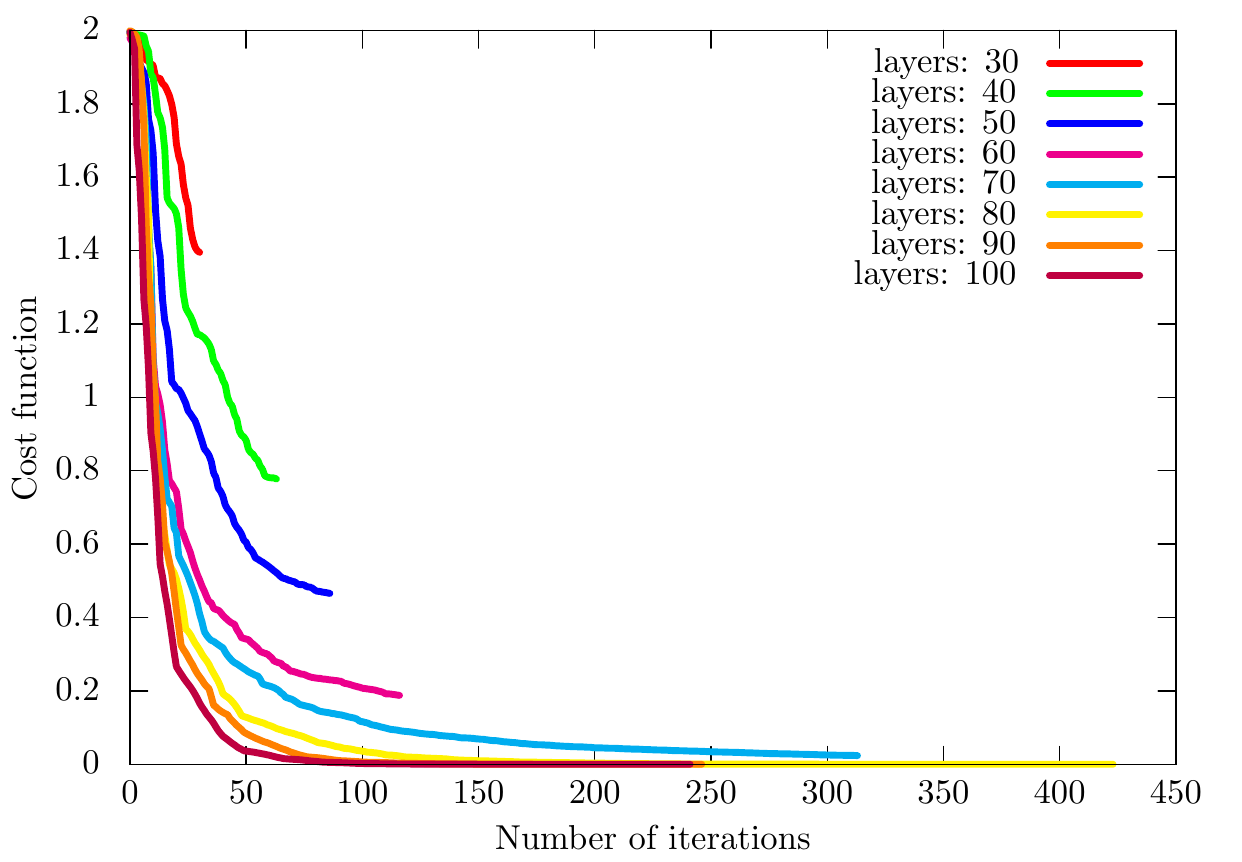}
\caption{Values of the cost function $J_\mathrm{cost} (\theta, \lambda; L, 2, \hat{U})$, Eq.~\eqref{main-arXiv-theta-lambda-VCR-001-001} with $p = 2$, for the cancer dataset ($0$ or $1$). We set $L = 30, 40, 50, 60, 70, 80, 90, 100$.}
\label{main-arXiv-numerical-result-VCR-UCI-cancer-0-1-001}
\end{figure}

Let us consider the cancer dataset ($0$ or $1$) and minimizing Eq.~\eqref{main-arXiv-theta-lambda-VCR-001-001} with $p = 2$.
As $\hat{U}$, we use the unitary operator that gives the success rate for the training dataset $0.7565$ and that for the test dataset $0.8286$.
In Fig.~\ref{main-arXiv-numerical-result-VCR-UCI-cancer-0-1-001}, we show the values of the cost function in the right-hand side of Eq.~\eqref{main-arXiv-theta-lambda-VCR-001-001} with different numbers of layers.
In Table~\ref{main-arXiv-table-VCR-UCI-cancer-0-1-001}, we summarize the performance of the input unitary operator, QCL, and the circuit geometry computed by the VCR.
\begin{table}[tb]
  \begin{tabular}{cc|ccc}
    \hline \hline
    Algo. & Condition & Cost & Training & Test \\
    \hline
    Input & UKM, $\hat{P}$, real, w/o bias & --- & 0.9139 & 0.9483 \\
    \hline
    VCR & \# of layers: 10 & 1.9694 & 0.3929 & 0.2931 \\
    VCR & \# of layers: 20 & 1.9734 & 0.6071 & 0.7069 \\
    VCR & \# of layers: 30 & 1.3950 & 0.6071 & 0.7069 \\
    VCR & \# of layers: 40 & 0.7777 & 0.6909 & 0.7586 \\
    VCR & \# of layers: 50 & 0.4657 & 0.8499 & 0.9224 \\
    VCR & \# of layers: 60 & 0.1877 & 0.9073 & 0.9483 \\
    VCR & \# of layers: 70 & 0.0236 & 0.9073 & 0.9483 \\
    VCR & \# of layers: 80 & 0.0000 & 0.9139 & 0.9483 \\
    VCR & \# of layers: 90 & 0.0000 & 0.9139 & 0.9483 \\
    VCR & \# of layers: 100 & 0.0000 & 0.9139 & 0.9483 \\
    \hline
    UKM & $\hat{P}$, real, w/o bias & --- & 0.9194 & 0.9131 \\
    QCL & \# of layers: 5 & --- & 0.8798 & 0.8768 \\
    QCL & \# of layers: 10 & --- & 0.7814 & 0.7767 \\
    \hline \hline
  \end{tabular}
\caption{Performance of the VCR for the cancer dataset ($0$ or $1$). We show the success rates for the training and test datasets and the value of the cost function for the VCR. The input for the VCR is $\hat{P}$ created by the UKM under the condition of real matrices without the bias term with $r = 0.010$. For reference, we add the last three rows that show the results of 5-fold CV. The table shows that around 50 layers, by combining UKM with VCR one can get a better performance than that of QCL.}
\label{main-arXiv-table-VCR-UCI-cancer-0-1-001}
\end{table}

Fig.~\ref{main-arXiv-numerical-result-VCR-UCI-cancer-0-1-001} shows the values of the cost function, Eq.~\eqref{main-arXiv-theta-lambda-VCR-001-001}, over iterations with different number of layers $L$.
Table~\ref{main-arXiv-table-VCR-UCI-cancer-0-1-001} shows that $\hat{U}_\mathrm{c} (\theta)$ gives high performance.
We have $L_{0.001} = 80$.

\textit{Discussions on the VCR.}---
Recall that $M$ and $L$ are the dimension of the data points and the number of layers in an ansatz adopted in QCL, respectively.
Note that we use the amplitude encoding in this letter. 
Then circuits in QCL have $\lceil \ln M \rceil$ qubits and have $\mathcal{O} (L \ln M)$ gates.
The number of parameters to estimate is of the same order since we use the three-dimensional rotation gate as a parametrized gate.
The UKM also has the same number of qubits $\lceil \ln M \rceil$; so it retains the qubit efficiency, but it optimizes over $\mathcal{O} (M^2)$ parameters.
Moreover, circuits obtained by the combination of the UKM and the VCR are still of complexity $\mathcal{O} (L \ln M)$, except that now $L$ is not a constant, as in QCL.
For the datasets used in this letter, VCR yields much more compact circuits than traditional methods for obtaining circuits for unitary operators, such as the QSD, where the number of gates will be $\mathcal{O} (M^2)$.
Thus, VCR yields better performance than the traditional methods.

Also, we show using the VCR that we can realize the unitary operator obtained by the UKM using the same ansatz used in QCL and for some cases we can beat the gate counts of QCL as well.
In other cases, we have bigger circuits (i.e., $L$ is larger) but with better performance.
If data has very high dimensions, i.e. $M$ is very large, the computational time and circuit size might be very large, $\mathcal{O} (M^2)$. But we still have the $\lceil \ln M \rceil$ advantage in the number of qubits $n$.

However, QCL also has two major potential problems, when $M$ is very large.
First, the dataset size has to be very large due to the curse of dimensionality.
So the training time and convergence complexity will be a problem no matter what the parameter size is.
Second, there is no guarantee that a kernel function with $\mathcal{O} (L \ln M)$ parameters will do well, especially for small $L$. The performance for small $L$ and large $M$ could be poor.
There is no theoretical proof that for large $M$ the QCL will do well with small $L$.
We both use the same number of qubits $\lceil \ln M \rceil$; so in terms of intermediate-scale quantum computers, we both have the same advantage.
And the computations of the VCR are in the order of $M^2$; so it is doable for any reasonable size dimension.

\textit{Concluding Remarks.}---
In this letter, we have first discussed the mathematical relationship between VQC/QCL and the kernel method.
This relationship implies that VQC including QCL is a subset of the classical kernel method and cannot outperform the kernel method.

Then we have proposed the UKM for classification problems.
Mathematically the UKM lies between the kernel method and the QCL, and thus it is expected to provide us an upper bound on the performance of QCL.
By numerical simulation, we have shown that the UKM is better than QCL, as expected.

We also have proposed the VCR to find a circuit geometry that realizes a given unitary operator.
By combining the UKM and the VCR, we have shown that we can find a circuit geometry that shows high performance in classification.

Also, we have assumed amplitude encoding, and in future work, we plan to explore the performance of VQC for other forms of encoding the classical data.
For example, one straightforward extension would be to  embed the feature vector $x_i \in \mathbb{R}^M$ into a higher dimensional vector $\phi(x_i) \in \mathbb{R}^L$ with $L= \mathcal{O} (M^c)$ and then use the rest of the framework; the number of qubits $n$ will still be $\mathcal{O} (\log M)$, thus retaining any potential quantum advantage. Such extensions can extend the power of VQC and QCL.


\bibliographystyle{apsrev4-1}
\bibliography{paper-circuit-learning-999-001-bib}

\begin{thebibliography}{66}%
\makeatletter
\providecommand \@ifxundefined [1]{%
 \@ifx{#1\undefined}
}%
\providecommand \@ifnum [1]{%
 \ifnum #1\expandafter \@firstoftwo
 \else \expandafter \@secondoftwo
 \fi
}%
\providecommand \@ifx [1]{%
 \ifx #1\expandafter \@firstoftwo
 \else \expandafter \@secondoftwo
 \fi
}%
\providecommand \natexlab [1]{#1}%
\providecommand \enquote  [1]{``#1''}%
\providecommand \bibnamefont  [1]{#1}%
\providecommand \bibfnamefont [1]{#1}%
\providecommand \citenamefont [1]{#1}%
\providecommand \href@noop [0]{\@secondoftwo}%
\providecommand \href [0]{\begingroup \@sanitize@url \@href}%
\providecommand \@href[1]{\@@startlink{#1}\@@href}%
\providecommand \@@href[1]{\endgroup#1\@@endlink}%
\providecommand \@sanitize@url [0]{\catcode `\\12\catcode `\$12\catcode
  `\&12\catcode `\#12\catcode `\^12\catcode `\_12\catcode `\%12\relax}%
\providecommand \@@startlink[1]{}%
\providecommand \@@endlink[0]{}%
\providecommand \url  [0]{\begingroup\@sanitize@url \@url }%
\providecommand \@url [1]{\endgroup\@href {#1}{\urlprefix }}%
\providecommand \urlprefix  [0]{URL }%
\providecommand \Eprint [0]{\href }%
\providecommand \doibase [0]{http://dx.doi.org/}%
\providecommand \selectlanguage [0]{\@gobble}%
\providecommand \bibinfo  [0]{\@secondoftwo}%
\providecommand \bibfield  [0]{\@secondoftwo}%
\providecommand \translation [1]{[#1]}%
\providecommand \BibitemOpen [0]{}%
\providecommand \bibitemStop [0]{}%
\providecommand \bibitemNoStop [0]{.\EOS\space}%
\providecommand \EOS [0]{\spacefactor3000\relax}%
\providecommand \BibitemShut  [1]{\csname bibitem#1\endcsname}%
\let\auto@bib@innerbib\@empty
\bibitem [{\citenamefont {Shor}(1999)}]{Shor001}%
  \BibitemOpen
  \bibfield  {author} {\bibinfo {author} {\bibfnamefont {P.~W.}\ \bibnamefont
  {Shor}},\ }\href@noop {} {\bibfield  {journal} {\bibinfo  {journal} {SIAM
  review}\ }\textbf {\bibinfo {volume} {41}},\ \bibinfo {pages} {303} (\bibinfo
  {year} {1999})}\BibitemShut {NoStop}%
\bibitem [{\citenamefont {Arute}\ \emph {et~al.}(2019)\citenamefont {Arute},
  \citenamefont {Arya}, \citenamefont {Babbush}, \citenamefont {Bacon},
  \citenamefont {Bardin}, \citenamefont {Barends}, \citenamefont {Biswas},
  \citenamefont {Boixo}, \citenamefont {Brandao}, \citenamefont {Buell} \emph
  {et~al.}}]{Arute001}%
  \BibitemOpen
  \bibfield  {author} {\bibinfo {author} {\bibfnamefont {F.}~\bibnamefont
  {Arute}}, \bibinfo {author} {\bibfnamefont {K.}~\bibnamefont {Arya}},
  \bibinfo {author} {\bibfnamefont {R.}~\bibnamefont {Babbush}}, \bibinfo
  {author} {\bibfnamefont {D.}~\bibnamefont {Bacon}}, \bibinfo {author}
  {\bibfnamefont {J.~C.}\ \bibnamefont {Bardin}}, \bibinfo {author}
  {\bibfnamefont {R.}~\bibnamefont {Barends}}, \bibinfo {author} {\bibfnamefont
  {R.}~\bibnamefont {Biswas}}, \bibinfo {author} {\bibfnamefont
  {S.}~\bibnamefont {Boixo}}, \bibinfo {author} {\bibfnamefont {F.~G.}\
  \bibnamefont {Brandao}}, \bibinfo {author} {\bibfnamefont {D.~A.}\
  \bibnamefont {Buell}},  \emph {et~al.},\ }\href@noop {} {\bibfield  {journal}
  {\bibinfo  {journal} {Nature}\ }\textbf {\bibinfo {volume} {574}},\ \bibinfo
  {pages} {505} (\bibinfo {year} {2019})}\BibitemShut {NoStop}%
\bibitem [{\citenamefont {Cerezo}\ \emph {et~al.}(2020)\citenamefont {Cerezo},
  \citenamefont {Arrasmith}, \citenamefont {Babbush}, \citenamefont {Benjamin},
  \citenamefont {Endo}, \citenamefont {Fujii}, \citenamefont {McClean},
  \citenamefont {Mitarai}, \citenamefont {Yuan}, \citenamefont {Cincio} \emph
  {et~al.}}]{Cerezo01}%
  \BibitemOpen
  \bibfield  {author} {\bibinfo {author} {\bibfnamefont {M.}~\bibnamefont
  {Cerezo}}, \bibinfo {author} {\bibfnamefont {A.}~\bibnamefont {Arrasmith}},
  \bibinfo {author} {\bibfnamefont {R.}~\bibnamefont {Babbush}}, \bibinfo
  {author} {\bibfnamefont {S.~C.}\ \bibnamefont {Benjamin}}, \bibinfo {author}
  {\bibfnamefont {S.}~\bibnamefont {Endo}}, \bibinfo {author} {\bibfnamefont
  {K.}~\bibnamefont {Fujii}}, \bibinfo {author} {\bibfnamefont {J.~R.}\
  \bibnamefont {McClean}}, \bibinfo {author} {\bibfnamefont {K.}~\bibnamefont
  {Mitarai}}, \bibinfo {author} {\bibfnamefont {X.}~\bibnamefont {Yuan}},
  \bibinfo {author} {\bibfnamefont {L.}~\bibnamefont {Cincio}},  \emph
  {et~al.},\ }\href@noop {} {\bibfield  {journal} {\bibinfo  {journal} {arXiv
  preprint arXiv:2012.09265}\ } (\bibinfo {year} {2020})}\BibitemShut {NoStop}%
\bibitem [{\citenamefont {Farhi}\ \emph {et~al.}(2014)\citenamefont {Farhi},
  \citenamefont {Goldstone},\ and\ \citenamefont {Gutmann}}]{Farhi001}%
  \BibitemOpen
  \bibfield  {author} {\bibinfo {author} {\bibfnamefont {E.}~\bibnamefont
  {Farhi}}, \bibinfo {author} {\bibfnamefont {J.}~\bibnamefont {Goldstone}}, \
  and\ \bibinfo {author} {\bibfnamefont {S.}~\bibnamefont {Gutmann}},\
  }\href@noop {} {\bibfield  {journal} {\bibinfo  {journal} {arXiv preprint
  arXiv:1411.4028}\ } (\bibinfo {year} {2014})}\BibitemShut {NoStop}%
\bibitem [{\citenamefont {McClean}\ \emph {et~al.}(2016)\citenamefont
  {McClean}, \citenamefont {Romero}, \citenamefont {Babbush},\ and\
  \citenamefont {Aspuru-Guzik}}]{McClean001}%
  \BibitemOpen
  \bibfield  {author} {\bibinfo {author} {\bibfnamefont {J.~R.}\ \bibnamefont
  {McClean}}, \bibinfo {author} {\bibfnamefont {J.}~\bibnamefont {Romero}},
  \bibinfo {author} {\bibfnamefont {R.}~\bibnamefont {Babbush}}, \ and\
  \bibinfo {author} {\bibfnamefont {A.}~\bibnamefont {Aspuru-Guzik}},\
  }\href@noop {} {\bibfield  {journal} {\bibinfo  {journal} {New Journal of
  Physics}\ }\textbf {\bibinfo {volume} {18}},\ \bibinfo {pages} {023023}
  (\bibinfo {year} {2016})}\BibitemShut {NoStop}%
\bibitem [{\citenamefont {Mitarai}\ \emph {et~al.}(2018)\citenamefont
  {Mitarai}, \citenamefont {Negoro}, \citenamefont {Kitagawa},\ and\
  \citenamefont {Fujii}}]{Mitarai001}%
  \BibitemOpen
  \bibfield  {author} {\bibinfo {author} {\bibfnamefont {K.}~\bibnamefont
  {Mitarai}}, \bibinfo {author} {\bibfnamefont {M.}~\bibnamefont {Negoro}},
  \bibinfo {author} {\bibfnamefont {M.}~\bibnamefont {Kitagawa}}, \ and\
  \bibinfo {author} {\bibfnamefont {K.}~\bibnamefont {Fujii}},\ }\href@noop {}
  {\bibfield  {journal} {\bibinfo  {journal} {Physical Review A}\ }\textbf
  {\bibinfo {volume} {98}},\ \bibinfo {pages} {032309} (\bibinfo {year}
  {2018})}\BibitemShut {NoStop}%
\bibitem [{\citenamefont {Schuld}\ \emph
  {et~al.}(2020{\natexlab{a}})\citenamefont {Schuld}, \citenamefont {Bocharov},
  \citenamefont {Svore},\ and\ \citenamefont {Wiebe}}]{Schuld001}%
  \BibitemOpen
  \bibfield  {author} {\bibinfo {author} {\bibfnamefont {M.}~\bibnamefont
  {Schuld}}, \bibinfo {author} {\bibfnamefont {A.}~\bibnamefont {Bocharov}},
  \bibinfo {author} {\bibfnamefont {K.~M.}\ \bibnamefont {Svore}}, \ and\
  \bibinfo {author} {\bibfnamefont {N.}~\bibnamefont {Wiebe}},\ }\href@noop {}
  {\bibfield  {journal} {\bibinfo  {journal} {Physical Review A}\ }\textbf
  {\bibinfo {volume} {101}},\ \bibinfo {pages} {032308} (\bibinfo {year}
  {2020}{\natexlab{a}})}\BibitemShut {NoStop}%
\bibitem [{\citenamefont {Bishop}(2006)}]{Bishop001}%
  \BibitemOpen
  \bibfield  {author} {\bibinfo {author} {\bibfnamefont {C.~M.}\ \bibnamefont
  {Bishop}},\ }\href@noop {} {\emph {\bibinfo {title} {Pattern recognition and
  machine learning}}}\ (\bibinfo  {publisher} {springer},\ \bibinfo {year}
  {2006})\BibitemShut {NoStop}%
\bibitem [{\citenamefont {Murphy}(2012)}]{Murphy001}%
  \BibitemOpen
  \bibfield  {author} {\bibinfo {author} {\bibfnamefont {K.~P.}\ \bibnamefont
  {Murphy}},\ }\href@noop {} {\emph {\bibinfo {title} {Machine learning: a
  probabilistic perspective}}}\ (\bibinfo  {publisher} {MIT press},\ \bibinfo
  {year} {2012})\BibitemShut {NoStop}%
\bibitem [{\citenamefont {Boyd}\ \emph {et~al.}(2004)\citenamefont {Boyd},
  \citenamefont {Boyd},\ and\ \citenamefont {Vandenberghe}}]{Boyd001}%
  \BibitemOpen
  \bibfield  {author} {\bibinfo {author} {\bibfnamefont {S.}~\bibnamefont
  {Boyd}}, \bibinfo {author} {\bibfnamefont {S.~P.}\ \bibnamefont {Boyd}}, \
  and\ \bibinfo {author} {\bibfnamefont {L.}~\bibnamefont {Vandenberghe}},\
  }\href@noop {} {\emph {\bibinfo {title} {Convex optimization}}}\ (\bibinfo
  {publisher} {Cambridge university press},\ \bibinfo {year}
  {2004})\BibitemShut {NoStop}%
\bibitem [{\citenamefont {Fletcher}(2013)}]{Fletcher001}%
  \BibitemOpen
  \bibfield  {author} {\bibinfo {author} {\bibfnamefont {R.}~\bibnamefont
  {Fletcher}},\ }\href@noop {} {\emph {\bibinfo {title} {Practical methods of
  optimization}}}\ (\bibinfo  {publisher} {John Wiley \& Sons},\ \bibinfo
  {year} {2013})\BibitemShut {NoStop}%
\bibitem [{Note1()}]{Note1}%
  \BibitemOpen
  \bibinfo {note} {The kernel method is discussed in Sec.~\ref
  {supp-arXiv-sec-kernel-001} of the SI and the relationship between a VQC and
  the kernel method is discussed in Sec.~\ref
  {supp-arXiv-sec-correpondence-QCL-kernel-001} of the SI in
  detail.}\BibitemShut {Stop}%
\bibitem [{Note2()}]{Note2}%
  \BibitemOpen
  \bibinfo {note} {Refer to Sec.~\ref {supp-arXiv-sec-quantum-circuit-001-001}
  of the SI for the details of quantum circuits.}\BibitemShut {Stop}%
\bibitem [{Note3()}]{Note3}%
  \BibitemOpen
  \bibinfo {note} {For details, refer to Sec.~\ref {supp-arXiv-sec-QCL-001-001}
  of the SI.}\BibitemShut {Stop}%
\bibitem [{\citenamefont {Nelder}\ and\ \citenamefont
  {Mead}(1965)}]{Nelder001}%
  \BibitemOpen
  \bibfield  {author} {\bibinfo {author} {\bibfnamefont {J.~A.}\ \bibnamefont
  {Nelder}}\ and\ \bibinfo {author} {\bibfnamefont {R.}~\bibnamefont {Mead}},\
  }\href@noop {} {\bibfield  {journal} {\bibinfo  {journal} {The computer
  journal}\ }\textbf {\bibinfo {volume} {7}},\ \bibinfo {pages} {308} (\bibinfo
  {year} {1965})}\BibitemShut {NoStop}%
\bibitem [{\citenamefont {McClean}\ \emph {et~al.}(2018)\citenamefont
  {McClean}, \citenamefont {Boixo}, \citenamefont {Smelyanskiy}, \citenamefont
  {Babbush},\ and\ \citenamefont {Neven}}]{McClean002}%
  \BibitemOpen
  \bibfield  {author} {\bibinfo {author} {\bibfnamefont {J.~R.}\ \bibnamefont
  {McClean}}, \bibinfo {author} {\bibfnamefont {S.}~\bibnamefont {Boixo}},
  \bibinfo {author} {\bibfnamefont {V.~N.}\ \bibnamefont {Smelyanskiy}},
  \bibinfo {author} {\bibfnamefont {R.}~\bibnamefont {Babbush}}, \ and\
  \bibinfo {author} {\bibfnamefont {H.}~\bibnamefont {Neven}},\ }\href@noop {}
  {\bibfield  {journal} {\bibinfo  {journal} {Nature communications}\ }\textbf
  {\bibinfo {volume} {9}},\ \bibinfo {pages} {1} (\bibinfo {year}
  {2018})}\BibitemShut {NoStop}%
\bibitem [{\citenamefont {Lai}\ and\ \citenamefont {Osher}(2014)}]{Lai001}%
  \BibitemOpen
  \bibfield  {author} {\bibinfo {author} {\bibfnamefont {R.}~\bibnamefont
  {Lai}}\ and\ \bibinfo {author} {\bibfnamefont {S.}~\bibnamefont {Osher}},\
  }\href@noop {} {\bibfield  {journal} {\bibinfo  {journal} {Journal of
  Scientific Computing}\ }\textbf {\bibinfo {volume} {58}},\ \bibinfo {pages}
  {431} (\bibinfo {year} {2014})}\BibitemShut {NoStop}%
\bibitem [{\citenamefont {Osher}\ \emph {et~al.}(2005)\citenamefont {Osher},
  \citenamefont {Burger}, \citenamefont {Goldfarb}, \citenamefont {Xu},\ and\
  \citenamefont {Yin}}]{Osher001}%
  \BibitemOpen
  \bibfield  {author} {\bibinfo {author} {\bibfnamefont {S.}~\bibnamefont
  {Osher}}, \bibinfo {author} {\bibfnamefont {M.}~\bibnamefont {Burger}},
  \bibinfo {author} {\bibfnamefont {D.}~\bibnamefont {Goldfarb}}, \bibinfo
  {author} {\bibfnamefont {J.}~\bibnamefont {Xu}}, \ and\ \bibinfo {author}
  {\bibfnamefont {W.}~\bibnamefont {Yin}},\ }\href@noop {} {\bibfield
  {journal} {\bibinfo  {journal} {Multiscale Modeling \& Simulation}\ }\textbf
  {\bibinfo {volume} {4}},\ \bibinfo {pages} {460} (\bibinfo {year}
  {2005})}\BibitemShut {NoStop}%
\bibitem [{\citenamefont {Yin}\ \emph {et~al.}(2008)\citenamefont {Yin},
  \citenamefont {Osher}, \citenamefont {Goldfarb},\ and\ \citenamefont
  {Darbon}}]{Yin001}%
  \BibitemOpen
  \bibfield  {author} {\bibinfo {author} {\bibfnamefont {W.}~\bibnamefont
  {Yin}}, \bibinfo {author} {\bibfnamefont {S.}~\bibnamefont {Osher}}, \bibinfo
  {author} {\bibfnamefont {D.}~\bibnamefont {Goldfarb}}, \ and\ \bibinfo
  {author} {\bibfnamefont {J.}~\bibnamefont {Darbon}},\ }\href@noop {}
  {\bibfield  {journal} {\bibinfo  {journal} {SIAM Journal on Imaging
  sciences}\ }\textbf {\bibinfo {volume} {1}},\ \bibinfo {pages} {143}
  (\bibinfo {year} {2008})}\BibitemShut {NoStop}%
\bibitem [{Note4()}]{Note4}%
  \BibitemOpen
  \bibinfo {note} {See Sec.~\ref {supp-arXiv-sec-BIR-001} in the SI for
  details.}\BibitemShut {Stop}%
\bibitem [{Note5()}]{Note5}%
  \BibitemOpen
  \bibinfo {note} {See Sec.~\ref {supp-arXiv-sec-SOC-001-001} for
  details.}\BibitemShut {Stop}%
\bibitem [{Note6()}]{Note6}%
  \BibitemOpen
  \bibinfo {note} {See Sec.~\ref {supp-arXiv-sec-UKM-001-002} for
  details.}\BibitemShut {Stop}%
\bibitem [{Note7()}]{Note7}%
  \BibitemOpen
  \bibinfo {note} {See Secs.~\ref {supp-arXiv-sec-SOC-001-001} and \ref
  {supp-arXiv-sec-UKM-001-001} of the SI for details.}\BibitemShut {Stop}%
\bibitem [{Note8()}]{Note8}%
  \BibitemOpen
  \bibinfo {note} {For the details of the UKM, refer to Sec.~\ref
  {supp-arXiv-sec-UKM-001-001} of the SI.}\BibitemShut {Stop}%
\bibitem [{Note9()}]{Note9}%
  \BibitemOpen
  \bibinfo {note} {OU is explained in Sec.~\ref {supp-arXiv-sec-UKM-001-001} of
  the SI.}\BibitemShut {Stop}%
\bibitem [{\citenamefont {Shende}\ \emph {et~al.}(2006)\citenamefont {Shende},
  \citenamefont {Bullock},\ and\ \citenamefont {Markov}}]{Shende001}%
  \BibitemOpen
  \bibfield  {author} {\bibinfo {author} {\bibfnamefont {V.~V.}\ \bibnamefont
  {Shende}}, \bibinfo {author} {\bibfnamefont {S.~S.}\ \bibnamefont {Bullock}},
  \ and\ \bibinfo {author} {\bibfnamefont {I.~L.}\ \bibnamefont {Markov}},\
  }\href@noop {} {\bibfield  {journal} {\bibinfo  {journal} {IEEE Transactions
  on Computer-Aided Design of Integrated Circuits and Systems}\ }\textbf
  {\bibinfo {volume} {25}},\ \bibinfo {pages} {1000} (\bibinfo {year}
  {2006})}\BibitemShut {NoStop}%
\bibitem [{\citenamefont {Nielsen}\ and\ \citenamefont
  {Chuang}(2002)}]{Nielsen001}%
  \BibitemOpen
  \bibfield  {author} {\bibinfo {author} {\bibfnamefont {M.~A.}\ \bibnamefont
  {Nielsen}}\ and\ \bibinfo {author} {\bibfnamefont {I.}~\bibnamefont
  {Chuang}},\ }\href@noop {} {\enquote {\bibinfo {title} {Quantum computation
  and quantum information},}\ } (\bibinfo {year} {2002})\BibitemShut {NoStop}%
\bibitem [{\citenamefont {Plesch}\ and\ \citenamefont
  {Brukner}(2011)}]{Plesch001}%
  \BibitemOpen
  \bibfield  {author} {\bibinfo {author} {\bibfnamefont {M.}~\bibnamefont
  {Plesch}}\ and\ \bibinfo {author} {\bibfnamefont {{\v{C}}.}~\bibnamefont
  {Brukner}},\ }\href@noop {} {\bibfield  {journal} {\bibinfo  {journal}
  {Physical Review A}\ }\textbf {\bibinfo {volume} {83}},\ \bibinfo {pages}
  {032302} (\bibinfo {year} {2011})}\BibitemShut {NoStop}%
\bibitem [{Note10()}]{Note10}%
  \BibitemOpen
  \bibinfo {note} {The iris dataset in the UCI repository~\cite {Dua001} has
  two labels: (0) `B' and (1) `M.' In the cancer dataset ($0$ or $1$), we
  consider the classification problem between the 0 label and the 1 label.
  Furthermore, we relabel $0$ with $-1$ to adjust labels with the eigenvalues
  of $\protect \cc@accent {"705E}{\sigma }_z$. For the numerical results for
  other datasets, refer to Sec.~\ref {supp-arXiv-sec-datasets-UKM-001-001} of
  the SI.}\BibitemShut {Stop}%
\bibitem [{\citenamefont {Dua}\ and\ \citenamefont {Graff}(2017)}]{Dua001}%
  \BibitemOpen
  \bibfield  {author} {\bibinfo {author} {\bibfnamefont {D.}~\bibnamefont
  {Dua}}\ and\ \bibinfo {author} {\bibfnamefont {C.}~\bibnamefont {Graff}},\
  }\href {http://archive.ics.uci.edu/ml} {\enquote {\bibinfo {title} {{UCI}
  machine learning repository},}\ } (\bibinfo {year} {2017})\BibitemShut
  {NoStop}%
\bibitem [{Note11()}]{Note11}%
  \BibitemOpen
  \bibinfo {note} {For the details of numerical settings, refer to Sec.~\ref
  {supp-arXiv-sec-numerical-settings-UKM-001} of the SI.}\BibitemShut {Stop}%
\bibitem [{Note12()}]{Note12}%
  \BibitemOpen
  \bibinfo {note} {Refer to Sec.~\ref {supp-arXiv-sec-CG-001-001} of the SI for
  the details of the CG method and Sec.~\ref {supp-arXiv-sec-UKM-CG-001-001} of
  the SI for the details of the UKM with the CG method.}\BibitemShut {Stop}%
\bibitem [{Note13()}]{Note13}%
  \BibitemOpen
  \bibinfo {note} {The definitions of the CNOT-based circuit, the CRot-based
  circuit, the 1d Heisenberg circuit, and the FC Heisenberg circuit are given
  in Sec.~\ref {supp-arXiv-sec-quantum-circuit-001-001} of the SI.}\BibitemShut
  {Stop}%
\bibitem [{Note14()}]{Note14}%
  \BibitemOpen
  \bibinfo {note} {Particularly, we use Ridge classification as the kernel
  method. In Ridge classification, we make prediction on the label of $x_i$ by
  $f_\protect \mathrm {pred} (x_i; v) \mathrel {\mathop :}\mathrel {\mkern
  -1.2mu}=\protect \frac {1}{N} \DOTSB \tsum \slimits@ _j v_j \phi _j (x_i)$
  where $v \mathrel {\mathop :}\mathrel {\mkern -1.2mu}=[v_1, v_2, \protect
  \dots v_G]^\intercal $ is a parameter and use the squared error function. For
  the details of Ridge classification, see Sec.~\ref {supp-arXiv-sec-Ridge-001}
  of the SI. Furthermore, the kernel method is described in Sec.~\ref
  {supp-arXiv-sec-kernel-001} of the SI. Ref.~\cite {Bishop001, Murphy001} are
  also helpful.}\BibitemShut {Stop}%
\bibitem [{Note15()}]{Note15}%
  \BibitemOpen
  \bibinfo {note} {In Sec.~\ref {supp-arXiv-sec-numerical-result-UKM-001} of
  the SI, the numerical results for other datasets are shown.}\BibitemShut
  {Stop}%
\bibitem [{\citenamefont {DiVincenzo}(1995)}]{Divincenzo001}%
  \BibitemOpen
  \bibfield  {author} {\bibinfo {author} {\bibfnamefont {D.~P.}\ \bibnamefont
  {DiVincenzo}},\ }\href@noop {} {\bibfield  {journal} {\bibinfo  {journal}
  {Physical Review A}\ }\textbf {\bibinfo {volume} {51}},\ \bibinfo {pages}
  {1015} (\bibinfo {year} {1995})}\BibitemShut {NoStop}%
\bibitem [{\citenamefont {Barenco}\ \emph {et~al.}(1995)\citenamefont
  {Barenco}, \citenamefont {Bennett}, \citenamefont {Cleve}, \citenamefont
  {DiVincenzo}, \citenamefont {Margolus}, \citenamefont {Shor}, \citenamefont
  {Sleator}, \citenamefont {Smolin},\ and\ \citenamefont
  {Weinfurter}}]{Barenco001}%
  \BibitemOpen
  \bibfield  {author} {\bibinfo {author} {\bibfnamefont {A.}~\bibnamefont
  {Barenco}}, \bibinfo {author} {\bibfnamefont {C.~H.}\ \bibnamefont
  {Bennett}}, \bibinfo {author} {\bibfnamefont {R.}~\bibnamefont {Cleve}},
  \bibinfo {author} {\bibfnamefont {D.~P.}\ \bibnamefont {DiVincenzo}},
  \bibinfo {author} {\bibfnamefont {N.}~\bibnamefont {Margolus}}, \bibinfo
  {author} {\bibfnamefont {P.}~\bibnamefont {Shor}}, \bibinfo {author}
  {\bibfnamefont {T.}~\bibnamefont {Sleator}}, \bibinfo {author} {\bibfnamefont
  {J.~A.}\ \bibnamefont {Smolin}}, \ and\ \bibinfo {author} {\bibfnamefont
  {H.}~\bibnamefont {Weinfurter}},\ }\href@noop {} {\bibfield  {journal}
  {\bibinfo  {journal} {Physical review A}\ }\textbf {\bibinfo {volume} {52}},\
  \bibinfo {pages} {3457} (\bibinfo {year} {1995})}\BibitemShut {NoStop}%
\bibitem [{\citenamefont {Crooks}(2019)}]{Crooks001}%
  \BibitemOpen
  \bibfield  {author} {\bibinfo {author} {\bibfnamefont {G.~E.}\ \bibnamefont
  {Crooks}},\ }\href@noop {} {\bibfield  {journal} {\bibinfo  {journal} {arXiv
  preprint arXiv:1905.13311}\ } (\bibinfo {year} {2019})}\BibitemShut {NoStop}%
\bibitem [{\citenamefont {Petersen}\ and\ \citenamefont
  {Pedersen}(2008)}]{Petersen001}%
  \BibitemOpen
  \bibfield  {author} {\bibinfo {author} {\bibfnamefont {K.}~\bibnamefont
  {Petersen}}\ and\ \bibinfo {author} {\bibfnamefont {M.}~\bibnamefont
  {Pedersen}},\ }\href@noop {} {\bibfield  {journal} {\bibinfo  {journal}
  {Technical University of Denmark}\ }\textbf {\bibinfo {volume} {15}}
  (\bibinfo {year} {2008})}\BibitemShut {NoStop}%
\bibitem [{\citenamefont {Kreutz-Delgado}(2009)}]{Kreutz001}%
  \BibitemOpen
  \bibfield  {author} {\bibinfo {author} {\bibfnamefont {K.}~\bibnamefont
  {Kreutz-Delgado}},\ }\href@noop {} {\bibfield  {journal} {\bibinfo  {journal}
  {arXiv preprint arXiv:0906.4835}\ } (\bibinfo {year} {2009})}\BibitemShut
  {NoStop}%
\bibitem [{\citenamefont {Fletcher}\ and\ \citenamefont
  {Reeves}(1964)}]{Fletcher002}%
  \BibitemOpen
  \bibfield  {author} {\bibinfo {author} {\bibfnamefont {R.}~\bibnamefont
  {Fletcher}}\ and\ \bibinfo {author} {\bibfnamefont {C.~M.}\ \bibnamefont
  {Reeves}},\ }\href@noop {} {\bibfield  {journal} {\bibinfo  {journal} {The
  computer journal}\ }\textbf {\bibinfo {volume} {7}},\ \bibinfo {pages} {149}
  (\bibinfo {year} {1964})}\BibitemShut {NoStop}%
\bibitem [{\citenamefont {Shewchuk}\ \emph {et~al.}(1994)\citenamefont
  {Shewchuk} \emph {et~al.}}]{Shewchuk001}%
  \BibitemOpen
  \bibfield  {author} {\bibinfo {author} {\bibfnamefont {J.~R.}\ \bibnamefont
  {Shewchuk}} \emph {et~al.},\ }\href@noop {} {\enquote {\bibinfo {title} {An
  introduction to the conjugate gradient method without the agonizing pain},}\
  } (\bibinfo {year} {1994})\BibitemShut {NoStop}%
\bibitem [{\citenamefont {Broyden}(1970)}]{Broyden001}%
  \BibitemOpen
  \bibfield  {author} {\bibinfo {author} {\bibfnamefont {C.~G.}\ \bibnamefont
  {Broyden}},\ }\href@noop {} {\bibfield  {journal} {\bibinfo  {journal} {IMA
  Journal of Applied Mathematics}\ }\textbf {\bibinfo {volume} {6}},\ \bibinfo
  {pages} {76} (\bibinfo {year} {1970})}\BibitemShut {NoStop}%
\bibitem [{\citenamefont {Fletcher}(1970)}]{Fletcher003}%
  \BibitemOpen
  \bibfield  {author} {\bibinfo {author} {\bibfnamefont {R.}~\bibnamefont
  {Fletcher}},\ }\href@noop {} {\bibfield  {journal} {\bibinfo  {journal} {The
  computer journal}\ }\textbf {\bibinfo {volume} {13}},\ \bibinfo {pages} {317}
  (\bibinfo {year} {1970})}\BibitemShut {NoStop}%
\bibitem [{\citenamefont {Goldfarb}(1970)}]{Goldfarb001}%
  \BibitemOpen
  \bibfield  {author} {\bibinfo {author} {\bibfnamefont {D.}~\bibnamefont
  {Goldfarb}},\ }\href@noop {} {\bibfield  {journal} {\bibinfo  {journal}
  {Mathematics of computation}\ }\textbf {\bibinfo {volume} {24}},\ \bibinfo
  {pages} {23} (\bibinfo {year} {1970})}\BibitemShut {NoStop}%
\bibitem [{\citenamefont {Shanno}(1970)}]{Shanno001}%
  \BibitemOpen
  \bibfield  {author} {\bibinfo {author} {\bibfnamefont {D.~F.}\ \bibnamefont
  {Shanno}},\ }\href@noop {} {\bibfield  {journal} {\bibinfo  {journal}
  {Mathematics of computation}\ }\textbf {\bibinfo {volume} {24}},\ \bibinfo
  {pages} {647} (\bibinfo {year} {1970})}\BibitemShut {NoStop}%
\bibitem [{\citenamefont {Armijo}(1966)}]{Armijo001}%
  \BibitemOpen
  \bibfield  {author} {\bibinfo {author} {\bibfnamefont {L.}~\bibnamefont
  {Armijo}},\ }\href@noop {} {\bibfield  {journal} {\bibinfo  {journal}
  {Pacific Journal of mathematics}\ }\textbf {\bibinfo {volume} {16}},\
  \bibinfo {pages} {1} (\bibinfo {year} {1966})}\BibitemShut {NoStop}%
\bibitem [{\citenamefont {Wolfe}(1969)}]{Wolfe001}%
  \BibitemOpen
  \bibfield  {author} {\bibinfo {author} {\bibfnamefont {P.}~\bibnamefont
  {Wolfe}},\ }\href@noop {} {\bibfield  {journal} {\bibinfo  {journal} {SIAM
  review}\ }\textbf {\bibinfo {volume} {11}},\ \bibinfo {pages} {226} (\bibinfo
  {year} {1969})}\BibitemShut {NoStop}%
\bibitem [{\citenamefont {Wolfe}(1971)}]{Wolfe002}%
  \BibitemOpen
  \bibfield  {author} {\bibinfo {author} {\bibfnamefont {P.}~\bibnamefont
  {Wolfe}},\ }\href@noop {} {\bibfield  {journal} {\bibinfo  {journal} {SIAM
  review}\ }\textbf {\bibinfo {volume} {13}},\ \bibinfo {pages} {185} (\bibinfo
  {year} {1971})}\BibitemShut {NoStop}%
\bibitem [{\citenamefont {Polak}\ and\ \citenamefont
  {Ribiere}(1969)}]{Polak001}%
  \BibitemOpen
  \bibfield  {author} {\bibinfo {author} {\bibfnamefont {E.}~\bibnamefont
  {Polak}}\ and\ \bibinfo {author} {\bibfnamefont {G.}~\bibnamefont
  {Ribiere}},\ }\href@noop {} {\bibfield  {journal} {\bibinfo  {journal}
  {ESAIM: Mathematical Modelling and Numerical Analysis-Mod{\'e}lisation
  Math{\'e}matique et Analyse Num{\'e}rique}\ }\textbf {\bibinfo {volume}
  {3}},\ \bibinfo {pages} {35} (\bibinfo {year} {1969})}\BibitemShut {NoStop}%
\bibitem [{\citenamefont {Hestenes}\ \emph {et~al.}(1952)\citenamefont
  {Hestenes}, \citenamefont {Stiefel} \emph {et~al.}}]{Hestenes001}%
  \BibitemOpen
  \bibfield  {author} {\bibinfo {author} {\bibfnamefont {M.~R.}\ \bibnamefont
  {Hestenes}}, \bibinfo {author} {\bibfnamefont {E.}~\bibnamefont {Stiefel}},
  \emph {et~al.},\ }\href@noop {} {\bibfield  {journal} {\bibinfo  {journal}
  {Journal of research of the National Bureau of Standards}\ }\textbf {\bibinfo
  {volume} {49}},\ \bibinfo {pages} {409} (\bibinfo {year} {1952})}\BibitemShut
  {NoStop}%
\bibitem [{\citenamefont {Dai}\ and\ \citenamefont {Yuan}(2001)}]{Dai001}%
  \BibitemOpen
  \bibfield  {author} {\bibinfo {author} {\bibfnamefont {Y.-H.}\ \bibnamefont
  {Dai}}\ and\ \bibinfo {author} {\bibfnamefont {Y.-x.}\ \bibnamefont {Yuan}},\
  }\href@noop {} {\bibfield  {journal} {\bibinfo  {journal} {Annals of
  Operations Research}\ }\textbf {\bibinfo {volume} {103}},\ \bibinfo {pages}
  {33} (\bibinfo {year} {2001})}\BibitemShut {NoStop}%
\bibitem [{\citenamefont {Schuld}\ \emph
  {et~al.}(2020{\natexlab{b}})\citenamefont {Schuld}, \citenamefont {Sweke},\
  and\ \citenamefont {Meyer}}]{Schuld002}%
  \BibitemOpen
  \bibfield  {author} {\bibinfo {author} {\bibfnamefont {M.}~\bibnamefont
  {Schuld}}, \bibinfo {author} {\bibfnamefont {R.}~\bibnamefont {Sweke}}, \
  and\ \bibinfo {author} {\bibfnamefont {J.~J.}\ \bibnamefont {Meyer}},\
  }\href@noop {} {\bibfield  {journal} {\bibinfo  {journal} {arXiv preprint
  arXiv:2008.08605}\ } (\bibinfo {year} {2020}{\natexlab{b}})}\BibitemShut
  {NoStop}%
\bibitem [{\citenamefont {Fujii}\ and\ \citenamefont
  {Nakajima}(2017)}]{Fujii001}%
  \BibitemOpen
  \bibfield  {author} {\bibinfo {author} {\bibfnamefont {K.}~\bibnamefont
  {Fujii}}\ and\ \bibinfo {author} {\bibfnamefont {K.}~\bibnamefont
  {Nakajima}},\ }\href@noop {} {\bibfield  {journal} {\bibinfo  {journal}
  {Physical Review Applied}\ }\textbf {\bibinfo {volume} {8}},\ \bibinfo
  {pages} {024030} (\bibinfo {year} {2017})}\BibitemShut {NoStop}%
\bibitem [{\citenamefont {Nakajima}\ \emph {et~al.}(2019)\citenamefont
  {Nakajima}, \citenamefont {Fujii}, \citenamefont {Negoro}, \citenamefont
  {Mitarai},\ and\ \citenamefont {Kitagawa}}]{Nakajima001}%
  \BibitemOpen
  \bibfield  {author} {\bibinfo {author} {\bibfnamefont {K.}~\bibnamefont
  {Nakajima}}, \bibinfo {author} {\bibfnamefont {K.}~\bibnamefont {Fujii}},
  \bibinfo {author} {\bibfnamefont {M.}~\bibnamefont {Negoro}}, \bibinfo
  {author} {\bibfnamefont {K.}~\bibnamefont {Mitarai}}, \ and\ \bibinfo
  {author} {\bibfnamefont {M.}~\bibnamefont {Kitagawa}},\ }\href@noop {}
  {\bibfield  {journal} {\bibinfo  {journal} {Physical Review Applied}\
  }\textbf {\bibinfo {volume} {11}},\ \bibinfo {pages} {034021} (\bibinfo
  {year} {2019})}\BibitemShut {NoStop}%
\bibitem [{\citenamefont {Rockafellar}(1970)}]{Rockafellar001}%
  \BibitemOpen
  \bibfield  {author} {\bibinfo {author} {\bibfnamefont {R.~T.}\ \bibnamefont
  {Rockafellar}},\ }\href@noop {} {\emph {\bibinfo {title} {Convex
  analysis}}},\ \bibinfo {number} {28}\ (\bibinfo  {publisher} {Princeton
  university press},\ \bibinfo {year} {1970})\BibitemShut {NoStop}%
\bibitem [{\citenamefont {Bregman}(1967)}]{Bregman001}%
  \BibitemOpen
  \bibfield  {author} {\bibinfo {author} {\bibfnamefont {L.~M.}\ \bibnamefont
  {Bregman}},\ }\href@noop {} {\bibfield  {journal} {\bibinfo  {journal} {USSR
  computational mathematics and mathematical physics}\ }\textbf {\bibinfo
  {volume} {7}},\ \bibinfo {pages} {200} (\bibinfo {year} {1967})}\BibitemShut
  {NoStop}%
\bibitem [{\citenamefont {Hale}\ \emph {et~al.}(2007)\citenamefont {Hale},
  \citenamefont {Yin},\ and\ \citenamefont {Zhang}}]{Hale001}%
  \BibitemOpen
  \bibfield  {author} {\bibinfo {author} {\bibfnamefont {E.~T.}\ \bibnamefont
  {Hale}}, \bibinfo {author} {\bibfnamefont {W.}~\bibnamefont {Yin}}, \ and\
  \bibinfo {author} {\bibfnamefont {Y.}~\bibnamefont {Zhang}},\ }\href@noop {}
  {\emph {\bibinfo {title} {A fixed-point continuation method for
  L\_1-regularization with application to compressed sensing}}},\ \bibinfo
  {type} {Tech. Rep.}\ (\bibinfo {year} {2007})\BibitemShut {NoStop}%
\bibitem [{\citenamefont {Manton}(2002)}]{Manton001}%
  \BibitemOpen
  \bibfield  {author} {\bibinfo {author} {\bibfnamefont {J.~H.}\ \bibnamefont
  {Manton}},\ }\href@noop {} {\bibfield  {journal} {\bibinfo  {journal} {IEEE
  Transactions on Signal Processing}\ }\textbf {\bibinfo {volume} {50}},\
  \bibinfo {pages} {635} (\bibinfo {year} {2002})}\BibitemShut {NoStop}%
\bibitem [{\citenamefont {Gibson}(1962)}]{Gibson001}%
  \BibitemOpen
  \bibfield  {author} {\bibinfo {author} {\bibfnamefont {W.}~\bibnamefont
  {Gibson}},\ }\href@noop {} {\bibfield  {journal} {\bibinfo  {journal}
  {Psychometrika}\ }\textbf {\bibinfo {volume} {27}},\ \bibinfo {pages} {193}
  (\bibinfo {year} {1962})}\BibitemShut {NoStop}%
\bibitem [{\citenamefont {Schuld}\ and\ \citenamefont
  {Killoran}(2019)}]{Schuld003}%
  \BibitemOpen
  \bibfield  {author} {\bibinfo {author} {\bibfnamefont {M.}~\bibnamefont
  {Schuld}}\ and\ \bibinfo {author} {\bibfnamefont {N.}~\bibnamefont
  {Killoran}},\ }\href@noop {} {\bibfield  {journal} {\bibinfo  {journal}
  {Physical review letters}\ }\textbf {\bibinfo {volume} {122}},\ \bibinfo
  {pages} {040504} (\bibinfo {year} {2019})}\BibitemShut {NoStop}%
\bibitem [{Note16()}]{Note16}%
  \BibitemOpen
  \bibinfo {note} {The SciPy package~\cite {Virtanen001} provides the \protect
  \textbf {optimize} function, in which the CG method, the BFGS method, and
  other optimization methods are implemented.}\BibitemShut {Stop}%
\bibitem [{\citenamefont {Virtanen}\ \emph {et~al.}(2020)\citenamefont
  {Virtanen}, \citenamefont {Gommers}, \citenamefont {Oliphant}, \citenamefont
  {Haberland}, \citenamefont {Reddy}, \citenamefont {Cournapeau}, \citenamefont
  {Burovski}, \citenamefont {Peterson}, \citenamefont {Weckesser},
  \citenamefont {Bright}, \citenamefont {{van der Walt}}, \citenamefont
  {Brett}, \citenamefont {Wilson}, \citenamefont {Millman}, \citenamefont
  {Mayorov}, \citenamefont {Nelson}, \citenamefont {Jones}, \citenamefont
  {Kern}, \citenamefont {Larson}, \citenamefont {Carey}, \citenamefont {Polat},
  \citenamefont {Feng}, \citenamefont {Moore}, \citenamefont {{VanderPlas}},
  \citenamefont {Laxalde}, \citenamefont {Perktold}, \citenamefont {Cimrman},
  \citenamefont {Henriksen}, \citenamefont {Quintero}, \citenamefont {Harris},
  \citenamefont {Archibald}, \citenamefont {Ribeiro}, \citenamefont
  {Pedregosa}, \citenamefont {{van Mulbregt}},\ and\ \citenamefont {{SciPy 1.0
  Contributors}}}]{Virtanen001}%
  \BibitemOpen
  \bibfield  {author} {\bibinfo {author} {\bibfnamefont {P.}~\bibnamefont
  {Virtanen}}, \bibinfo {author} {\bibfnamefont {R.}~\bibnamefont {Gommers}},
  \bibinfo {author} {\bibfnamefont {T.~E.}\ \bibnamefont {Oliphant}}, \bibinfo
  {author} {\bibfnamefont {M.}~\bibnamefont {Haberland}}, \bibinfo {author}
  {\bibfnamefont {T.}~\bibnamefont {Reddy}}, \bibinfo {author} {\bibfnamefont
  {D.}~\bibnamefont {Cournapeau}}, \bibinfo {author} {\bibfnamefont
  {E.}~\bibnamefont {Burovski}}, \bibinfo {author} {\bibfnamefont
  {P.}~\bibnamefont {Peterson}}, \bibinfo {author} {\bibfnamefont
  {W.}~\bibnamefont {Weckesser}}, \bibinfo {author} {\bibfnamefont
  {J.}~\bibnamefont {Bright}}, \bibinfo {author} {\bibfnamefont {S.~J.}\
  \bibnamefont {{van der Walt}}}, \bibinfo {author} {\bibfnamefont
  {M.}~\bibnamefont {Brett}}, \bibinfo {author} {\bibfnamefont
  {J.}~\bibnamefont {Wilson}}, \bibinfo {author} {\bibfnamefont {K.~J.}\
  \bibnamefont {Millman}}, \bibinfo {author} {\bibfnamefont {N.}~\bibnamefont
  {Mayorov}}, \bibinfo {author} {\bibfnamefont {A.~R.~J.}\ \bibnamefont
  {Nelson}}, \bibinfo {author} {\bibfnamefont {E.}~\bibnamefont {Jones}},
  \bibinfo {author} {\bibfnamefont {R.}~\bibnamefont {Kern}}, \bibinfo {author}
  {\bibfnamefont {E.}~\bibnamefont {Larson}}, \bibinfo {author} {\bibfnamefont
  {C.~J.}\ \bibnamefont {Carey}}, \bibinfo {author} {\bibfnamefont
  {{\.I}.}~\bibnamefont {Polat}}, \bibinfo {author} {\bibfnamefont
  {Y.}~\bibnamefont {Feng}}, \bibinfo {author} {\bibfnamefont {E.~W.}\
  \bibnamefont {Moore}}, \bibinfo {author} {\bibfnamefont {J.}~\bibnamefont
  {{VanderPlas}}}, \bibinfo {author} {\bibfnamefont {D.}~\bibnamefont
  {Laxalde}}, \bibinfo {author} {\bibfnamefont {J.}~\bibnamefont {Perktold}},
  \bibinfo {author} {\bibfnamefont {R.}~\bibnamefont {Cimrman}}, \bibinfo
  {author} {\bibfnamefont {I.}~\bibnamefont {Henriksen}}, \bibinfo {author}
  {\bibfnamefont {E.~A.}\ \bibnamefont {Quintero}}, \bibinfo {author}
  {\bibfnamefont {C.~R.}\ \bibnamefont {Harris}}, \bibinfo {author}
  {\bibfnamefont {A.~M.}\ \bibnamefont {Archibald}}, \bibinfo {author}
  {\bibfnamefont {A.~H.}\ \bibnamefont {Ribeiro}}, \bibinfo {author}
  {\bibfnamefont {F.}~\bibnamefont {Pedregosa}}, \bibinfo {author}
  {\bibfnamefont {P.}~\bibnamefont {{van Mulbregt}}}, \ and\ \bibinfo {author}
  {\bibnamefont {{SciPy 1.0 Contributors}}},\ }\href {\doibase
  10.1038/s41592-019-0686-2} {\bibfield  {journal} {\bibinfo  {journal} {Nature
  Methods}\ }\textbf {\bibinfo {volume} {17}},\ \bibinfo {pages} {261}
  (\bibinfo {year} {2020})}\BibitemShut {NoStop}%
\bibitem [{\citenamefont {Iten}\ \emph {et~al.}(2016)\citenamefont {Iten},
  \citenamefont {Colbeck}, \citenamefont {Kukuljan}, \citenamefont {Home},\
  and\ \citenamefont {Christandl}}]{Iten001}%
  \BibitemOpen
  \bibfield  {author} {\bibinfo {author} {\bibfnamefont {R.}~\bibnamefont
  {Iten}}, \bibinfo {author} {\bibfnamefont {R.}~\bibnamefont {Colbeck}},
  \bibinfo {author} {\bibfnamefont {I.}~\bibnamefont {Kukuljan}}, \bibinfo
  {author} {\bibfnamefont {J.}~\bibnamefont {Home}}, \ and\ \bibinfo {author}
  {\bibfnamefont {M.}~\bibnamefont {Christandl}},\ }\href@noop {} {\bibfield
  {journal} {\bibinfo  {journal} {Physical Review A}\ }\textbf {\bibinfo
  {volume} {93}},\ \bibinfo {pages} {032318} (\bibinfo {year}
  {2016})}\BibitemShut {NoStop}%
\bibitem [{\citenamefont {Knill}(1995)}]{Knill001}%
  \BibitemOpen
  \bibfield  {author} {\bibinfo {author} {\bibfnamefont {E.}~\bibnamefont
  {Knill}},\ }\href@noop {} {\bibfield  {journal} {\bibinfo  {journal} {arXiv
  preprint quant-ph/9508006}\ } (\bibinfo {year} {1995})}\BibitemShut {NoStop}%
\bibitem [{\citenamefont {LeCun}\ and\ \citenamefont
  {Cortes}(2010)}]{LeCun001}%
  \BibitemOpen
  \bibfield  {author} {\bibinfo {author} {\bibfnamefont {Y.}~\bibnamefont
  {LeCun}}\ and\ \bibinfo {author} {\bibfnamefont {C.}~\bibnamefont {Cortes}},\
  }\href {http://yann.lecun.com/exdb/mnist/} {\  (\bibinfo {year}
  {2010})}\BibitemShut {NoStop}%
\end{thebibliography}%

\onecolumngrid
\appendix

\section{Introduction}

This is the supplemental information (SI) for the paper entitled ``Ansatz-Independent Variational Quantum Classifier."

The SI has two major purposes.
The first one is to provide the detailed descriptions of the unitary kernel method (UKM) and the variational circuit realization (VCR).
The second one is to show additional numerical results of the UKM and the VCR to support the statements in the main text.
Furthermore, we explain some useful formulas and related algorithms.

\section{Notation}

We introduce some symbols: product operators and matrix norms.

\subsection{Product operator} \label{supp-arXiv-sec-product-operator-001}

We define
\begin{align}
  \sideset{}{^\uparrow} \prod_{j=k}^{k+l} \hat{A}_j &\coloneqq \hat{A}_k \hat{A}_{k+1} \dots \hat{A}_{k+l}, \label{supp-arXiv-def-product-operator-uparrow-001} \\
  \sideset{}{^\downarrow} \prod_{j=k}^{k+l} \hat{A}_j &\coloneqq \hat{A}_{k+l} \hat{A}_{k+l-1} \dots \hat{A}_k. \label{supp-arXiv-def-product-operator-downarrow-001}
\end{align}
When Eq.~\eqref{supp-arXiv-def-product-operator-uparrow-001} and Eq.~\eqref{supp-arXiv-def-product-operator-downarrow-001} are identical, we denote them by $\prod_{j=k}^{k+l} \hat{A}_j$.

\subsection{Matrix norms}

Let $A$ be a $M \times N$ matrix.
We define $a_{i, j} \coloneqq [ A ]_{i,j}$ for $i = 1, 2, \dots, M$ and $j = 1, 2, \dots, N$.
For $p > 0$, we then define the matrix norm by
\begin{align}
  \| A \|_p &\coloneqq \bigg( \sum_i \sum_j | a_{i, j} |^p \bigg)^\frac{1}{p}. \label{supp-arXiv-def-matrix-norm-001}
\end{align}

From Eq.~\eqref{supp-arXiv-def-matrix-norm-001}, we have
\begin{align}
  \| A \|_\mathrm{F} &\coloneqq \| A \|_2 \\
  &= ( \mathrm{Tr} [A^\mathrm{H} A])^\frac{1}{2} \label{supp-arXiv-another-expression-matrix-norm-001} \\
  &= \bigg( \sum_{i=1}^M \sum_{j=1}^N | a_{i, j} |^2 \bigg)^\frac{1}{2} \\
  &= \bigg( \sum_{i=1}^{\min (M, N)} \sigma_i^2 \bigg)^\frac{1}{2},
\end{align}
where $(\cdot)^\mathrm{H}$ denotes the Hermitian conjugate of a matrix.

\section{Rotation operators and their controlled versions}

In QCL, rotation gates and their controlled versions play an important role.
In this section, we then review them.

\subsection{Many-body operator}

When we consider a $n$-qubit system, it is important to introduce many-body operators.
Let $\hat{\sigma}$ be a single-qubit operator.
We denote, by $\hat{\sigma}_i$, $\hat{\sigma}$ acting on the $i$-th qubit of a $n$-qubit system:
\begin{align}
  \hat{\sigma}_i &\coloneqq \underbrace{\hat{1} \otimes \dots \otimes \hat{1}}_{i-1} \otimes \hat{\sigma} \otimes \underbrace{\hat{1} \otimes \dots \otimes \hat{1}}_{n-i}.
\end{align}

For practical computations, we use the Kronecker product.
We give its definition:
\begin{align}
A \otimes B &\coloneqq
\begin{bmatrix}
a_{1, 1} B & a_{1, 2} B & \dots \\
a_{2, 1} B & a_{2, 2} B & \dots \\
\vdots & \vdots & \ddots
\end{bmatrix},
\end{align}
where $a_{i, j} \coloneqq [A]_{i, j}$ is the element of $A$ in the $i$-th row and $j$-th column.

The Kronecker product satisfies the following relation:
\begin{align}
  \hat{A} \otimes \hat{B} &= (\hat{A} \otimes \hat{1}) (\hat{1} \otimes \hat{B}) \\
  &= \hat{A}_1 \hat{B}_2,
\end{align}
where
\begin{align}
  \hat{A}_1 &\coloneqq \hat{A} \otimes \hat{1}, \\
  \hat{B}_2 &\coloneqq \hat{1} \otimes \hat{B}.
\end{align}
In the case of projection operators, we have
\begin{align}
  \hat{P}^0 \otimes \hat{A} + \hat{P}^1 \otimes \hat{B} &= \hat{P}_1^0 \hat{A}_2 + \hat{P}_1^1 \hat{B}_2,
\end{align}
where
\begin{align}
  \hat{P}^0 &\coloneqq | 0 \rangle \langle 0 |, \\
  \hat{P}^1 &\coloneqq | 1 \rangle \langle 1 |.
\end{align}

\subsection{Rotation gates} \label{supp-arXiv-sec-def-rotation-qubit-001}

Two-qubit gates are known to be universal~\cite{Divincenzo001, Nielsen001}, and we often use single-qubit rotation gates and CNOT gates in practice~\cite{Barenco001, Nielsen001}.
Then, we review single-qubit rotation operators $\hat{R}^x (\phi)$, $\hat{R}^y (\phi)$, and $\hat{R}^z (\phi)$.

\subsubsection{One-qubit case}

We first introduce Pauli-matrices:
\begin{align}
\hat{X} &\coloneqq \hat{\sigma}^x \\
&=
\begin{bmatrix}
0 & 1 \\
1 & 0
\end{bmatrix}, \\
\hat{Y} &\coloneqq \hat{\sigma}^y \\
&=
\begin{bmatrix}
0 & -i \\
i & 0
\end{bmatrix}, \\
\hat{Z} &\coloneqq \hat{\sigma}^z \\
&=
\begin{bmatrix}
1 & 0 \\
0 & -1
\end{bmatrix}.
\end{align}
Rotation gates with respect to $\hat{X}$, $\hat{Y}$, and $\hat{Z}$ are written as
\begin{align}
\hat{R}^x (\phi) &\coloneqq e^{-i \frac{\phi}{2} \hat{X}} \\
&= \hat{1}_2 \cos (\phi / 2) - i \hat{X} \sin (\phi / 2) \\
&=
\begin{bmatrix}
\cos (\phi / 2) & -i \sin (\phi / 2) \\
-i \sin (\phi / 2) & \cos (\phi / 2)
\end{bmatrix}, \label{supp-arXiv-def-rotation-X-001} \\
\hat{R}^y (\phi) &\coloneqq e^{-i \frac{\phi}{2} \hat{Y}} \\
&= \hat{1}_2 \cos (\phi / 2) - i \hat{Y} \sin (\phi / 2) \\
&=
\begin{bmatrix}
\cos (\phi / 2) & - \sin (\phi / 2) \\
\sin (\phi / 2) & \cos (\phi / 2)
\end{bmatrix}, \label{supp-arXiv-def-rotation-Y-001} \\
\hat{R}^z (\phi) &\coloneqq e^{-i \frac{\phi}{2} \hat{Z}} \\
&= \hat{1}_2 \cos (\phi / 2) - i \hat{Z} \sin (\phi / 2) \\
&=
\begin{bmatrix}
e^{-i \phi / 2} & 0 \\
0 & e^{i \phi / 2}
\end{bmatrix}. \label{supp-arXiv-def-rotation-Z-001}
\end{align}

As shown in Ref.~\cite{Barenco001}, the $3$-dimensional rotation gate can be expressed as
\begin{align}
\hat{R}^\mathrm{3d} (\phi, \theta, \omega) &\coloneqq \hat{R}^z (\omega) \hat{R}^y (\theta) \hat{R}^z (\phi) \\
&=
\begin{bmatrix}
e^{-i \frac{\phi + \omega}{2}} \cos (\theta / 2) & - e^{i \frac{\phi - \omega}{2}} \sin (\theta / 2) \\
e^{-i \frac{\phi - \omega}{2}} \sin (\theta / 2) & e^{i \frac{\phi + \omega}{2}} \cos (\theta / 2)
\end{bmatrix}. \label{supp-arXiv-def-3d-rotation-001}
\end{align}
Eq.~\eqref{supp-arXiv-def-3d-rotation-001} is called the $ZYZ$ decomposition and the combination of this decomposition and the CNOT gate is very useful.

By applying the global phase operator to Eq.~\eqref{supp-arXiv-def-3d-rotation-001}, we can express any unitary operator:
\begin{align}
  \hat{R}^\mathrm{3d} (\phi, \theta, \omega) \hat{\Phi}_2 (\lambda) &=
  \begin{bmatrix}
  e^{-i (\lambda + \frac{\phi + \omega}{2})} \cos (\theta / 2) & - e^{-i (\lambda - \frac{\phi - \omega}{2})} \sin (\theta / 2) \\
  e^{-i (\lambda + \frac{\phi - \omega}{2})} \sin (\theta / 2) & e^{-i (\lambda - \frac{\phi + \omega}{2})} \cos (\theta / 2)
  \end{bmatrix}, \label{supp-arXiv-ZYZ-global-phase-001-001}
\end{align}
where, for $n = 1, 2, \dots$,
\begin{align}
  \hat{\Phi}_{2^n} (\lambda) &\coloneqq e^{- i \lambda} \hat{1}_{2^n}. \label{supp-arXiv-def-global-phase-unitary-001-001}
\end{align}

\subsubsection{Many-qubit case}

We denote $\hat{X}$, $\hat{Y}$, and $\hat{Z}$ acting on the $i$-th qubit of a $n$-qubit system by $\hat{X}_i$, $\hat{Y}_i$, and $\hat{Z}_i$, respectively.
Furthermore, $\hat{R}_j^\mathrm{3d} (\phi, \theta, \omega) \coloneqq \hat{R}_j^z (\omega) \hat{R}_j^y (\theta) \hat{R}_j^z (\phi)$ is the 3-dimensional rotation gate defined in Eq.~\eqref{supp-arXiv-def-3d-rotation-001} on the $j$-th qubit of a $n$-qubit system~\cite{Barenco001, Nielsen001}.

\subsection{Controlled gates}

\subsubsection{Many-qubit case}

Suppose that we have a $n$-qubit system.
We denote, by $\mathrm{Ct}_i [\cdot]$, the controlled two-qubit gate of a given single-qubit gate controlled by the $i$-th qubit of a $n$-qubit system.
Then, we have
\begin{align}
  \mathrm{Ct}_i [\hat{G}_j] &= \hat{P}_i^0 + \hat{P}_i^1 \hat{G}_j,
\end{align}
where $\hat{G}_j$ is $\hat{G}$ acting on the $j$-th qubit for $j=1, 2, \dots, n$, and
\begin{align}
  \hat{P}_i^0 &\coloneqq \underbrace{\hat{1}_2 \otimes \dots \otimes \hat{1}_2}_{i-1} \otimes | 0 \rangle \langle 0 | \otimes \underbrace{\hat{1}_2 \otimes \dots \otimes \hat{1}_2}_{n-i} \\
  &= \frac{1}{2} (\hat{1}_{2^n} + \hat{Z}_i), \\
  \hat{P}_i^1 &\coloneqq \underbrace{\hat{1}_2 \otimes \dots \otimes \hat{1}_2}_{i-1} \otimes | 1 \rangle \langle 1 | \otimes \underbrace{\hat{1}_2 \otimes \dots \otimes \hat{1}_2}_{n-i} \\
  &= \frac{1}{2} (\hat{1}_{2^n} - \hat{Z}_i).
\end{align}
Here, $\hat{Z}_i$ is the Pauli-$Z$ operator acting on the $i$-th qubit of a $n$-qubit system.

\subsubsection{Two-qubit case}

Let us consider a controlled gate where the first qubit is the control qubit and the second qubit is the target qubit.
For any single-qubit operator $\hat{A}$, its controlled version is defined as
\begin{align}
\mathrm{Ct}_1 [\hat{A}_2] &\coloneqq | 0 \rangle \langle 0 | \otimes \hat{1}_2 + | 1 \rangle \langle 1 | \otimes \hat{A} \\
&=
\begin{bmatrix}
\hat{1}_2 & \hat{0}_2  \\
\hat{0}_2 & \hat{A}
\end{bmatrix},
\end{align}
where $\hat{1}_2$ and $\hat{0}_2$ are the two-dimensional identity operator and the two-dimensional zero operator, respectively.
Note that $| 0 \rangle = | \uparrow \rangle = [1, 0]^\intercal$ and $| 1 \rangle = | \downarrow \rangle = [0, 1]^\intercal$.

For example, we have
\begin{align}
\mathrm{Ct}_1 [\hat{X}_2] &= | 0 \rangle \langle 0 | \otimes \hat{1}_2 + | 1 \rangle \langle 1 | \otimes \hat{X} \\
&=
\begin{bmatrix}
1 & 0 & 0 & 0 \\
0 & 1 & 0 & 0 \\
0 & 0 & 0 & 1 \\
0 & 0 & 1 & 0
\end{bmatrix}, \\
\mathrm{Ct}_2 [\hat{X}_1] &= \hat{1}_2 \otimes | 0 \rangle \langle 0 | + \hat{X} \otimes | 1 \rangle \langle 1 | \\
&=
\begin{bmatrix}
1 & 0 & 0 & 0 \\
0 & 0 & 0 & 1 \\
0 & 0 & 1 & 0 \\
0 & 1 & 0 & 0
\end{bmatrix}, \\
\mathrm{Ct}_1 [\hat{R}_2^z (\phi)] &= \mathrm{Ct}_1 [e^{-i \frac{\phi}{2} \hat{Z}_2}] \\
&=
\begin{bmatrix}
1 & 0 & 0 & 0 \\
0 & 1 & 0 & 0 \\
0 & 0 & e^{-i \phi / 2} & 0 \\
0 & 0 & 0 & e^{i \phi / 2}
\end{bmatrix},
\end{align}
and
\begin{align}
\mathrm{Ct}_1 [\hat{R}_2^\mathrm{3d} (\phi, \theta, \omega)] &= \mathrm{Ct}_1 [\hat{R}_2^z (\omega)] \mathrm{Ct}_1 [\hat{R}_2^y (\theta)] \mathrm{Ct}_1 [\hat{R}_2^z (\phi)] \\
&=
\begin{bmatrix}
1 & 0 & 0 & 0 \\
0 & 1 & 0 & 0 \\
0 & 0 & e^{-i(\phi + \omega)/2} \cos (\theta / 2) & - e^{-i(\phi - \omega)/2} \sin (\theta / 2) \\
0 & 0 & e^{-i(\phi - \omega)/2} \sin (\theta / 2) & e^{i(\phi + \omega)/2} \cos (\theta / 2)
\end{bmatrix}.
\end{align}

\subsection{Gradients on gate operators} \label{supp-arXiv-sec-gradient-circuit-parameters-001}

In the case of QCL, we often estimate the gradients of expectation values.
Here, we review it by following Ref.~\cite{Crooks001}.

\subsubsection{Multiple gates}

The unitary operator in QCL takes the following form:
\begin{align}
  \hat{U}_\mathrm{c} (\theta) &\coloneqq \sideset{}{^\downarrow}\prod_{i=1}^L \hat{U}_{\mathrm{c}, i} (\theta_i), \\
  \theta &\coloneqq [\theta_1, \theta_2, \dots, \theta_L]^\intercal.
\end{align}
Then, its derivative becomes
\begin{align}
  \frac{d}{d \theta_j} \hat{U}_\mathrm{c} (\theta) &= \sum_{i=1}^L \Bigg[ \sideset{}{^\downarrow}\prod_{i'=i+1}^L \hat{U}_{\mathrm{c}, i'} (\theta_{i'}) \Bigg] \Bigg( \frac{d}{d \theta_j} \hat{U}_{\mathrm{c}, i} (\theta_i) \Bigg) \Bigg[ \sideset{}{^\downarrow}\prod_{i'=1}^{i-1} \hat{U}_{\mathrm{c}, i'} (\theta_{i'}) \Bigg]. \label{supp-arXiv-derivative-circuit-geometry-001}
\end{align}

\subsubsection{Expectation values}

Formula for computing the derivative of parameters of the rotation gates, Eqs.~\eqref{supp-arXiv-def-rotation-X-001}, \eqref{supp-arXiv-def-rotation-Y-001}, and \eqref{supp-arXiv-def-rotation-Z-001}, are very useful.
Here, we describe them by following Ref.~\cite{Crooks001}.

Let us consider the following function:
\begin{align}
  f_{\hat{O}} (\theta) &\coloneqq \langle \psi | \hat{U}_\mathrm{c}^\dagger (\theta) \hat{O} \hat{U}_\mathrm{c} (\theta) | \psi \rangle, \label{supp-arXiv-def-function-f-expectation-001}
\end{align}
where $a$ is a constant, $\hat{G}$ is a unitary and Hermitian operator that has two unique eigenvalues, and
\begin{align}
  \hat{U}_\mathrm{c} (\theta) &\coloneqq \exp ( - i a \theta \hat{G}).
\end{align}
Note that $\hat{X}_i$, $\hat{Y}_i$, and $\hat{Z}_i$ all satisfy the conditions of $\hat{G}$.

The derivative of Eq.~\eqref{supp-arXiv-def-function-f-expectation-001} is given by
\begin{align}
  \frac{d}{d \theta} f_{\hat{O}} (\theta) &= a [ f_{\hat{O}} (\theta + \pi / 4 a) - f_{\hat{O}} (\theta - \pi / 4 a) ]. \label{supp-arXiv-formula-gradient-f-001-001}
\end{align}

The proof of Eq.~\eqref{supp-arXiv-formula-gradient-f-001-001} is shown as follows.
\begin{proof}

Since $\hat{G}$ is unitary and Hermitian, we have
\begin{align}
  \hat{G} \hat{G} &= \hat{G}^\dagger \hat{G} \\
  &= \hat{1}.
\end{align}
Then, it leads to
\begin{align}
  \hat{U}_\mathrm{c} (\theta) &= \exp (- i a \theta \hat{G}) \\
  &= \sum_{j=0}^\infty \frac{1}{j!} (-i a \theta \hat{G})^j \\
  &= \sum_{j=0}^\infty \frac{1}{(2j)!} (-i a \theta \hat{G})^{(2j)} + \sum_{j=0}^\infty \frac{1}{(2j+1)!} (-i a \theta \hat{G})^{(2j+1)} \\
  &= \hat{1} \sum_{j=0}^\infty \frac{(-1)^j}{(2j)!} (a \theta)^{(2j)} - i \hat{G} \sum_{j=0}^\infty \frac{(-1)^j}{(2j+1)!} (a \theta)^{(2j+1)} \\
  &= \hat{1} \cos (a \theta) - i \hat{G} \sin (a \theta),
\end{align}
and
\begin{align}
  \frac{d}{d \theta} \hat{U}_\mathrm{c} (\theta) &= \frac{d}{d \theta} \exp (- i a \theta \hat{G}) \\
  &= - i a \hat{G} \exp (- i a \theta \hat{G}) \\
  &= - i a \hat{G} \hat{U}_\mathrm{c} (\theta). \label{supp-arXiv-derivative-Unitary-gate-001-001}
\end{align}

The derivative of Eq.~\eqref{supp-arXiv-def-function-f-expectation-001} becomes
\begin{align}
  \frac{d}{d \theta} f_{\hat{O}} (\theta) &= \frac{d}{d \theta} \langle \psi | \hat{U}_\mathrm{c}^\dagger (\theta) \hat{O} \hat{U}_\mathrm{c} (\theta) | \psi \rangle \\
  &= \langle \psi | \bigg[ \frac{d}{d \theta} \hat{U}_\mathrm{c}^\dagger (\theta) \bigg] \hat{O} \hat{U}_\mathrm{c} (\theta) | \psi \rangle + \langle \psi | \hat{U}_\mathrm{c}^\dagger (\theta) \hat{O} \bigg[ \frac{d}{d \theta} \hat{U}_\mathrm{c} (\theta) \bigg] | \psi \rangle \\
  &= \langle \psi | [ i a \hat{G} \hat{U}_\mathrm{c}^\dagger (\theta) ] \hat{O} \hat{U}_\mathrm{c} (\theta) | \psi \rangle + \langle \psi | \hat{U}_\mathrm{c}^\dagger (\theta) \hat{O} [ - i a \hat{G} \hat{U}_\mathrm{c} (\theta) ] | \psi \rangle \\
  &= a \langle \psi | \hat{U}_\mathrm{c}^\dagger (\theta) \bigg[ \frac{\hat{1} + i \hat{G}}{\sqrt{2}} \bigg] \hat{O} \bigg[ \frac{\hat{1} - i \hat{G}}{\sqrt{2}} \bigg] \hat{U}_\mathrm{c} (\theta) | \psi \rangle - a \langle \psi | \hat{U}_\mathrm{c}^\dagger (\theta) \bigg[ \frac{\hat{1} - i \hat{G}}{\sqrt{2}} \bigg] \hat{O} \bigg[ \frac{\hat{1} + i \hat{G}}{\sqrt{2}} \bigg] \hat{U}_\mathrm{c} (\theta) | \psi \rangle \\
  &= a \langle \psi | \hat{U}_\mathrm{c}^\dagger (\theta + \pi / 4 a) \hat{O} \hat{U}_\mathrm{c} (\theta + \pi / 4 a) | \psi \rangle - a \langle \psi | \hat{U}_\mathrm{c}^\dagger (\theta - \pi / 4 a) \hat{O} \hat{U}_\mathrm{c} (\theta - \pi / 4 a) | \psi \rangle \\
  &= a [ f_{\hat{O}} (\theta + \pi / 4 a) - f_{\hat{O}} (\theta - \pi / 4 a) ],
\end{align}
where we have used
\begin{align}
  \hat{U}_\mathrm{c} (\pm \pi / 4 a) &\coloneqq \frac{\hat{1} \mp i \hat{G}}{\sqrt{2}}.
\end{align}
Then we have obtained Eq.~\eqref{supp-arXiv-formula-gradient-f-001-001}.

\end{proof}

By setting $a = 1 / 2$, the expectation value of the derivatives of the rotation gates, Eqs.~\eqref{supp-arXiv-def-rotation-X-001}, \eqref{supp-arXiv-def-rotation-Y-001}, and \eqref{supp-arXiv-def-rotation-Z-001}, are analytically computed as
\begin{align}
\frac{d}{d\phi} \mathbb{E}_\psi [\hat{R}^w (\phi)] &= \frac{1}{2} (\mathbb{E}_\psi [\hat{R}^w (\phi + \pi / 2)] - \mathbb{E}_\psi [\hat{R}^w (\phi - \pi / 2)]), \label{supp-arXiv-derivative-expectation-rotation-001-001}
\end{align}
for $w = x, y, z$.
Here, $\mathbb{E}_\psi [\cdot] \coloneqq \langle \psi | \cdot | \psi \rangle$.

\section{Matrix derivative}

In the UKM, we optimize the cost function of a unitary matrix.
Then, we provide the formula of matrix derivatives.

\subsection{Matrix derivative}

We review the derivative of the trace with respect to a matrix.
For details, refer to Ref.~\cite{Petersen001, Kreutz001}.

\subsubsection{Real case}

Let $A$ and $X$ be square matrices and define $a_{i, j}$ and $x_{i, j}$ by
\begin{align}
  a_{i, j} &\coloneqq [A]_{i, j}, \\
  x_{i, j} &\coloneqq [X]_{i, j},
\end{align}
where $[\cdot]_{i, j}$ is the element in the $i$-th row and the $j$-th column of a matrix.
Then, we have
\begin{align}
  \frac{d}{d x_{i, j}} \mathrm{Tr} [A X^\intercal] &= a_{i, j}, \label{supp-arXiv-formula-derivative-real-matrix-trace-001-001} \\
  \frac{d}{d x_{i, j}} \mathrm{Tr} [A X] &= - a_{j, i}. \label{supp-arXiv-formula-derivative-real-matrix-trace-001-002}
\end{align}
The matrix representations of Eqs.~\eqref{supp-arXiv-formula-derivative-real-matrix-trace-001-001} and \eqref{supp-arXiv-formula-derivative-real-matrix-trace-001-002} are, respectively,
\begin{align}
  \frac{d}{dX} \mathrm{Tr} [A X^\intercal] &= A, \label{supp-arXiv-formula-derivative-real-matrix-trace-002-001} \\
  \frac{d}{dX} \mathrm{Tr} [A X] &= A^\intercal. \label{supp-arXiv-formula-derivative-real-matrix-trace-002-002}
\end{align}

By using Eq.~\eqref{supp-arXiv-formula-derivative-real-matrix-trace-002-001} and \eqref{supp-arXiv-formula-derivative-real-matrix-trace-002-002}, we have
\begin{align}
  \frac{d}{dX} \mathrm{Tr} [A X B X^\intercal C] &= A^\intercal C^\intercal X B^\intercal + C A X B.
\end{align}

\subsubsection{Complex case}

So far, we have considered the matrix derivatives of the trace in the case of real matrices.
Here, we present the matrix derivatives in the case of complex matrices.

Let $(\cdot)^*$, $(\cdot)^\intercal$, and $(\cdot)^\mathrm{H}$ denote the complex conjugate, the transpose, and the Hermitian conjugate of a matrix.
Then we have
\begin{align}
A^\mathrm{H} &= (A^*)^\intercal \\
&= (A^\intercal)^*.
\end{align}
For a complex matrix $X$, we denote, by $\Re [X]$ and $\Im [X]$, the real and complex parts of $X$:
\begin{align}
  \Re [X] &\coloneqq \frac{X + X^*}{2}, \label{supp-arXiv-complex-matrix-real-and-complex-parts-001-001} \\
  \Im [X] &\coloneqq \frac{X - X^*}{2i}. \label{supp-arXiv-complex-matrix-real-and-complex-parts-001-002}
\end{align}
Furthermore, we define
\begin{align}
  x_{i, j} &\coloneqq [X]_{i, j},
\end{align}
and
\begin{align}
  x_{i, j}^\Re &\coloneqq \Re [x_{i, j}], \\
  x_{i, j}^\Im &\coloneqq \Im [x_{i, j}].
\end{align}

From Eq.~\eqref{supp-arXiv-complex-matrix-real-and-complex-parts-001-001}, we have
\begin{align}
  \mathrm{Tr} \big[ A X^\mathrm{H} \big] &= \mathrm{Tr} \big[ A \Re \big[ X^\mathrm{H} \big] \big] + i \mathrm{Tr} \big[ A \Im \big[ X^\mathrm{H} \big] \big] \\
  &= \mathrm{Tr} \big[ A \Re \big[ X^\intercal \big] \big] - i \mathrm{Tr} \big[ A \Im \big[ X ^\intercal \big] \big].
\end{align}
Then, the complex versions of Eqs.~\eqref{supp-arXiv-formula-derivative-real-matrix-trace-001-001} and \eqref{supp-arXiv-formula-derivative-real-matrix-trace-001-002} are, respectively,
\begin{align}
  \frac{d}{d x_{i, j}^\Re} \mathrm{Tr} [A X^\mathrm{H}] &= a_{i, j}, \label{supp-arXiv-formula-derivative-complex-matrix-trace-003-001} \\
  i \frac{d}{d x_{i, j}^\Im} \mathrm{Tr} [A X^\mathrm{H}] &= a_{i, j}, \label{supp-arXiv-formula-derivative-complex-matrix-trace-003-002}
\end{align}
and
\begin{align}
  \frac{d}{d x_{i, j}^\Re} \mathrm{Tr} [A X] &= a_{j, i}, \\
  i \frac{d}{d x_{i, j}^\Im} \mathrm{Tr} [A X] &= a_{j, i}.
\end{align}
Furthermore, from Eqs.~\eqref{supp-arXiv-formula-derivative-complex-matrix-trace-003-001} and \eqref{supp-arXiv-formula-derivative-complex-matrix-trace-003-002}, the complex versions of Eqs.~\eqref{supp-arXiv-formula-derivative-real-matrix-trace-002-001} and \eqref{supp-arXiv-formula-derivative-real-matrix-trace-002-002} are, respectively,
\begin{align}
  \frac{d}{d\Re [X]} \mathrm{Tr} [A X^\mathrm{H}] &= A, \label{supp-arXiv-formula-derivative-complex-matrix-trace-004-001} \\
  i \frac{d}{d\Im [X]} \mathrm{Tr} [A X^\mathrm{H}] &= A, \label{supp-arXiv-formula-derivative-complex-matrix-trace-004-002}
\end{align}
and
\begin{align}
  \frac{d}{d\Re [X]} \mathrm{Tr} [A X] &= A^\intercal, \label{supp-arXiv-formula-derivative-complex-matrix-trace-004-011} \\
  i \frac{d}{d\Im [X]} \mathrm{Tr} [A X] &= - A^\intercal. \label{supp-arXiv-formula-derivative-complex-matrix-trace-004-012}
\end{align}

\subsection{Derivative of the expectation value with respect to a unitary operator}

Let $(\cdot)^\dagger$ denote the Hermitian conjugate of an operator.
For an operator $\hat{A}$, we have
\begin{align}
  \hat{A}^\dagger &= (\hat{A}^*)^\intercal \\
  &= (\hat{A}^\intercal)^*.
\end{align}

As defined in Eqs.~\eqref{supp-arXiv-complex-matrix-real-and-complex-parts-001-001} and \eqref{supp-arXiv-complex-matrix-real-and-complex-parts-001-002}, we denote, by $\Re [\hat{U}]$ and $\Im [\hat{U}]$, the real and imaginary parts of $\hat{U}$:
\begin{align}
  \Re [\hat{U}] &\coloneqq \frac{\hat{U} + \hat{U}^*}{2}, \\
  \Im [\hat{U}] &\coloneqq \frac{\hat{U} - \hat{U}^*}{2i}.
\end{align}

The expectation value of $\hat{O}$ with respect to the state that evolves from $\hat{\rho}$ by $\hat{U}$ is given by $\mathrm{Tr} [\hat{U}^\dagger \hat{O} \hat{U} \hat{\rho}]$.
In the UKM, it is the central problem to find $\hat{U}$ that minimizes a given cost function.
Then, it is helpful if we have the derivative of $\mathrm{Tr} [\hat{U}^\dagger \hat{O} \hat{U} \hat{\rho}]$ with respect to $\hat{U}$.
By using Eqs.~\eqref{supp-arXiv-formula-derivative-complex-matrix-trace-004-001} and \eqref{supp-arXiv-formula-derivative-complex-matrix-trace-004-002}, we have
\begin{align}
  \frac{d}{d\Re [\hat{U}]} \mathrm{Tr} [\hat{U}^\dagger \hat{O} \hat{U} \hat{\rho}] &= \frac{d}{d\Re [\hat{U}]} \mathrm{Tr} [\hat{U}^\dagger \hat{O} \hat{V} \hat{\rho}]|_{\hat{V}=\hat{U}} + \frac{d}{d\Re [\hat{U}]} \mathrm{Tr} [\hat{V}^\dagger \hat{O} \hat{U} \hat{\rho}]|_{\hat{V}^\dagger=\hat{U}^\dagger} \\
  &= \hat{O} \hat{U} \hat{\rho} + \hat{O}^\intercal \hat{U}^* \hat{\rho}^\intercal. \label{supp-arXiv-derivative-UBUrho-ReU-001}
\end{align}
Here, we have used
\begin{align}
  \frac{d}{d\Re [\hat{U}]} \mathrm{Tr} [\hat{U}^\dagger \hat{O} \hat{V} \hat{\rho}]|_{\hat{V}=\hat{U}} &= \frac{d}{d\Re [\hat{U}]} \mathrm{Tr} [\hat{O} \hat{V} \hat{\rho} \hat{U}^\dagger]|_{\hat{V}=\hat{U}} \\
  &= \hat{O} \hat{U} \hat{\rho}, \\
  \frac{d}{d\Re [\hat{U}]} \mathrm{Tr} [\hat{V}^\dagger \hat{O} \hat{U} \hat{\rho}]|_{\hat{V}^\dagger=\hat{U}^\dagger} &= \frac{d}{d\Re [\hat{U}]} \mathrm{Tr} [\hat{\rho} \hat{V}^\dagger \hat{O} \hat{U}]|_{\hat{V}^\dagger=\hat{U}^\dagger} \\
  &= (\hat{\rho} \hat{U}^\dagger \hat{O})^\intercal \\
  &= \hat{O}^\intercal \hat{U}^* \hat{\rho}^\intercal.
\end{align}
Similarly, we also have
\begin{align}
  \frac{d}{d\Im [\hat{U}]} \mathrm{Tr} [\hat{U}^\dagger \hat{O} \hat{U} \hat{\rho}] &= \frac{d}{d\Im [\hat{U}]} \mathrm{Tr} [\hat{U}^\dagger \hat{O} \hat{V} \hat{\rho}]|_{\hat{V}=\hat{U}} + \frac{d}{d\Im [\hat{U}]} \mathrm{Tr} [\hat{V}^\dagger \hat{O} \hat{U} \hat{\rho}]|_{\hat{V}^\dagger=\hat{U}^\dagger} \\
  &= - i \hat{O} \hat{U} \hat{\rho} + i \hat{O}^\intercal \hat{U}^* \hat{\rho}^\intercal. \label{supp-arXiv-derivative-UBUrho-ImU-001}
\end{align}
Here, we have used
\begin{align}
  i \frac{d}{d\Im [\hat{U}]} \mathrm{Tr} [\hat{U}^\dagger \hat{O} \hat{V} \hat{\rho}]|_{\hat{V}=\hat{U}} &= i \frac{d}{d\Im [\hat{U}]} \mathrm{Tr} [\hat{O} \hat{V} \hat{\rho} \hat{U}^\dagger]|_{\hat{V}=\hat{U}} \\
  &= \hat{O} \hat{U} \hat{\rho}, \\
  i \frac{d}{d\Im [\hat{U}]} [\hat{V}^\dagger \hat{O} \hat{U} \hat{\rho}]|_{\hat{V}^\dagger=\hat{U}^\dagger} &= i \frac{d}{d\Im [\hat{U}]} [\hat{\rho} \hat{V}^\dagger \hat{O} \hat{U}]|_{\hat{V}^\dagger=\hat{U}^\dagger} \\
  &= - (\hat{\rho} \hat{U}^\dagger \hat{O})^\intercal \\
  &= - \hat{O}^\intercal \hat{U}^* \hat{\rho}^\intercal.
\end{align}

We utilize Eqs.~\eqref{supp-arXiv-derivative-UBUrho-ReU-001} and \eqref{supp-arXiv-derivative-UBUrho-ImU-001} in the UKM.

\section{Review of optimization methods}

In the UKM and the VCR, optimization algorithms play an important role.
There are many approaches to solve minimization problems, such as the gradient method, the Newton method, the quasi-Newton method~\cite{Boyd001, Fletcher001}.
Among them, the nonlinear conjugate gradient (CG) method~\cite{Fletcher002, Shewchuk001} is very simple and work well for optimization problems associated with a high dimensional vector~\cite{Fletcher002, Shewchuk001} and the Broyden-Fletcher-Goldfarb-Shanno (BFGS) method~\cite{Broyden001, Fletcher003, Goldfarb001, Shanno001, Fletcher002} is one of the most sophisticated methods.
In this paper, the CG method and the UKM and the BFGS method are used for the UKM and the VCR, respectively.

In this section, we review the CG method and the BFGS method as optimization methods.
For details, refer to Refs.~\cite{Boyd001, Fletcher001}.

\subsection{Basics of the gradient method}

Before getting into the CG method the BFGS method, we review the basics of the gradient method.

\subsubsection{Iteration method based on Linear search}

Let us consider the following optimization problem:
\begin{align}
  \min_x \mathcal{J}_\mathrm{cost} (x). \label{supp-arXiv-optimization-problem-001-001}
\end{align}
Let $d_k$ and $\alpha_k$ be the gradient descent direction and the step size at the $k$-th iteration, respectively.
Then we update $x_k$ by
\begin{align}
  x_k &= x_{k-1} + \alpha_k d_k. \label{supp-arXiv-iterative-method-001-001}
\end{align}
To decrease $\mathcal{J}_\mathrm{cost} (x)$ in Eq.~\eqref{supp-arXiv-optimization-problem-001-001}, the gradient direction $d_k$ must satisfy the following condition:
\begin{align}
  \nabla \mathcal{J}_\mathrm{cost} (x_k)^\intercal d_k &< 0. \label{supp-arXiv-condition-gradient-direction-001-001}
\end{align}
By using the Taylor expansion, we have
\begin{align}
  \mathcal{J}_\mathrm{cost} (x_k) - \mathcal{J}_\mathrm{cost} (x_{k-1}) &= \alpha_k \bigg( \nabla \mathcal{J}_\mathrm{cost} (x_{k-1})^\intercal d_k + \frac{o (\alpha_k)}{\alpha_k} \bigg).
\end{align}
Then, $\alpha_k$ is sufficiently small, Eq.~\eqref{supp-arXiv-iterative-method-001-001} is expected to work well.
We summarize the iterative method based on linear search in Algo.~\ref{supp-arXiv-iterative-method-based-on-linear-search-001-001}.
\begin{algorithm}[t]
\caption{Iterative method based on linear search} \label{supp-arXiv-iterative-method-based-on-linear-search-001-001}
\begin{algorithmic}[1]
\STATE initialize $x_0$
\FOR{$k = 1, 2, \dots, K$}
\STATE compute $d_k$ that satisfies Eq.~\eqref{supp-arXiv-condition-gradient-direction-001-001}
\STATE compute $\alpha_k$
\STATE compute $x_k$ by Eq.~\eqref{supp-arXiv-iterative-method-001-001}
\ENDFOR
\end{algorithmic}
\end{algorithm}

\subsubsection{Linear search}

It is quite important to compute appropriate $\alpha_k$, and there exist well-known conditions for good $\alpha_k$.
Here we review the conditions for $\alpha_k$.

To derive the conditions for $\alpha_k$, let us consider the minimization problem of the following function:
\begin{align}
  \phi (\alpha) &\coloneqq \mathcal{J}_\mathrm{cost} (x + \alpha d).
\end{align}

The simplest condition for $\alpha_k$ is the Armijo condition~\cite{Armijo001} given by
\begin{align}
  \phi (\alpha) &\le \phi (0) + c_1 \alpha \phi' (0), \label{supp-arXiv-condition-Armijo-001}
\end{align}
for $0 < c_1 < 1$.
When we use Eq.~\eqref{supp-arXiv-condition-Armijo-001}, $\alpha_k$ is likely to be very small.
Another well-used condition for $\alpha_k$ is the Wolfe condition~\cite{Wolfe001, Wolfe002} given by
\begin{subequations}
\begin{align}
  \phi (\alpha) &\le \phi (0) + c_1 \alpha \phi' (0), \\
  \phi' (\alpha) &\ge c_2 \phi' (0), \label{supp-arXiv-condition-Wolfe-002-002}
\end{align}
\label{supp-arXiv-condition-Wolfe-001-001}%
\end{subequations}
for $0 < c_1 < c_2 < 1$.
In this case, $\phi' (\alpha)$ may take a large positive number.
To avoid this problem, the following condition for $\alpha$, which is called the strong Wolfe condition, is used:
\begin{subequations}
\begin{align}
  \phi (\alpha) &\le \phi (0) + c_1 \alpha \phi' (0), \\
  | \phi' (\alpha) | &\ge c_2 | \phi' (0) |,
\end{align}
\end{subequations}
for $0 < c_1 < c_2 < 1$.
Finally, we also mention the Goldstein condition for $\alpha$ given by
\begin{align}
  \phi (0) + (1 - c) \alpha \phi' (0) &\le \phi (\alpha) \\
  &\le \phi (0) + c \alpha \phi' (0),
\end{align}
for $0 < c < 1/2$.

To find $\alpha_k$ that satisfies one of the above conditions, the backtracking line search is often used.
In the backtracking line search, we set $\alpha_0 > 0$ and iterate the following equation until Eq.~\eqref{supp-arXiv-condition-Armijo-001} is satisfied:
\begin{align}
  \alpha_k &= \rho \alpha_{k-1},
\end{align}
where $\rho \in (0, 1)$.

\subsection{Nonlinear CG method} \label{supp-arXiv-sec-CG-001-001}

In the UKM, the gradient method is utilized to $\hat{U}_k$ in Eq.~\eqref{supp-arXiv-quantum-kernel-method-001-011}.
The BFGS method is known as a sophisticated method but requires relatively large memory space.
Then, we use the CG method in the UKM.
Here, we explain the CG method.

\subsubsection{Linear CG method}

We begin with the CG method.
Let us consider minimizing
\begin{align}
  \mathcal{J}_\mathrm{cost} (x) &\coloneqq \frac{1}{2} (x - x_*)^\intercal A (x - x_*),
\end{align}
where $A \succ O$.

In the CG method, we set $x_0$ and $d_0 = - \nabla \mathcal{J}_\mathrm{cost} (x_0)$, and then iterate, until convergence,
\begin{subequations}
\begin{align}
  x_k &= x_{k-1} + \alpha_k d_{k-1}, \label{supp-arXiv-CG-001-011} \\
  d_k &= - \nabla \mathcal{J}_\mathrm{cost} (x_k) + \beta_k d_{k-1}, \label{supp-arXiv-CG-001-012}
\end{align}
\label{supp-arXiv-CG-001-001}%
\end{subequations}
where
\begin{subequations}
\begin{align}
  \alpha_k &= - \frac{\nabla \mathcal{J}_\mathrm{cost} (x_{k-1})^\intercal d_{k-1}}{d_{k-1}^\intercal A d_{k-1}}, \label{supp-arXiv-alpha-001-001} \\
  \beta_k &= \frac{\nabla \mathcal{J}_\mathrm{cost} (x_k)^\intercal A d_{k-1}}{d_{k-1}^\intercal A d_{k-1}}. \label{supp-arXiv-beta-001-001}
\end{align}
\label{supp-arXiv-CG-002-001}%
\end{subequations}

The CG method is summrized in Algo.~\ref{supp-arXiv-conjugate-gradient-method-001-001}.
\begin{algorithm}[t]
\caption{Conjugate gradient (CG) method} \label{supp-arXiv-conjugate-gradient-method-001-001}
\begin{algorithmic}[1]
\STATE initialize $x_0$
\STATE set $d_0 = - \nabla \mathcal{J}_\mathrm{cost} (x_0)$
\FOR{$k = 1, 2, \dots, K$}
\STATE compute $x_k$ by Eq.~\eqref{supp-arXiv-CG-001-011} with Eq.~\eqref{supp-arXiv-alpha-001-001}
\STATE compute $d_k$ by Eq.~\eqref{supp-arXiv-CG-001-012} with Eq.~\eqref{supp-arXiv-beta-001-001}
\ENDFOR
\end{algorithmic}
\end{algorithm}

\subsubsection{Nonlinier CG method}

Eq.~\eqref{supp-arXiv-CG-002-001} requires the Hessian matrix $A$.
When we do not know the Hessian, we need to replace Eq.~\eqref{supp-arXiv-CG-002-001} with different approaches.
These approaches are called the nonlinear CG method~\cite{Fletcher002, Polak001, Hestenes001, Dai001, Shewchuk001}.

First, let us focus on how to compute $\alpha_k$ because Eq.~\eqref{supp-arXiv-alpha-001-001} is not available when we do not know the Hessian.
It is very simple because we can use the line search method:
\begin{align}
  \alpha_k &= \argmin_\alpha \mathcal{J}_\mathrm{cost} (x_{k-1} + \alpha d_{k-1}). \label{supp-arXiv-alpha-002-001}
\end{align}
Here, Eqs.~\eqref{supp-arXiv-condition-Armijo-001} and \eqref{supp-arXiv-condition-Wolfe-001-001} are often used.

Next, let us turn our attention to $\beta_k$.
While we do not elaborate on them, there are several methods to replace $\beta_k$ in Eq.~\eqref{supp-arXiv-beta-001-001}:
\begin{align}
  \beta_k &= \frac{\| \nabla \mathcal{J}_\mathrm{cost} (x_k) \|_\mathrm{F}^2}{\| \nabla \mathcal{J}_\mathrm{cost} (x_{k-1}) \|_\mathrm{F}^2}, \label{supp-arXiv-beta-002-001} \\
  \beta_k &= \frac{\nabla \mathcal{J}_\mathrm{cost} (x_k)^\intercal y_k}{\| \nabla \mathcal{J}_\mathrm{cost} (x_{k-1}) \|_\mathrm{F}^2}, \label{supp-arXiv-beta-002-002} \\
  \beta_k &= \frac{\nabla \mathcal{J}_\mathrm{cost} (x_k)^\intercal y_k}{d_k^\intercal y_k}, \label{supp-arXiv-beta-002-003} \\
  \beta_k &= \frac{\| \nabla \mathcal{J}_\mathrm{cost} (x_k) \|_\mathrm{F}^2}{d_k^\intercal y_k}, \label{supp-arXiv-beta-002-004}
\end{align}
where $y_k \coloneqq \nabla \mathcal{J}_\mathrm{cost} (x_k) - \nabla \mathcal{J}_\mathrm{cost} (x_{k-1})$.
In this paper, we utilize Eq.~\eqref{supp-arXiv-beta-002-001} called the Fletcher-Reeves method~\cite{Fletcher002}.
Eq.~\eqref{supp-arXiv-beta-002-002}, Eq.~\eqref{supp-arXiv-beta-002-003}, and Eq.~\eqref{supp-arXiv-beta-002-004} are called the Polak-Ribi\`{e}re method~\cite{Polak001}, the Hestenes-Stiefel method~\cite{Hestenes001}, and the Dai-Yuan method~\cite{Dai001}, respectively.

The nonlinear CG method is summrized in Algo.~\ref{supp-arXiv-nonlinear-conjugate-gradient-method-001-001}.
\begin{algorithm}[t]
\caption{Nonlinear conjugate gradient (CG) method} \label{supp-arXiv-nonlinear-conjugate-gradient-method-001-001}
\begin{algorithmic}[1]
\STATE initialize $x_0$
\STATE set $d_0 = - \nabla \mathcal{J}_\mathrm{cost} (x_0)$
\FOR{$k = 1, 2, \dots, K$}
\STATE compute $\alpha_k$ by Eq.~\eqref{supp-arXiv-alpha-002-001}
\STATE compute $x_k$ by Eq.~\eqref{supp-arXiv-CG-001-011}
\STATE compute $\beta_k$ by Eq.~\eqref{supp-arXiv-beta-002-001}
\STATE compute $d_k$ by Eq.~\eqref{supp-arXiv-CG-001-012}
\ENDFOR
\end{algorithmic}
\end{algorithm}

\subsection{BFGS method}

In this paper, the BFGS method is used for the VCR.
The aim of this subsection is to describe the BFGS method~\cite{Broyden001, Fletcher003, Goldfarb001, Shanno001, Fletcher002}.

\subsubsection{Algorithmic details of the quasi-Newton method}

Let us consider the optimization problem, Eq.~\eqref{supp-arXiv-optimization-problem-001-001}.
We begin with Newton's method.
In Newton's method, we set $x_0$ and iterate the following equation until convergence:
\begin{align}
  x_k &= x_{k-1} - (\nabla^2 \mathcal{J}_\mathrm{cost} (x_{k-1}))^{-1} \nabla \mathcal{J}_\mathrm{cost} (x_{k-1}). \label{supp-arXiv-Newtons-method-001-001}
\end{align}
Newton's method is summarized in Algo.~\ref{supp-arXiv-Newton-method-001-001}
\begin{algorithm}[t]
\caption{Newton's method} \label{supp-arXiv-Newton-method-001-001}
\begin{algorithmic}[1]
\STATE initialize $x_0$
\FOR{$k = 1, 2, \dots, K$}
\STATE compute $x_k$ by Eq.~\eqref{supp-arXiv-Newtons-method-001-001}
\ENDFOR
\end{algorithmic}
\end{algorithm}
The biggest problem of Newton's method is the difficulty of computing $\nabla^2 \mathcal{J}_\mathrm{cost} (x_{k-1}))^{-1}$ in Eq.~\eqref{supp-arXiv-Newtons-method-001-001}.
Then some alternative methods, called the quasi-Newton method, were proposed~\cite{Fletcher001}.

Let us turn our attention to the quasi-Newton method.
In the quasi-Newton method, the gradient descent direction is computed as
\begin{align}
  d_k &= - B_k^{-1} \nabla \mathcal{J}_\mathrm{cost} (x_k), \label{supp-arXiv-BFGS-001-012}
\end{align}
where $B_k$ is an approximation of $\nabla^2 \mathcal{J}_\mathrm{cost} (x)$.
Similarly, Eq.~\eqref{supp-arXiv-BFGS-001-012} is equivalent to
\begin{align}
  d_k &= - H_k \nabla \mathcal{J}_\mathrm{cost} (x_k),
\end{align}
where $H_k$ is an approximation of $[\nabla^2 \mathcal{J}_\mathrm{cost} (x)]^{-1}$.
Next, by using Eqs.~\eqref{supp-arXiv-condition-Armijo-001} and \eqref{supp-arXiv-condition-Wolfe-001-001}, we compute $\alpha_k$ by
\begin{align}
  \alpha_k &= \argmin_\alpha \mathcal{J}_\mathrm{cost} (x_{k-1} + \alpha d_k). \label{supp-arXiv-alpha-003-001}
\end{align}
Then, we update $x_k$, in the quasi-Newton method, by
\begin{align}
  x_k &= x_{k-1} + \alpha_k d_k. \label{supp-arXiv-BFGS-001-032}
\end{align}

\subsubsection{Condition for $B_k$}

Next, we consider the condition for $B_k$.
By the Taylor expansion, we have
\begin{align}
  \nabla^2 \mathcal{J}_\mathrm{cost} (x_k) \times (x_k - x_{k-1}) &= \nabla \mathcal{J}_\mathrm{cost} (x_k) - \nabla \mathcal{J}_\mathrm{cost} (x_{k-1}) + \mathcal{O} (\| x_k - x_{k-1} \|^2). \label{supp-arXiv-Taylor-expansion-001}
\end{align}
By introducing
\begin{align}
  y_k &\coloneqq \nabla \mathcal{J}_\mathrm{cost} (x_k) - \nabla \mathcal{J}_\mathrm{cost} (x_{k-1}), \\
  s_k &\coloneqq x_k - x_{k-1},
\end{align}
Eq.~\eqref{supp-arXiv-Taylor-expansion-001} is also expressed as
\begin{align}
  \nabla^2 \mathcal{J}_\mathrm{cost} (x_k) s_k &= y_k + \mathcal{O} (\| s_k \|^2).
\end{align}
Thus, in the quasi-Newton method, the following secant condition is imposed:
\begin{align}
  B_k s_k &= y_k. \label{supp-arXiv-secant-condition-001}
\end{align}
Eq.~\eqref{supp-arXiv-secant-condition-001} is equivalent to
\begin{align}
  H_k y_k &= s_k.
\end{align}

The necessary condition for $B_{k}$ that satisfies the secant condition, Eq.~\eqref{supp-arXiv-secant-condition-001}, to exist, is
\begin{align}
  s_k^\intercal y_k &> 0, \label{supp-arXiv-condition-secant-001}
\end{align}

The proof of Eq.~\eqref{supp-arXiv-condition-secant-001} is as follows.
\begin{proof}

From Eq.~\eqref{supp-arXiv-condition-Wolfe-002-002} and $\alpha_k > 0$, we have
\begin{align}
  \nabla \mathcal{J}_\mathrm{cost} (x_k)^\intercal s_k &\ge c_2 \nabla \mathcal{J}_\mathrm{cost} (x_{k-1})^\intercal s_k,
\end{align}
where $c_2$ is a constant that appears in the Wolfe condition, Eq.~\eqref{supp-arXiv-condition-Wolfe-001-001}.
Then, we have
\begin{align}
  y_k^\intercal s_k &= \nabla \mathcal{J}_\mathrm{cost} (x_k)^\intercal s_k - \nabla \mathcal{J}_\mathrm{cost} (x_{k-1})^\intercal s_k \\
  &\ge - (1 - c_2) \nabla \mathcal{J}_\mathrm{cost} (x_{k-1})^\intercal s_k.
\end{align}
Since $c_2 < 1$ and $s_k$ is a gradient descent direction, we obtain Eq.~\eqref{supp-arXiv-condition-secant-001} as a necessary condition for the secant condition, Eq.~\eqref{supp-arXiv-secant-condition-001}.

\end{proof}

\subsubsection{BFGS formula}

We do not elaborate on the BFGS method, we here give the BFGS formula.
The BFGS formula is given by
\begin{align}
  B_k &= B_{k-1} - \frac{B_{k-1} s_k s_k^\intercal B_{k-1}}{s_k^\intercal B_{k-1} s_k} + \frac{y_k y_k^\intercal}{s_k^\intercal y_k}. \label{supp-arXiv-BGFS-001-021}
\end{align}
In the case of BFGS formula, we have
\begin{align}
  H_k &= \bigg( I - \frac{s_k y_k^\intercal}{s_k^\intercal y_k} \bigg) H_{k-1} \bigg( I - \frac{y_k s_k^\intercal}{s_k^\intercal y_k} \bigg) + \frac{s_k s_k^\intercal}{s_k^\intercal y_k}.
\end{align}
The quasi-Newton method is summarized in Algo.~\ref{supp-arXiv-BFGS-method-001-001}
\begin{algorithm}[t]
\caption{BFGS method} \label{supp-arXiv-BFGS-method-001-001}
\begin{algorithmic}[1]
\STATE initialize $x_0$
\STATE set $B_0 = I$
\FOR{$k = 1, 2, \dots, K$}
\STATE compute $d_k$ by Eq.~\eqref{supp-arXiv-BFGS-001-012}
\STATE compute $\alpha_k$ by Eq.~\eqref{supp-arXiv-alpha-003-001}
\STATE compute $x_k$ by Eq.~\eqref{supp-arXiv-BFGS-001-032}
\STATE compute $B_k$ by Eq.~\eqref{supp-arXiv-BGFS-001-021}
\ENDFOR
\end{algorithmic}
\end{algorithm}

So far, we explained Eq.~\eqref{supp-arXiv-BGFS-001-021}, but another formula for $B_k$ is also known.
The Davidon-Fletcher-Powell (DFP) formula is given by~\cite{Fletcher001}
\begin{align}
  B_k &= \bigg( I - \frac{y_k s_k^\intercal}{s_k^\intercal y_k} \bigg) B_{k-1} \bigg( I - \frac{s_k y_k^\intercal}{s_k^\intercal y_k} \bigg) + \frac{y_k y_k^\intercal}{s_k^\intercal y_k}.
\end{align}

\section{Quantum circuit learning} \label{supp-arXiv-sec-QCL-001-001}

QCL was proposed in Refs.~\cite{Schuld001, Mitarai001} as an application of the QAOA and the VQE for machine learning.
In this section, we review Refs.~\cite{Schuld001, Mitarai001}.
We first explain the ingredients of QCL and then describe the whole procedure of QCL.

\subsection{Encoding} \label{supp-arXiv-sec-encoding-001}

Suppose that we have $\mathcal{D} \coloneqq \{ x_i, y_i \}_{i=1}^N$ and let $y_i$ be the label of $x_i$.
It is an important thing to encode $x_i \in \mathbb{R}^M$ on qubits efficiently.

Suppose that we have a $n$-qubit system and that the initial state is written as $| \mathrm{init} \rangle$.
A typical example of $| \mathrm{init} \rangle$ is given by
\begin{align}
  | \mathrm{init} \rangle &\coloneqq | \underbrace{00 \dots 0}_n \rangle.
\end{align}
We then encode $x_i$ by applying $\hat{S} (x_i)$ on $| \mathrm{init} \rangle$:
\begin{align}
  | \psi^\mathrm{in} (x_i) \rangle &\coloneqq \hat{S} (x_i) | \mathrm{init} \rangle. \label{supp-arXiv-encoding-QCL-001-001}
\end{align}

There exist some novel quantum encoding methods; in Refs.~\cite{Schuld001, Plesch001}, the amplitude encoding is used.
In the amplitude encoding, we construct the quantum state for $x_i$:
\begin{align}
  | \psi^\mathrm{in} (x_i) \rangle &= \sum_{j=1}^{2^n} \tilde{x}_i^j | j \rangle, \label{supp-arXiv-amplitude-encoding-001-001}
\end{align}
where
\begin{subequations}
\begin{align}
  \tilde{x}_i &\coloneqq \frac{1}{\chi_i} [x_i^1, x_i^2, \dots, x_i^M, c_i^1, c_i^2, \dots, c_i^{M'}], \\
  \chi_i &\coloneqq \sqrt{\sum_{j=1}^M (x_i^j)^2 + \sum_{j=1}^{M'} (c_i^j)^2}.
\end{align}
\label{supp-arXiv-def-tilde-x-001-001}%
\end{subequations}
Here, $\{ c_i^j \}_{j=1}^{M'}$ are certain constant values for padding and $M + M' = 2^n$.

\subsection{Quantum circuit} \label{supp-arXiv-sec-quantum-circuit-001-001}

A near-term quantum device is believed to realize a quantum circuit~\cite{Arute001}.
Suppose that we have a $n$-qubit system and a quantum circuit that is parameterized by $\theta$.
By applying $\hat{U}_\mathrm{c} (\theta)$ realized by the quantum circuit, $| \psi^\mathrm{in} (x_i) \rangle$ evolves into
\begin{align}
  | \psi^\mathrm{out} (x_i; \theta) \rangle &\coloneqq \hat{U}_\mathrm{c} (\theta) | \psi^\mathrm{in} (x_i) \rangle, \label{supp-arXiv-state-output-001-001}
\end{align}
where
\begin{align}
  \hat{U}_\mathrm{c} (\theta) &\coloneqq \sideset{}{^\downarrow}\prod_{i=1}^L \hat{U}_{\mathrm{c}, i} (\theta_i), \label{supp-arXiv-form-U-theta-001-001}
\end{align}
$\theta \coloneqq \{ \theta_i \}_{i=1}^L$, and the definition of $\sideset{}{^\downarrow}\prod$ is given by Eq.~\eqref{supp-arXiv-def-product-operator-downarrow-001}.

The functional form of $\hat{U}_{\mathrm{c}, i} (\theta_i)$ in Eq.~\eqref{supp-arXiv-form-U-theta-001-001}, depends on quantum circuits.
In Refs.~\cite{Schuld001, Schuld002, Mitarai001}, the following form is often assumed for $\hat{U}_{\mathrm{c}, i} (\theta_i)$ in Eq.~\eqref{supp-arXiv-form-U-theta-001-001}:
\begin{align}
\hat{U}_{\mathrm{c}, i} (\theta_i) &\coloneqq \bigg[ \prod_{j=1}^n \hat{R}_j^\mathrm{3d} (\theta_{i, j}) \bigg] \hat{U}_\mathrm{ent}. \label{supp-arXiv-form-U-i-theta-i-001-001}
\end{align}
Graphically, $\hat{U}_{\mathrm{c}, i} (\theta_i)$, Eq.~\eqref{supp-arXiv-form-U-i-theta-i-001-001}, for $n = 4$ is expressed as
\begin{align}
\hat{U}_{\mathrm{c}, i} (\theta_i)
&=
\begin{tikzcd}[row sep = 0.2cm, column sep = 0.2cm]
& \gate[wires=4]{\hat{U}_\mathrm{ent}} & \gate{\hat{R}^\mathrm{3d}} & \qw \\
&                                      & \gate{\hat{R}^\mathrm{3d}} & \qw \\
&                                      & \gate{\hat{R}^\mathrm{3d}} & \qw \\
&                                      & \gate{\hat{R}^\mathrm{3d}} & \qw
\end{tikzcd}.
\end{align}
Here, $\theta_i \coloneqq \{ \theta_{i, j} \}_{j=1}^n$, $\theta_{i, j} \coloneqq \{ \theta_{i, j, k} \}_{k=1}^3$ and $\hat{R}_j^\mathrm{3d} (\theta_{i, j}) \coloneqq \hat{R}_j^z (\theta_{i, j, 3}) \hat{R}_j^y (\theta_{i, j, 2}) \hat{R}_j^z (\theta_{i, j, 1})$ is the 3-dimensional rotation gate defined in Eq.~\eqref{supp-arXiv-def-3d-rotation-001} on the $j$-th qubit~\cite{Barenco001, Nielsen001}.
Note that the 3-dimensional rotation can be, in general, decomposed into the product of the $z$-rotation gate and the $y$-rotation gate as shown in Refs.~\cite{Barenco001, Nielsen001}.
For details, see Sec.~\ref{supp-arXiv-sec-def-rotation-qubit-001}.

Two-qubit gates are known to be universal~\cite{Divincenzo001}.
In Refs.~\cite{Schuld001, Schuld002}, CNOT gates and controlled-rotation gates are used for $\hat{U}_\mathrm{ent}$ in Eq.~\eqref{supp-arXiv-form-U-i-theta-i-001-001}.
For example, we can use the following for $\hat{U}_\mathrm{ent}$ in Eq.~\eqref{supp-arXiv-form-U-i-theta-i-001-001}:
\begin{align}
  \hat{U}_\mathrm{ent} &= \sideset{}{^\downarrow}\prod_{j=1}^n \mathrm{Ct}_j [\hat{X}_{j+1}], \label{supp-arXiv-gate-CNOT-001-001}
\end{align}
where $\hat{X}_{n+1} \coloneqq \hat{X}_1$.
Graphically $\hat{U}_{\mathrm{c}, i} (\theta_i)$, Eq.~\eqref{supp-arXiv-gate-CNOT-001-001}, for $n = 4$ is expressed as
\begin{align}
\hat{U}_\mathrm{ent}
&=
\begin{tikzcd}[row sep = 0.2cm, column sep = 0.2cm]
& \ctrl{1} & \qw      & \qw      & \targ{}   & \qw \\
& \targ{}  & \ctrl{1} & \qw      & \qw       & \qw \\
& \qw      & \targ{}  & \ctrl{1} & \qw       & \qw \\
& \qw      & \qw      & \targ{}  & \ctrl{-3} & \qw
\end{tikzcd}.
\end{align}
Then, $\hat{U}_{\mathrm{c}, i} (\theta_i)$ in Eq.~\eqref{supp-arXiv-form-U-i-theta-i-001-001} with Eq.~\eqref{supp-arXiv-gate-CNOT-001-001} for $n = 4$ graphically becomes
\begin{align}
\hat{U}_{\mathrm{c}, i} (\theta_i)
&=
\begin{tikzcd}[row sep = 0.2cm, column sep = 0.2cm]
& \ctrl{1} & \qw      & \qw      & \targ{}   & \gate{\hat{R}^\mathrm{3d}} & \qw \\
& \targ{}  & \ctrl{1} & \qw      & \qw       & \gate{\hat{R}^\mathrm{3d}} & \qw \\
& \qw      & \targ{}  & \ctrl{1} & \qw       & \gate{\hat{R}^\mathrm{3d}} & \qw \\
& \qw      & \qw      & \targ{}  & \ctrl{-3} & \gate{\hat{R}^\mathrm{3d}} & \qw
\end{tikzcd}.
\end{align}

If only CNOT gates are used to construct $\hat{U}_\mathrm{ent}$ like the above example, $\hat{U}_\mathrm{ent}$ is fixed and free from parameters.
Instead of Eq.~\eqref{supp-arXiv-form-U-i-theta-i-001-001}, we can also use
\begin{align}
\hat{U}_{\mathrm{c}, i} (\theta_i) &\coloneqq \bigg[ \prod_{j=1}^n \hat{R}_j^\mathrm{3d} (\theta_{i, j}^1) \bigg] \hat{U}_\mathrm{ent} (\theta_{i, j}^2), \label{supp-arXiv-form-U-i-theta-i-001-002}
\end{align}
where
\begin{align}
  \hat{U}_\mathrm{ent} (\theta_i^2) &\coloneqq \sideset{}{^\downarrow}\prod_{j=1}^n \mathrm{Ct}_j [\hat{R}_{j+1}^\mathrm{3d} (\theta_{i, j}^2)], \label{supp-arXiv-gate-CRot-001-001}
\end{align}
and $\hat{R}_{n+1}^\mathrm{3d} (\theta_{i, n+1}^2) \coloneqq \hat{R}_1^\mathrm{3d} (\theta_{i, 1}^2)$.
In the case of Eq.~\eqref{supp-arXiv-form-U-i-theta-i-001-002}, the definition of $\theta_i$ is repaced by $\theta_i \coloneqq \{ \theta_i^1, \theta_i^2 \}$, and we set $\theta_i^1 \coloneqq \{ \theta_{i, j}^1 \}_{j=1}^n$, $\theta_i^2 \coloneqq \{ \theta_{i, j}^2 \}_{j=1}^n$, $\theta_{i, j}^1 \coloneqq \{ \theta_{i, j, k}^1 \}_{k=1}^3$, and $\theta_{i, j}^2 \coloneqq \{ \theta_{i, j, k}^2 \}_{k=1}^3$.
Graphically, $\hat{U}_{\mathrm{c}, i} (\theta_i)$ in Eq.~\eqref{supp-arXiv-form-U-i-theta-i-001-002} for $n = 4$ becomes
\begin{align}
\hat{U}_{\mathrm{c}, i} (\theta_i)
&=
\begin{tikzcd}[row sep = 0.2cm, column sep = 0.2cm]
& \ctrl{1} & \qw      & \qw      & \gate{\hat{R}^\mathrm{3d}}  & \gate{\hat{R}^\mathrm{3d}}   & \qw \\
& \gate{\hat{R}^\mathrm{3d}} & \ctrl{1} & \qw      & \qw       & \gate{\hat{R}^\mathrm{3d}}   & \qw \\
& \qw      & \gate{\hat{R}^\mathrm{3d}} & \ctrl{1} & \qw       & \gate{\hat{R}^\mathrm{3d}}   & \qw \\
& \qw      & \qw      & \gate{\hat{R}^\mathrm{3d}} & \ctrl{-3} & \gate{\hat{R}^\mathrm{3d}}   & \qw
\end{tikzcd}.
\end{align}

In Ref.~\cite{Mitarai001}, the following operator is used for $\hat{U}_\mathrm{ent}$ in Eq.~\eqref{supp-arXiv-form-U-i-theta-i-001-001}:
\begin{align}
\hat{U}_\mathrm{ent} &= e^{-i \Delta t \hat{H}_\mathrm{H}}, \label{supp-arXiv-gate-Mitarai-001-001}
\end{align}
where $\hat{H}_\mathrm{H}$ is the Heisenberg model given by
\begin{align}
\hat{H}_\mathrm{H} &\coloneqq \sum_{i, j} J_{i, j} \hat{\sigma}_i \cdot \hat{\sigma}_j + \sum_i \sum_{w = x, y ,z} h_i^w \hat{\sigma}_i^w, \\
\hat{\sigma}_i &\coloneqq [\hat{\sigma}_i^x, \hat{\sigma}_i^y, \hat{\sigma}_i^z]^\intercal.
\end{align}
More generally, $\hat{H}_\mathrm{H}$ can involve $n$-body terms with $n \ge 3$.
Graphically, $\hat{U}_{\mathrm{c}, i} (\theta_i)$, Eq.~\eqref{supp-arXiv-form-U-i-theta-i-001-001}, with Eq.~\eqref{supp-arXiv-gate-Mitarai-001-001} for $n = 4$ is expressed as
\begin{align}
\hat{U}_{\mathrm{c}, i} (\theta_i)
&=
\begin{tikzcd}[row sep = 0.2cm, column sep = 0.2cm]
& \gate[wires=4]{e^{-i \Delta t \hat{H}_\mathrm{H}}} & \gate{\hat{R}^\mathrm{3d}} & \qw \\
&                                      & \gate{\hat{R}^\mathrm{3d}} & \qw \\
&                                      & \gate{\hat{R}^\mathrm{3d}} & \qw \\
&                                      & \gate{\hat{R}^\mathrm{3d}} & \qw
\end{tikzcd}.
\end{align}

In numerical simulation, we compare Eqs.~\eqref{supp-arXiv-gate-CNOT-001-001}, \eqref{supp-arXiv-gate-CRot-001-001}, and \eqref{supp-arXiv-gate-Mitarai-001-001}.

\subsection{Measurement and prediction}

Next, we turn our attention to measurements on $| \psi^\mathrm{out} (x_i; \theta) \rangle$ and prediction based on measurements.
We obtain the measurement of $\hat{O}_j$ with respect to $| \psi^\mathrm{out} (x_i; \theta) \rangle$:
\begin{align}
  \langle \hat{O}_j \rangle_{x_i, \theta} \coloneqq \langle \psi^\mathrm{out} (x_i; \theta) | \hat{O}_j | \psi^\mathrm{out} (x_i; \theta) \rangle. \label{supp-arXiv-measurement-j-001-001}
\end{align}
In QCL, the label of $x_i$ is predicted by
\begin{align}
  f_\mathrm{pred} (x_i; \theta, \theta_\mathrm{b}) &\coloneqq \sum_{j=1}^Q \xi_j \langle \hat{O}_j \rangle_{x_i, \theta} + \theta_\mathrm{b}, \label{supp-arXiv-prediction-QCL-001-001}
\end{align}
where $\xi \coloneqq [\xi_1, \xi_2, \dots, \xi_Q]$ are fixed parameters.
Here $\theta_\mathrm{b}$ is a bias term to be estimated, but we may set $\theta_\mathrm{b} = 0$ for simplicity.

So far, we have focused on quantum circuit leaning~\cite{Schuld001, Mitarai001}, in which $\theta$ is optimized.
We here mention the difference between QCL and related algorithms.
In the cases of the QAOA and the VQE, $\{ \xi_j \}_{j=1}^Q$ are fixed as in case case of QCL; buts $\{ \xi_j \}_{j=1}^Q$ in QAOA and VQE encode the problem of interest, though $\{ \xi_j \}_{j=1}^Q$ in QCL are just fixed parameters.

For the almost same purpose with QCL, quantum reservoir computing (QRC) was proposed in Ref.~\cite{Fujii001} and Ref.~\cite{Nakajima001}.
In QRC, $\theta$ is fixed and $\xi$ is optimized.
Then, in the above problem setting, $\{ \xi_j \}_{j=1}^Q$ are optimized.

\subsection{Cost function and loss functions} \label{supp-arXiv-sec-QCL-cost-function-001-001}

To find a good $\theta$, we formulate the optimization problem with respect to $\theta$.
In QCL, we consider the optimization problem of minimizing the following cost function:
\begin{align}
  \mathcal{J}_\mathrm{cost} (\theta, \theta_\mathrm{b}) &\coloneqq \frac{1}{N} \sum_{i=1}^N \ell (y_i, f_\mathrm{pred} (x_i; \theta, \theta_\mathrm{b})), \label{supp-arXiv-cost-function-QCL-001-001}
\end{align}
where $\ell (\cdot, \cdot)$ is a loss function~\cite{Bishop001, Murphy001}.
For example, $\ell (\cdot, \cdot)$ takes the form
\begin{align}
  \ell_\mathrm{SE} (y, \tilde{y}) &\coloneqq \frac{1}{2} | y - \tilde{y} |^2, \label{supp-arXiv-squared-error-function-001-001}
\end{align}
which is called the squared error function.
Letting $y$ and $\tilde{y}$ be a label that takes $\pm 1$ and a prediction on $y$, respectively, another example is
\begin{align}
  \ell_\mathrm{hinge} (y, \tilde{y}) &\coloneqq \max(0, 1 - y \tilde{y}), \label{supp-arXiv-hinge-function-001-001}
\end{align}
which is called the hinge function.
Letting $y$ and $\tilde{y}$ be a label that takes $0$ or $1$ and a prediction on $y$, respectively, we can use
\begin{align}
  \ell_\mathrm{XE} (y, \tilde{y}) &\coloneqq - y \ln \tilde{y} - (1 - y) \ln (1 - \tilde{y}), \label{supp-arXiv-XE-function-001-001}
\end{align}
which is called the cross-entropy loss function.
In Eq.~\eqref{supp-arXiv-XE-function-001-001}, $y$ and $\tilde{y}$ are not symmetric though, in Eqs.~\eqref{supp-arXiv-XE-function-001-001} and \eqref{supp-arXiv-hinge-function-001-001}, $y$ and $\tilde{y}$ are symmetric.
Note that Eqs.~\eqref{supp-arXiv-squared-error-function-001-001}, \eqref{supp-arXiv-hinge-function-001-001}, and \eqref{supp-arXiv-XE-function-001-001} are not specific to QCL but also used for many machine learning algorithms~\cite{Bishop001, Murphy001} and the UKM, which is one of the main algorithms of this paper.
In Ref.~\cite{Schuld001}, a different cost function is introduced.
However, we do not use it in this paper.

\subsection{Algorithmic procedure of QCL}

So far, we have explained the ingredients of QCL.
Here we explain the whole procedure of QCL.
In QCL, we estimate $\theta$ in Eq.~\eqref{supp-arXiv-form-U-theta-001-001} by iterating the following two steps.
The first step is compute $\langle \hat{O}_j \rangle_{x_i, \theta}$ in Eq.~\eqref{supp-arXiv-measurement-j-001-001}.
The first step is done in a quantum device.
The second step is update $\theta$ and $\theta_\mathrm{b}$ by minimizing $\mathcal{J}_\mathrm{cost} (\theta, \theta_\mathrm{b})$ in Eq.~\eqref{supp-arXiv-cost-function-QCL-001-001}.
To update $\theta$, the derivative of $\mathcal{J}_\mathrm{cost} (\theta, \theta_\mathrm{b})$ is required; thus, Eq.~\eqref{supp-arXiv-derivative-circuit-geometry-001} and Eq.~\eqref{supp-arXiv-derivative-expectation-rotation-001-001} are useful.
The second step is expected done in a classical device.
We repeat these steps until convergence.

Finally, we summarize QCL in Algo.~\ref{supp-arXiv-QCL-without-SGD-001-001}.
\begin{algorithm}[t]
\caption{Quantum circuit learning (QCL)} \label{supp-arXiv-QCL-without-SGD-001-001}
\begin{algorithmic}[1]
\WHILE{termination condition is not satisfied}
\STATE compute $| \psi^\mathrm{in} (x_i) \rangle$ in Eq.~\eqref{supp-arXiv-encoding-QCL-001-001} for $i = 1, 2, \dots, N$
\STATE compute $| \psi^\mathrm{out} (x_i; \theta) \rangle$ in Eq.~\eqref{supp-arXiv-state-output-001-001} for $i = 1, 2, \dots, N$
\STATE compute $\langle \hat{O}_j \rangle_{x_i, \theta}$ in Eq.~\eqref{supp-arXiv-measurement-j-001-001} for $j = 1, 2, \dots, Q$ and $i = 1, 2, \dots, N$
\STATE compute $f_\mathrm{pred} (x_i; \theta, \theta_\mathrm{b}) = \sum_j \xi_j \langle \hat{O}_j \rangle_{x_i, \theta} + \theta_\mathrm{b}$ in Eq.~\eqref{supp-arXiv-prediction-QCL-001-001} for $i = 1, 2, \dots, N$
\STATE update $\theta$ and $\theta_\mathrm{b}$ by minimizing $\mathcal{J}_\mathrm{cost} (\theta, \theta_\mathrm{b})$ in Eq.~\eqref{supp-arXiv-cost-function-QCL-001-001}
\ENDWHILE
\end{algorithmic}
\end{algorithm}
In practice, stochastic gradient descent (SGD) is also used for QCL.
Then, the procedure of QCL with SGD is summarized in Algo.~\ref{supp-arXiv-QCL-with-SGD-001-001}.
\begin{algorithm}[t]
\caption{Quantum circuit learning (QCL) with stochastic gradient descent (SGD)} \label{supp-arXiv-QCL-with-SGD-001-001}
\begin{algorithmic}[1]
\STATE set $N' < N$
\WHILE{termination condition is not satisfied}
\STATE sample $i_l$ from $[1, 2, \dots, N]$ for $l = 1, 2, \dots, N'$
\STATE compute $| \psi^\mathrm{in} (x_{i_l}) \rangle$ in Eq.~\eqref{supp-arXiv-encoding-QCL-001-001} for $l = 1, 2, \dots, N'$
\STATE compute $| \psi^\mathrm{out} (x_{i_l}; \theta) \rangle$ in Eq.~\eqref{supp-arXiv-state-output-001-001} for $l = 1, 2, \dots, N'$
\STATE compute $\langle \hat{O}_j \rangle_{x_{i_l}, \theta}$ in Eq.~\eqref{supp-arXiv-measurement-j-001-001} for $j = 1, 2, \dots, Q$ and $l = 1, 2, \dots, N'$
\STATE compute $f_\mathrm{pred} (x_{i_l}; \theta, \theta_\mathrm{b}) = \sum_j \xi_j \langle \hat{O}_j \rangle_{x_{i_l}, \theta} + \theta_\mathrm{b}$ in Eq.~\eqref{supp-arXiv-prediction-QCL-001-001} for $l = 1, 2, \dots, N'$
\STATE update $\theta$ and $\theta_\mathrm{b}$ by minimizing $\sum_l \ell (y_{i_l}, f_\mathrm{pred} (x_{i_l}; \theta, \theta_\mathrm{b}))$
\ENDWHILE
\end{algorithmic}
\end{algorithm}

\section{Kernel method} \label{supp-arXiv-sec-kernel-001}

Machine learning provides many kinds of tool kits to analyze a wide range of datasets~\cite{Bishop001, Murphy001}.
Among them, the kernel method is one of the most important approaches because it is very powerful and mathematically simple compared with a mixture model, neural networks, etc~\cite{Bishop001, Murphy001}.
This section is devoted to the kernel method~\cite{Bishop001, Murphy001}.

\subsection{Algorithmic details of the kernel method}

Let us assume that we have
\begin{align}
  \mathcal{D} &\coloneqq [(x_1, y_1), (x_2, y_2), \dots, (x_N, y_N)],
\end{align}
where $x_i$ is a data point and $y_i$ is the label of $x_i$ for $i = 1, 2, \dots, N$.
We also assume that each data point $x_i$ is a $M$-dimensional vector:
\begin{align}
  x_i &\coloneqq [x_i^1, x_i^2, \dots, x_i^M]^\intercal,
\end{align}
where $(\cdot)^\intercal$ represents the transpose.

We first define the feature map $\phi (\cdot)$ by
\begin{align}
  \phi (x_i) &\coloneqq [\phi_1 (x_i), \phi_2 (x_i), \dots, \phi_G (x_i)]^\intercal. \label{supp-arXiv-def-feature-map-001-001}
\end{align}
Then, in the kernel method, by using the feature map~\eqref{supp-arXiv-def-feature-map-001-001}, the prediction on $x_i$ is done by
\begin{align}
  f_\mathrm{pred} (x_i; v) &= \sum_{j=1}^G v_j \phi_j (x_i), \label{supp-arXiv-f-pred-kernel-method-001-001}
\end{align}
where $v \coloneqq [v_1, v_2, \dots, v_G]^\intercal$ is the parameter to be estimated.
The performance of $f_\mathrm{pred} (\cdot; v)$ heavily depends on $v$; then it is quite important to estimate $v$.

To find a good $v$, the following optimization problem is often considered:
\begin{align}
  \min_v \mathcal{J}_\mathrm{cost} (v), \label{supp-arXiv-optimization-problem-kernel-001-001}
\end{align}
where
\begin{align}
  \mathcal{J}_\mathrm{cost} (v) &\coloneqq \frac{1}{N} \sum_{i=1}^N \ell (y_i, f_\mathrm{pred} (x_i; v)) + \frac{\lambda}{2} \| v \|_\mathrm{F}^2, \label{supp-arXiv-cost-function-kernel-method-001-001}
\end{align}
and $\| \cdot \|_\mathrm{F}$ is the Frobenius norm.
For example, the squared error function, Eq.~\eqref{supp-arXiv-squared-error-function-001-001}, the hinge function, Eq.~\eqref{supp-arXiv-hinge-function-001-001}, and the cross-entropy loss function, Eq.~\eqref{supp-arXiv-XE-function-001-001}, are often used for $\ell (\cdot, \cdot)$ in Eq.~\eqref{supp-arXiv-cost-function-kernel-method-001-001}.
Then, the optimal $v$ is given by
\begin{align}
  v_* &= \argmin_v \mathcal{J}_\mathrm{cost} (v). \label{supp-arXiv-optimization-problem-kernel-method-001-001}
\end{align}

\subsection{Ridge classification} \label{supp-arXiv-sec-Ridge-001}

As Eq.~\eqref{supp-arXiv-f-pred-kernel-method-001-001}, we consider the following function to predict the label of $x_i$:
\begin{align}
  f_\mathrm{pred} (x_i; v) &= \sum_{j=1}^G v_j \phi_j (x_i). \label{supp-arXiv-f-pred-kernel-method-001-002}
\end{align}

Furthermore, let us consider the squared error function, Eq.~\eqref{supp-arXiv-squared-error-function-001-001} for Eq.~\eqref{supp-arXiv-cost-function-kernel-method-001-001}:
In this case, Eq.~\eqref{supp-arXiv-cost-function-kernel-method-001-001} becomes
\begin{align}
  \mathcal{J}_\mathrm{cost} (v) &= \frac{1}{N} \sum_{i=1}^N \ell_\mathrm{SE} (y_i, f_\mathrm{pred} (x_i; v)) + \frac{\lambda}{2} \| v \|_\mathrm{F}^2 \\
  &= \frac{1}{N} \sum_{i=1}^N \ell_\mathrm{SE} \bigg( y_i, \sum_{j=1}^G v_j \phi_j (x_i) \bigg) + \frac{\lambda}{2} \sum_{j=1}^G | v_j |^2 \\
  &= \frac{1}{2 N} \sum_{i=1}^N \bigg[ y_i - \sum_{j=1}^G v_j \phi_j (x_i) \bigg]^2 + \frac{\lambda}{2} \sum_{j=1}^G | v_j |^2. \label{supp-arXiv-cost-function-kernel-method-001-002}
\end{align}

The classification method based on Eq.~\eqref{supp-arXiv-optimization-problem-kernel-method-001-001} with Eq.~\eqref{supp-arXiv-cost-function-kernel-method-001-002} is called Ridge classification.

\subsection{Optimal solution in the case of Ridge classification}

Here, we consider the optimal solution of Eq.~\eqref{supp-arXiv-optimization-problem-kernel-method-001-001} in the case of Eq.~\eqref{supp-arXiv-cost-function-kernel-method-001-002}.
By solving $\frac{d}{dv} \mathcal{J}_\mathrm{cost} (v) = 0$, we have
\begin{align}
  v &= \frac{1}{\lambda N} \sum_{i=1}^N \bigg[ y_i - \sum_{j=1}^G v_j \phi_j (x_i) \bigg] \phi (x_i). \label{supp-arXiv-v-min-J-001-001}
\end{align}
Defining, for $i = 1, 2, \dots, N$,
\begin{align}
  a_i &\coloneqq \frac{1}{\lambda N} \bigg[ y_i - \sum_{j=1}^G v_j \phi_j (x_i) \bigg],
\end{align}
Eq.~\eqref{supp-arXiv-v-min-J-001-001} is rewritten as
\begin{align}
  v &= \sum_{i=1}^N a_i \phi (x_i). \label{supp-arXiv-relation-v-a-001-001}
\end{align}
Introducing
\begin{align}
  a &\coloneqq [a_1, a_2, \dots, a_N]^\intercal, \label{supp-arXiv-def-a-001-001}
\end{align}
Eq.~\eqref{supp-arXiv-cost-function-kernel-method-001-002} is transformed into
\begin{align}
  \mathcal{J}_\mathrm{cost} (v) &= \frac{1}{2 N} \sum_{i=1}^N \bigg[ y_i - \sum_{l=1}^G \bigg( \sum_{j=1}^N a_j \phi_l (x_j) \bigg) \phi_l (x_i) \bigg]^2 + \frac{\lambda}{2} \sum_{l=1}^G \bigg[ \sum_{i=1}^N a_i \phi_l (x_i) \bigg]^2 \\
  &= \frac{1}{2 N} \sum_{i=1}^N \bigg[ y_i - \sum_{j=1}^N a_j k (x_j, x_i) \bigg]^2 + \frac{\lambda}{2} \sum_{i, j=1}^N a_i k (x_i, x_j) a_j \\
  &= \frac{1}{2 N} [ y - K a ]^\intercal [ y - K a ] + \frac{\lambda}{2} a^\intercal K a. \label{supp-arXiv-cost-function-kernel-method-001-003}
\end{align}
Here we have used
\begin{align}
  k (x_i, x_j) &\coloneqq \phi^\intercal (x_i) \phi (x_j), \\
  y &\coloneqq [y_1, y_2, \dots, y_N]^\intercal, \\
  K &\coloneqq \Phi \Phi^\intercal, \label{supp-arXiv-def-K-001-001} \\
  \Phi &\coloneqq [\phi (x_1), \phi (x_2), \dots, \phi (x_N)]^\intercal. \label{supp-arXiv-def-Phi-001-001}
\end{align}
Each element of the matrix $K$ is also expressed as
\begin{align}
 [K]_{i, j} &= k (x_i, x_j),
\end{align}
where $[\cdot]_{i, j}$ is the element in the $i$-th row and $j$-th column.
Thus, Eq.~\eqref{supp-arXiv-optimization-problem-kernel-001-001} is equivalent to
\begin{align}
  \min_a \mathcal{J}_\mathrm{cost} (a), \label{supp-arXiv-optimization-problem-kernel-001-002}
\end{align}
where, as shown in Eq.~\eqref{supp-arXiv-cost-function-kernel-method-001-003},
\begin{align}
  \mathcal{J}_\mathrm{cost} (a) &\coloneqq \frac{1}{2 N} [ y - K a ]^\intercal [ y - K a ] + \frac{\lambda}{2} a^\intercal K a. \label{supp-arXiv-cost-function-kernel-method-001-004}
\end{align}
As the solution of Eq.~\eqref{supp-arXiv-optimization-problem-kernel-001-002}, we have
\begin{align}
  a_* &= \frac{1}{N} (K + \lambda N I_N)^{-1} y. \label{supp-arXiv-solution-kernel-001-001}
\end{align}
where that $I_N$ is the $N \times N$ identity matrix.
The proof of Eq.~\eqref{supp-arXiv-solution-kernel-001-001} is as follows.
\begin{proof}

From Eq.~\eqref{supp-arXiv-cost-function-kernel-method-001-004}, the derivative of $\mathcal{J}_\mathrm{cost} (a)$ with respect to $a$ is given by
\begin{align}
  \frac{d}{da} \mathcal{J}_\mathrm{cost} (a) &= - \frac{1}{N} K [y - K a] + \lambda K a \\
  &= - \frac{1}{N} K [y - (K + \lambda N I_N) a]. \label{supp-arXiv-derivative-optimization-problem-kernel-001-001}
\end{align}

By setting the derivative of $\mathcal{J}_\mathrm{cost} (a)$, Eq.~\eqref{supp-arXiv-derivative-optimization-problem-kernel-001-001}, equal to zero and solving it, we have
\begin{align}
  \frac{d}{da} \mathcal{J}_\mathrm{cost} (a_*) = 0 &\Leftrightarrow - \frac{1}{N} K [y - (K + \lambda N I_N) a_*] = 0 \\
  &\Leftrightarrow (K + \lambda N I_N) a_* = y \\
  &\Leftrightarrow a_* = (K + \lambda N I_N)^{-1} y.
\end{align}
Therefore, we have obtained Eq.~\eqref{supp-arXiv-solution-kernel-001-001}.

\end{proof}

From Eq.~\eqref{supp-arXiv-def-a-001-001} and \eqref{supp-arXiv-def-Phi-001-001}, we have
\begin{align}
  v &= a^\intercal \Phi.
\end{align}
Then, the optimal $f_\mathrm{pred} (x; v_*)$ is written as
\begin{align}
  f_\mathrm{pred} (x; v_*) &= v_*^\intercal \phi (x) \\
  &= a_*^\intercal \Phi \phi (x) \\
  &= y^\intercal (K + \lambda N I_N)^{-1} k (x),
\end{align}
where
\begin{align}
  k (x) &\coloneqq \Phi \phi (x) \\
  &= [k (x_1, x), k (x_2, x), \dots, k (x_N, x)]^\intercal.
\end{align}

\section{Correspondence between QCL and the kernel method} \label{supp-arXiv-sec-correpondence-QCL-kernel-001}

To clarify the properties of QCL~\cite{Schuld001, Mitarai001}, we discuss the relationship between quantum computing and the kernel method~\cite{Bishop001, Murphy001} in this section.

\subsection{Rewriting QCL}

Here, we consider the correspondence between QCL~\cite{Schuld001, Mitarai001} and the kernel method~\cite{Bishop001, Murphy001}.
As discussed in Sec.~\ref{supp-arXiv-sec-QCL-cost-function-001-001}, the optimization of QCL is understood as the minimization problem:
\begin{align}
  \min_{\theta, \theta_\mathrm{b}} \mathcal{J}_\mathrm{cost} (\theta, \theta_\mathrm{b}), \label{supp-arXiv-optimization-problem-QCL-theta-001-001}
\end{align}
where
\begin{align}
  \mathcal{J}_\mathrm{cost} (\theta, \theta_\mathrm{b}) &\coloneqq \frac{1}{N} \sum_{i=1}^N \ell (y_i, f_\mathrm{pred} (x_i; \theta, \theta_\mathrm{b})). \label{supp-arXiv-cost-function-QCL-001-002}
\end{align}
Note that Eq.~\eqref{supp-arXiv-cost-function-QCL-001-002} is already given in Eq.~\eqref{supp-arXiv-cost-function-QCL-001-001},
Here, we have used
\begin{align}
  f_\mathrm{pred} (x_i; \theta, \theta_\mathrm{b}) &\coloneqq \sum_{j=1}^Q \xi_j \langle \hat{O}_j \rangle_{x_i, \theta} + \theta_\mathrm{b}, \label{supp-arXiv-prediction-QCL-001-002} \\
  \langle \hat{O}_j \rangle_{x_i, \theta} &\coloneqq \langle \psi^\mathrm{out} (x_i; \theta) | \hat{O}_j | \psi^\mathrm{out} (x_i; \theta) \rangle, \label{supp-arXiv-measurement-j-001-002} \\
  | \psi^\mathrm{out} (x_i; \theta) \rangle &\coloneqq \hat{U}_\mathrm{c} (\theta) | \psi^\mathrm{in} (x_i) \rangle, \label{supp-arXiv-state-output-001-002} \\
  | \psi^\mathrm{in} (x_i) \rangle &\coloneqq \hat{S} (x_i) | \mathrm{init} \rangle. \label{supp-arXiv-encoding-QCL-001-002}
\end{align}
Note that Eqs.~\eqref{supp-arXiv-prediction-QCL-001-002}, \eqref{supp-arXiv-measurement-j-001-002}, \eqref{supp-arXiv-state-output-001-002}, and \eqref{supp-arXiv-encoding-QCL-001-002} are already given in Eqs.~\eqref{supp-arXiv-prediction-QCL-001-001}, \eqref{supp-arXiv-measurement-j-001-001}, \eqref{supp-arXiv-state-output-001-001}, and \eqref{supp-arXiv-encoding-QCL-001-001}.

To rewrite Eq.~\eqref{supp-arXiv-optimization-problem-QCL-theta-001-001}, we define
\begin{align}
  \tilde{O}_j (\{ u_{k, l} \}_{k, l = 1}^{2^n}) &\coloneqq \sum_{\substack{k_1, k_2, \\ k_3, k_4}} [ \psi_{k_1}^\mathrm{in} (x_i) ]^* u_{k_1, k_2}^* O_j^{k_2, k_3} u_{k_3, k_4} \psi_{k_4}^\mathrm{in} (x_i), \label{supp-arXiv-def-O-tilde-001-001}
\end{align}
where
\begin{align}
  \psi_l^\mathrm{in} (x_i) &\coloneqq \langle l | \psi^\mathrm{in} (x_i) \rangle, \\
  O_j^{k, l} &\coloneqq \langle k | \hat{O}_j | l \rangle,
\end{align}
for $k, l = 1, 2, \dots, 2^n$, and $(\cdot)^*$ represents the complex conjugate for scalars.
Thus, we have the following relation on Eqs.~\eqref{supp-arXiv-measurement-j-001-002} and \eqref{supp-arXiv-def-O-tilde-001-001}:
\begin{align}
  \langle \hat{O}_j \rangle_{x_i, \theta} &= \tilde{O}_j (\{ u_{k, l} \}_{k, l = 1}^{2^n}).
\end{align}
So far, we have assumed that $\hat{U}_\mathrm{c} (\theta)$ is a parametrized universal quantum circuit.
By regarding $\{ u_{k, l} \}_{k,l = 1}^{2^n}$ as variables to be estimated instead of $\theta$ in $\hat{U}_\mathrm{c} (\theta)$, Eq.~\eqref{supp-arXiv-optimization-problem-QCL-theta-001-001} is rewritten as
\begin{subequations}
\begin{align}
  \min_{\{ u_{k, l} \}_{k, l = 1}^{2^n}, \theta_\mathrm{b}} \ & \mathcal{J}_\mathrm{cost} (\{ u_{k, l} \}_{k, l = 1}^{2^n}, \theta_\mathrm{b}), \\
  \mathrm{subject \ to} \ & u_k^\mathrm{H} u_l = \delta_{k, l} \ (k, l = 1, 2, \dots, 2^n), \label{supp-arXiv-constraint-optimization-problem-QCL-u-001-001}
\end{align}
\label{supp-arXiv-optimization-problem-QCL-u-001-001}%
\end{subequations}
where
\begin{align}
  \mathcal{J}_\mathrm{cost} (\{ u_{k, l} \}_{k, l = 1}^{2^n}, \theta_\mathrm{b}) &\coloneqq \frac{1}{N} \sum_{i=1}^N \ell \bigg( y_i, \sum_{j=1}^Q \xi_j \tilde{O}_j (\{ u_{k, l} \}_{k, l = 1}^{2^n}) + \theta_\mathrm{b} \bigg).
\end{align}

Furthermore, by using the penalty method~\cite{Boyd001, Fletcher001}, Eq.~\eqref{supp-arXiv-optimization-problem-QCL-u-001-001} can be expressed as
\begin{align}
  \lim_{\lambda \to \infty} \min_{\{ u_{k, l} \}_{k, l = 1}^{2^n}, \theta_\mathrm{b}} \mathcal{J}_\mathrm{pm} (\{ u_{k, l} \}_{k, l = 1}^{2^n}, \theta_\mathrm{b}; \lambda), \label{supp-arXiv-optimization-problem-QCL-u-002-001}
\end{align}
where
\begin{align}
  \mathcal{J}_\mathrm{pm} (\{ u_{k, l} \}_{k, l = 1}^{2^n}, \theta_\mathrm{b}; \lambda) &\coloneqq \mathcal{J}_\mathrm{cost} (\{ u_{k, l} \}_{k, l = 1}^{2^n}, \theta_\mathrm{b}) + \frac{\lambda}{2} \sum_{k, l = 1, 2, \dots, 2^n} \big[ u_k^\mathrm{H} u_l - \delta_{k, l} \big]^2. \label{supp-arXiv-def-J-pm-QCL-001-001}
\end{align}
Here we define, for $k = 1, 2, \dots, 2^n$,
\begin{align}
  u_k \coloneqq [u_{1, k}, u_{2, k}, \dots, u_{2^n, k}]^\mathrm{H}, \label{supp-arXiv-def-u-vector-001-001}
\end{align}
where $(\cdot)^\mathrm{H} \coloneqq ((\cdot)^*)^\intercal = ((\cdot)^\intercal)^*$ is the Hermitian conjugate for vectors and matrices.
Furthermore, $\delta_{k, l}$ is the Kronecker delta function:
\begin{align}
  \delta_{k, l} &=
  \begin{cases}
    1 \ (k = l), \\
    0 \ (k \ne l).
  \end{cases}
\end{align}
As shown in Eq.~\eqref{supp-arXiv-def-u-vector-001-001}, $\{ u_k \}_{k=1}^{2^n}$ are the $k$-th column vectors, but we can also use the row vectors for $\{ u_k \}_{k=1}^{2^n}$ by replacing Eq.~\eqref{supp-arXiv-constraint-optimization-problem-QCL-u-001-001} with $u_k u_l^\mathrm{H} = \delta_{k, l}$.

\subsection{Rewriting the kernel method: Quadratic formulation}

Here, to consider the correspondence between QCL and the kernel method, we consider a slightly different formulation of the kernel method.
Instead of Eq.~\eqref{supp-arXiv-f-pred-kernel-method-001-001}, let us make a prediction on $y_i$ by
\begin{align}
  f_\mathrm{pred} (x; w) &\coloneqq \sum_{k, l = 1, 2, \dots, G} w_{k, l} \tilde{\phi}_k (x) \tilde{\phi}_l (x),
\end{align}
where $[w]_{i, j} \coloneqq w_{i, j}$.
Here, $[\cdot]_{i, j}$ is the element in the $i$-th row and the $j$-th column.
Then, Eq.~\eqref{supp-arXiv-optimization-problem-kernel-001-001} is transformed into
\begin{align}
  \min_w \mathcal{J}_\mathrm{cost} (w), \label{supp-arXiv-optimization-problem-kernel-quad-001-001}
\end{align}
where
\begin{align}
  \mathcal{J}_\mathrm{cost} (w) &\coloneqq \frac{1}{N} \sum_{i=1}^N \ell (y_i,f_\mathrm{pred} (x_i; w)) + \frac{\lambda}{2} \| w \|_\mathrm{F}^2 \\
  &= \frac{1}{N} \sum_{i=1}^N \ell \bigg( y_i, \sum_{k, l = 1, 2, \dots, G} w_{k, l} \tilde{\phi}_k (x_i) \tilde{\phi}_l (x_i) \bigg) + \frac{\lambda}{2} \sum_{k, l = 1, 2, \dots, G} | w_{k, l} |^2. \label{supp-arXiv-def-J-quad-001-001}
\end{align}

At the end of this subsection, we also introduce
\begin{align}
  w_{k, l}^\delta &\coloneqq w_{k, l} + \delta_{k, l}.
\end{align}
Then we can rewrite Eq.~\eqref{supp-arXiv-def-J-quad-001-001} as
\begin{align}
  \mathcal{J}_\mathrm{cost} (w^\delta) &= \frac{1}{N} \sum_{i=1}^N \ell \bigg( y_i, \sum_{k, l = 1, 2, \dots, G} (w_{k, l}^\delta - \delta_{k, l}) \tilde{\phi}_k (x_i) \tilde{\phi}_l (x_i) \bigg) + \frac{\lambda}{2} \sum_{k, l = 1, 2, \dots, G} | w_{k, l}^\delta - \delta_{k, l} |^2. \label{supp-arXiv-def-J-quad-002-001}
\end{align}
In the rest of this section, we clarify the correspondence between QCL and the kernel method by using Eqs.~\eqref{supp-arXiv-def-J-quad-001-001} and \eqref{supp-arXiv-def-J-quad-002-001}.

\subsection{Correspondence with QCL}

So far, we have finished the preparation.
We discuss the correspondence between QCL and the kernel method.
For simplicity, we set $\theta_\mathrm{b} = 0$ in Eq.~\eqref{supp-arXiv-prediction-QCL-001-001} in this subsection.

\subsubsection{Cost functions}

We here compare QCL and the kernel method from the viewpoint of cost functions.
Eq.~\eqref{supp-arXiv-def-J-quad-001-001} and Eq.~\eqref{supp-arXiv-def-J-pm-QCL-001-001} have the following correspondence:
\begin{subequations}
\begin{align}
  \psi_l^\mathrm{in} (x_i) &\Leftrightarrow \tilde{\phi}_l (x_i), \\
  \sum_j \sum_{k_1, k_2} \xi_j u_{k, k_1}^* O_j^{k_1, k_2} u_{k_2, l} &\Leftrightarrow w_{k, l}, \\
  \frac{\lambda}{2} \sum_{k, l} | u_k^\mathrm{H} u_l - \delta_{k, l} |^2 &\Leftrightarrow \frac{\lambda}{2} \sum_{k, l} | w_{k, l} |^2.
\end{align}
\label{supp-arXiv-correspondence-QCL-kernel-001-001}%
\end{subequations}
By using Eq.~\eqref{supp-arXiv-def-J-quad-002-001}, we can rewrite Eq.~\eqref{supp-arXiv-correspondence-QCL-kernel-001-001} as
\begin{align}
  \psi_l^\mathrm{in} (x_i) &\Leftrightarrow \tilde{\phi}_l (x_i), \\
  \sum_j \sum_{k_1, k_2} \xi_j u_{k, k_1}^* O_j^{k_1, k_2} u_{k_2, l} &\Leftrightarrow w_{k, l}^\delta - \delta_{k, l}, \\
  \frac{\lambda}{2} \sum_{k, l} | u_k^\mathrm{H} u_l - \delta_{k, l} |^2 &\Leftrightarrow \frac{\lambda}{2} \sum_{k, l} | w_{k, l}^\delta - \delta_{k, l} |^2.
\end{align}

\subsubsection{Feature map and encoding}

At the end of this section, we mention the relationship between quantum encoding and the feature map.
Depending on $\hat{S}_x$, $| \psi^\mathrm{in} (x_i) \rangle$ varies.
When we employ amplitude encoding, described by Eq.~\eqref{supp-arXiv-amplitude-encoding-001-001}, we have
\begin{align}
  \psi_l^\mathrm{in} (x_i)&= \langle l | \psi^\mathrm{in} (x_i) \rangle \nonumber \\
  &= \langle l | \bigg[ \sum_{k=1}^{M + M'} \tilde{x}_i^k \bigg] | k \rangle \nonumber \\
  &= \tilde{x}_i^l, \label{supp-arXiv-amplitude-encoding-011-001}
\end{align}
for $l = 1, 2, \dots, 2^n$.
But, we can extend Eq.~\eqref{supp-arXiv-amplitude-encoding-011-001}:
\begin{align}
  | \psi^\mathrm{in} (x_i) \rangle &\coloneqq \frac{1}{\chi} \sum_{k_1, k_2, \dots} (x_i^{k_1} x_i^{k_2} \dots) | k_1, k_2, \dots, k_M \rangle, \label{supp-arXiv-extended-amplitude-encoding-001-001}
\end{align}
where
\begin{align}
  \chi &\coloneqq \sqrt{\sum_{k_1, k_2, \dots, k_M} (x_i^{k_1} x_i^{k_2} \dots x_i^{k_M})},
\end{align}
for $k_m = 1, 2, \dots$ and $m = 1, 2, \dots, M$.
Note that $M$ is the dimensions of each $x_i$.

On the other hand, in the kernel method, we need to select the functional form of the feature map.
For example, we may adopt the linear map given by
\begin{align}
  \tilde{\phi} (x_i) &= [x_i^1, x_i^2, \dots], \label{supp-arXiv-phi-example-001-001}
\end{align}
and
the nonlinear map $\tilde{\phi} (x_i)$ given by
\begin{align}
  \tilde{\phi} (x_i) &= [x_i^1, x_i^2, x_i^1 x_i^2, \dots]. \label{supp-arXiv-phi-example-002-001}
\end{align}
Thus, by tuning parameters, we can associate Eqs.~\eqref{supp-arXiv-phi-example-001-001} and \eqref{supp-arXiv-phi-example-002-001} to Eqs.~\eqref{supp-arXiv-amplitude-encoding-011-001} and \eqref{supp-arXiv-extended-amplitude-encoding-001-001}, respectively.

\section{Splitting method for orthogonality constrained problems}

The main purpose of this section is to review the method of splitting orthogonality constraints (SOC) proposed in Ref.~\cite{Lai001}.
The method of SOC is based on Bregman iterative regularization proposed in Refs.~\cite{Osher001, Yin001}.
We begin with Bregman iterative regularization and discuss its relation with the augmented Lagrange method.
Then, we review the method of SOC.

\subsection{Bregman iterative regularization} \label{supp-arXiv-sec-BIR-001}

The method of SOC proposed in Ref.~\cite{Lai001} is based on Bregman iterative regularization proposed in Refs.~\cite{Osher001, Yin001}.
We then explain Bregman iterative regularization.
Let us consider the following optimization problem:
\begin{subequations}
\begin{align}
  \min_x \ & \mathcal{J}_\mathrm{cost} (x), \\
  \mathrm{subject \ to} \ & A x = g,
\end{align}
\label{supp-arXiv-problem-Bregman-iterative-regularization-001-001}%
\end{subequations}
where $A$ is a matrix.

Bregman iterative regularization solves Eq.~\eqref{supp-arXiv-problem-Bregman-iterative-regularization-001-001} by the following approach:
\begin{subequations}
\begin{align}
  x_{k+1} &= \argmin_x \mathcal{J}_\mathrm{BIR} (x; x_k, p_k), \\
  p_{k+1} &= p_k - r A^\intercal (A x_{k+1} - g), \label{supp-arXiv-algorithm-Bregman-iterative-regularization-001-012}
\end{align}
\label{supp-arXiv-algorithm-Bregman-iterative-regularization-001-001}%
\end{subequations}
where
\begin{align}
  \mathcal{J}_\mathrm{BIR} (x; y, p) &\coloneqq B_{\mathcal{J}_\mathrm{cost} (\cdot)}^p (x, y) + \frac{r}{2} \| A x - g \|_\mathrm{F}^2, \label{supp-arXiv-def-J-BIR-001-001} \\
  B_{\mathcal{J}_\mathrm{cost} (\cdot)}^p (x, y) &\coloneqq \mathcal{J}_\mathrm{cost} (x) - \mathcal{J}_\mathrm{cost} (y) - \langle p, x - y \rangle. \label{supp-arXiv-def-Bregman-divergence-subderivative-001-001}
\end{align}
Here, $r$ is a positive constant.
The performance of Bregman iterative regularization depends on $r$; thus we need to find an appropriate value of $r$.
Letting $x_*$ and $p_*$ be $x$ and $p$ at convergence, we have $p_* \in \partial \mathcal{J}_\mathrm{cost} (x_*)$ where $\partial \mathcal{J}_\mathrm{cost} (x)$ is the subderivative of $\mathcal{J}_\mathrm{cost} (\cdot)$ at $x$~\cite{Rockafellar001}.
Suppose that $\mathcal{J}_\mathrm{cost} (x)$ is convex for $\Omega$; the definition of the subderivative is
\begin{align}
  \partial \mathcal{J}_\mathrm{cost} (y) &= \Big\{ c \Big| \forall x \in \Omega, \mathcal{J}_\mathrm{cost} (x) - \mathcal{J}_\mathrm{cost} (y) \ge c (x - y) \Big\}.
\end{align}
Note that the conventional Bregman divergence is~\cite{Bregman001}
\begin{align}
  B_{\mathcal{J}_\mathrm{cost} (\cdot)} (x, y) &\coloneqq \mathcal{J}_\mathrm{cost} (x) - \mathcal{J}_\mathrm{cost} (y) - \langle \nabla \mathcal{J}_\mathrm{cost} (y), x - y \rangle; \label{supp-arXiv-def-Bregman-divergence-001-001}
\end{align}
then Eq.~\eqref{supp-arXiv-def-Bregman-divergence-subderivative-001-001} can be regarded as the subderivative extension of Eq.~\eqref{supp-arXiv-def-Bregman-divergence-001-001}.

Eq.~\eqref{supp-arXiv-algorithm-Bregman-iterative-regularization-001-001} can be rewritten as~\cite{Lai001}
\begin{subequations}
\begin{align}
  x_{k+1} &= \argmin_x \bigg[ \mathcal{J}_\mathrm{cost} (x) + \frac{r}{2} \| A x - g + b_k \|_\mathrm{F}^2 \bigg], \label{supp-arXiv-algorithm-Bregman-iterative-regularization-002-011} \\
  b_{k+1} &= b_k + A x_{k+1} - g.
\end{align}
\label{supp-arXiv-algorithm-Bregman-iterative-regularization-002-001}%
\end{subequations}
By setting $g_k \coloneqq g - b_k$, Eq.~\eqref{supp-arXiv-algorithm-Bregman-iterative-regularization-002-001} is also expressed as~\cite{Yin001}
\begin{subequations}
\begin{align}
  x_{k+1} &= \argmin_x \bigg[ \mathcal{J}_\mathrm{cost} (x) + \frac{r}{2} \| A x - g_k \|_\mathrm{F}^2 \bigg], \\
  g_{k+1} &= g_k - (A x_{k+1} - g).
\end{align}
\label{supp-arXiv-algorithm-Bregman-iterative-regularization-003-001}%
\end{subequations}

The equivalence between Eqs.~\eqref{supp-arXiv-algorithm-Bregman-iterative-regularization-001-001} with $r = 1$ and \eqref{supp-arXiv-algorithm-Bregman-iterative-regularization-003-001} with $r = 1$ is shown as follows.
\begin{proof}

For clarity, when we consider the algorithm described by \eqref{supp-arXiv-algorithm-Bregman-iterative-regularization-003-001}, we use $\tilde{x}_k$ instead of $x_k$.

We first assume, for a certain $k$, $p_k = A^\intercal (g_k - A \tilde{x}_k)$.
From Eq.~\eqref{supp-arXiv-def-Bregman-divergence-subderivative-001-001}, we have
\begin{align}
  B_{\mathcal{J}_\mathrm{cost} (\cdot)}^{p_k} (x, x_k) + \frac{1}{2} \| A x - g \|_\mathrm{F}^2 &= \mathcal{J}_\mathrm{cost} (x) - \mathcal{J}_\mathrm{cost} (x_k) - \langle p_k, x - x_k \rangle + \frac{1}{2} \| A x - g \|_\mathrm{F}^2 \\
  &= \mathcal{J}_\mathrm{cost} (x) - \langle (g_k - A \tilde{x}_k), A x \rangle + \frac{1}{2} \| A x - g \|_\mathrm{F}^2 + C_1 \\
  &= \mathcal{J}_\mathrm{cost} (x) + \frac{1}{2} \| A x - g - (g_k - A \tilde{x}_k) \|_\mathrm{F}^2 + C_2 \\
  &= \mathcal{J}_\mathrm{cost} (x) + \frac{1}{2} \| A x - g_{k+1} \|_\mathrm{F}^2 + C_2,
\end{align}
where
\begin{align}
  C_1 &\coloneqq - \mathcal{J}_\mathrm{cost} (x_k) - \langle p_k, - x_k \rangle, \\
  C_2 &\coloneqq C_1 - \frac{1}{2} \| g_k - A \tilde{x}_k \|_\mathrm{F}^2.
\end{align}
Thus, we have shown that Eqs.~\eqref{supp-arXiv-algorithm-Bregman-iterative-regularization-001-001} and \eqref{supp-arXiv-algorithm-Bregman-iterative-regularization-003-001} are identical.
For details, refer to Ref.~\cite{Hale001}.

\end{proof}

\subsection{Correspondence between Bregman iterative regularization and the augmented Lagrange method}

It seemsf meaningful to clarify the relationship between Bregman iterative regularization and a well-known algorithm.
Here we explain the penalty method and the augmented Lagrangian method~\cite{Boyd001, Fletcher001}.
Then we see the correspondence between Bregman iterative regularization and the augmented Lagrange method.
To this end, let us consider the following optimization problem:
\begin{subequations}
\begin{align}
  \min_x \ & \mathcal{J}_\mathrm{cost} (x), \\
  \mathrm{subject \ to} \ & c_i (x) = 0 \ (i = 1, 2, \dots, m).
\end{align}
\label{supp-arXiv-problem-optimization-correspondence-001-001}%
\end{subequations}

To find the solution of Eq.~\eqref{supp-arXiv-problem-optimization-correspondence-001-001}, the penalty method solves
\begin{align}
  \lim_{\lambda \to \infty} \min_x \mathcal{J}_\mathrm{pm} (x; \lambda),
\end{align}
where
\begin{align}
  \mathcal{J}_\mathrm{pm} (x; \lambda) &\coloneqq \mathcal{J}_\mathrm{cost} (x) + \frac{\lambda}{2} \sum_i \| c_i (x) \|_\mathrm{F}^2.
\end{align}

Introducing $\lambda \coloneqq [\lambda_1, \lambda_2, \dots, \lambda_m]^\intercal$ and $c (x) \coloneqq [c_1 (x), c_2 (x), \dots, c_m (x)]^\intercal$, the augmented Lagrange method for Eq.~\eqref{supp-arXiv-problem-optimization-correspondence-001-001} solves, for $k = 1, 2, \dots$,
\begin{align}
  \lim_{k \to \infty} \min_x \mathcal{J}_\mathrm{alm} (x; \lambda_k),
\end{align}
where
\begin{align}
  \mathcal{J}_\mathrm{alm} (x; \lambda_k) &= \mathcal{J}_\mathrm{cost} (x) + \lambda_k^\intercal c (x) + \frac{r}{2} \| c (x) \|_\mathrm{F}^2, \label{supp-arXiv-problem-optimization-correspondence-alm-001-001} \\
  \lambda_{k+1} &= \lambda_k + r c (x_{k+1}), \label{supp-arXiv-problem-optimization-correspondence-alm-001-002}
\end{align}
and $r > 0$ is a parameter.

We note that Eq.~\eqref{supp-arXiv-def-J-BIR-001-001} with $r = 1$ and Eq.~\eqref{supp-arXiv-problem-optimization-correspondence-alm-001-001} with $r = 1$ are equivalent when $c (x)$ is linear.
The proof is shown below.
\begin{proof}

From the assumption, we can set
\begin{align}
  c (x) &= A x - g.
\end{align}
For a certain $k$, we impose the following relation:
\begin{align}
  p_k &= - A^\intercal \lambda_k.
\end{align}
Then we have
\begin{align}
  \mathcal{J}_\mathrm{alm} (x; \lambda_k) &= \mathcal{J}_\mathrm{cost} (x) + \langle \lambda_k, (A x - g) \rangle + \frac{1}{2} \| A x - g \|_\mathrm{F}^2 \\
  &= \mathcal{J}_\mathrm{cost} (x) + \langle \lambda_k, A x \rangle + \frac{1}{2} \| A x - g \|_\mathrm{F}^2 + C_1 \\
  &= \mathcal{J}_\mathrm{cost} (x) - \langle p_k , x \rangle + \frac{1}{2} \| A x - g \|_\mathrm{F}^2 + C_1 \\
  &= \mathcal{J}_\mathrm{cost} (x) - \mathcal{J}_\mathrm{cost} (x_k) - \langle p_k , (x - x_k) \rangle + \frac{1}{2} \| A x - g \|_\mathrm{F}^2 + \mathcal{J}_\mathrm{cost} (x_k) - \langle p_k, x_k \rangle + C_1 \\
  &= B_{\mathcal{J}_\mathrm{cost} (\cdot)}^{p_k} (x, x_k) + \frac{1}{2} \| A x - g \|_\mathrm{F}^2 + C_2.
\end{align}
where $C_1$ and $C_2$ are constant terms with respect to $x$ given by
\begin{align}
  C_1 &\coloneqq - \langle \lambda_k, g \rangle, \\
  C_2 &\coloneqq C_1 + \mathcal{J}_\mathrm{cost} (x_k) - \langle p_k, x_k \rangle.
\end{align}
Thus we have shown the equivalence between Eq.~\eqref{supp-arXiv-def-J-BIR-001-001} with $r = 1$ and Eq.~\eqref{supp-arXiv-problem-optimization-correspondence-alm-001-001} with $r = 1$.

\end{proof}

Furthermore, Eq.~\eqref{supp-arXiv-problem-optimization-correspondence-alm-001-002} leads to Eq.~\eqref{supp-arXiv-algorithm-Bregman-iterative-regularization-001-012}.
Therefore, we have confirmed the equivalence between Bregman iterative regularization and the augmented Lagrange method.

\subsection{Penalty method and augmented Lagrangian method for optimization problems with orthogonality constraints}

Before getting into the method of SOC, we explain the penalty method and the augmented Lagrangian method for optimization problems with orthogonality constraints~\cite{Boyd001, Fletcher001}.
Then let us consider the following optimization problem:
\begin{subequations}
\begin{align}
  \min_X \ & \mathcal{J}_\mathrm{cost} (X), \\
  \mathrm{subject \ to} \ & X^\intercal A X = I,
\end{align}
\label{supp-arXiv-problem-SOC-matrix-001-001}%
\end{subequations}
where $A \succ O$.
Here, $I$ and $O$ are the identity matrix and the zero matrix, respectively.

The penalty method for Eq.~\eqref{supp-arXiv-problem-SOC-matrix-001-001} can be formulated as follows:
\begin{align}
  \lim_{\lambda \to \infty} \min_X \mathcal{J}_\mathrm{pm} (X; \lambda),
\end{align}
where
\begin{align}
  \mathcal{J}_\mathrm{pm} (X; \lambda) &\coloneqq \mathcal{J}_\mathrm{cost} (X) + \frac{\lambda}{2} \| X^\intercal A X - I \|_\mathrm{F}^2.
\end{align}

The augmented Lagrange method for Eq.~\eqref{supp-arXiv-problem-SOC-matrix-001-001} solves, for $k = 1, 2, \dots$,
\begin{align}
  \lim_{k \to \infty} \min_X \mathcal{J}_\mathrm{alm} (X; \Lambda_k),
\end{align}
where
\begin{align}
  \mathcal{J}_\mathrm{alm} (X; \Lambda_k) &\coloneqq \mathcal{J}_\mathrm{cost} (X) + \mathrm{Tr} [(\Lambda_k)^\intercal (X^\intercal A X - I)] + \frac{r}{2} \| X^\intercal A X - I \|_\mathrm{F}^2, \\
  \Lambda_{k+1} &= \Lambda_k + r (X^\intercal A X - I),
\end{align}
and $r > 0$ is a parameter.

The main problem with the above approaches is that they are very slow.
Then, we review the method of SOC proposed in Ref.~\cite{Lai001} in the next subsection.

\subsection{Algorithmic details of the method of SOC} \label{supp-arXiv-sec-SOC-001-001}

Let us consider the optimization problem given in Eq.~\eqref{supp-arXiv-problem-SOC-matrix-001-001}.
By splitting the constraints, we first rewrite Eq.~\eqref{supp-arXiv-problem-SOC-matrix-001-001} as
\begin{subequations}
\begin{align}
  \min_{X, P} \ & \mathcal{J}_\mathrm{cost} (X), \\
  \mathrm{subject \ to} \ & L X = P, \\
  & P^\intercal P = I,
\end{align}
\label{supp-arXiv-problem-SOC-matrix-002-001}%
\end{subequations}
where $L$ is a matrix that satisfies
\begin{align}
  A &= L^\intercal L.
\end{align}
Next we apply Bregman iterative regularization to Eq.~\eqref{supp-arXiv-problem-SOC-matrix-002-001}; then, we solve it by iterating the following equations:
\begin{subequations}
\begin{align}
  \{ X_k, P_k \} &= \argmin_{X, P} \mathcal{J}_\mathrm{SOC} (X; P, D_{k-1}), \nonumber \\
  & \quad \mathrm{subject \ to} \ P^\intercal P = I, \label{supp-arXiv-algorithm-SOC-001-011} \\
  D_k &= D_{k-1} + L X_k - P_k, \\
\end{align}
\label{supp-arXiv-algorithm-SOC-001-001}%
\end{subequations}
where
\begin{align}
 \mathcal{J}_\mathrm{SOC} (X; P, D) &\coloneqq \mathcal{J}_\mathrm{cost} (X) + \frac{r}{2} \| L X - P + D \|_\mathrm{F}^2.
\end{align}
Here, $r$ is a positive constant.
The performance of the method of SOC depends on $r$; thus, we need to find a good value of $r$.
Note that Eq.~\eqref{supp-arXiv-algorithm-SOC-001-011} comes from Eq.~\eqref{supp-arXiv-algorithm-Bregman-iterative-regularization-002-011}.

By dividing Eq.~\eqref{supp-arXiv-algorithm-SOC-001-011} into two equations, we rewrite Eq.~\eqref{supp-arXiv-algorithm-SOC-001-001} as follows:
\begin{subequations}
\begin{align}
  X_k &= \argmin_X \mathcal{J}_\mathrm{SOC} (X; P_{k-1}, D_{k-1}), \label{supp-arXiv-algorithm-SOC-002-011} \\
  P_k &= \argmin_P \frac{r}{2} \| P - (L X_k + D_{k-1}) \|_\mathrm{F}^2, \nonumber \\
  & \quad \mathrm{subject \ to} \ P^\intercal P = I, \label{supp-arXiv-algorithm-SOC-002-012} \\
  D_k &= D_{k-1} + L X_k - P_k. \label{supp-arXiv-algorithm-SOC-002-013}
\end{align}
\label{supp-arXiv-algorithm-SOC-002-001}%
\end{subequations}
Note that Eq.~\eqref{supp-arXiv-algorithm-SOC-002-012} comes form the fact that $\mathcal{J}_\mathrm{cost} (X)$ does not depend on $P$:
\begin{align}
  \argmin_P \frac{r}{2} \| P - (L X_k + D_{k-1}) \|_\mathrm{F}^2 &= \argmin_P \mathcal{J}_\mathrm{SOC} (X_k; P, D_{k-1}).
\end{align}
Furthermore, the closed form solution of Eq.~\eqref{supp-arXiv-algorithm-SOC-002-012} is already known as shown in Refs.~\cite{Lai001, Manton001, Gibson001}.
We first compute
\begin{align}
  K_{1, k} \Sigma_k K_{2, k}^\intercal = L X_k + D_{k-1},
\end{align}
where $K_{1, k}$ and $K_{2, k}^\intercal$ are orthogonal matrices and $\Sigma_k$ is a $p \times q$ matrix, the elements of which in the $i$-th row and the $j$-th column for $i \ne j$ are zero, and then we can compute $P_k$ in Eq.~\eqref{supp-arXiv-algorithm-SOC-002-012} in a closed form by
\begin{align}
  P_k &= K_{1, k} I_{p, q} K_{2, k}^\intercal. \label{supp-arXiv-algorithm-SOC-002-022}
\end{align}
We call Eq.~\eqref{supp-arXiv-algorithm-SOC-002-001} the method of SOC and it is summarized in Algo.~\ref{supp-arXiv-SOC-001-001}.
\begin{algorithm}[t]
\caption{Method of splitting orthogonality constraints (SOC)} \label{supp-arXiv-SOC-001-001}
\begin{algorithmic}[1]
\STATE set $P_0$ and $D_0$
\FOR{$k = 1, 2, \dots, K$}
\STATE compute $X_k$ by Eq.~\eqref{supp-arXiv-algorithm-SOC-002-011}
\STATE compute $P_k$ by Eq.~\eqref{supp-arXiv-algorithm-SOC-002-022}
\STATE compute $D_k$ by Eq.~\eqref{supp-arXiv-algorithm-SOC-002-013}
\ENDFOR
\end{algorithmic}
\end{algorithm}

To explain Eq.~\eqref{supp-arXiv-algorithm-SOC-002-022}, we then state the related theorem by following Ref.~\cite{Lai001}.
\begin{theorem} \label{supp-arXiv-theorem-algorithm-SOC-002-001}

Suppose that we have a $p \times q$ matrix $Y$ and let us consider the following problem:
\begin{align}
  P_Y &= \argmin_P \frac{1}{2} \| P - Y \|, \nonumber \\
  & \quad \mathrm{subject \ to} \ P^\intercal P = I. \label{supp-arXiv-minimization-on-P-Y-real-001-001}
\end{align}
Then, the solution of Eq.~\eqref{supp-arXiv-minimization-on-P-Y-real-001-001} is given by
\begin{align}
  P_Y &= K_1 I_{p, q} K_2^\intercal. \label{supp-arXiv-update-P-Y-real-001-001}
\end{align}
where $K_1$ and $K_2^\intercal$ satisfy $K_1 \Sigma K_2^\intercal = Y$ and $\Sigma$ is a diagonal matrix whose diagonal elements are the singular values of $Y$.

\end{theorem}

By setting $Y = L X_k + D_{k-1}$ in Eq.~\eqref{supp-arXiv-minimization-on-P-Y-real-001-001}, Eq.~\eqref{supp-arXiv-update-P-Y-real-001-001} leads to Eq.~\eqref{supp-arXiv-algorithm-SOC-002-022}.
The proof of Thm.~\ref{supp-arXiv-theorem-algorithm-SOC-002-001} is given as follows.
\begin{proof}

Let us consider the singular value decomposition of $Y$:
\begin{align}
  Y &= K_1 \Sigma K_2^\intercal, \label{supp-arXiv-SVD-Y-001}
\end{align}
where $K_1$ and $K_2^\intercal$ are $p \times p$ and $q \times q$ orthogonal matrices, respectively, and $\Sigma$ is a $p \times q$ diagonal matrix in the sense that $\{ [\Sigma]_{i, i} \}_i$ are the singular values of $Y$ and $[\Sigma]_{i, j} = 0$ for $i \ne j$.

By using $K_1$ and $K_2^\intercal$ in Eq.~\eqref{supp-arXiv-SVD-Y-001}, we define
\begin{align}
  \tilde{P} &\coloneqq K_1^\intercal P K_2. \label{supp-arXiv-def-tilde-P-001}
\end{align}
Then, we have
\begin{align}
  \| P - Y \|_\mathrm{F}^2 &= \| K_1 (K_1^\intercal P K_2 - \Sigma) K_2^\intercal \|_\mathrm{F}^2 \\
  &= \| K_1 (\tilde{P} - \Sigma) K_2^\intercal \|_\mathrm{F}^2 \\
  &= \| \tilde{P} - \Sigma \|_\mathrm{F}^2.
\end{align}
Furthermore, from Eq.~\eqref{supp-arXiv-def-tilde-P-001}, we also have
\begin{align}
  P^\intercal P = I &\Leftrightarrow \tilde{P}^\intercal \tilde{P} = I.
\end{align}
Then, Eq.~\eqref{supp-arXiv-minimization-on-P-Y-real-001-001} is almost equivalent to
\begin{align}
  P_\Sigma &= \argmin_{\tilde{P}} \frac{1}{2} \| \tilde{P} - \Sigma \|_\mathrm{F}^2, \nonumber \\
  & \quad \mathrm{subject \ to} \ \tilde{P}^\intercal \tilde{P} = I. \label{supp-arXiv-minimization-on-P-Sigma-001}
\end{align}
It is almost trivial that the solution of Eq.~\eqref{supp-arXiv-minimization-on-P-Sigma-001} takes the following form:
\begin{align}
  P_\Sigma &= I_{p, q},
\end{align}
where
\begin{align}
  [I_{p, q}]_{k, l} &=
  \begin{cases}
    1 & (k = l) \\
    0 & (k \ne l)
  \end{cases}. \label{supp-arXiv-Ipq-001-001}
\end{align}
Here $[\cdot]_{k, l}$ is the element in the $k$-th row and $l$-th column.
Thus, the colosed form solution of Eq.~\eqref{supp-arXiv-minimization-on-P-Y-real-001-001} is given by
\begin{align}
  P_Y &= K_1 P_\Sigma K_2^\intercal \\
  &= K_1 I_{p, q} K_2^\intercal. \label{supp-arXiv-PY-001-001}
\end{align}
Thus we have obtained Eq.~\eqref{supp-arXiv-update-P-Y-real-001-001}.

\end{proof}

The UKM described in Sec.~\ref{sec-UKM-001} is based on the method of SOC.
More precisely, the UKM is based on the complex version of the method of SOC.

\section{Unitary kernel method} \label{sec-UKM-001}

This section describes the UKM, which is one of the main algorithms in this paper.
We first provide an overview of the UKM and then discuss how to implement it by using the CG method.

\subsection{Algorithmic details of the UKM} \label{supp-arXiv-sec-UKM-001-001}

We explain the UKM in this section.
We first describe the optimization problem and then state how to solve it.

In the UKM, we make a prediction on $y_i$ by $f_\mathrm{pred} (x_i; \hat{U}, \theta_\mathrm{b})$.
The problem of the UKM is to estimate $\hat{U}$.
To do so, similarly to Eq.~\eqref{supp-arXiv-prediction-QCL-001-001}, we consider the following cost function:
\begin{align}
  \mathcal{J}_\mathrm{cost} (\hat{U}, \theta_\mathrm{b}) &\coloneqq \frac{1}{N} \sum_{i=1}^N \ell (y_i, f_\mathrm{pred} (x_i; \hat{U}, \theta_\mathrm{b})), \label{supp-arXiv-cost-UKM-001-001}
\end{align}
where
\begin{align}
  f_\mathrm{pred} (x_i; \hat{U}, \theta_\mathrm{b}) &\coloneqq \sum_{j=1}^Q \xi_j \langle \hat{O}_j \rangle_{x_i, \hat{U}} + \theta_\mathrm{b}, \label{supp-arXiv-prediction-UKM-001-001} \\
  \langle \hat{O}_j \rangle_{x_i, \hat{U}} &\coloneqq \langle \psi^\mathrm{out} (x_i; \hat{U}) | \hat{O}_j | \psi^\mathrm{out} (x_i; \hat{U}) \rangle, \\
  | \psi^\mathrm{out} (x_i; \hat{U}) \rangle &\coloneqq \hat{U} | \psi^\mathrm{in} (x_i) \rangle, \\
  | \psi^\mathrm{in} (x_i) \rangle &\coloneqq \hat{S} (x_i) | \mathrm{init} \rangle. \label{supp-arXiv-encoding-UKM-001-001}
\end{align}
Here, $| \mathrm{init} \rangle$ is a initial quantum state, $\{ \xi_j \}_{j=1}^Q$ are fixed parameters, $\{ \hat{O}_j \}_{j=1}^Q$ are the set of measurements, $\hat{S} (x_i)$ is the operator to encode $x_i$ on qubits, and $\ell (\cdot, \cdot)$ is a loss function.
For example, the squared error function, Eq.~\eqref{supp-arXiv-squared-error-function-001-001}, the hinge function, Eq.~\eqref{supp-arXiv-hinge-function-001-001}, and the cross-entropy loss function, Eq.~\eqref{supp-arXiv-XE-function-001-001}, are often used for $\ell (\cdot, \cdot)$~\cite{Bishop001, Murphy001}, and the amplitude encoding is used for $\hat{S} (x_i)$~\cite{Schuld001, Schuld003}.
Note that the dimension of $\hat{U}$ can take any positive integer, but to consider the correspondence with quantum computing, we consider a $n$-qubit system and then the dimension of $\hat{U}$ is $2^n$.
Here, $\theta_\mathrm{b}$ is a bias term to be estimated, but we may set $\theta_\mathrm{b} = 0$ for simplicity.
Due to the nature of quantum mechanics, $\hat{U}$ in Eq.~\eqref{supp-arXiv-cost-UKM-001-001} is a unitary operator and, to find a unitary operator, the method of SOC discussed in Sec.~\ref{supp-arXiv-sec-SOC-001-001} is applicable.
Similarly to Eq.~\eqref{supp-arXiv-problem-SOC-matrix-001-001}, we estimate $\hat{U}$ and $\theta_\mathrm{b}$ by solving the following optimization problem:
\begin{subequations}
\begin{align}
  \min_{\hat{U}, \theta_\mathrm{b}} \ & \mathcal{J}_\mathrm{cost} (\hat{U}, \theta_\mathrm{b}), \\
  \mathrm{subject \ to} \ & \hat{U}^\dagger \hat{U} = \hat{1}_{2^n}.
\end{align}
\label{supp-arXiv-problem-quantum-kernel-method-001-001}%
\end{subequations}

To solve Eq.~\eqref{supp-arXiv-problem-quantum-kernel-method-001-001}, we use the method of SOC, Eq.~\eqref{supp-arXiv-problem-SOC-matrix-002-001}.
Hereafter we denote, by $\hat{X}$, an operator obtained by the method of SOC since it does not strictly satisfy the unitarity condition.
Then, we rewrite Eq.~\eqref{supp-arXiv-problem-quantum-kernel-method-001-001} as
\begin{subequations}
\begin{align}
  \min_{\hat{X}, \theta_\mathrm{b}} \ & \mathcal{J}_\mathrm{cost} (\hat{X}, \theta_\mathrm{b}), \\
  \mathrm{subject \ to} \ & \hat{X} = \hat{P}, \\
  & \hat{P}^\dagger \hat{P} = \hat{1}_{2^n}.
\end{align}
\label{supp-arXiv-problem-quantum-kernel-method-002-001}%
\end{subequations}
Note that $L$ in Eq.~\eqref{supp-arXiv-problem-SOC-matrix-002-001} is replaced by the identity operator and the UKM deals with unitary operators though the method of SOC considers the real matrices.
Then, we try to explicitly write the update equations for Eq.~\eqref{supp-arXiv-problem-quantum-kernel-method-002-001} by following the method of SOC.
From now on, we denote $\hat{X}$, $\hat{P}$, $\hat{D}$, and $\theta_\mathrm{b}$ at the $k$-th iteration by $\hat{X}_k$, $\hat{P}_k$, $\hat{D}_k$, and $\theta_{\mathrm{b}, k}$, respectively.
Like Eqs.~\eqref{supp-arXiv-algorithm-SOC-002-011} and \eqref{supp-arXiv-algorithm-SOC-002-012}, by introducing $\hat{D}$, we split Eq.~\eqref{supp-arXiv-problem-quantum-kernel-method-002-001} into
\begin{align}
\{ \hat{X}_k, \theta_{\mathrm{b}, k} \} &= \argmin_{\hat{X}, \theta_\mathrm{b}} \mathcal{J}_\mathrm{SOC} (\hat{X}, \theta_\mathrm{b}; \hat{P}_{k-1}, \hat{D}_{k-1}), \label{supp-arXiv-quantum-kernel-method-001-011}
\end{align}
and
\begin{align}
  \hat{P}_k &= \argmin_{\hat{P}} \frac{r}{2} \| \hat{P} - (\hat{X}_k + \hat{D}_{k-1}) \|_\mathrm{F}^2, \nonumber \\
  & \quad \mathrm{subject \ to} \ \hat{P}^\dagger \hat{P} = \hat{1}_{2^n}, \label{supp-arXiv-quantum-kernel-method-001-012}
\end{align}
where
\begin{align}
  \mathcal{J}_\mathrm{SOC} (\hat{X}, \theta_\mathrm{b}; \hat{P}, \hat{D}) &\coloneqq \mathcal{J}_\mathrm{cost} (\hat{X}, \theta_\mathrm{b}) + \frac{r}{2} \| \hat{X} - \hat{P} + \hat{D} \|_\mathrm{F}^2. \label{supp-arXiv-cost-UKM-Frobenius-norm-001-001}
\end{align}

Then, we turn our attention to how to solve Eq.~\eqref{supp-arXiv-quantum-kernel-method-001-012}.
In Thm.~\ref{supp-arXiv-theorem-algorithm-SOC-002-001}, we limit ourselves to real matrices.
Note that, while Thm.~\ref{supp-arXiv-theorem-algorithm-SOC-002-001} in the previous section and Ref.~\cite{Lai001} focus on an orthogonal case, Refs.~\cite{Manton001, Gibson001} discuss a unitary case.
In the case of operators, Eq.~\eqref{supp-arXiv-minimization-on-P-Y-real-001-001} is transformed into
\begin{align}
  \hat{P}_* &= \argmin_{\hat{P}} \frac{1}{2} \| \hat{P} - \hat{Y} \|, \nonumber \\
  & \quad \mathrm{subject \ to} \ \hat{P}^\dagger \hat{P} = \hat{1}. \label{supp-arXiv-minimization-on-P-Y-complex-001-001}
\end{align}
Then, the solution of Eq.~\eqref{supp-arXiv-minimization-on-P-Y-complex-001-001} becomes
\begin{align}
\hat{P}_* &= \hat{K}_1 \hat{K}_2^\dagger, \label{supp-arXiv-OU-main-001-001}
\end{align}
where $\hat{K}_1$ and $\hat{K}_2$ are unitary matrices that satisfy
\begin{align}
  \hat{Y} = \hat{K}_1 \hat{\Sigma} \hat{K}_2^\dagger, \label{supp-arXiv-OU-sub-001-001}
\end{align}
and $\hat{\Sigma}$ is a diagonal operator in the sense that $\{ \langle i | \hat{\Sigma} | i \rangle \}_i$ are the singular values of $\hat{Y}$ and $\langle i | \hat{\Sigma} | j \rangle = 0$ for $i \ne j$.
In this paper, we call Eq.~\eqref{supp-arXiv-OU-main-001-001} with Eq.~\eqref{supp-arXiv-OU-sub-001-001} operator unitarization (OU) and in the numerical sections, Secs.~\ref{supp-arXiv-sec-numerical-result-UKM-001} and \ref{supp-arXiv-sec-numerical-result-VCR-001}, we use OU to obtain matrices that exactly satisfy the unitarity constraint.
Thus, to find the solution of Eq.~\eqref{supp-arXiv-quantum-kernel-method-001-012}, we first apply the SVD to $\hat{X}_k + \hat{D}_{k-1}$ and compute $\hat{K}_{1, k}$, $\hat{\Sigma}_k$, and $\hat{K}_{2, k}^\dagger$ such that
\begin{align}
  \hat{K}_{1, k} \hat{\Sigma}_k \hat{K}_{2, k}^\dagger &= \hat{X}_k + \hat{D}_{k-1},
\end{align}
where $\hat{K}_{1, k}$ and $\hat{K}_{2, k}^\dagger$ are unitary operators and $\hat{\Sigma}_k$ is a diagonal operator whose diagonal elements are the singular values of $\hat{X}_k + \hat{D}_{k-1}$.
We then compute
\begin{align}
  \hat{P}_k &= \hat{K}_{1, k} \hat{K}_{2, k}^\dagger. \label{supp-arXiv-quantum-kernel-method-001-013}
\end{align}
Note that the dimensions of $\hat{K}_{1, k}$ and $\hat{K}_{2, k}^\dagger$ are identical and thus nothing need to be inserted between $\hat{K}_{1, k}$ and $\hat{K}_{2, k}^\dagger$ though $I_{p, q}$ in Eq.~\eqref{supp-arXiv-Ipq-001-001} is inserted between $K_1$ and $K_2^\intercal$ in Eq.~\eqref{supp-arXiv-update-P-Y-real-001-001}.

At the end of the $k$-th iteration, like Eq.~\eqref{supp-arXiv-algorithm-SOC-002-013}, we update $\hat{D}_k$ by
\begin{align}
  \hat{D}_k &= \hat{D}_{k-1} + \hat{X}_k - \hat{P}_k. \label{supp-arXiv-quantum-kernel-method-001-014}
\end{align}
In the UKM, we iterate Eq.~\eqref{supp-arXiv-quantum-kernel-method-001-011}, Eq.~\eqref{supp-arXiv-quantum-kernel-method-001-013}, and Eq.~\eqref{supp-arXiv-quantum-kernel-method-001-014} until convergence.
The UKM is summarized in Algo.~\ref{supp-arXiv-quantum-kernel-method-001-001}
\begin{algorithm}[t]
\caption{Unitary kernel method (UKM)} \label{supp-arXiv-quantum-kernel-method-001-001}
\begin{algorithmic}[1]
\STATE set $\hat{P}_0$ and $\hat{D}_0$
\FOR{$k = 1, 2, \dots, K$}
\STATE compute $\hat{X}_k$ and $\theta_{\mathrm{b}, k}$ by Eq.~\eqref{supp-arXiv-quantum-kernel-method-001-011}
\STATE compute $\hat{P}_k$ by Eq.~\eqref{supp-arXiv-quantum-kernel-method-001-013}
\STATE compute $\hat{D}_k$ by Eq.~\eqref{supp-arXiv-quantum-kernel-method-001-014}
\ENDFOR
\end{algorithmic}
\end{algorithm}
In the next subsection, we elaborate on how to solve Eq.~\eqref{supp-arXiv-quantum-kernel-method-001-011}.

\subsection{Derivatives of the cost function} \label{supp-arXiv-sec-UKM-001-002}

Let us consider the derivatives of Eq.~\eqref{supp-arXiv-cost-UKM-001-001} and \eqref{supp-arXiv-cost-UKM-Frobenius-norm-001-001}.
For simplicity, we set $Q = 1$ and $\xi_1 = 1$ in Eq.~\eqref{supp-arXiv-prediction-UKM-001-001}; we denote the measurement by $\hat{O}$.
Furthermore, we define $\hat{\rho}_i \coloneqq | \psi^\mathrm{in} (x_i) \rangle \langle \psi^\mathrm{in} (x_i) |$.

First, we consider the derivatives of Eq.~\eqref{supp-arXiv-cost-UKM-001-001}.
By using Eq.~\eqref{supp-arXiv-derivative-UBUrho-ReU-001}, we have
\begin{align}
  \frac{d}{d \Re [\hat{X}]} \mathcal{J}_\mathrm{cost} (\hat{X}, \theta_\mathrm{b}) &= \frac{1}{N} \sum_{i=1}^N \Bigg[ \frac{d}{d z} \ell (y_i, z) \bigg|_{z = f_\mathrm{pred} (x_i; \hat{X}, \theta_\mathrm{b})} \Bigg] \frac{d}{d \Re [\hat{X}]} f_\mathrm{pred} (x_i; \hat{X}, \theta_\mathrm{b}) \\
  &= \frac{1}{N} \sum_{i=1}^N \Bigg[ \frac{d}{d z} \ell (y_i, z) \bigg|_{z = f_\mathrm{pred} (x_i; \hat{X}, \theta_\mathrm{b})} \Bigg] (\hat{O} \hat{X} \hat{\rho}_i + \hat{O}^\intercal \hat{X}^* \hat{\rho}_i^\intercal). \label{supp-arXiv-derivative-cost-Re-UKM-cost-001-001}
\end{align}
Similarly, by using Eq.~\eqref{supp-arXiv-derivative-UBUrho-ImU-001}, we get
\begin{align}
  \frac{d}{d \Im [\hat{X}]} \mathcal{J}_\mathrm{cost} (\hat{X}, \theta_\mathrm{b}) &= \frac{1}{N} \sum_{i=1}^N \Bigg[ \frac{d}{d z} \ell (y_i, z) \bigg|_{z = f_\mathrm{pred} (x_i; \hat{X}, \theta_\mathrm{b})} \Bigg] \frac{d}{d \Im [\hat{X}]} f_\mathrm{pred} (x_i; \hat{X}, \theta_\mathrm{b}) \\
  &= \frac{1}{N} \sum_{i=1}^N \Bigg[ \frac{d}{d z} \ell (y_i, z) \bigg|_{z = f_\mathrm{pred} (x_i; \hat{X}, \theta_\mathrm{b})} \Bigg] (-i \hat{O} \hat{X} \hat{\rho}_i + i \hat{O}^\intercal \hat{X}^* \hat{\rho}_i^\intercal). \label{supp-arXiv-derivative-cost-Im-UKM-cost-001-001}
\end{align}
Finally, we have
\begin{align}
  \frac{d}{d \theta_\mathrm{b}} \mathcal{J}_\mathrm{cost} (\hat{X}, \theta_\mathrm{b}) &= \frac{1}{N} \sum_{i=1}^N \Bigg[ \frac{d}{d z} \ell (y_i, z) \bigg|_{z = f_\mathrm{pred} (x_i; \hat{X}, \theta_\mathrm{b})} \Bigg] \frac{d}{d \theta_\mathrm{b}} f_\mathrm{pred} (x_i; \hat{X}, \theta_\mathrm{b}) \\
  &= \frac{1}{N} \sum_{i=1}^N \Bigg[ \frac{d}{d z} \ell (y_i, z) \bigg|_{z = f_\mathrm{pred} (x_i; \hat{X}, \theta_\mathrm{b})} \Bigg]. \label{supp-arXiv-derivative-cost-theta-bias-cost-001-001}
\end{align}
Thus, we have obtained the derivatives of Eq.~\eqref{supp-arXiv-cost-UKM-001-001}.

We also give the derivatives of the square of the Frobenius norm since they appear in Eq.~\eqref{supp-arXiv-quantum-kernel-method-001-011}.
By using Eqs.~\eqref{supp-arXiv-formula-derivative-complex-matrix-trace-004-001} and \eqref{supp-arXiv-formula-derivative-complex-matrix-trace-004-011}, we have
\begin{align}
  \frac{d}{d \Re [\hat{X}]} \| \hat{X} - \hat{Y} \|_\mathrm{F}^2 &= \frac{d}{d \Re [\hat{X}]} \mathrm{Tr} [(\hat{X} - \hat{Y})^\dagger (\hat{X} - \hat{Y})] \\
  &= (\hat{X} - \hat{Y}) + (\hat{X} - \hat{Y})^* \\
  &= 2 \Re [\hat{X} - \hat{Y}]. \label{supp-arXiv-derivative-Frobenius-Re-001-001}
\end{align}
Similarly, by using Eqs.~\eqref{supp-arXiv-formula-derivative-complex-matrix-trace-004-002} and \eqref{supp-arXiv-formula-derivative-complex-matrix-trace-004-012}, we have
\begin{align}
  \frac{d}{d \Im [\hat{X}]} \| \hat{X} - \hat{Y} \|_\mathrm{F}^2 &= \frac{d}{d \Im [\hat{X}]} \mathrm{Tr} [(\hat{X} - \hat{Y})^\dagger (\hat{X} - \hat{Y})] \\
  &= -i (\hat{X} - \hat{Y}) + i (\hat{X} - \hat{Y})^* \\
  &= 2 \Im [\hat{X} - \hat{Y}]. \label{supp-arXiv-derivative-Frobenius-Im-001-001}
\end{align}

Finally, we consider the derivatives of Eq.~\eqref{supp-arXiv-cost-UKM-Frobenius-norm-001-001}.
Eqs.~\eqref{supp-arXiv-derivative-cost-Re-UKM-cost-001-001} and \eqref{supp-arXiv-derivative-Frobenius-Re-001-001} lead to
\begin{align}
  \frac{d}{d \Re [\hat{X}]} \mathcal{J}_\mathrm{SOC} (\hat{X}, \theta_\mathrm{b}; \hat{P}, \hat{D}) &= \frac{1}{N} \Bigg[ \frac{d}{d z} \ell (y_i, z) \bigg|_{z = f_\mathrm{pred} (x_i; \hat{X}, \theta_\mathrm{b})} \Bigg] (\hat{O} \hat{X} \hat{\rho}_i + \hat{O}^\intercal \hat{X}^* \hat{\rho}_i^\intercal) + r \Re [\hat{X} - \hat{P} + \hat{D}], \label{supp-arXiv-derivative-cost-Re-UKM-SOC-001-001}
\end{align}
and Eqs.~\eqref{supp-arXiv-derivative-cost-Im-UKM-cost-001-001} and \eqref{supp-arXiv-derivative-Frobenius-Im-001-001} also lead to
\begin{align}
  \frac{d}{d \Im [\hat{X}]} \mathcal{J}_\mathrm{SOC} (\hat{X}, \theta_\mathrm{b}; \hat{P}, \hat{D}) &= \frac{1}{N} \sum_{i=1}^N \Bigg[ \frac{d}{d z} \ell (y_i, z) \bigg|_{z = f_\mathrm{pred} (x_i; \hat{X}, \theta_\mathrm{b})} \Bigg] (-i \hat{O} \hat{X} \hat{\rho}_i + i \hat{O}^\intercal \hat{X}^* \hat{\rho}_i^\intercal) + r \Im [\hat{X} - \hat{P} + \hat{D}]. \label{supp-arXiv-derivative-cost-Im-UKM-SOC-001-001}
\end{align}
The second term of Eq.~\eqref{supp-arXiv-cost-UKM-Frobenius-norm-001-001} does not depend on $\theta_\mathrm{b}$.
Thus, similarly to Eq.~\eqref{supp-arXiv-derivative-cost-theta-bias-cost-001-001}, we have
\begin{align}
  \frac{d}{d \theta_\mathrm{b}} \mathcal{J}_\mathrm{SOC} (\hat{X}, \theta_\mathrm{b}) &= \frac{1}{N} \sum_{i=1}^N \Bigg[ \frac{d}{d z} \ell (y_i, z) \bigg|_{z = f_\mathrm{pred} (x_i; \hat{X}, \theta_\mathrm{b})} \Bigg]. \label{supp-arXiv-derivative-cost-theta-bias-SOC-001-001}
\end{align}
Thus, we have obtained the derivatives of Eq.~\eqref{supp-arXiv-cost-UKM-Frobenius-norm-001-001}.

Without specifying $\ell (\cdot, \cdot)$ in Eq.~\eqref{supp-arXiv-cost-UKM-001-001}, we cannot go further.
Then let us consider the squared error function, Eq.~\eqref{supp-arXiv-squared-error-function-001-001}, for $\ell (\cdot, \cdot)$ in Eq.~\eqref{supp-arXiv-cost-UKM-001-001}; the derivative of the loss function takes the form
\begin{align}
  \frac{d}{d z} \ell_\mathrm{SE} (y_i, z) \bigg|_{z = f_\mathrm{pred} (x_i; \hat{X}, \theta_\mathrm{b})} &= - [y_i - f_\mathrm{pred} (x_i; \hat{X}, \theta_\mathrm{b})]. \label{supp-arXiv-derivative-loss-SE-001-001}
\end{align}
Substituting Eq.~\eqref{supp-arXiv-derivative-loss-SE-001-001} into Eqs.~\eqref{supp-arXiv-derivative-cost-Re-UKM-SOC-001-001} and \eqref{supp-arXiv-derivative-cost-Im-UKM-SOC-001-001}, we obtain the derivatives of Eq.~\eqref{supp-arXiv-cost-UKM-Frobenius-norm-001-001} in the case of the squared error function, Eq.~\eqref{supp-arXiv-squared-error-function-001-001}.

We employ the \textbf{optimize} function~\footnote{The SciPy package~\cite{Virtanen001} provides the \textbf{optimize} function, in which the CG method, the BFGS method, and other optimization methods are implemented.} in the SciPy package~\cite{Virtanen001} to solve Eq.~\eqref{supp-arXiv-quantum-kernel-method-001-011}.
Eqs.~\eqref{supp-arXiv-derivative-cost-Re-UKM-SOC-001-001} and \eqref{supp-arXiv-derivative-cost-Im-UKM-SOC-001-001} help it run faster.

\subsection{UKM with the nonlinear CG method} \label{supp-arXiv-sec-UKM-CG-001-001}

We have explained the UKM in the previous subsection.
Here we state how to solve Eq.~\eqref{supp-arXiv-quantum-kernel-method-001-011}.

We first define the $2 N^2 + 1$-dimensional real vector $\tilde{x}$ whose element is given by
\begin{subequations}
\begin{align}
  [\tilde{x}]_{2 (N(i-1) + (j-1)) + 1} &\coloneqq \Re [\langle i | \hat{X} | j \rangle], \\
  [\tilde{x}]_{2 (N(i-1) + (j-1)) + 2} &\coloneqq \Im [\langle i | \hat{X} | j \rangle], \\
  [\tilde{x}]_{2 N^2 + 1} &\coloneqq \theta_\mathrm{b},
\end{align}
\label{supp-arXiv-def-tilde-u-vector-001-001}%
\end{subequations}
for $i, j = 1, 2, \dots, N$, where $N$ is the dimension of $\hat{X}$.
From Eq.~\eqref{supp-arXiv-cost-UKM-001-001} and Eq.~\eqref{supp-arXiv-quantum-kernel-method-001-011}, we then define
\begin{align}
  \mathcal{J}_\mathrm{CG} (\tilde{x}) &\coloneqq \mathcal{J}_\mathrm{cost} (\hat{X}, \theta_\mathrm{b}) + \frac{r}{2} \| \hat{X} - \hat{P}_{k-1} + \hat{D}_{k-1} \|_\mathrm{F}^2.
\end{align}

Then, similarly to Eq.~\eqref{supp-arXiv-CG-001-001}, we solve Eq.~\eqref{supp-arXiv-quantum-kernel-method-001-011} by
\begin{subequations}
\begin{align}
  \tilde{x}_k &= \tilde{x}_{k-1} + \alpha_k d_{k-1}, \label{supp-arXiv-CG-quantum-kernel-method-001-011} \\
  d_k &= - \nabla \mathcal{J}_\mathrm{CG} (\tilde{x}_k) + \beta_k d_{k-1}. \label{supp-arXiv-CG-quantum-kernel-method-001-012}
\end{align}
\label{supp-arXiv-CG-quantum-kernel-method-001-001}%
\end{subequations}
Here we use the CG method: Eq.~\eqref{supp-arXiv-alpha-002-001} for $\alpha_k$ and Eq.~\eqref{supp-arXiv-beta-002-001} for $\beta_k$, since the BFGS method requires a large memory space then the CG method and for the case of a large number of qubits, the memory space problem becomes severe.
Note that Eqs.~\eqref{supp-arXiv-derivative-UBUrho-ReU-001} and \eqref{supp-arXiv-derivative-UBUrho-ImU-001} are available to compute $\nabla \mathcal{J}_\mathrm{cost} (\hat{X}, \theta_\mathrm{b})$ and $\nabla \mathcal{J}_\mathrm{CG} (\tilde{x})$.
Then, we compute $\hat{X}_k$ from $\tilde{x}$ by using by Eq.~\eqref{supp-arXiv-def-tilde-u-vector-001-001}.
Furthermore, we repeat the SOC update $K$ times and the CG update $K'$ times.

We summarize the UKM with the CG method in Algo.~\ref{supp-arXiv-quantum-kernel-method-002-001}.
In this paper, We perform the UKM with the CG method, Algo.~\ref{supp-arXiv-quantum-kernel-method-002-001}, and see its performance in Sec.~\ref{supp-arXiv-sec-numerical-result-UKM-001}.
\begin{algorithm}[t]
\caption{Unitary kernel method (UKM) with the conjugate gradient (CG) method} \label{supp-arXiv-quantum-kernel-method-002-001}
\begin{algorithmic}[1]
\STATE set $\hat{P}_0$ and $\hat{D}_0$
\FOR{$k = 1, 2, \dots, K$}
\STATE initialize $\tilde{x}_0$ by Eq.~\eqref{supp-arXiv-def-tilde-u-vector-001-001}
\STATE set $d_0 = - \nabla \mathcal{J}_\mathrm{CG} (\tilde{x}_0)$
\FOR{$k' = 1, 2, \dots, K'$}
\STATE compute $\tilde{x}_{k'}$ by Eq.~\eqref{supp-arXiv-CG-quantum-kernel-method-001-011}
\STATE compute $d_{k'}$ by Eq.~\eqref{supp-arXiv-CG-quantum-kernel-method-001-012}
\ENDFOR
\STATE compute $\hat{X}_k$ and $\theta_{\mathrm{b}, k}$ from $\tilde{x}_{K'}$ by Eq.~\eqref{supp-arXiv-def-tilde-u-vector-001-001}
\STATE compute $\hat{P}_k$ by Eq.~\eqref{supp-arXiv-quantum-kernel-method-001-012}
\STATE compute $\hat{D}_k$ by Eq.~\eqref{supp-arXiv-quantum-kernel-method-001-013}
\ENDFOR
\end{algorithmic}
\end{algorithm}

\section{Variational circuit realization}

In the literature, some novel unitary decomposition methods were proposed: the quantum Shannon decomposition (QSD)~\cite{Shende001}, the column-by-column decomposition~\cite{Iten001}, Knill's decomposition~\cite{Knill001, Nielsen001}, etc.
Despite their usefulness, they require an exponentially large number of CNOT gates when the number of qubits $n$ increases.

In this section, we propose a variational method to obtain a quantum circuit that realizes a given unitary operator, which we call the VCR.

\subsection{Algorithmic details of the VCR}

We first assume that a target unitary and a quantum circuit that has the set of parameters.
Then, let $\hat{U}$ and $\hat{U}_\mathrm{c} (\theta; L)$ be the target unitary operator and the unitary operator composed of gates that are parametrized by $\theta \coloneqq [\theta_1, \theta_2, \dots]^\intercal$, respectively.
So far, we have denoted, by $\hat{U}_\mathrm{c} (\theta)$, the unitary realized by a quantum circuit, but we denote it by $\hat{U}_\mathrm{c} (\theta; L)$ to emphasize the number of layers $L$ in this section.

In the VCR, we construct a cost function and estimate $\theta$ by minimizing the cost function.
For example, we can estimate $\theta$ by
\begin{align}
  \{ \theta_*, \lambda_* \} &= \argmin_{\theta, \lambda} \tilde{\mathcal{J}}_\mathrm{cost} (\theta, \lambda; L, p, \hat{U}),
\end{align}
where, for arbitrary $p > 0$,
\begin{align}
  \tilde{\mathcal{J}}_\mathrm{cost} (\theta, \lambda; L, p, \hat{U}) &\coloneqq \| \hat{U} - \hat{U}_\mathrm{c+p} (\theta, \lambda; L) \|_\mathrm{F}^p. \label{supp-arXiv-def-g1-001}
\end{align}
The global phase does not matter physically, but it matters for approximating a unitary operator; then we have used, in Eq.~\eqref{supp-arXiv-def-g1-001}, the unitary operator $\hat{U}_\mathrm{c+p} (\theta, \lambda; L)$ defined by
\begin{align}
  \hat{U}_\mathrm{c+p} (\theta, \lambda; L) &\coloneqq \hat{\Phi}_{2^n} (\lambda) \hat{U}_\mathrm{c} (\theta; L),
\end{align}
where $\hat{\Phi}_{2^n} (\lambda) \coloneqq e^{- i \lambda} \hat{1}_{2^n}$ is defined in Eq.~\eqref{supp-arXiv-def-global-phase-unitary-001-001}.

We give another example of a cost function.
When $\hat{U}$ and $\hat{U}_\mathrm{c+p} (\theta, \lambda; L)$ are identical, we have
\begin{align}
  \hat{U}^\dagger \hat{U}_\mathrm{c+p} (\theta, \lambda; L) &= \hat{1}_{2^n}.
\end{align}
Then, we can estimate $\theta$ by
\begin{align}
  \{ \theta_*, \lambda_* \} &= \argmin_{\theta, \lambda} \mathcal{J}_\mathrm{cost} (\theta, \lambda; L, p, \hat{U}),
\end{align}
where, for any $p > 0$,
\begin{align}
  \mathcal{J}_\mathrm{cost} (\theta, \lambda; L, p, \hat{U}) &\coloneqq \| \hat{U}^\dagger \hat{U}_\mathrm{c+p} (\theta, \lambda; L) - \hat{1}_{2^n} \|_\mathrm{F}^p. \label{supp-arXiv-def-g2-001}
\end{align}

In a circuit realization, the complexity of a circuit is of great interest.
In this letter, we assume a layered structure for a quantum circuit.
Thus, given an error threshold $\delta$, it is convenient to define $L_\delta$:
\begin{align}
  L_\delta &= \argmin_L \epsilon_L, \nonumber \\
  & \quad \mathrm{subject \ to} \ \epsilon_L \le \delta,
\end{align}
where
\begin{align}
  \epsilon_L &\coloneqq \min_{\theta, \lambda} \mathcal{J}_\mathrm{cost} (\theta, \lambda; L, p, \hat{U}).
\end{align}

The advantage of the VCR is that the number of CNOT gates is expected to be small though the number of CNOT gates exponentially grows in the QSD.
In Sec.~\ref{supp-arXiv-sec-numerical-result-VCR-001}, we perform the VCR and see its performance.

\subsection{Derivatives of the cost functions}

To employ the \textbf{optimize} function in the SciPy package~\cite{Virtanen001}, the derivative of a cost function is required for fast computing.
We explain the derivatives of Eqs.~\eqref{supp-arXiv-def-g1-001} and \eqref{supp-arXiv-def-g2-001}.

From Eq.~\eqref{supp-arXiv-another-expression-matrix-norm-001}, we can rewrite the Frobenius norm by using the trace operator; then Eqs.~\eqref{supp-arXiv-def-g1-001} and \eqref{supp-arXiv-def-g1-001} for $p = 2$ become, respectively,
\begin{align}
  \tilde{\mathcal{J}}_\mathrm{cost} (\theta, \lambda; L, 2, \hat{U}) &= \mathrm{Tr} \Big[ \Big(\hat{U} - \hat{U}_\mathrm{c+p} (\theta, \lambda; L) \Big)^\dagger \Big( \hat{U} - \hat{U}_\mathrm{c+p} (\theta, \lambda; L) \Big) \Big], \\
  \mathcal{J}_\mathrm{cost} (\theta, \lambda; L, 2, \hat{U}) &= \mathrm{Tr} \Big[ \Big( \hat{U}^\dagger \hat{U}_\mathrm{c+p} (\theta, \lambda; L) - \hat{1}_{2^n} \Big)^\dagger \Big( \hat{U}^\dagger \hat{U}_\mathrm{c+p} (\theta, \lambda; L) - \hat{1}_{2^n} \Big) \Big].
\end{align}

Due to the linearity of the trace operation, the following relation holds:
\begin{align}
  \frac{d}{d \theta_i} \mathrm{Tr} [\hat{A} (\theta)] &= \mathrm{Tr} \bigg[ \frac{d}{d \theta_i} \hat{A} (\theta) \bigg].
\end{align}
Then we have
\begin{align}
  \frac{d}{d \theta_i} \tilde{\mathcal{J}}_\mathrm{cost} (\theta, \lambda; L, 2, \hat{U}) &= - \mathrm{Tr} \bigg[ \frac{d}{d \theta_i} \hat{U}_\mathrm{c+p}^\dagger (\theta, \lambda; L) \bigg( \hat{U} - \hat{U}_\mathrm{c+p} (\theta, \lambda; L) \bigg) \bigg] - \mathrm{Tr} \bigg[ \bigg( \hat{U} - \hat{U}_\mathrm{c+p} (\theta, \lambda; L) \bigg)^\dagger \frac{d}{d \theta_i} \hat{U}_\mathrm{c+p} (\theta, \lambda; L) \bigg], \label{supp-arXiv-derivative-g1-001}
\end{align}
and
\begin{align}
  & \frac{d}{d \theta_i} \mathcal{J}_\mathrm{cost} (\theta, \lambda; L, 2, \hat{U}) \nonumber \\
  & \quad = \mathrm{Tr} \bigg[ \bigg( \hat{U}^\dagger \frac{d}{d \theta_i} \hat{U}_\mathrm{c+p} (\theta, \lambda; L) \bigg)^\dagger \bigg( \hat{U}^\dagger \hat{U}_\mathrm{c+p} (\theta, \lambda; L) - \hat{1}_{2^n} \bigg) \bigg] + \mathrm{Tr} \bigg[ \bigg( \hat{U}^\dagger \hat{U}_\mathrm{c+p} (\theta, \lambda; L) - \hat{1}_{2^n} \bigg)^\dagger \bigg( \hat{U}^\dagger \frac{d}{d \theta_i} \hat{U}_\mathrm{c+p} (\theta, \lambda; L) \bigg) \bigg]. \label{supp-arXiv-derivative-g2-001}
\end{align}
By using Eq.~\eqref{supp-arXiv-derivative-circuit-geometry-001}, we compute $\frac{d}{d \theta_i} \hat{U}_\mathrm{c+p} (\theta, \lambda; L)$ in Eqs.~\eqref{supp-arXiv-derivative-g1-001} and \eqref{supp-arXiv-derivative-g2-001}.

\section{Numerical simulation of the UKM} \label{supp-arXiv-sec-numerical-result-UKM-001}

In this section, we show numerical simulations of QCL and the UKM for several datasets.
We first describe the numerical setting and the properties of datasets.
Then, we show the results.

\subsection{Numerical setting} \label{supp-arXiv-sec-numerical-settings-UKM-001}

To evaluate the performance of QCL and the UKM, we use 5-fold cross-validation (CV) with 5 different random initial conditions.
For each method, we select the best model for the training dataset over iterations to compute the performance.

For the UKM, we show the performance of $\hat{X}$, $\hat{P}$, and OU of $\hat{X}$.
Refer to Sec.~\ref{supp-arXiv-sec-SOC-001-001} for the details of OU.
Furthermore, we consider real and complex matrices as initial inputs.

For QCL, we show the performance of the CNOT-based circuit geometry~\eqref{supp-arXiv-gate-CNOT-001-001}, the CRot-based circuit geometry~\eqref{supp-arXiv-gate-CRot-001-001}, and the Heisenberg circuit geometry~\eqref{supp-arXiv-gate-Mitarai-001-001}.
In the Heisenberg circuit geometry, Eq.\eqref{supp-arXiv-gate-Mitarai-001-001}, we put $h_i^w = 0$ for $i = 1, 2, \dots, n$ and $w = x, y, z$, and we consider two cases for $J_{i, j}$:
\begin{align}
  J_{i, j} &= \delta_{j, i+1}, \label{supp-arXiv-Heisenberg-1d-001} \\
  J_{i, j} &= \frac{1}{N}, \label{supp-arXiv-Heisenberg-full-001}
\end{align}
for $i, j = 1, 2, \dots, n$.
Note that Eq.~\eqref{supp-arXiv-Heisenberg-1d-001} is the 1-dimensional (1d) Heisenberg model and Eq.~\eqref{supp-arXiv-Heisenberg-1d-001} is the fully-connected (FC) Heisenberg model.
In the case of the Heisenberg circuits, we set $\Delta t = 0.1$ in Eq.~\eqref{supp-arXiv-gate-Mitarai-001-001}.
For QCL, we utilize the stochastic gradient descent method~\cite{Murphy001}.

To create $| \psi^\mathrm{in} (x_i) \rangle$ in Eq.~\eqref{supp-arXiv-encoding-QCL-001-001} and \eqref{supp-arXiv-encoding-UKM-001-001}, we need to fix $\hat{S} (x_i)$.
For $\hat{S} (x_i)$ for $i = 1, 2, \dots, N$, we use the amplitude encoding with $c_i^k = 0$ in Eq.~\eqref{supp-arXiv-def-tilde-x-001-001} for all $i$ and $k$, described in Sec.~\ref{supp-arXiv-sec-encoding-001}.
For $\ell (\cdot, \cdot)$ in Eqs.~\eqref{supp-arXiv-cost-function-QCL-001-001} and \eqref{supp-arXiv-cost-UKM-001-001}, we utilize the squared error function, Eq.~\eqref{supp-arXiv-squared-error-function-001-001}.
Furthermore, we set $Q = 1$ and $\xi_1 = 1$ in Eqs.~\eqref{supp-arXiv-prediction-QCL-001-001} and \eqref{supp-arXiv-prediction-UKM-001-001}, and consider two cases on $\theta_\mathrm{b}$ in Eqs.~\eqref{supp-arXiv-prediction-QCL-001-001} and \eqref{supp-arXiv-prediction-UKM-001-001}: the cases of fixing $\theta_\mathrm{b} = 0$ and estimating $\theta_\mathrm{b}$.

\subsection{Datasets} \label{supp-arXiv-sec-datasets-UKM-001-001}

In this section, the following six datasets are used to compare the performance of QCL and the UKM: the iris dataset, the cancer dataset, the wine dataset, the sonar dataset, the semeion dataset in the UCI repository~\cite{Dua001}, and the MNIST dataset~\cite{LeCun001}.

The original MNIST dataset has $28 \times 28$ dimensions.
We create the MNIST256 dataset by reducing its dimensions to $16 \times 16$ using coarse-graining.

We consider the binary classification problem.
The cancer dataset has two labels, (0) ``B" and (1) ``M", and the wine dataset has two labels, (0) ``R" and (1) ``M".

On the other hand, other datasets have more labels.
The iris dataset has three labels: (0) Iris-setosa, (1) Iris-versicolor, and (2) Iris-virginica.
When we deal with the classification problem between (0) Iris-setosa and (1) Iris-versicolor, we write the dataset by Iris ($0$ or $1$).
When we deal with the classification problem between (0) Iris-setosa and the others, we write it by Iris ($0$ or non-$0$).
Similarly, when we deal with the classification problem between (1) Iris-setosa and the others, we write the dataset by Iris ($1$ or non-$1$).

The wine dataset has three labels: (0) class 1, (1) class 2, and (3) class 3.
We consider the classification problem between (0) class 1 and others, we write it as the wine dataset ($0$ or non-$0$).

The semeion dataset and the MNIST256 dataset has ten labels from 0 to 9.
We write it as ($0$ or $1$) when we consider the classification problem between $0$ and $1$, and we write it as ($0$ or non-$0$) when we consider the classification problem between $0$ and the others.

At the end of this subsection, the numbers of datapoints $N$ and dimensions $M$ of the datasets are shown in Table~\ref{supp-arXiv-table-datasets-001}.
\begin{table}[htb]
  \begin{tabular}{c|cccc}
    \hline \hline
    ID & $N$ & $M$ & $n$ \\
    \hline
    Iris ($0$ or $1$)         &  100 &   4 & 2 \\
    Iris ($0$ or non-$0$)     &  150 &   4 & 2 \\
    Iris ($1$ or non-$1$)     &  150 &   4 & 2 \\
    Cancer ($0$ or $1$)       &  569 &  30 & 5 \\
    Sonar ($0$ or $1$)        &  208 &  60 & 6 \\
    Wine ($0$ or non-$0$)     &  178 &  14 & 4 \\
    Semeion ($0$ or $1$)      &  323 & 256 & 8 \\
    Semeion ($0$ or non-$0$)  & 1593 & 256 & 8 \\
    MNIST256 ($0$ or $1$)     &  569 & 256 & 8 \\
    MNIST256 ($0$ or non-$0$) & 2766 & 256 & 8 \\
    \hline \hline
  \end{tabular}
\caption{Numbers of datapoints $N$ dimensions $M$ of the datasets and the number of qubits $n$ required for amplitude encoding. Note that $n = \lceil \log_2 M \rceil$.}
\label{supp-arXiv-table-datasets-001}
\end{table}

\subsection{Iris dataset ($0$ or $1$)}

We here show the numerical result for the iris dataset ($0$ or $1$).
For the UKM, we put $r = 0.010$ and set $K = 30$ and $K' = 10$ in Algo.~\ref{supp-arXiv-quantum-kernel-method-002-001}.
For QCL, we run iterations $300$ times.

\begin{figure*}[htb]
\centering
\includegraphics[scale=0.25]{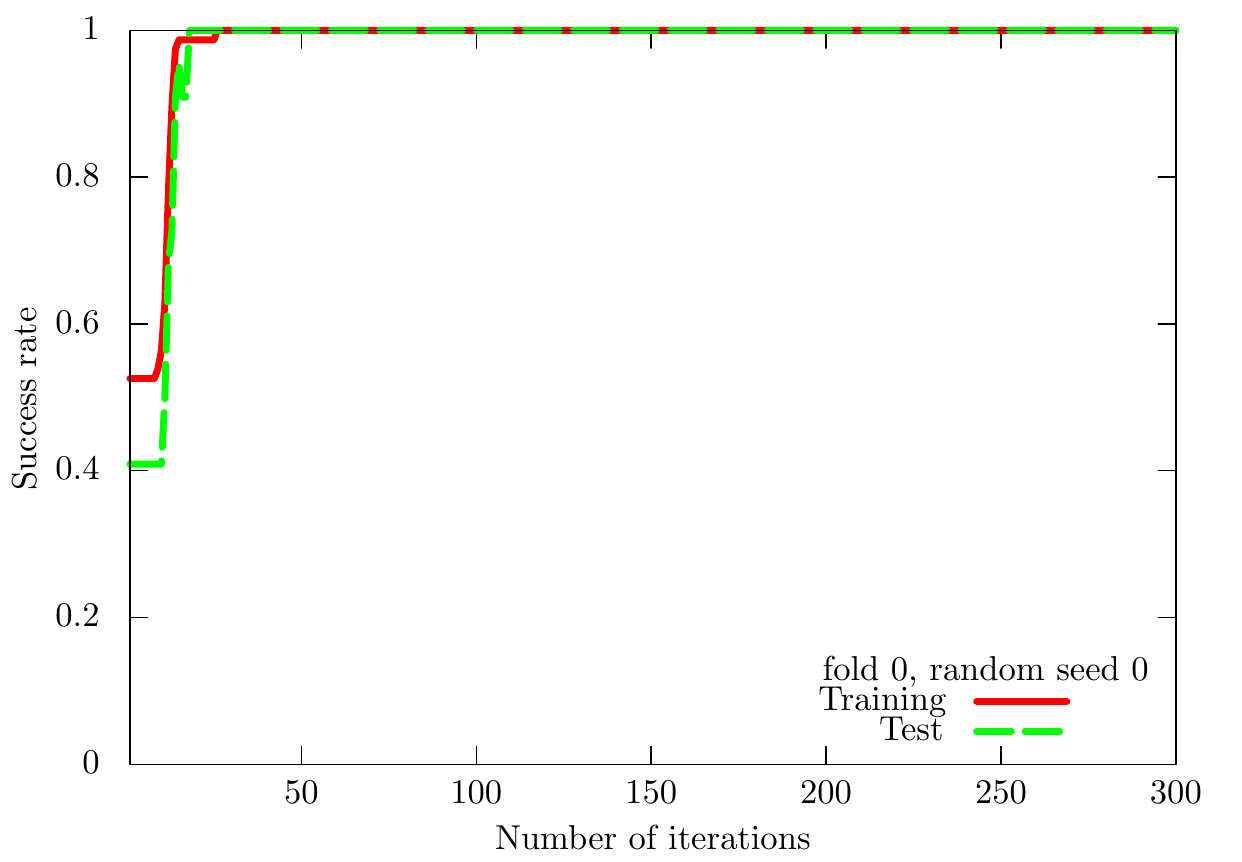}
\includegraphics[scale=0.25]{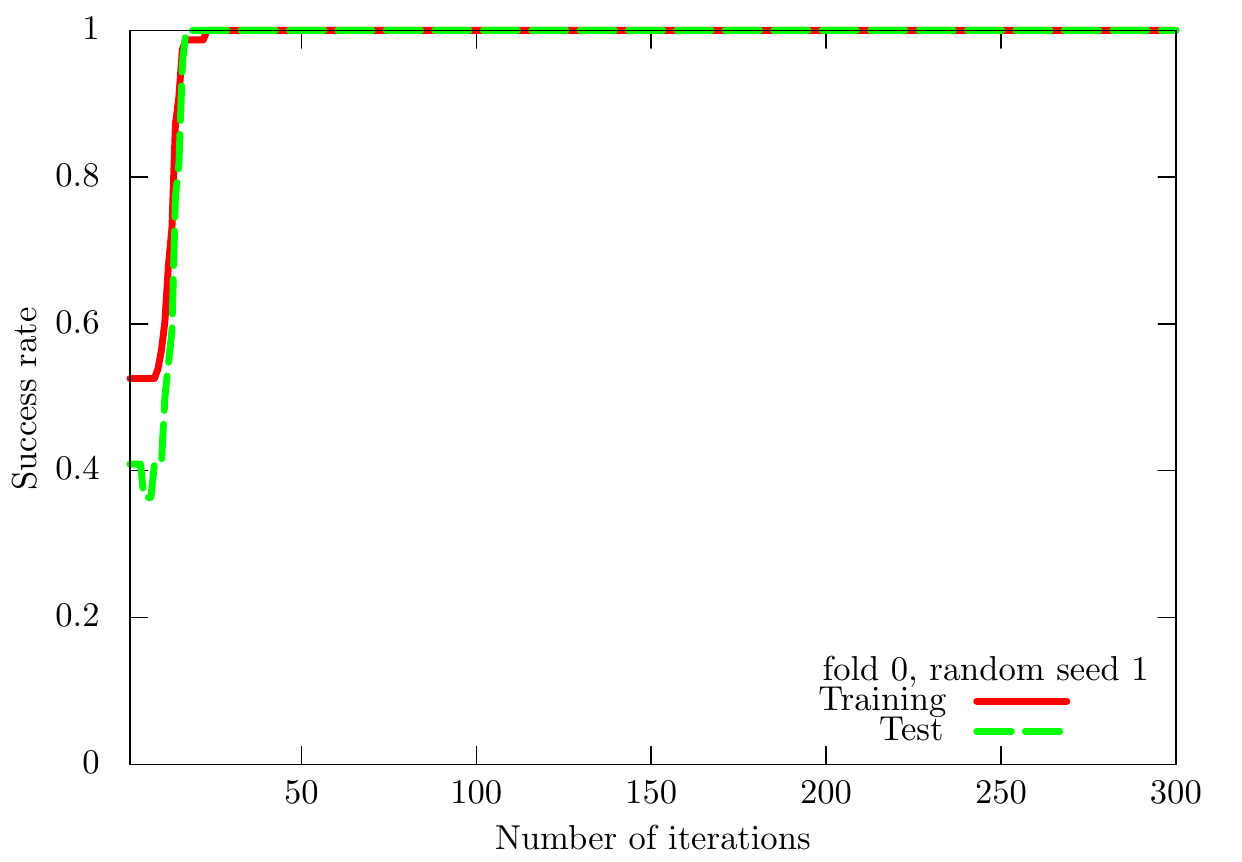}
\includegraphics[scale=0.25]{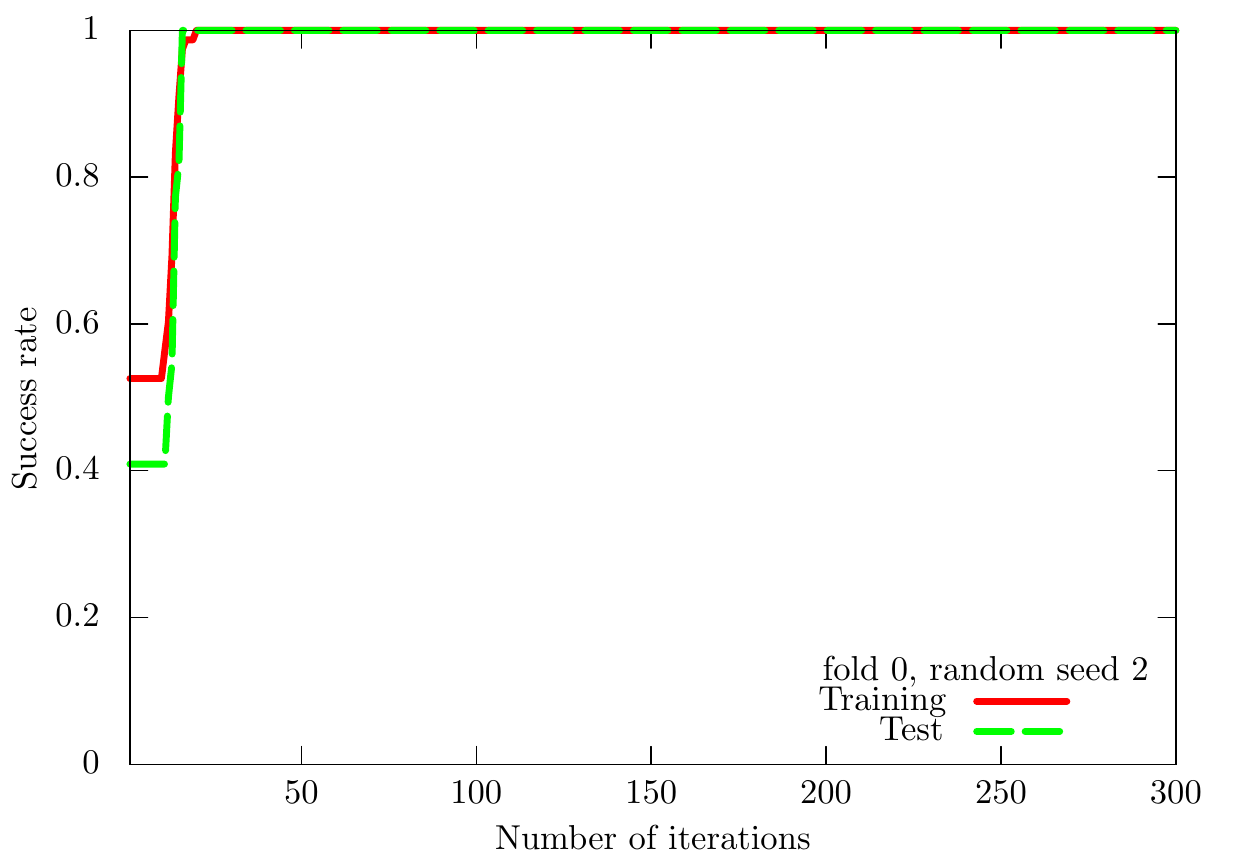}
\includegraphics[scale=0.25]{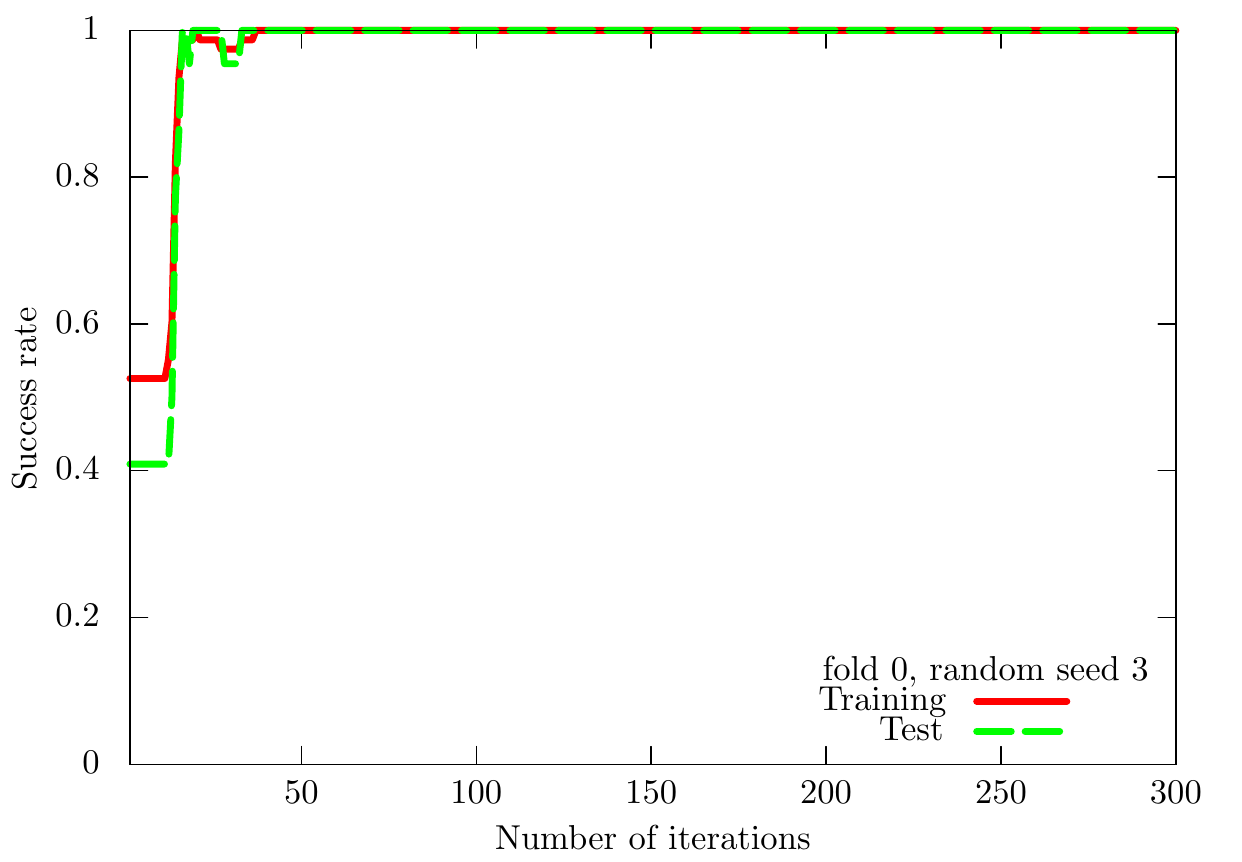}
\includegraphics[scale=0.25]{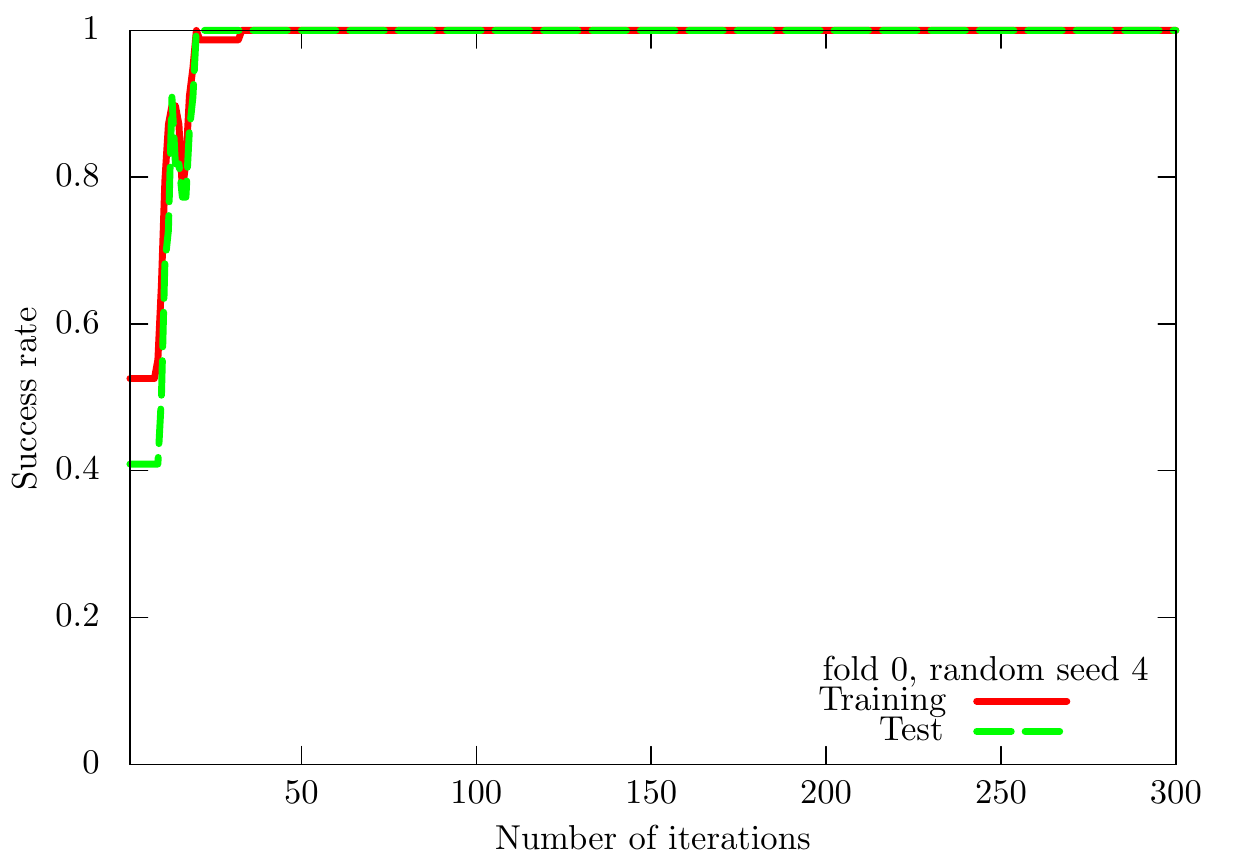}
\includegraphics[scale=0.25]{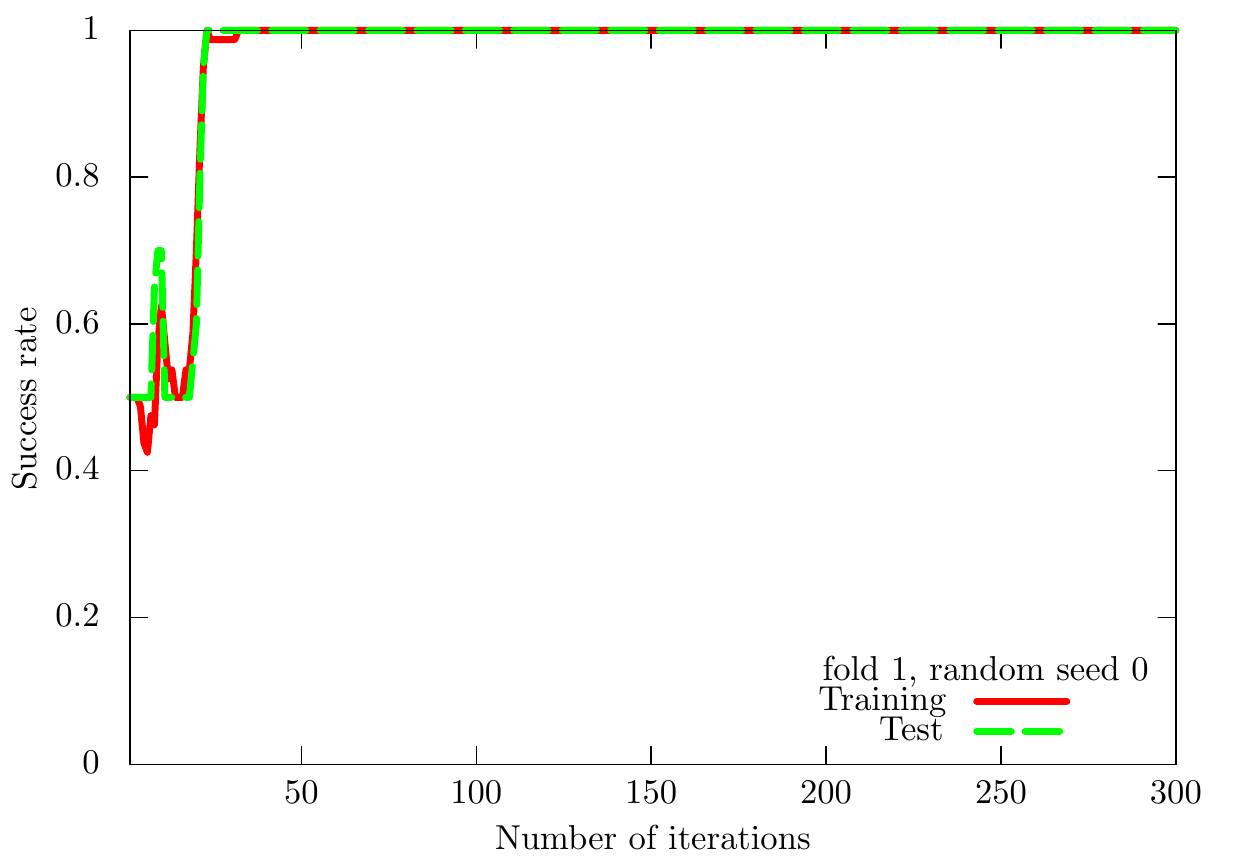}
\includegraphics[scale=0.25]{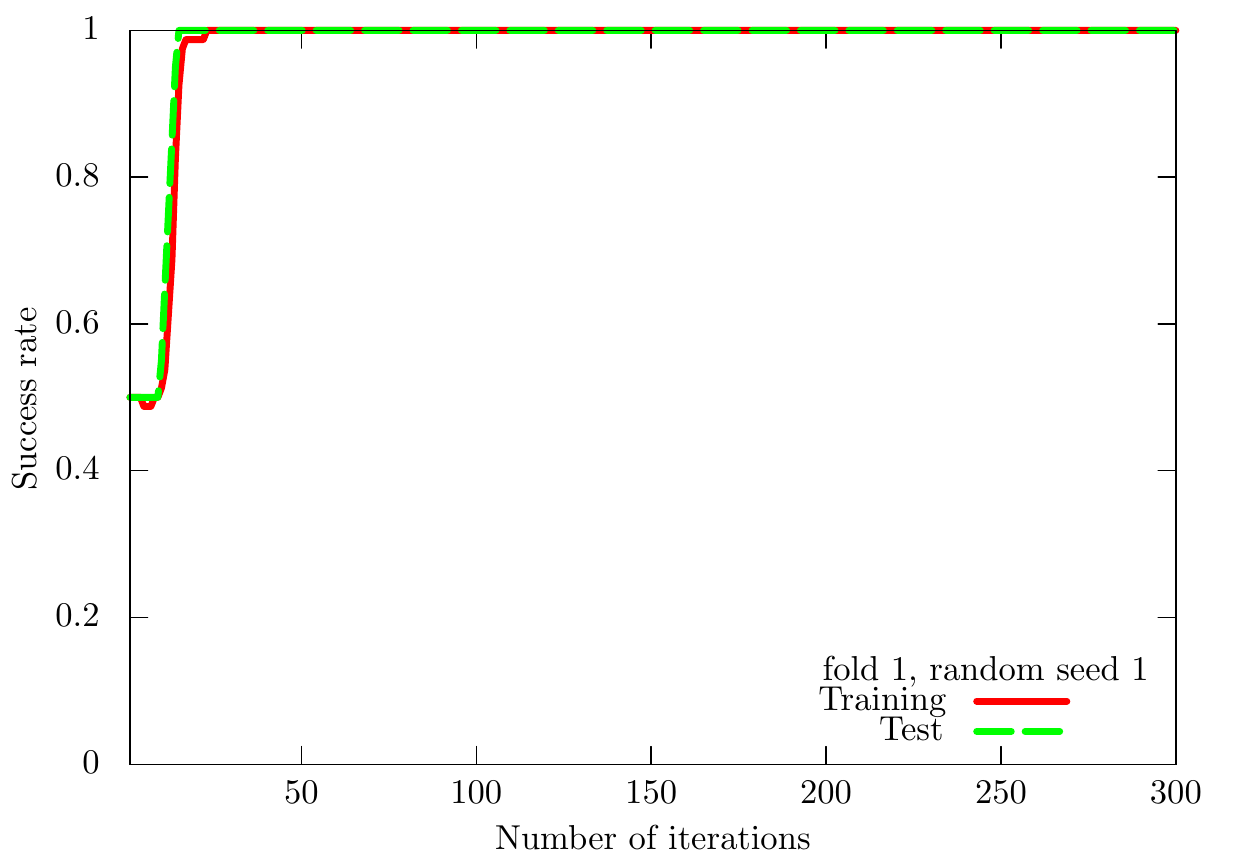}
\includegraphics[scale=0.25]{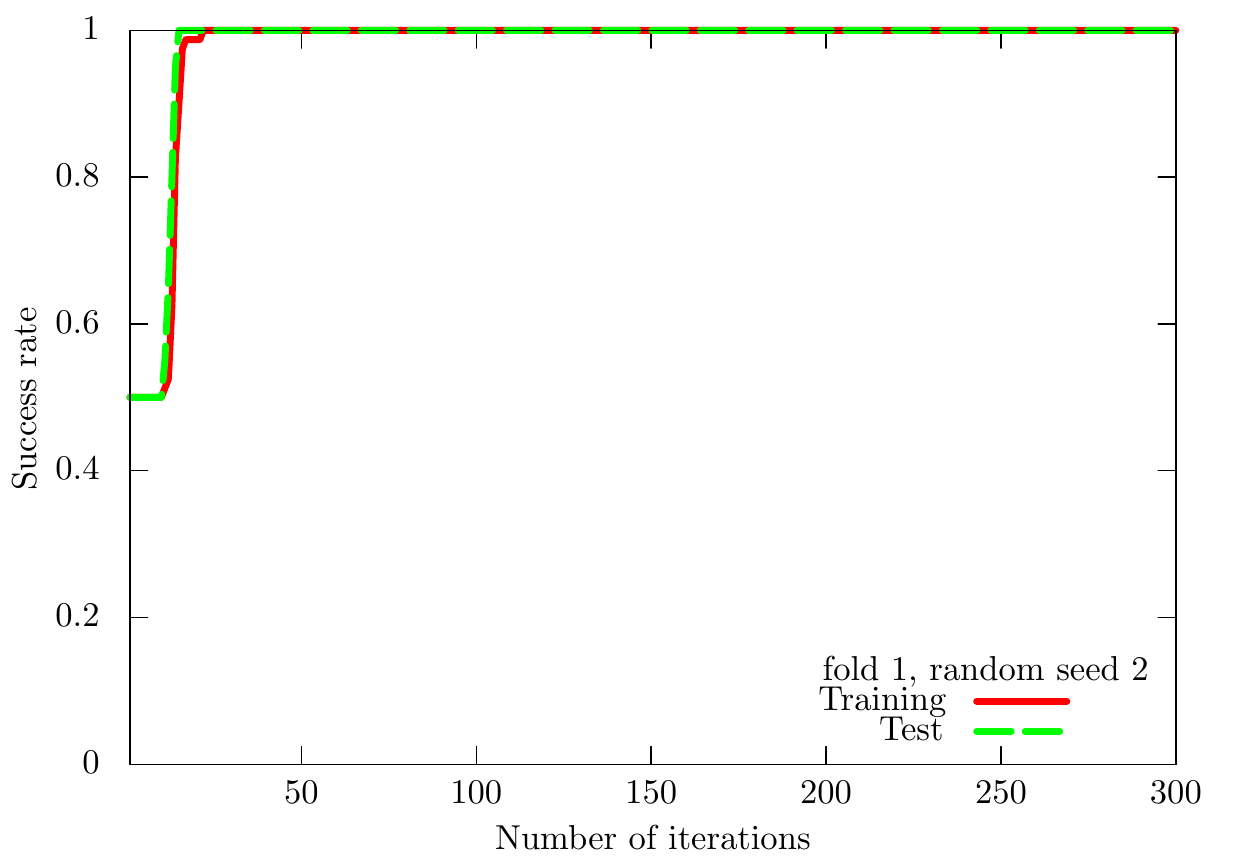}
\includegraphics[scale=0.25]{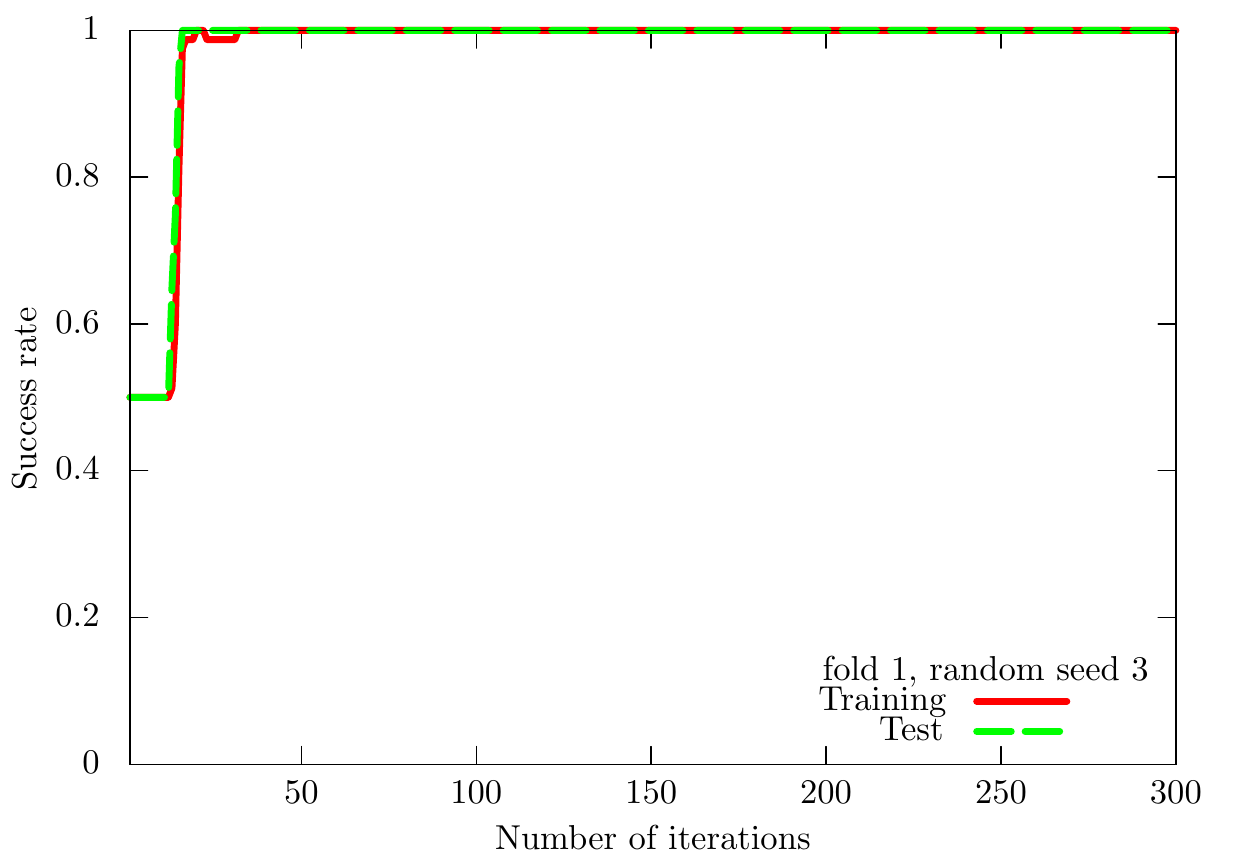}
\includegraphics[scale=0.25]{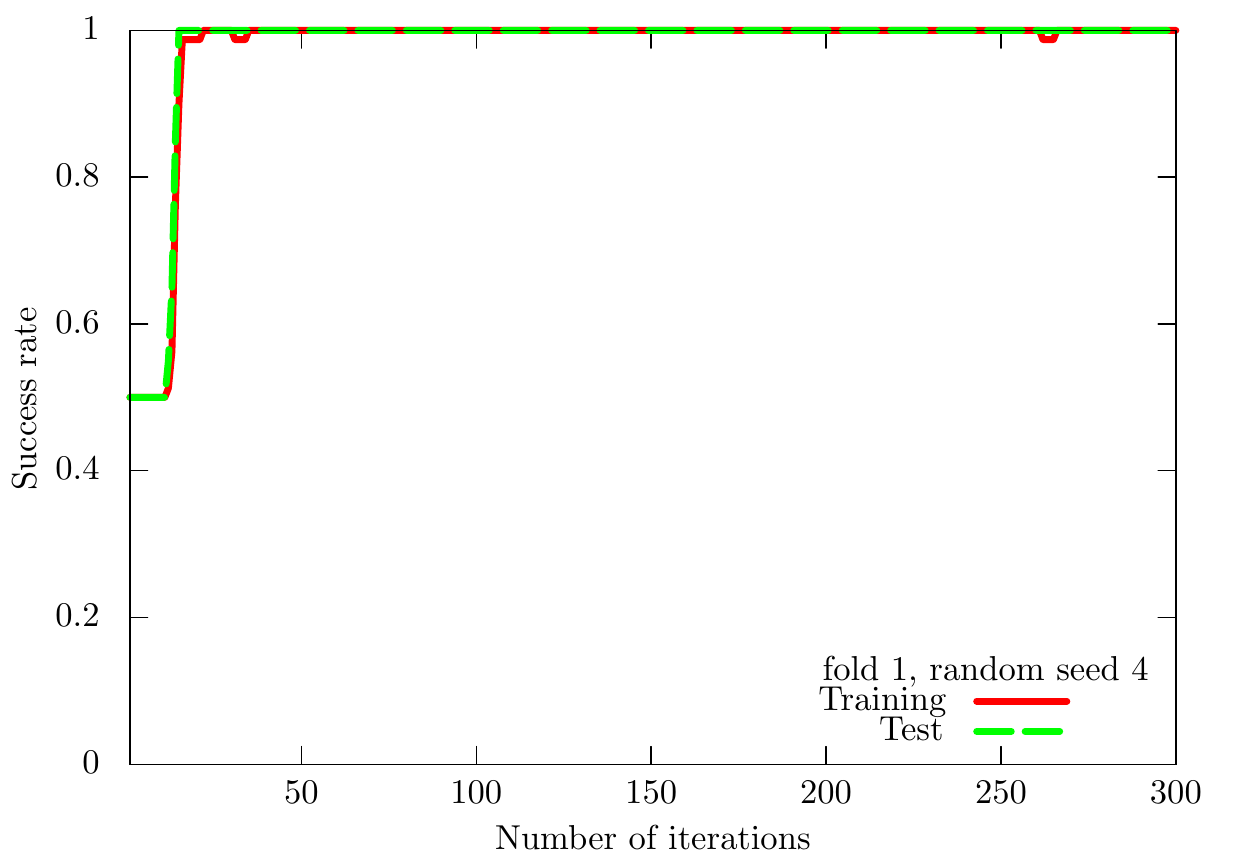}
\includegraphics[scale=0.25]{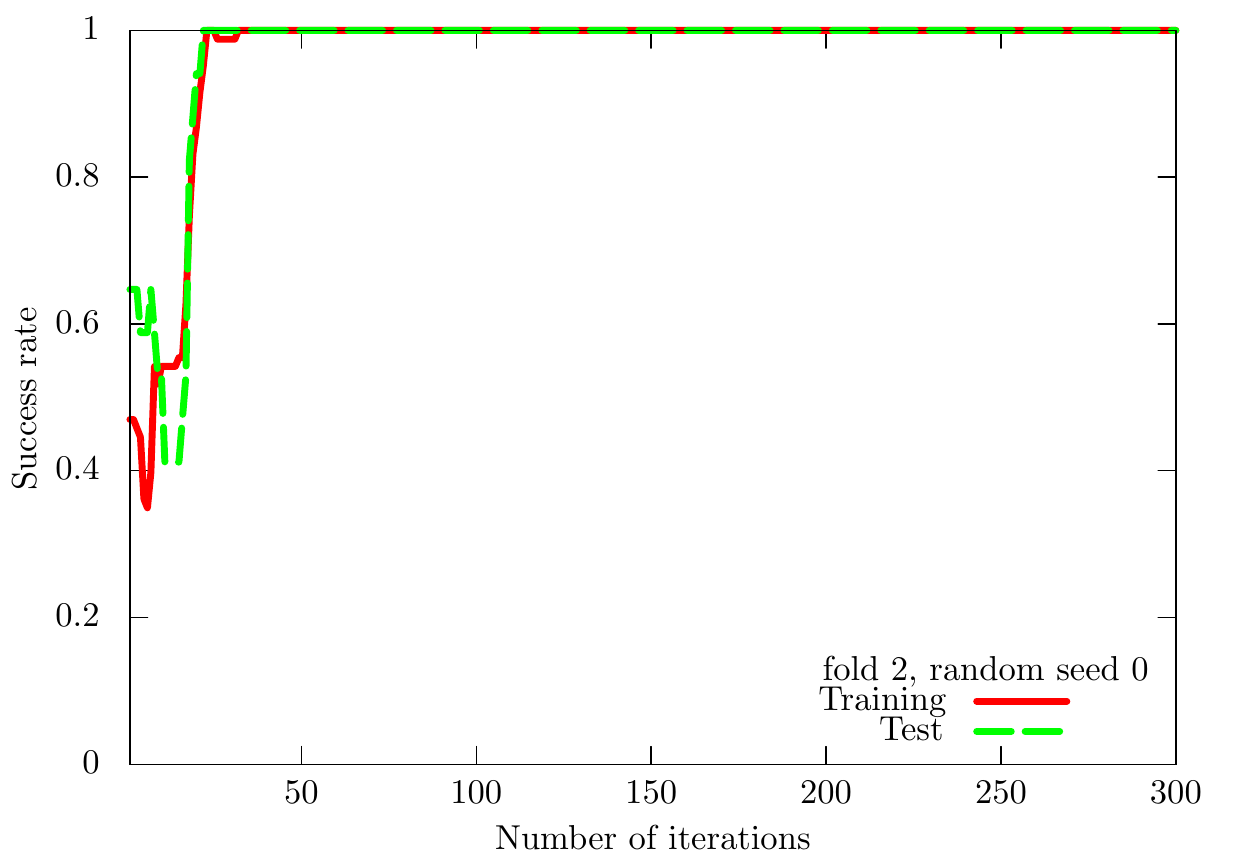}
\includegraphics[scale=0.25]{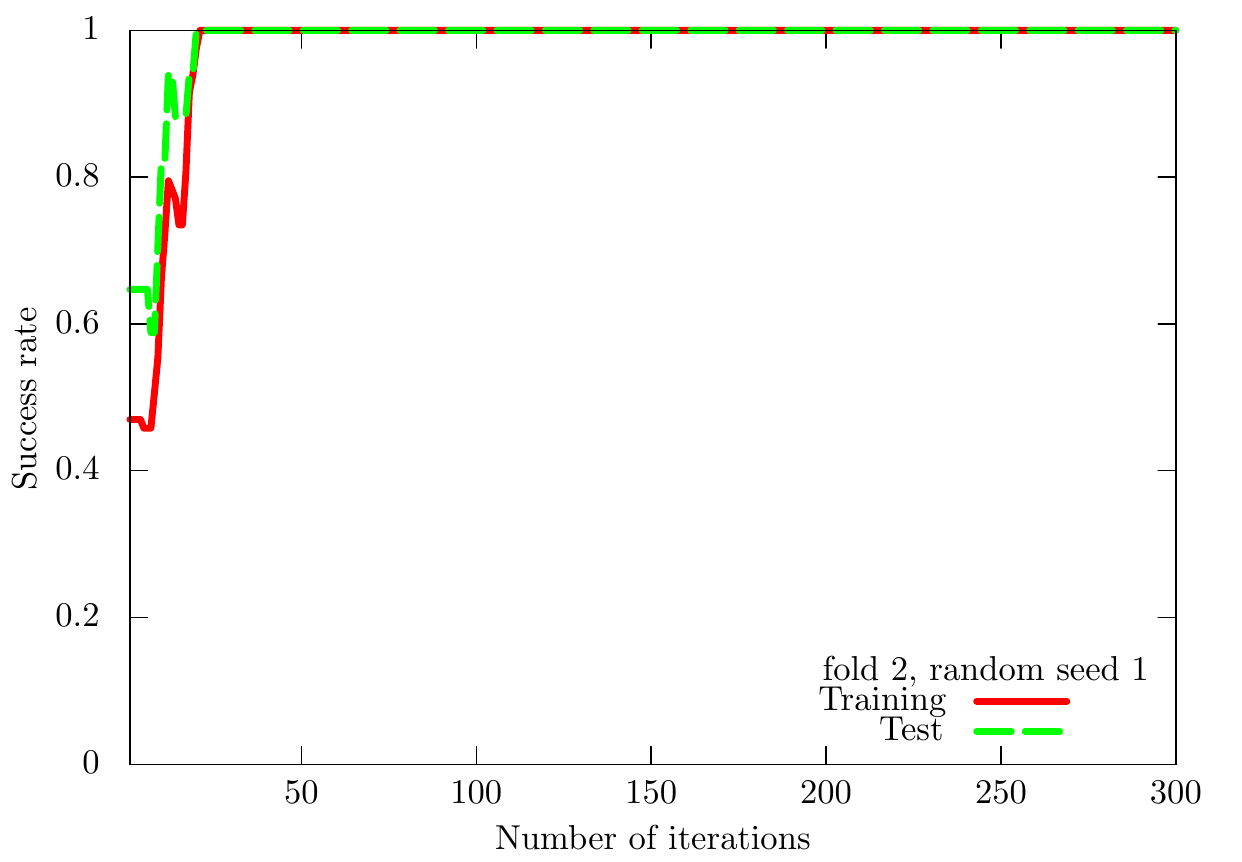}
\includegraphics[scale=0.25]{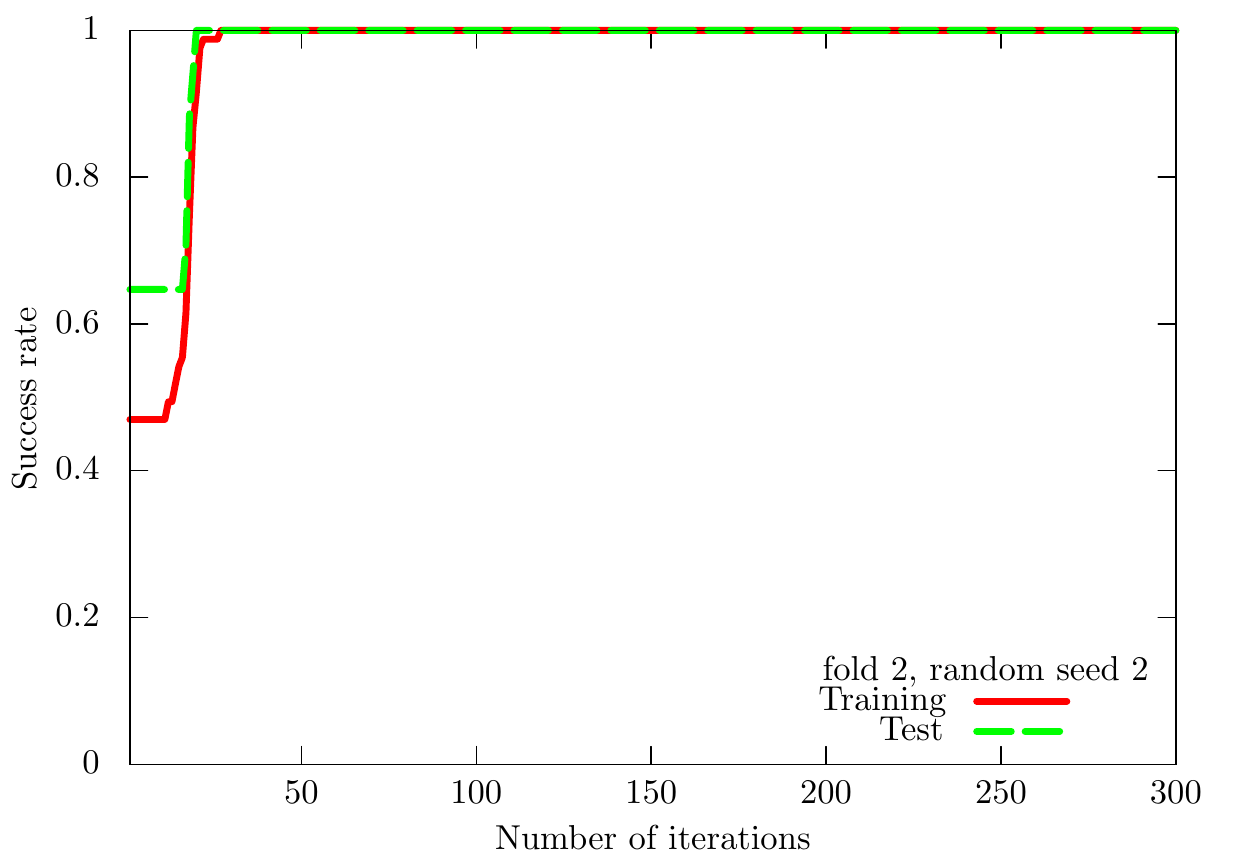}
\includegraphics[scale=0.25]{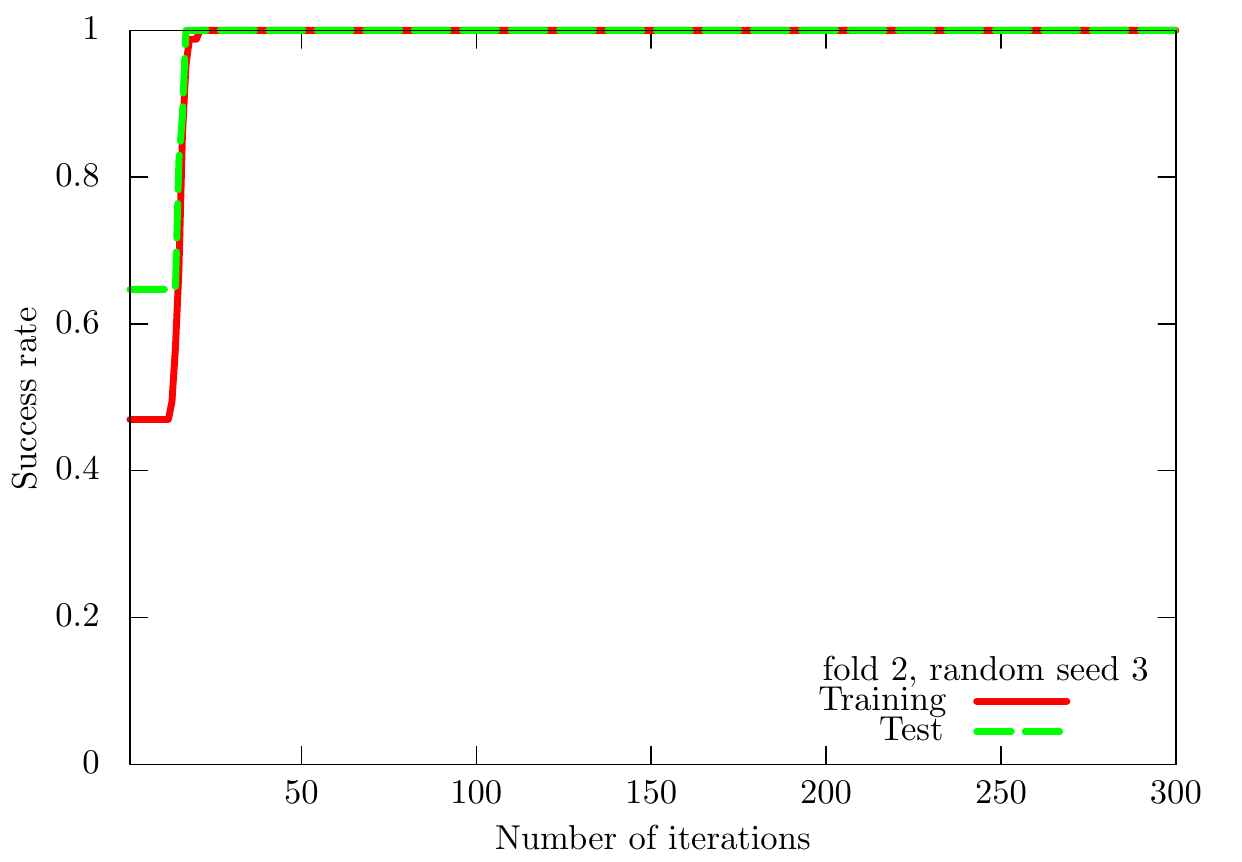}
\includegraphics[scale=0.25]{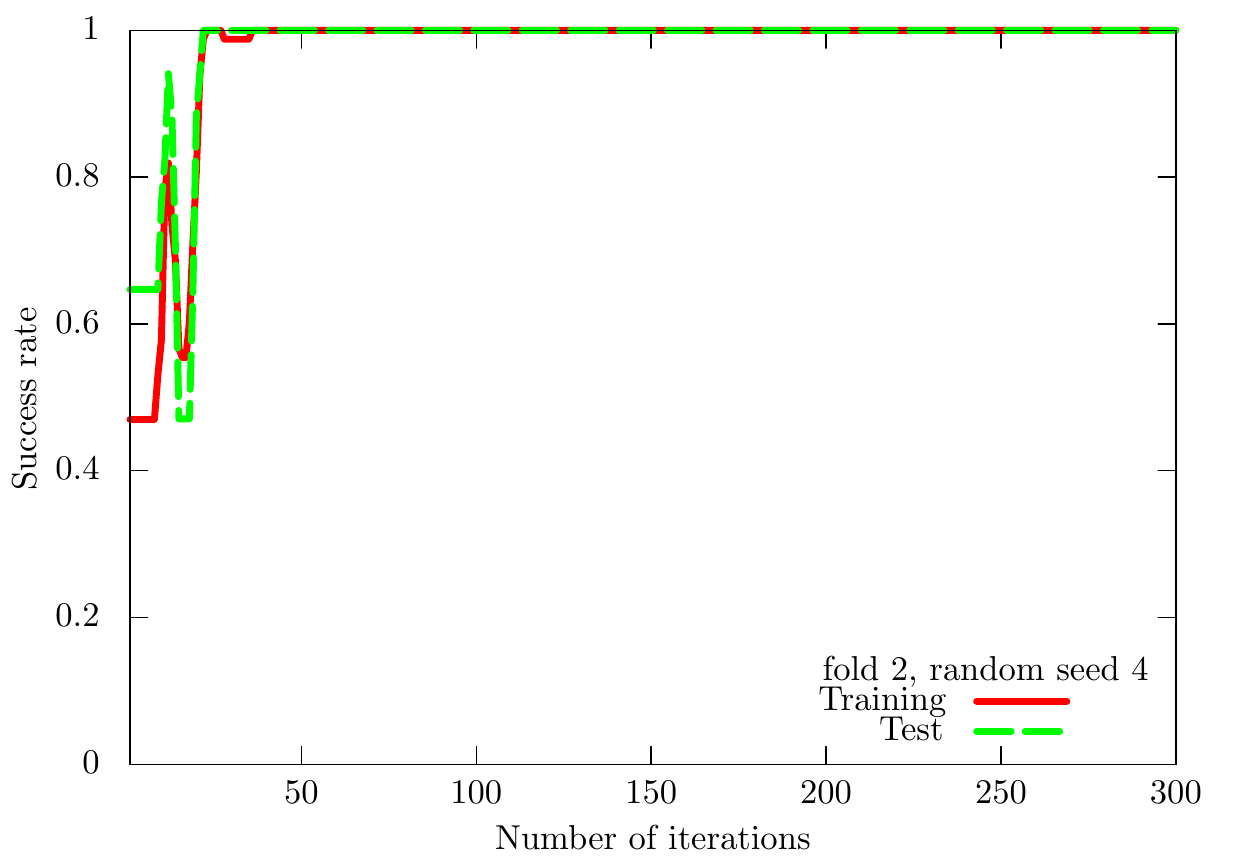}
\includegraphics[scale=0.25]{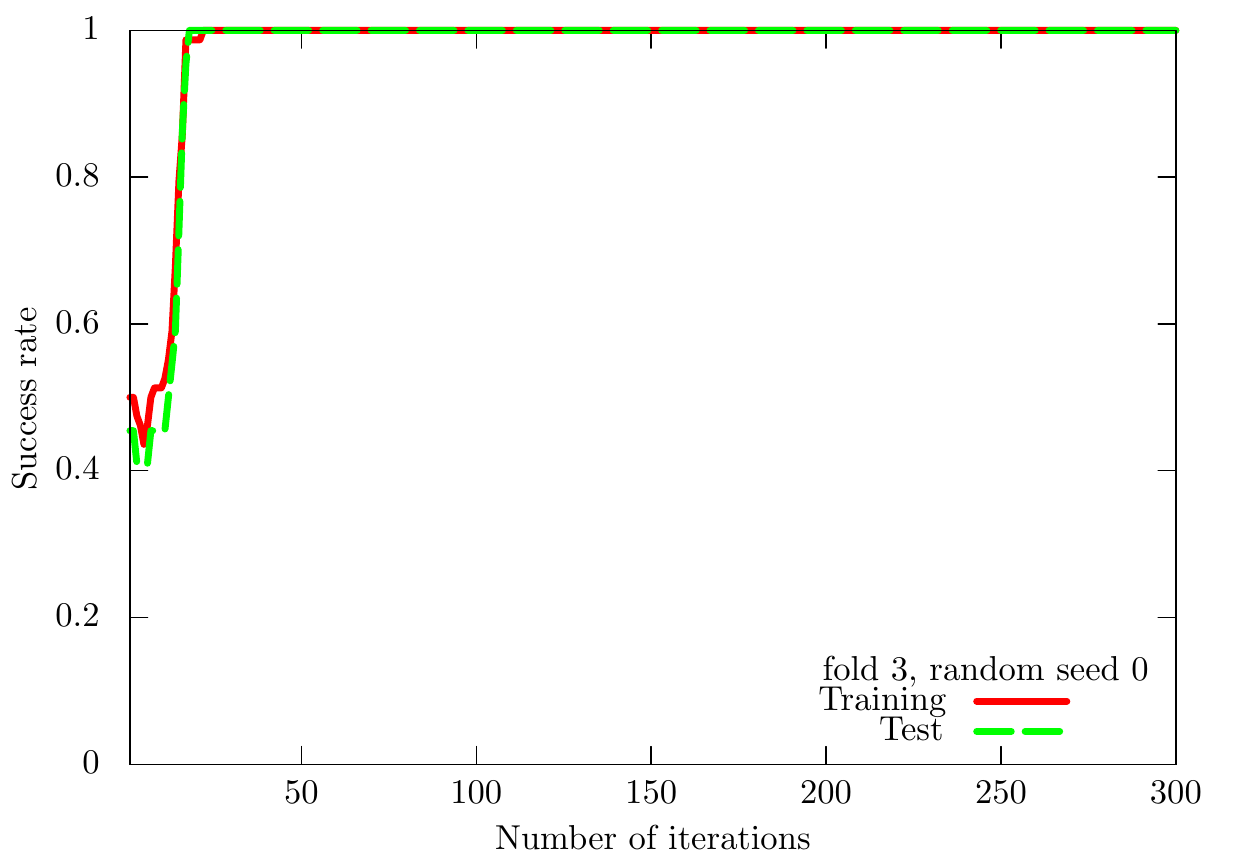}
\includegraphics[scale=0.25]{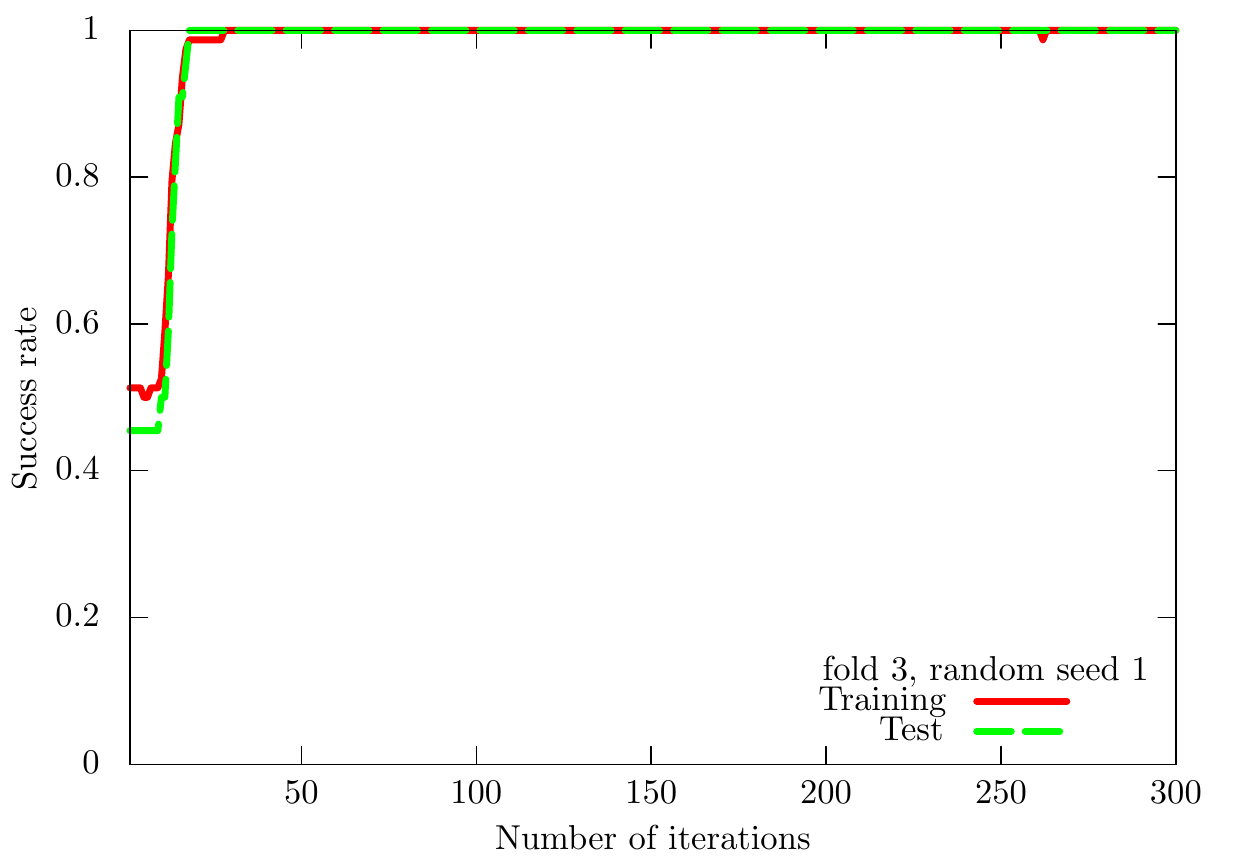}
\includegraphics[scale=0.25]{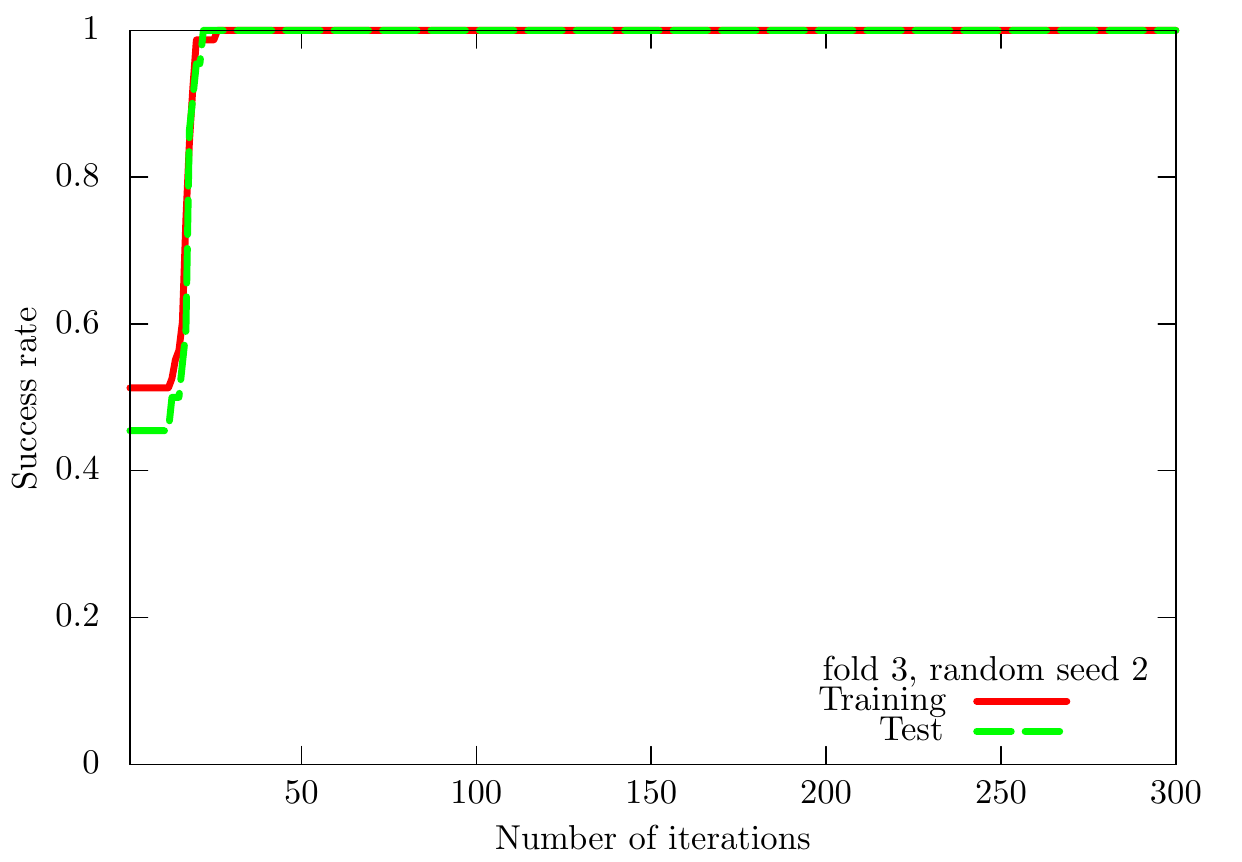}
\includegraphics[scale=0.25]{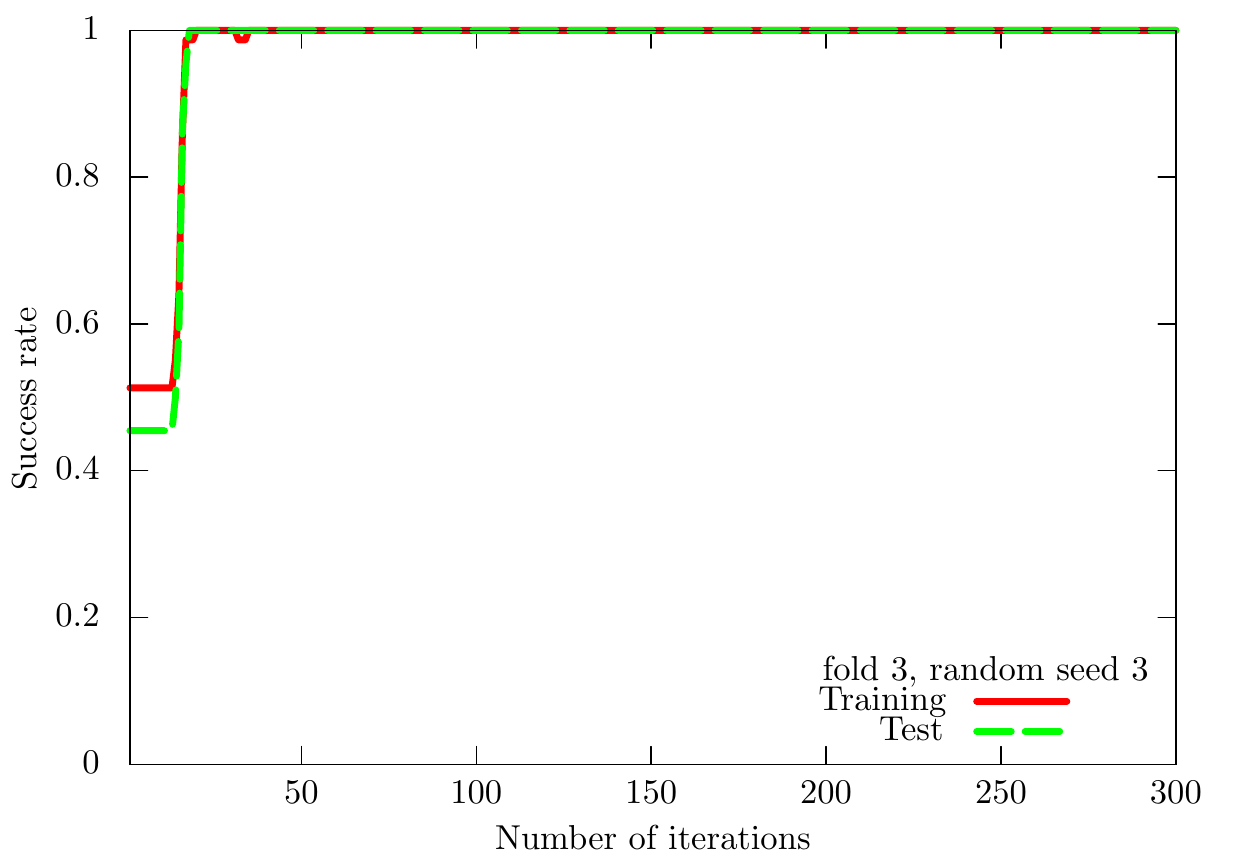}
\includegraphics[scale=0.25]{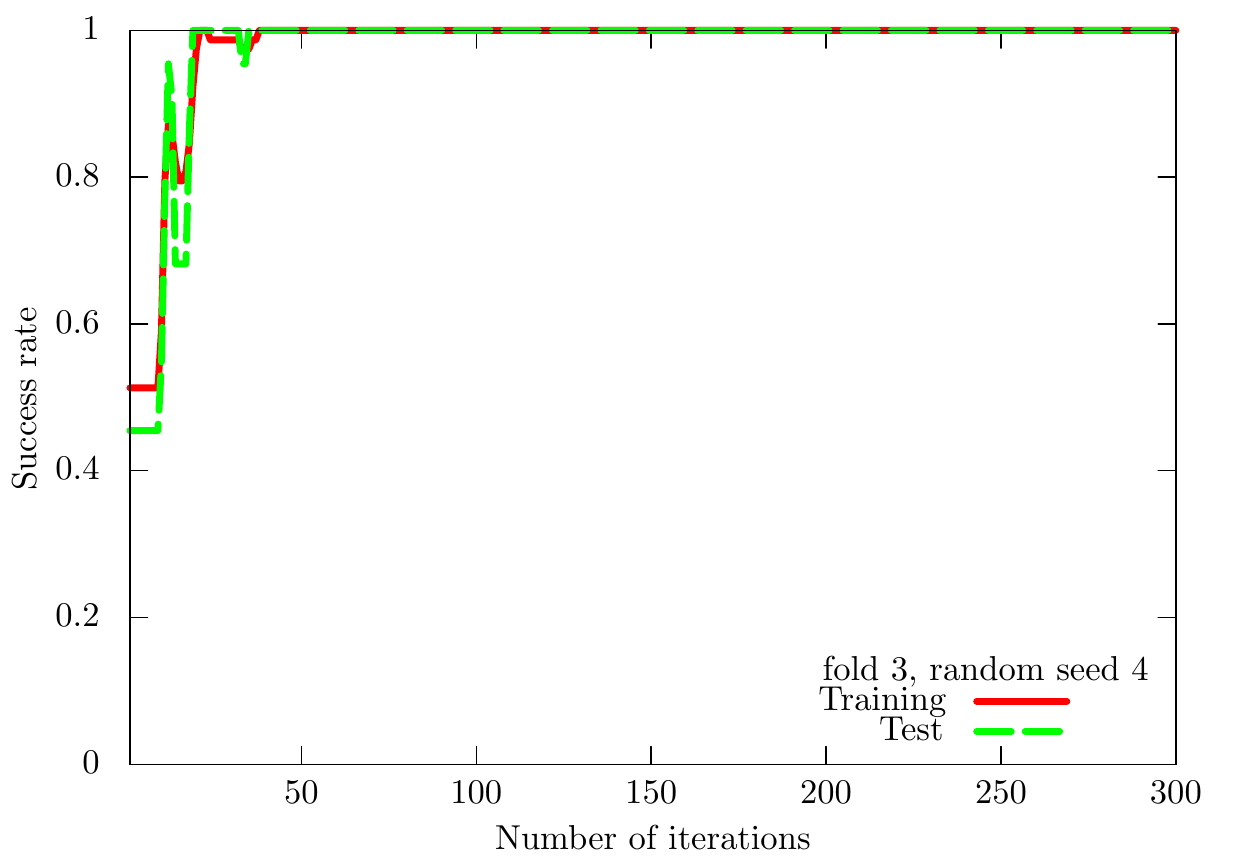}
\includegraphics[scale=0.25]{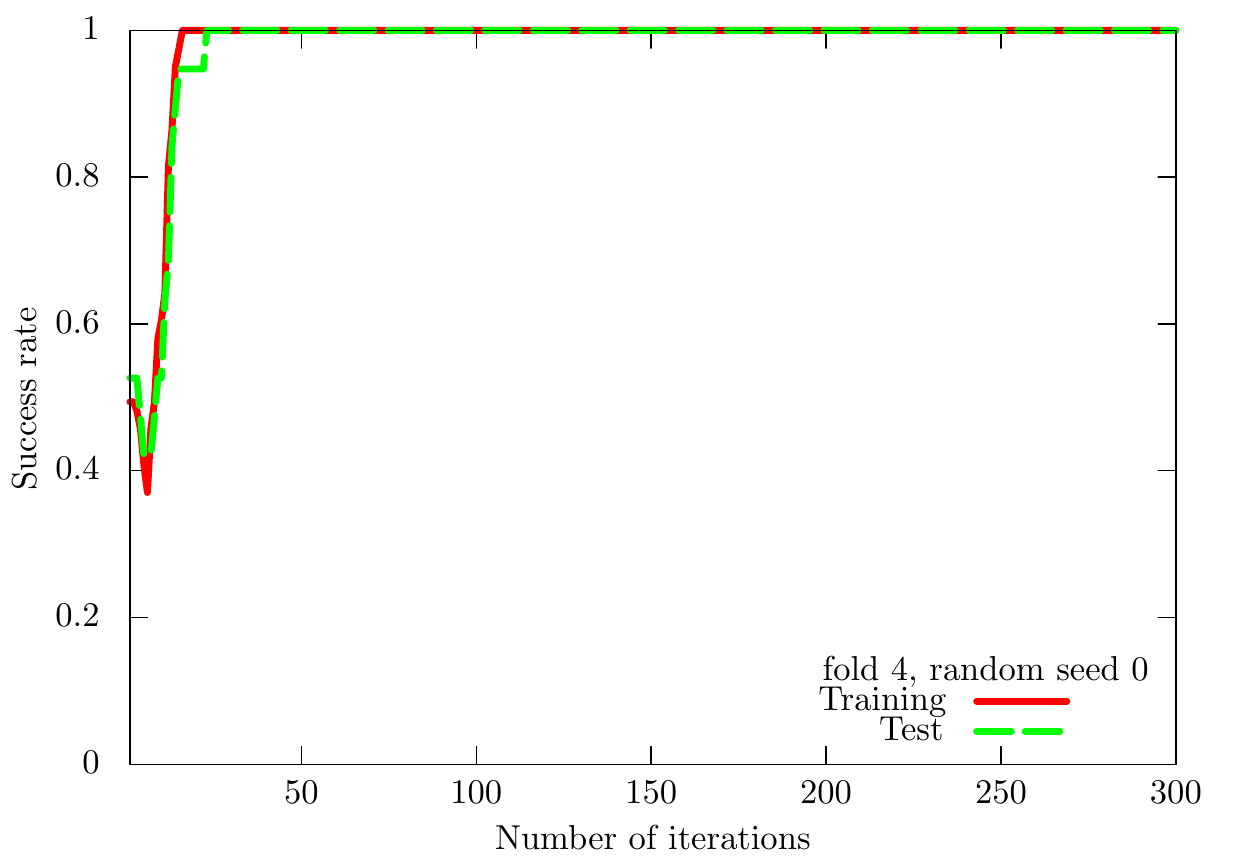}
\includegraphics[scale=0.25]{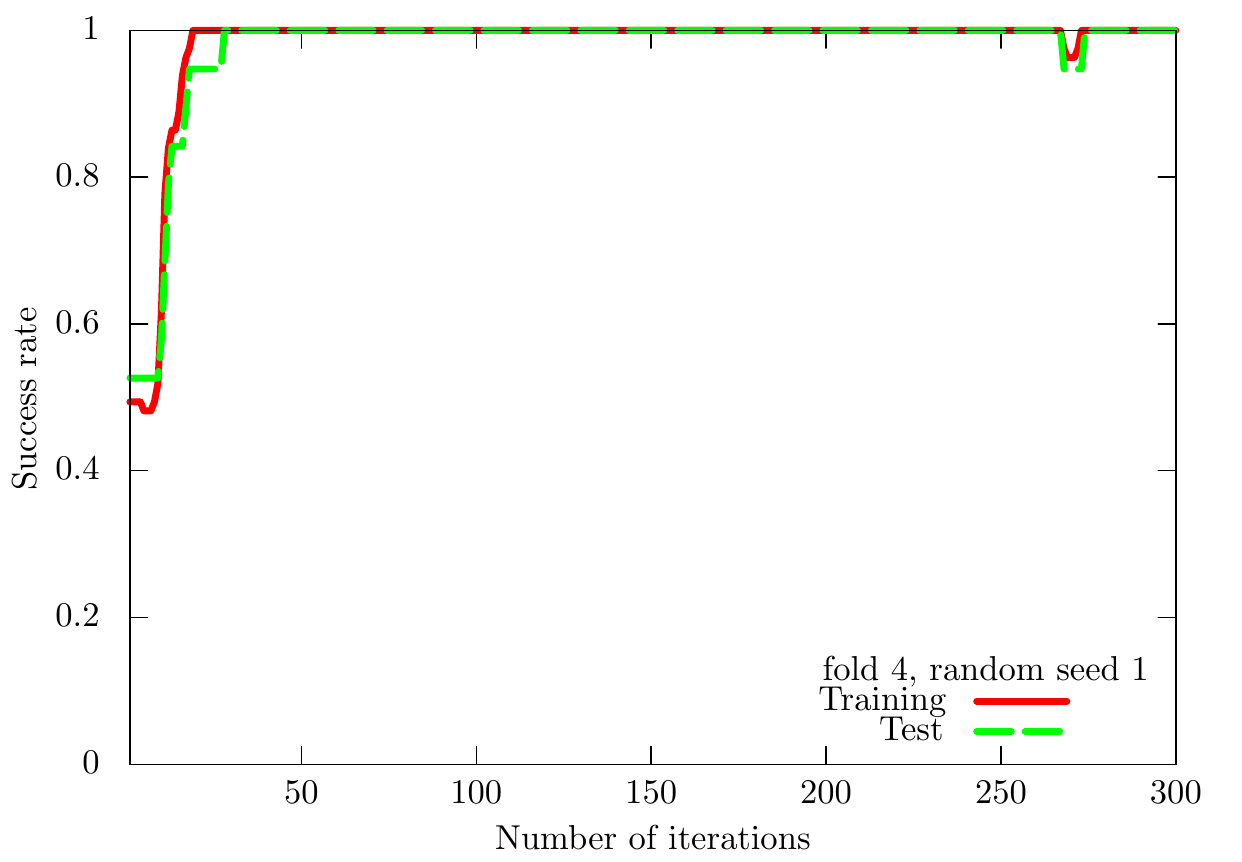}
\includegraphics[scale=0.25]{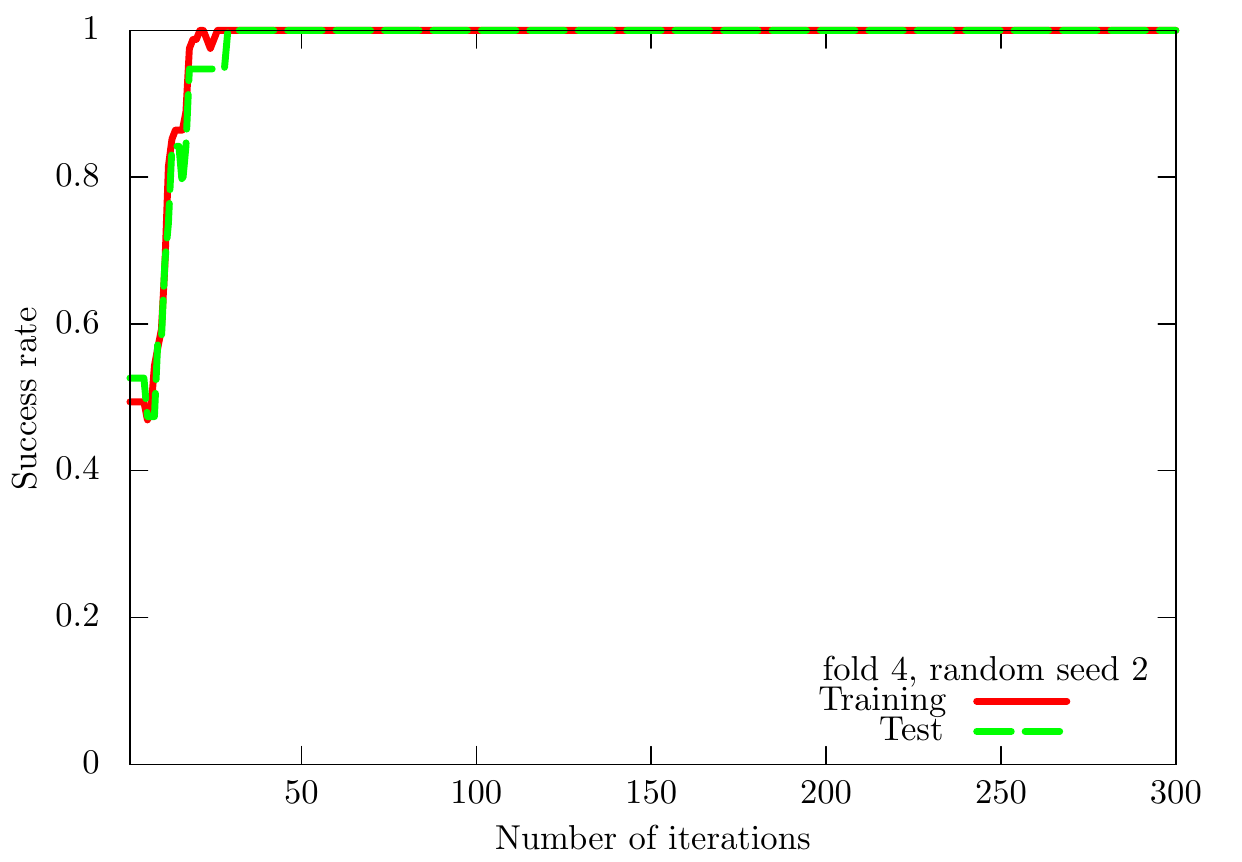}
\includegraphics[scale=0.25]{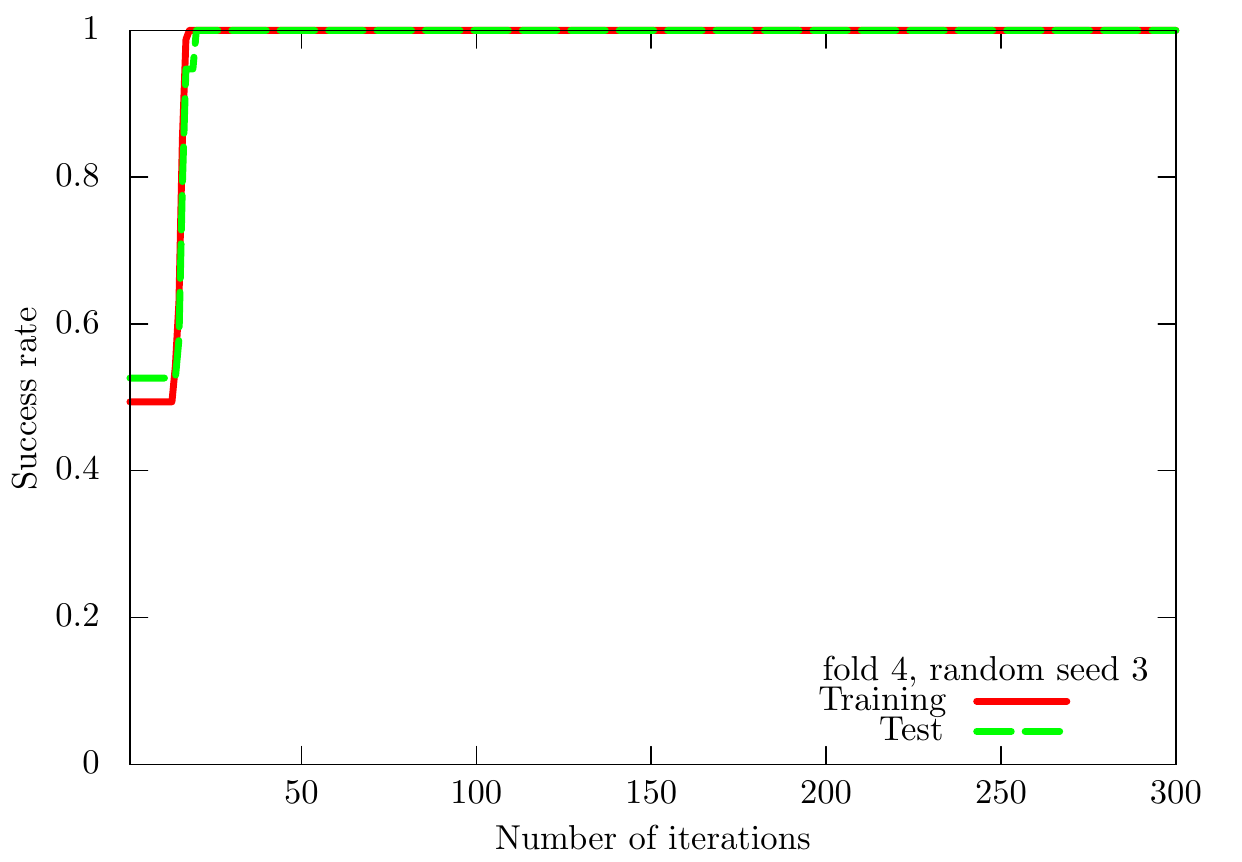}
\includegraphics[scale=0.25]{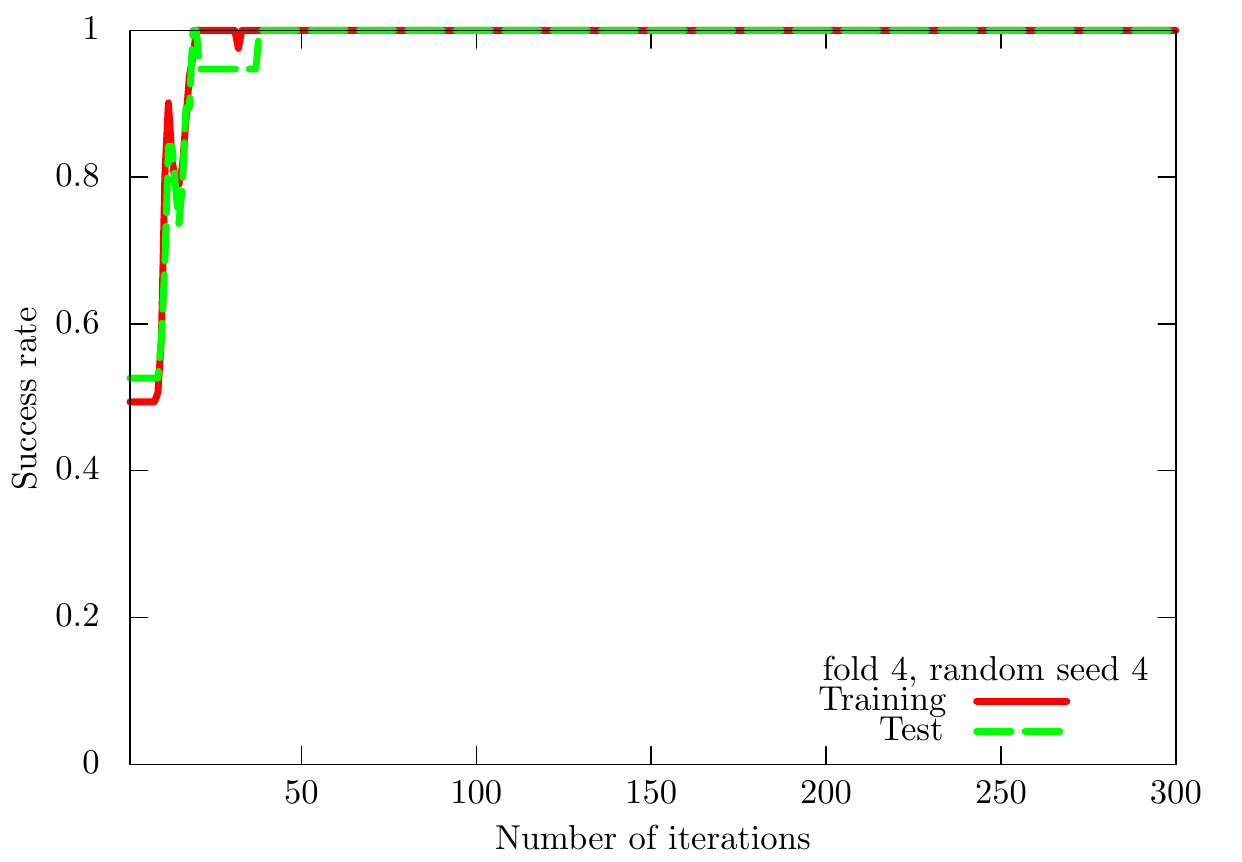}
\caption{Results of QCL on the $5$-fold datasets with $5$ different random seeds for the iris dataset ($0$ or $1$). We use the CNOT-based circuit and set $\theta_\mathrm{bias} = 0$. The number of layers $L$ is set to $5$.}
\label{supp-arXiv-numerical-result-raw-data-fold-001-rand-001-QCL-UCI-iris-0-1}
\end{figure*}

\begin{figure*}[htb]
\centering
\includegraphics[scale=0.25]{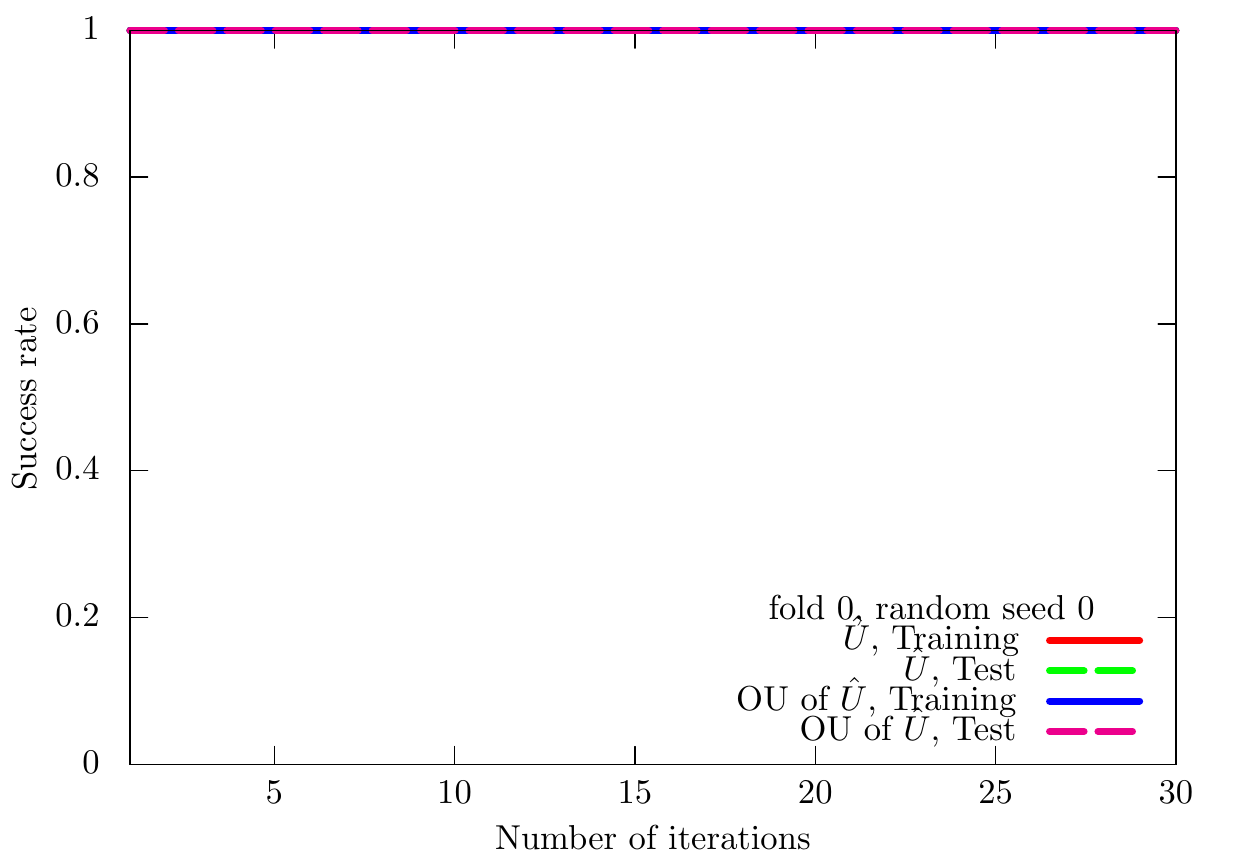}
\includegraphics[scale=0.25]{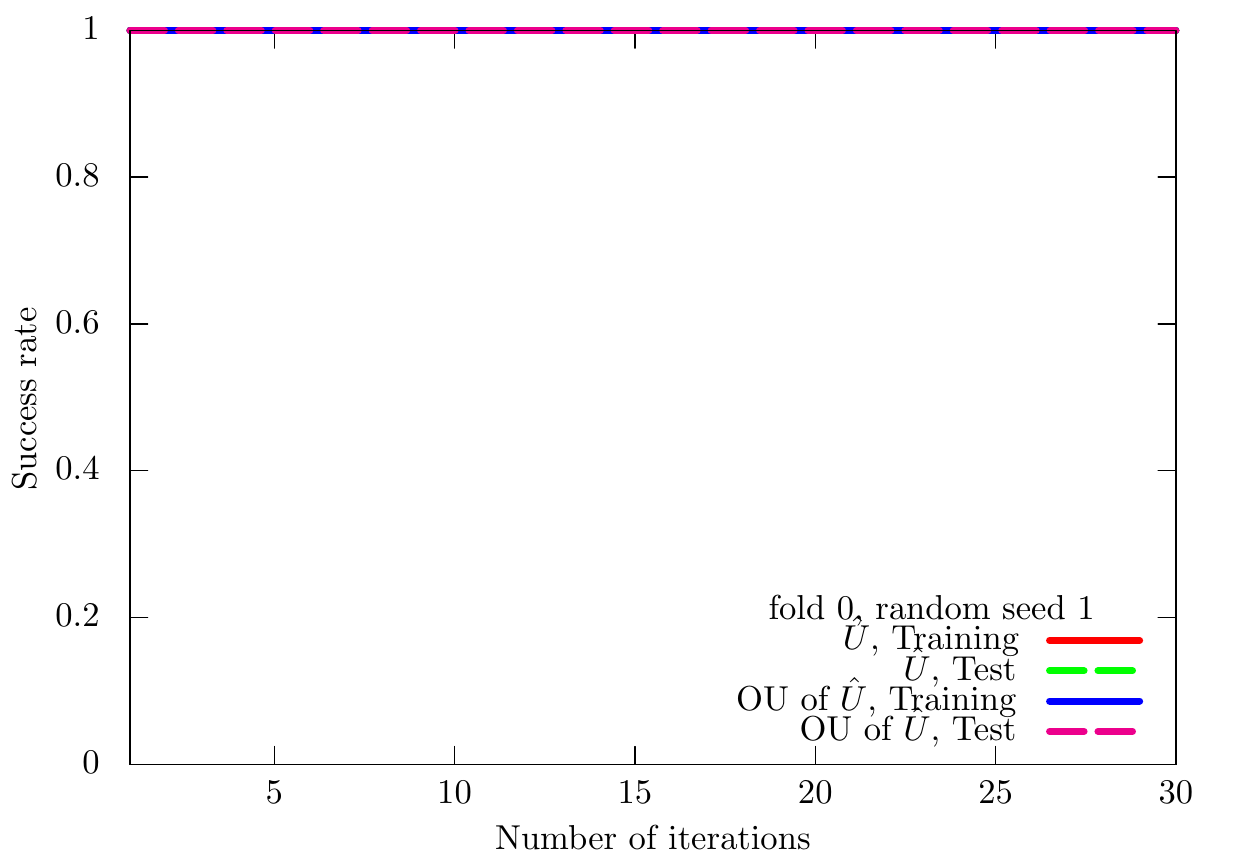}
\includegraphics[scale=0.25]{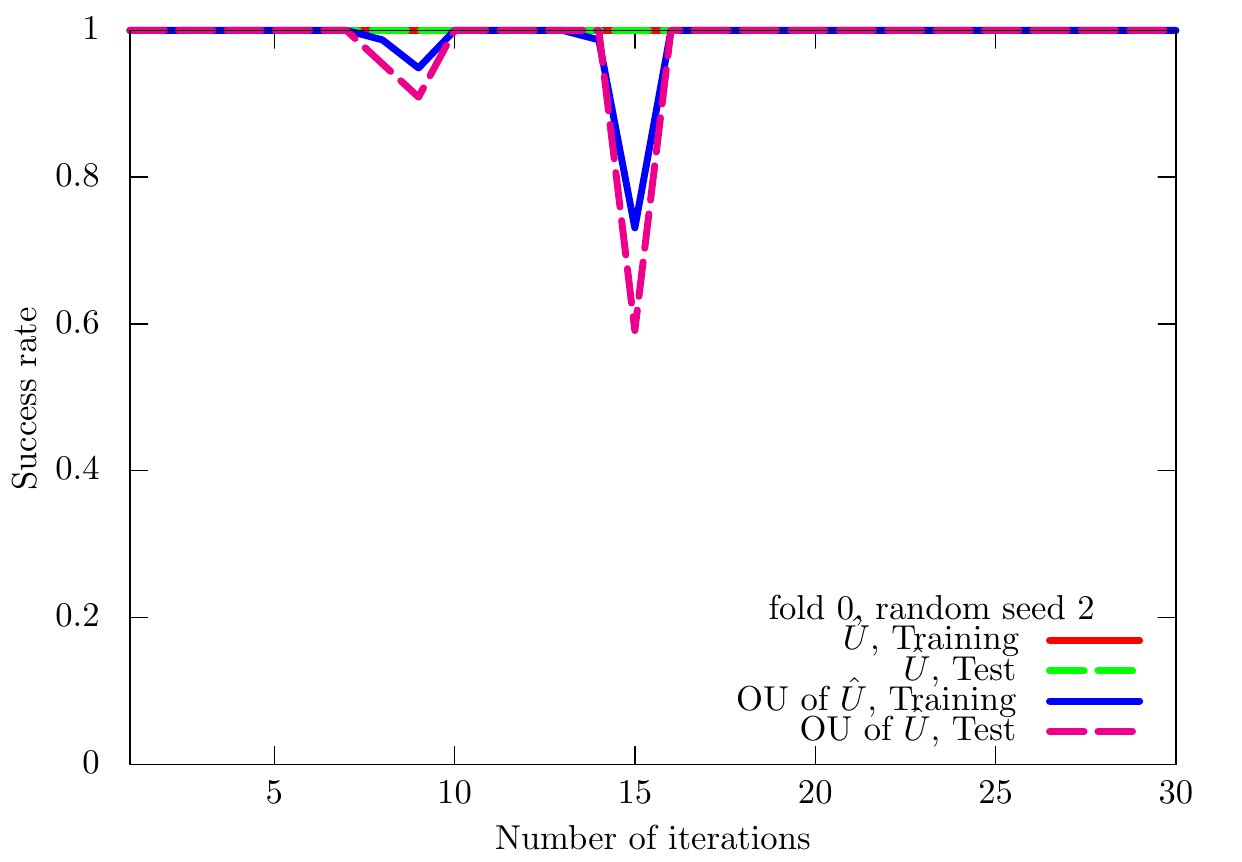}
\includegraphics[scale=0.25]{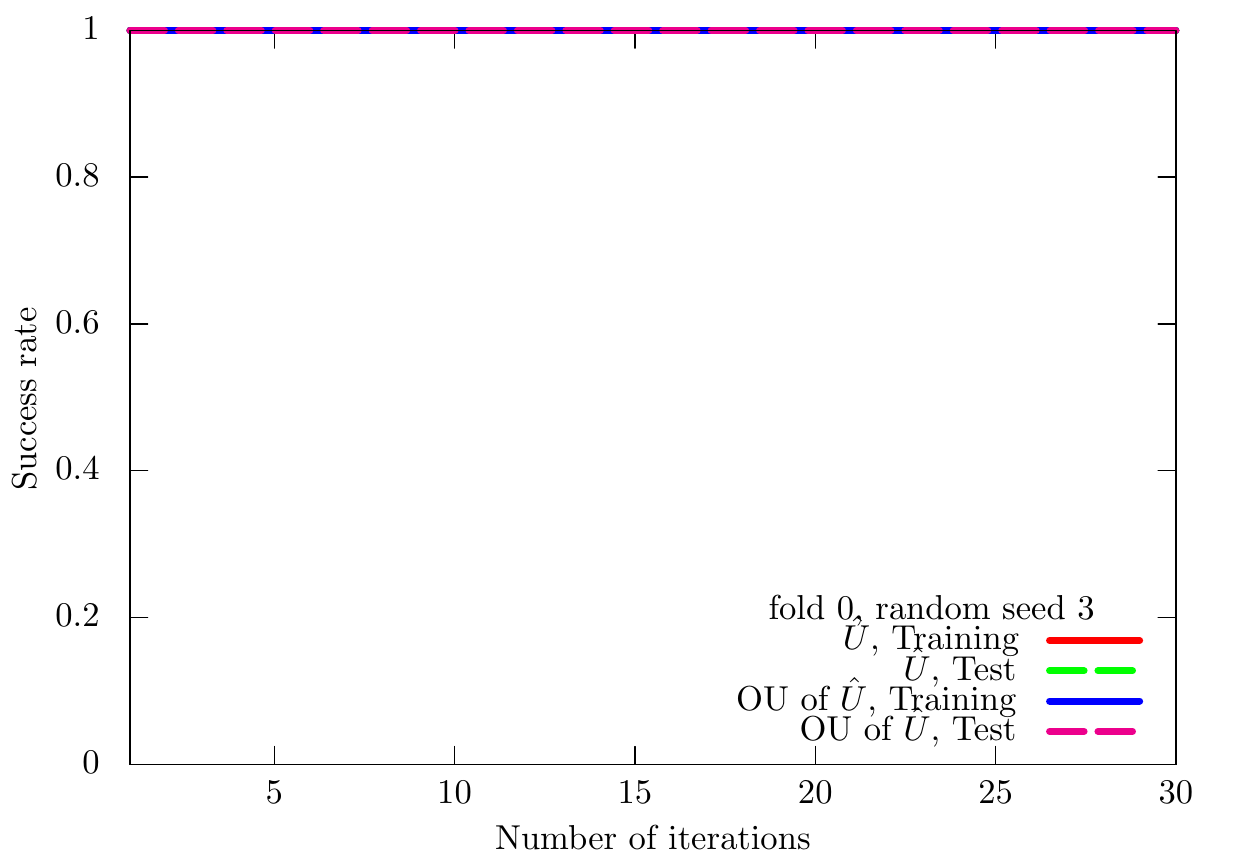}
\includegraphics[scale=0.25]{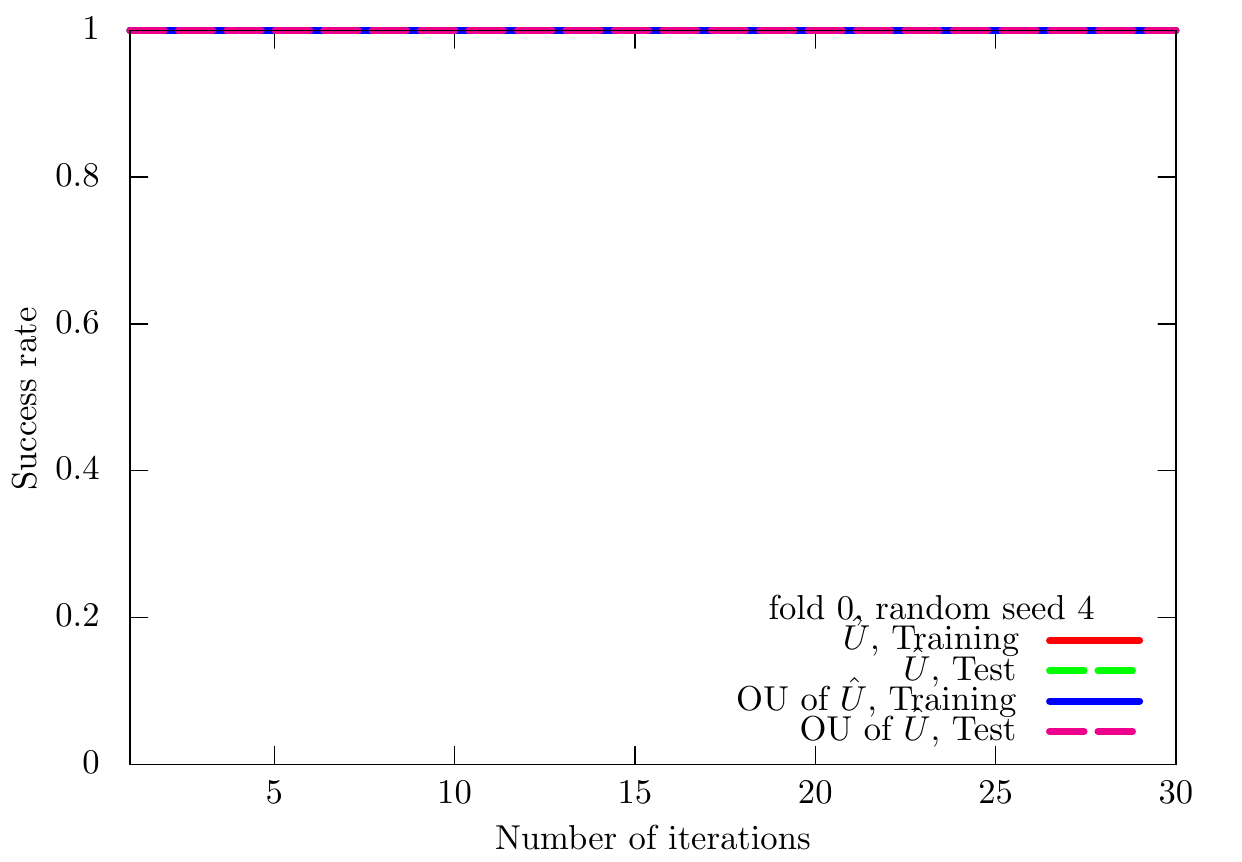}
\includegraphics[scale=0.25]{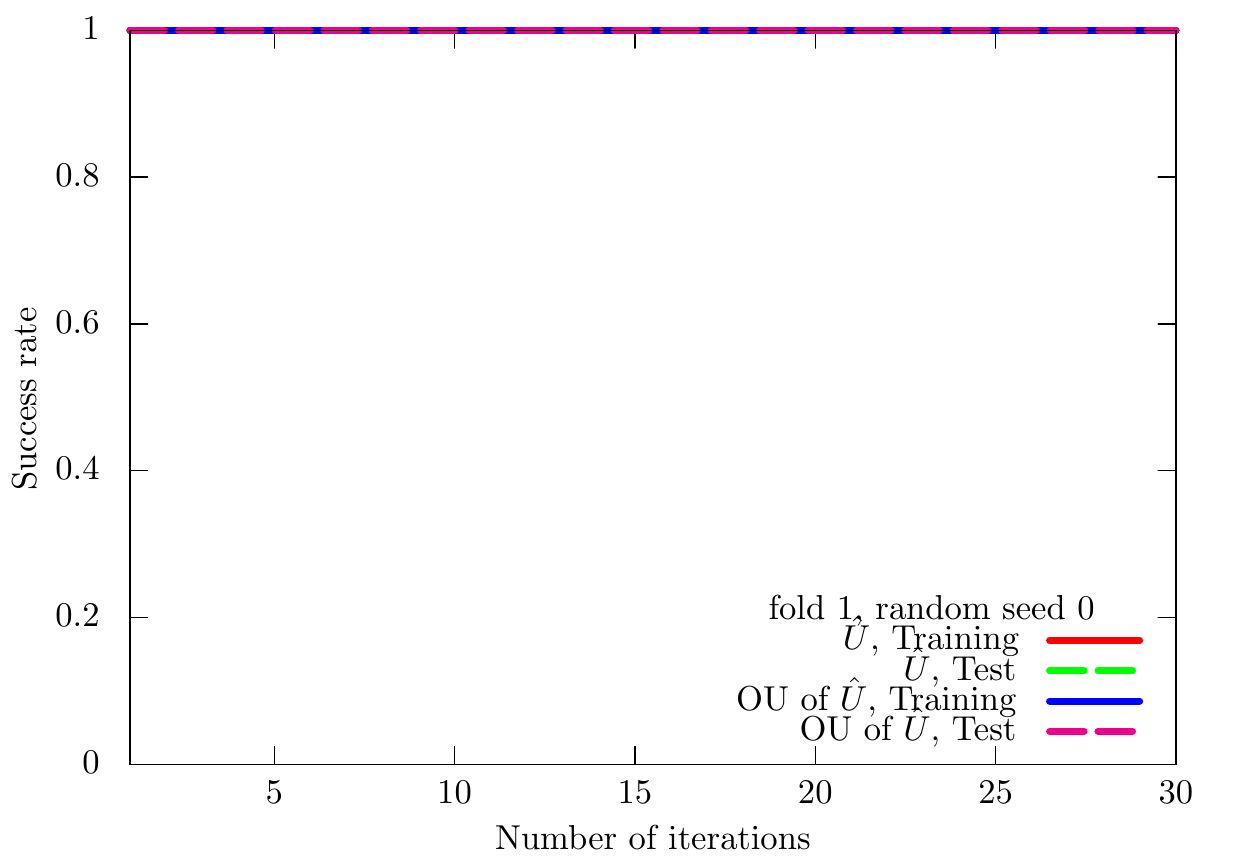}
\includegraphics[scale=0.25]{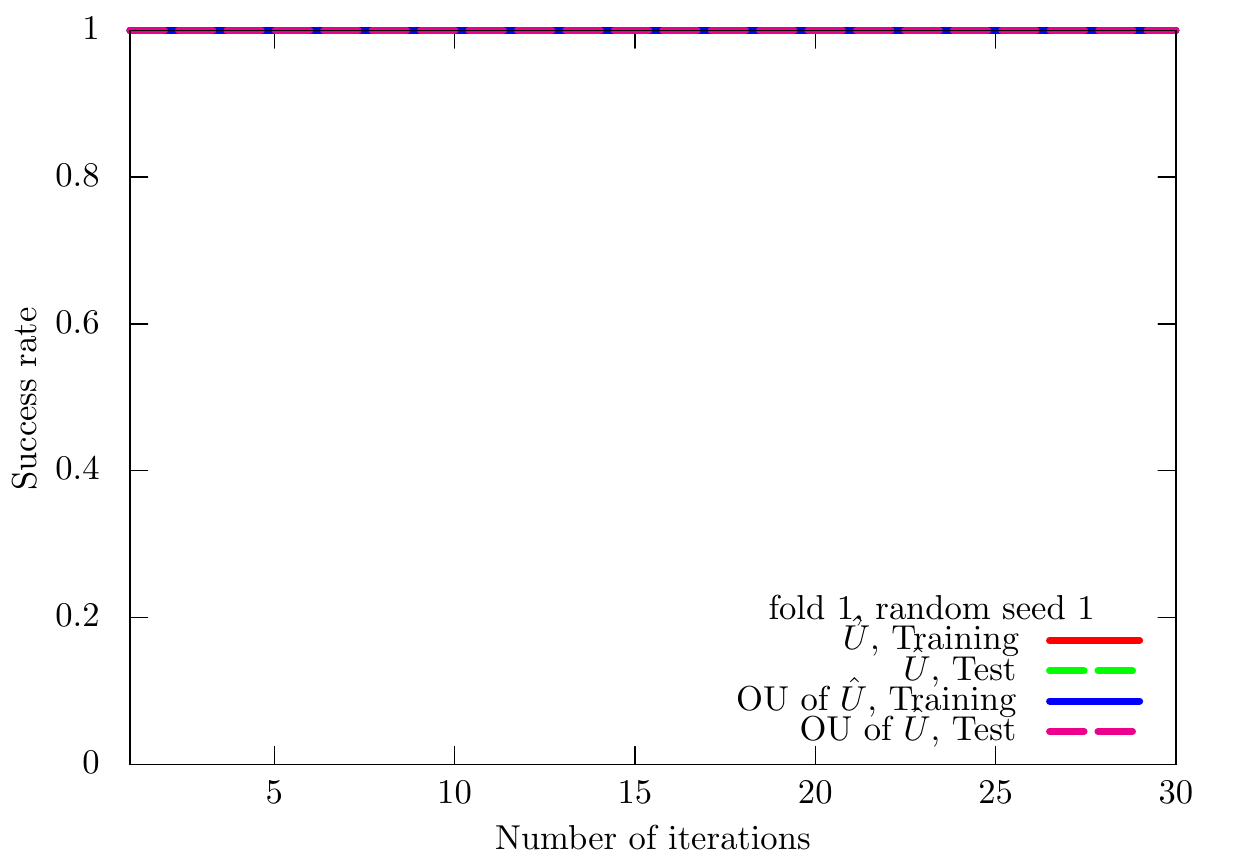}
\includegraphics[scale=0.25]{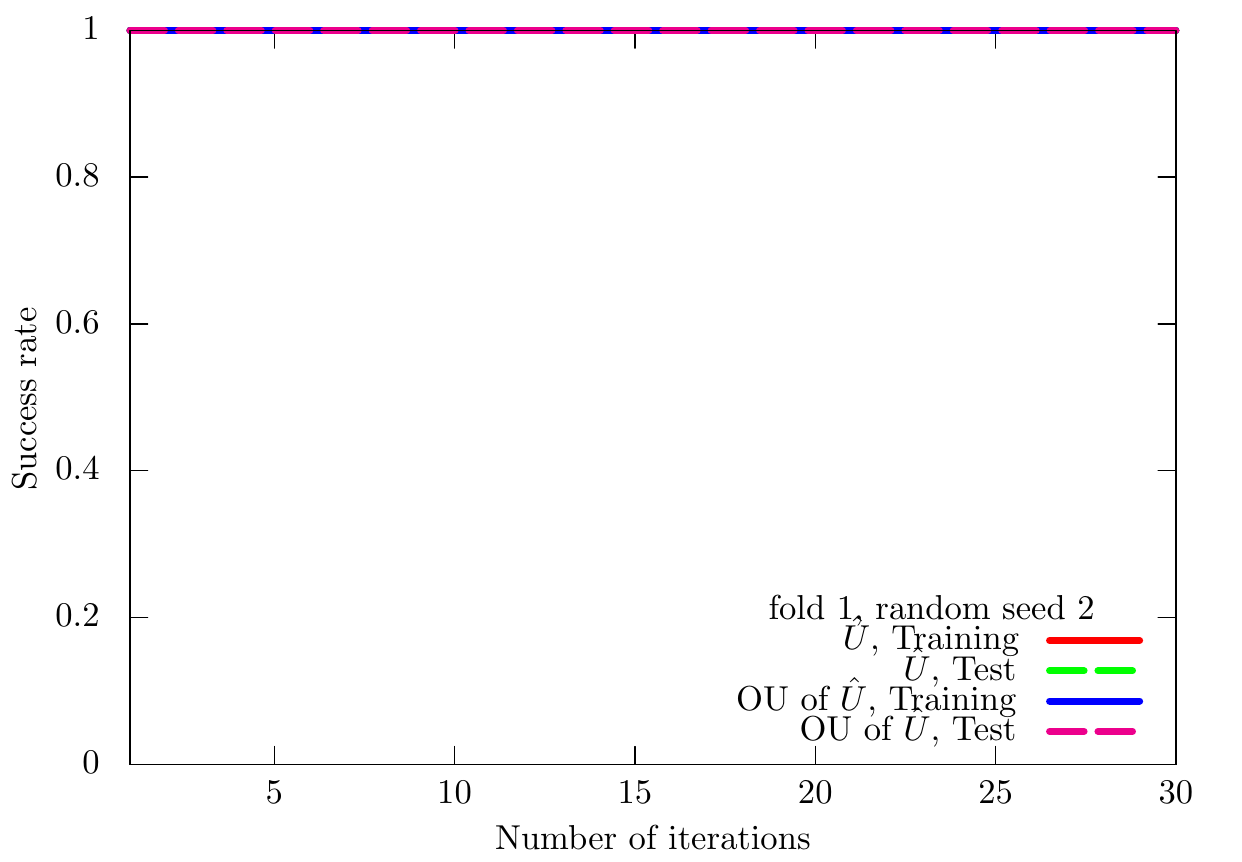}
\includegraphics[scale=0.25]{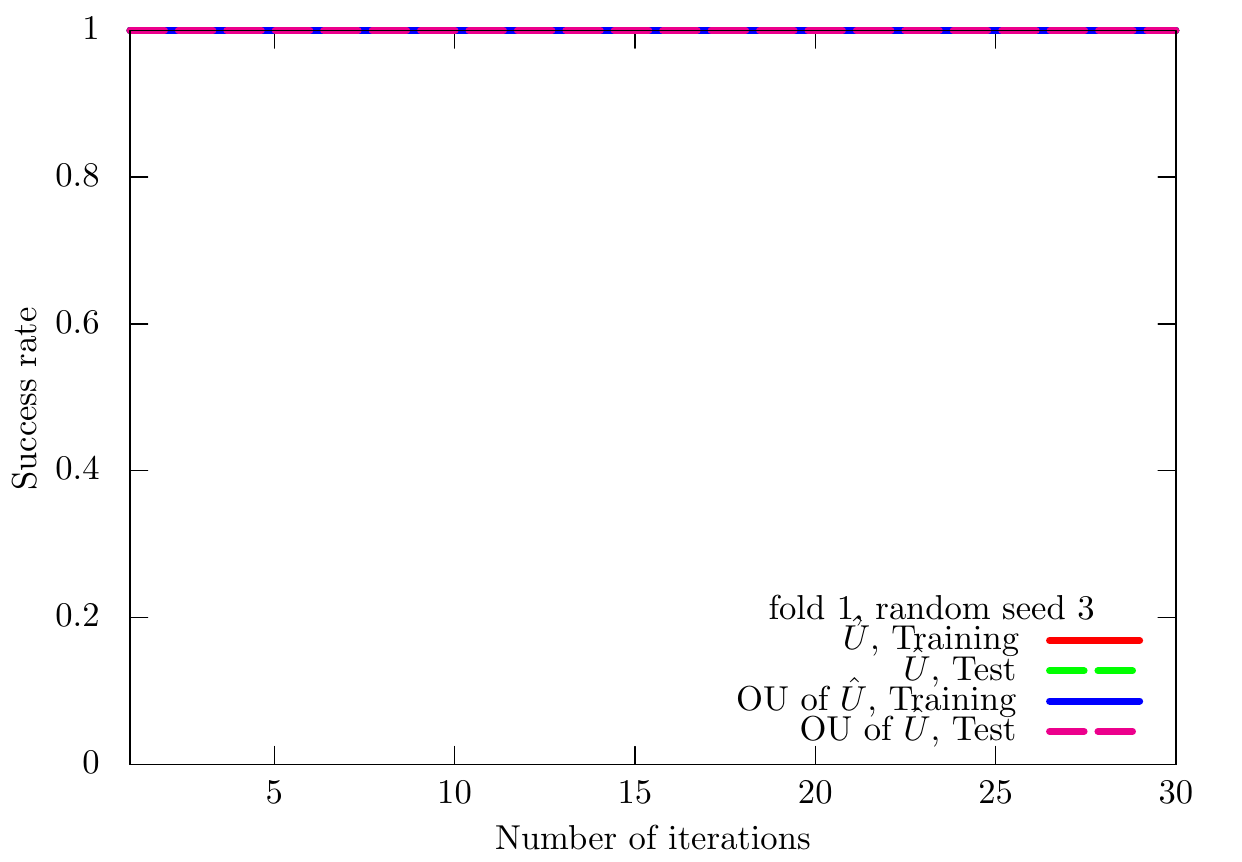}
\includegraphics[scale=0.25]{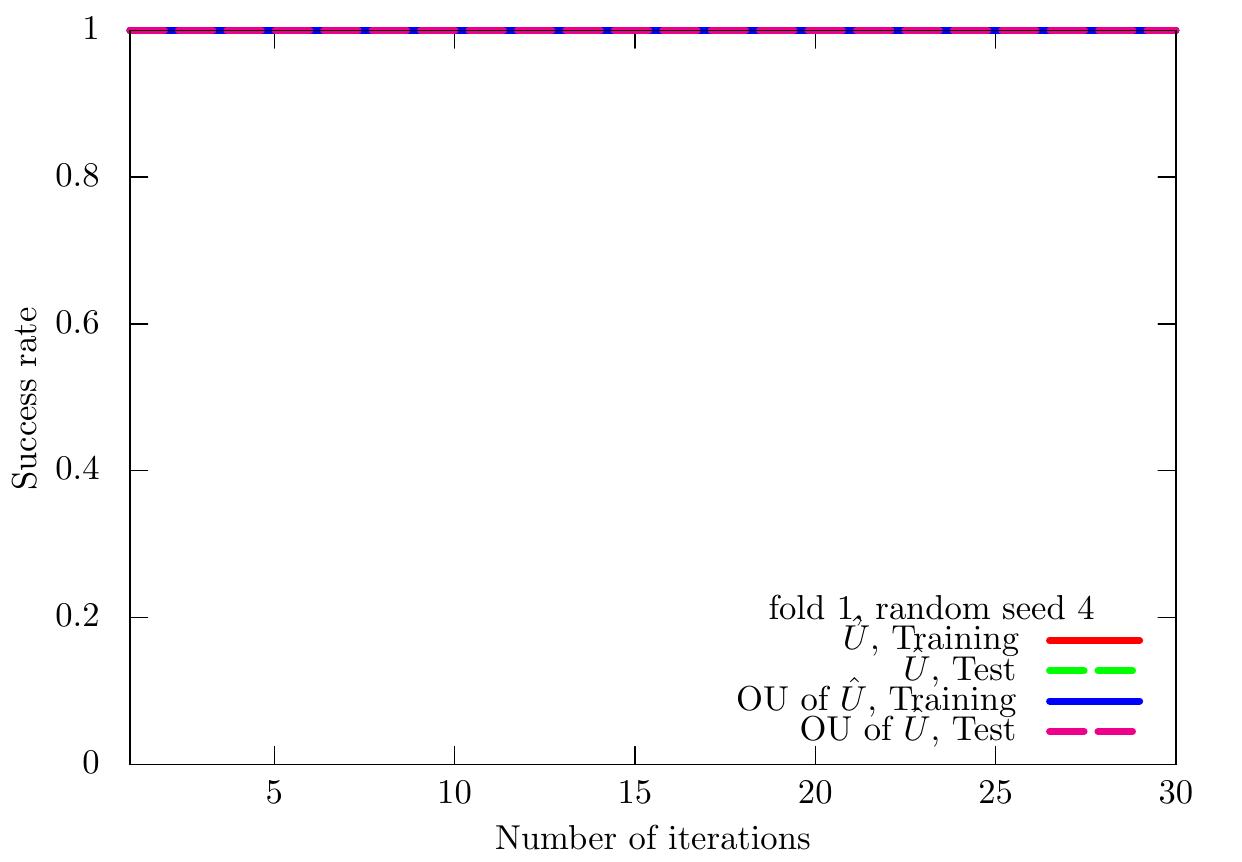}
\includegraphics[scale=0.25]{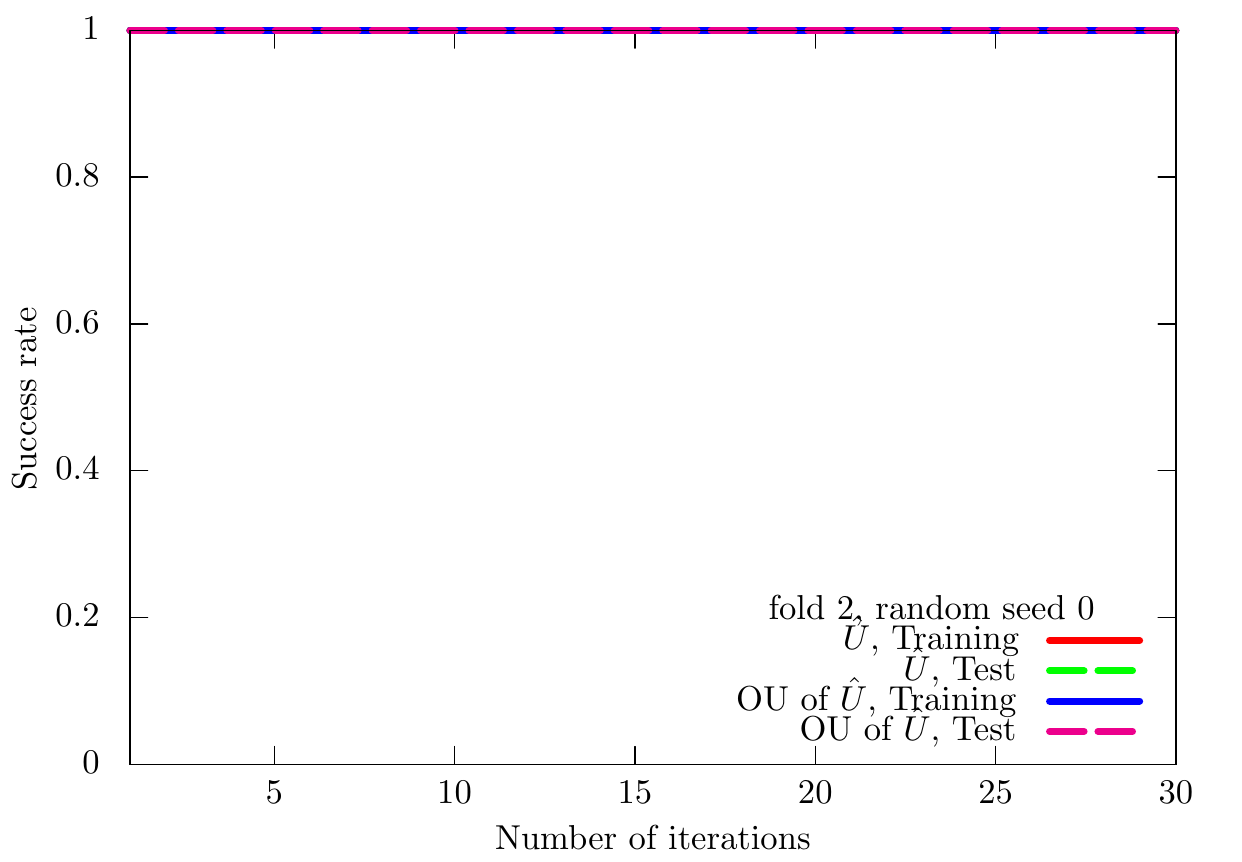}
\includegraphics[scale=0.25]{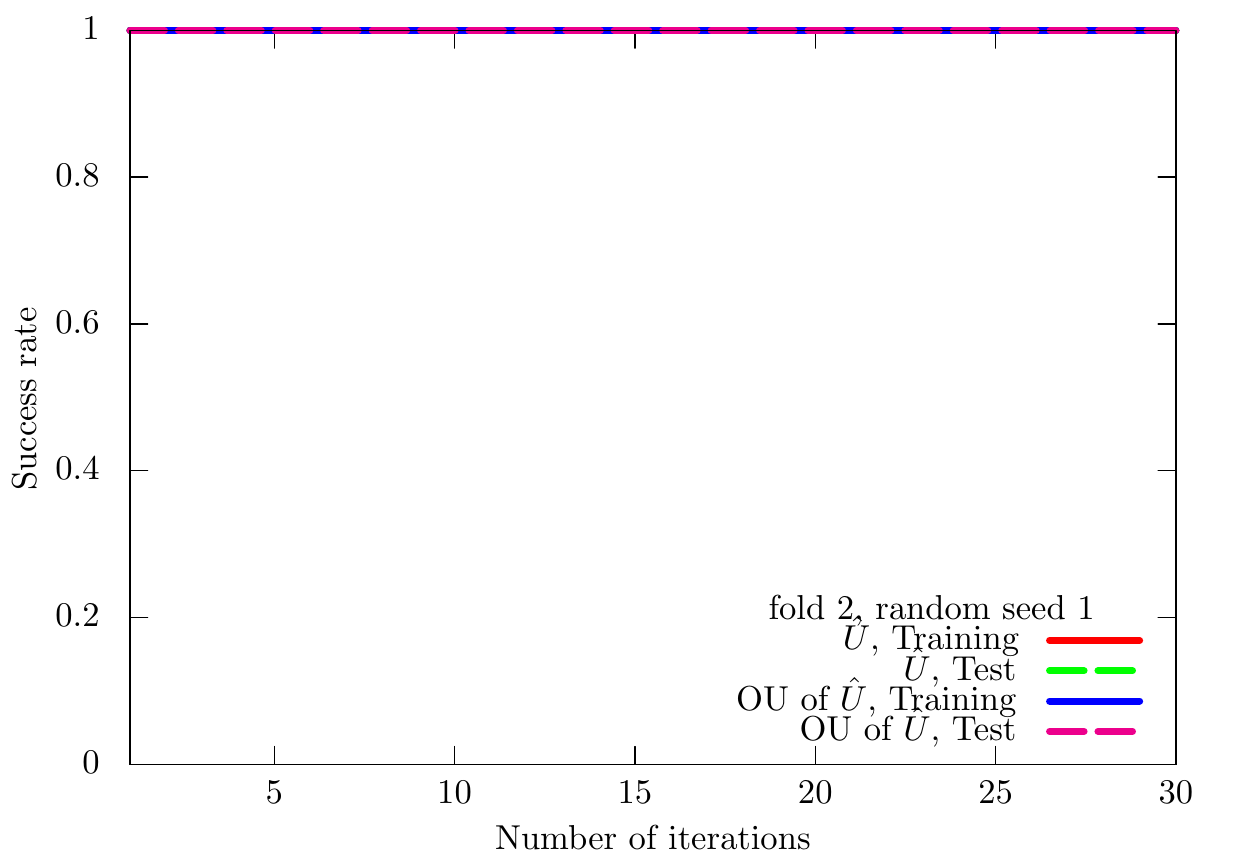}
\includegraphics[scale=0.25]{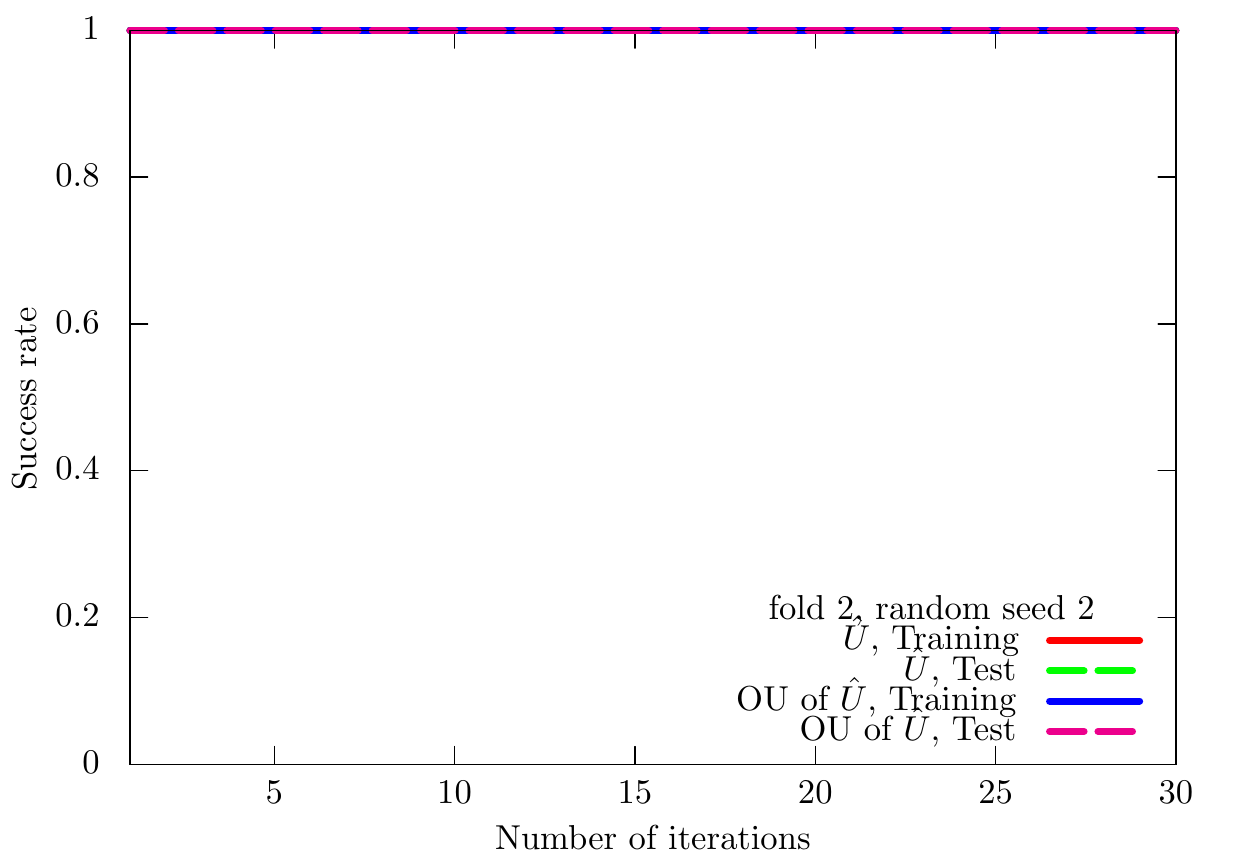}
\includegraphics[scale=0.25]{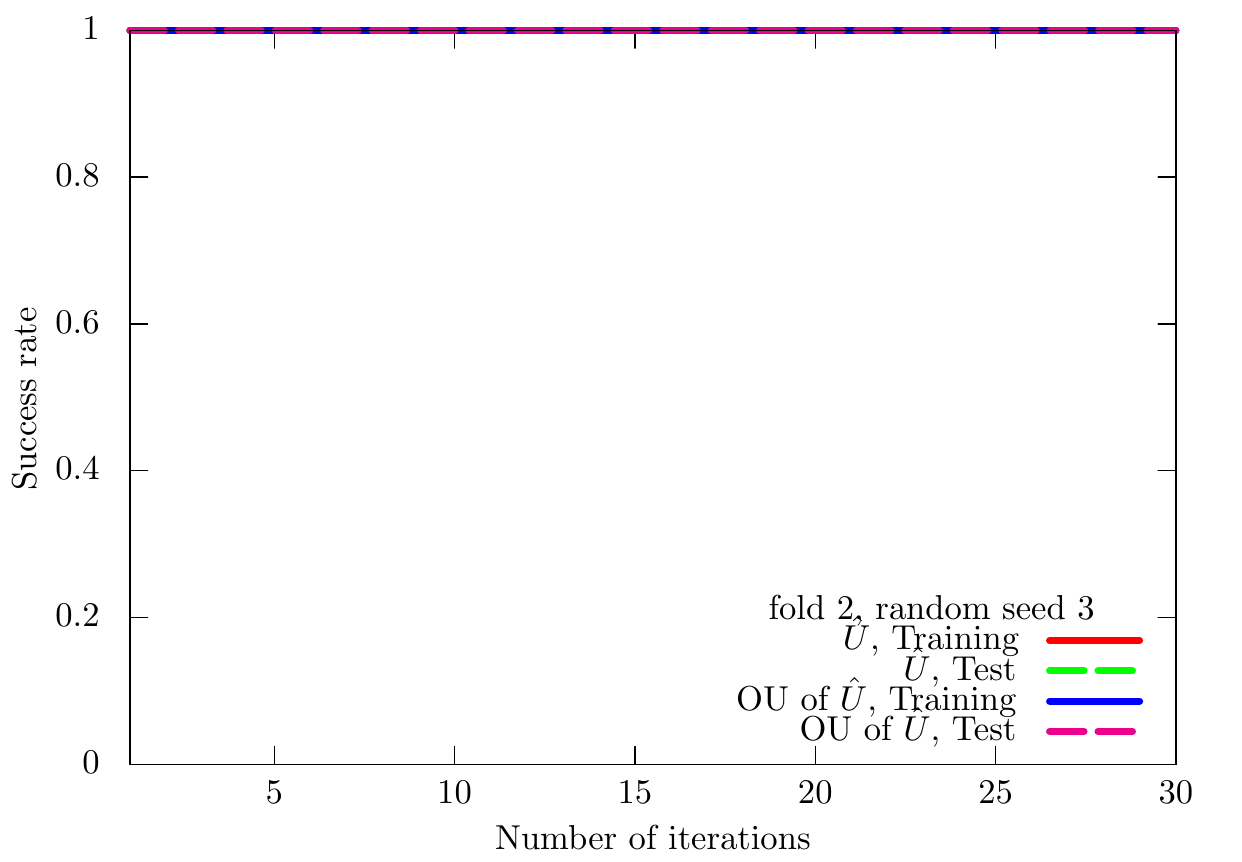}
\includegraphics[scale=0.25]{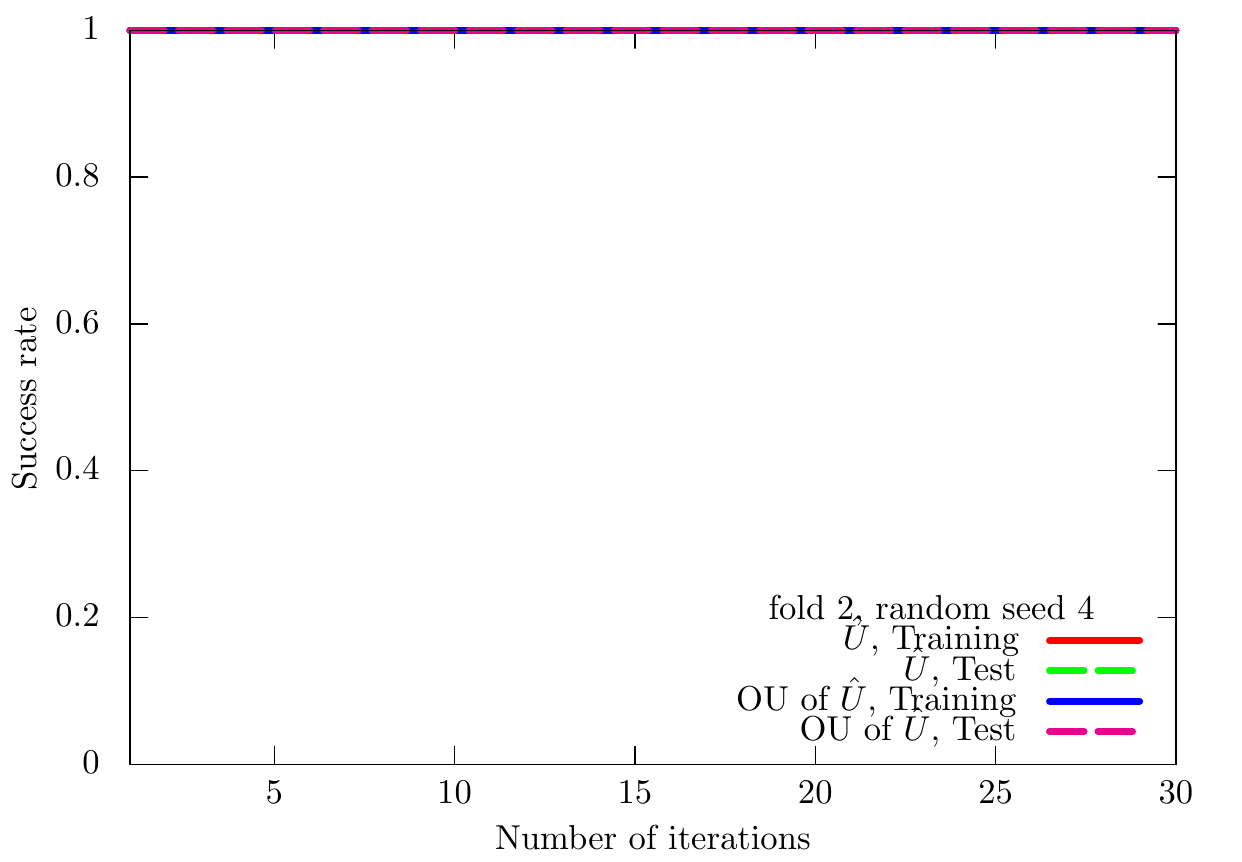}
\includegraphics[scale=0.25]{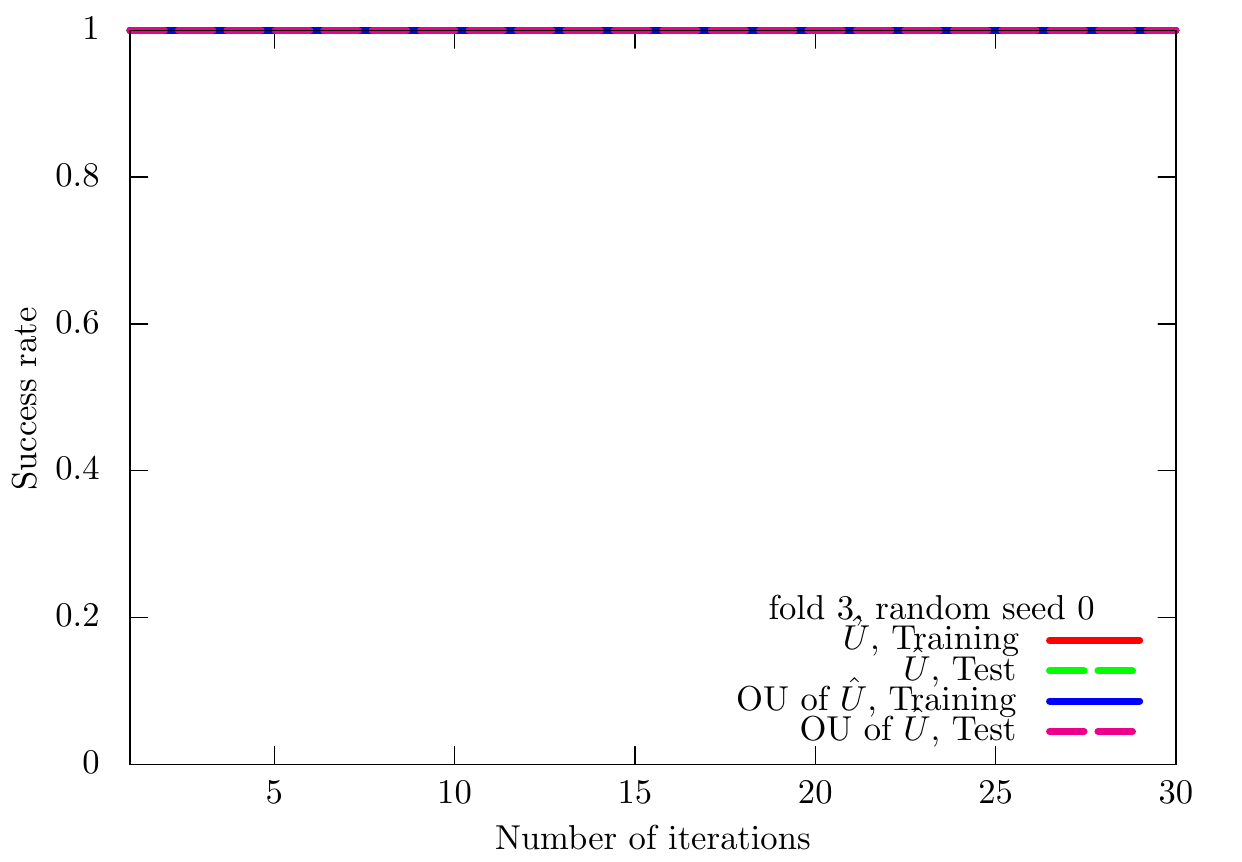}
\includegraphics[scale=0.25]{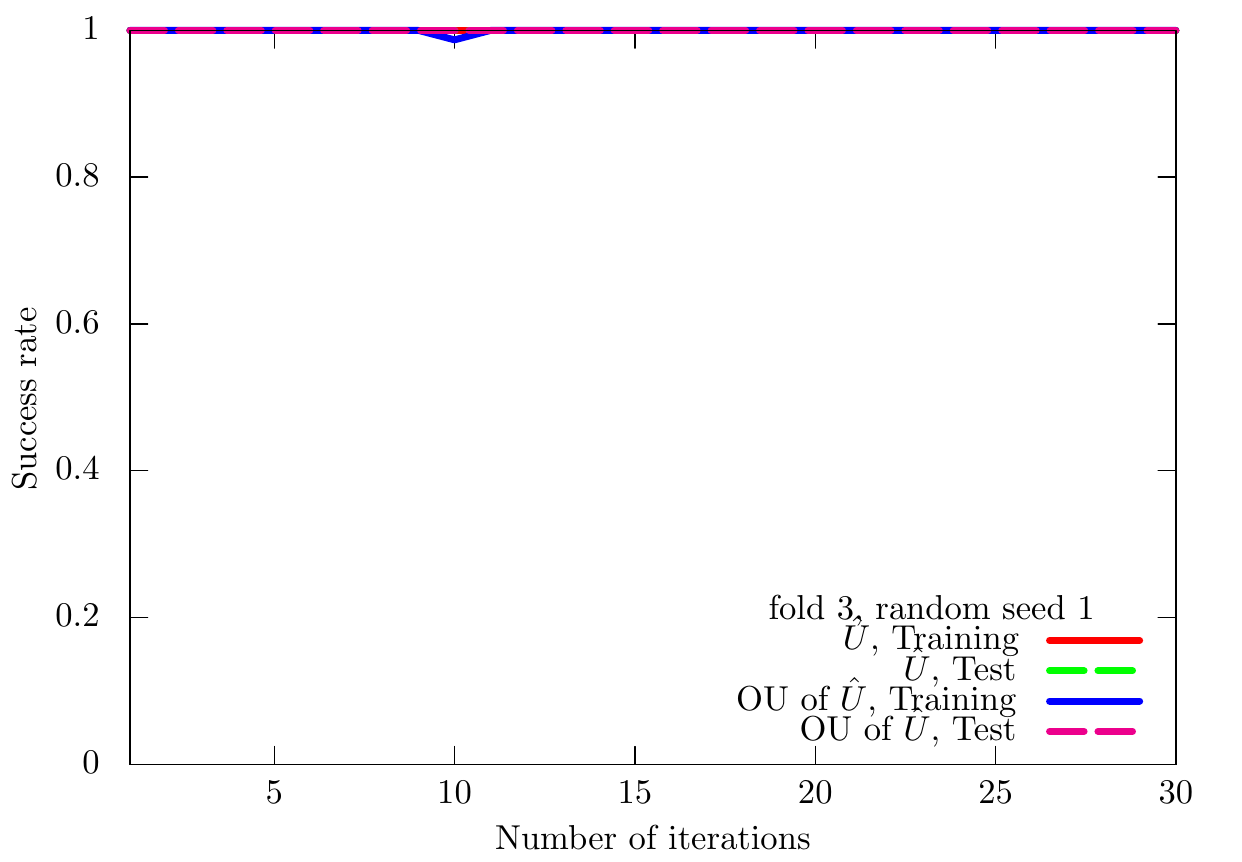}
\includegraphics[scale=0.25]{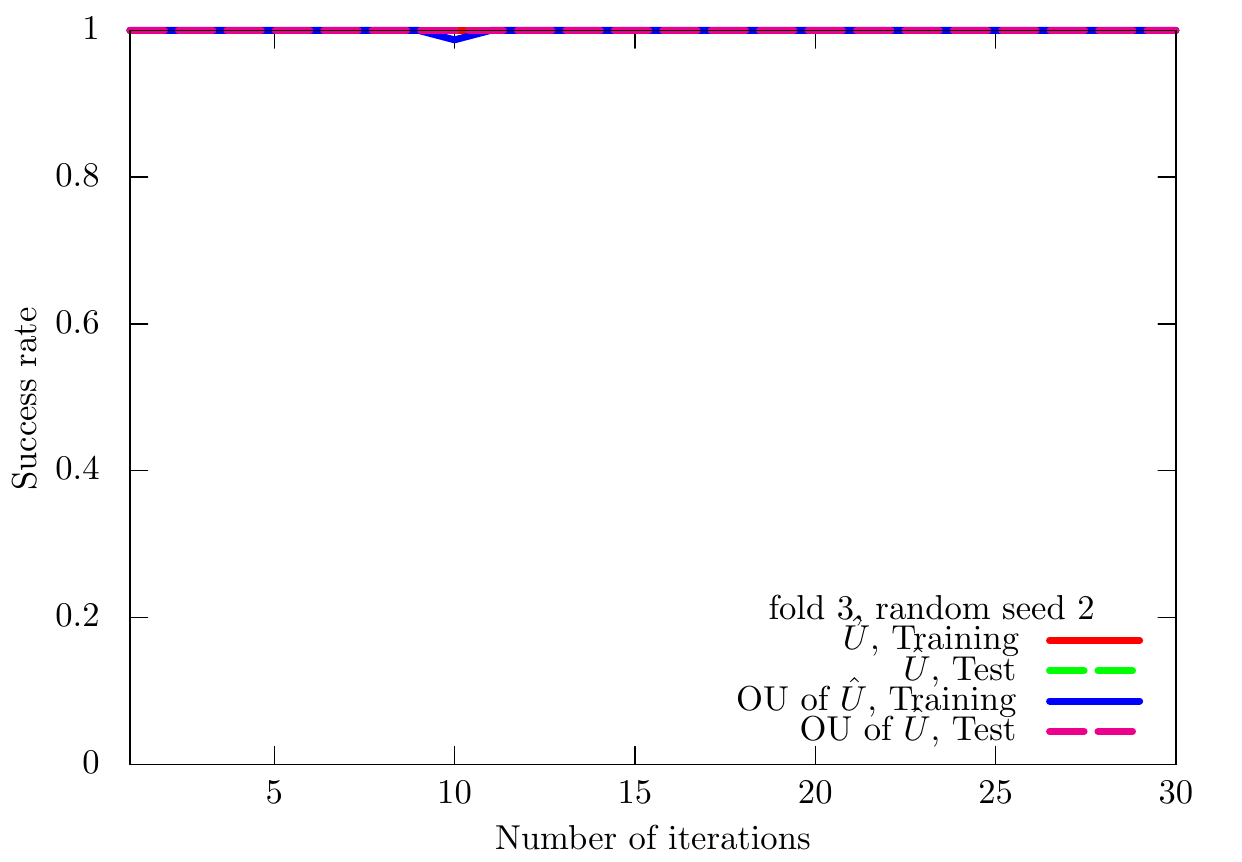}
\includegraphics[scale=0.25]{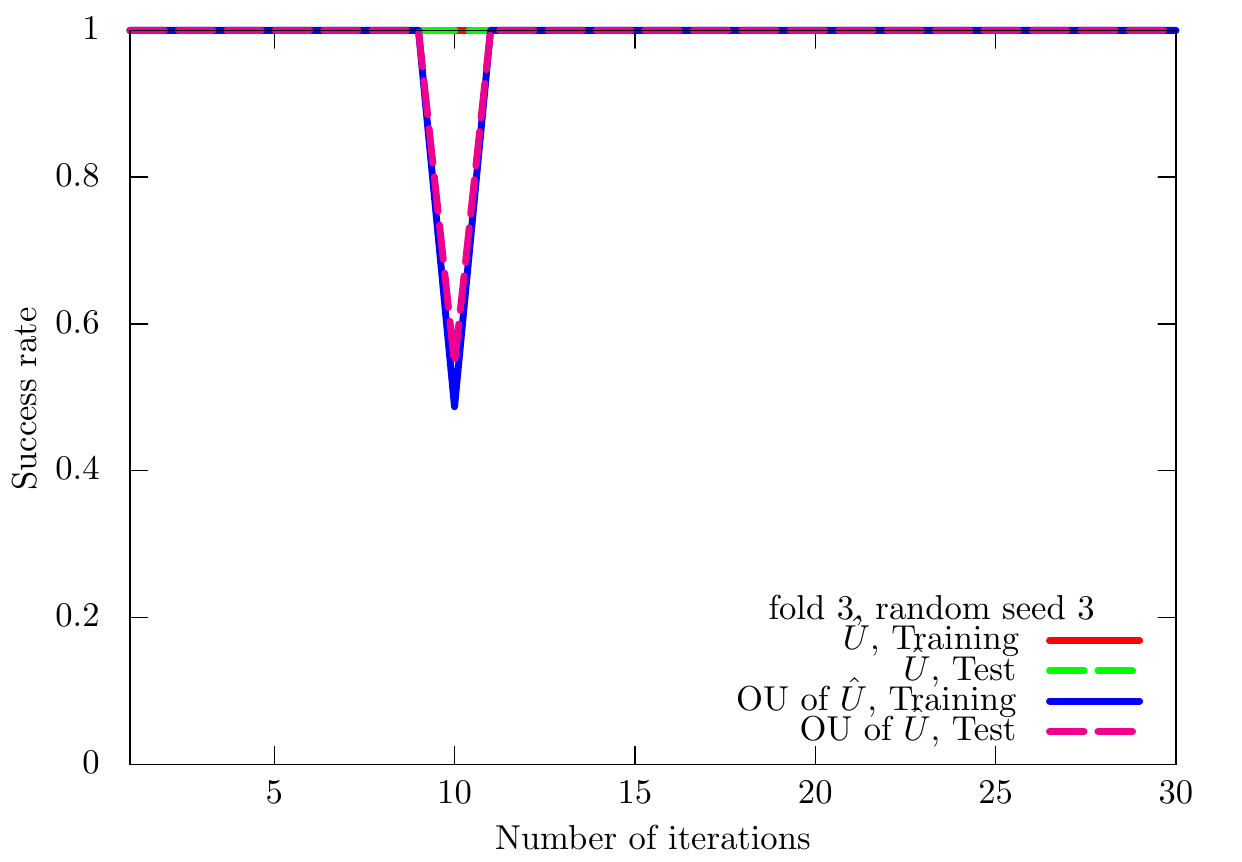}
\includegraphics[scale=0.25]{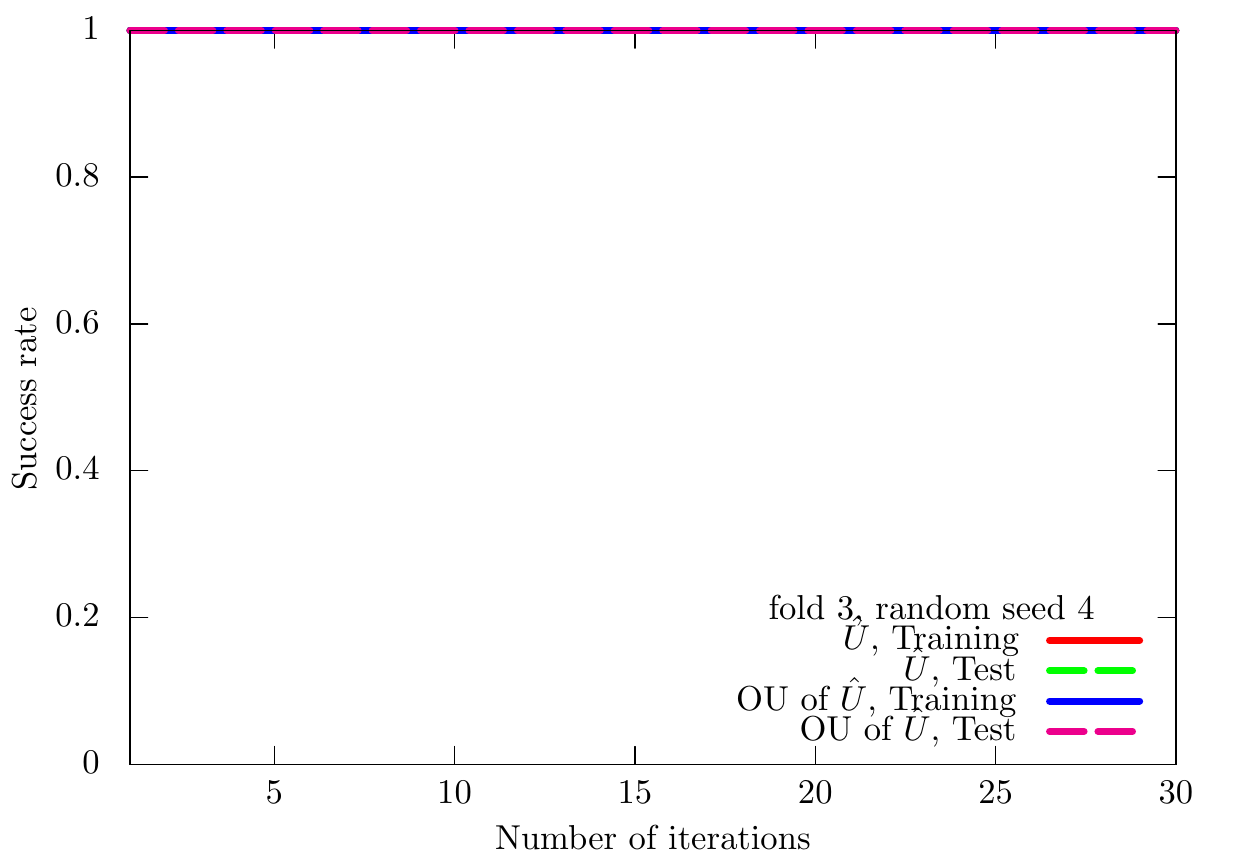}
\includegraphics[scale=0.25]{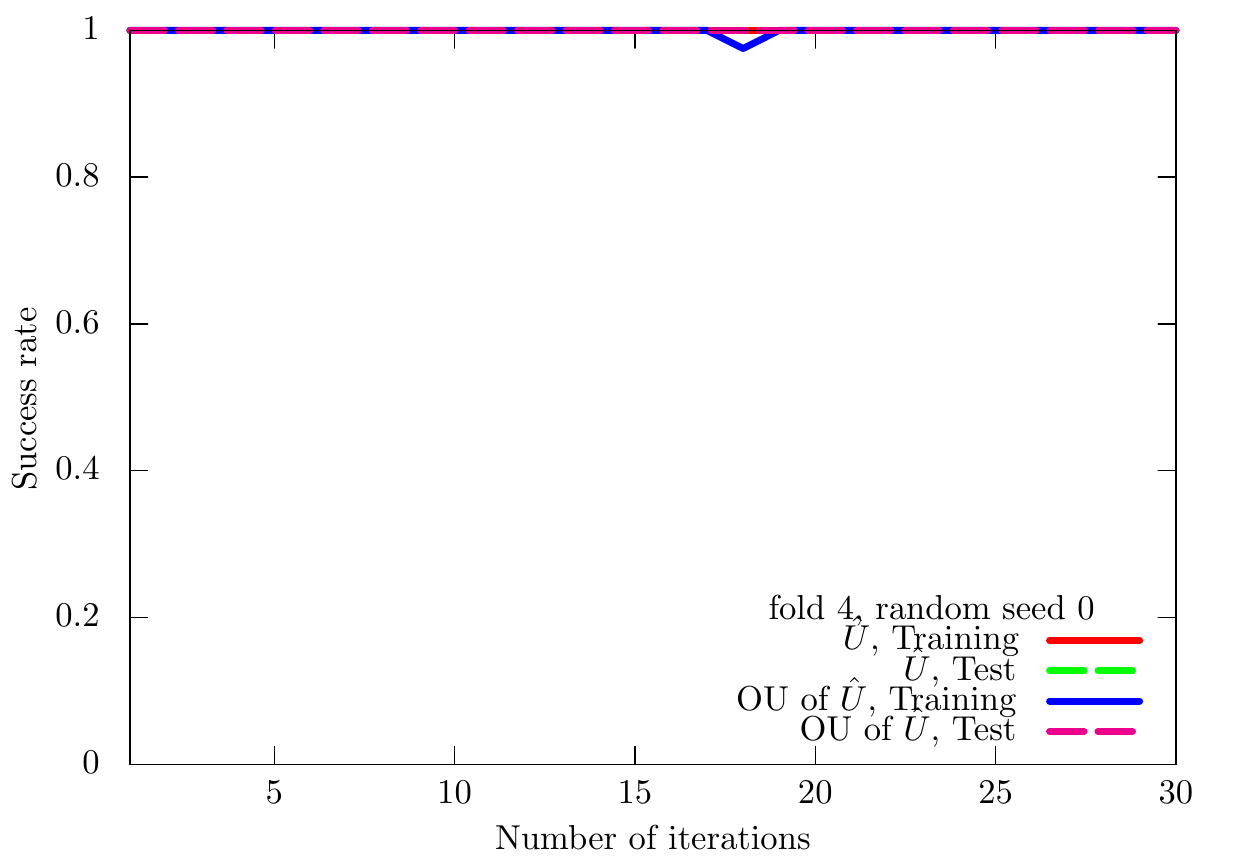}
\includegraphics[scale=0.25]{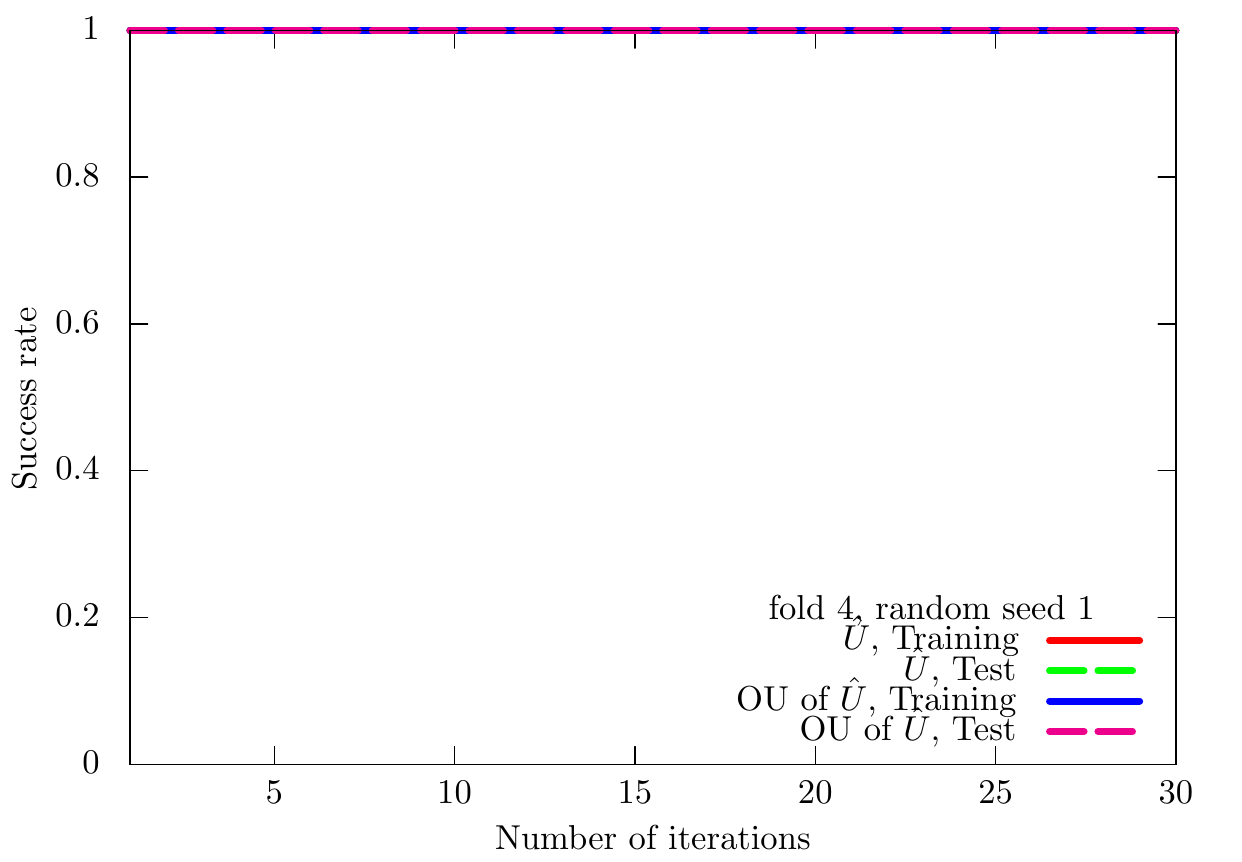}
\includegraphics[scale=0.25]{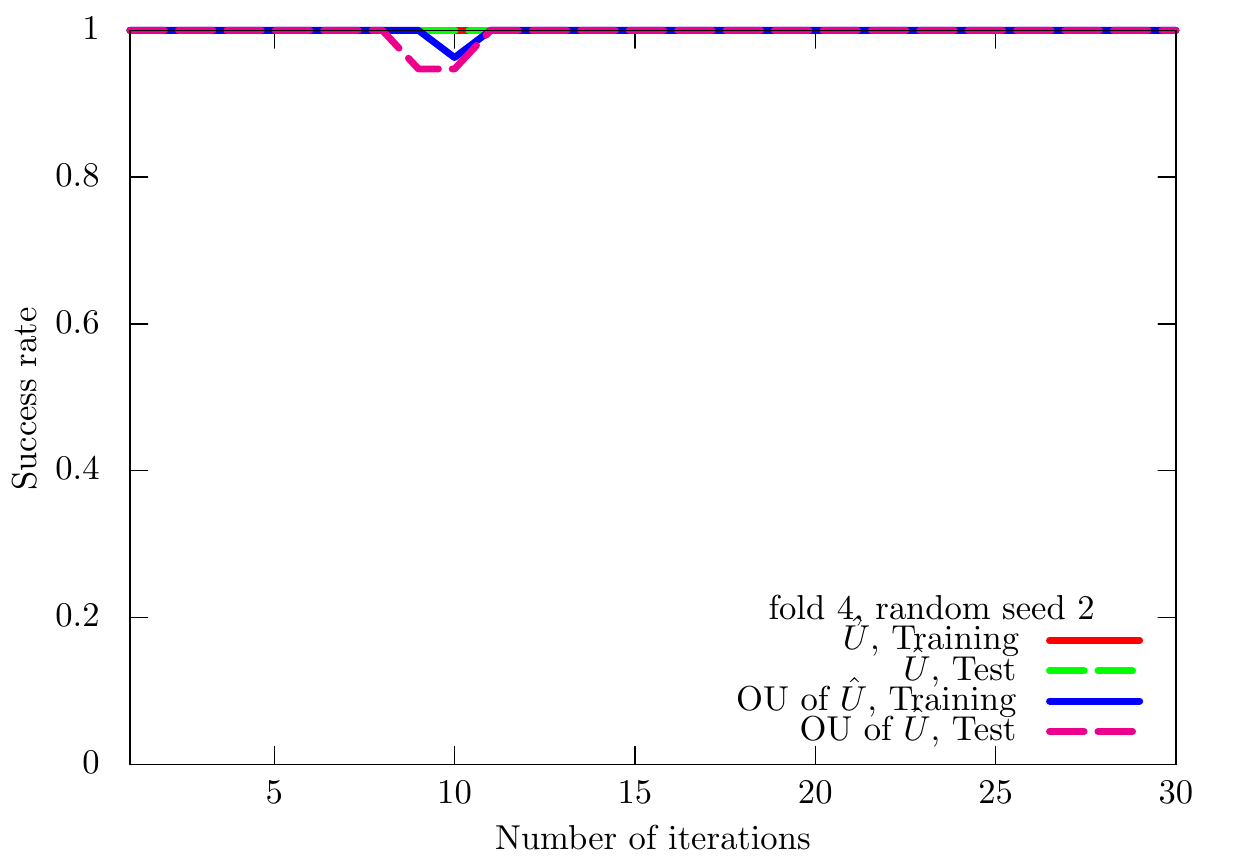}
\includegraphics[scale=0.25]{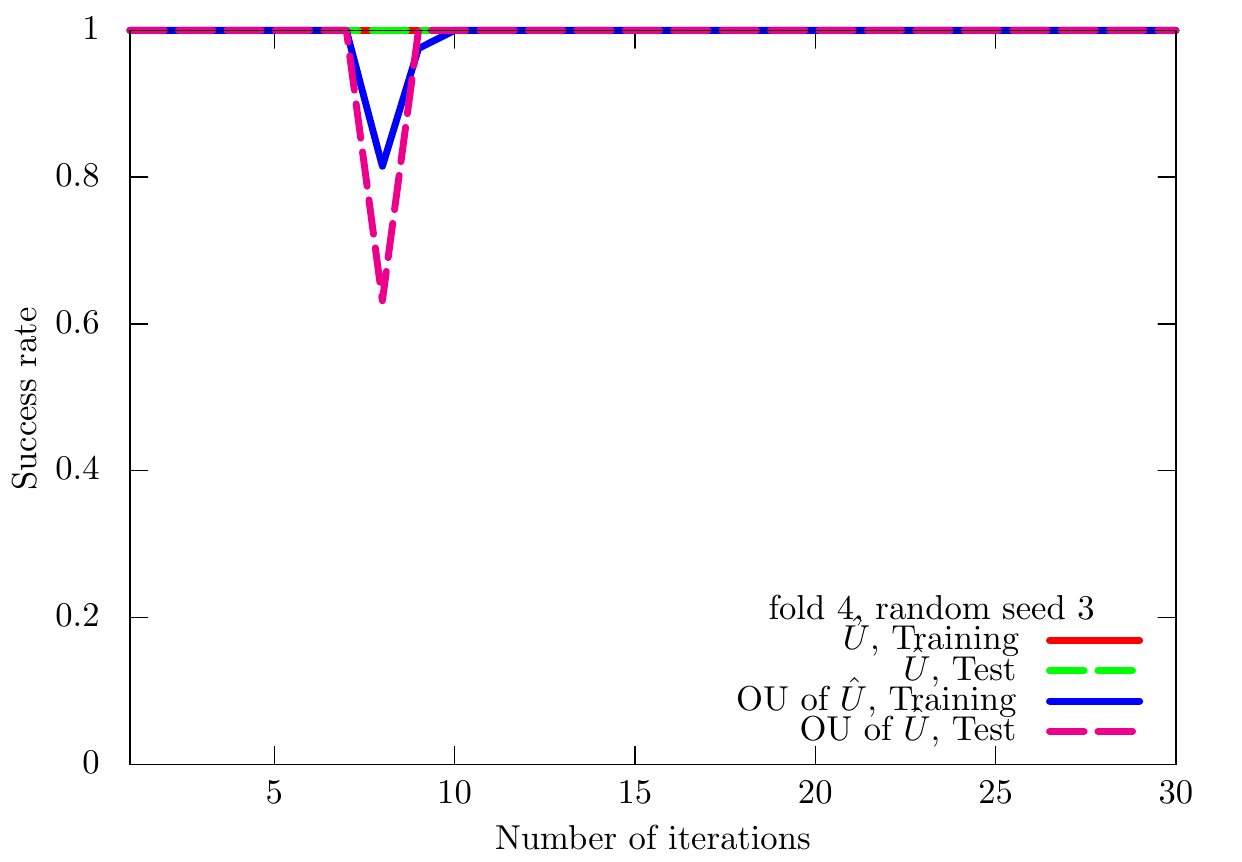}
\includegraphics[scale=0.25]{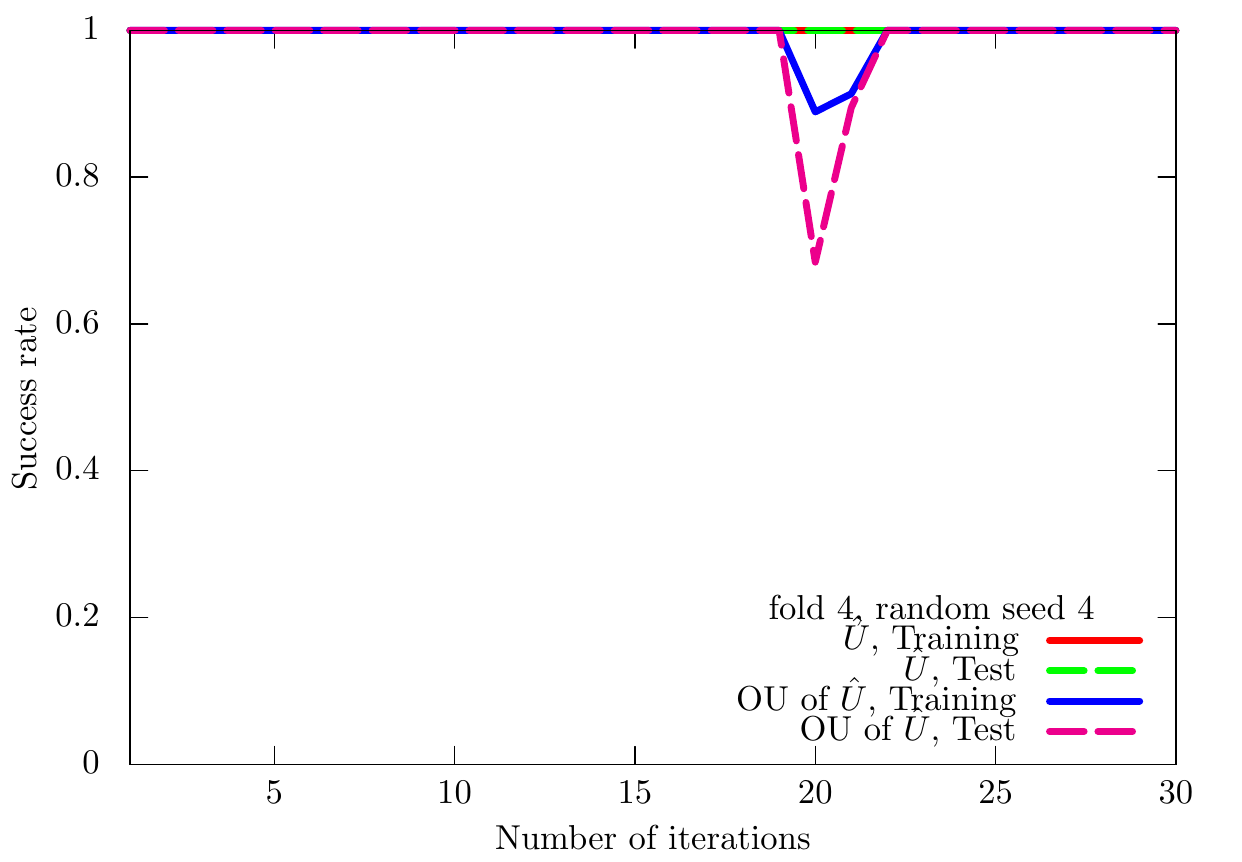}
\caption{Results of the UKM ($\hat{X}$ and $\hat{P}$) on the $5$-fold datasets with $5$ different random seeds for the iris dataset ($0$ or $1$). We use complex matrices and set $\theta_\mathrm{bias} = 0$. We set $r = 0.010$.}
\label{supp-arXiv-numerical-result-raw-data-fold-001-rand-001-UKM-P-UCI-iris-0-1}
\end{figure*}

\begin{figure*}[htb]
\centering
\includegraphics[scale=0.25]{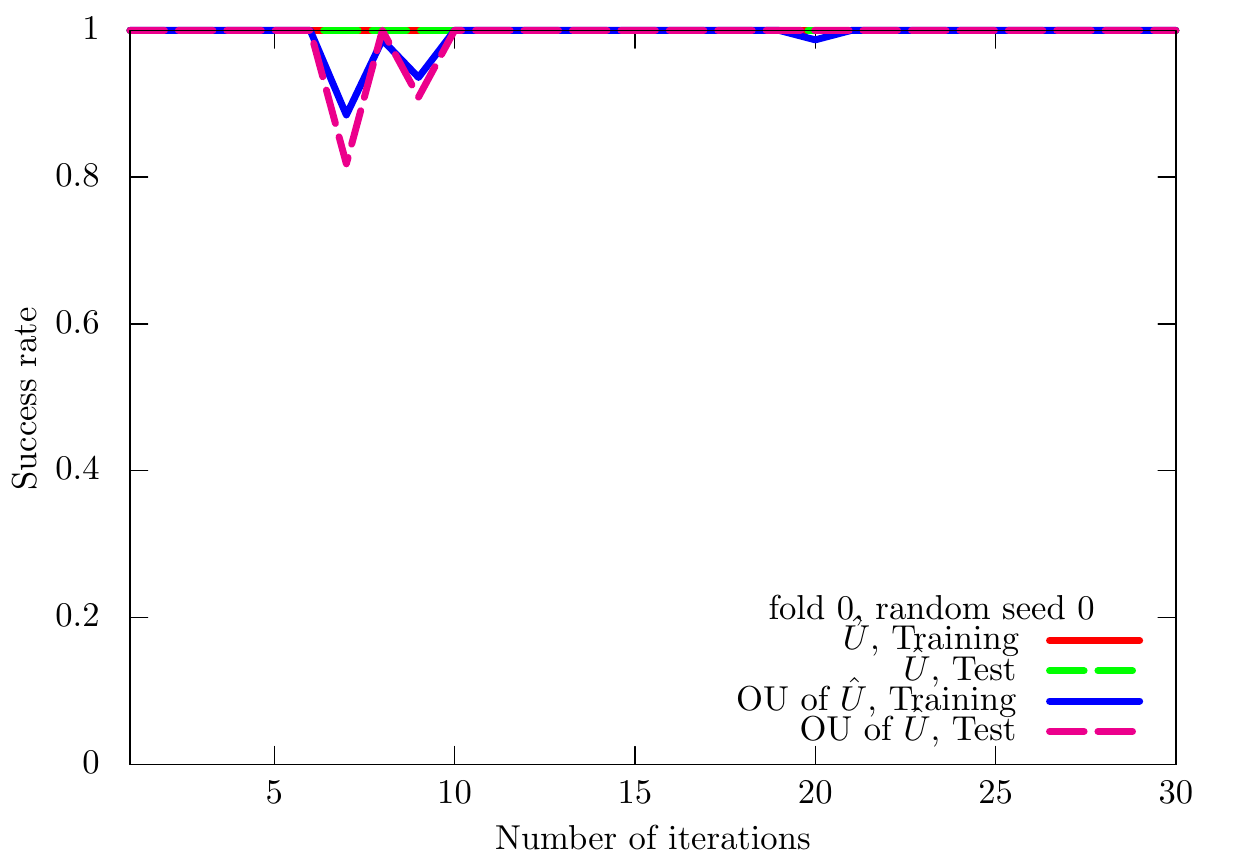}
\includegraphics[scale=0.25]{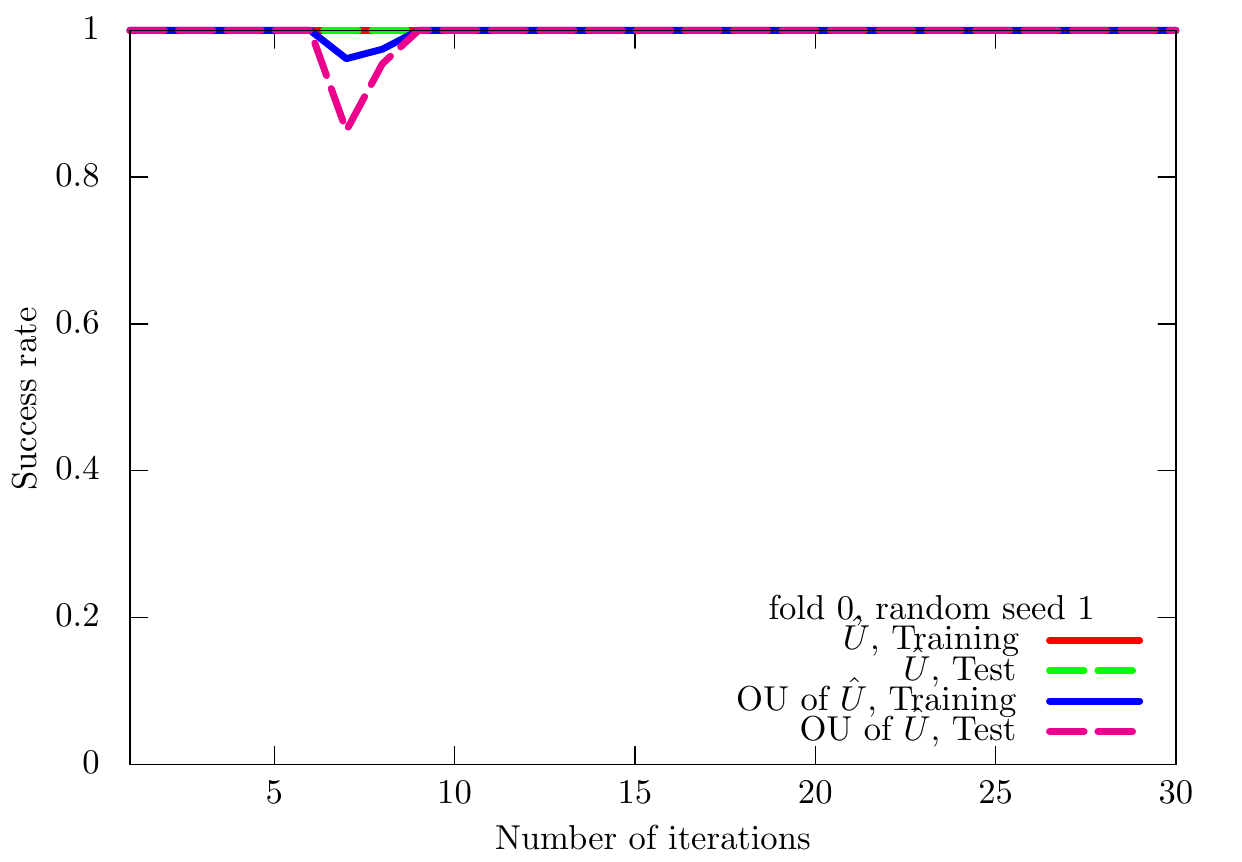}
\includegraphics[scale=0.25]{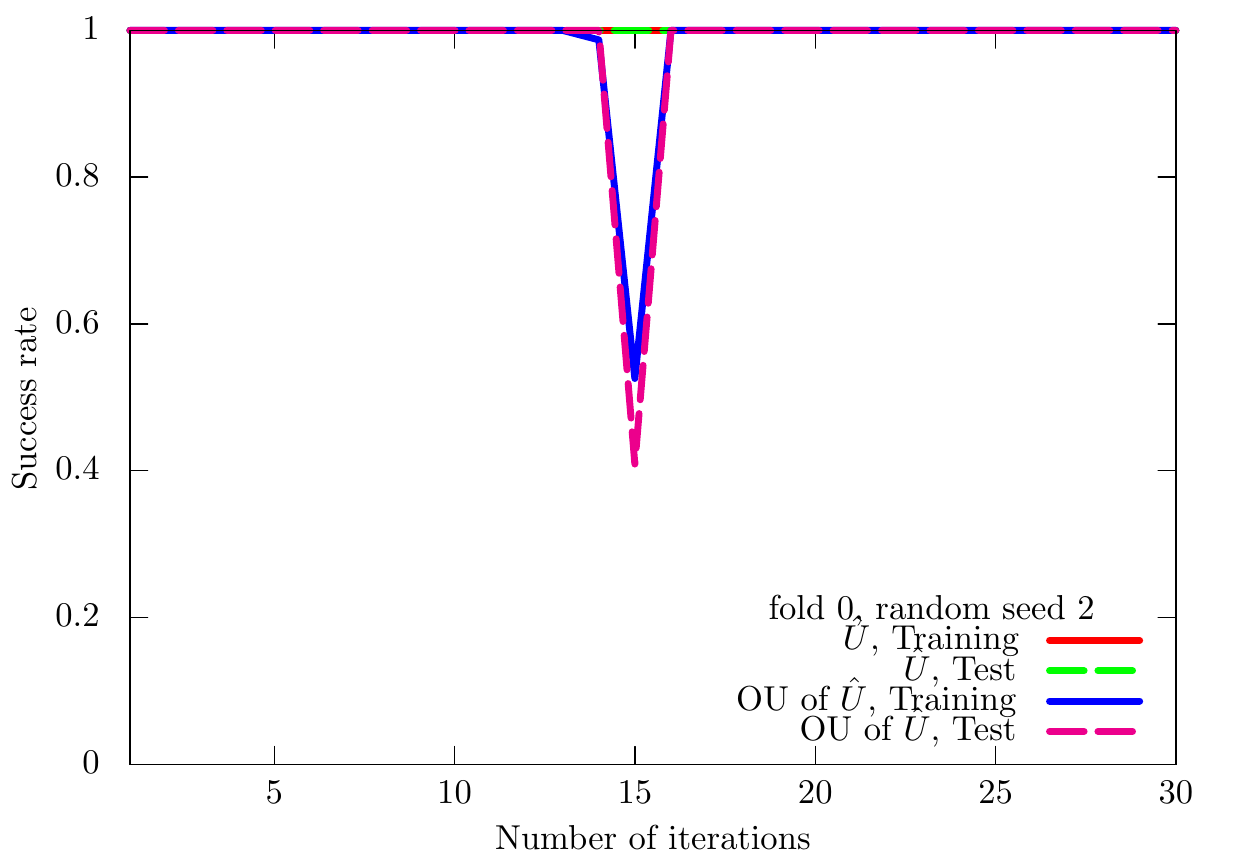}
\includegraphics[scale=0.25]{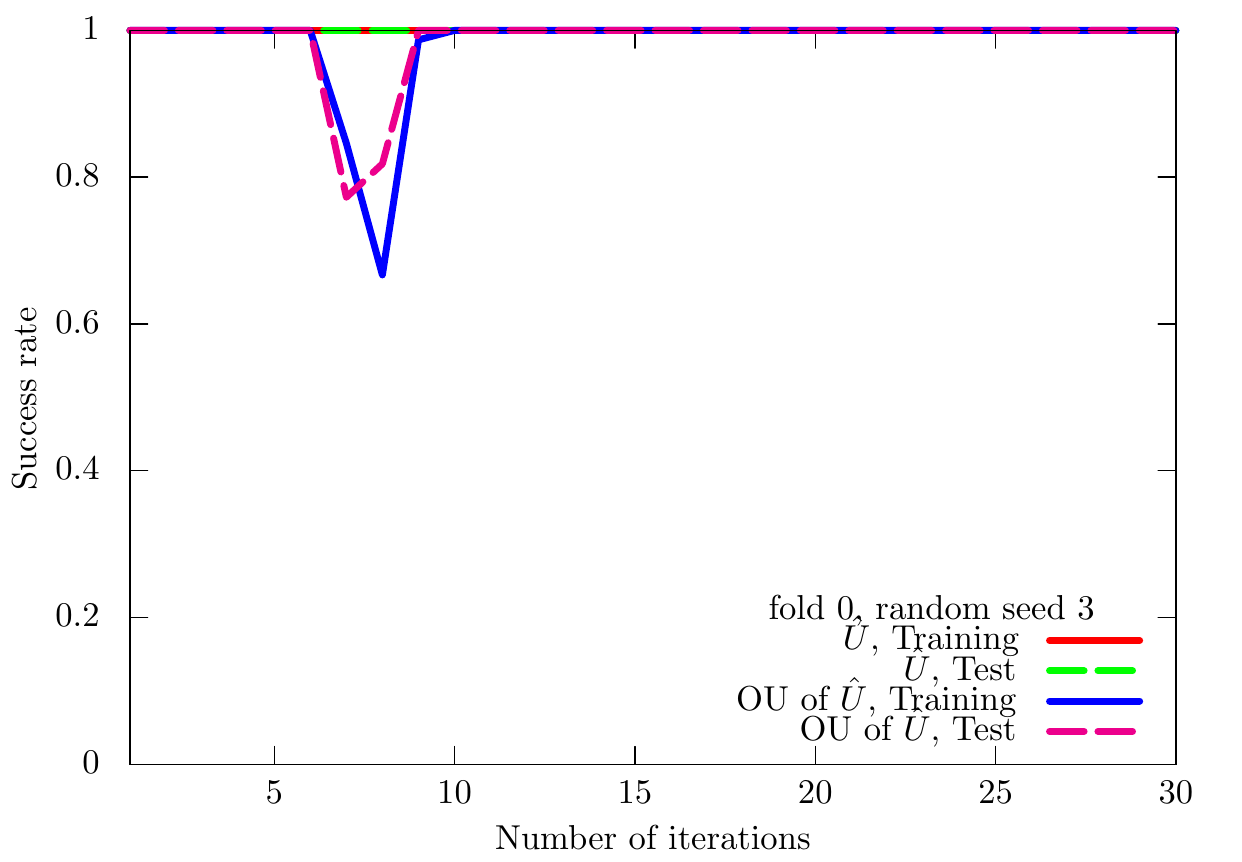}
\includegraphics[scale=0.25]{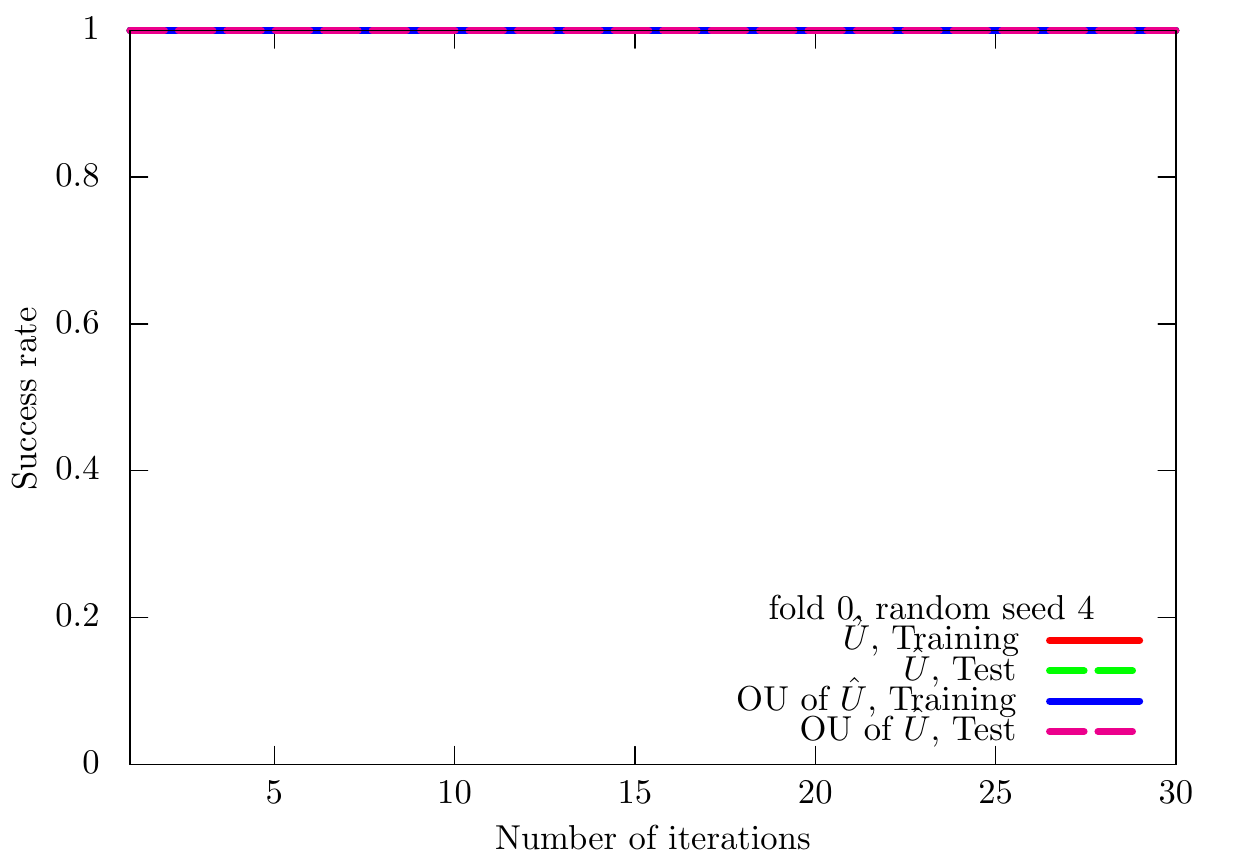}
\includegraphics[scale=0.25]{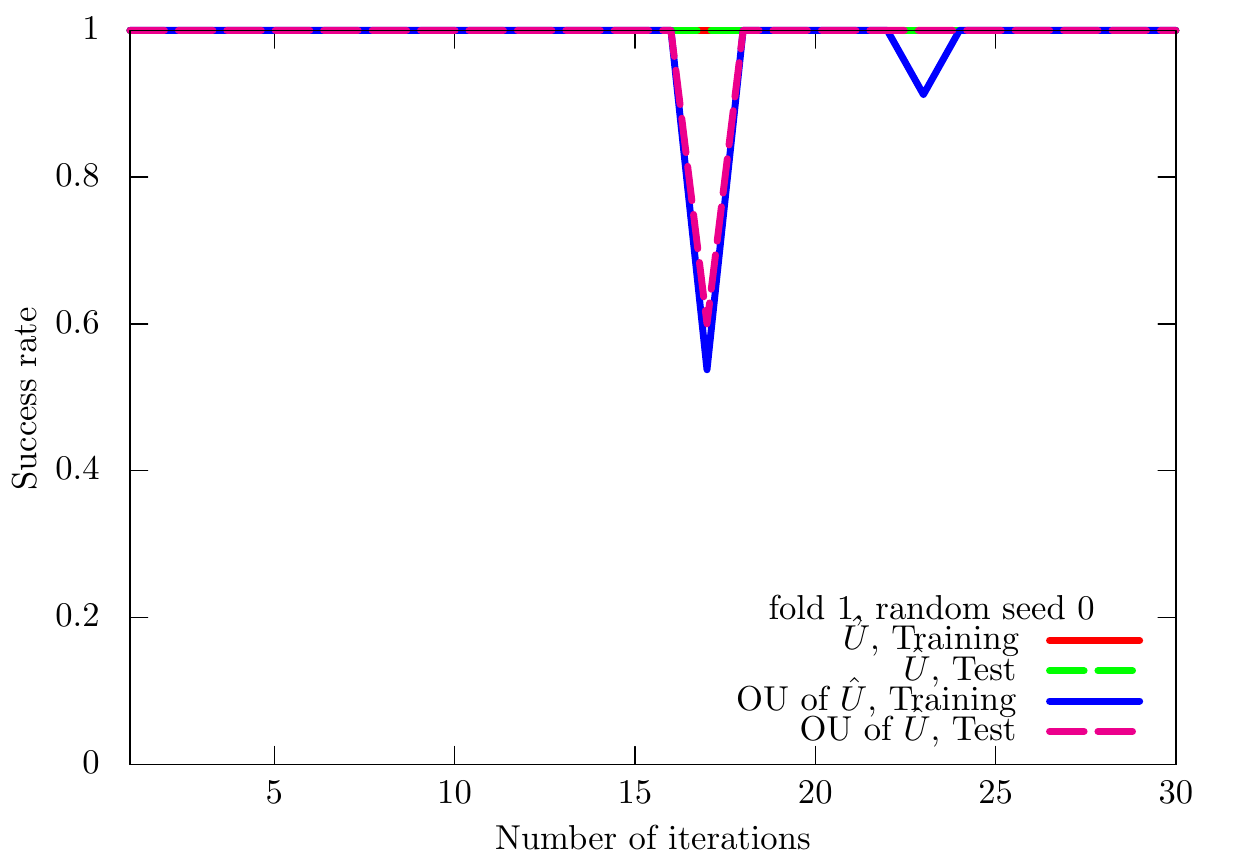}
\includegraphics[scale=0.25]{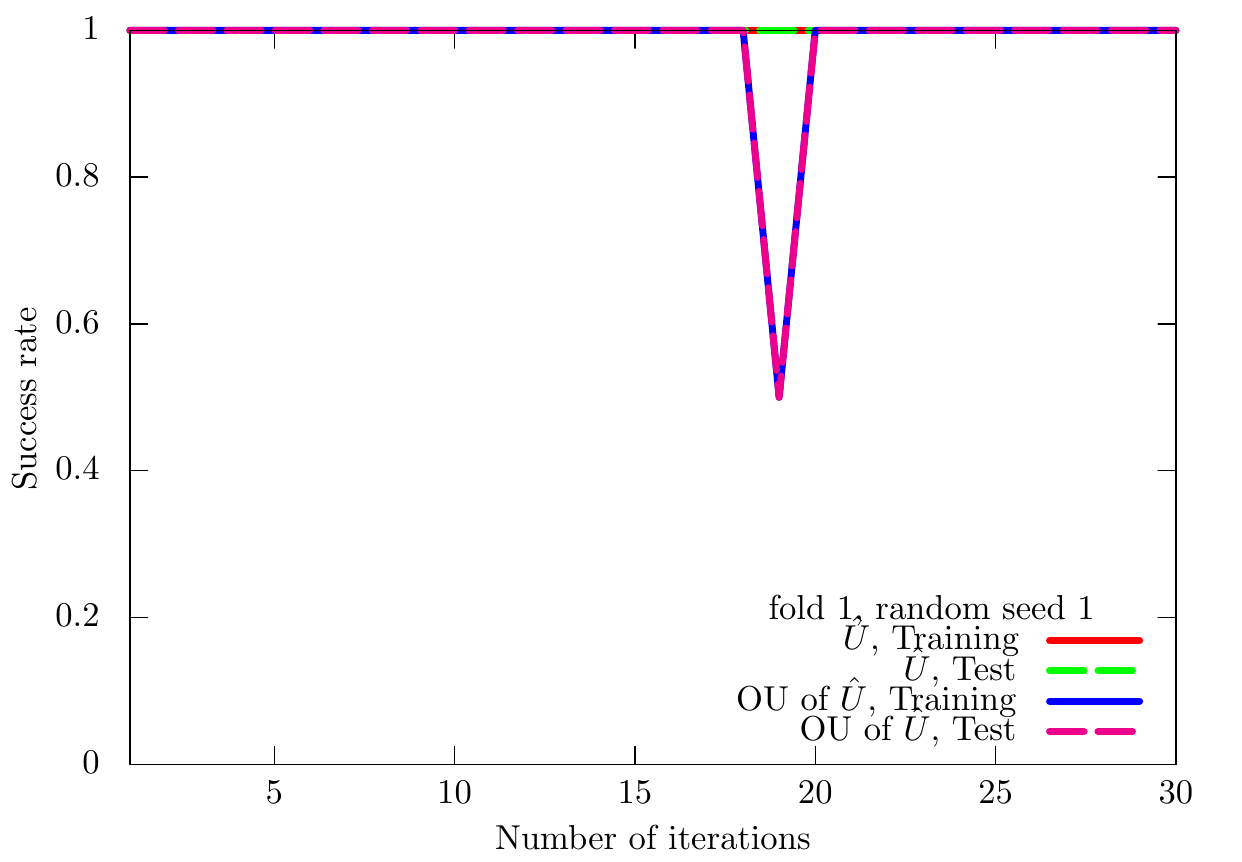}
\includegraphics[scale=0.25]{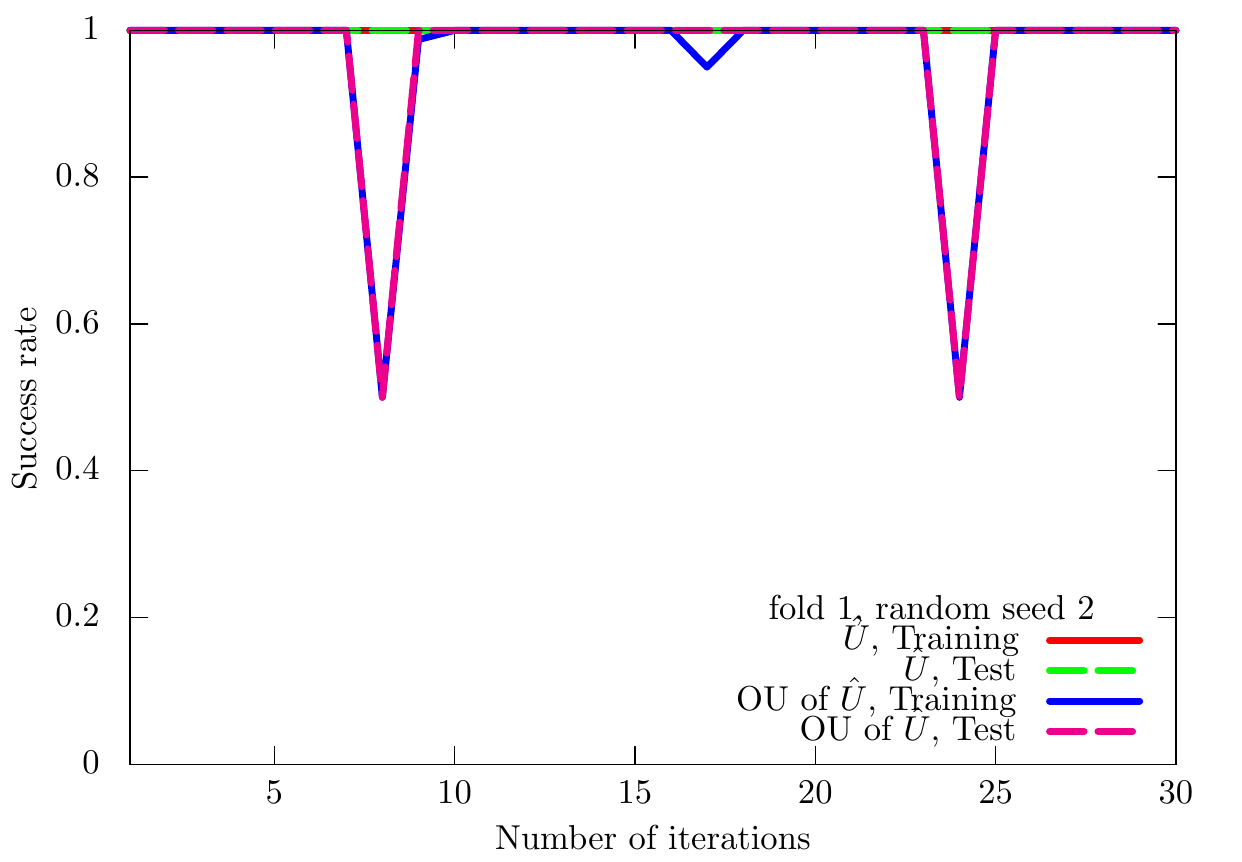}
\includegraphics[scale=0.25]{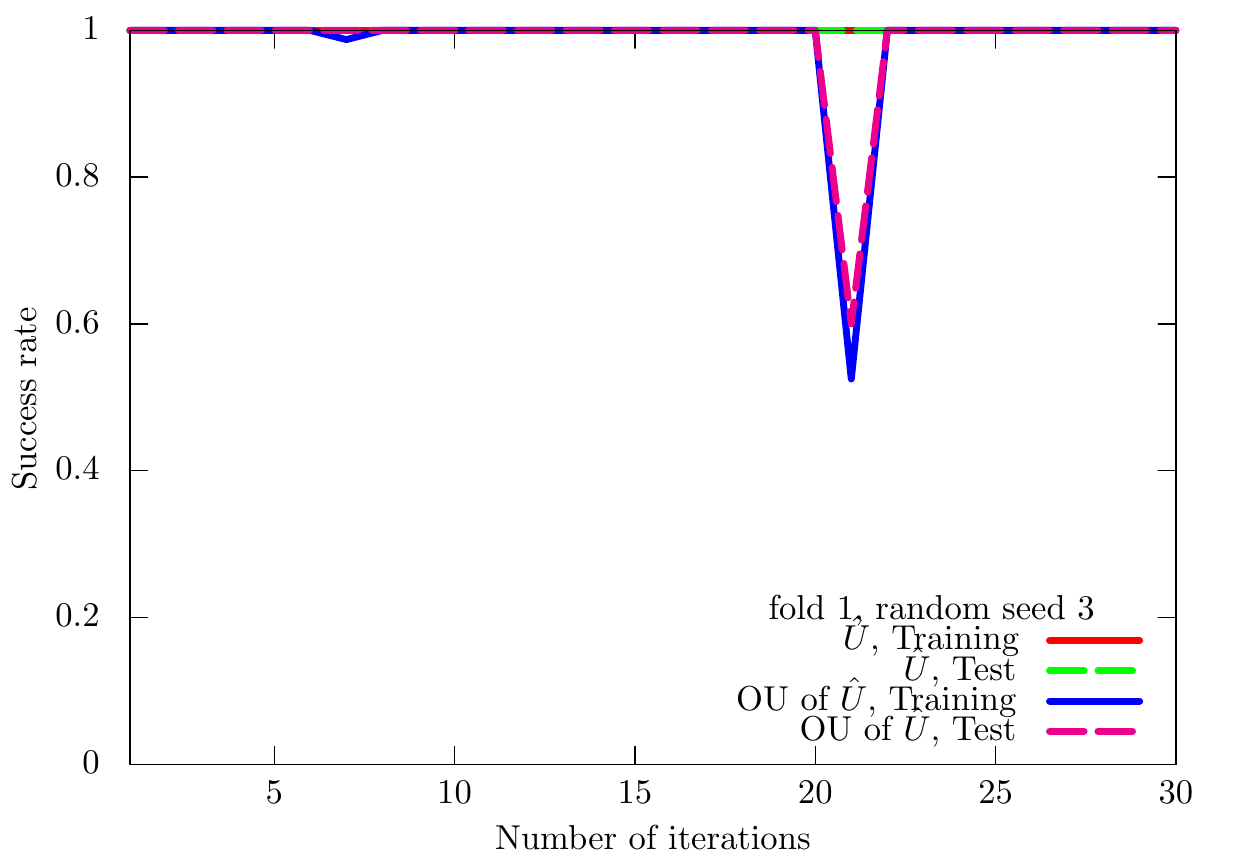}
\includegraphics[scale=0.25]{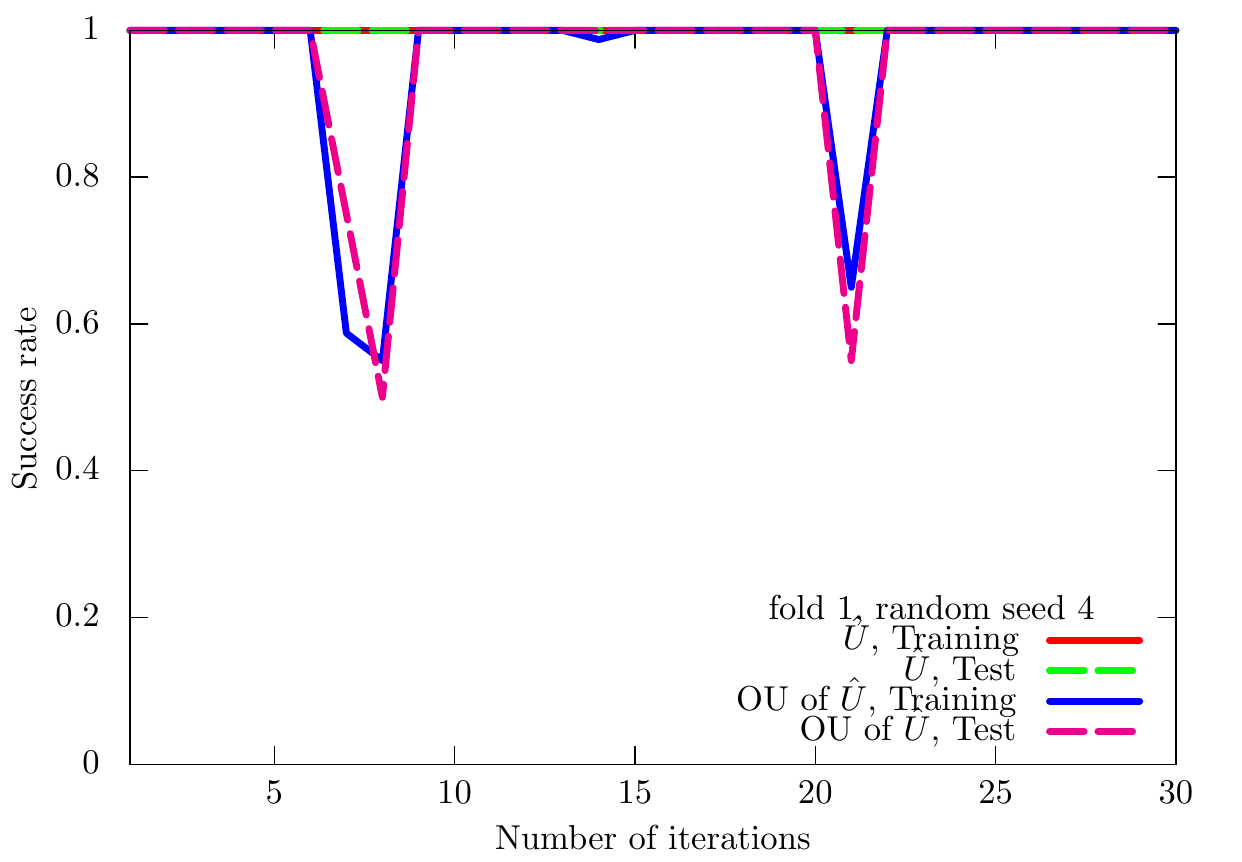}
\includegraphics[scale=0.25]{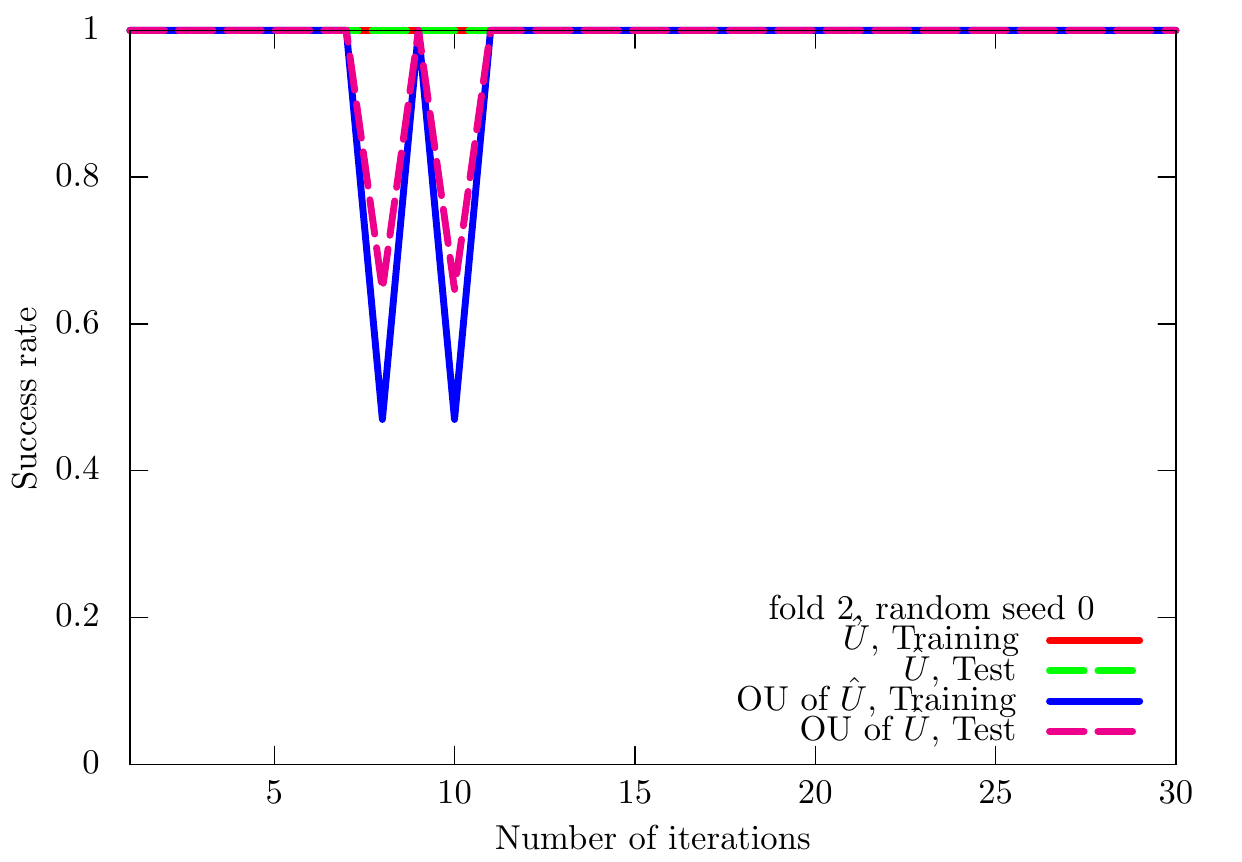}
\includegraphics[scale=0.25]{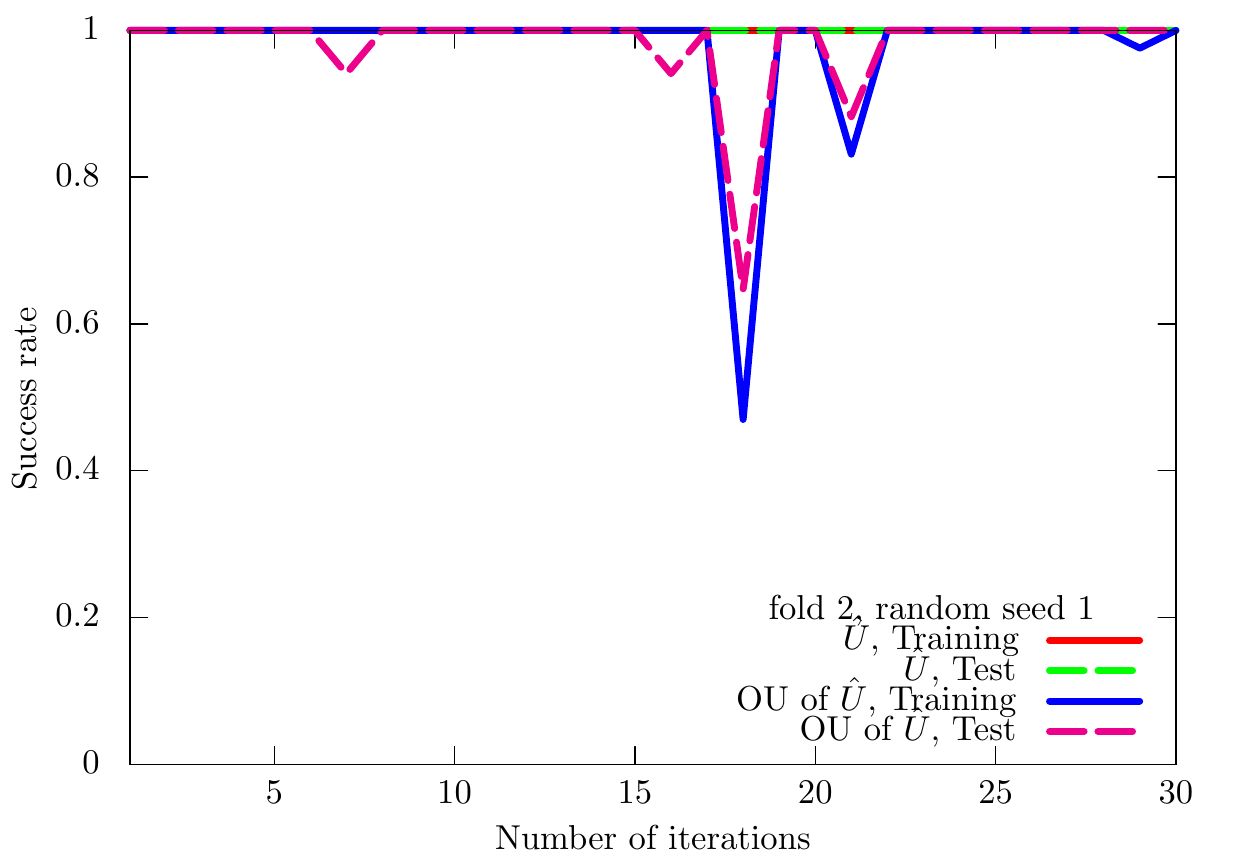}
\includegraphics[scale=0.25]{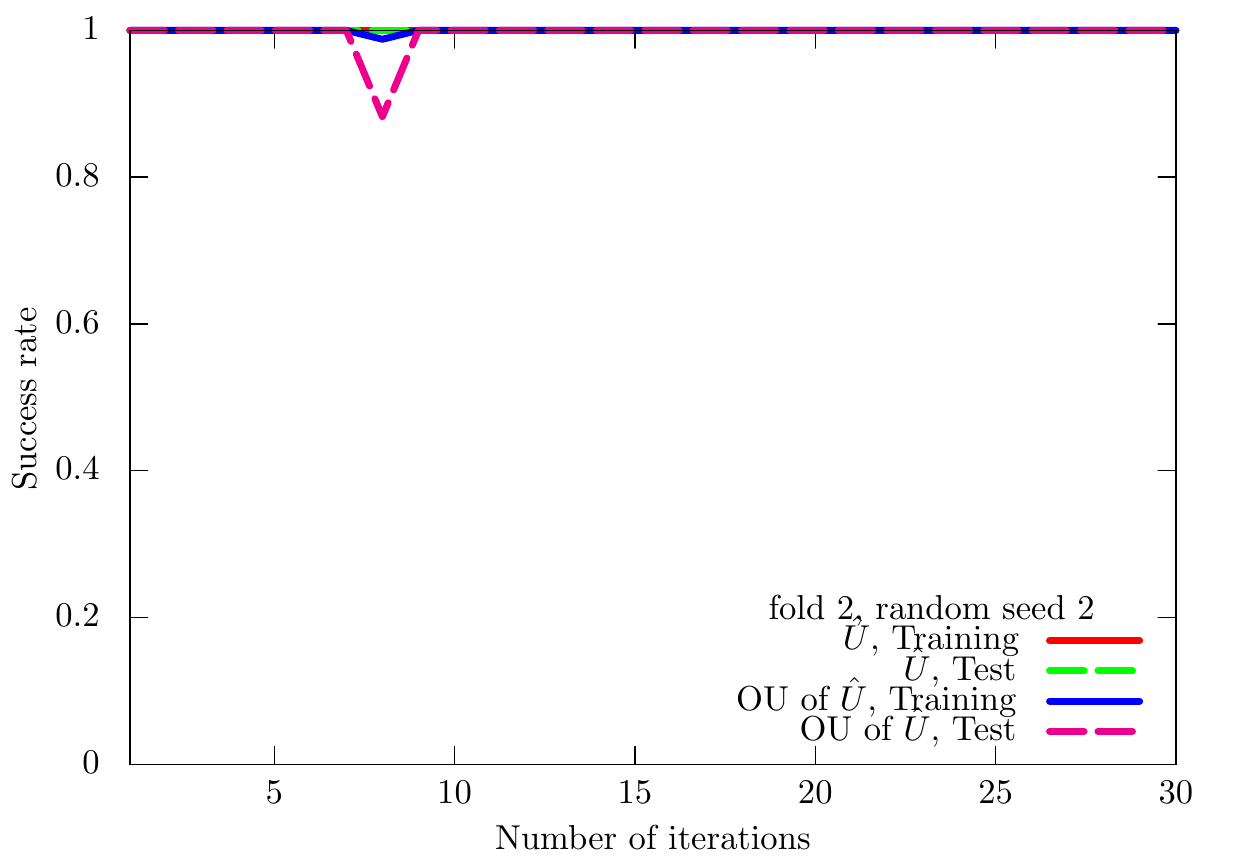}
\includegraphics[scale=0.25]{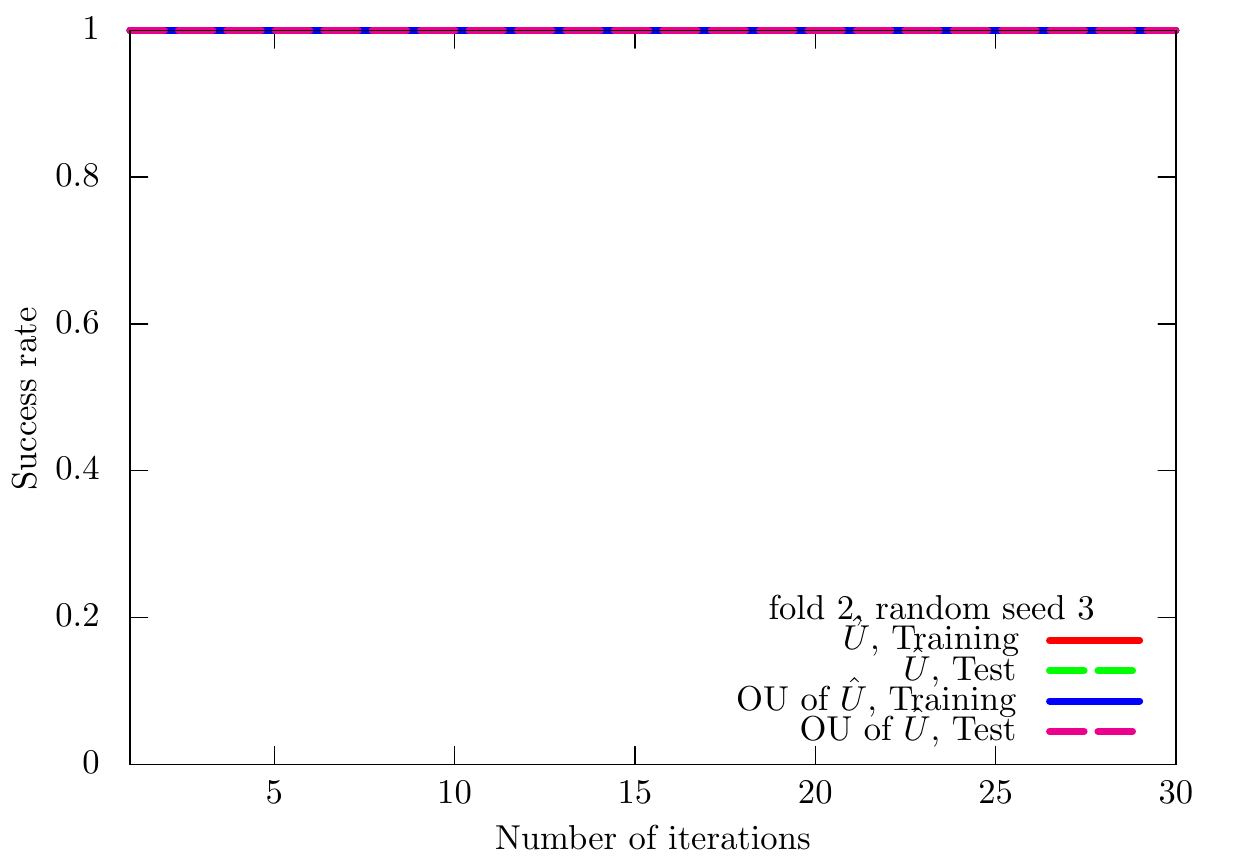}
\includegraphics[scale=0.25]{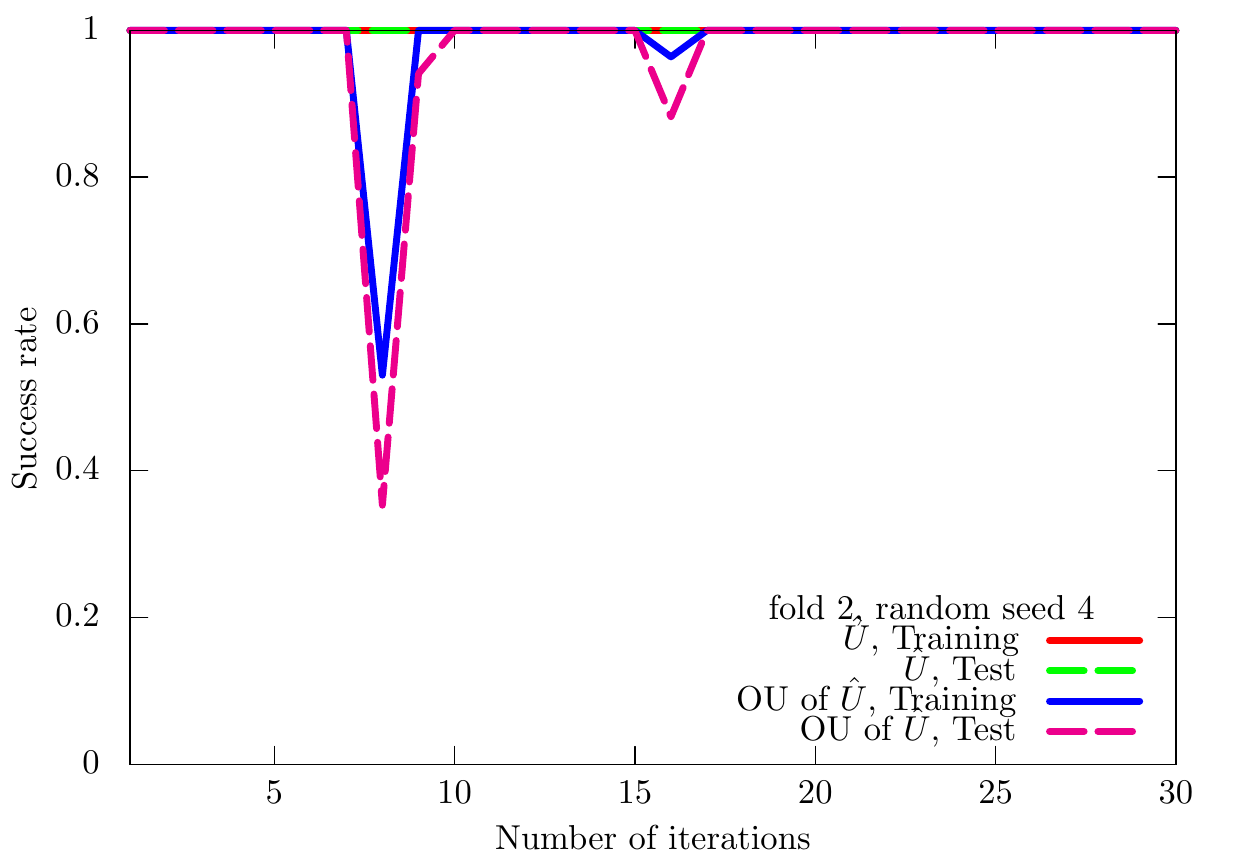}
\includegraphics[scale=0.25]{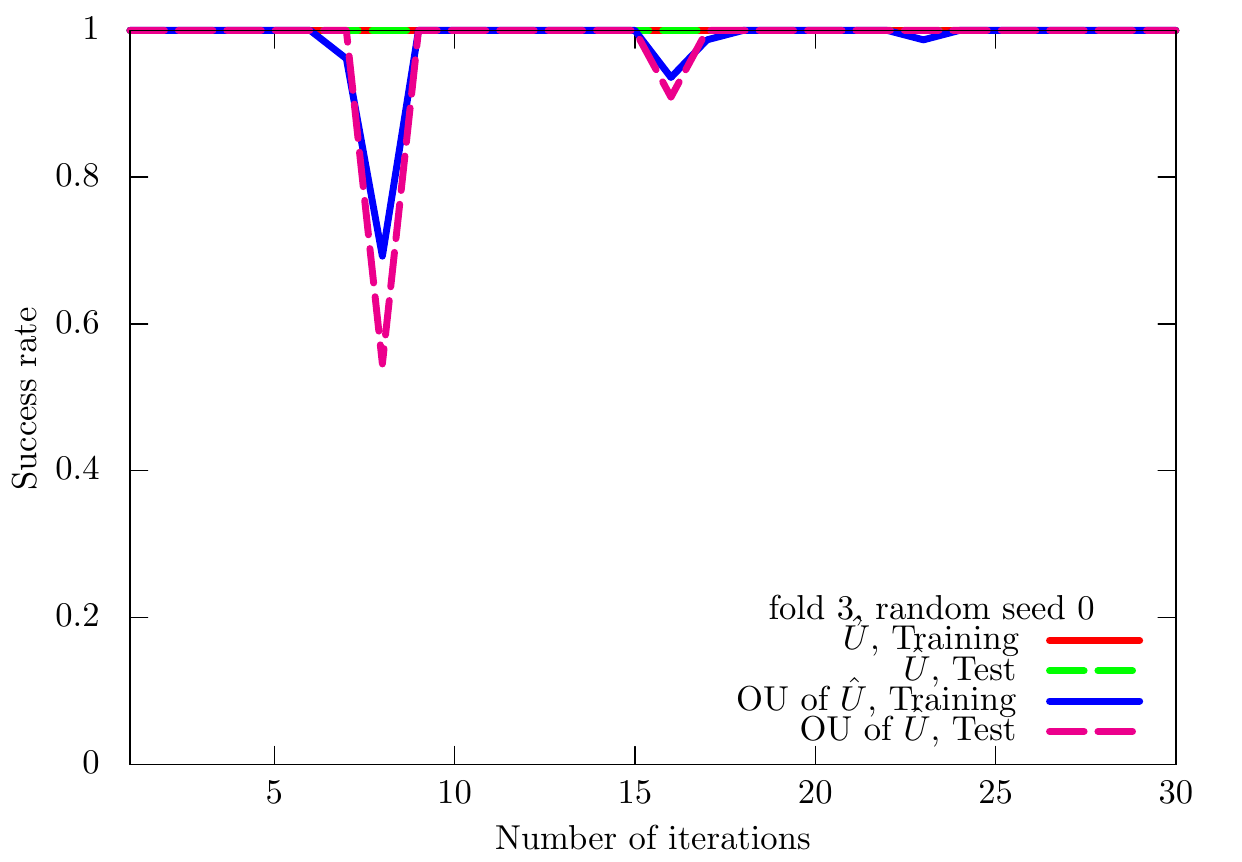}
\includegraphics[scale=0.25]{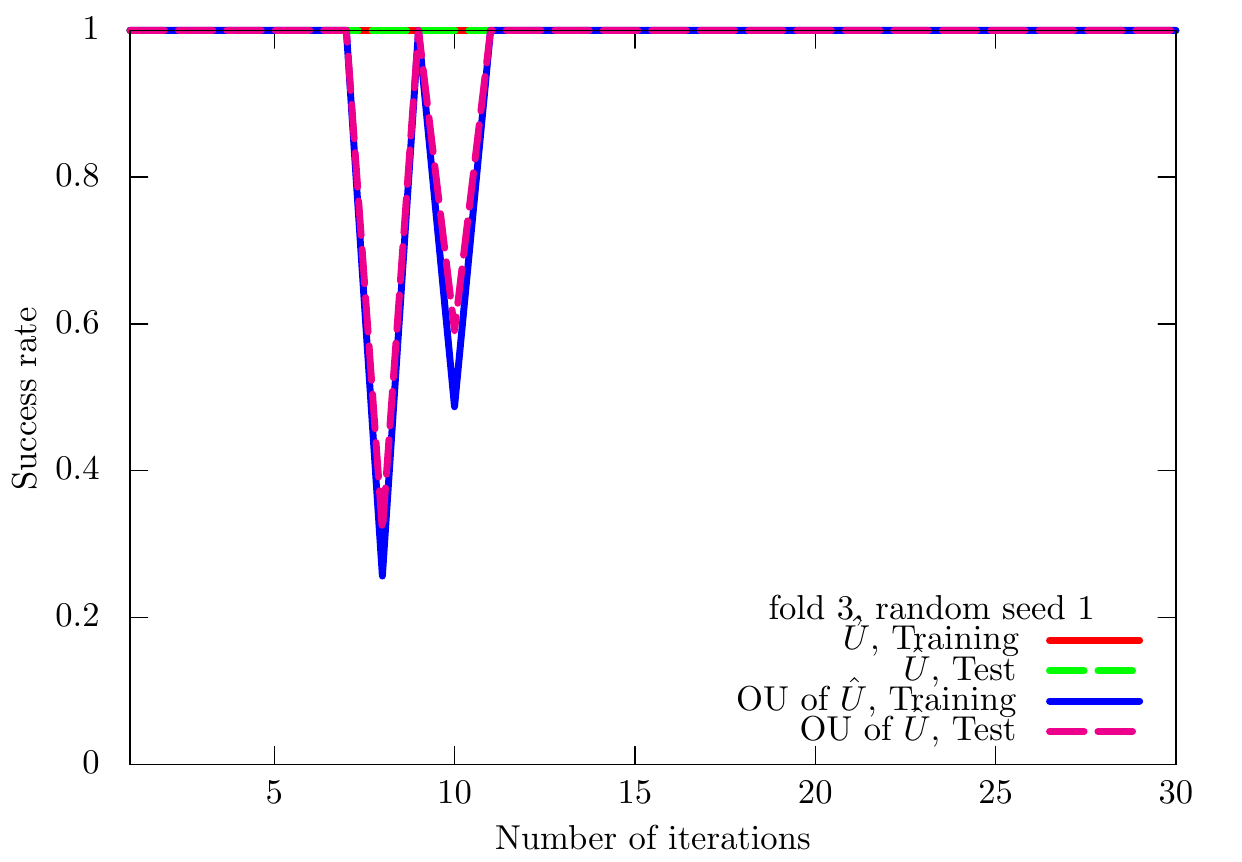}
\includegraphics[scale=0.25]{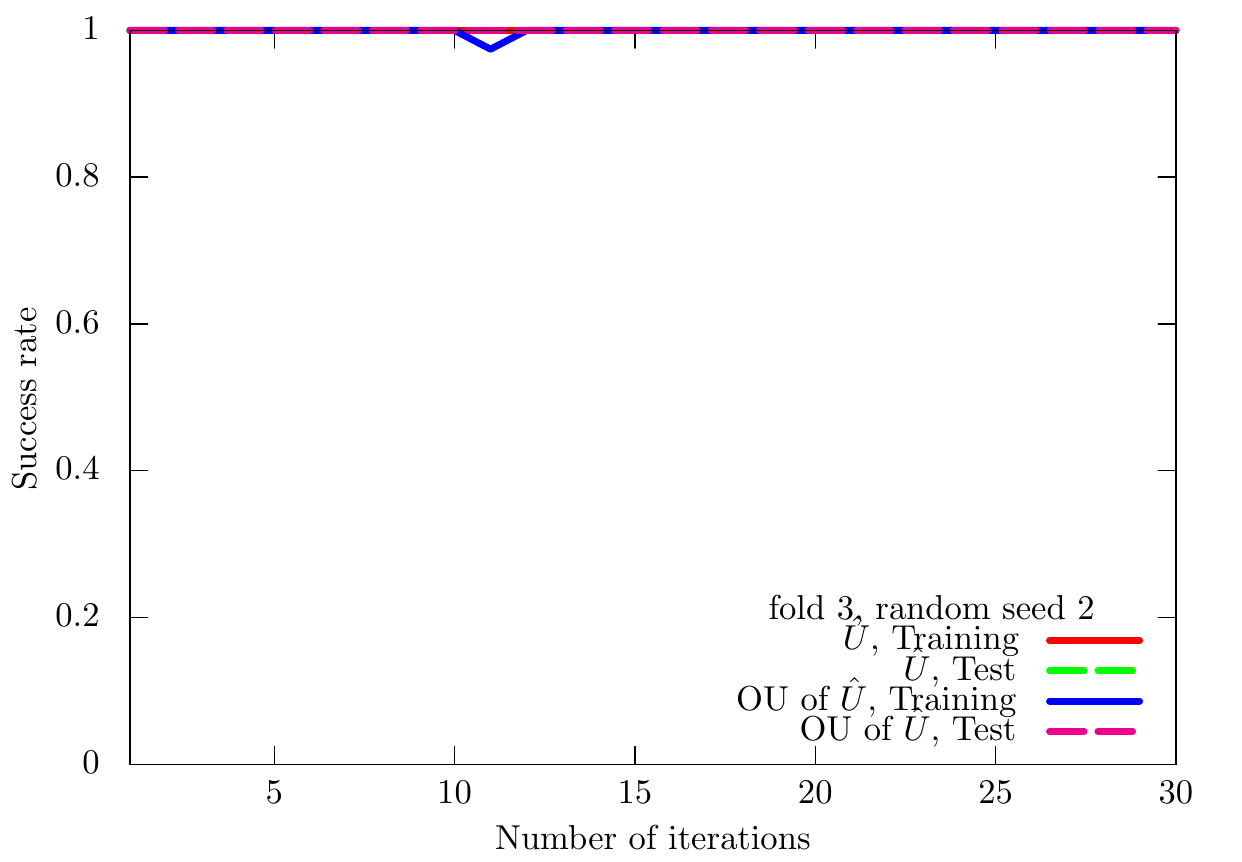}
\includegraphics[scale=0.25]{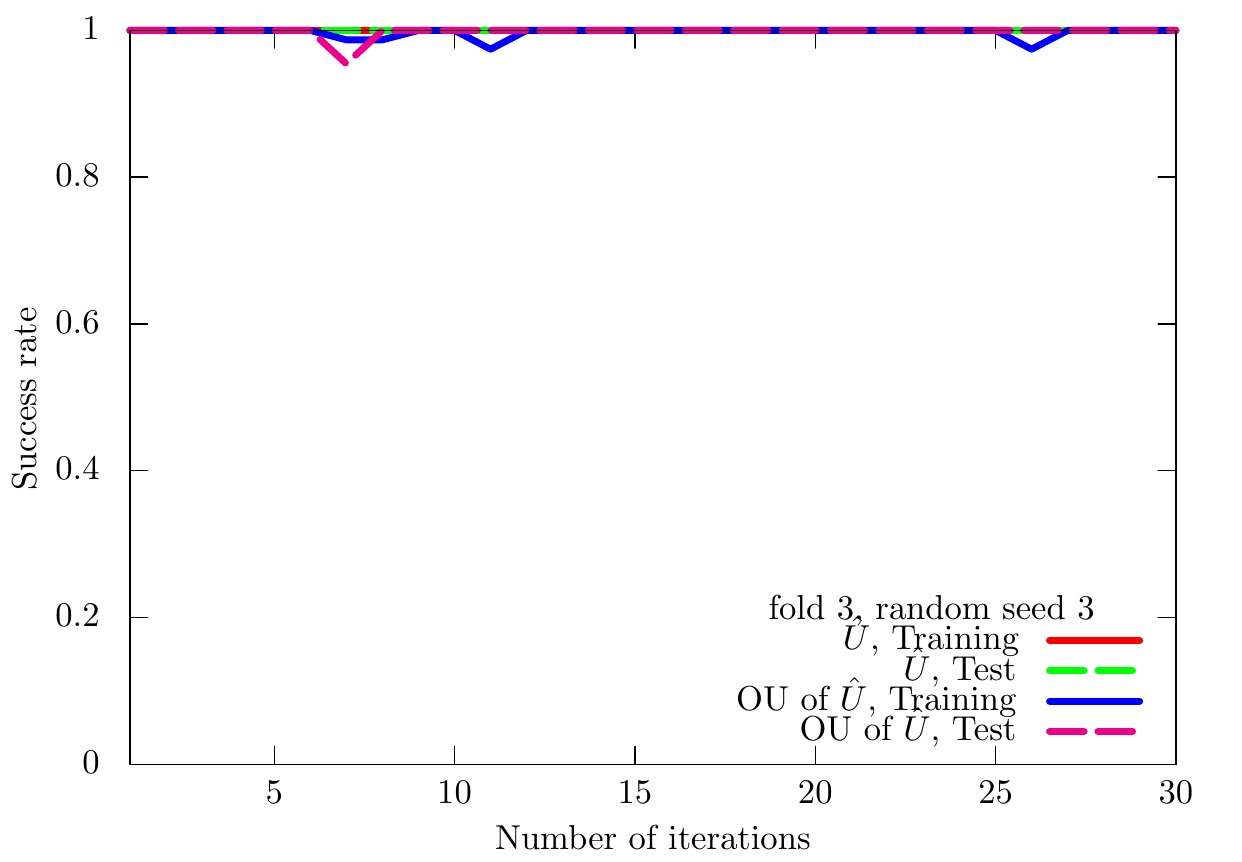}
\includegraphics[scale=0.25]{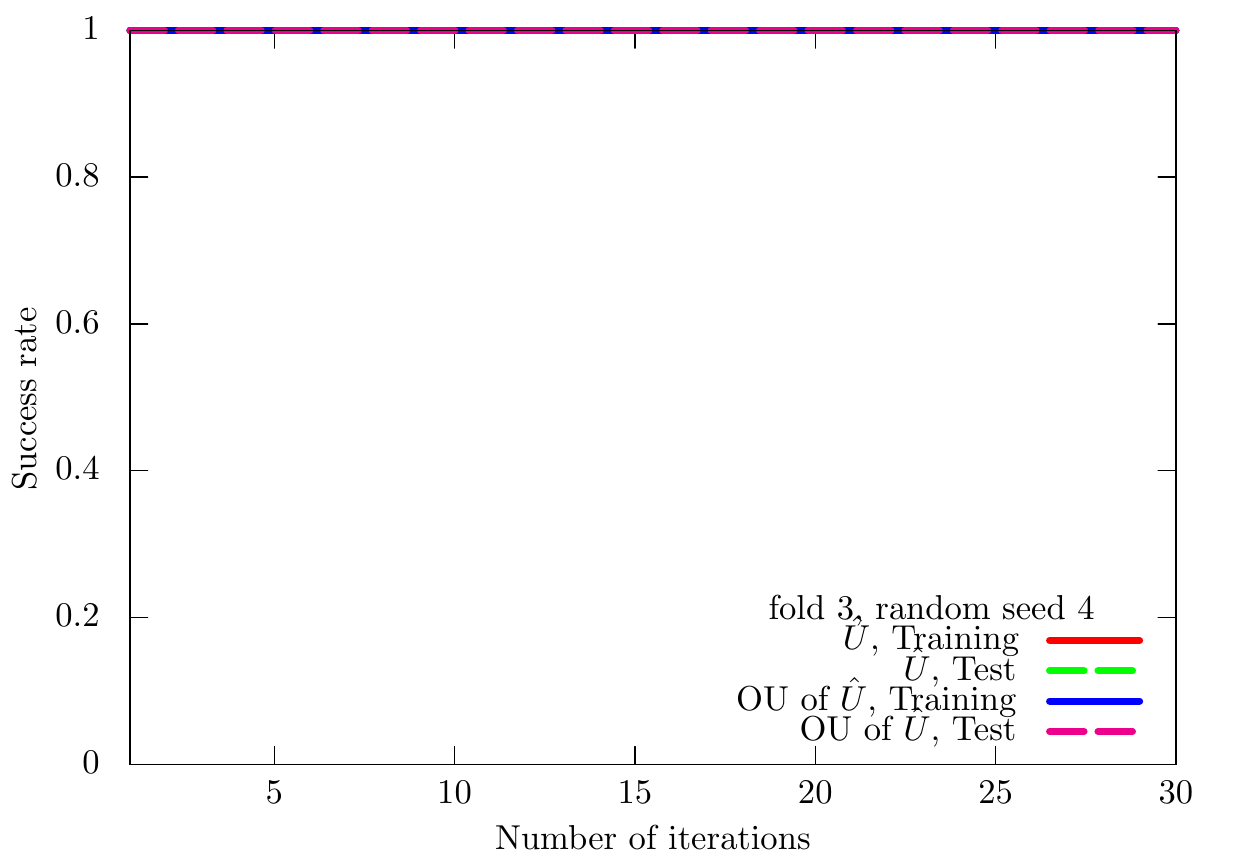}
\includegraphics[scale=0.25]{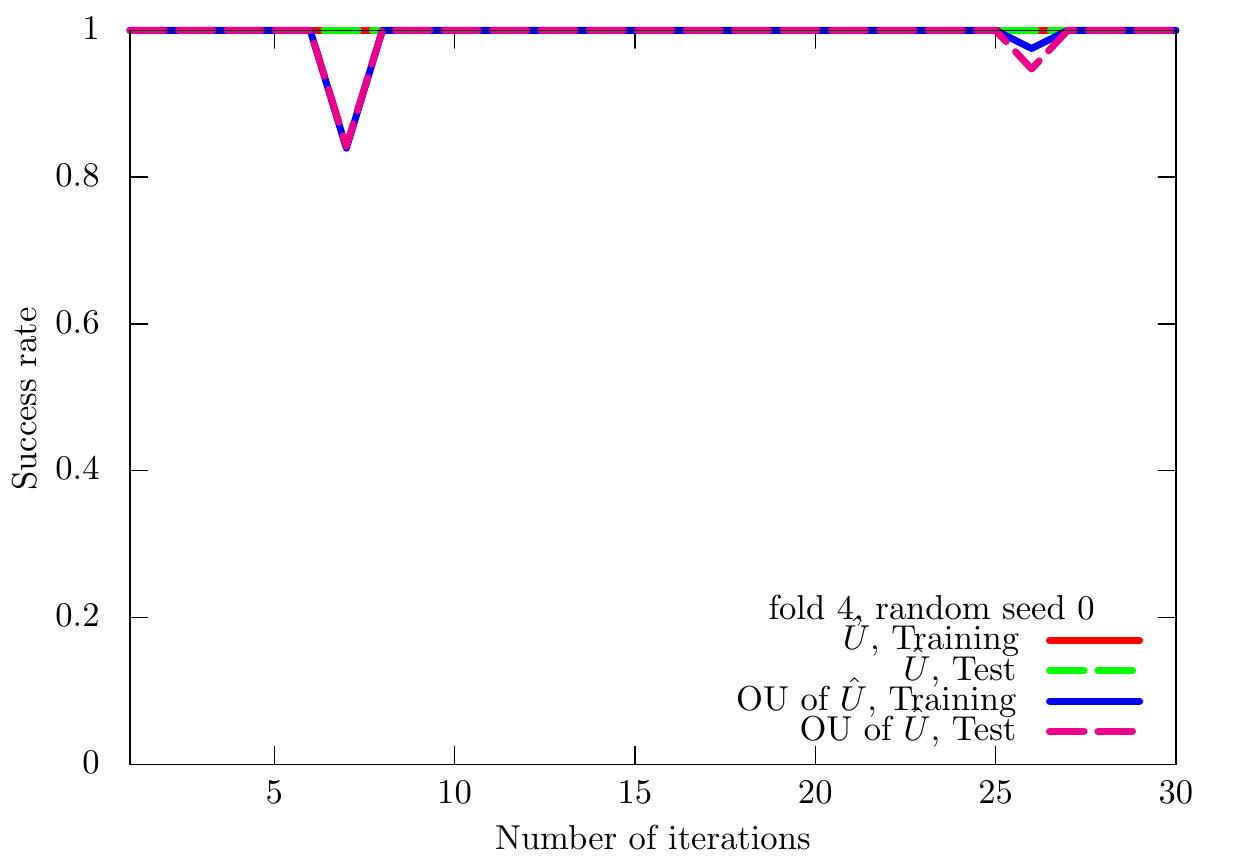}
\includegraphics[scale=0.25]{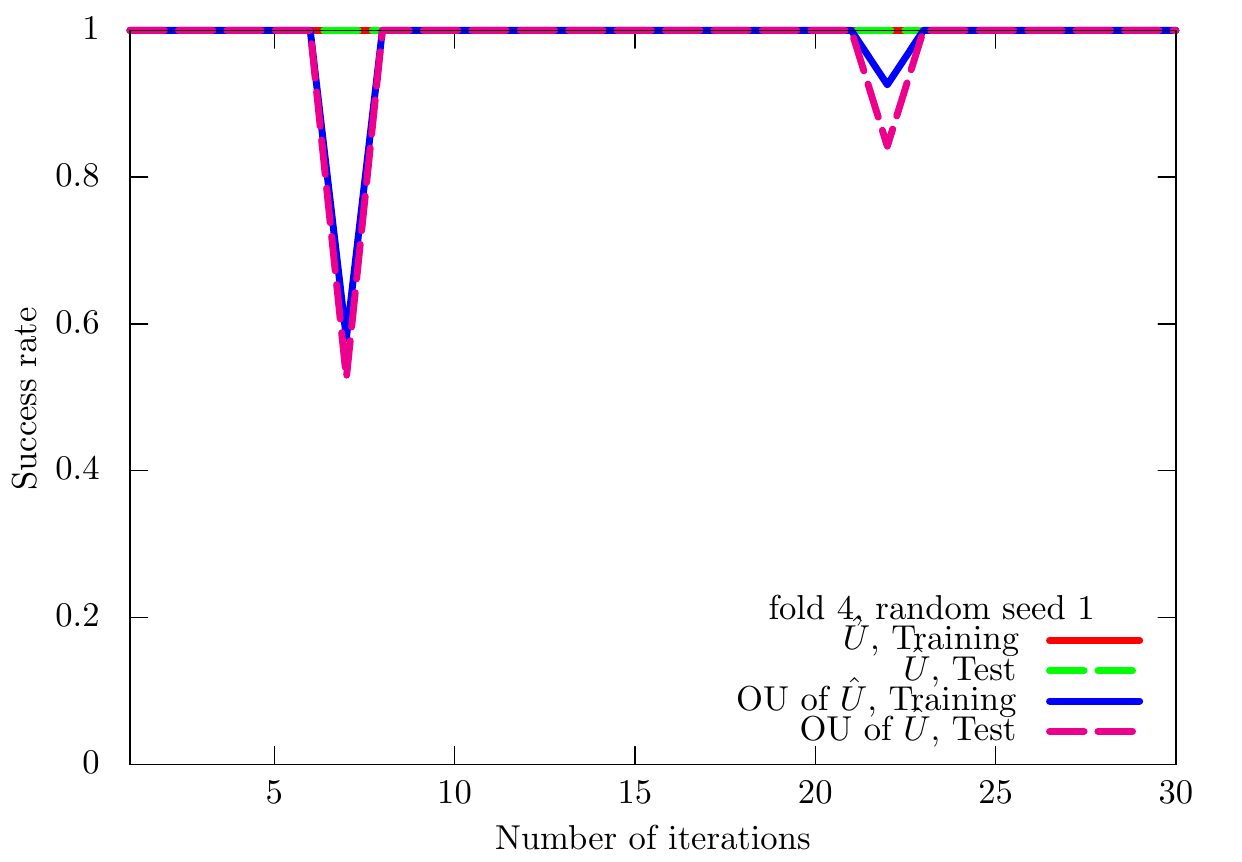}
\includegraphics[scale=0.25]{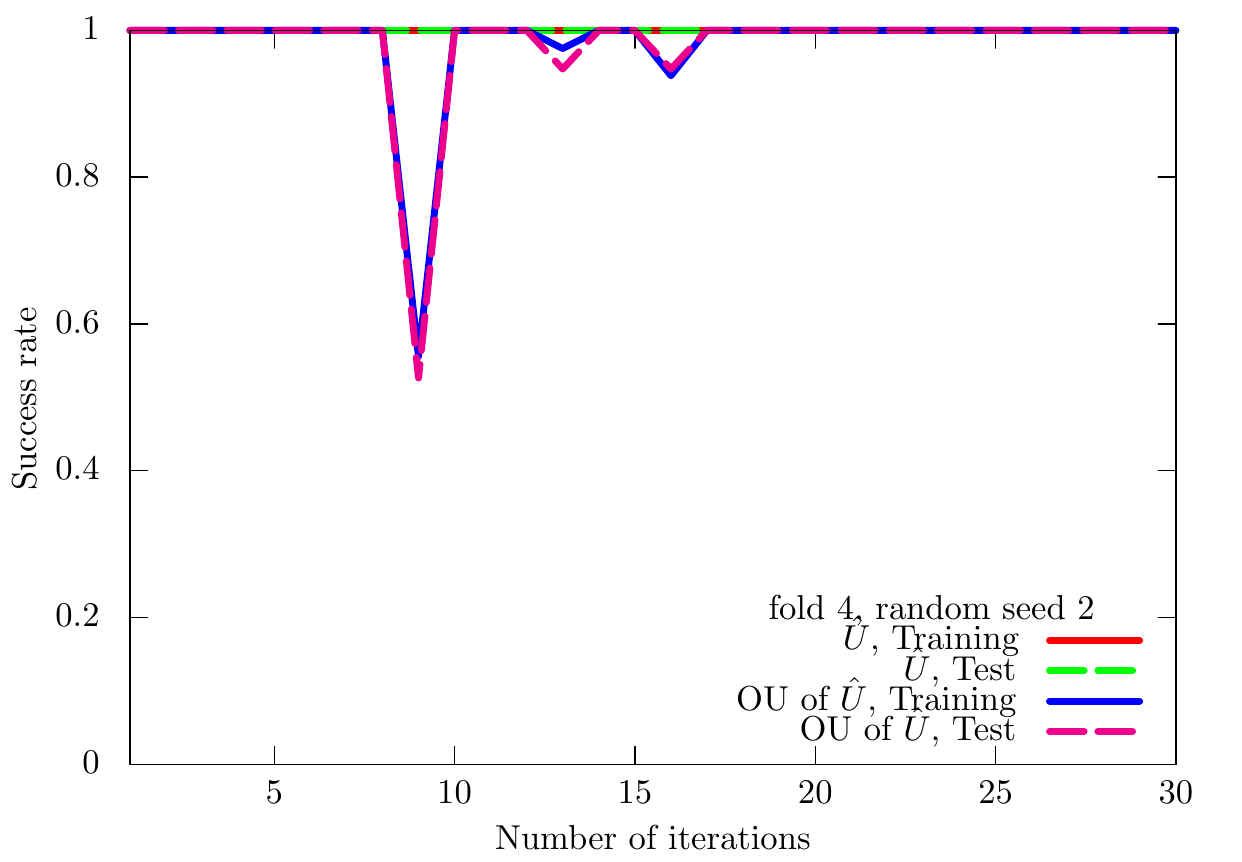}
\includegraphics[scale=0.25]{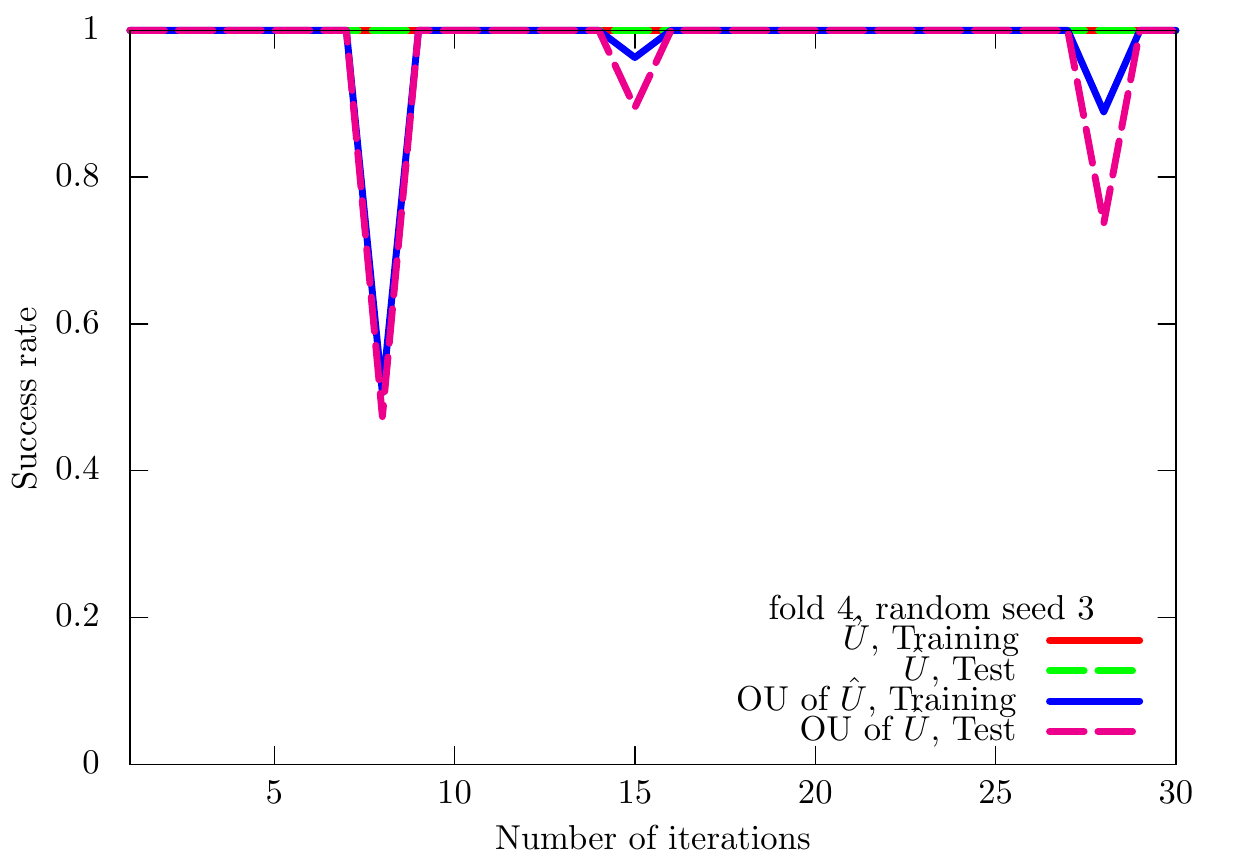}
\includegraphics[scale=0.25]{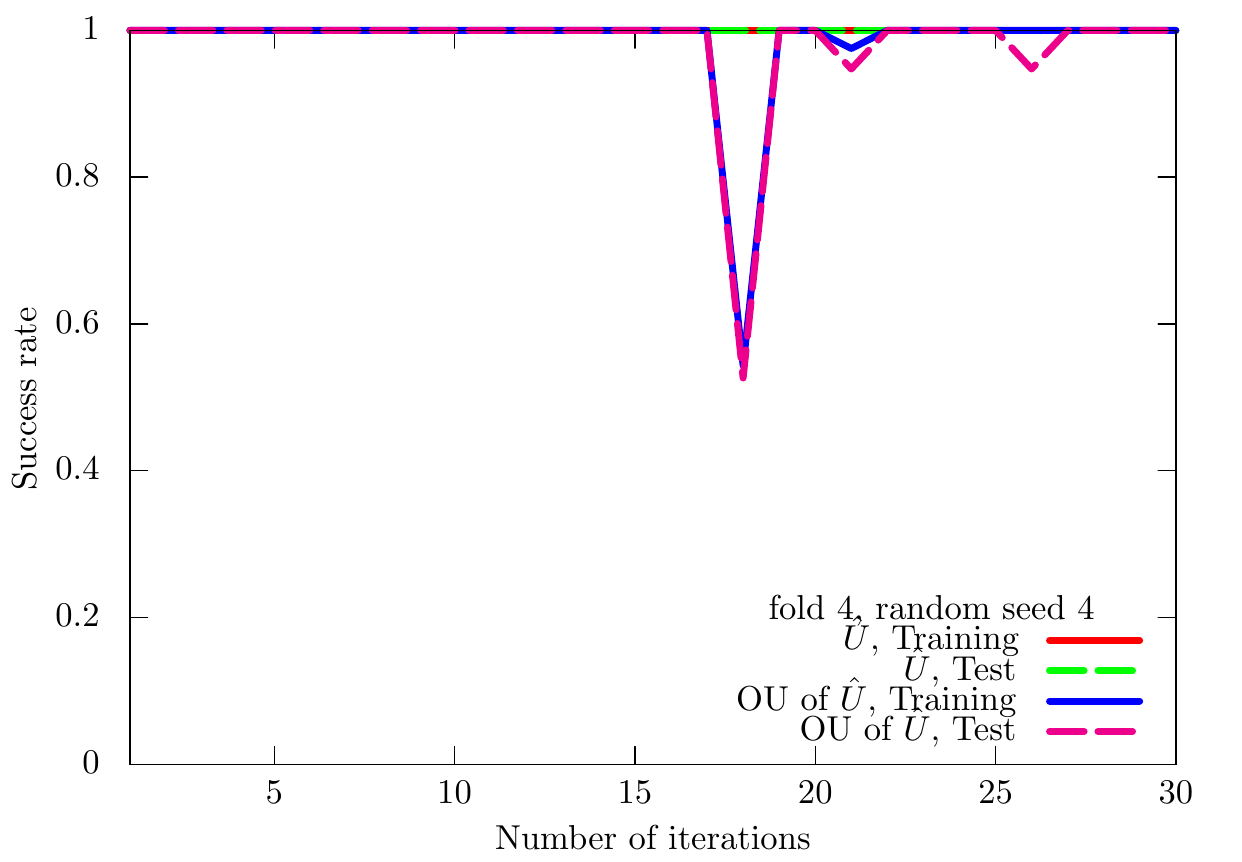}
\caption{Results of the UKM ($\hat{X}$ and OU of $\hat{X}$) on the $5$-fold datasets with $5$ different random seeds for the iris dataset ($0$ or $1$). We use complex matrices and set $\theta_\mathrm{bias} = 0$. We set $r = 0.010$.}
\label{supp-arXiv-numerical-result-raw-data-fold-001-rand-001-UKM-OUU-UCI-iris-0-1}
\end{figure*}

In Fig.~\ref{supp-arXiv-numerical-result-raw-data-fold-001-rand-001-QCL-UCI-iris-0-1}, we show the numerical results of QCL for the $5$-fold datasets with $5$ different random seeds.
In Fig.~\ref{supp-arXiv-numerical-result-raw-data-fold-001-rand-001-UKM-P-UCI-iris-0-1}, we show the numerical results of $\hat{P}$ of the UKM for the $5$-fold datasets with $5$ different random seeds.
In Fig.~\ref{supp-arXiv-numerical-result-raw-data-fold-001-rand-001-UKM-OUU-UCI-iris-0-1}, we also show the numerical results of OU of $\hat{X}$ of the UKM for the $5$-fold datasets with $5$ different random seeds.

\begin{figure}[htb]
\centering
\includegraphics[scale=0.45]{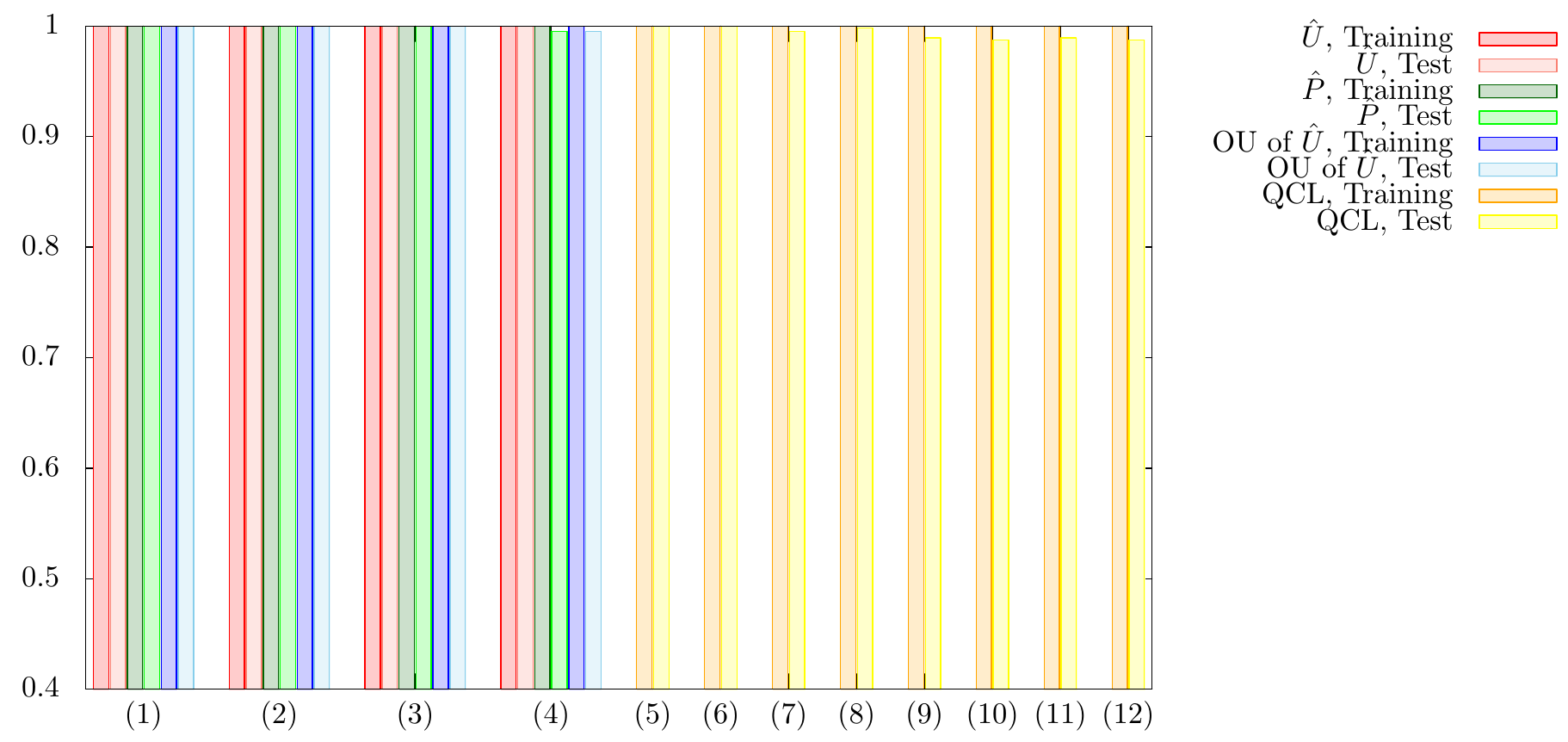}
\caption{Results of $5$-fold CV with $5$ different random seeds for the iris dataset ($0$ or $1$). For the UKM, we put $r = 0.010$ and set $K = 30$ and $K' = 10$. For QCL, the number of layers $L$ is $5$ and the number of iterations is $300$. The numerical settings are as follows: (1) complex matrices without the bias term, (2) complex matrices with the bias term, (3) real matrices without the bias term, (4) real matrices with the bias term, (5) CNOT-based circuit without the bias term, (6) CNOT-based circuit with the bias term, (7) CRot-based circuit without the bias term, (8) CRot-based circuit with the bias term, (9) 1d Heisenberg circuit without the bias term, (10) 1d Heisenberg circuit with the bias term, (11) FC Heisenberg circuit without the bias term, and (12) FC Heisenberg circuit with the bias term.}
\label{supp-arXiv-numerical-result-performance-UKM-QCL-UCI-iris-0-1}
\end{figure}

We summarize the results of 5-fold CV with 5 different random seeds of QCL and the UKM in Tables~\ref{supp-arXiv-table-UCI-iris-0-1-002} and \ref{supp-arXiv-table-UCI-iris-0-1-001}, respectively.
For QCL and the UKM, we select the best model for the training dataset over iterations to compute the performance.
\begin{table}[htb]
  \begin{tabular}{cc|cc}
    \hline \hline
    Algo. & Condition & Training & Test \\
    \hline
    QCL & CNOT-based, w/o bias & 1.0 & 1.0 \\
    QCL & CNOT-based, w/ bias & 1.0 & 1.0 \\
    \hline
    QCL & CRot-based, w/o bias & 1.0 & 0.9953 \\
    QCL & CRot-based, w/ bias & 1.0 & 0.9982 \\
    \hline
    QCL & 1d Heisenberg, w/o bias & 1.0 & 0.9895 \\
    QCL & 1d Heisenberg, w/ bias & 1.0 & 0.9874 \\
    \hline
    QCL & FC Heisenberg, w/o bias & 1.0 & 0.9895 \\
    QCL & FC Heisenberg, w/ bias & 1.0 & 0.9874 \\
    \hline \hline
  \end{tabular}
\caption{Results of $5$-fold CV with $5$ different random seeds of QCL for the iris dataset ($0$ or $1$). The number of layers $L$ is set to $5$ and the number of iterations is set to $300$.}
\label{supp-arXiv-table-UCI-iris-0-1-002}
\end{table}
\begin{table}[htb]
  \begin{tabular}{cc|cc}
    \hline \hline
    Algo. & Condition & Training & Test \\
    \hline
    UKM & $\hat{X}$, complex, w/o bias & 1.0 & 1.0 \\
    UKM & $\hat{P}$, complex, w/o bias & 1.0 & 1.0 \\
    UKM & OU of $\hat{X}$, complex, w/ bias & 1.0 & 1.0 \\
    \hline
    UKM & $\hat{X}$, complex, w/ bias & 1.0 & 1.0 \\
    UKM & $\hat{P}$, complex, w/ bias & 1.0 & 1.0 \\
    UKM & OU of $\hat{X}$, real, w/o bias & 1.0 & 1.0 \\
    \hline
    UKM & $\hat{X}$, real, w/o bias & 1.0 & 1.0 \\
    UKM & $\hat{P}$, real, w/o bias & 1.0 & 1.0 \\
    UKM & OU of $\hat{X}$, real w/o bias & 1.0 & 1.0 \\
    \hline
    UKM & $\hat{X}$, real, w/ bias & 1.0 & 1.0 \\
    UKM & $\hat{P}$, real, w/ bias & 1.0 & 0.9953 \\
    UKM & OU of $\hat{X}$, real, w/ bias & 1.0 & 0.9953 \\
    \hline \hline
  \end{tabular}
\caption{Results of $5$-fold CV with $5$ different random seeds of the UKM for the iris dataset ($0$ or $1$). We put $r = 0.010$ and set $K = 30$ and $K' = 10$.}
\label{supp-arXiv-table-UCI-iris-0-1-001}
\end{table}
In Fig.~\ref{supp-arXiv-numerical-result-performance-UKM-QCL-UCI-iris-0-1}, we plot the data shown in Tables~\ref{supp-arXiv-table-UCI-iris-0-1-002} and \ref{supp-arXiv-table-UCI-iris-0-1-001}.
We also summarize the results of 5-fold CV with 5 different random seeds of the kernel method in Table~\ref{supp-arXiv-table-UCI-iris-0-1-003}.
More specifically, we use Ridge classification in Sec.~\ref{supp-arXiv-sec-Ridge-001}.
We consider the linear functions and the second-order polynomial functions for $\phi (\cdot)$ in Eq.~\eqref{supp-arXiv-f-pred-kernel-method-001-002} with and without normalization.
We set $\lambda = 10^{-2}, 10^{-1}, 1$ where $\lambda$ is the coefficient of the regularization term.
\begin{table}[htb]
  \begin{tabular}{cc|cc}
    \hline \hline
    Algo. & Condition & Training & Test \\
    \hline
  Kernel method & Linear, w/o normalization, $\lambda = 10^{-2}$ & 1.0000 & 1.0000 \\
  Kernel method & Linear, w/o normalization, $\lambda = 10^{-1}$ & 1.0000 & 1.0000 \\
  Kernel method & Linear, w/o normalization, $\lambda = 1$ & 1.0000 & 1.0000 \\
    \hline
  Kernel method & Linear, w/ normalization, $\lambda = 10^{-2}$ & 1.0000 & 1.0000 \\
  Kernel method & Linear, w/ normalization, $\lambda = 10^{-1}$ & 1.0000 & 1.0000 \\
  Kernel method & Linear, w/ normalization, $\lambda = 1$ & 1.0000 & 1.0000 \\
    \hline
  Kernel method & Poly-2, w/o normalization, $\lambda = 10^{-2}$ & 1.0000 & 1.0000 \\
  Kernel method & Poly-2, w/o normalization, $\lambda = 10^{-1}$ & 1.0000 & 1.0000 \\
  Kernel method & Poly-2, w/o normalization, $\lambda = 1$ & 1.0000 & 1.0000 \\
    \hline
  Kernel method & Poly-2, w/ normalization, $\lambda = 10^{-2}$ & 1.0000 & 1.0000 \\
  Kernel method & Poly-2, w/ normalization, $\lambda = 10^{-1}$ & 1.0000 & 1.0000 \\
  Kernel method & Poly-2, w/ normalization, $\lambda = 1$ & 1.0000 & 1.0000 \\
    \hline \hline
  \end{tabular}
\caption{Results of 5-fold CV with 5 different random seeds of the kernel method for the iris dataset ($0$ or $1$).}
\label{supp-arXiv-table-UCI-iris-0-1-003}
\end{table}

\begin{figure}[htb]
\centering
\includegraphics[scale=0.45]{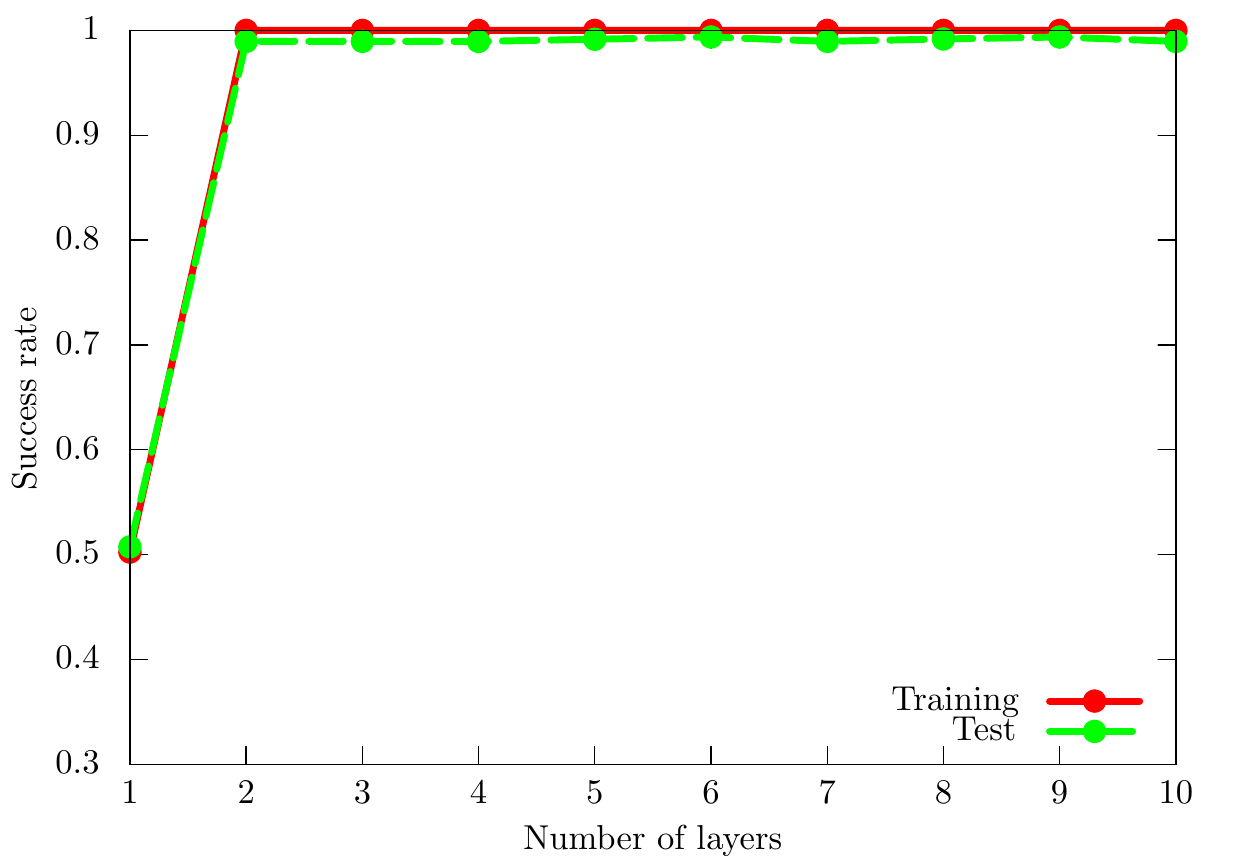}
\caption{Performance dependence of QCL on the number of layers $L$ for the iris dataset ($0$ or $1$). We use the CNOT-based circuit geometry and set $\theta_\mathrm{bias} = 0$. We iterate the computation $300$ times.}
\label{supp-arXiv-numerical-result-layers-dependence-QCL-UCI-iris-0-1}
\end{figure}

\begin{figure}[htb]
\centering
\includegraphics[scale=0.45]{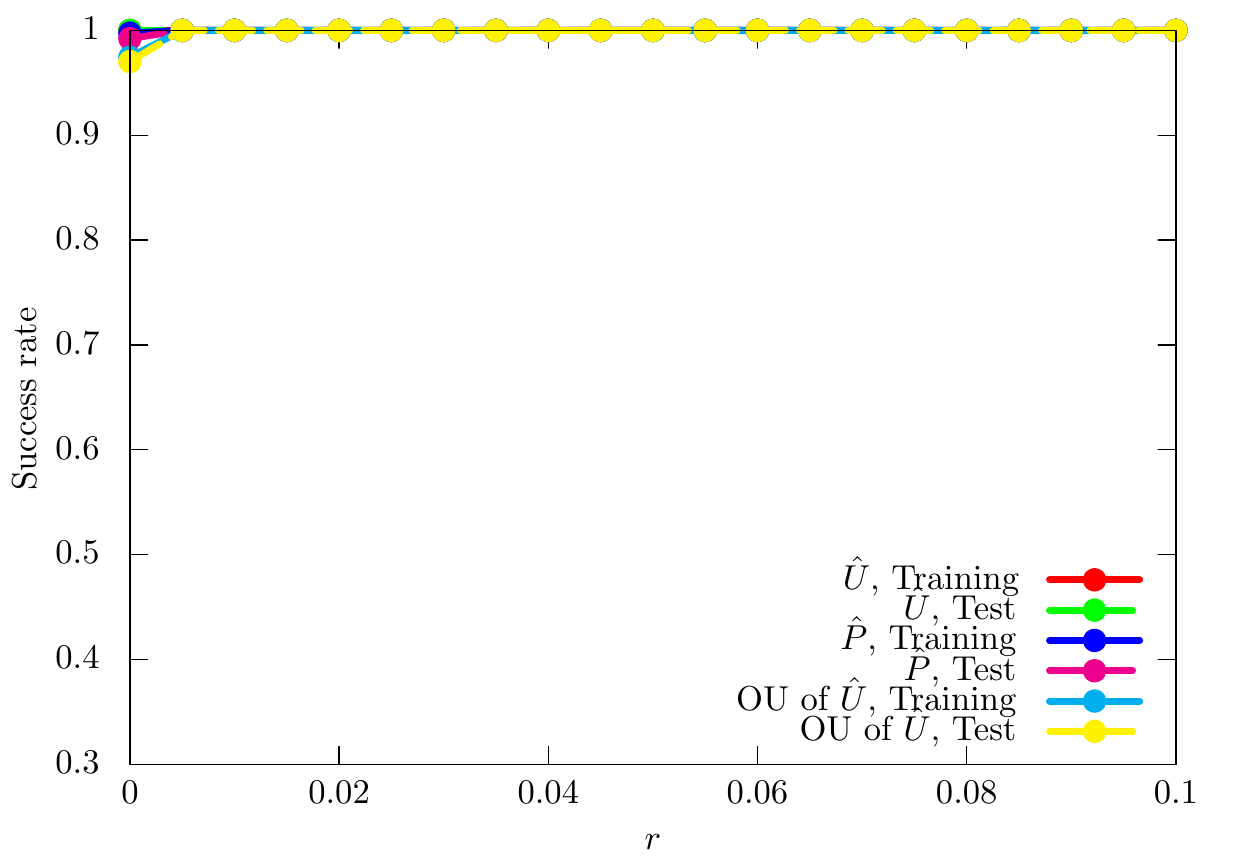}
\includegraphics[scale=0.45]{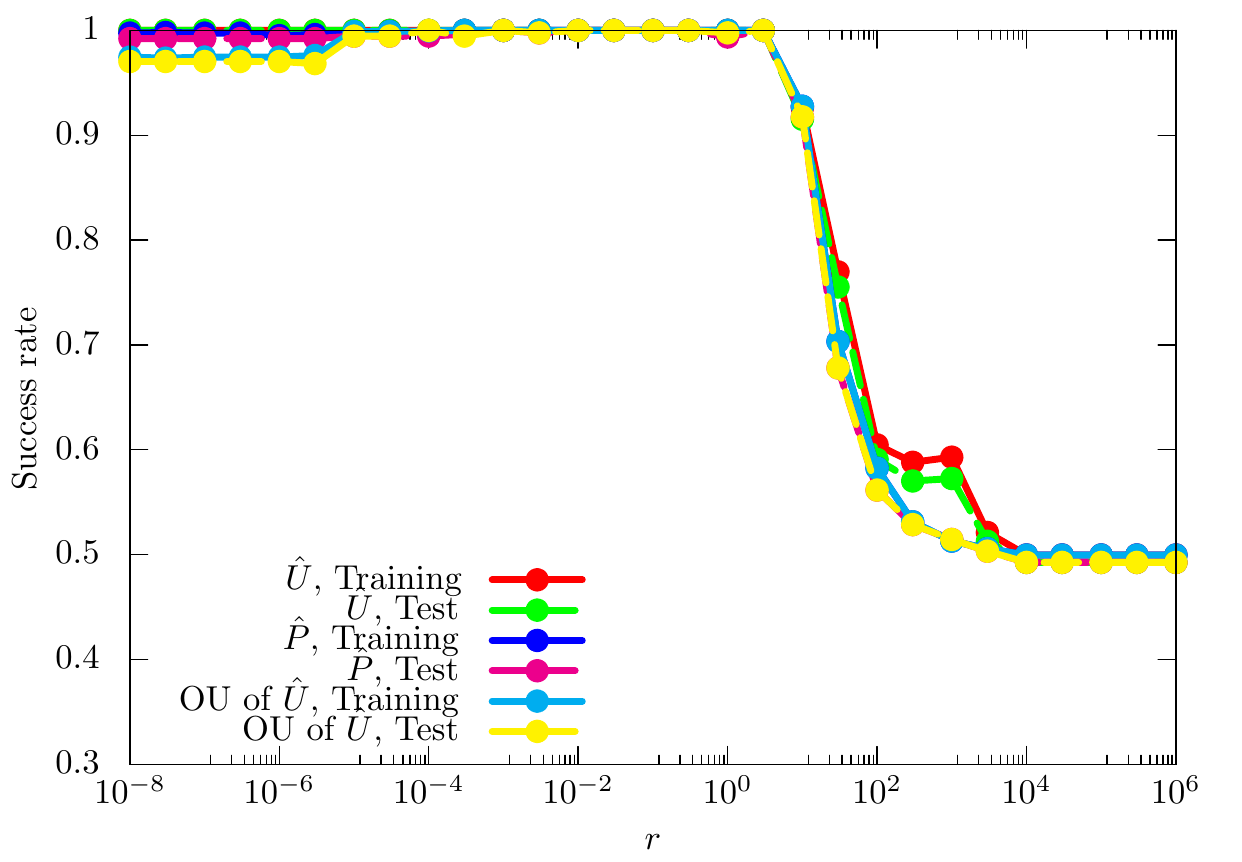}
\caption{Performance dependence of the UKM on $r$, which is the coefficient of the second term in the right-hand side of Eq.~\eqref{supp-arXiv-quantum-kernel-method-001-011} for the iris dataset ($0$ or $1$). We use complex matrices and set $\theta_\mathrm{bias} = 0$. We set $K = 30$ and $K' = 10$.}
\label{supp-arXiv-numerical-result-r-dependence-UKM-UCI-iris-0-1}
\end{figure}

\begin{figure}[htb]
\centering
\includegraphics[scale=0.45]{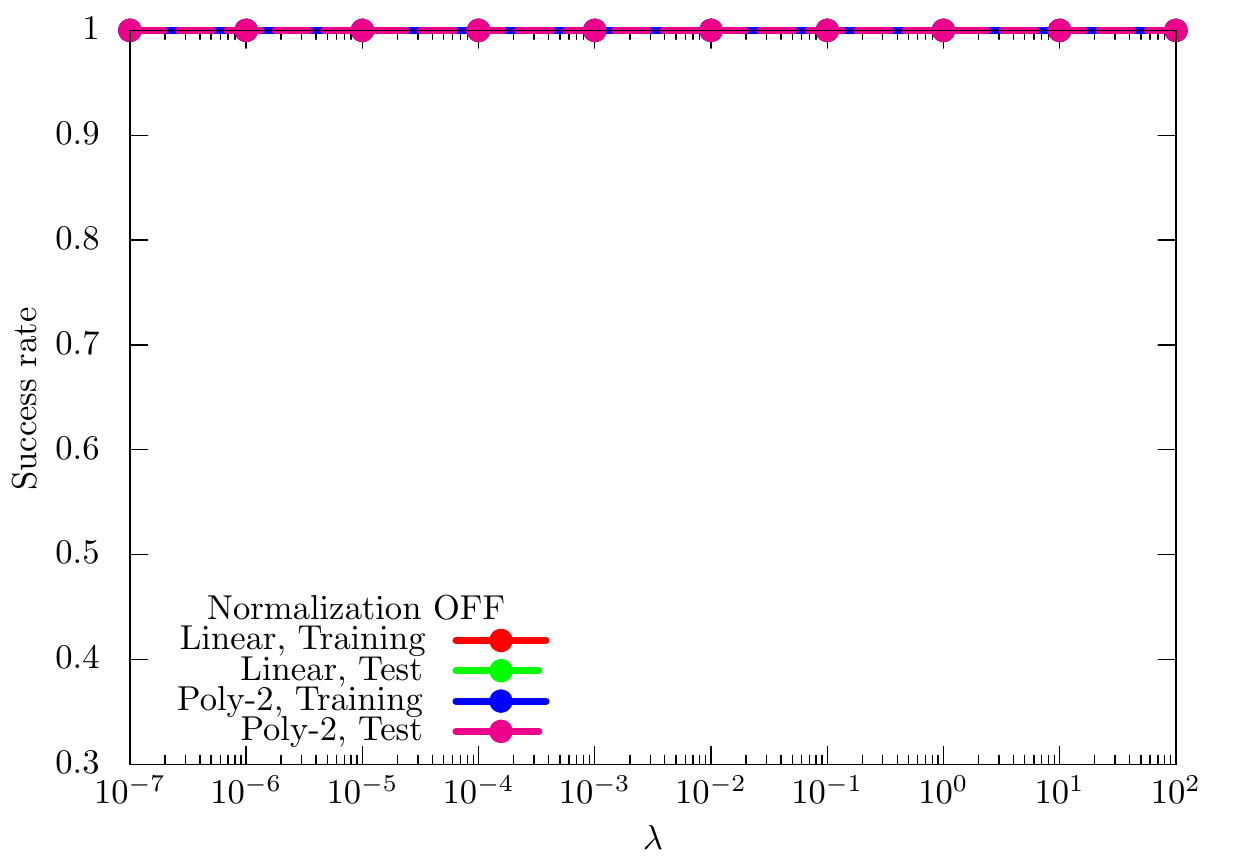}
\includegraphics[scale=0.45]{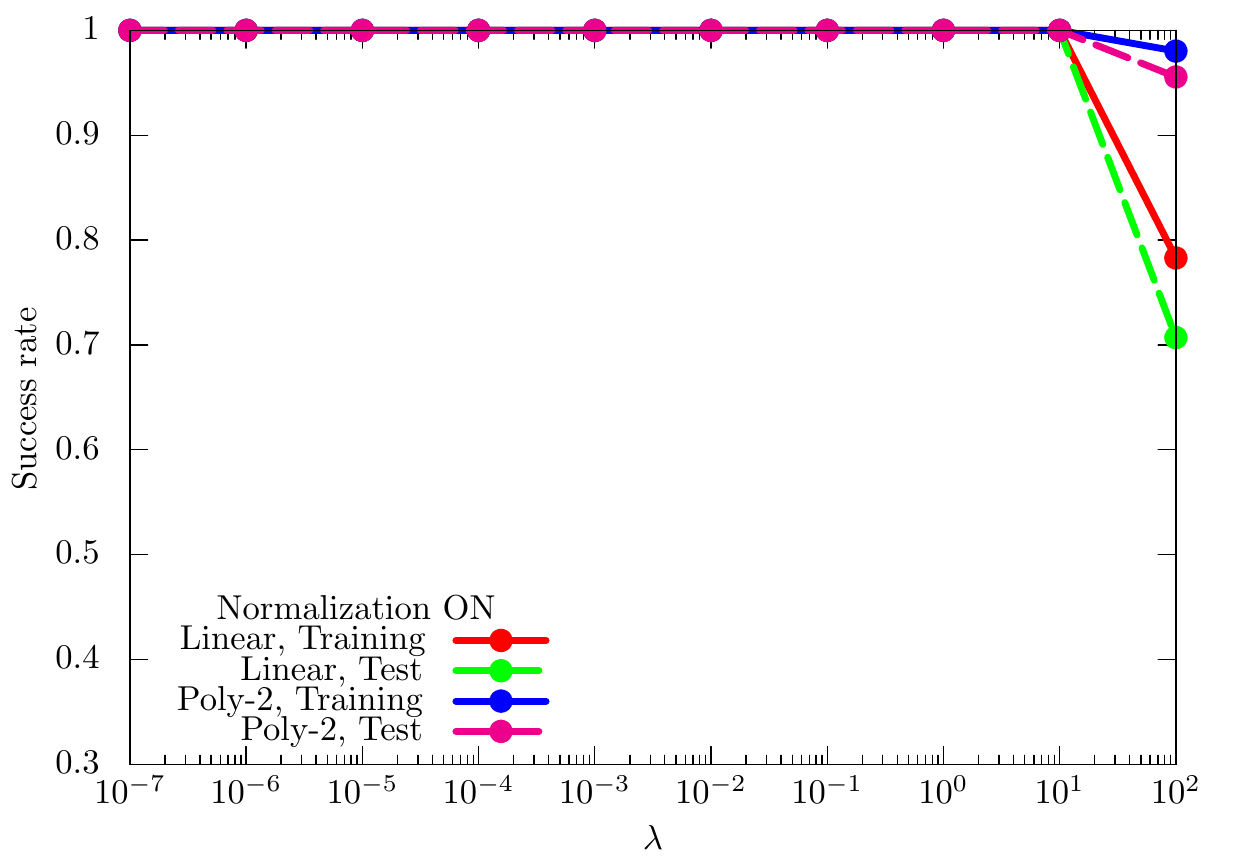}
\caption{Performance dependence of the kernel method on $\lambda$, which is the coefficient of the second term in the right-hand side of Eq.~\eqref{supp-arXiv-cost-function-kernel-method-001-002} for the iris dataset ($0$ or $1$). For $\phi (\cdot)$ in Eq.~\eqref{supp-arXiv-f-pred-kernel-method-001-002}, we use the linear functions and the second-degree polynomial functions with and without normalization.}
\label{supp-arXiv-numerical-result-lambda-dependence-kernel-method-iris-0-1}
\end{figure}

Next, we show the performance dependence of the three algorithms on their key parameters.
We see the performance dependence of QCL on the number of layers $L$.
The result is shown in Fig.~\ref{supp-arXiv-numerical-result-layers-dependence-QCL-UCI-iris-0-1}.
We then see the performance dependence of the UKM on $r$, which is the coefficient of the second term in the right-hand side of Eq.~\eqref{supp-arXiv-quantum-kernel-method-001-011}.
The result is shown in Fig.~\ref{supp-arXiv-numerical-result-r-dependence-UKM-UCI-iris-0-1}.
In Fig.~\ref{supp-arXiv-numerical-result-lambda-dependence-kernel-method-iris-0-1}, we show the performance dependence of the kernel method on $\lambda$, which is the coefficient of the second term in the right-hand side of Eq.~\eqref{supp-arXiv-cost-function-kernel-method-001-002}.

\begin{figure}[htb]
\centering
\includegraphics[scale=0.45]{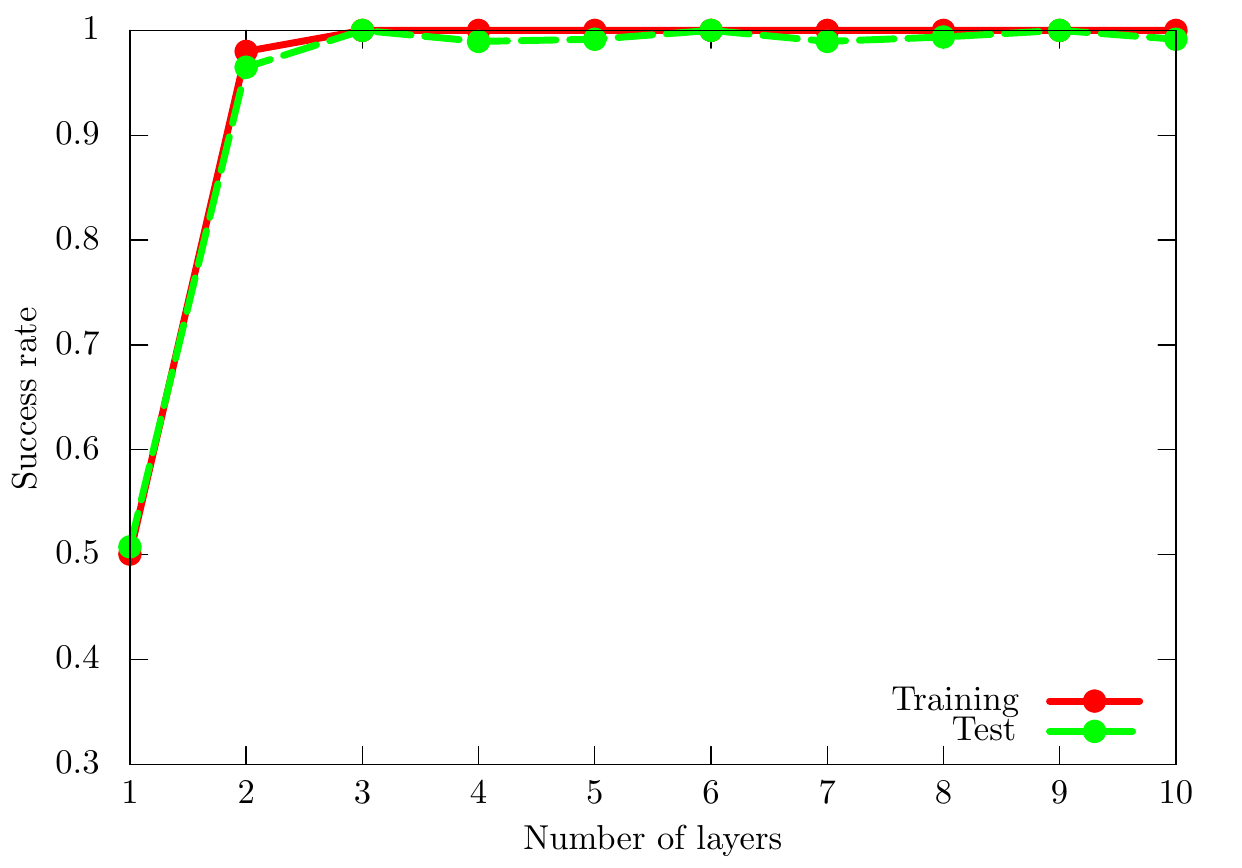}
\caption{Performance dependence of QCL on the number of layers $L$ for the iris dataset ($0$ or $1$) in the case of the hinge function, Eq.~\eqref{supp-arXiv-hinge-function-001-001}. We use the CNOT-based circuit geometry and set $\theta_\mathrm{bias} = 0$. We iterate the computation $300$ times.}
\label{supp-arXiv-numerical-result-layers-dependence-QCL-UCI-iris-0-1-hinge}
\end{figure}

\begin{figure}[htb]
\centering
\includegraphics[scale=0.45]{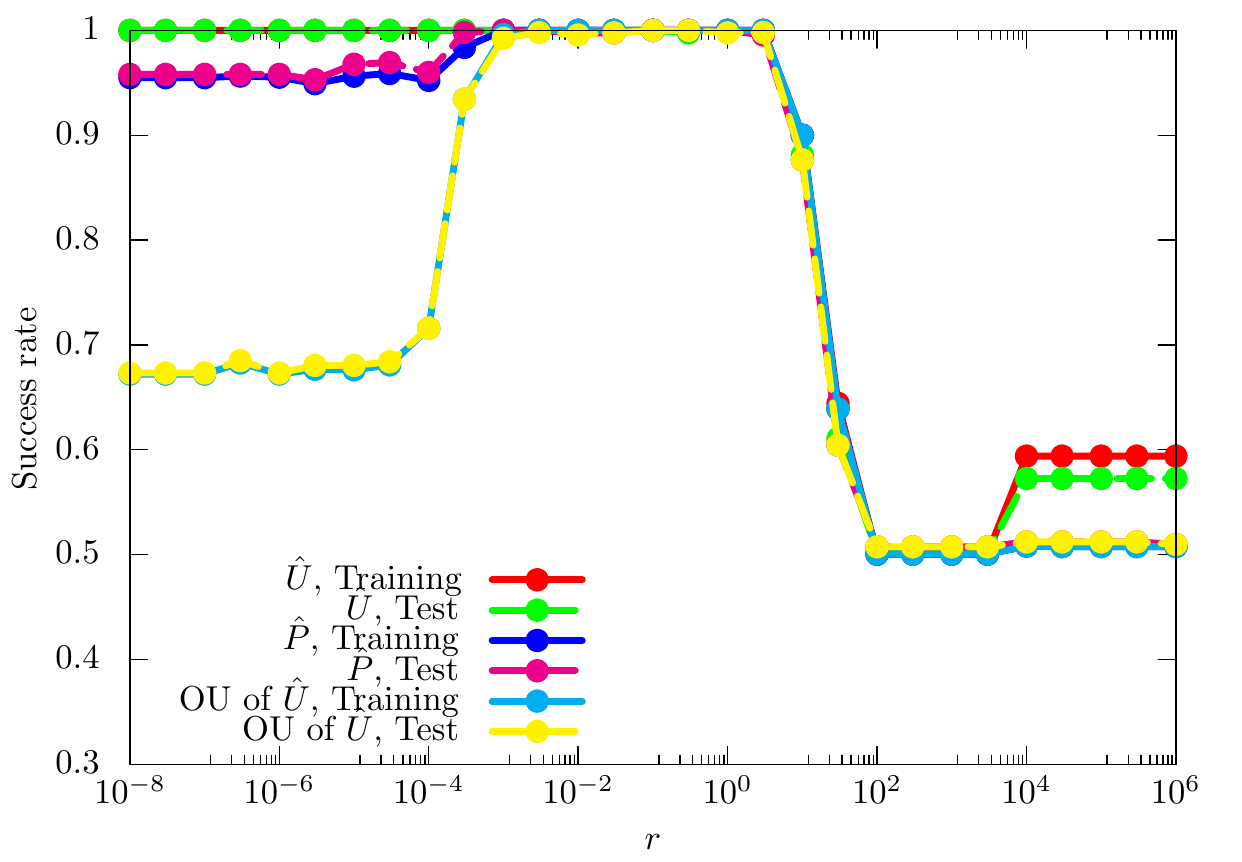}
\caption{Performance dependence of the UKM on $r$, which is the coefficient of the second term in the right-hand side of Eq.~\eqref{supp-arXiv-quantum-kernel-method-001-011} for the iris dataset ($0$ or $1$) in the case of the hinge function, Eq.~\eqref{supp-arXiv-hinge-function-001-001}. We use complex matrices and set $\theta_\mathrm{bias} = 0$. We set $K = 30$ and $K' = 10$.}
\label{supp-arXiv-numerical-result-r-dependence-UKM-UCI-iris-0-1-hinge}
\end{figure}

So far, we have used the squared error function, Eq.~\eqref{supp-arXiv-squared-error-function-001-001}.
In Fig.~\ref{supp-arXiv-numerical-result-layers-dependence-QCL-UCI-iris-0-1-hinge}, we show the performance dependence of QCL on the number of layers $L$ in the case of the hinge function, Eq.~\eqref{supp-arXiv-hinge-function-001-001}.
In Fig.~\ref{supp-arXiv-numerical-result-r-dependence-UKM-UCI-iris-0-1-hinge}, we show the performance dependence of the UKM on $r$, which is the coefficient of the second term in the right-hand side of Eq.~\eqref{supp-arXiv-quantum-kernel-method-001-011}, in the case of the hinge function, Eq.~\eqref{supp-arXiv-hinge-function-001-001}.

\clearpage

\subsection{Iris dataset ($0$ or non-$0$)}

We here show the numerical result for the iris dataset ($0$ or non-$0$).
For the UKM, we put $r = 0.010$ and set $K = 30$ and $K' = 10$ in Algo.~\ref{supp-arXiv-quantum-kernel-method-002-001}.
For QCL, we run iterations $300$ times.

\begin{figure*}[htb]
\centering
\includegraphics[scale=0.25]{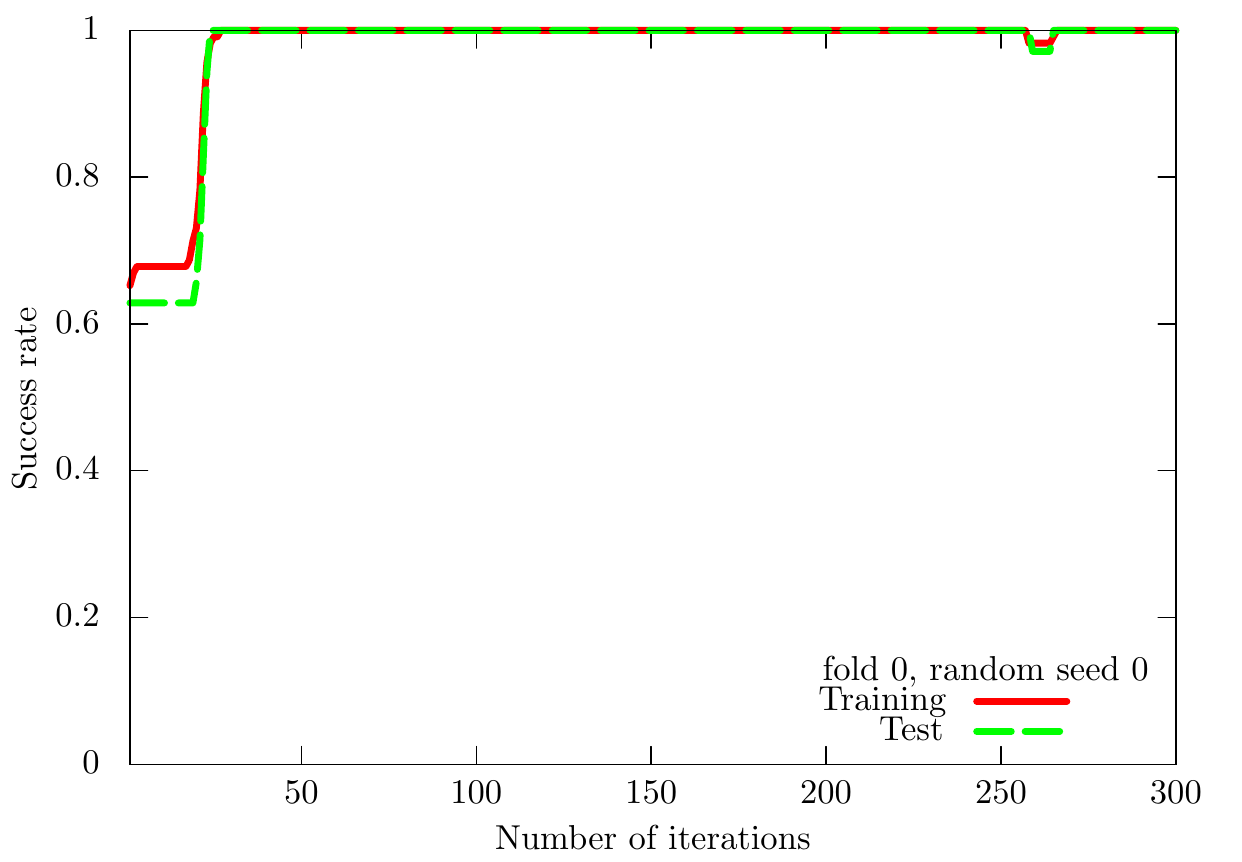}
\includegraphics[scale=0.25]{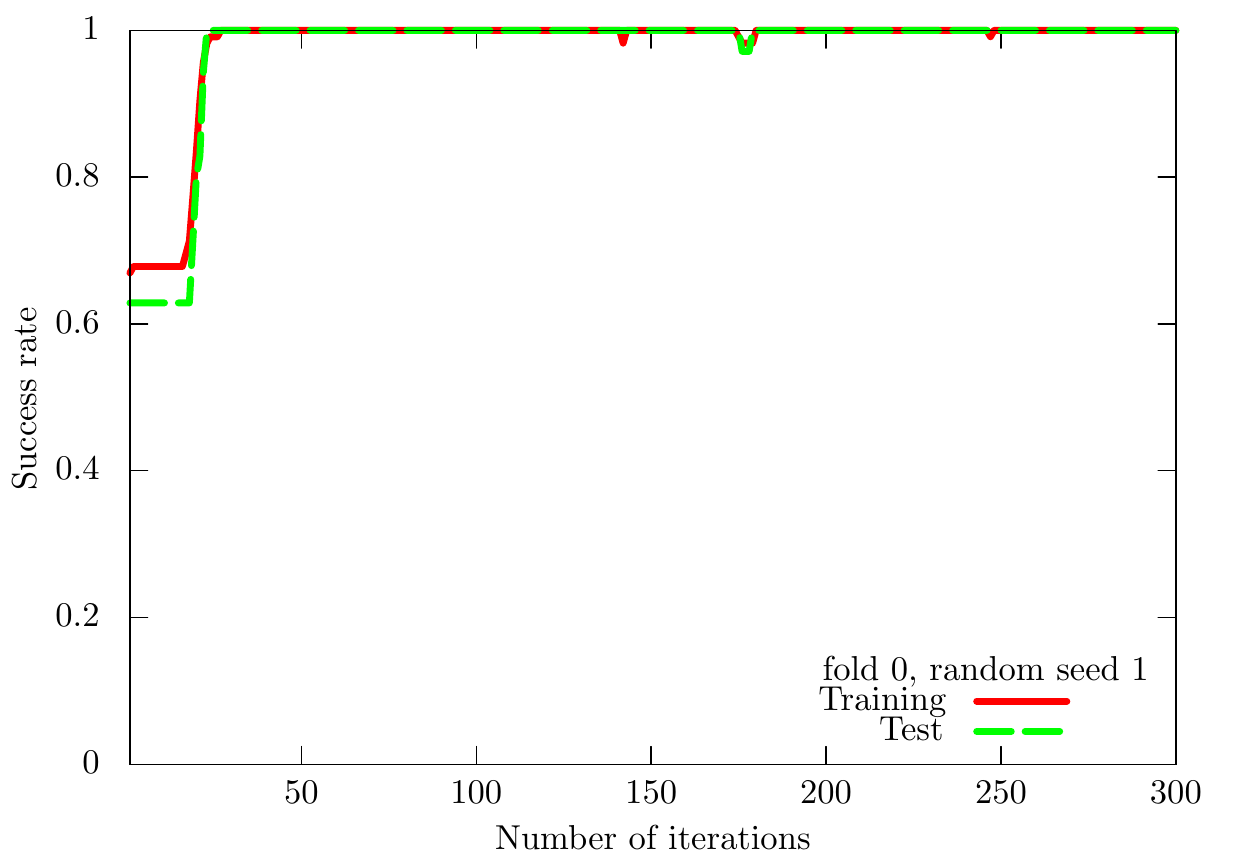}
\includegraphics[scale=0.25]{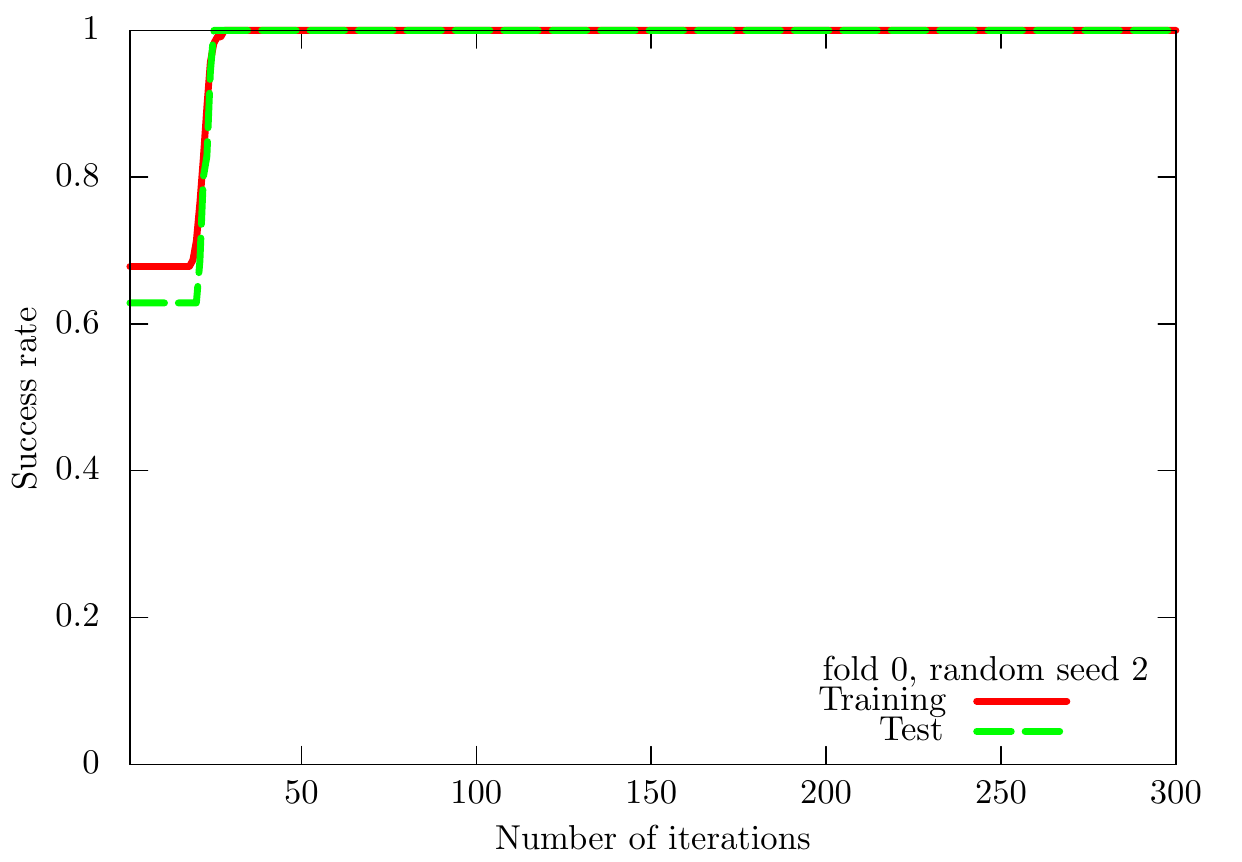}
\includegraphics[scale=0.25]{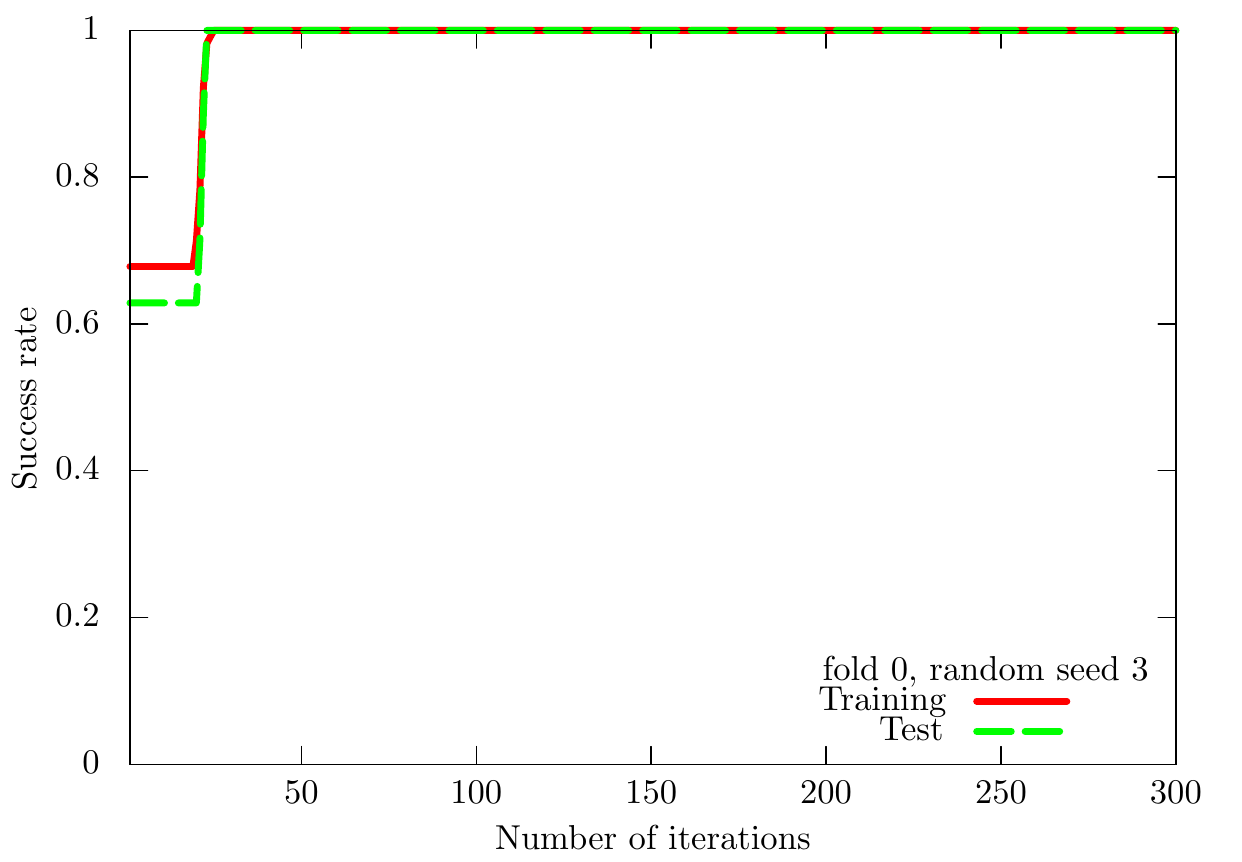}
\includegraphics[scale=0.25]{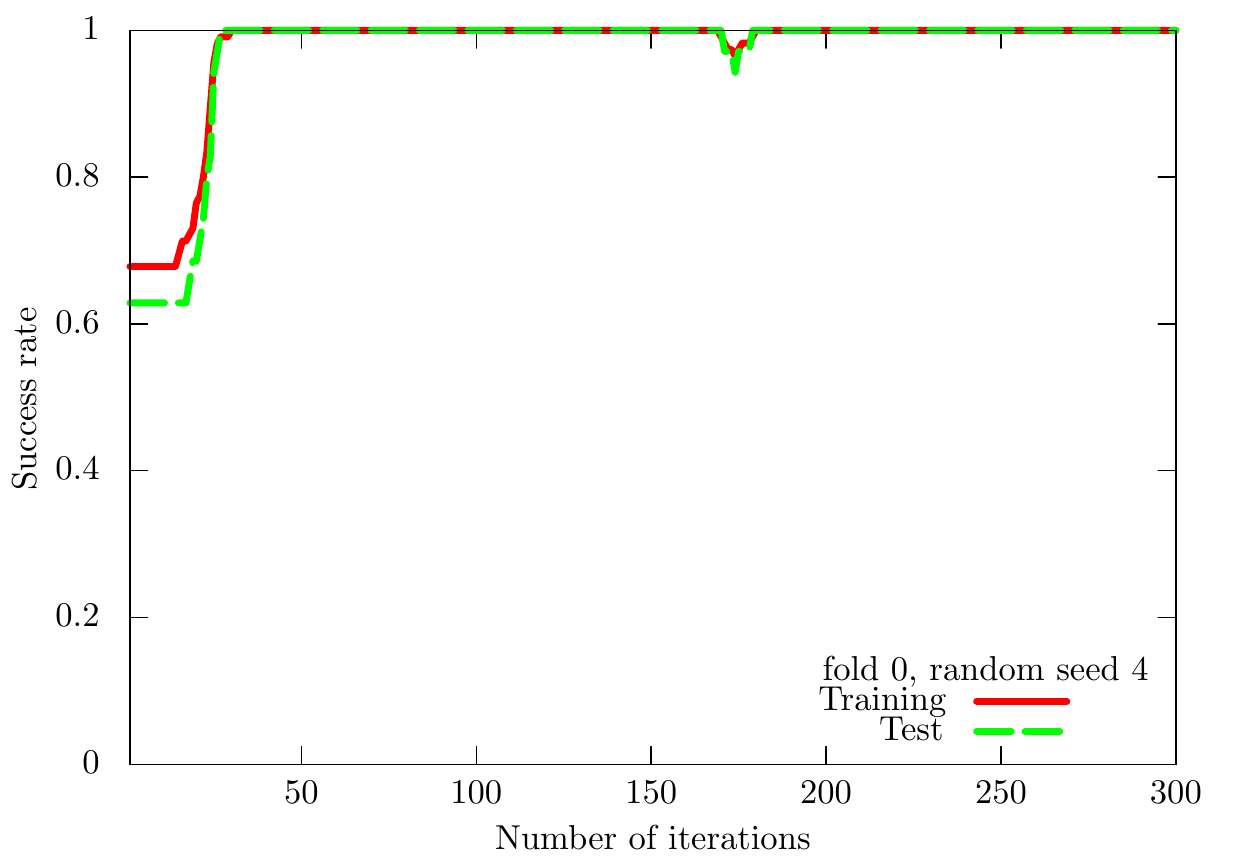}
\includegraphics[scale=0.25]{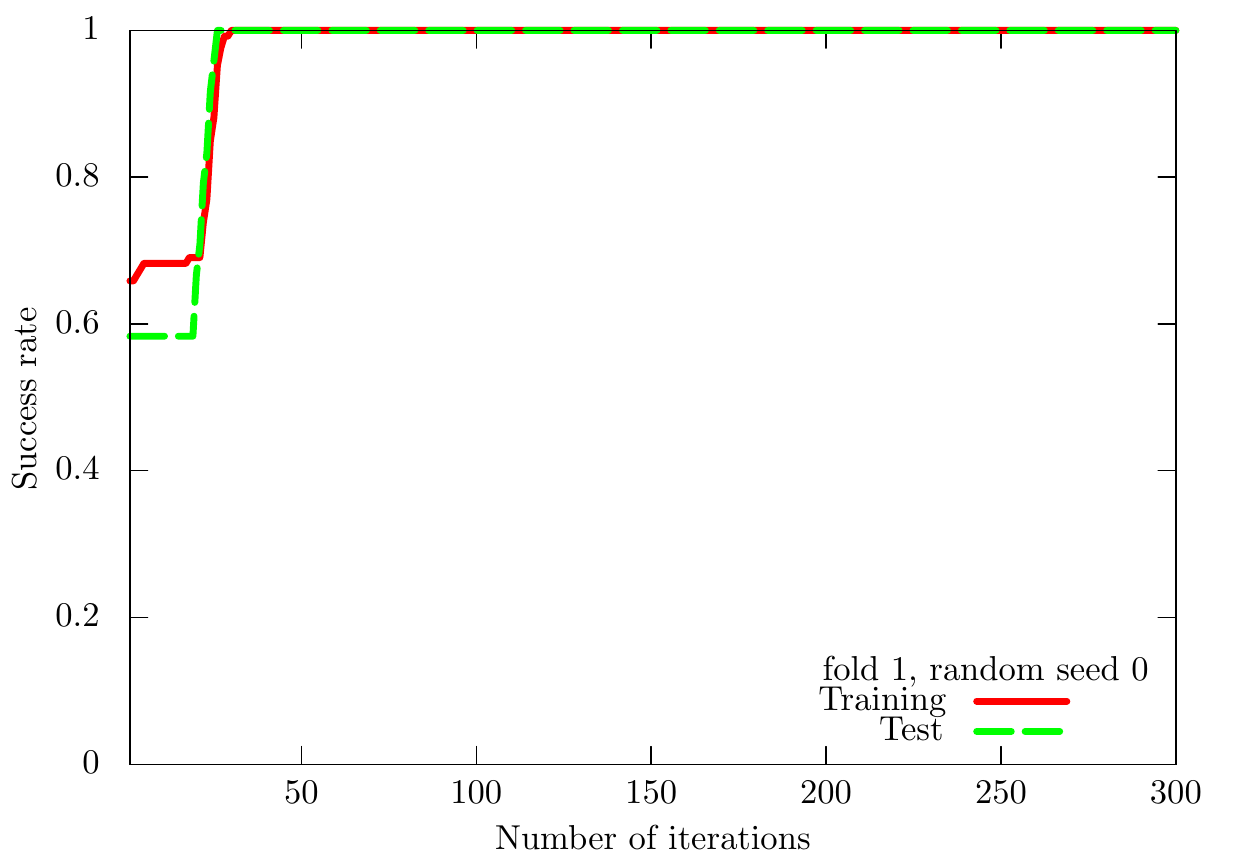}
\includegraphics[scale=0.25]{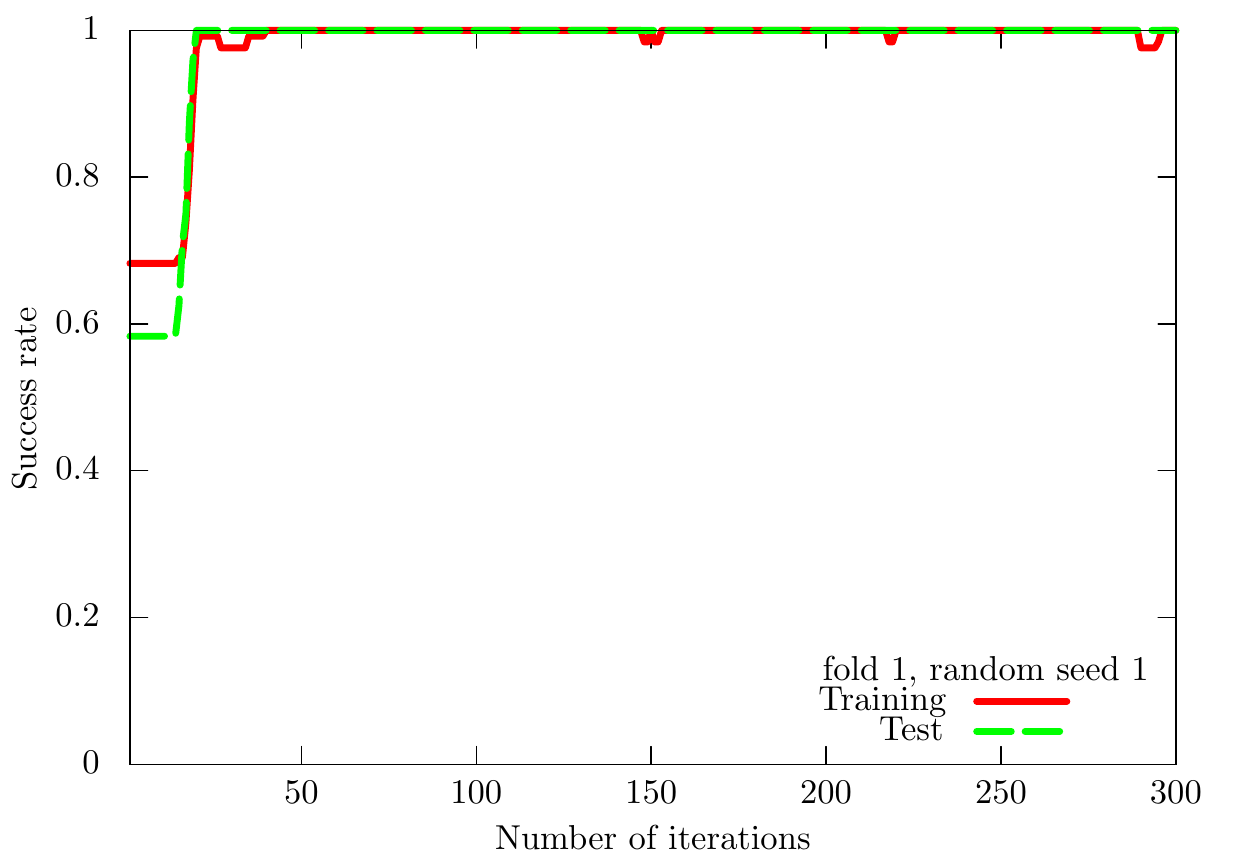}
\includegraphics[scale=0.25]{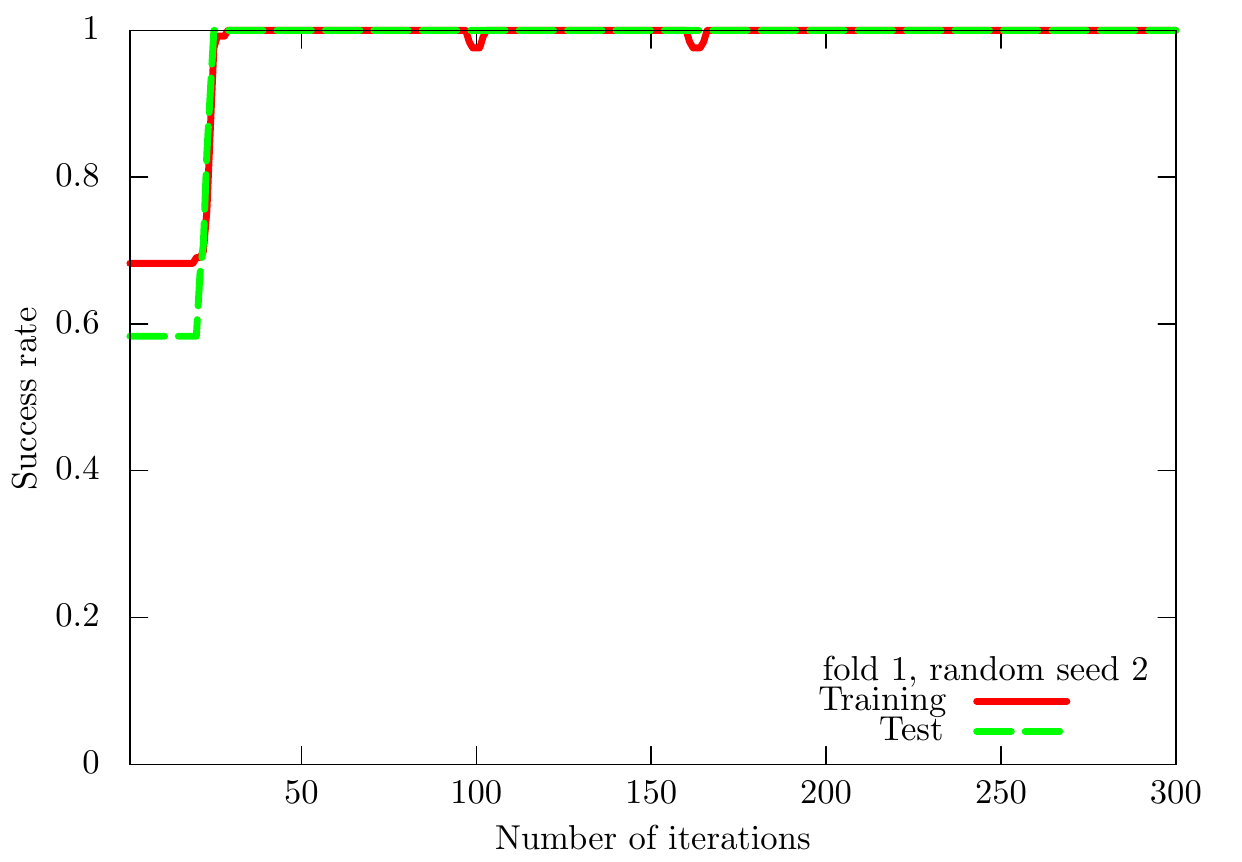}
\includegraphics[scale=0.25]{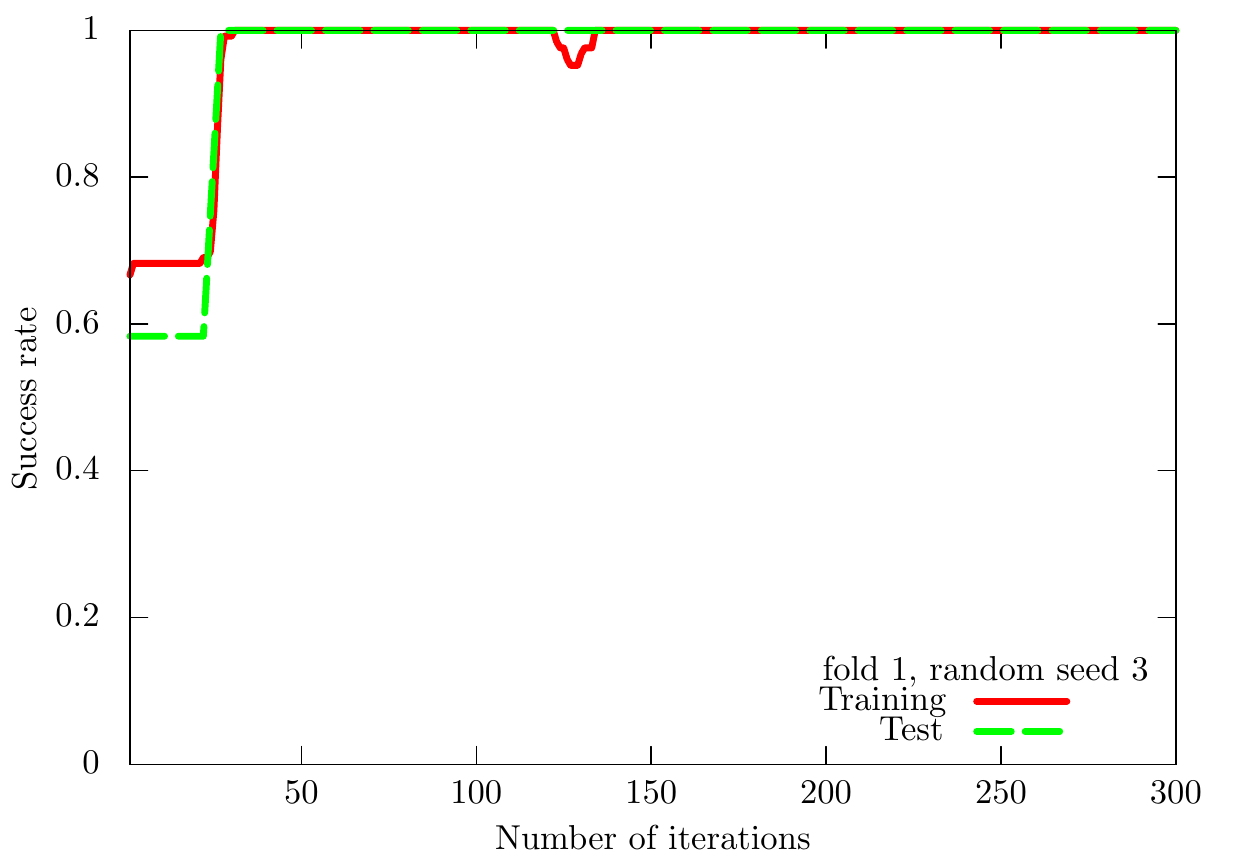}
\includegraphics[scale=0.25]{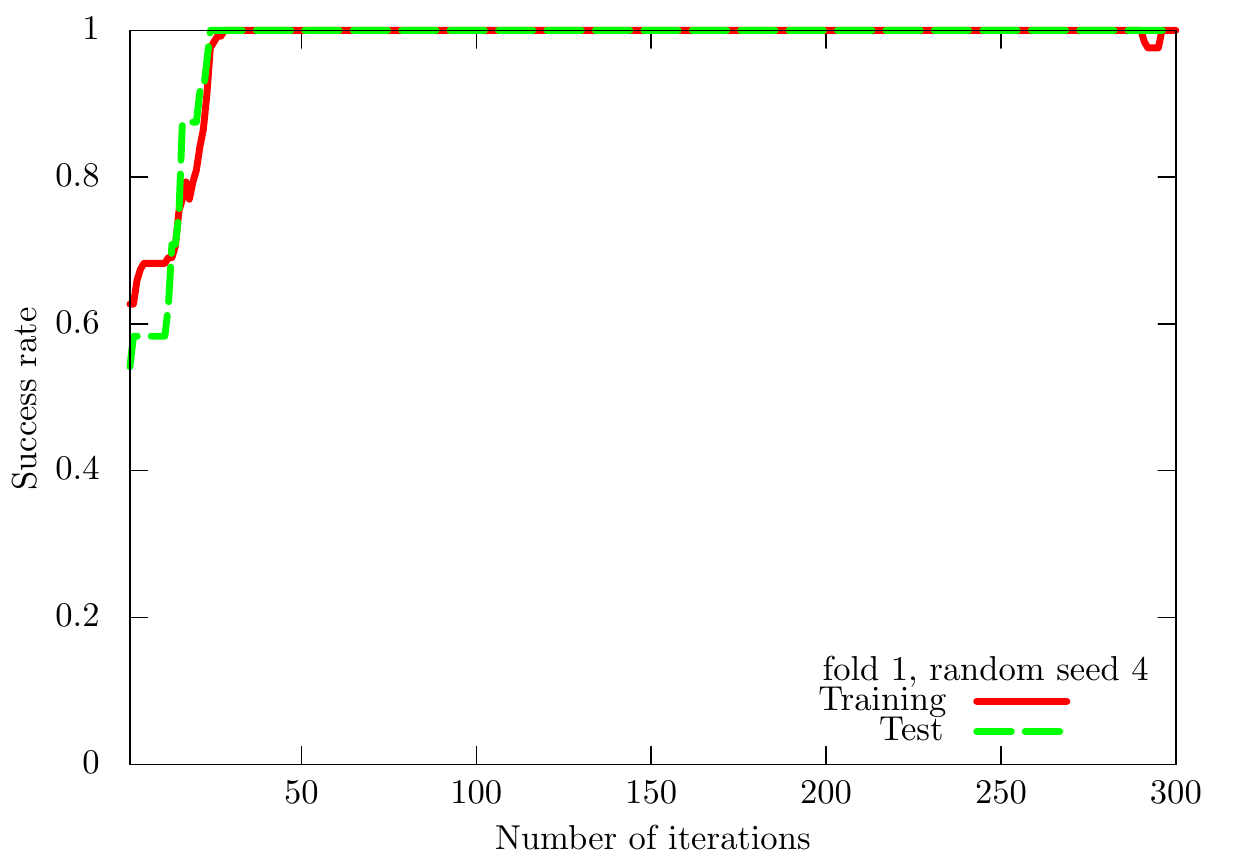}
\includegraphics[scale=0.25]{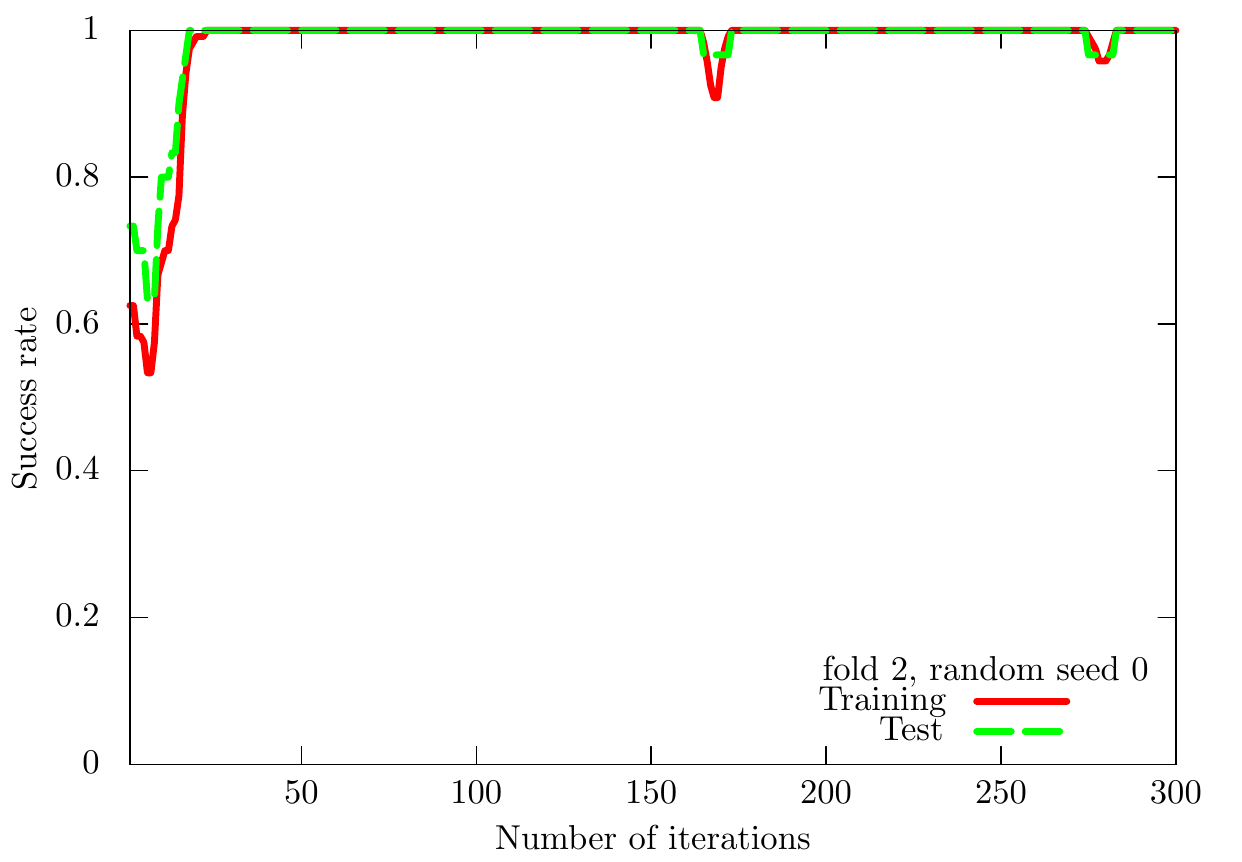}
\includegraphics[scale=0.25]{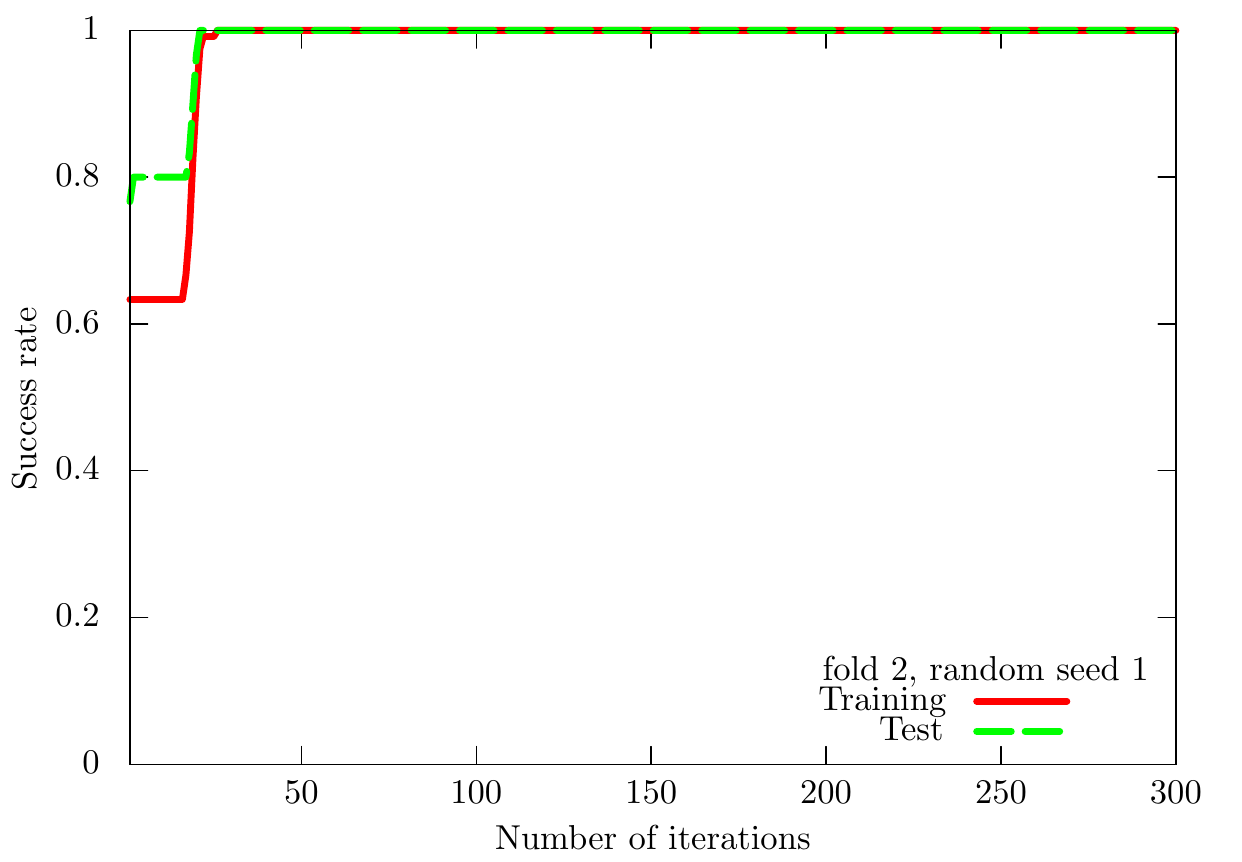}
\includegraphics[scale=0.25]{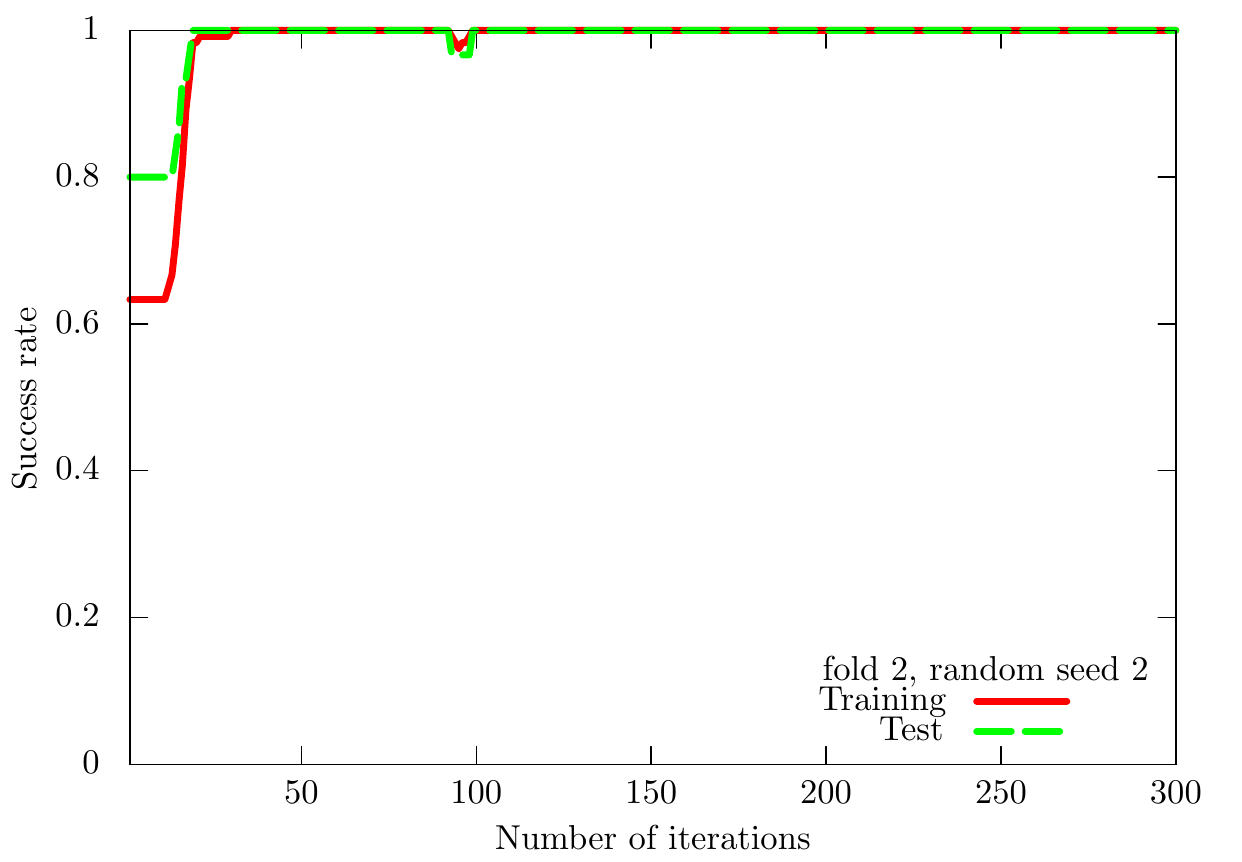}
\includegraphics[scale=0.25]{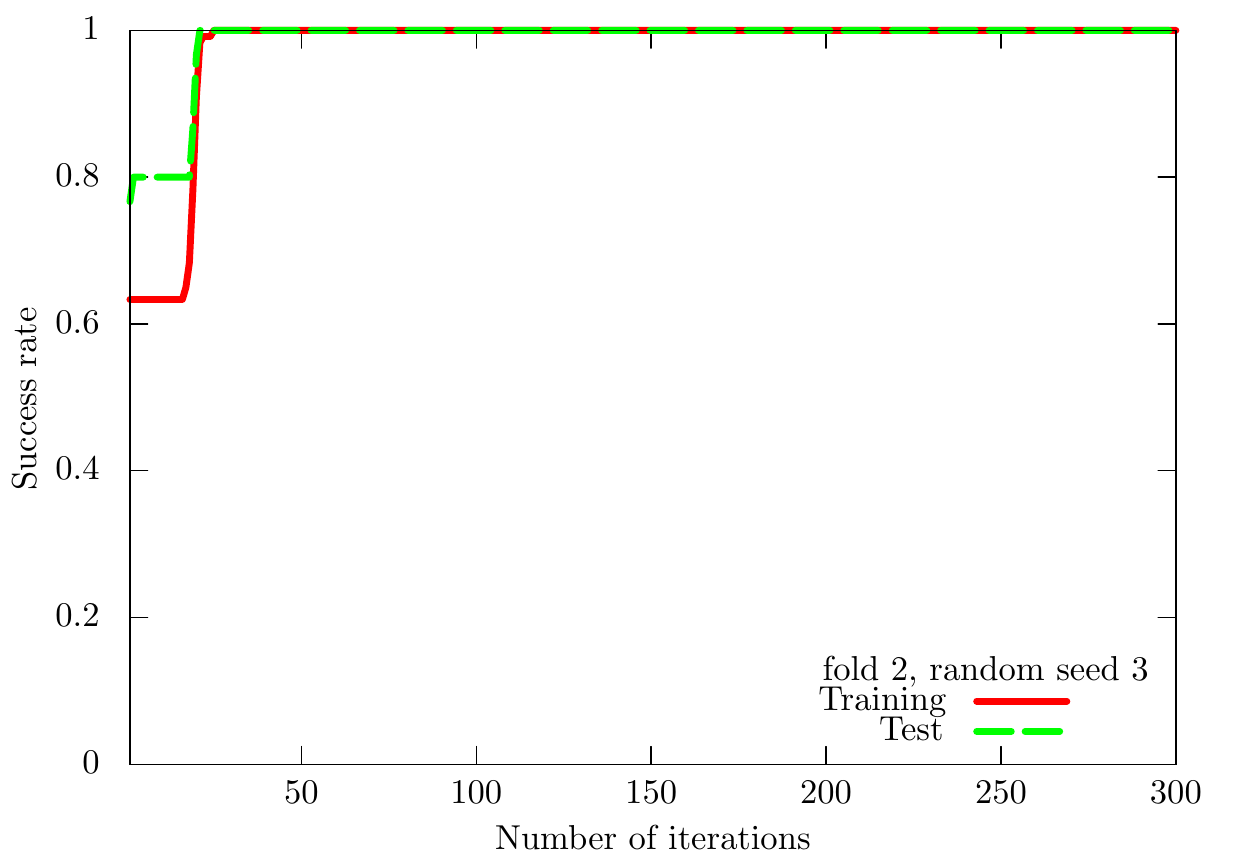}
\includegraphics[scale=0.25]{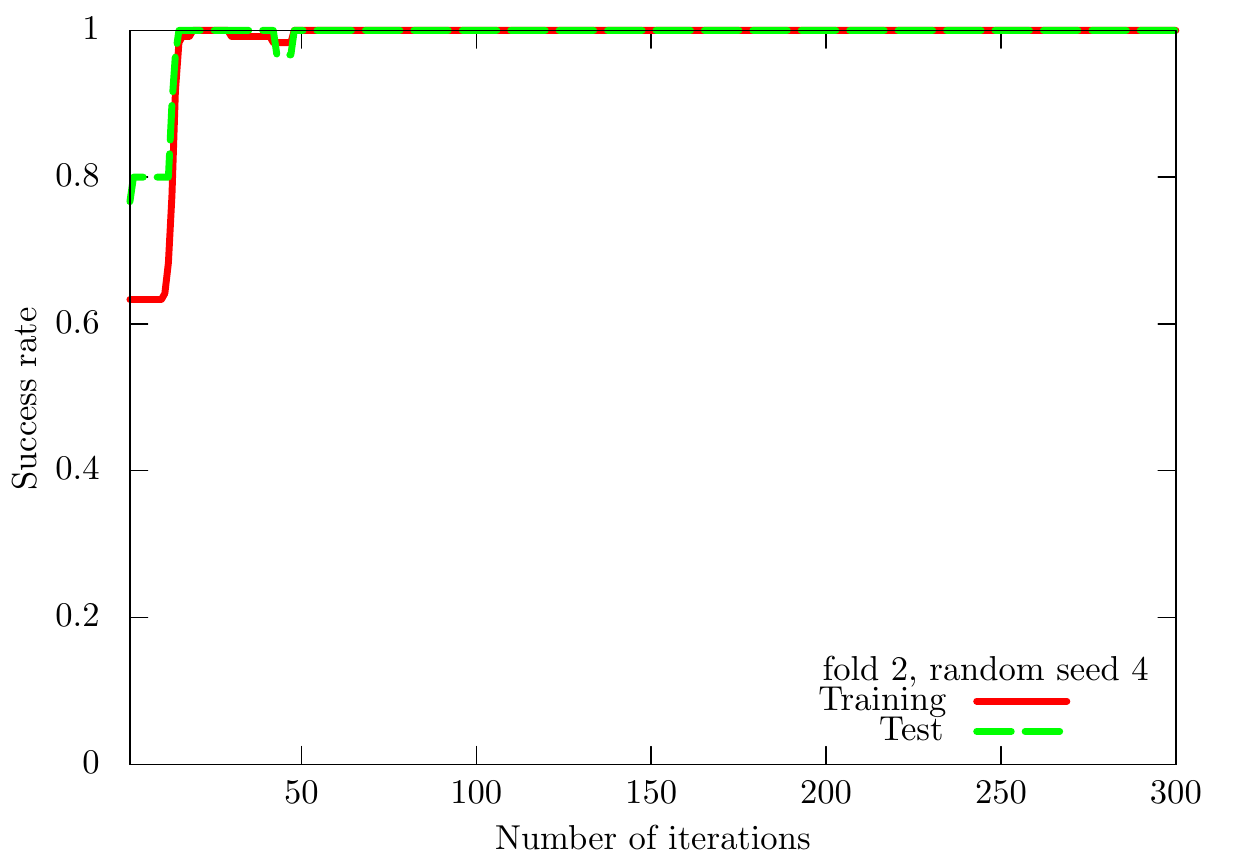}
\includegraphics[scale=0.25]{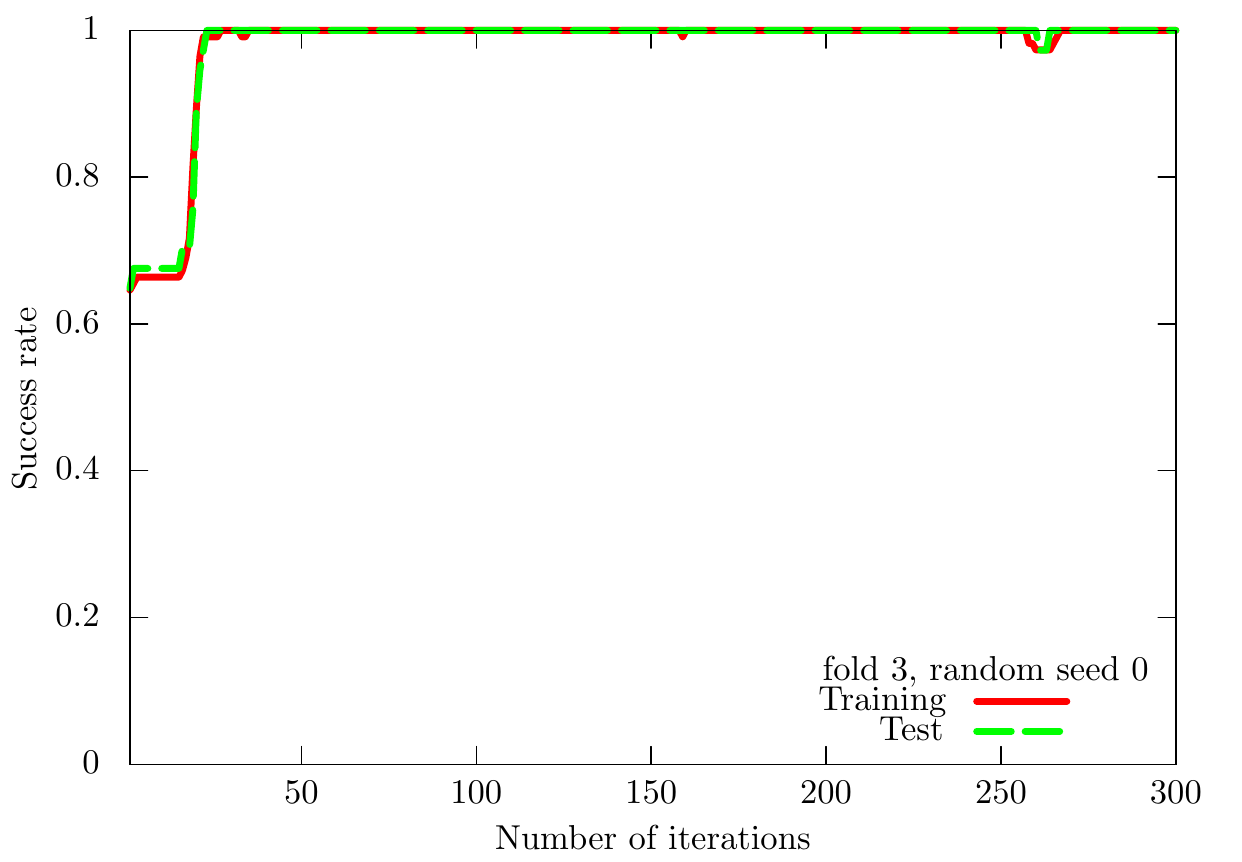}
\includegraphics[scale=0.25]{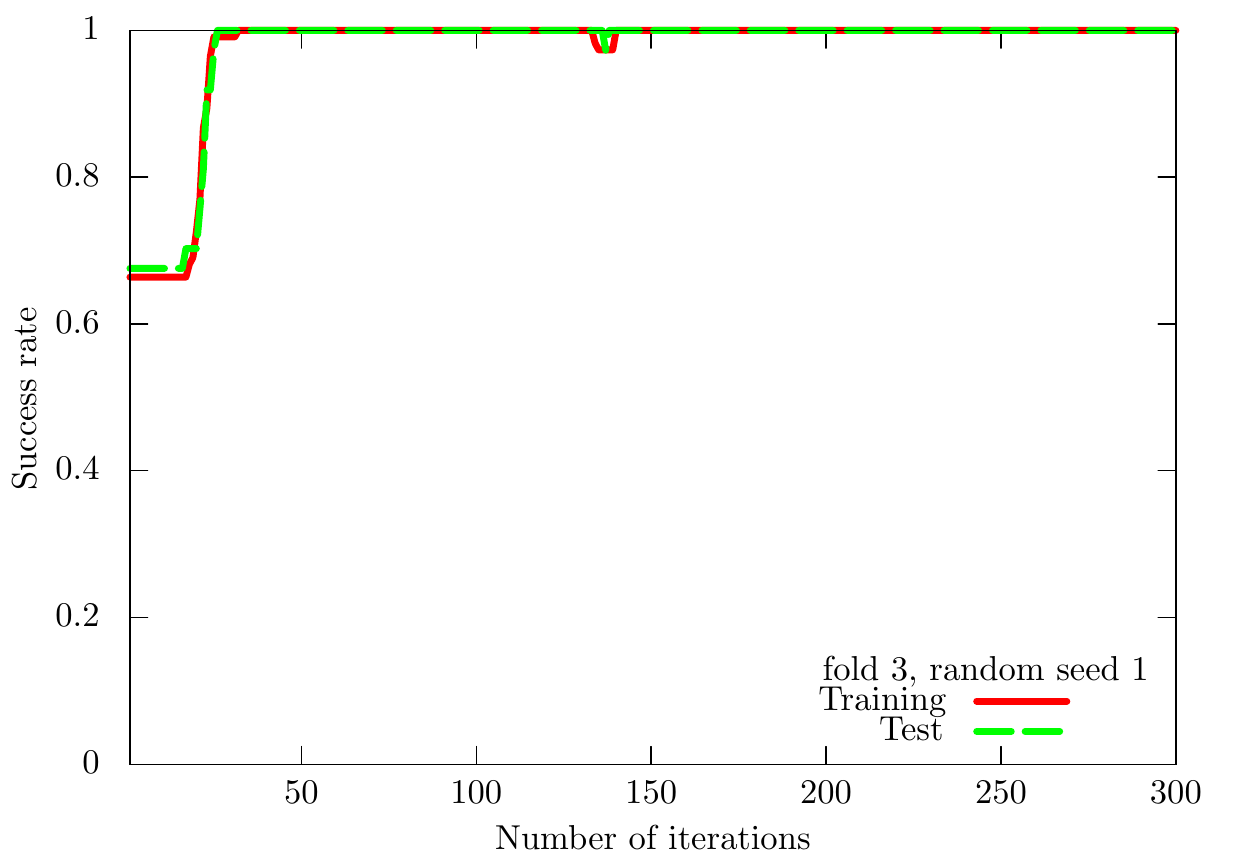}
\includegraphics[scale=0.25]{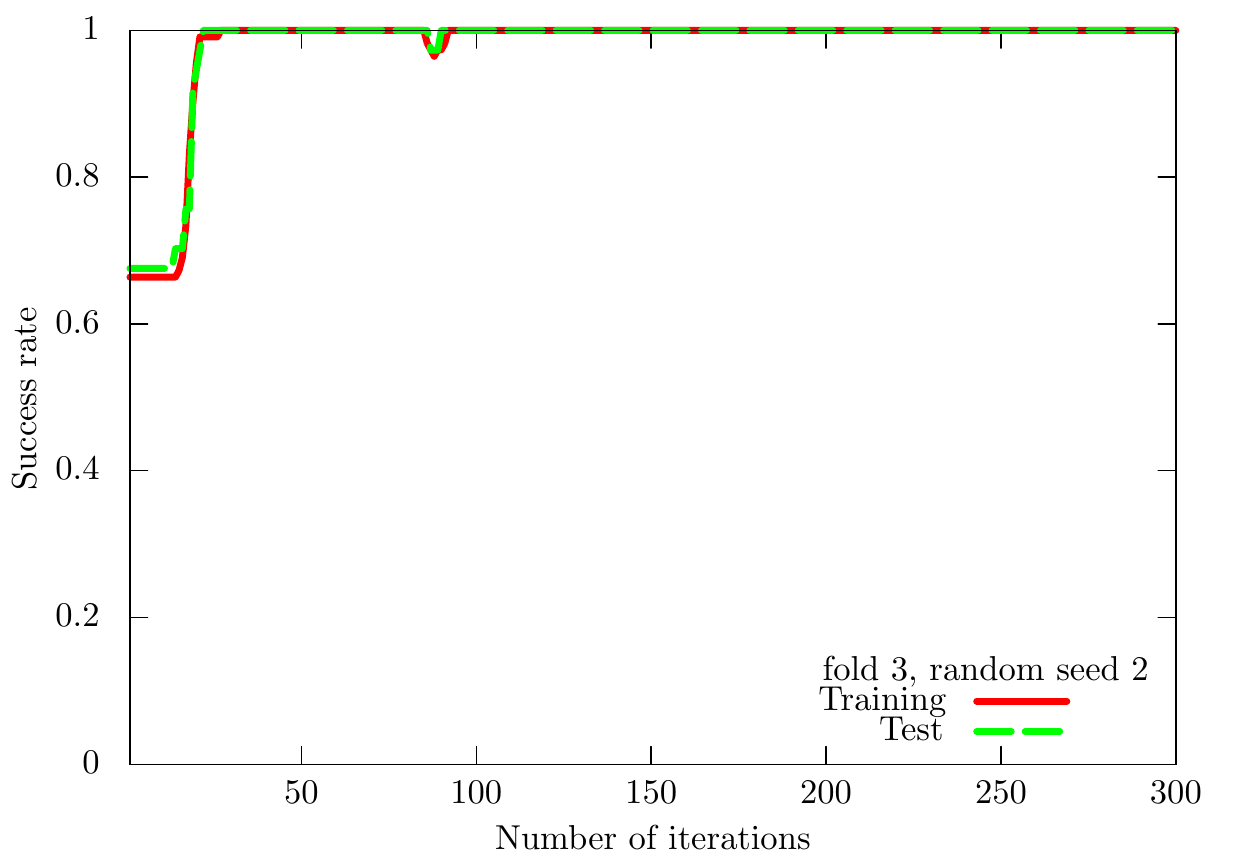}
\includegraphics[scale=0.25]{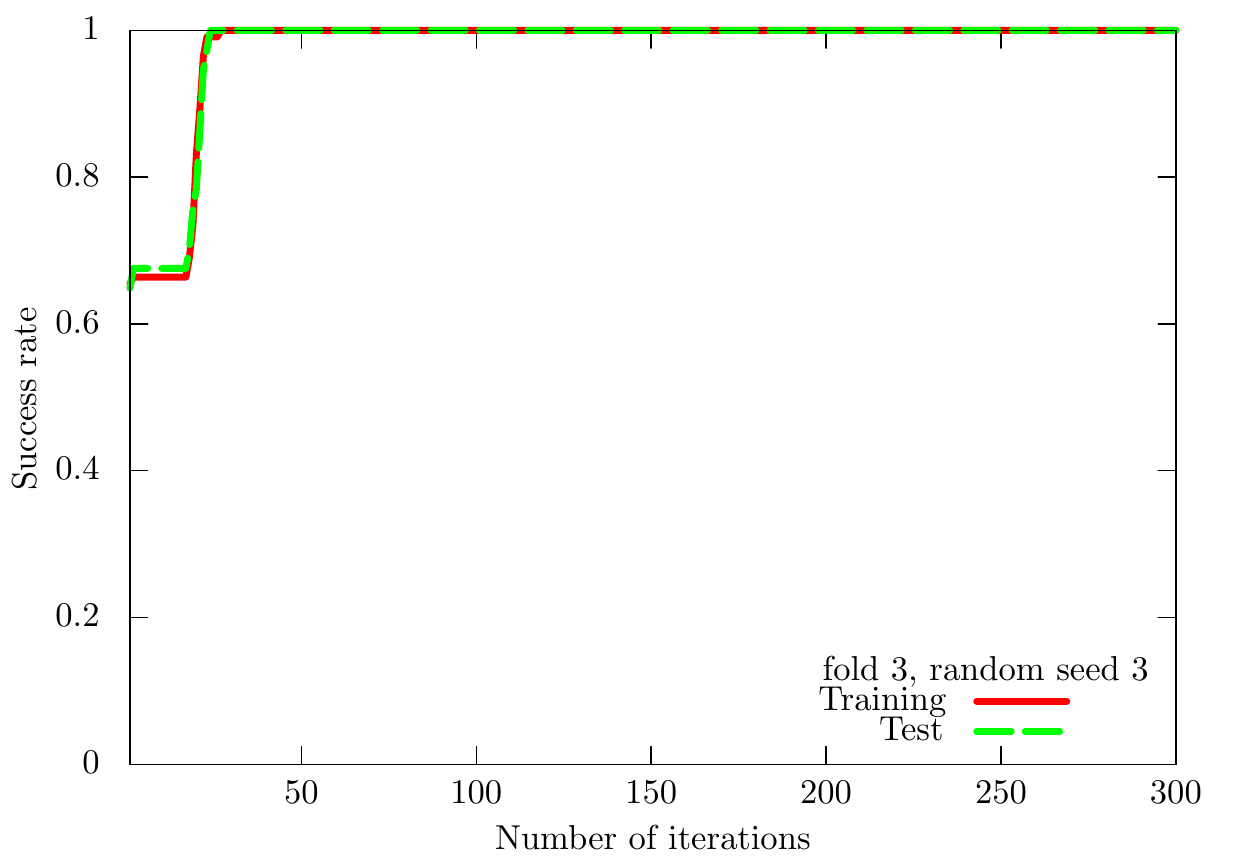}
\includegraphics[scale=0.25]{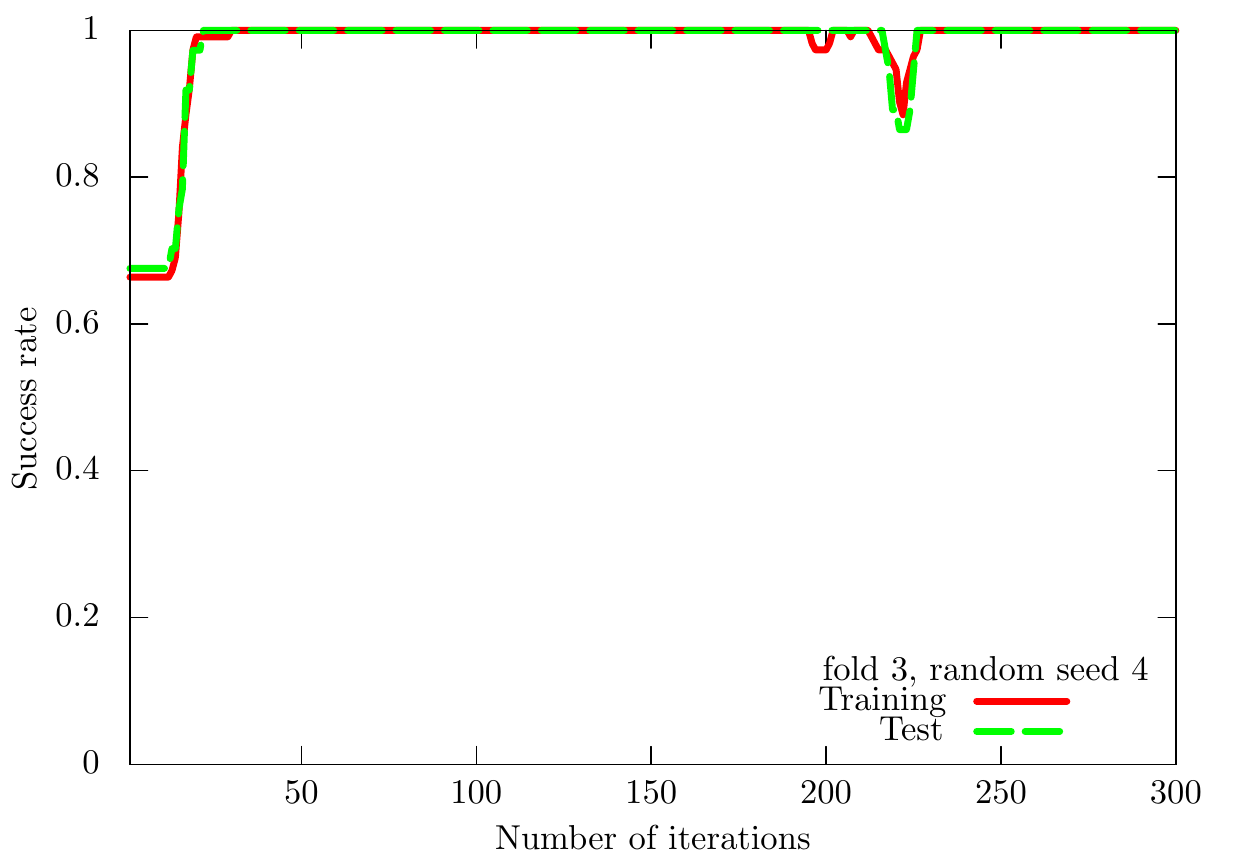}
\includegraphics[scale=0.25]{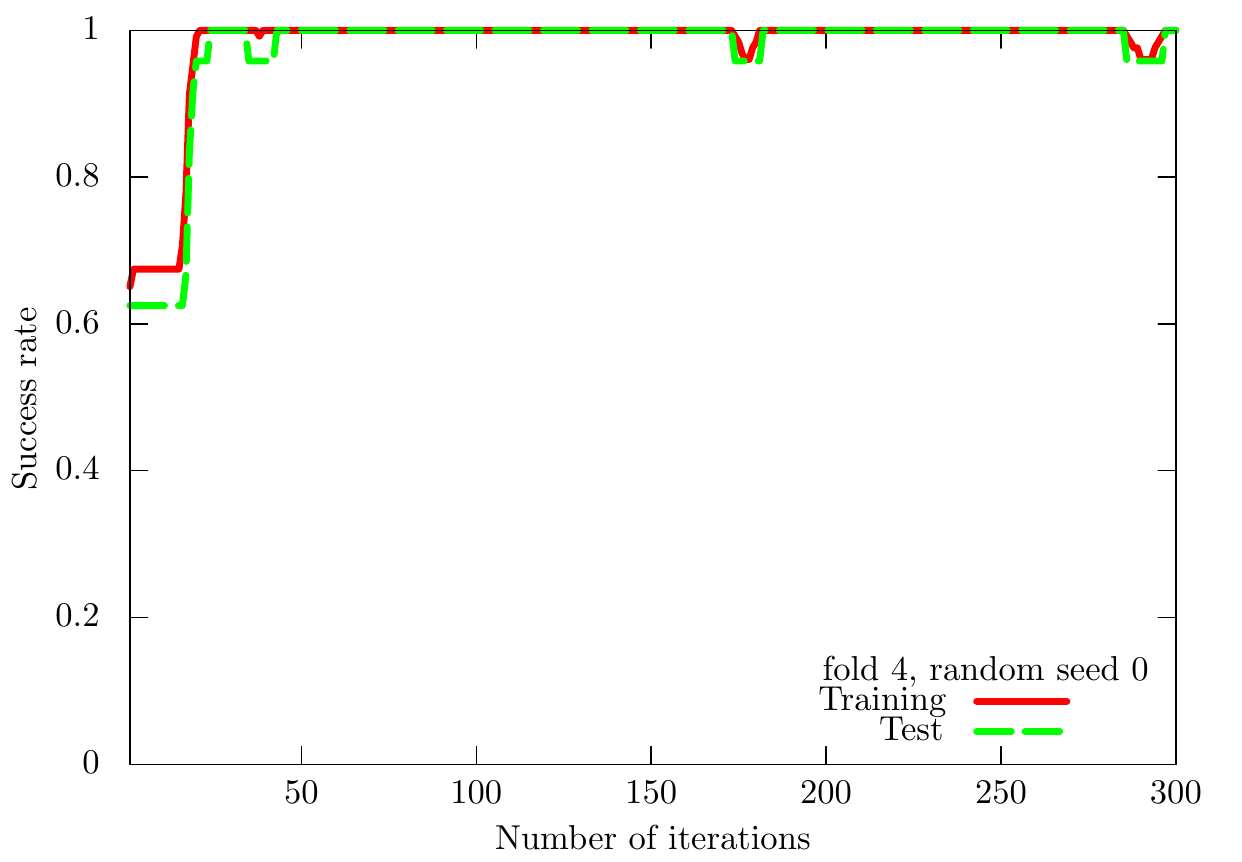}
\includegraphics[scale=0.25]{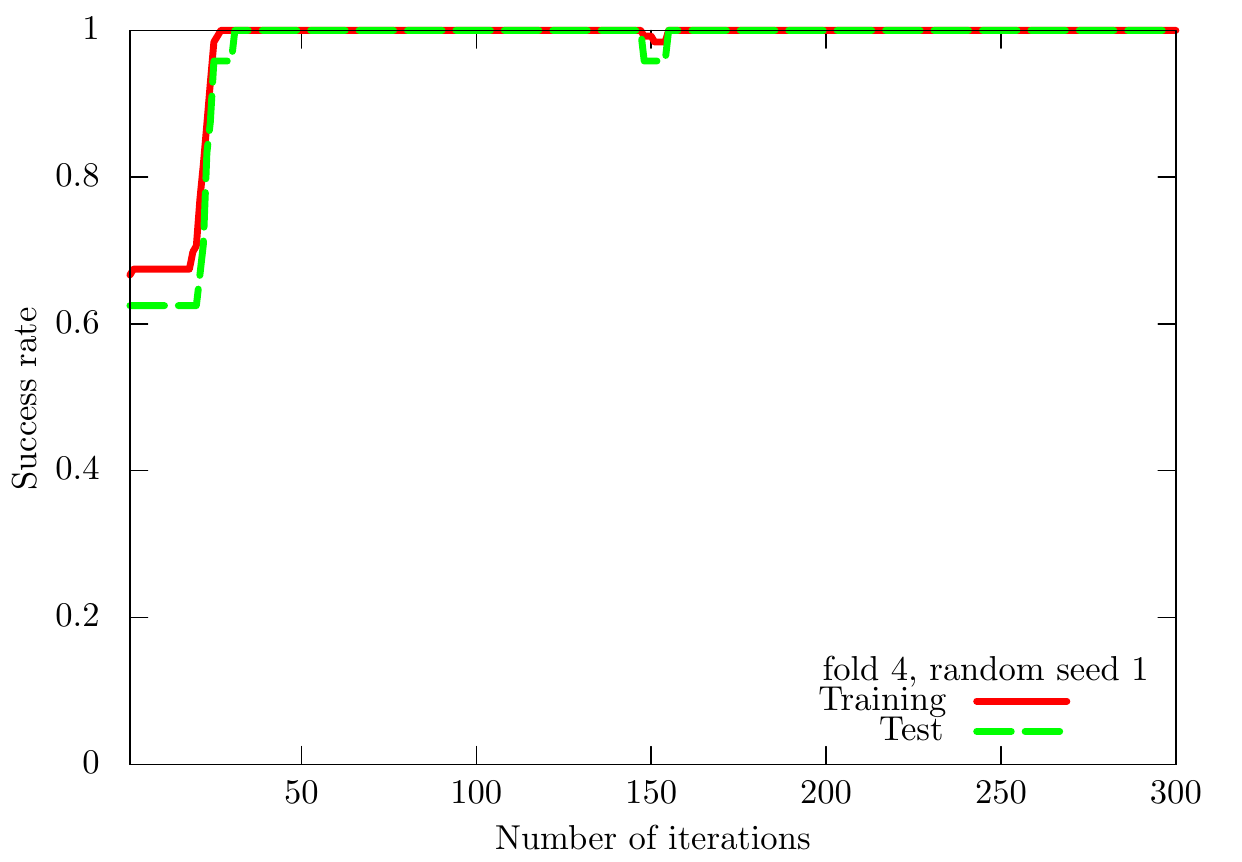}
\includegraphics[scale=0.25]{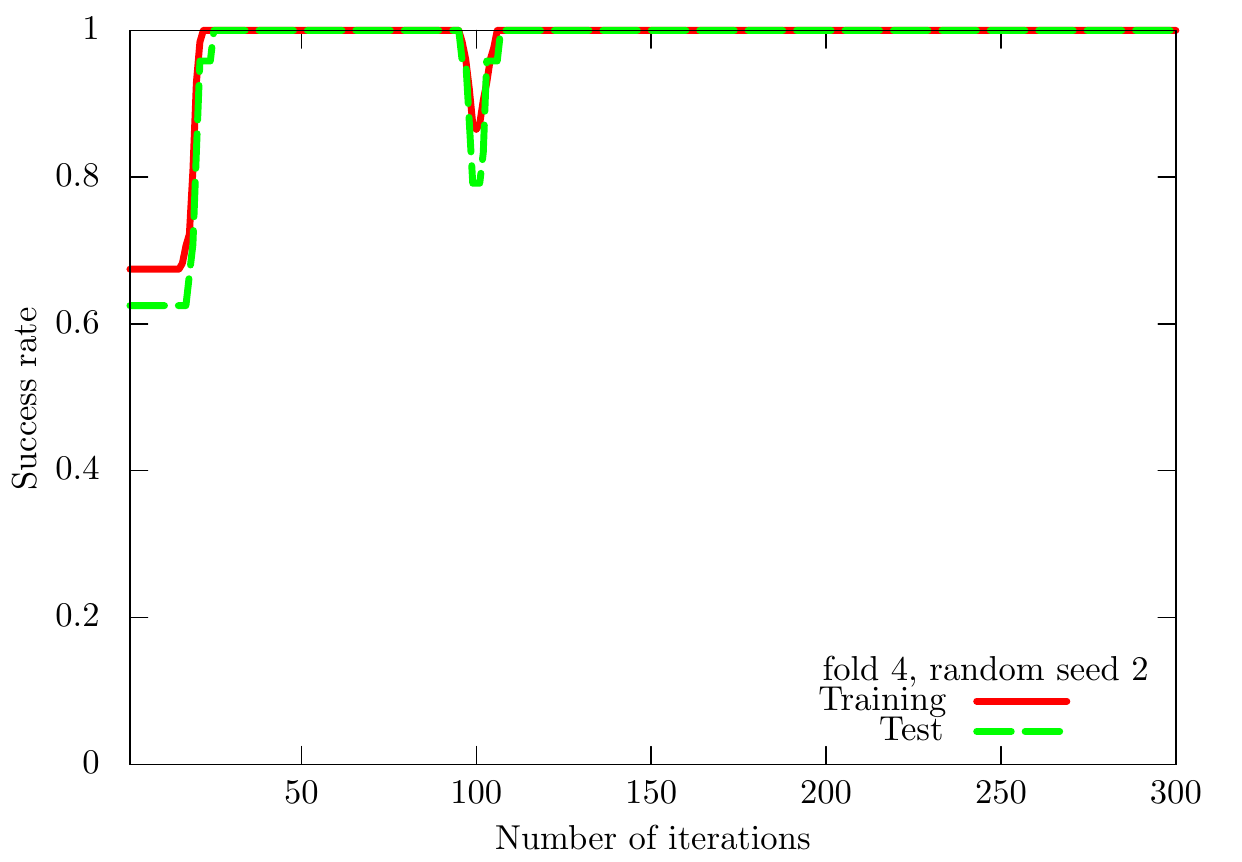}
\includegraphics[scale=0.25]{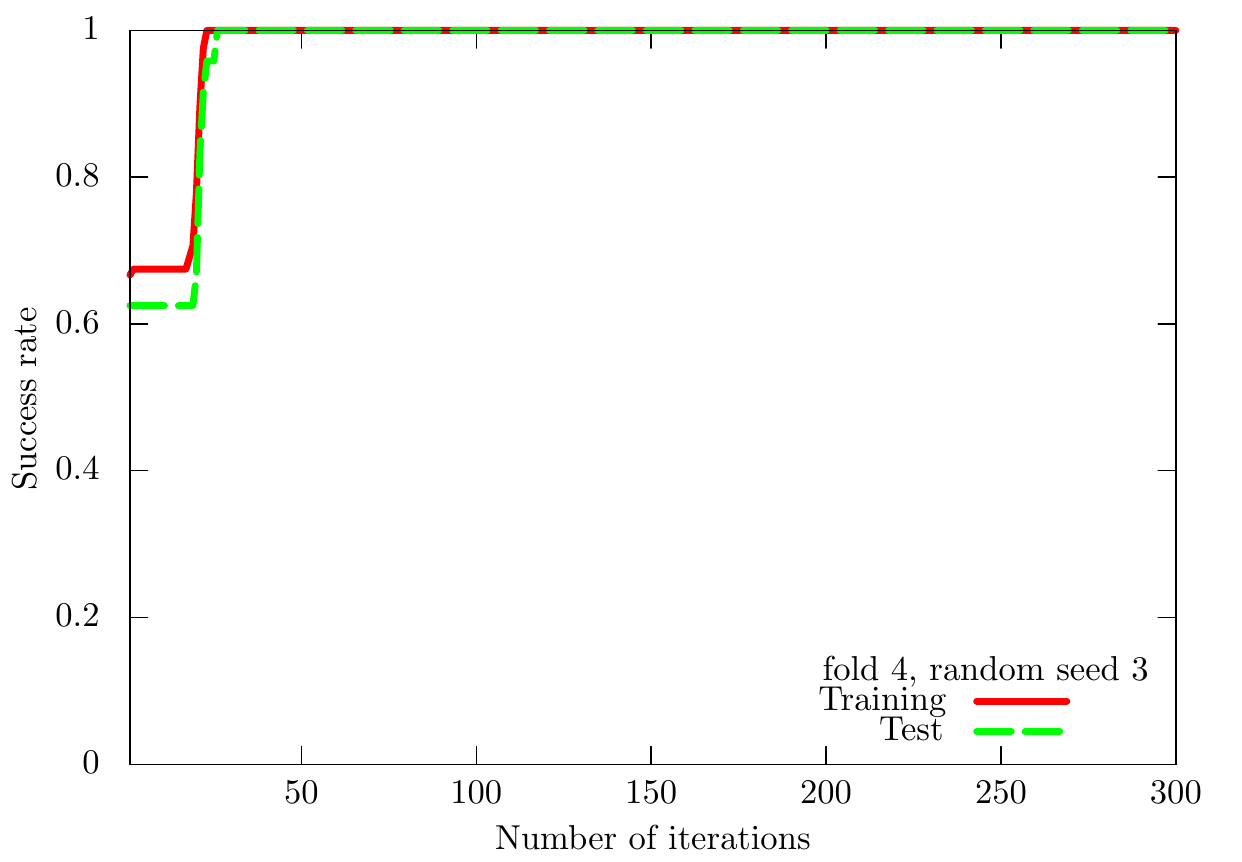}
\includegraphics[scale=0.25]{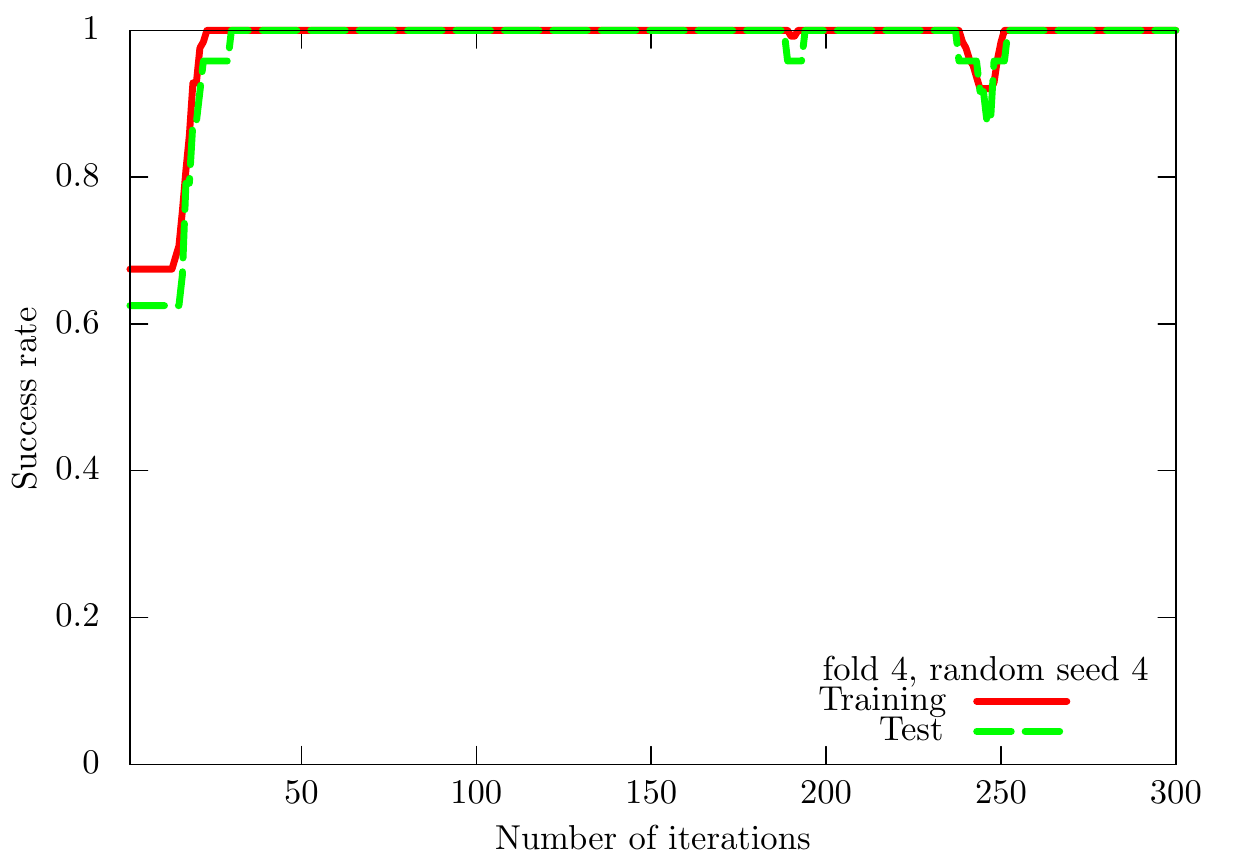}
\caption{Results of QCL on the $5$-fold datasets with $5$ different random seeds for the iris dataset ($0$ or non-$0$). We use the CNOT-based circuit and set $\theta_\mathrm{bias} = 0$. The number of layers $L$ is set to $5$.}
\label{supp-arXiv-numerical-result-raw-data-fold-001-rand-001-QCL-UCI-iris-0-non0}
\end{figure*}

\begin{figure*}[htb]
\centering
\includegraphics[scale=0.25]{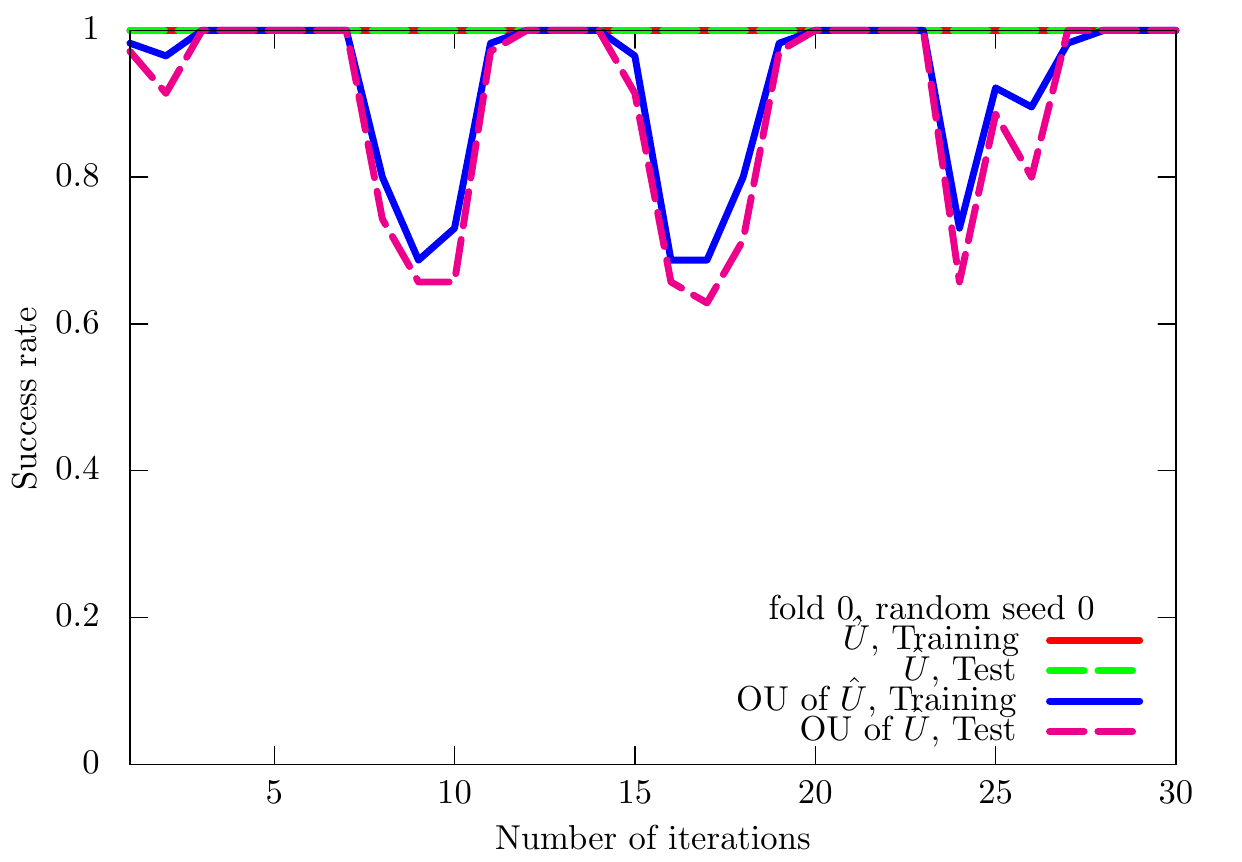}
\includegraphics[scale=0.25]{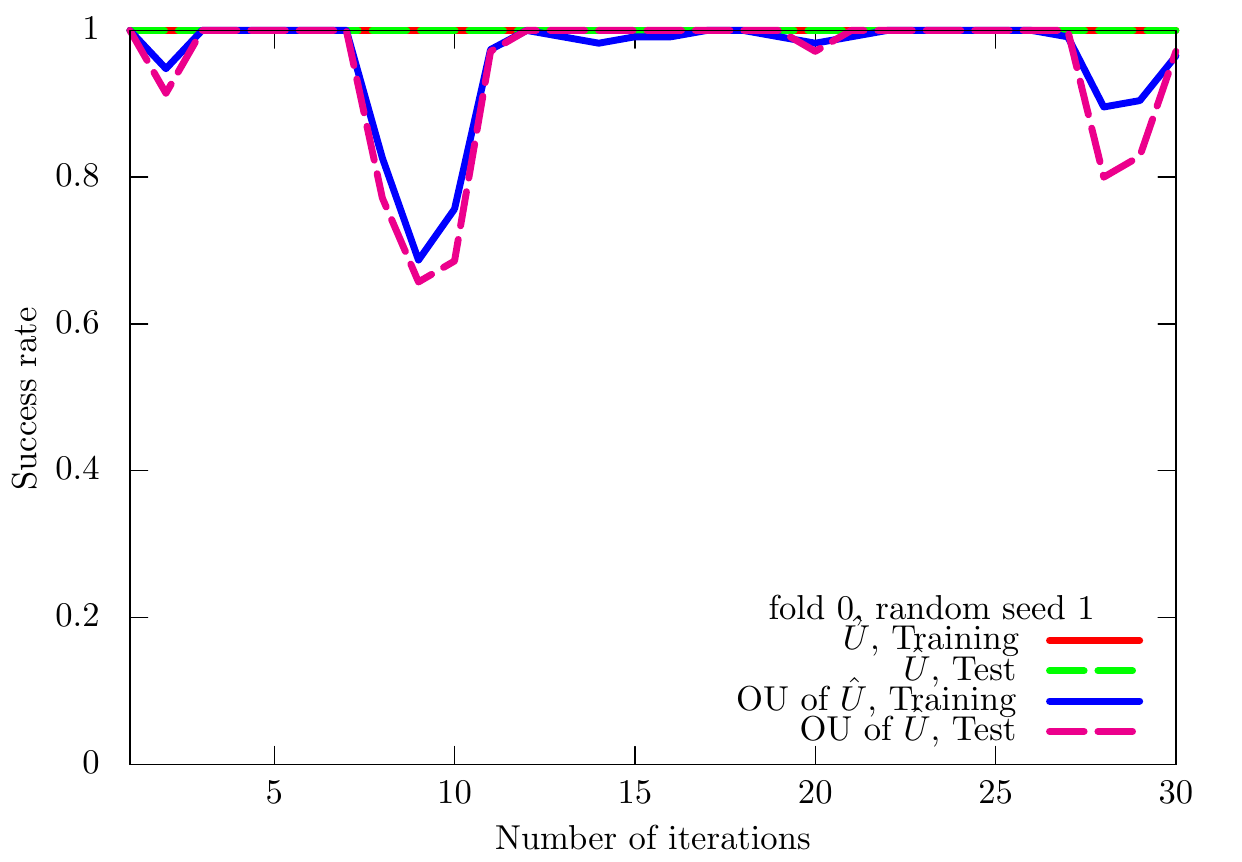}
\includegraphics[scale=0.25]{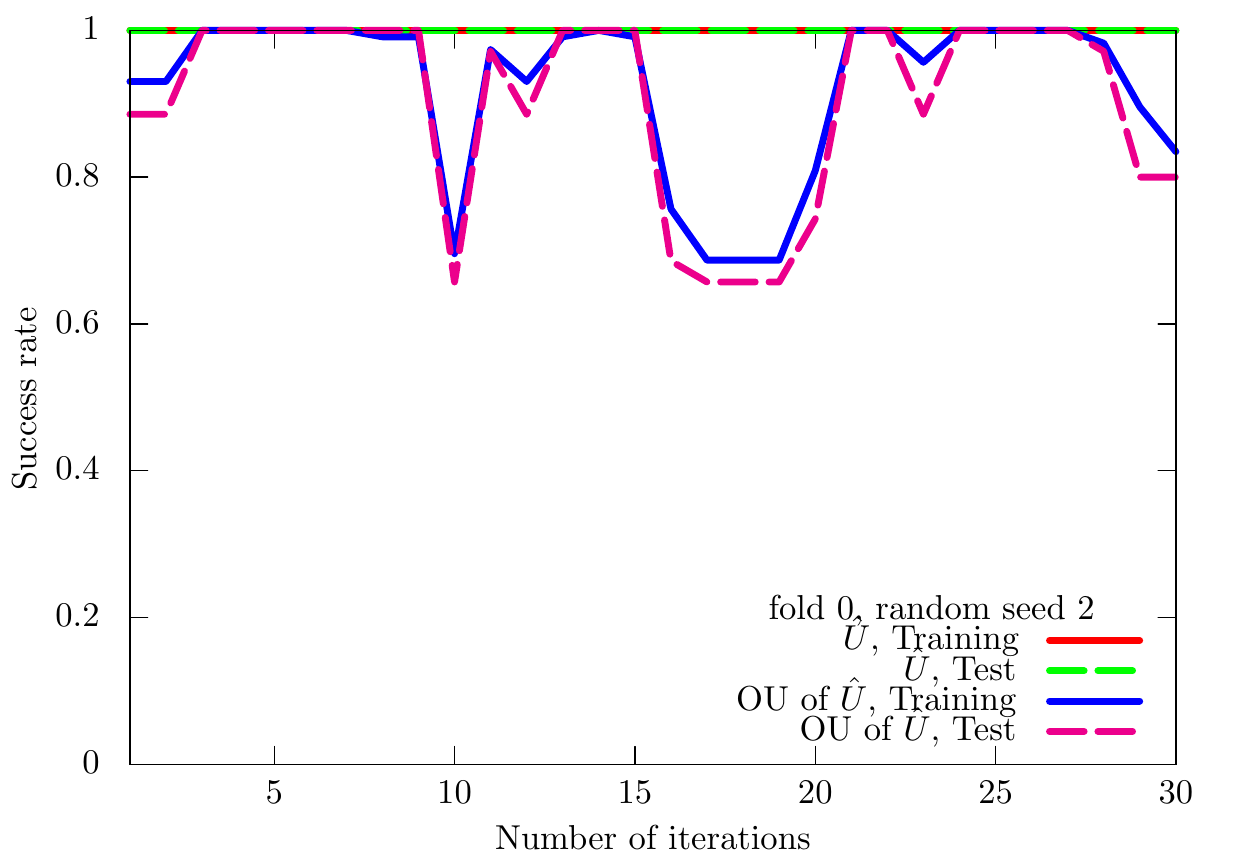}
\includegraphics[scale=0.25]{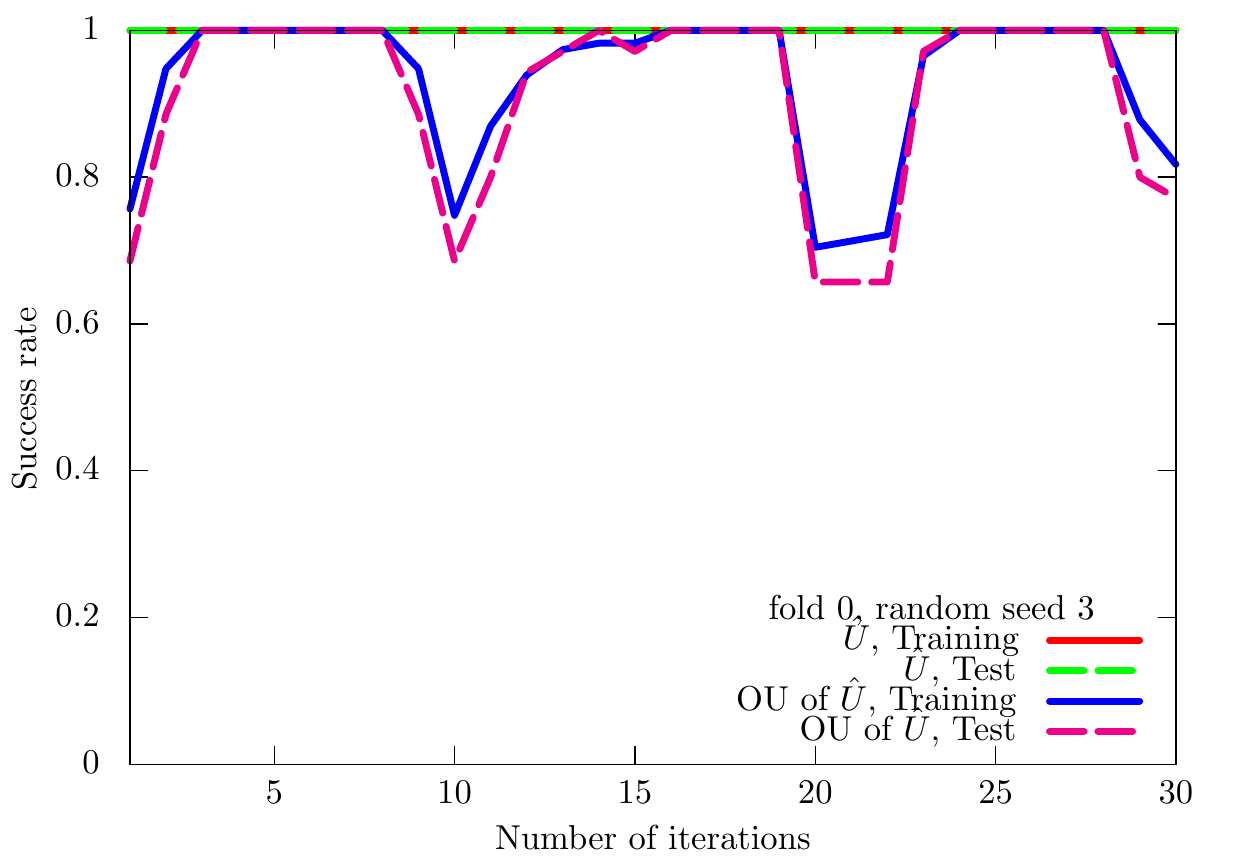}
\includegraphics[scale=0.25]{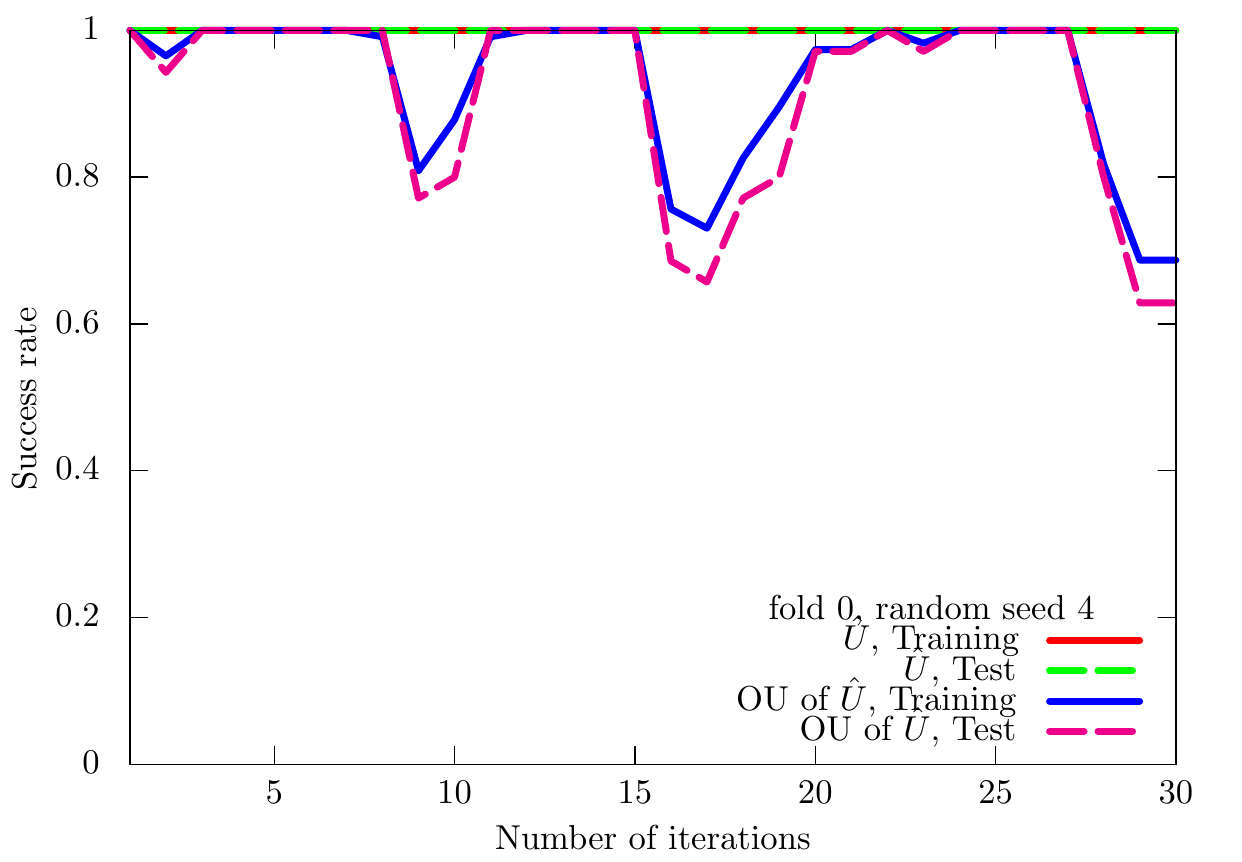}
\includegraphics[scale=0.25]{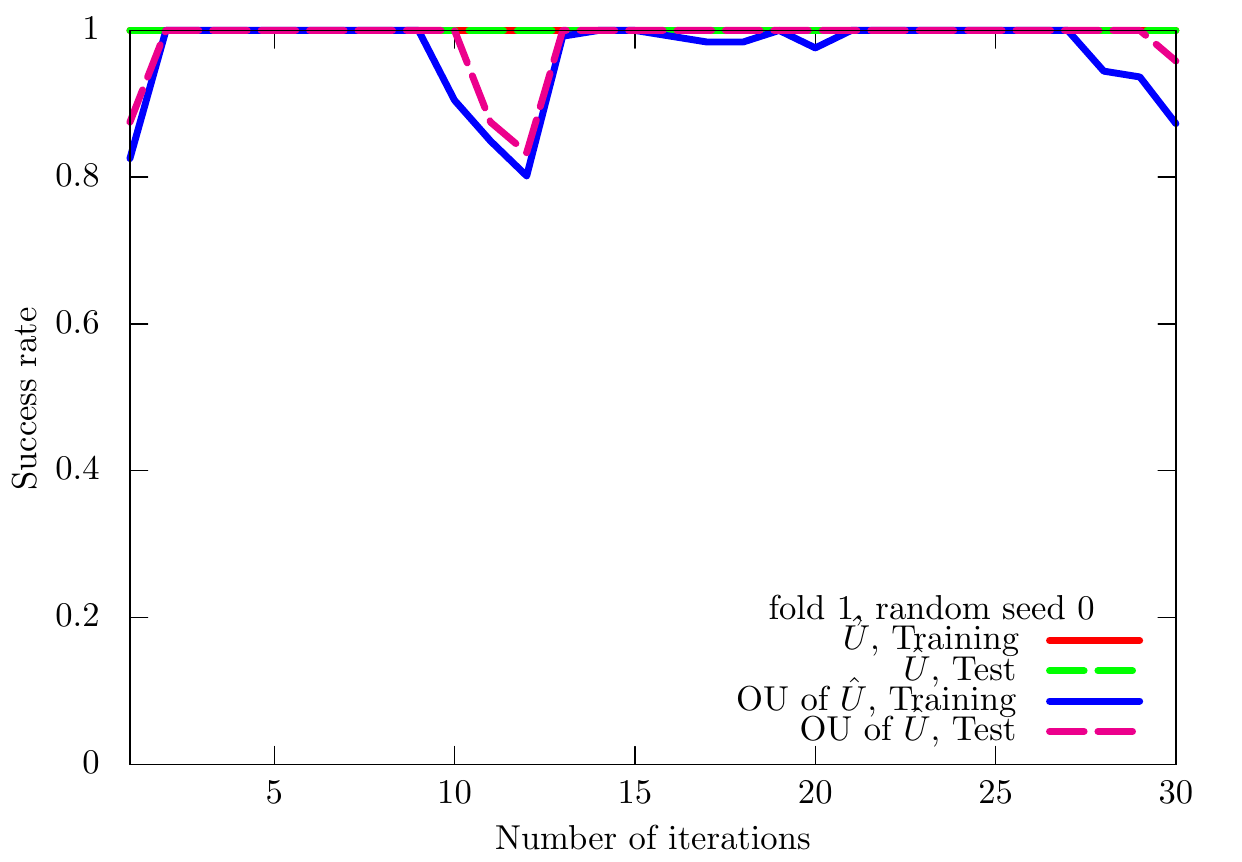}
\includegraphics[scale=0.25]{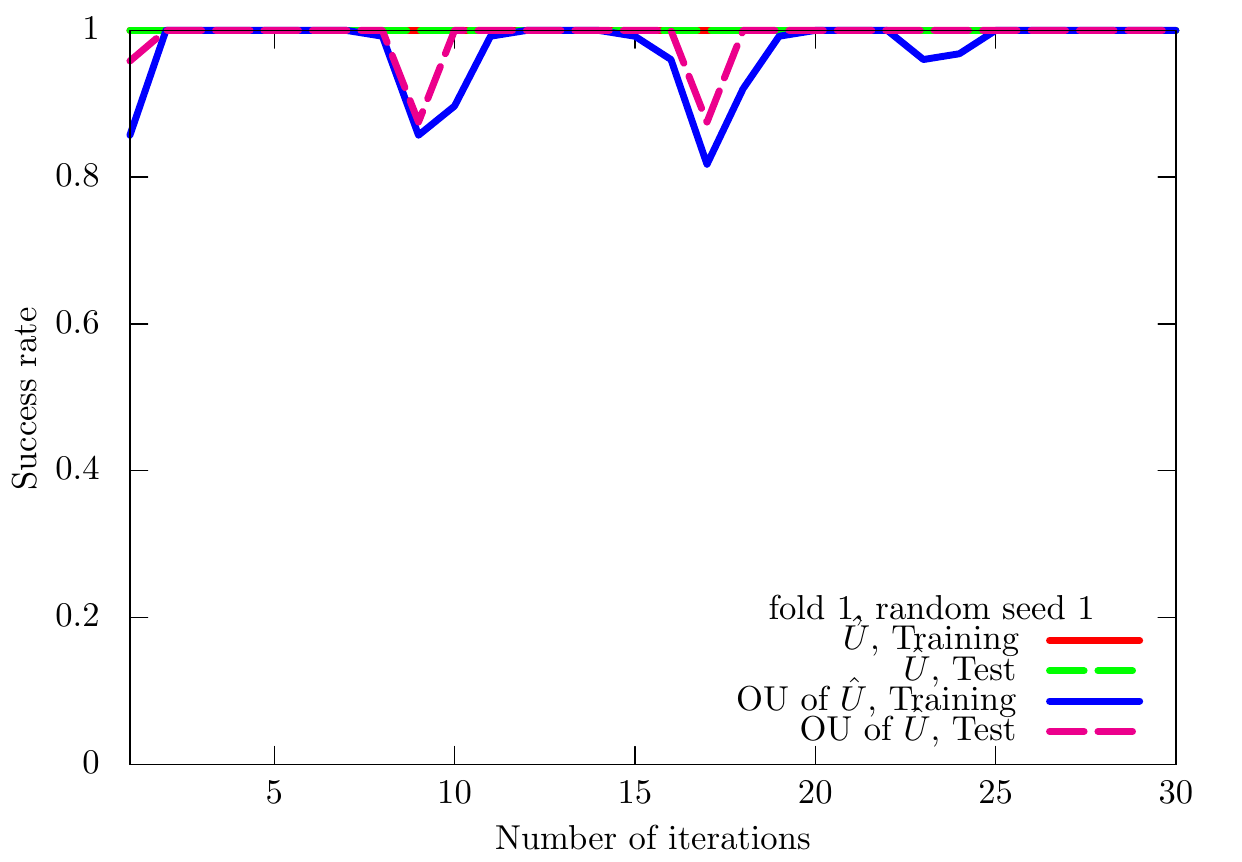}
\includegraphics[scale=0.25]{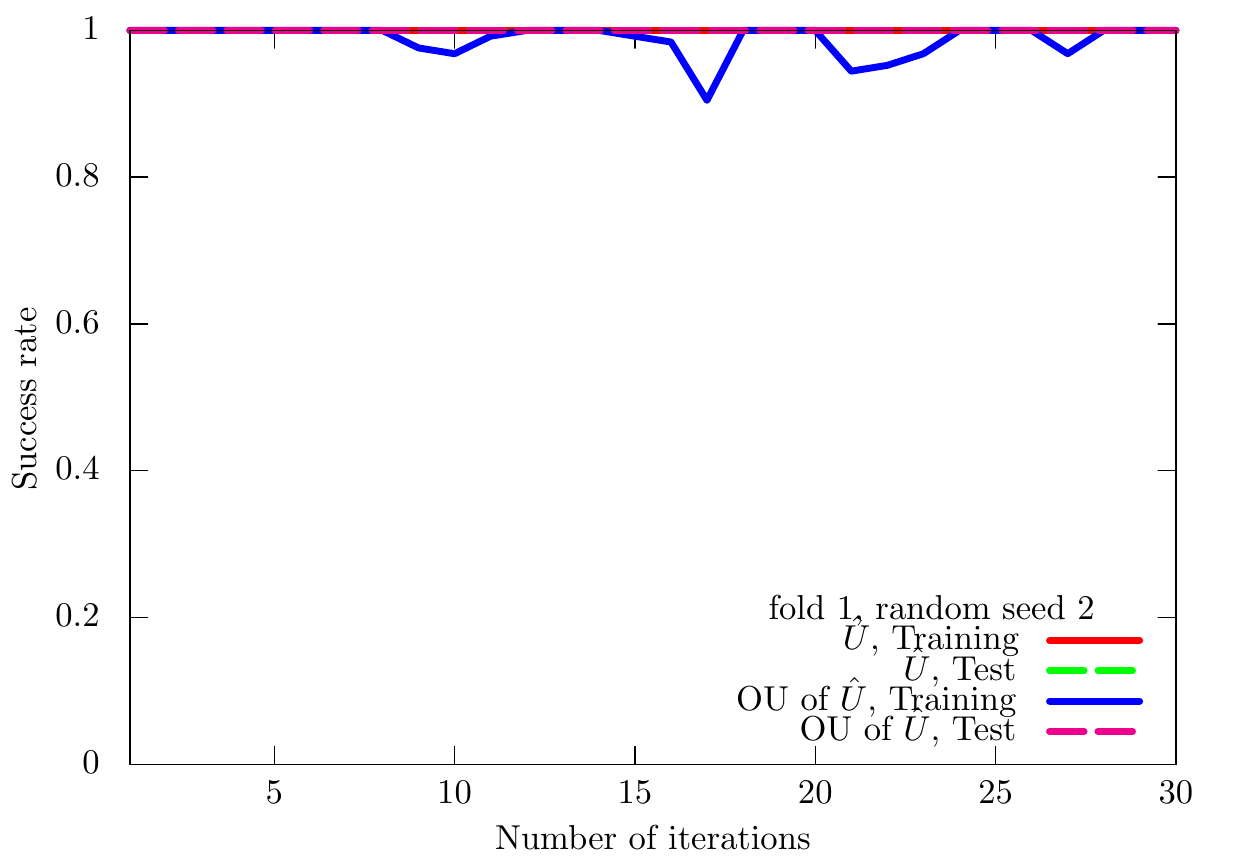}
\includegraphics[scale=0.25]{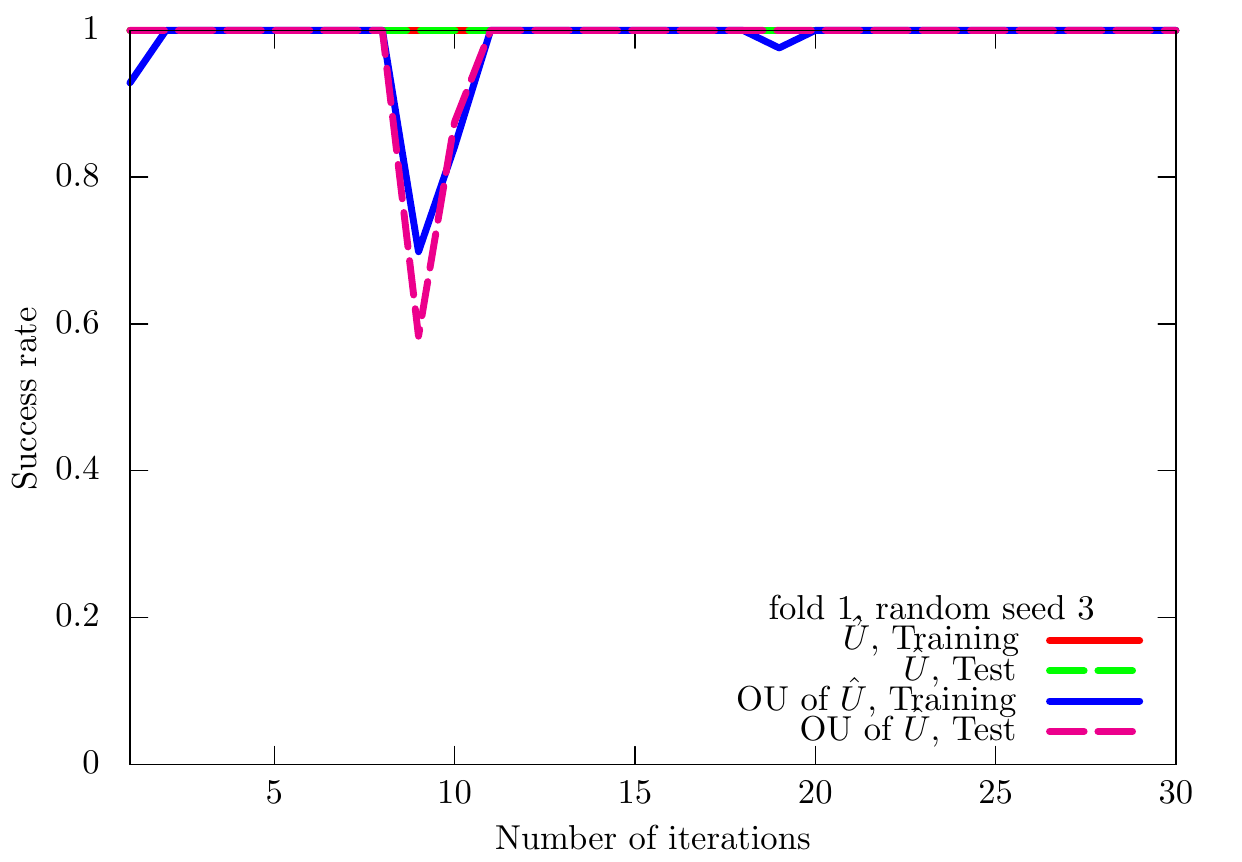}
\includegraphics[scale=0.25]{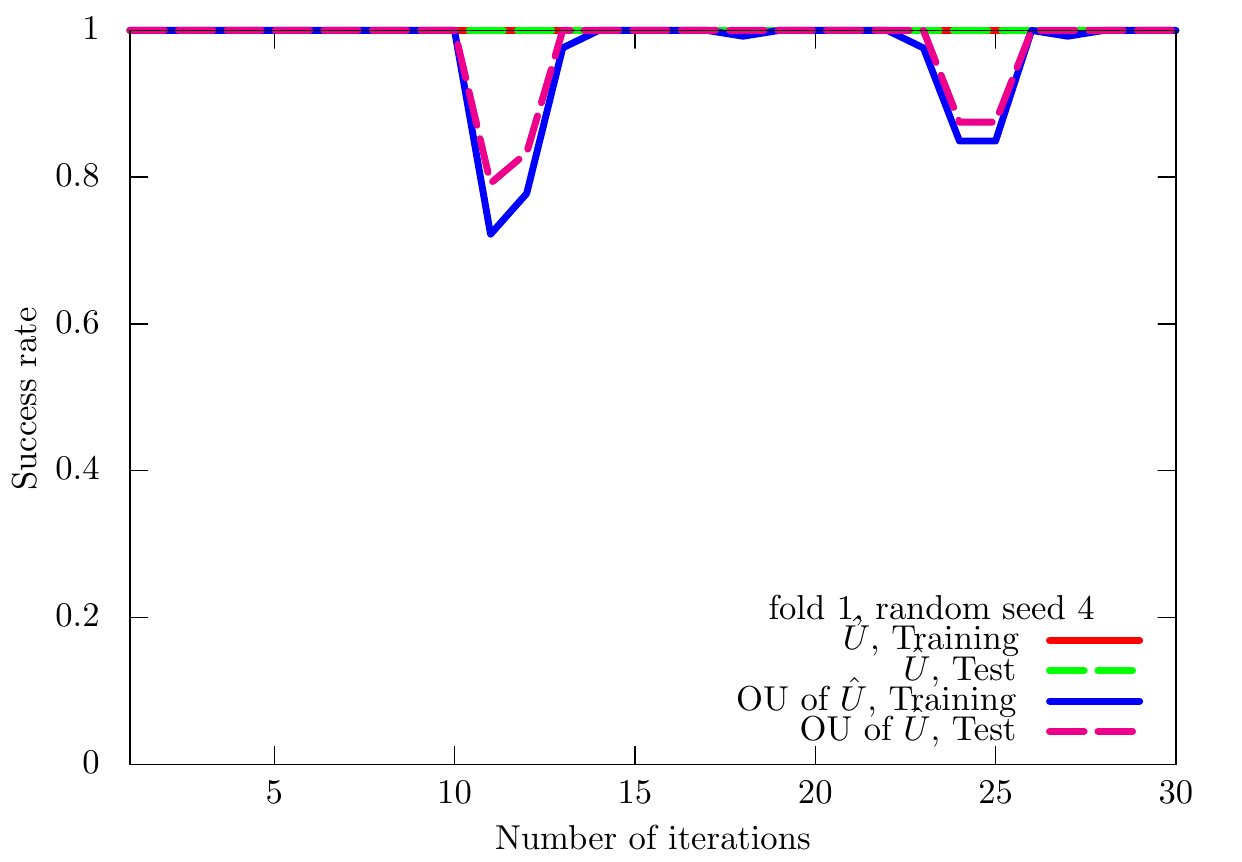}
\includegraphics[scale=0.25]{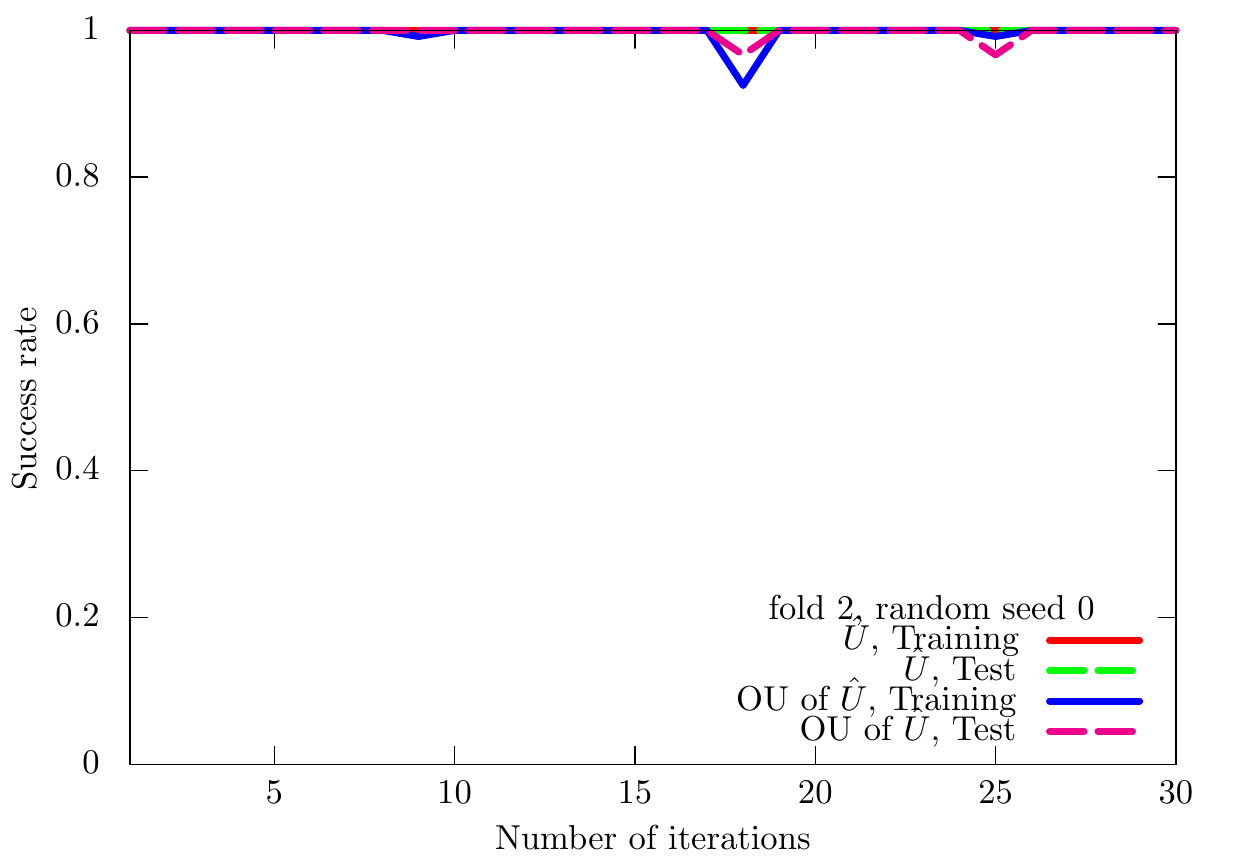}
\includegraphics[scale=0.25]{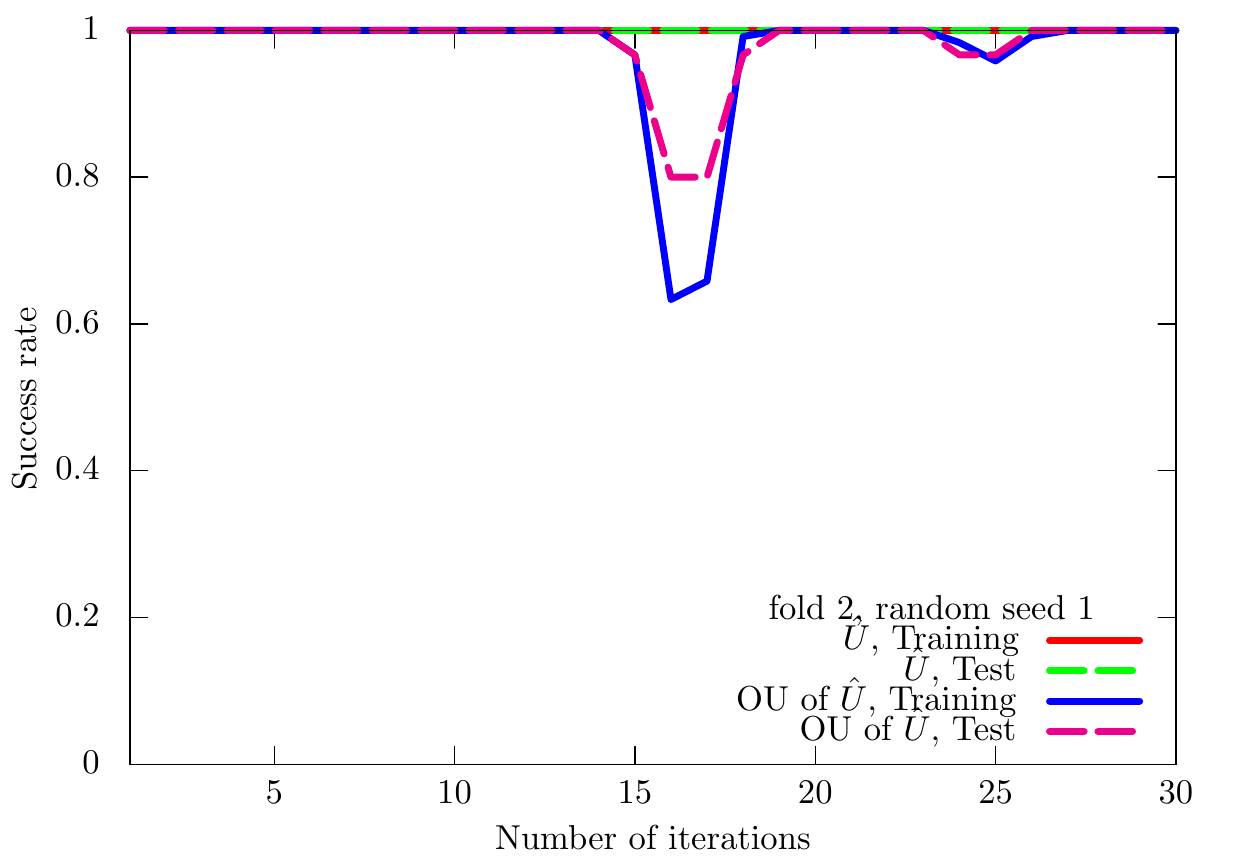}
\includegraphics[scale=0.25]{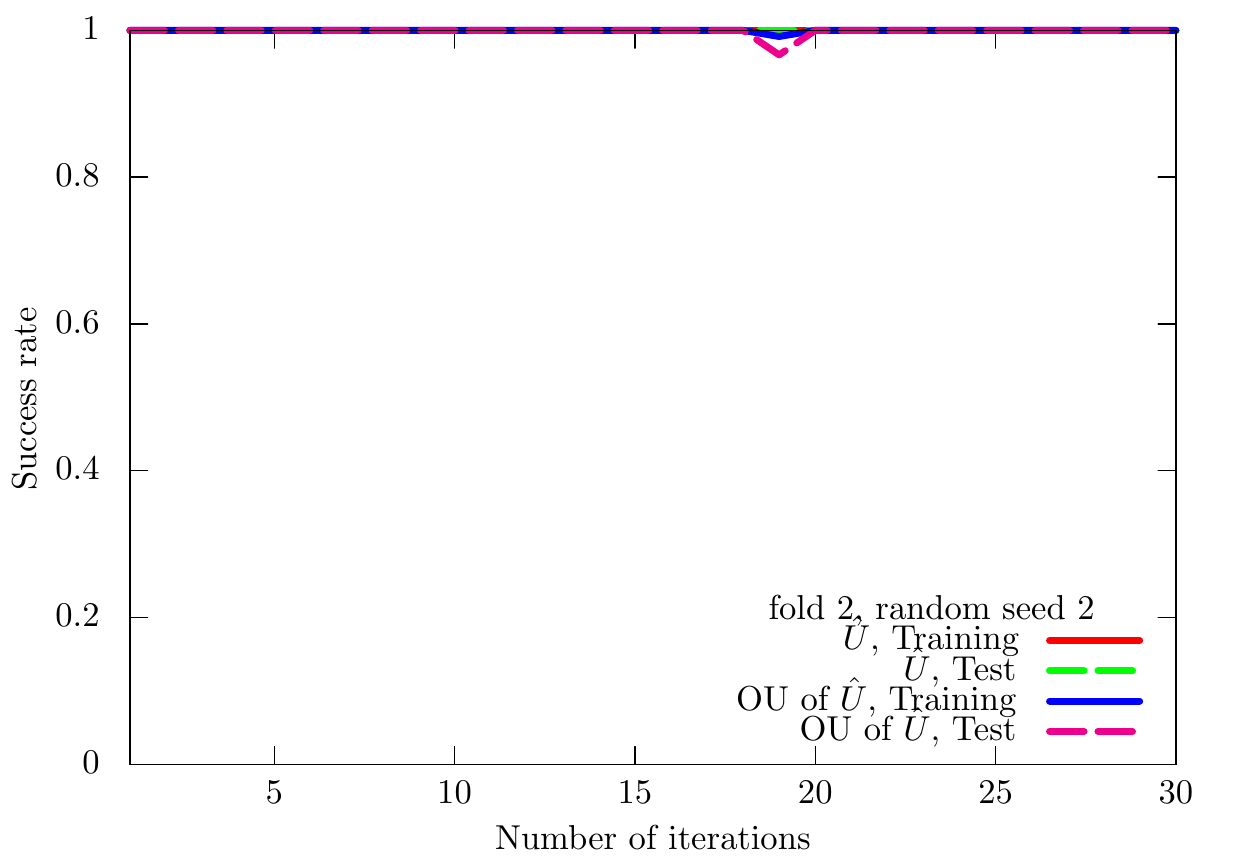}
\includegraphics[scale=0.25]{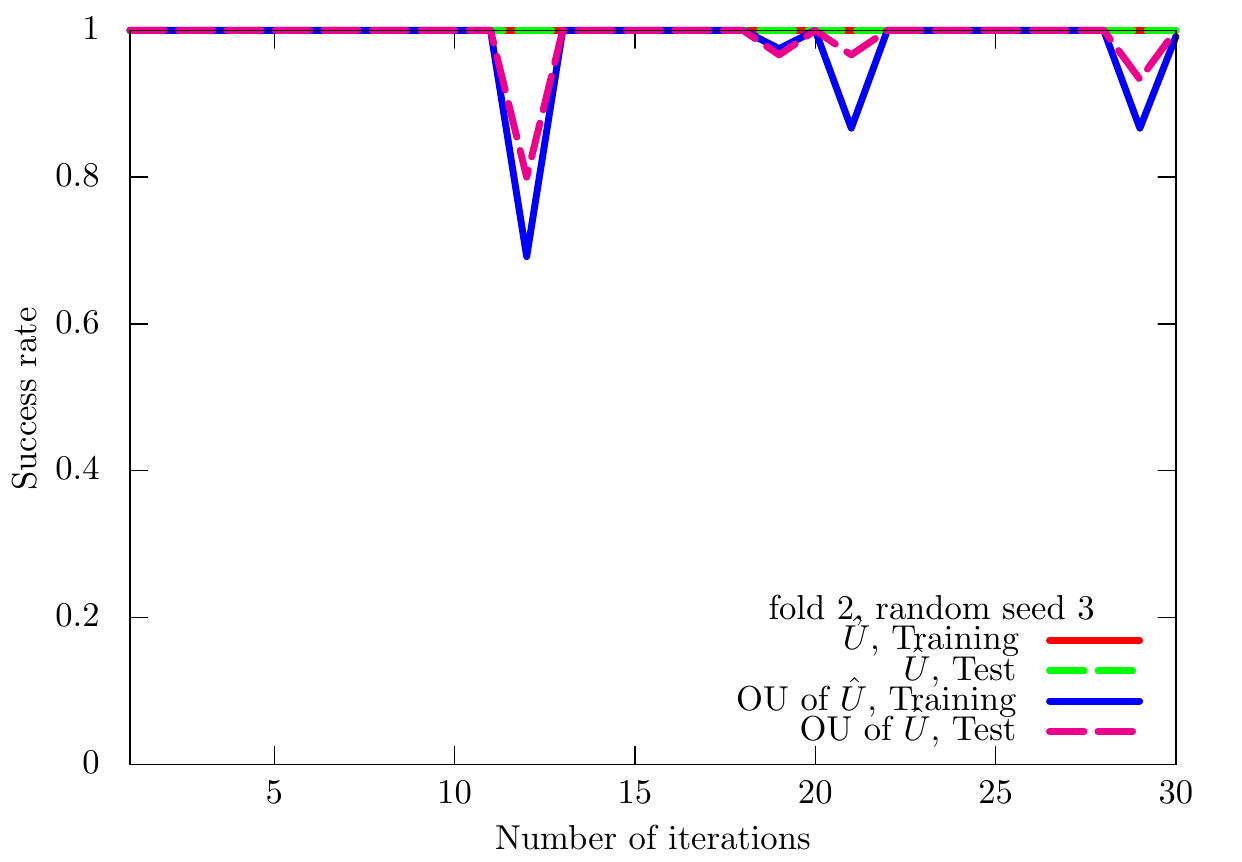}
\includegraphics[scale=0.25]{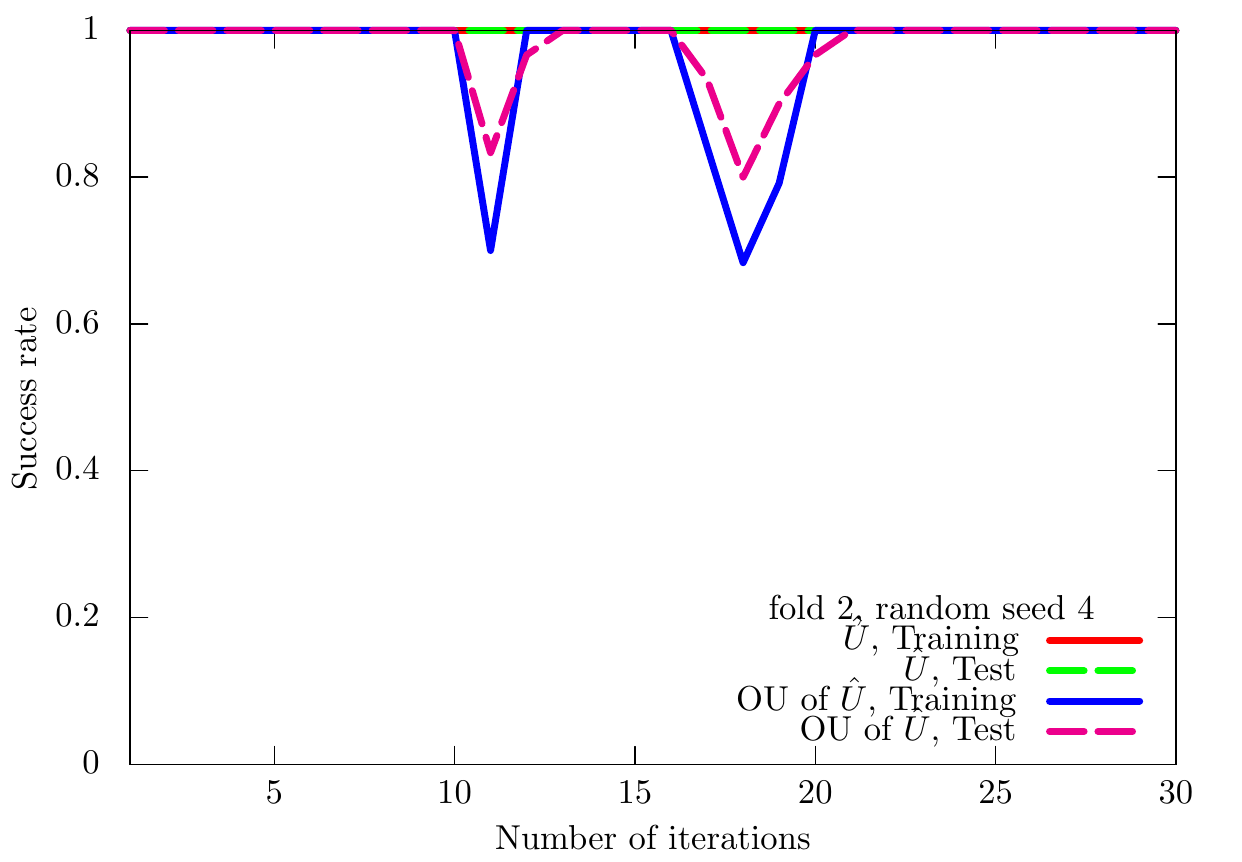}
\includegraphics[scale=0.25]{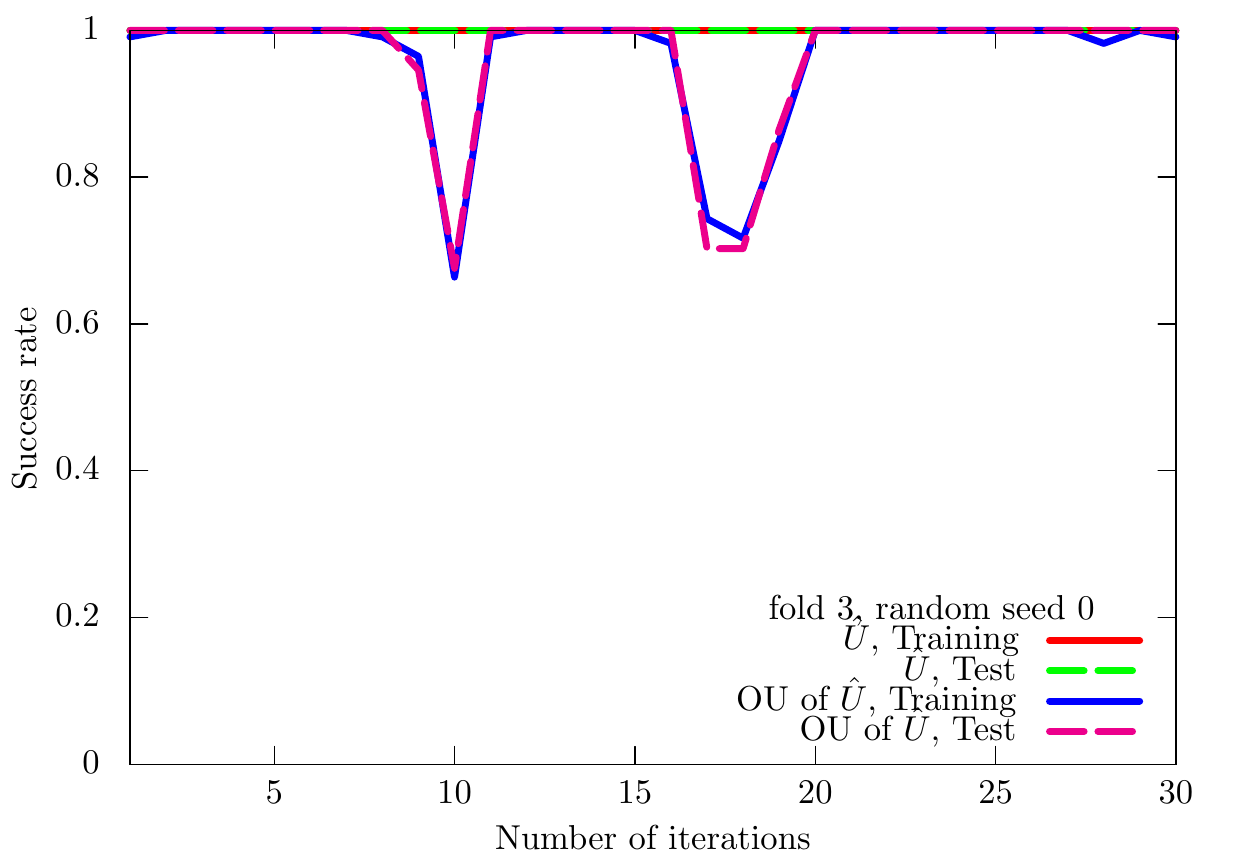}
\includegraphics[scale=0.25]{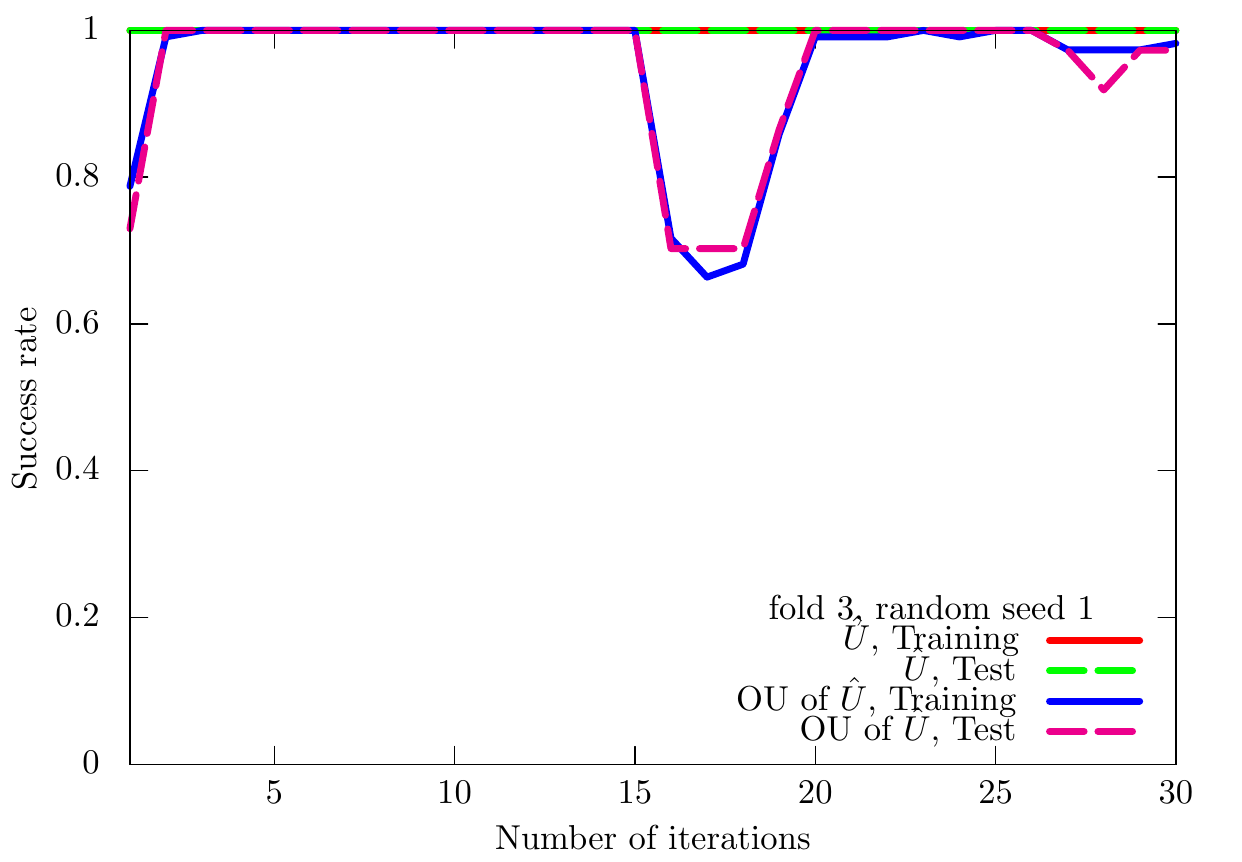}
\includegraphics[scale=0.25]{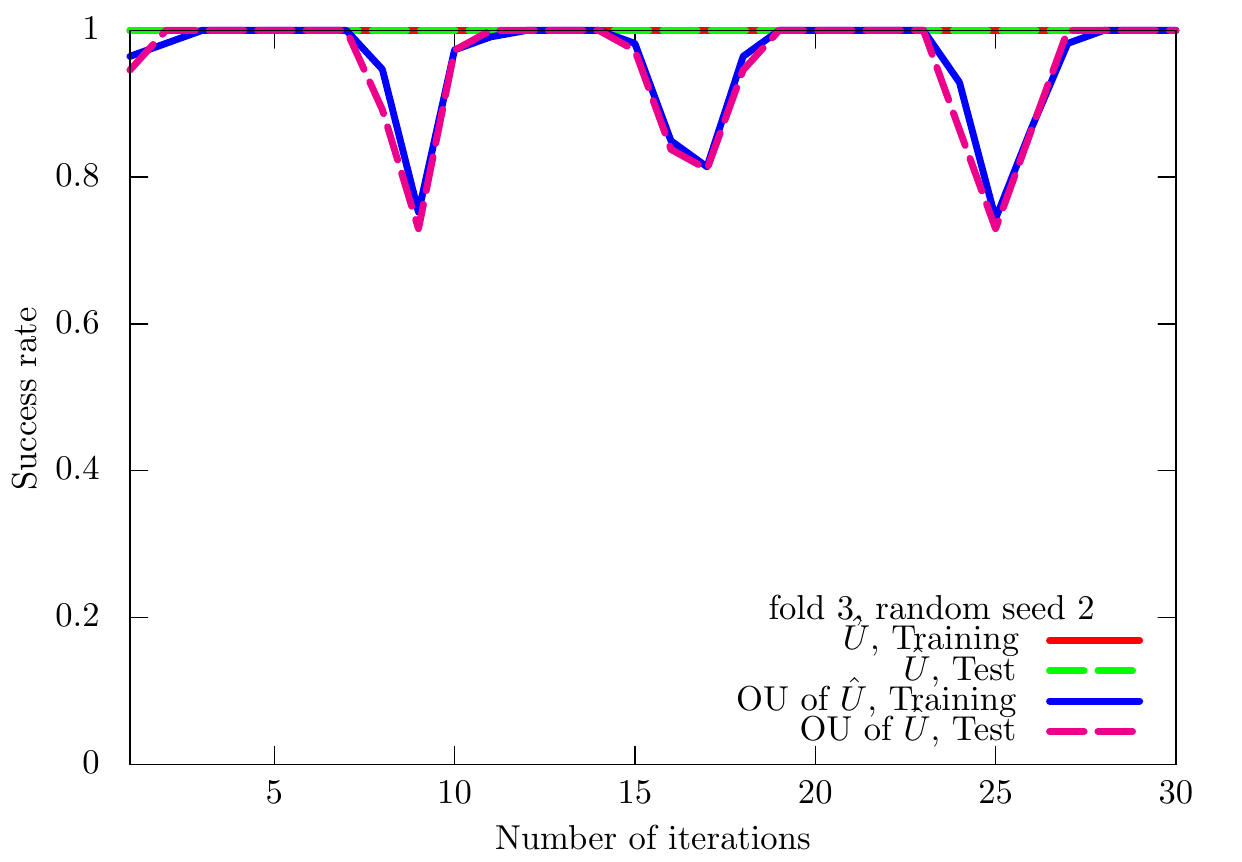}
\includegraphics[scale=0.25]{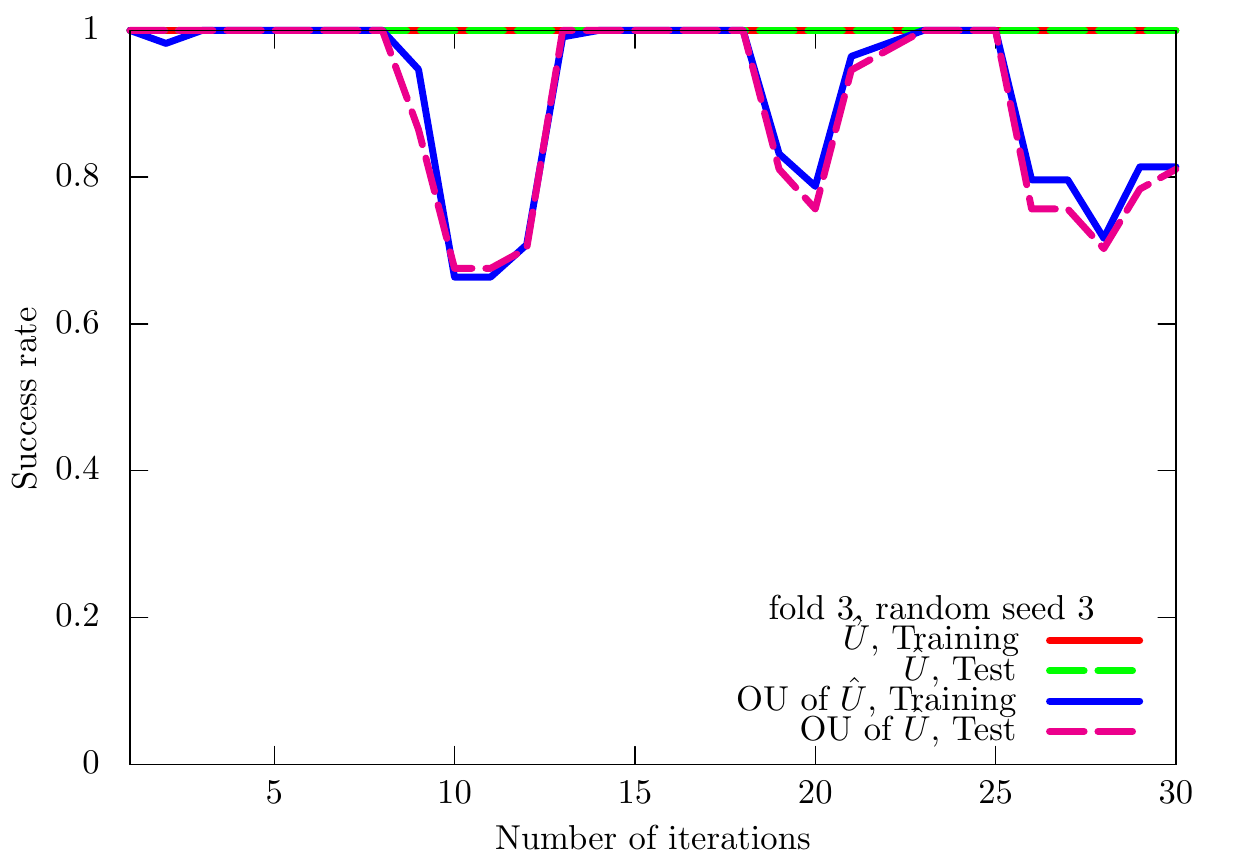}
\includegraphics[scale=0.25]{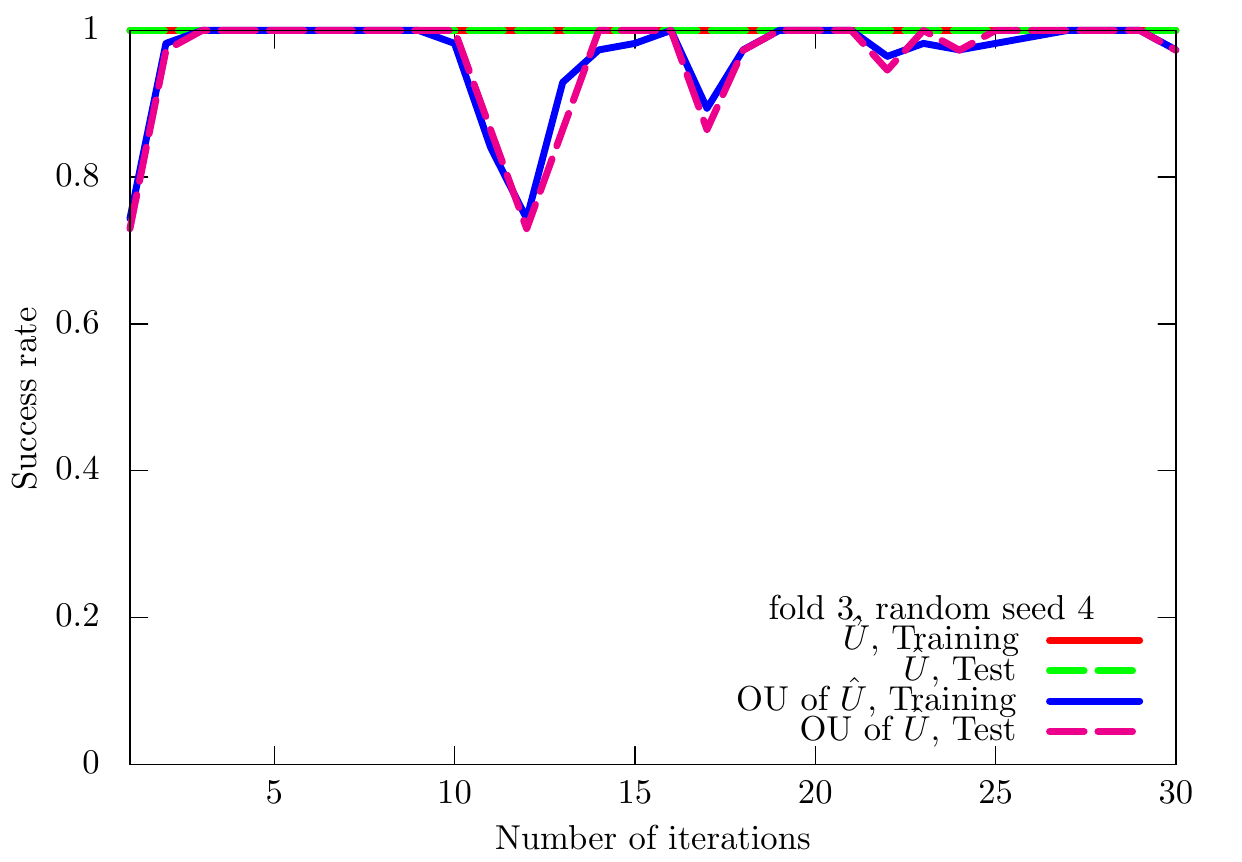}
\includegraphics[scale=0.25]{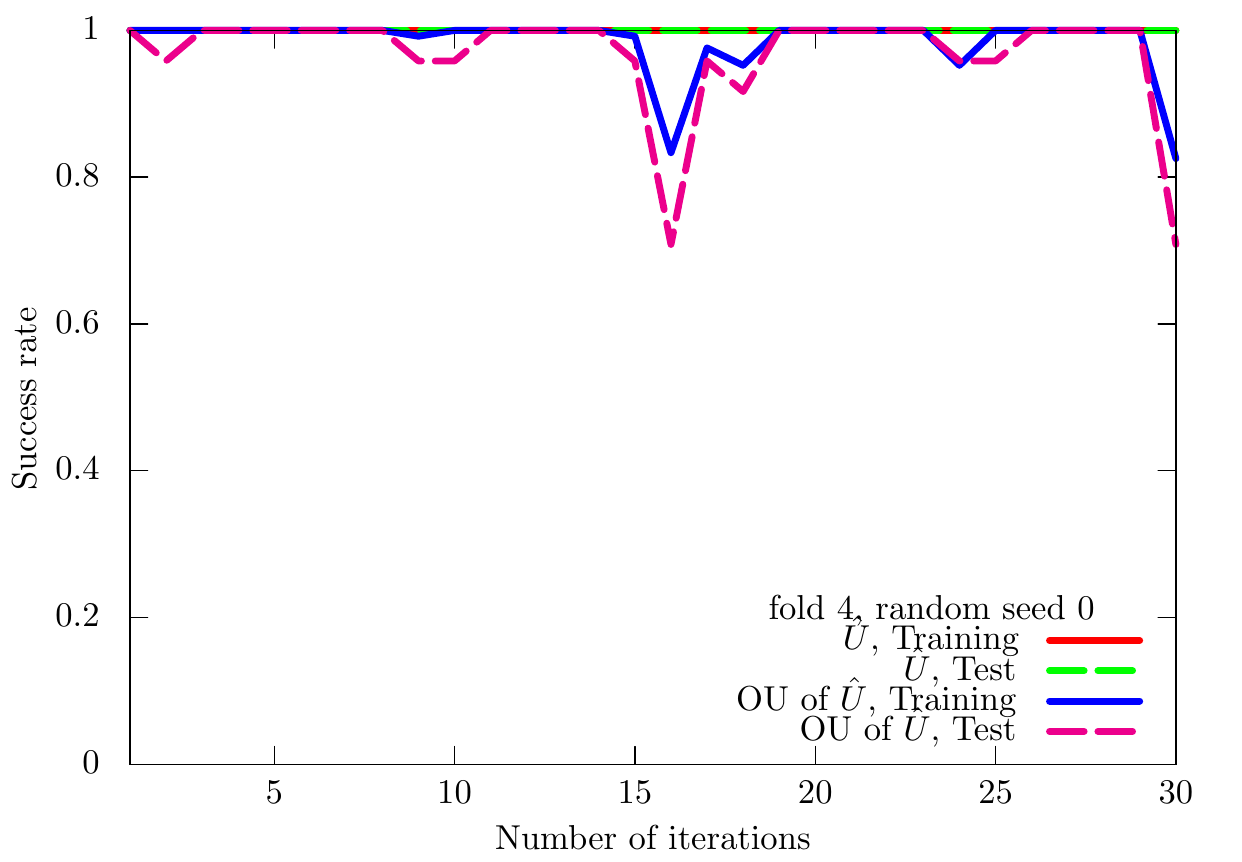}
\includegraphics[scale=0.25]{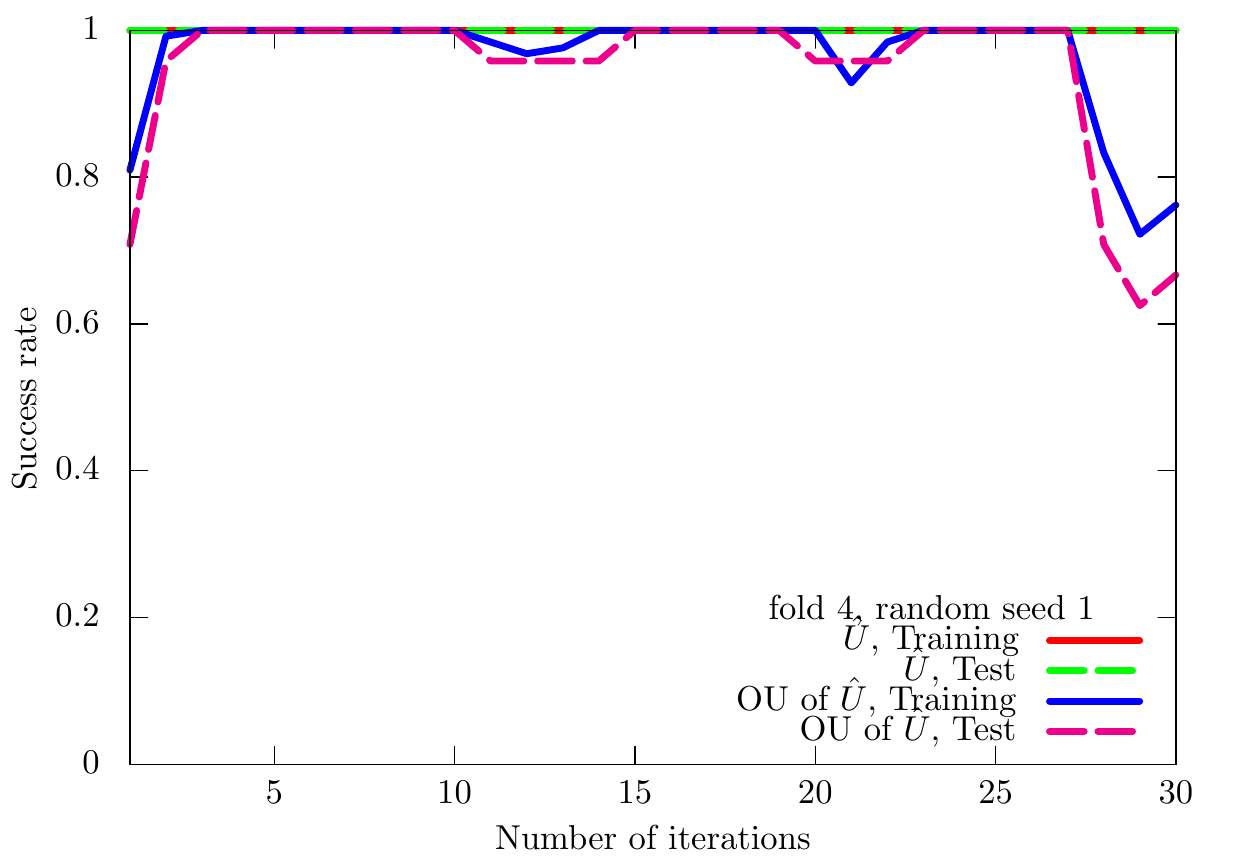}
\includegraphics[scale=0.25]{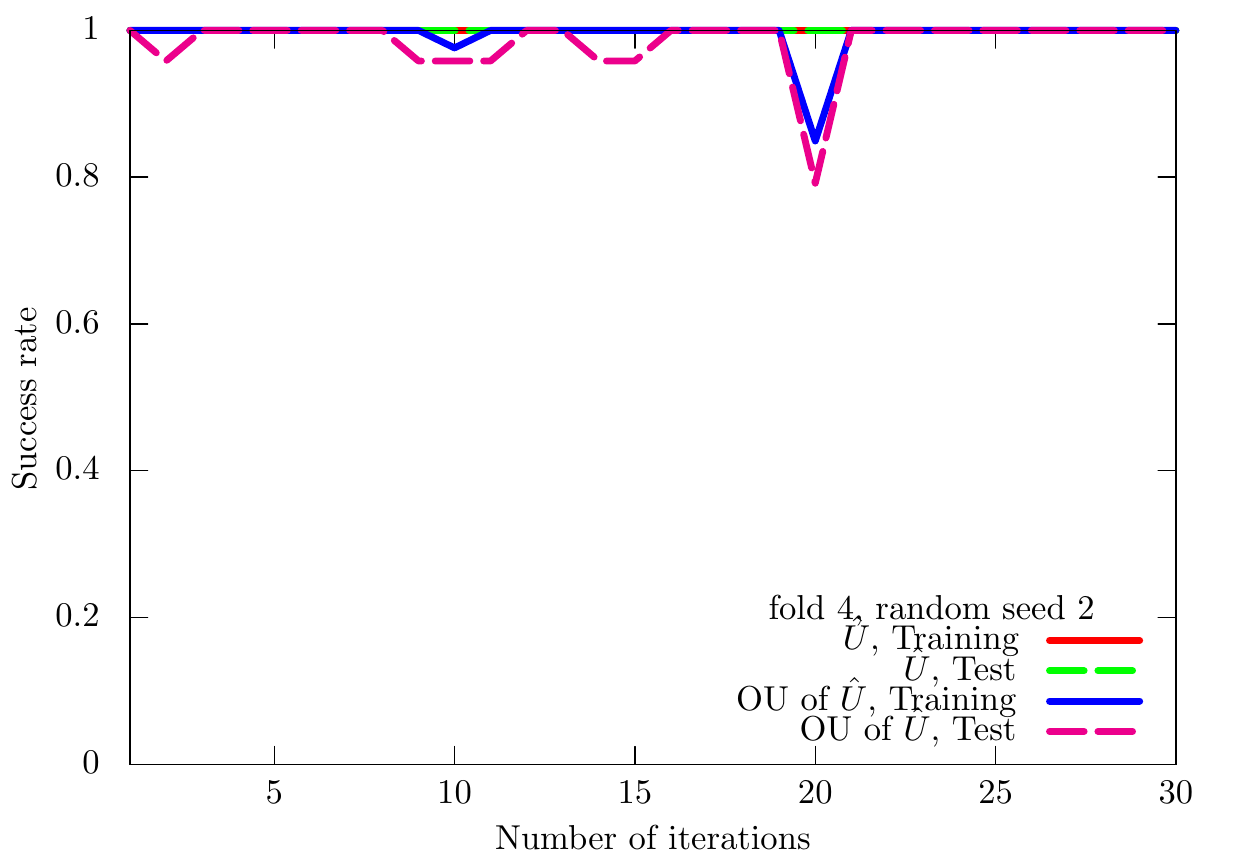}
\includegraphics[scale=0.25]{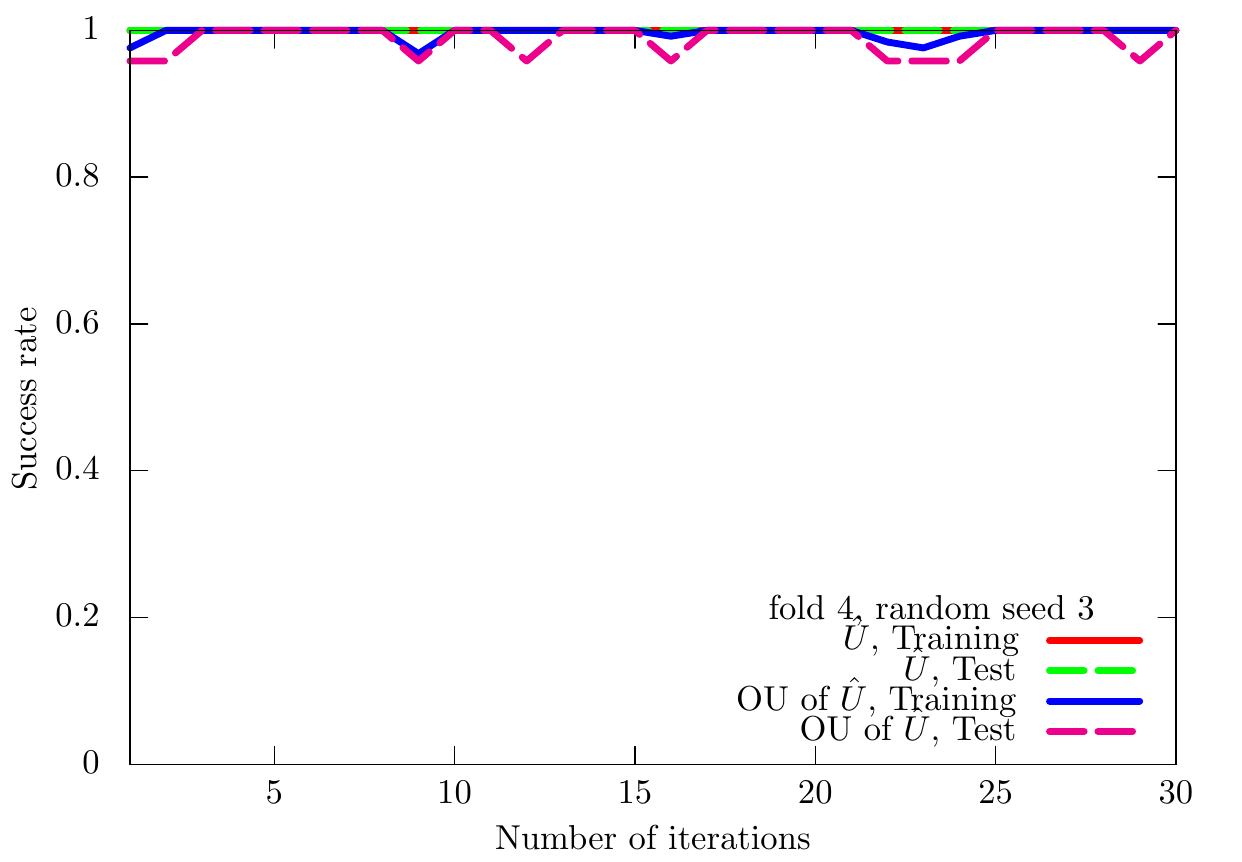}
\includegraphics[scale=0.25]{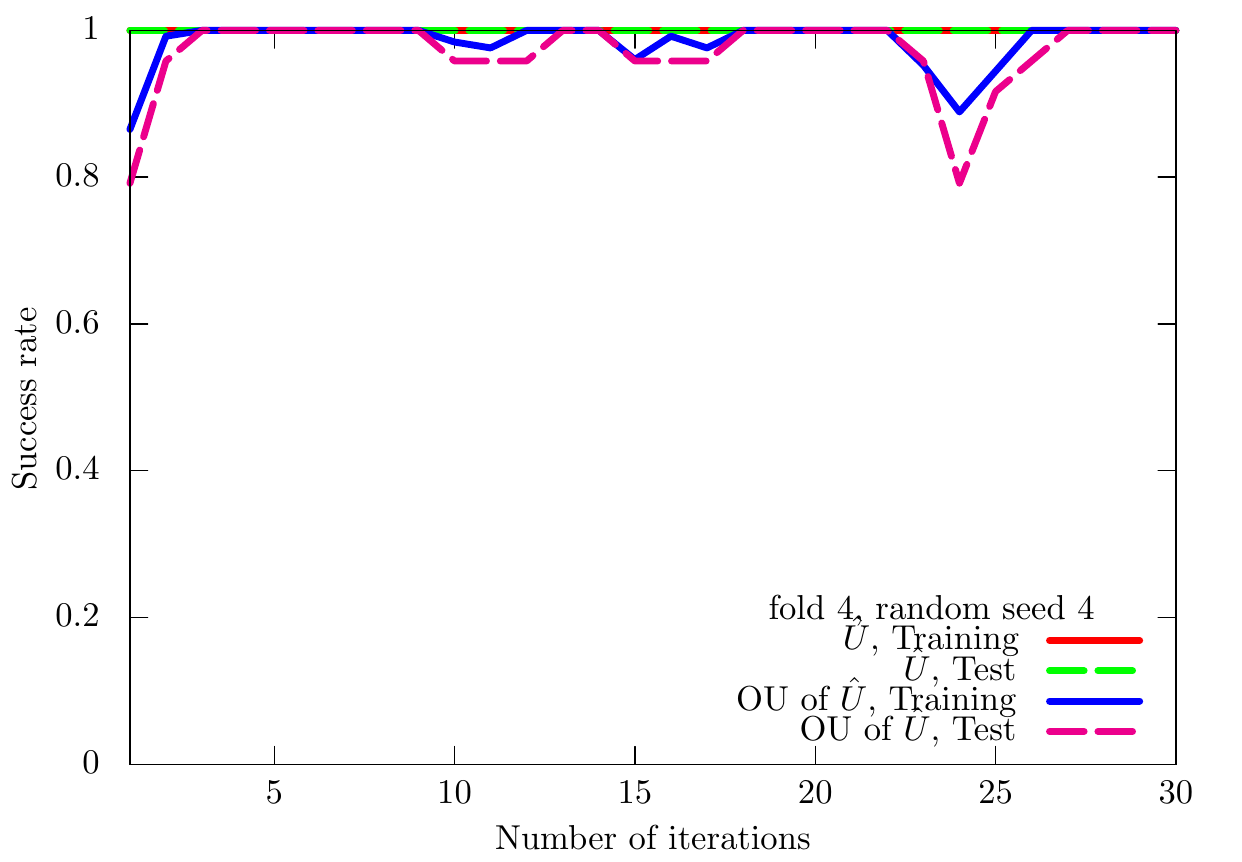}
\caption{Results of the UKM ($\hat{X}$ and $\hat{P}$) on the $5$-fold datasets with $5$ different random seeds for the iris dataset ($0$ or non-$0$). We use complex matrices and set $\theta_\mathrm{bias} = 0$. We set $r = 0.010$.}
\label{supp-arXiv-numerical-result-raw-data-fold-001-rand-001-UKM-P-UCI-iris-0-non0}
\end{figure*}

\begin{figure*}[htb]
\centering
\includegraphics[scale=0.25]{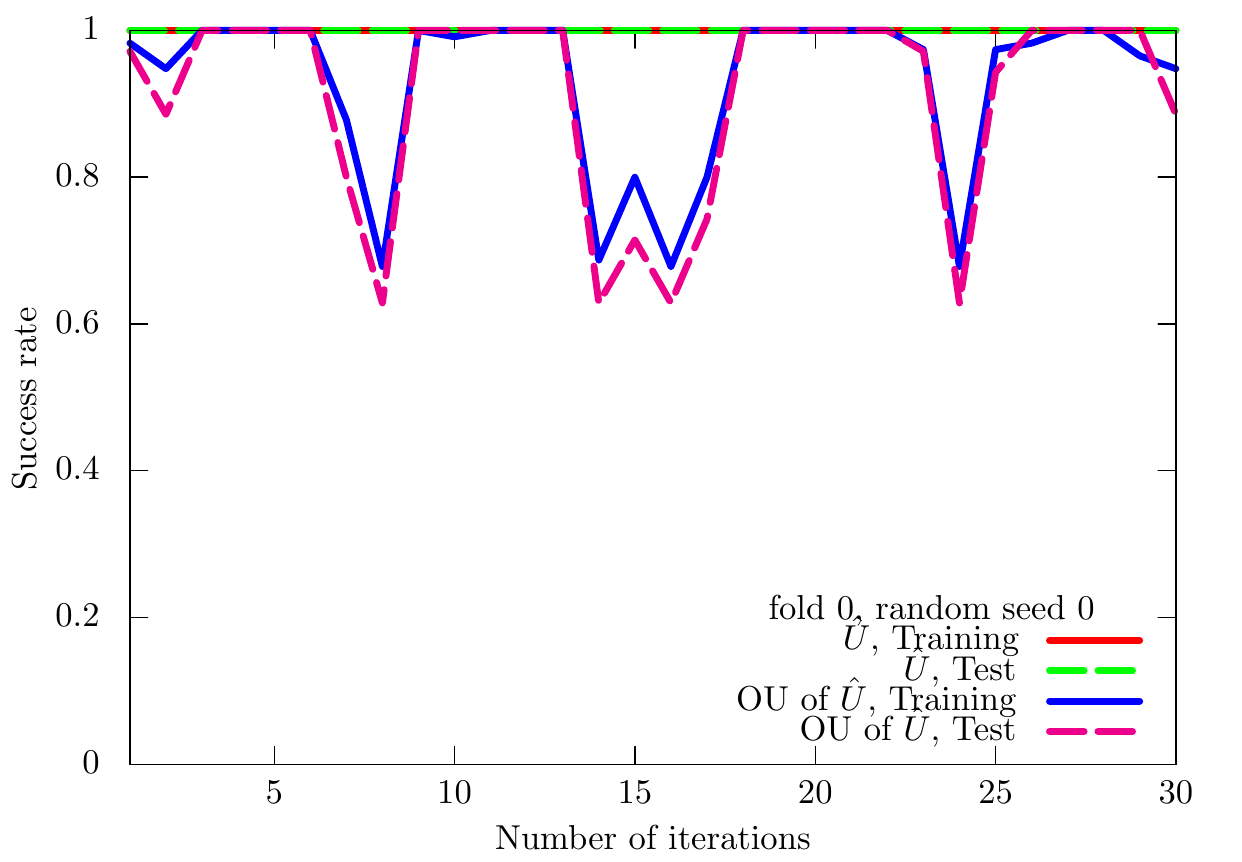}
\includegraphics[scale=0.25]{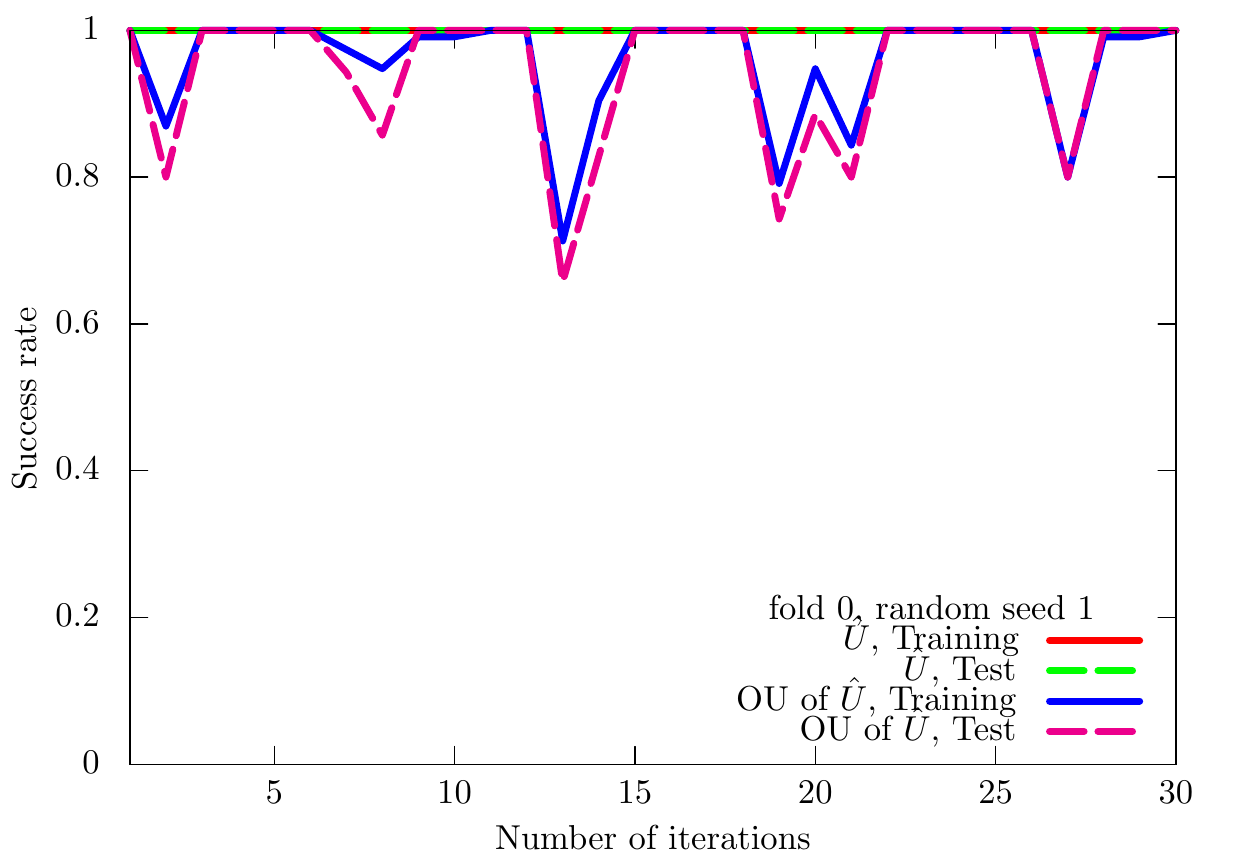}
\includegraphics[scale=0.25]{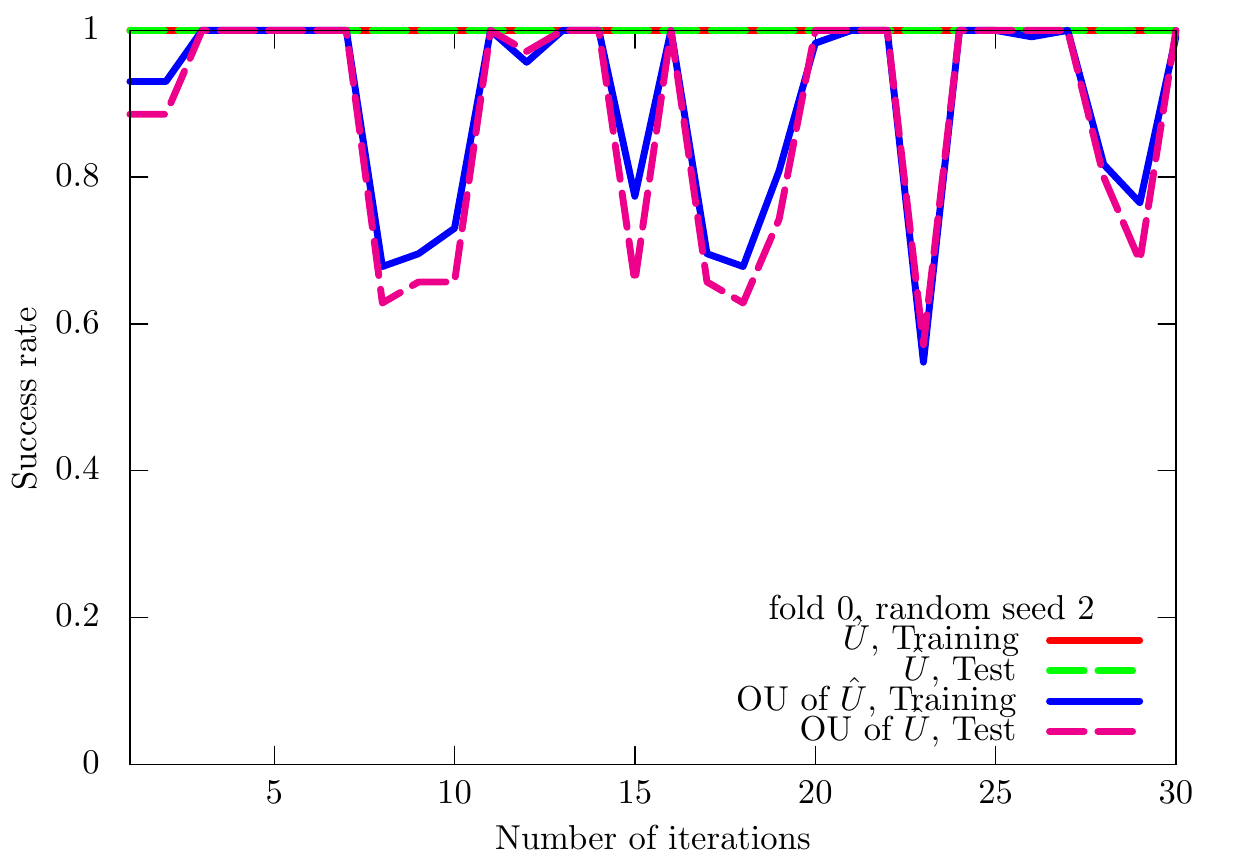}
\includegraphics[scale=0.25]{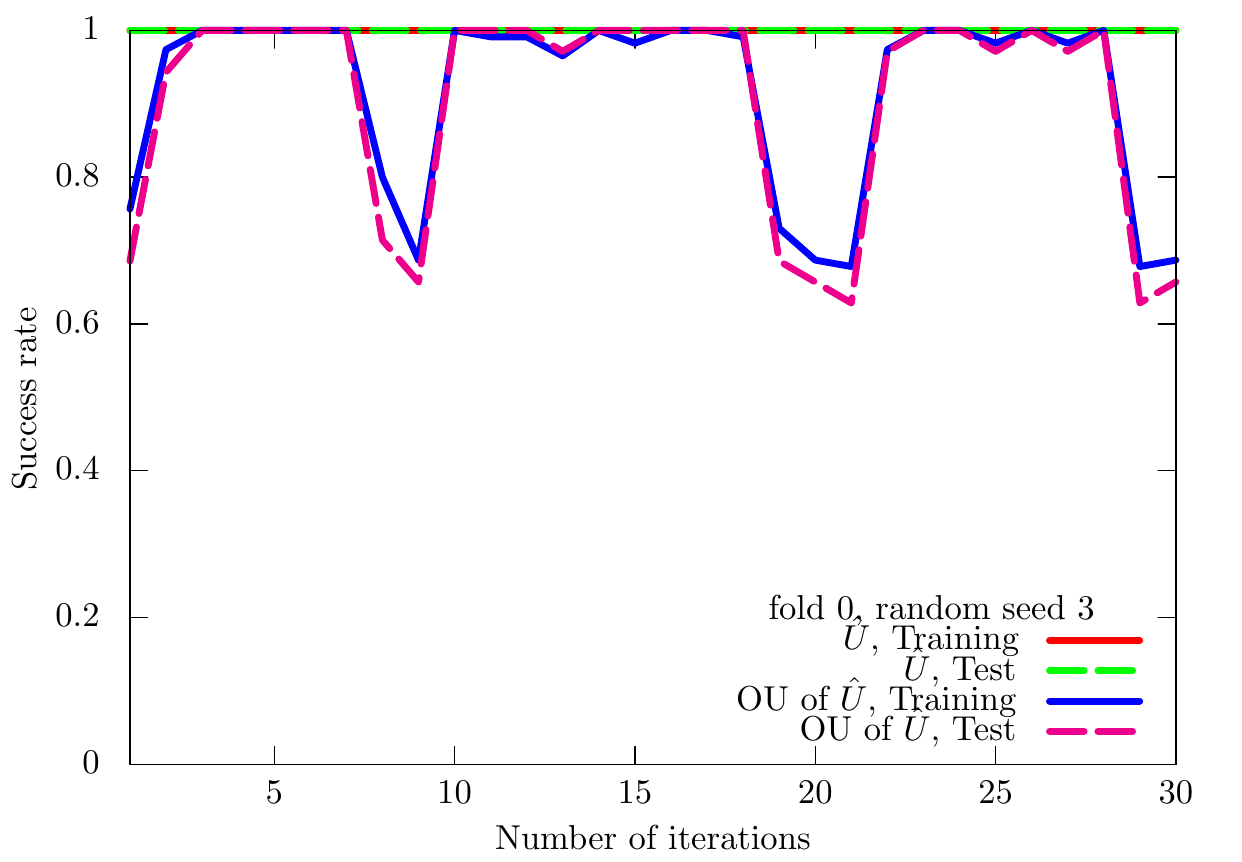}
\includegraphics[scale=0.25]{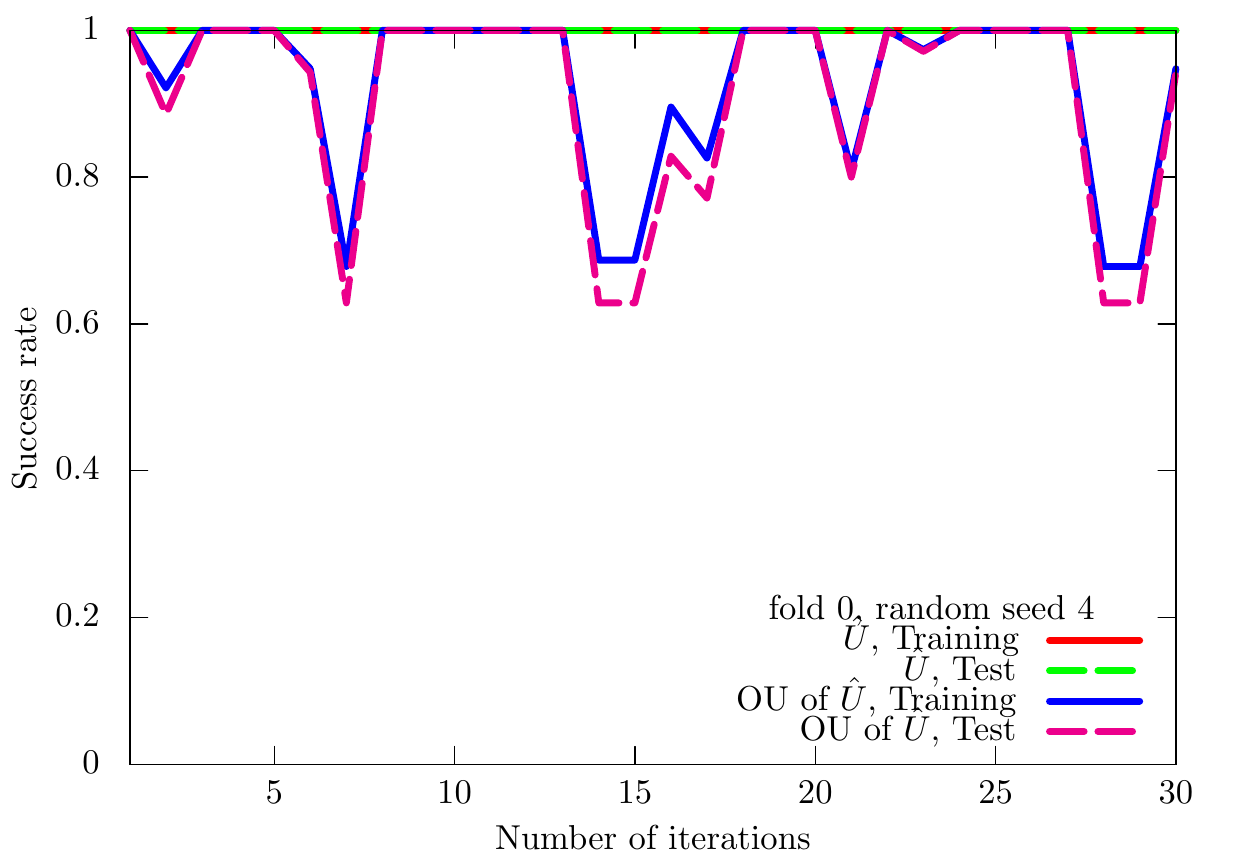}
\includegraphics[scale=0.25]{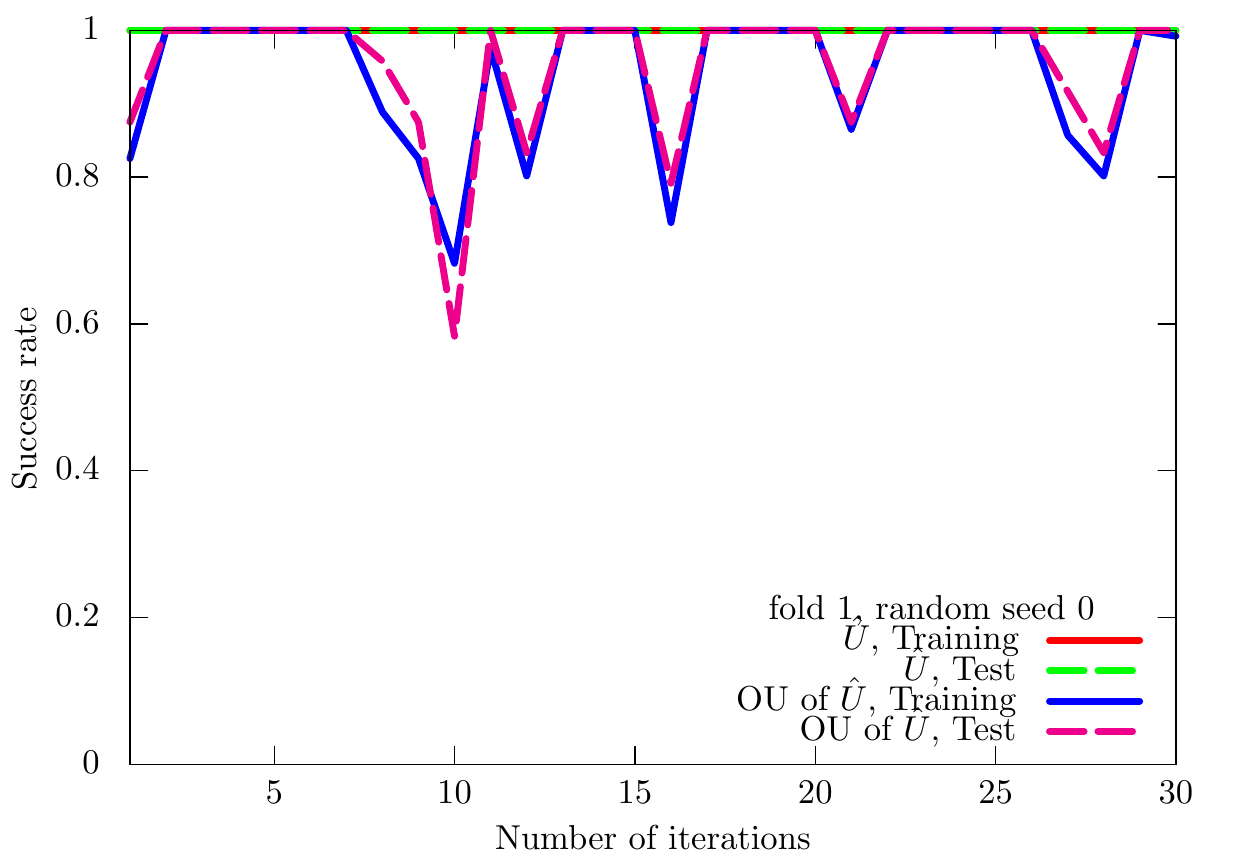}
\includegraphics[scale=0.25]{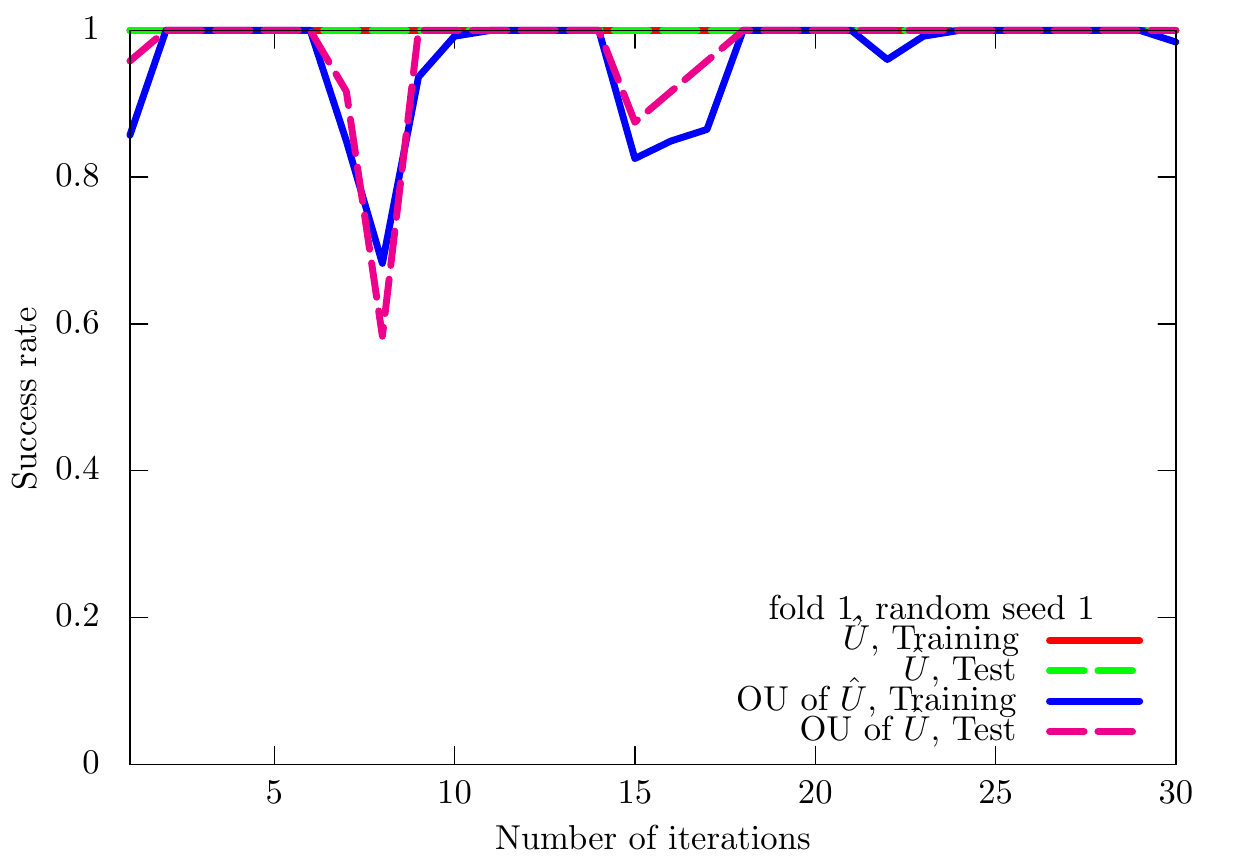}
\includegraphics[scale=0.25]{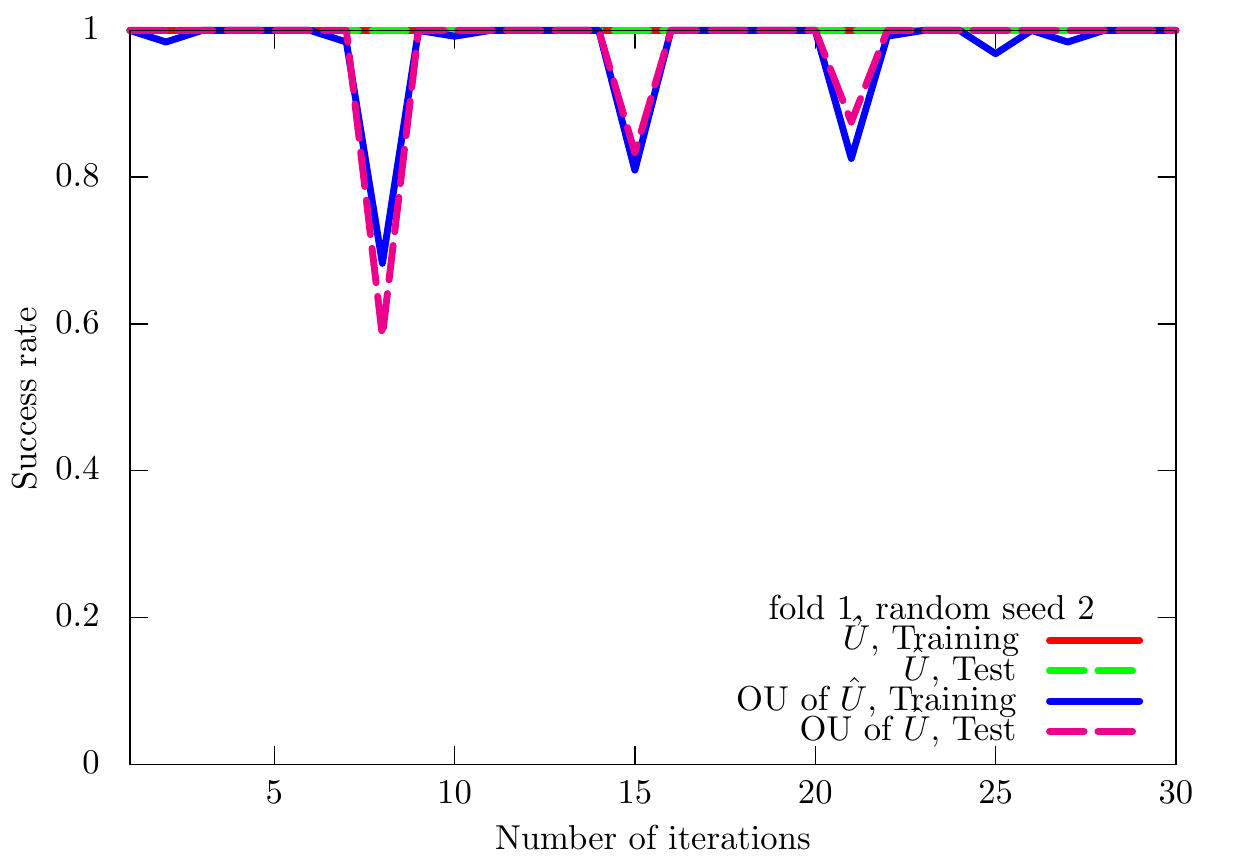}
\includegraphics[scale=0.25]{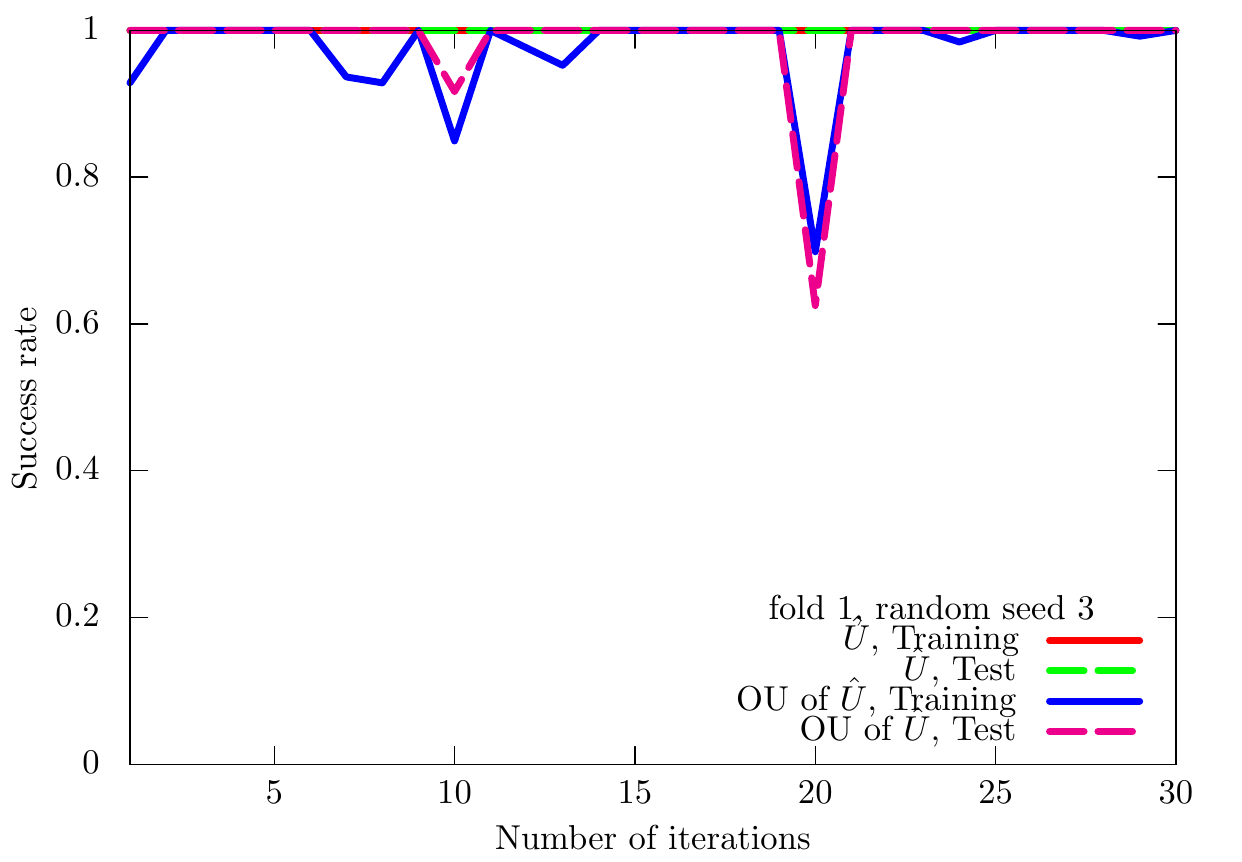}
\includegraphics[scale=0.25]{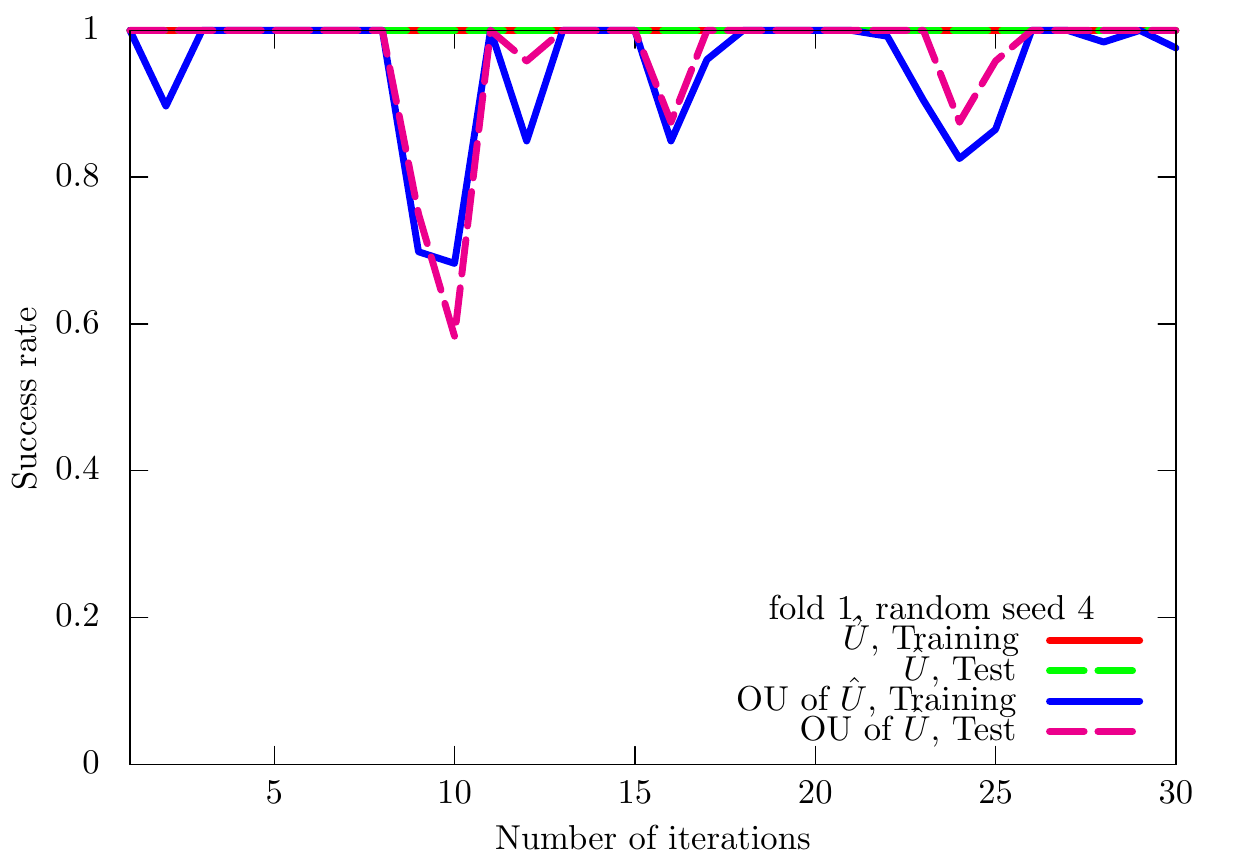}
\includegraphics[scale=0.25]{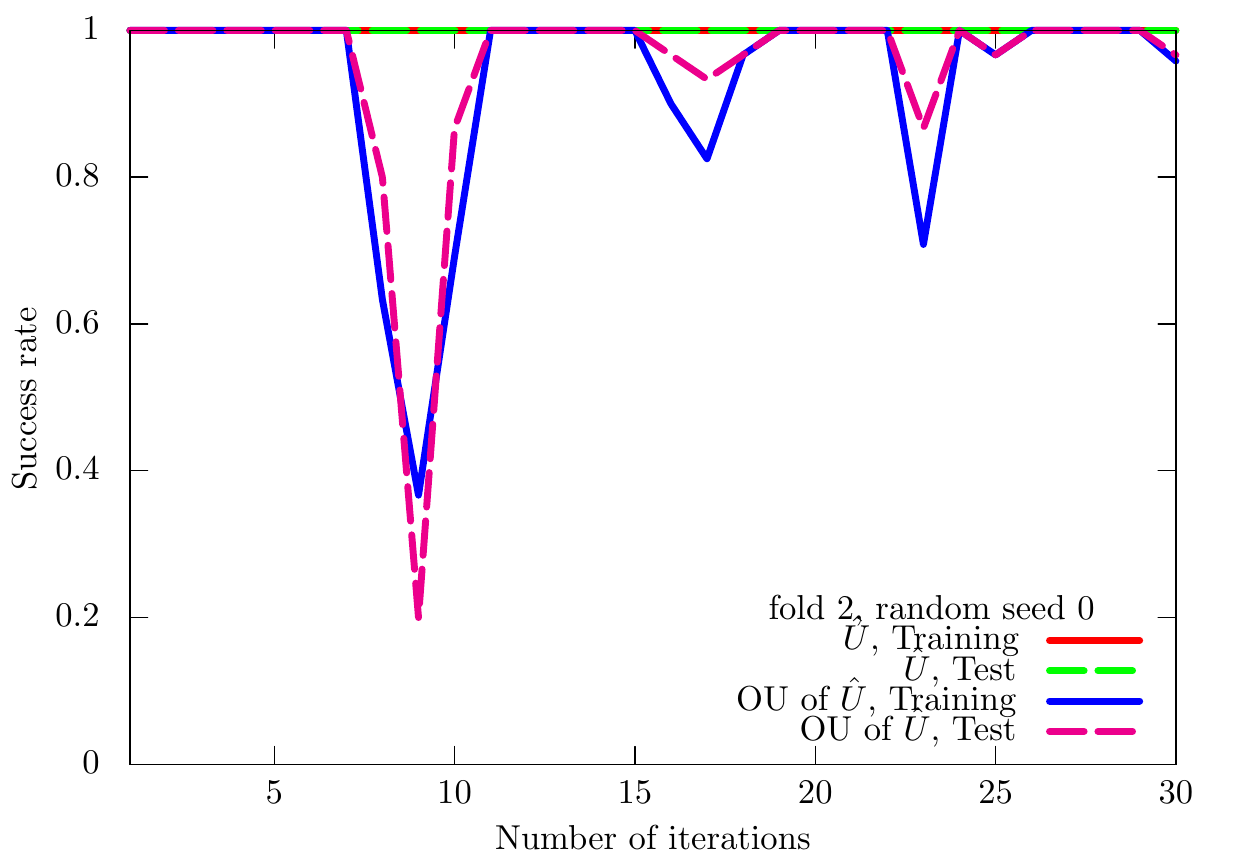}
\includegraphics[scale=0.25]{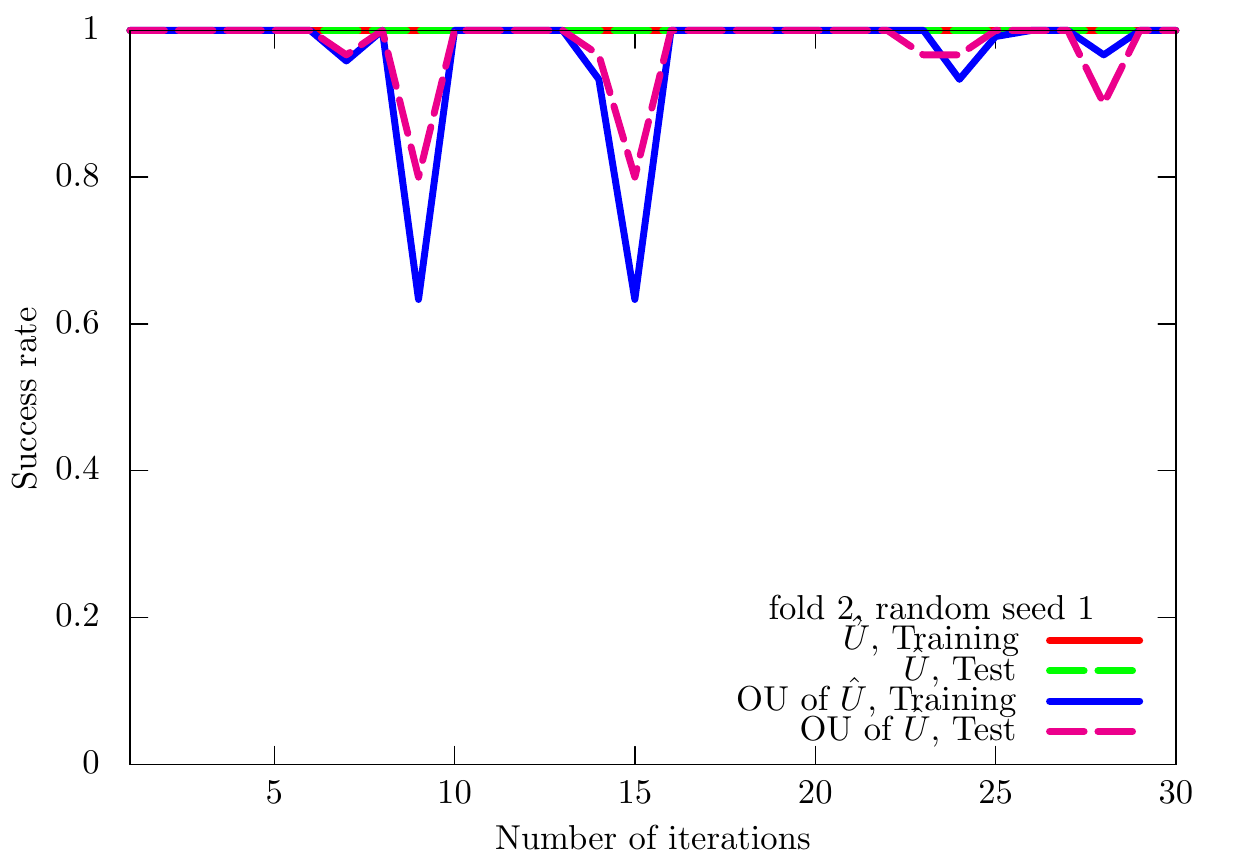}
\includegraphics[scale=0.25]{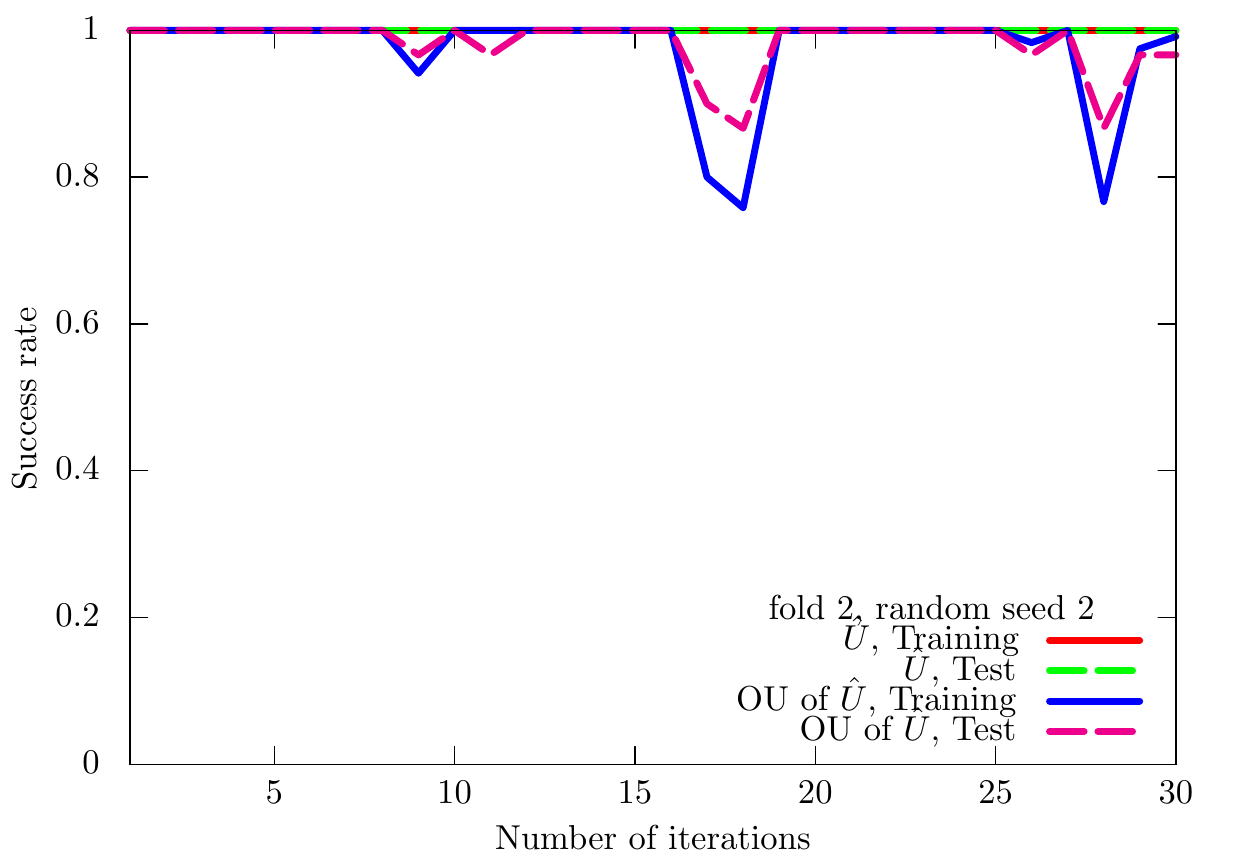}
\includegraphics[scale=0.25]{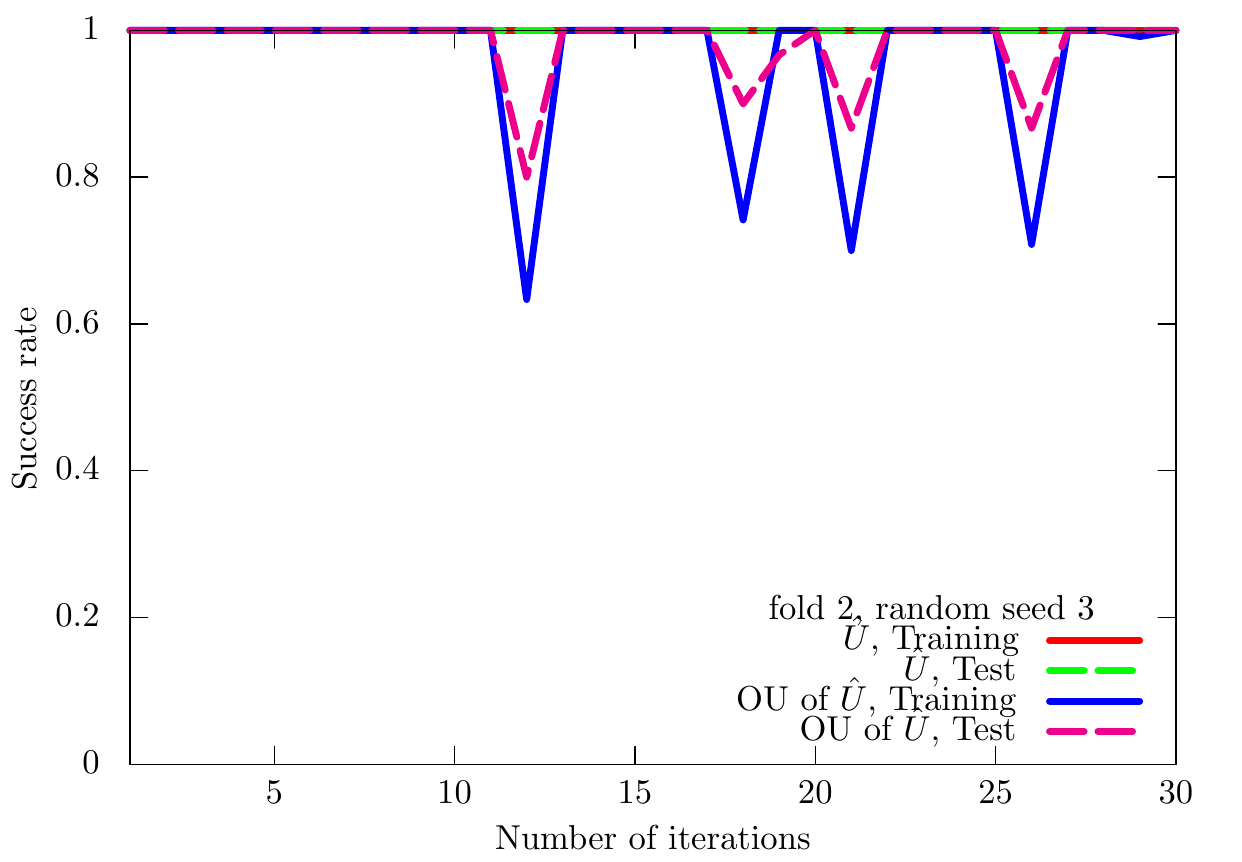}
\includegraphics[scale=0.25]{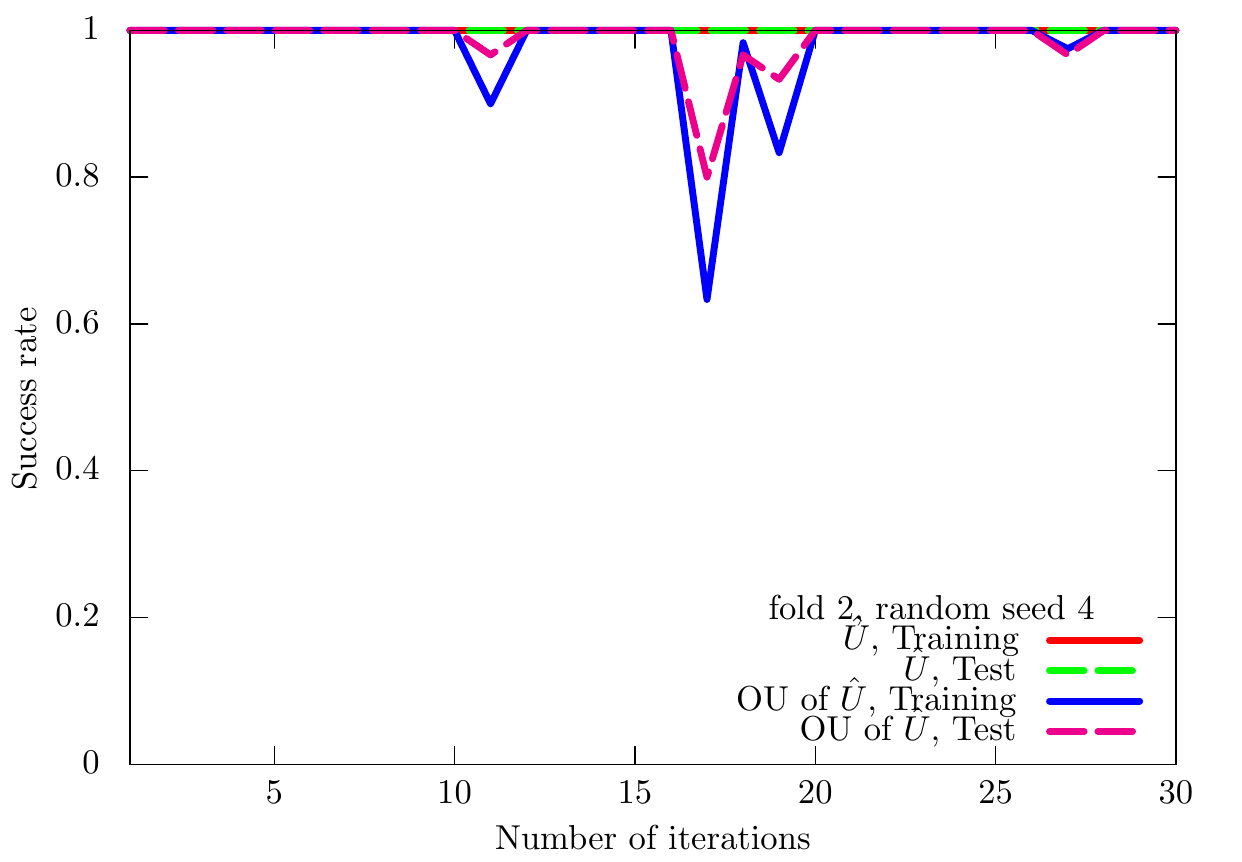}
\includegraphics[scale=0.25]{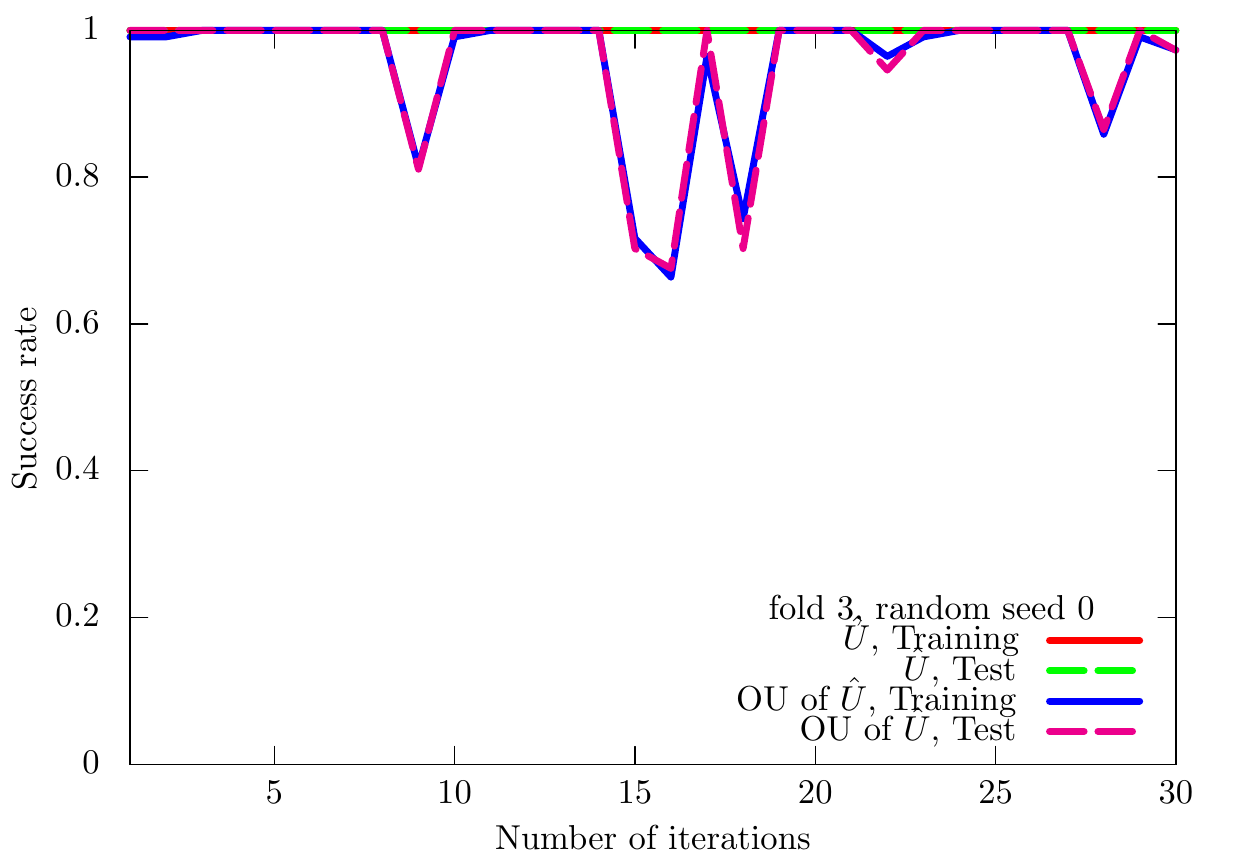}
\includegraphics[scale=0.25]{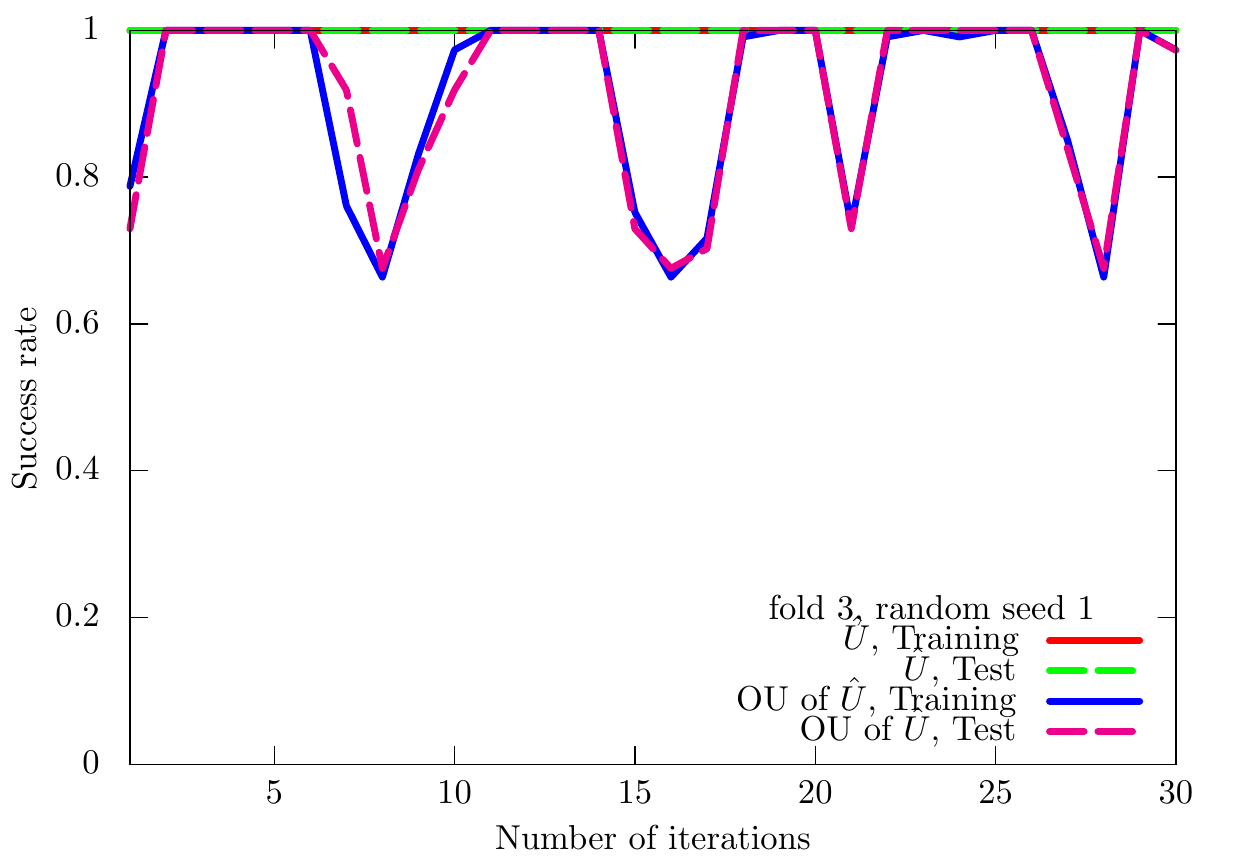}
\includegraphics[scale=0.25]{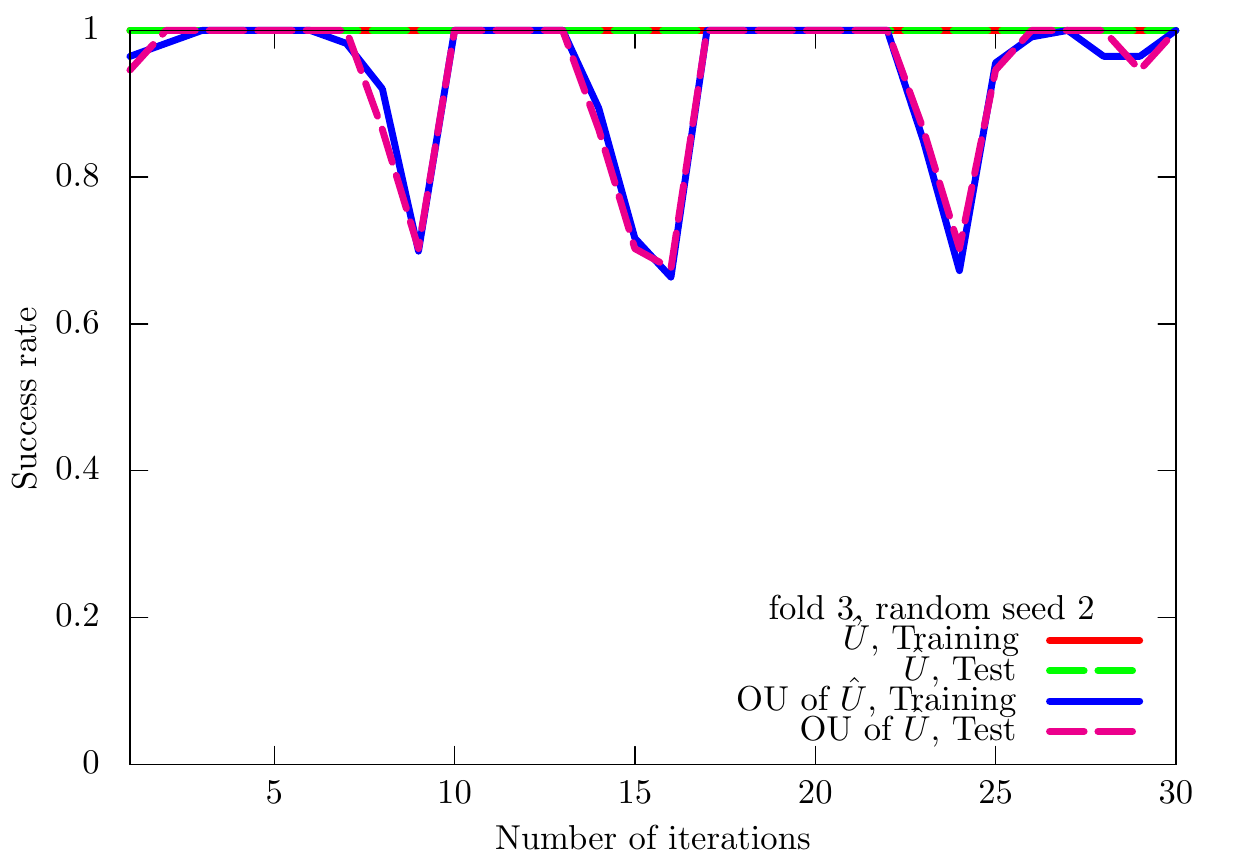}
\includegraphics[scale=0.25]{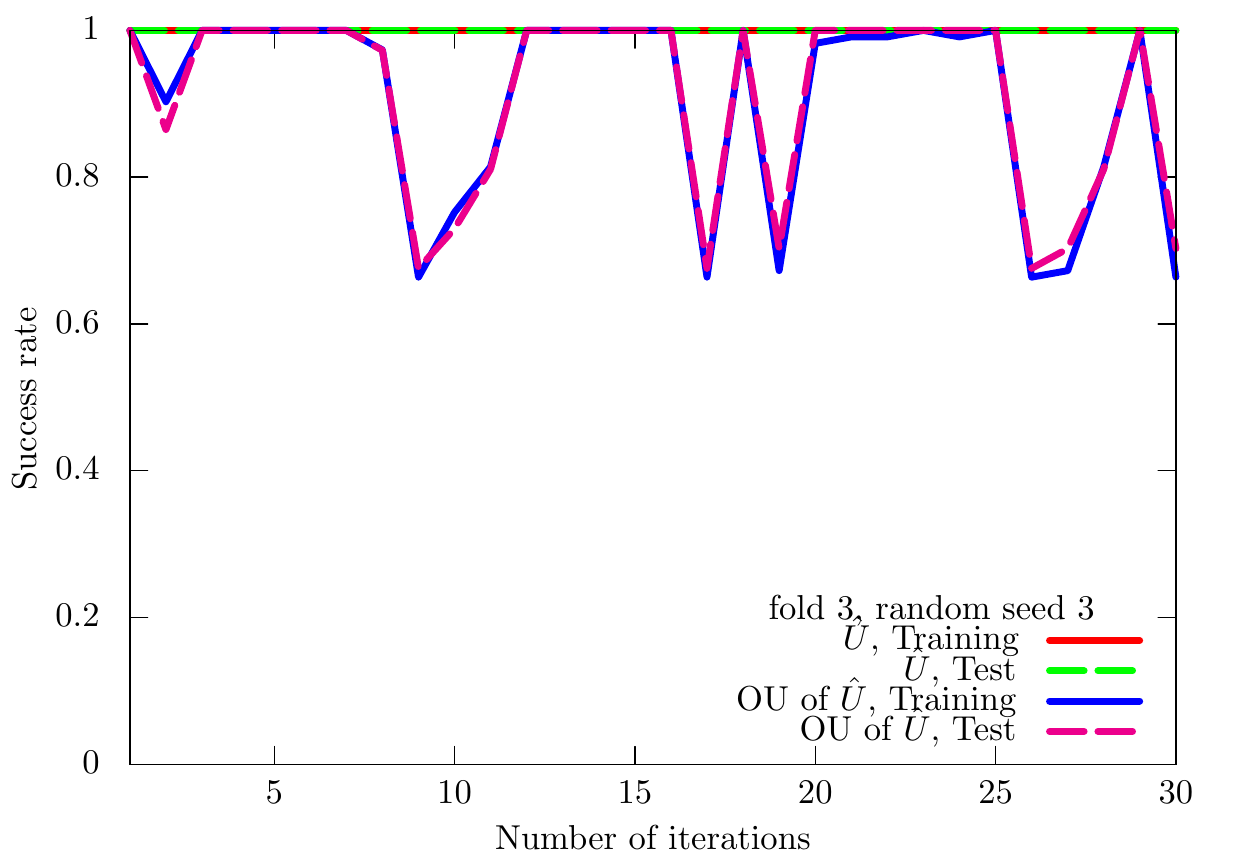}
\includegraphics[scale=0.25]{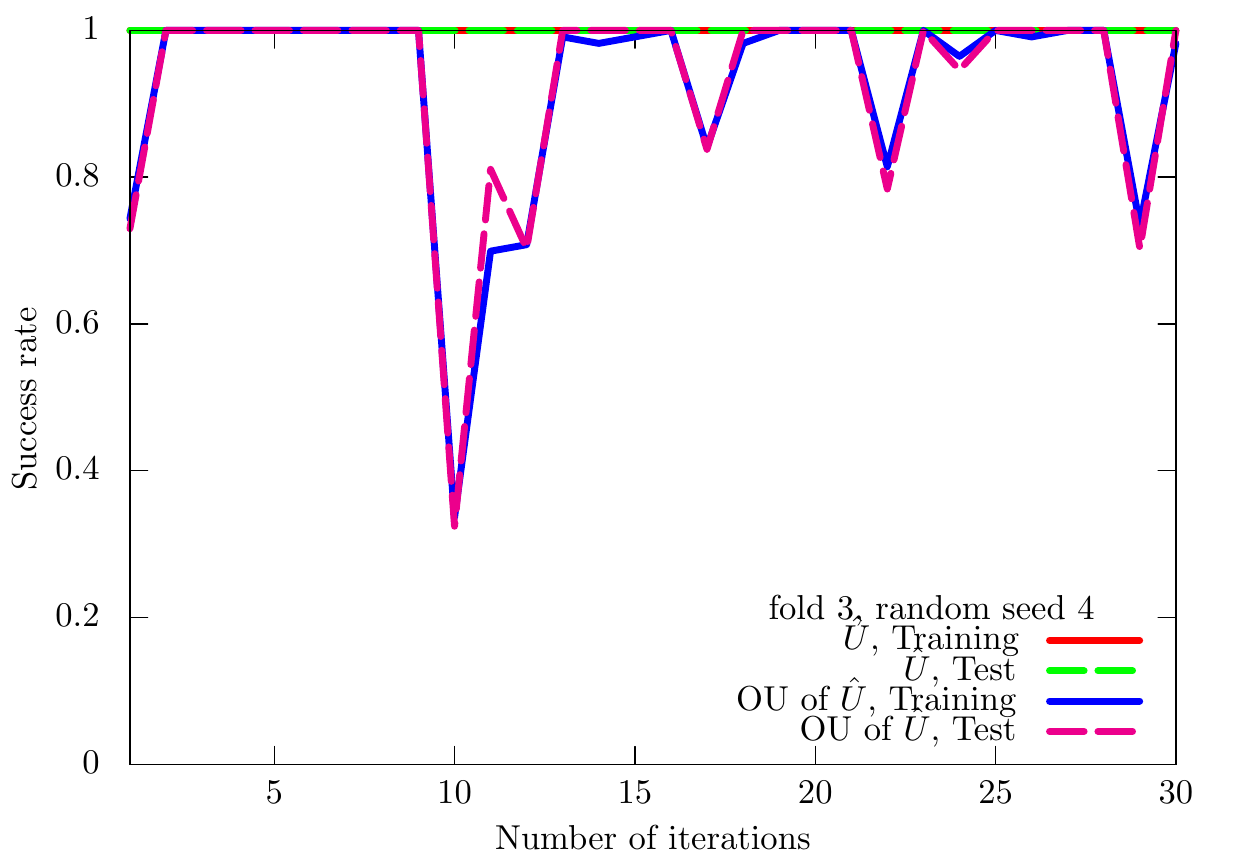}
\includegraphics[scale=0.25]{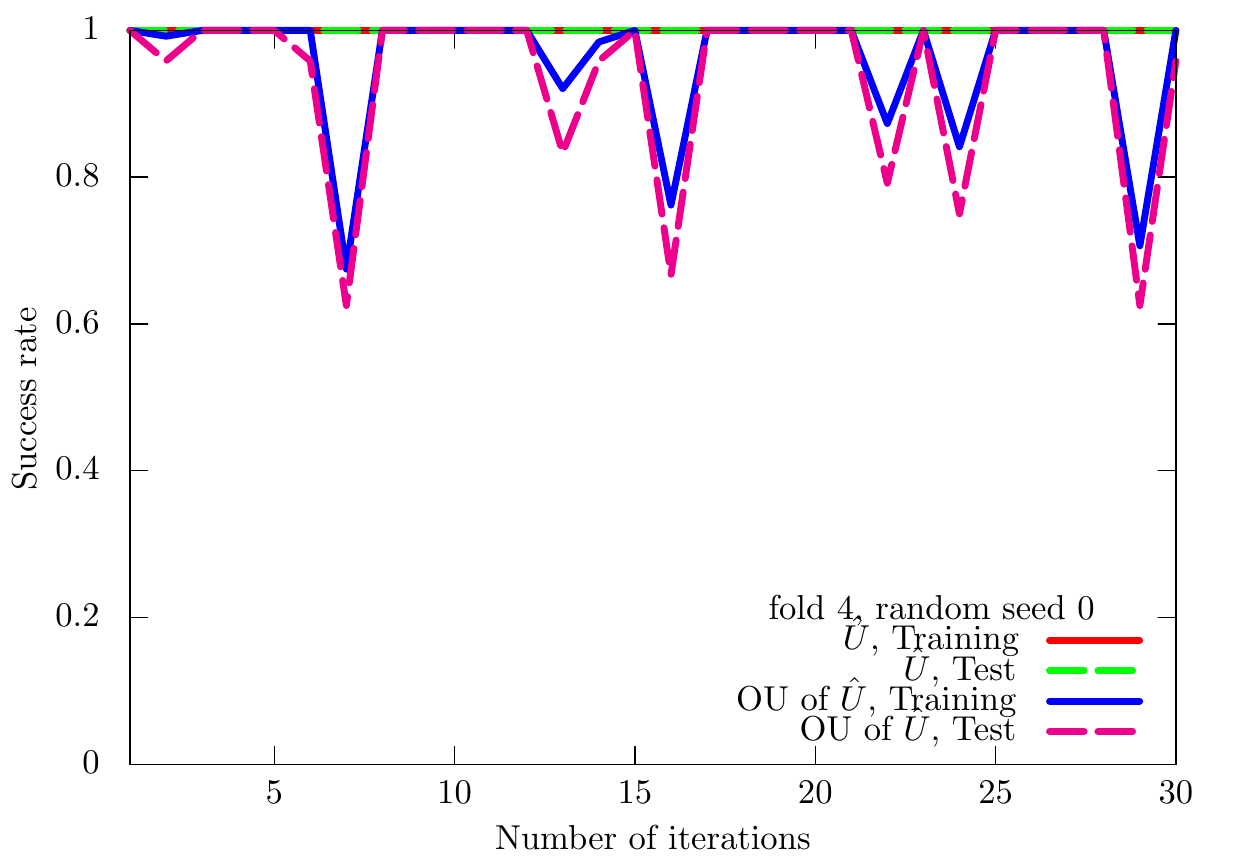}
\includegraphics[scale=0.25]{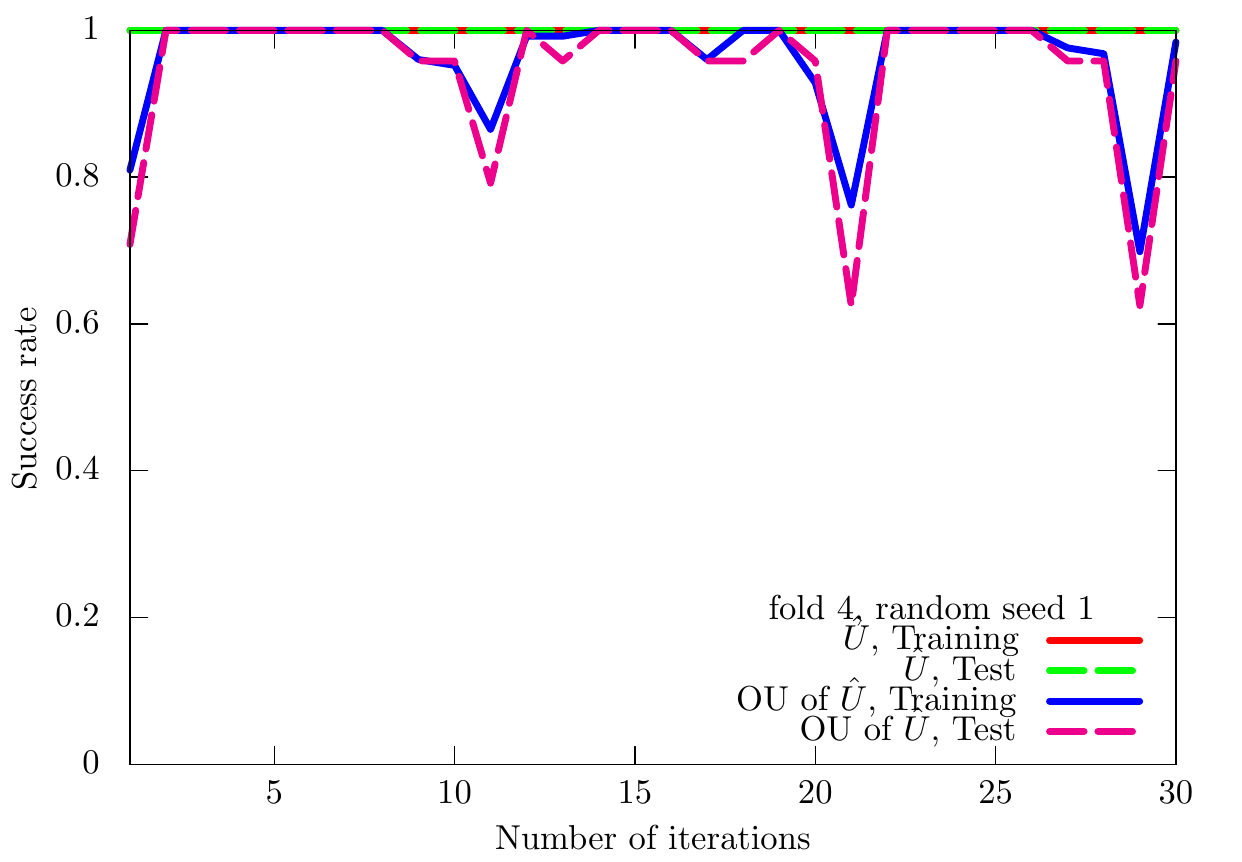}
\includegraphics[scale=0.25]{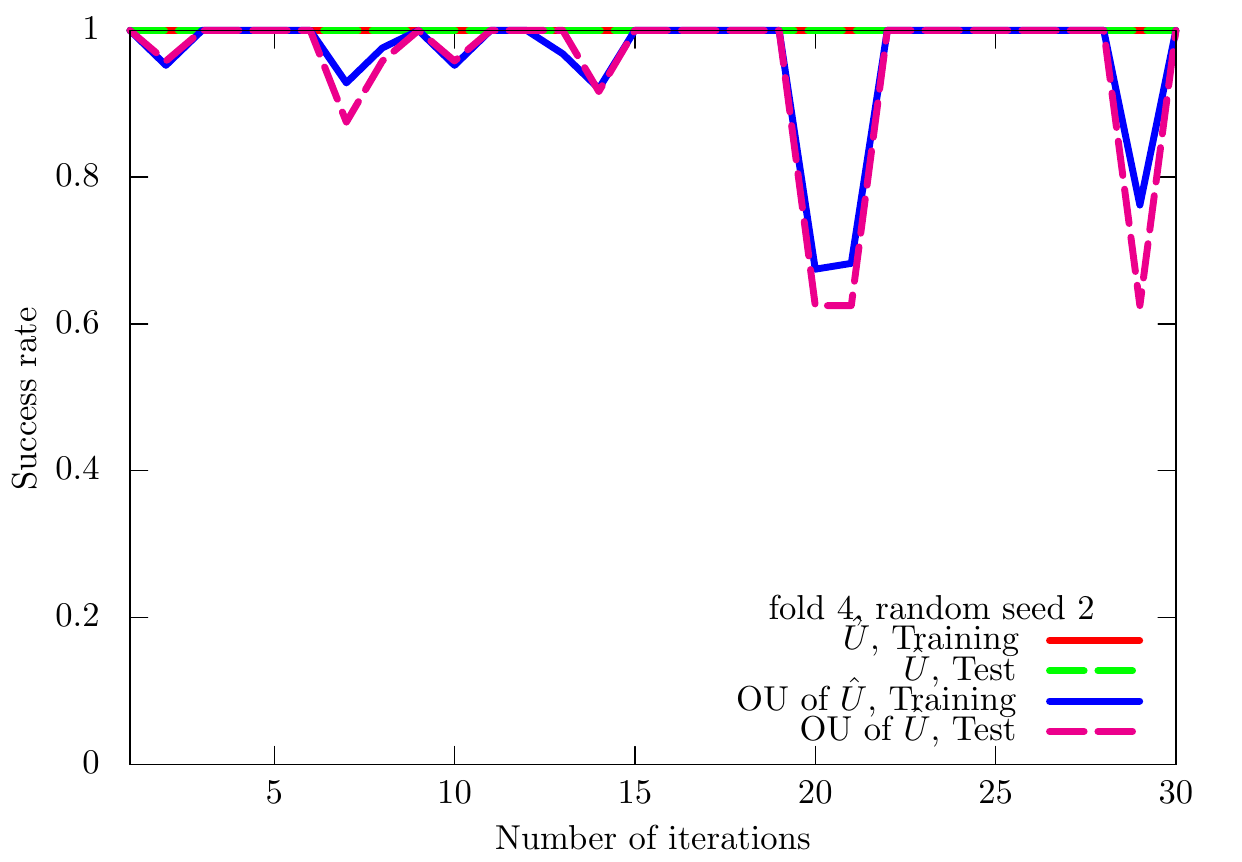}
\includegraphics[scale=0.25]{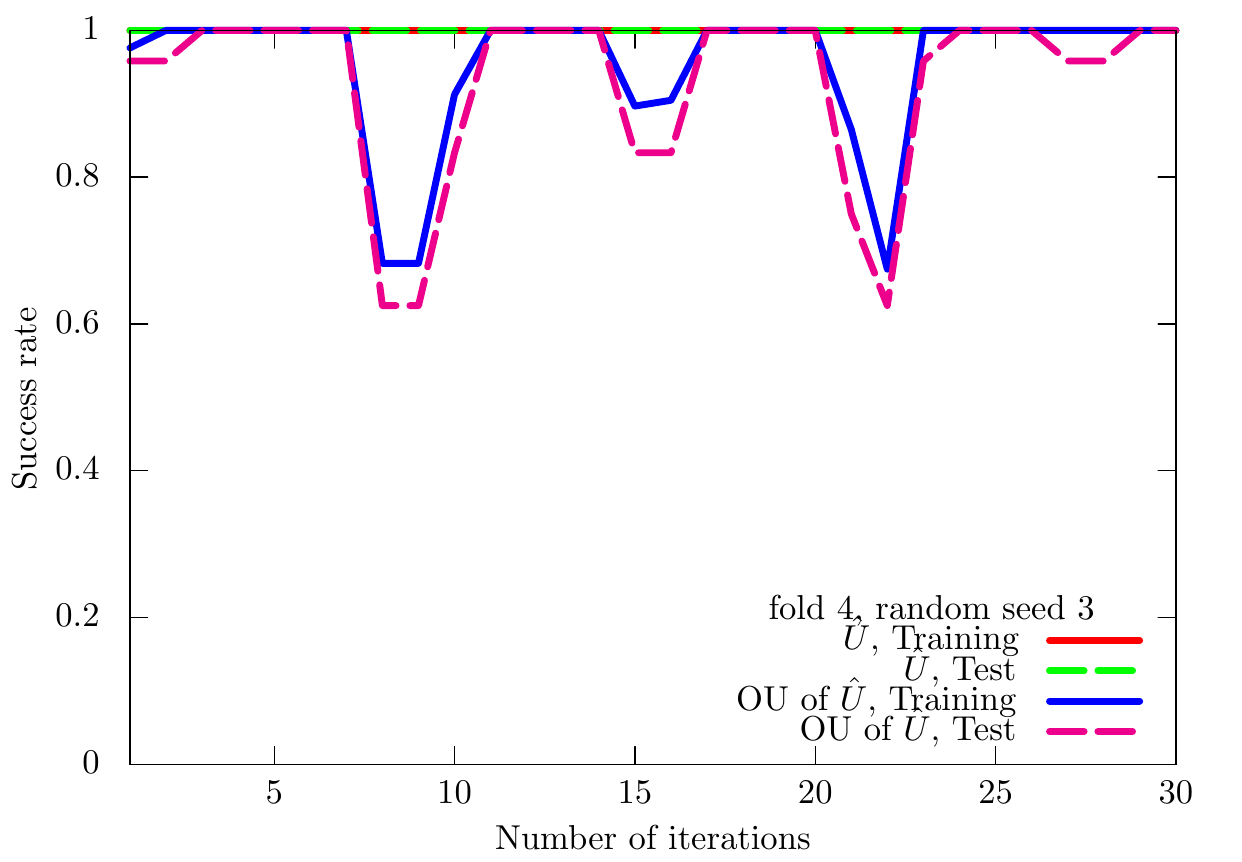}
\includegraphics[scale=0.25]{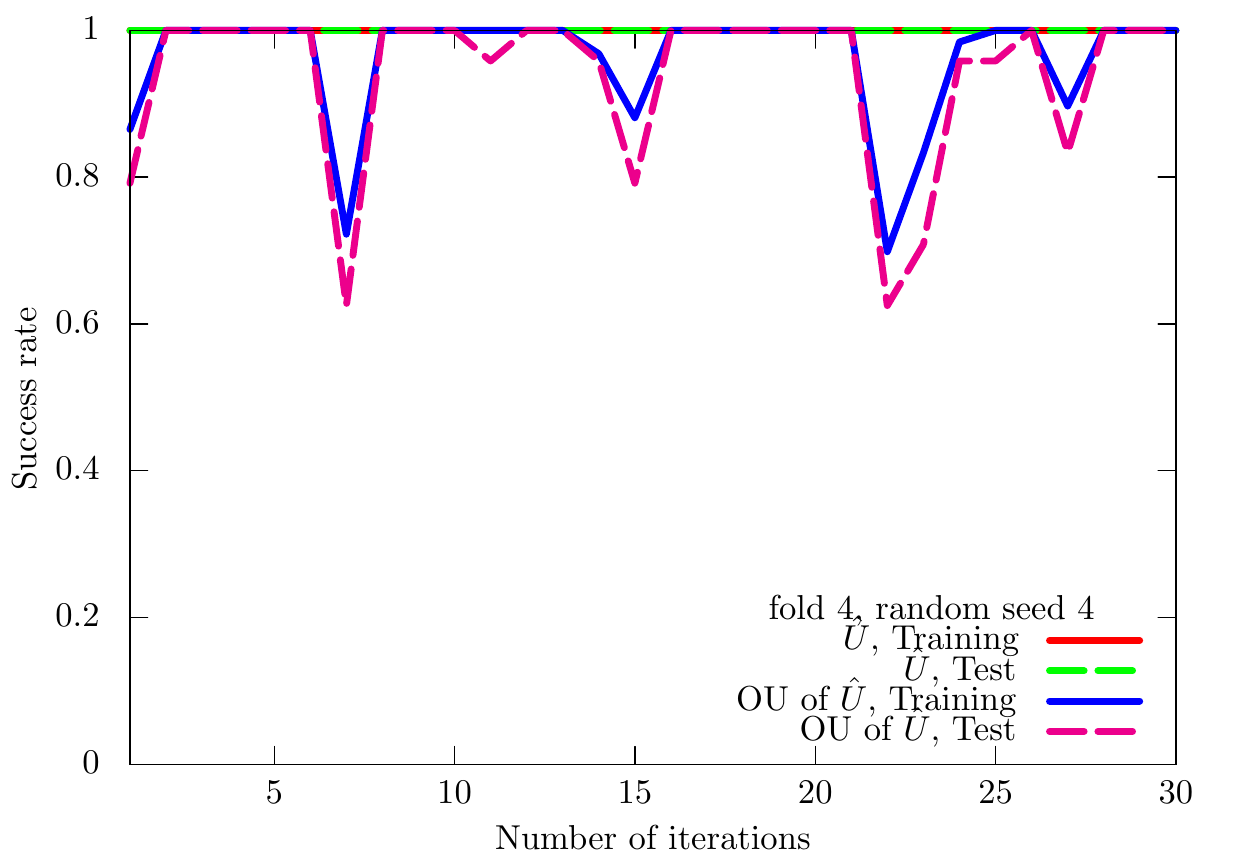}
\caption{Results of the UKM ($\hat{X}$ and OU of $\hat{X}$) on the $5$-fold datasets with $5$ different random seeds for the iris dataset ($0$ or non-$0$). We use complex matrices and set $\theta_\mathrm{bias} = 0$. We set $r = 0.010$.}
\label{supp-arXiv-numerical-result-raw-data-fold-001-rand-001-UKM-OUU-UCI-iris-0-non0}
\end{figure*}

In Fig.~\ref{supp-arXiv-numerical-result-raw-data-fold-001-rand-001-QCL-UCI-iris-0-non0}, we show the numerical results of QCL for the $5$-fold datasets with $5$ different random seeds.
In Fig.~\ref{supp-arXiv-numerical-result-raw-data-fold-001-rand-001-UKM-P-UCI-iris-0-non0}, we show the numerical results of $\hat{P}$ of the UKM for the $5$-fold datasets with $5$ different random seeds.
In Fig.~\ref{supp-arXiv-numerical-result-raw-data-fold-001-rand-001-UKM-OUU-UCI-iris-0-non0}, we also show the numerical results of OU of $\hat{X}$ of the UKM for the $5$-fold datasets with $5$ different random seeds.

\begin{figure}[htb]
\centering
\includegraphics[scale=0.45]{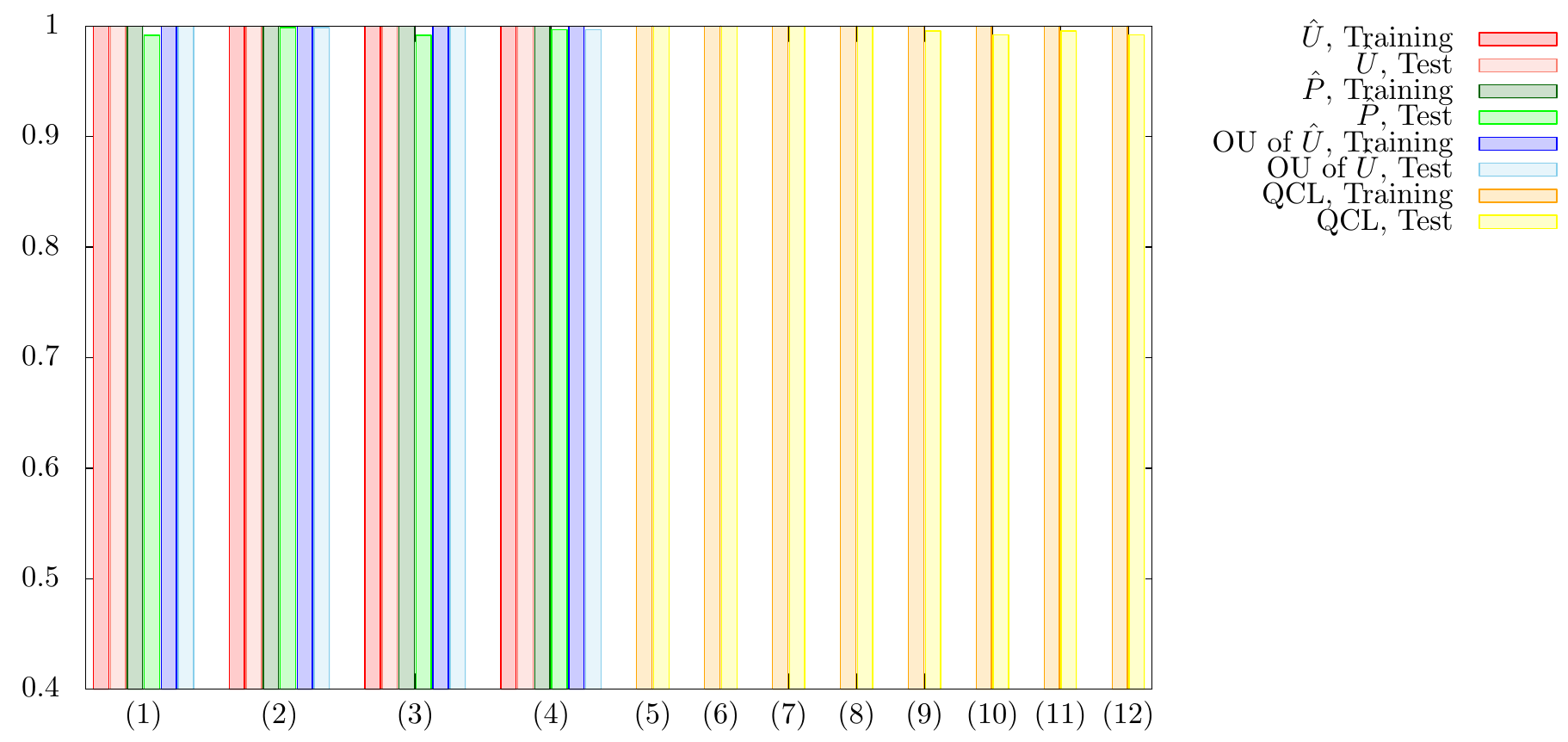}
\caption{Results of $5$-fold CV with $5$ different random seeds for the iris dataset ($0$ or non-$0$). For the UKM, we put $r = 0.010$ and set $K = 30$ and $K' = 10$. For QCL, the number of layers $L$ is $5$ and the number of iterations is $300$. The numerical settings are as follows: (1) complex matrices without the bias term, (2) complex matrices with the bias term, (3) real matrices without the bias term, (4) real matrices with the bias term, (5) CNOT-based circuit without the bias term, (6) CNOT-based circuit with the bias term, (7) CRot-based circuit without the bias term, (8) CRot-based circuit with the bias term, (9) 1d Heisenberg circuit without the bias term, (10) 1d Heisenberg circuit with the bias term, (11) FC Heisenberg circuit without the bias term, and (12) FC Heisenberg circuit with the bias term.}
\label{supp-arXiv-numerical-result-performance-UKM-QCL-UCI-iris-0-non0}
\end{figure}

We summarize the results of 5-fold CV with 5 different random seeds of QCL and the UKM in Tables~\ref{supp-arXiv-table-UCI-iris-0-non0-002} and \ref{supp-arXiv-table-UCI-iris-0-non0-001}.
For QCL and the UKM, we select the best model for the training dataset over iterations to compute the performance.
\begin{table}[htb]
  \begin{tabular}{cc|cc}
    \hline \hline
    Algo. & Condition & Training & Test \\
    \hline
    QCL & CNOT-based, w/o bias & 1.0 & 1.0 \\
    QCL & CNOT-based, w/ bias & 1.0 & 1.0 \\
    \hline
    QCL & CRot-based, w/o bias & 1.0 & 1.0 \\
    QCL & CRot-based, w/ bias & 1.0 & 1.0 \\
    \hline
    QCL & 1d Heisenberg, w/o bias & 1.0 & 0.9957 \\
    QCL & 1d Heisenberg, w/ bias & 1.0 & 0.9920 \\
    \hline
    QCL & FC Heisenberg, w/o bias & 1.0 & 0.9957 \\
    QCL & FC Heisenberg, w/ bias & 1.0 & 0.9920 \\
    \hline \hline
  \end{tabular}
\caption{Results of $5$-fold CV with $5$ different random seeds of QCL for the iris dataset ($0$ or non-$0$). The number of layers $L$ is set to $5$ and the number of iterations is set to $300$.}
\label{supp-arXiv-table-UCI-iris-0-non0-002}
\end{table}
\begin{table}[htb]
  \begin{tabular}{cc|cc}
    \hline \hline
    Algo. & Condition & Training & Test \\
    \hline
    UKM & $\hat{X}$, complex, w/o bias & 1.0 & 1.0 \\
    UKM & $\hat{P}$, complex, w/o bias & 1.0 & 0.9917 \\
    UKM & OU of $\hat{X}$, complex, w/o bias & 1.0 & 1.0 \\
    \hline
    UKM & $\hat{X}$, complex, w/ bias & 1.0 & 1.0 \\
    UKM & $\hat{P}$, complex, w/ bias & 1.0 & 0.9987 \\
    UKM & OU of $\hat{X}$, complex, w/ bias & 1.0 & 0.9987 \\
    \hline
    UKM & $\hat{X}$, real, w/o bias & 1.0 & 1.0 \\
    UKM & $\hat{P}$, real, w/o bias & 1.0 & 0.9917 \\
    UKM & OU of $\hat{X}$, real, w/o bias & 1.0 & 1.0 \\
    \hline
    UKM & $\hat{X}$, real, w/ bias & 1.0 & 1.0 \\
    UKM & $\hat{P}$, real, w/ bias & 1.0 &	0.9970 \\
    UKM & OU of $\hat{X}$, real, w/ bias & 1.0 & 0.9970 \\
    \hline \hline
  \end{tabular}
\caption{Results of $5$-fold CV with $5$ different random seeds of the UKM for the iris dataset ($0$ or non-$0$). We put $r = 0.010$ and set $K = 30$ and $K' = 10$.}
\label{supp-arXiv-table-UCI-iris-0-non0-001}
\end{table}
In Fig.~\ref{supp-arXiv-numerical-result-performance-UKM-QCL-UCI-iris-0-non0}, we plot the data shown in Tables~\ref{supp-arXiv-table-UCI-iris-0-non0-002} and \ref{supp-arXiv-table-UCI-iris-0-non0-001}.
We also summarize the results of 5-fold CV with 5 different random seeds of the kernel method in Table~\ref{supp-arXiv-table-UCI-iris-0-non0-003}.
More specifically, we use Ridge classification in Sec.~\ref{supp-arXiv-sec-Ridge-001}.
We consider the linear functions and the second-order polynomial functions for $\phi (\cdot)$ in Eq.~\eqref{supp-arXiv-f-pred-kernel-method-001-002} with and without normalization.
We set $\lambda = 10^{-2}, 10^{-1}, 1$ where $\lambda$ is the coefficient of the regularization term.
\begin{table}[htb]
  \begin{tabular}{cc|cc}
    \hline \hline
    Algo. & Condition & Training & Test \\
    \hline
  Kernel method & Linear, w/o normalization, $\lambda = 10^{-2}$ & 1.0000 & 1.0000 \\
  Kernel method & Linear, w/o normalization, $\lambda = 10^{-1}$ & 1.0000 & 1.0000 \\
  Kernel method & Linear, w/o normalization, $\lambda = 1$ & 1.0000 & 1.0000 \\
    \hline
  Kernel method & Linear, w/ normalization, $\lambda = 10^{-2}$ & 1.0000 & 1.0000 \\
  Kernel method & Linear, w/ normalization, $\lambda = 10^{-1}$ & 1.0000 & 1.0000 \\
  Kernel method & Linear, w/ normalization, $\lambda = 1$ & 1.0000 & 1.0000 \\
    \hline
  Kernel method & Poly-2, w/o normalization, $\lambda = 10^{-2}$ & 1.0000 & 1.0000 \\
  Kernel method & Poly-2, w/o normalization, $\lambda = 10^{-1}$ & 1.0000 & 1.0000 \\
  Kernel method & Poly-2, w/o normalization, $\lambda = 1$ & 1.0000 & 1.0000 \\
    \hline
  Kernel method & Poly-2, w/ normalization, $\lambda = 10^{-2}$ & 1.0000 & 1.0000 \\
  Kernel method & Poly-2, w/ normalization, $\lambda = 10^{-1}$ & 1.0000 & 1.0000 \\
  Kernel method & Poly-2, w/ normalization, $\lambda = 1$ & 1.0000 & 1.0000 \\
    \hline \hline
  \end{tabular}
\caption{Results of 5-fold CV with 5 different random seeds of the kernel method for the iris dataset ($0$ or non-$0$).}
\label{supp-arXiv-table-UCI-iris-0-non0-003}
\end{table}

\begin{figure}[htb]
\centering
\includegraphics[scale=0.45]{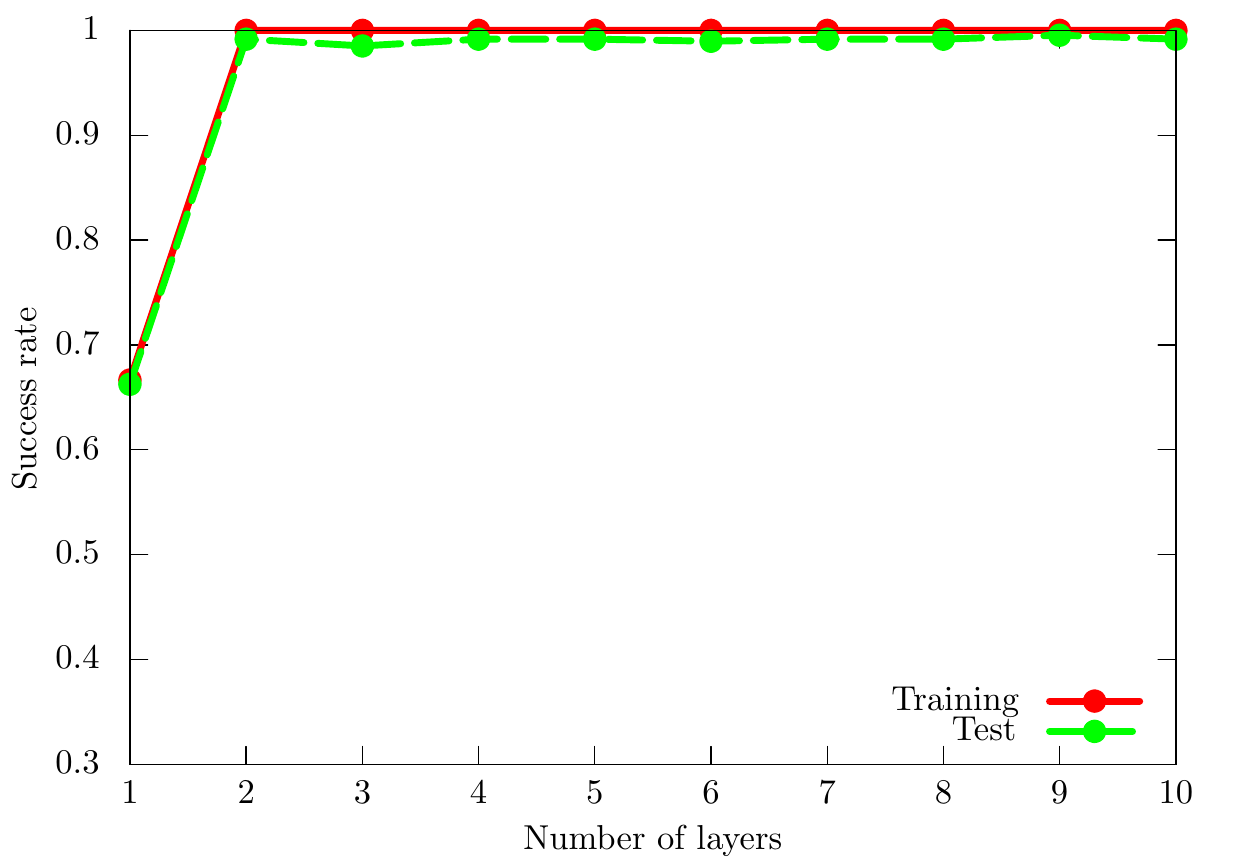}
\caption{Performance dependence of QCL on the number of layers $L$ for the iris dataset ($0$ or non-$0$). We use the CNOT-based circuit geometry and set $\theta_\mathrm{bias} = 0$. We iterate the computation $300$ times.}
\label{supp-arXiv-numerical-result-layers-dependence-QCL-UCI-iris-0-non0}
\end{figure}

\begin{figure}[htb]
\centering
\includegraphics[scale=0.45]{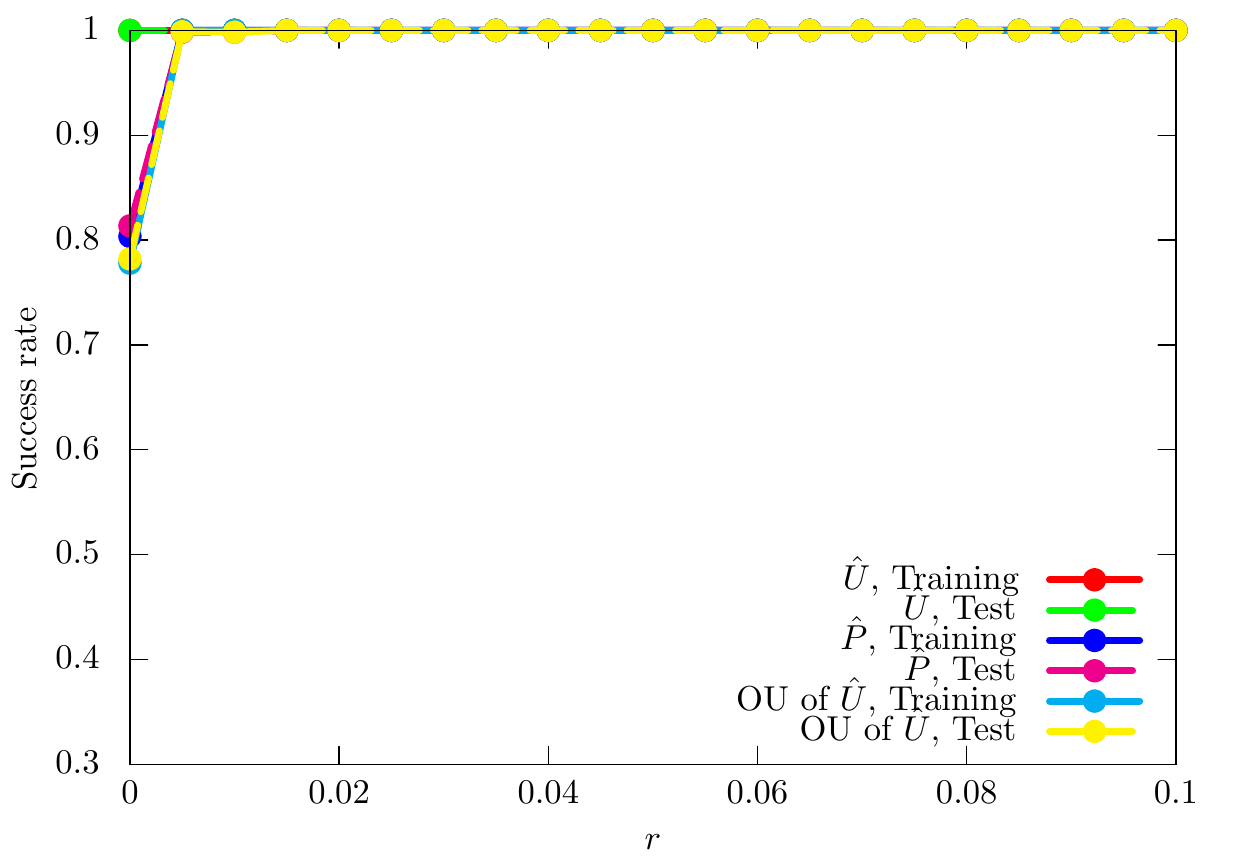}
\includegraphics[scale=0.45]{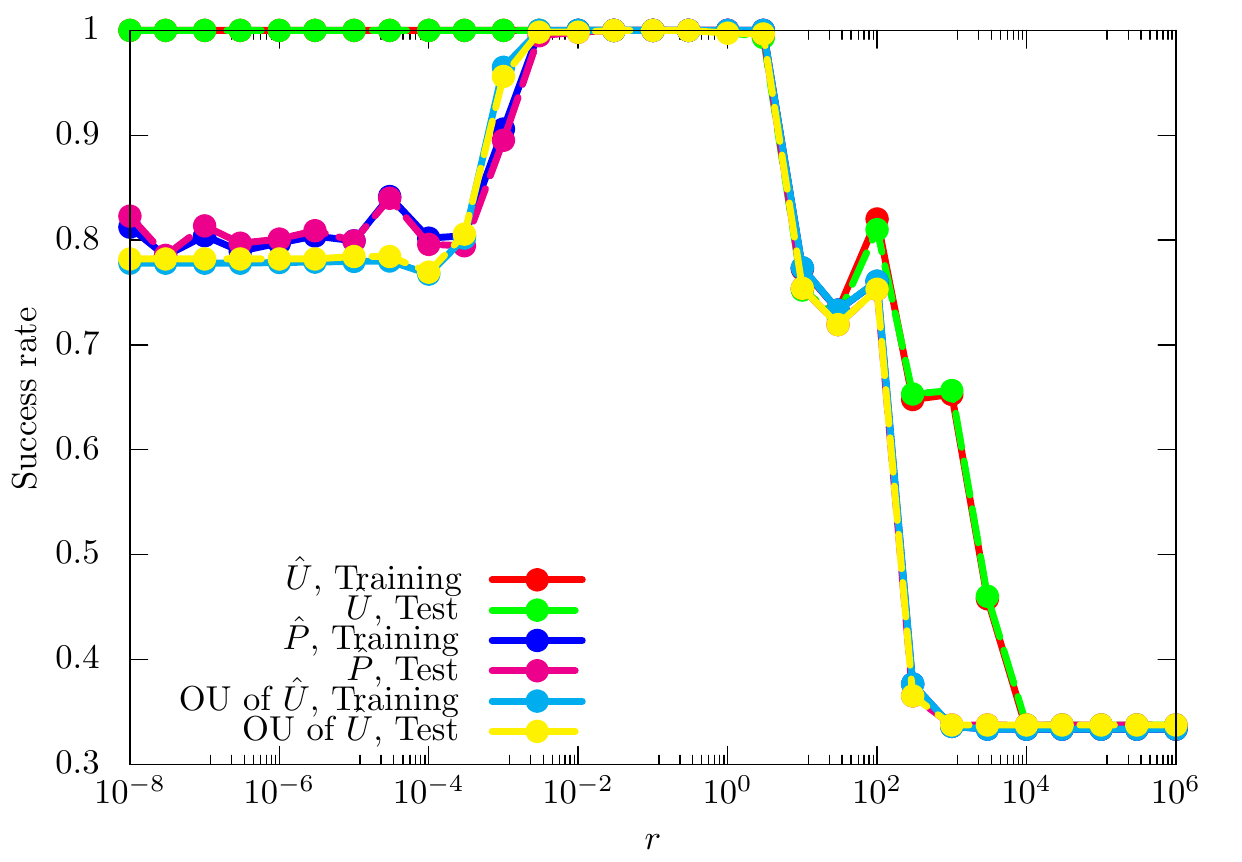}
\caption{Performance dependence of the UKM on $r$, which is the coefficient of the second term in the right-hand side of Eq.~\eqref{supp-arXiv-quantum-kernel-method-001-011} for the iris dataset ($0$ or non-$0$). We use complex matrices and set $\theta_\mathrm{bias} = 0$. We set $K = 30$ and $K' = 10$.}
\label{supp-arXiv-numerical-result-r-dependence-UKM-UCI-iris-0-non0}
\end{figure}

\begin{figure}[htb]
\centering
\includegraphics[scale=0.45]{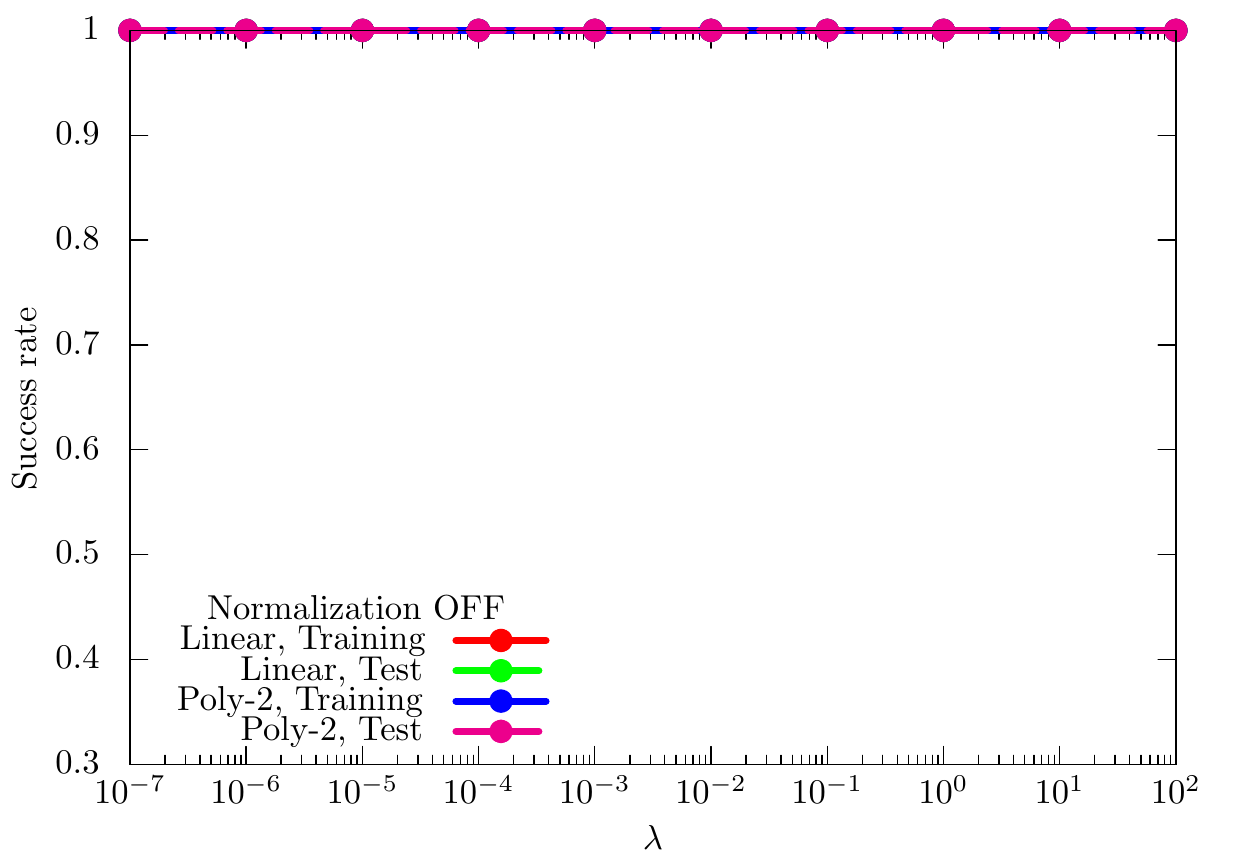}
\includegraphics[scale=0.45]{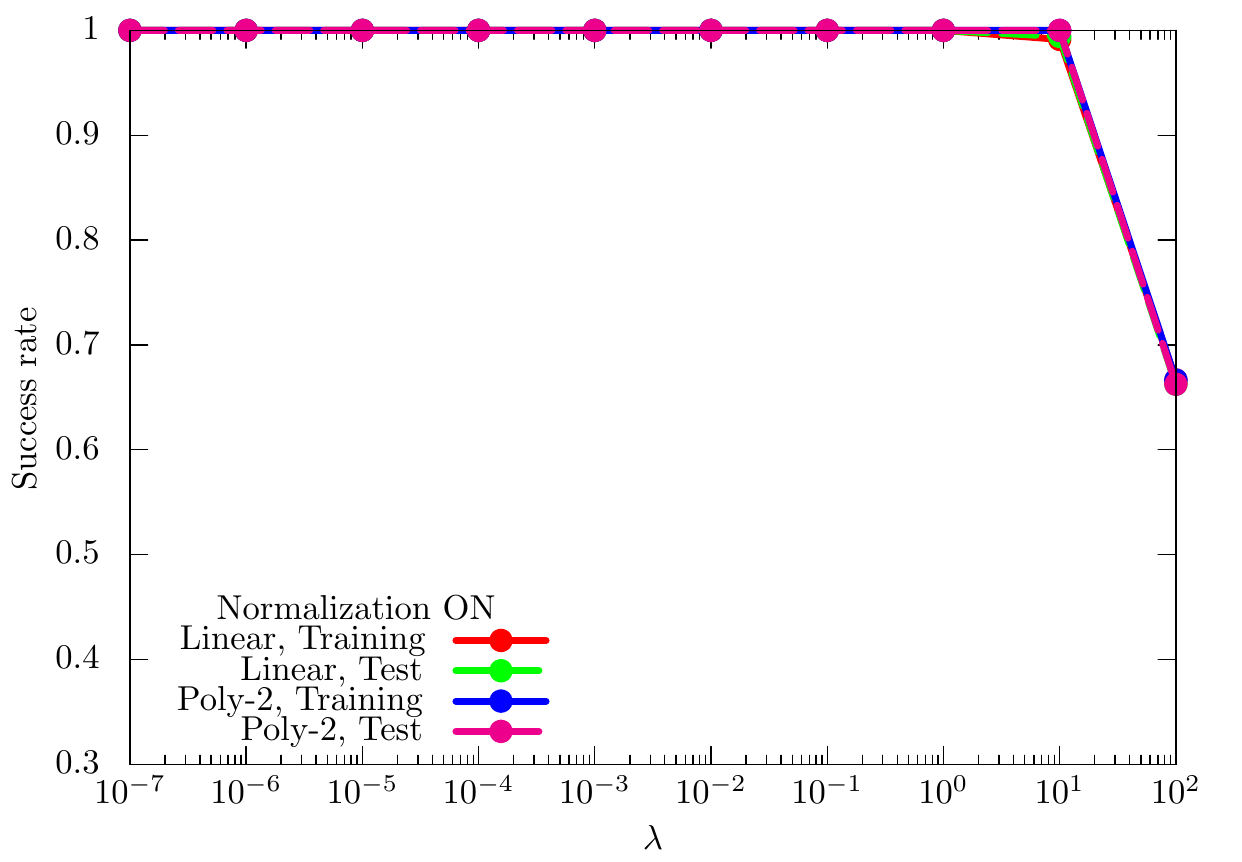}
\caption{Performance dependence of the kernel method on $\lambda$, which is the coefficient of the second term in the right-hand side of Eq.~\eqref{supp-arXiv-cost-function-kernel-method-001-002} for the iris dataset ($0$ or non-$0$). For $\phi (\cdot)$ in Eq.~\eqref{supp-arXiv-f-pred-kernel-method-001-002}, we use the linear functions and the second-degree polynomial functions with and without normalization.}
\label{supp-arXiv-numerical-result-lambda-dependence-kernel-method-iris-0-non0}
\end{figure}

Next, we show the performance dependence of the three algorithms on their key parameters.
We see the performance dependence of QCL on the number of layers $L$.
The result is shown in Fig.~\ref{supp-arXiv-numerical-result-layers-dependence-QCL-UCI-iris-0-non0}.
We then see the performance dependence of the UKM on $r$, which is the coefficient of the second term in the right-hand side of Eq.~\eqref{supp-arXiv-quantum-kernel-method-001-011}.
The result is shown in Fig.~\ref{supp-arXiv-numerical-result-r-dependence-UKM-UCI-iris-0-non0}.
In Fig.~\ref{supp-arXiv-numerical-result-lambda-dependence-kernel-method-iris-0-non0}, we show the performance dependence of the kernel method on $\lambda$, which is the coefficient of the second term in the right-hand side of Eq.~\eqref{supp-arXiv-cost-function-kernel-method-001-002}.

\begin{figure}[htb]
\centering
\includegraphics[scale=0.45]{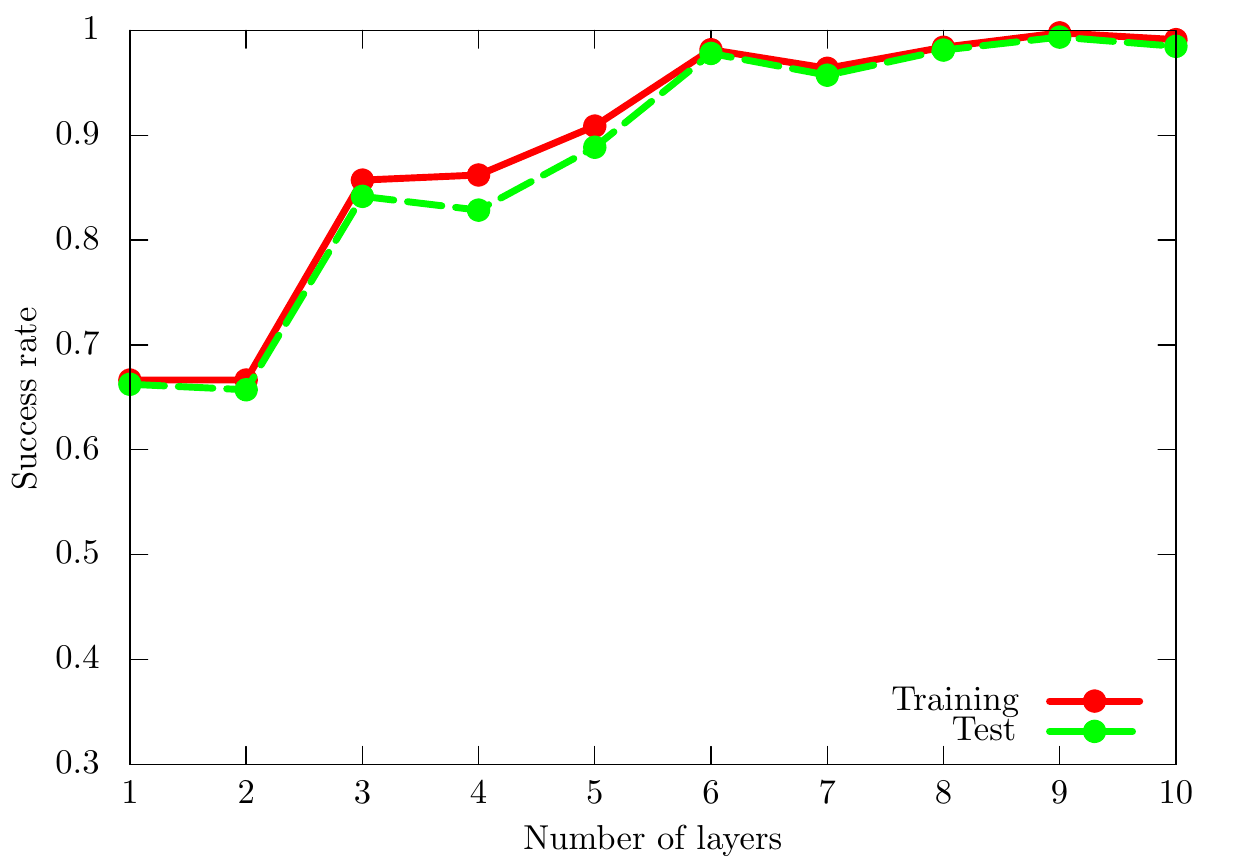}
\caption{Performance dependence of QCL on the number of layers $L$ for the iris dataset ($0$ or non-$0$) in the case of the hinge function, Eq.~\eqref{supp-arXiv-hinge-function-001-001}. We use the CNOT-based circuit geometry and set $\theta_\mathrm{bias} = 0$. We iterate the computation $300$ times.}
\label{supp-arXiv-numerical-result-layers-dependence-QCL-UCI-iris-0-non0-hinge}
\end{figure}

\begin{figure}[htb]
\centering
\includegraphics[scale=0.45]{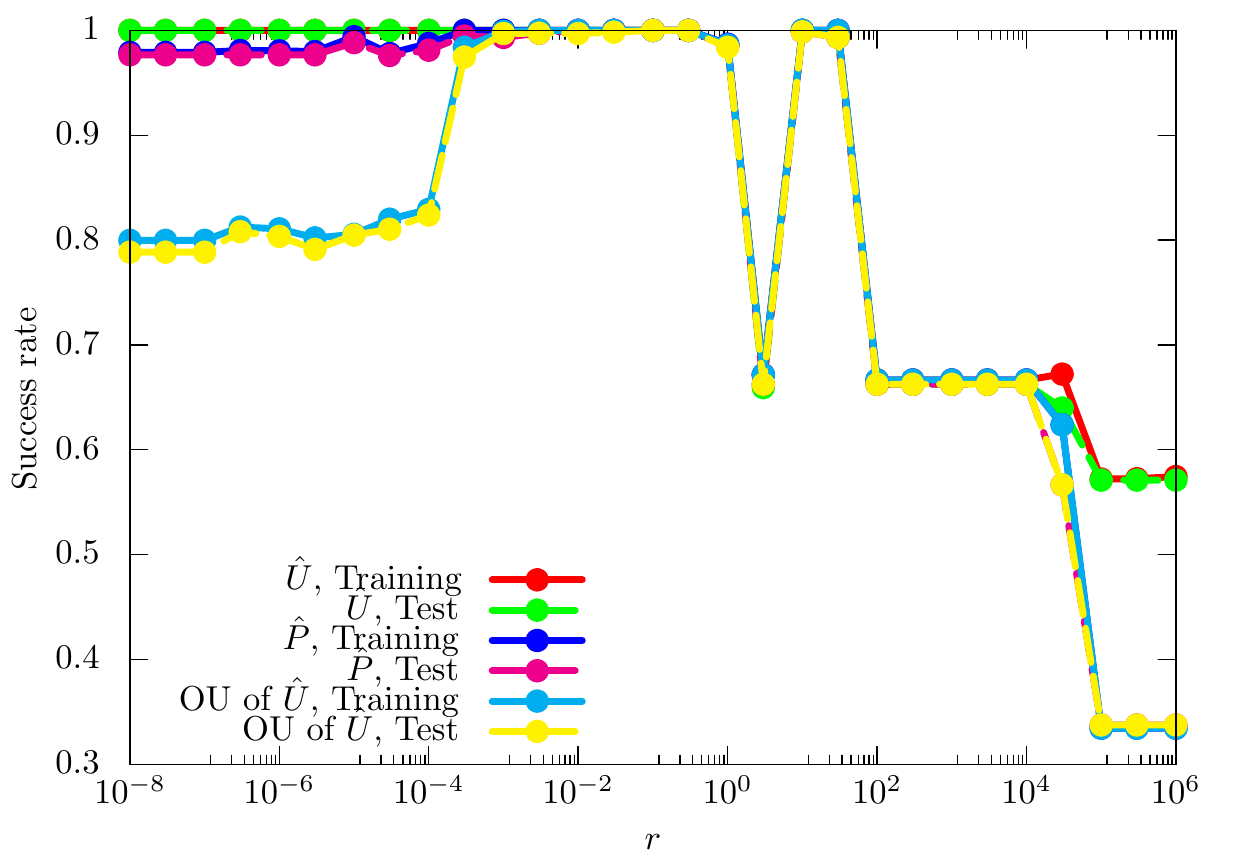}
\caption{Performance dependence of the UKM on $r$, which is the coefficient of the second term in the right-hand side of Eq.~\eqref{supp-arXiv-quantum-kernel-method-001-011} for the iris dataset ($0$ or non-$0$) in the case of the hinge function, Eq.~\eqref{supp-arXiv-hinge-function-001-001}. We use complex matrices and set $\theta_\mathrm{bias} = 0$. We set $K = 30$ and $K' = 10$.}
\label{supp-arXiv-numerical-result-r-dependence-UKM-UCI-iris-0-non0-hinge}
\end{figure}

So far, we have used the squared error function, Eq.~\eqref{supp-arXiv-squared-error-function-001-001}.
In Fig.~\ref{supp-arXiv-numerical-result-layers-dependence-QCL-UCI-iris-0-non0-hinge}, we show the performance dependence of QCL on the number of layers $L$ in the case of the hinge function, Eq.~\eqref{supp-arXiv-hinge-function-001-001}.
In Fig.~\ref{supp-arXiv-numerical-result-r-dependence-UKM-UCI-iris-0-non0-hinge}, we show the performance dependence of the UKM on $r$, which is the coefficient of the second term in the right-hand side of Eq.~\eqref{supp-arXiv-quantum-kernel-method-001-011}, in the case of the hinge function, Eq.~\eqref{supp-arXiv-hinge-function-001-001}.

\clearpage

\subsection{Iris dataset ($1$ or non-$1$)}

We here show the numerical result for the iris dataset ($1$ or non-$1$).
For the UKM, we put $r = 0.010$ and set $K = 30$ and $K' = 10$ in Algo.~\ref{supp-arXiv-quantum-kernel-method-002-001}.
For QCL, we run iterations $300$ times.

\begin{figure*}[htb]
\centering
\includegraphics[scale=0.25]{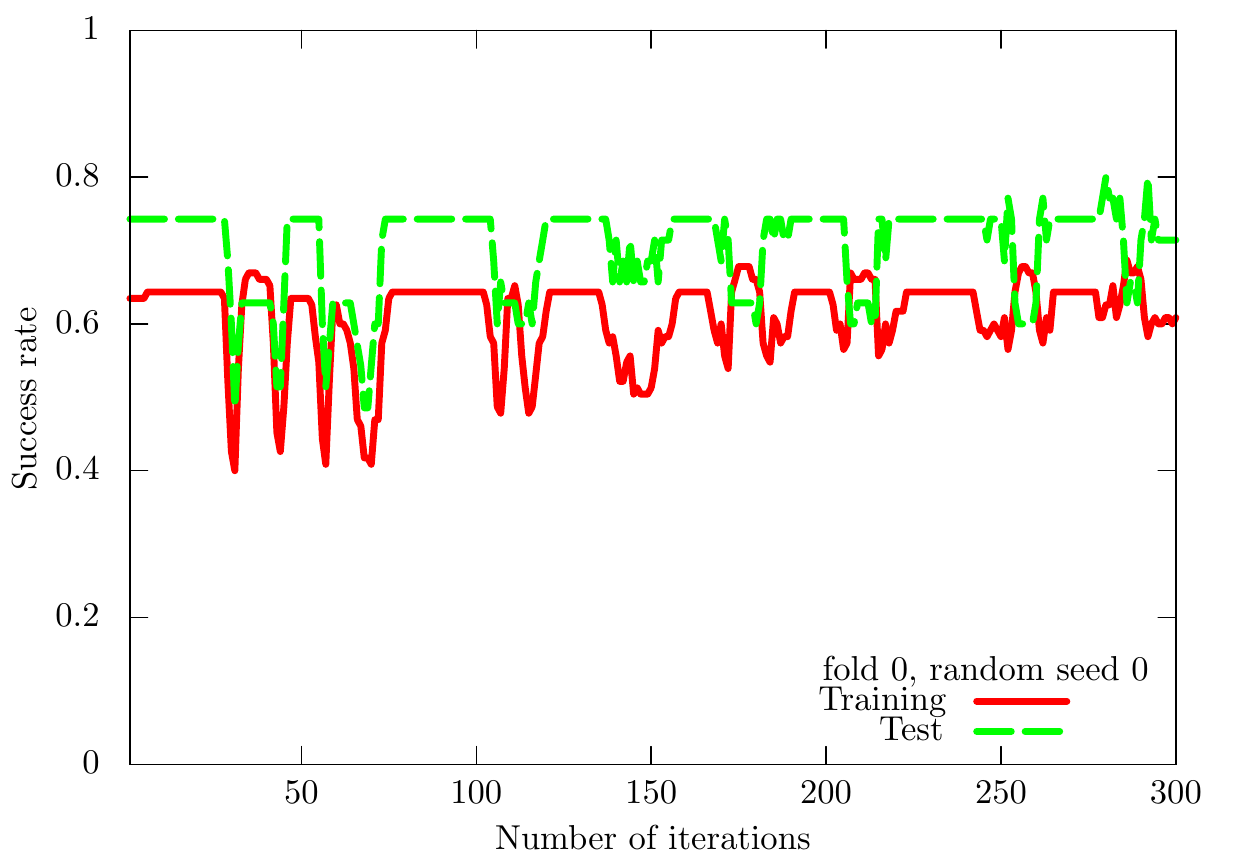}
\includegraphics[scale=0.25]{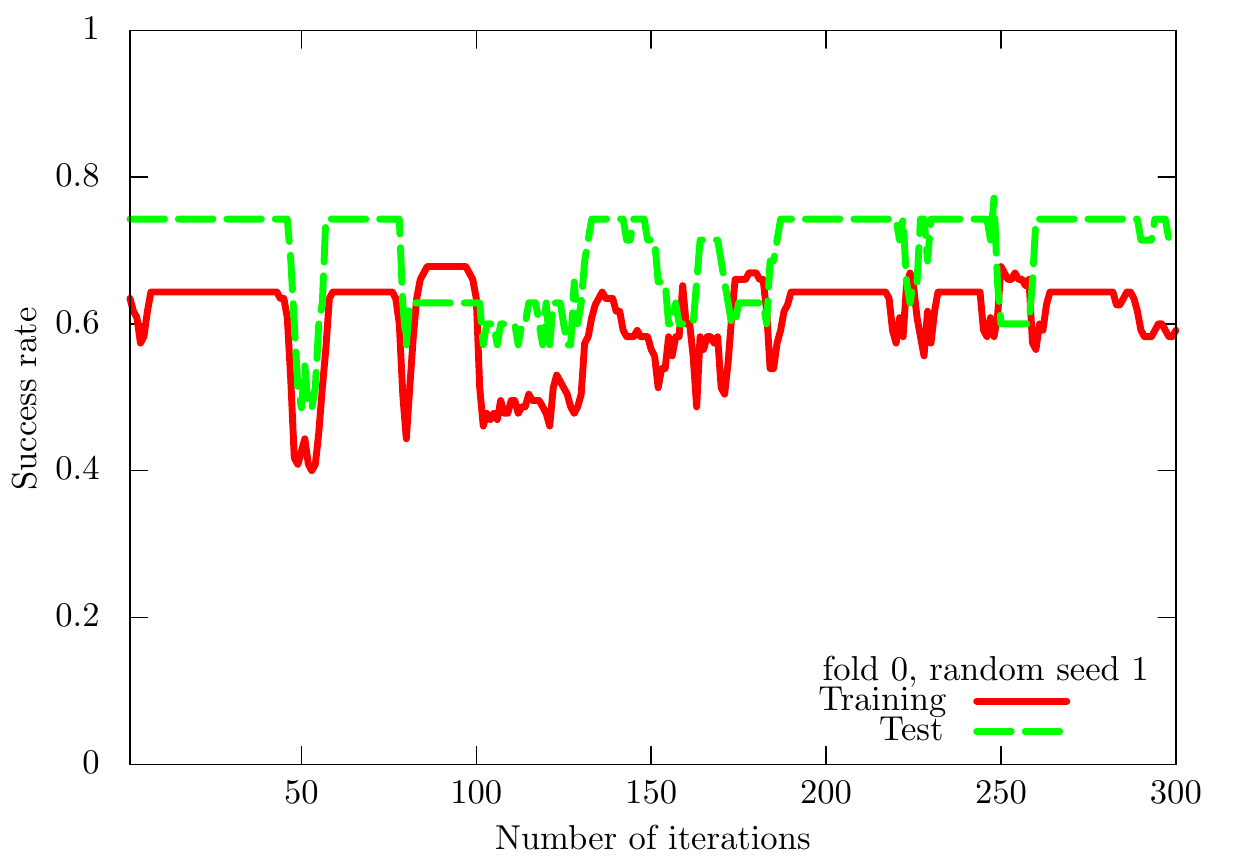}
\includegraphics[scale=0.25]{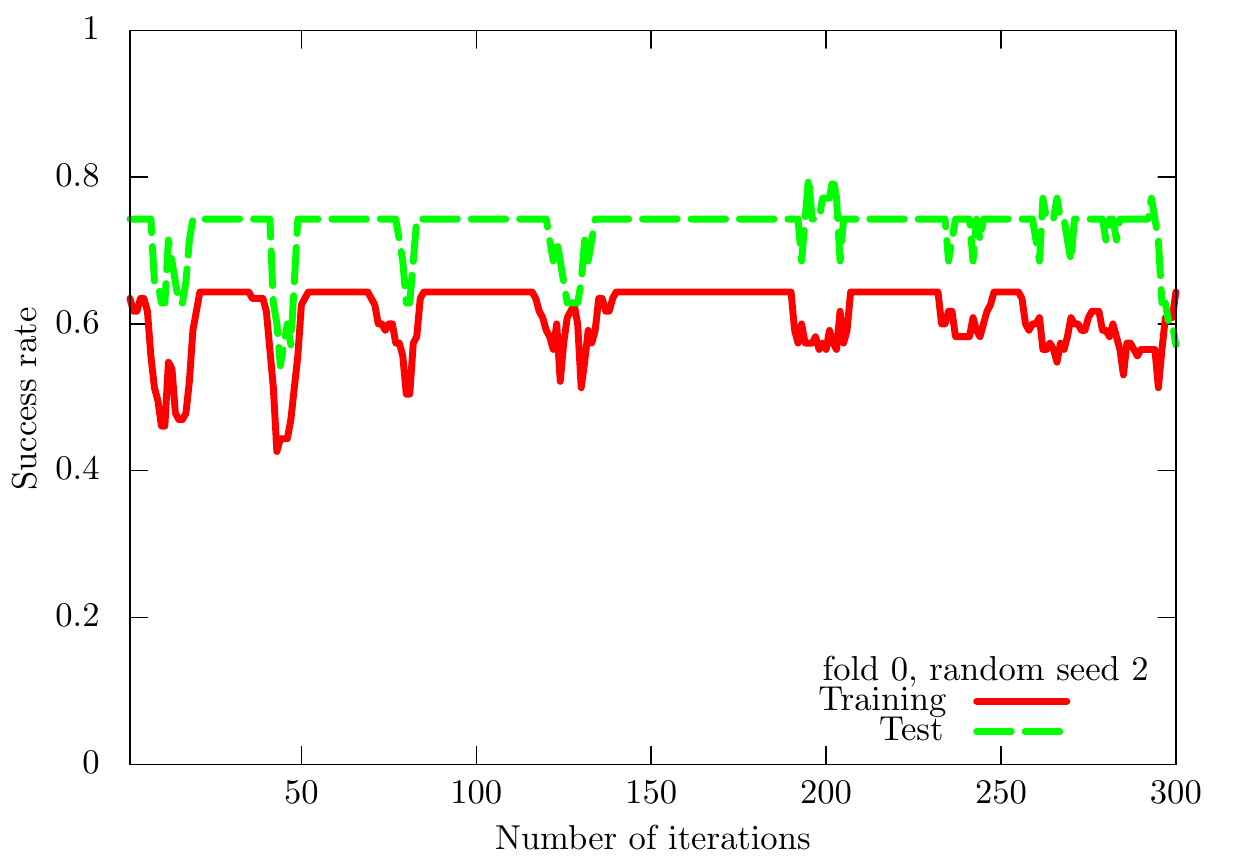}
\includegraphics[scale=0.25]{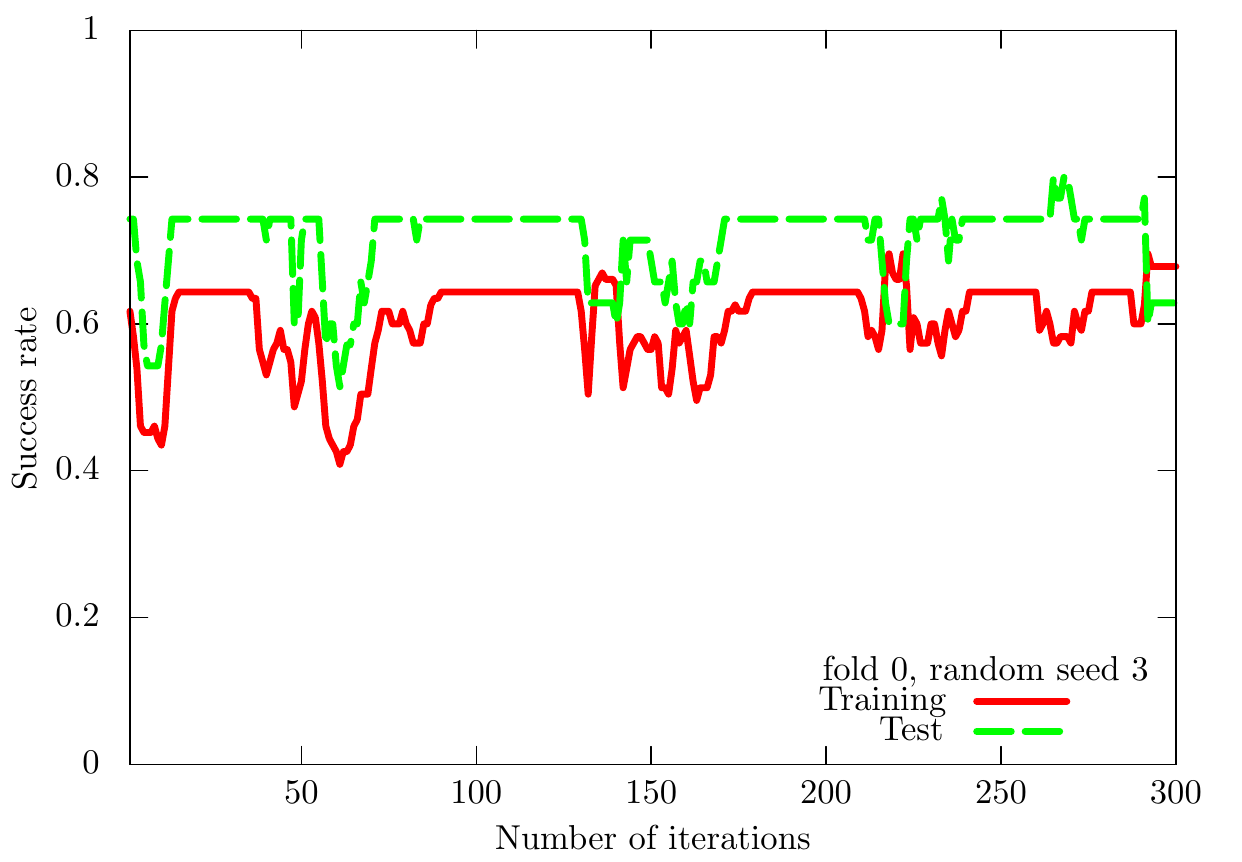}
\includegraphics[scale=0.25]{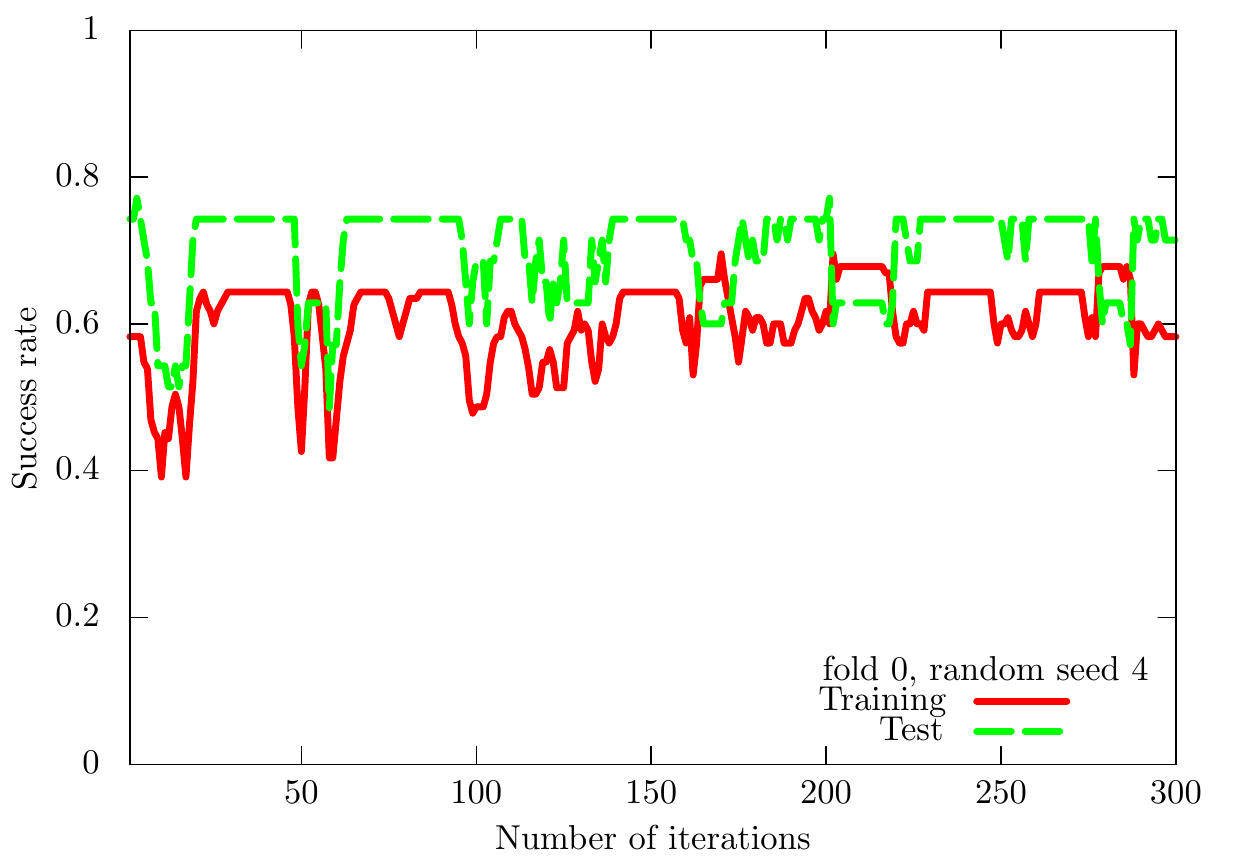}
\includegraphics[scale=0.25]{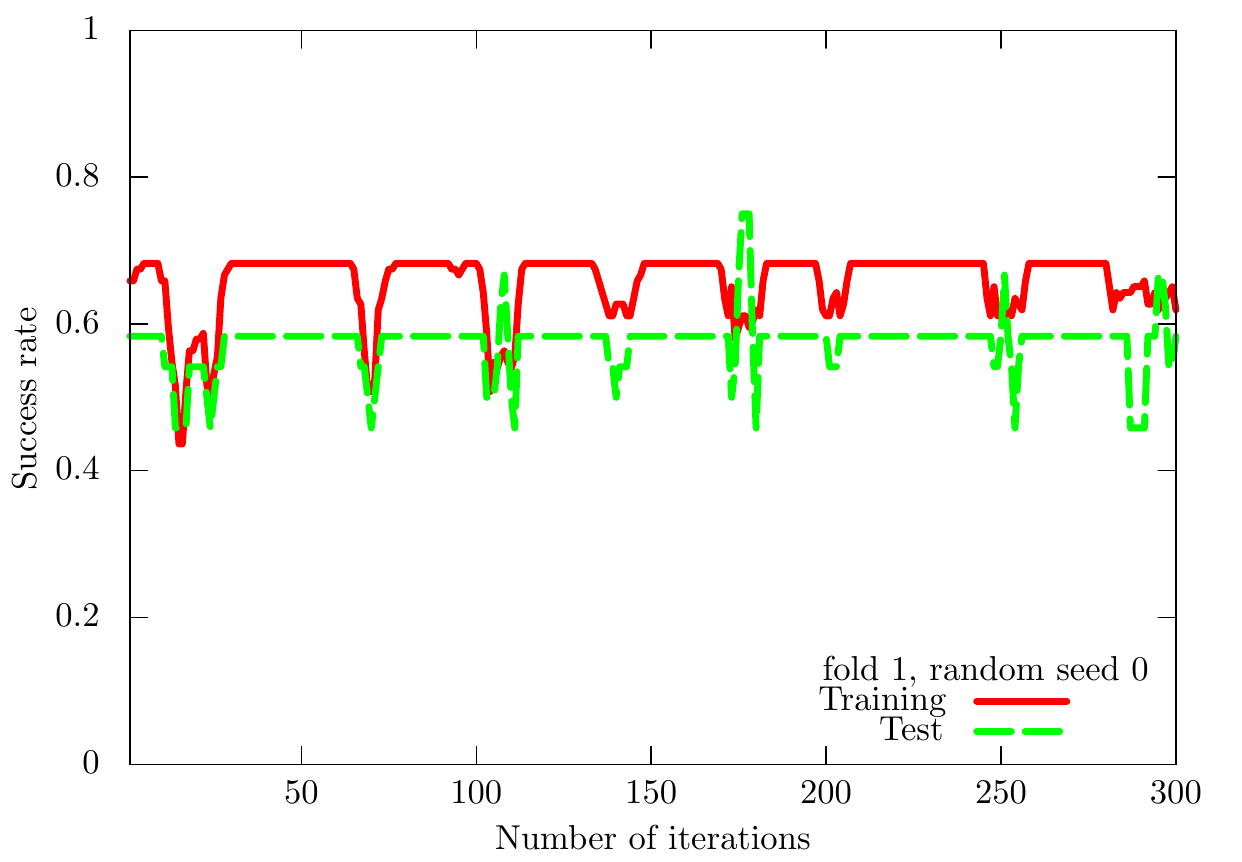}
\includegraphics[scale=0.25]{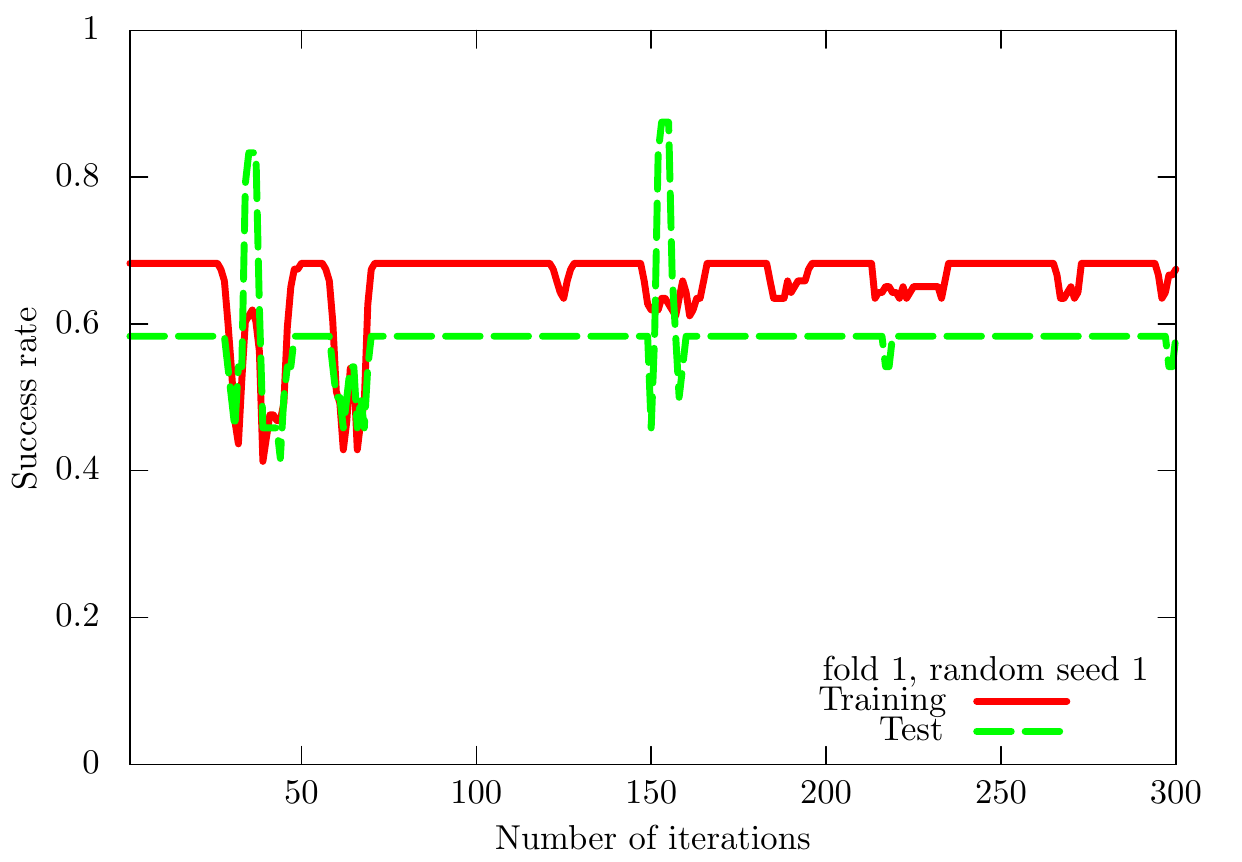}
\includegraphics[scale=0.25]{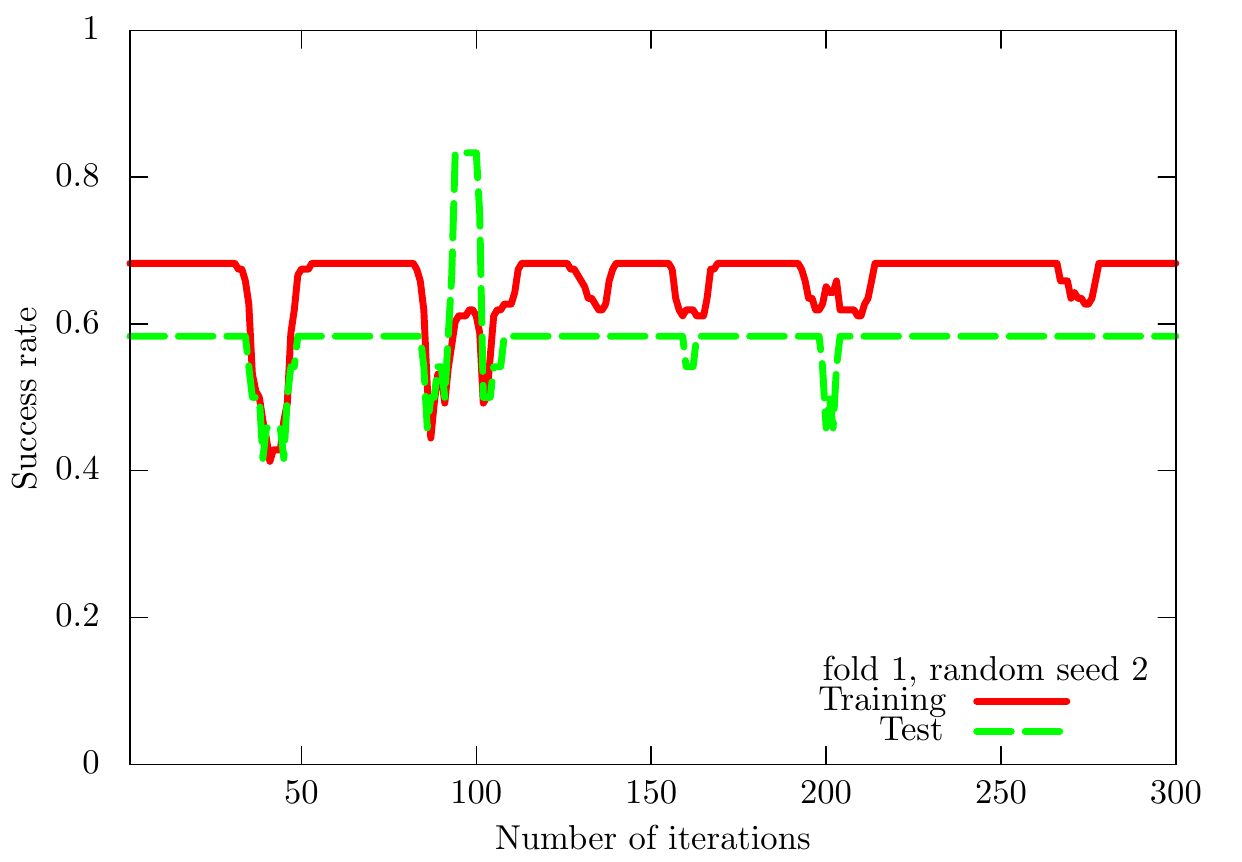}
\includegraphics[scale=0.25]{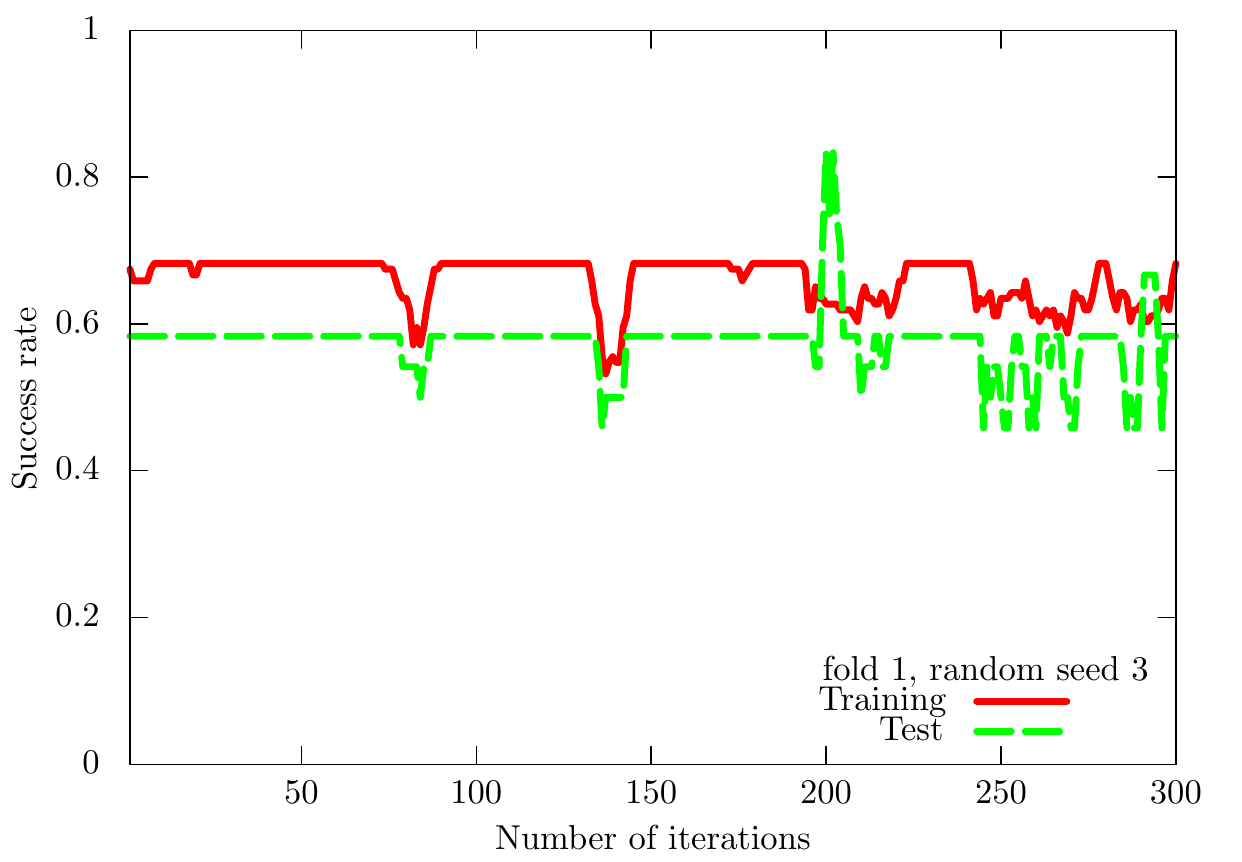}
\includegraphics[scale=0.25]{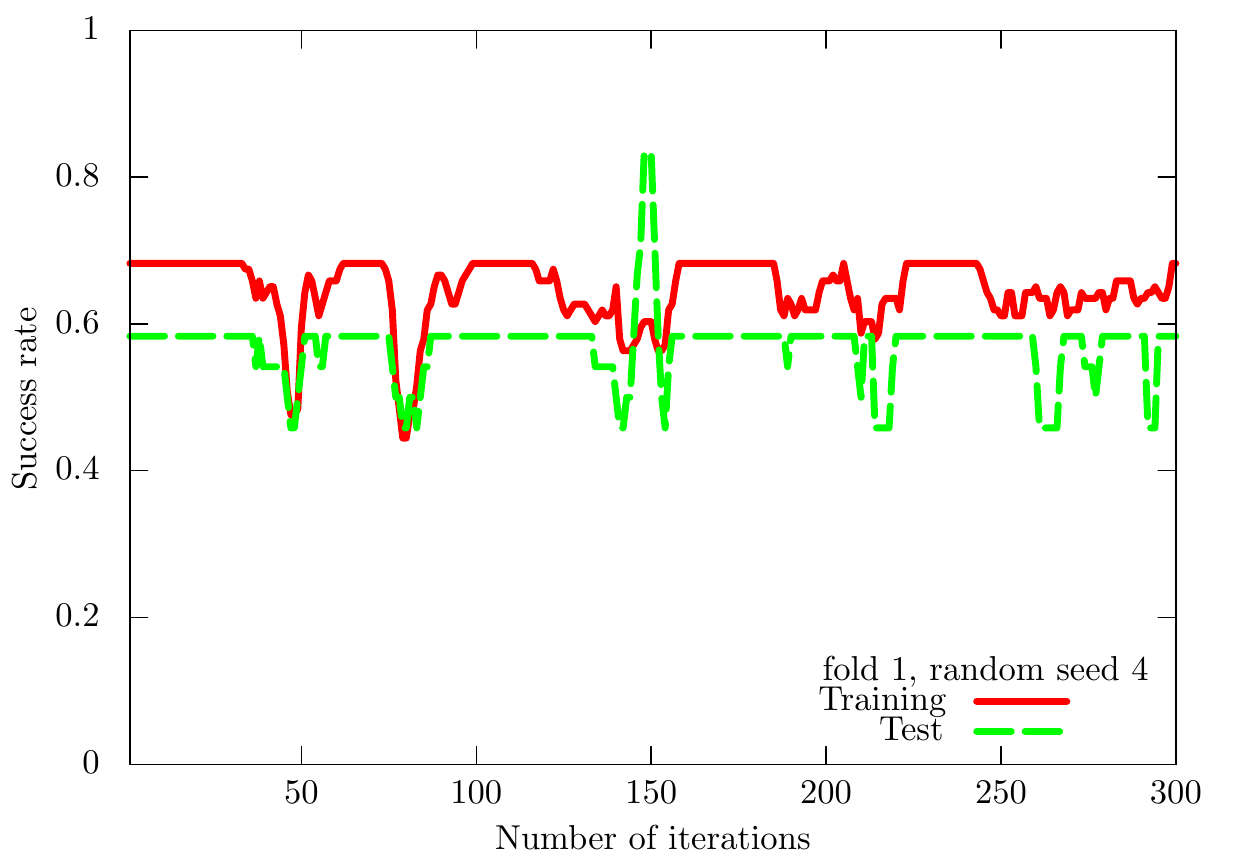}
\includegraphics[scale=0.25]{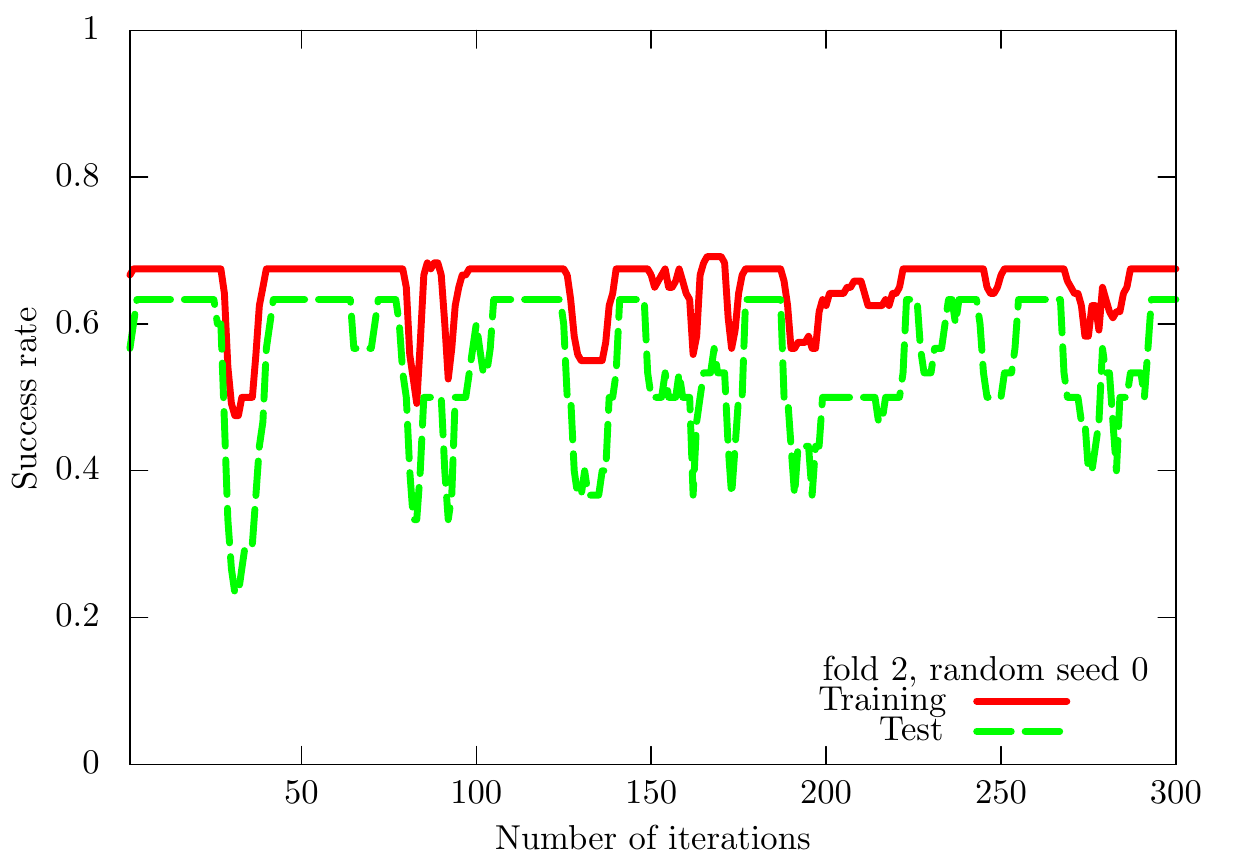}
\includegraphics[scale=0.25]{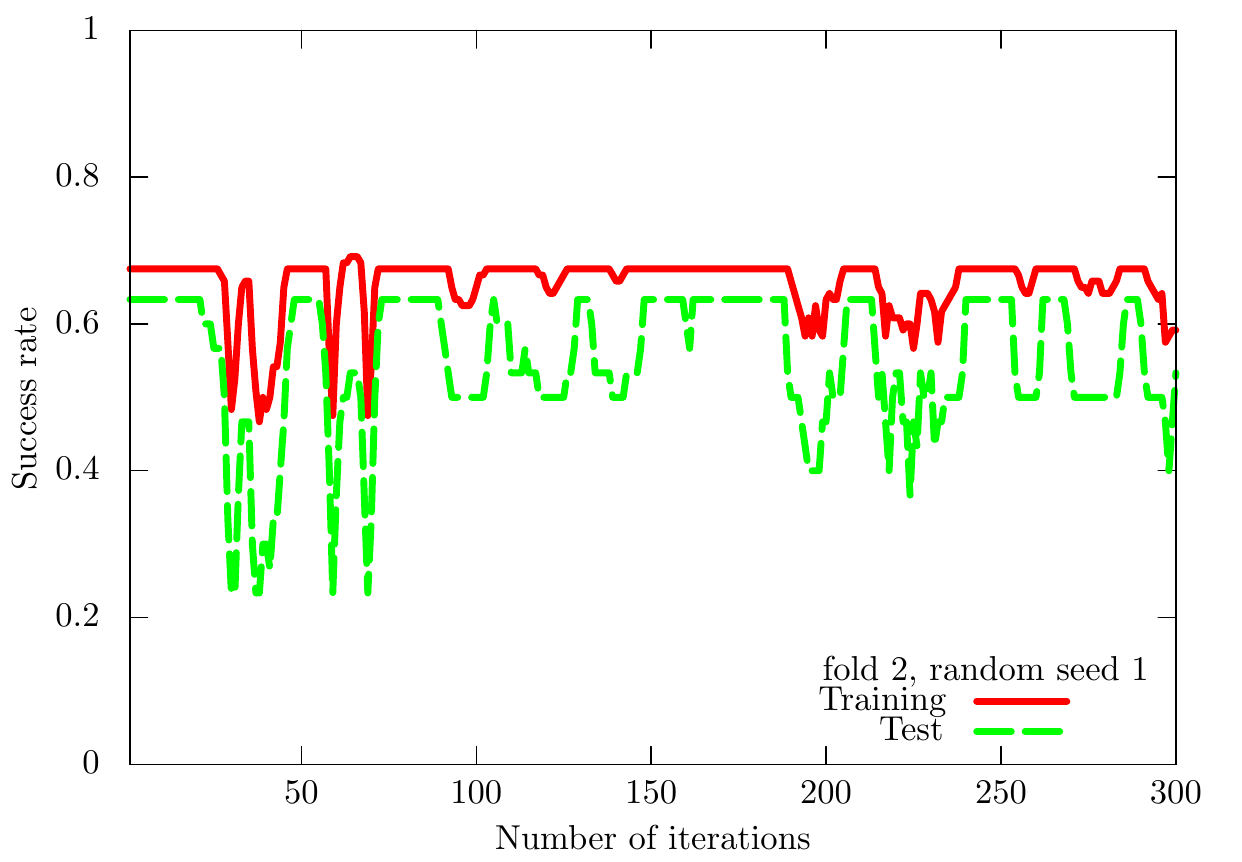}
\includegraphics[scale=0.25]{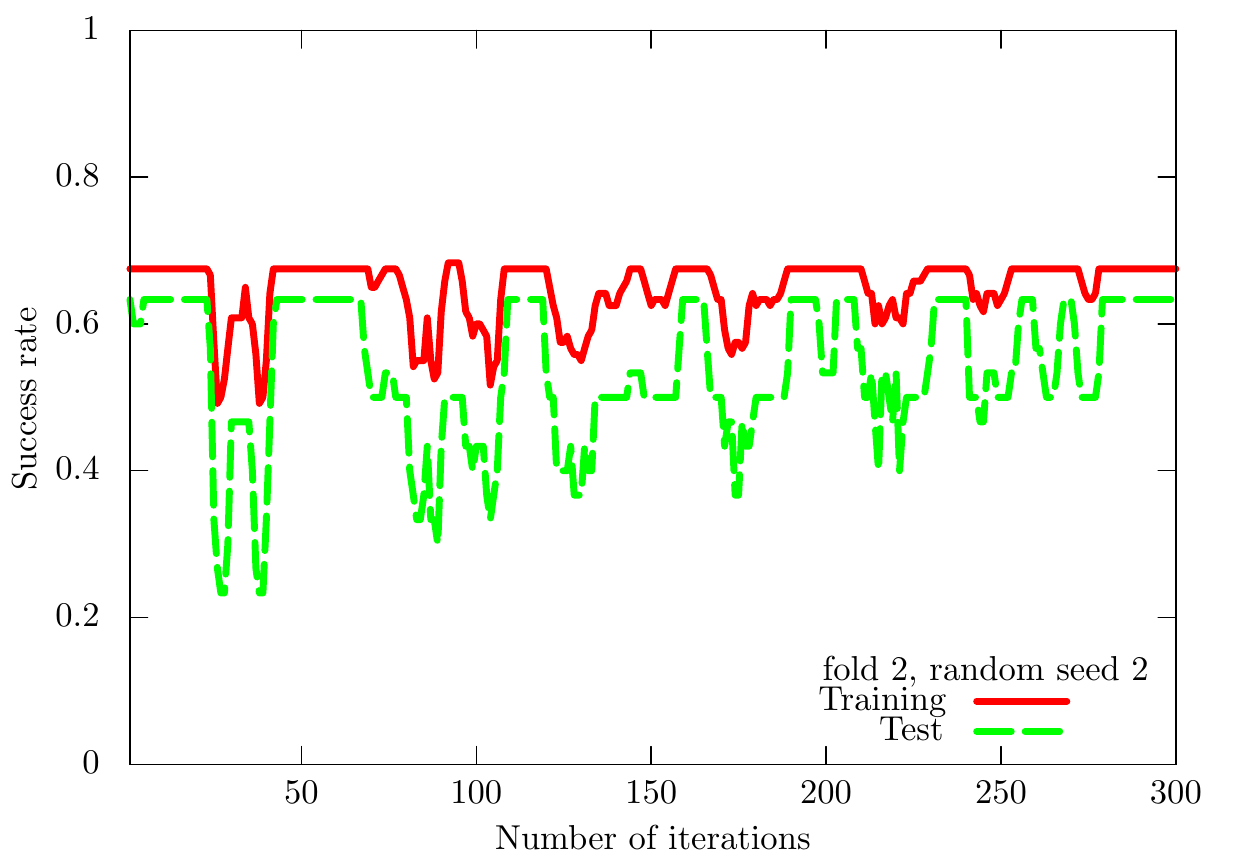}
\includegraphics[scale=0.25]{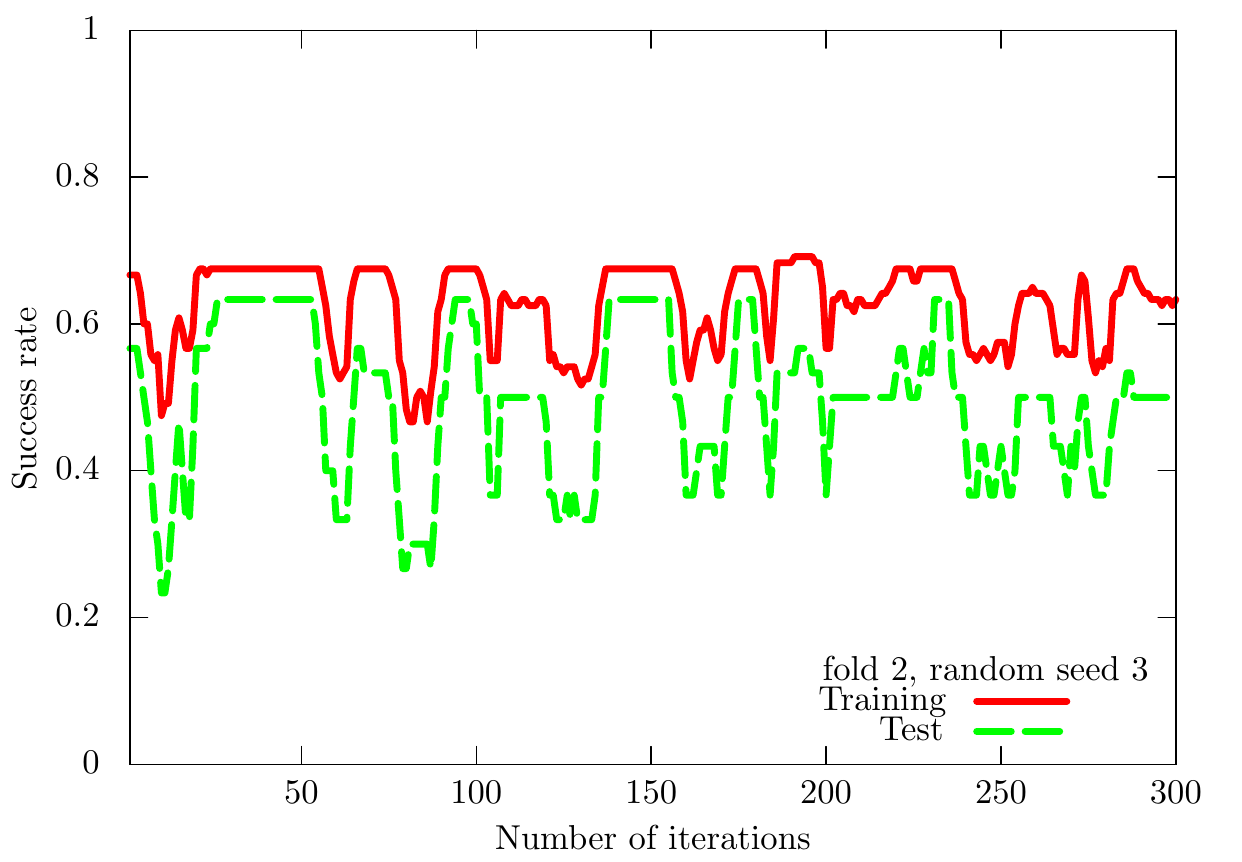}
\includegraphics[scale=0.25]{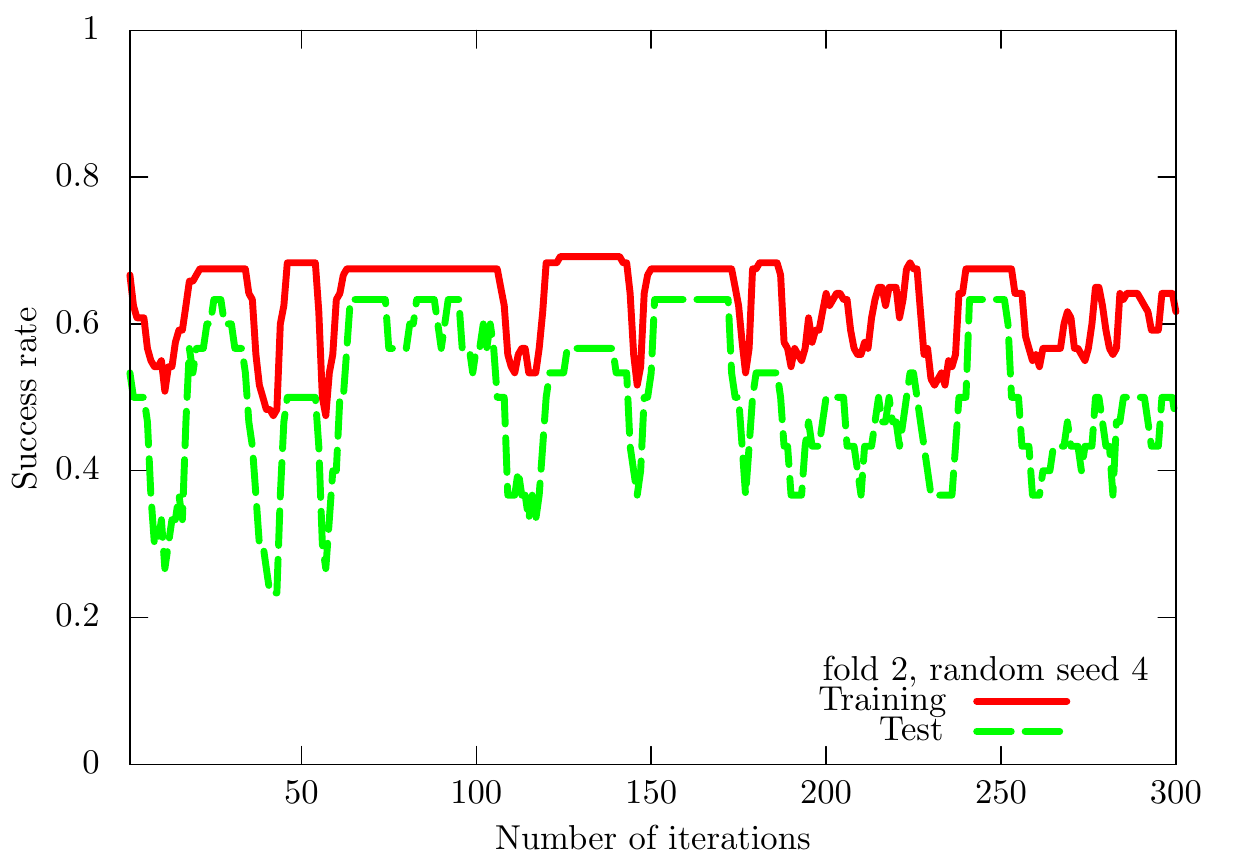}
\includegraphics[scale=0.25]{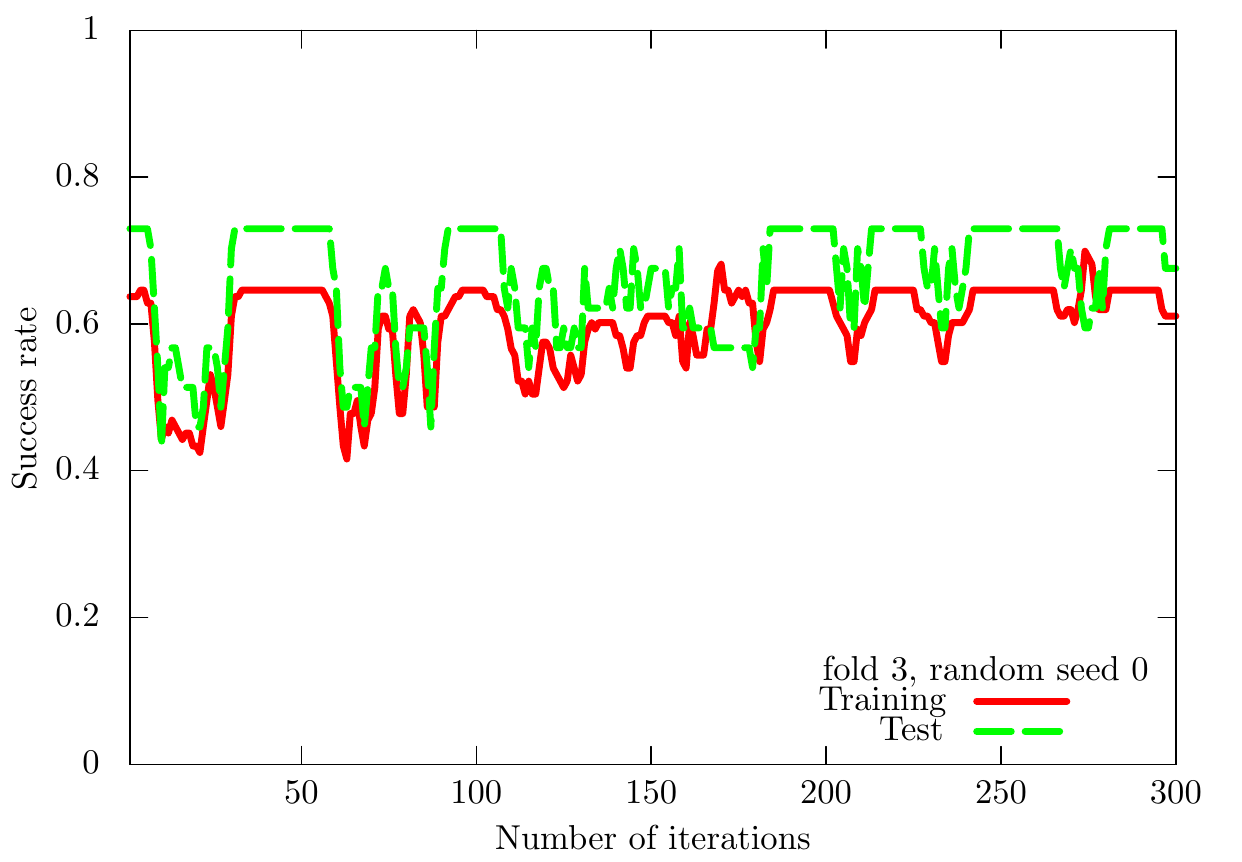}
\includegraphics[scale=0.25]{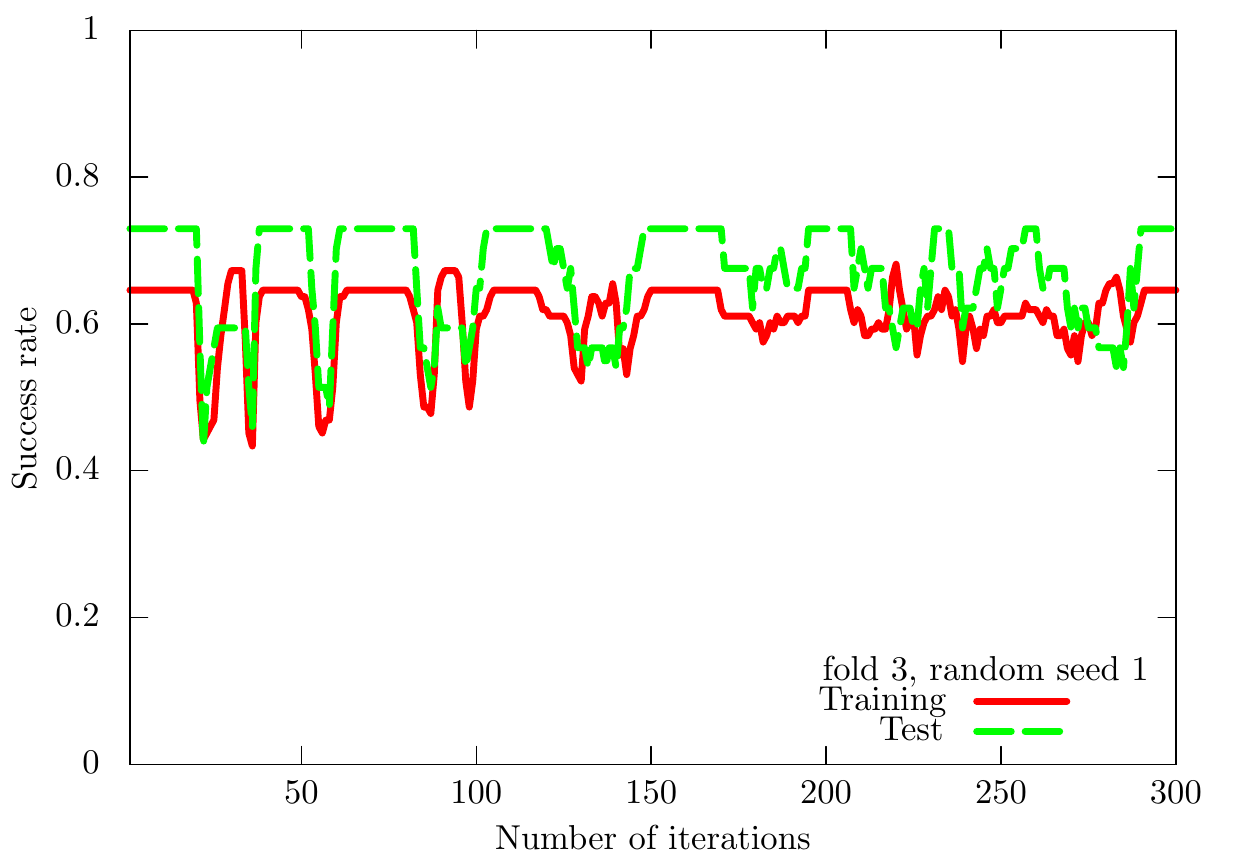}
\includegraphics[scale=0.25]{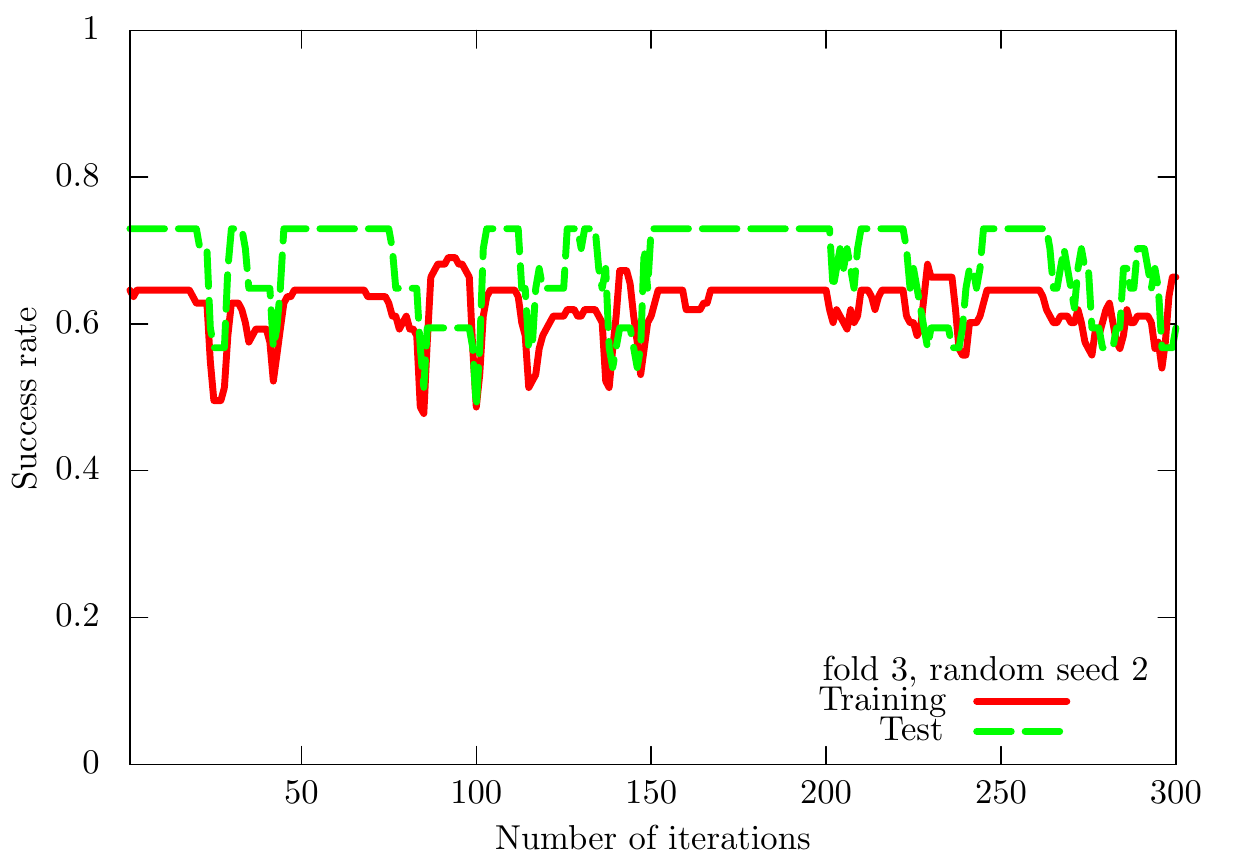}
\includegraphics[scale=0.25]{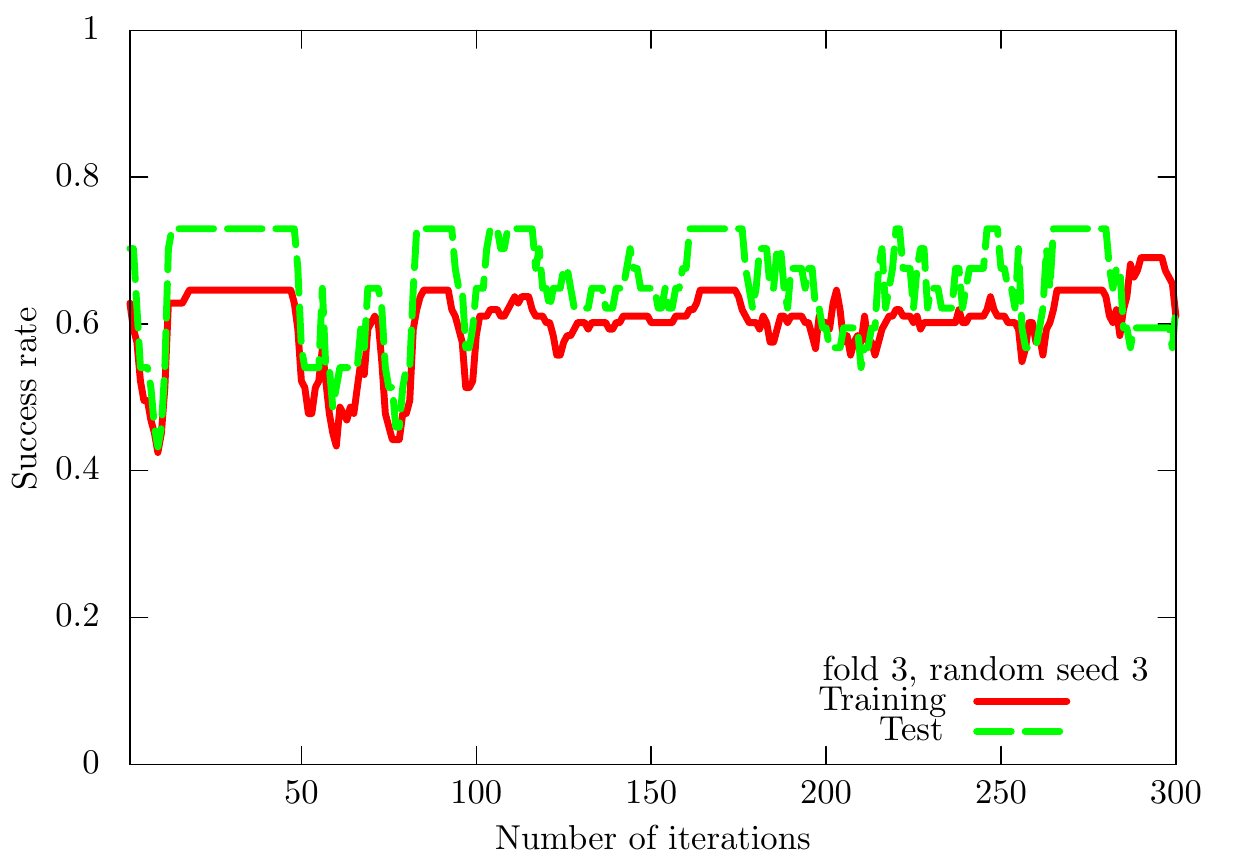}
\includegraphics[scale=0.25]{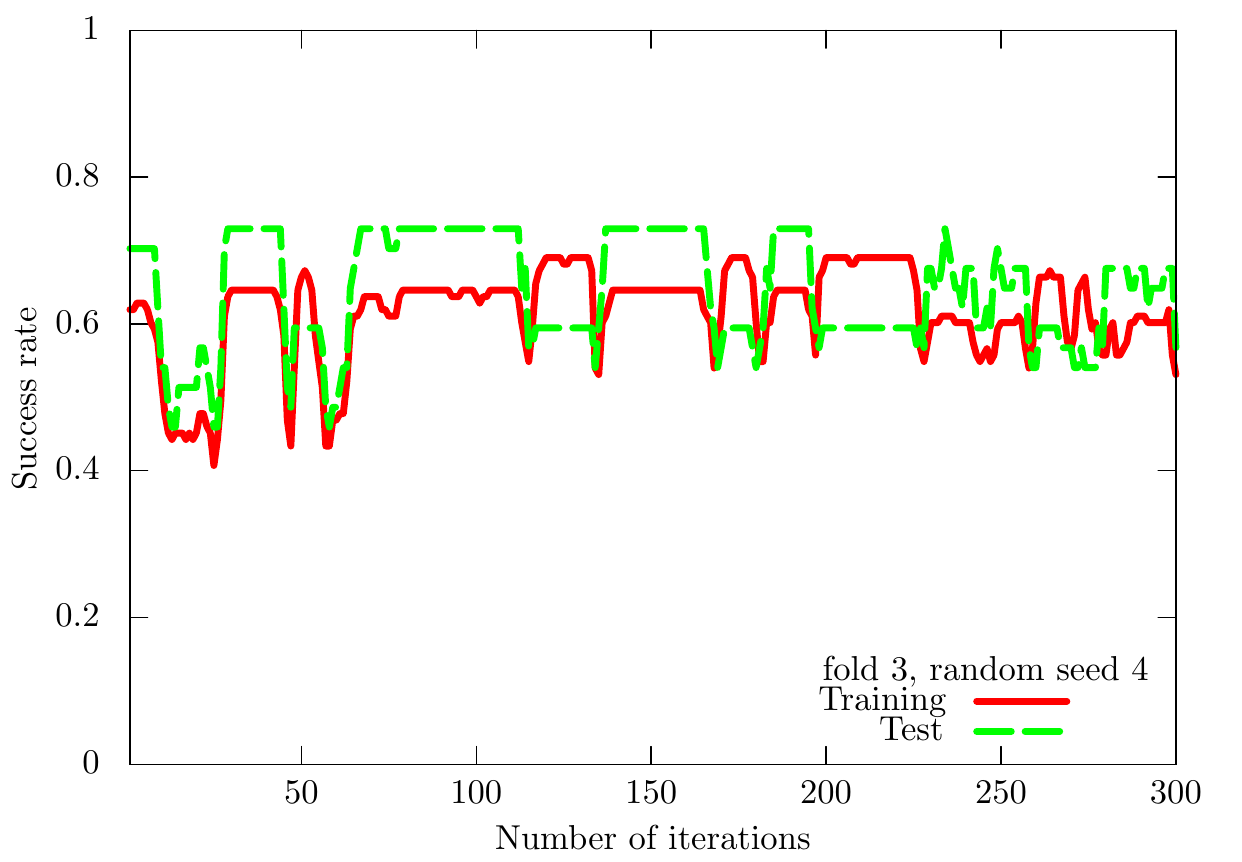}
\includegraphics[scale=0.25]{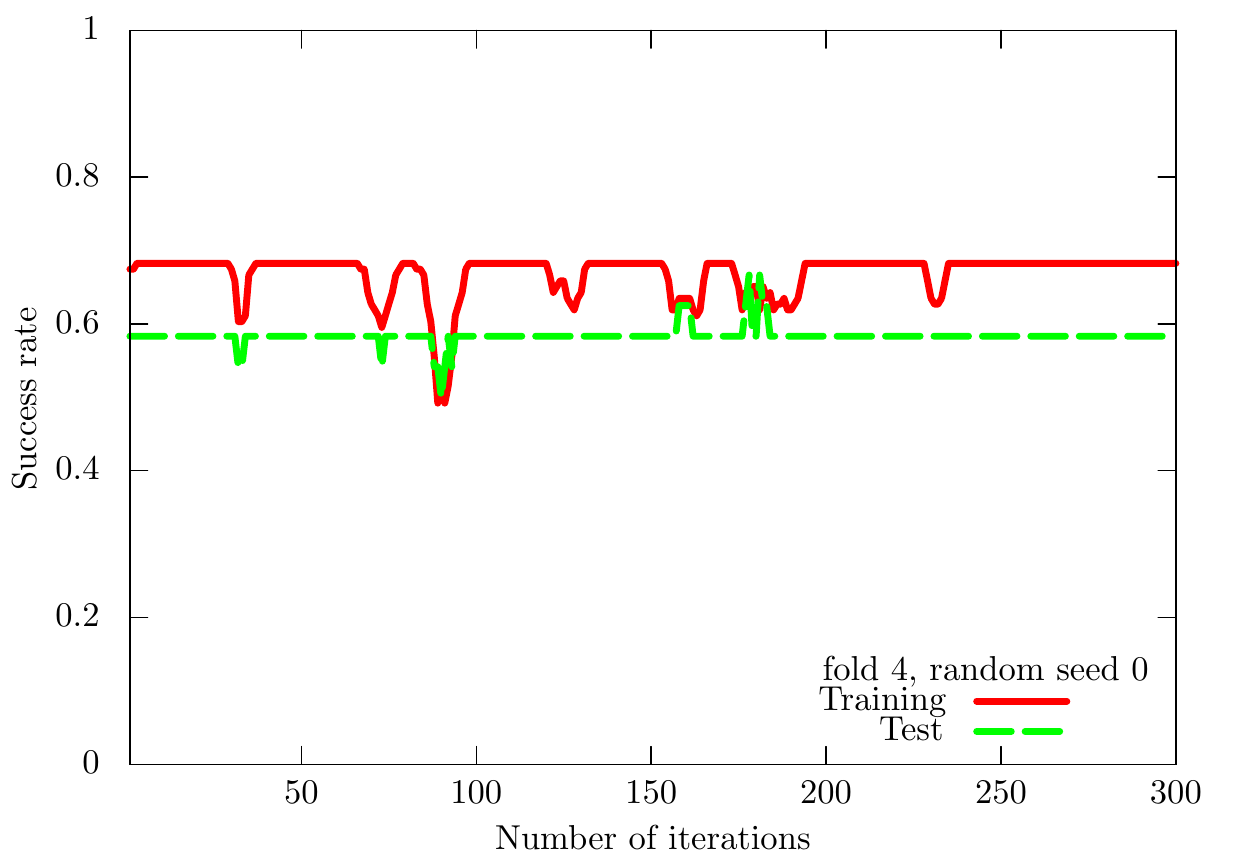}
\includegraphics[scale=0.25]{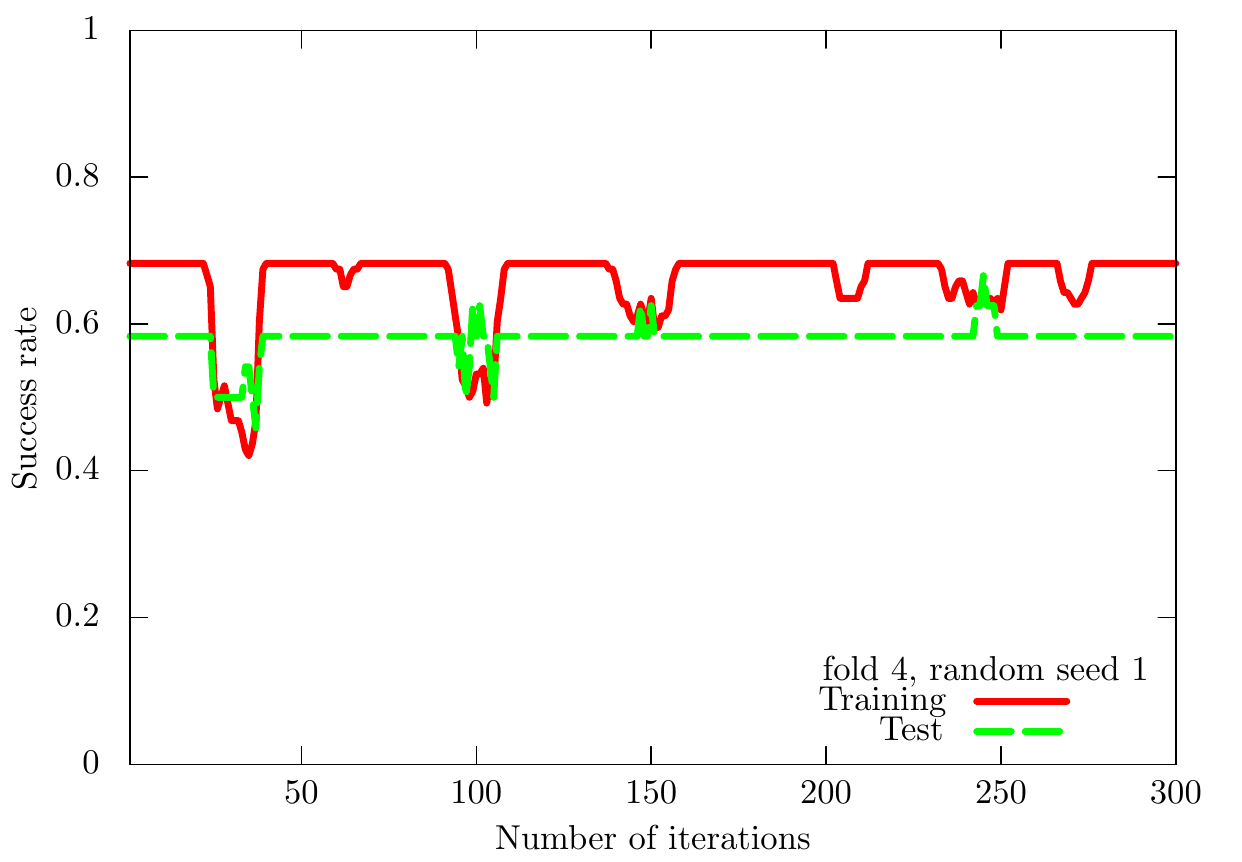}
\includegraphics[scale=0.25]{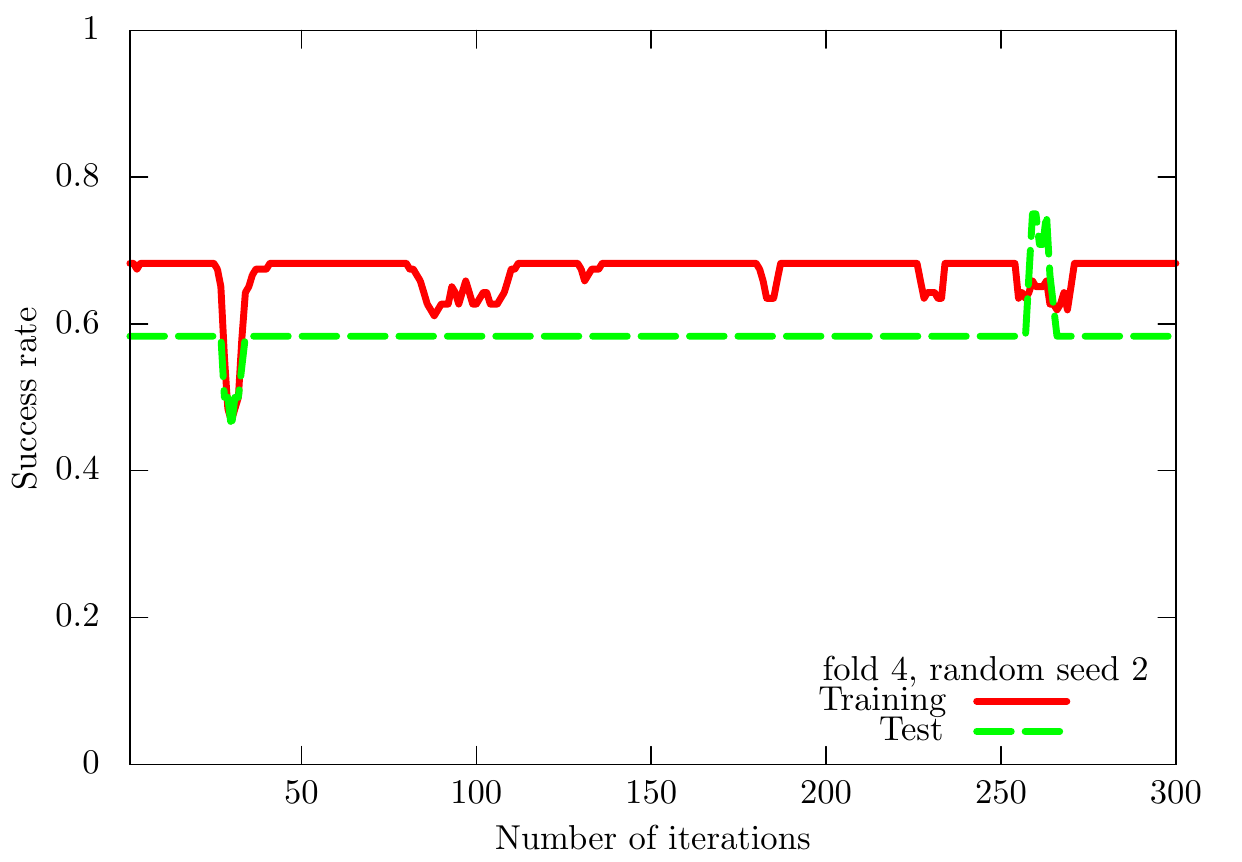}
\includegraphics[scale=0.25]{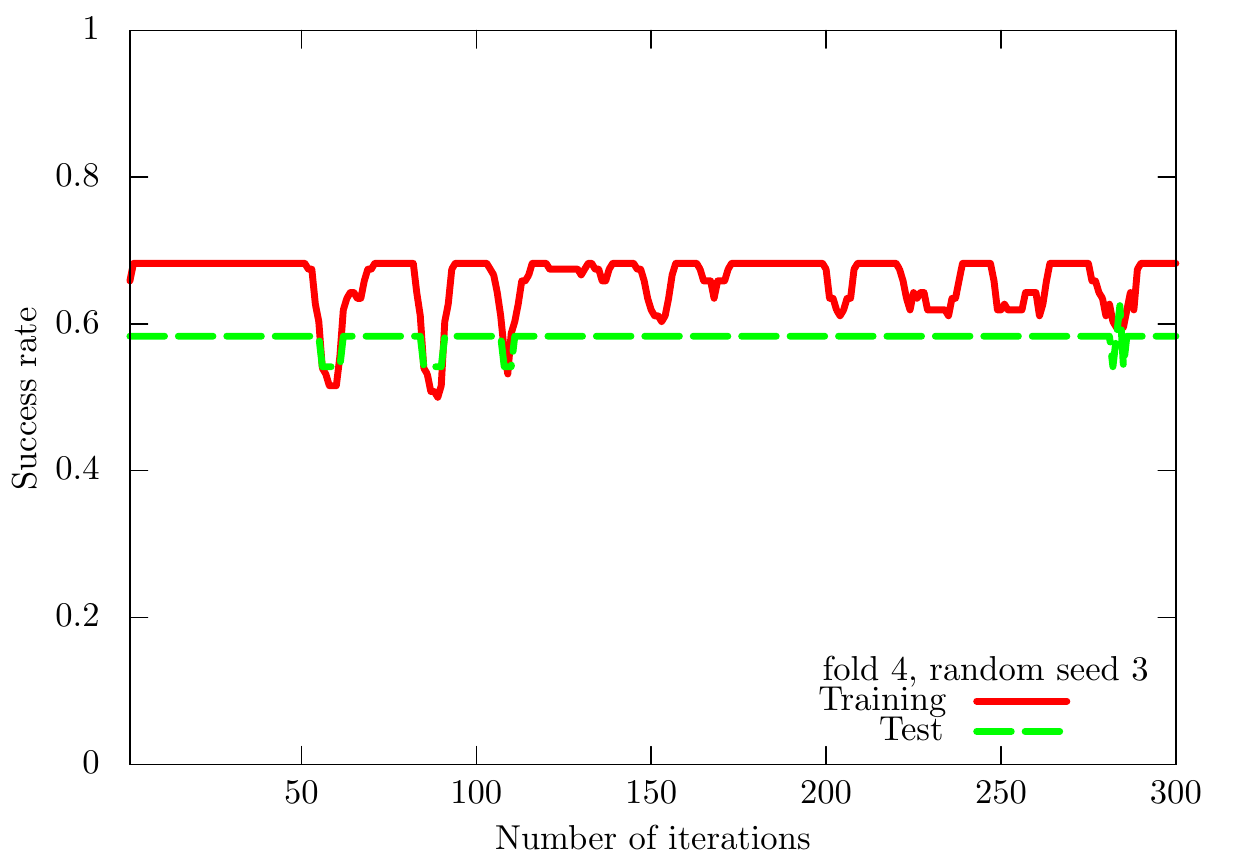}
\includegraphics[scale=0.25]{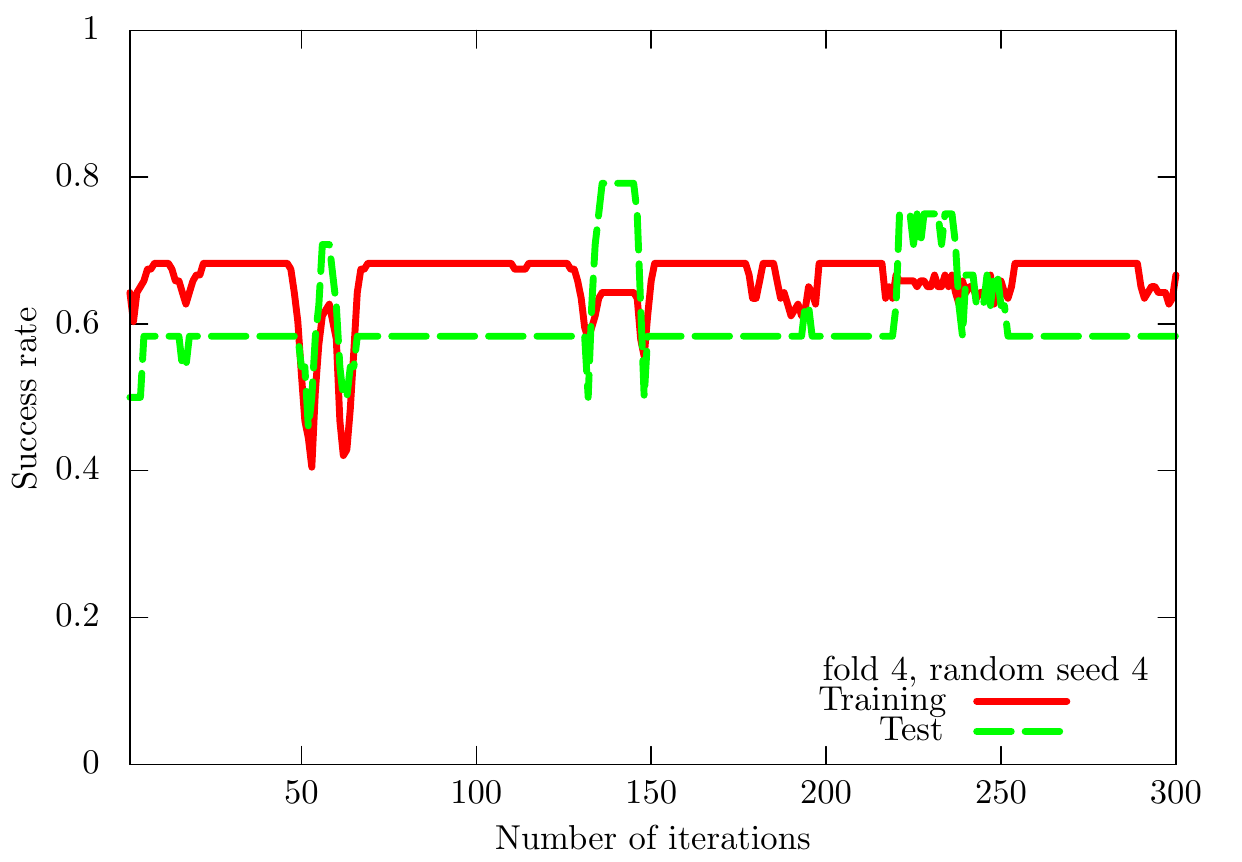}
\caption{Results of QCL on the $5$-fold datasets with $5$ different random seeds for the iris dataset ($1$ or non-$1$). We use the CNOT-based circuit and set $\theta_\mathrm{bias} = 0$. The number of layers $L$ is set to $5$.}
\label{supp-arXiv-numerical-result-raw-data-fold-001-rand-001-QCL-UCI-iris-1-non1}
\end{figure*}

\begin{figure*}[htb]
\centering
\includegraphics[scale=0.25]{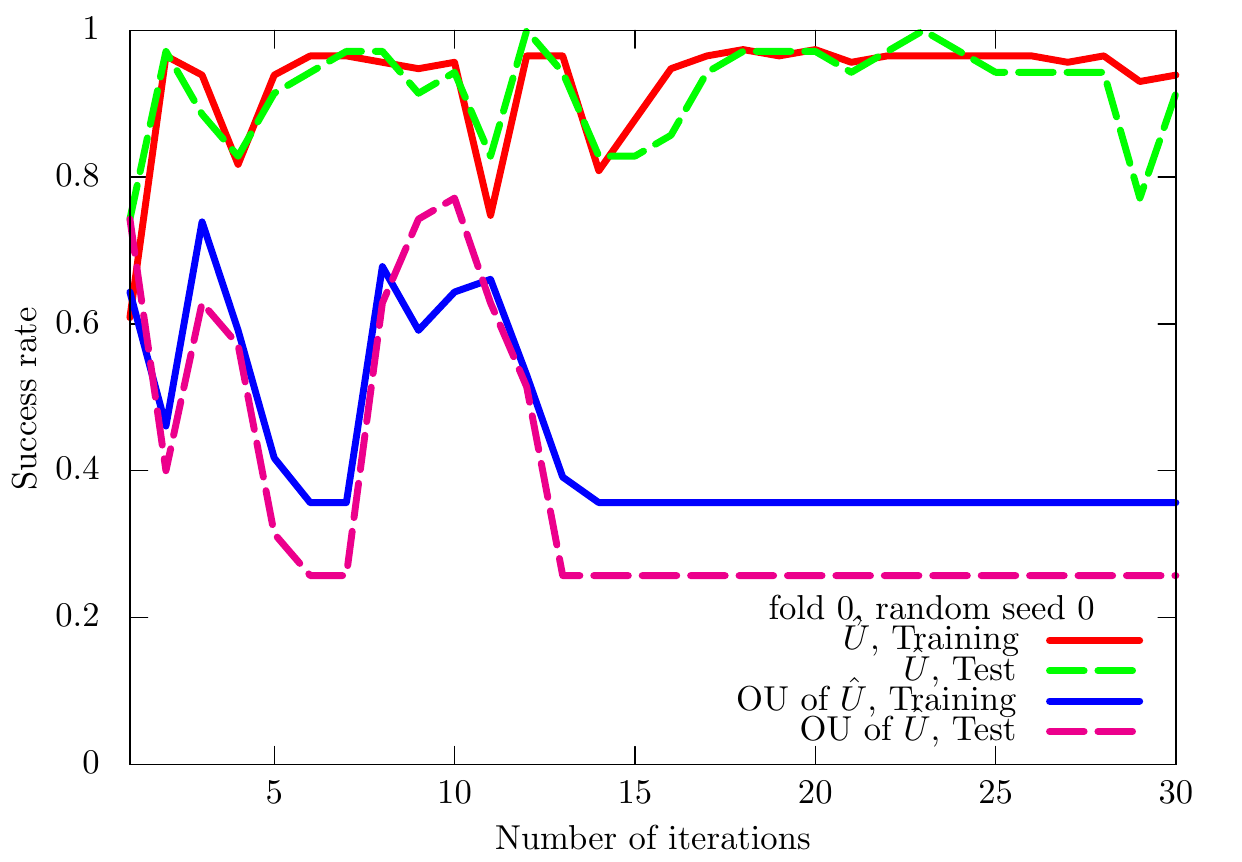}
\includegraphics[scale=0.25]{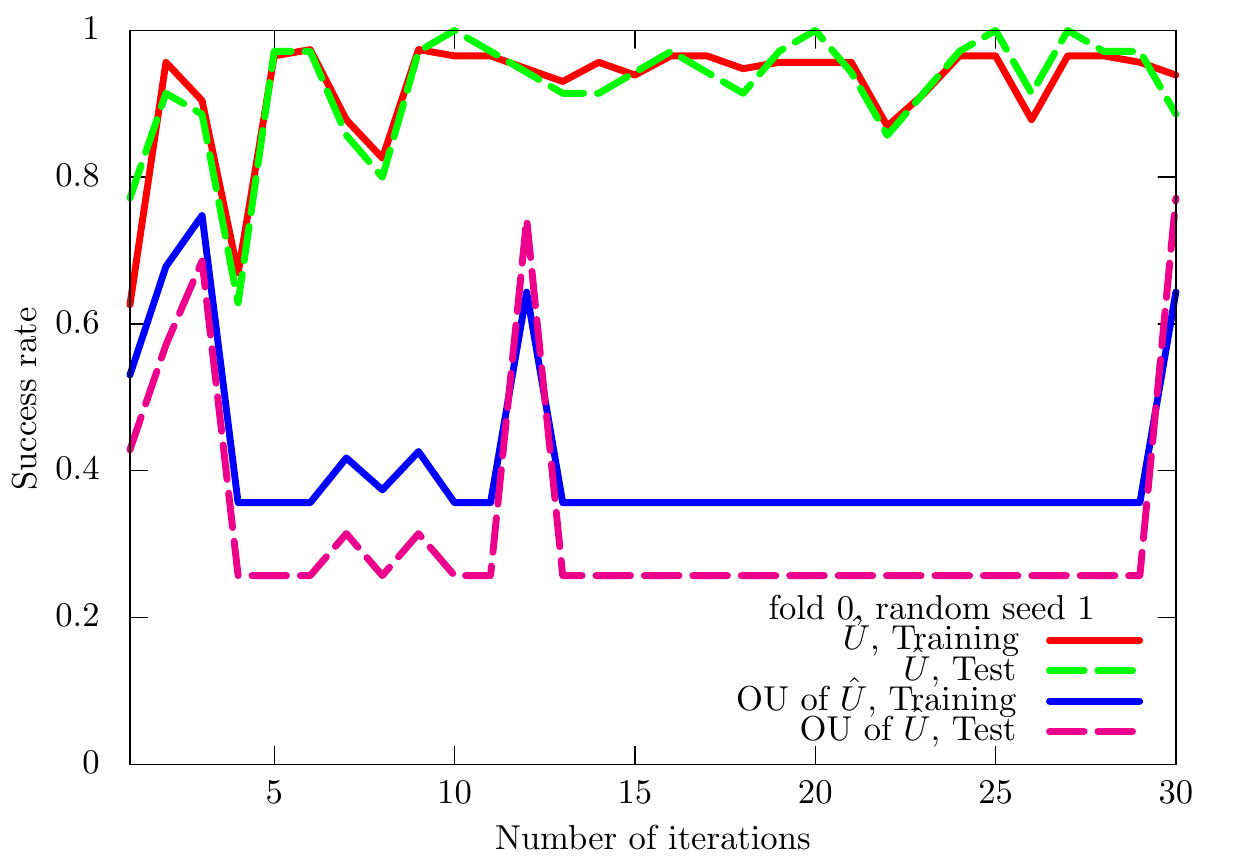}
\includegraphics[scale=0.25]{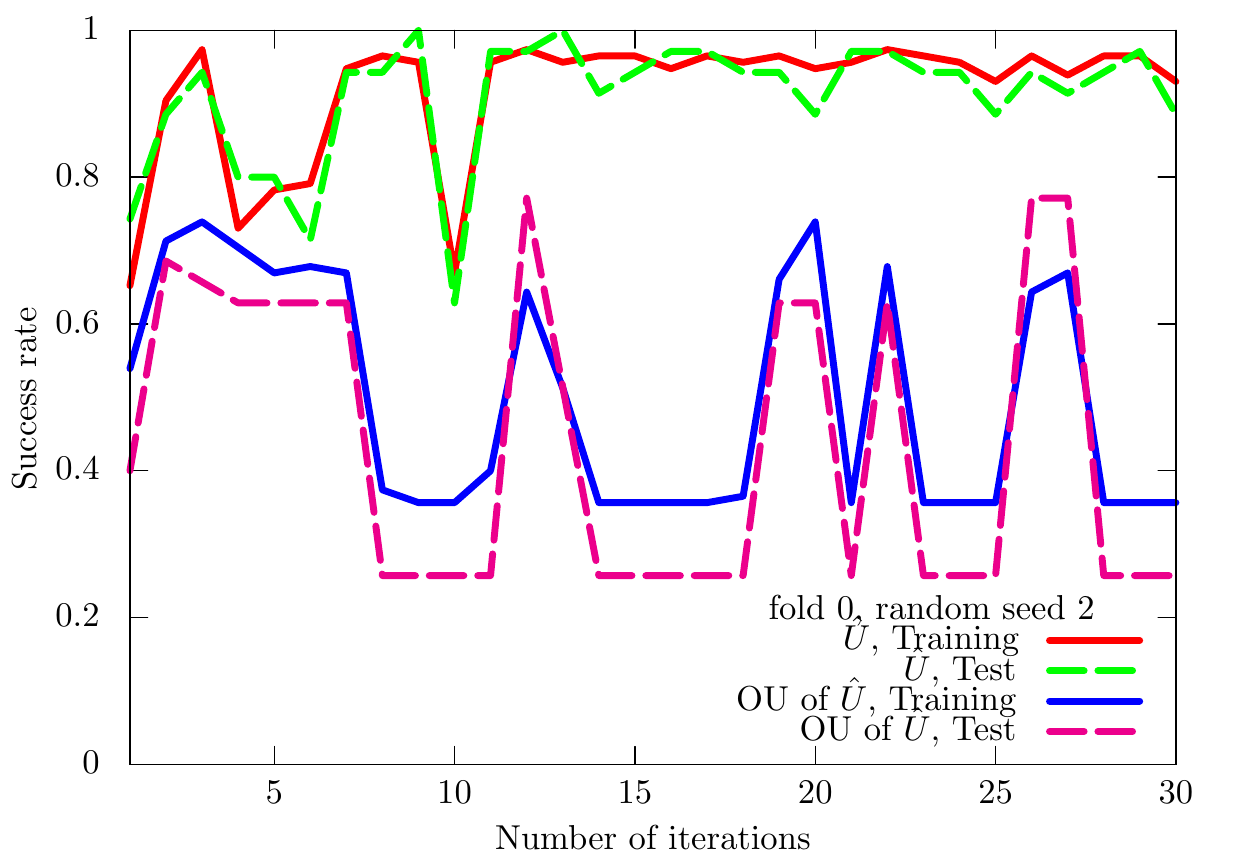}
\includegraphics[scale=0.25]{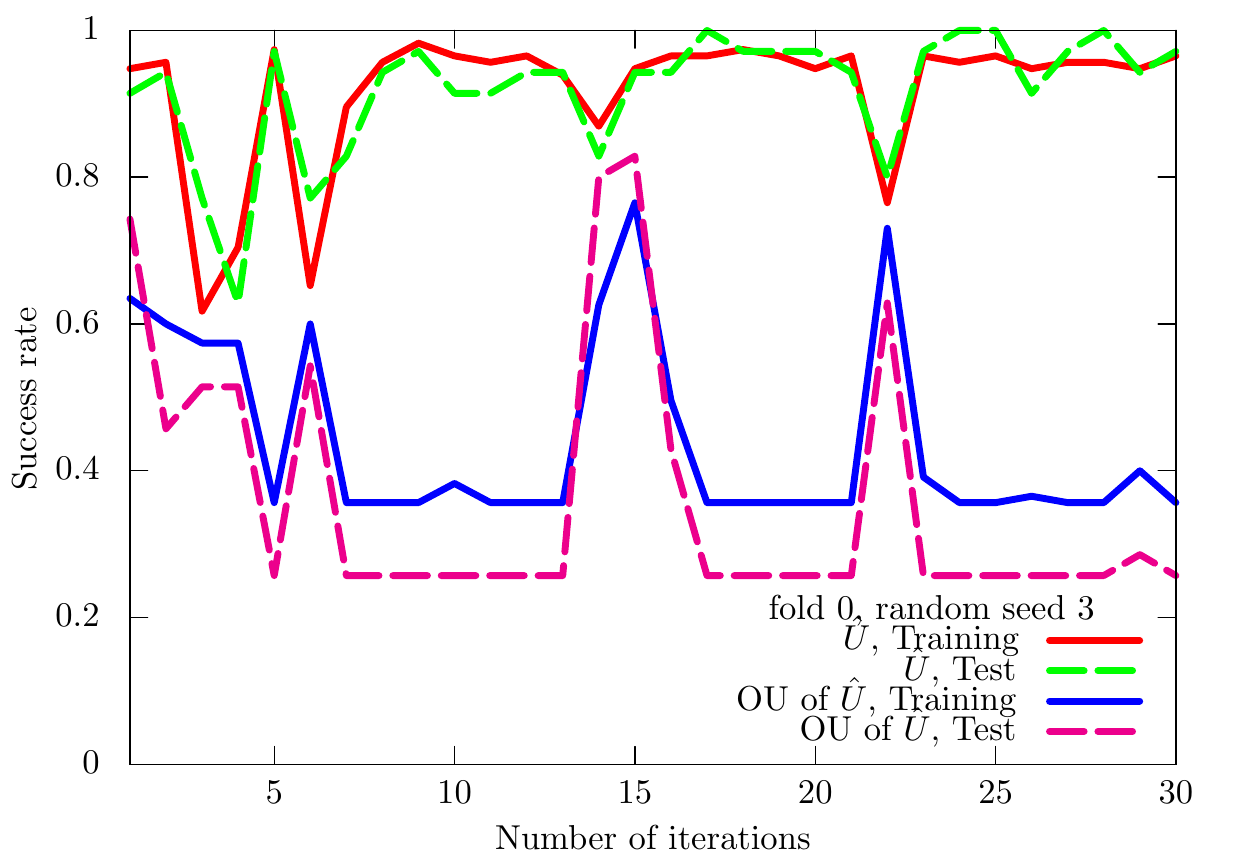}
\includegraphics[scale=0.25]{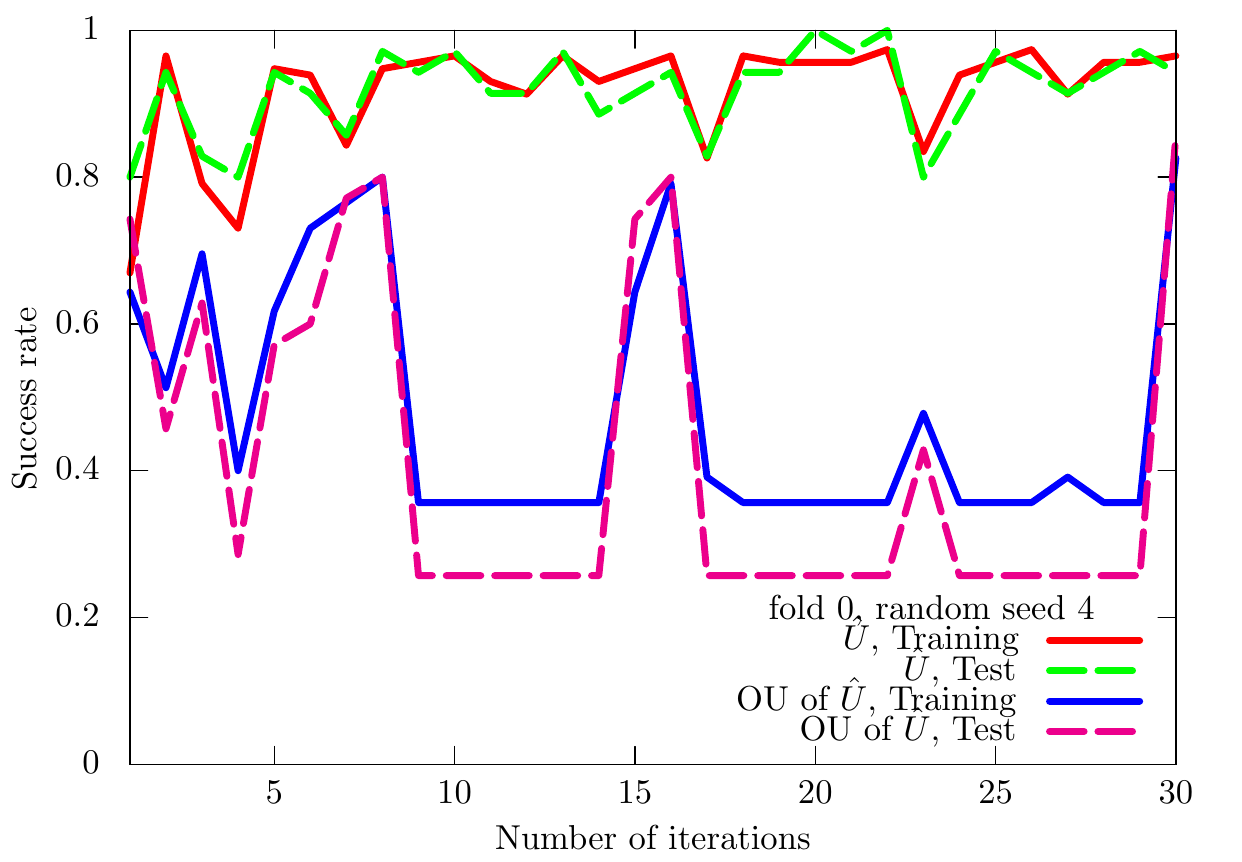}
\includegraphics[scale=0.25]{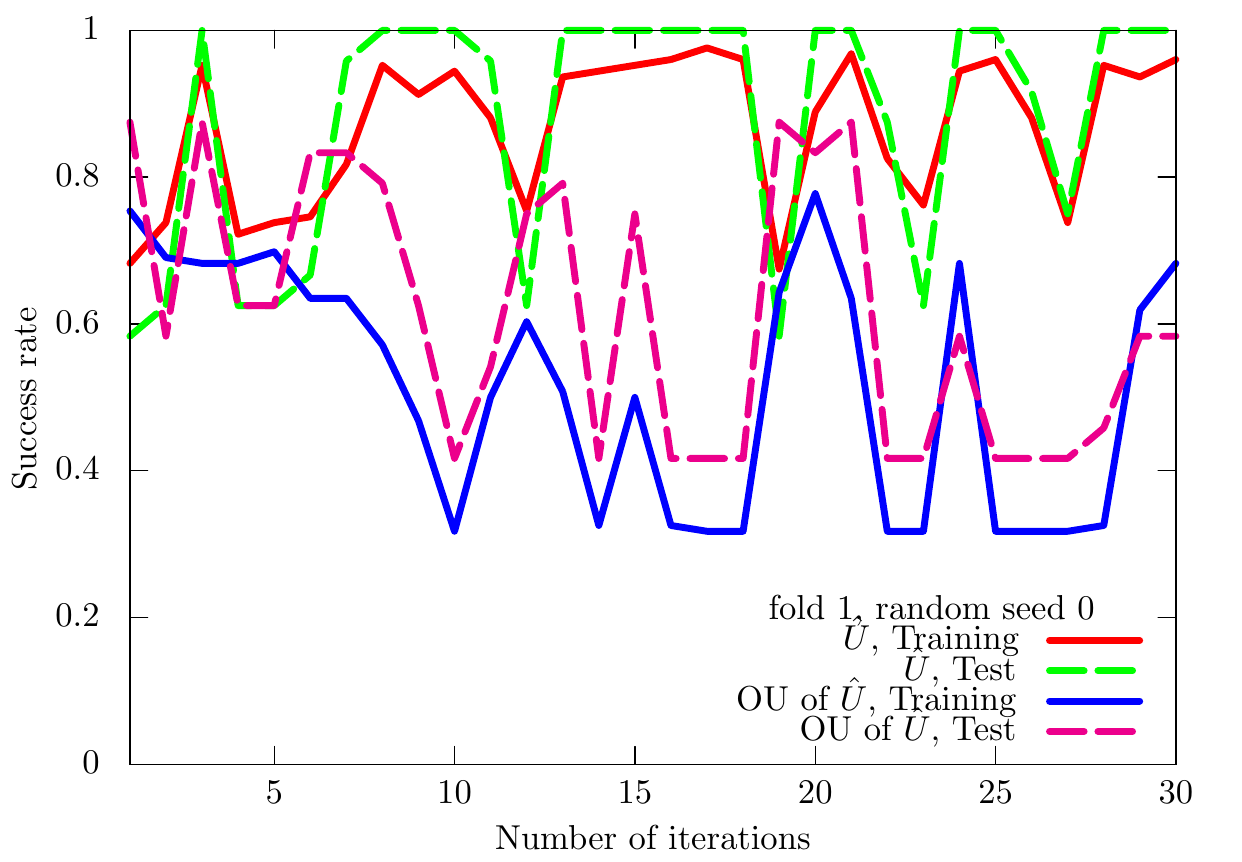}
\includegraphics[scale=0.25]{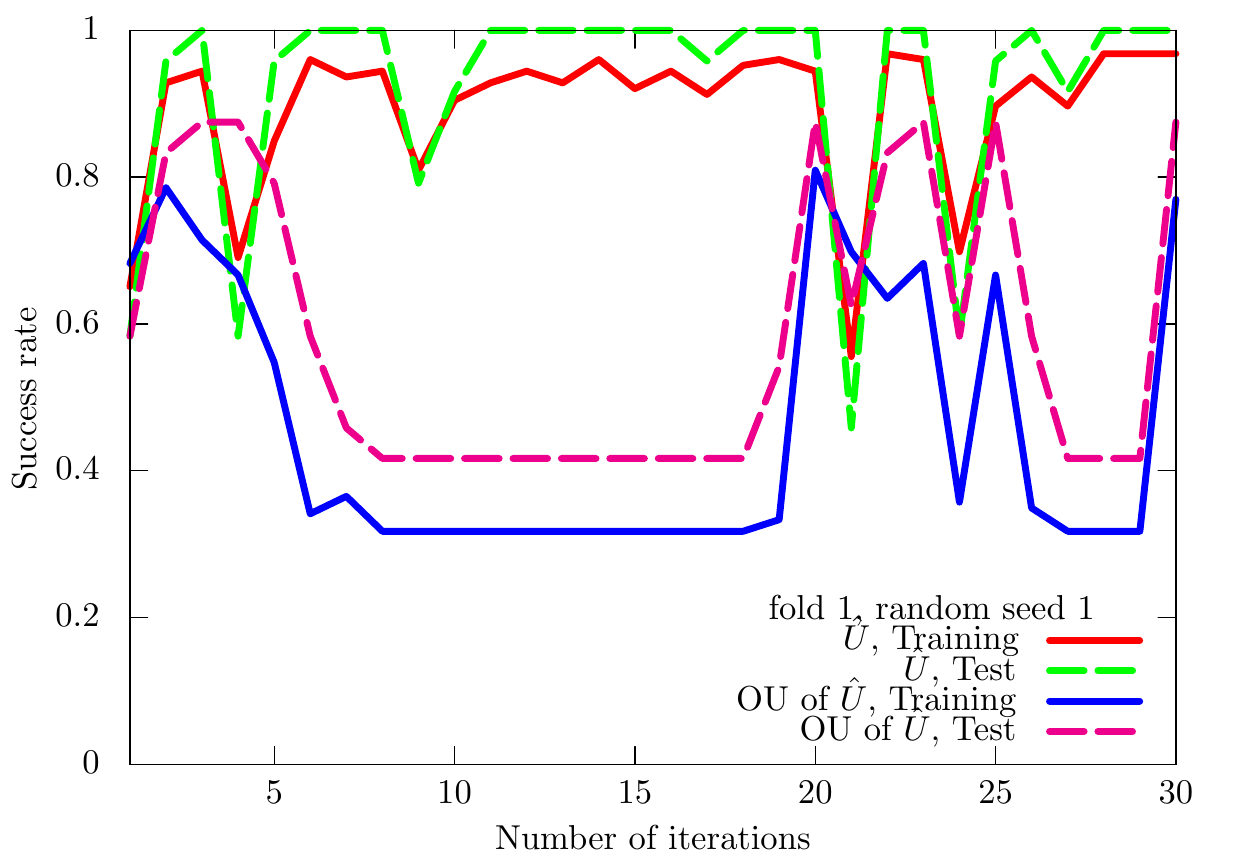}
\includegraphics[scale=0.25]{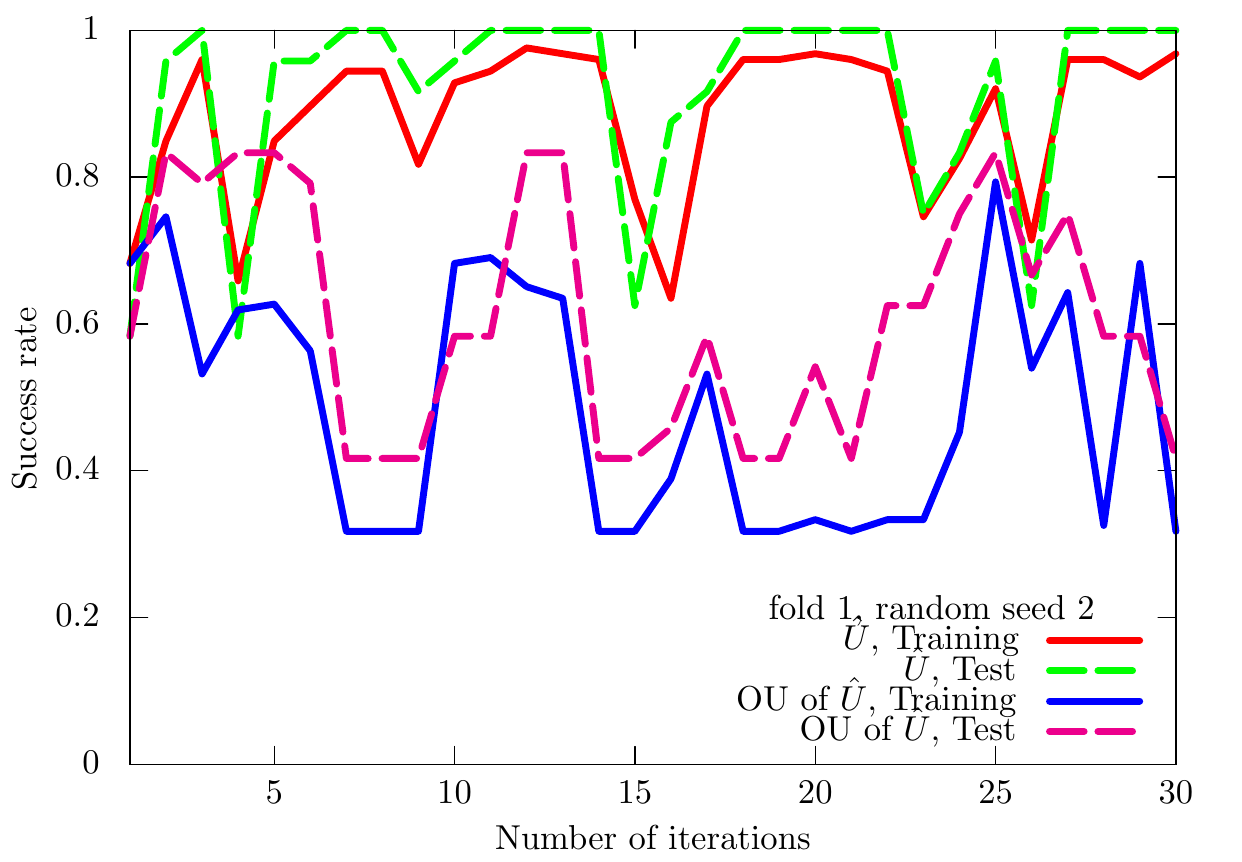}
\includegraphics[scale=0.25]{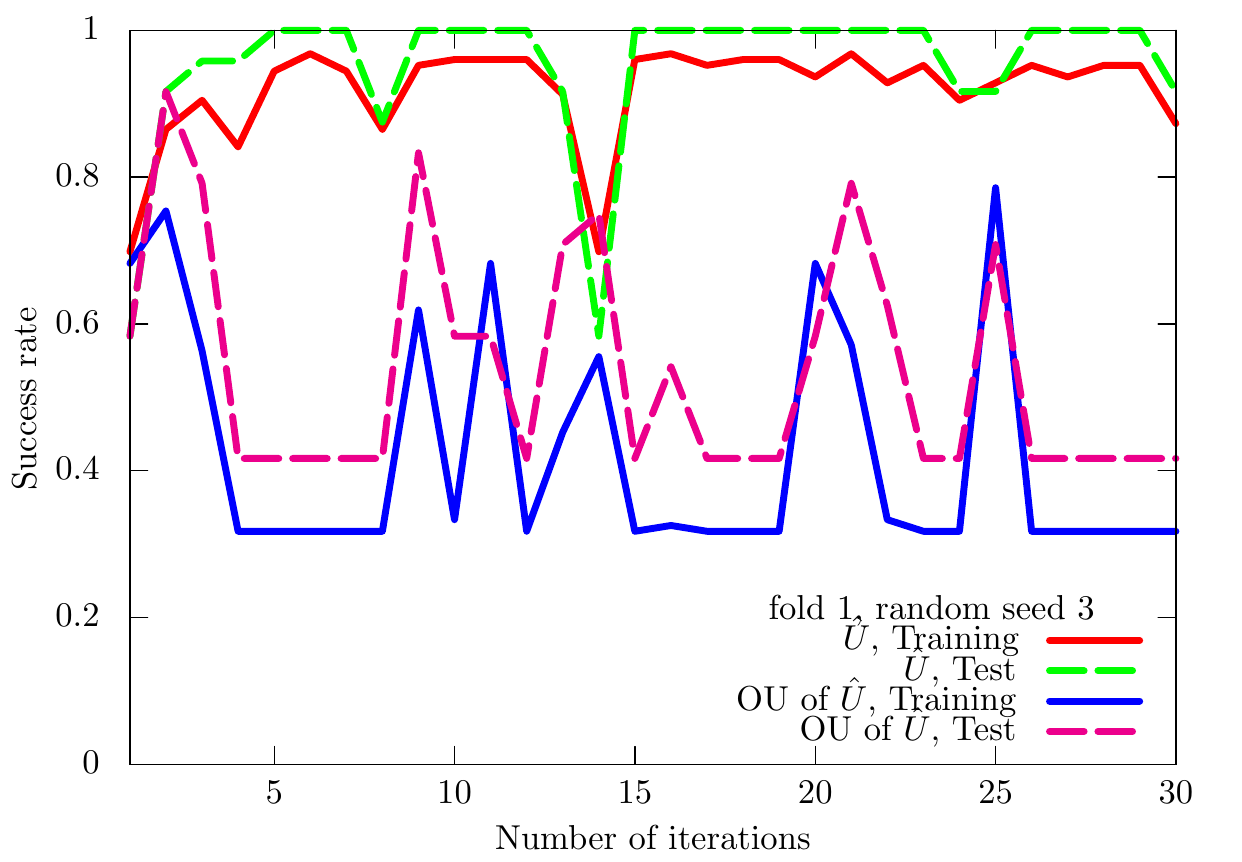}
\includegraphics[scale=0.25]{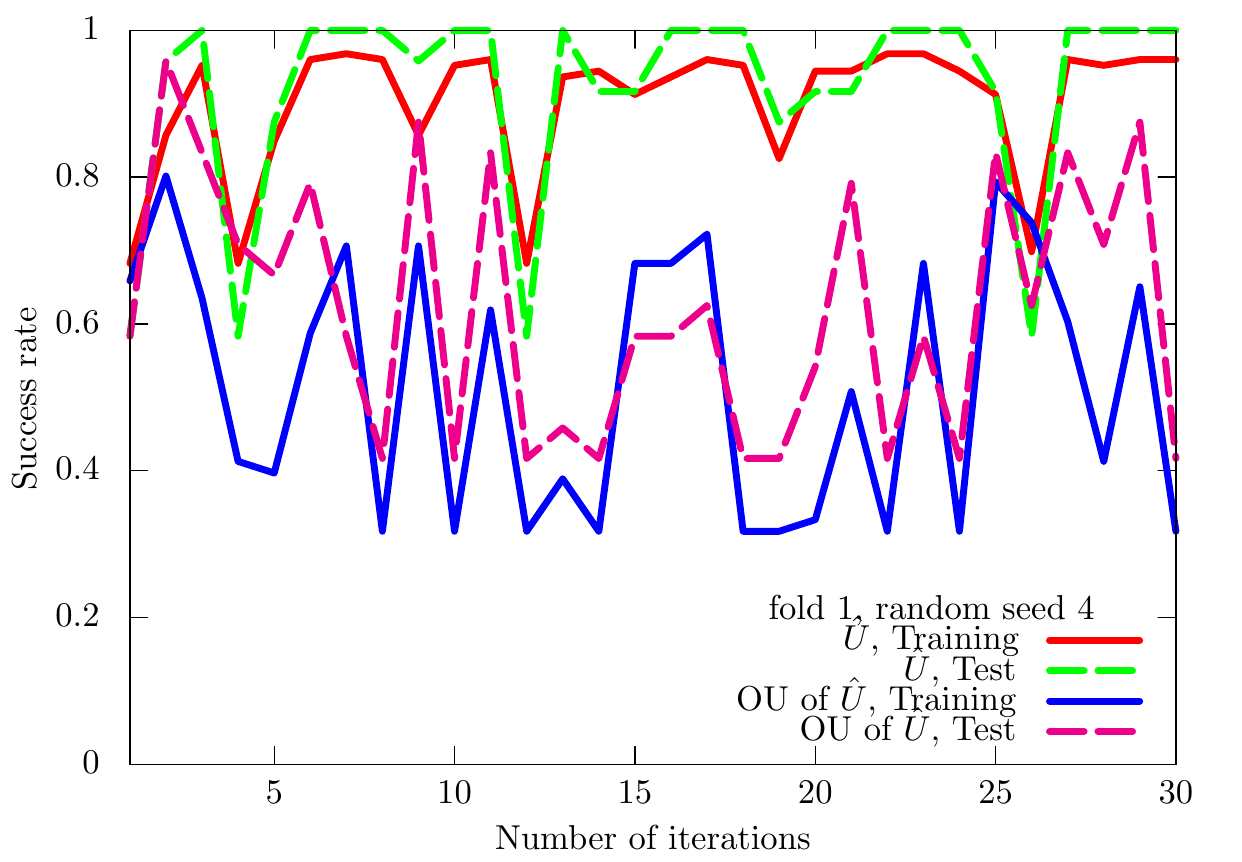}
\includegraphics[scale=0.25]{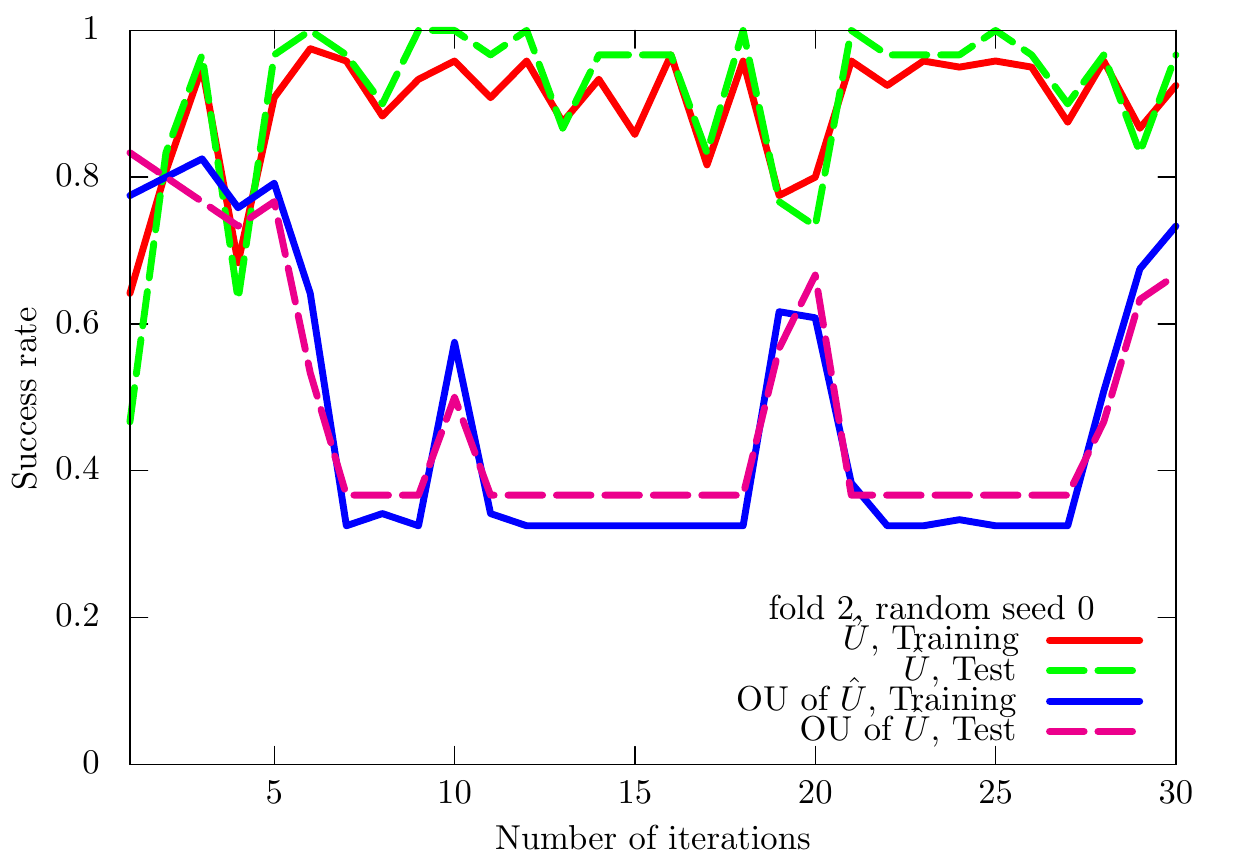}
\includegraphics[scale=0.25]{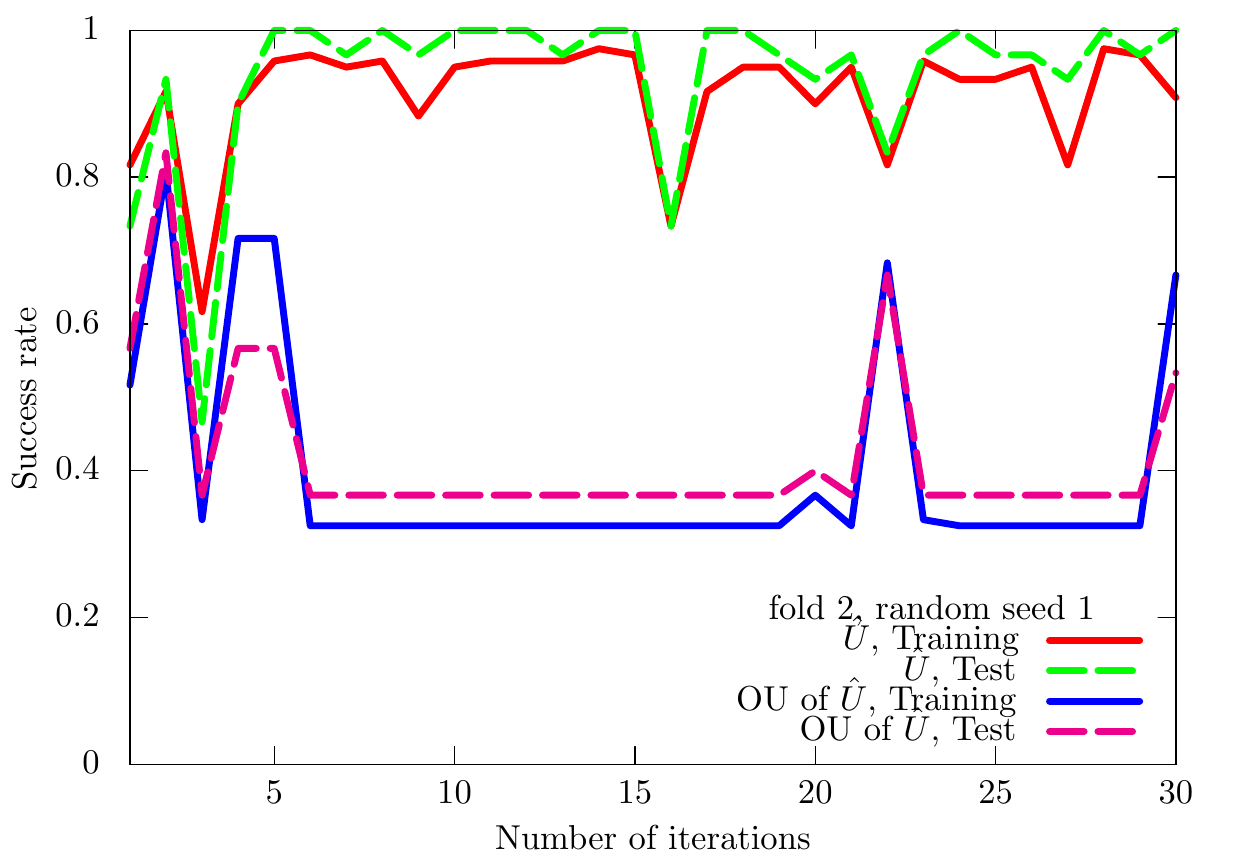}
\includegraphics[scale=0.25]{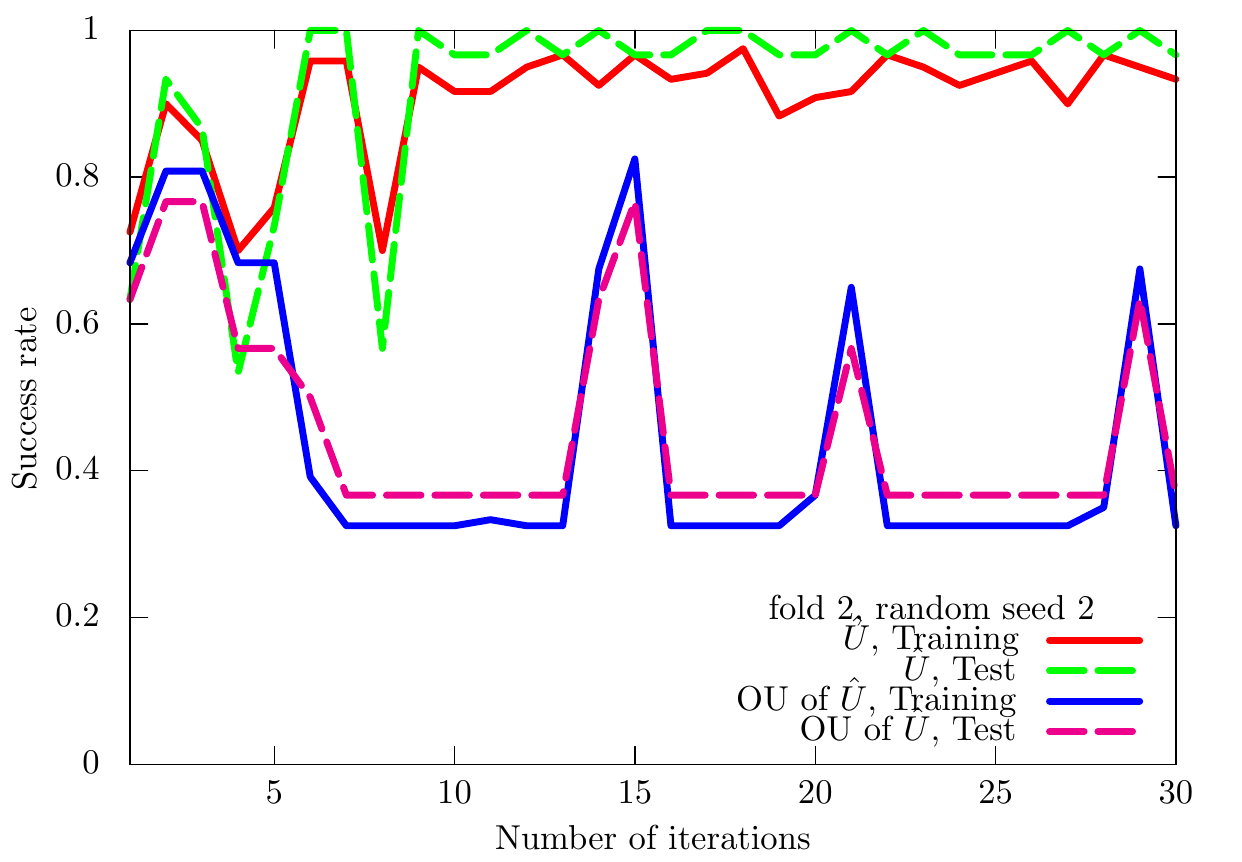}
\includegraphics[scale=0.25]{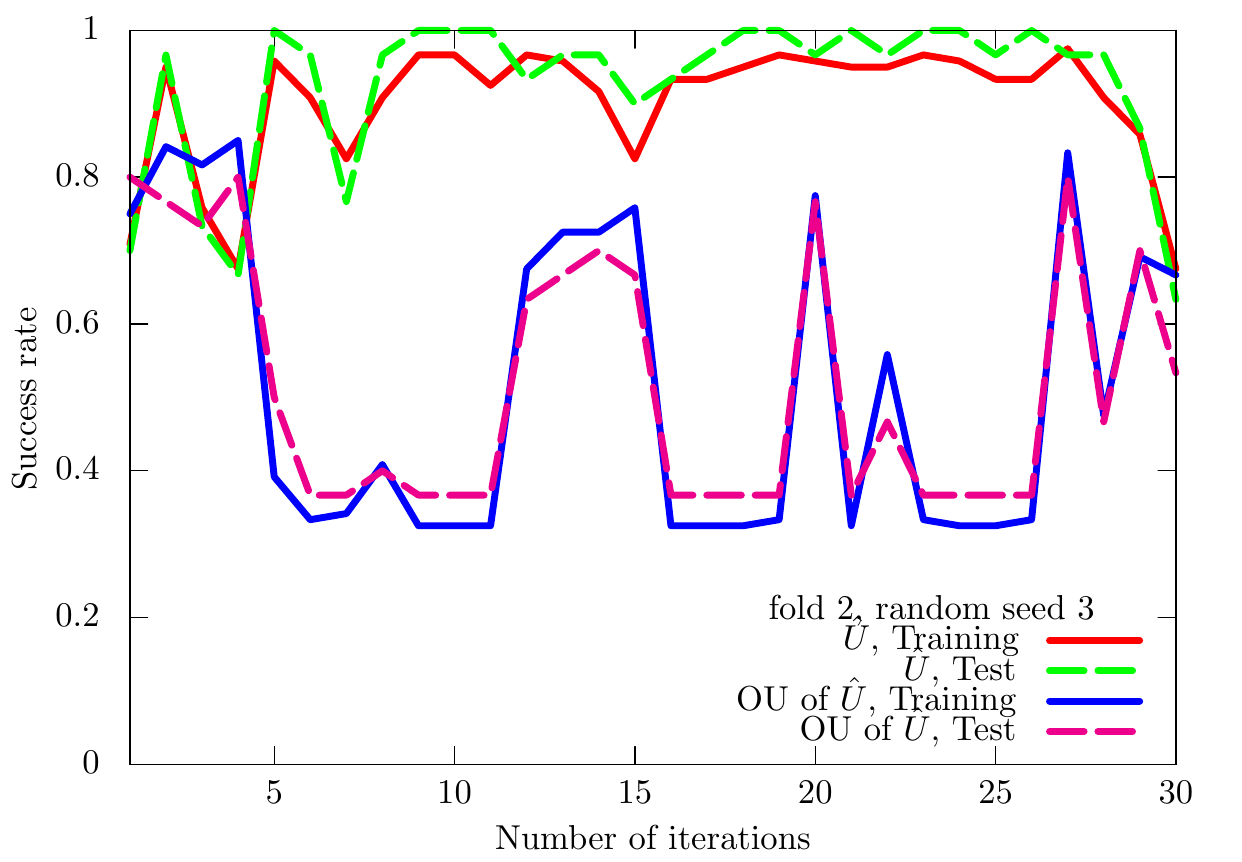}
\includegraphics[scale=0.25]{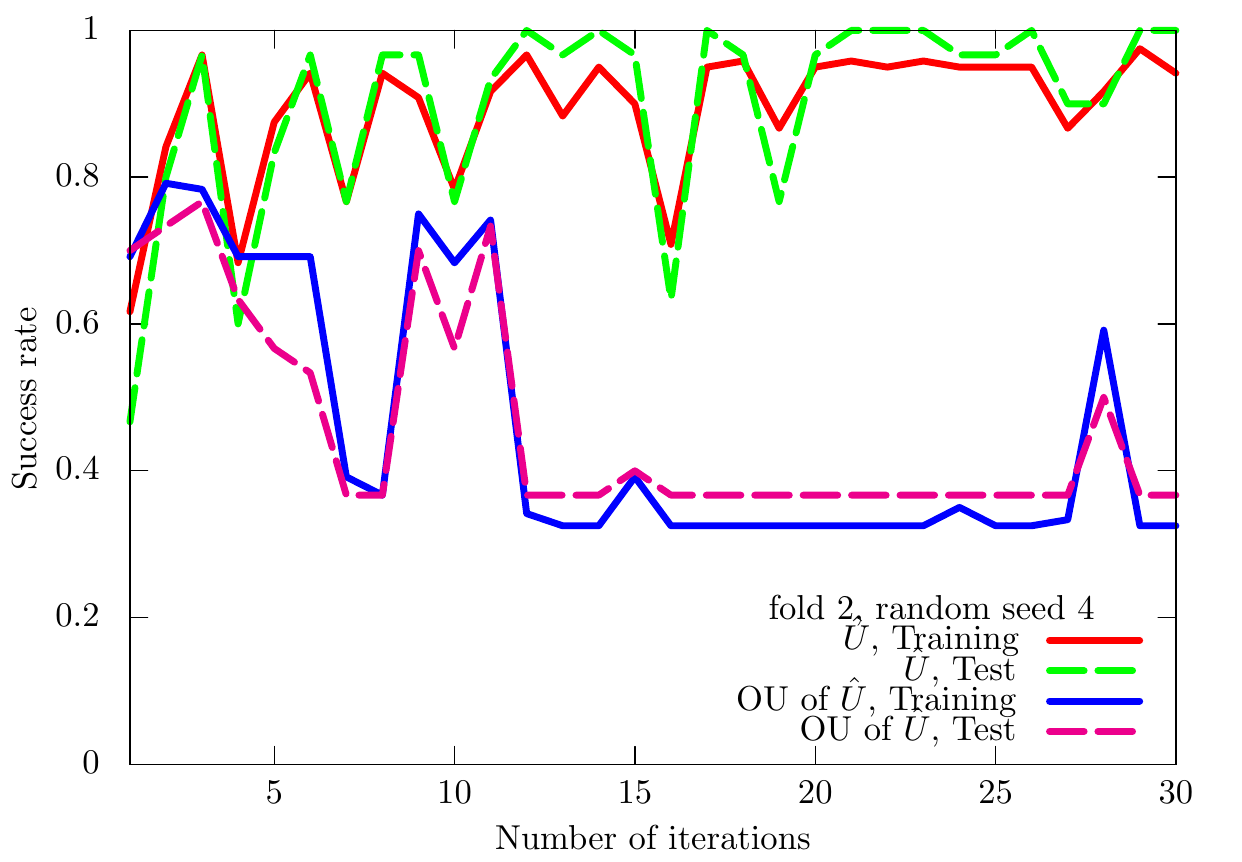}
\includegraphics[scale=0.25]{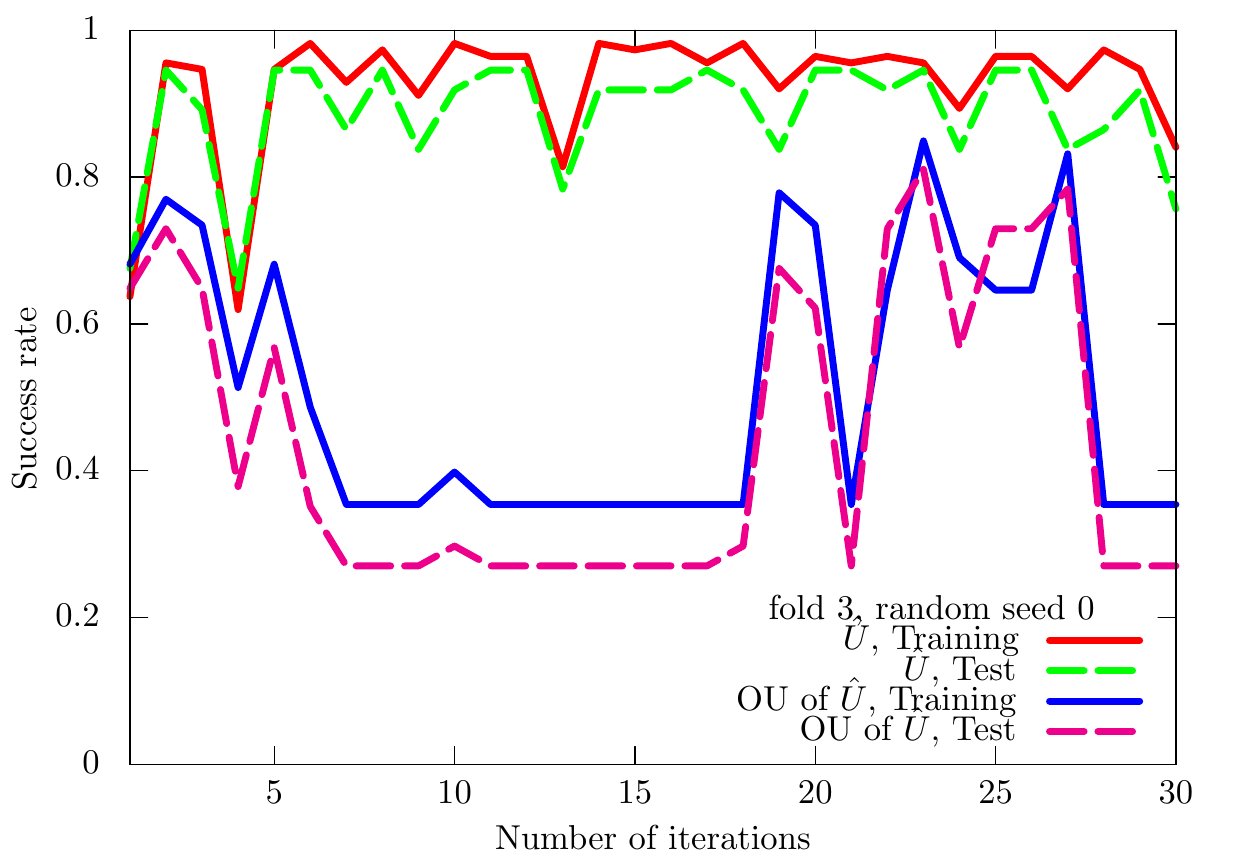}
\includegraphics[scale=0.25]{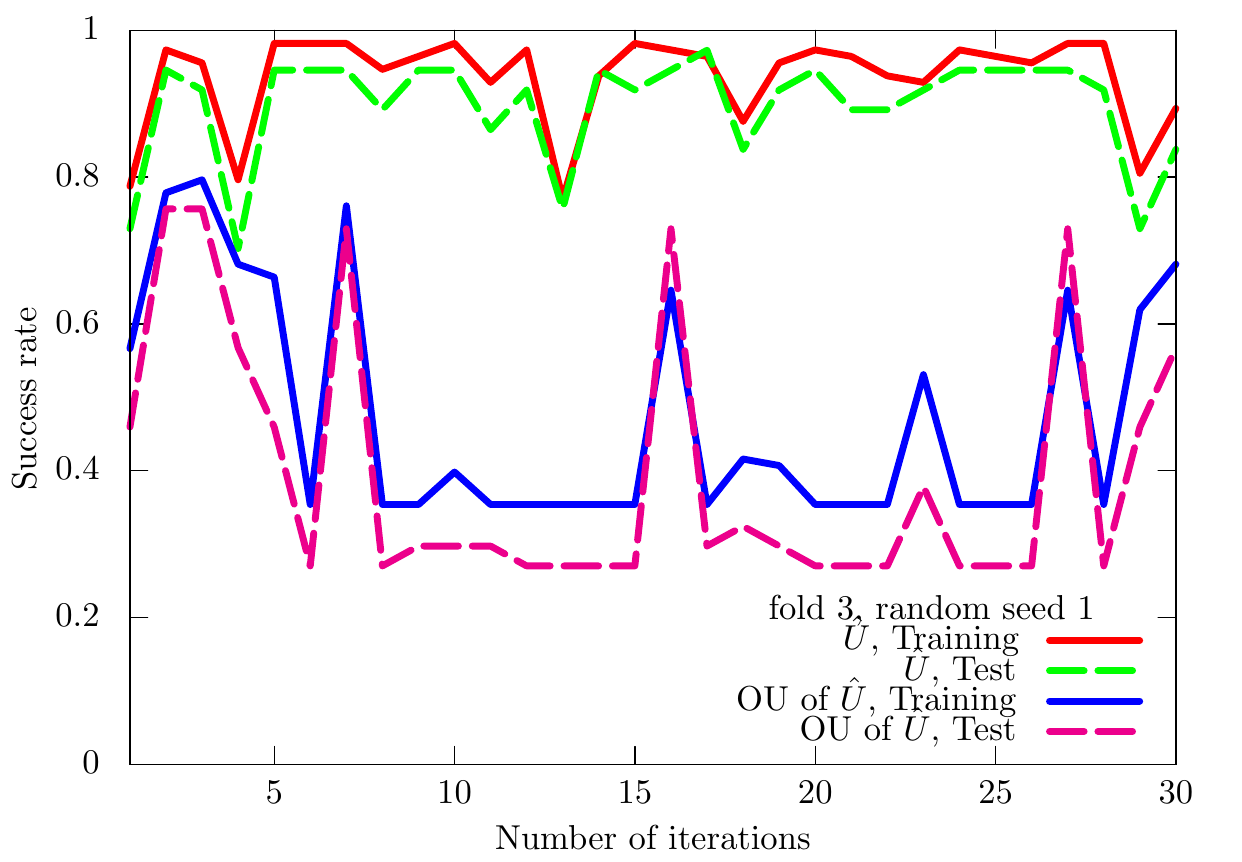}
\includegraphics[scale=0.25]{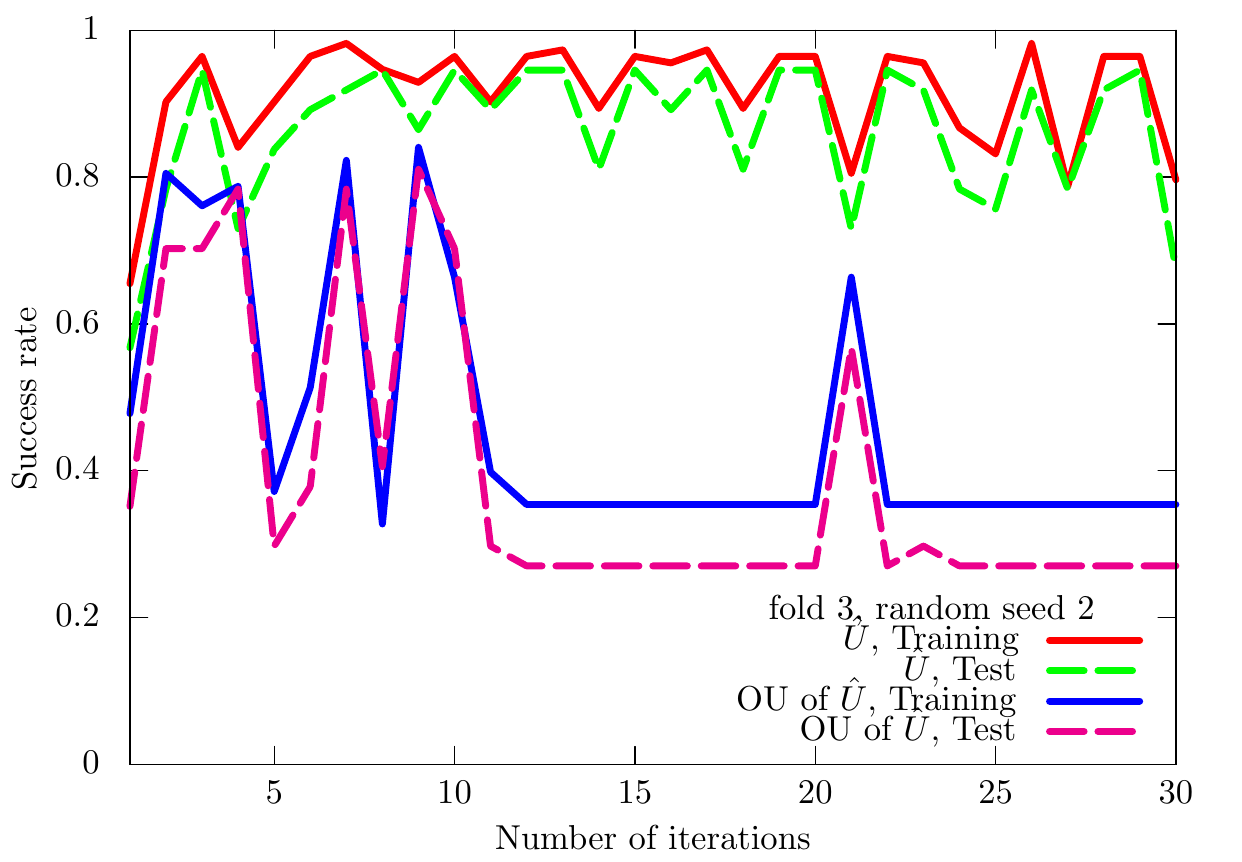}
\includegraphics[scale=0.25]{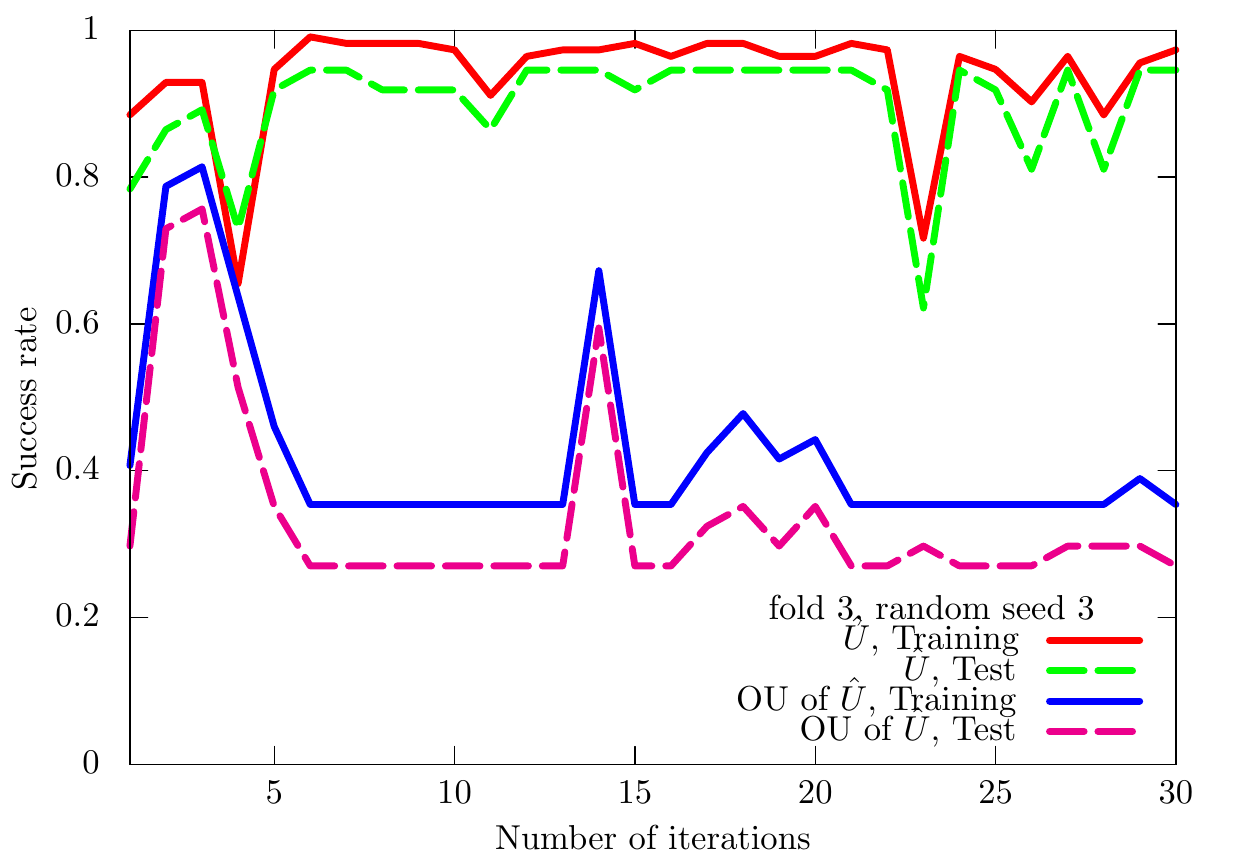}
\includegraphics[scale=0.25]{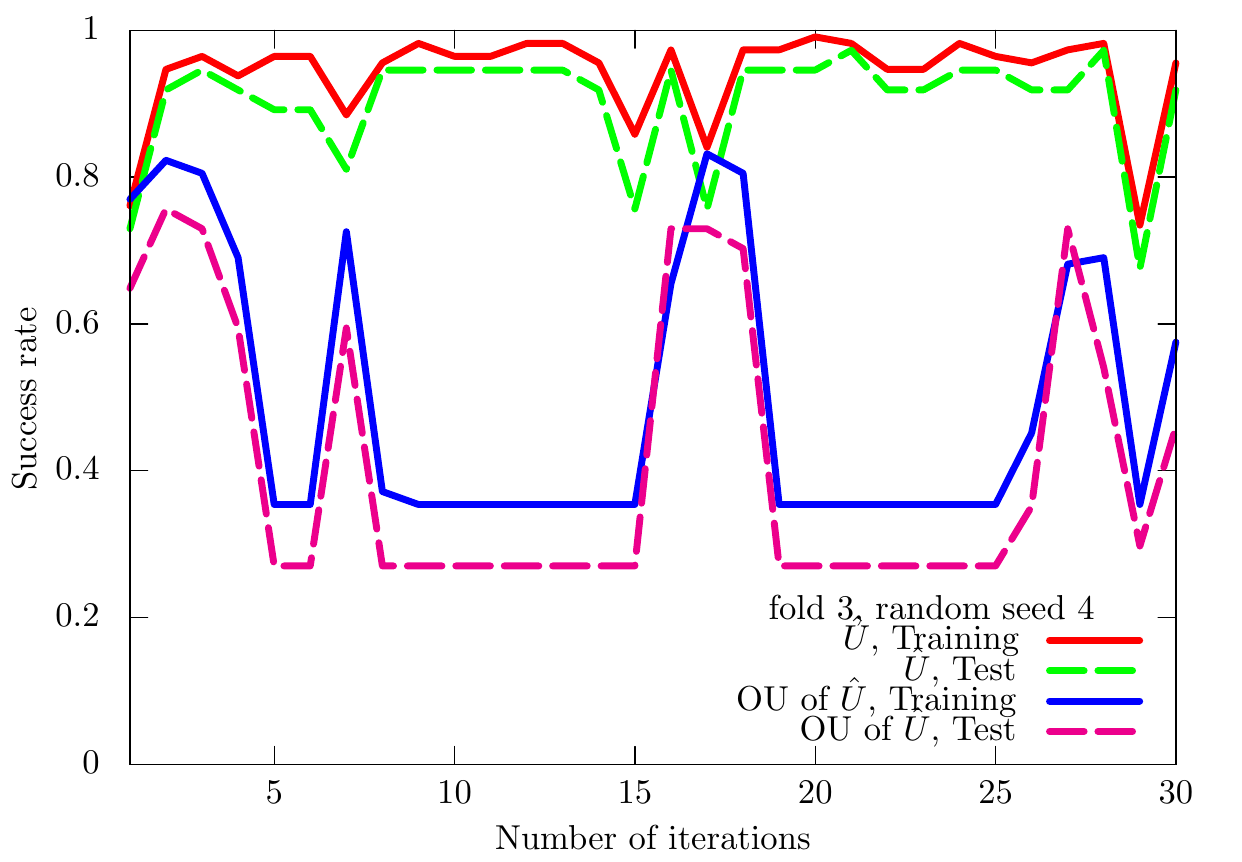}
\includegraphics[scale=0.25]{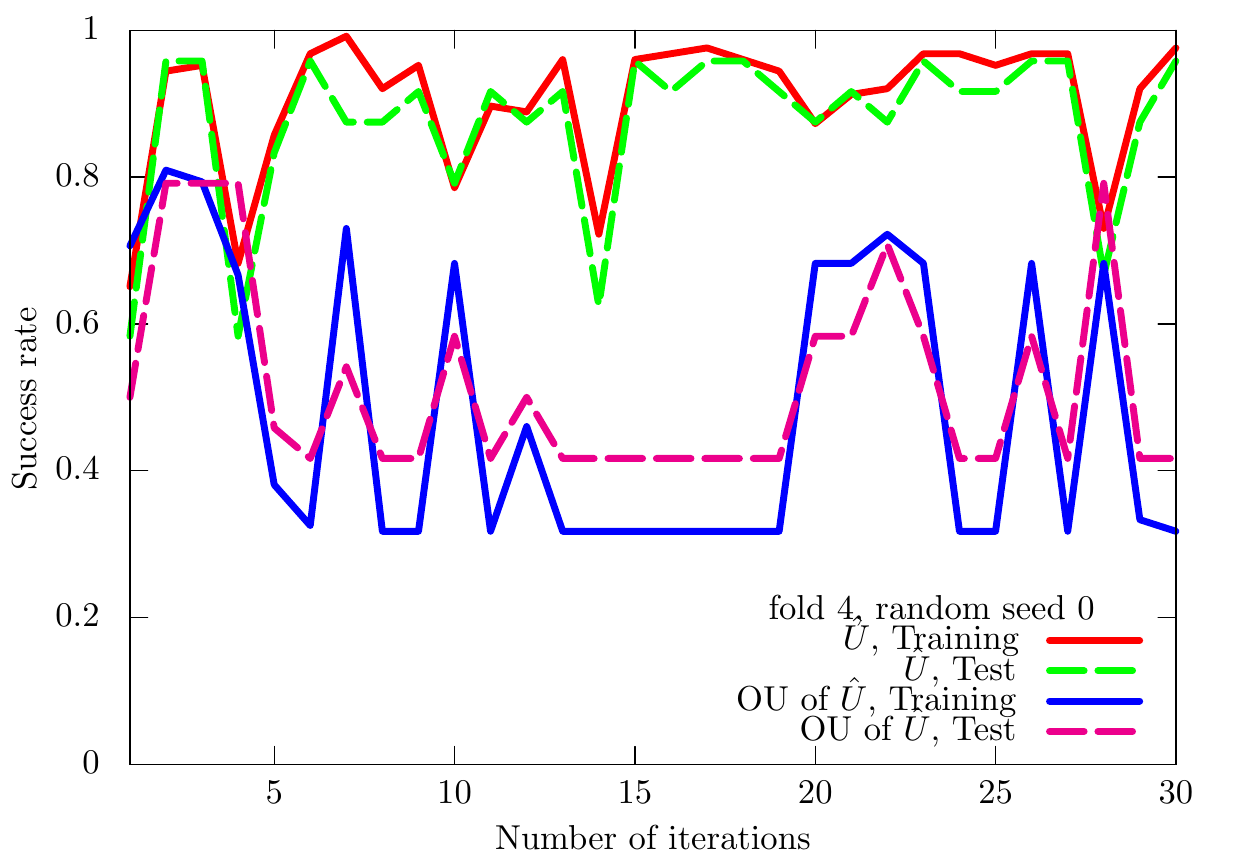}
\includegraphics[scale=0.25]{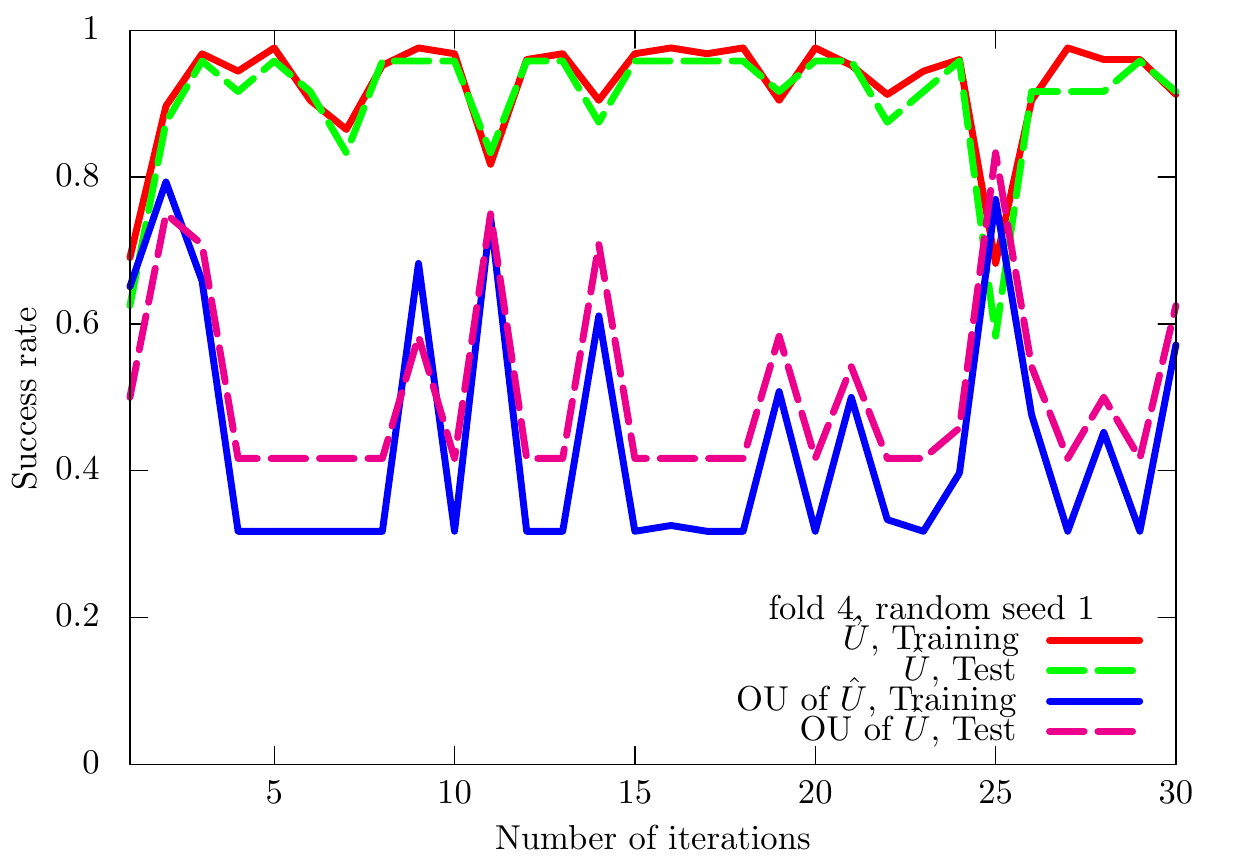}
\includegraphics[scale=0.25]{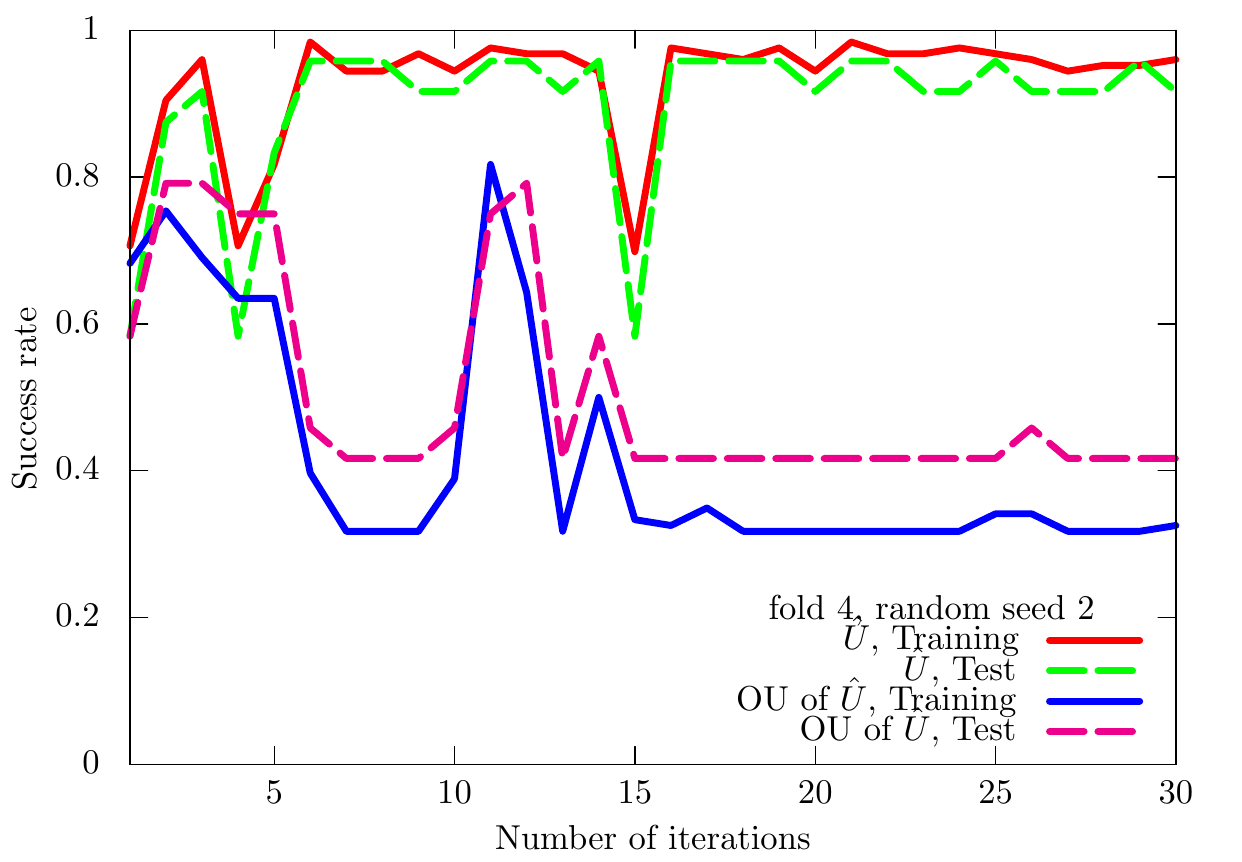}
\includegraphics[scale=0.25]{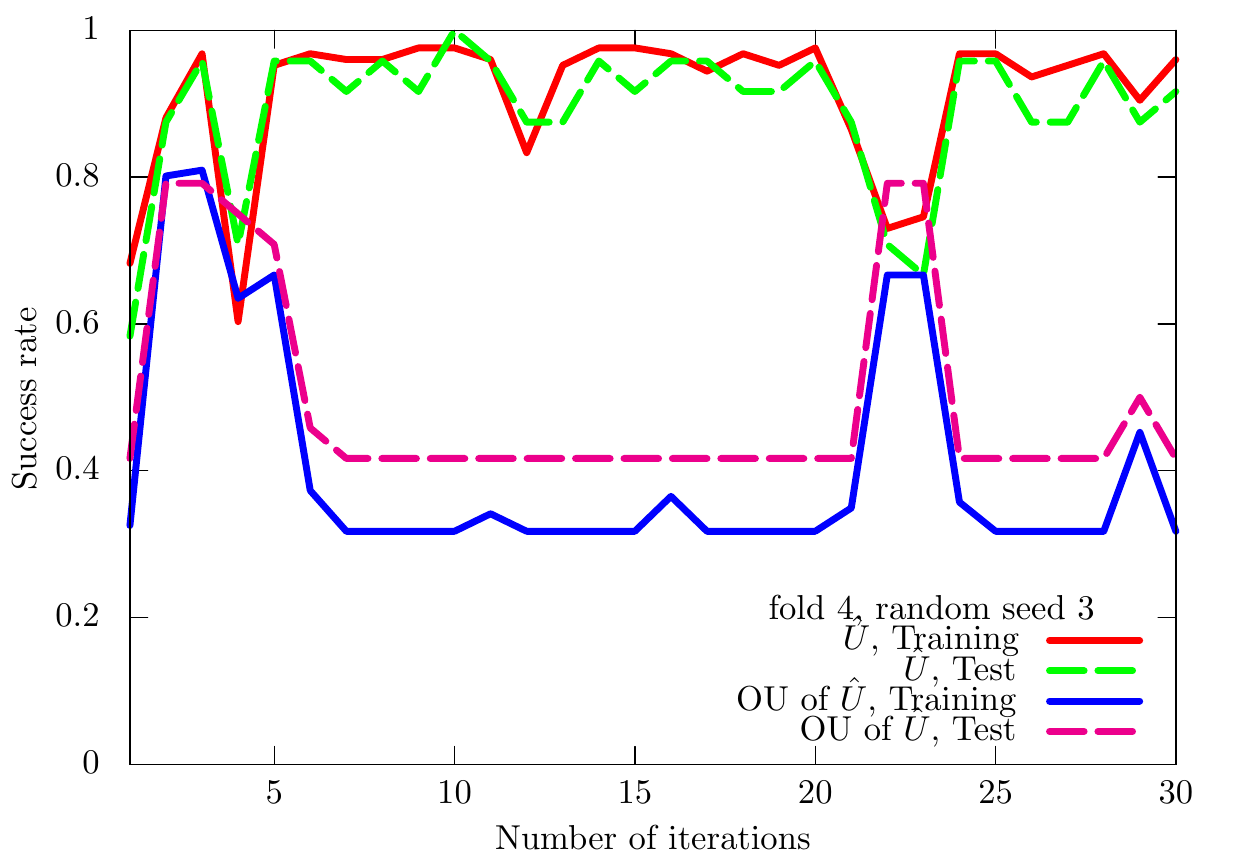}
\includegraphics[scale=0.25]{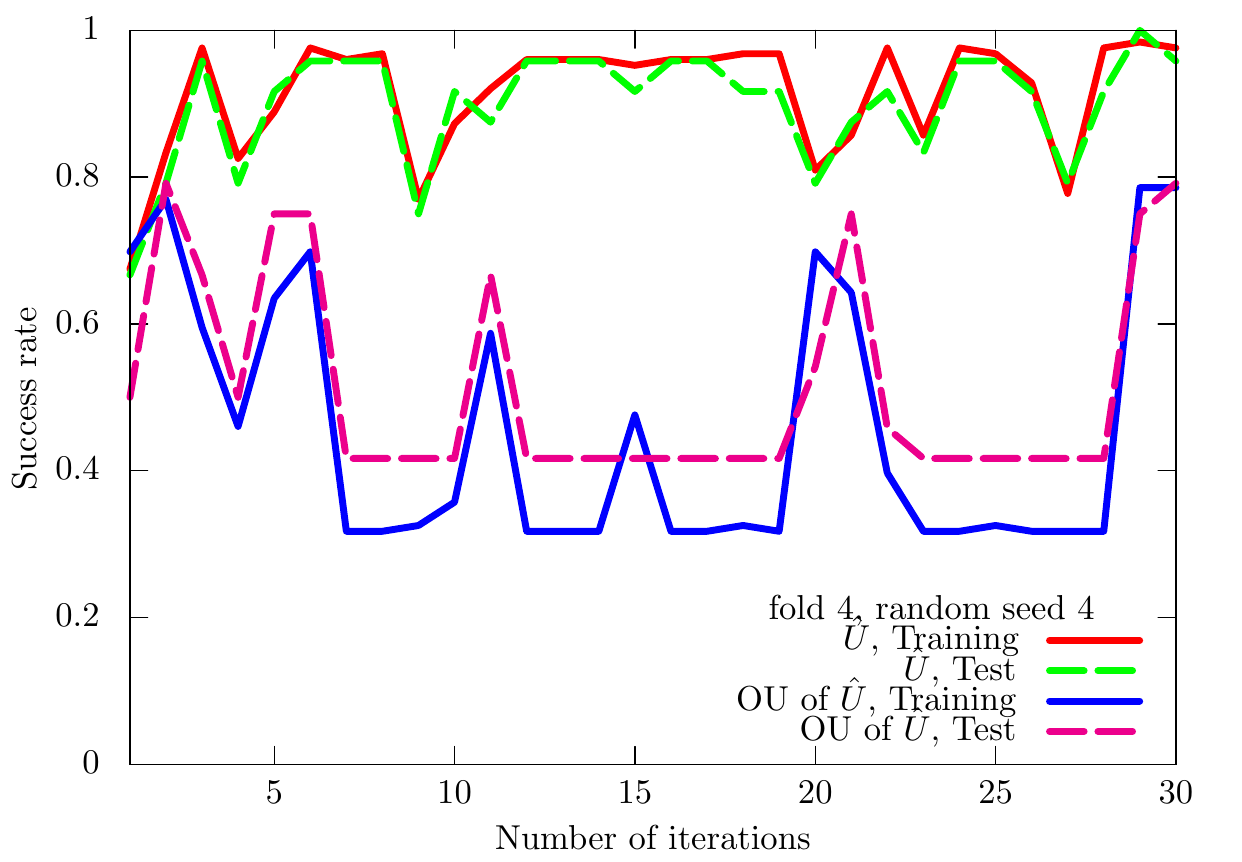}
\caption{Results of the UKM ($\hat{X}$ and $\hat{P}$) on the $5$-fold datasets with $5$ different random seeds for the iris dataset ($1$ or non-$1$). We use complex matrices and set $\theta_\mathrm{bias} = 0$. We set $r = 0.010$.}
\label{supp-arXiv-numerical-result-raw-data-fold-001-rand-001-UKM-P-UCI-iris-1-non1}
\end{figure*}

\begin{figure*}[htb]
\centering
\includegraphics[scale=0.25]{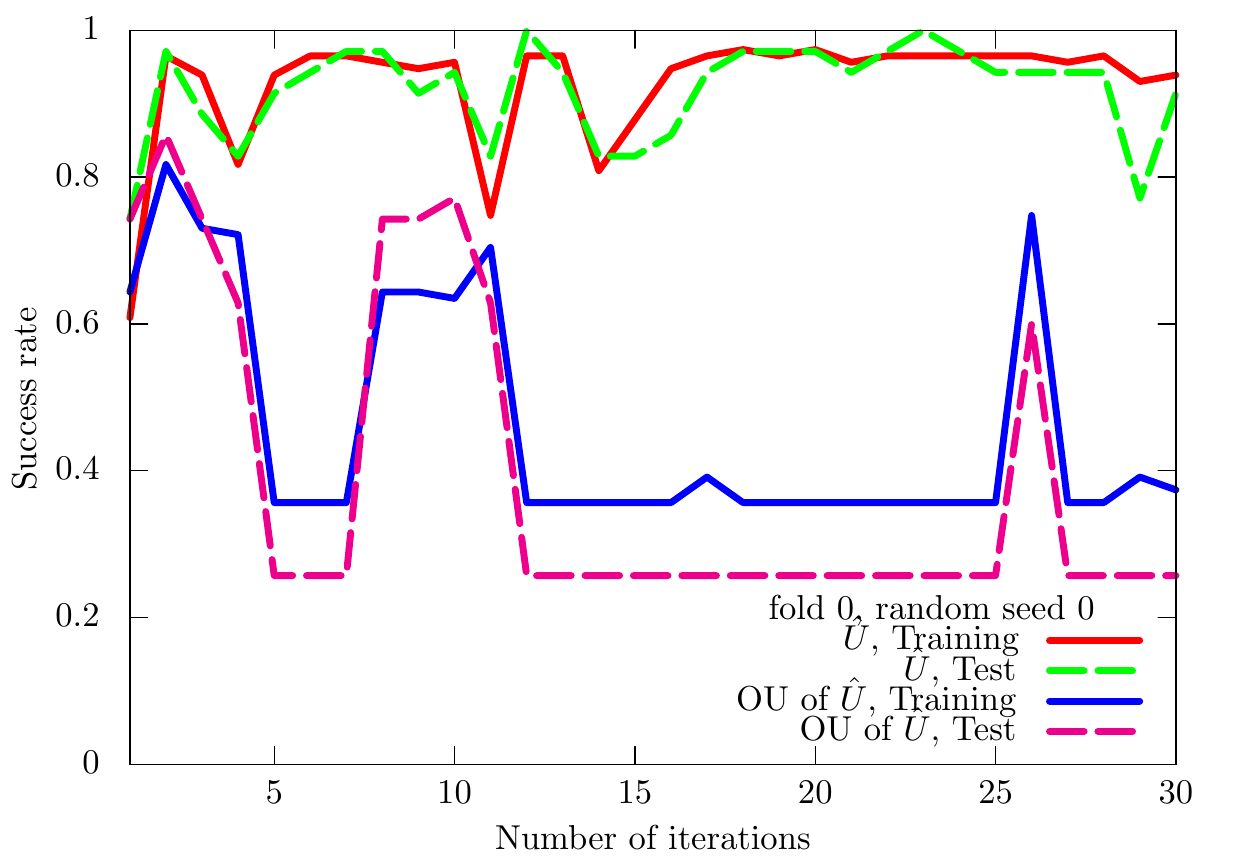}
\includegraphics[scale=0.25]{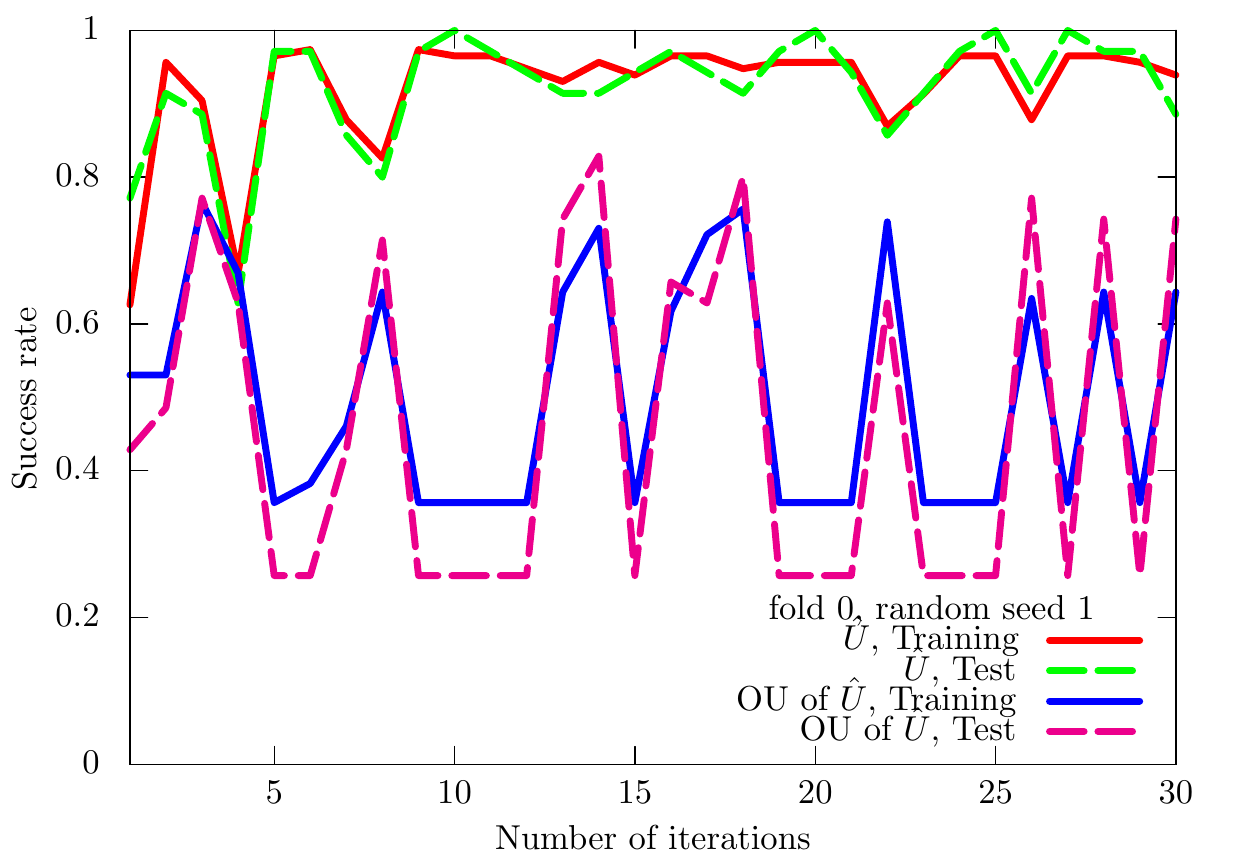}
\includegraphics[scale=0.25]{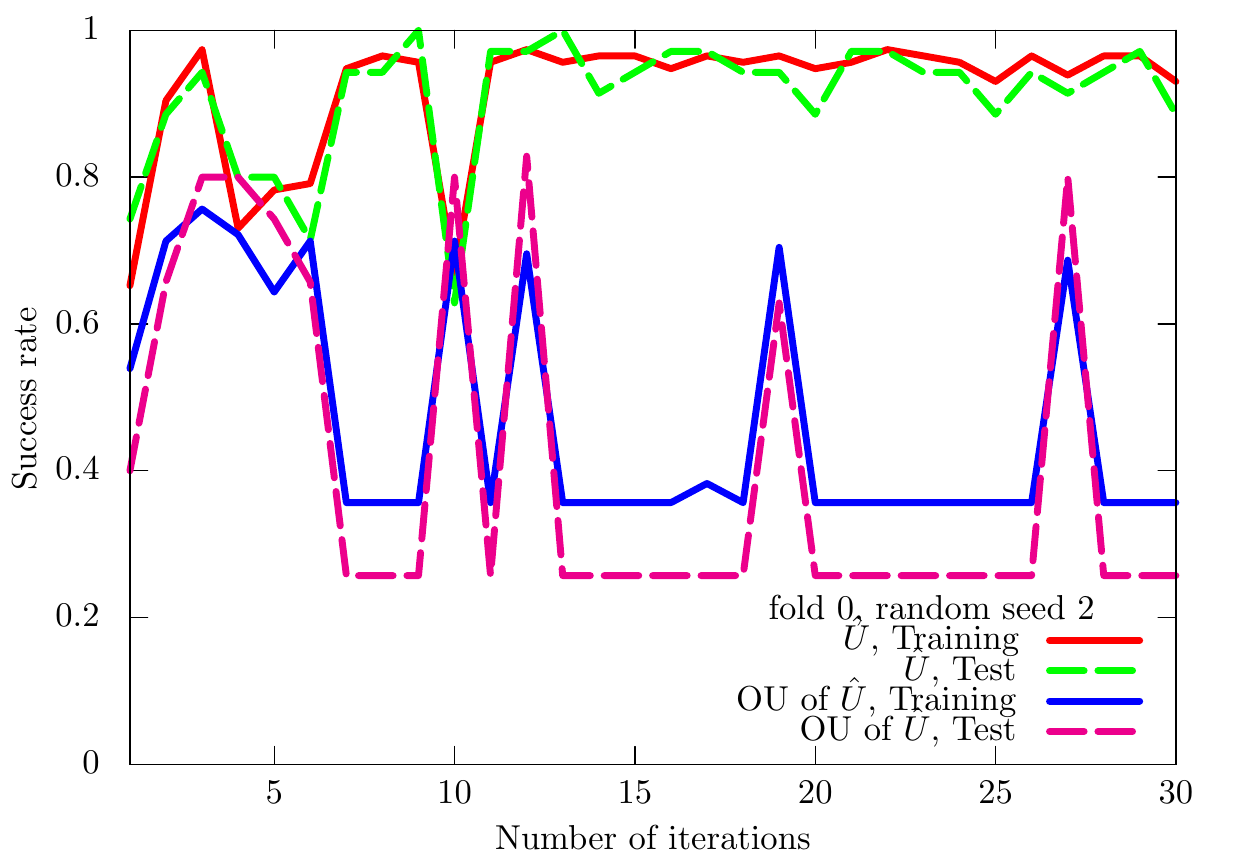}
\includegraphics[scale=0.25]{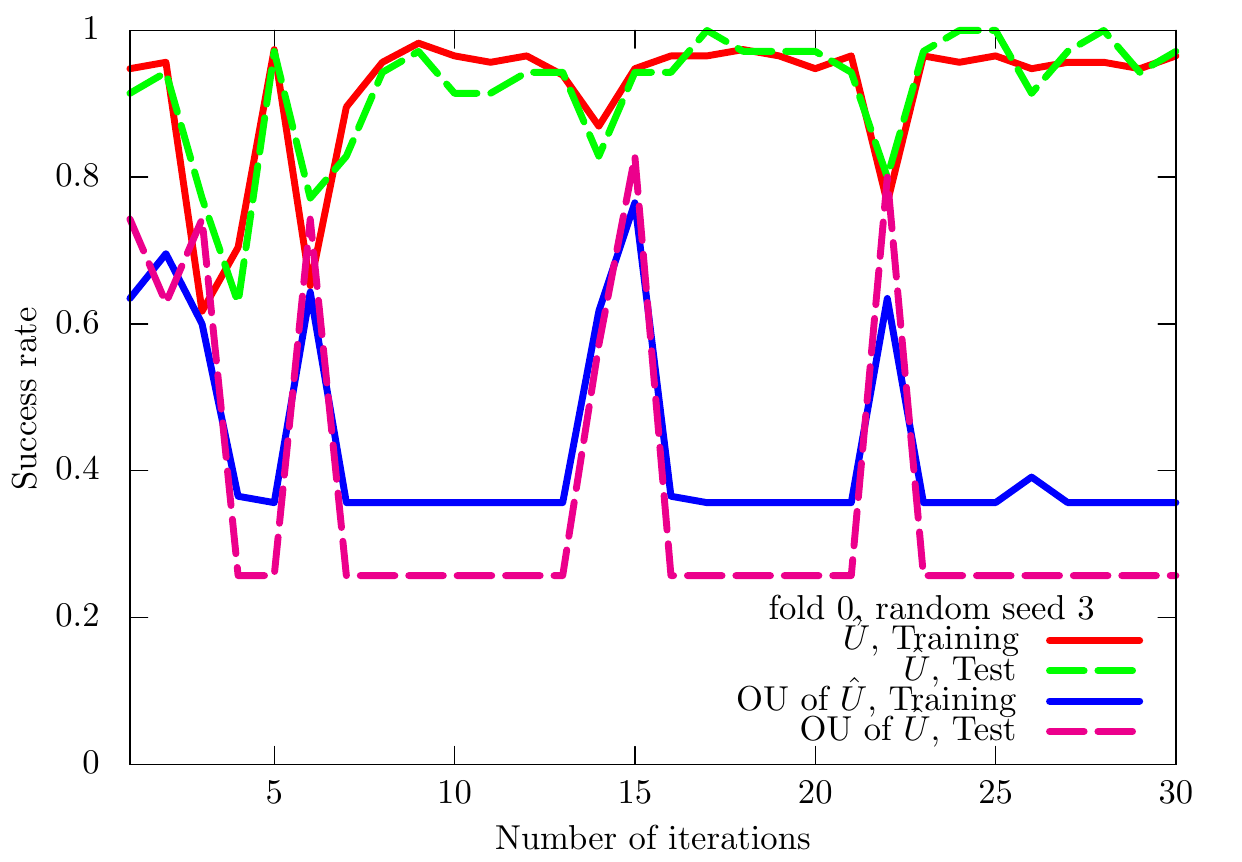}
\includegraphics[scale=0.25]{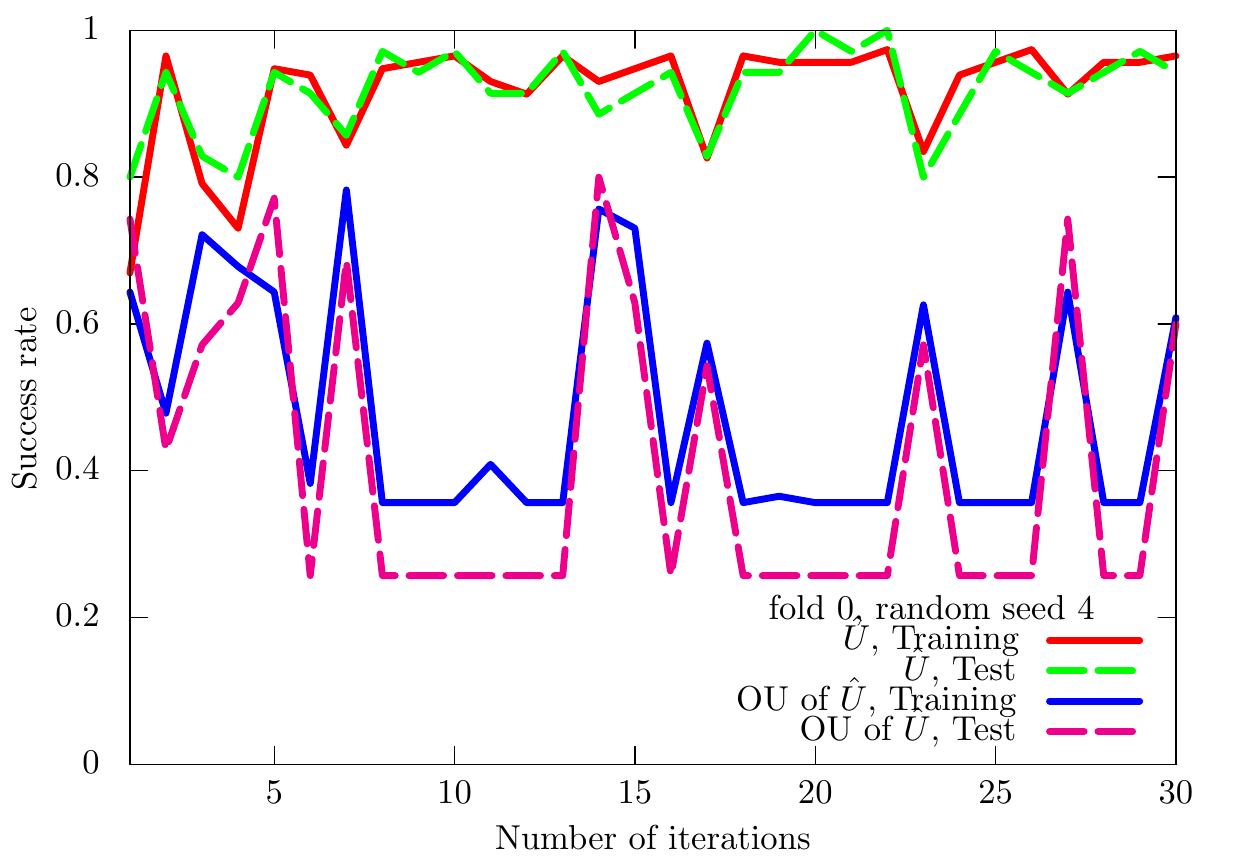}
\includegraphics[scale=0.25]{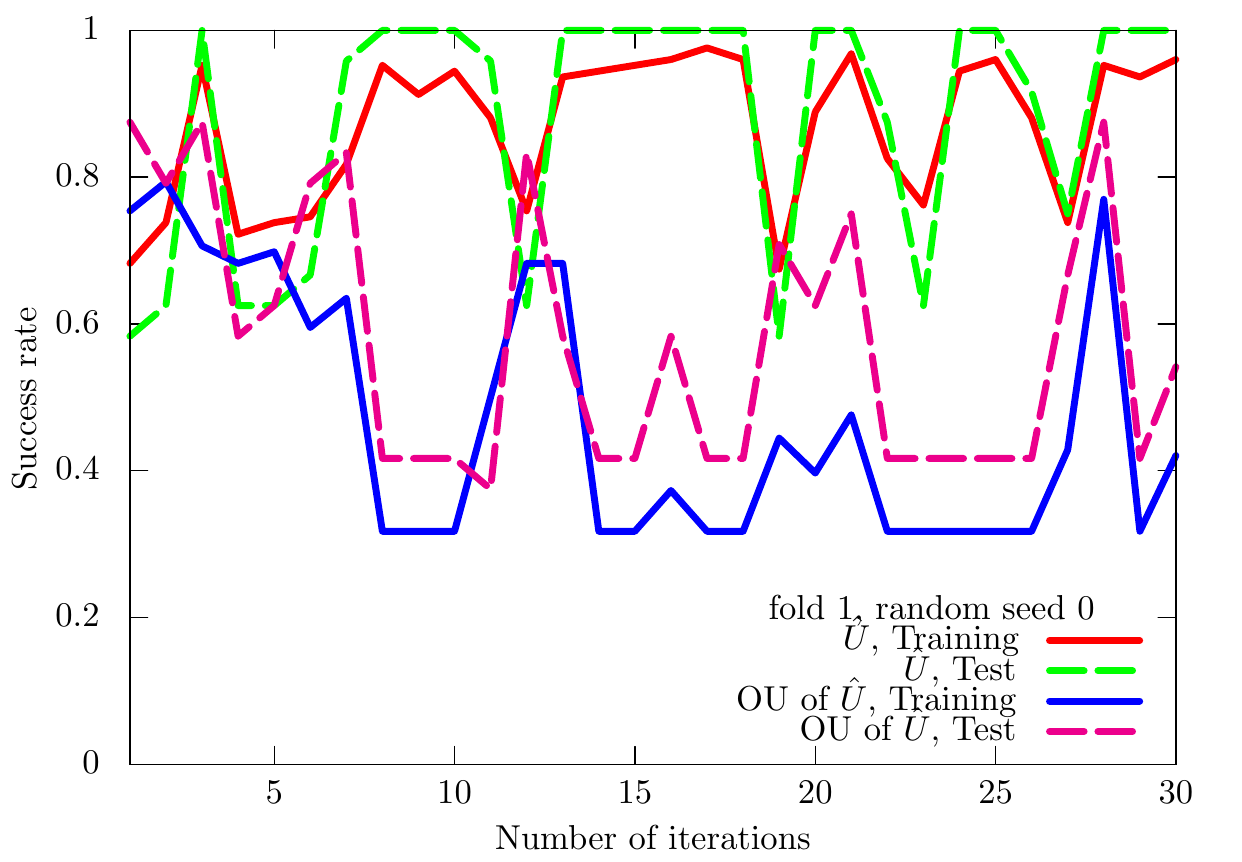}
\includegraphics[scale=0.25]{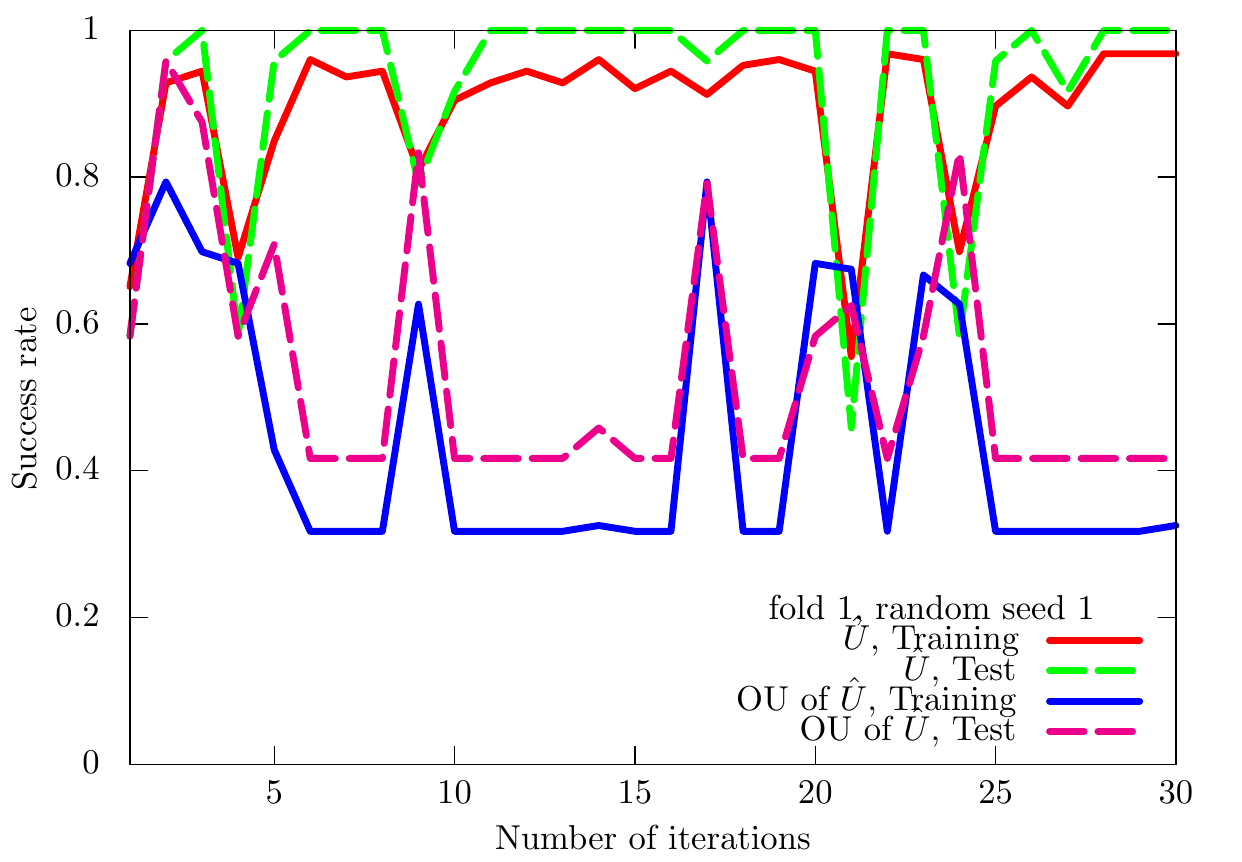}
\includegraphics[scale=0.25]{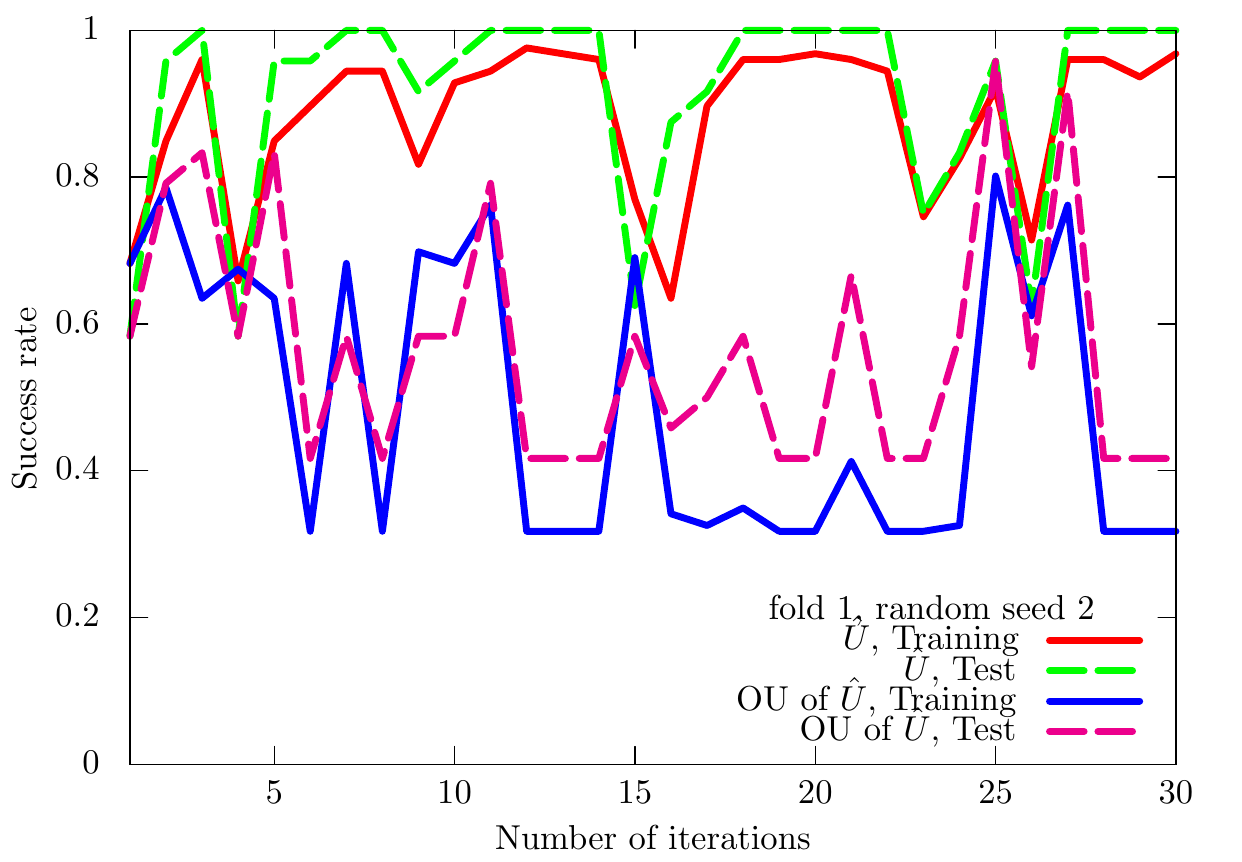}
\includegraphics[scale=0.25]{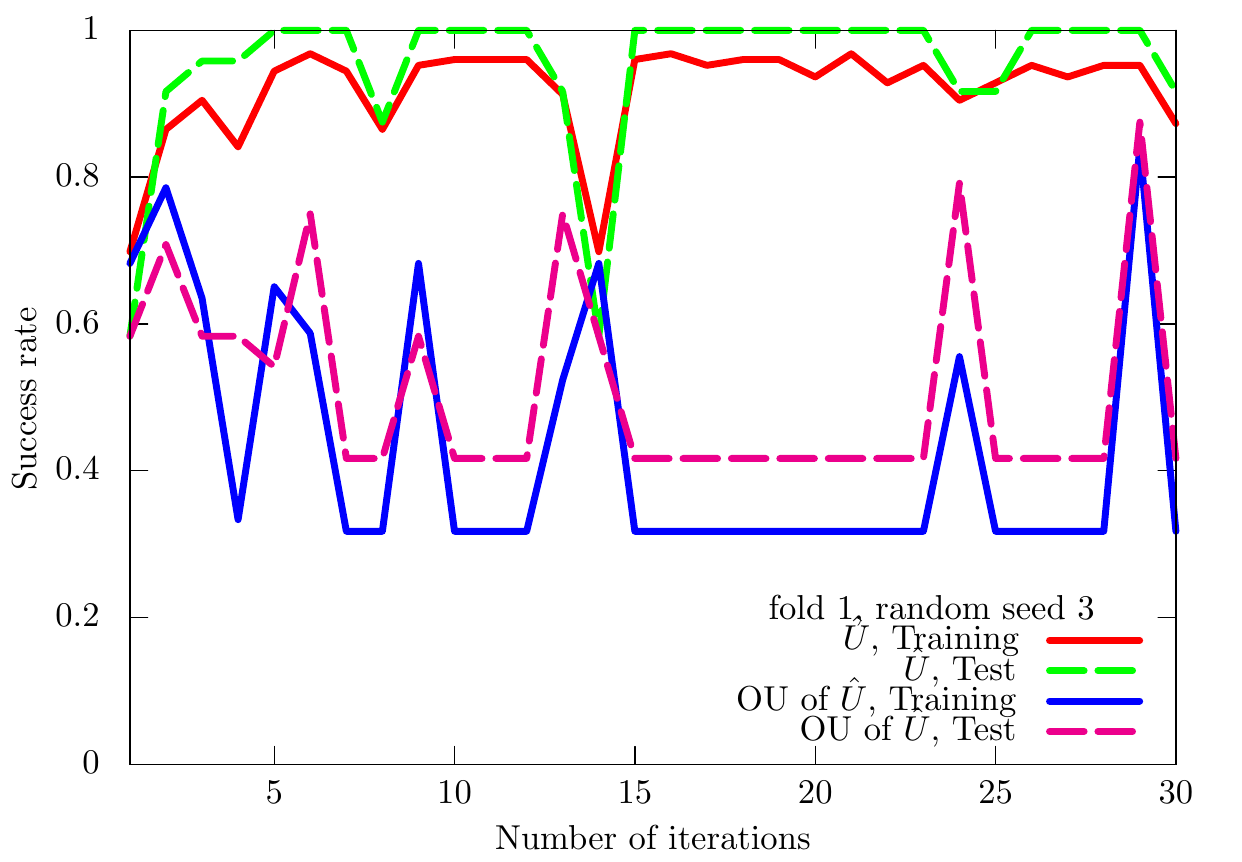}
\includegraphics[scale=0.25]{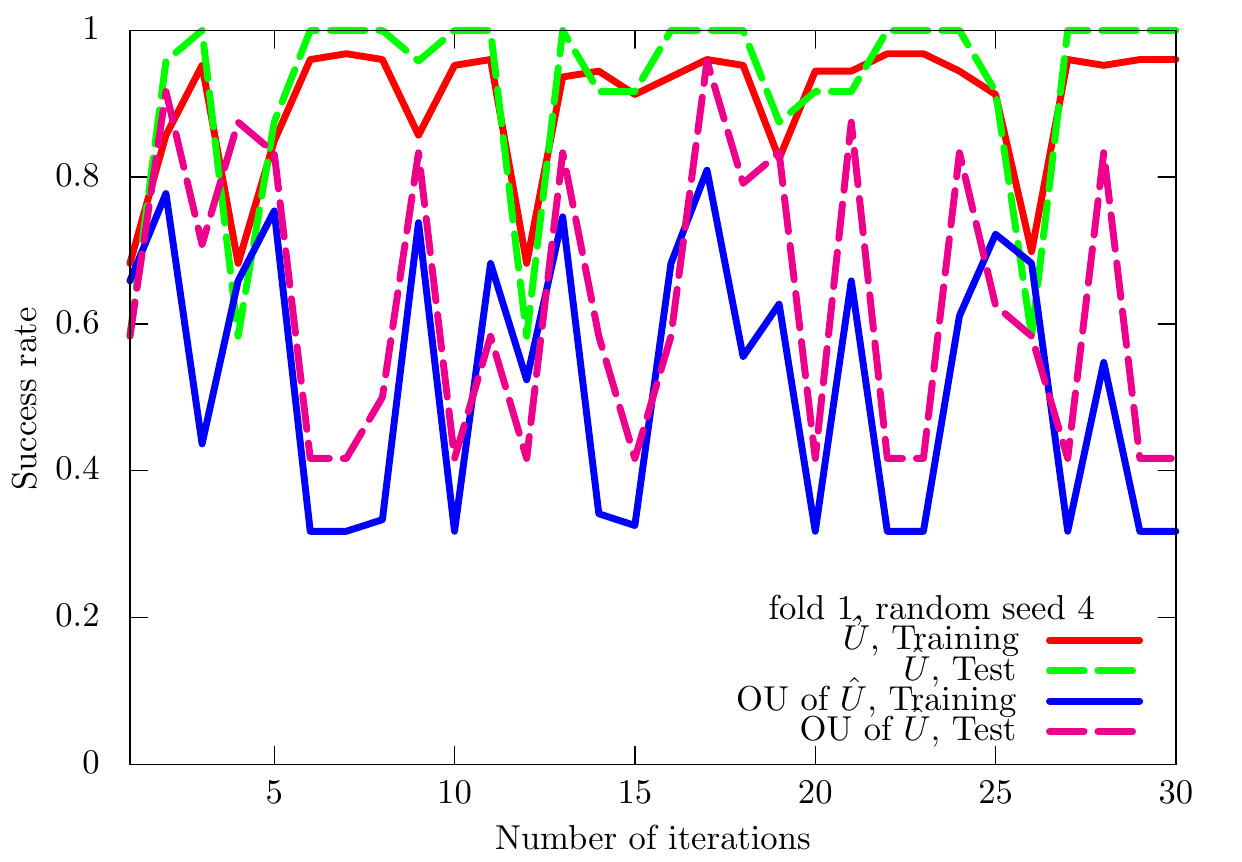}
\includegraphics[scale=0.25]{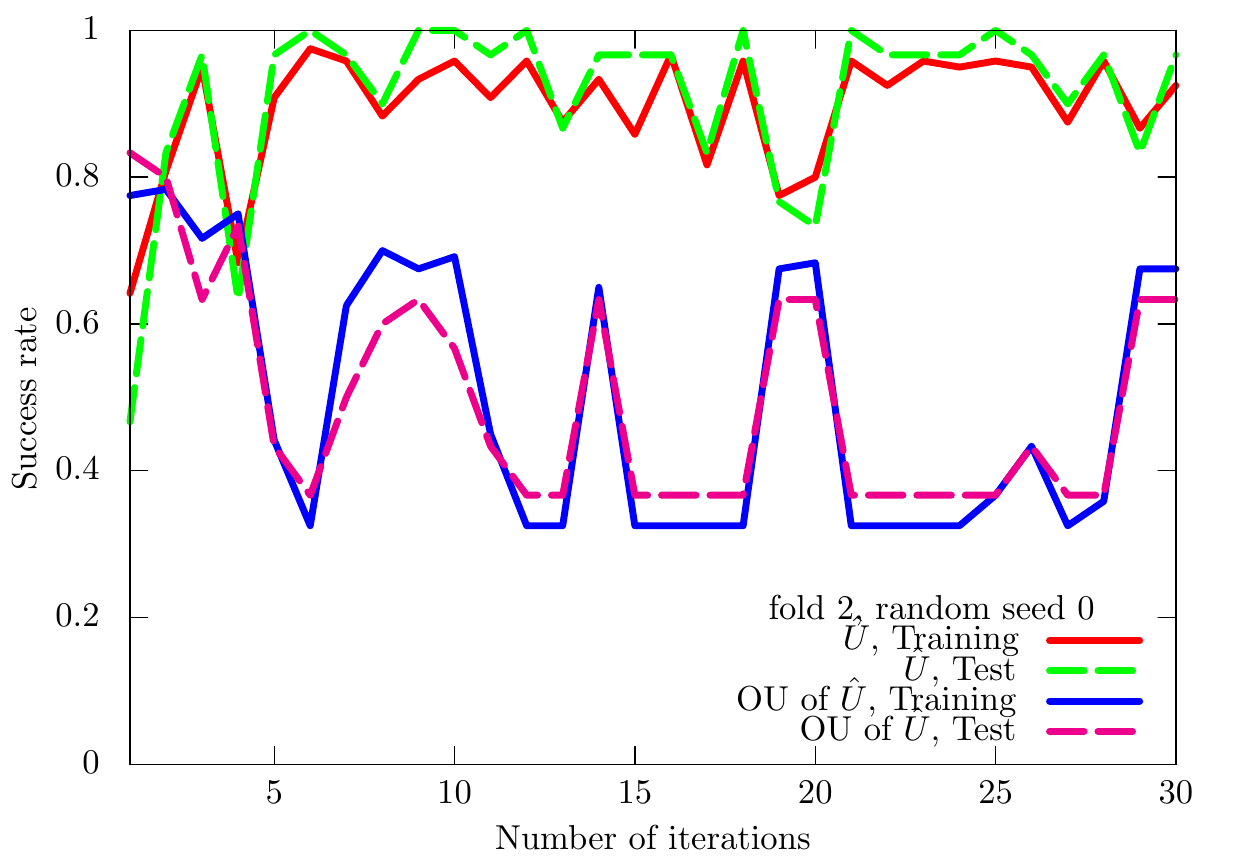}
\includegraphics[scale=0.25]{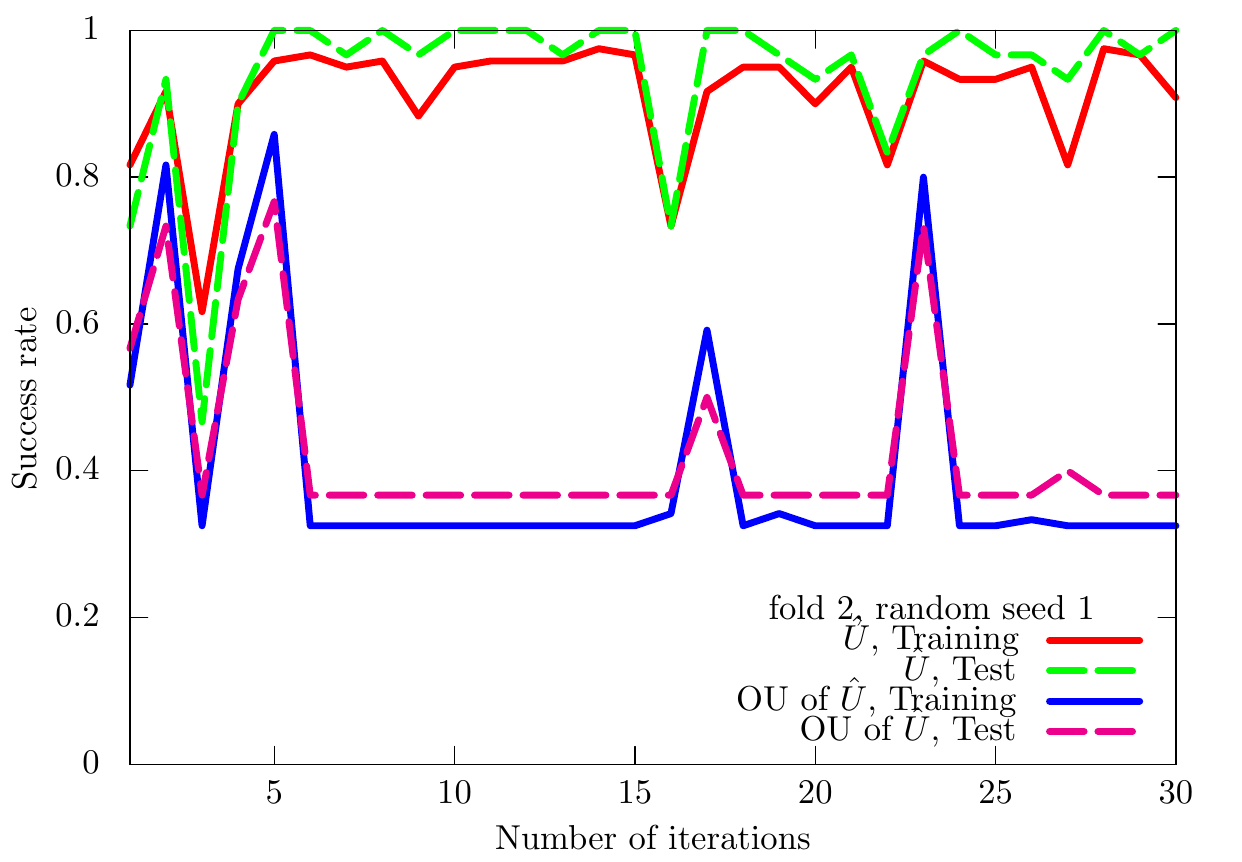}
\includegraphics[scale=0.25]{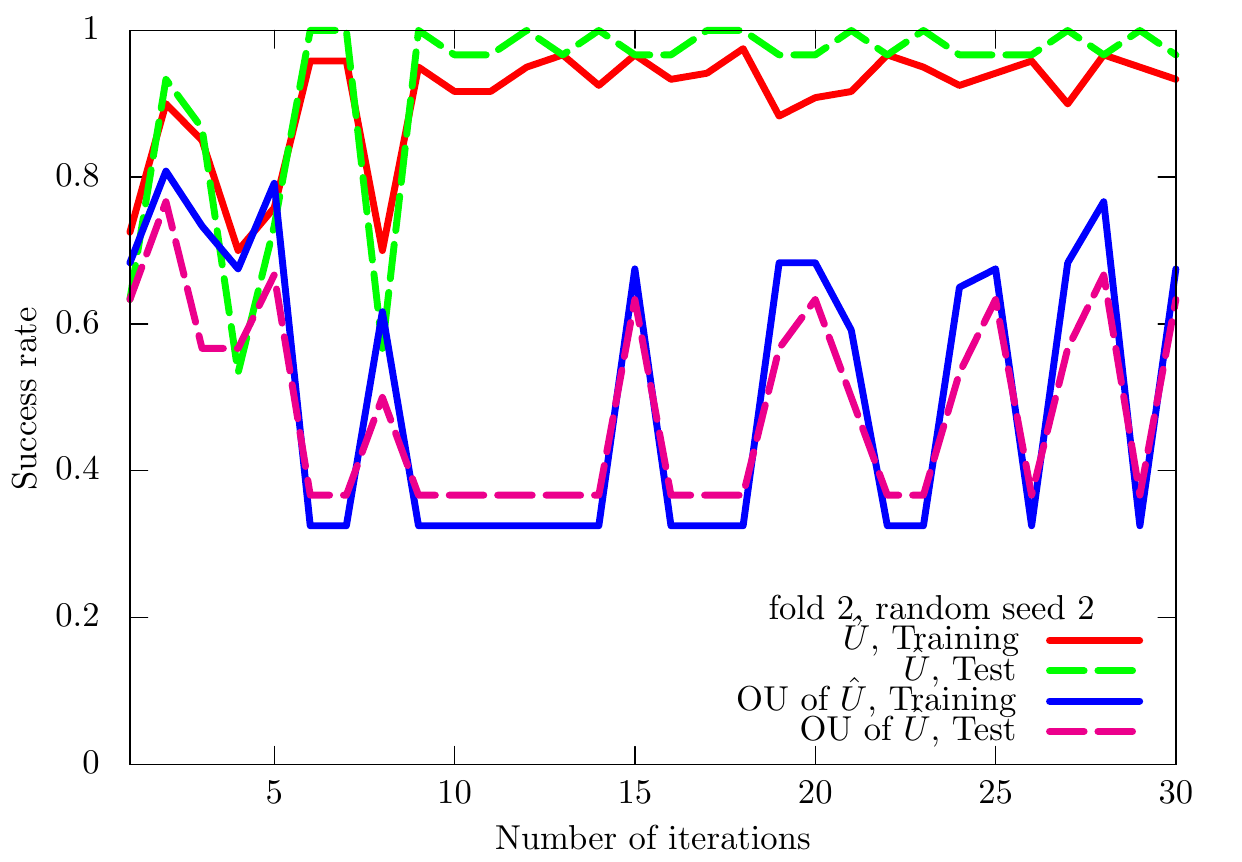}
\includegraphics[scale=0.25]{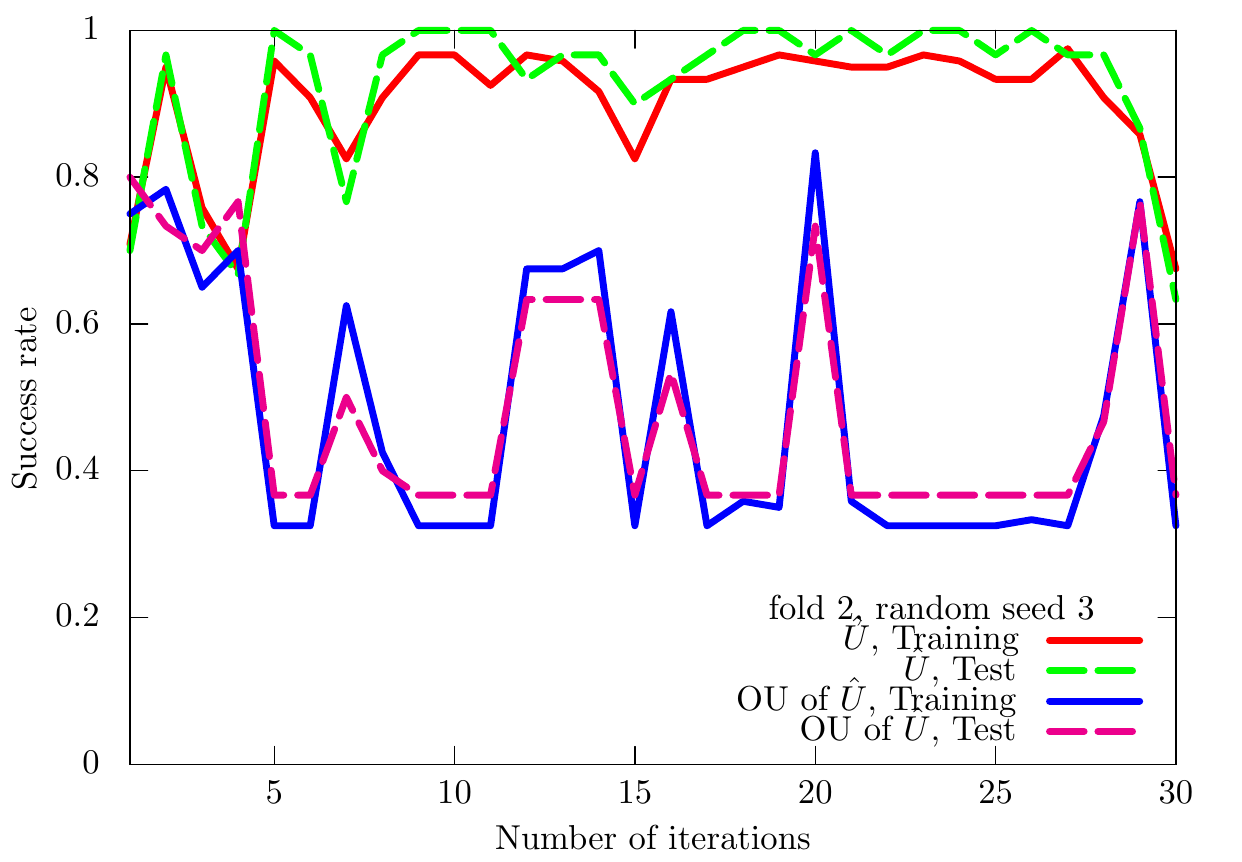}
\includegraphics[scale=0.25]{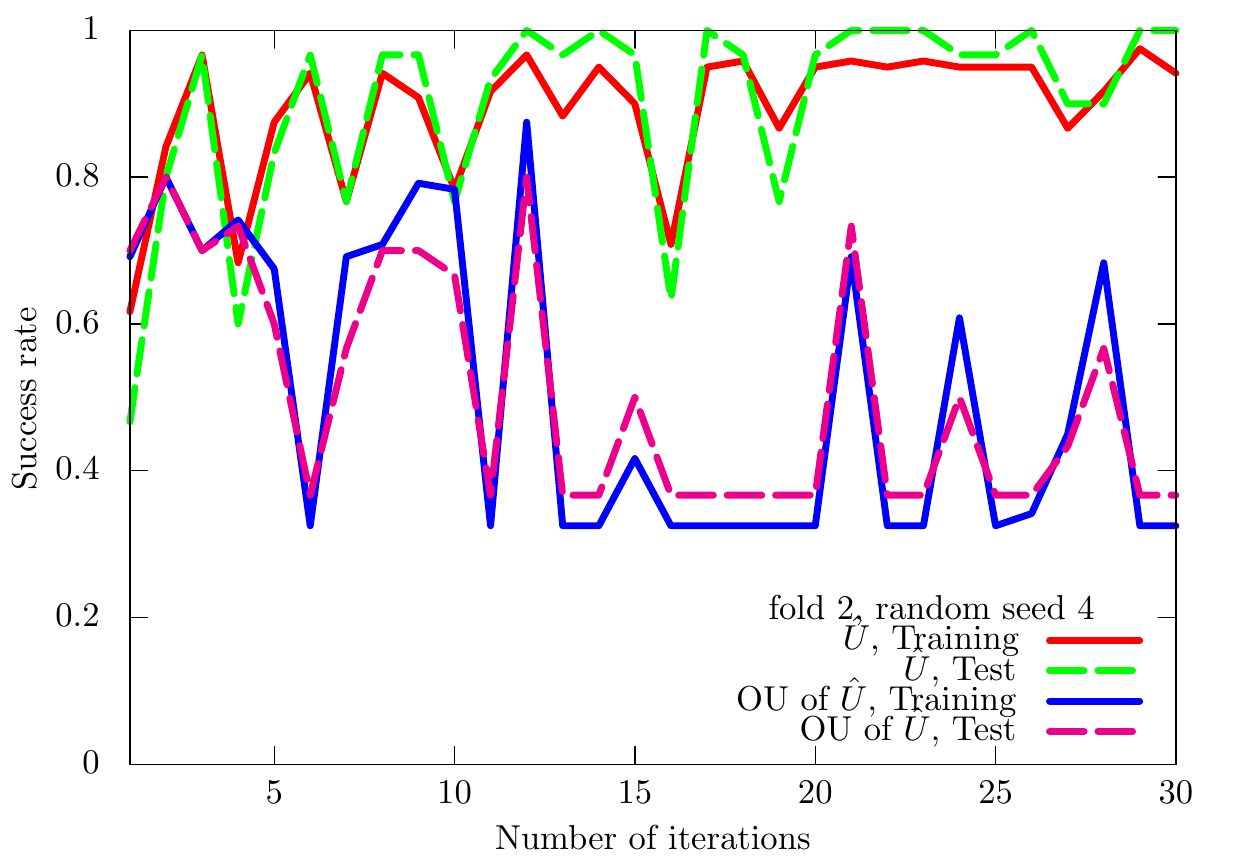}
\includegraphics[scale=0.25]{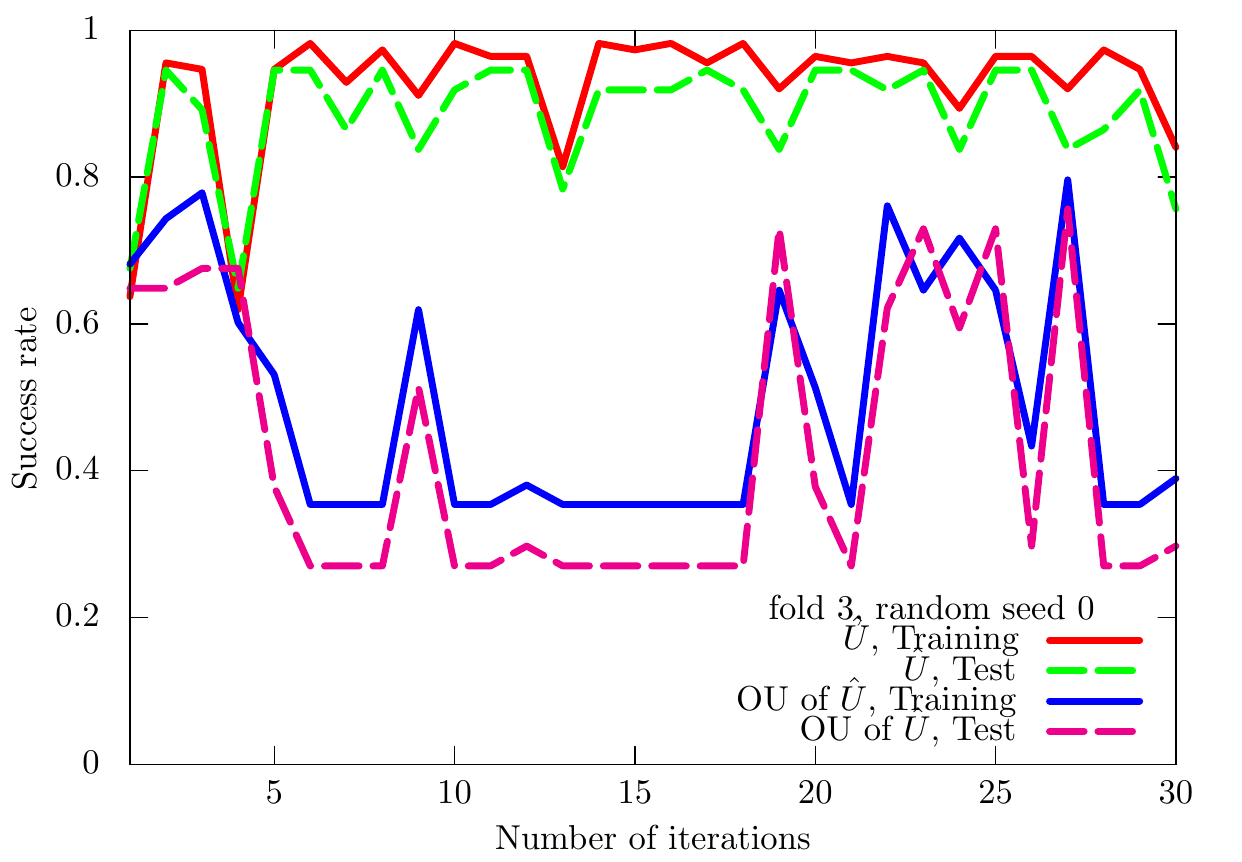}
\includegraphics[scale=0.25]{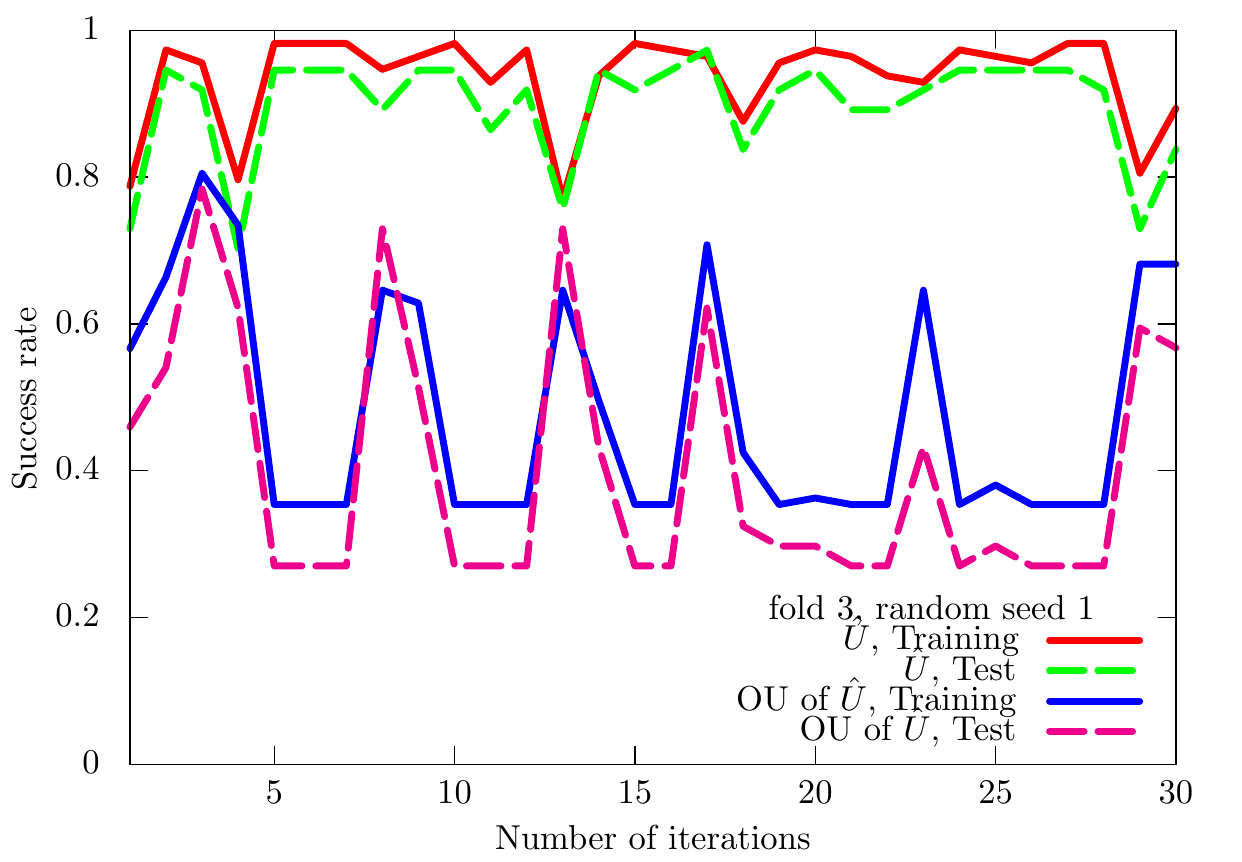}
\includegraphics[scale=0.25]{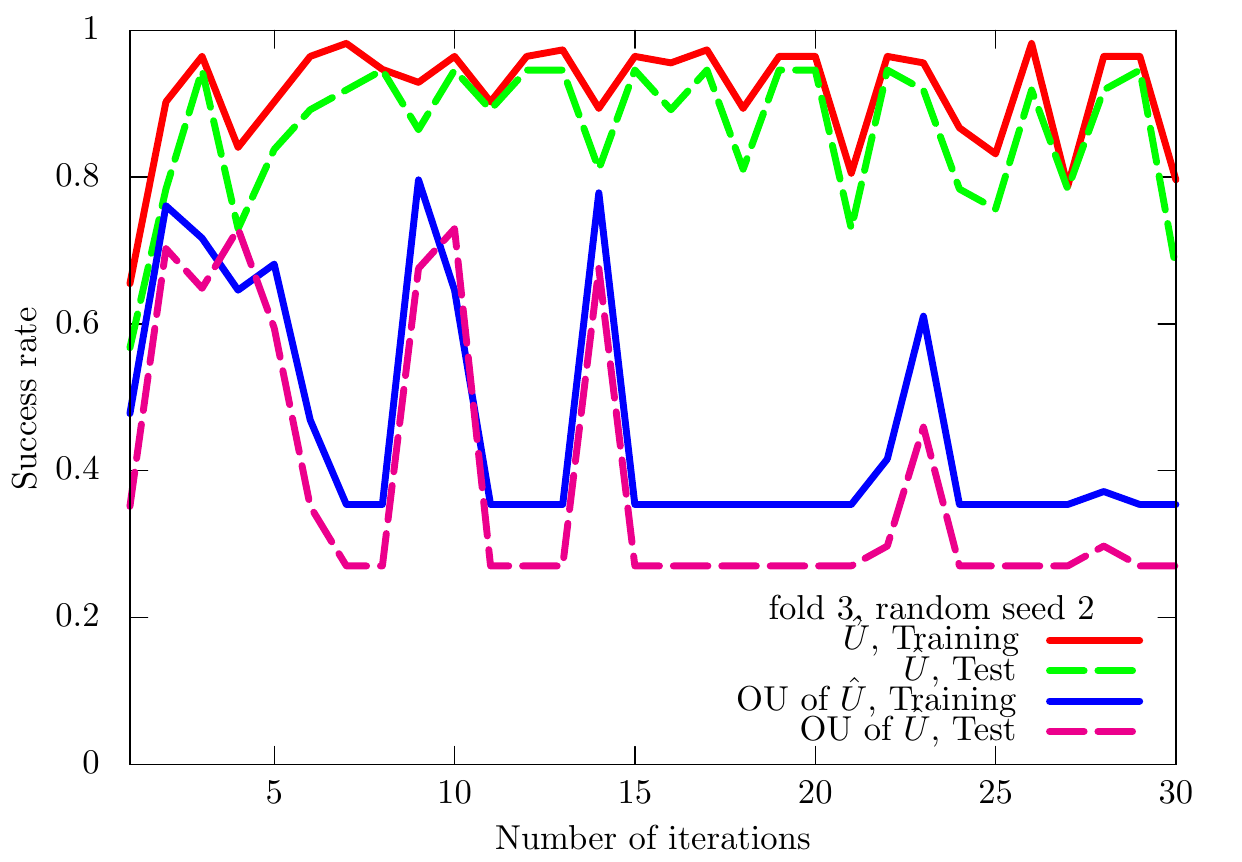}
\includegraphics[scale=0.25]{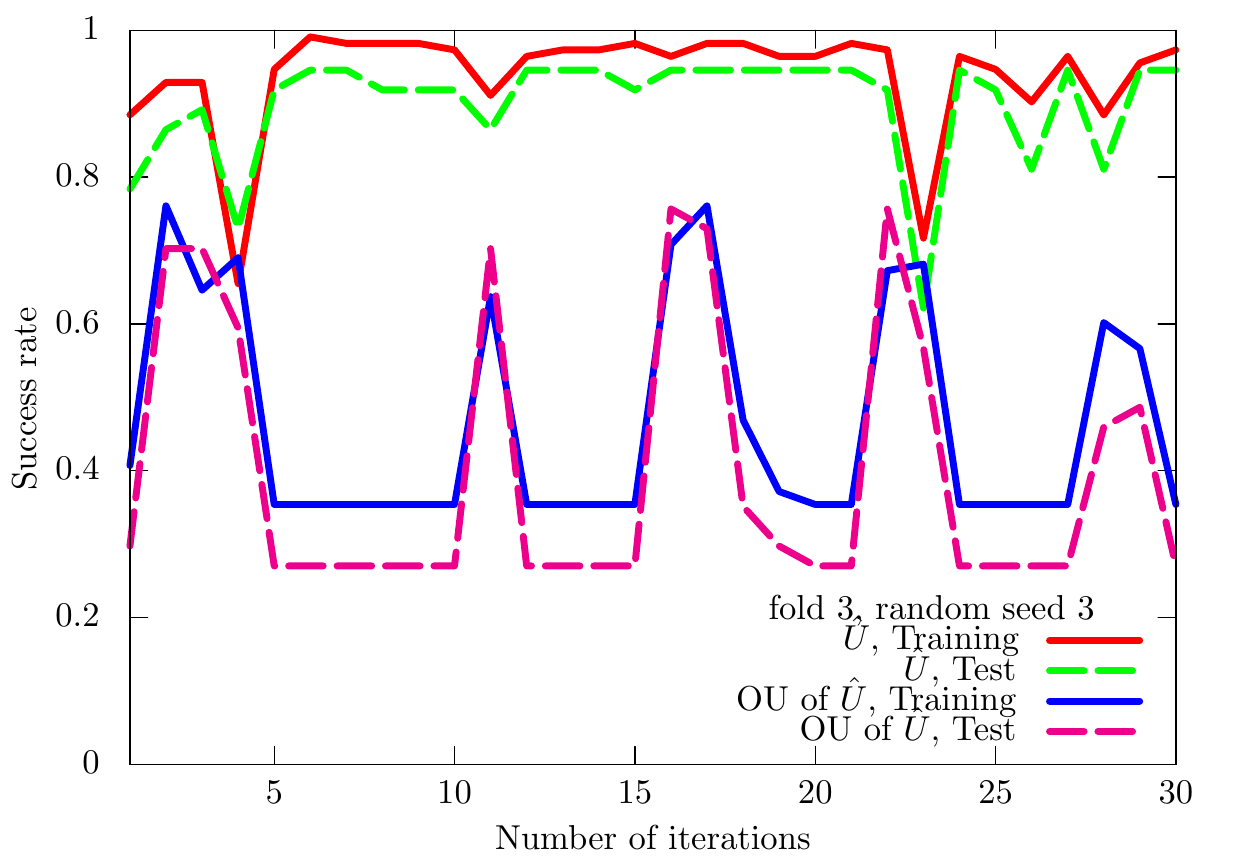}
\includegraphics[scale=0.25]{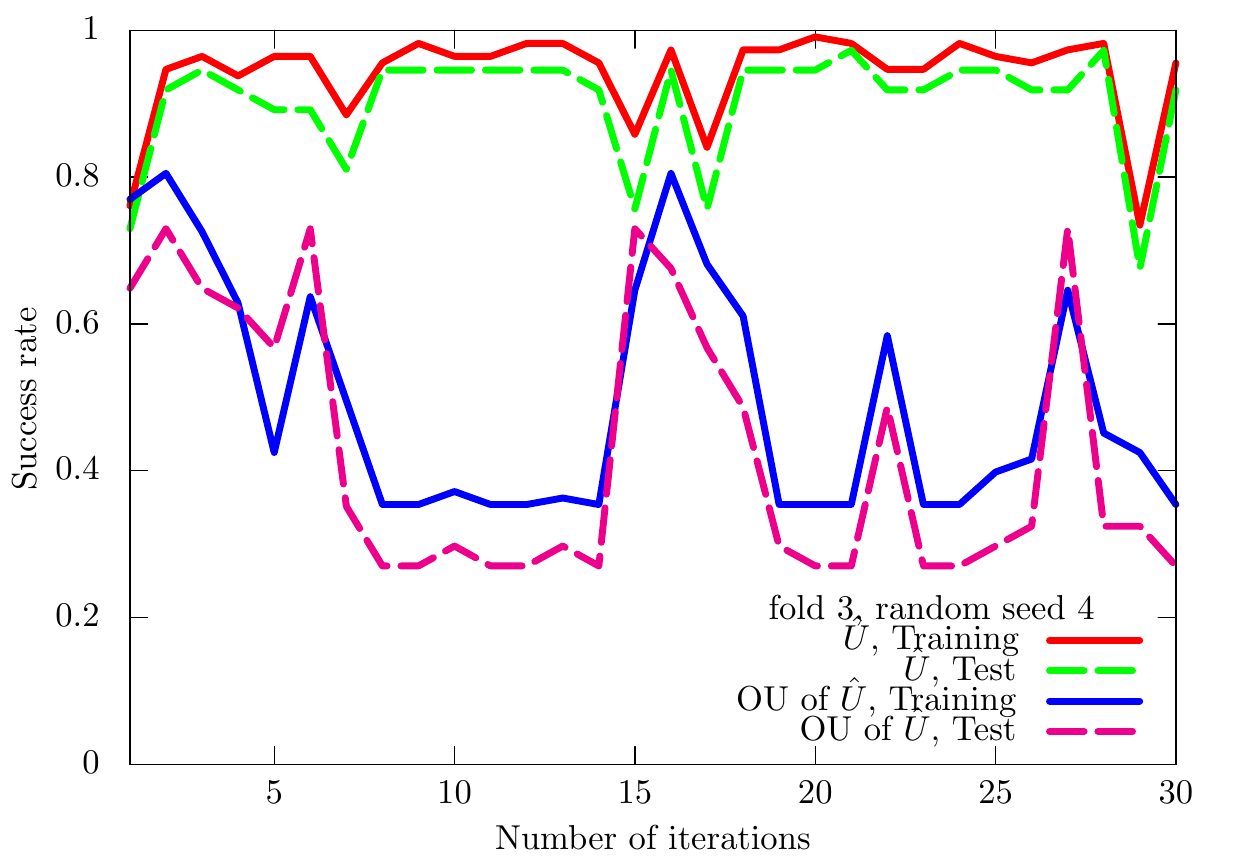}
\includegraphics[scale=0.25]{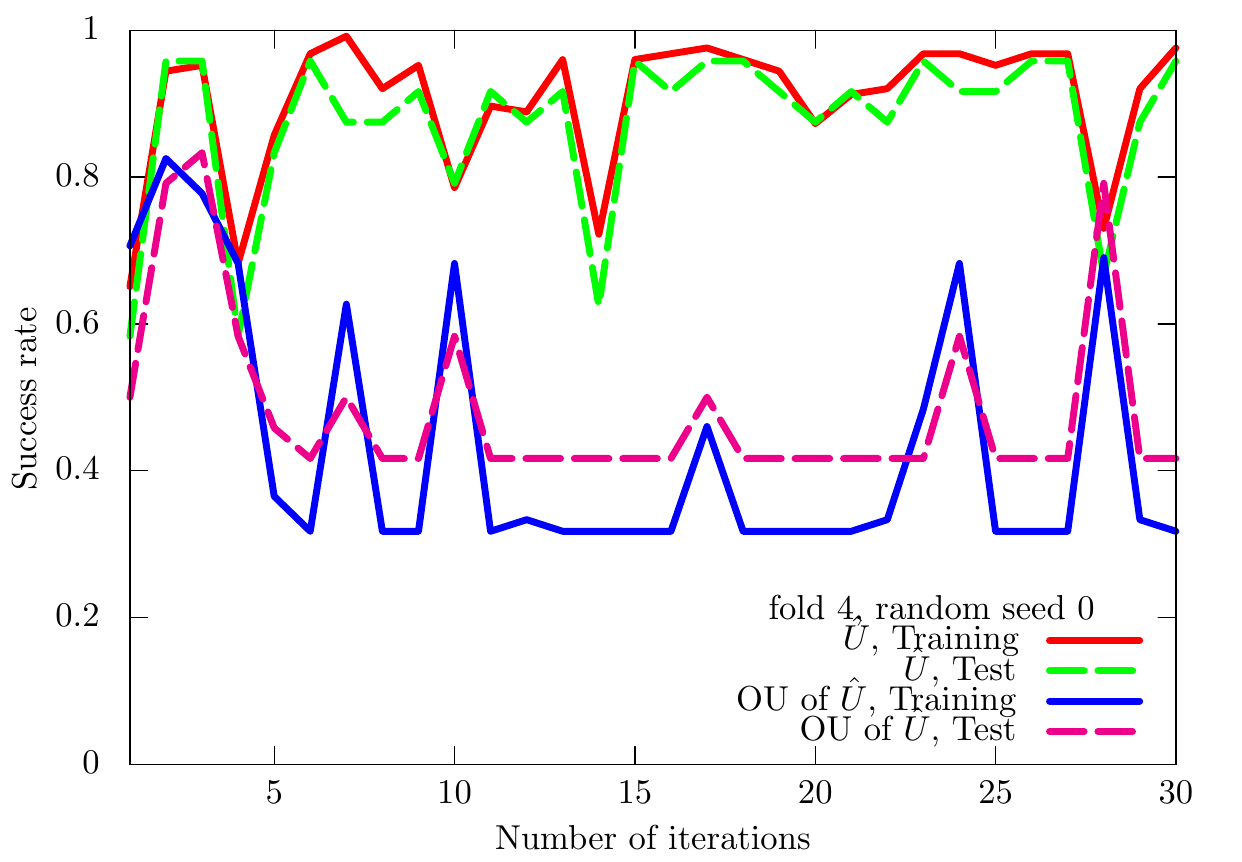}
\includegraphics[scale=0.25]{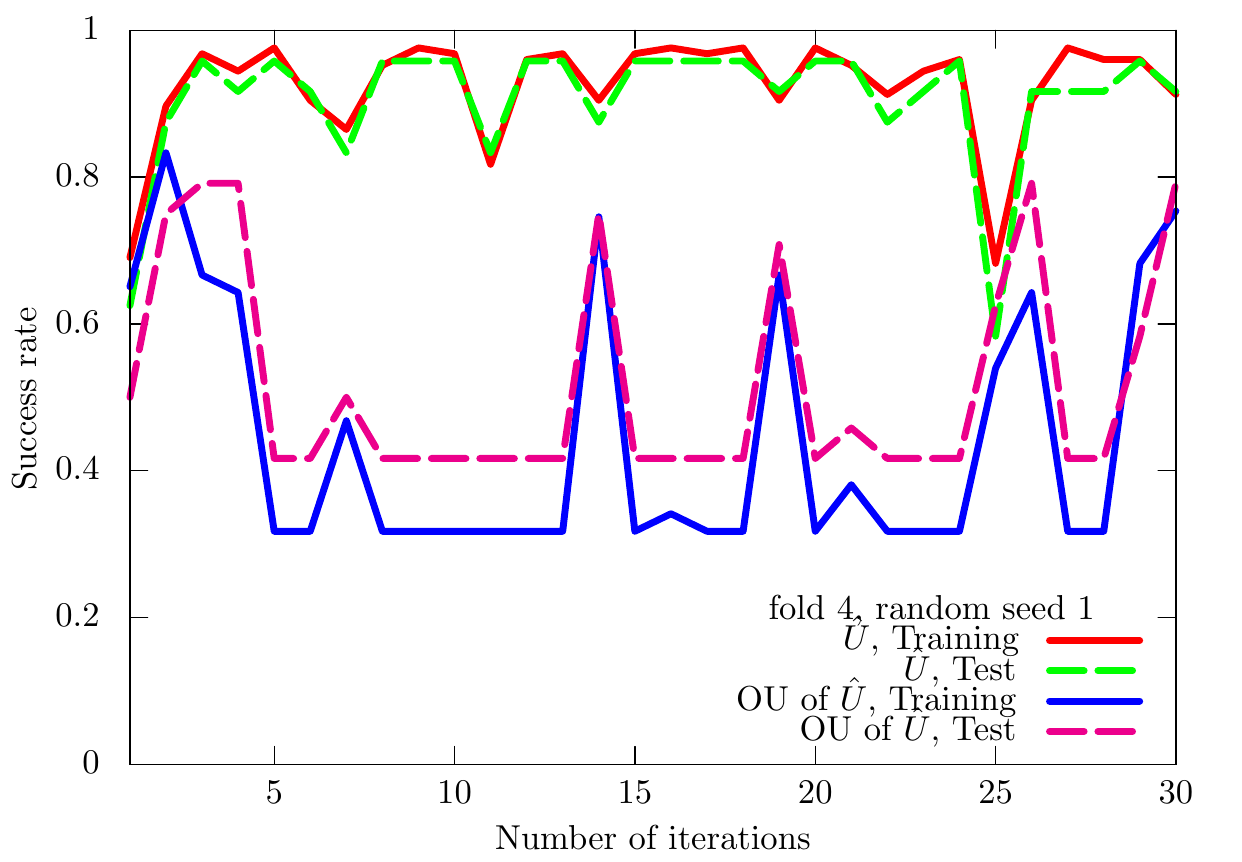}
\includegraphics[scale=0.25]{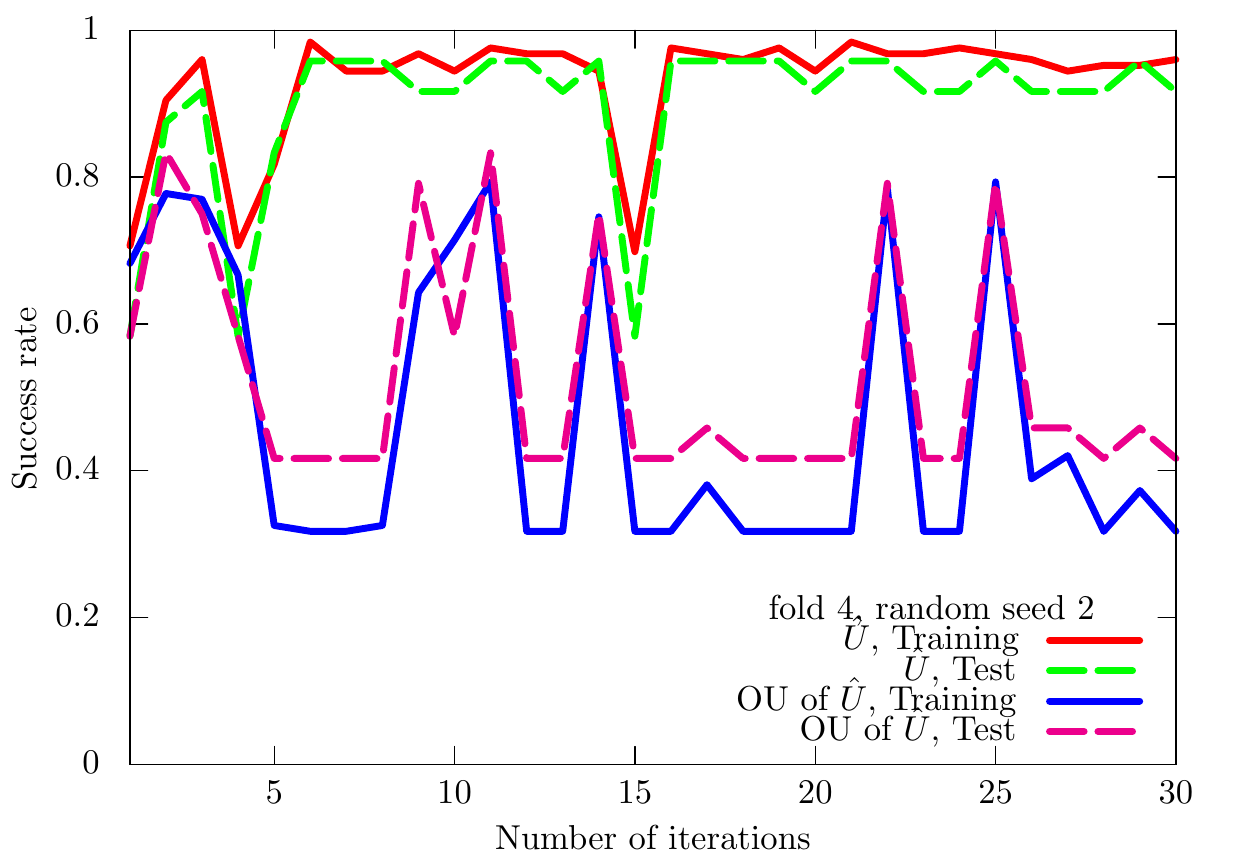}
\includegraphics[scale=0.25]{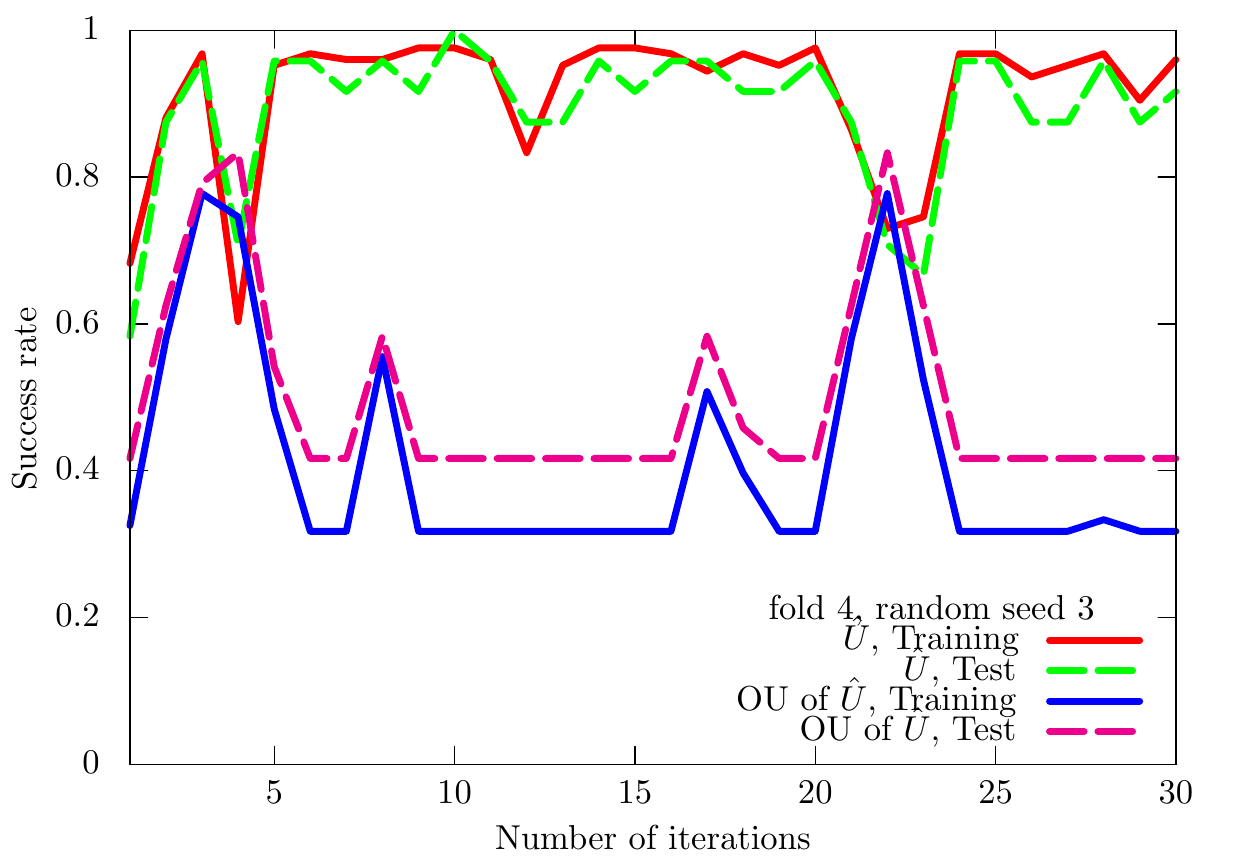}
\includegraphics[scale=0.25]{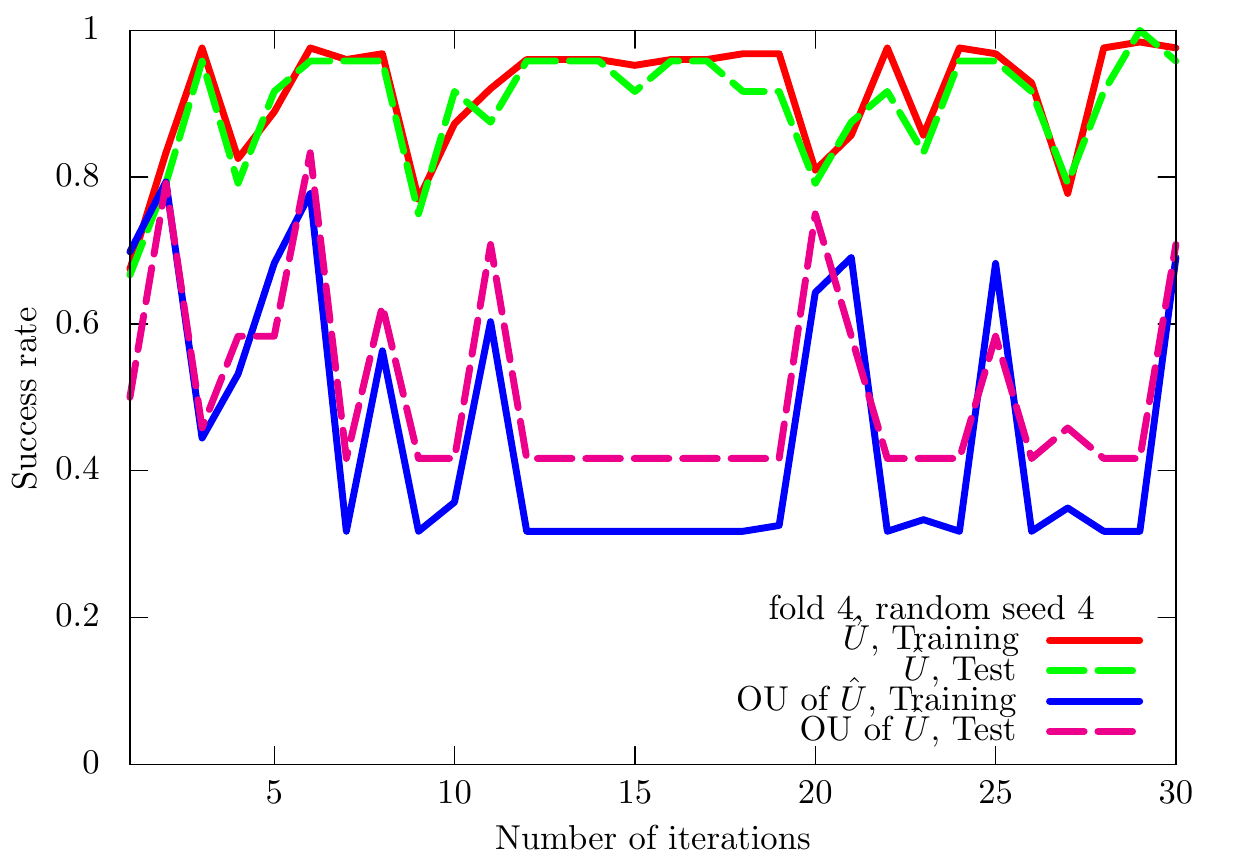}
\caption{Results of the UKM ($\hat{X}$ and OU of $\hat{X}$) on the $5$-fold datasets with $5$ different random seeds for the iris dataset ($1$ or non-$1$). We use complex matrices and set $\theta_\mathrm{bias} = 0$. We set $r = 0.010$.}
\label{supp-arXiv-numerical-result-raw-data-fold-001-rand-001-UKM-OUU-UCI-iris-1-non1}
\end{figure*}

In Fig.~\ref{supp-arXiv-numerical-result-raw-data-fold-001-rand-001-QCL-UCI-iris-1-non1}, we show the numerical results of QCL for the $5$-fold datasets with $5$ different random seeds.
In Fig.~\ref{supp-arXiv-numerical-result-raw-data-fold-001-rand-001-UKM-P-UCI-iris-1-non1}, we show the numerical results of $\hat{P}$ of the UKM for the $5$-fold datasets with $5$ different random seeds.
In Fig.~\ref{supp-arXiv-numerical-result-raw-data-fold-001-rand-001-UKM-OUU-UCI-iris-1-non1}, we also show the numerical results of OU of $\hat{X}$ of the UKM for the $5$-fold datasets with $5$ different random seeds.

\begin{figure}[htb]
\centering
\includegraphics[scale=0.45]{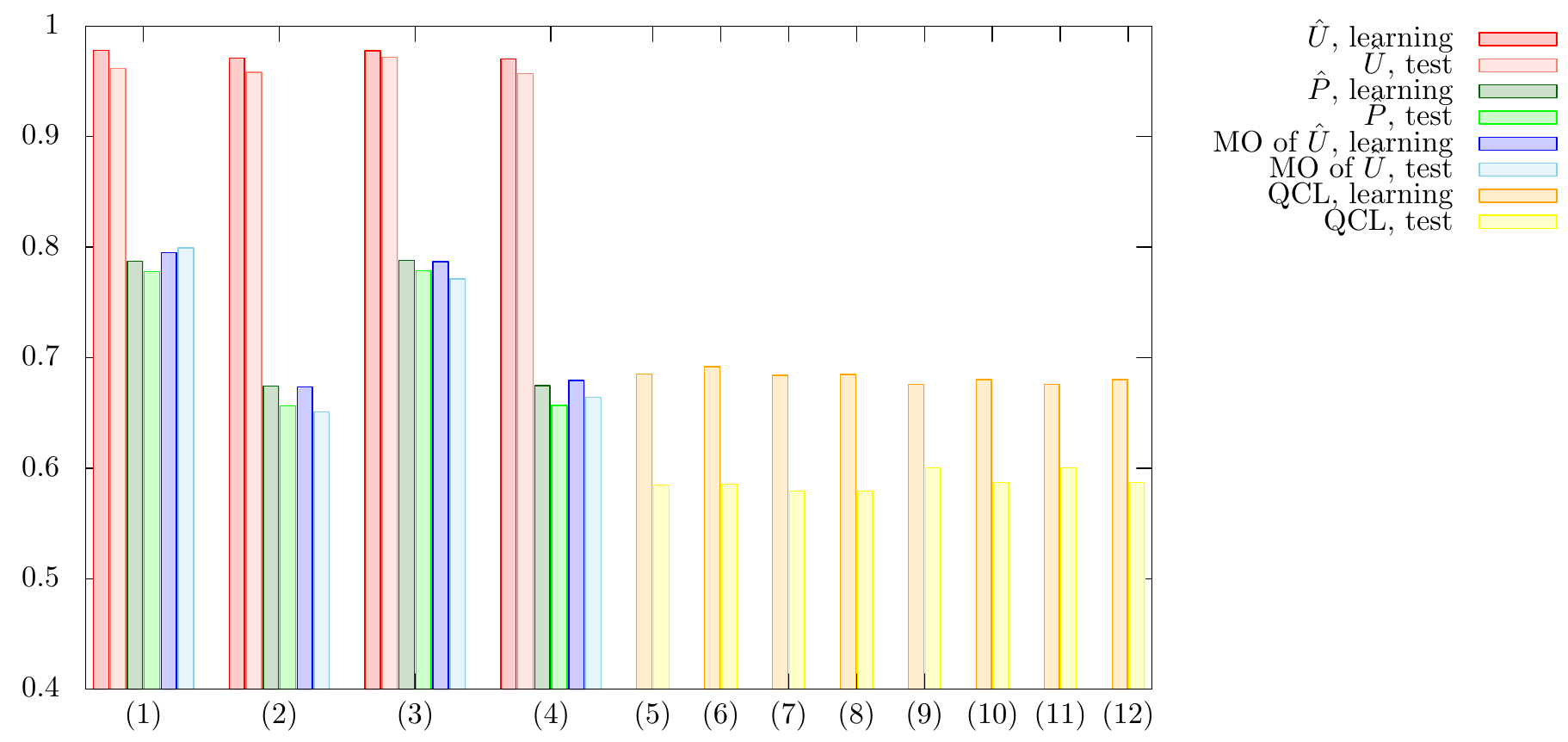}
\caption{Results of $5$-fold CV with $5$ different random seeds for the iris dataset ($1$ or non-$1$). For the UKM, we put $r = 0.010$ and set $K = 30$ and $K' = 10$. For QCL, the number of layers $L$ is $5$ and the number of iterations is $300$. The numerical settings are as follows: (1) complex matrices without the bias term, (2) complex matrices with the bias term, (3) real matrices without the bias term, (4) real matrices with the bias term, (5) CNOT-based circuit without the bias term, (6) CNOT-based circuit with the bias term, (7) CRot-based circuit without the bias term, (8) CRot-based circuit with the bias term, (9) 1d Heisenberg circuit without the bias term, (10) 1d Heisenberg circuit with the bias term, (11) FC Heisenberg circuit without the bias term, and (12) FC Heisenberg circuit with the bias term.}
\label{supp-arXiv-numerical-result-performance-UKM-QCL-UCI-iris-1-non1}
\end{figure}

We summarize the results of 5-fold CV with 5 different random seeds of QCL and the UKM in Tables~\ref{supp-arXiv-table-UCI-iris-1-non1-002} and \ref{supp-arXiv-table-UCI-iris-1-non1-001}, respectively.
For QCL and the UKM, we select the best model for the training dataset over iterations to compute the performance.
\begin{table}[htb]
  \begin{tabular}{cc|cc}
    \hline \hline
    Algo. & Condition & Training & Test \\
    \hline
  QCL & CNOT-based, w/o bias & 0.6851 & 0.5845 \\
  QCL & CNOT-based, w/ bias & 0.6919 & 0.5854 \\
    \hline
  QCL & CRot-based, w/o bias & 0.6840 & 0.5793 \\
  QCL & CRot-based, w/ bias & 0.6847 & 0.5793 \\
    \hline
  QCL & 1d Heisenberg, w/o bias & 0.6760 & 0.6004 \\
  QCL & 1d Heisenberg, w/ bias & 0.6801 & 0.5872 \\
    \hline
  QCL & FC Heisenberg, w/o bias & 0.6760 & 0.6004 \\
  QCL & FC Heisenberg, w/ bias & 0.6801 & 0.5872 \\
    \hline \hline
  \end{tabular}
\caption{Results of $5$-fold CV with $5$ different random seeds of QCL for the iris dataset ($1$ or non-$1$). The number of layers $L$ is set to $5$ and the number of iterations is set to $300$.}
\label{supp-arXiv-table-UCI-iris-1-non1-002}
\end{table}
\begin{table}[htb]
  \begin{tabular}{cc|cc}
    \hline \hline
    Algo. & Condition & Training & Test \\
    \hline
  UKM & $\hat{X}$, complex, w/o bias & 0.9781 & 0.9618 \\
  UKM & $\hat{P}$, complex, w/o bias & 0.7873 & 0.7781 \\
  UKM & OU of $\hat{X}$, complex, w/o bias & 0.7953 & 0.7994 \\
    \hline
  UKM & $\hat{X}$, complex, w/ bias & 0.9712 & 0.9581 \\
  UKM & $\hat{P}$, complex, w/ bias & 0.6744 & 0.6564 \\
  UKM & OU of $\hat{X}$, complex, w/ bias & 0.6734 & 0.6507 \\
    \hline
  UKM & $\hat{X}$, real, w/o bias & 0.9778 & 0.9717 \\
  UKM & $\hat{P}$, real, w/o bias & 0.7880 & 0.7789 \\
  UKM & OU of $\hat{X}$, real, w/o bias & 0.7869 & 0.7713 \\
    \hline
  UKM & $\hat{X}$, real, w/ bias & 0.9702 & 0.9569 \\
  UKM & $\hat{P}$, real, w/ bias & 0.6746 & 0.6568 \\
  UKM & OU of $\hat{X}$, real, w/ bias & 0.6794 & 0.6640 \\
    \hline \hline
  \end{tabular}
\caption{Results of $5$-fold CV with $5$ different random seeds of the UKM for the iris dataset ($1$ or non-$1$). We put $r = 0.010$ and set $K = 30$ and $K' = 10$.}
\label{supp-arXiv-table-UCI-iris-1-non1-001}
\end{table}
In Fig.~\ref{supp-arXiv-numerical-result-performance-UKM-QCL-UCI-iris-1-non1}, we plot the data shown in Tables~\ref{supp-arXiv-table-UCI-iris-1-non1-002} and \ref{supp-arXiv-table-UCI-iris-1-non1-001}.
We also summarize the results of 5-fold CV with 5 different random seeds of the kernel method in Table~\ref{supp-arXiv-table-UCI-iris-1-non1-003}.
More specifically, we use Ridge classification in Sec.~\ref{supp-arXiv-sec-Ridge-001}.
We consider the linear functions and the second-order polynomial functions for $\phi (\cdot)$ in Eq.~\eqref{supp-arXiv-f-pred-kernel-method-001-002} with and without normalization.
We set $\lambda = 10^{-2}, 10^{-1}, 1$ where $\lambda$ is the coefficient of the regularization term.
\begin{table}[htb]
  \begin{tabular}{cc|cc}
    \hline \hline
    Algo. & Condition & Training & Test \\
    \hline
  Kernel method & Linear, w/o normalization, $\lambda = 10^{-2}$ & 0.7463 & 0.7378 \\
  Kernel method & Linear, w/o normalization, $\lambda = 10^{-1}$ & 0.7445 & 0.7324 \\
  Kernel method & Linear, w/o normalization, $\lambda = 1$ & 0.7426 & 0.7240 \\
    \hline
  Kernel method & Linear, w/ normalization, $\lambda = 10^{-2}$ & 0.9401 & 0.9561 \\
  Kernel method & Linear, w/ normalization, $\lambda = 10^{-1}$ & 0.7697 & 0.7276 \\
  Kernel method & Linear, w/ normalization, $\lambda = 1$ & 0.6382 & 0.6113 \\
    \hline
  Kernel method & Poly-2, w/o normalization, $\lambda = 10^{-2}$ & 0.9751 & 0.9666 \\
  Kernel method & Poly-2, w/o normalization, $\lambda = 10^{-1}$ & 0.9733 & 0.9558 \\
  Kernel method & Poly-2, w/o normalization, $\lambda = 1$ & 0.9684 & 0.9558 \\
    \hline
  Kernel method & Poly-2, w/ normalization, $\lambda = 10^{-2}$ & 0.9601 & 0.9601 \\
  Kernel method & Poly-2, w/ normalization, $\lambda = 10^{-1}$ & 0.9253 & 0.9312 \\
  Kernel method & Poly-2, w/ normalization, $\lambda = 1$ & 0.6778 & 0.6475 \\
    \hline \hline
  \end{tabular}
\caption{Results of 5-fold CV with 5 different random seeds of the kernel method for the iris dataset ($1$ or non-$1$).}
\label{supp-arXiv-table-UCI-iris-1-non1-003}
\end{table}

\begin{figure}[htb]
\centering
\includegraphics[scale=0.45]{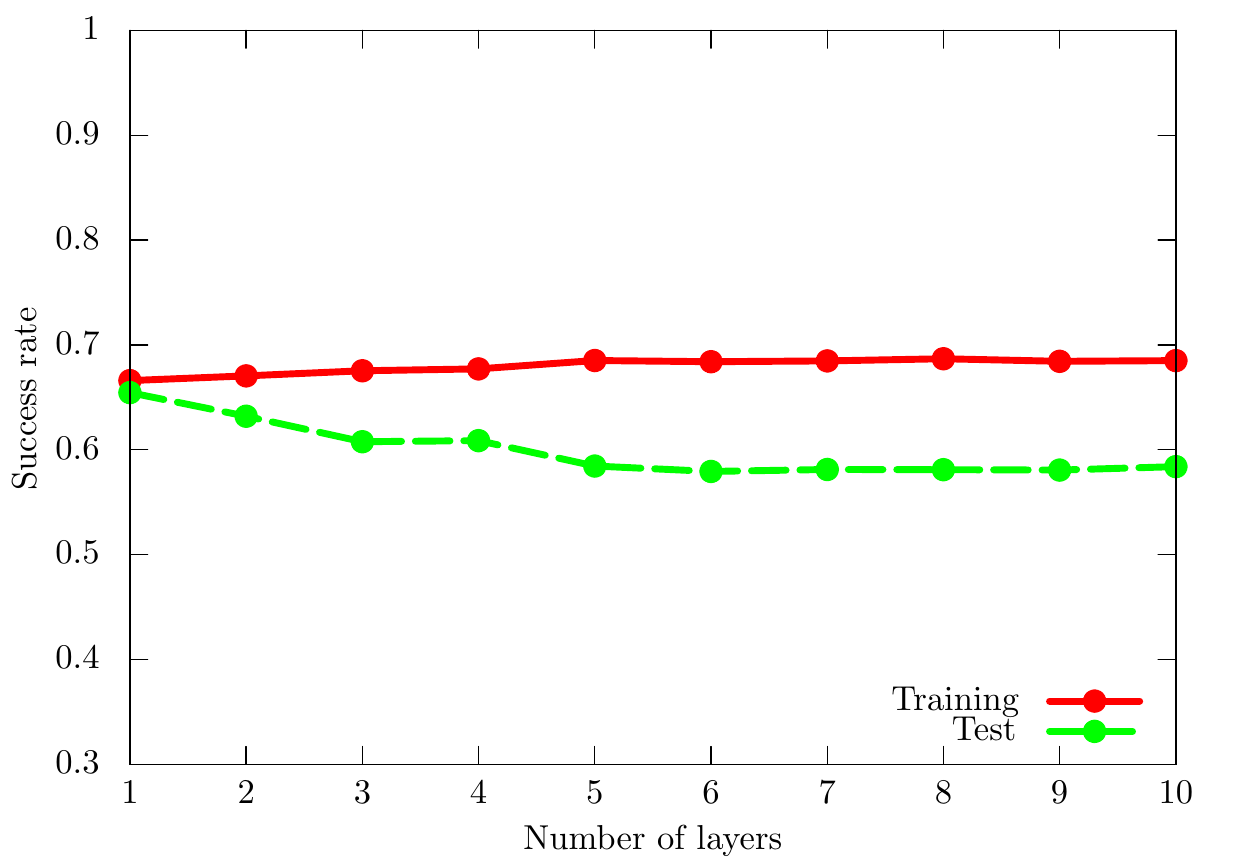}
\caption{Performance dependence of QCL on the number of layers $L$ for the iris dataset ($1$ or non-$1$). We use the CNOT-based circuit geometry and set $\theta_\mathrm{bias} = 0$. We iterate the computation $300$ times.}
\label{supp-arXiv-numerical-result-layers-dependence-QCL-UCI-iris-1-non1}
\end{figure}

\begin{figure}[htb]
\centering
\includegraphics[scale=0.45]{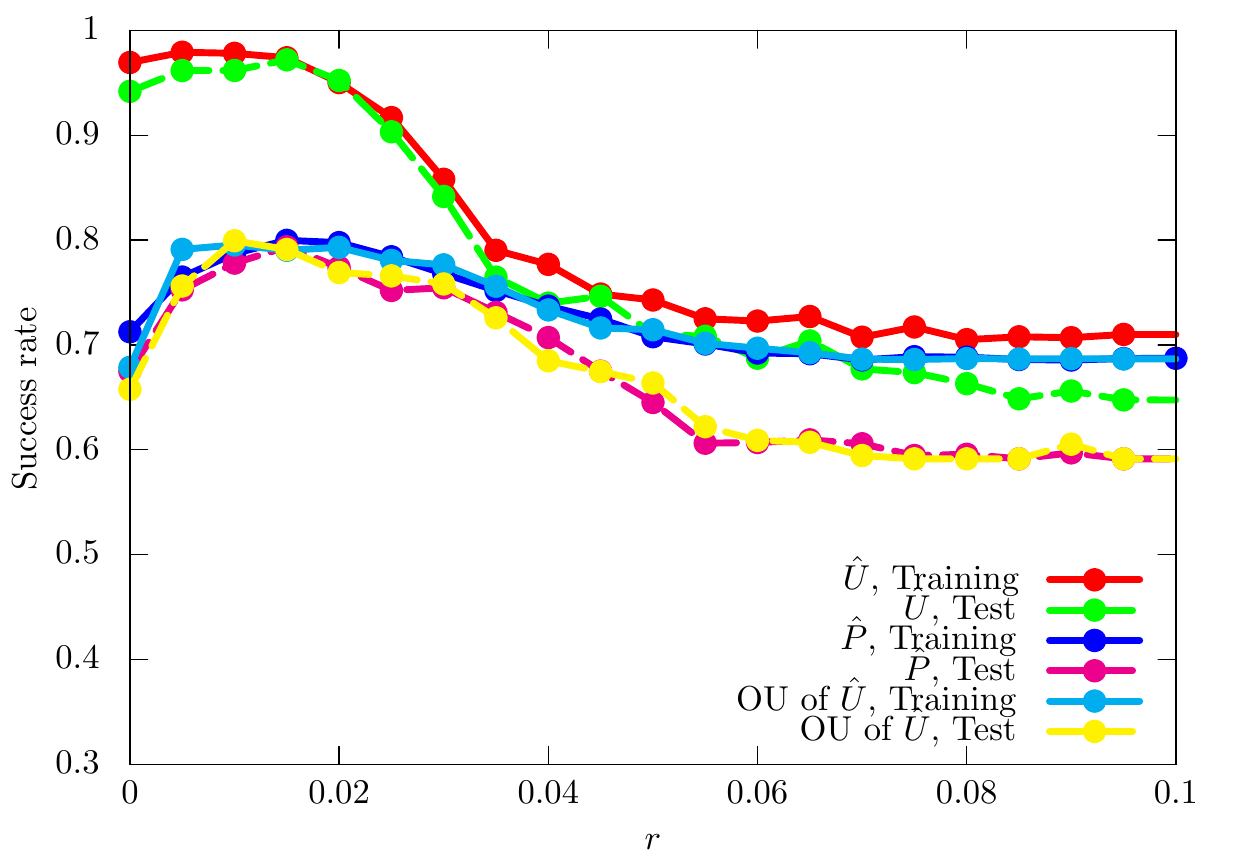}
\includegraphics[scale=0.45]{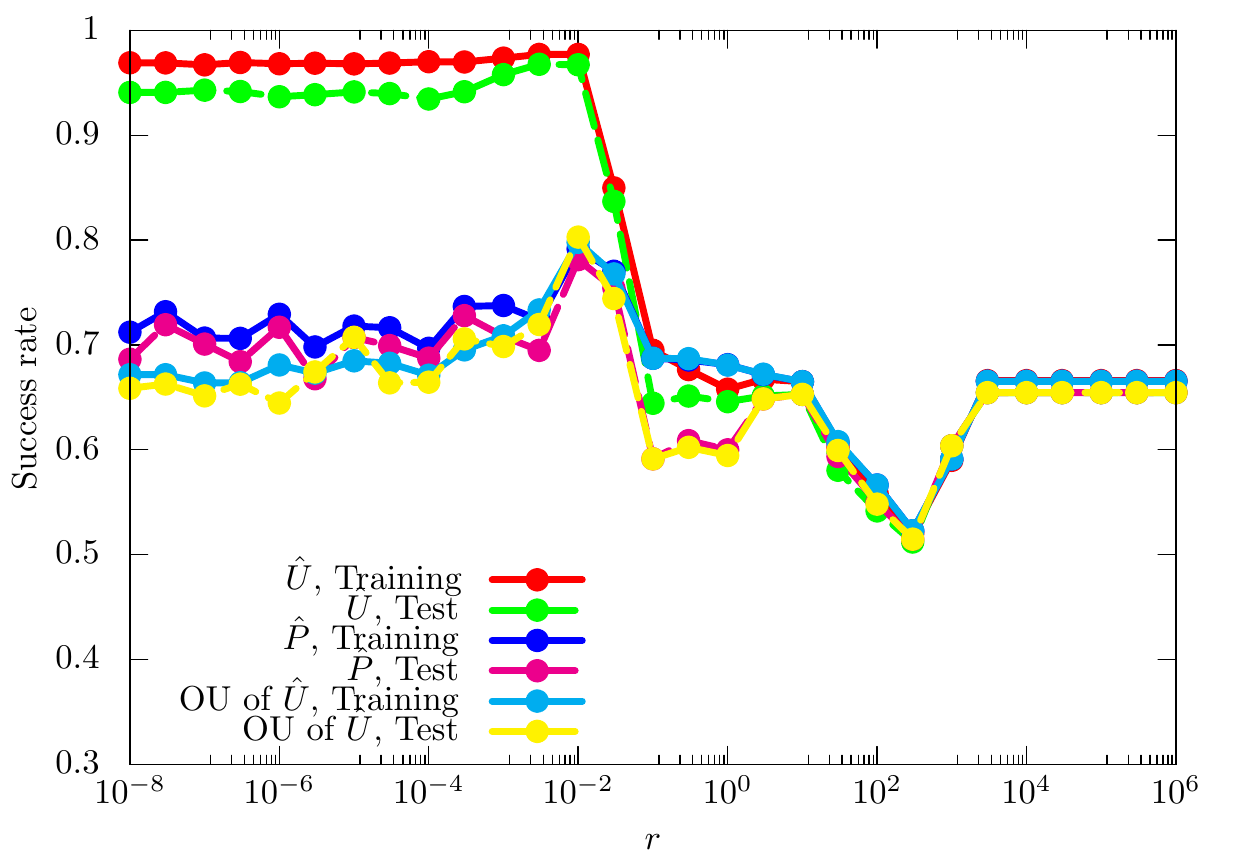}
\caption{Performance dependence of the UKM on $r$, which is the coefficient of the second term in the right-hand side of Eq.~\eqref{supp-arXiv-quantum-kernel-method-001-011} for the iris dataset ($1$ or non-$1$). We use complex matrices and set $\theta_\mathrm{bias} = 0$. We set $K = 30$ and $K' = 10$.}
\label{supp-arXiv-numerical-result-r-dependence-UKM-UCI-iris-1-non1}
\end{figure}

\begin{figure}[htb]
\centering
\includegraphics[scale=0.45]{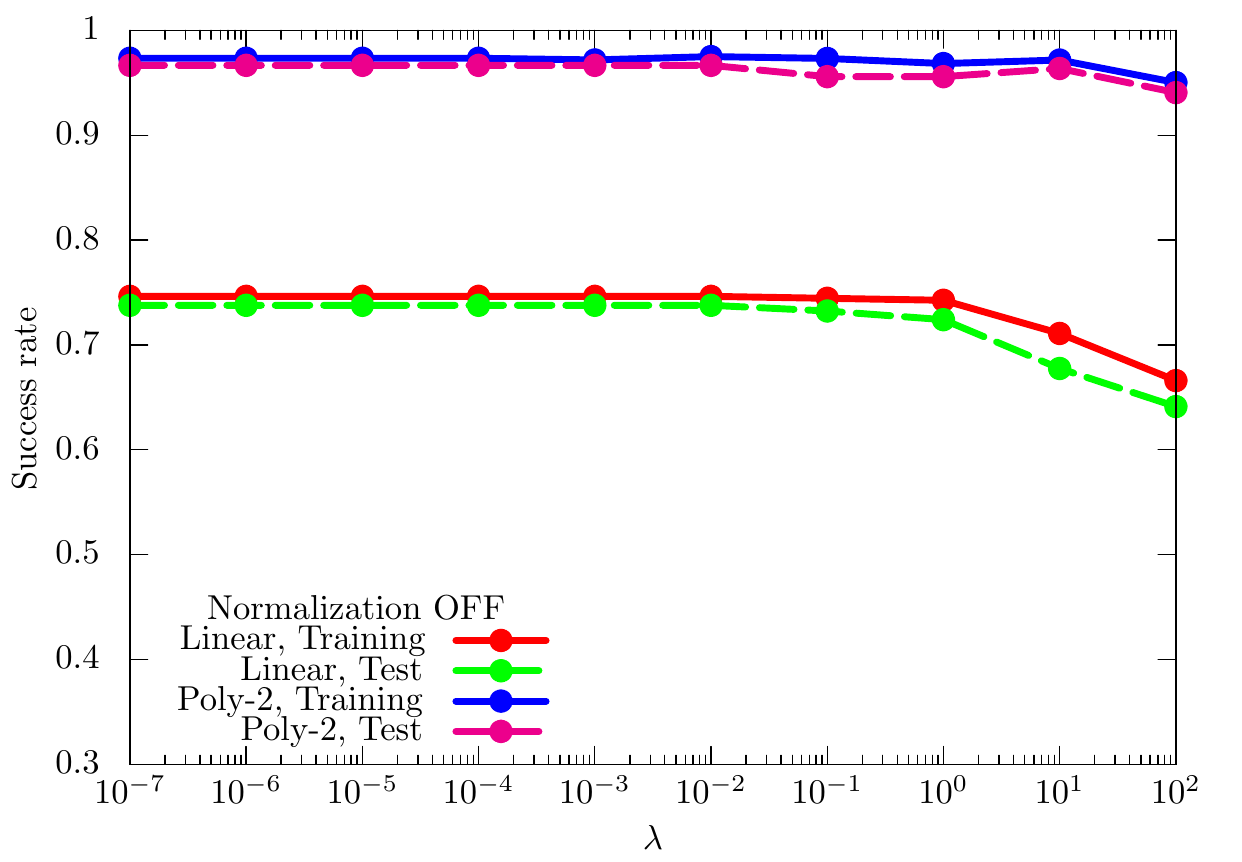}
\includegraphics[scale=0.45]{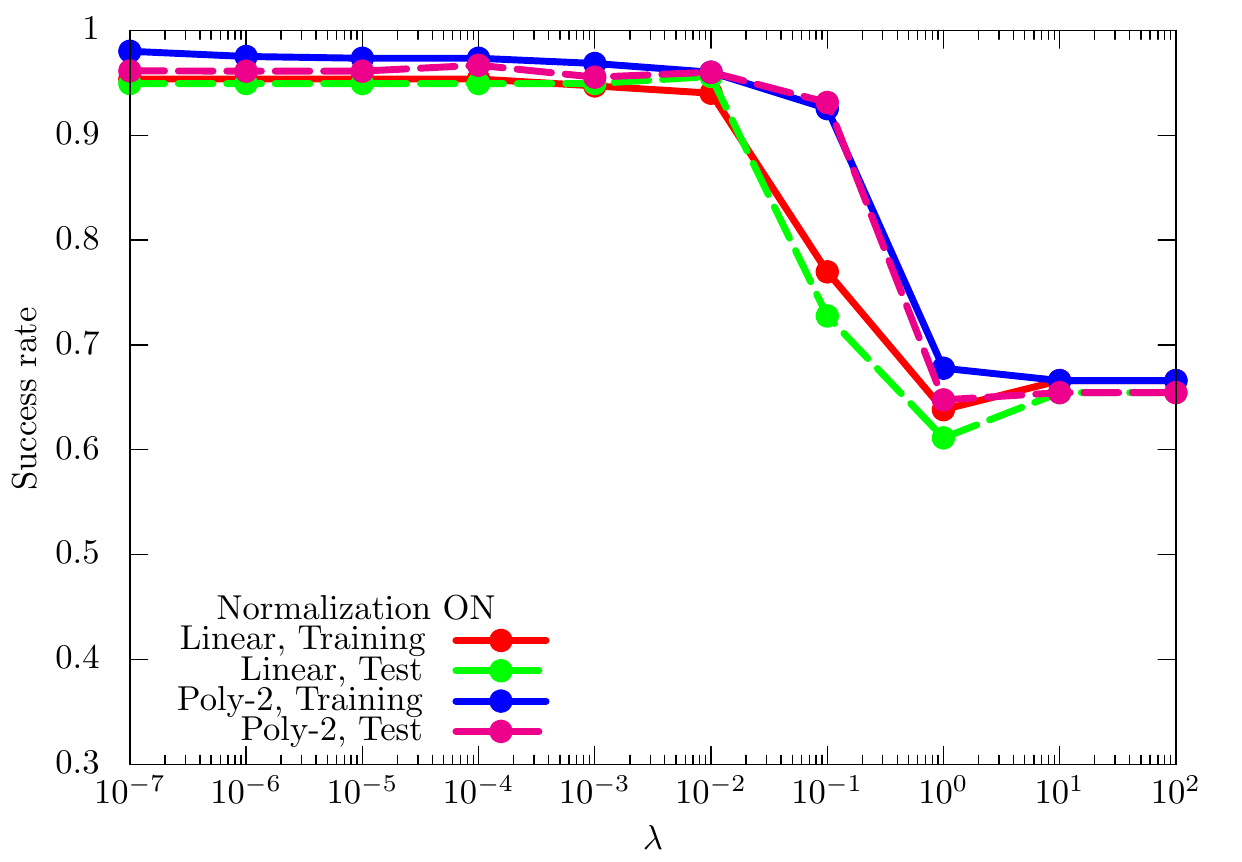}
\caption{Performance dependence of the kernel method on $\lambda$, which is the coefficient of the second term in the right-hand side of Eq.~\eqref{supp-arXiv-cost-function-kernel-method-001-002} for the iris dataset ($1$ or non-$1$). For $\phi (\cdot)$ in Eq.~\eqref{supp-arXiv-f-pred-kernel-method-001-002}, we use the linear functions and the second-degree polynomial functions with and without normalization.}
\label{supp-arXiv-numerical-result-lambda-dependence-kernel-method-iris-1-non1}
\end{figure}

Next, we show the performance dependence of the three algorithms on their key parameters.
We see the performance dependence of QCL on the number of layers $L$.
The result is shown in Fig.~\ref{supp-arXiv-numerical-result-layers-dependence-QCL-UCI-iris-1-non1}.
We then see the performance dependence of the UKM on $r$, which is the coefficient of the second term in the right-hand side of Eq.~\eqref{supp-arXiv-quantum-kernel-method-001-011}.
The result is shown in Fig.~\ref{supp-arXiv-numerical-result-r-dependence-UKM-UCI-iris-1-non1}.
In Fig.~\ref{supp-arXiv-numerical-result-lambda-dependence-kernel-method-iris-1-non1}, we show the performance dependence of the kernel method on $\lambda$, which is the coefficient of the second term in the right-hand side of Eq.~\eqref{supp-arXiv-cost-function-kernel-method-001-002}.

\begin{figure}[htb]
\centering
\includegraphics[scale=0.45]{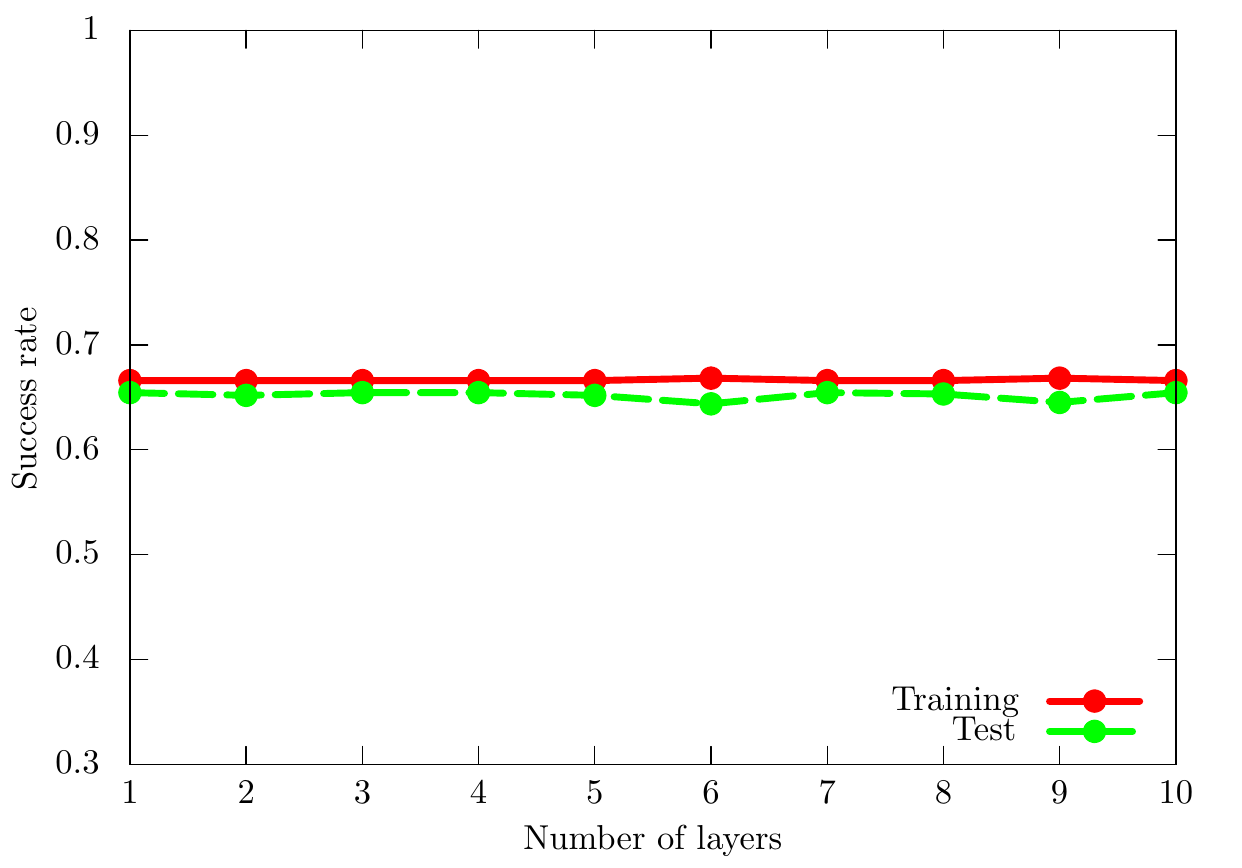}
\caption{Performance dependence of QCL on the number of layers $L$ for the iris dataset ($1$ or non-$1$) in the case of the hinge function, Eq.~\eqref{supp-arXiv-hinge-function-001-001}. We use the CNOT-based circuit geometry and set $\theta_\mathrm{bias} = 0$. We iterate the computation $300$ times.}
\label{supp-arXiv-numerical-result-layers-dependence-QCL-UCI-iris-1-non1-hinge}
\end{figure}

\begin{figure}[htb]
\centering
\includegraphics[scale=0.45]{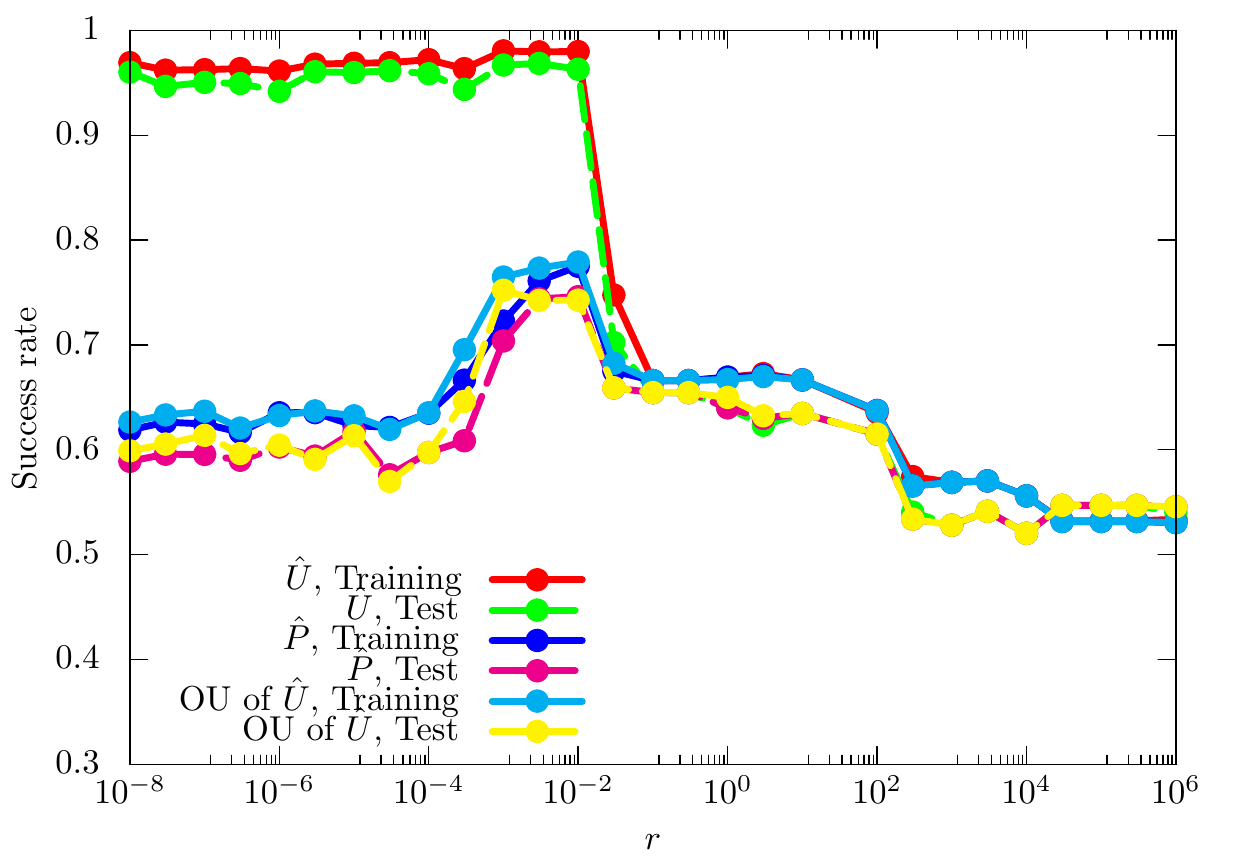}
\caption{Performance dependence of the UKM on $r$, which is the coefficient of the second term in the right-hand side of Eq.~\eqref{supp-arXiv-quantum-kernel-method-001-011} for the iris dataset ($1$ or non-$1$) in the case of the hinge function, Eq.~\eqref{supp-arXiv-hinge-function-001-001}. We use complex matrices and set $\theta_\mathrm{bias} = 0$. We set $K = 30$ and $K' = 10$.}
\label{supp-arXiv-numerical-result-r-dependence-UKM-UCI-iris-1-non1-hinge}
\end{figure}

So far, we have used the squared error function, Eq.~\eqref{supp-arXiv-squared-error-function-001-001}.
In Fig.~\ref{supp-arXiv-numerical-result-layers-dependence-QCL-UCI-iris-1-non1-hinge}, we show the performance dependence of QCL on the number of layers $L$ in the case of the hinge function, Eq.~\eqref{supp-arXiv-hinge-function-001-001}.
In Fig.~\ref{supp-arXiv-numerical-result-r-dependence-UKM-UCI-iris-1-non1-hinge}, we show the performance dependence of the UKM on $r$, which is the coefficient of the second term in the right-hand side of Eq.~\eqref{supp-arXiv-quantum-kernel-method-001-011}, in the case of the hinge function, Eq.~\eqref{supp-arXiv-hinge-function-001-001}.

\clearpage

\subsection{Cancer dataset ($0$ or $1$)}

We here show the numerical result for the cancer dataset ($0$ or $1$).
For the UKM, we put $r = 0.010$ and set $K = 30$ and $K' = 10$ in Algo.~\ref{supp-arXiv-quantum-kernel-method-002-001}.
For QCL, we run iterations $300$ times.

\begin{figure*}[htb]
\centering
\includegraphics[scale=0.25]{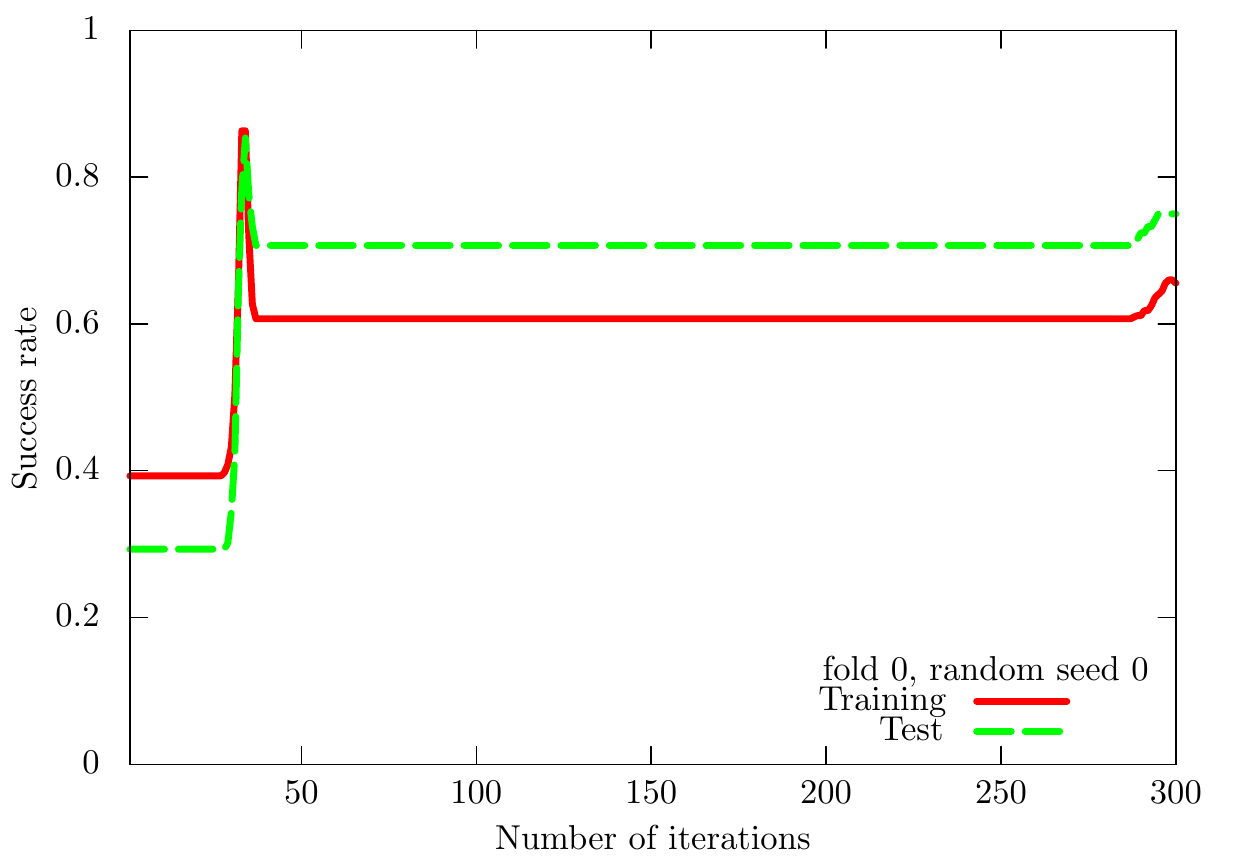}
\includegraphics[scale=0.25]{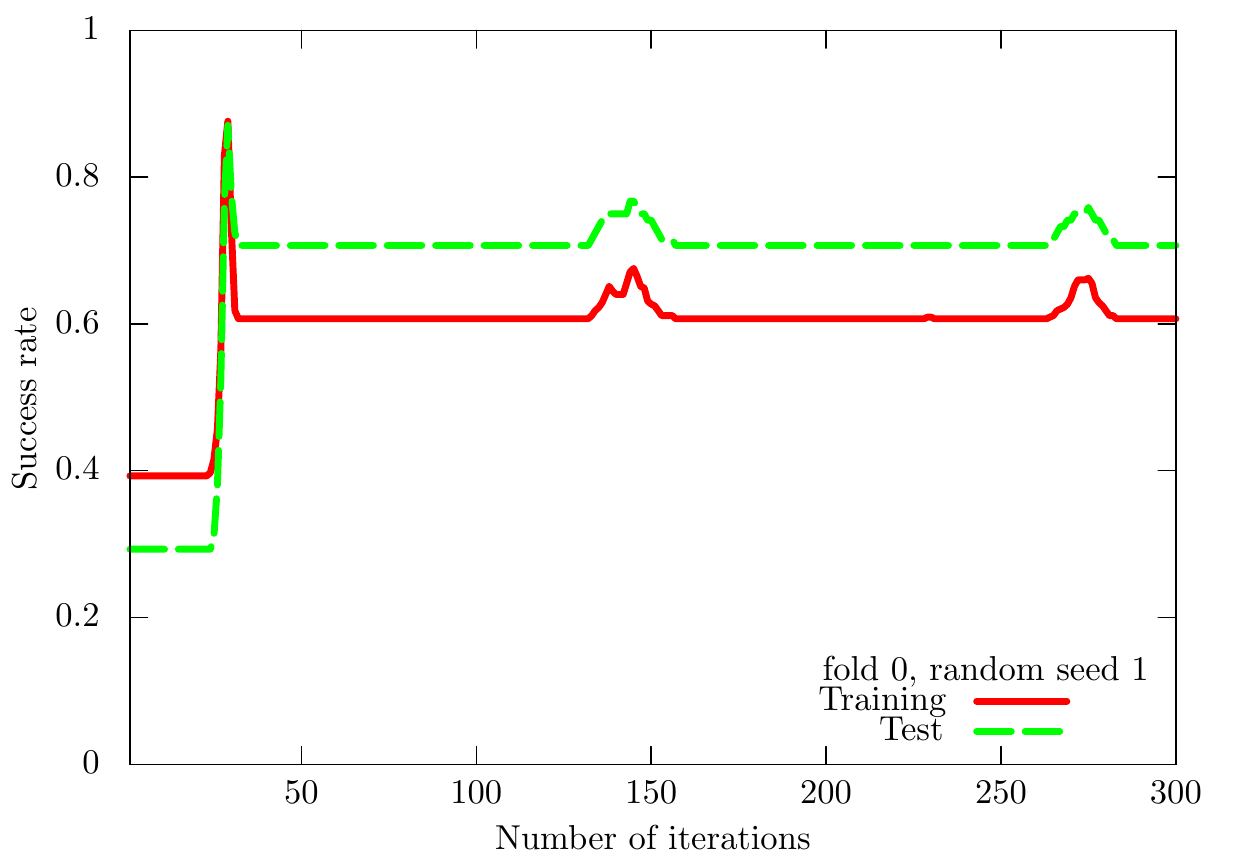}
\includegraphics[scale=0.25]{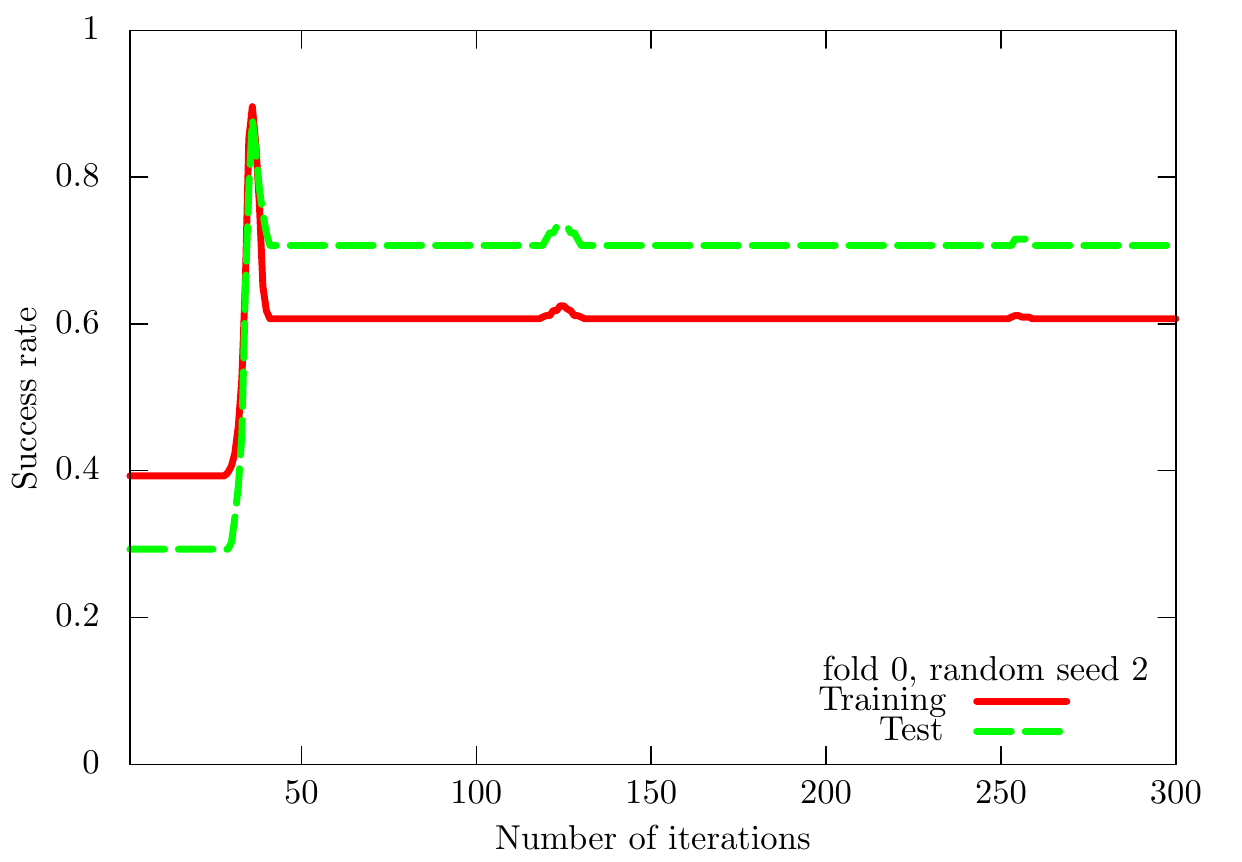}
\includegraphics[scale=0.25]{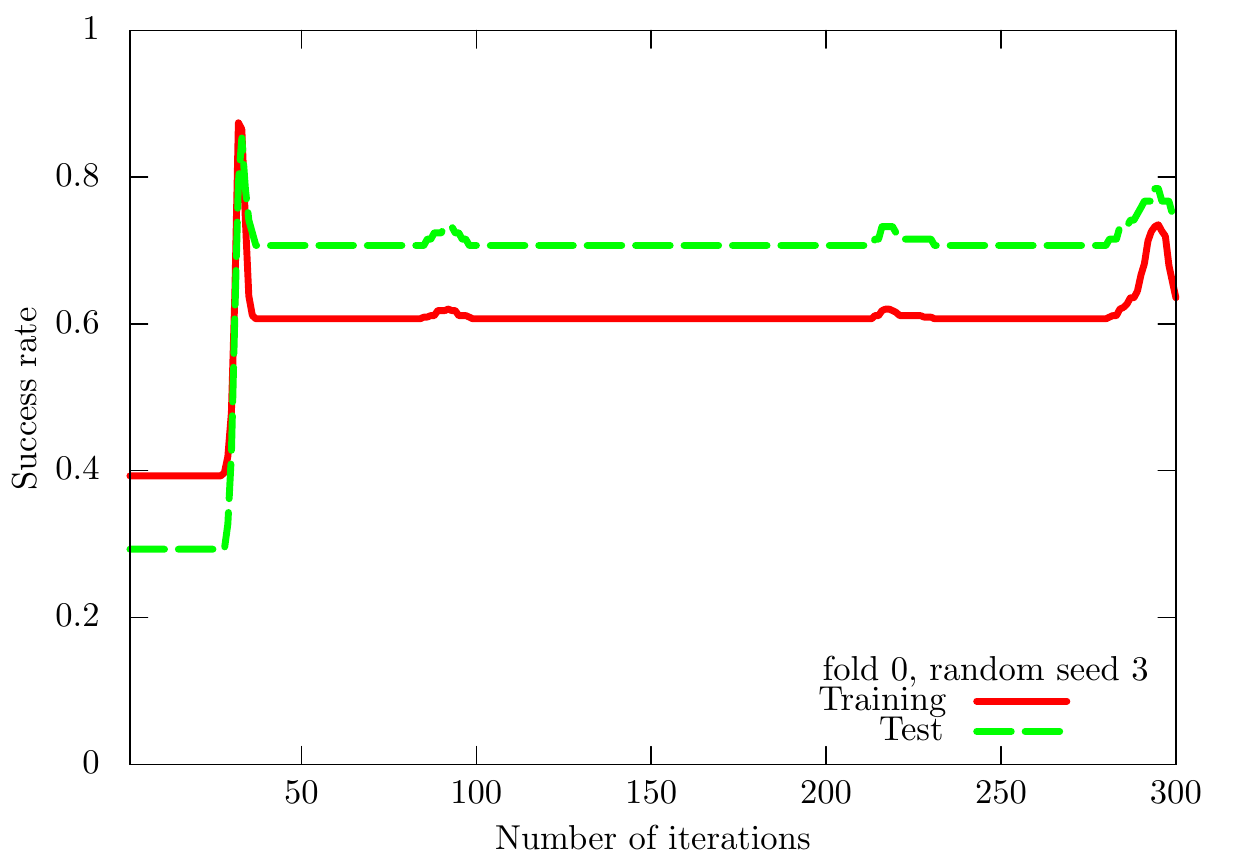}
\includegraphics[scale=0.25]{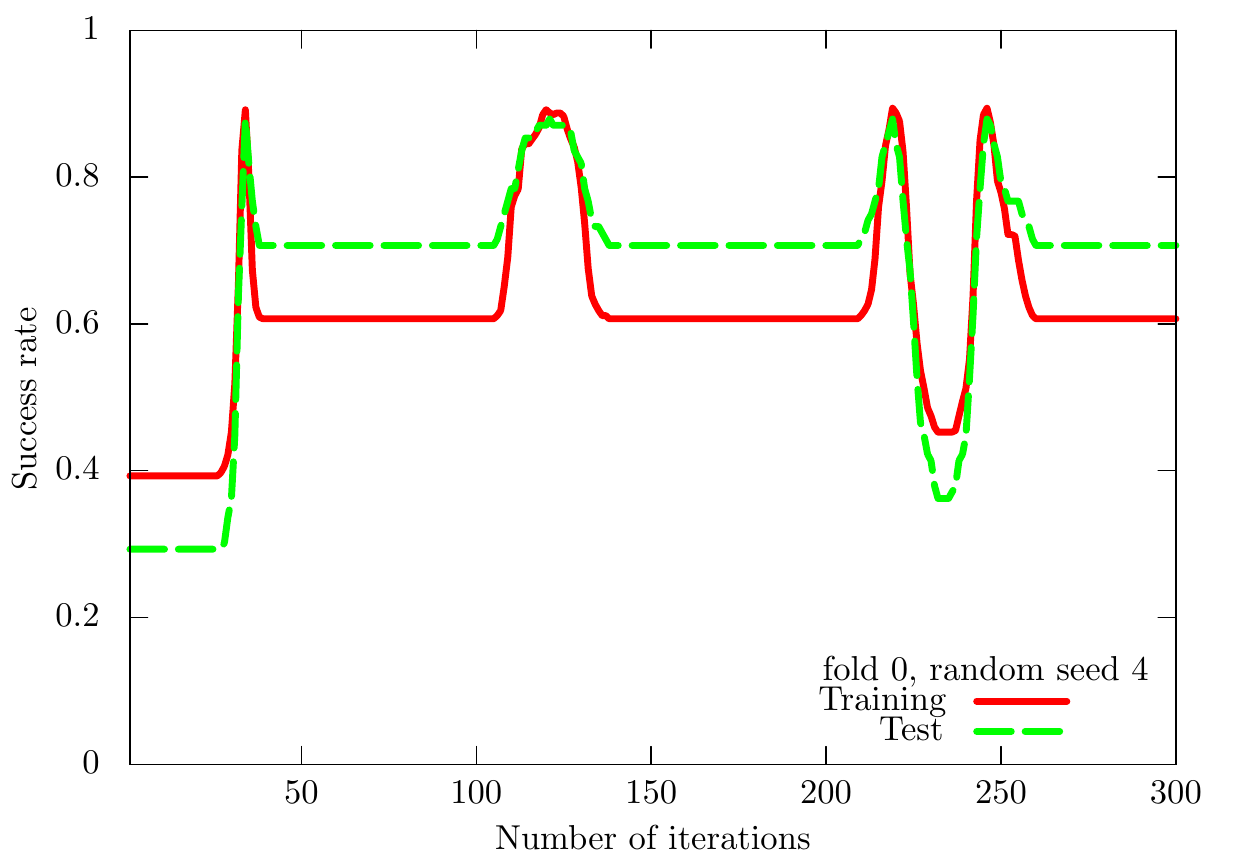}
\includegraphics[scale=0.25]{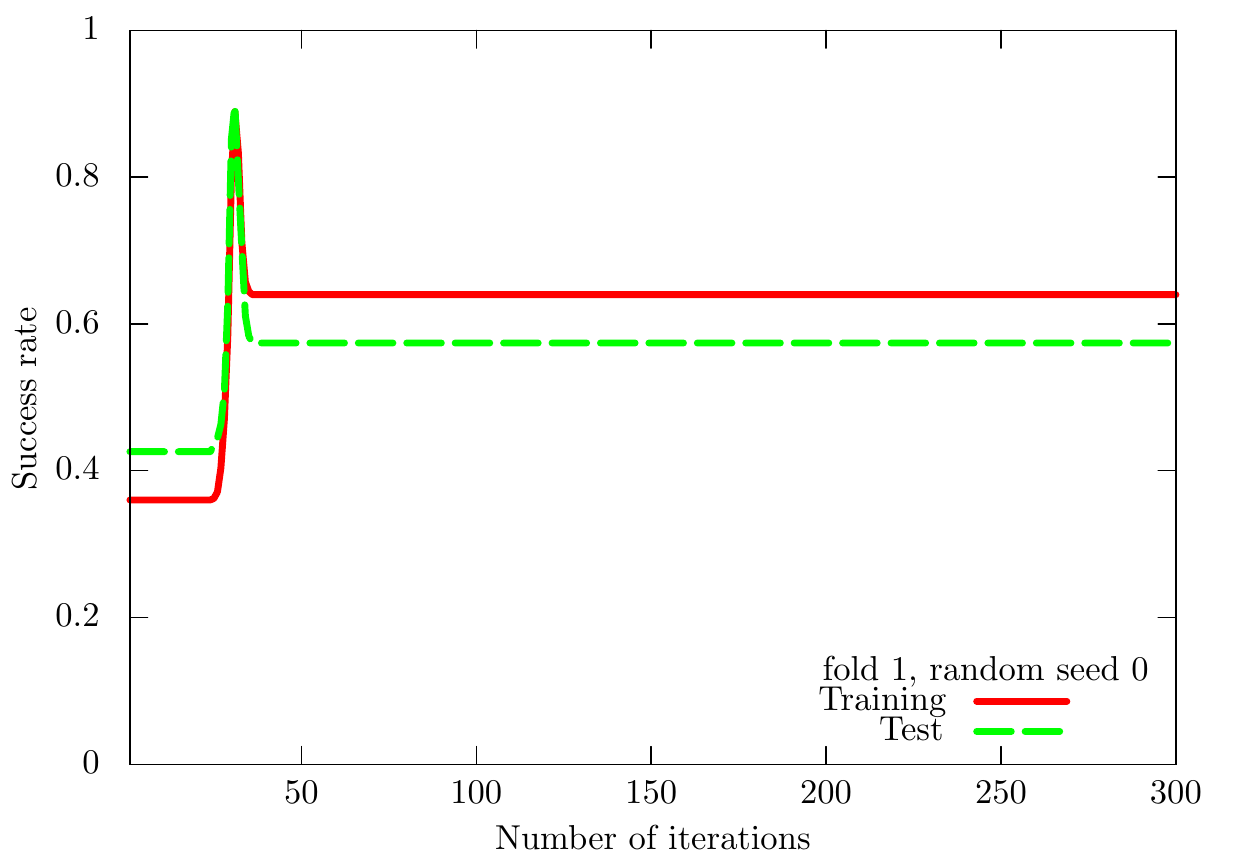}
\includegraphics[scale=0.25]{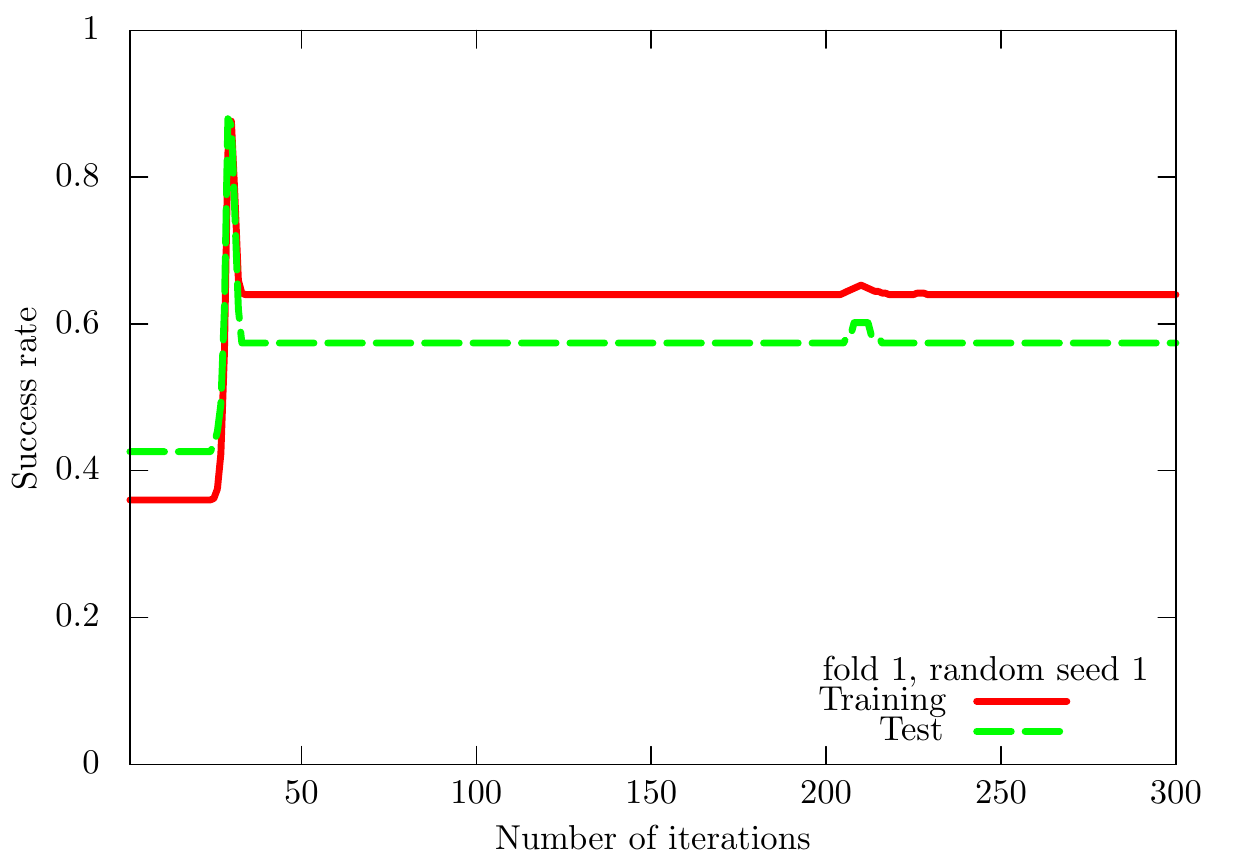}
\includegraphics[scale=0.25]{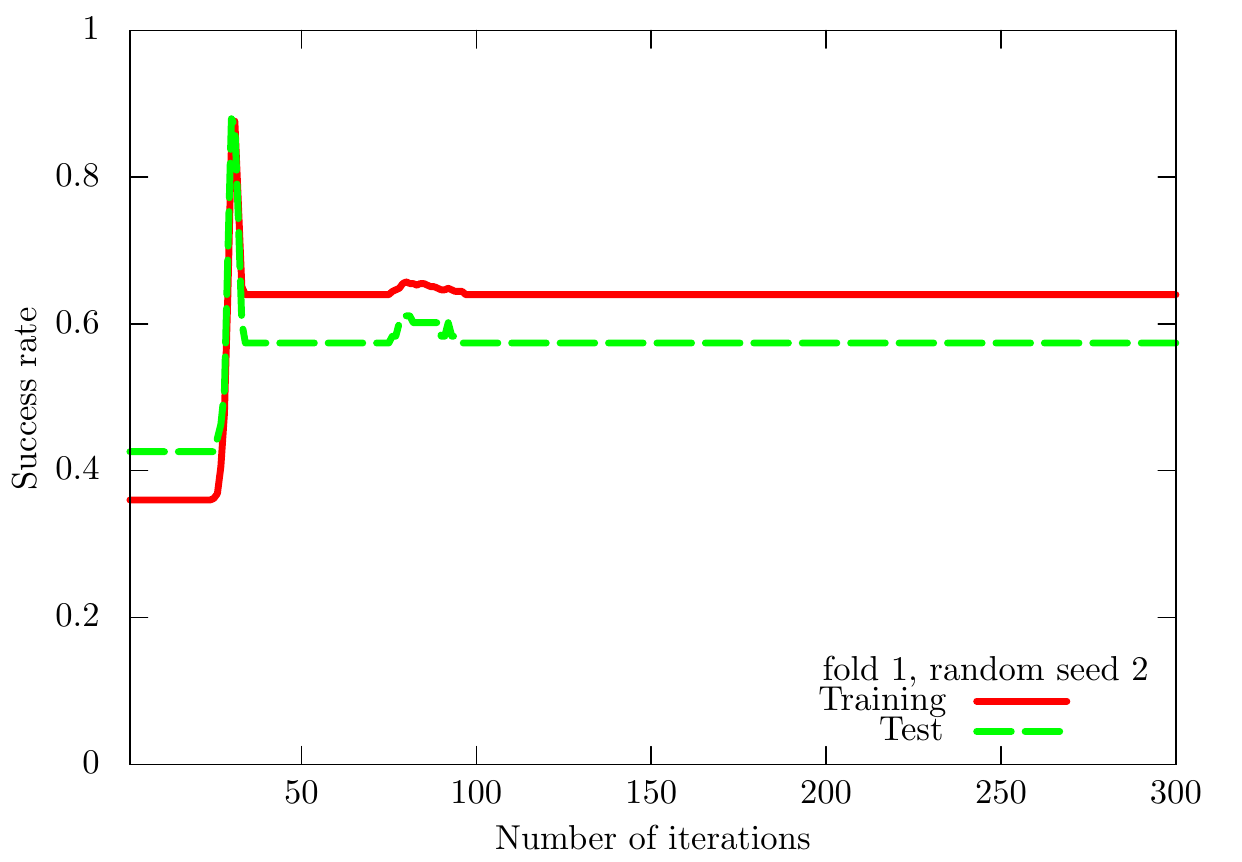}
\includegraphics[scale=0.25]{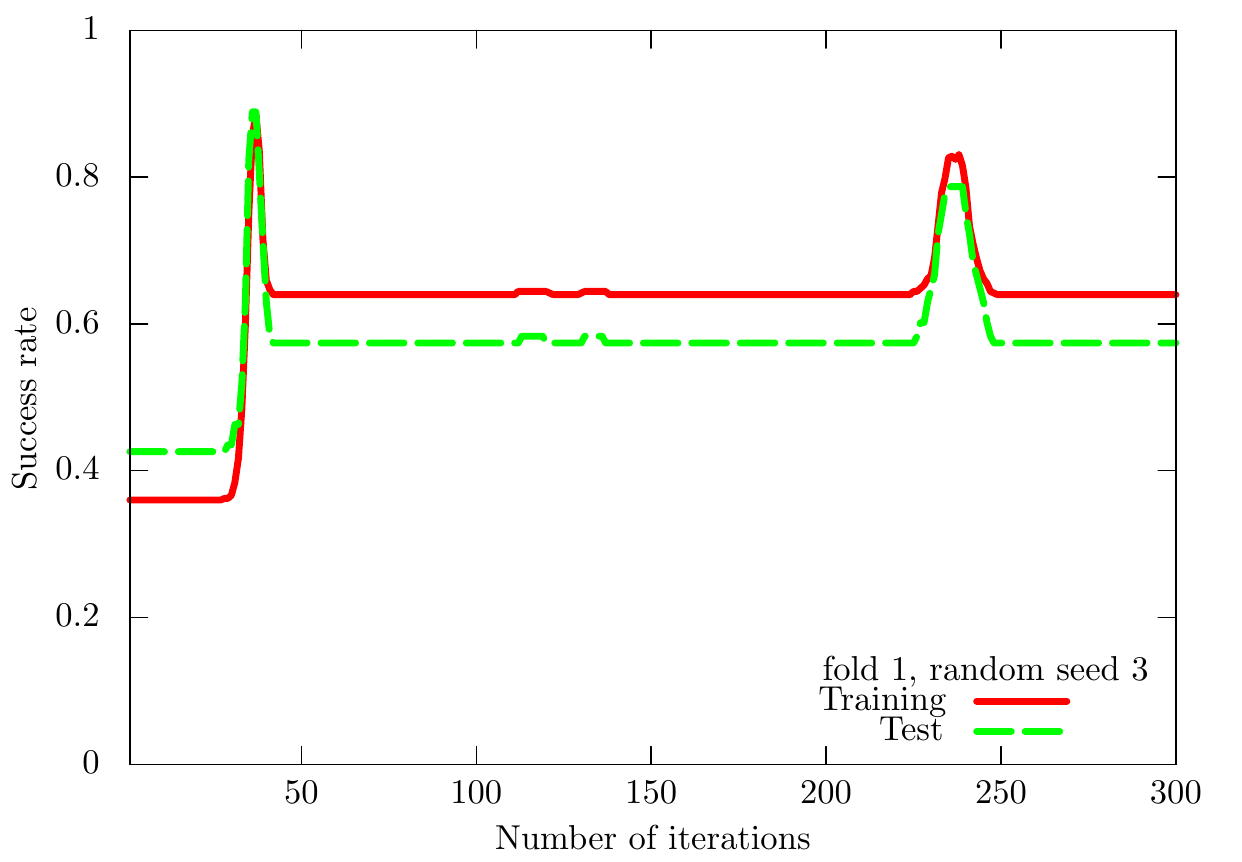}
\includegraphics[scale=0.25]{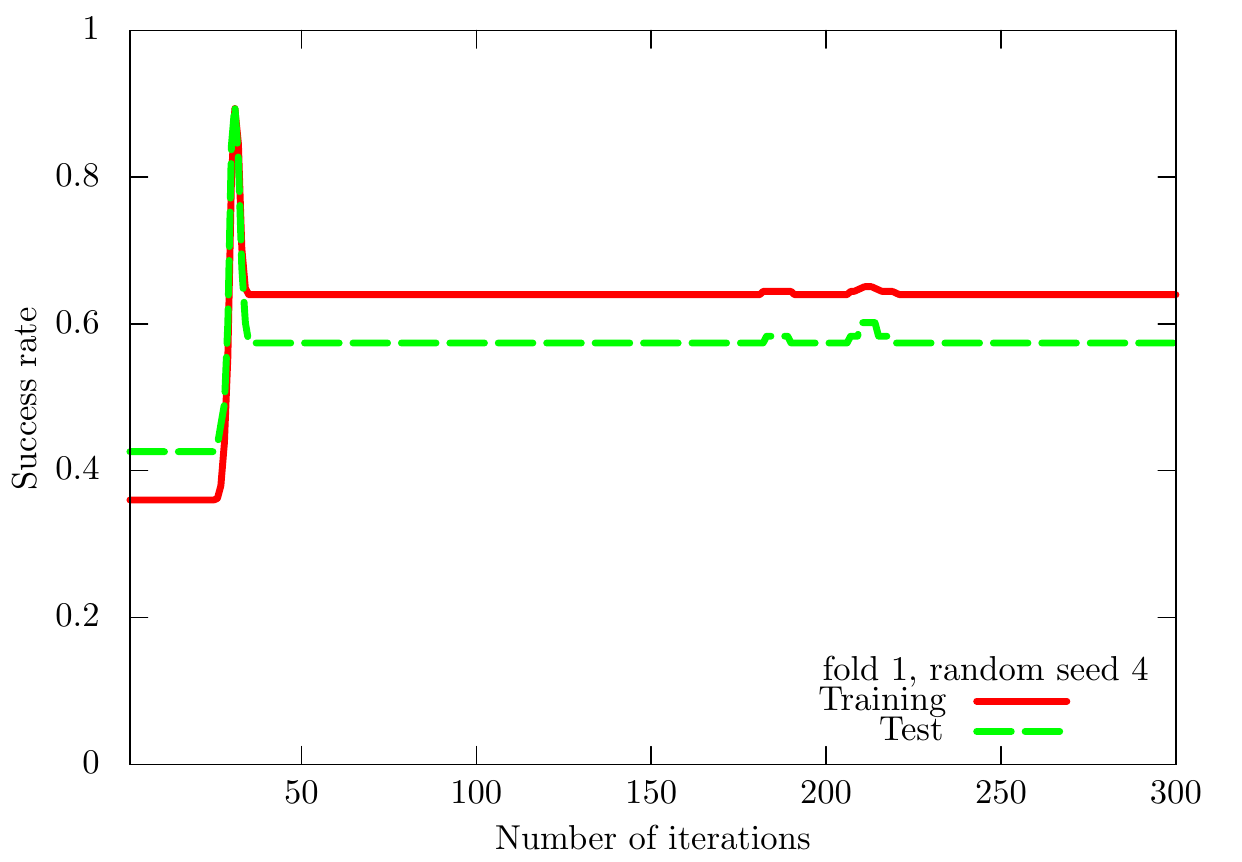}
\includegraphics[scale=0.25]{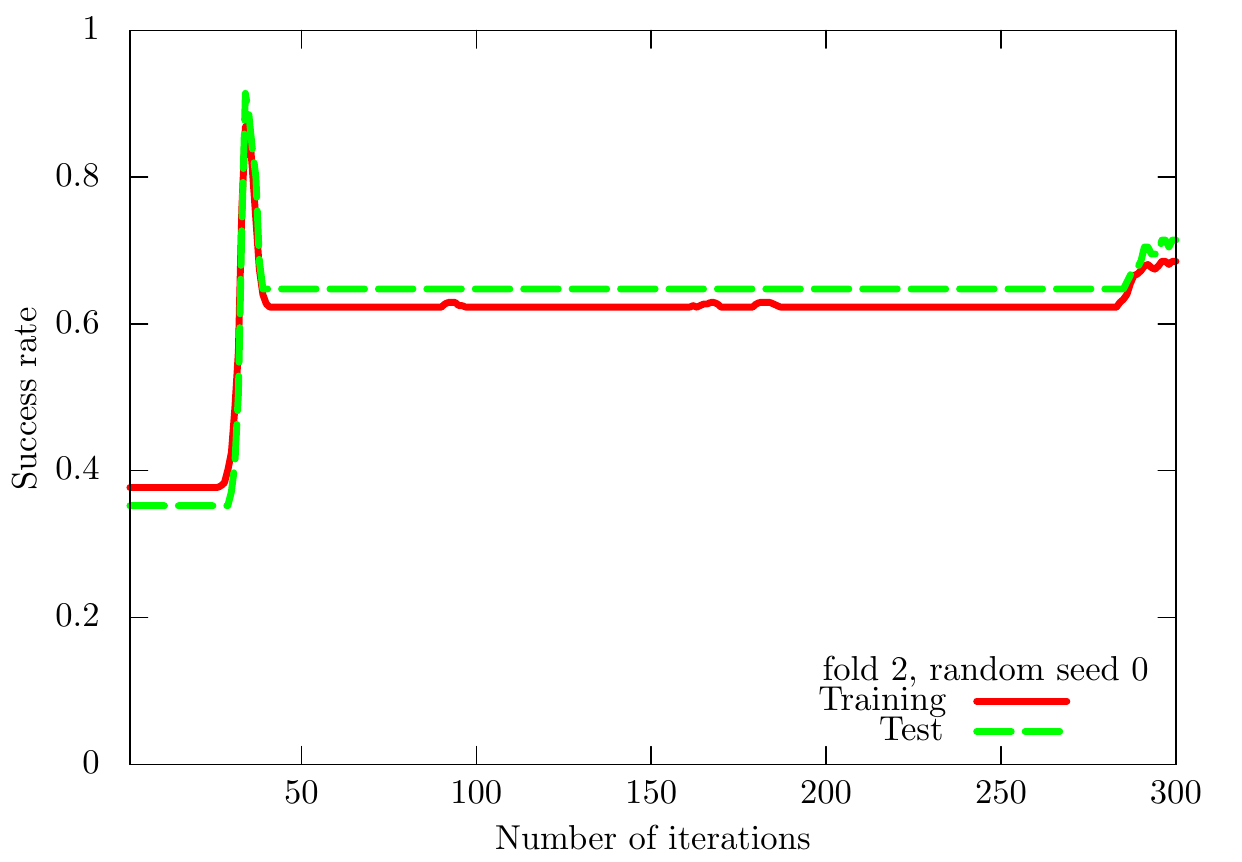}
\includegraphics[scale=0.25]{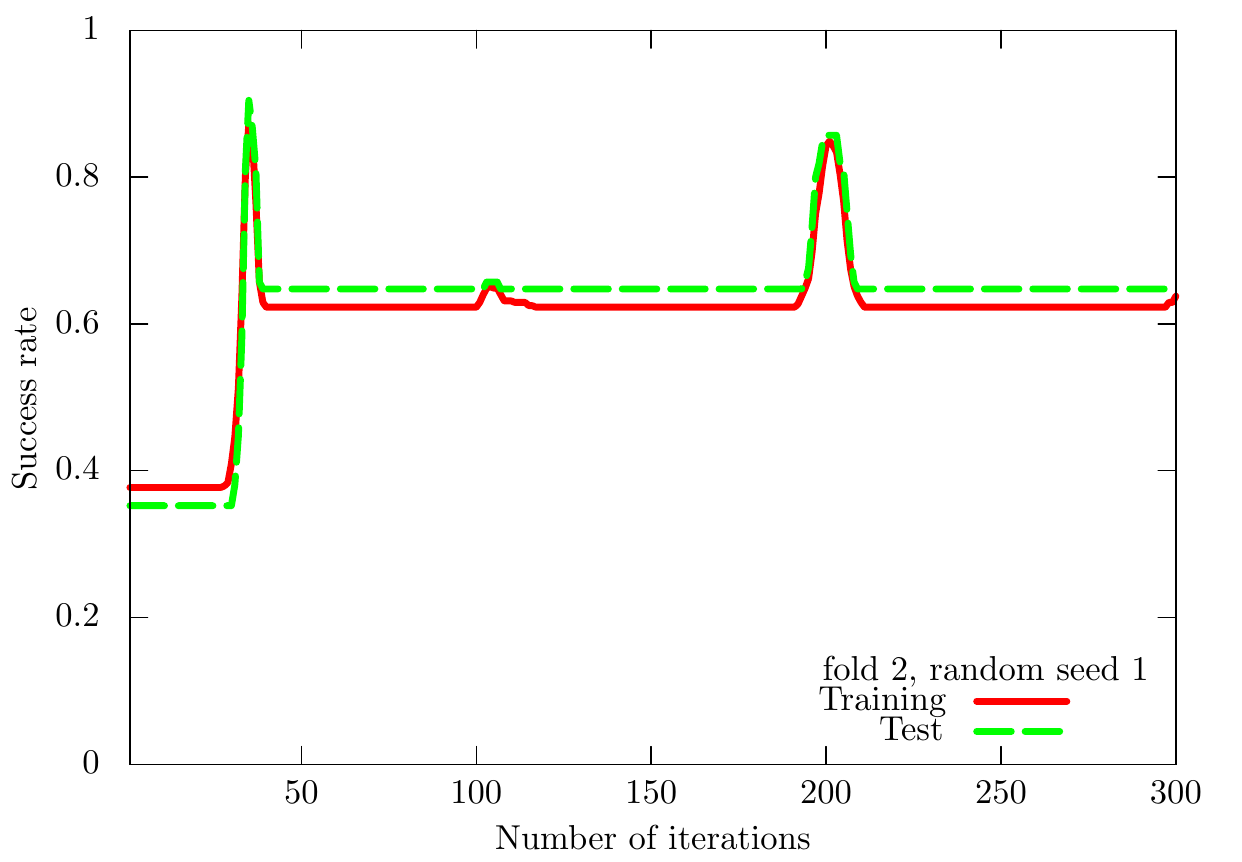}
\includegraphics[scale=0.25]{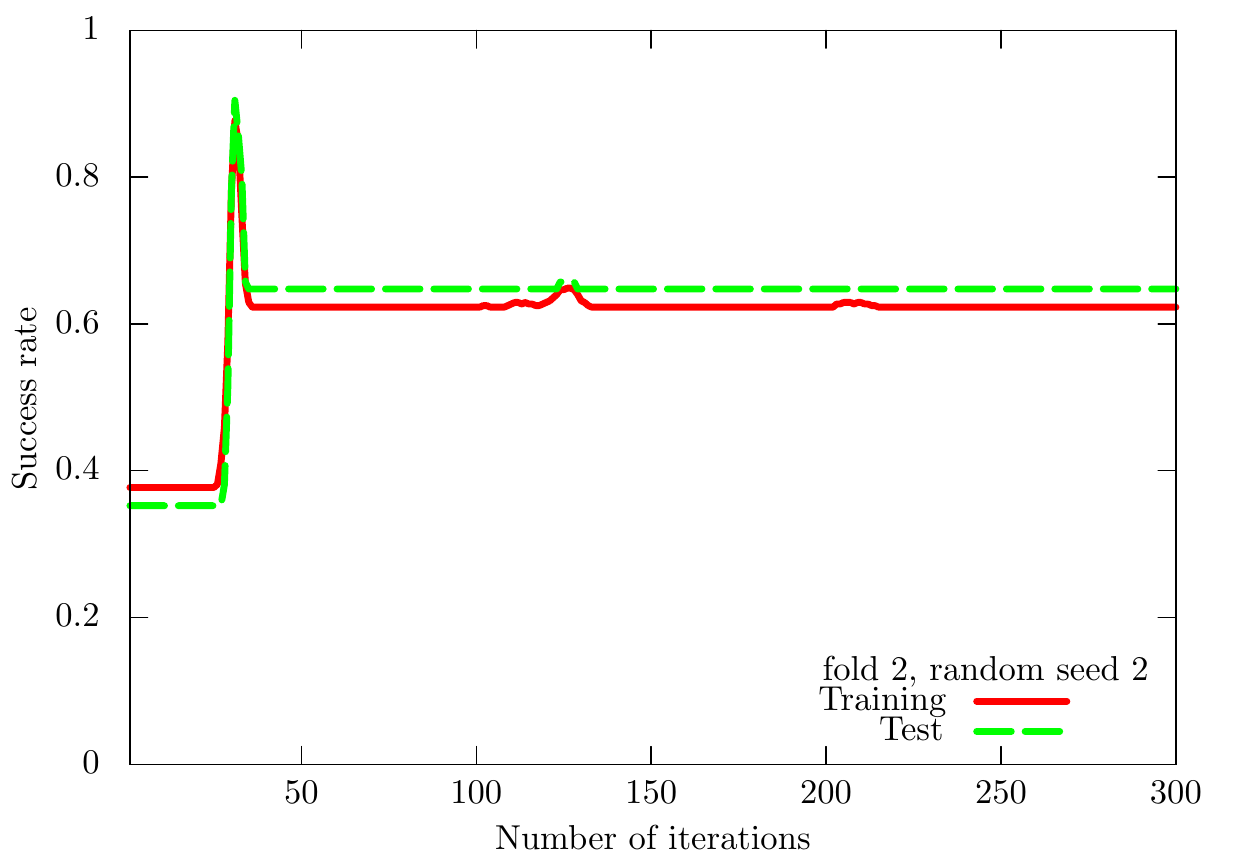}
\includegraphics[scale=0.25]{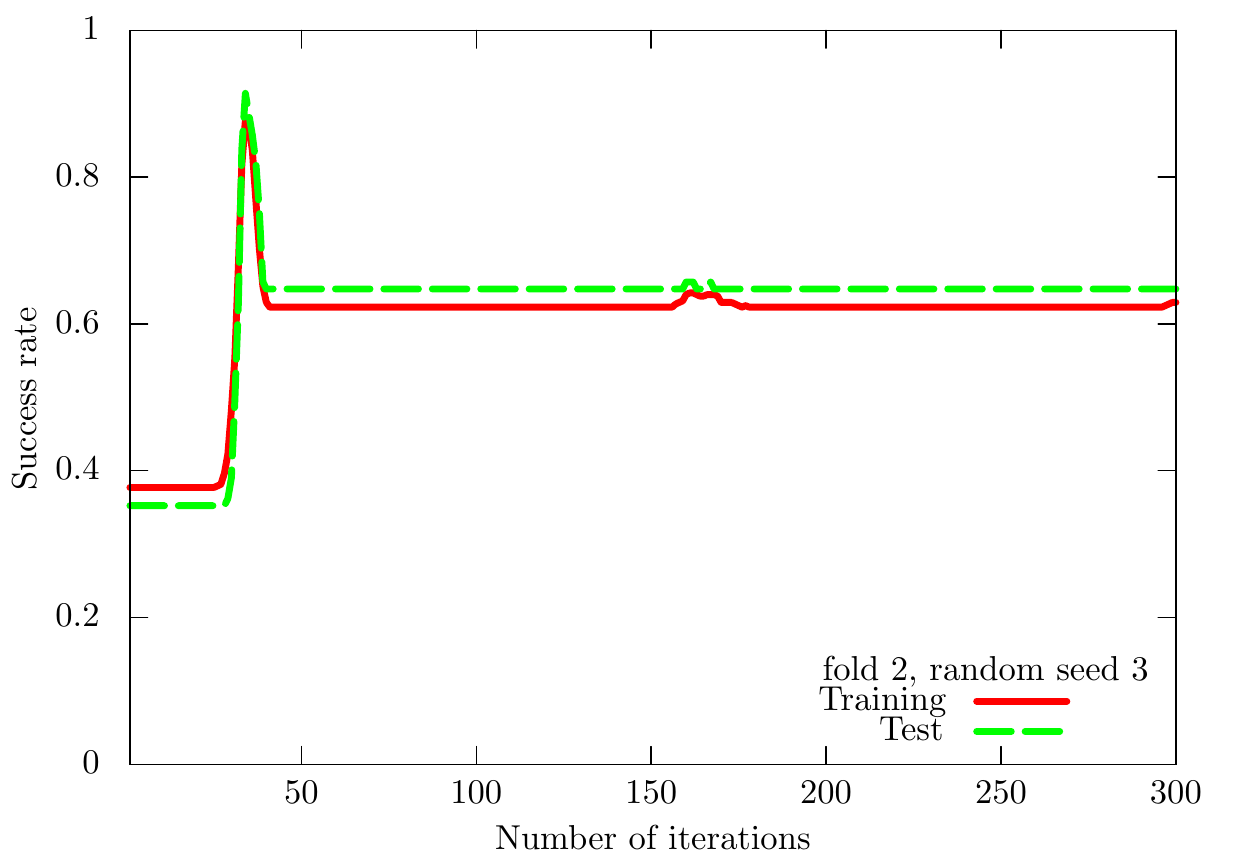}
\includegraphics[scale=0.25]{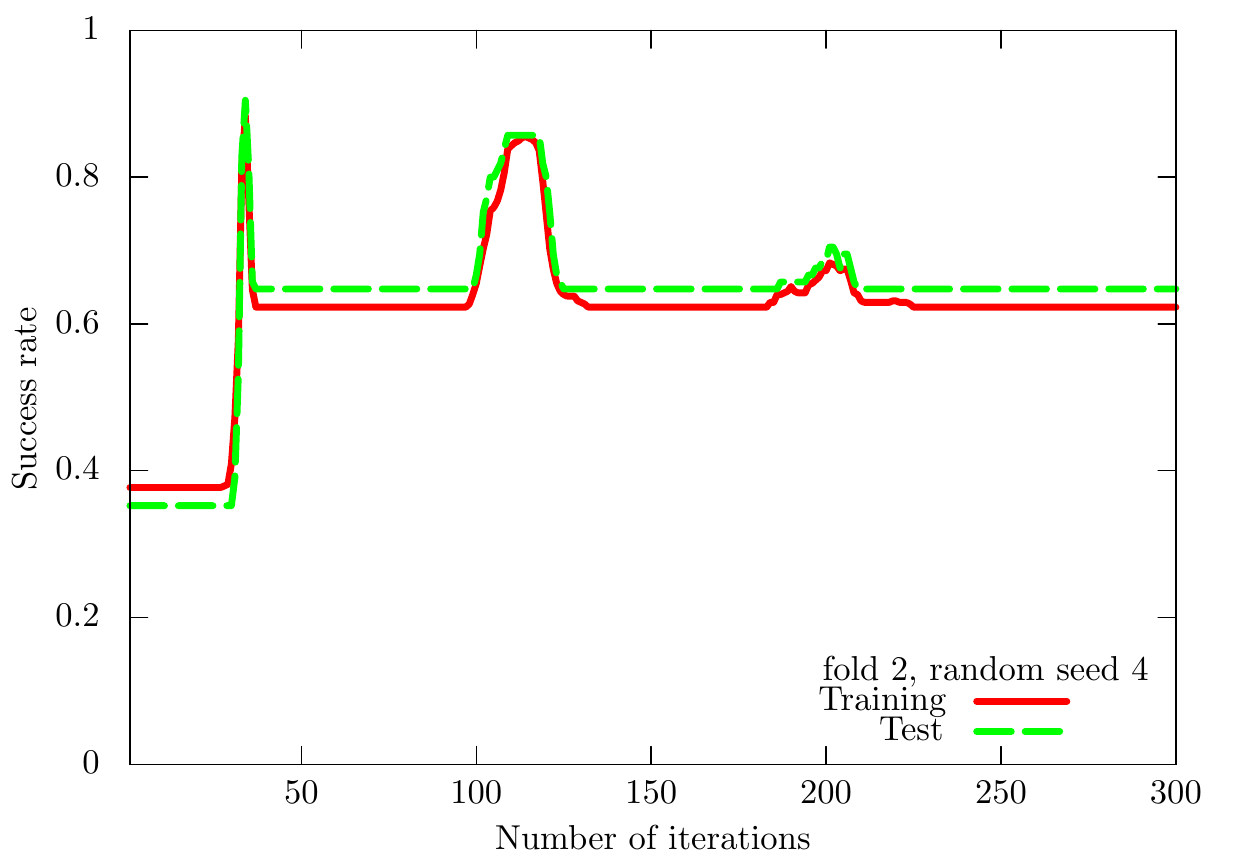}
\includegraphics[scale=0.25]{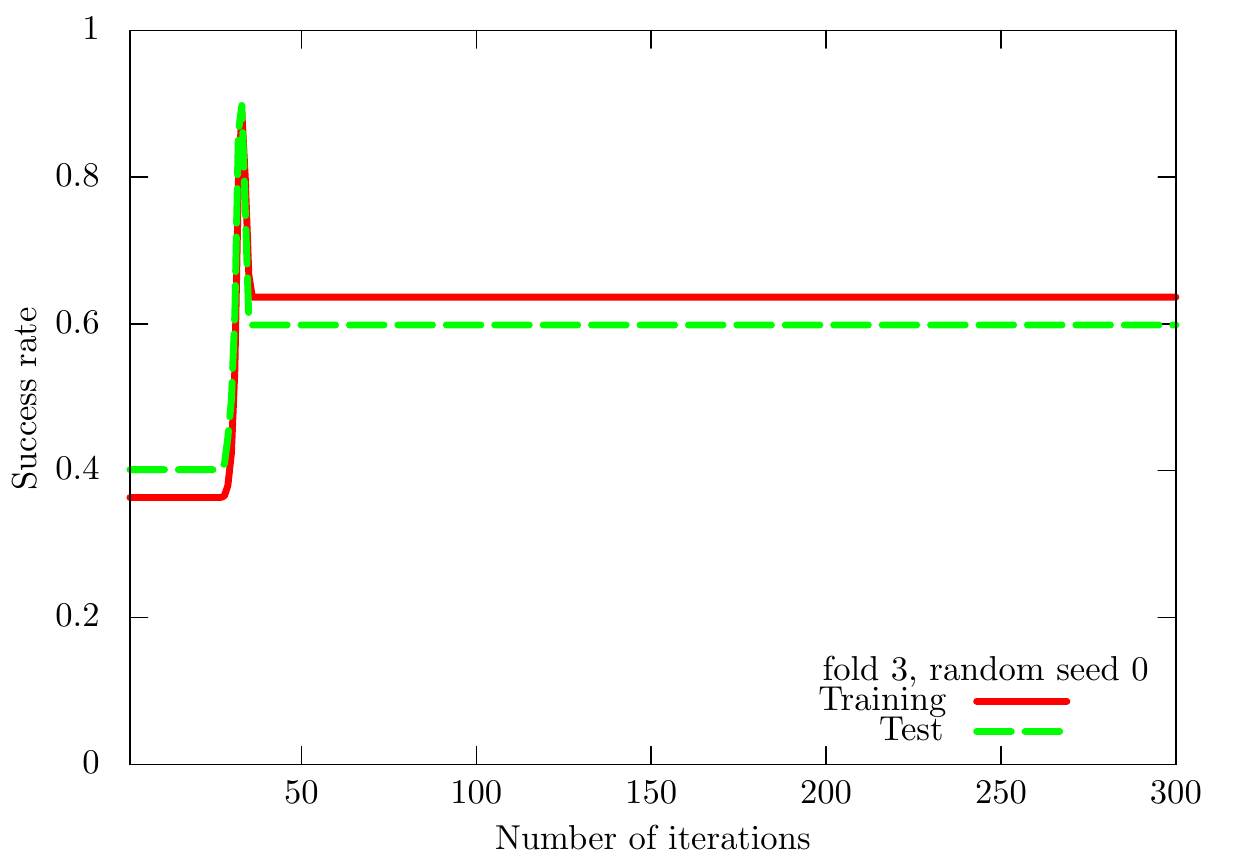}
\includegraphics[scale=0.25]{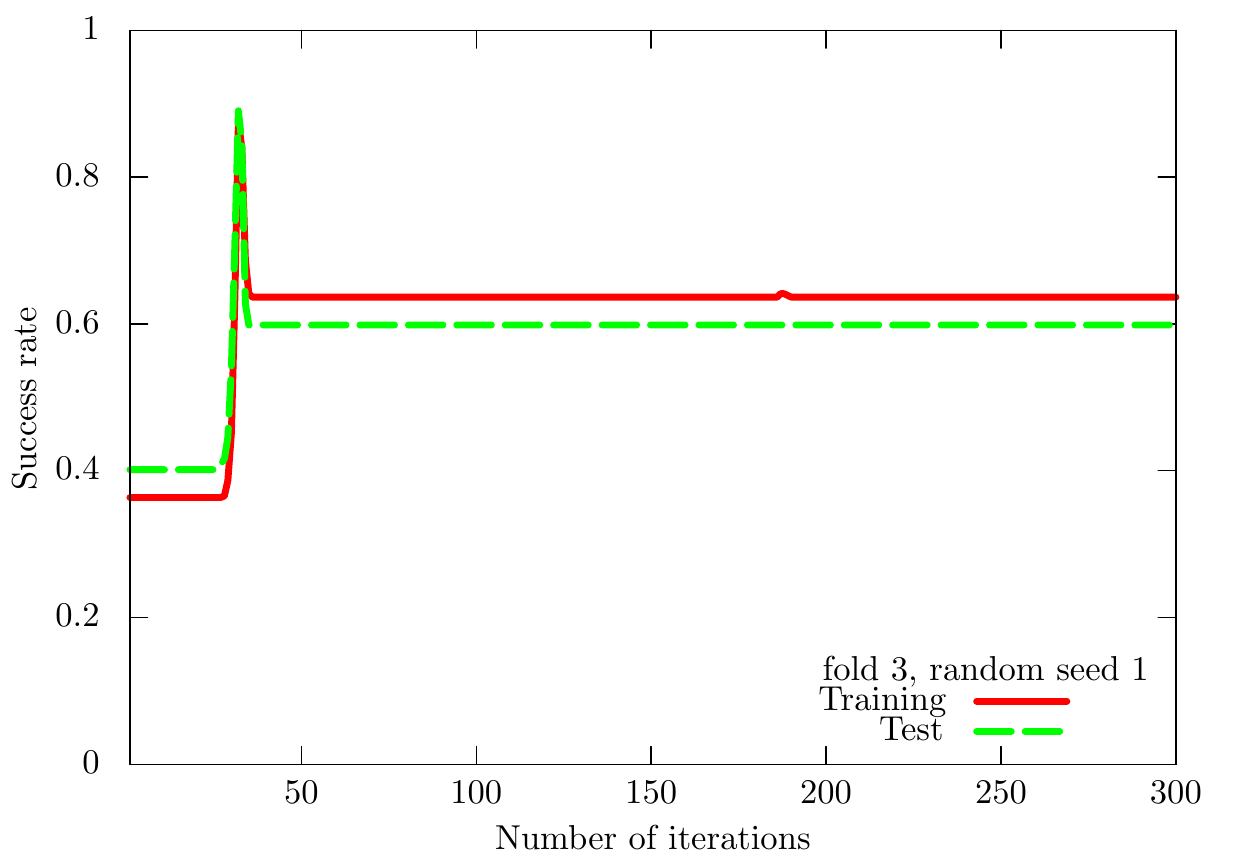}
\includegraphics[scale=0.25]{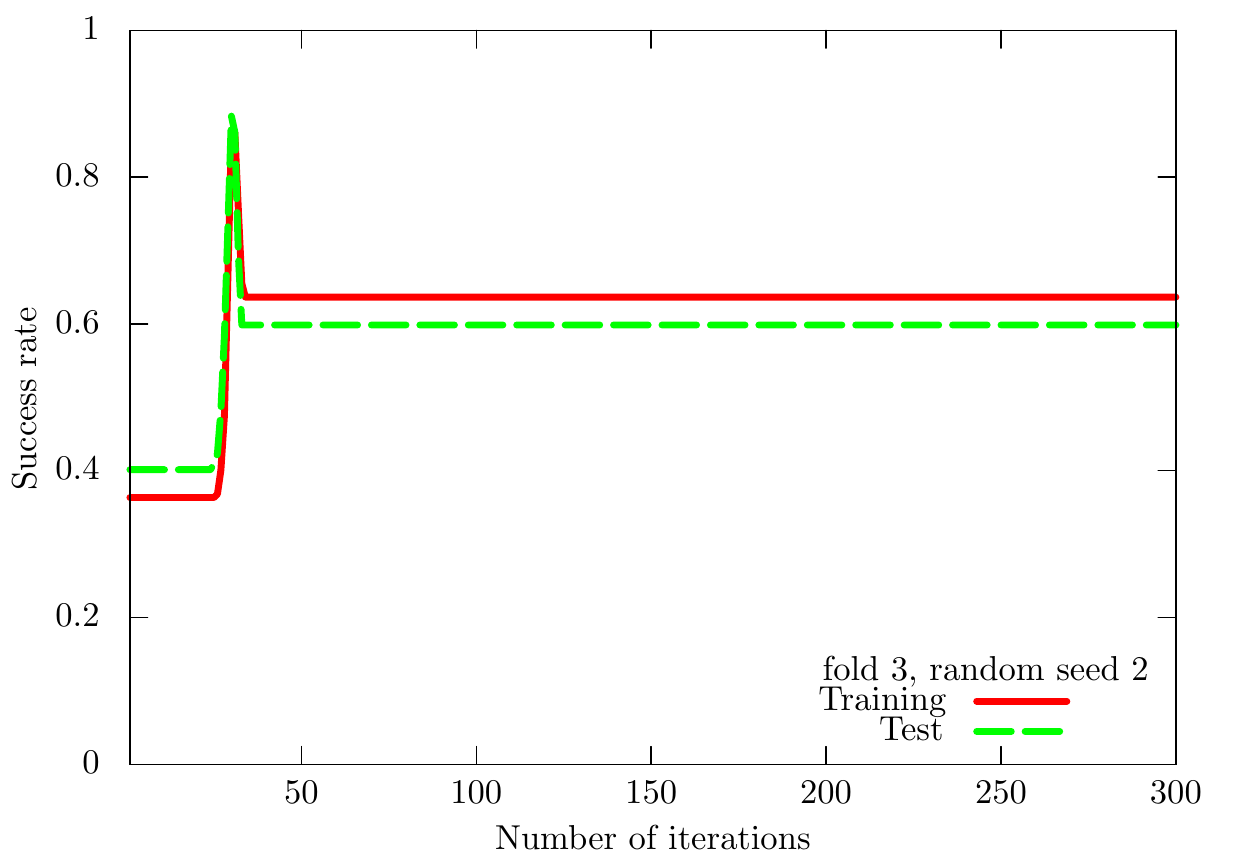}
\includegraphics[scale=0.25]{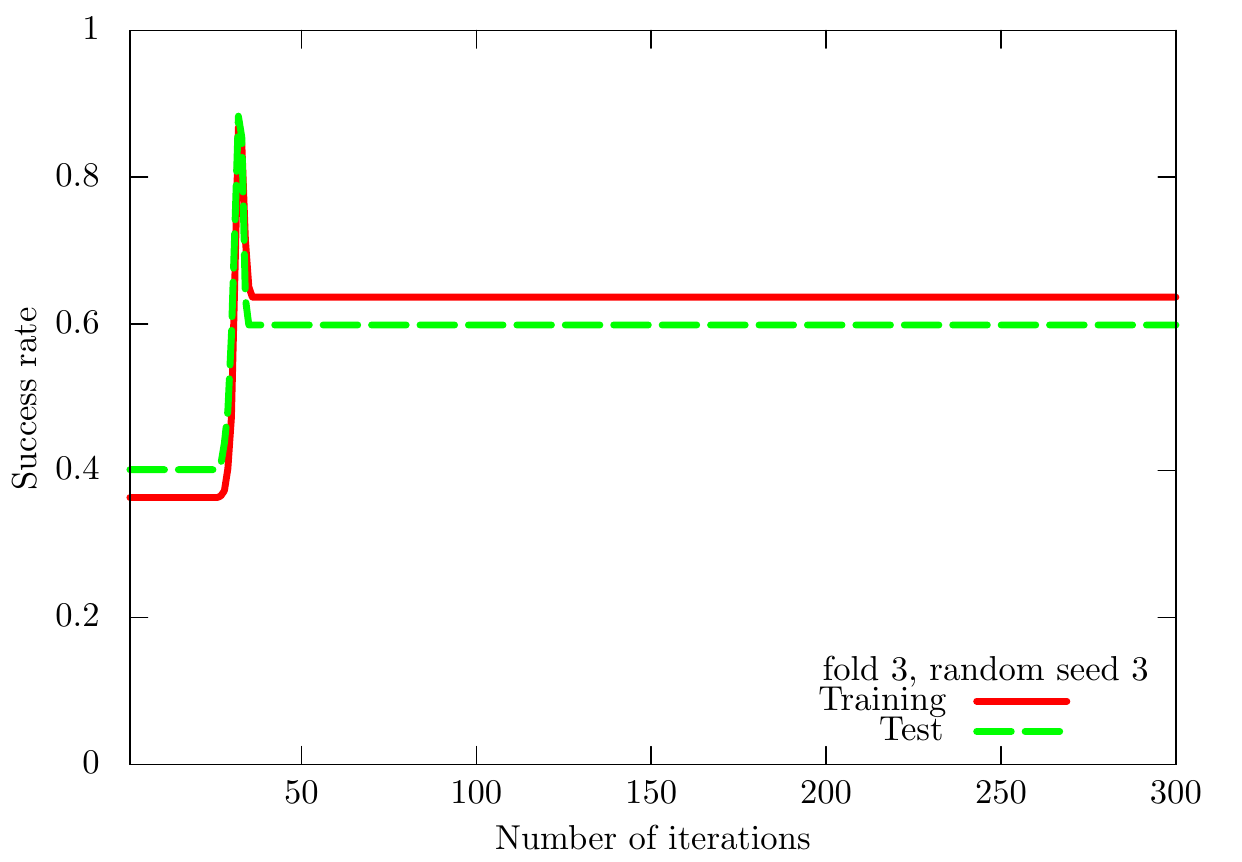}
\includegraphics[scale=0.25]{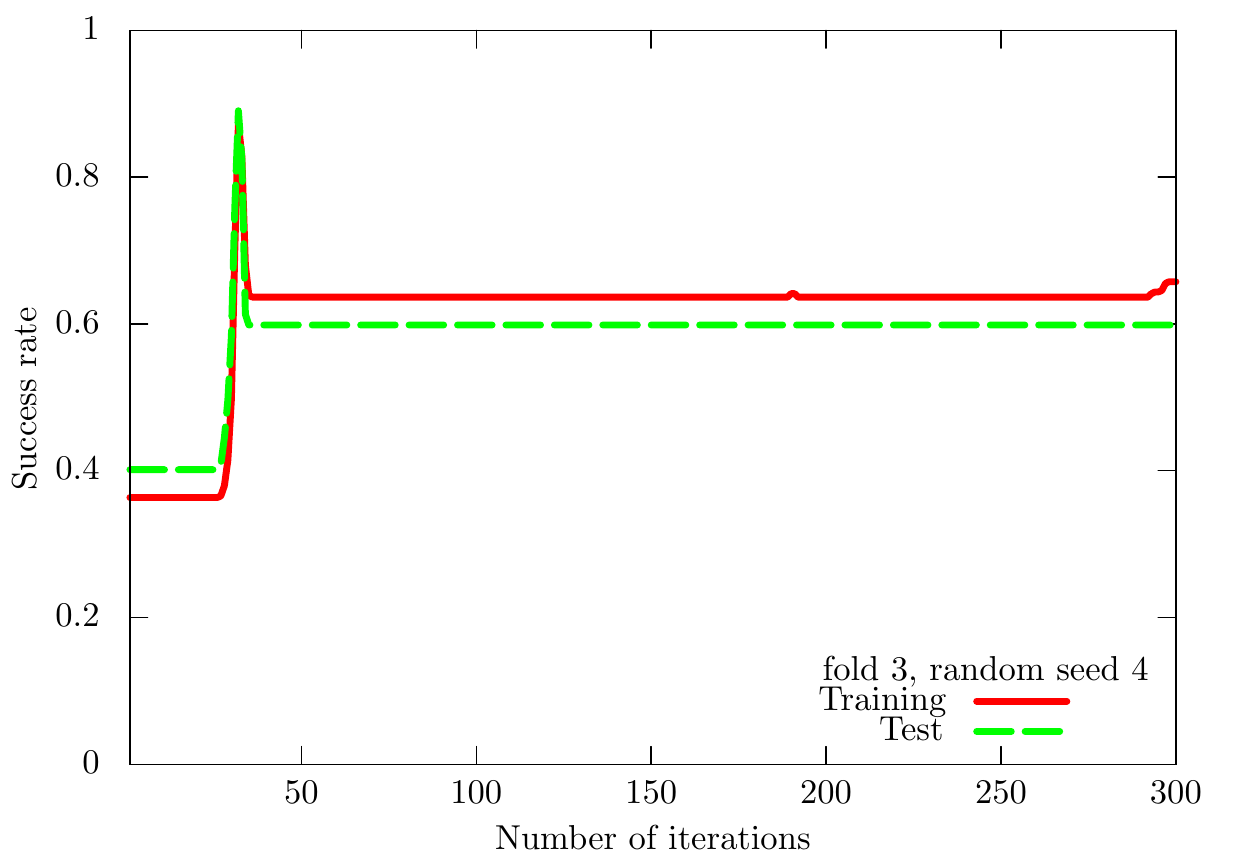}
\includegraphics[scale=0.25]{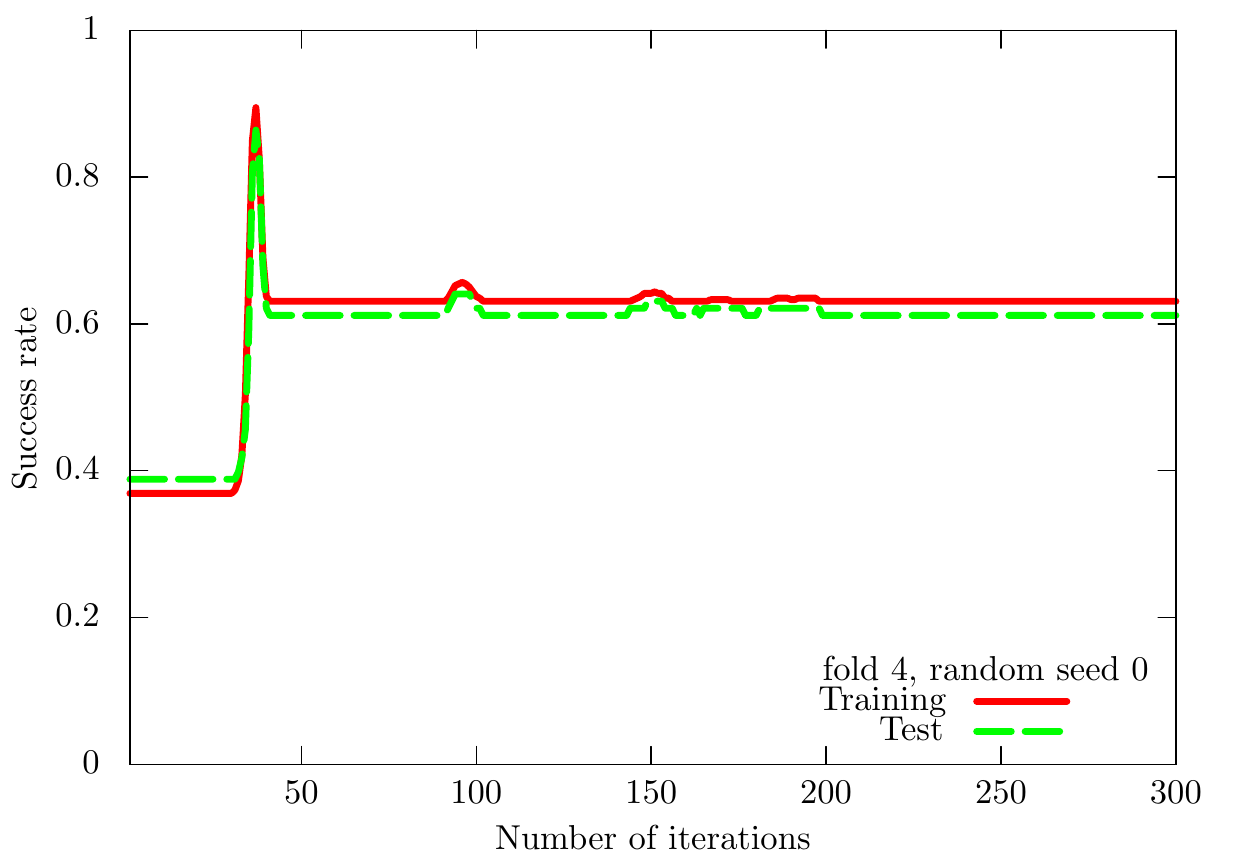}
\includegraphics[scale=0.25]{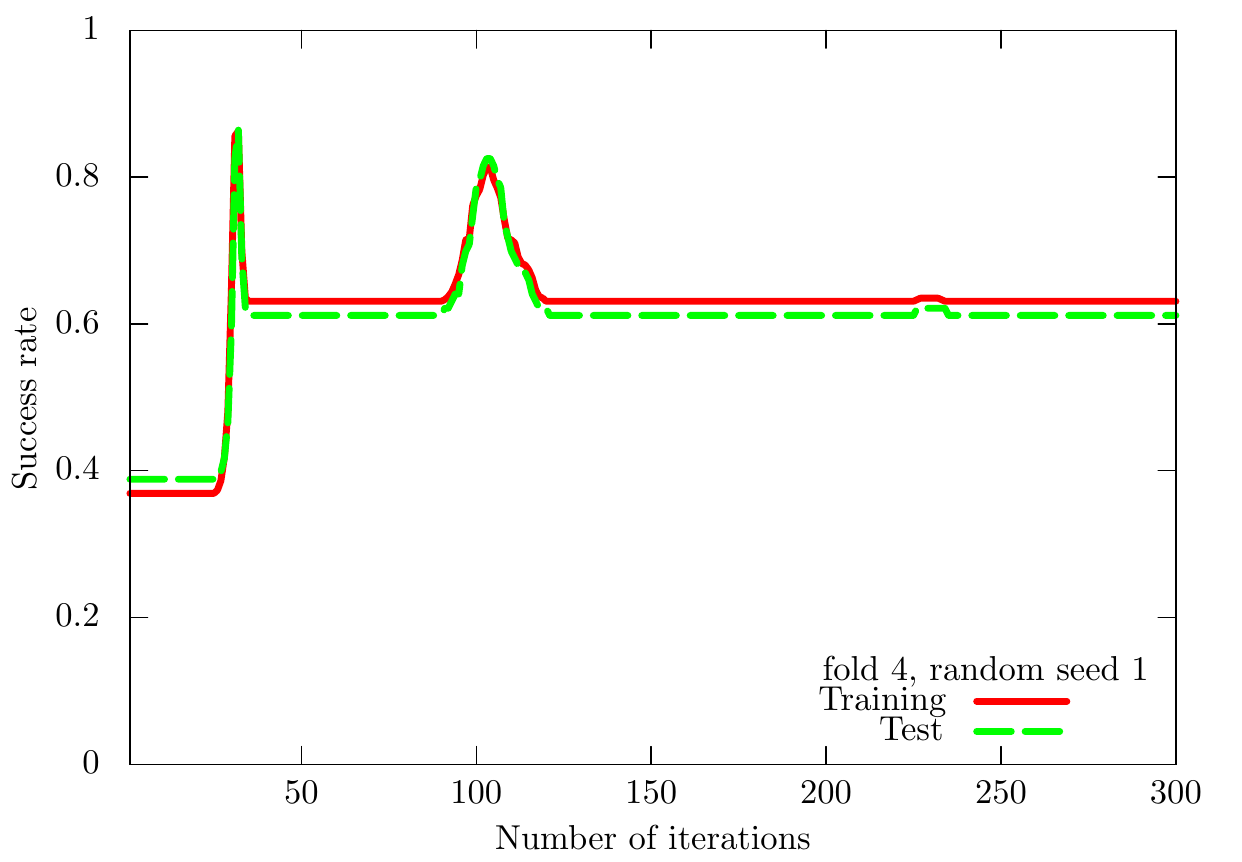}
\includegraphics[scale=0.25]{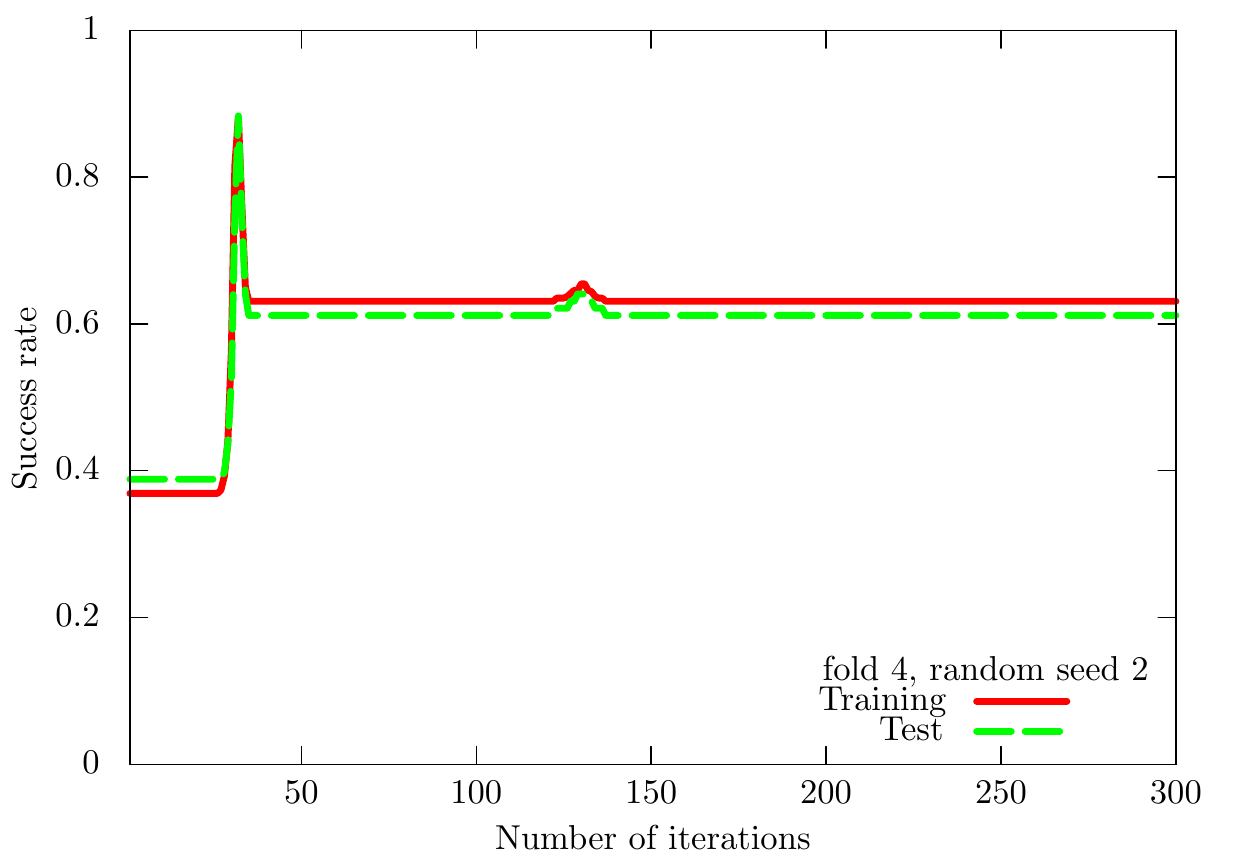}
\includegraphics[scale=0.25]{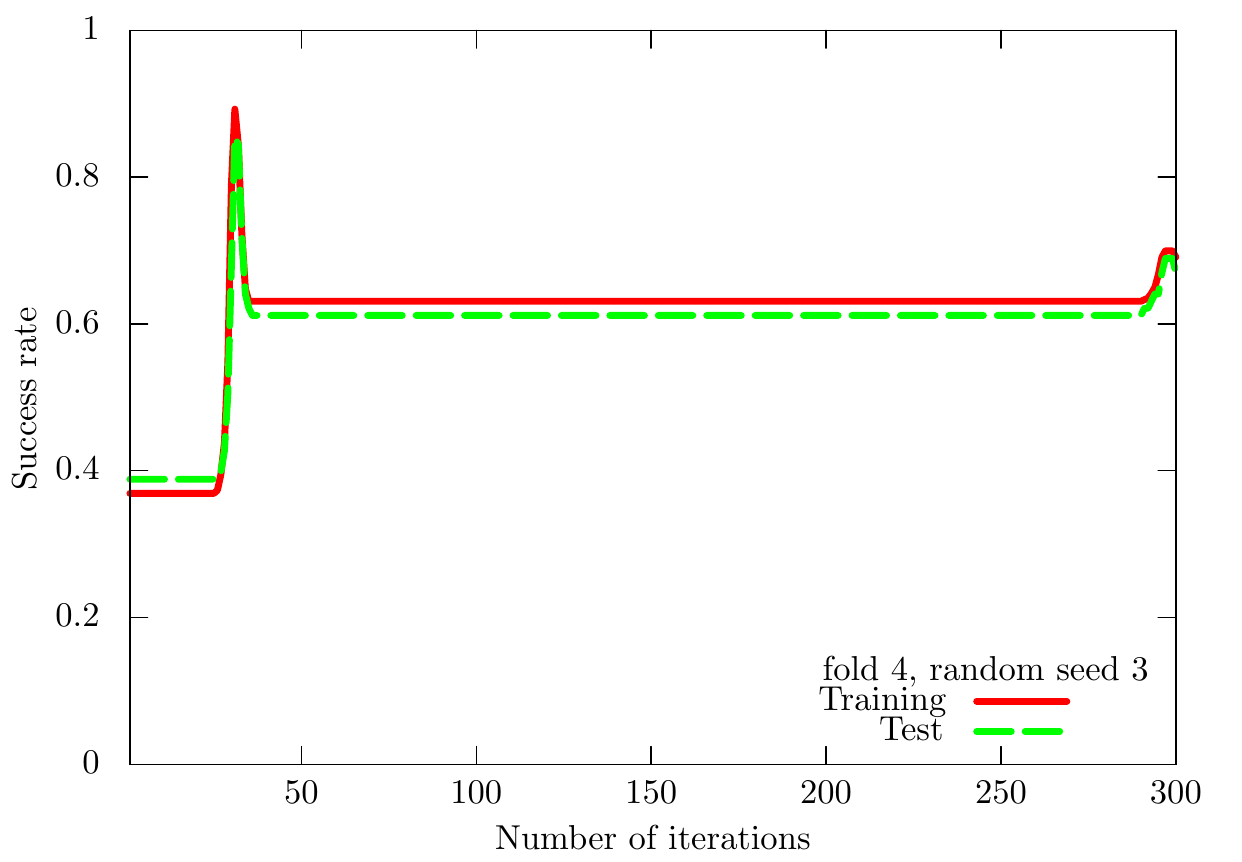}
\includegraphics[scale=0.25]{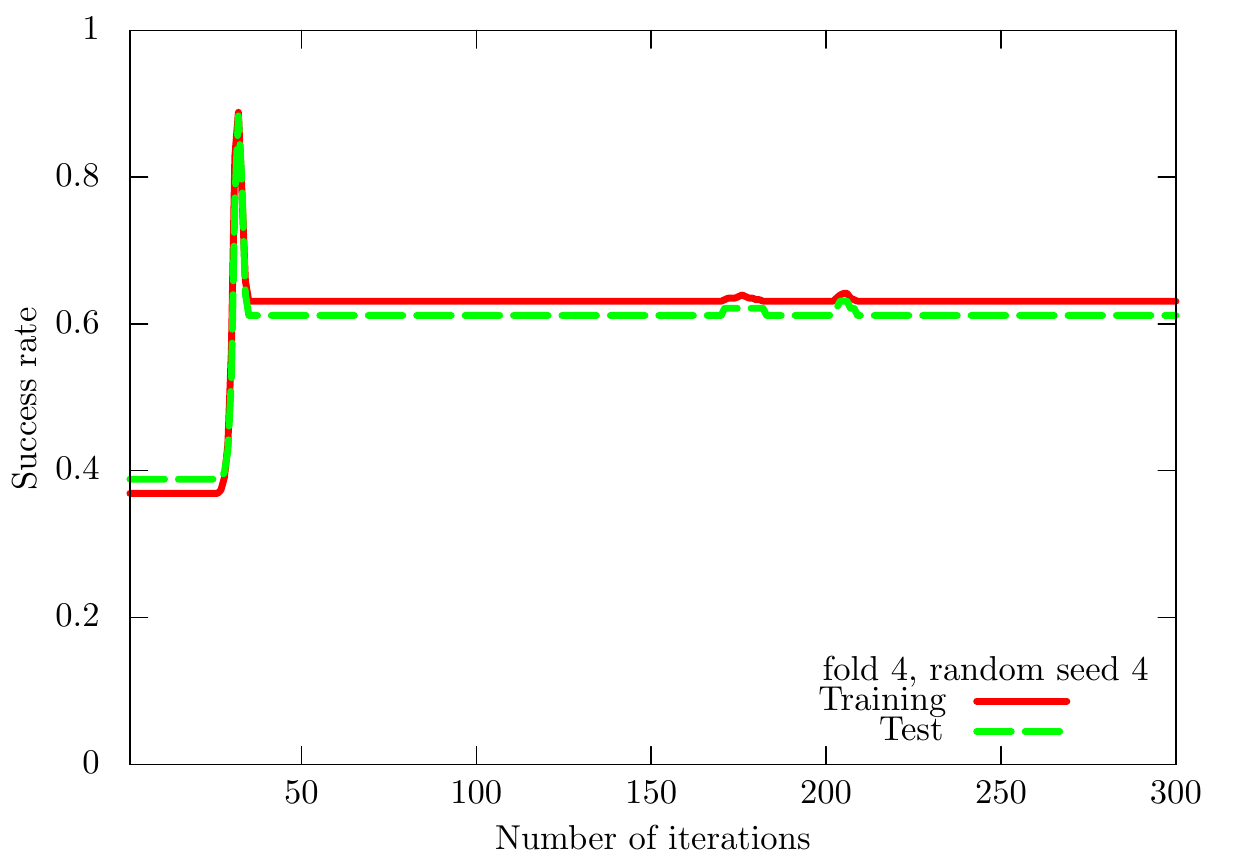}
\caption{Results of QCL on the $5$-fold datasets with $5$ different random seeds for the cancer dataset ($0$ or $1$). We use the CNOT-based circuit and set $\theta_\mathrm{bias} = 0$. The number of layers $L$ is set to $5$.}
\label{supp-arXiv-numerical-result-raw-data-fold-001-rand-001-QCL-UCI-cancer-0-1}
\end{figure*}

\begin{figure*}[htb]
\centering
\includegraphics[scale=0.25]{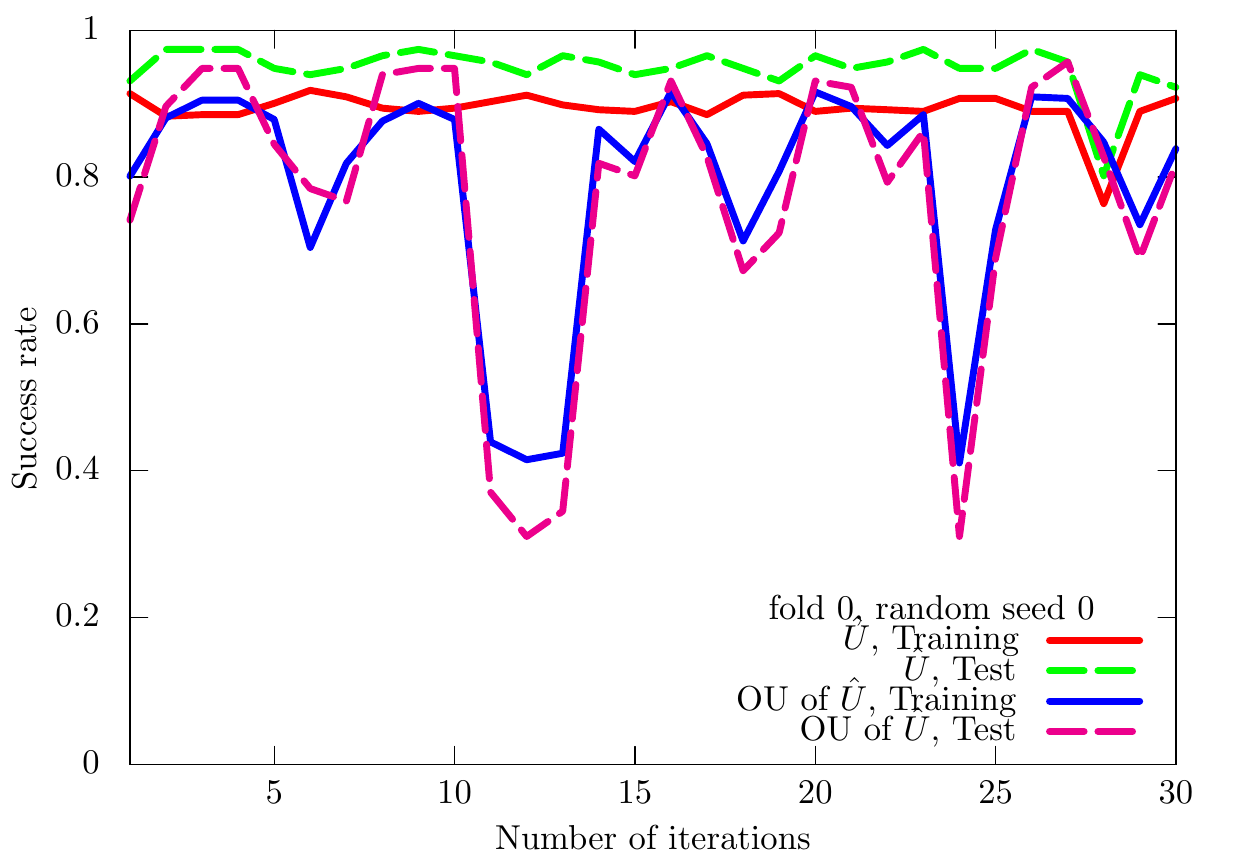}
\includegraphics[scale=0.25]{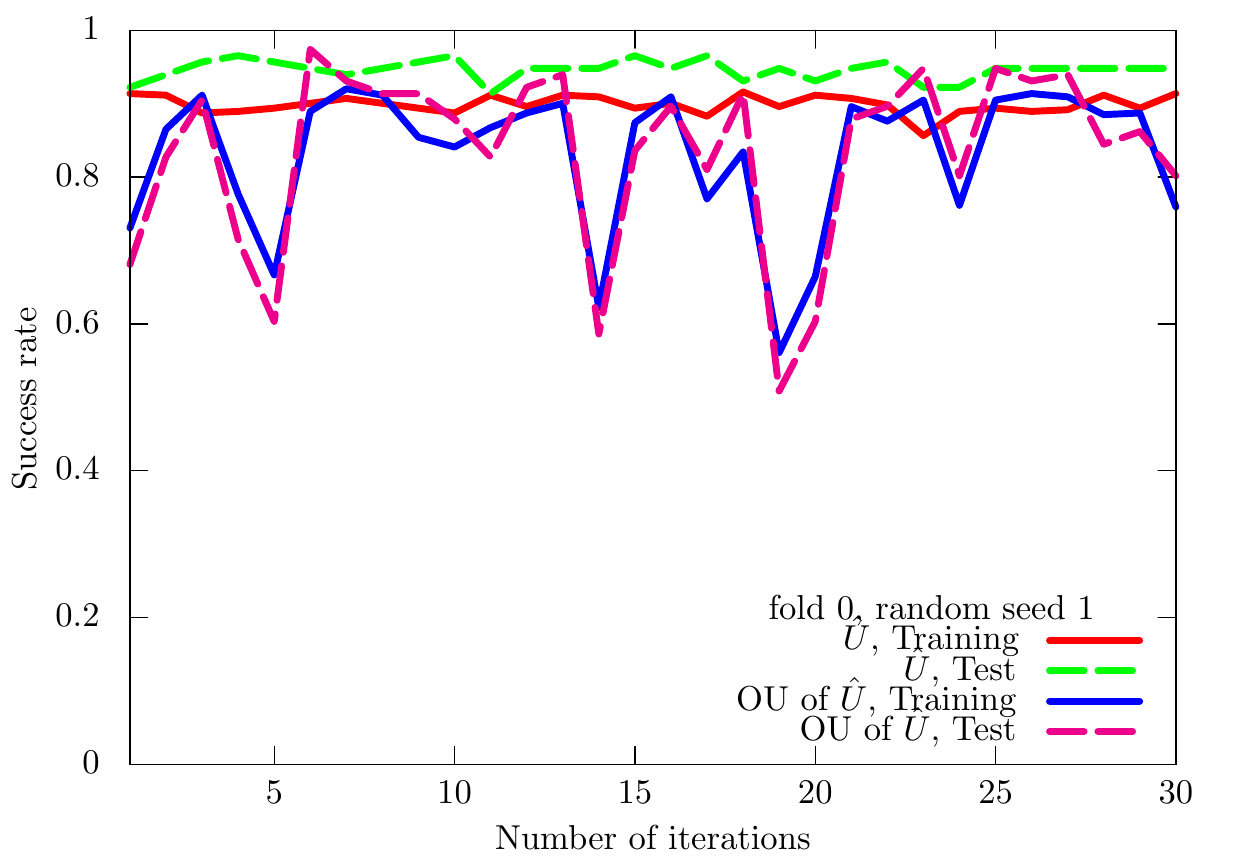}
\includegraphics[scale=0.25]{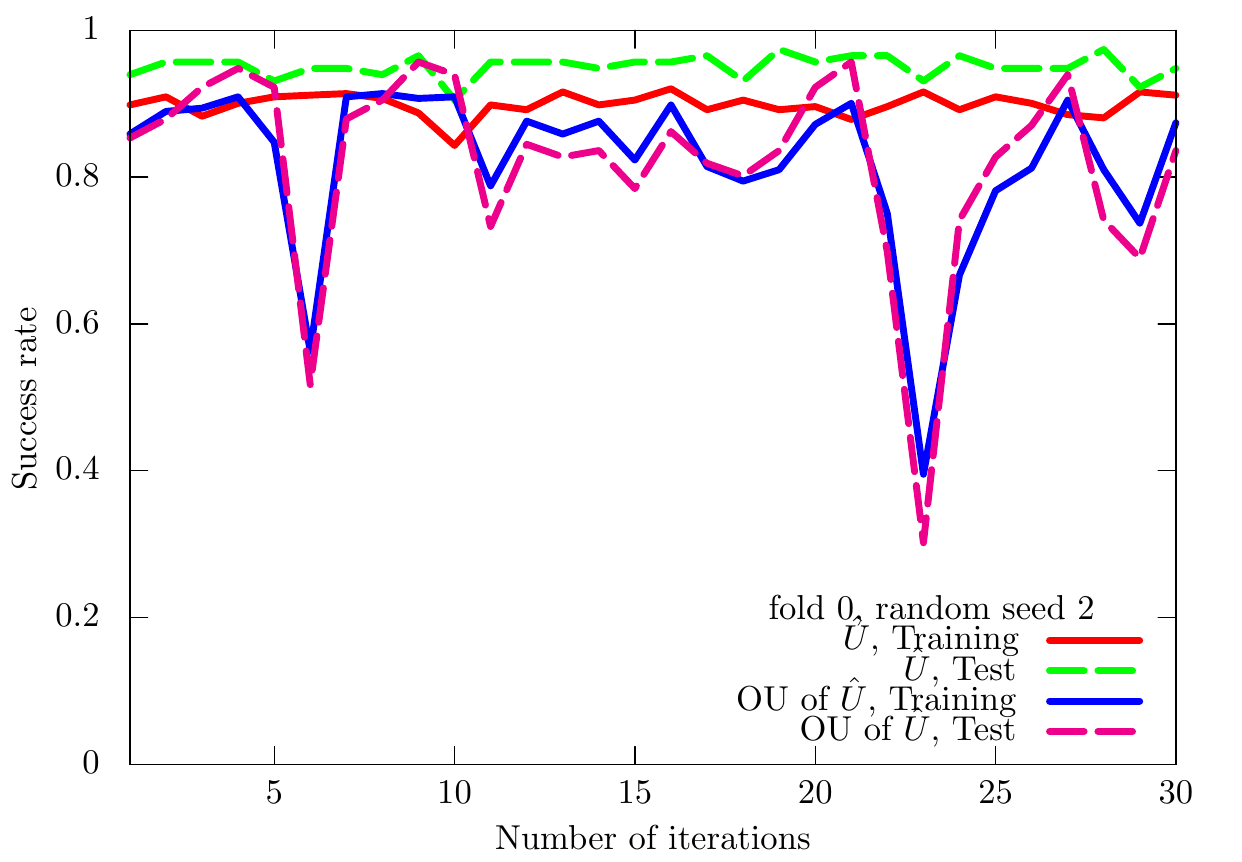}
\includegraphics[scale=0.25]{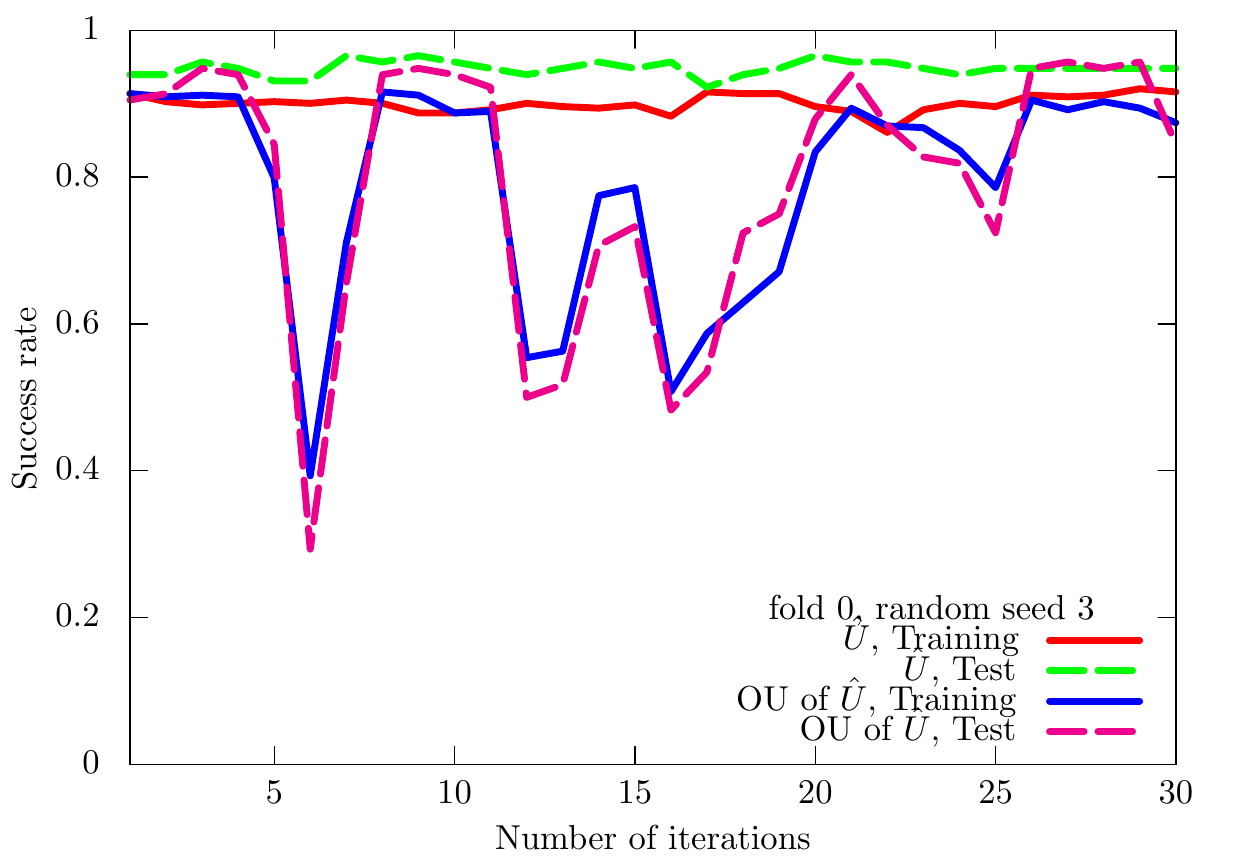}
\includegraphics[scale=0.25]{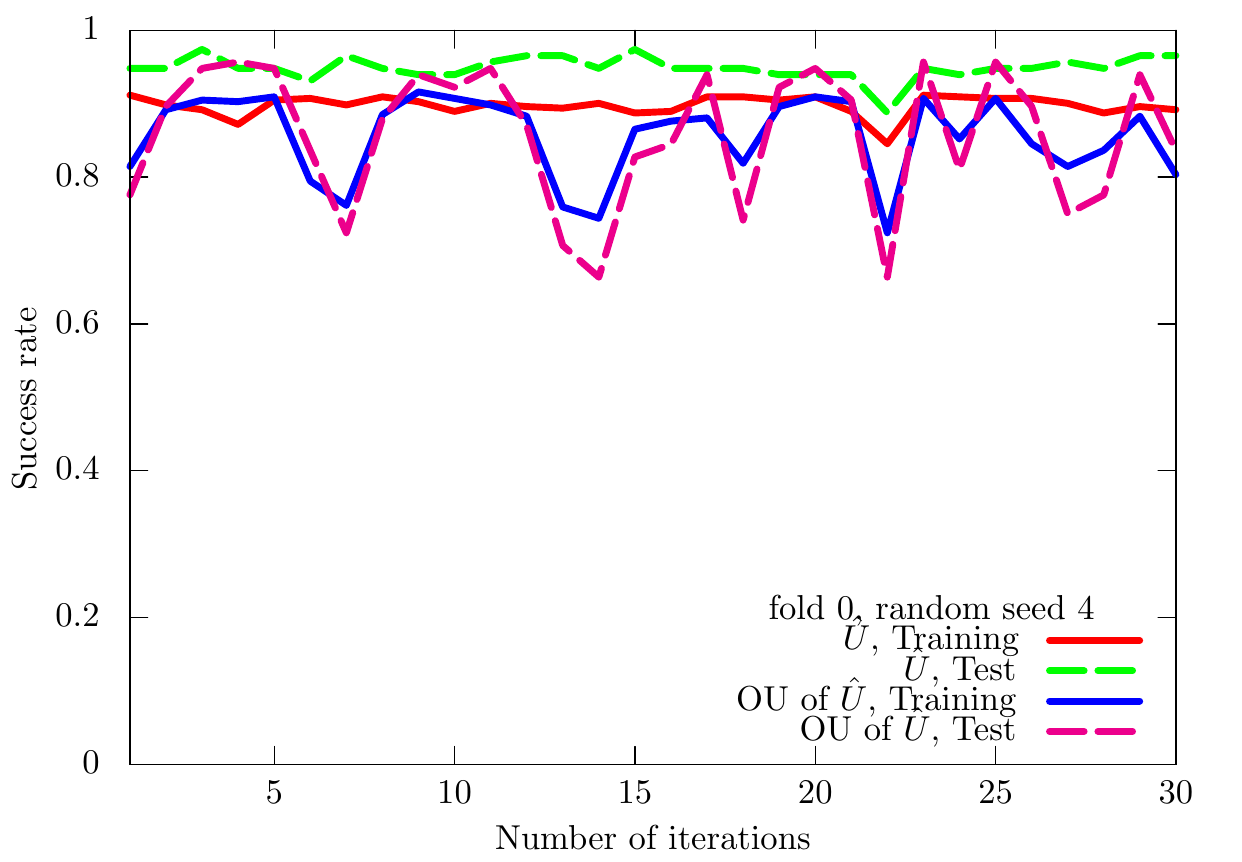}
\includegraphics[scale=0.25]{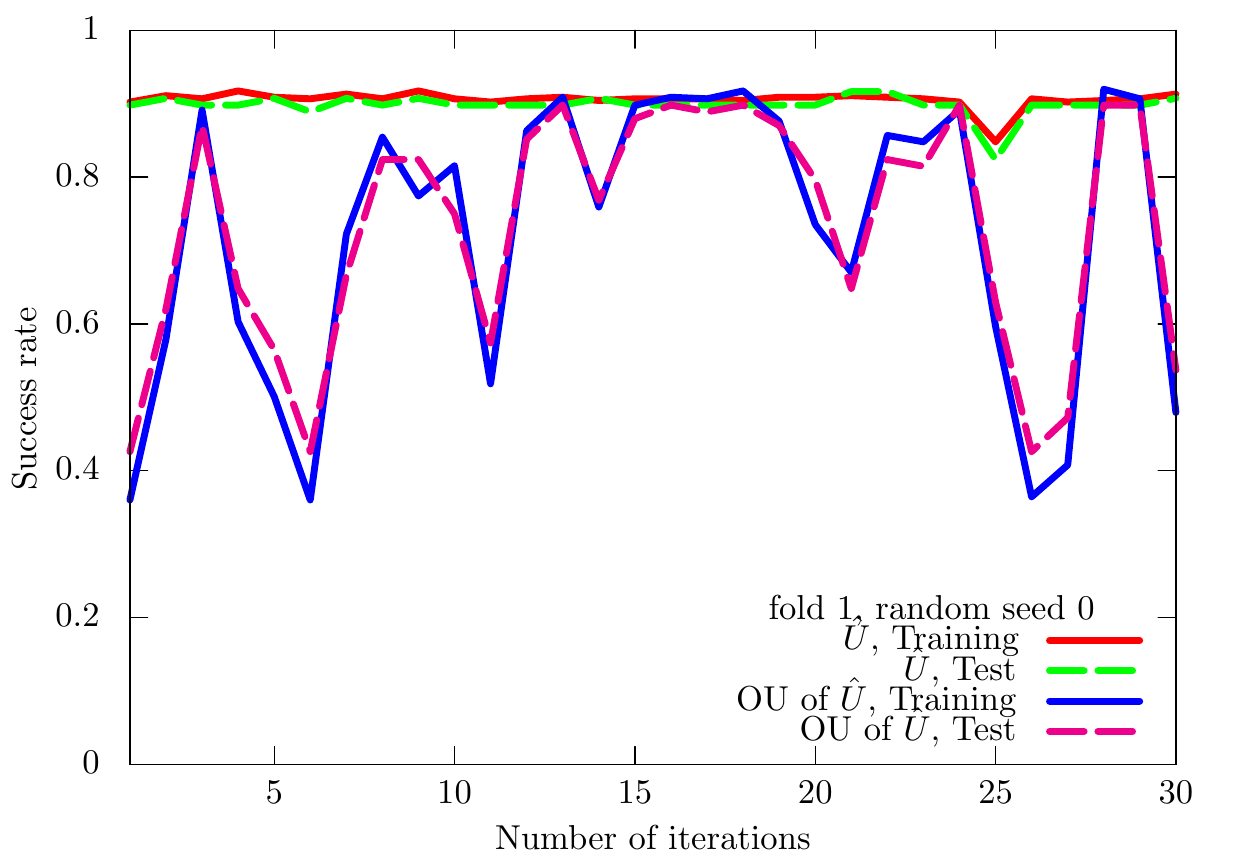}
\includegraphics[scale=0.25]{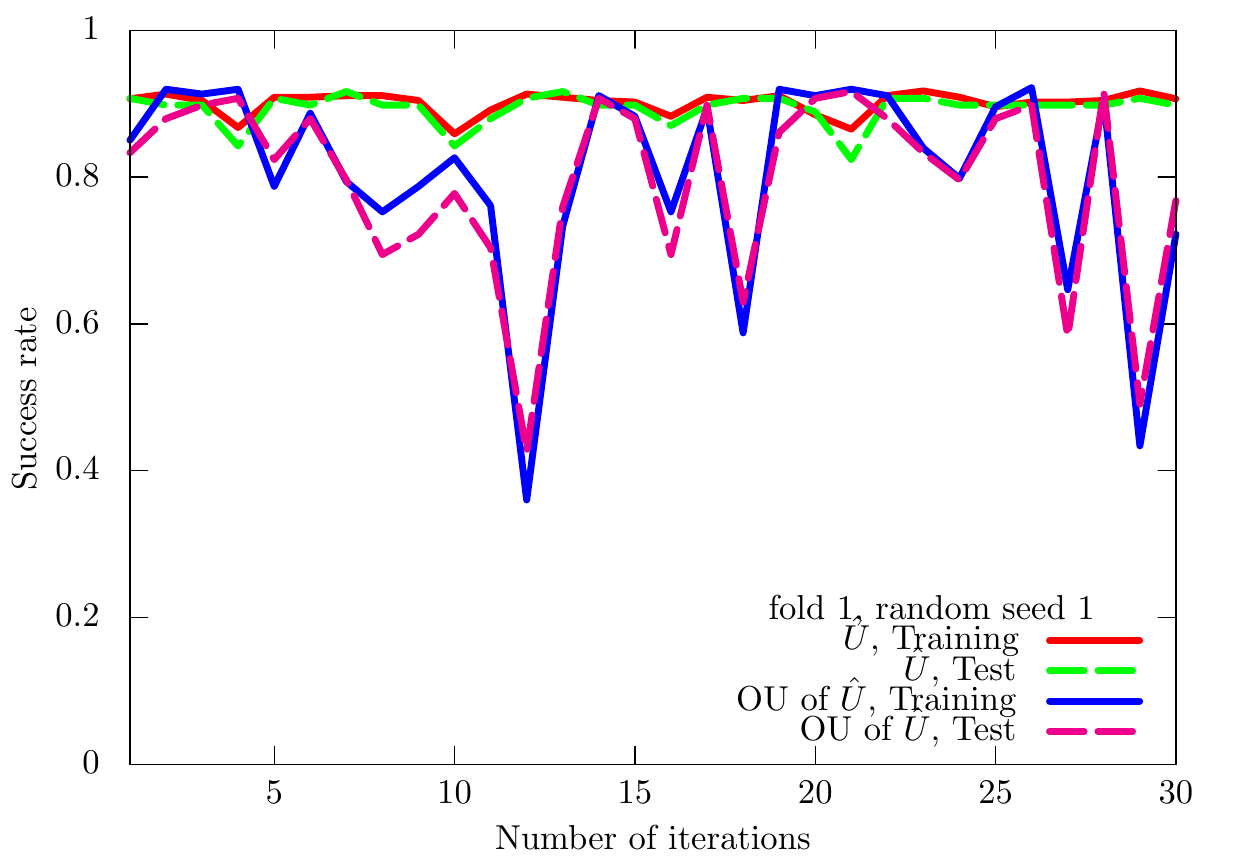}
\includegraphics[scale=0.25]{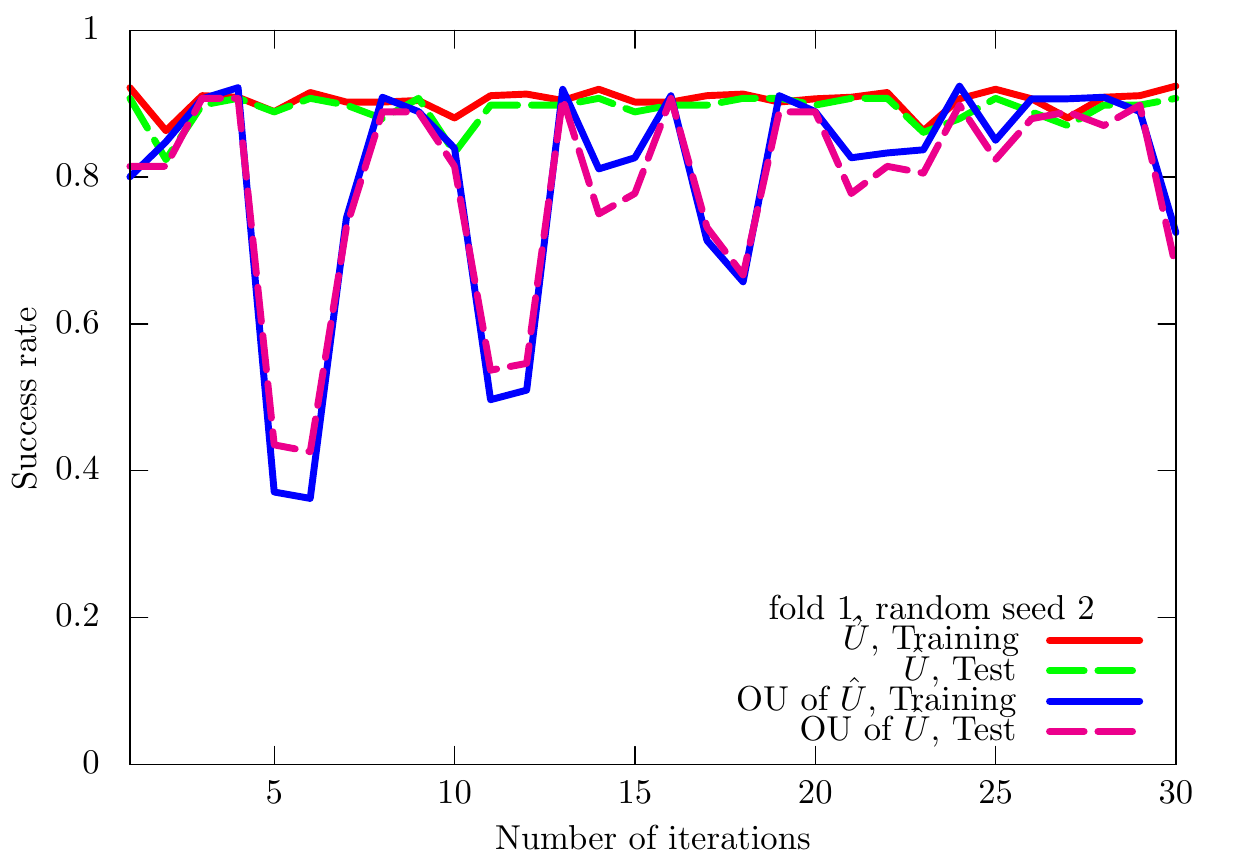}
\includegraphics[scale=0.25]{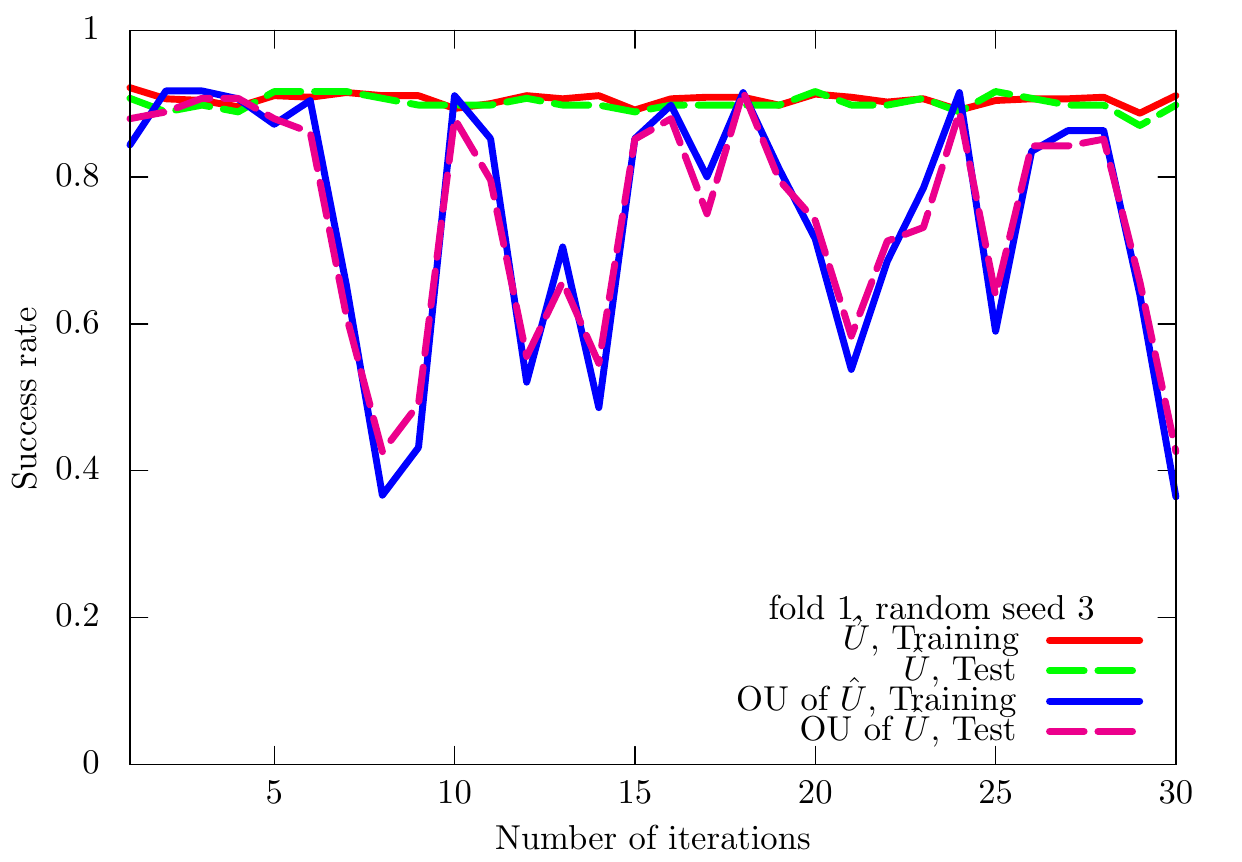}
\includegraphics[scale=0.25]{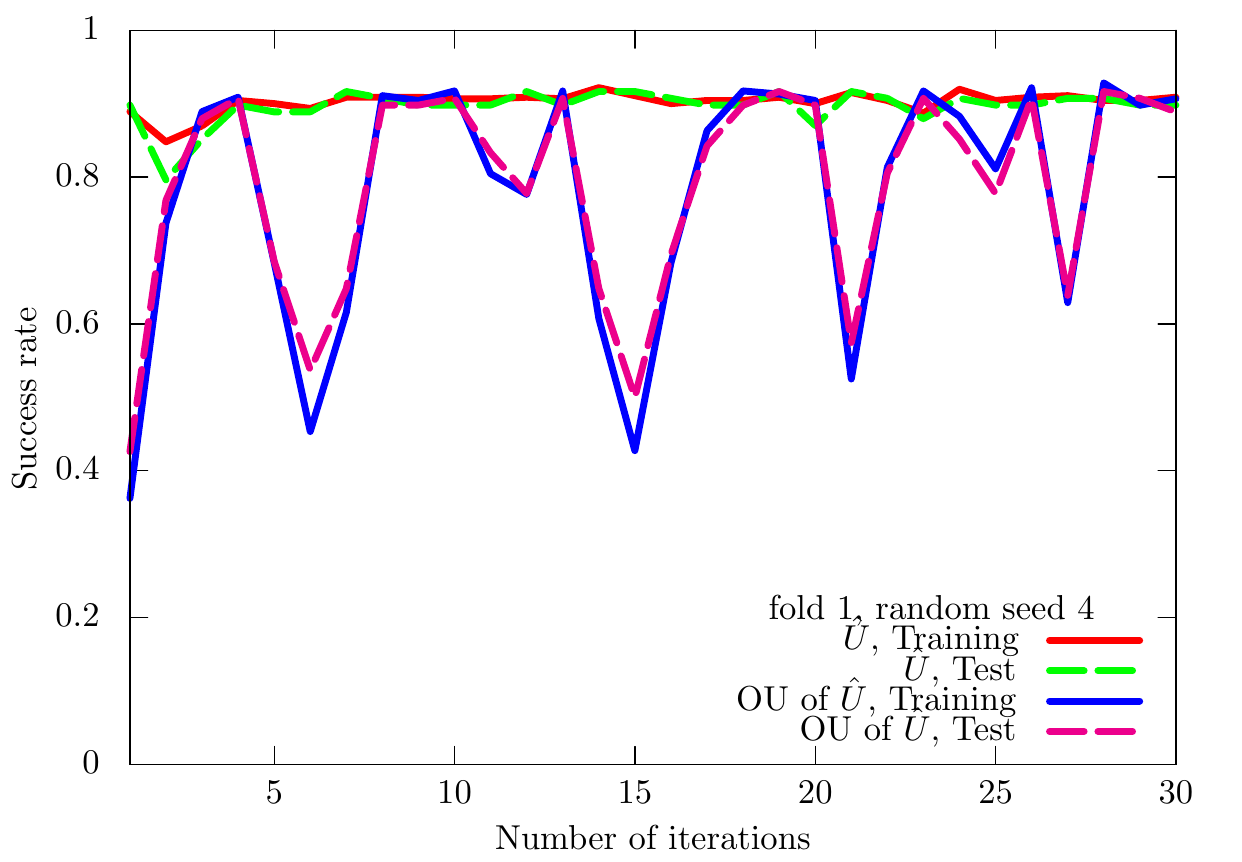}
\includegraphics[scale=0.25]{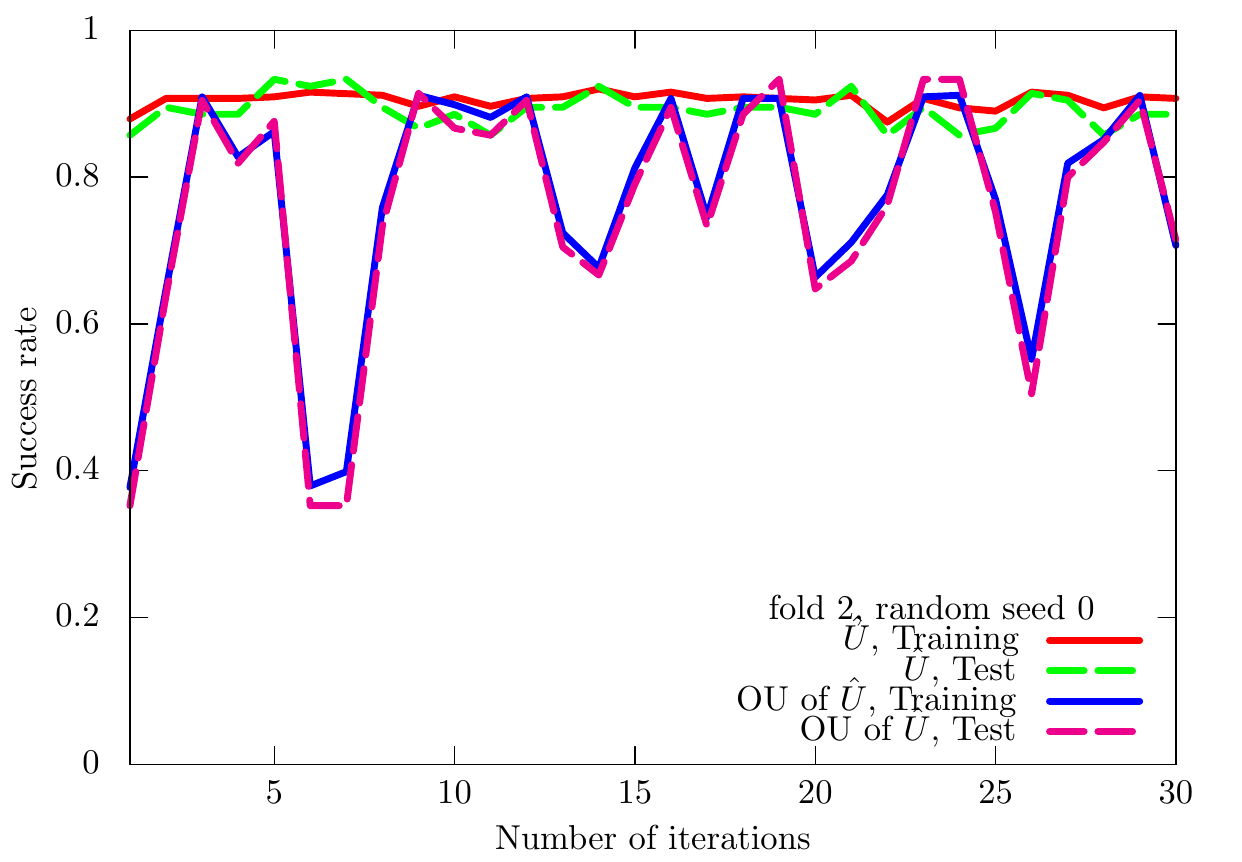}
\includegraphics[scale=0.25]{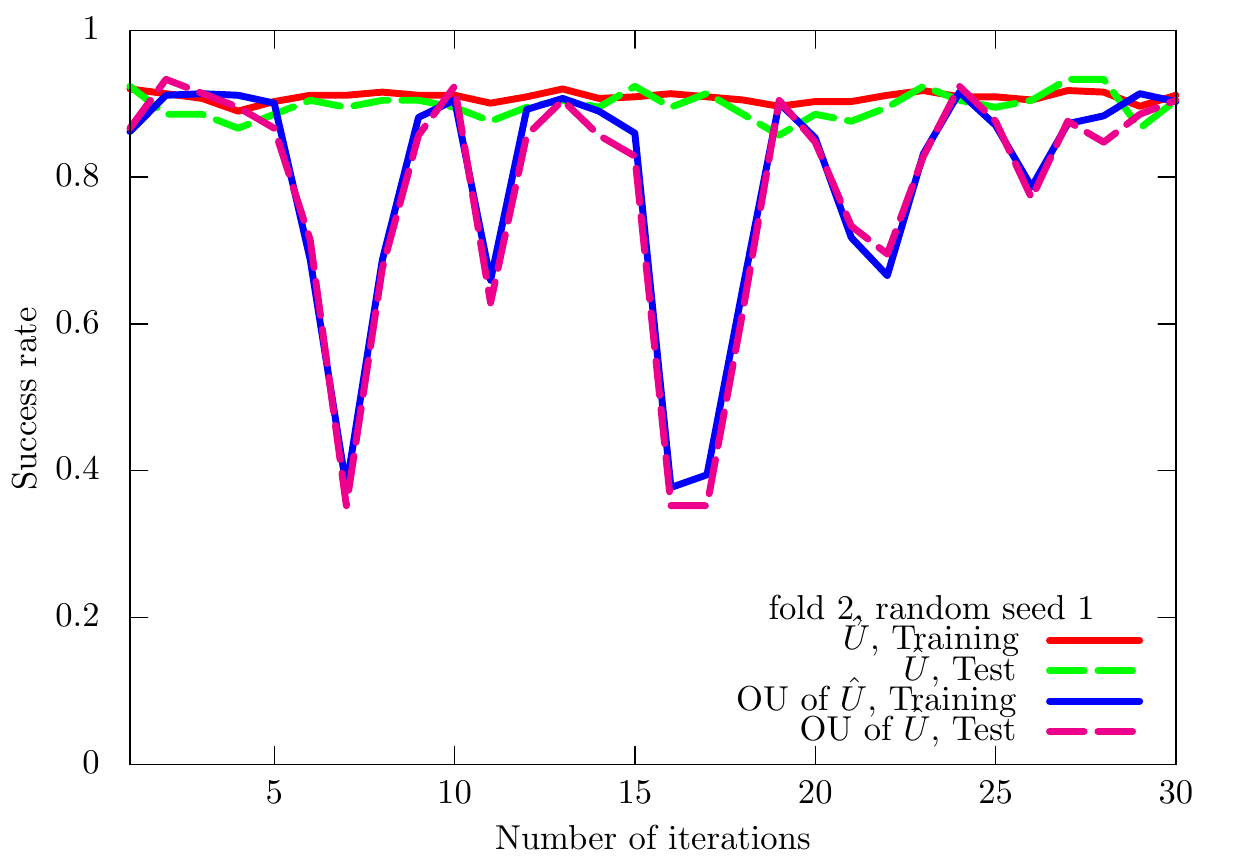}
\includegraphics[scale=0.25]{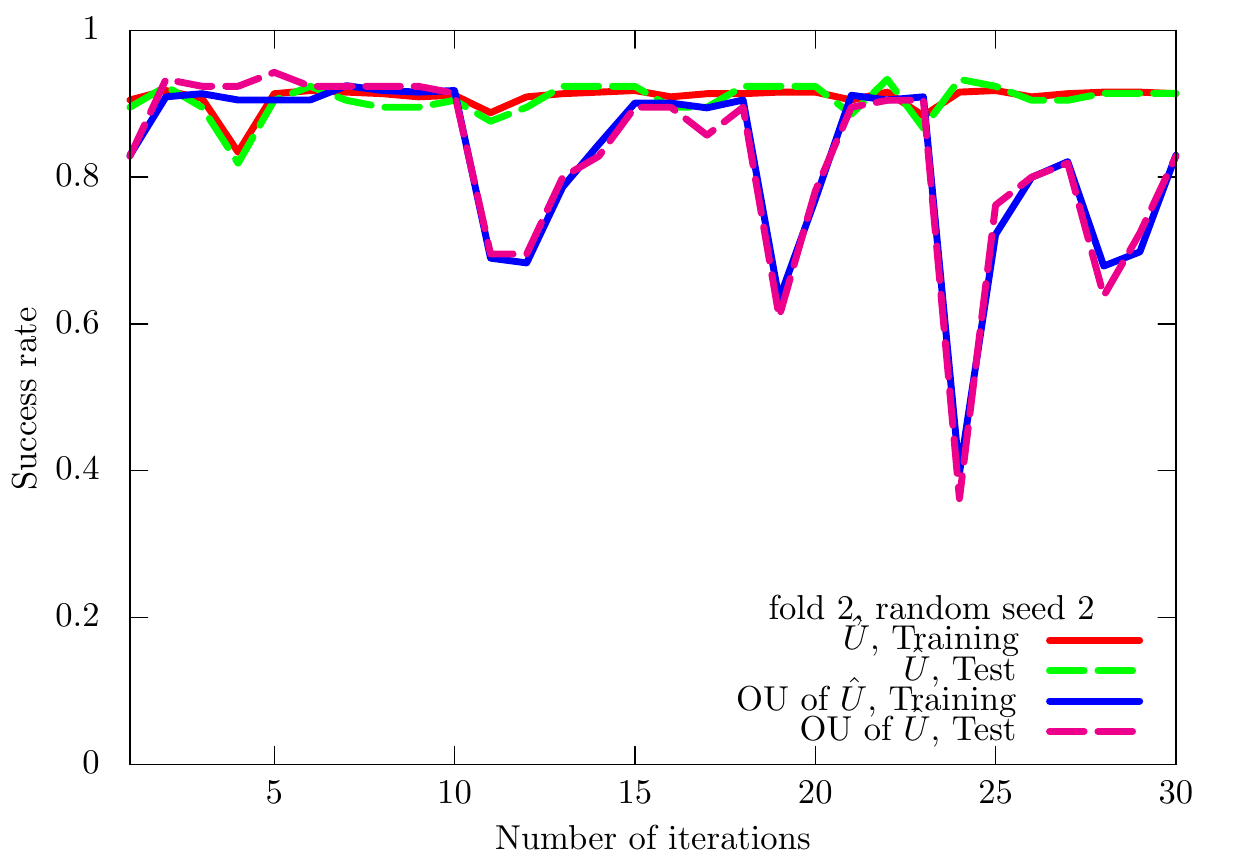}
\includegraphics[scale=0.25]{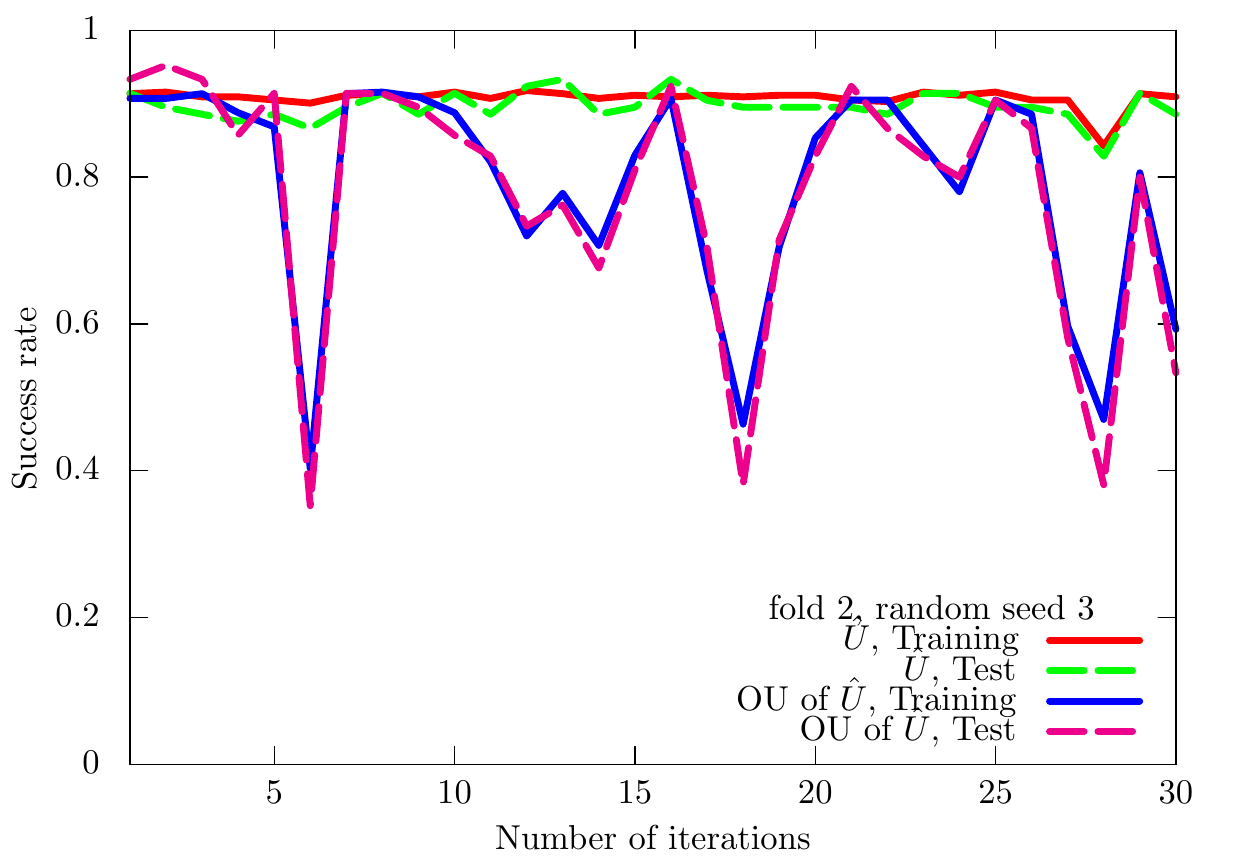}
\includegraphics[scale=0.25]{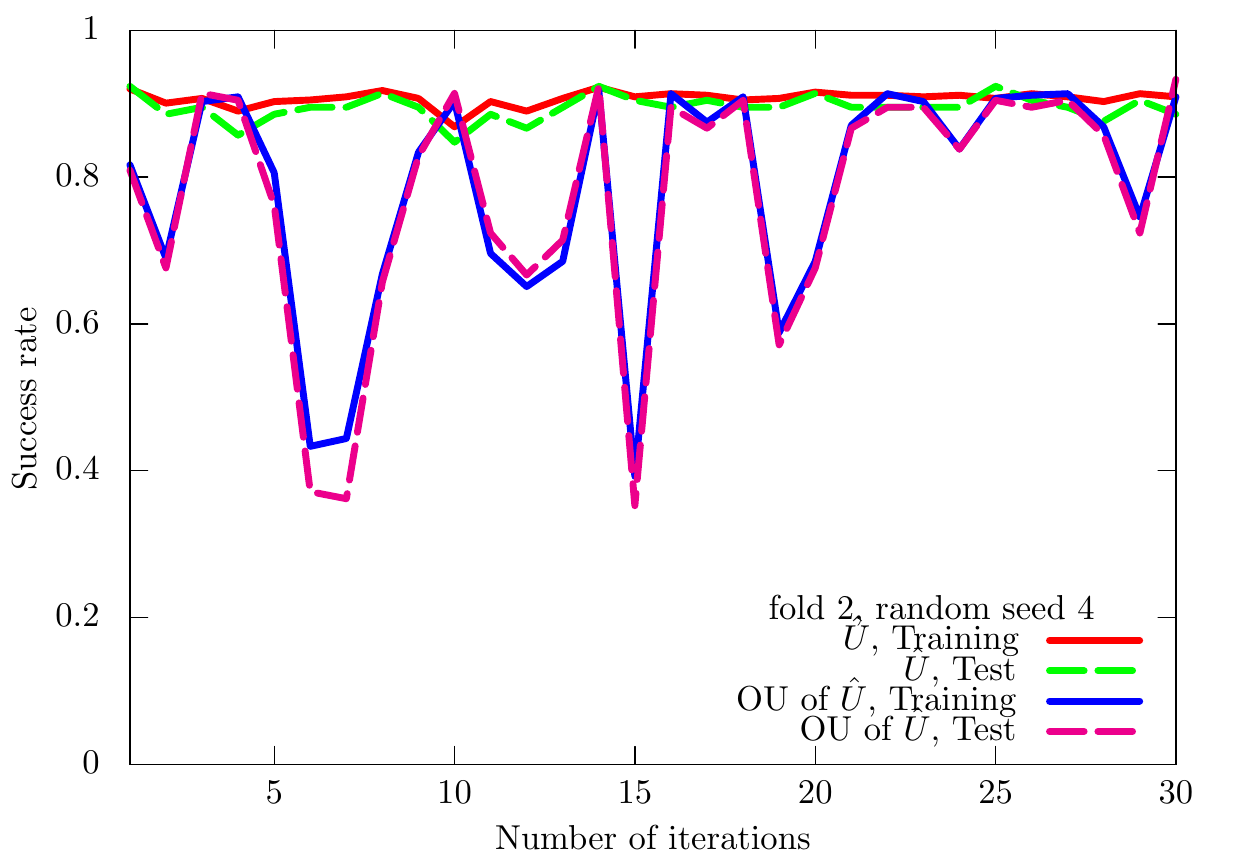}
\includegraphics[scale=0.25]{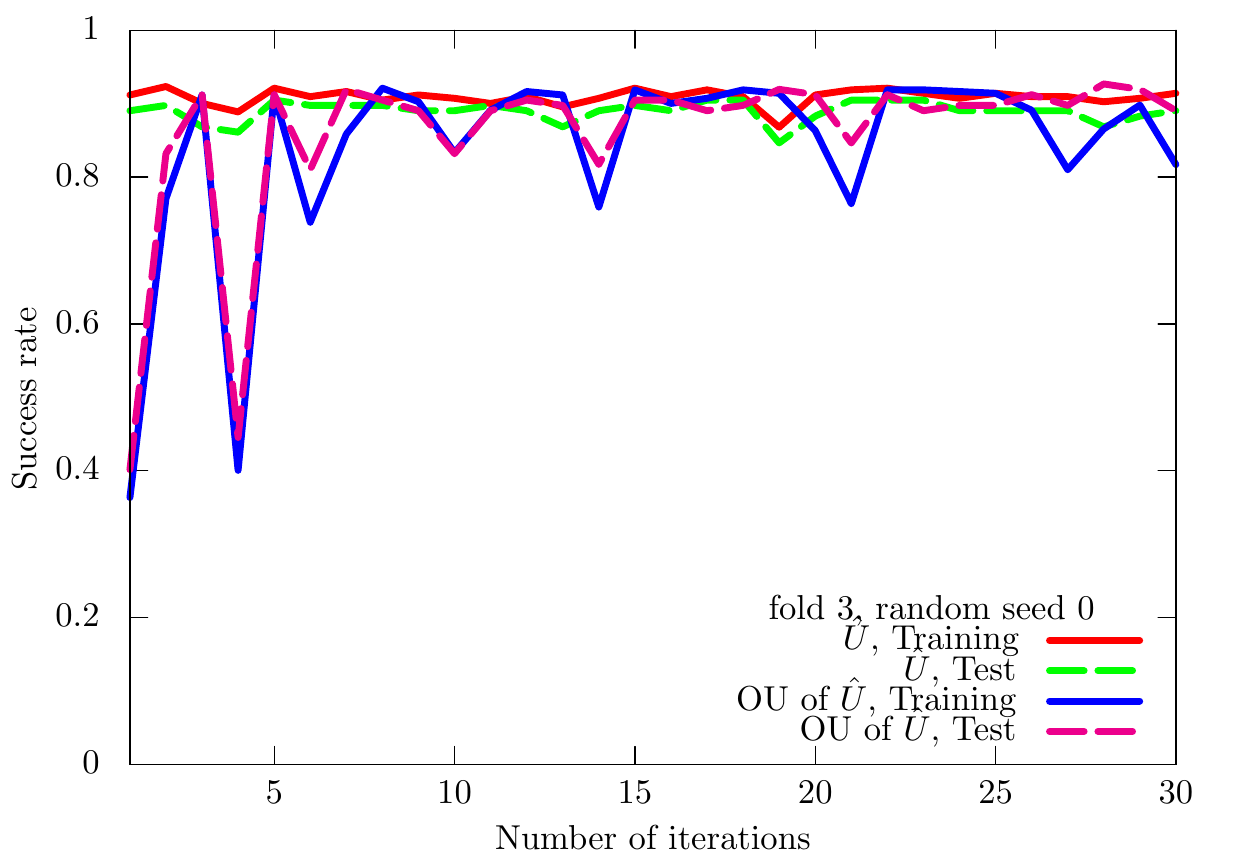}
\includegraphics[scale=0.25]{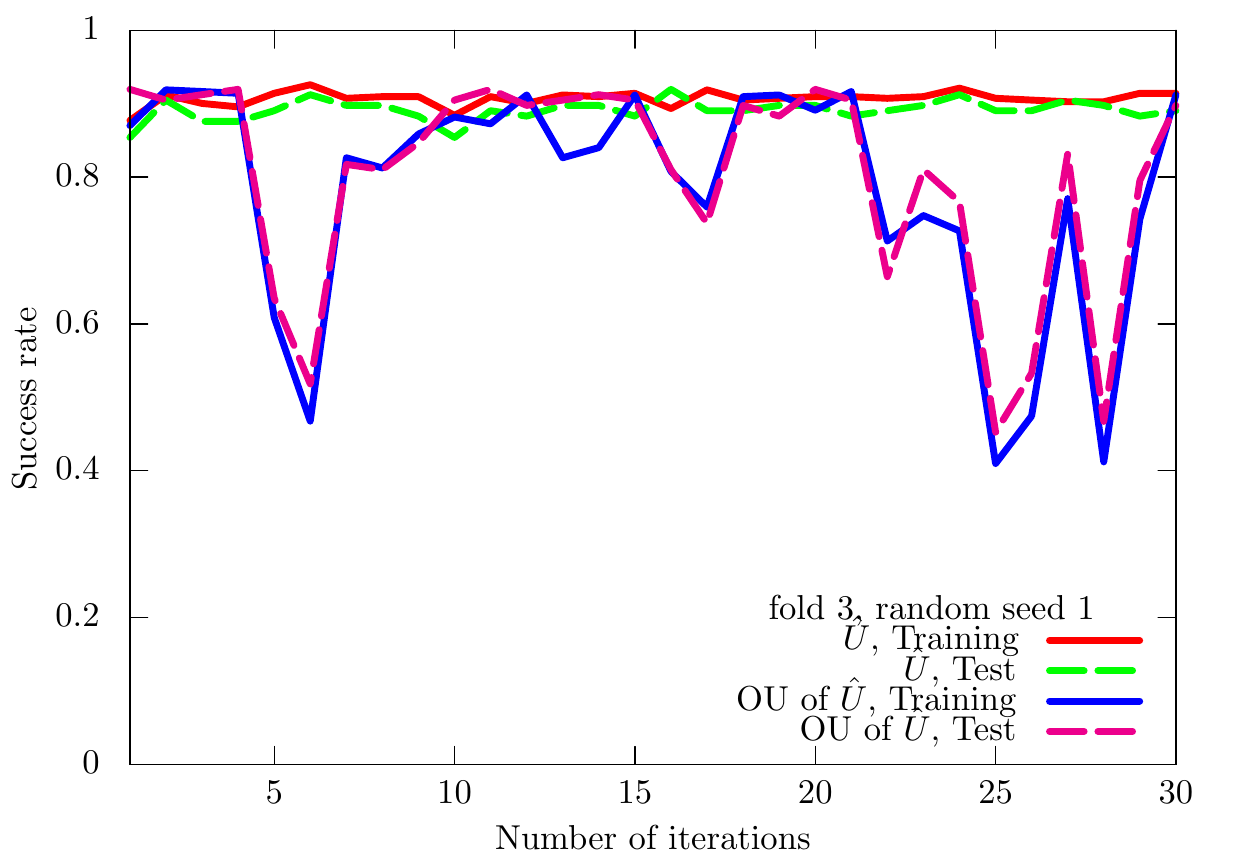}
\includegraphics[scale=0.25]{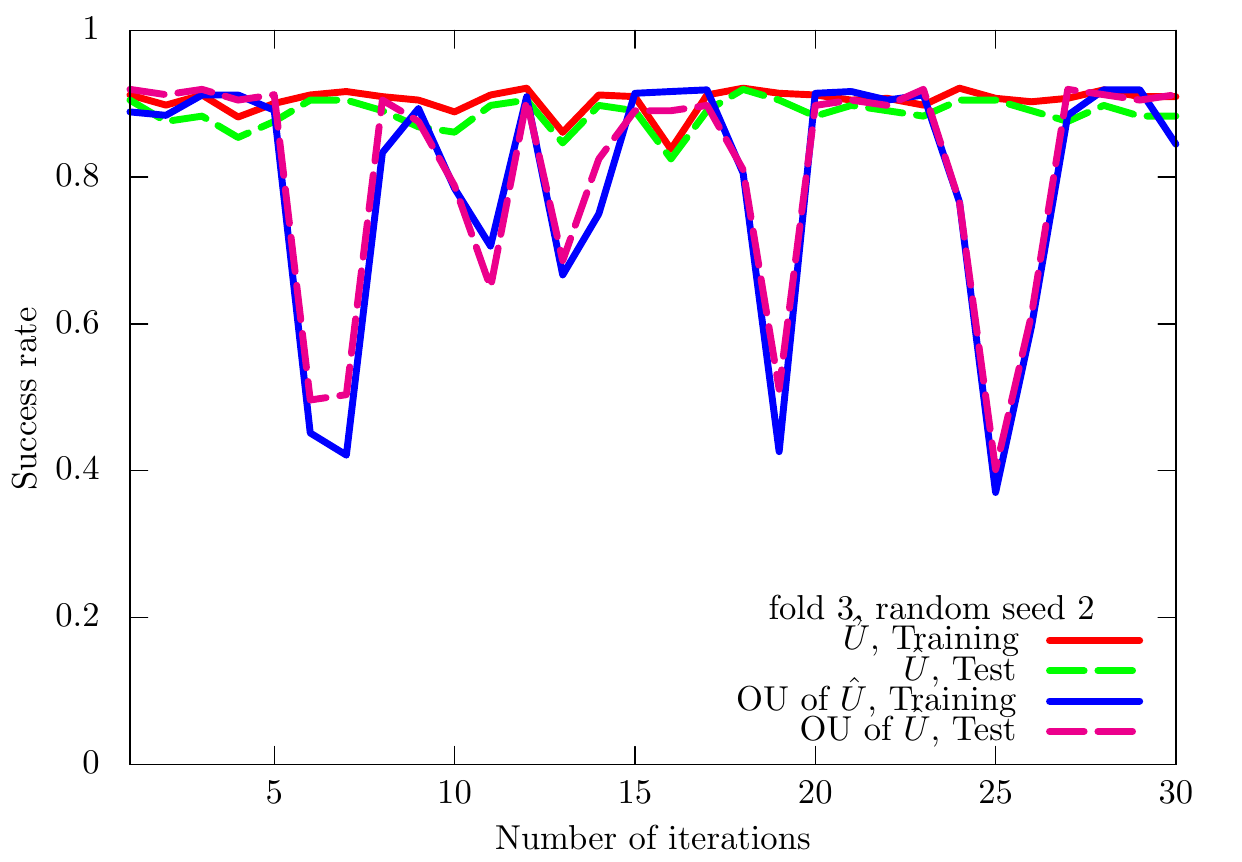}
\includegraphics[scale=0.25]{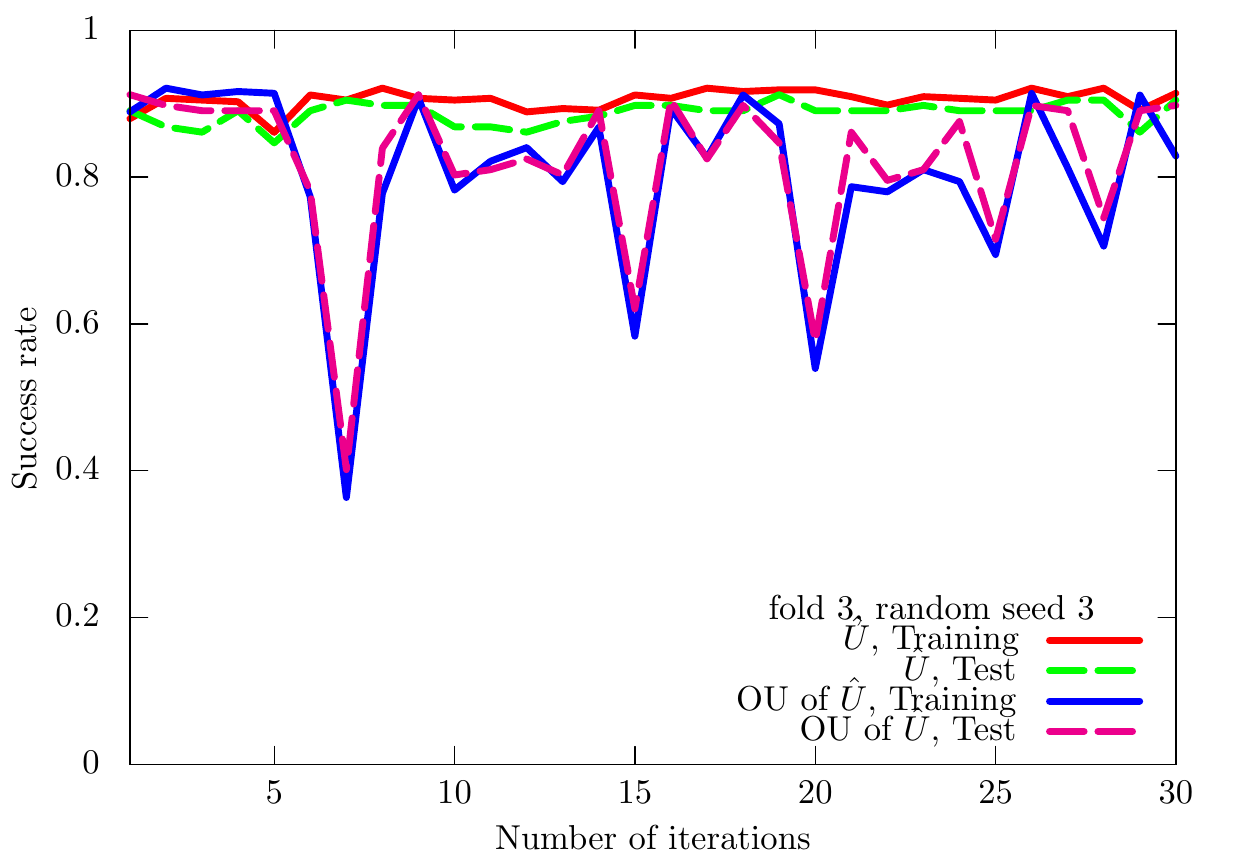}
\includegraphics[scale=0.25]{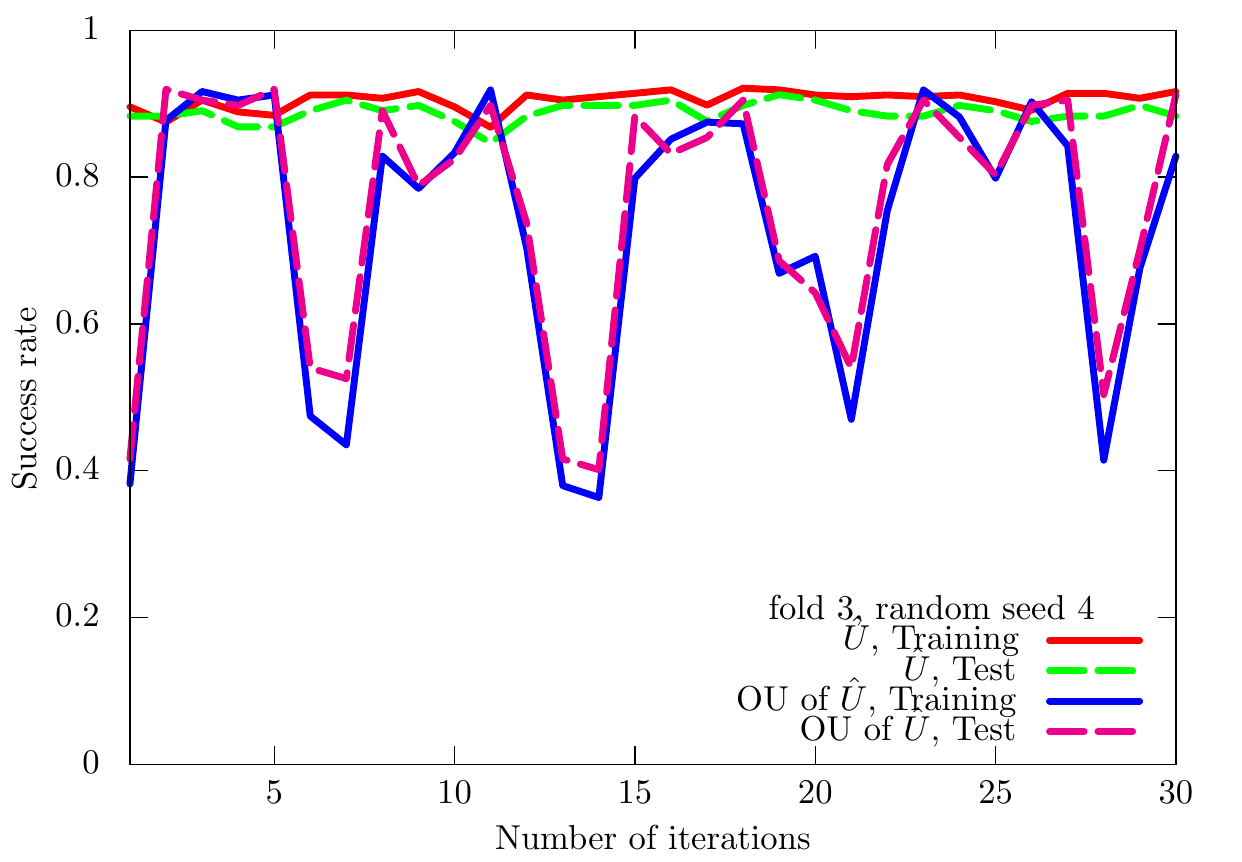}
\includegraphics[scale=0.25]{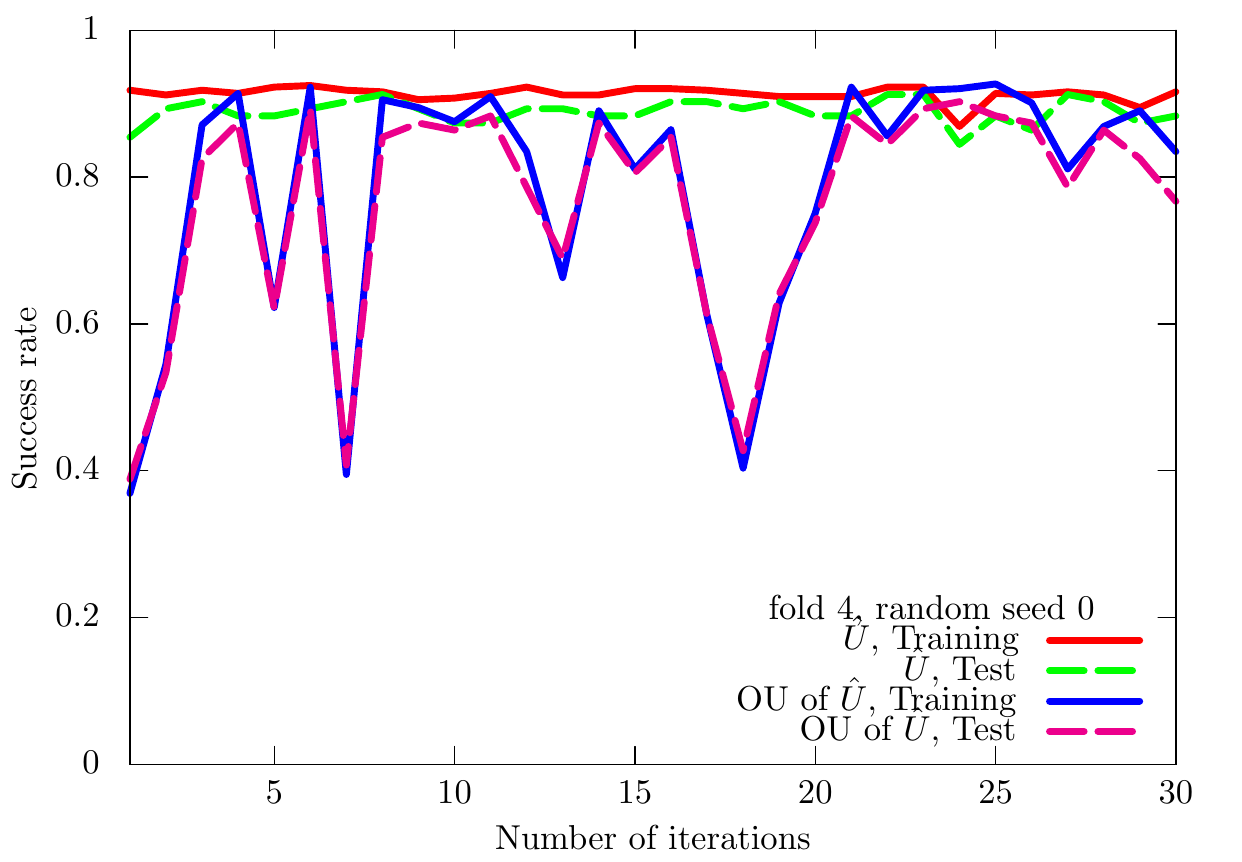}
\includegraphics[scale=0.25]{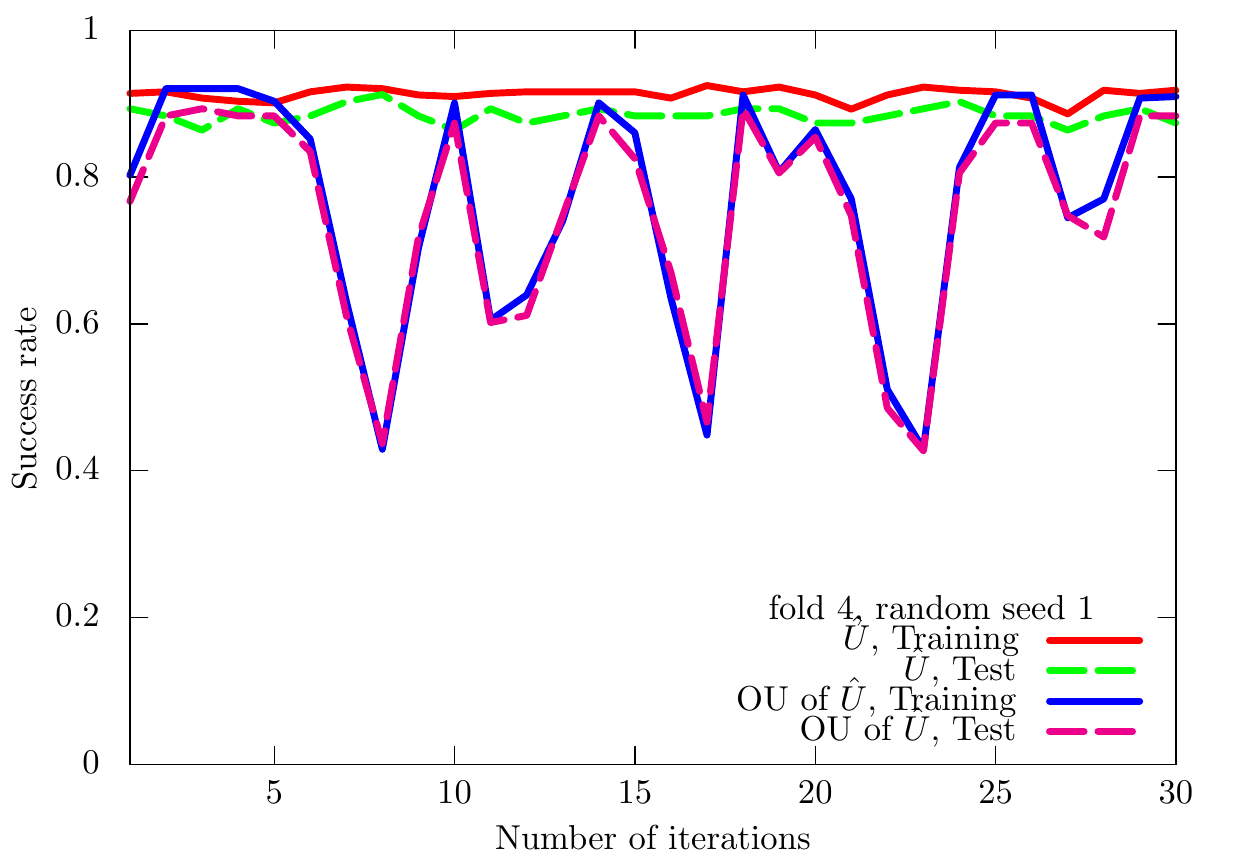}
\includegraphics[scale=0.25]{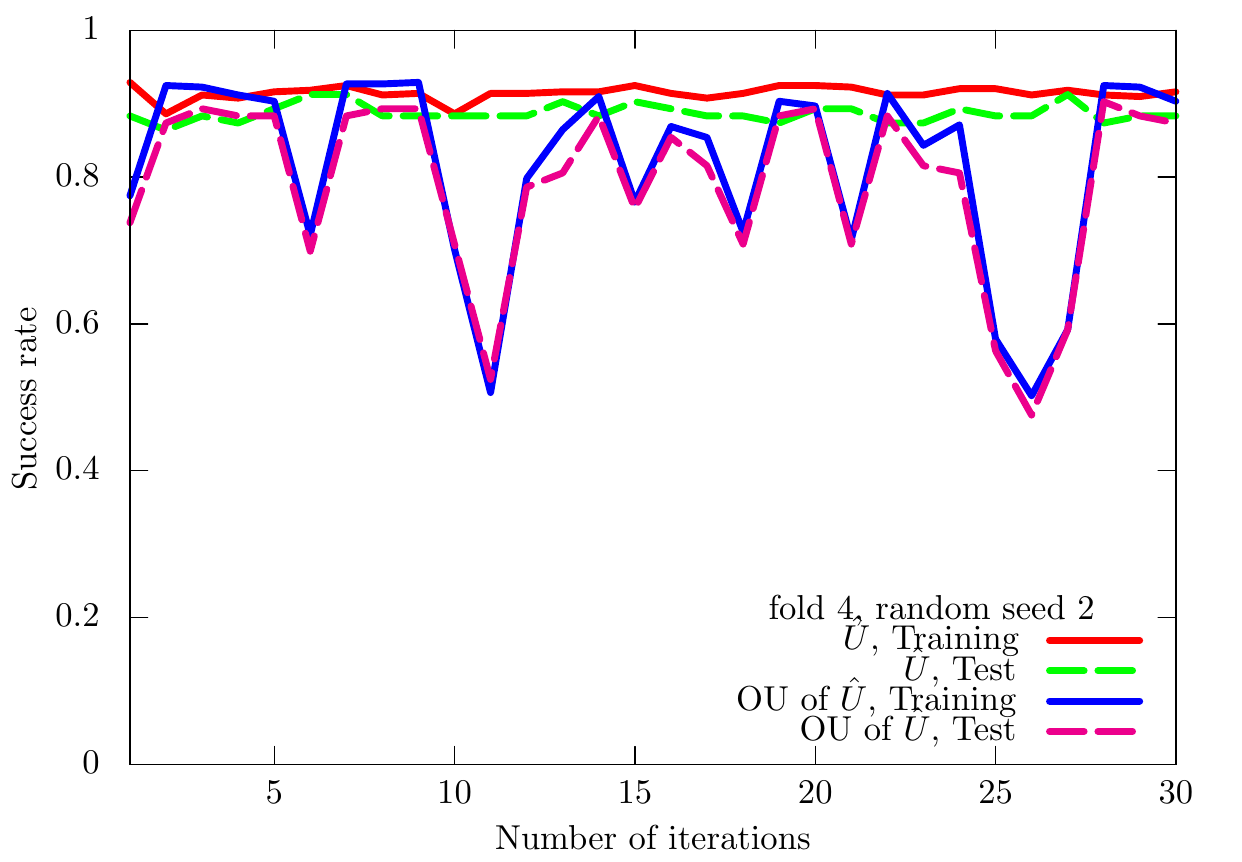}
\includegraphics[scale=0.25]{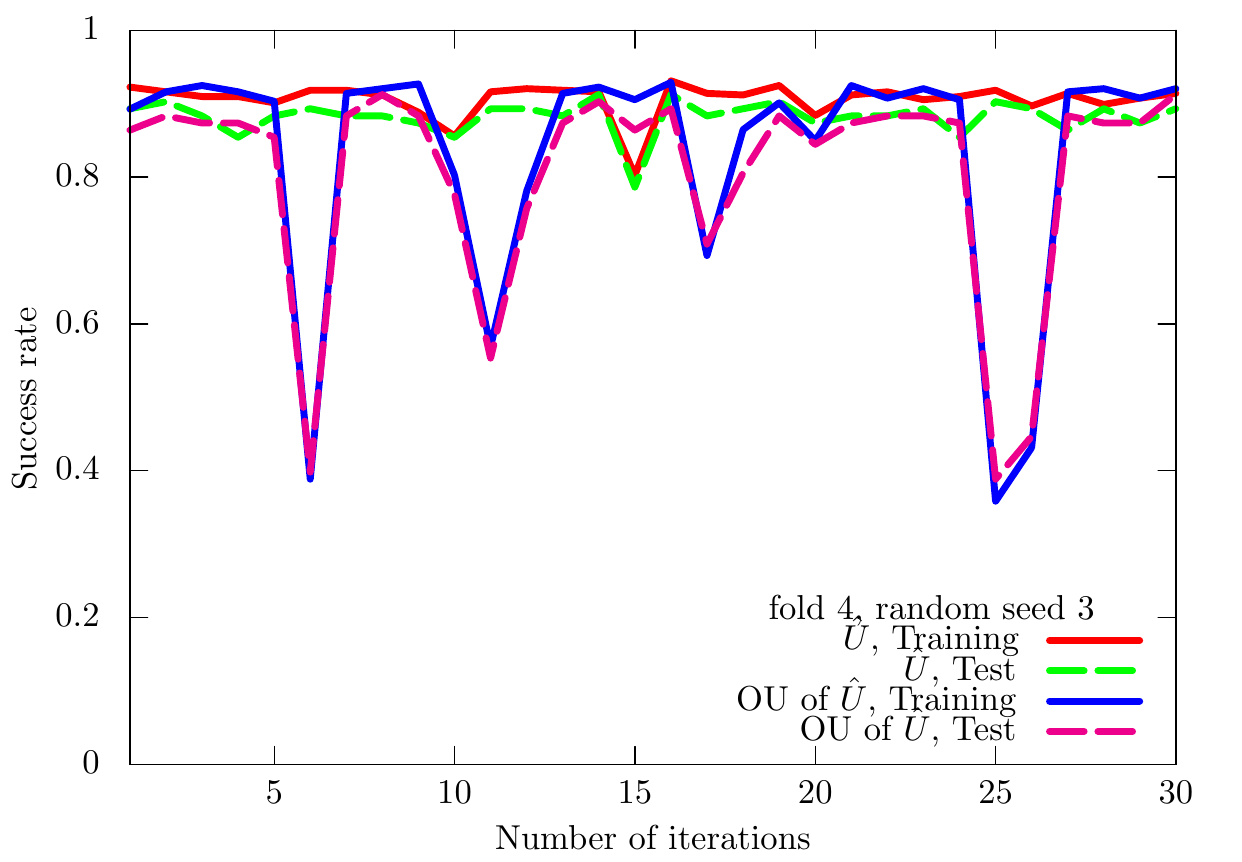}
\includegraphics[scale=0.25]{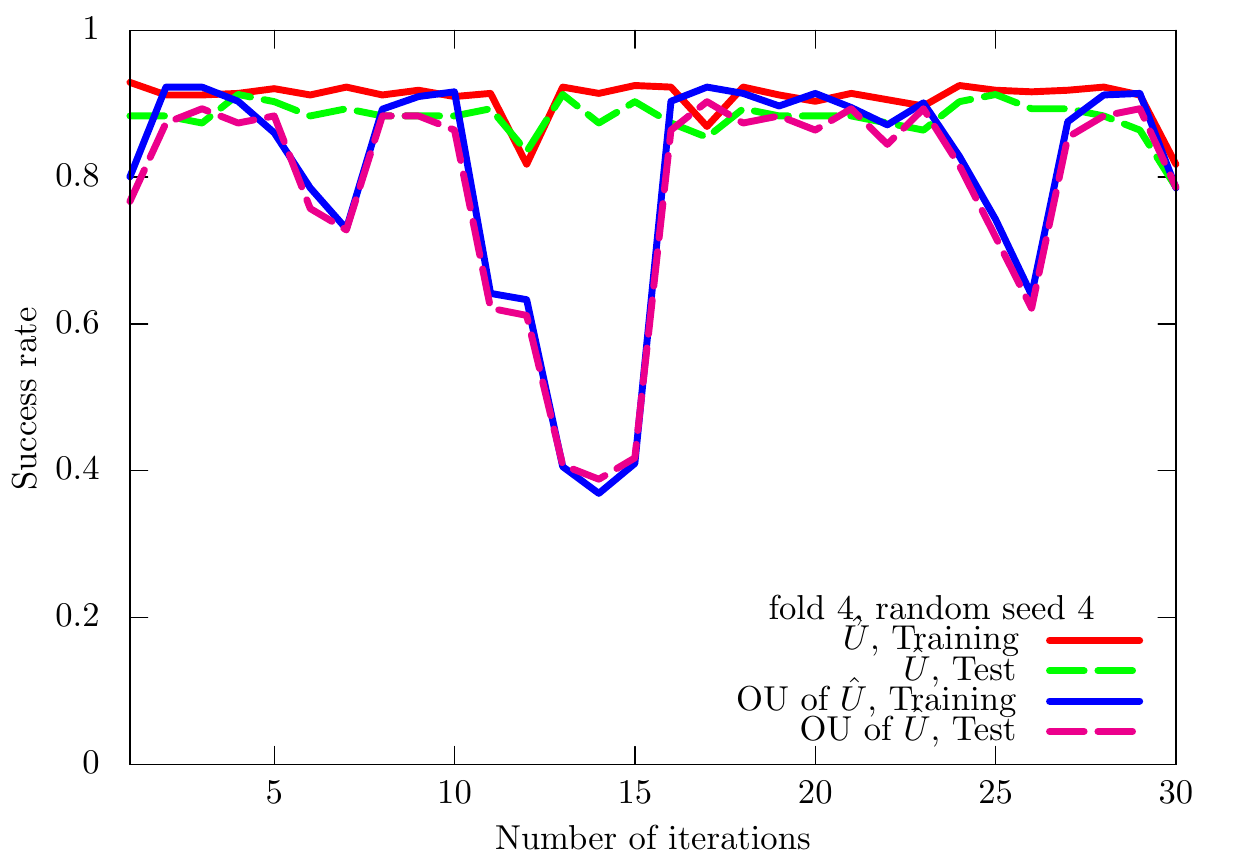}
\caption{Results of the UKM ($\hat{X}$ and $\hat{P}$) on the $5$-fold datasets with $5$ different random seeds for the cancer dataset ($0$ or $1$). We use complex matrices and set $\theta_\mathrm{bias} = 0$. We set $r = 0.010$.}
\label{supp-arXiv-numerical-result-raw-data-fold-001-rand-001-UKM-P-UCI-cancer-0-1}
\end{figure*}

\begin{figure*}[htb]
\centering
\includegraphics[scale=0.25]{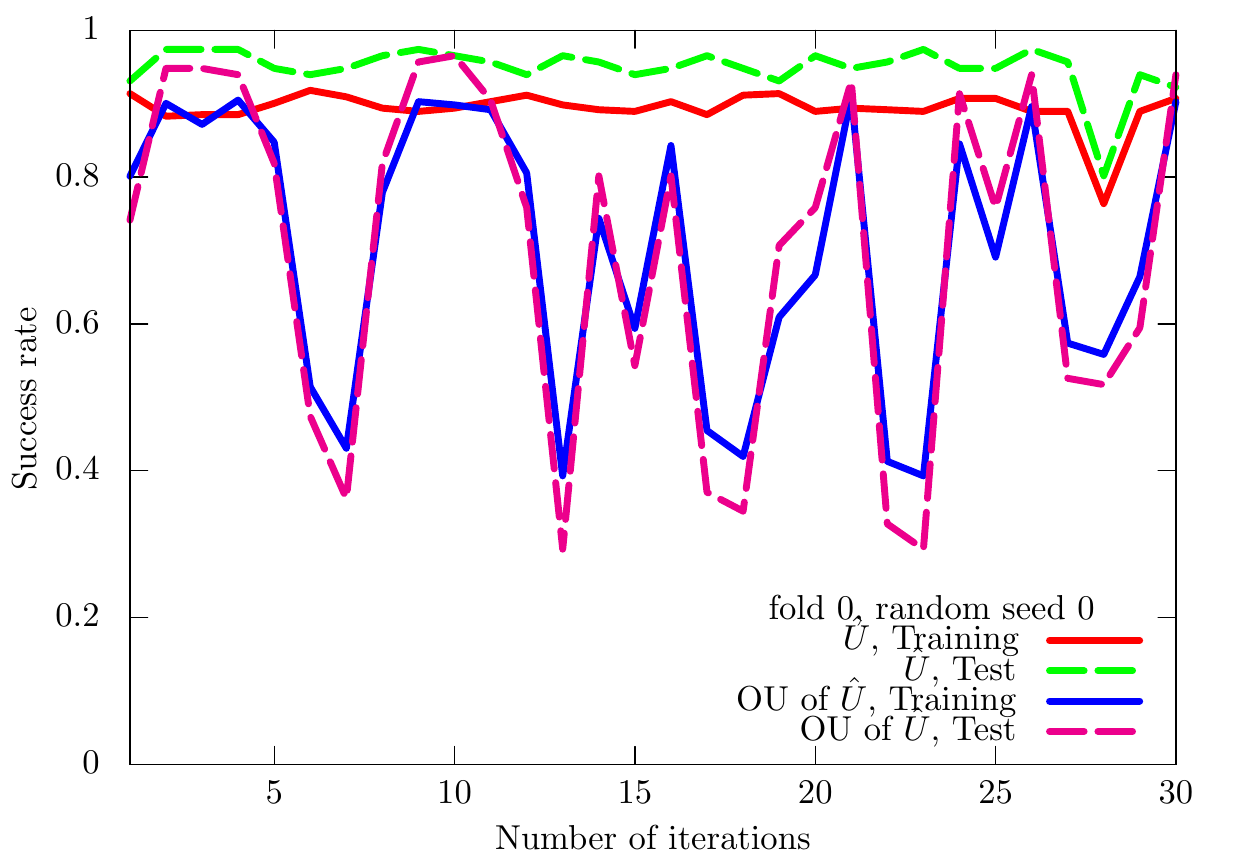}
\includegraphics[scale=0.25]{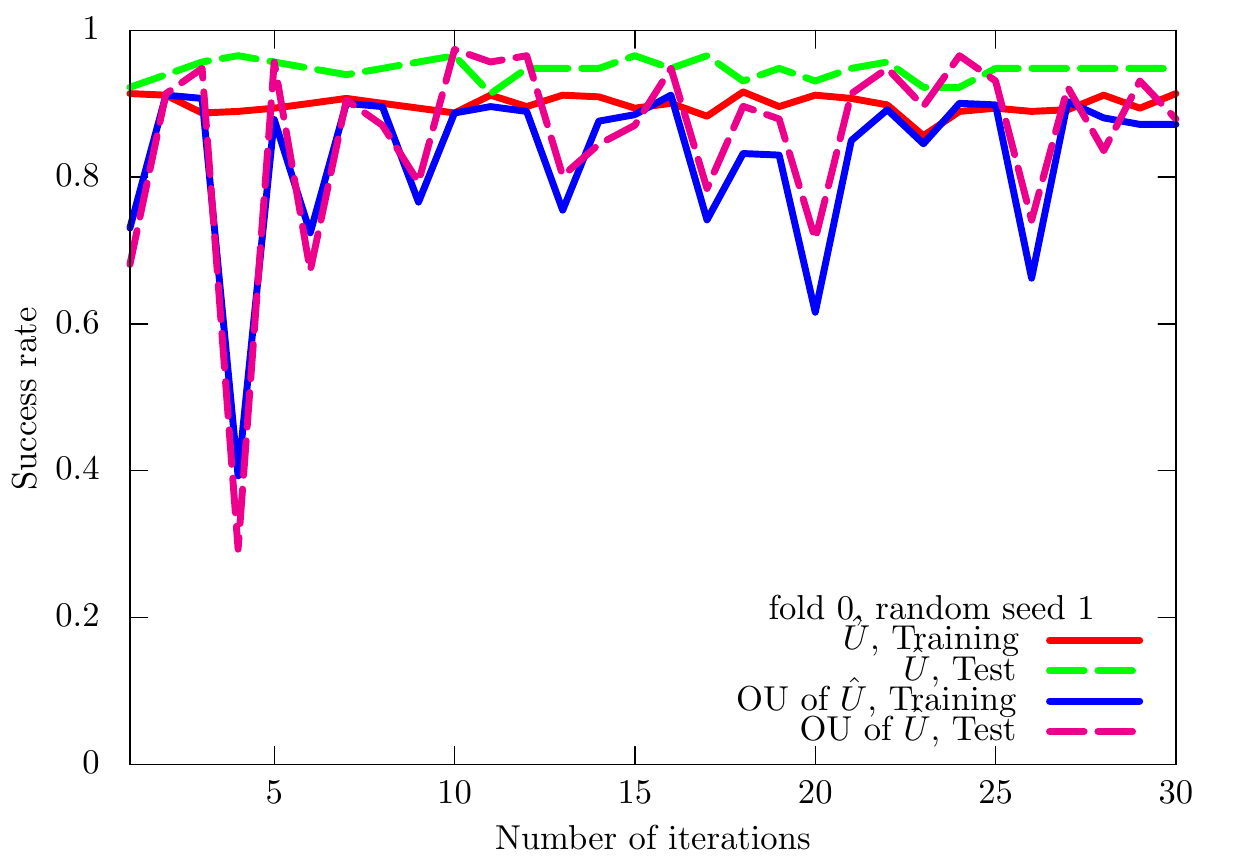}
\includegraphics[scale=0.25]{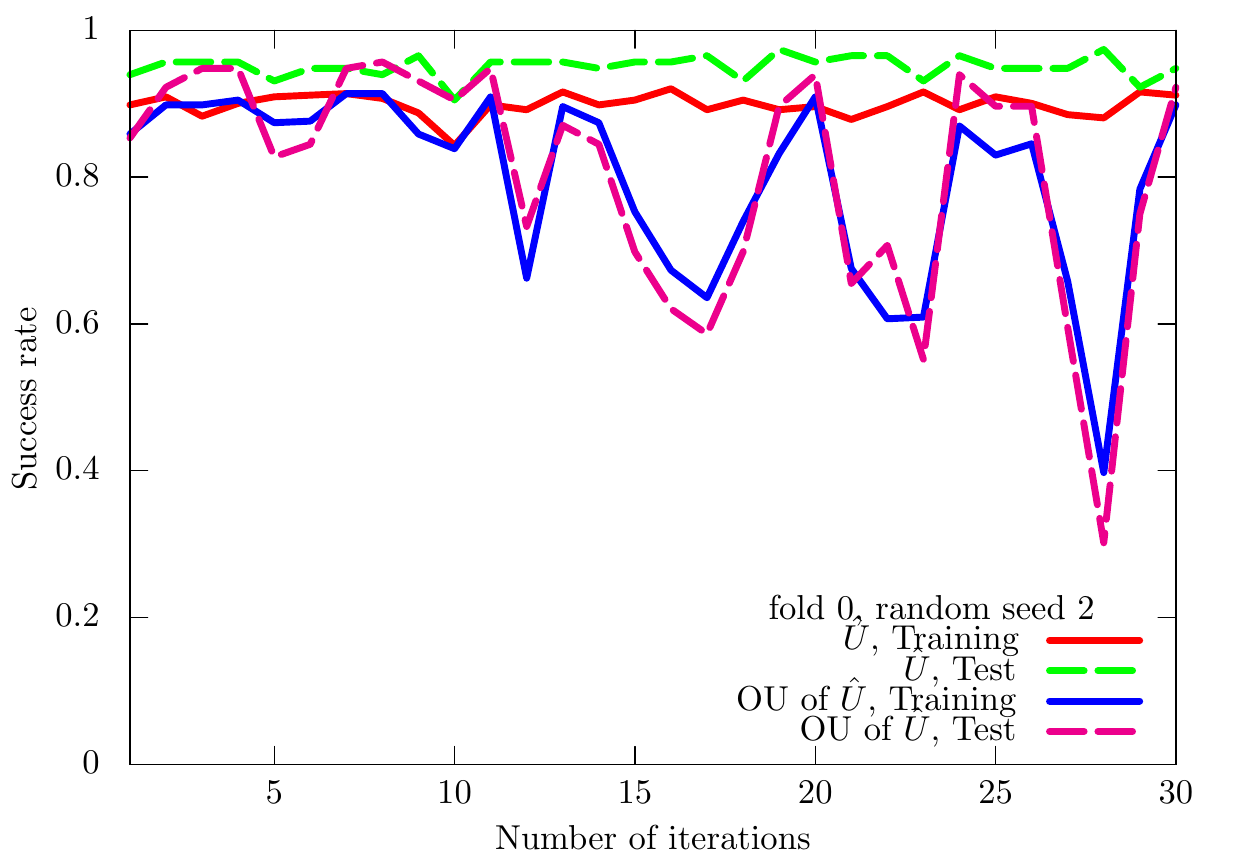}
\includegraphics[scale=0.25]{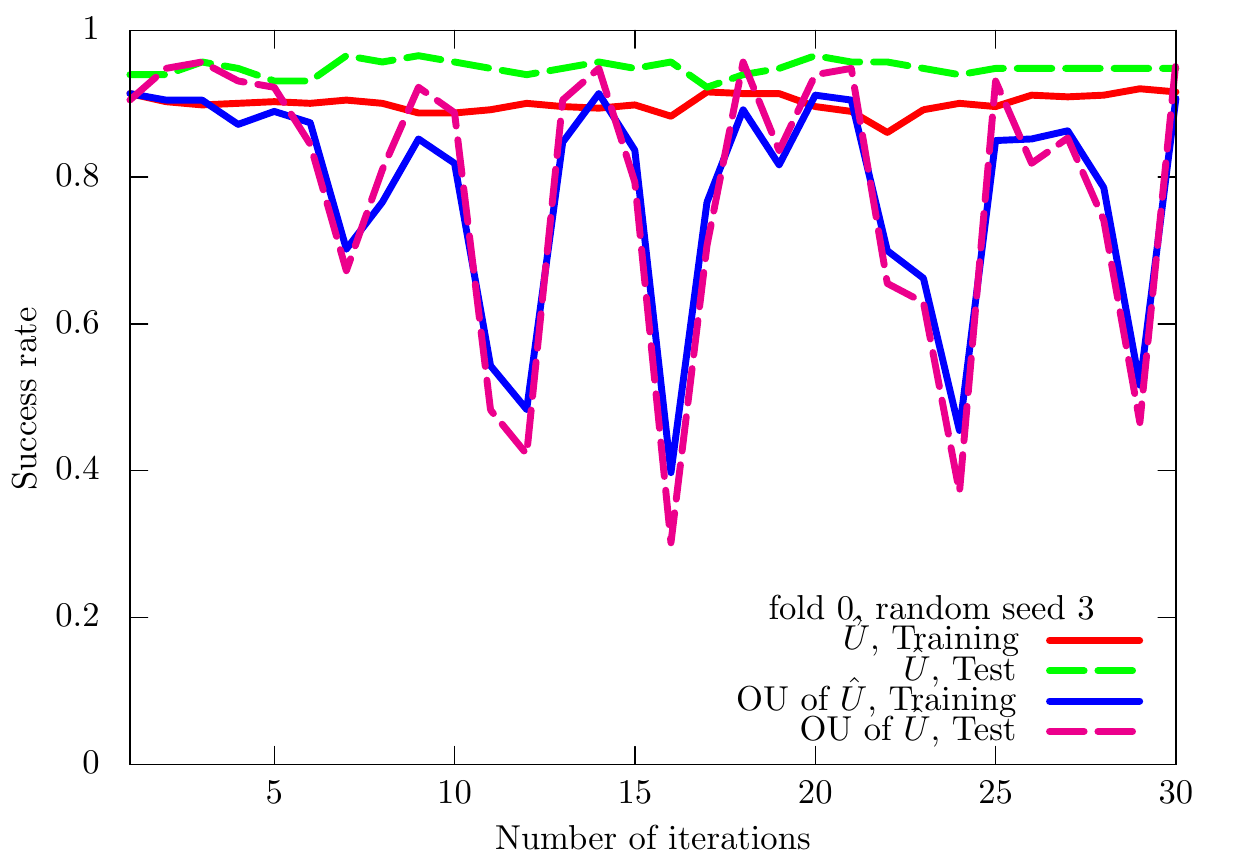}
\includegraphics[scale=0.25]{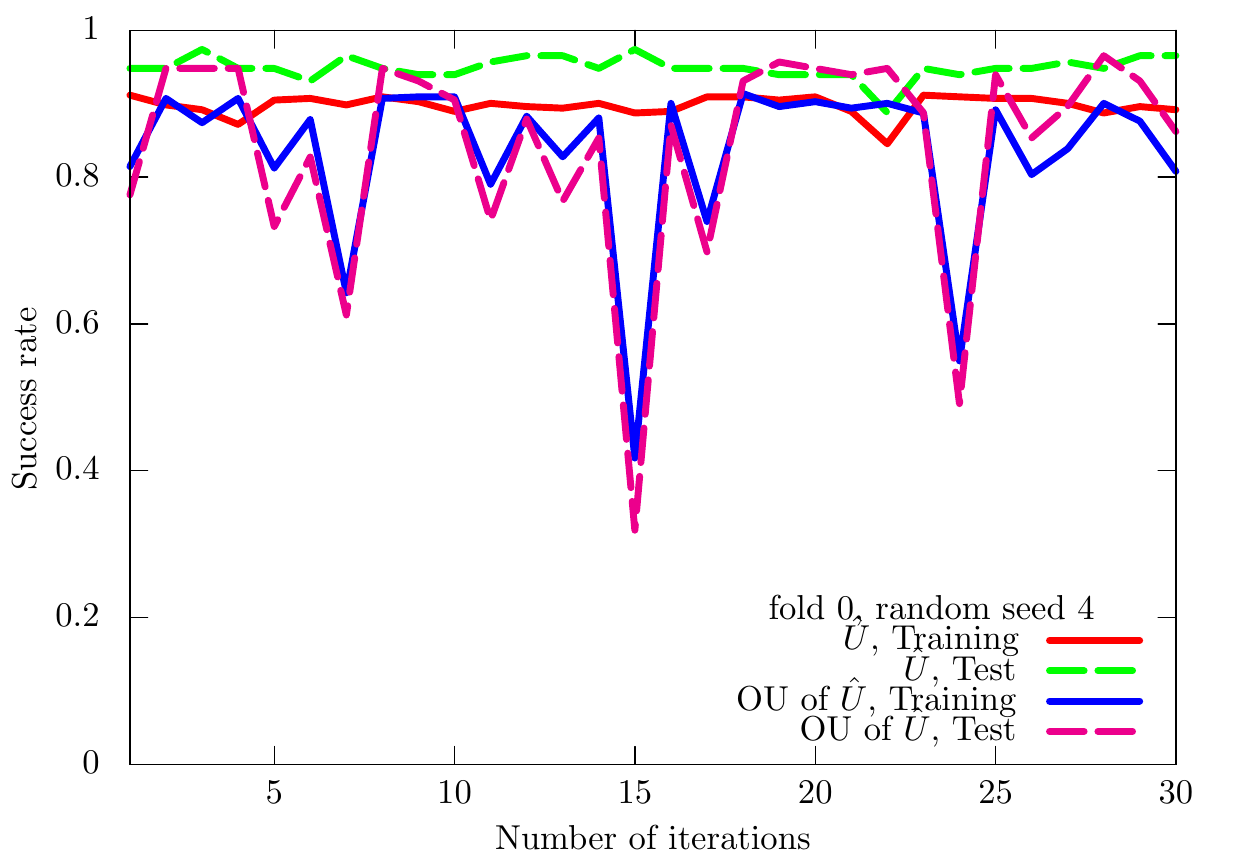}
\includegraphics[scale=0.25]{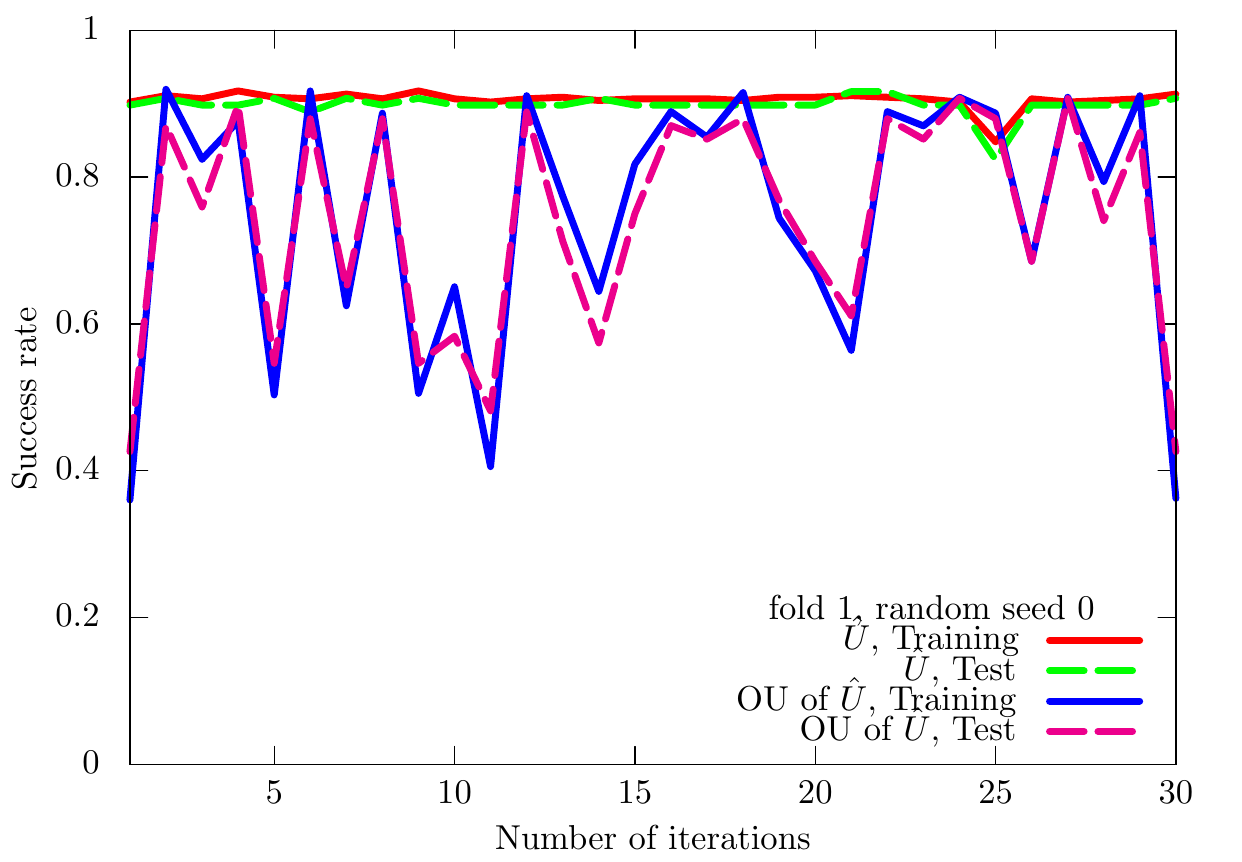}
\includegraphics[scale=0.25]{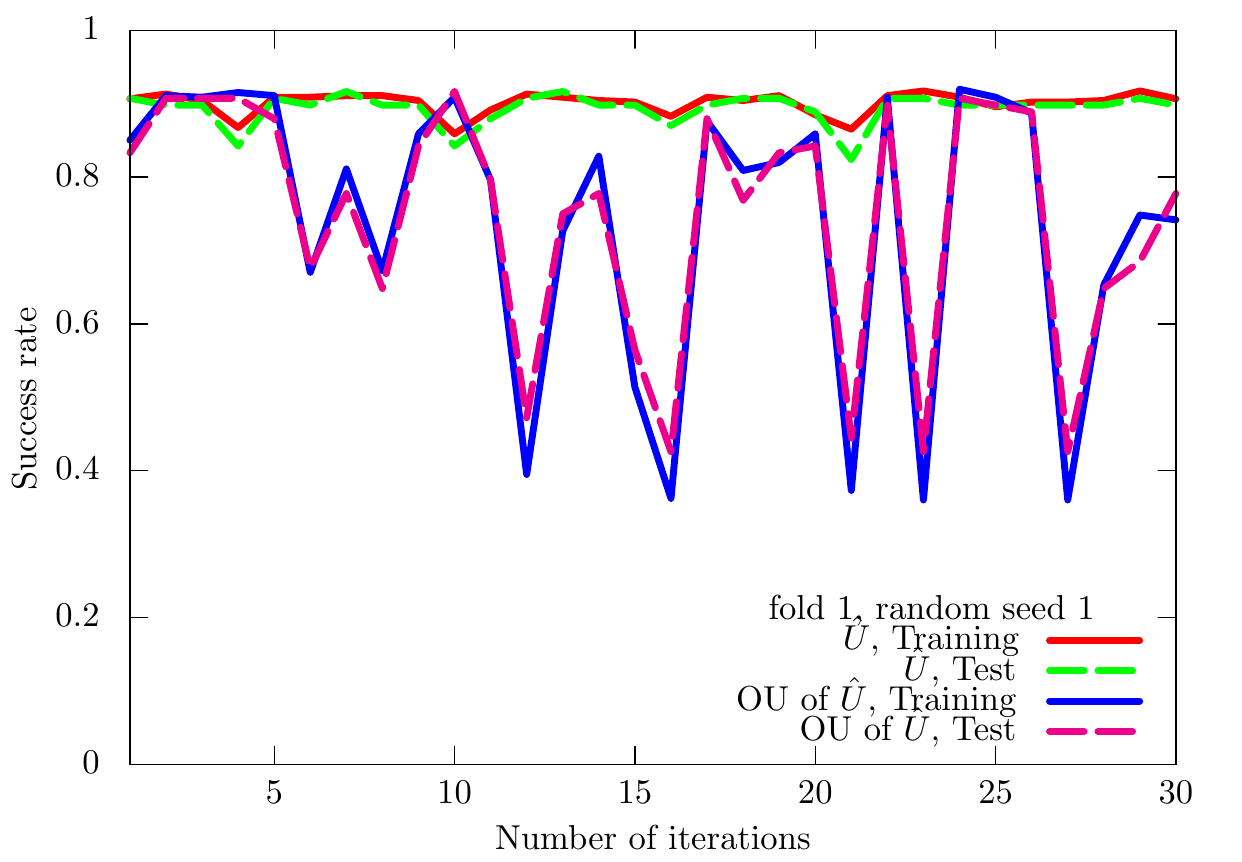}
\includegraphics[scale=0.25]{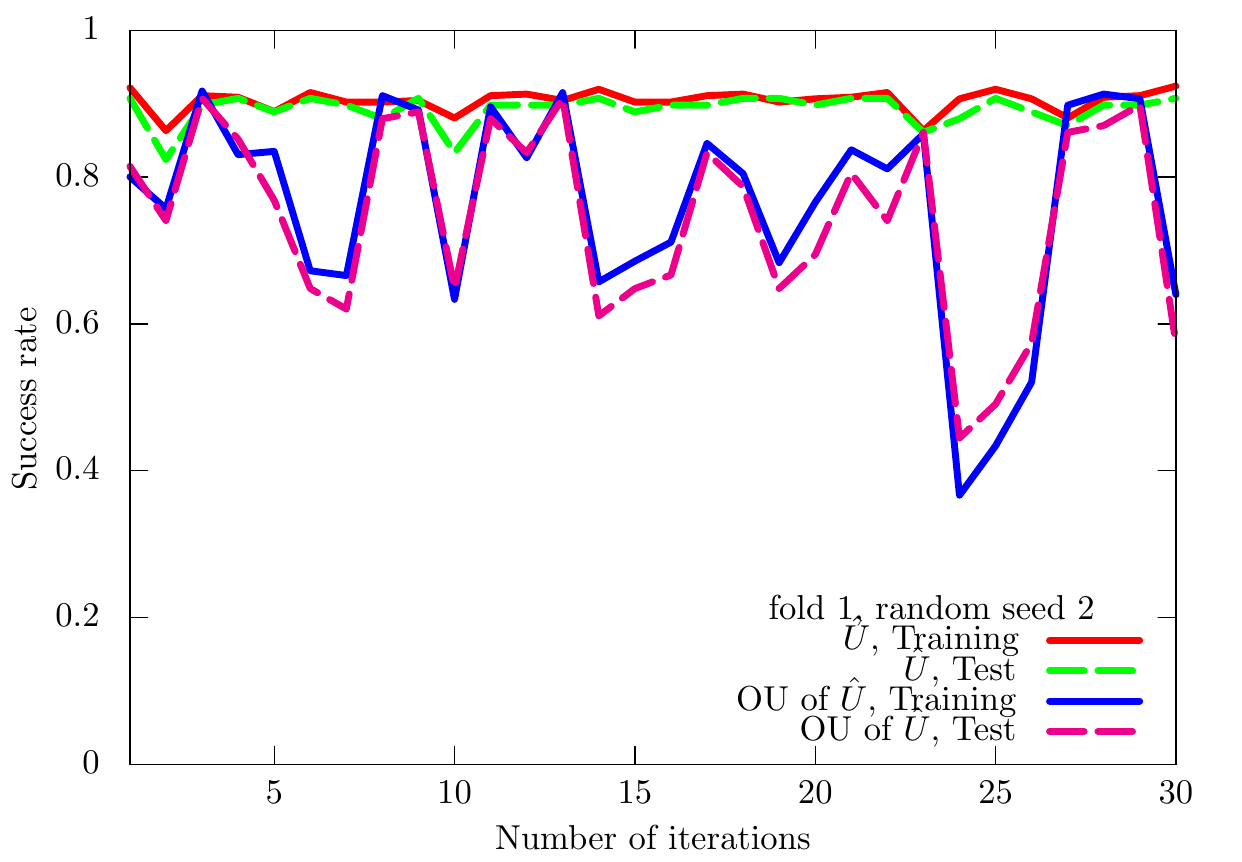}
\includegraphics[scale=0.25]{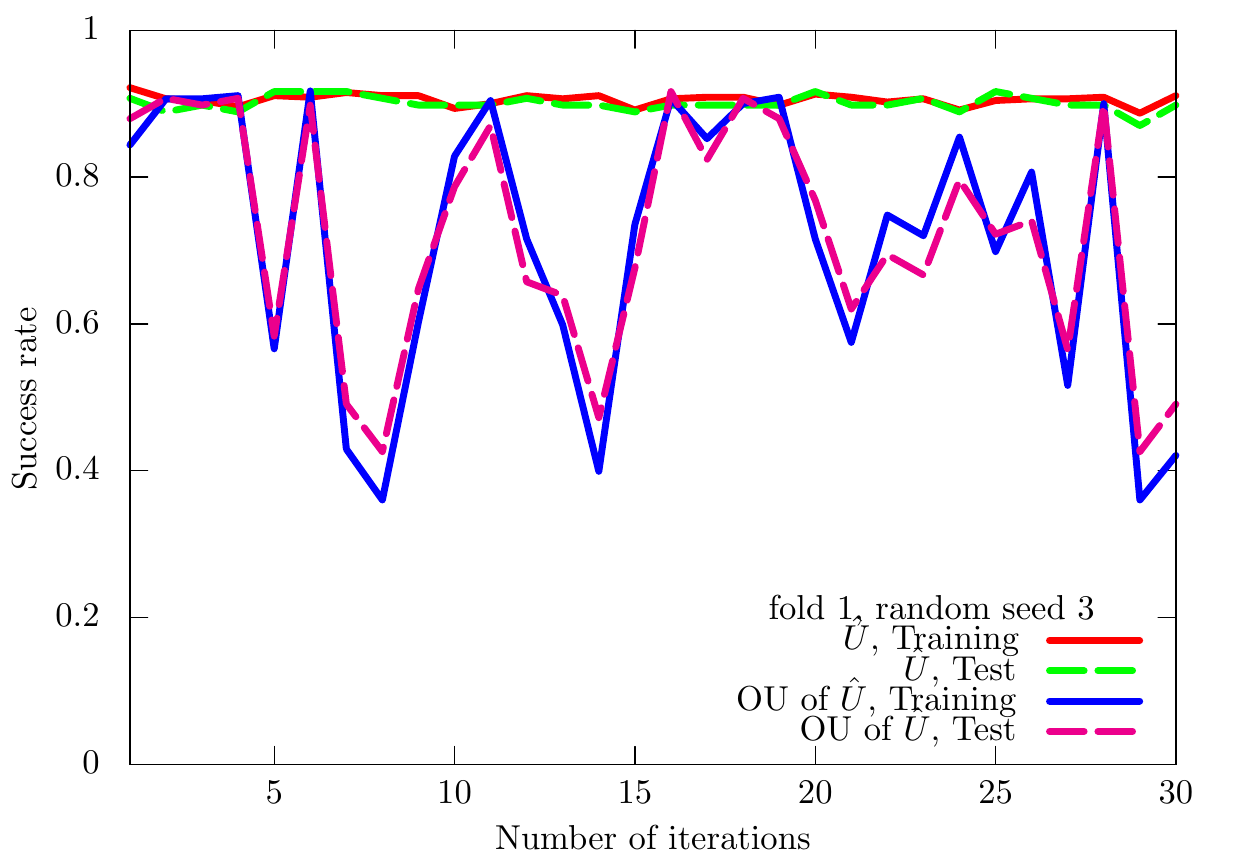}
\includegraphics[scale=0.25]{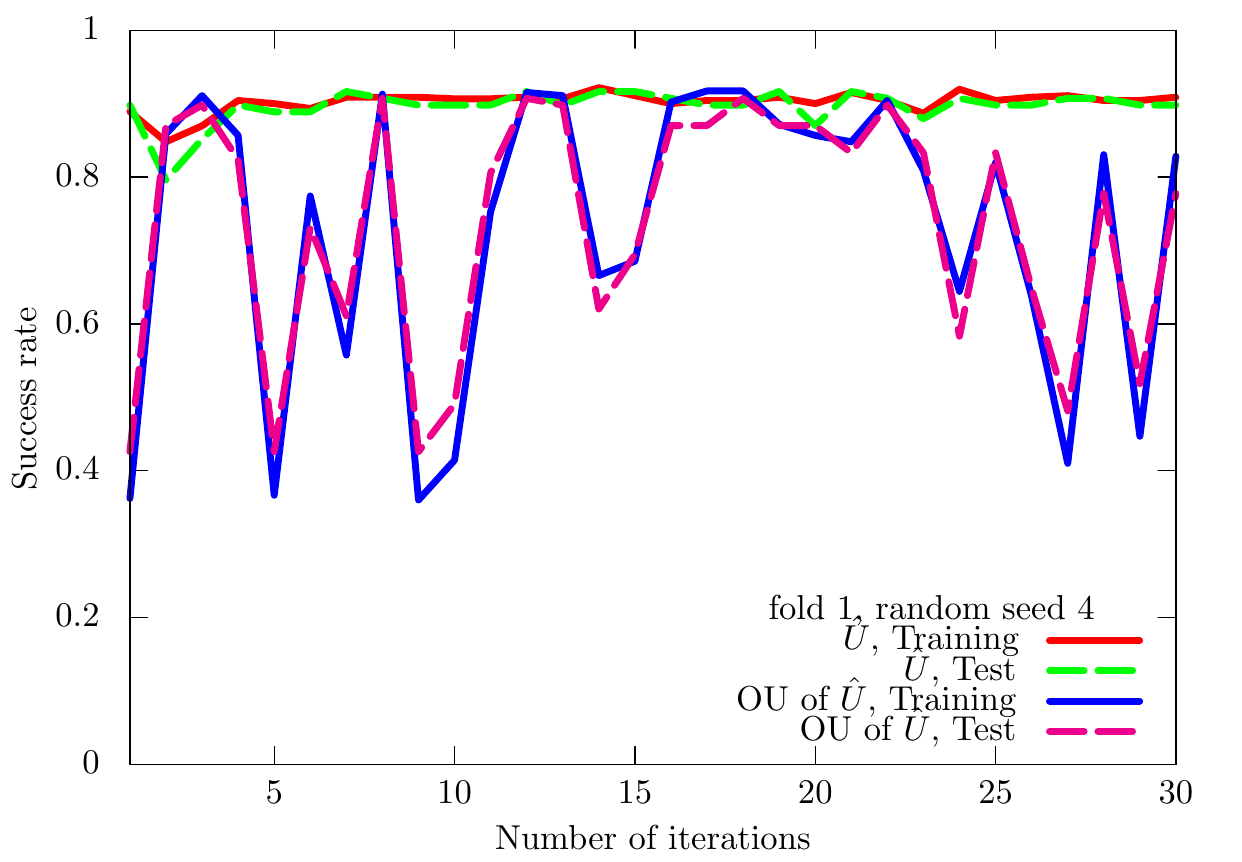}
\includegraphics[scale=0.25]{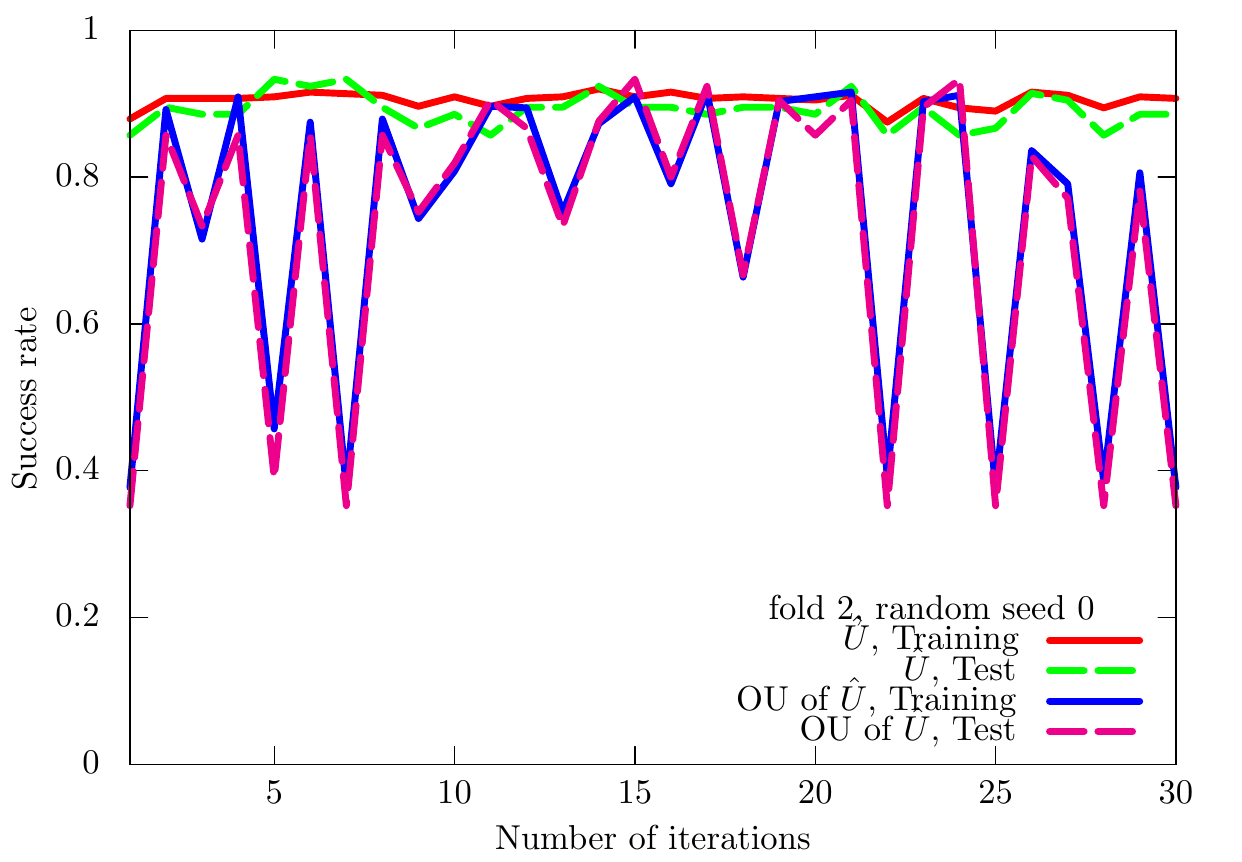}
\includegraphics[scale=0.25]{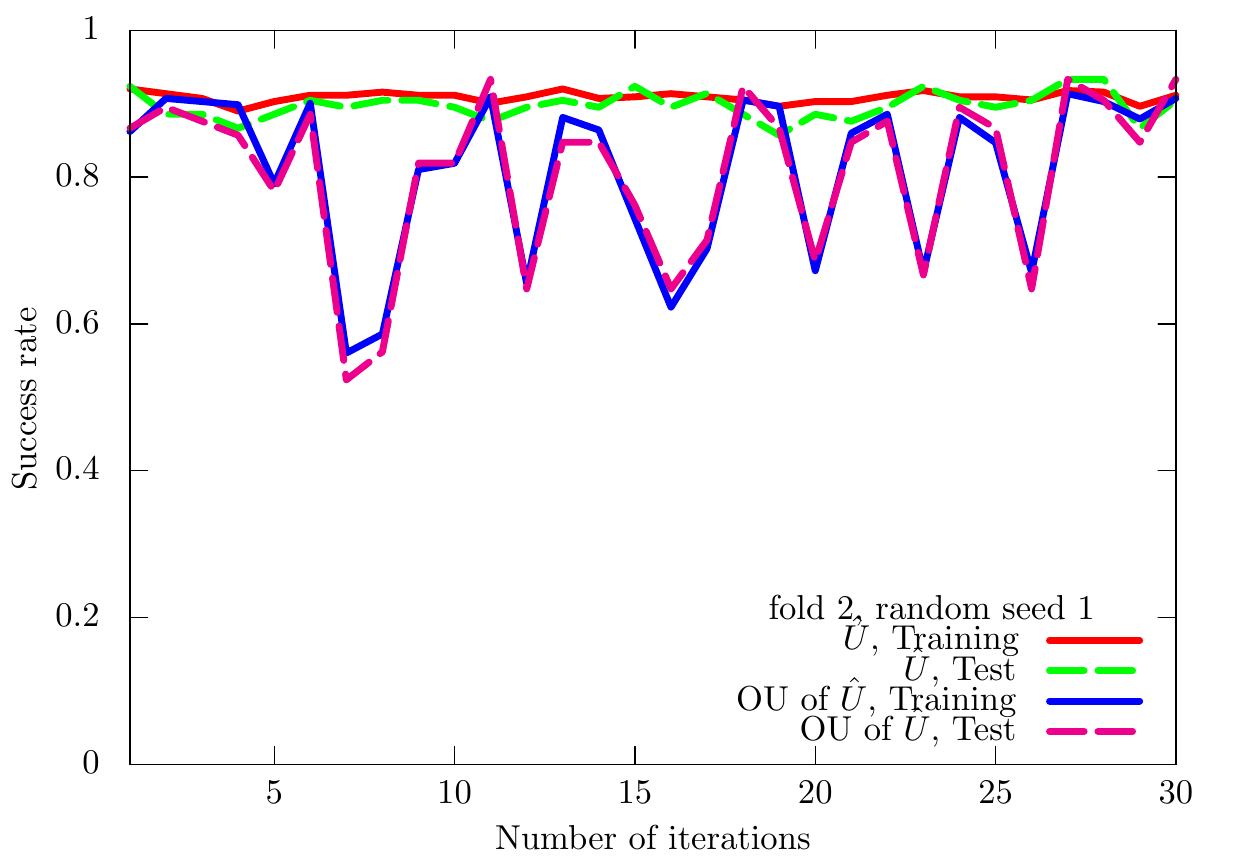}
\includegraphics[scale=0.25]{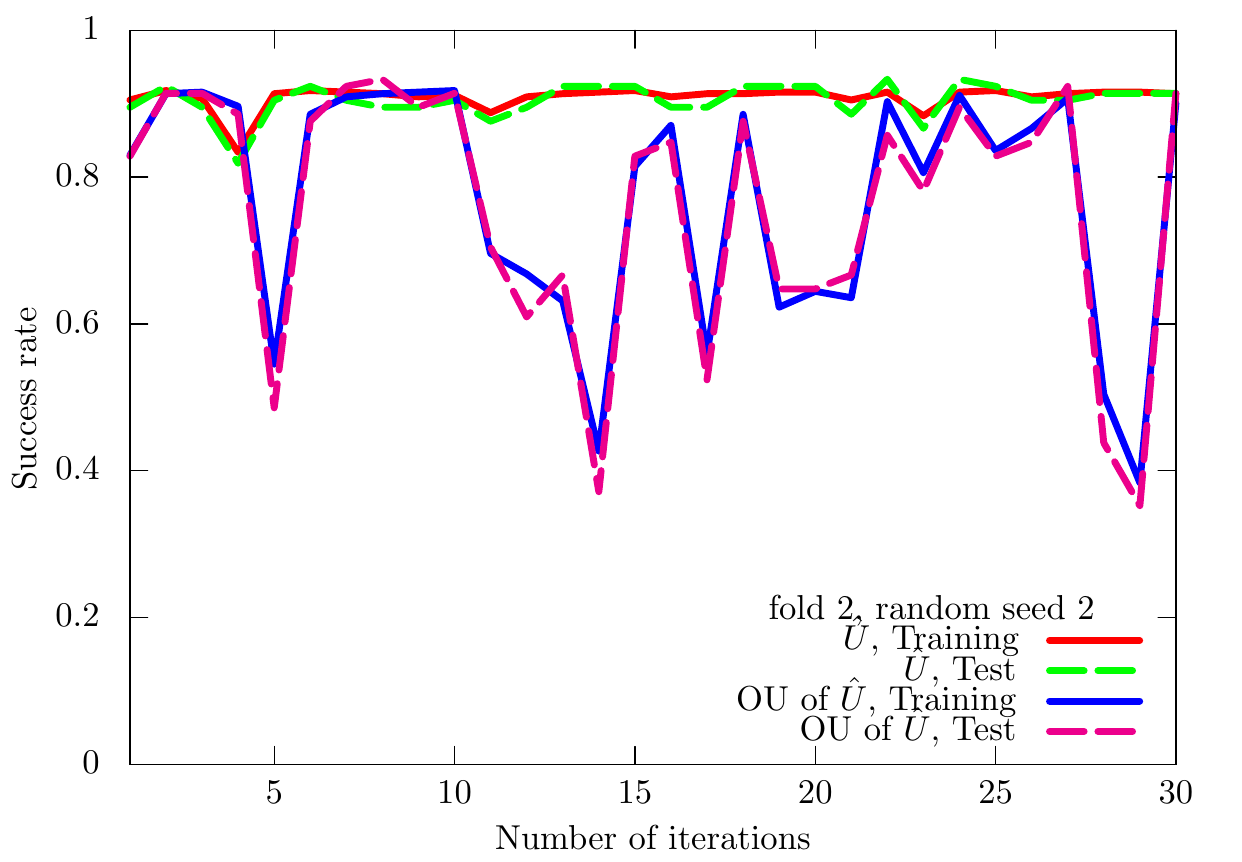}
\includegraphics[scale=0.25]{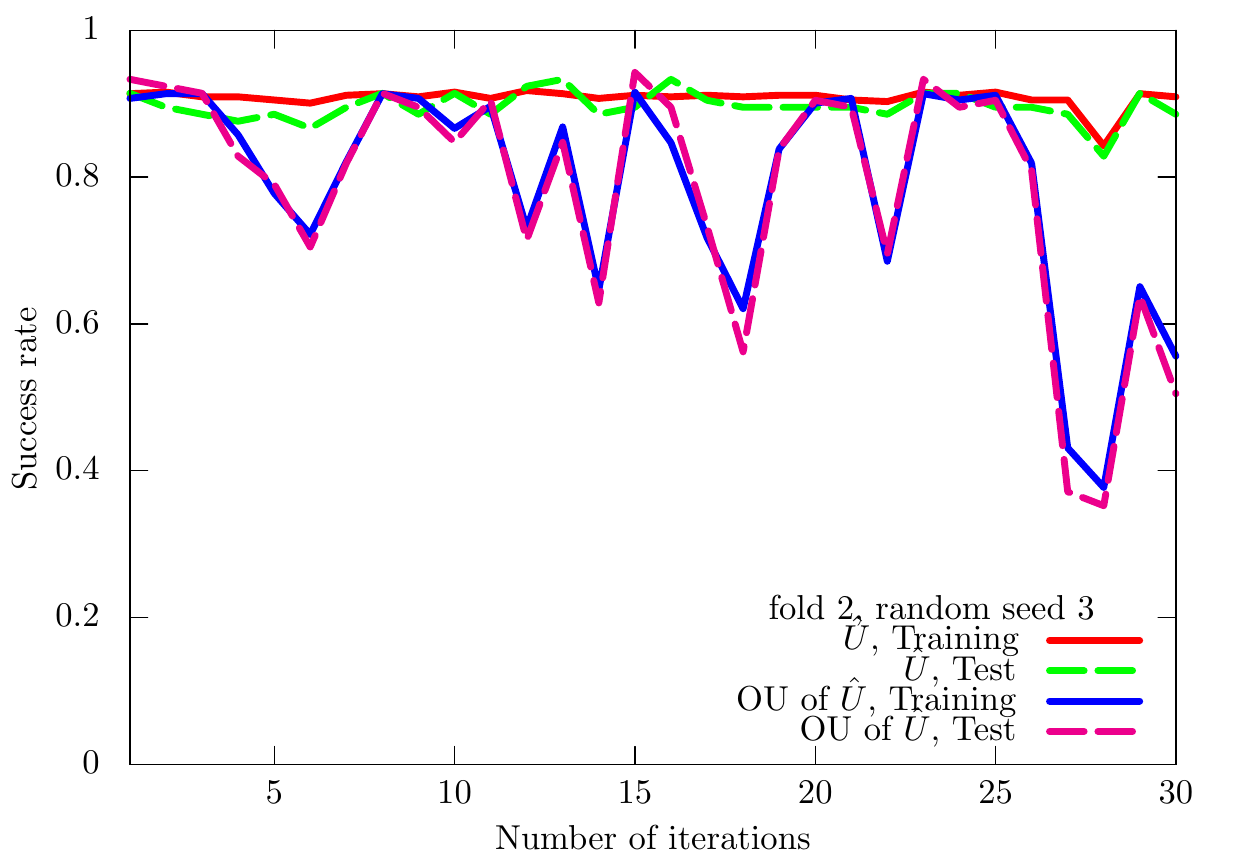}
\includegraphics[scale=0.25]{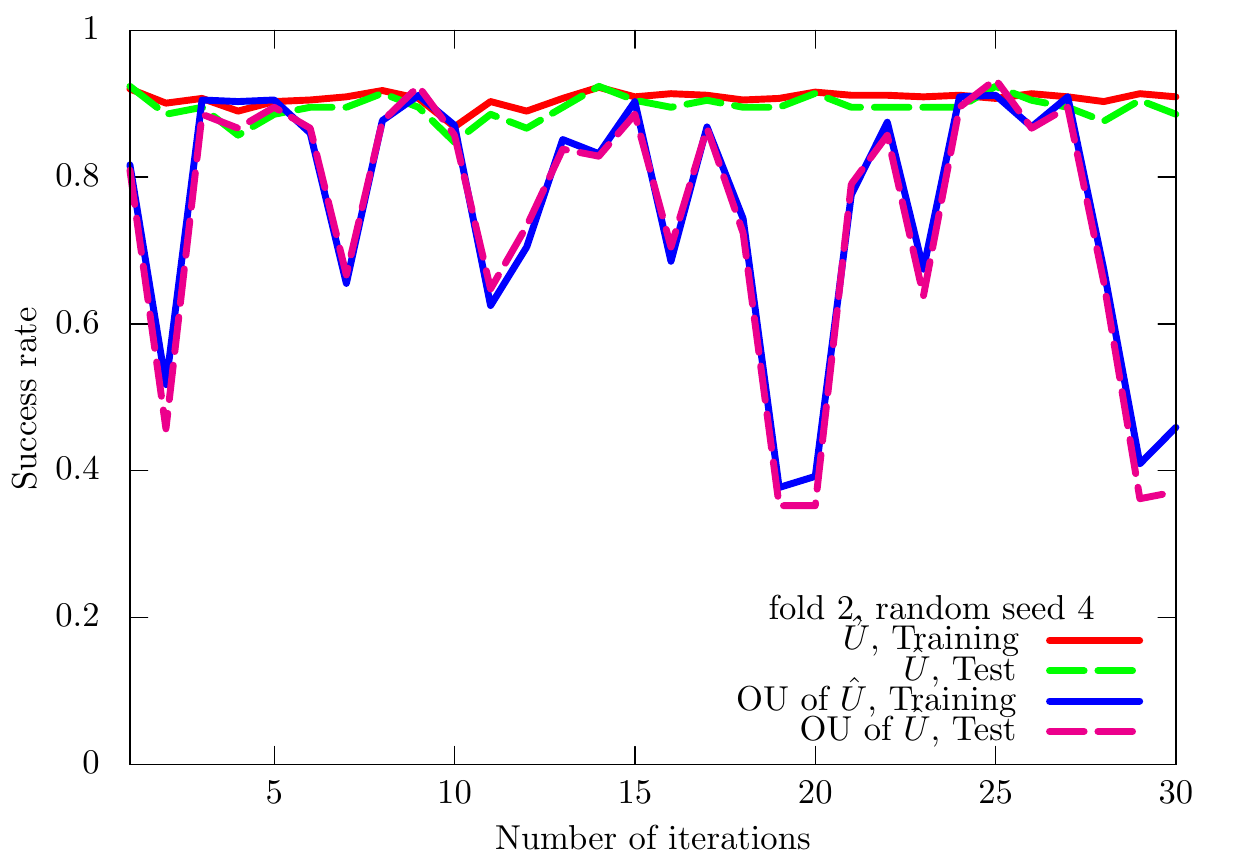}
\includegraphics[scale=0.25]{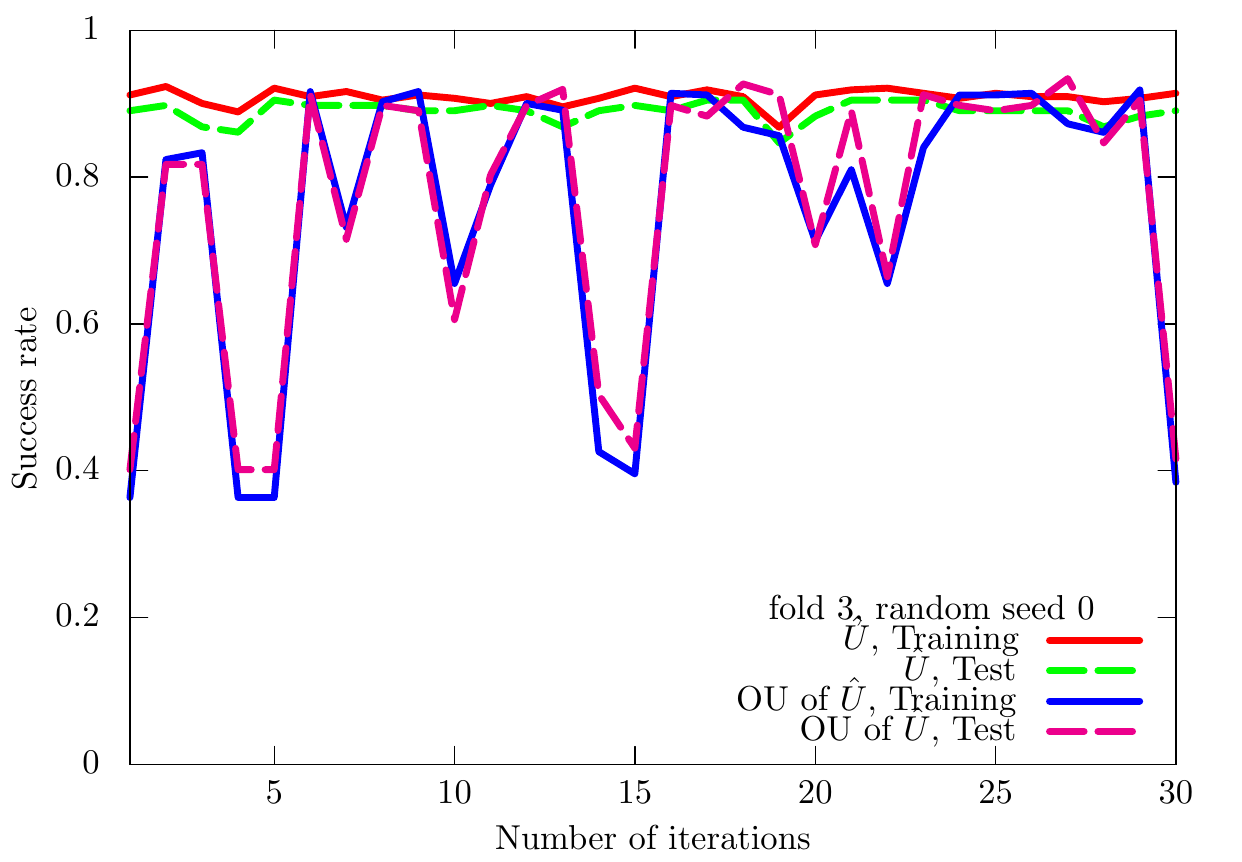}
\includegraphics[scale=0.25]{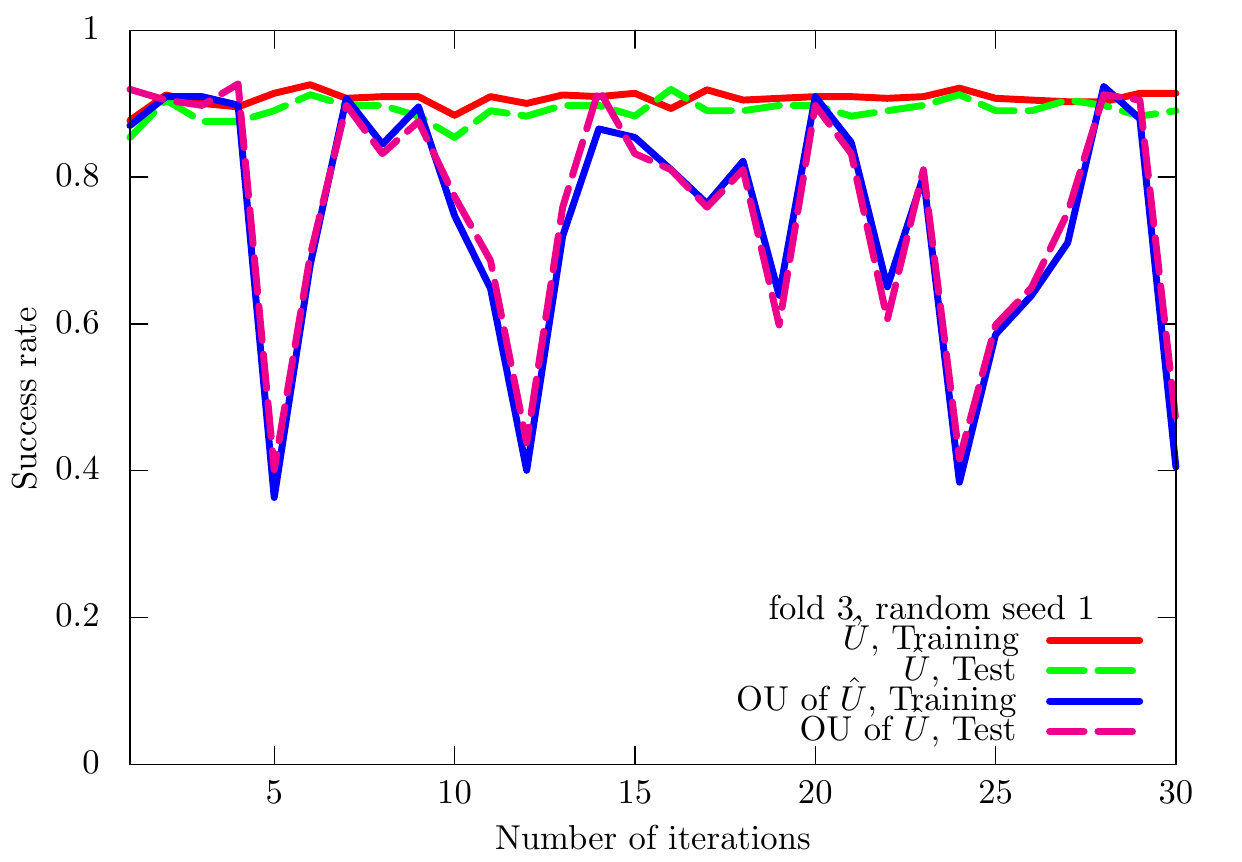}
\includegraphics[scale=0.25]{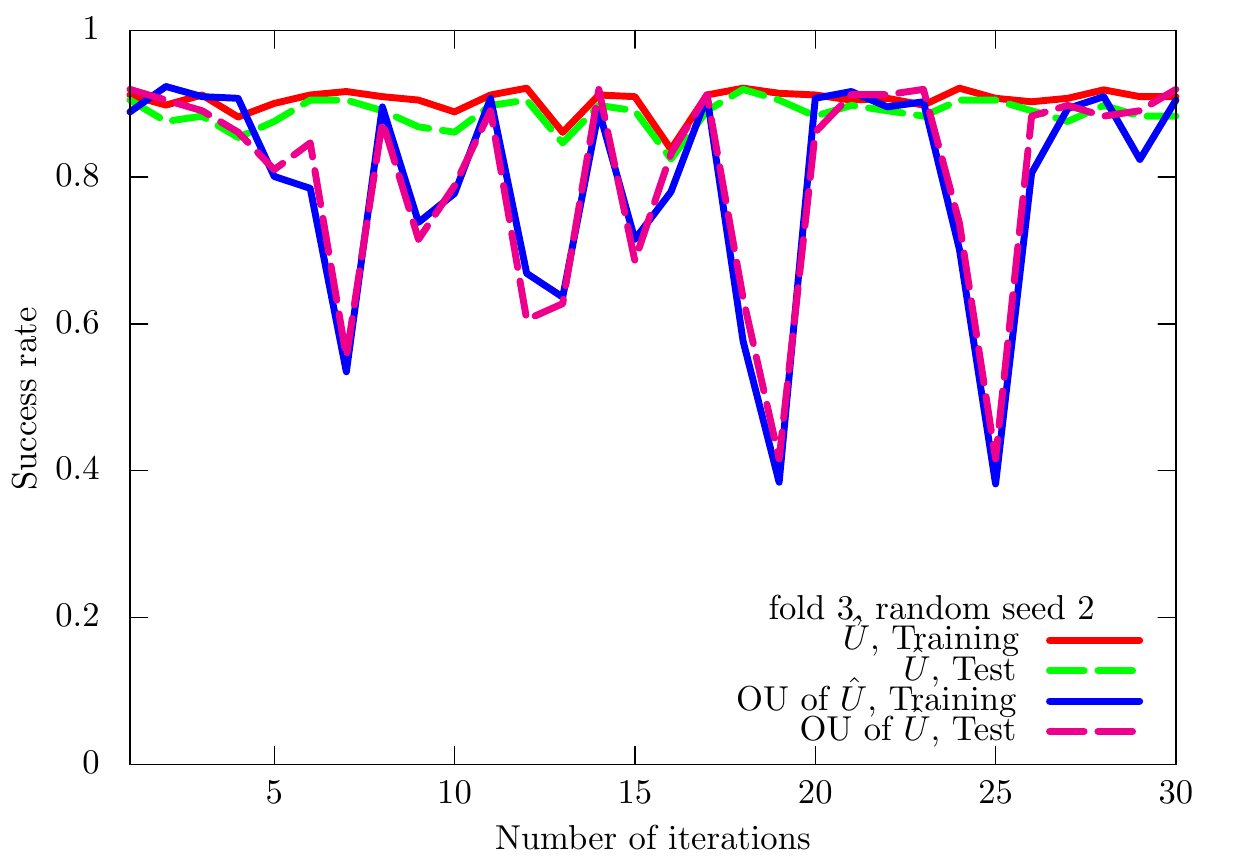}
\includegraphics[scale=0.25]{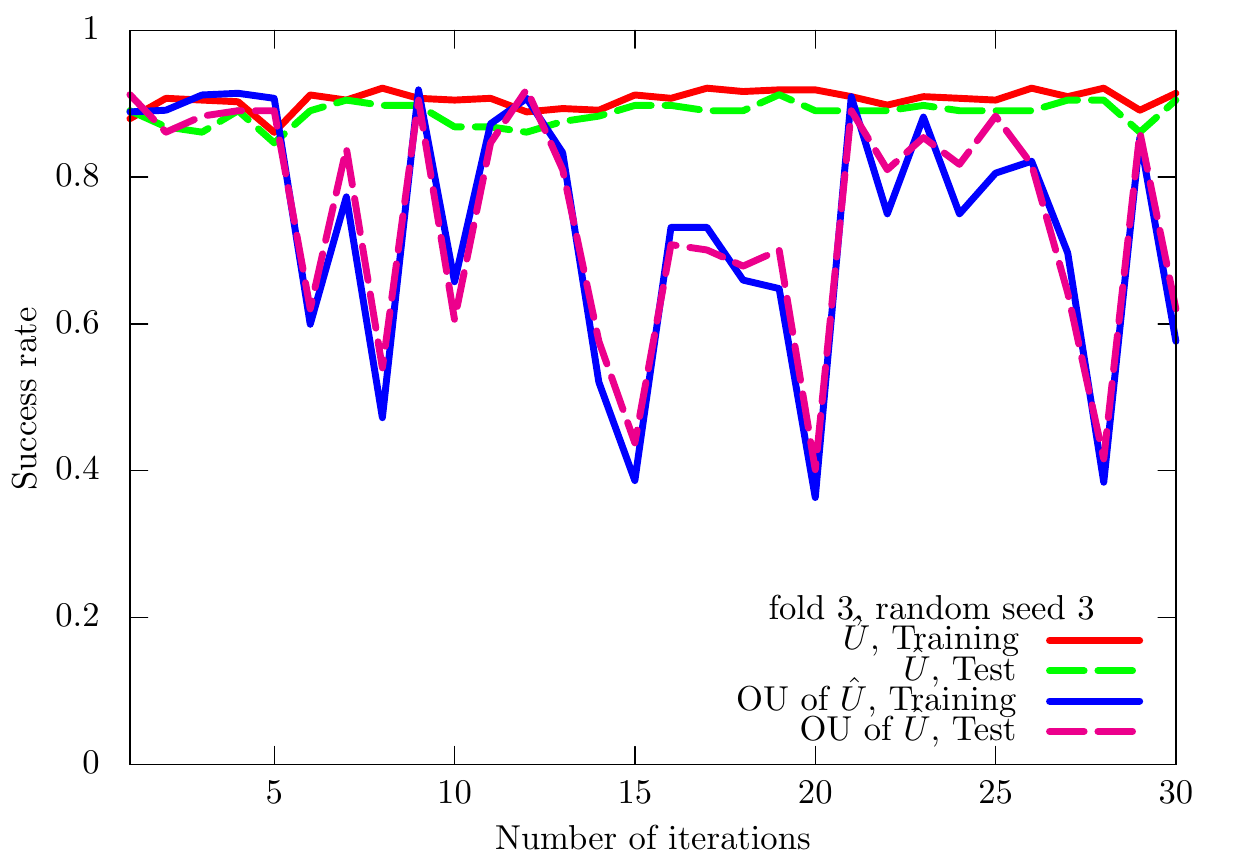}
\includegraphics[scale=0.25]{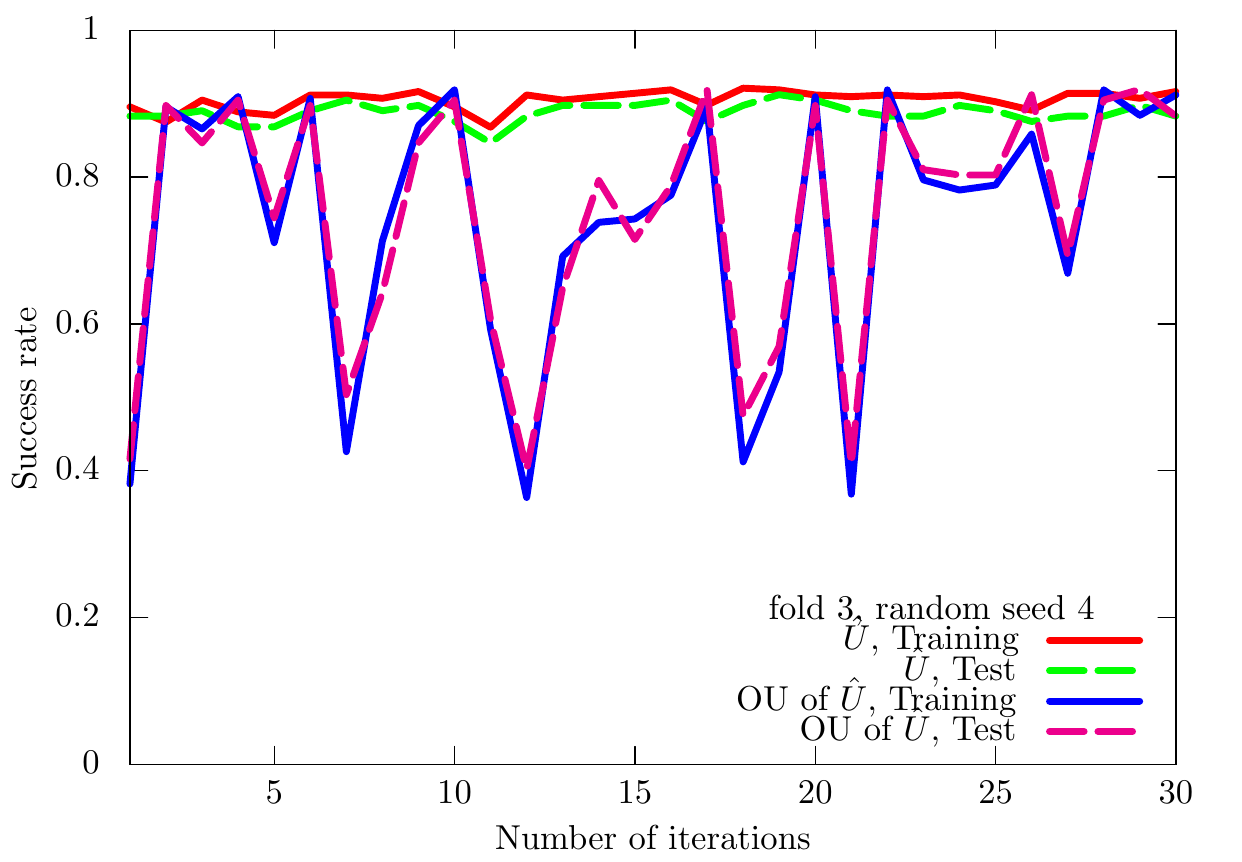}
\includegraphics[scale=0.25]{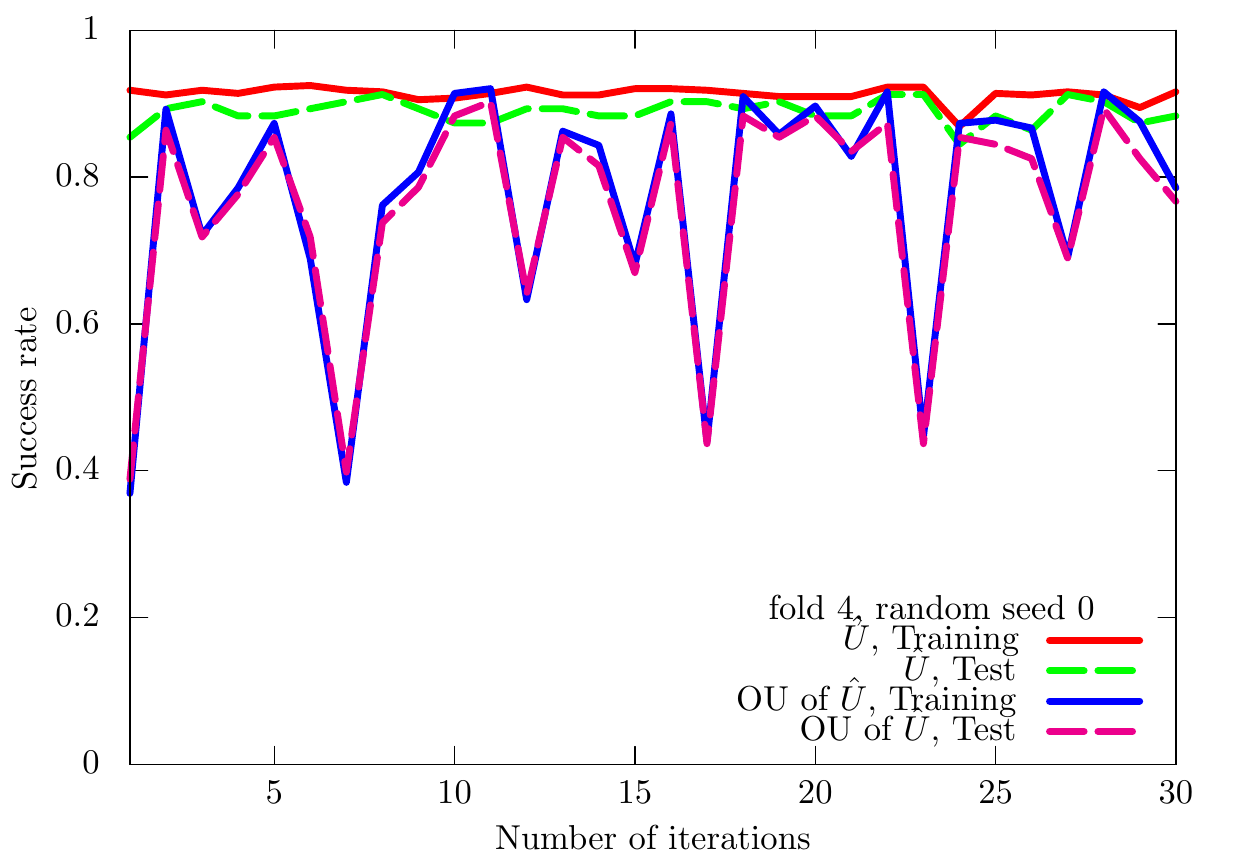}
\includegraphics[scale=0.25]{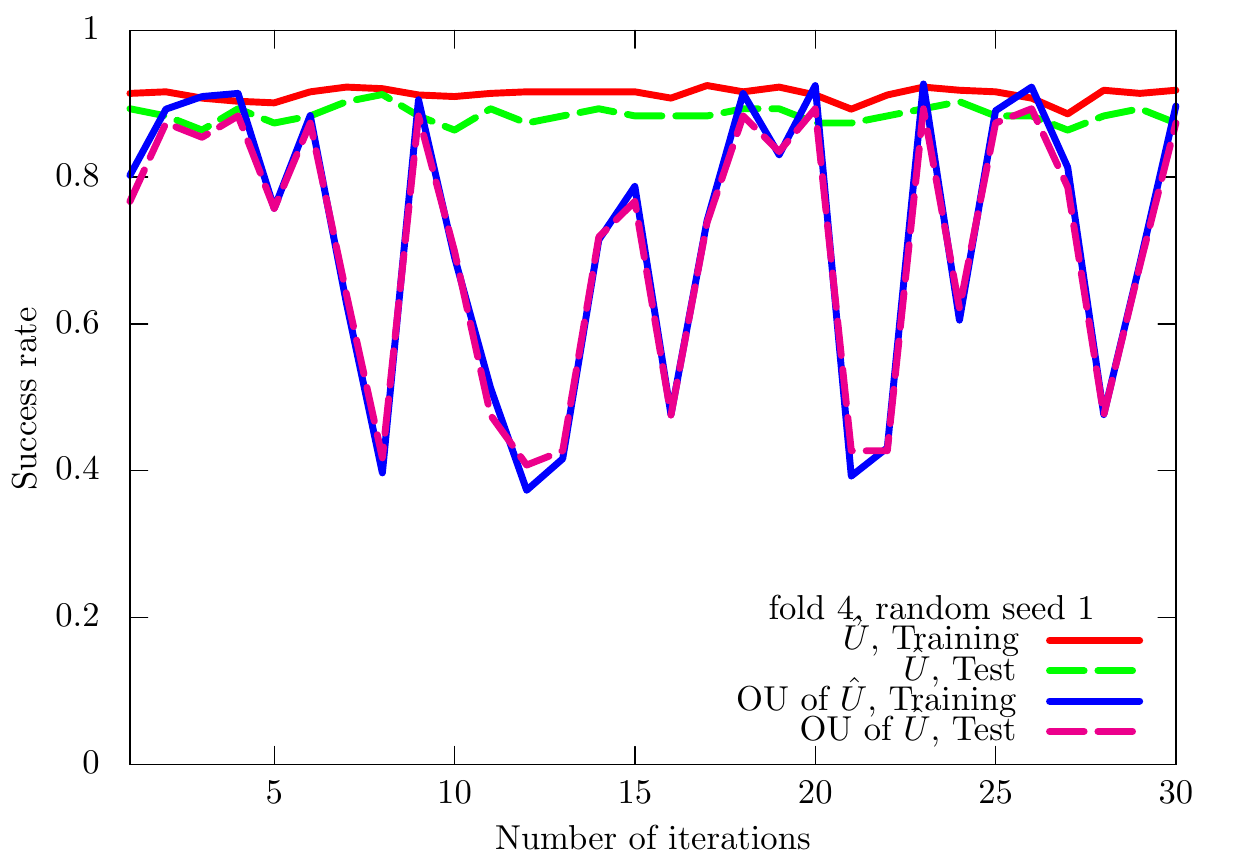}
\includegraphics[scale=0.25]{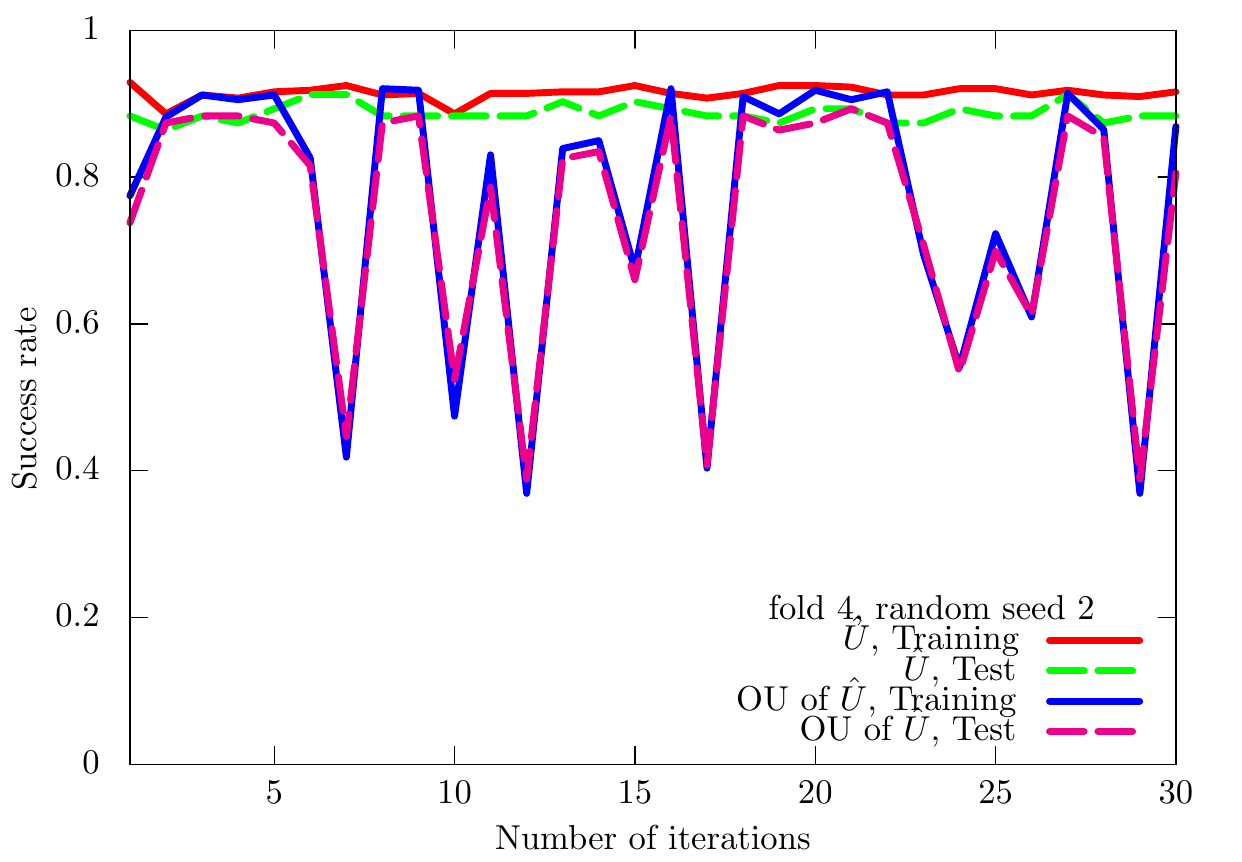}
\includegraphics[scale=0.25]{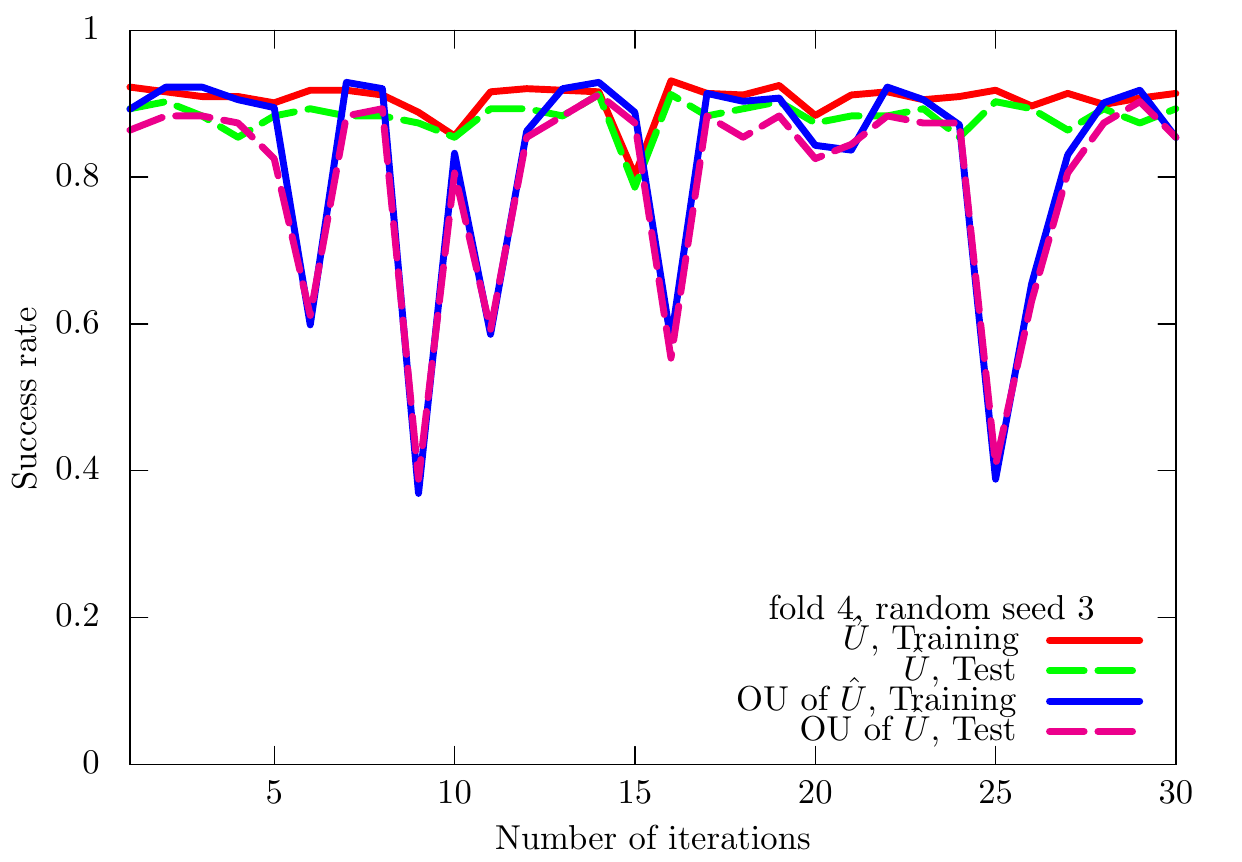}
\includegraphics[scale=0.25]{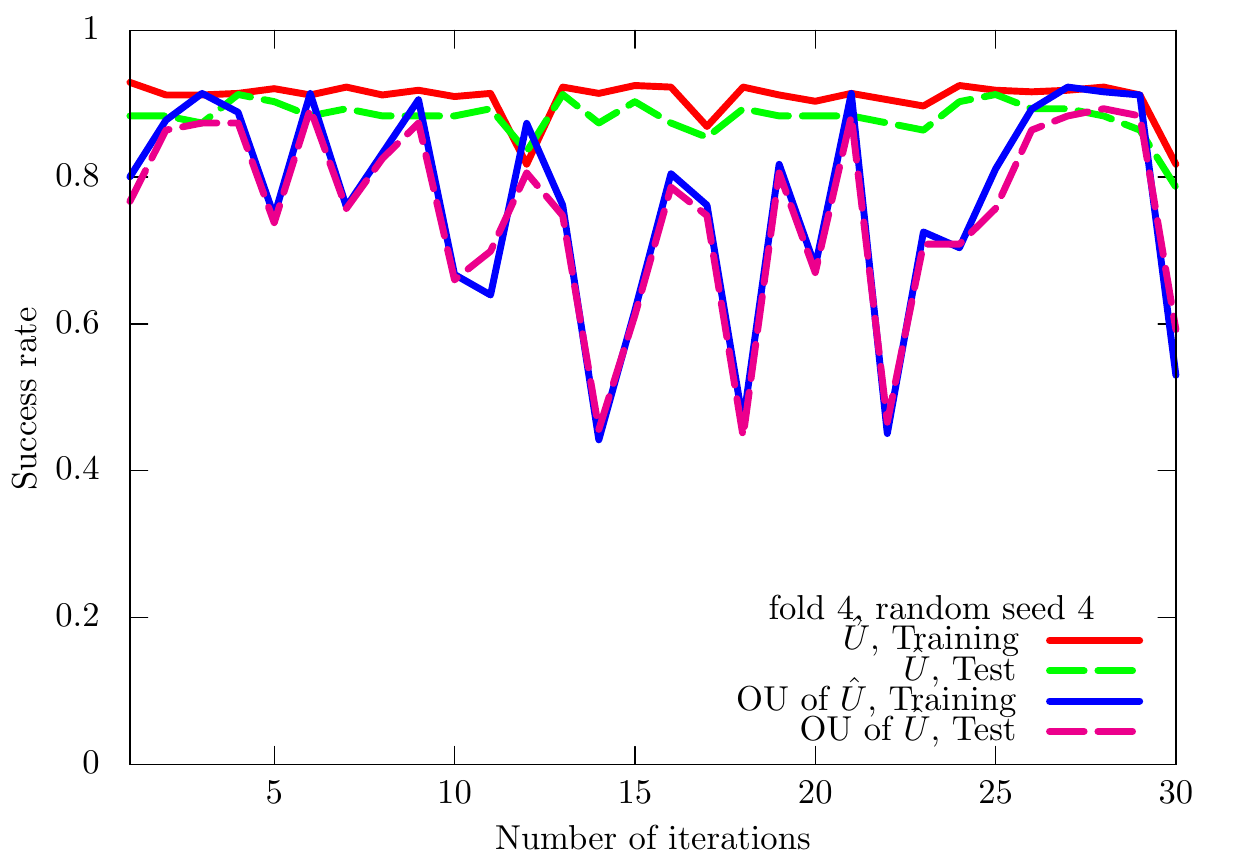}
\caption{Results of the UKM ($\hat{X}$ and OU of $\hat{X}$) on the $5$-fold datasets with $5$ different random seeds for the cancer dataset ($0$ or $1$). We use complex matrices and set $\theta_\mathrm{bias} = 0$. We set $r = 0.010$.}
\label{supp-arXiv-numerical-result-raw-data-fold-001-rand-001-UKM-OUU-UCI-cancer-0-1}
\end{figure*}

In Fig.~\ref{supp-arXiv-numerical-result-raw-data-fold-001-rand-001-QCL-UCI-cancer-0-1}, we show the numerical results of QCL for the $5$-fold datasets with $5$ different random seeds.
In Fig.~\ref{supp-arXiv-numerical-result-raw-data-fold-001-rand-001-UKM-P-UCI-cancer-0-1}, we show the numerical results of $\hat{P}$ of the UKM for the $5$-fold datasets with $5$ different random seeds.
In Fig.~\ref{supp-arXiv-numerical-result-raw-data-fold-001-rand-001-UKM-OUU-UCI-cancer-0-1}, we also show the numerical results of OU of $\hat{X}$ of the UKM for the $5$-fold datasets with $5$ different random seeds.

\begin{figure}[htb]
\centering
\includegraphics[scale=0.45]{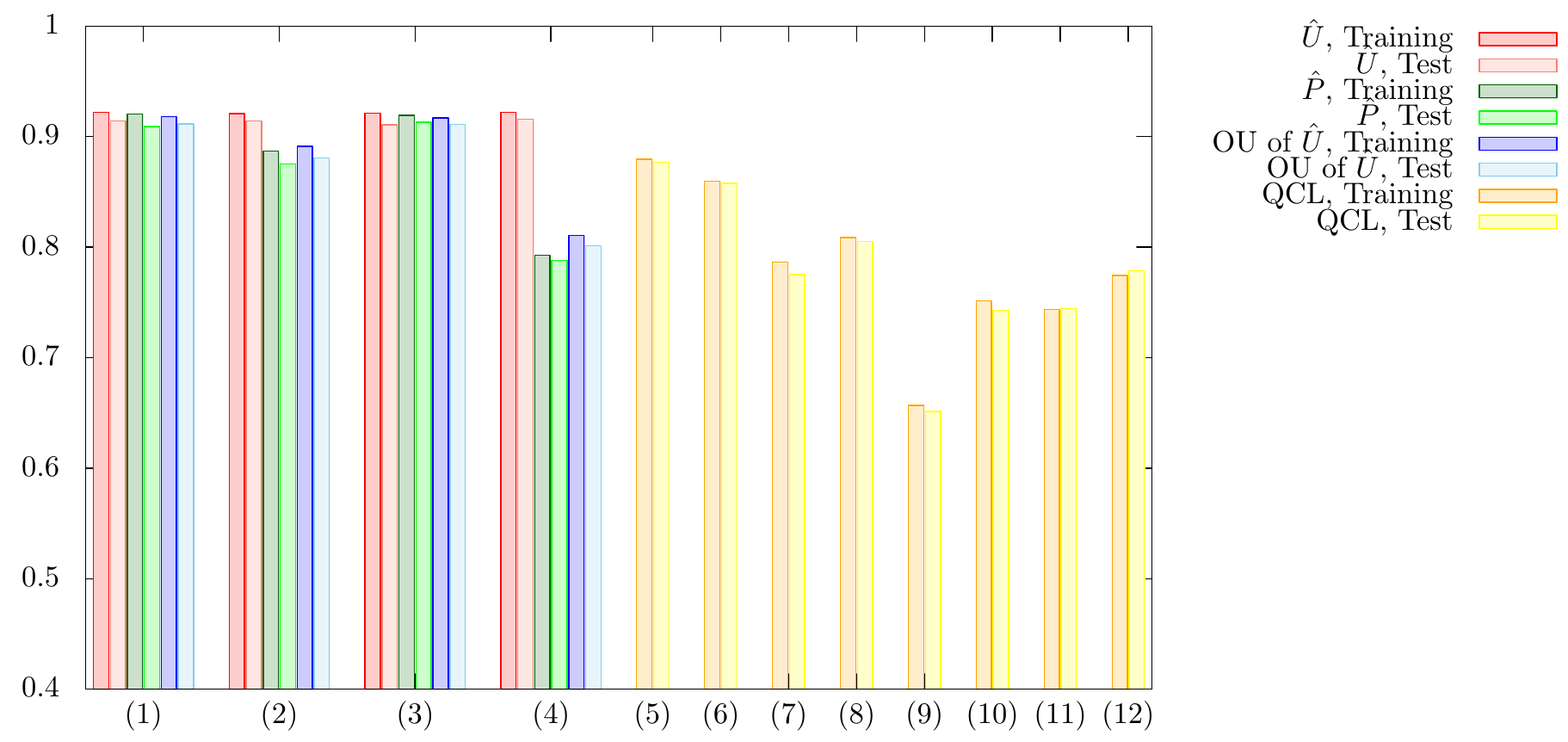}
\caption{Results of $5$-fold CV with $5$ different random seeds for the cancer dataset ($0$ or $1$). For the UKM, we put $r = 0.010$ and set $K = 30$ and $K' = 10$. For QCL, the number of layers $L$ is $5$ and the number of iterations is $300$. The numerical settings are as follows: (1) complex matrices without the bias term, (2) complex matrices with the bias term, (3) real matrices without the bias term, (4) real matrices with the bias term, (5) CNOT-based circuit without the bias term, (6) CNOT-based circuit with the bias term, (7) CRot-based circuit without the bias term, (8) CRot-based circuit with the bias term, (9) 1d Heisenberg circuit without the bias term, (10) 1d Heisenberg circuit with the bias term, (11) FC Heisenberg circuit without the bias term, and (12) FC Heisenberg circuit with the bias term.}
\label{supp-arXiv-numerical-result-performance-UKM-QCL-UCI-cancer-0-1}
\end{figure}

We summarize the results of 5-fold CV with 5 different random seeds of QCL and the UKM in Tables~\ref{supp-arXiv-table-UCI-cancer-0-1-002} and \ref{supp-arXiv-table-UCI-cancer-0-1-001}, respectively.
For QCL and the UKM, we select the best model for the training dataset over iterations to compute the performance.
\begin{table}[htb]
  \begin{tabular}{cc|cc}
    \hline \hline
    Algo. & Condition & Training & Test \\
    \hline
  QCL & CNOT-based, w/o bias & 0.8797 & 0.8768 \\
  QCL & CNOT-based, w/ bias & 0.8597 & 0.8577 \\
    \hline
  QCL & CRot-based, w/o bias & 0.7866 & 0.7752 \\
  QCL & CRot-based, w/ bias & 0.8085 & 0.8052 \\
    \hline
  QCL & 1d Heisenberg, w/o bias & 0.6568 & 0.6512 \\
  QCL & 1d Heisenberg, w/ bias & 0.7515 & 0.7427 \\
    \hline
  QCL & FC Heisenberg, w/o bias & 0.7435 & 0.7444 \\
  QCL & FC Heisenberg, w/ bias & 0.7744 & 0.7789 \\
    \hline \hline
  \end{tabular}
\caption{Results of $5$-fold CV with $5$ different random seeds of QCL for the cancer dataset ($0$ or $1$). The number of layers $L$ is set to $5$ and the number of iterations is set to $300$.}
\label{supp-arXiv-table-UCI-cancer-0-1-002}
\end{table}
\begin{table}[htb]
  \begin{tabular}{cc|cc}
    \hline \hline
    Algo. & Condition & Training & Test \\
    \hline
  UKM & $\hat{X}$, complex, w/o bias & 0.9219 & 0.9143 \\
  UKM & $\hat{P}$, complex, w/o bias & 0.9204 & 0.9093 \\
  UKM & OU of $\hat{X}$, complex, w/o bias & 0.9184 &	0.9115 \\
    \hline
  UKM & $\hat{X}$, complex, w/ bias & 0.9207 & 0.9143 \\
  UKM & $\hat{P}$, complex, w/ bias & 0.8870 & 0.8753 \\
  UKM & OU of $\hat{X}$, complex, w/ bias & 0.8912 & 0.8805 \\
    \hline
  UKM & $\hat{X}$, real, w/o bias & 0.9213 & 0.9107 \\
  UKM & $\hat{P}$, real, w/o bias & 0.9194 & 0.9131 \\
  UKM & OU of $\hat{X}$, real, w/o bias & 0.9170 & 0.9112 \\
    \hline
  UKM & $\hat{X}$, real, w/ bias & 0.9218 & 0.9160 \\
  UKM & $\hat{P}$, real, w/ bias & 0.7929 & 0.7879 \\
  UKM & OU of $\hat{X}$, real, w/ bias & 0.8107 & 0.8014 \\
    \hline \hline
  \end{tabular}
\caption{Results of $5$-fold CV with $5$ different random seeds of the UKM for the cancer dataset ($0$ or $1$). We put $r = 0.010$ and set $K = 30$ and $K' = 10$.}
\label{supp-arXiv-table-UCI-cancer-0-1-001}
\end{table}
In Fig.~\ref{supp-arXiv-numerical-result-performance-UKM-QCL-UCI-cancer-0-1}, we plot the data shown in Tables~\ref{supp-arXiv-table-UCI-cancer-0-1-002} and \ref{supp-arXiv-table-UCI-cancer-0-1-001}.
We also summarize the results of 5-fold CV with 5 different random seeds of the kernel method in Table~\ref{supp-arXiv-table-UCI-cancer-0-1-003}.
More specifically, we use Ridge classification in Sec.~\ref{supp-arXiv-sec-Ridge-001}.
We consider the linear functions and the second-order polynomial functions for $\phi (\cdot)$ in Eq.~\eqref{supp-arXiv-f-pred-kernel-method-001-002} with and without normalization.
We set $\lambda = 10^{-2}, 10^{-1}, 1$ where $\lambda$ is the coefficient of the regularization term.
\begin{table}[htb]
  \begin{tabular}{cc|cc}
    \hline \hline
    Algo. & Condition & Training & Test \\
    \hline
  Kernel method & Linear, w/o normalization, $\lambda = 10^{-2}$ & 0.9618 & 0.9568 \\
  Kernel method & Linear, w/o normalization, $\lambda = 10^{-1}$ & 0.9623 & 0.9549 \\
  Kernel method & Linear, w/o normalization, $\lambda = 1$ & 0.9591 & 0.9495 \\
    \hline
  Kernel method & Linear, w/ normalization, $\lambda = 10^{-2}$ & 0.9262 & 0.9247 \\
  Kernel method & Linear, w/ normalization, $\lambda = 10^{-1}$ & 0.9205 & 0.9176 \\
  Kernel method & Linear, w/ normalization, $\lambda = 1$ & 0.8830 & 0.8812 \\
    \hline
  Kernel method & Poly-2, w/o normalization, $\lambda = 10^{-2}$ & 0.9242 & 0.8491 \\
  Kernel method & Poly-2, w/o normalization, $\lambda = 10^{-1}$ & 0.9936 & 0.9361\\
  Kernel method & Poly-2, w/o normalization, $\lambda = 1$ & 0.9907 & 0.9454 \\
    \hline
  Kernel method & Poly-2, w/ normalization, $\lambda = 10^{-2}$ & 0.9298 & 0.9264 \\
  Kernel method & Poly-2, w/ normalization, $\lambda = 10^{-1}$ & 0.9210 & 0.9195 \\
  Kernel method & Poly-2, w/ normalization, $\lambda = 1$ & 0.9038 & 0.9052 \\
    \hline \hline
  \end{tabular}
\caption{Results of 5-fold CV with 5 different random seeds of the kernel method for the cancer dataset ($0$ or $1$).}
\label{supp-arXiv-table-UCI-cancer-0-1-003}
\end{table}

\begin{figure}[htb]
\centering
\includegraphics[scale=0.45]{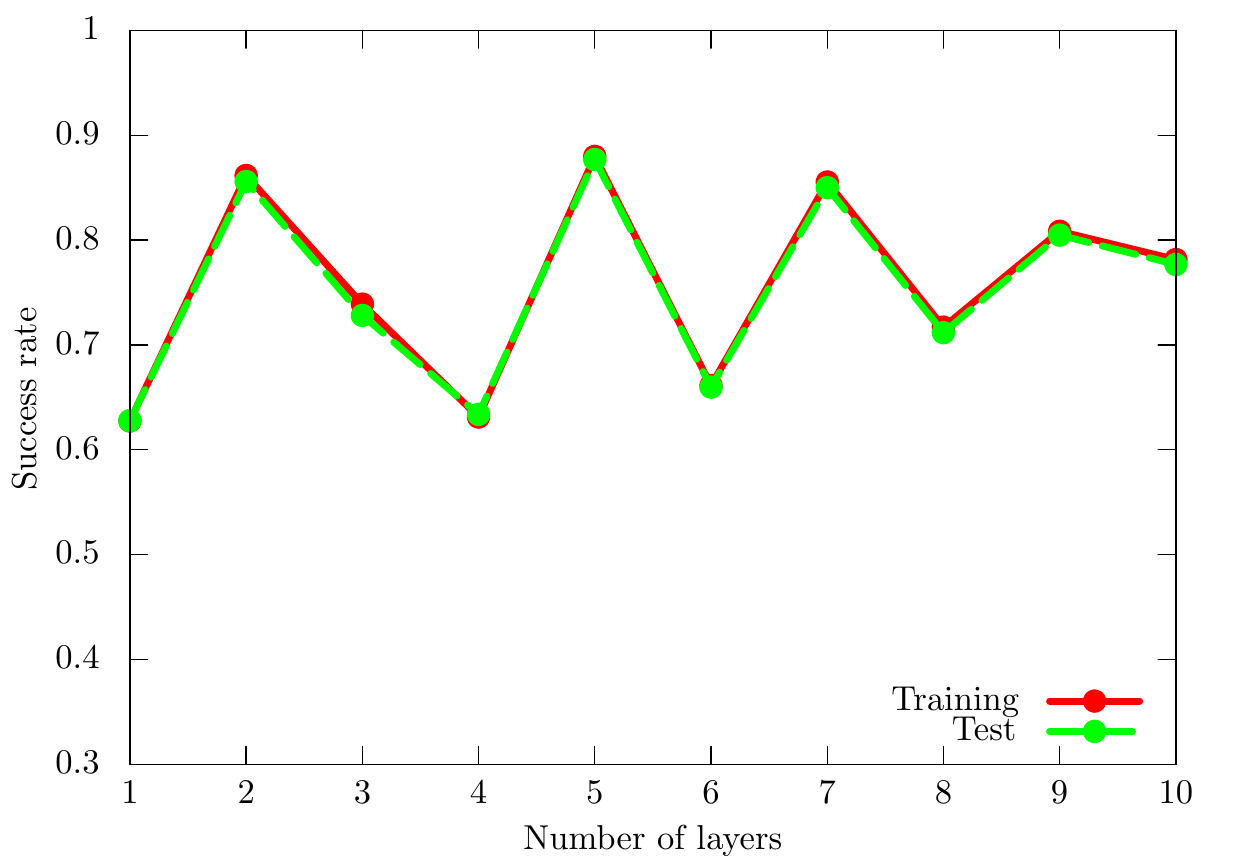}
\caption{Performance dependence of QCL on the number of layers $L$ for the cancer dataset ($0$ or $1$). We use the CNOT-based circuit geometry and set $\theta_\mathrm{bias} = 0$. We iterate the computation $300$ times.}
\label{supp-arXiv-numerical-result-layers-dependence-QCL-UCI-cancer-0-1}
\end{figure}

\begin{figure}[htb]
\centering
\includegraphics[scale=0.45]{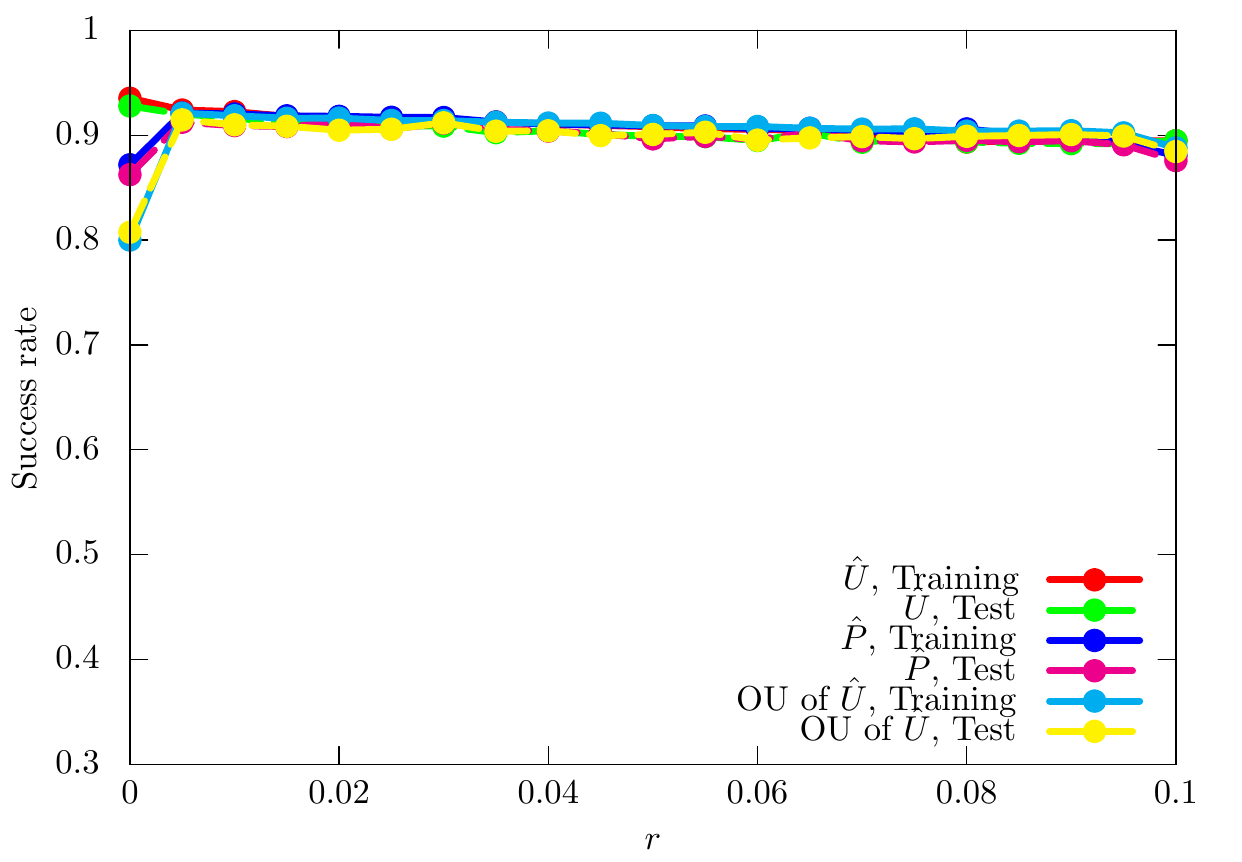}
\includegraphics[scale=0.45]{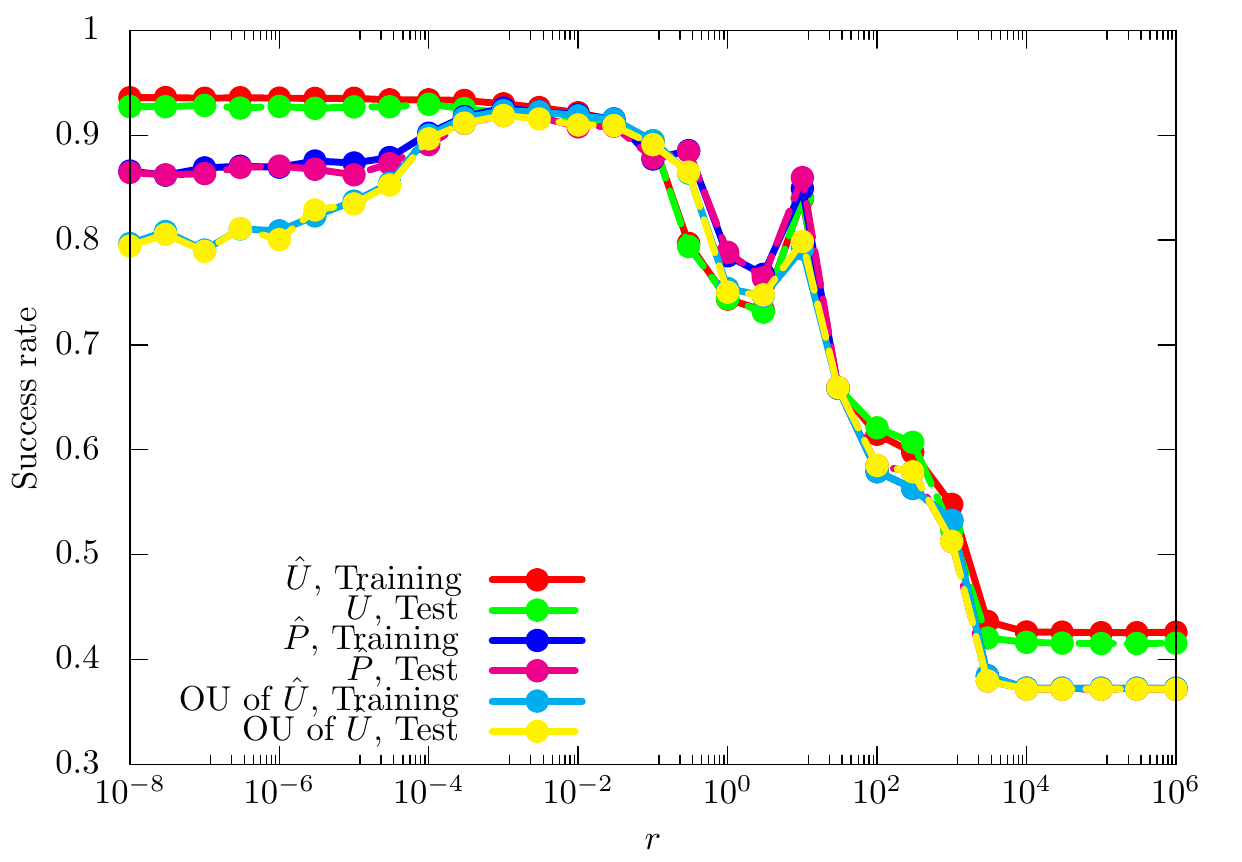}
\caption{Performance dependence of the UKM on $r$, which is the coefficient of the second term in the right-hand side of Eq.~\eqref{supp-arXiv-quantum-kernel-method-001-011} for the cancer dataset ($0$ or $1$). We use complex matrices and set $\theta_\mathrm{bias} = 0$. We set $K = 30$ and $K' = 10$.}
\label{supp-arXiv-numerical-result-r-dependence-UKM-UCI-cancer-0-1}
\end{figure}

\begin{figure}[htb]
\centering
\includegraphics[scale=0.45]{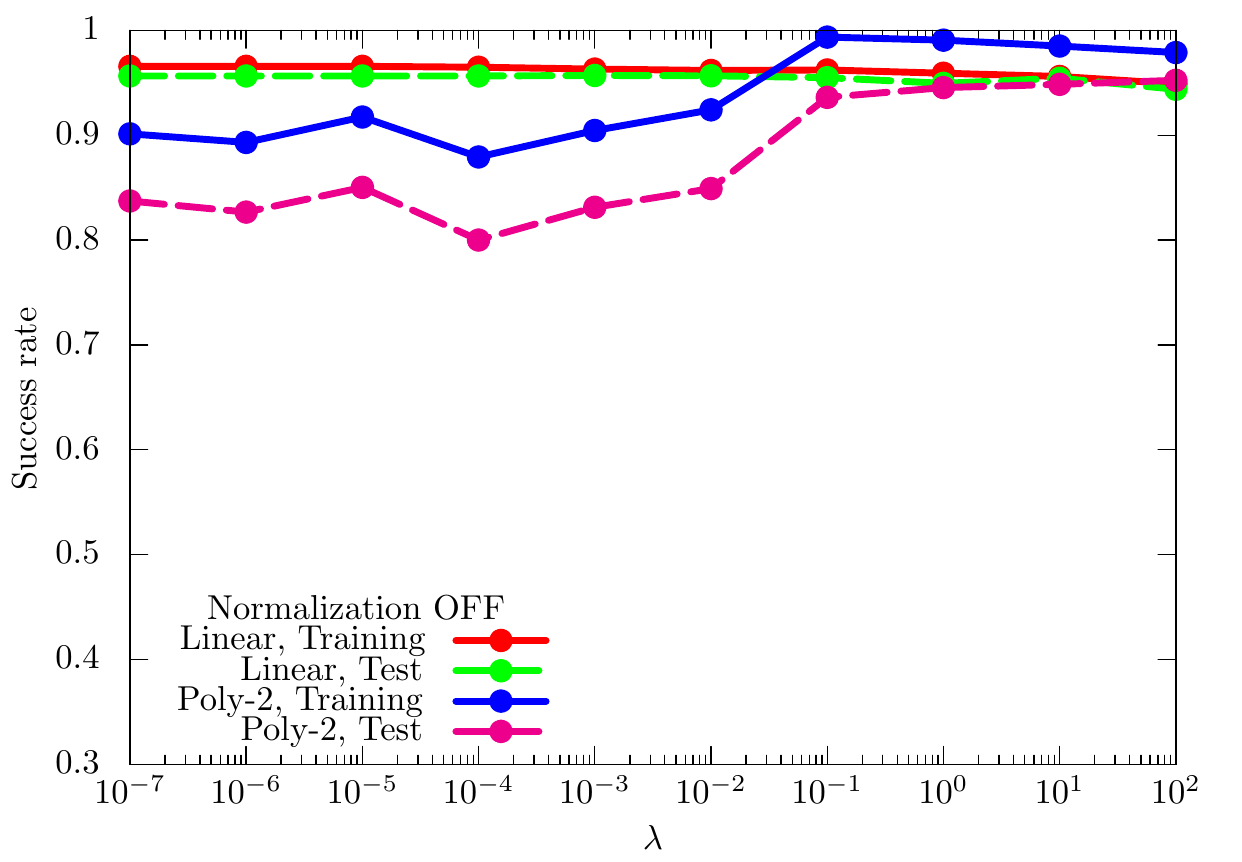}
\includegraphics[scale=0.45]{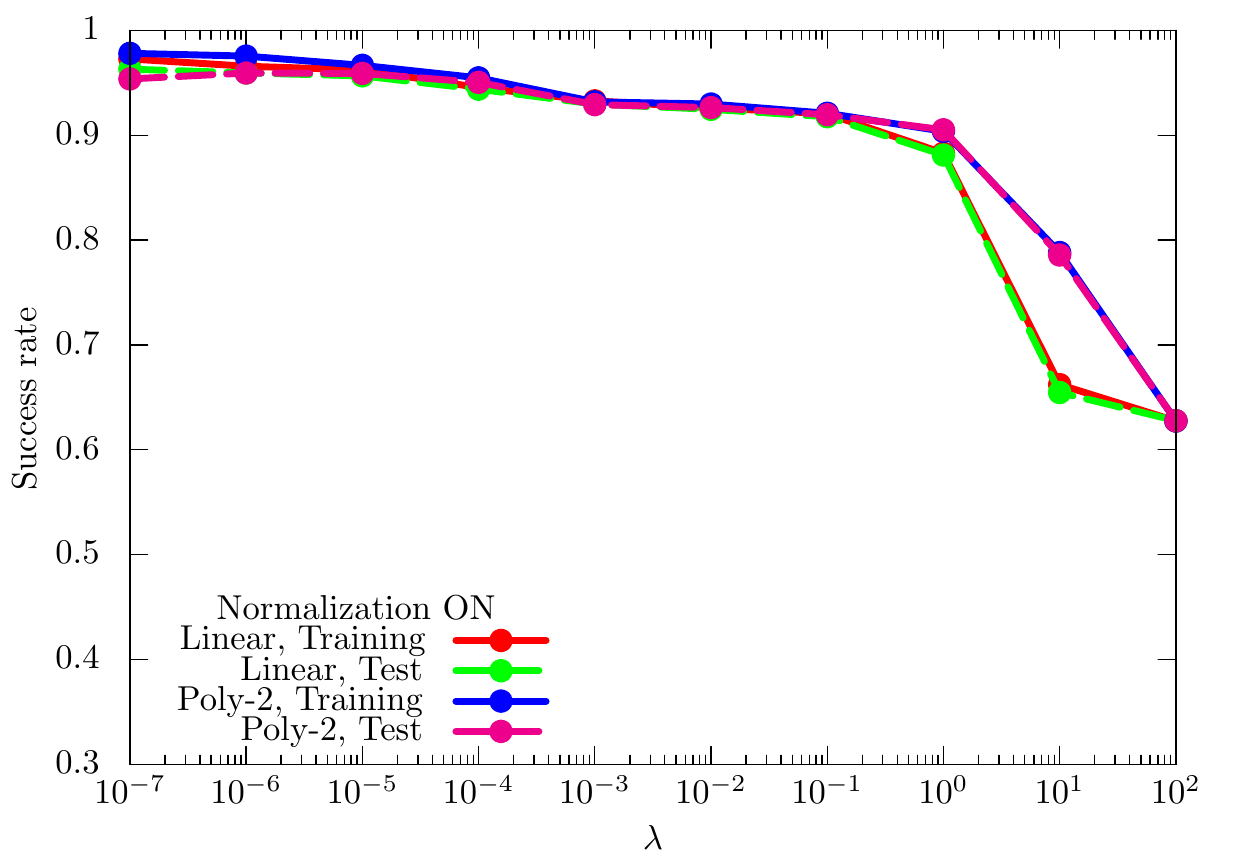}
\caption{Performance dependence of the kernel method on $\lambda$, which is the coefficient of the second term in the right-hand side of Eq.~\eqref{supp-arXiv-cost-function-kernel-method-001-002} for the cancer dataset ($0$ or $1$). For $\phi (\cdot)$ in Eq.~\eqref{supp-arXiv-f-pred-kernel-method-001-002}, we use the linear functions and the second-degree polynomial functions with and without normalization.}
\label{supp-arXiv-numerical-result-lambda-dependence-kernel-method-cancer-0-1}
\end{figure}

Next, we show the performance dependence of the three algorithms on their key parameters.
We see the performance dependence of QCL on the number of layers $L$.
The result is shown in Fig.~\ref{supp-arXiv-numerical-result-layers-dependence-QCL-UCI-cancer-0-1}.
We then ee the performance dependence of the UKM on $r$, which is the coefficient of the second term in the right-hand side of Eq.~\eqref{supp-arXiv-quantum-kernel-method-001-011}.
The result is shown in Fig.~\ref{supp-arXiv-numerical-result-r-dependence-UKM-UCI-cancer-0-1}.
In Fig.~\ref{supp-arXiv-numerical-result-lambda-dependence-kernel-method-cancer-0-1}, we show the performance dependence of the kernel method on $\lambda$, which is the coefficient of the second term in the right-hand side of Eq.~\eqref{supp-arXiv-cost-function-kernel-method-001-002}.

\begin{figure}[htb]
\centering
\includegraphics[scale=0.45]{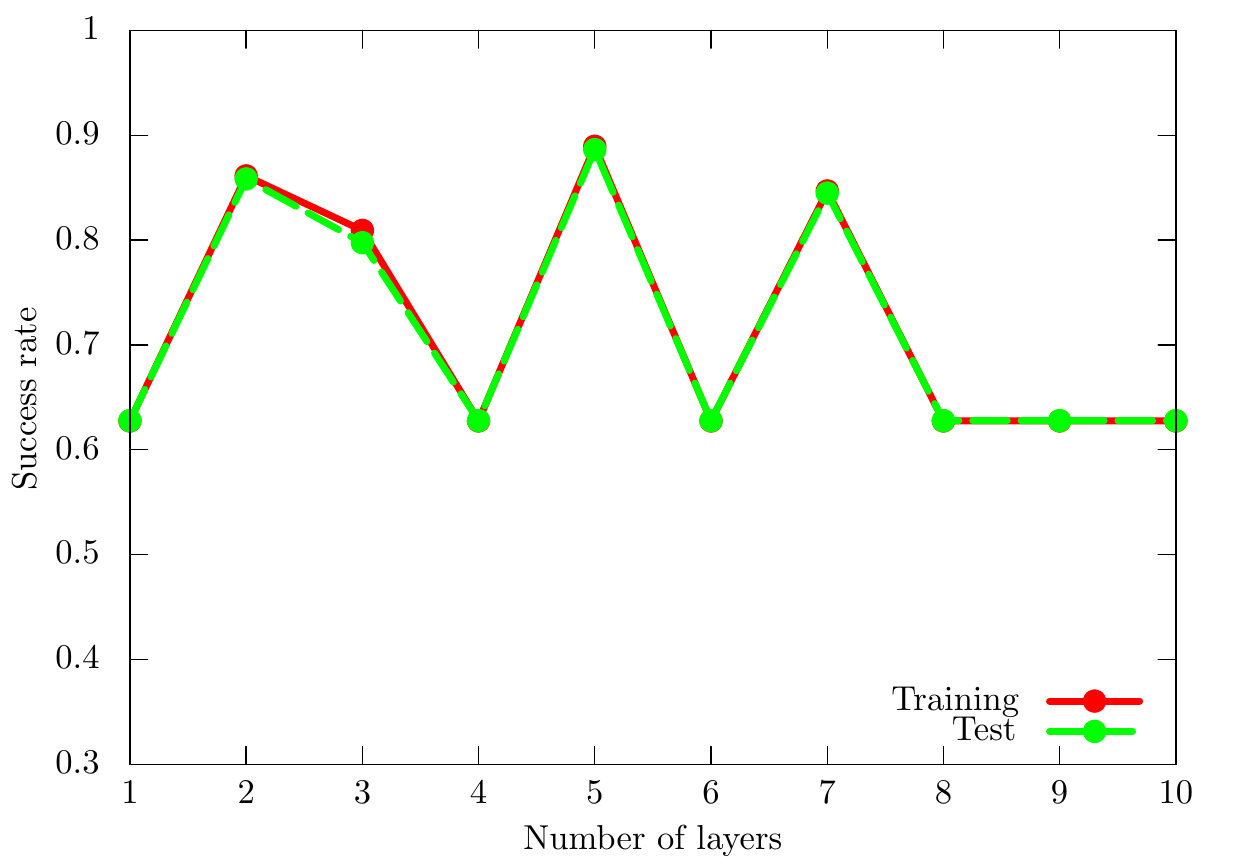}
\caption{Performance dependence of QCL on the number of layers $L$ for the cancer dataset ($0$ or $1$) in the case of the hinge function, Eq.~\eqref{supp-arXiv-hinge-function-001-001}. We use the CNOT-based circuit geometry and set $\theta_\mathrm{bias} = 0$. We iterate the computation $300$ times.}
\label{supp-arXiv-numerical-result-layers-dependence-QCL-UCI-cancer-0-1-hinge}
\end{figure}

\begin{figure}[htb]
\centering
\includegraphics[scale=0.45]{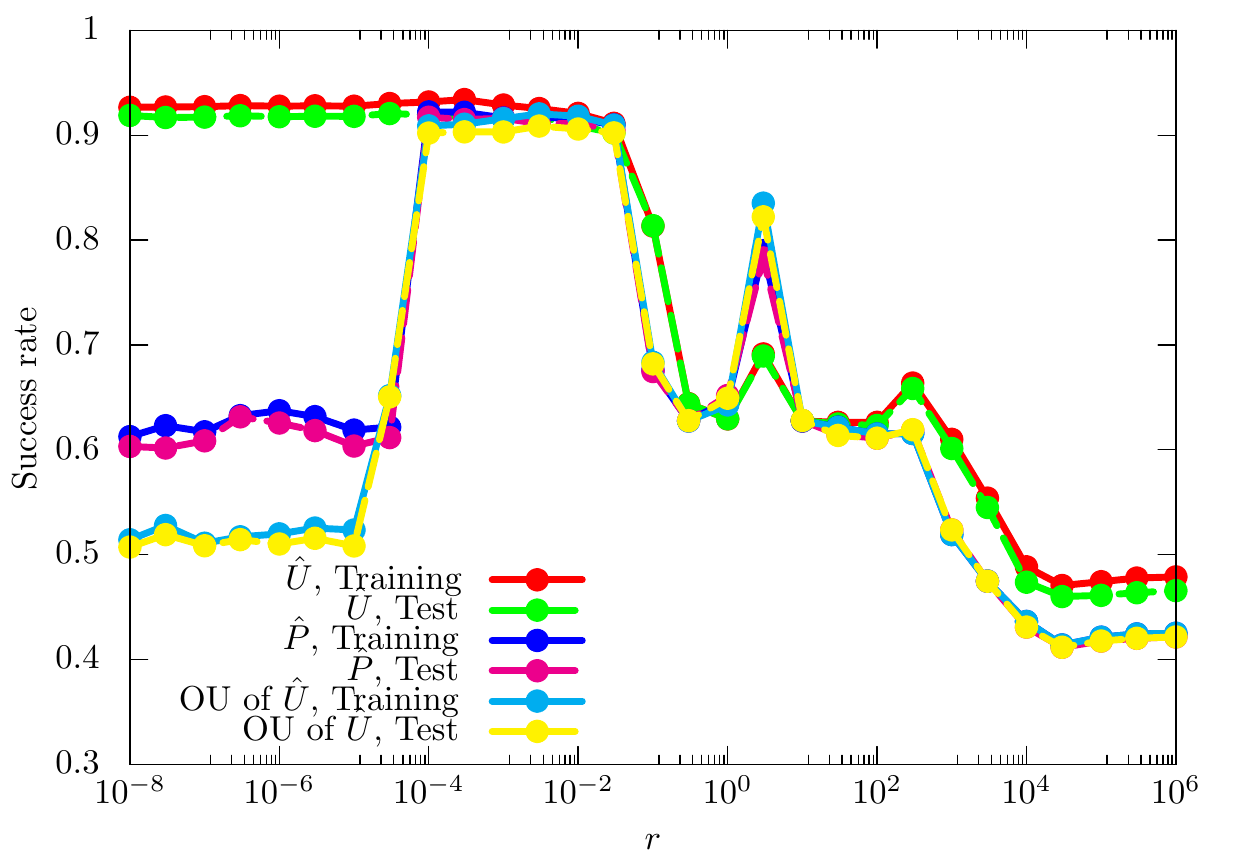}
\caption{Performance dependence of the UKM on $r$, which is the coefficient of the second term in the right-hand side of Eq.~\eqref{supp-arXiv-quantum-kernel-method-001-011} for the cancer dataset ($0$ or $1$) in the case of the hinge function, Eq.~\eqref{supp-arXiv-hinge-function-001-001}. We use complex matrices and set $\theta_\mathrm{bias} = 0$. We set $K = 30$ and $K' = 10$.}
\label{supp-arXiv-numerical-result-r-dependence-UKM-UCI-cancer-0-1-hinge}
\end{figure}

So far, we have used the squared error function, Eq.~\eqref{supp-arXiv-squared-error-function-001-001}.
In Fig.~\ref{supp-arXiv-numerical-result-layers-dependence-QCL-UCI-cancer-0-1-hinge}, we show the performance dependence of QCL on the number of layers $L$ in the case of the hinge function, Eq.~\eqref{supp-arXiv-hinge-function-001-001}.
In Fig.~\ref{supp-arXiv-numerical-result-r-dependence-UKM-UCI-cancer-0-1-hinge}, we show the performance dependence of the UKM on $r$, which is the coefficient of the second term in the right-hand side of Eq.~\eqref{supp-arXiv-quantum-kernel-method-001-011}, in the case of the hinge function, Eq.~\eqref{supp-arXiv-hinge-function-001-001}.

\clearpage

\subsection{Sonar dataset ($0$ or $1$)}

We here show the numerical result for the sonar dataset ($0$ or $1$).
For the UKM, we put $r = 0.010$ and set $K = 30$ and $K' = 10$ in Algo.~\ref{supp-arXiv-quantum-kernel-method-002-001}.
For QCL, we run iterations $300$ times.

\begin{figure*}[htb]
\centering
\includegraphics[scale=0.25]{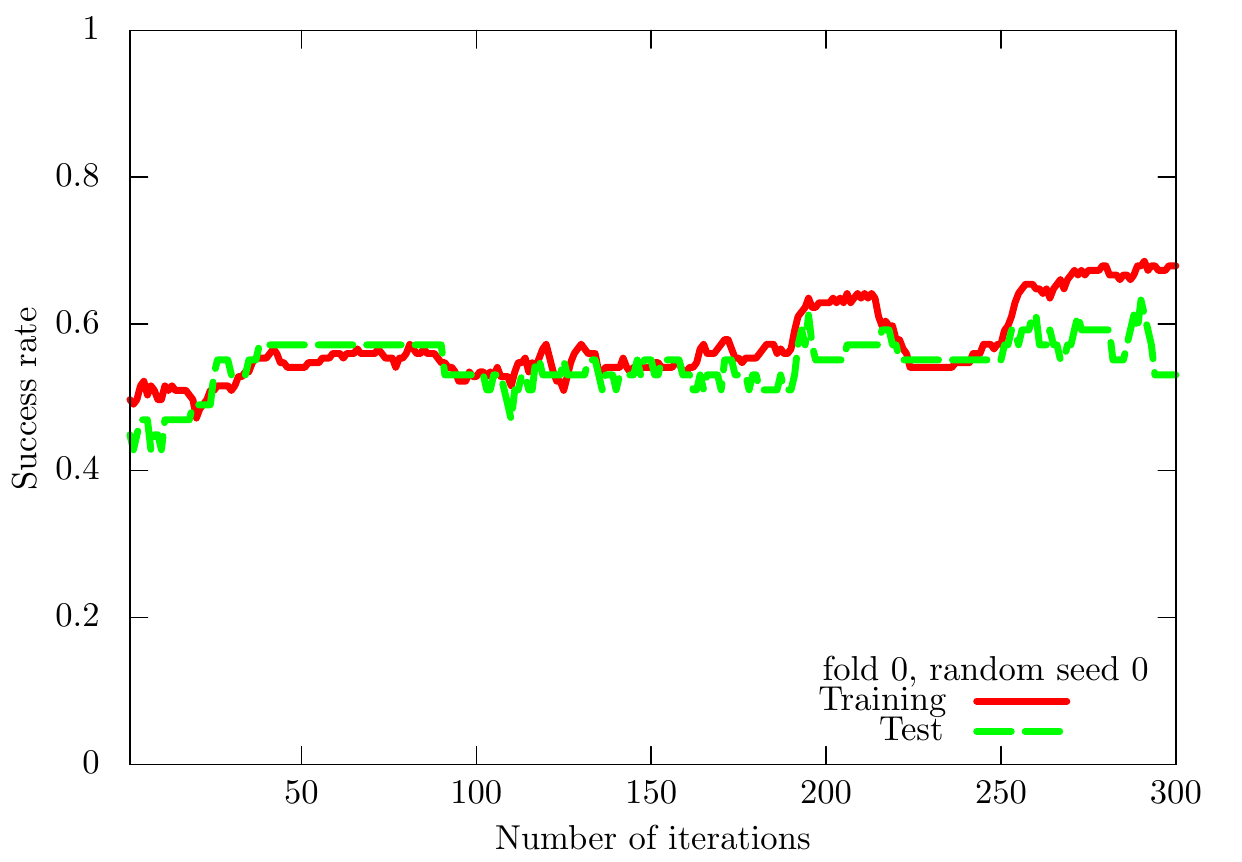}
\includegraphics[scale=0.25]{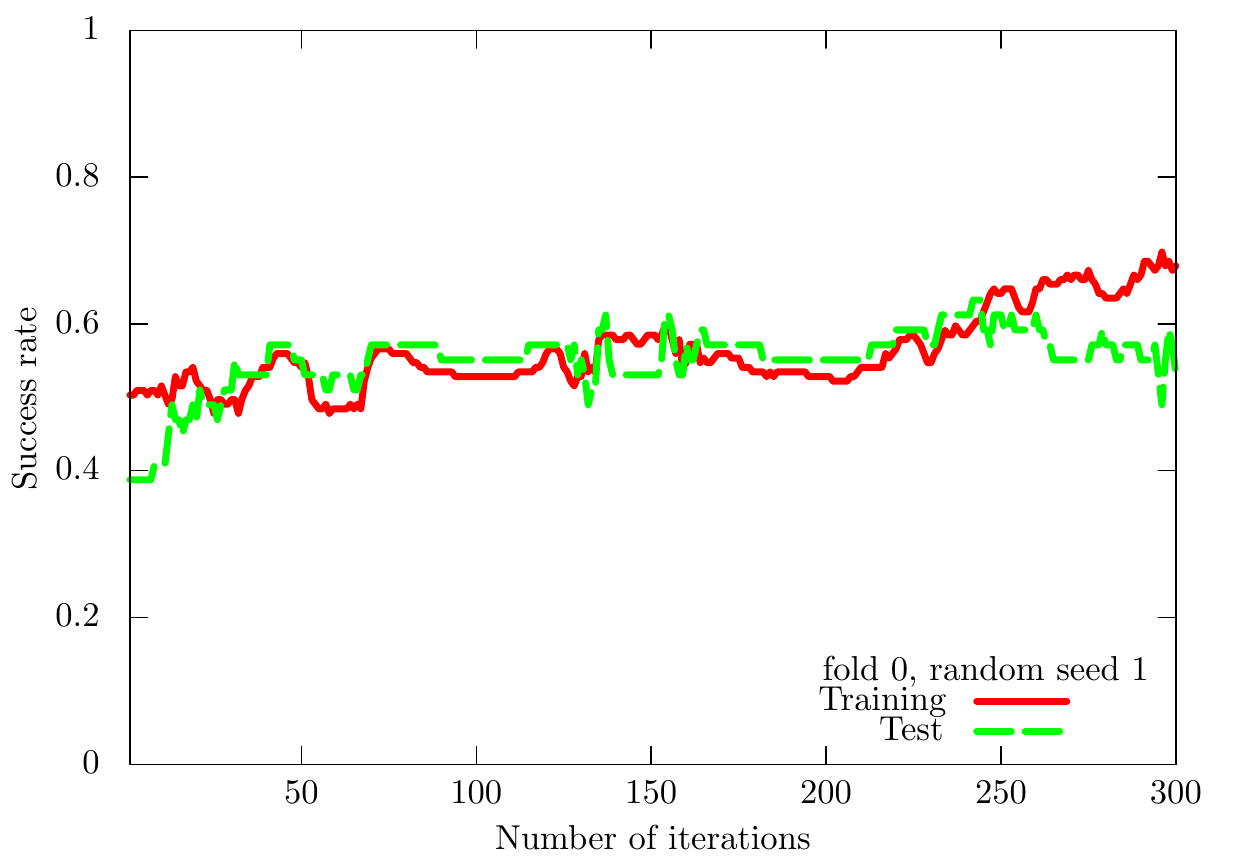}
\includegraphics[scale=0.25]{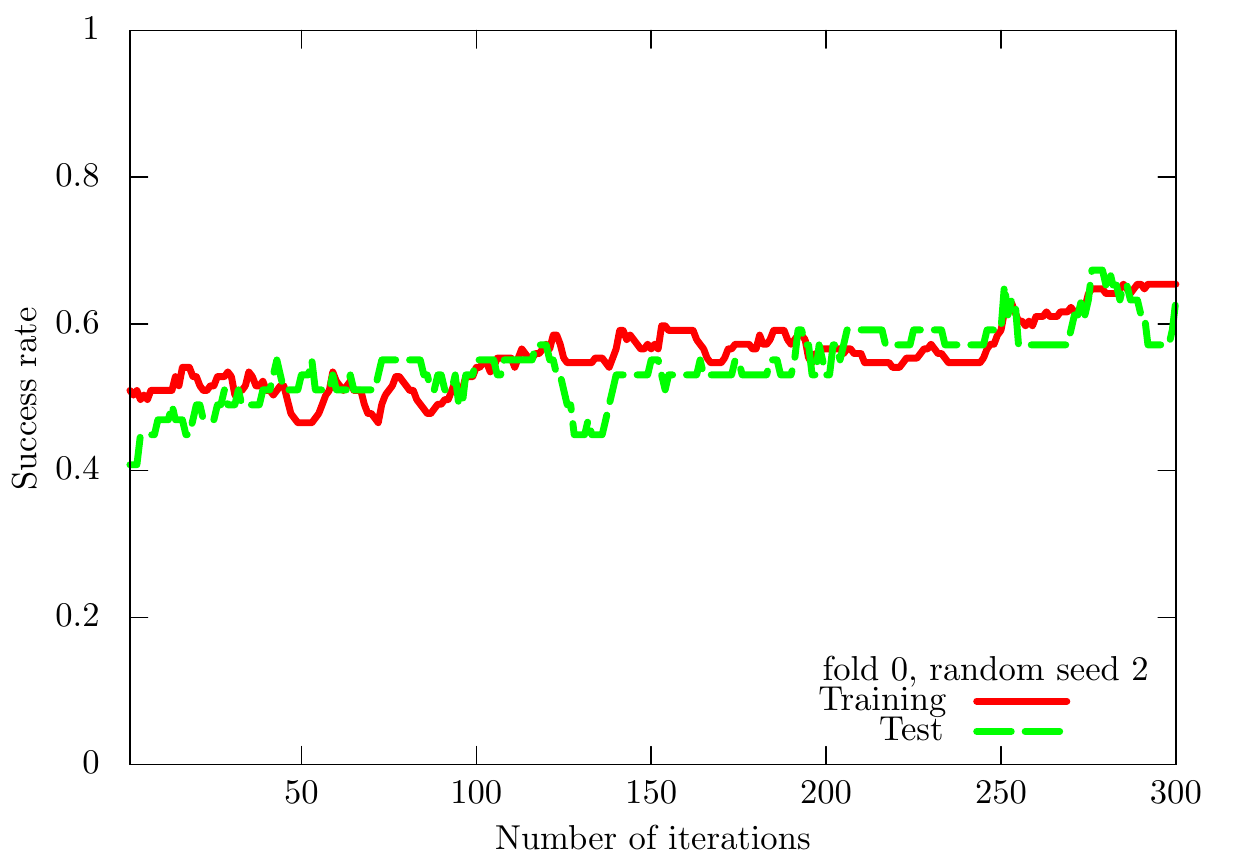}
\includegraphics[scale=0.25]{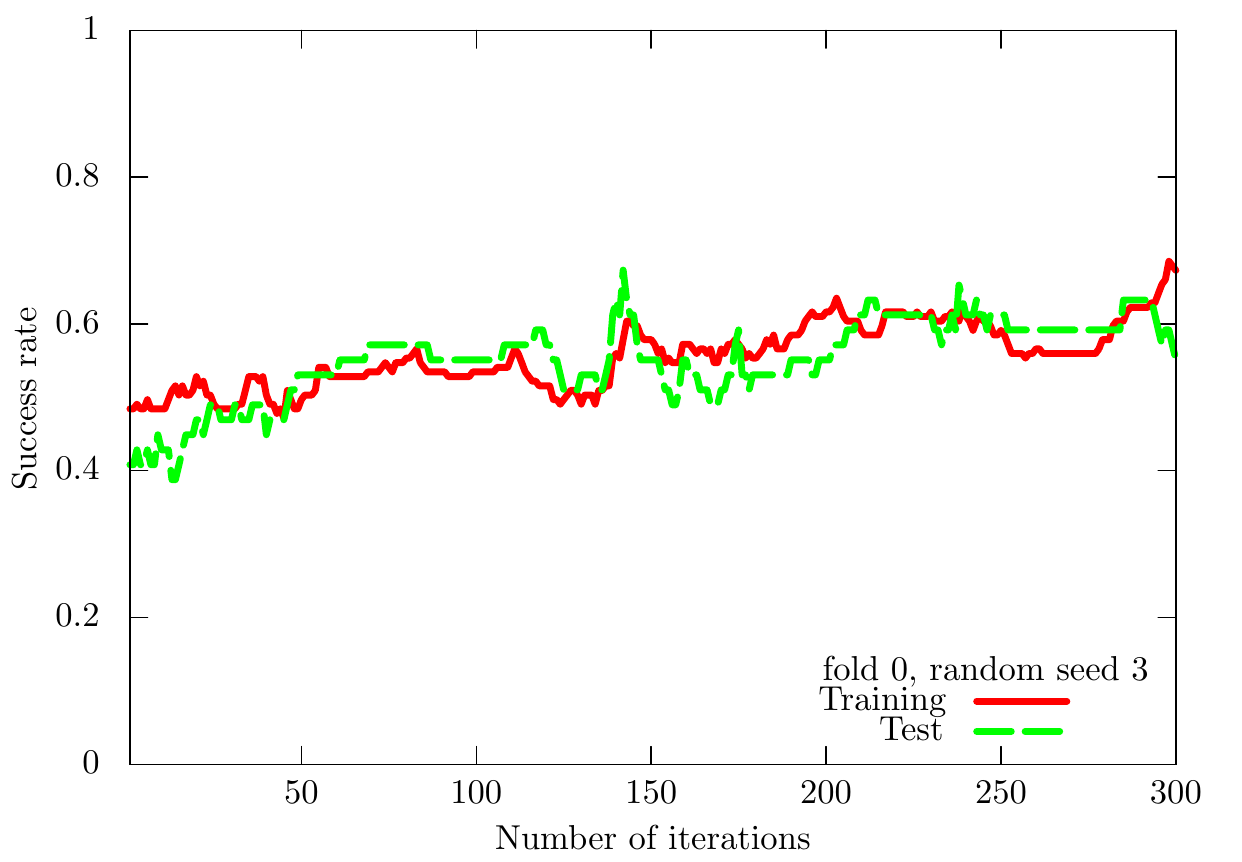}
\includegraphics[scale=0.25]{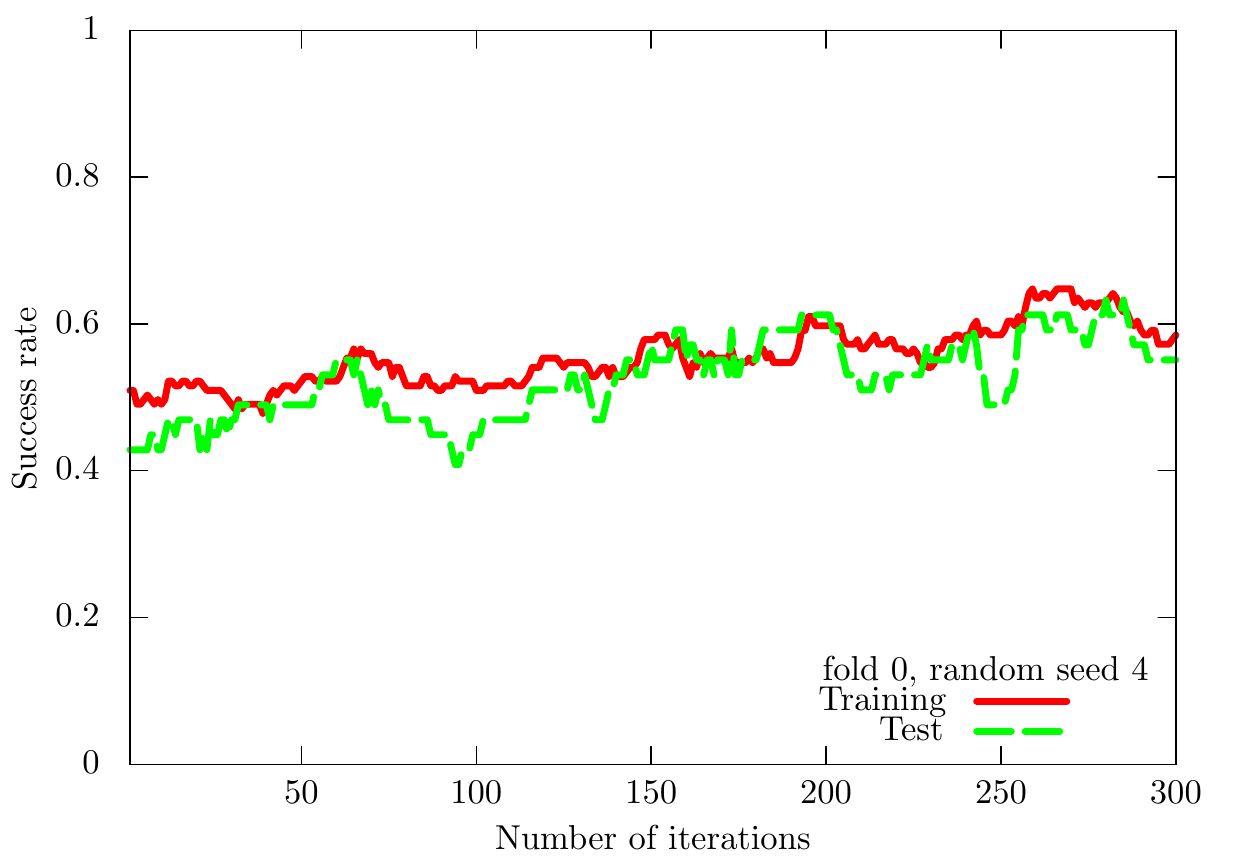}
\includegraphics[scale=0.25]{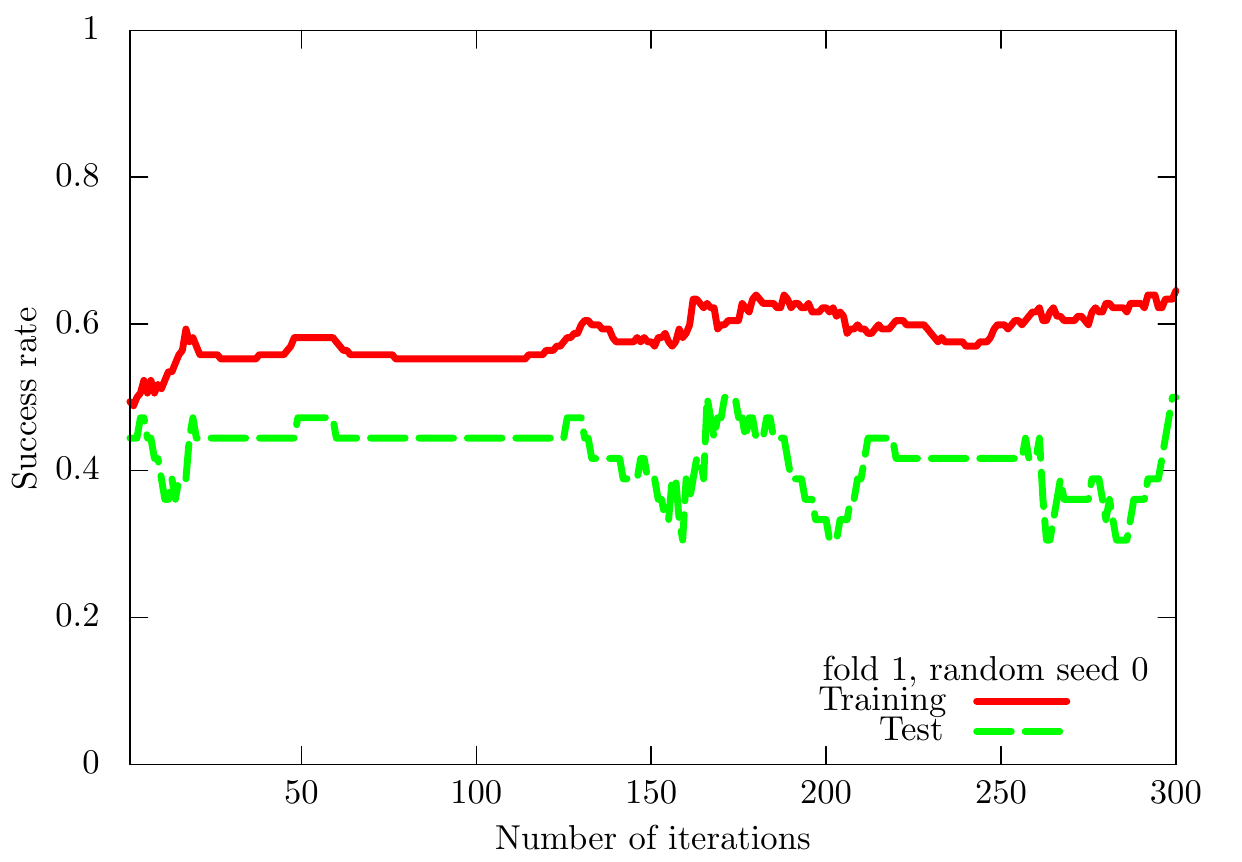}
\includegraphics[scale=0.25]{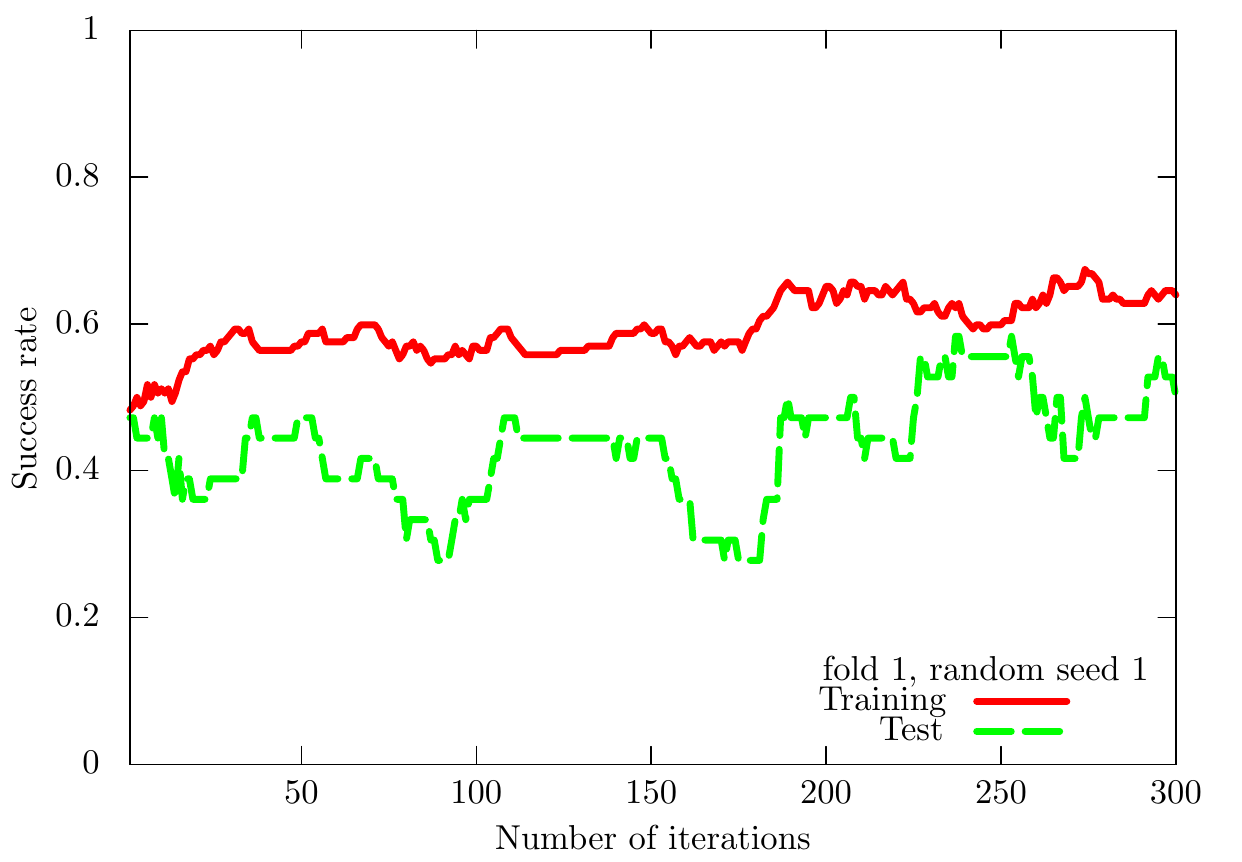}
\includegraphics[scale=0.25]{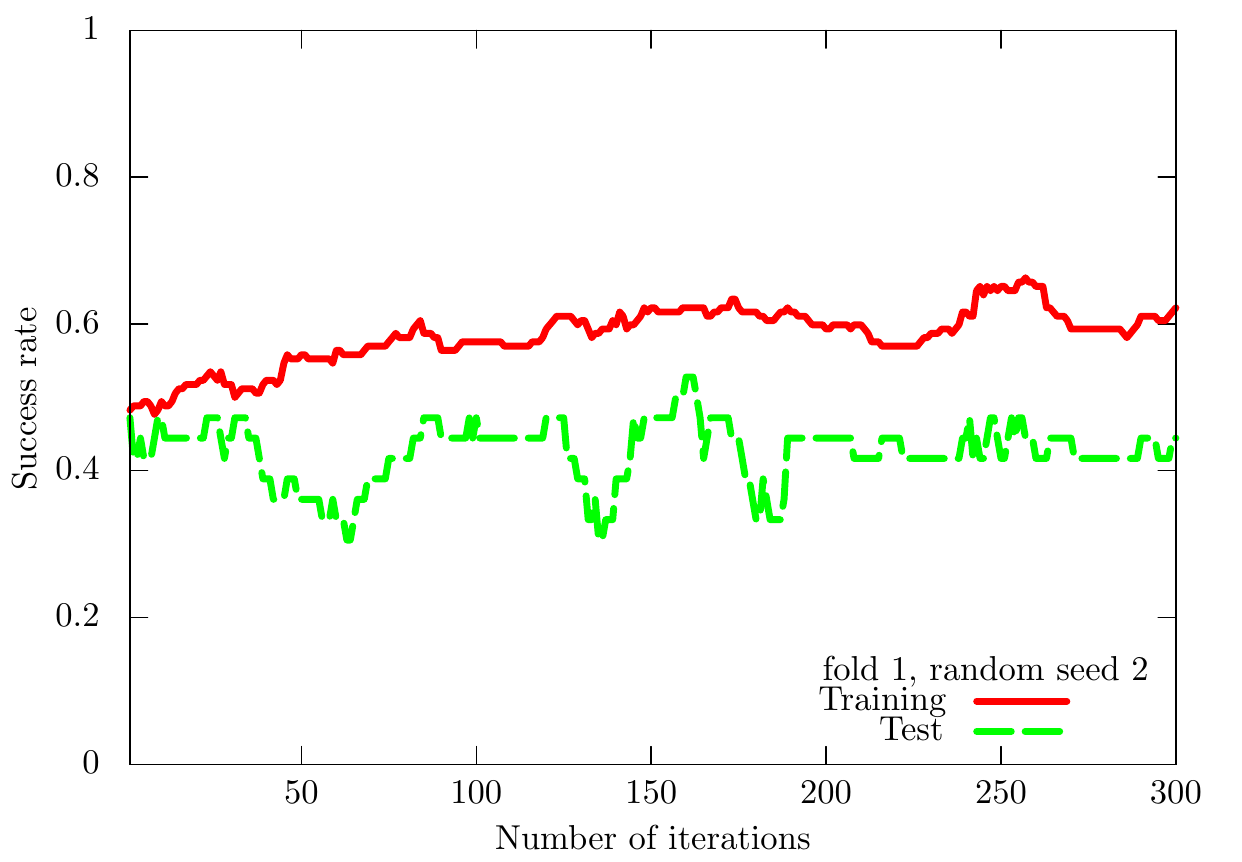}
\includegraphics[scale=0.25]{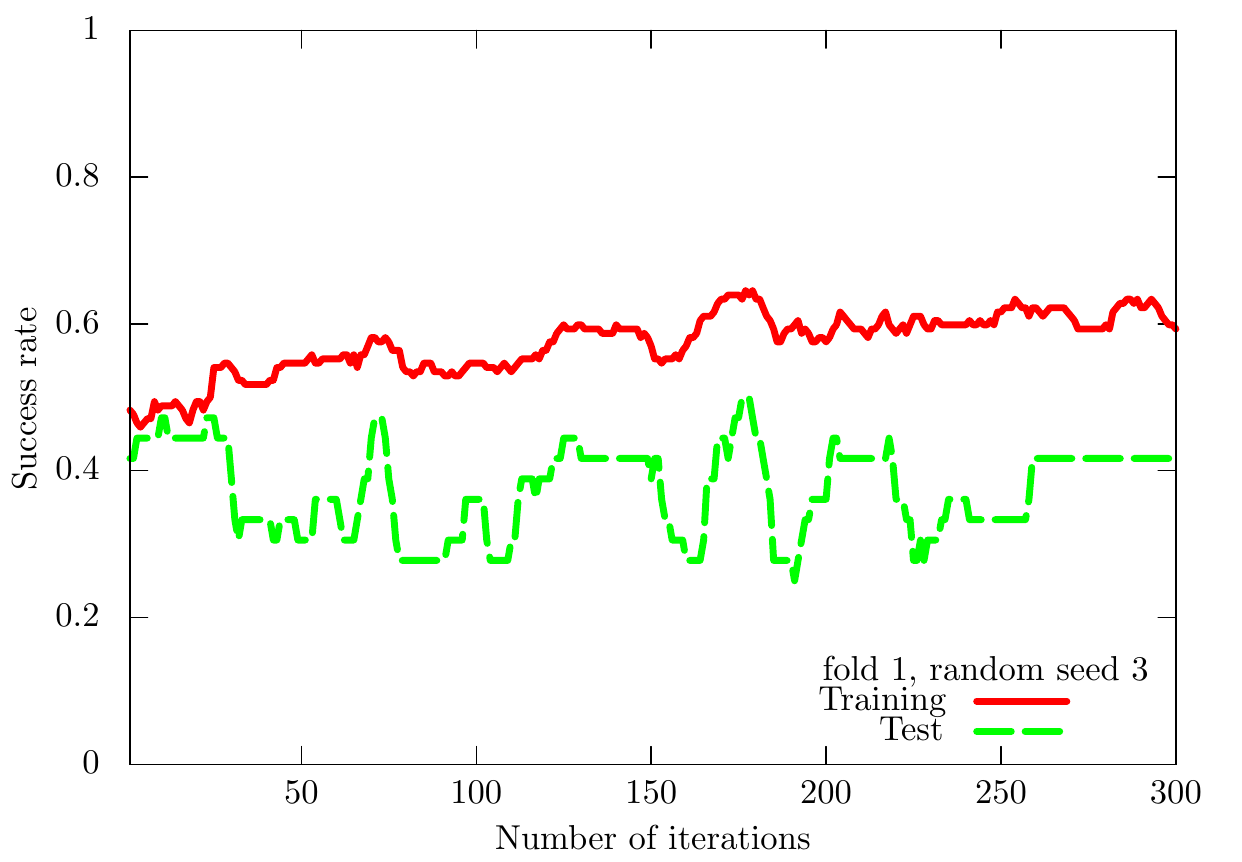}
\includegraphics[scale=0.25]{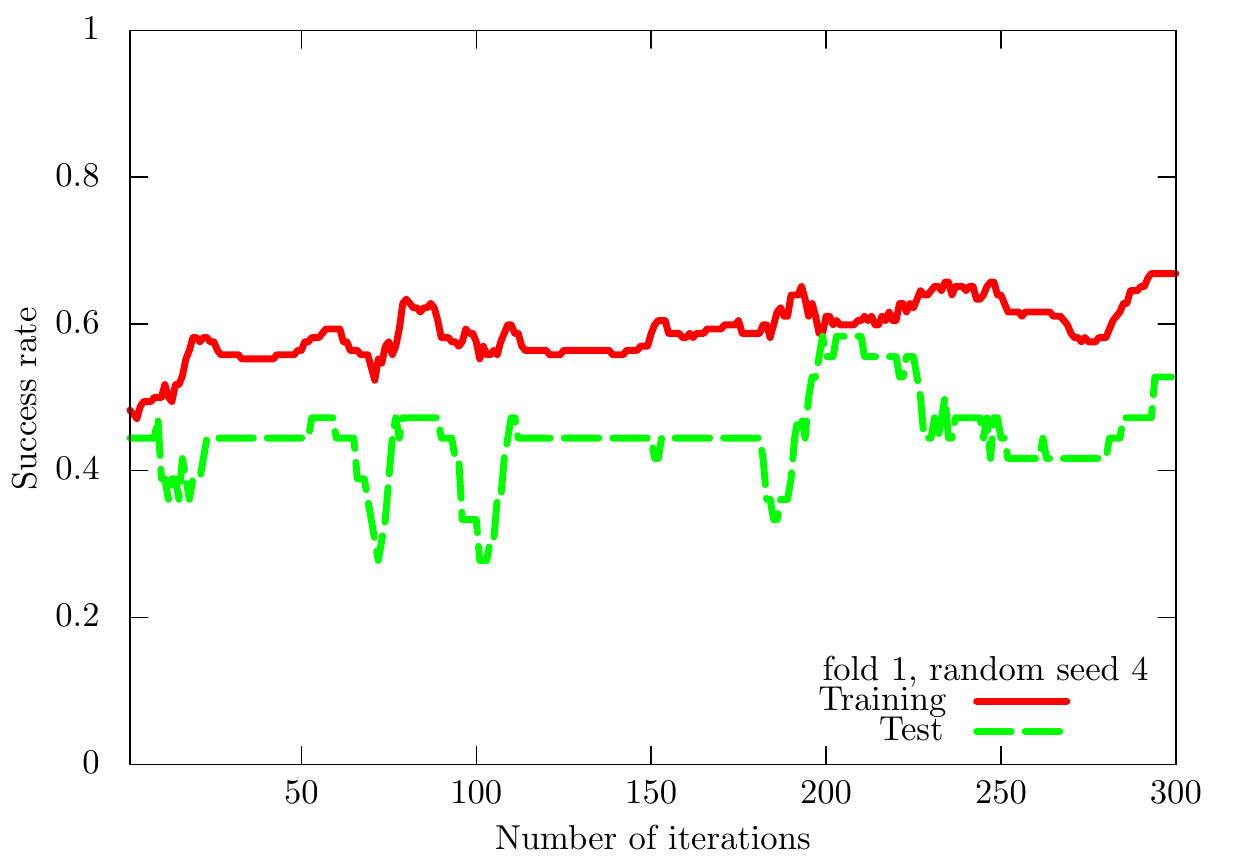}
\includegraphics[scale=0.25]{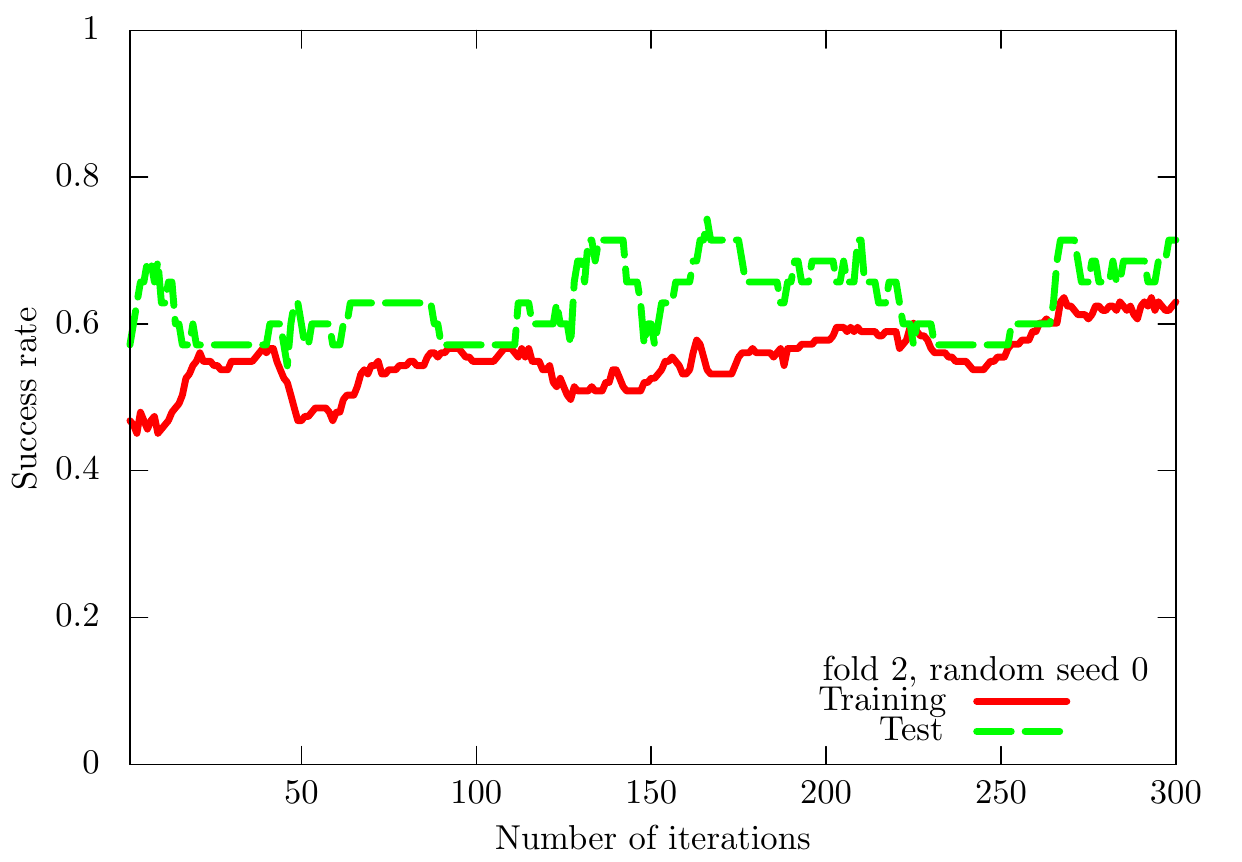}
\includegraphics[scale=0.25]{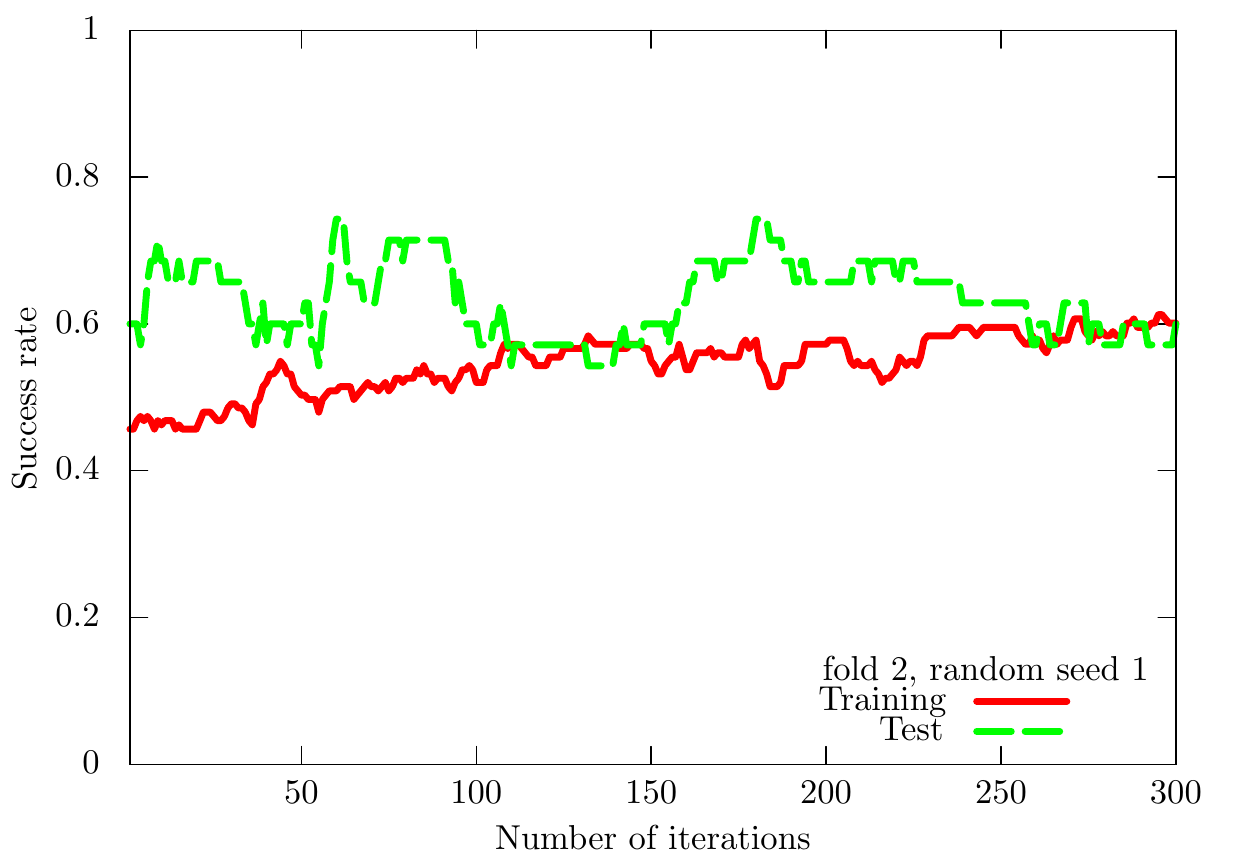}
\includegraphics[scale=0.25]{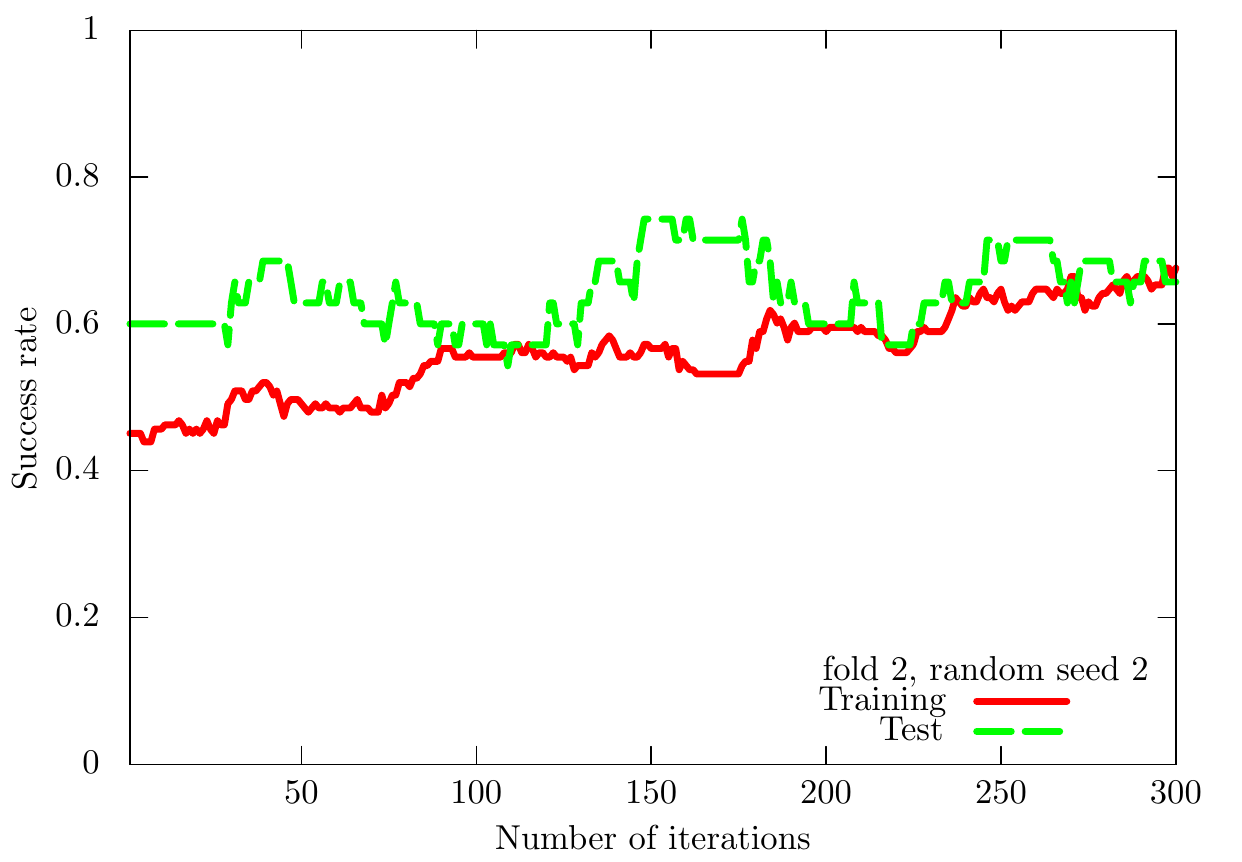}
\includegraphics[scale=0.25]{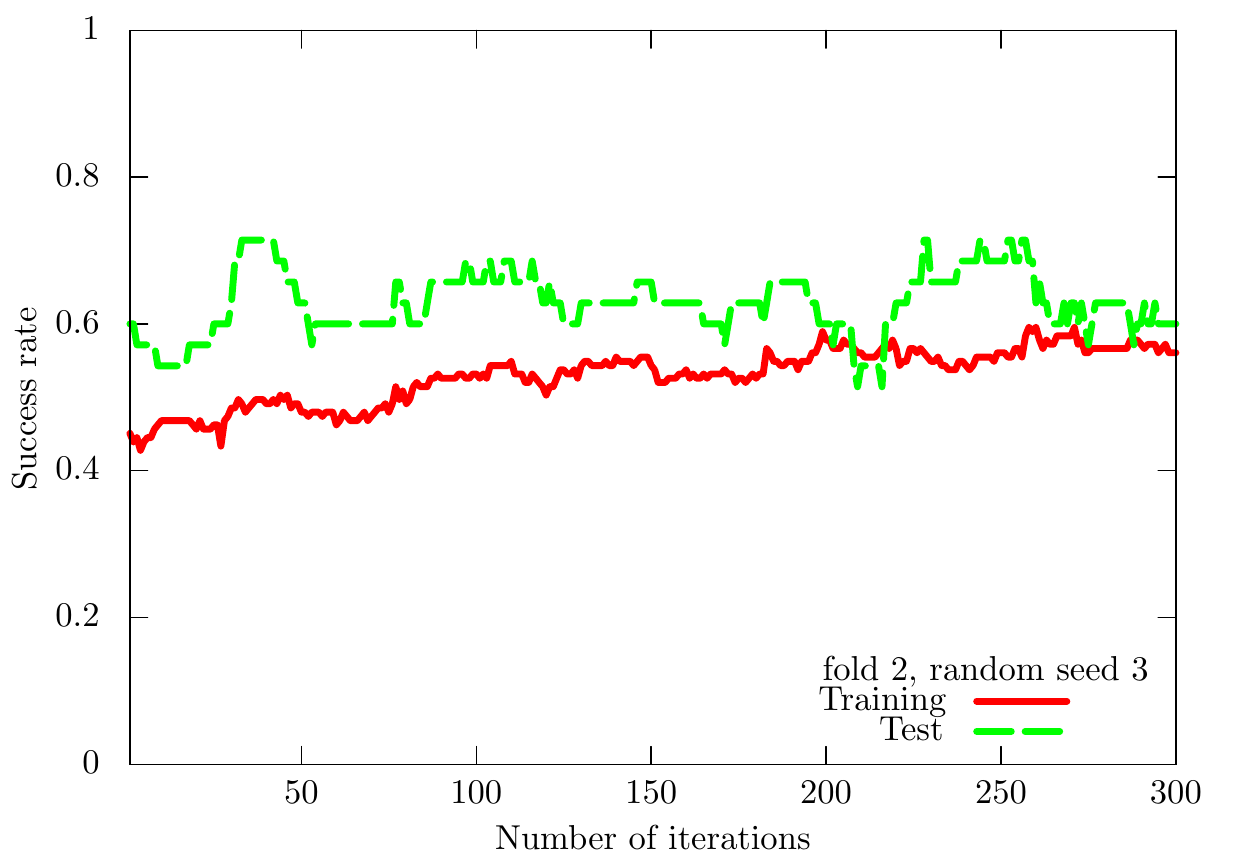}
\includegraphics[scale=0.25]{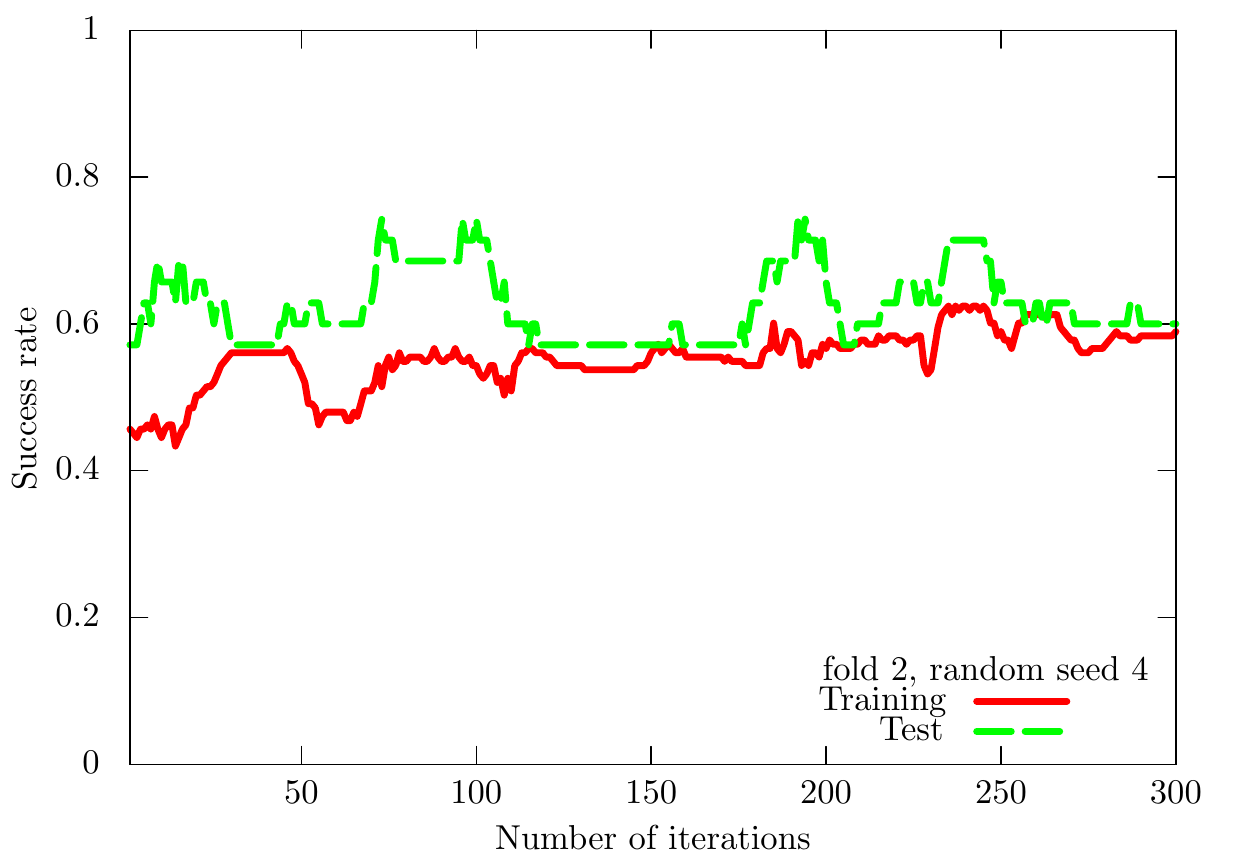}
\includegraphics[scale=0.25]{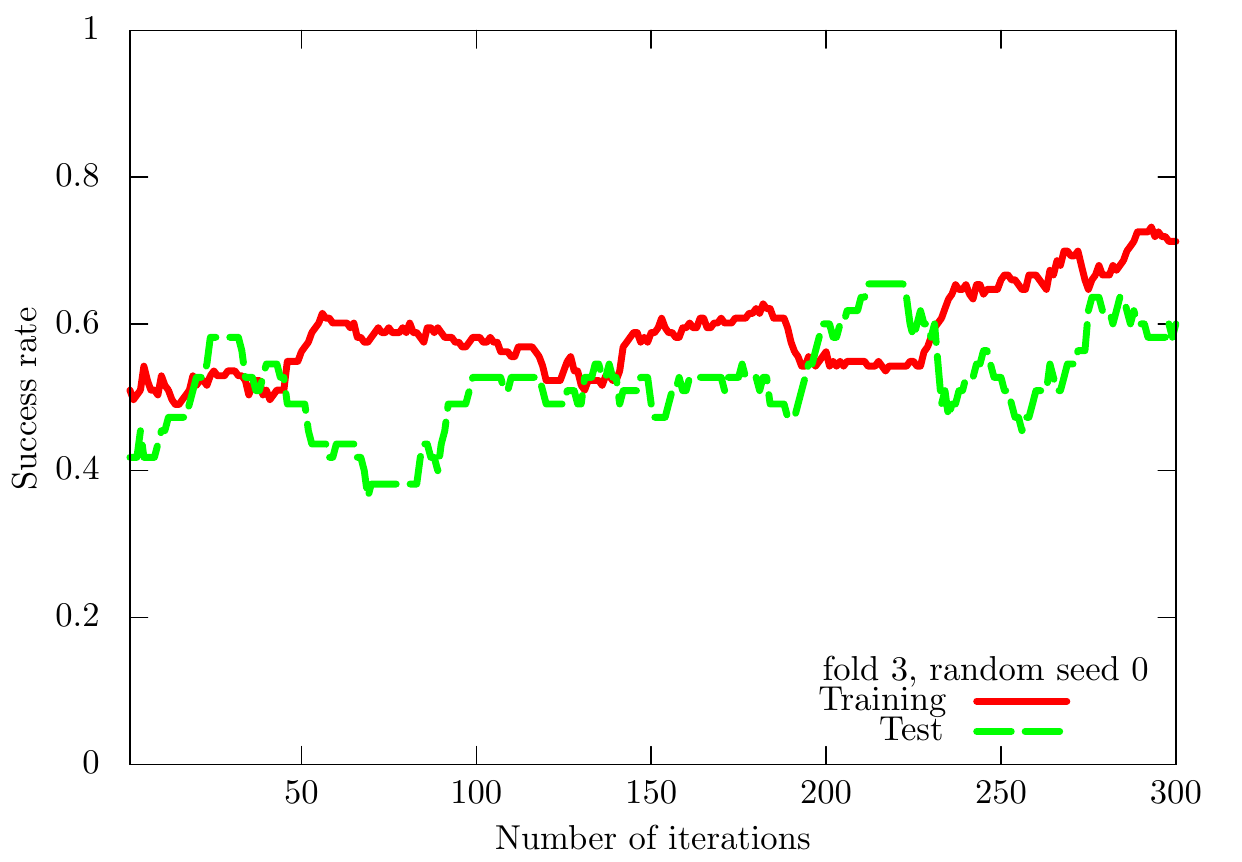}
\includegraphics[scale=0.25]{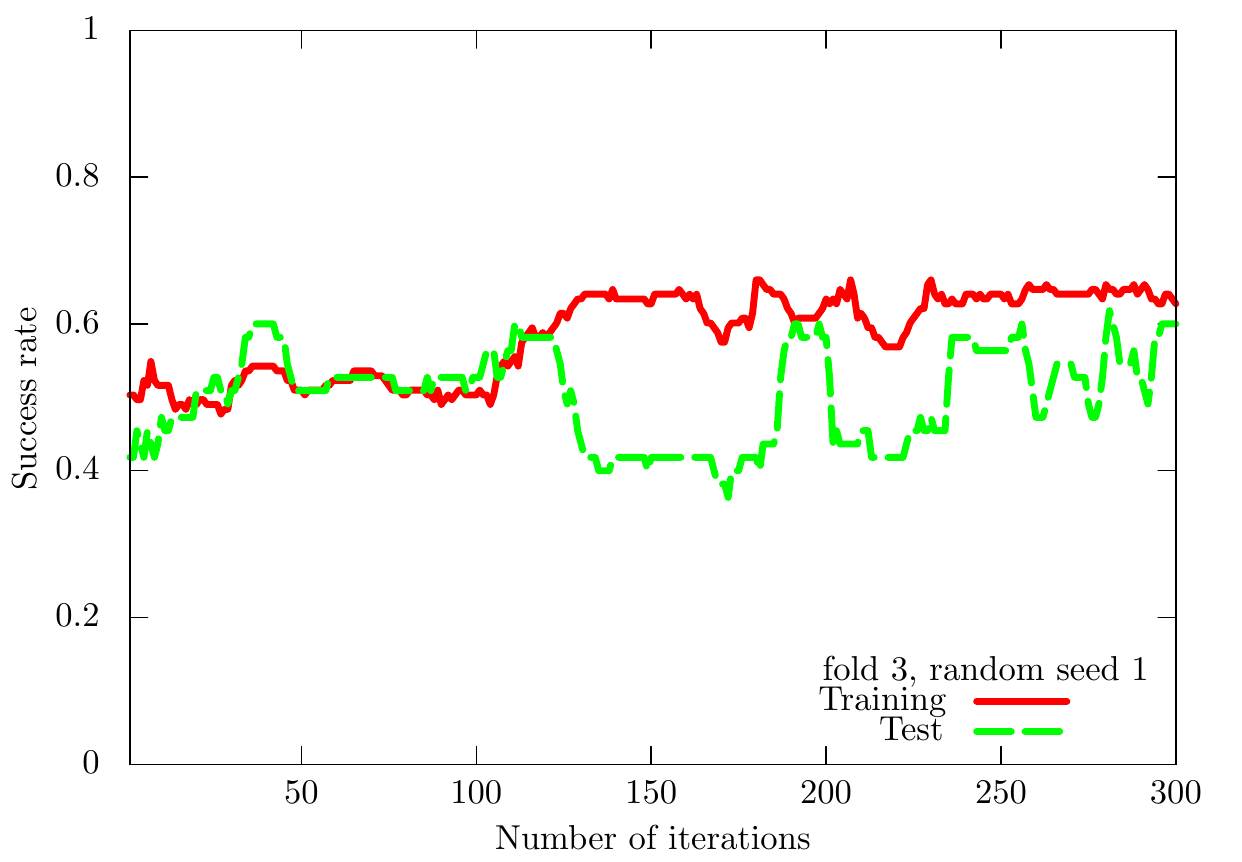}
\includegraphics[scale=0.25]{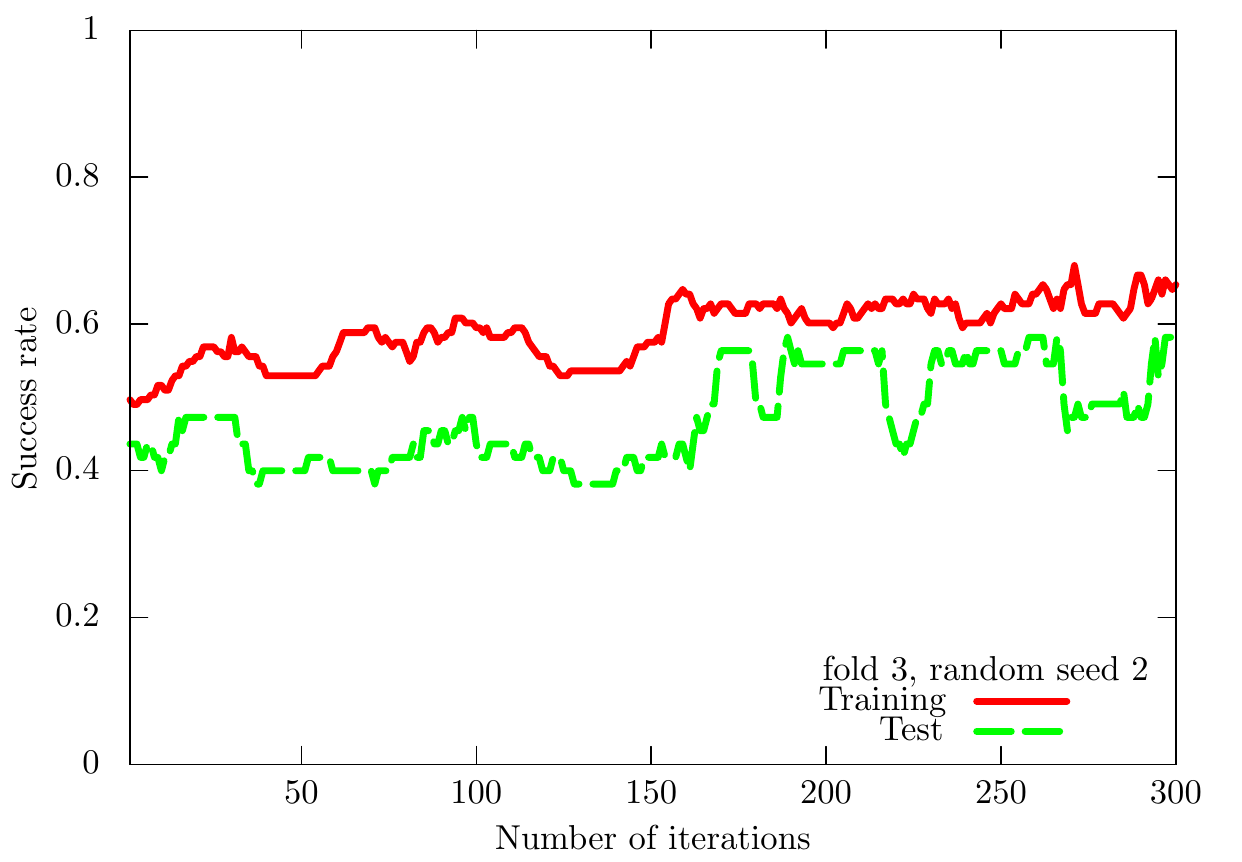}
\includegraphics[scale=0.25]{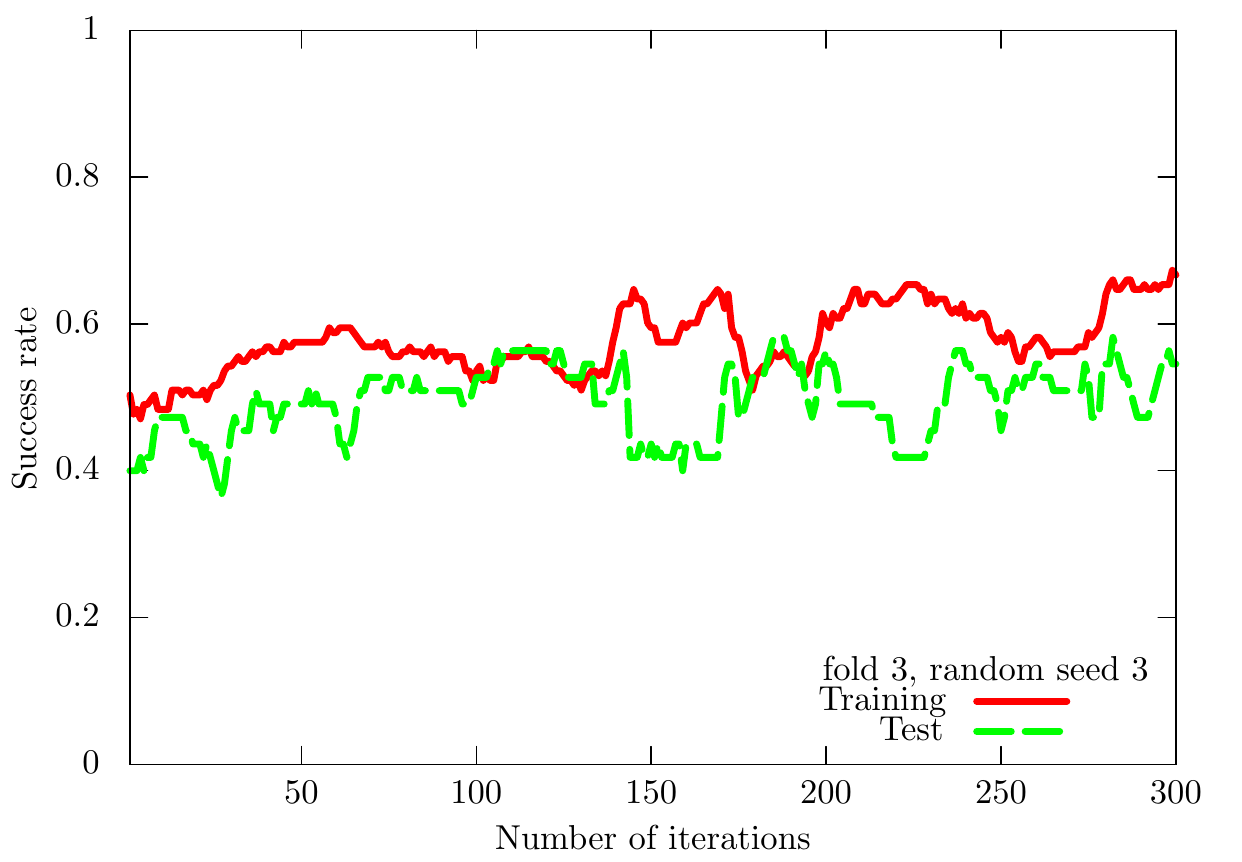}
\includegraphics[scale=0.25]{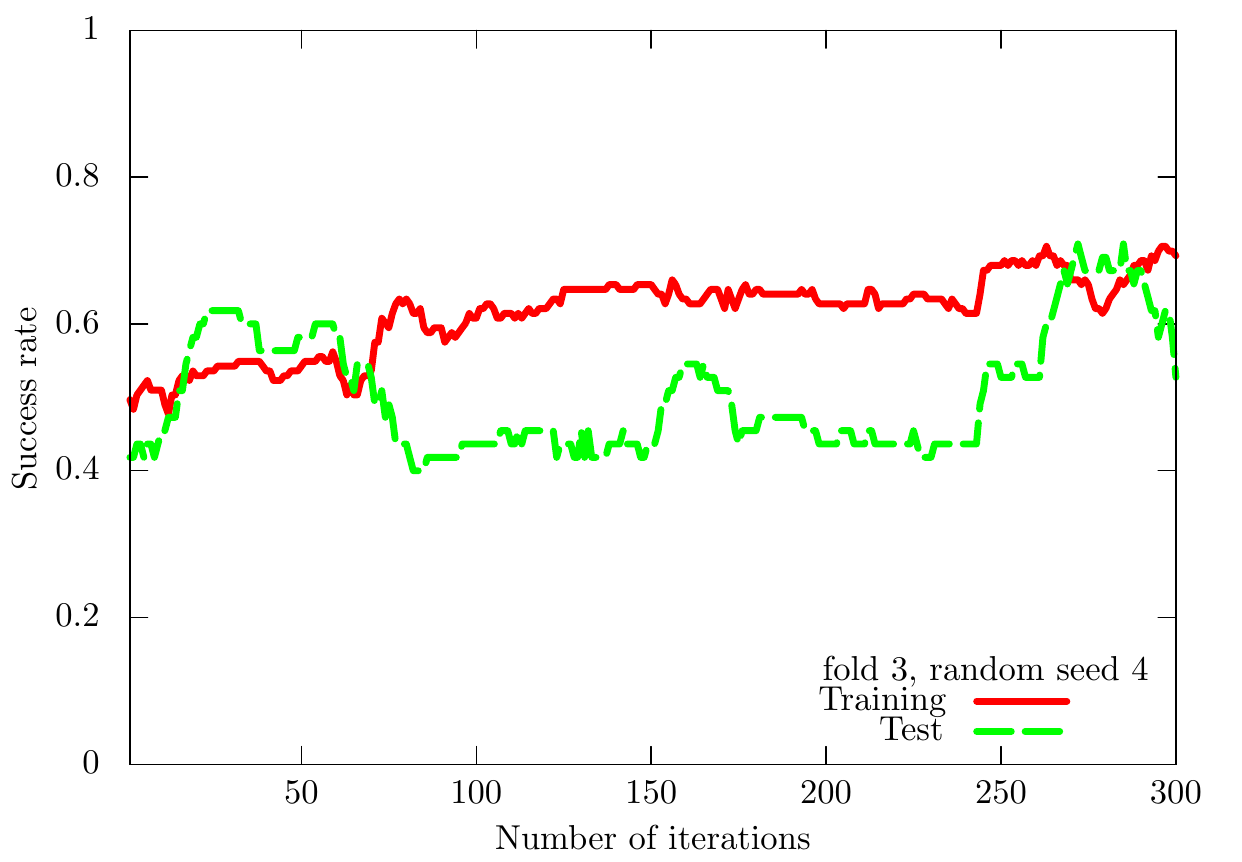}
\includegraphics[scale=0.25]{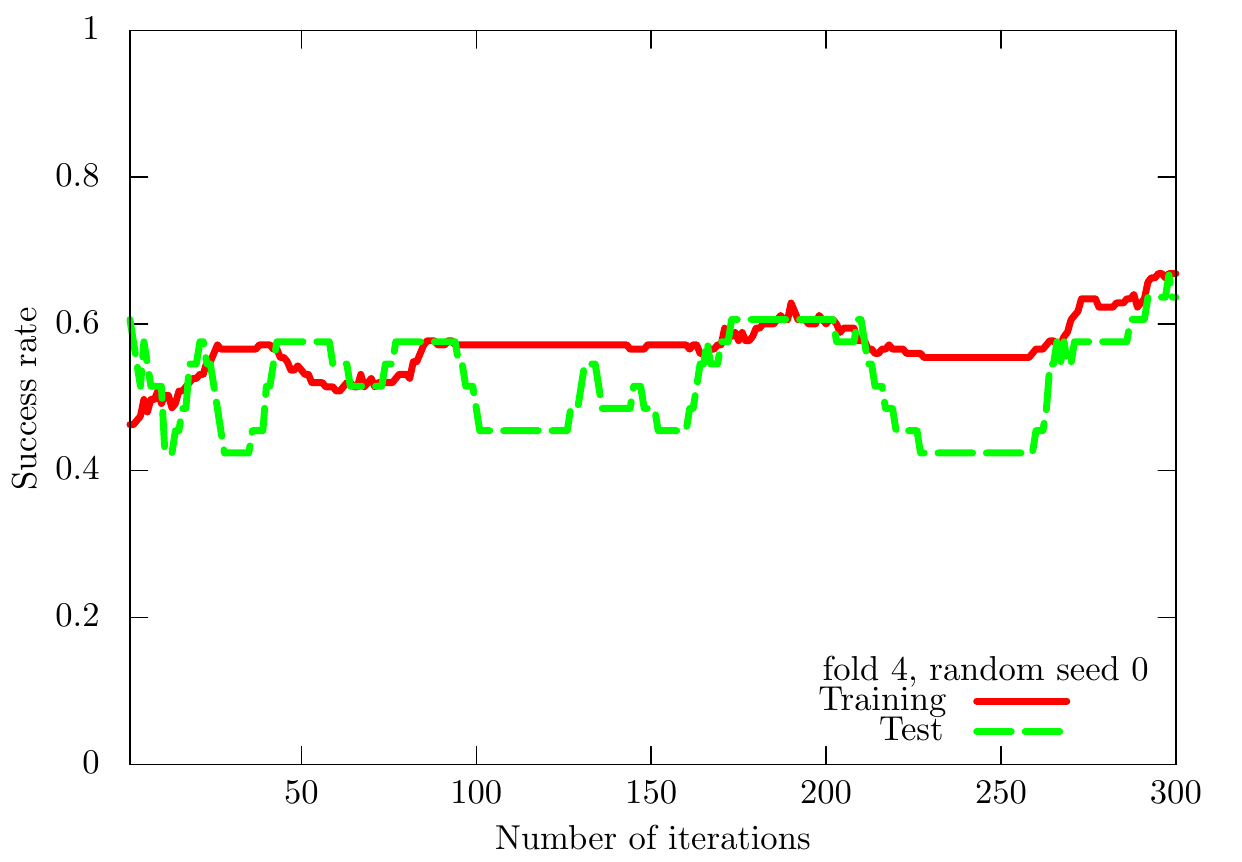}
\includegraphics[scale=0.25]{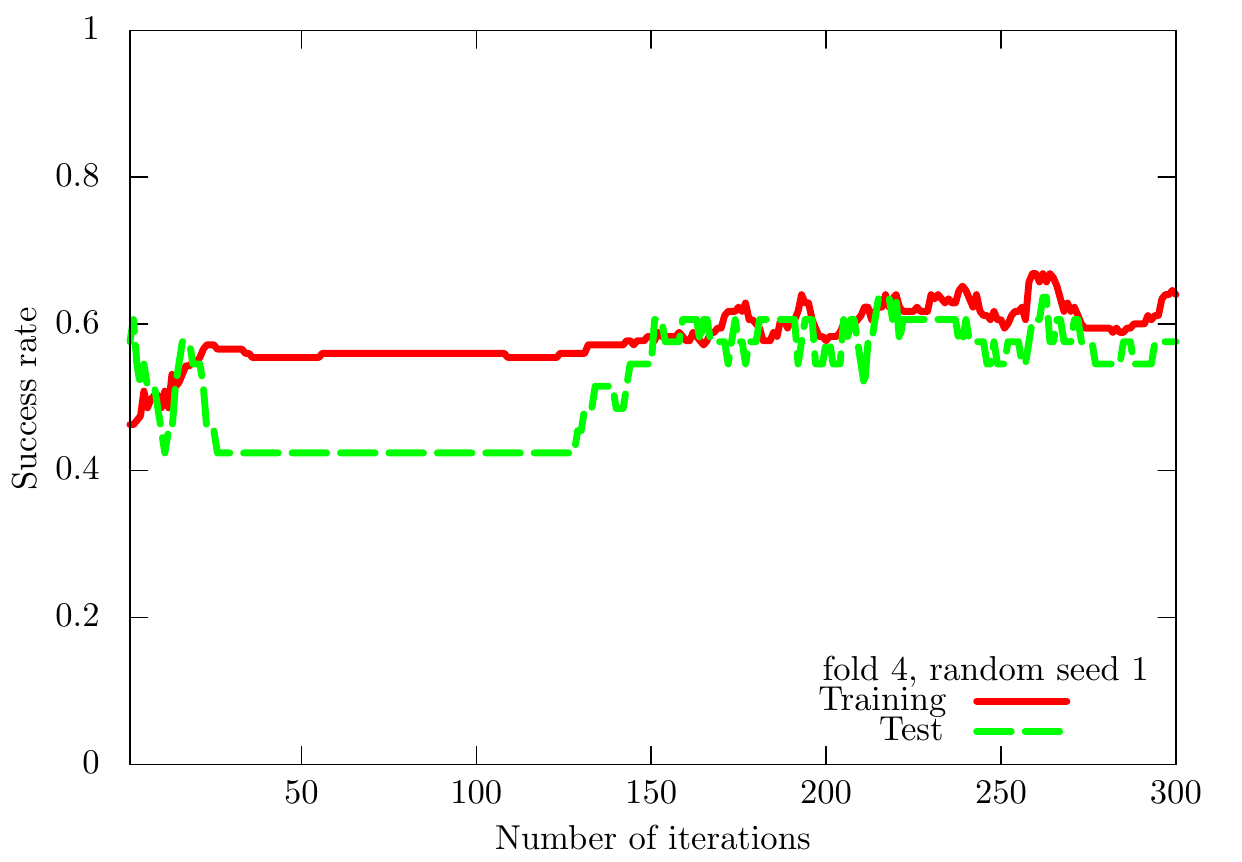}
\includegraphics[scale=0.25]{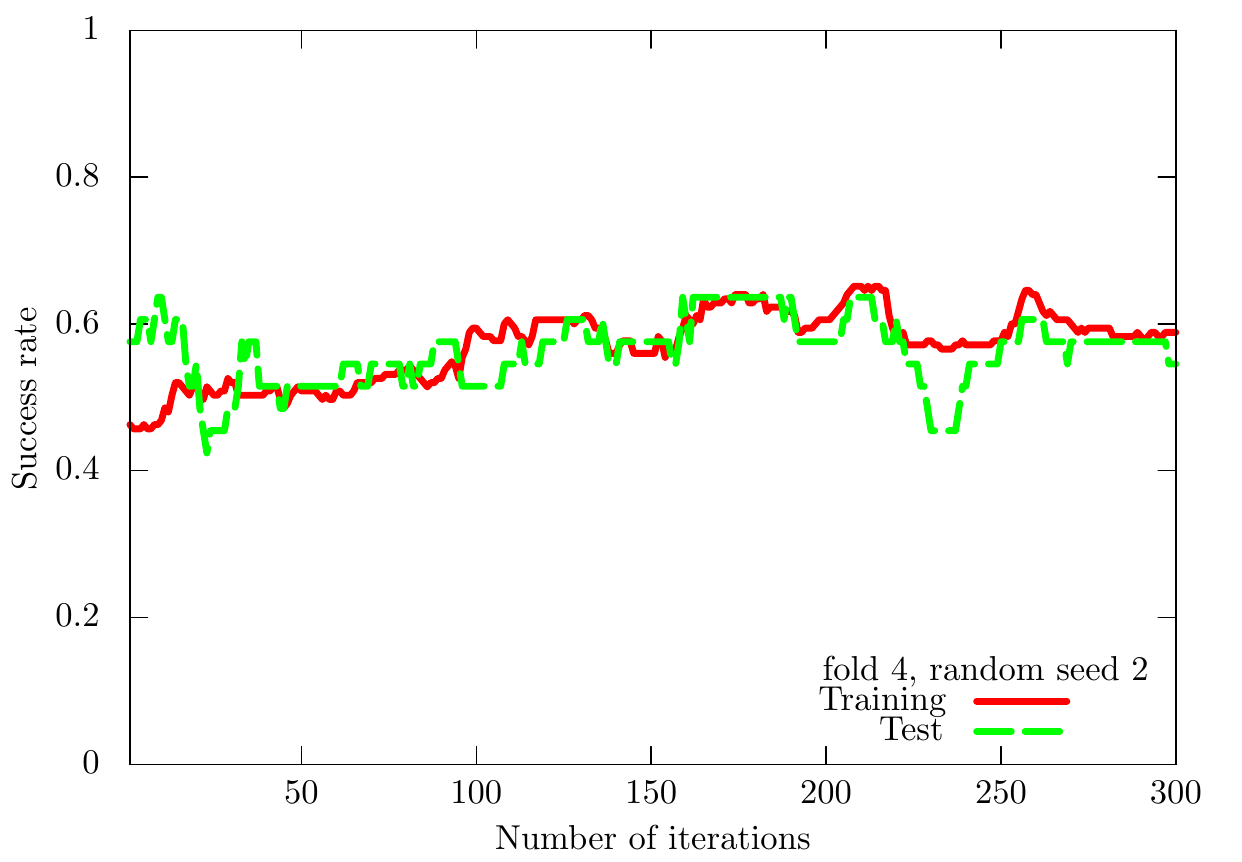}
\includegraphics[scale=0.25]{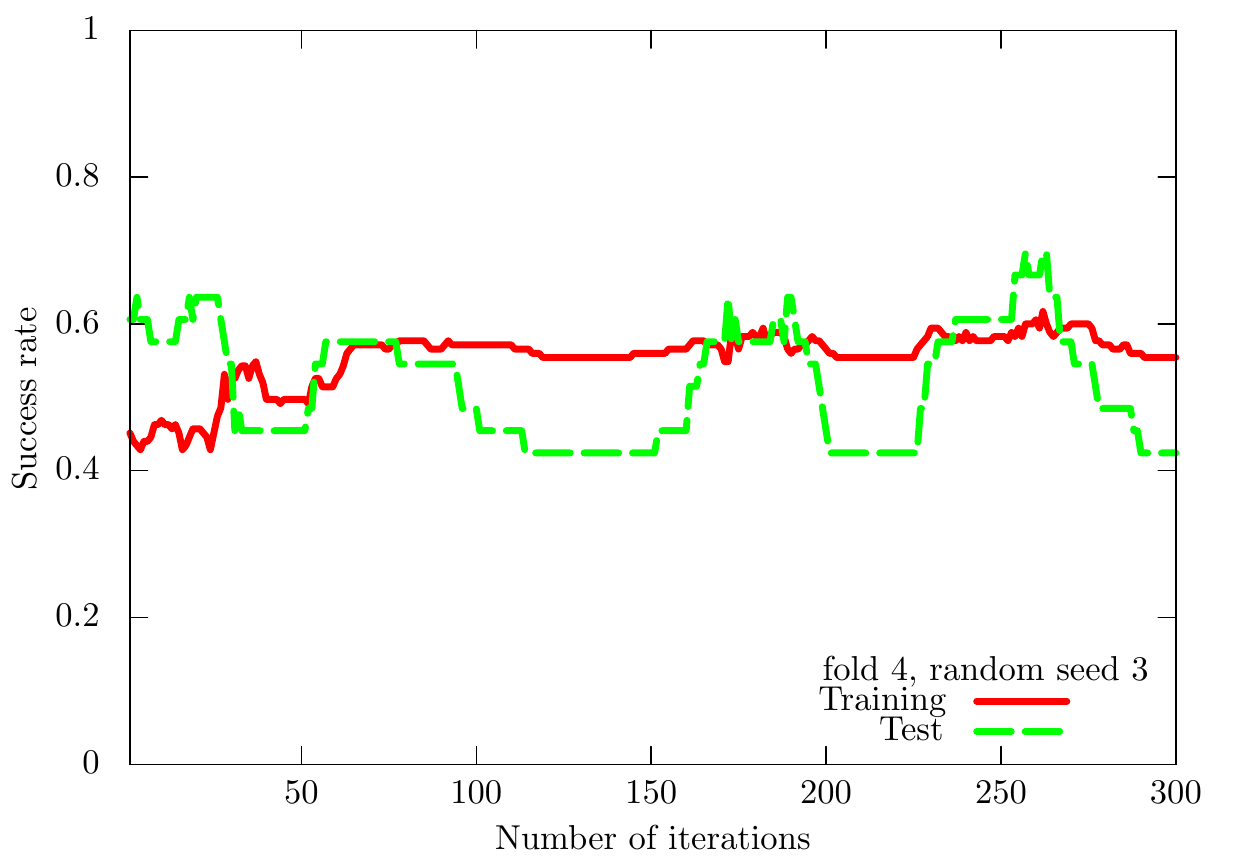}
\includegraphics[scale=0.25]{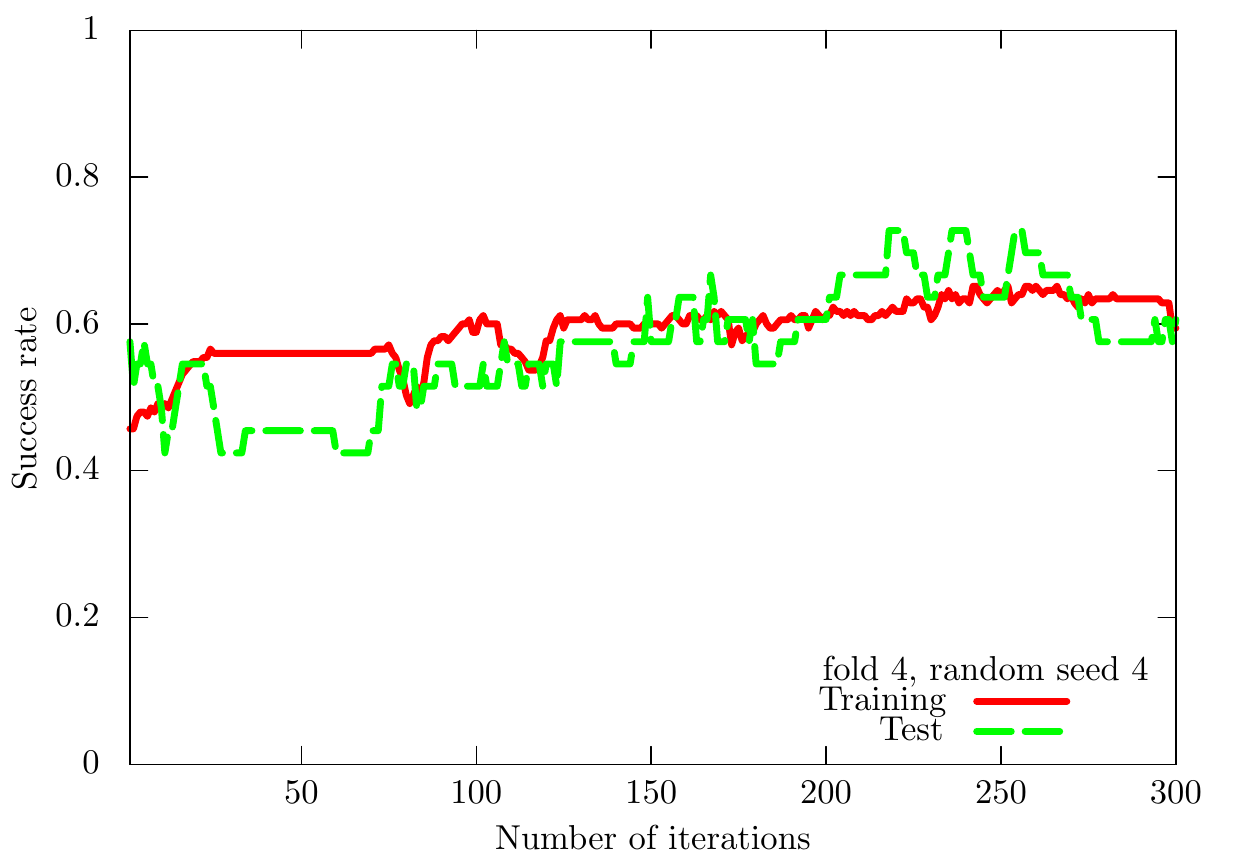}
\caption{Results of QCL on the $5$-fold datasets with $5$ different random seeds for the sonar dataset ($0$ or $1$). We use the CNOT-based circuit and set $\theta_\mathrm{bias} = 0$. The number of layers $L$ is set to $5$.}
\label{supp-arXiv-numerical-result-raw-data-fold-001-rand-001-QCL-UCI-sonar-0-1}
\end{figure*}

\begin{figure*}[htb]
\centering
\includegraphics[scale=0.25]{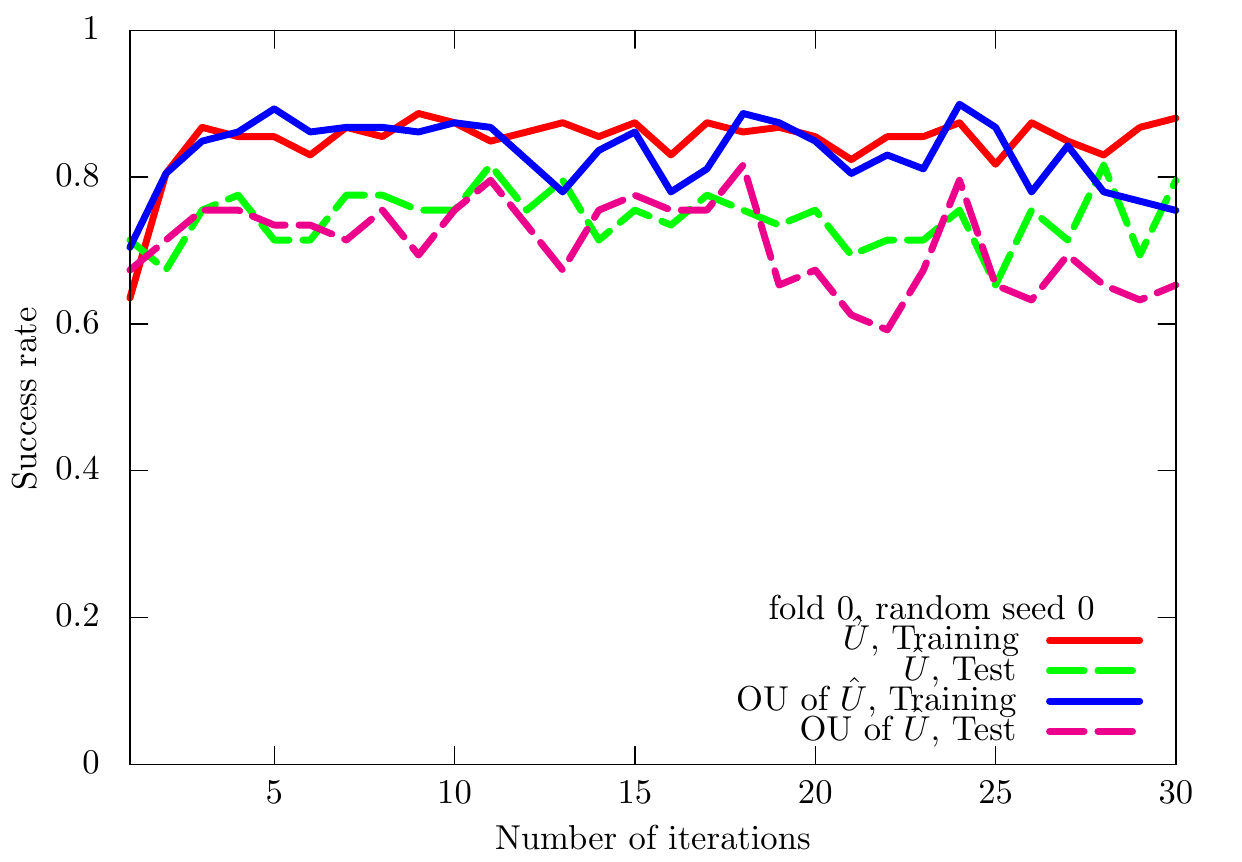}
\includegraphics[scale=0.25]{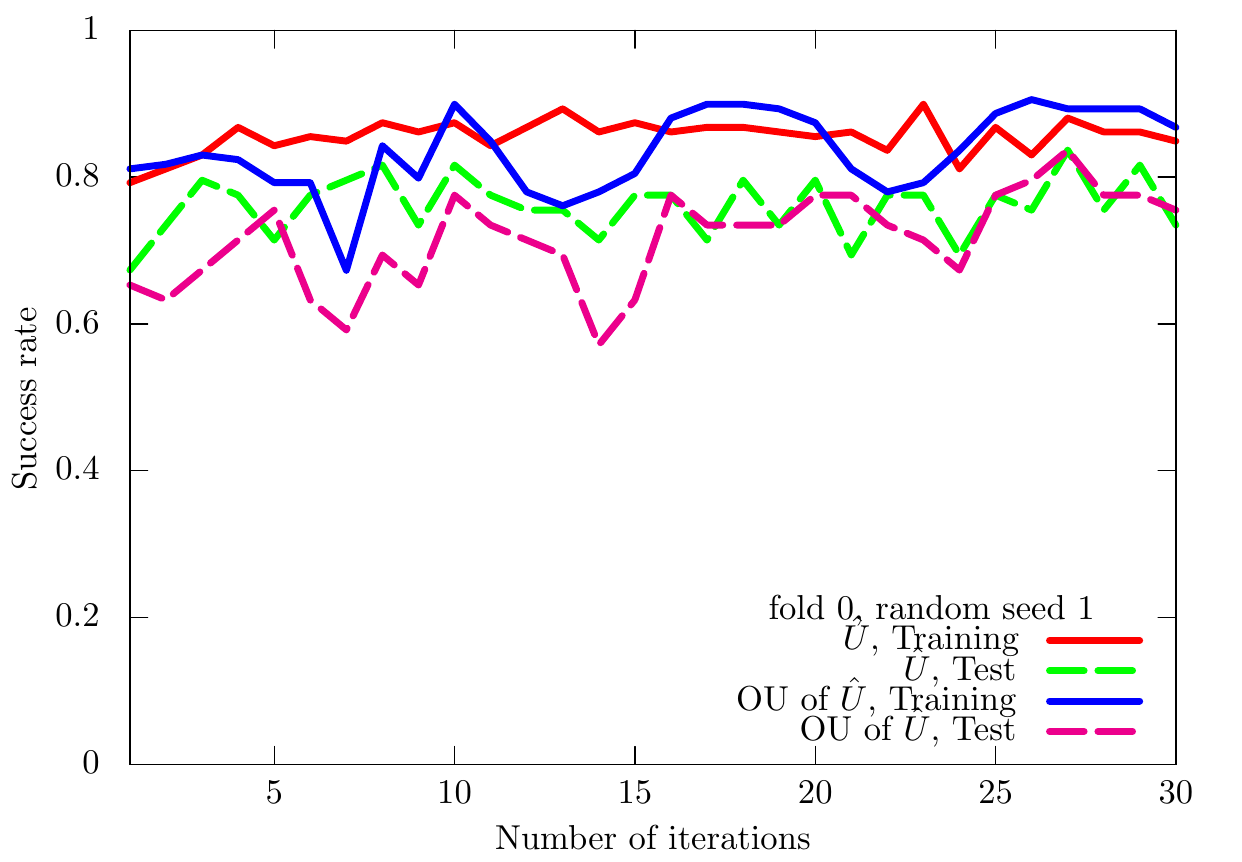}
\includegraphics[scale=0.25]{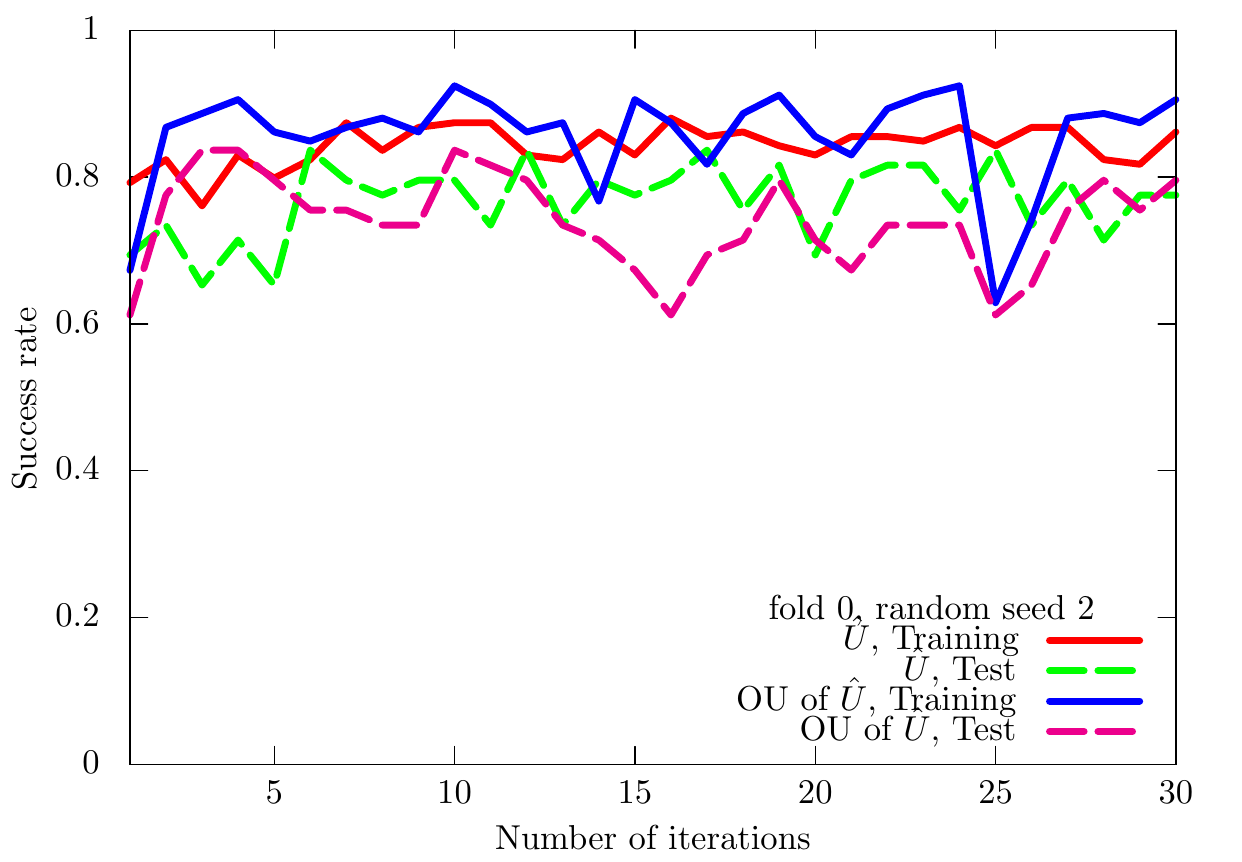}
\includegraphics[scale=0.25]{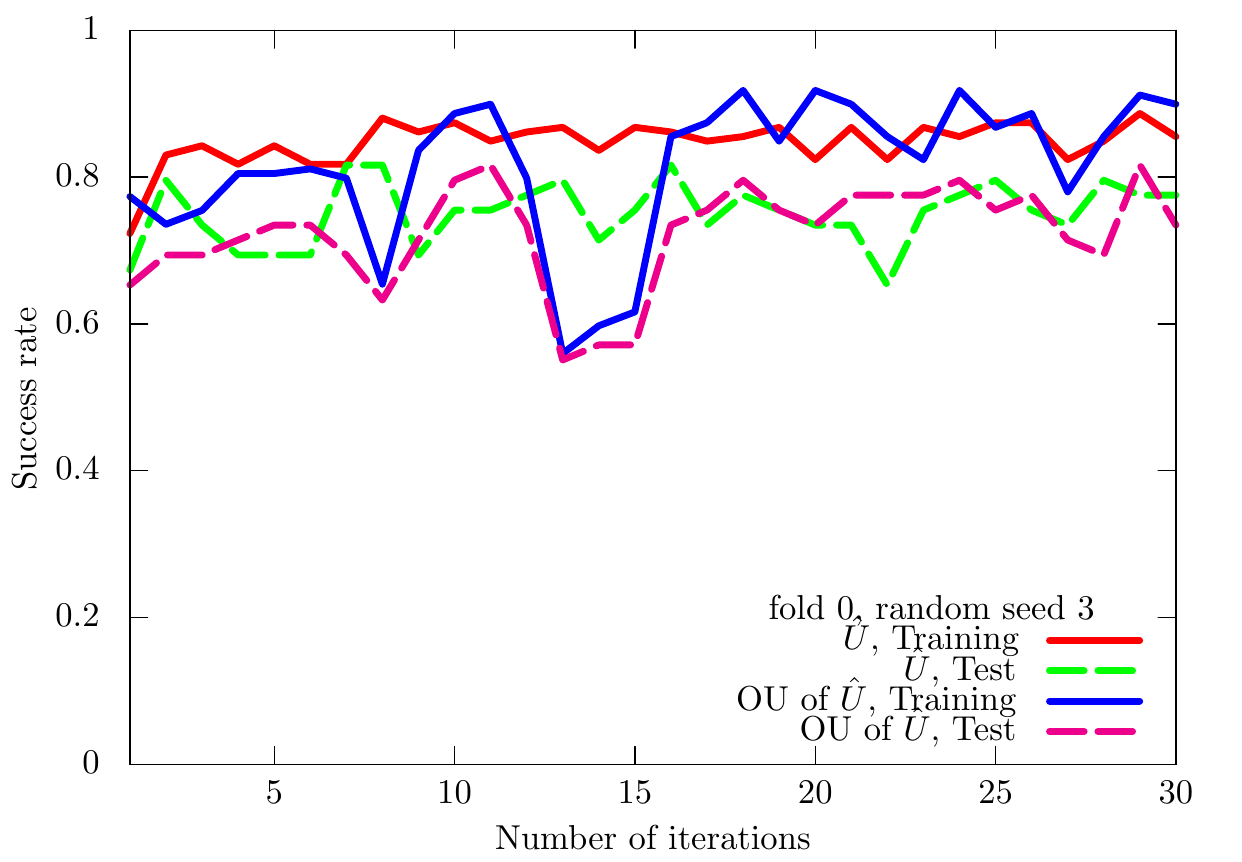}
\includegraphics[scale=0.25]{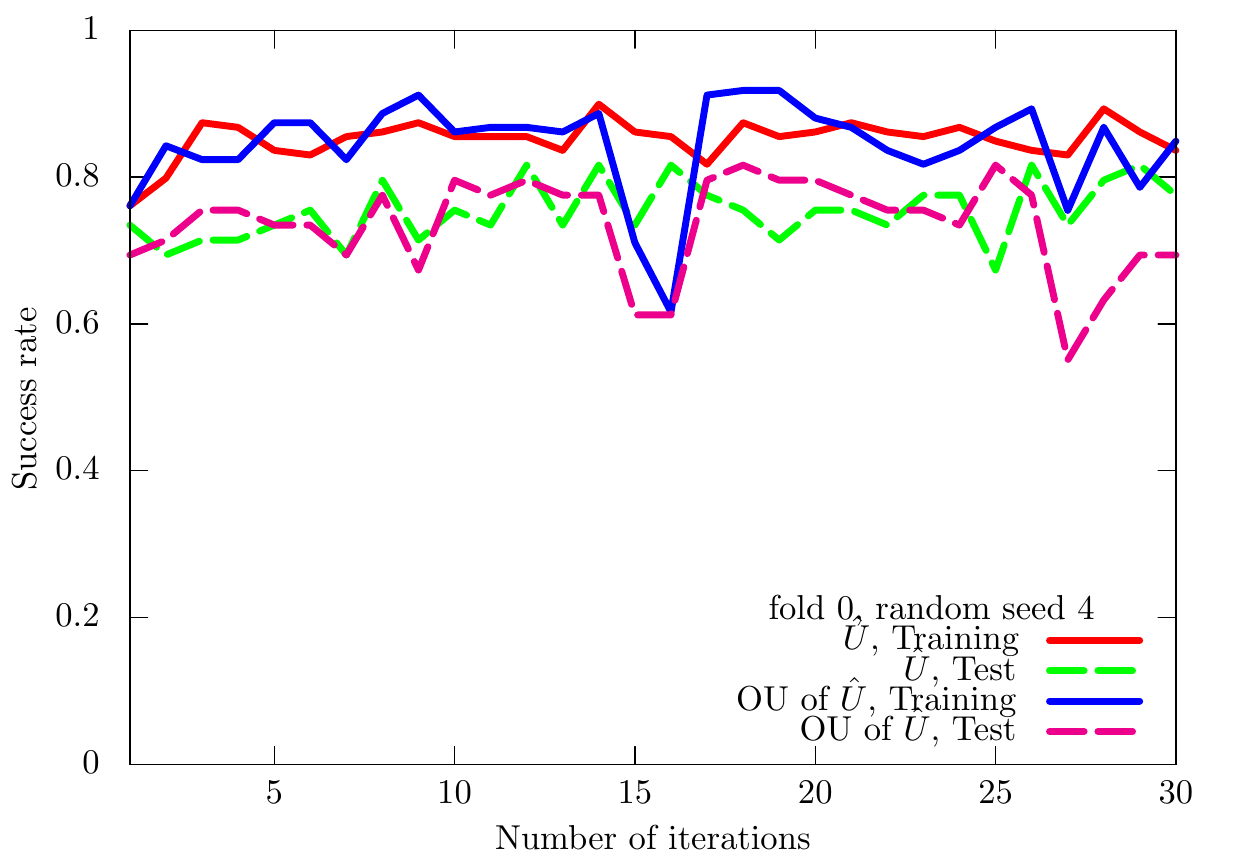}
\includegraphics[scale=0.25]{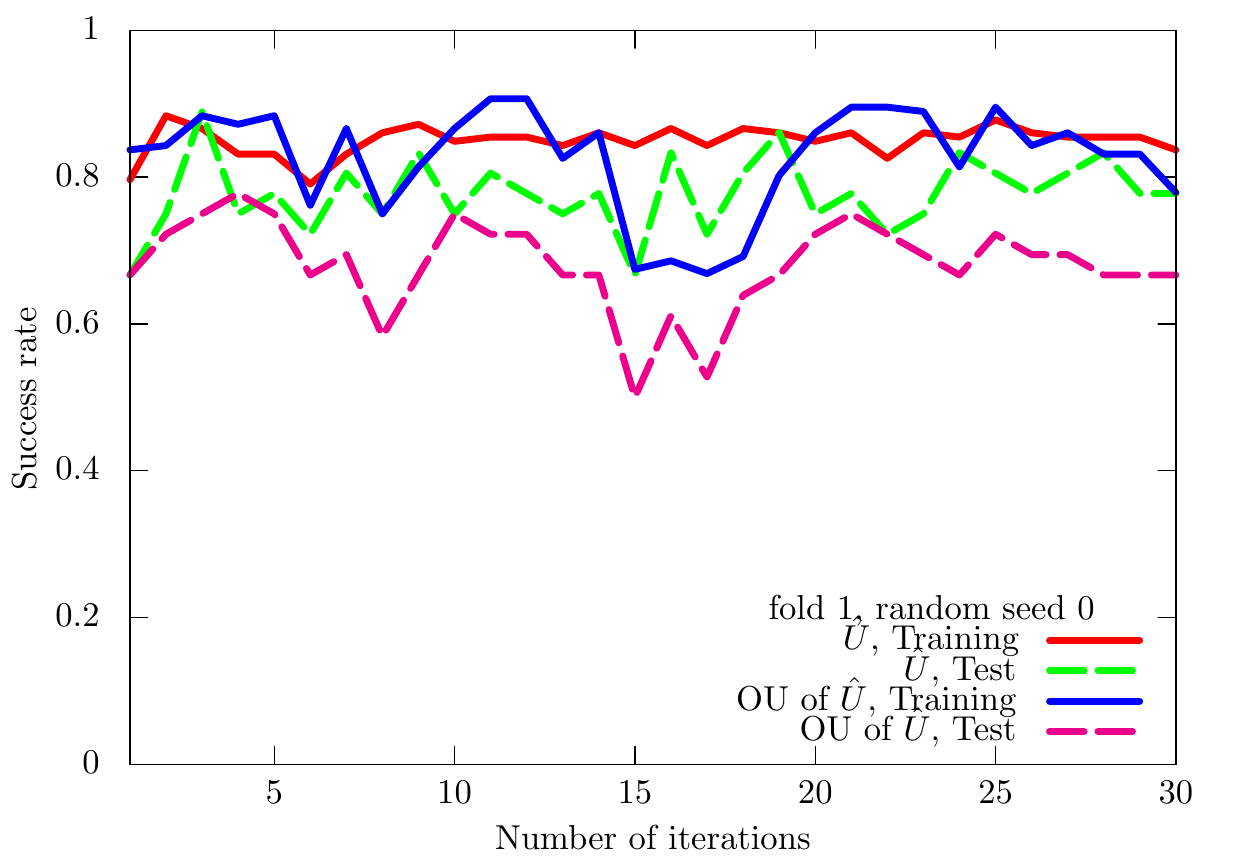}
\includegraphics[scale=0.25]{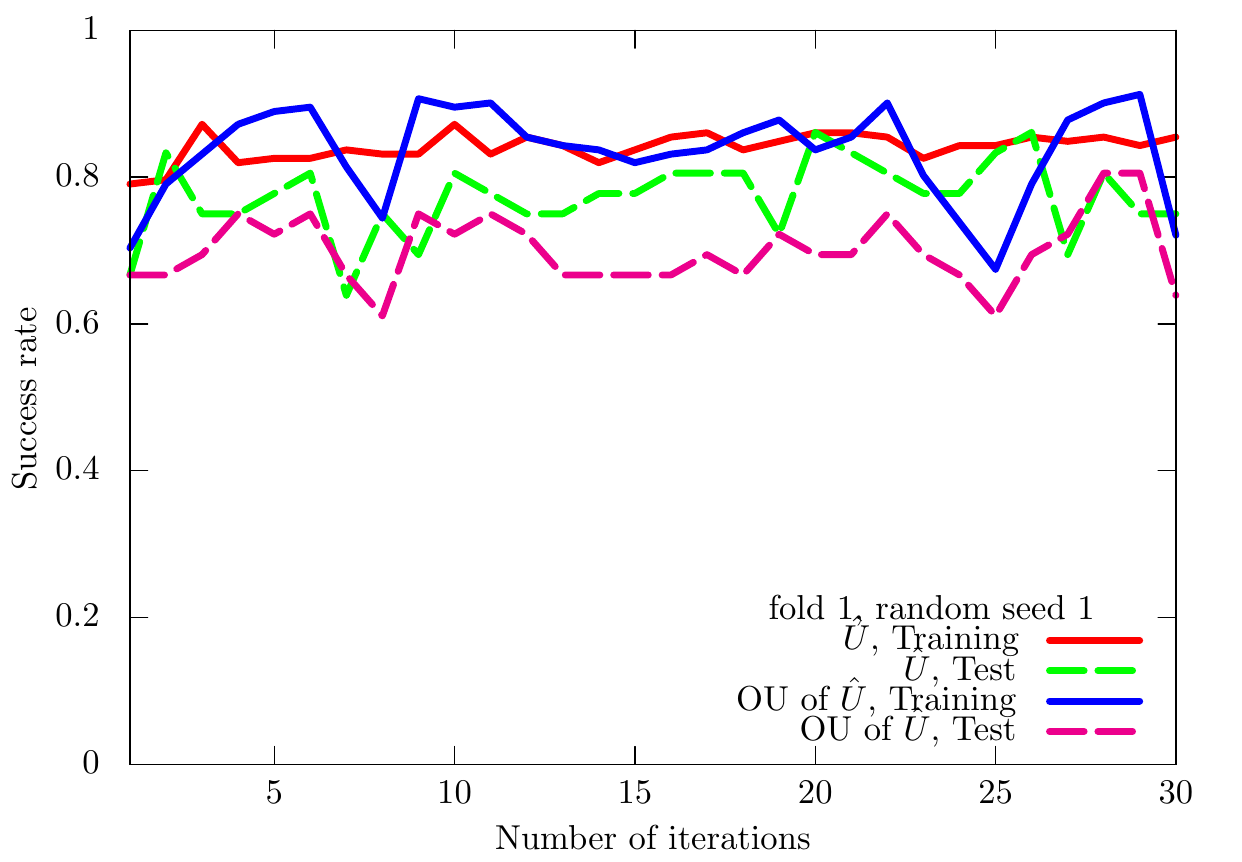}
\includegraphics[scale=0.25]{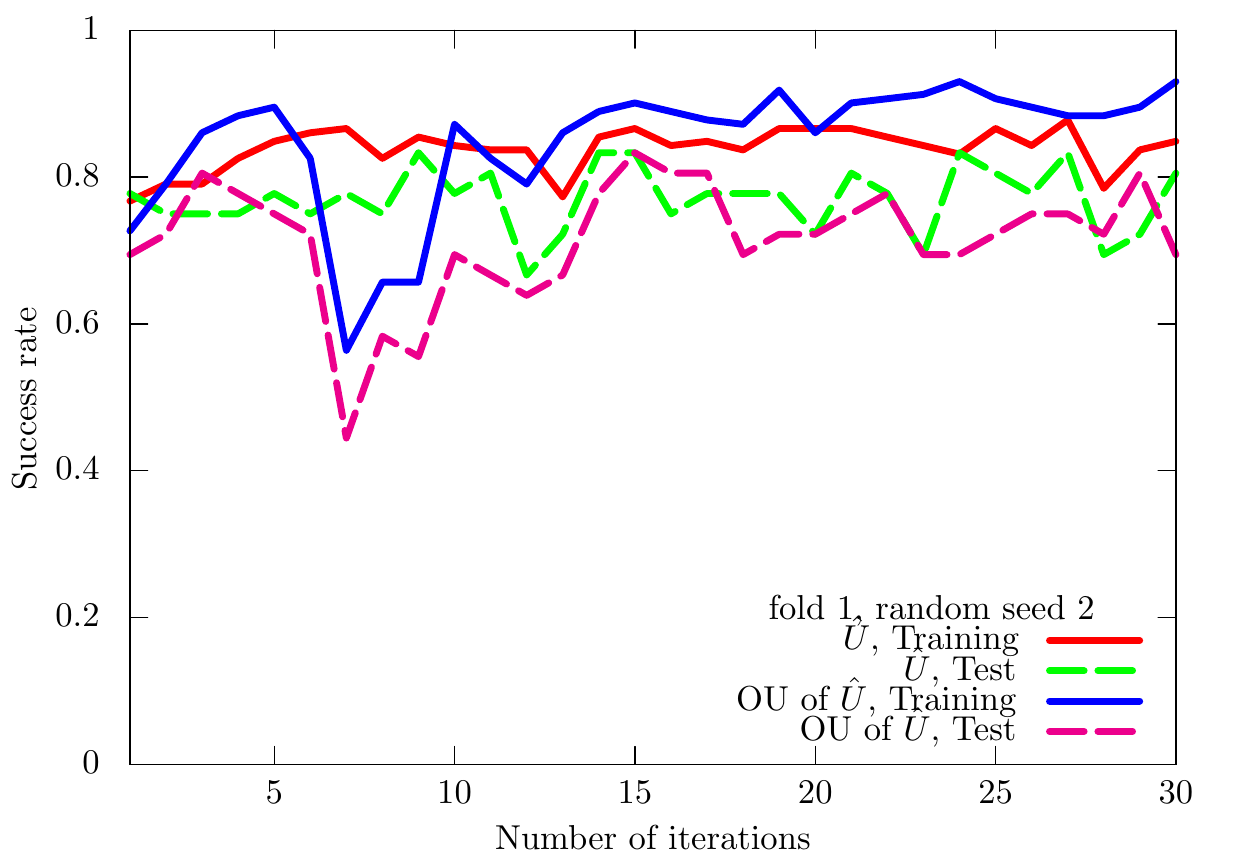}
\includegraphics[scale=0.25]{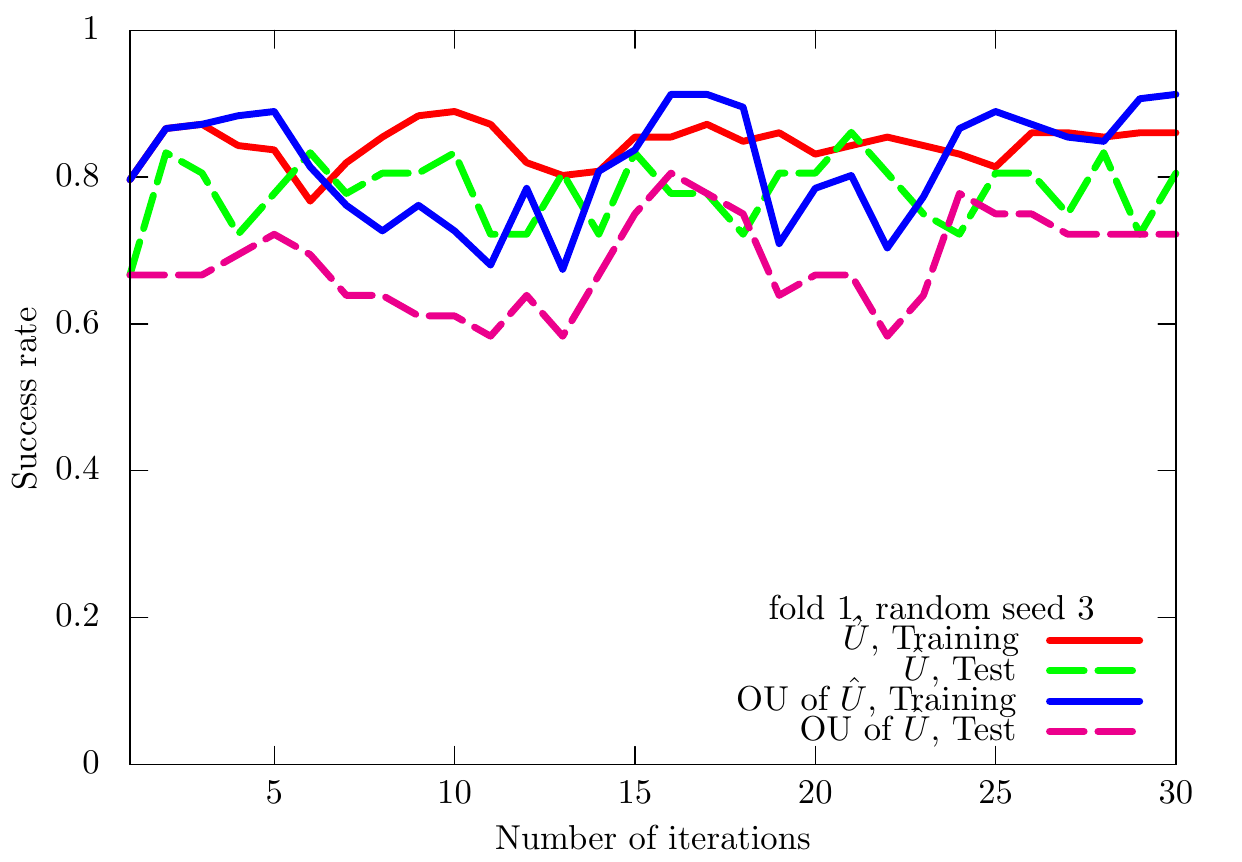}
\includegraphics[scale=0.25]{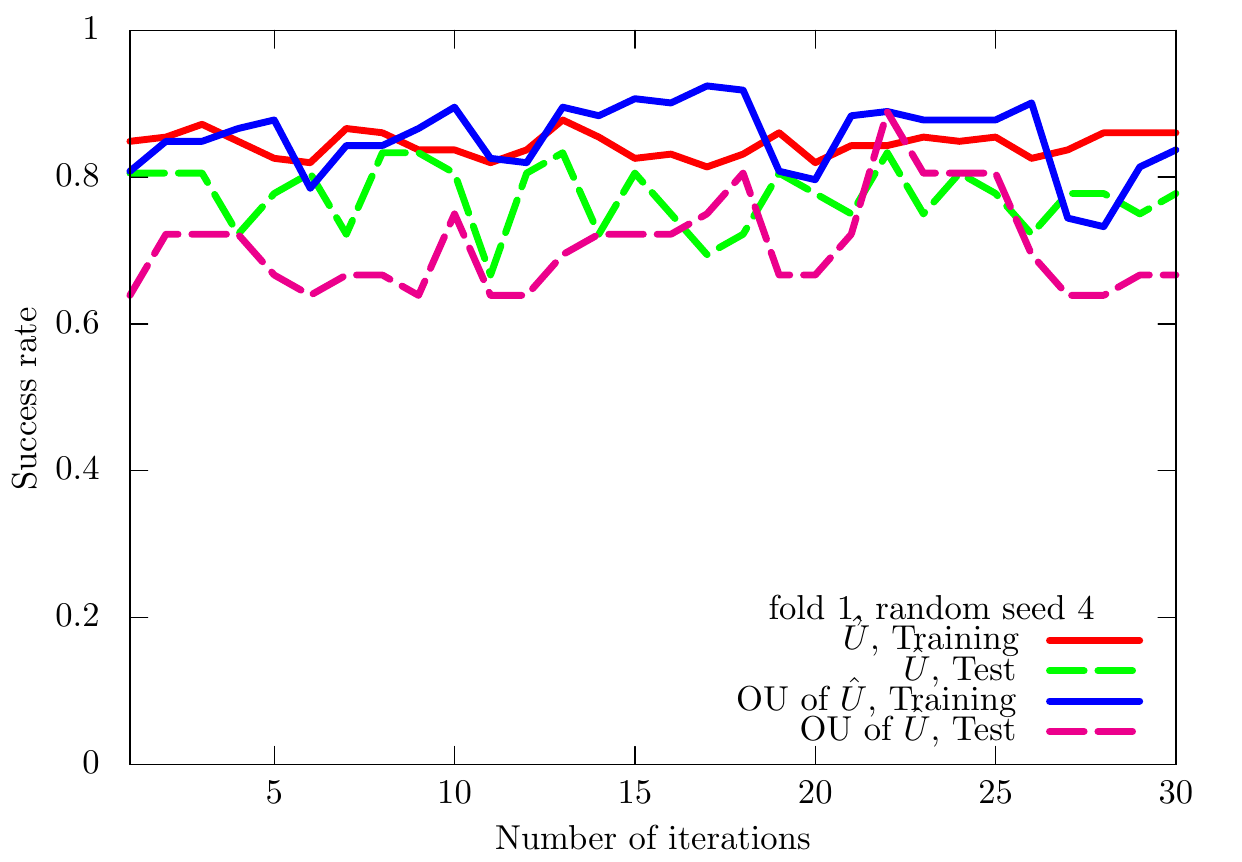}
\includegraphics[scale=0.25]{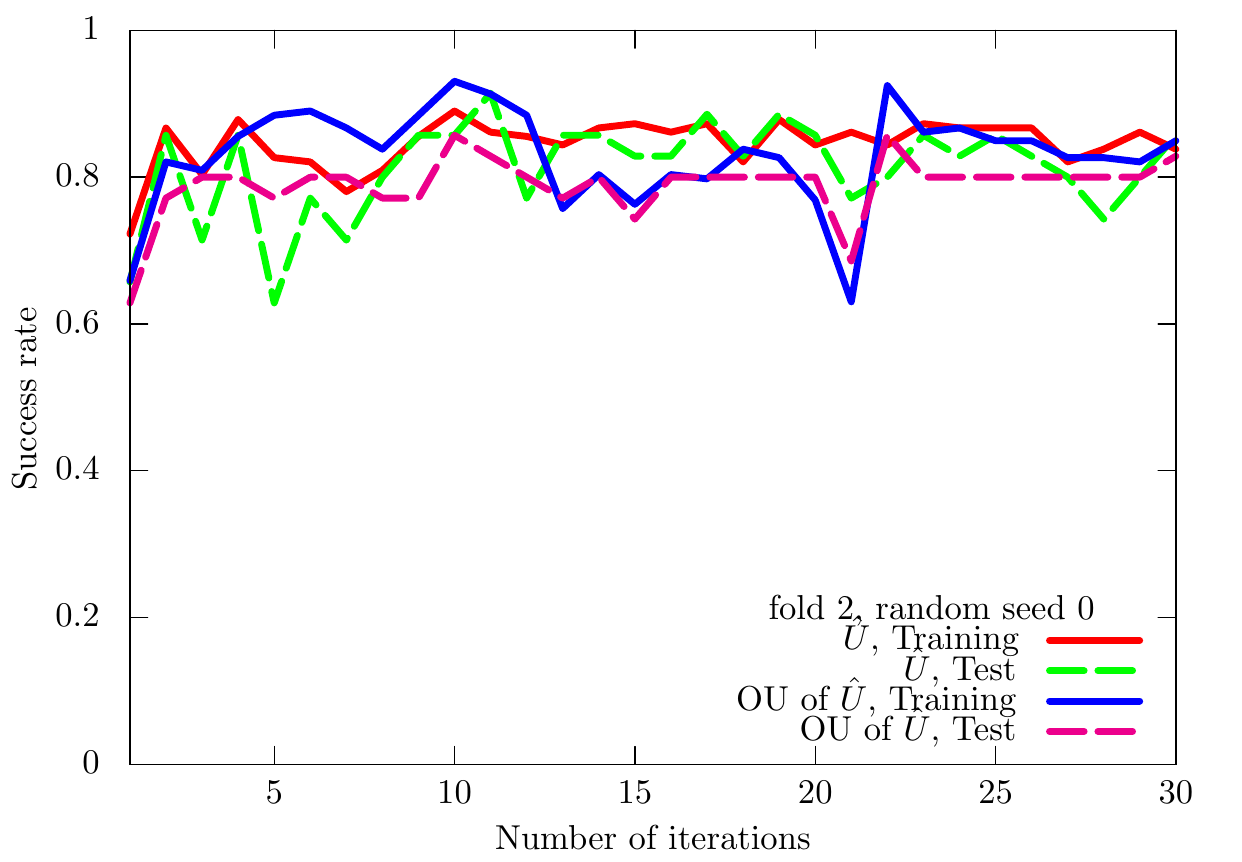}
\includegraphics[scale=0.25]{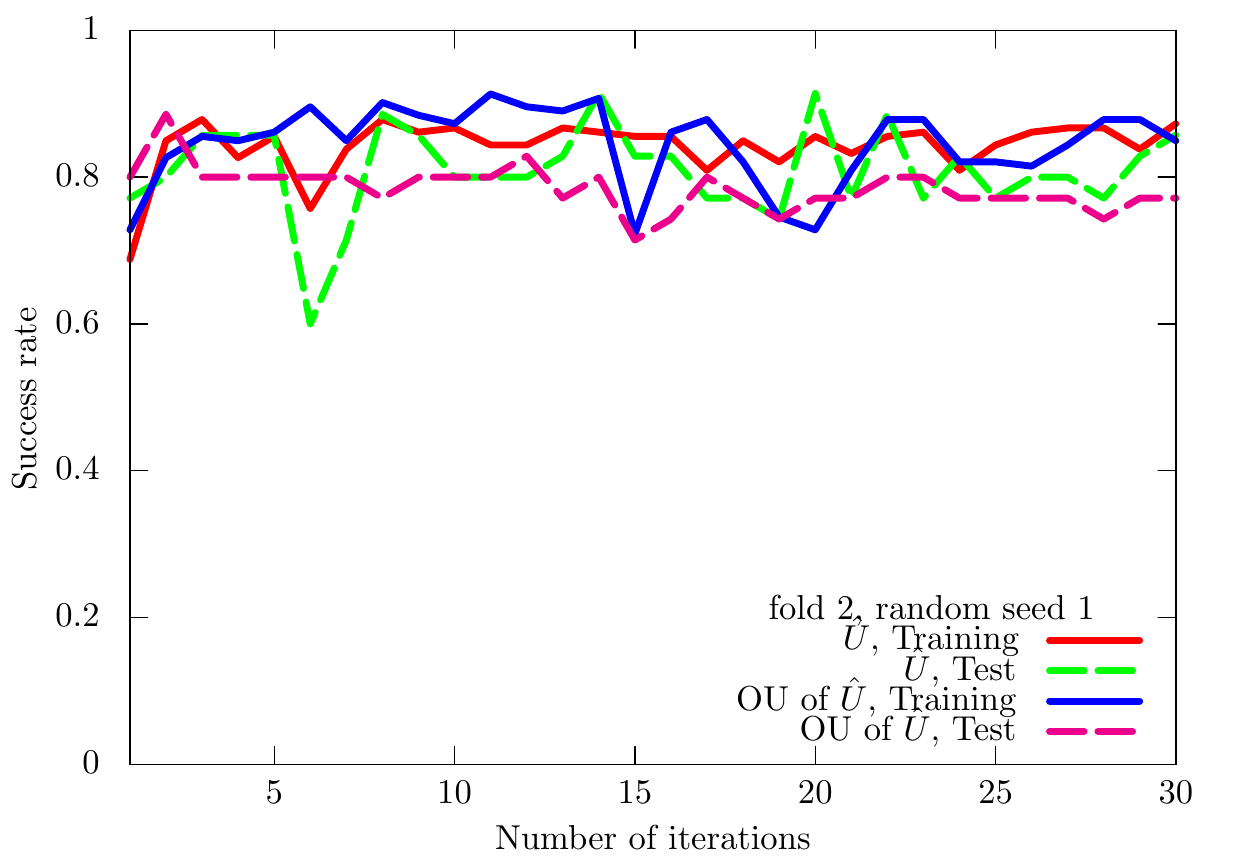}
\includegraphics[scale=0.25]{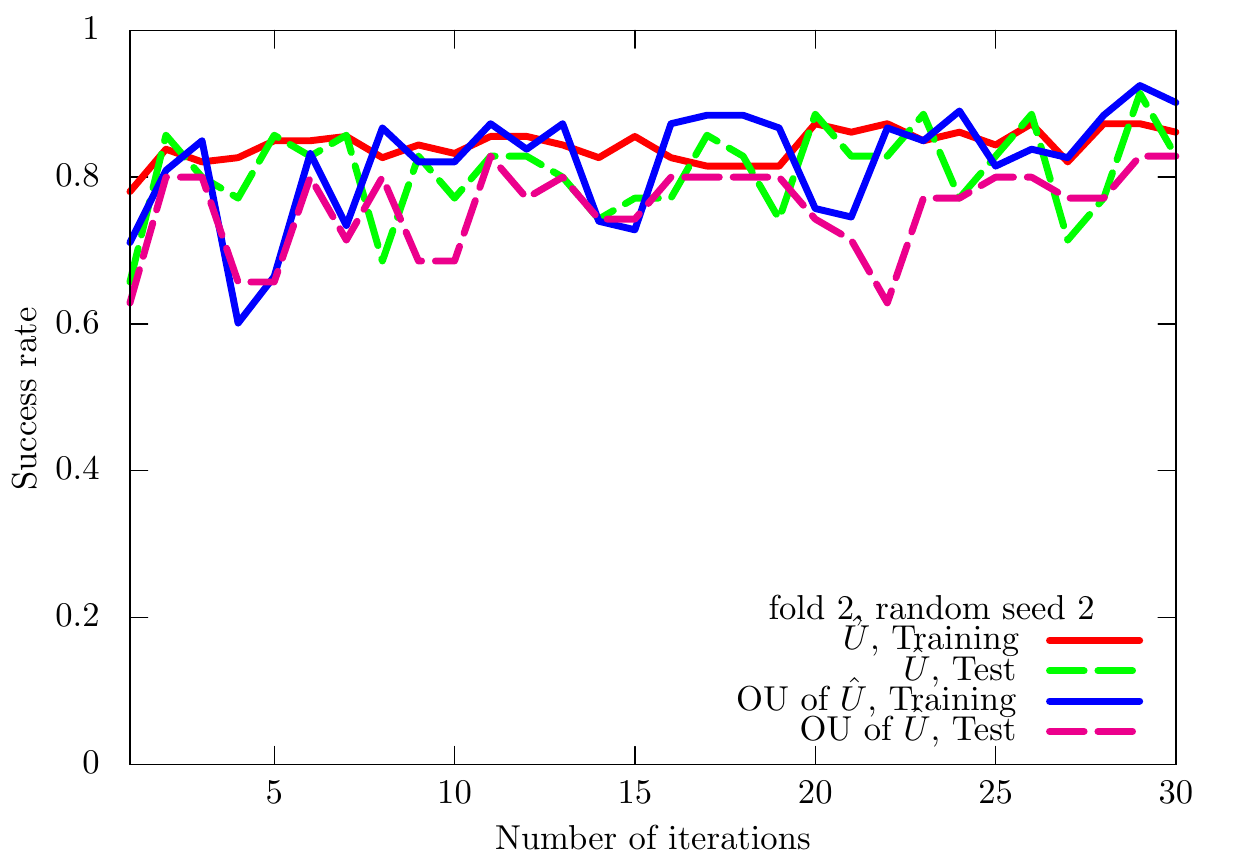}
\includegraphics[scale=0.25]{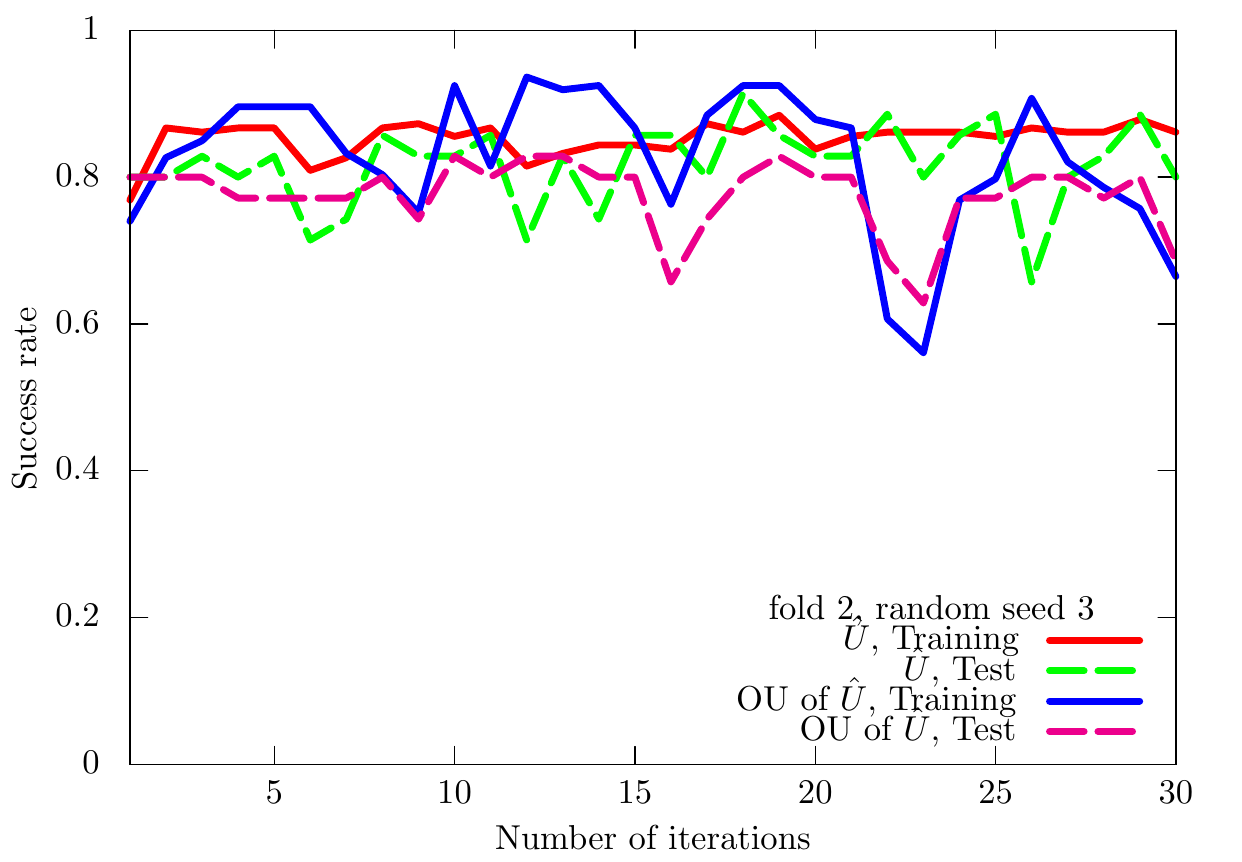}
\includegraphics[scale=0.25]{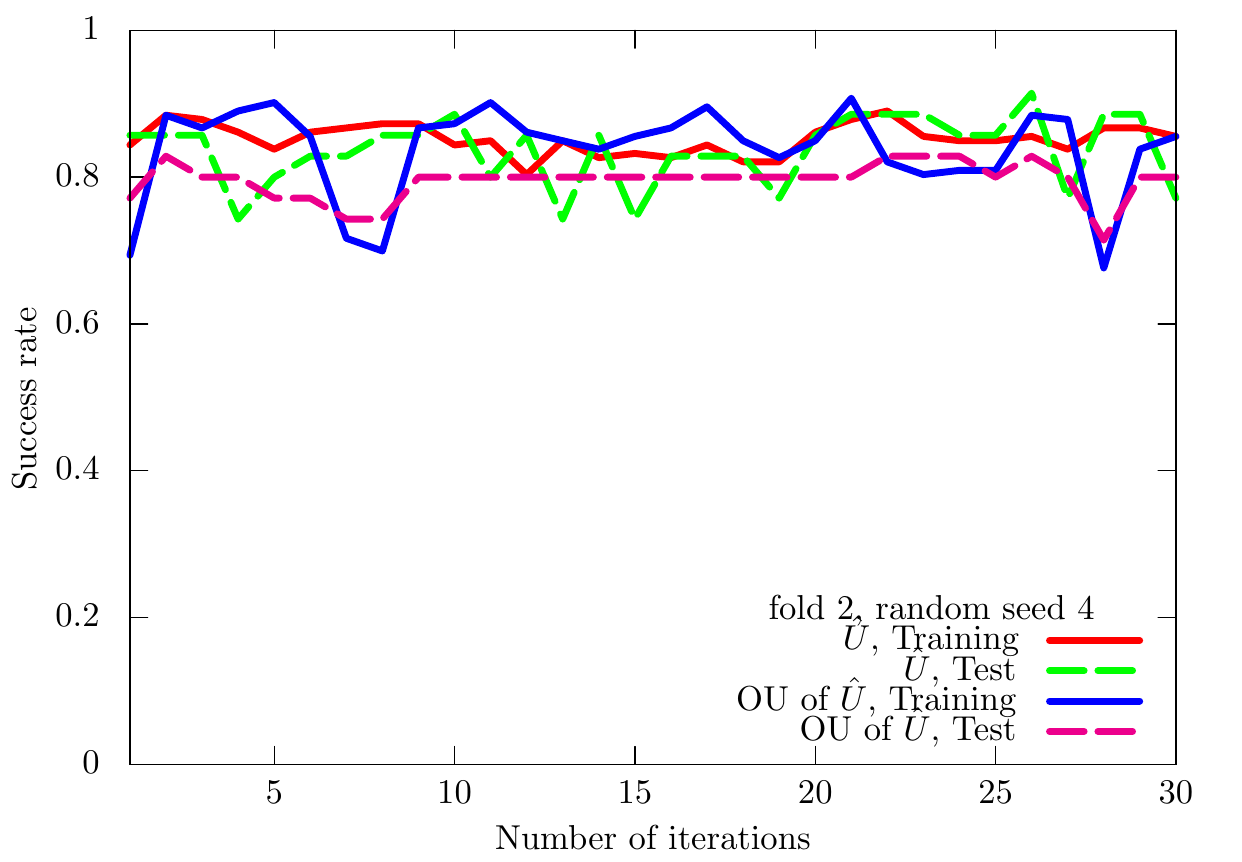}
\includegraphics[scale=0.25]{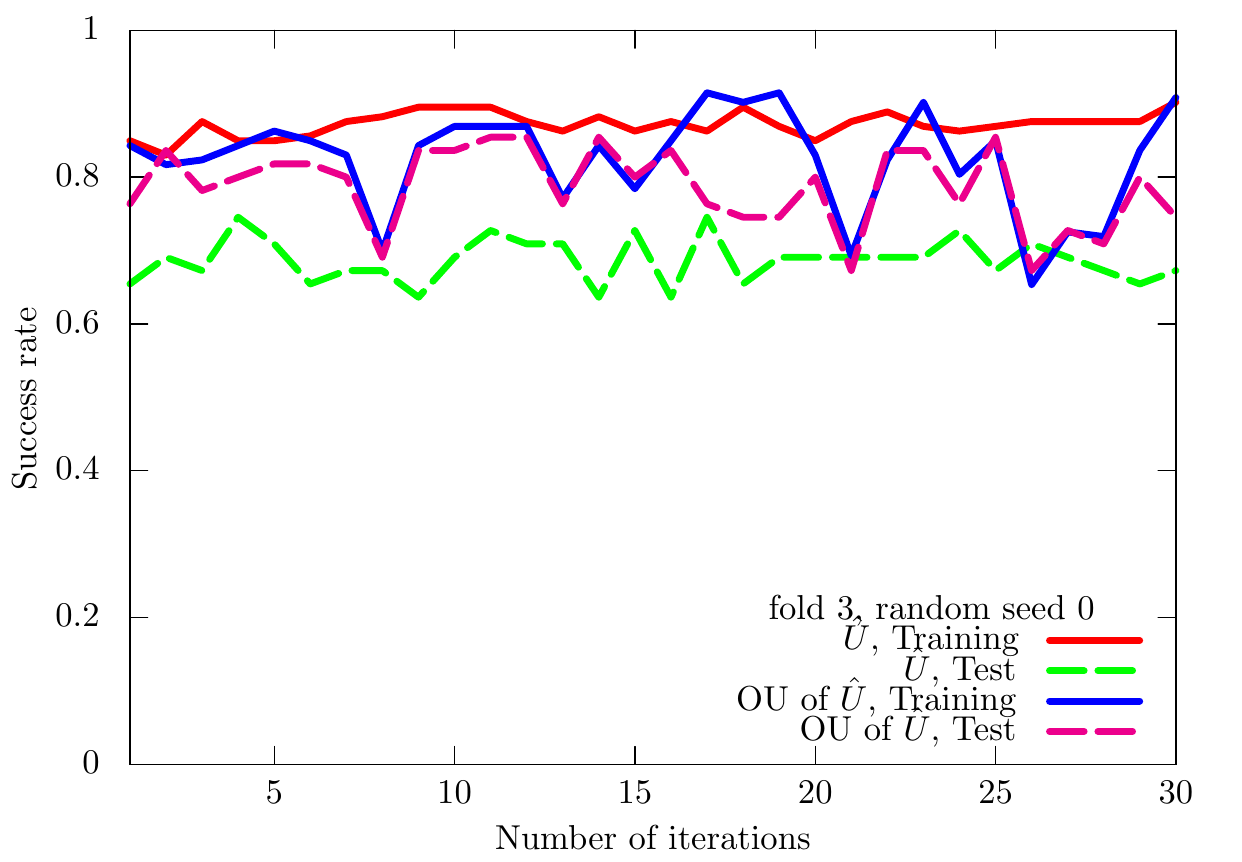}
\includegraphics[scale=0.25]{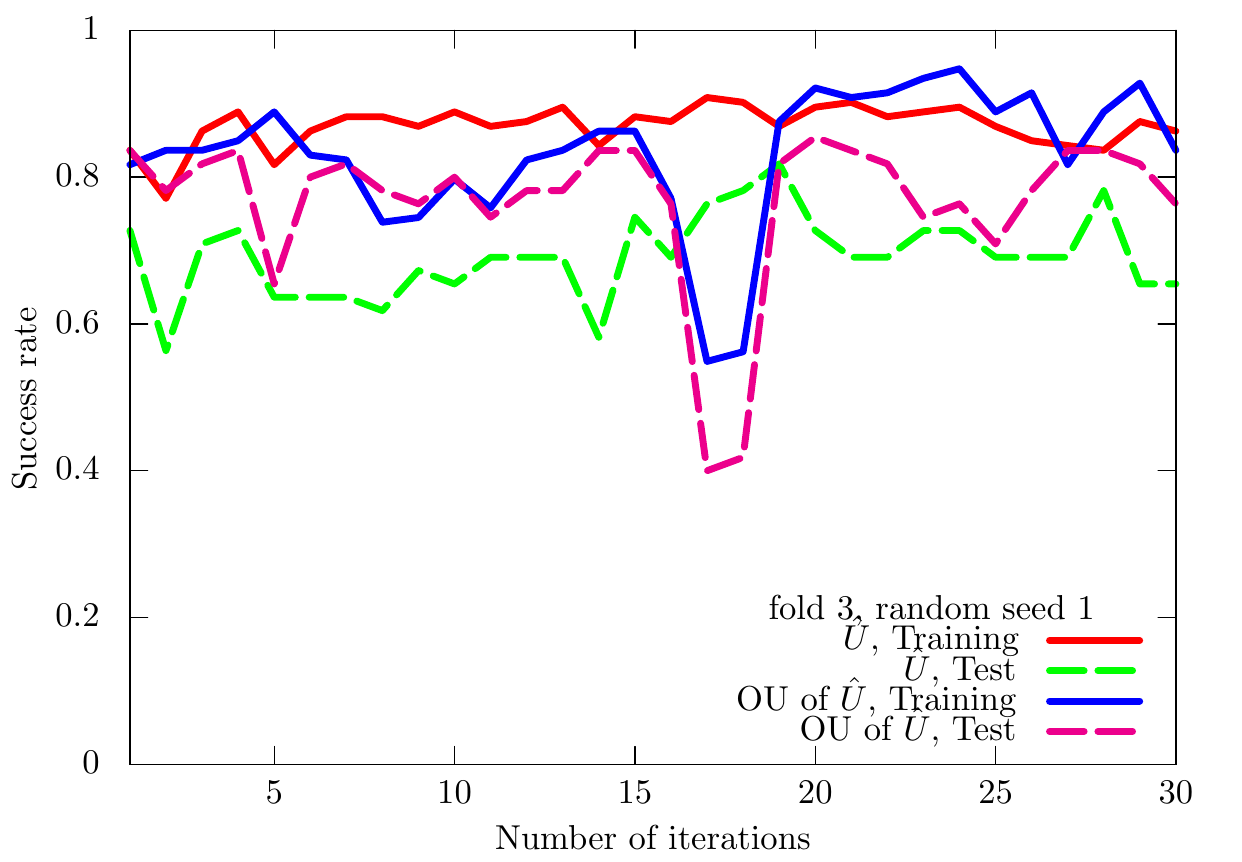}
\includegraphics[scale=0.25]{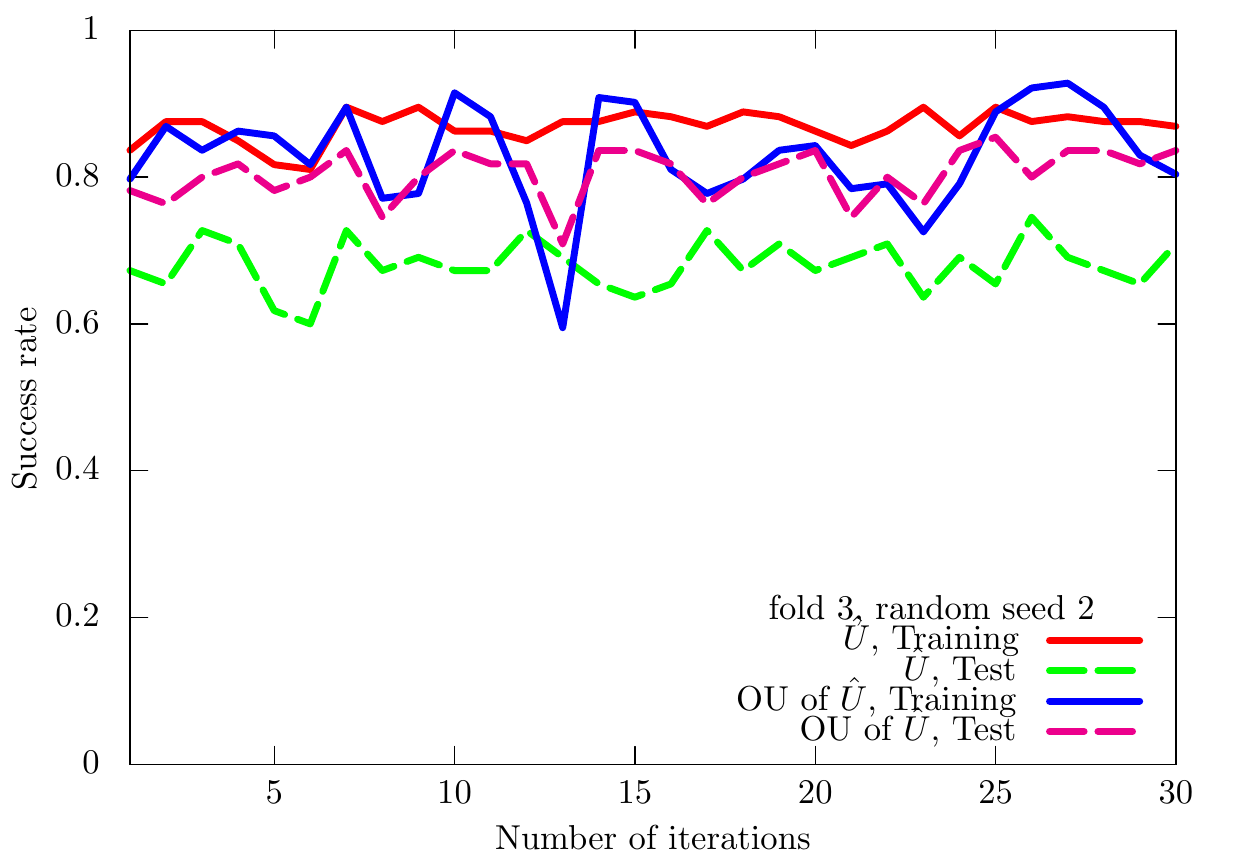}
\includegraphics[scale=0.25]{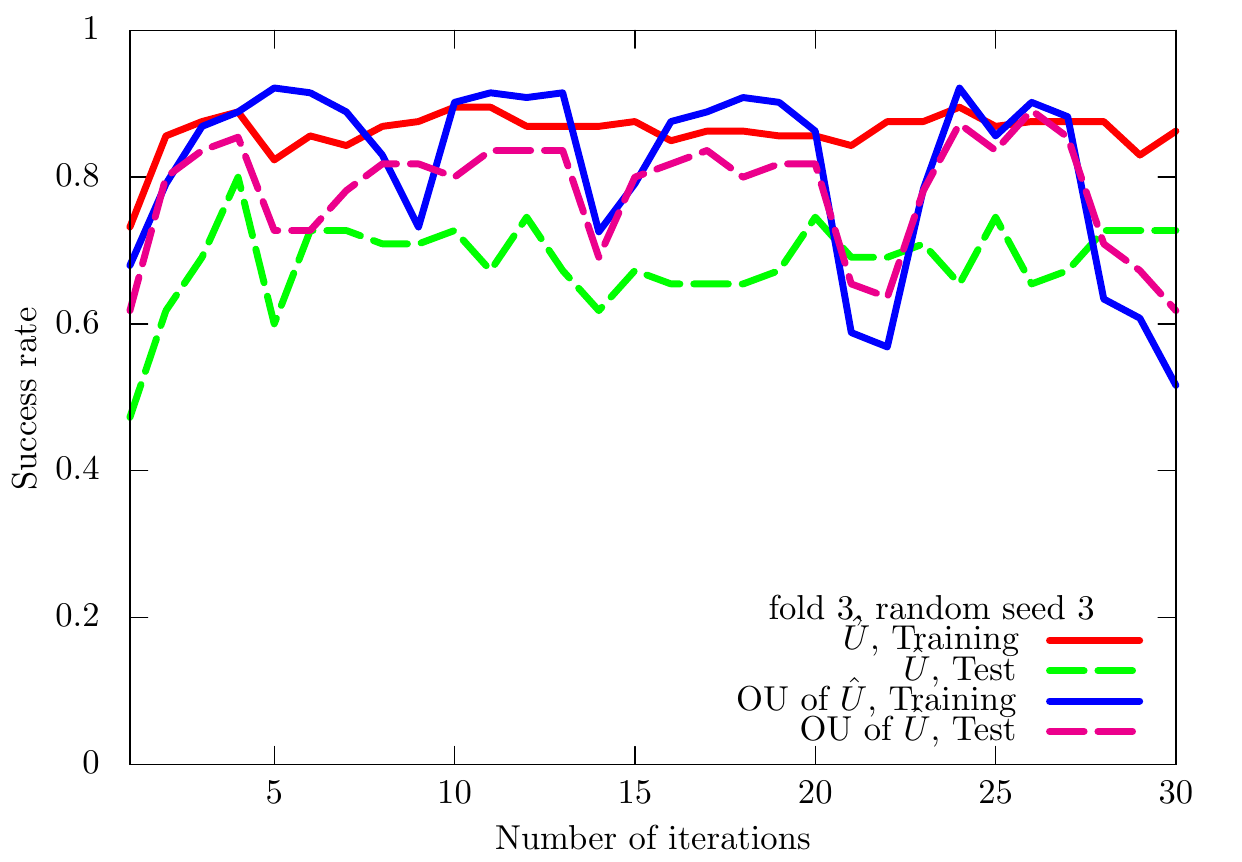}
\includegraphics[scale=0.25]{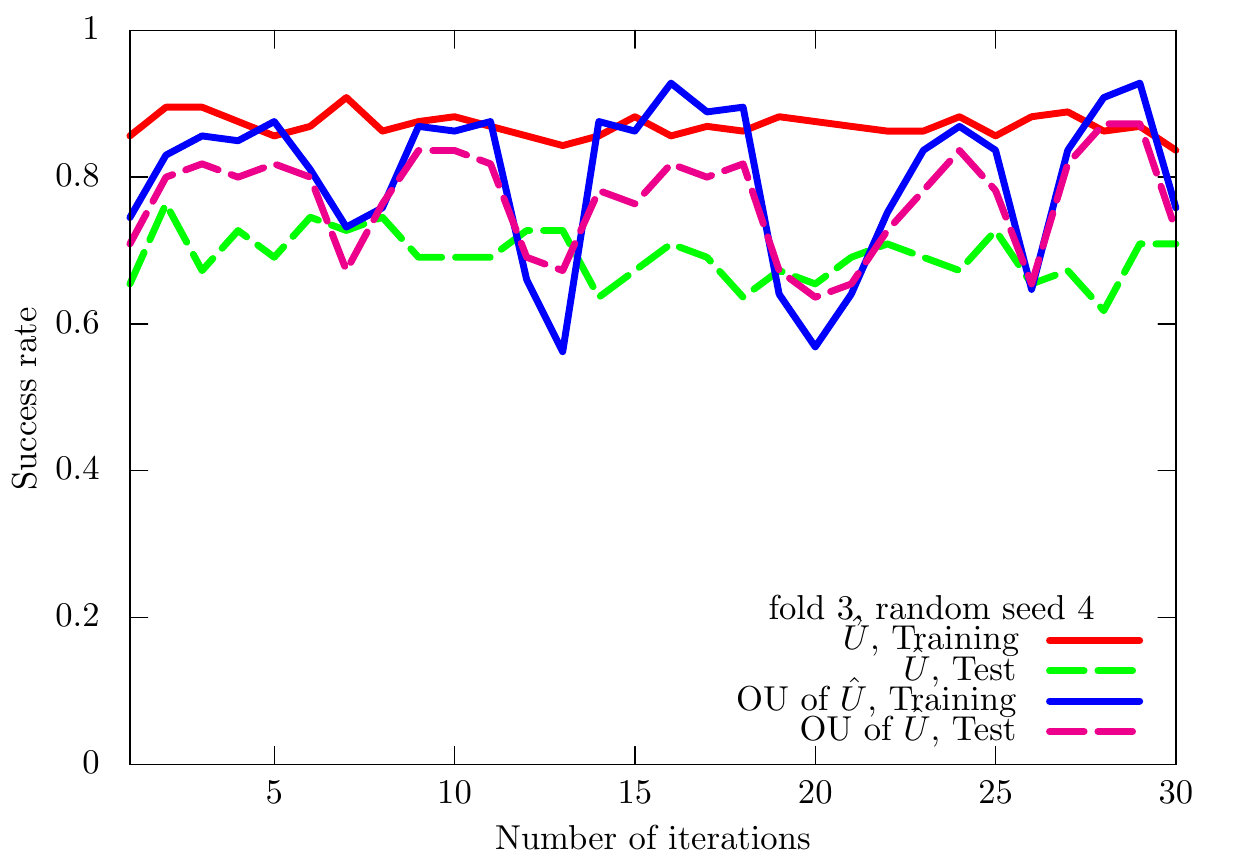}
\includegraphics[scale=0.25]{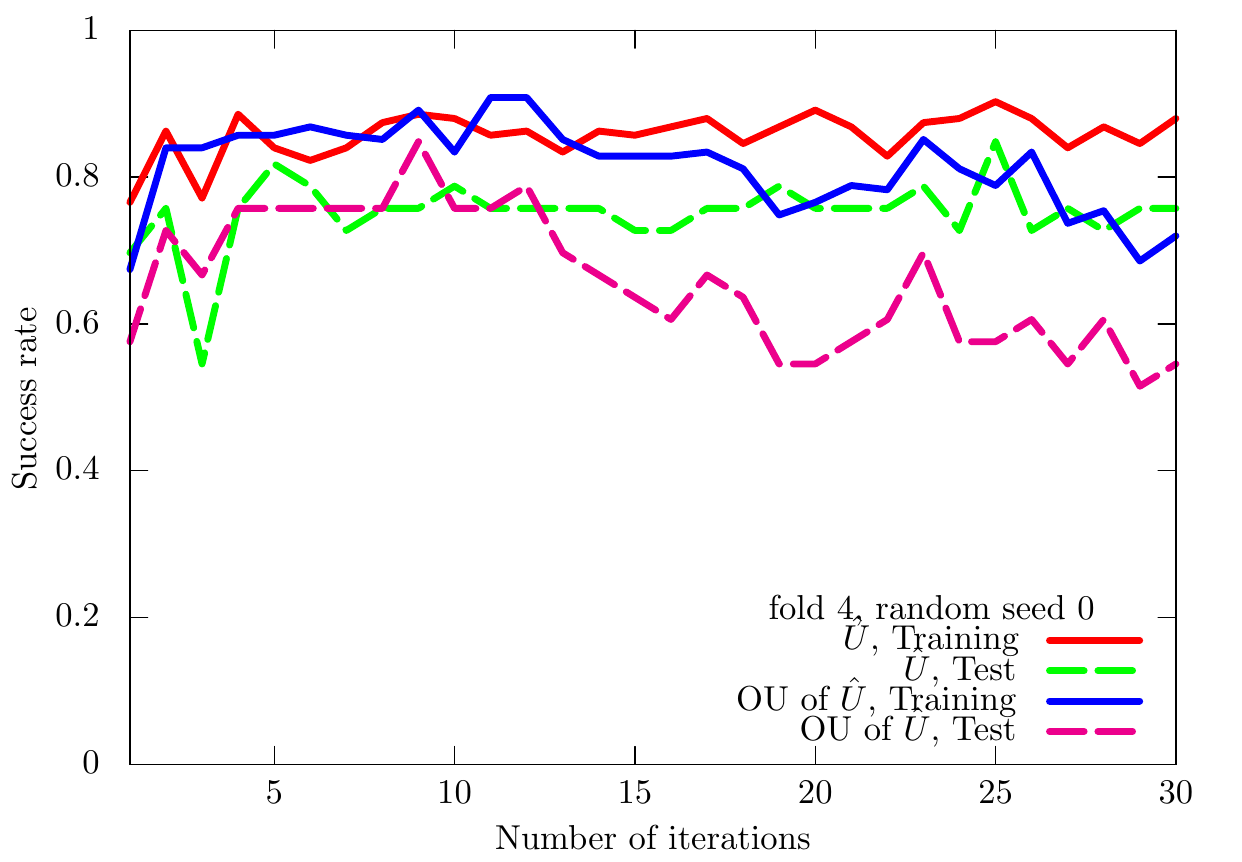}
\includegraphics[scale=0.25]{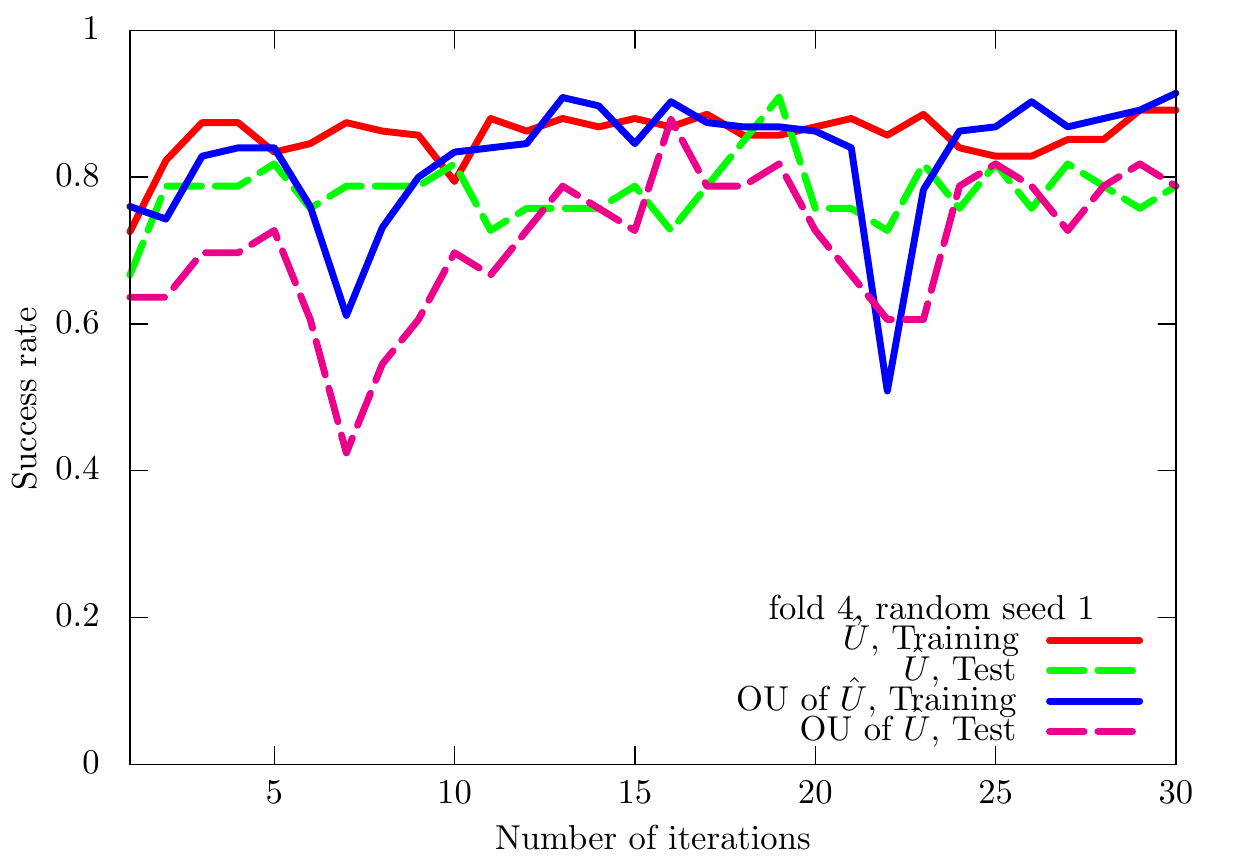}
\includegraphics[scale=0.25]{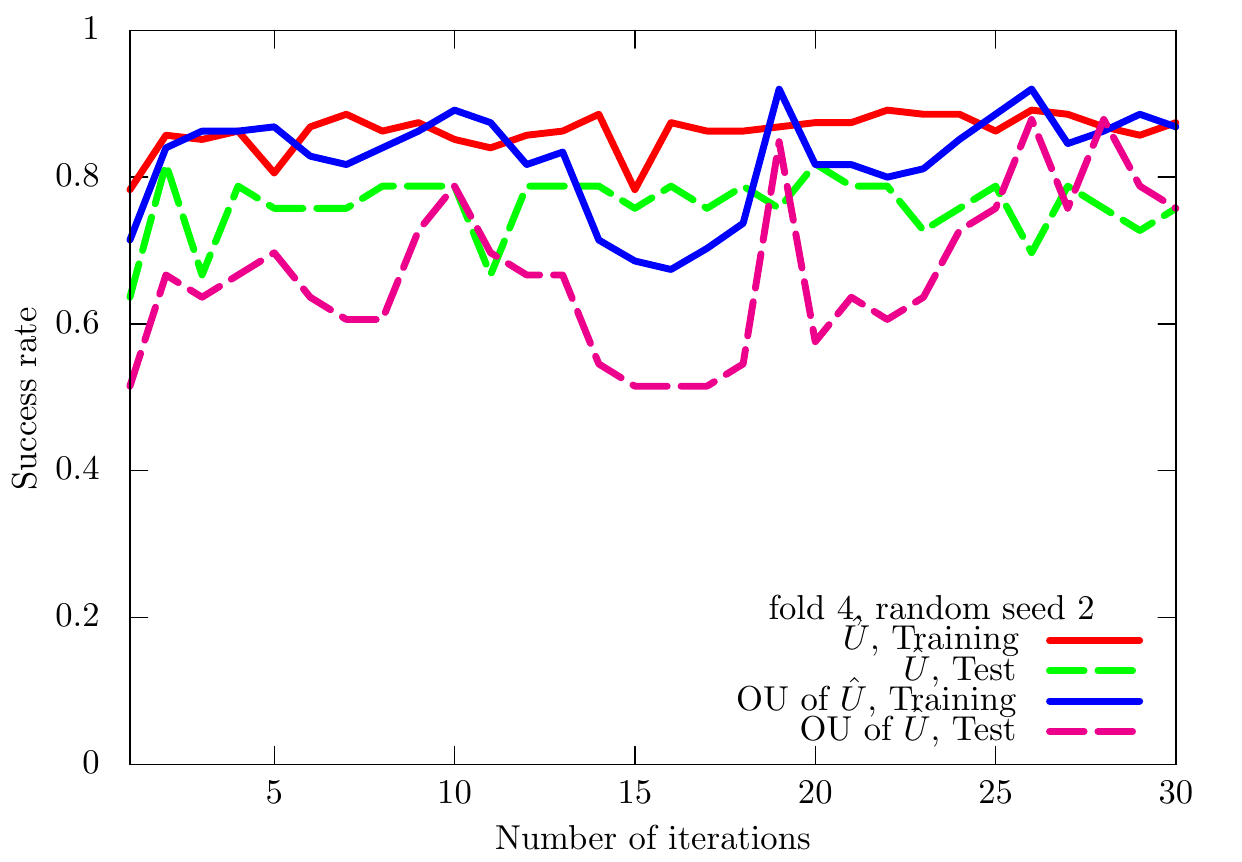}
\includegraphics[scale=0.25]{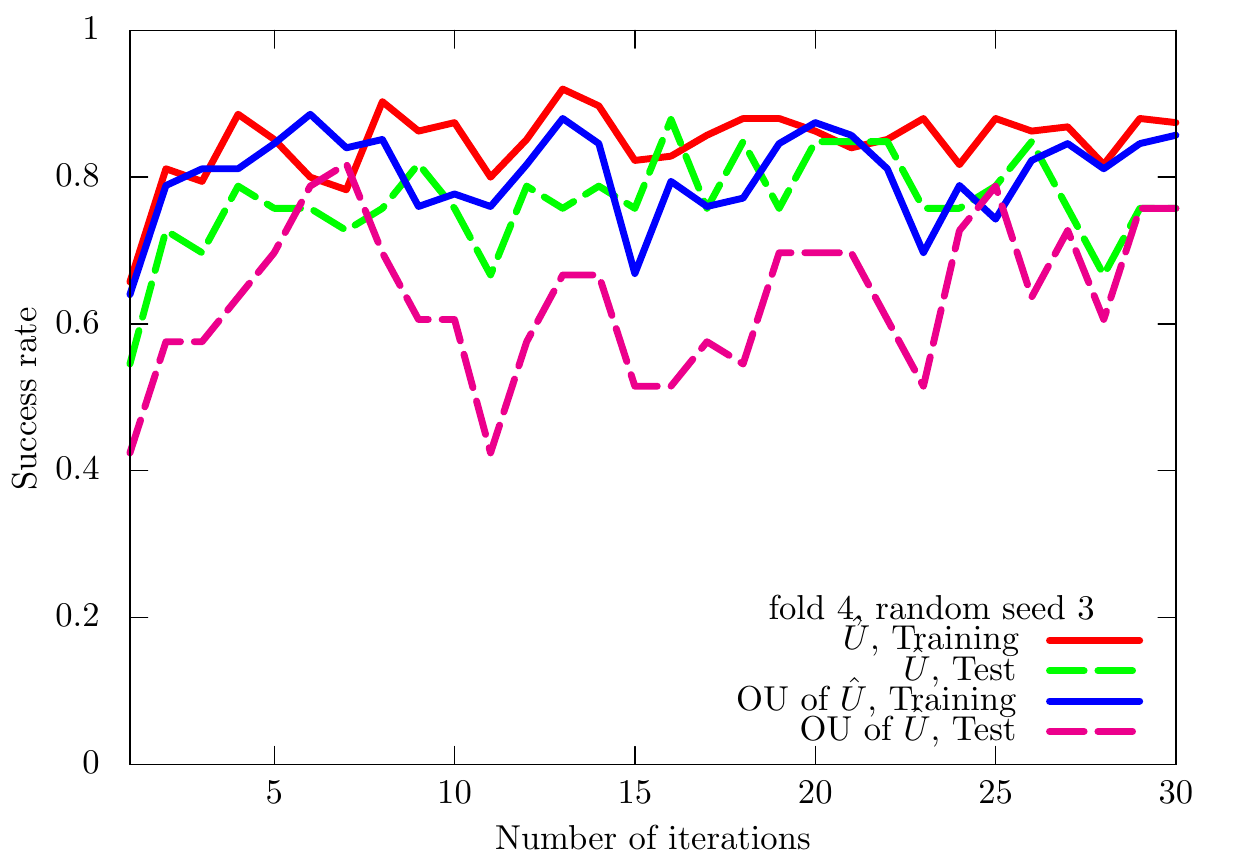}
\includegraphics[scale=0.25]{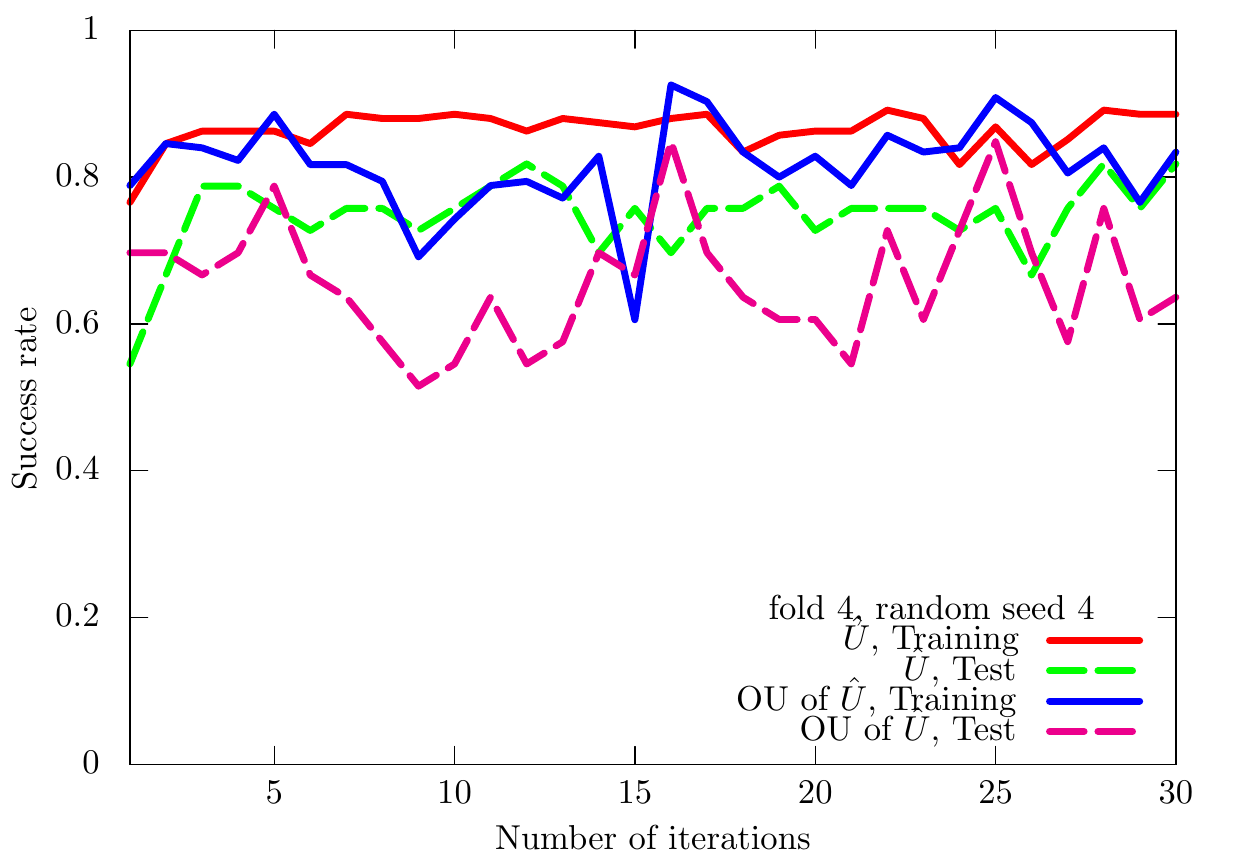}
\caption{Results of the UKM ($\hat{X}$ and $\hat{P}$) on the $5$-fold datasets with $5$ different random seeds for the sonar dataset ($0$ or $1$). We use complex matrices and set $\theta_\mathrm{bias} = 0$. We set $r = 0.010$.}
\label{supp-arXiv-numerical-result-raw-data-fold-001-rand-001-UKM-P-UCI-sonar-0-1}
\end{figure*}

\begin{figure*}[htb]
\centering
\includegraphics[scale=0.25]{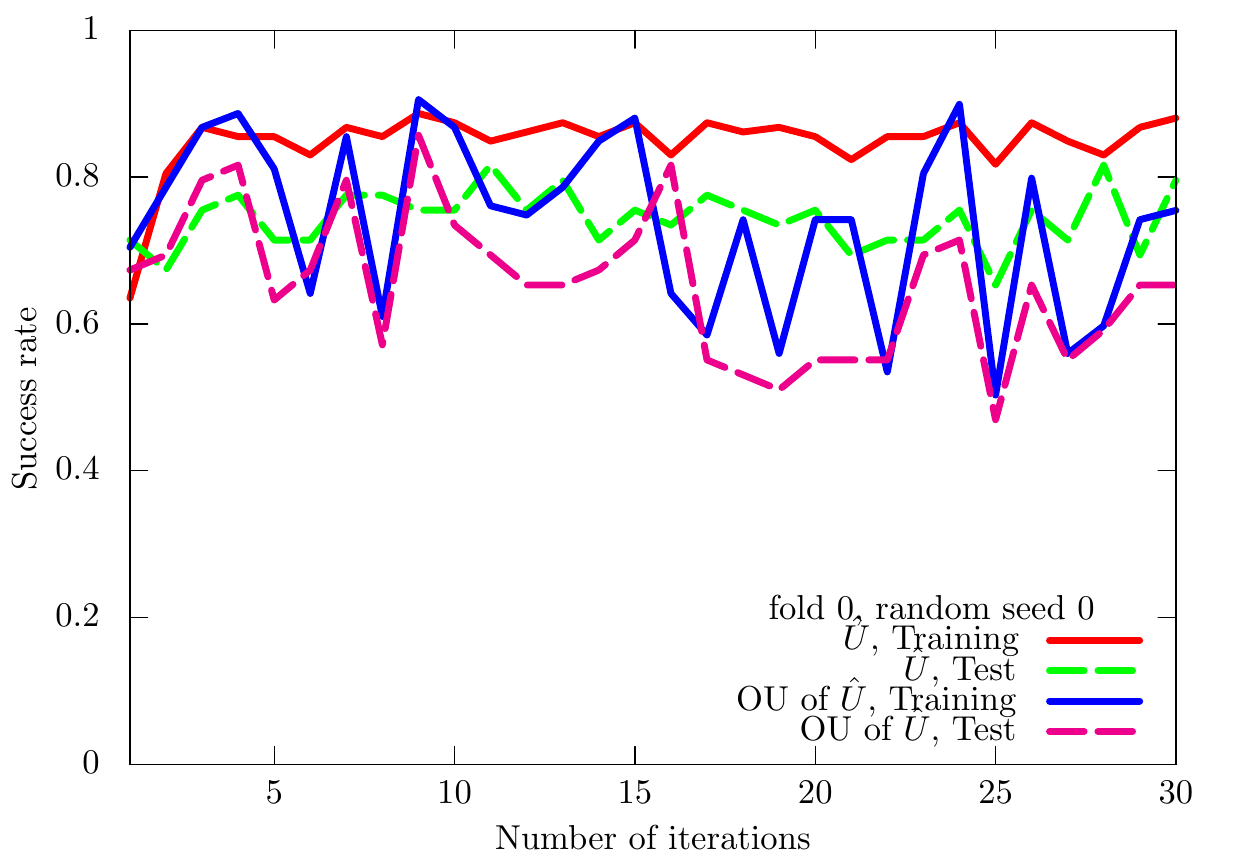}
\includegraphics[scale=0.25]{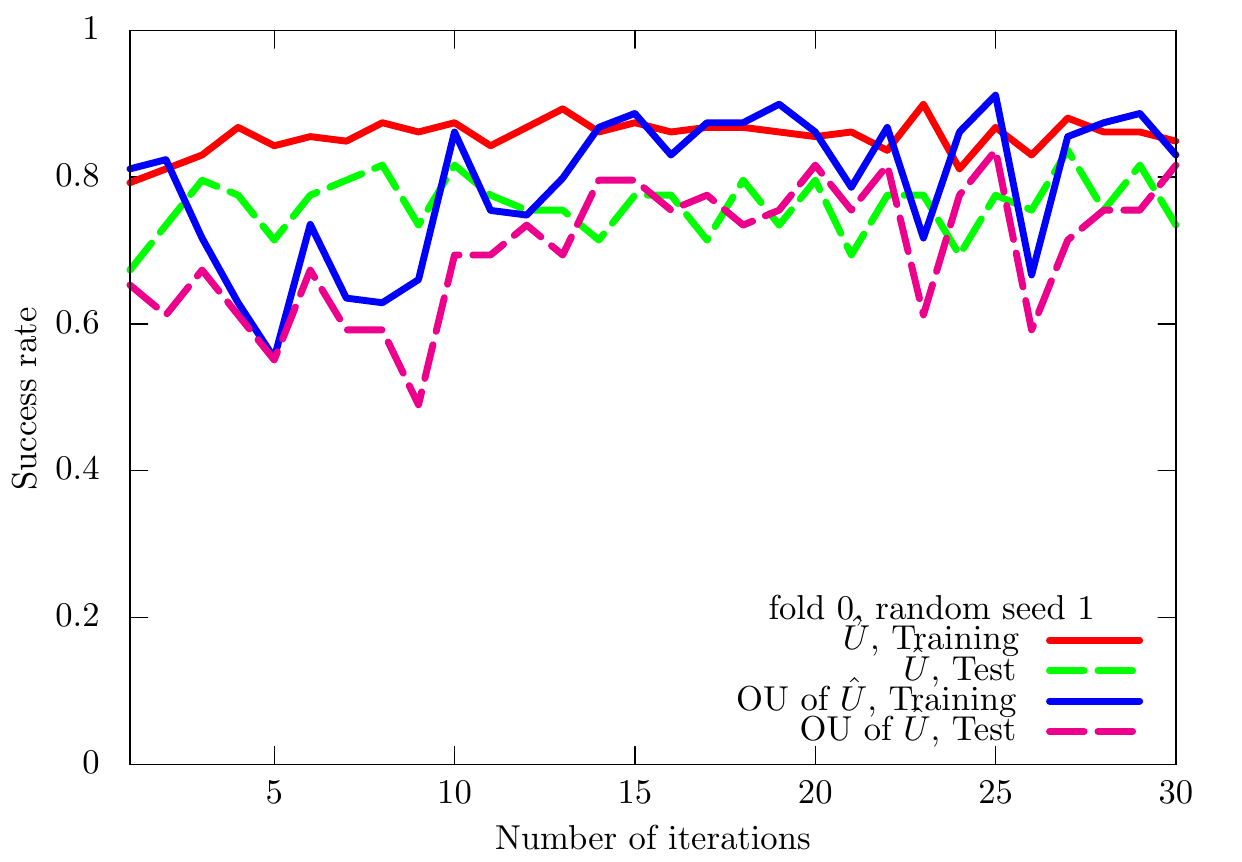}
\includegraphics[scale=0.25]{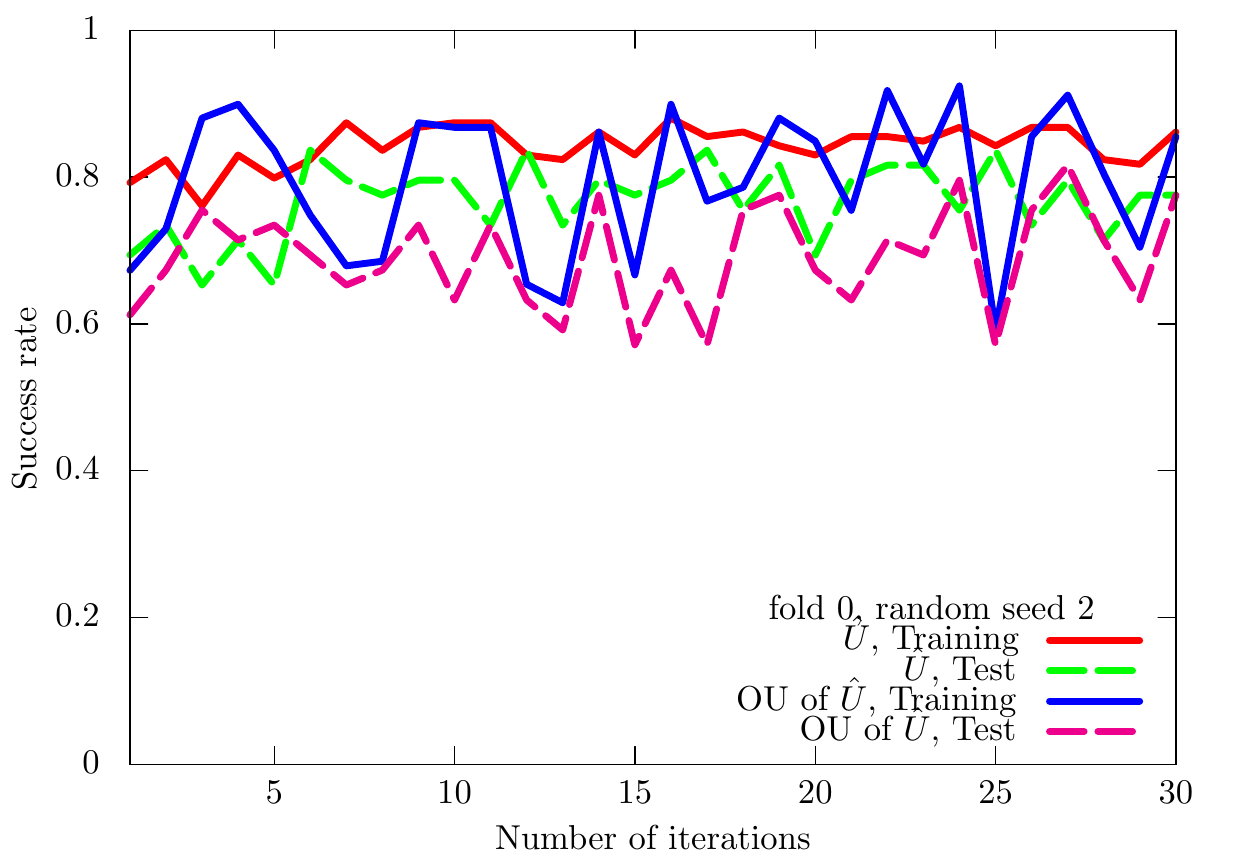}
\includegraphics[scale=0.25]{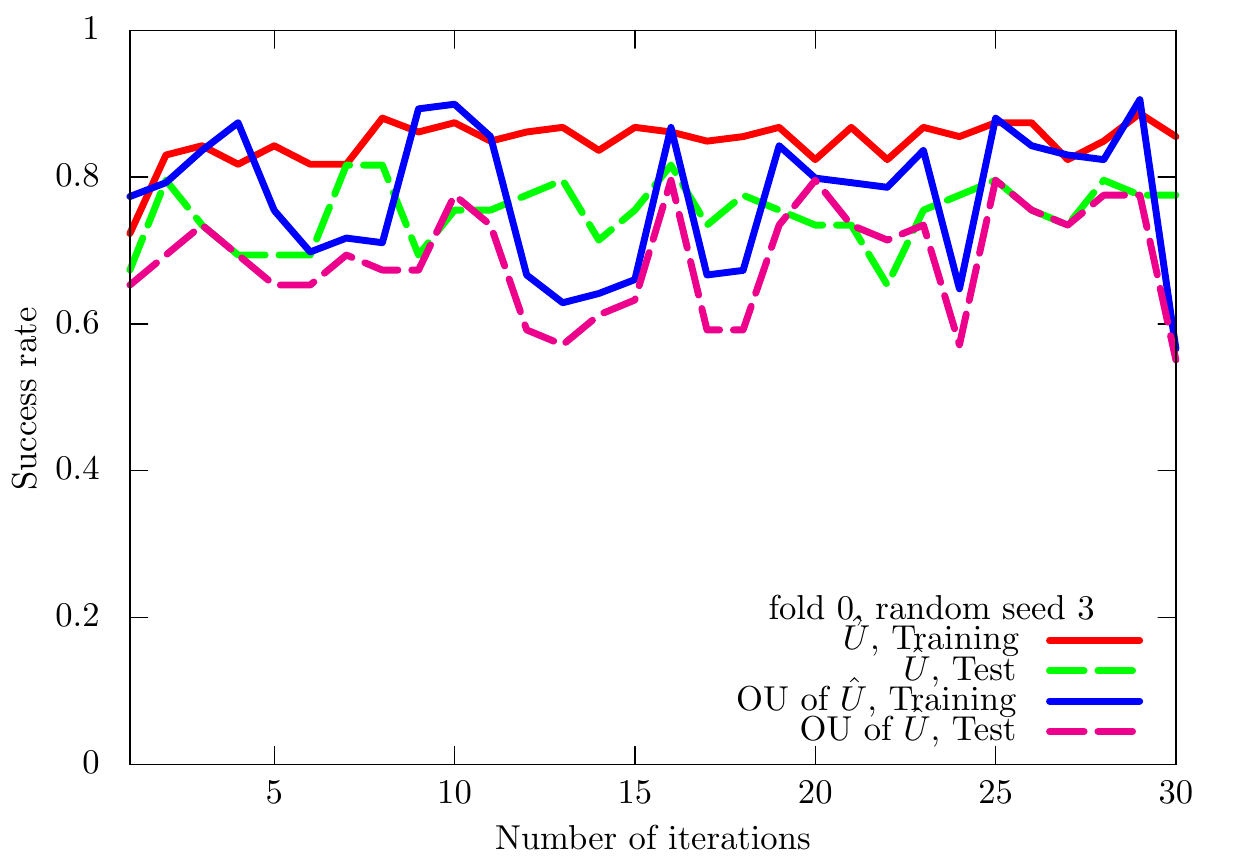}
\includegraphics[scale=0.25]{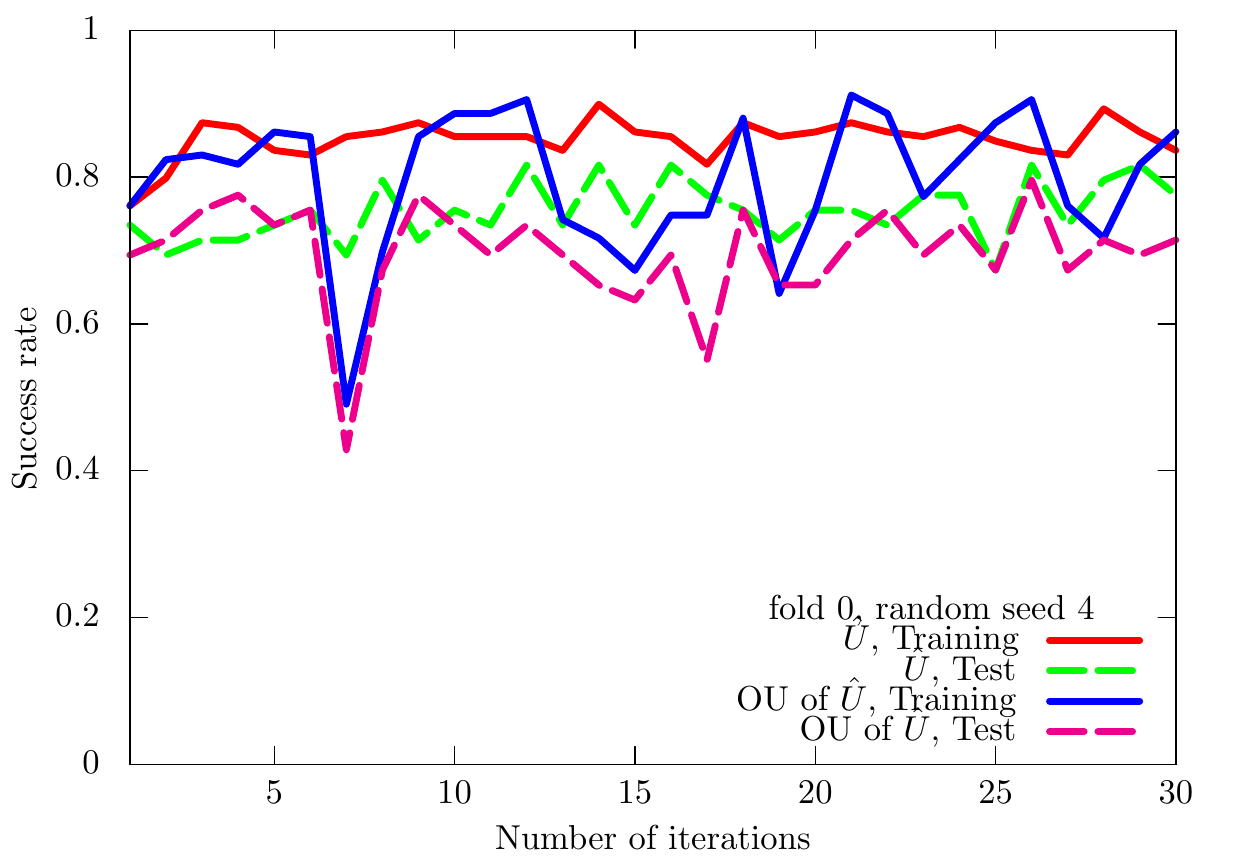}
\includegraphics[scale=0.25]{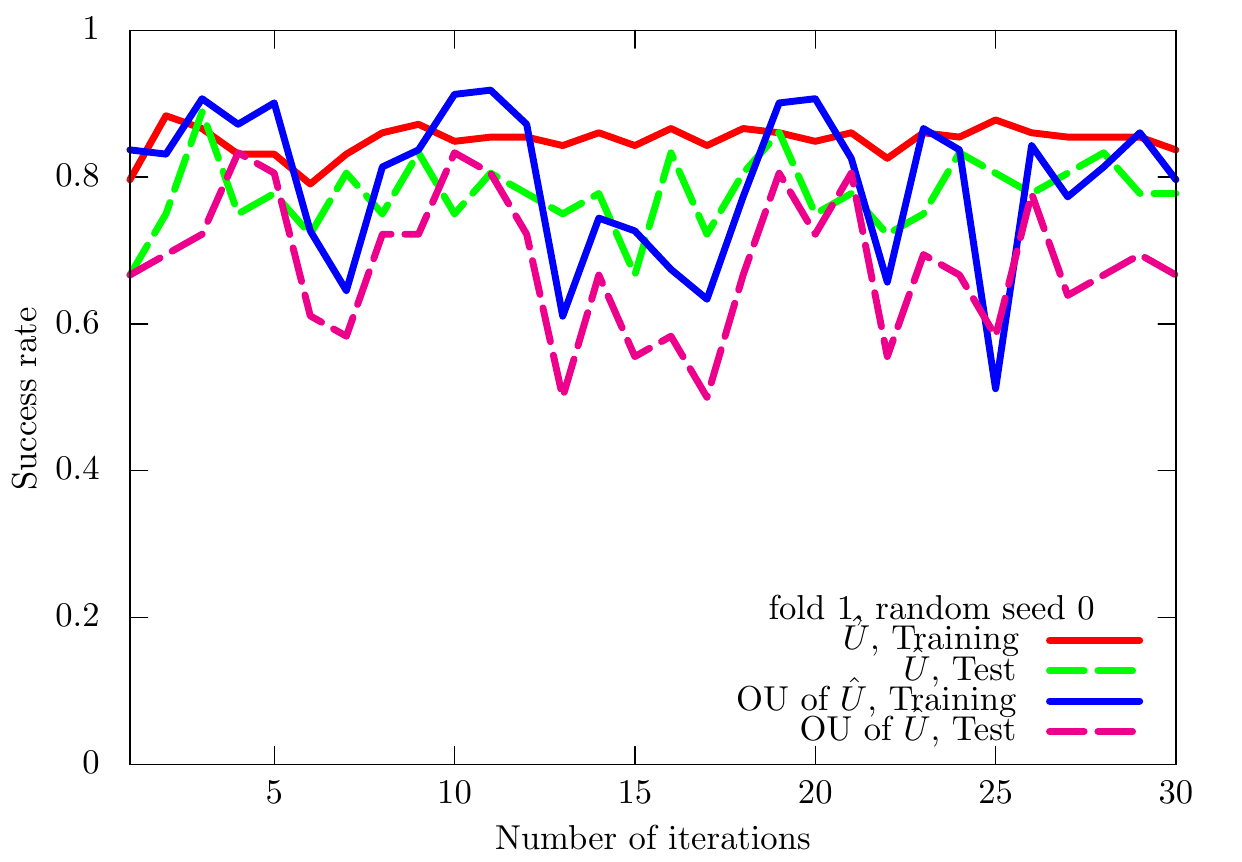}
\includegraphics[scale=0.25]{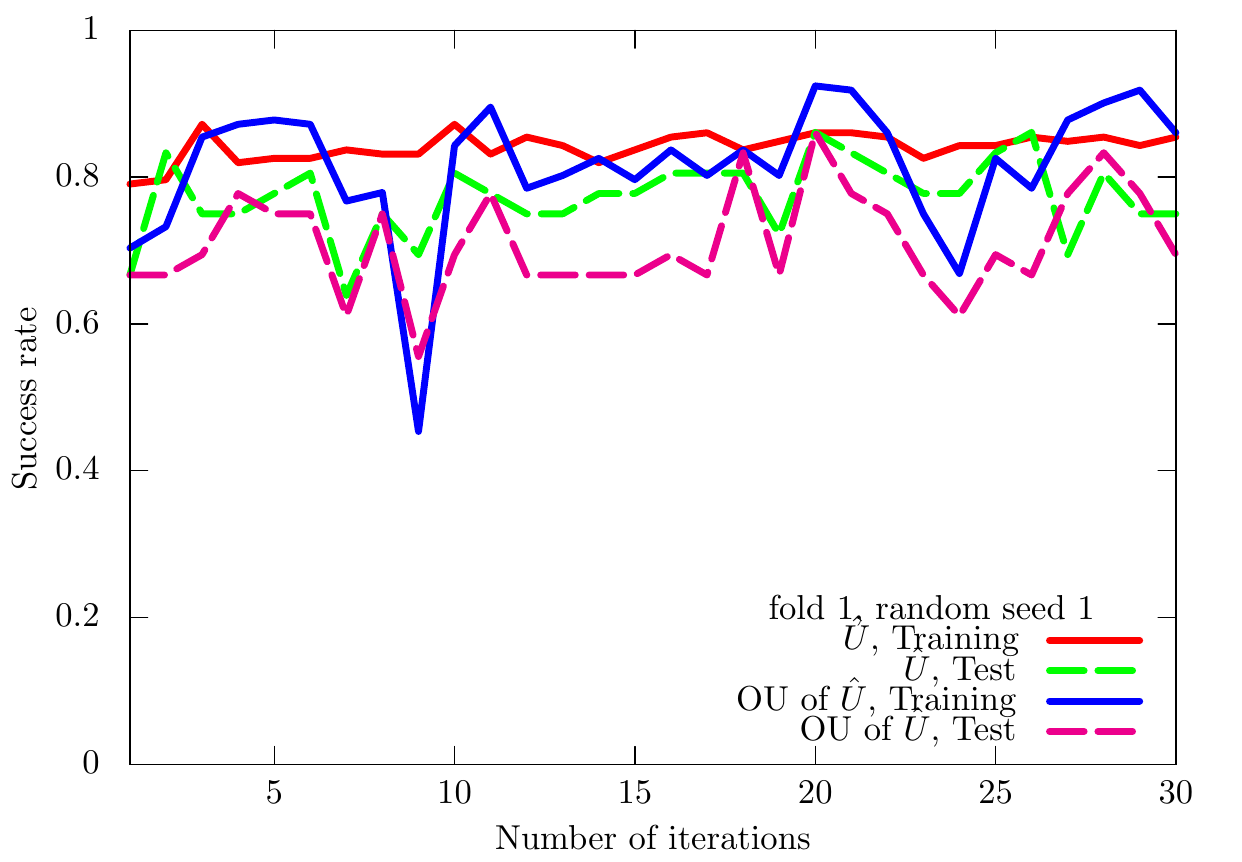}
\includegraphics[scale=0.25]{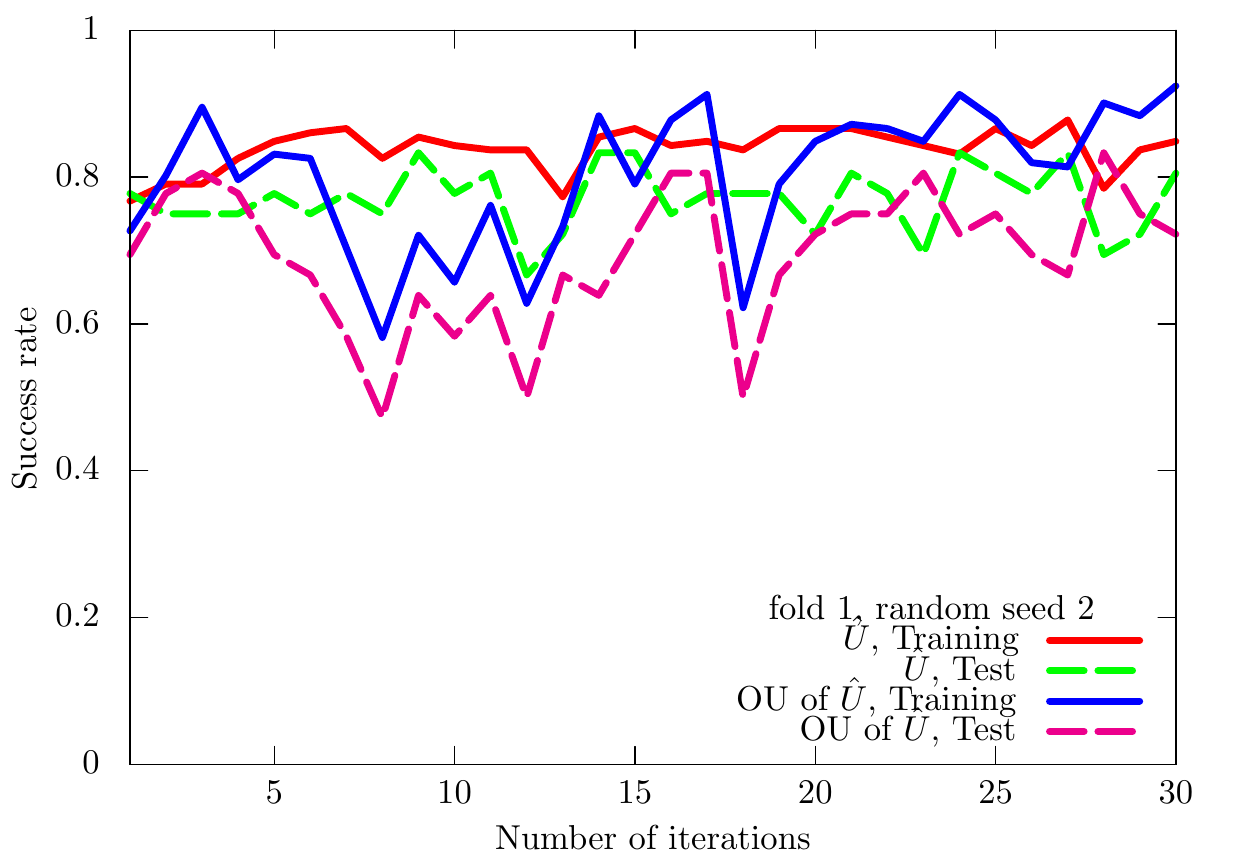}
\includegraphics[scale=0.25]{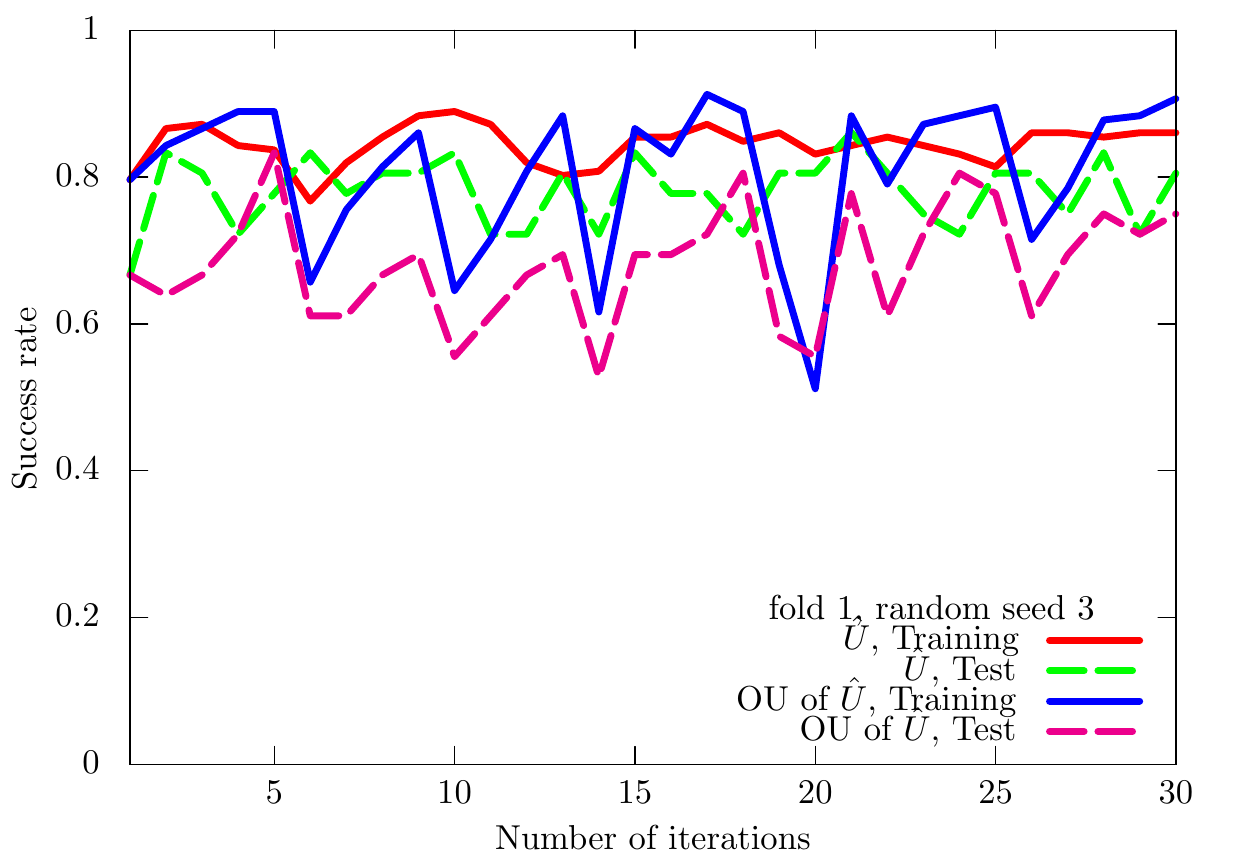}
\includegraphics[scale=0.25]{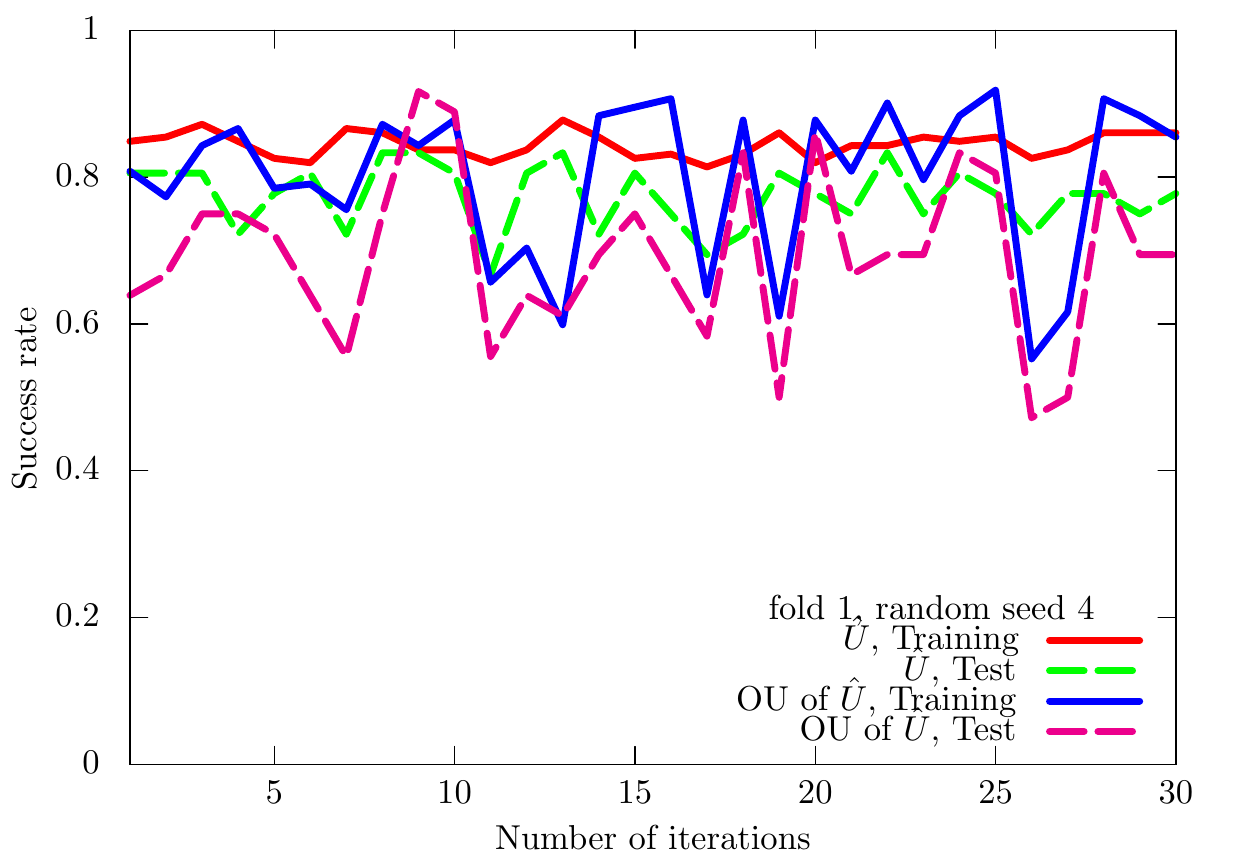}
\includegraphics[scale=0.25]{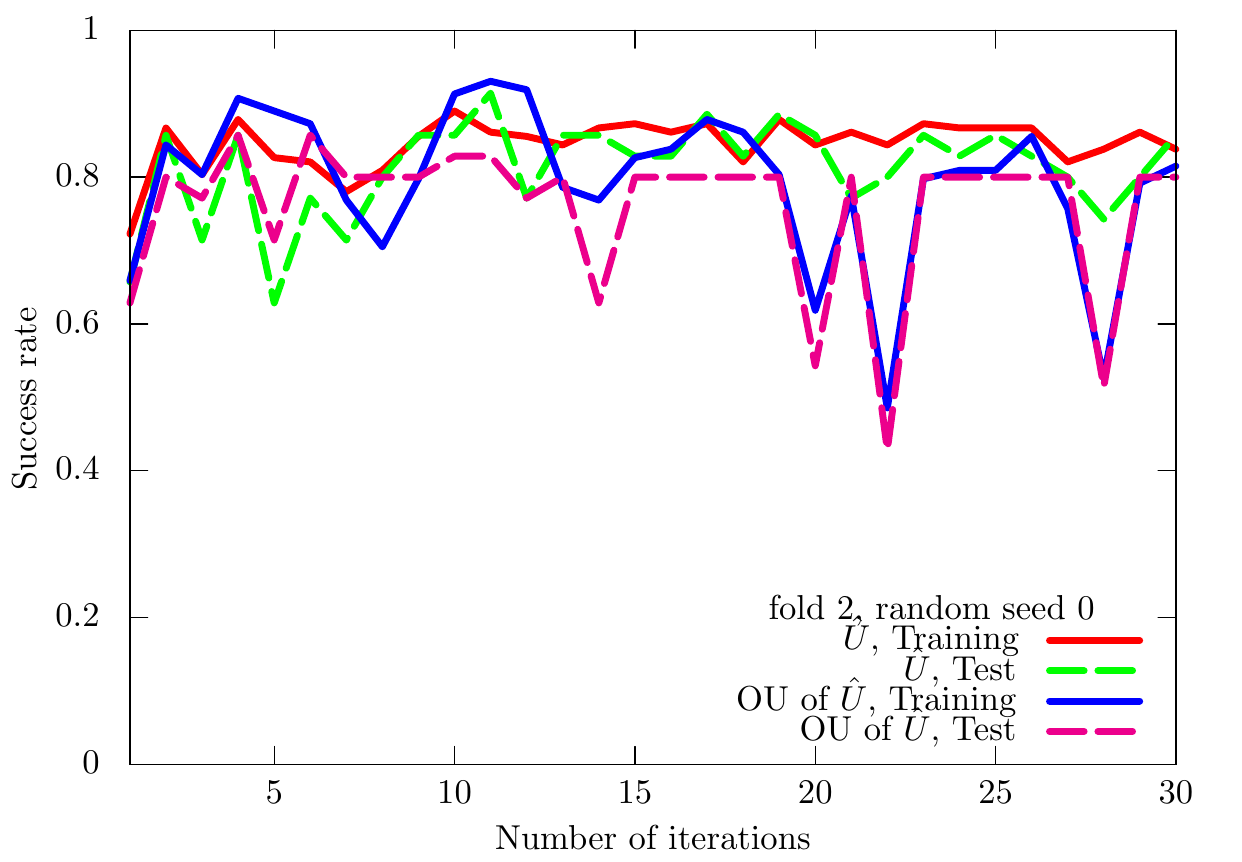}
\includegraphics[scale=0.25]{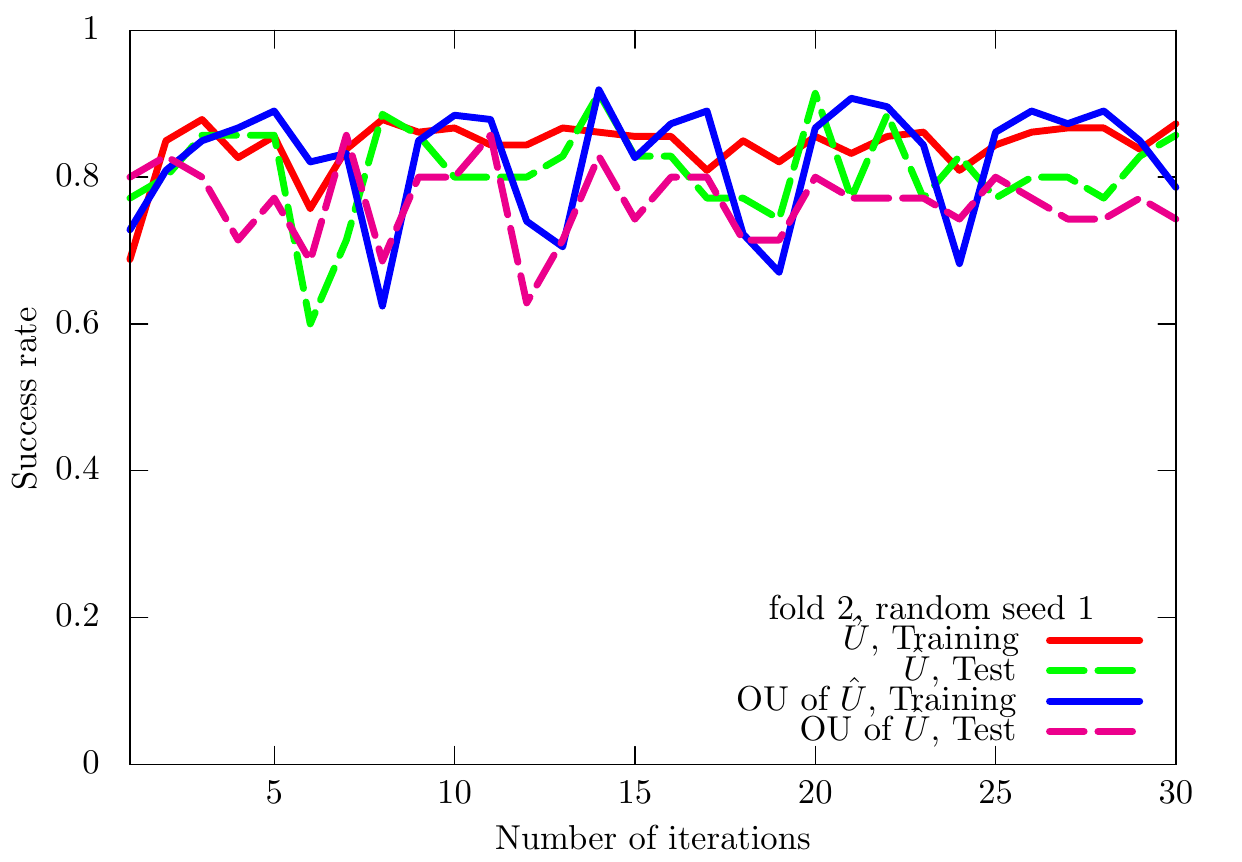}
\includegraphics[scale=0.25]{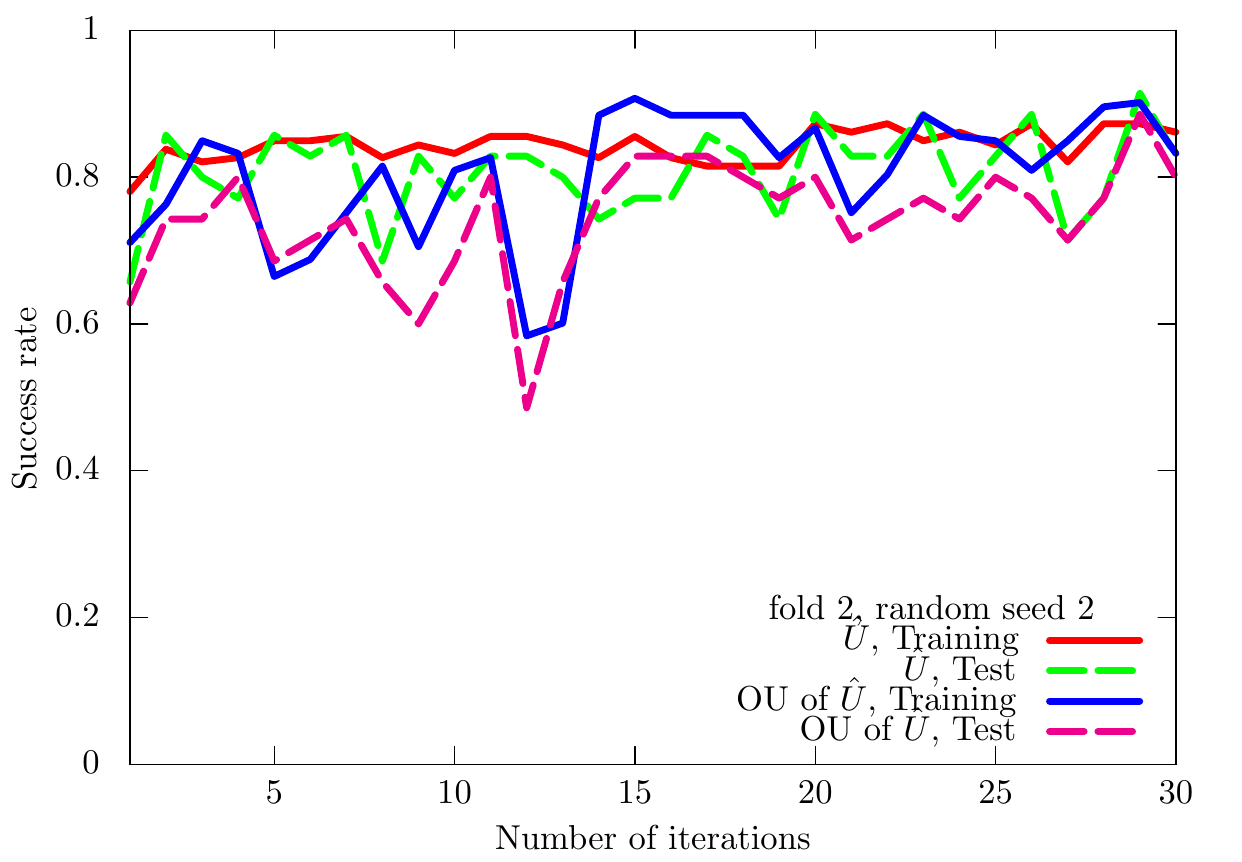}
\includegraphics[scale=0.25]{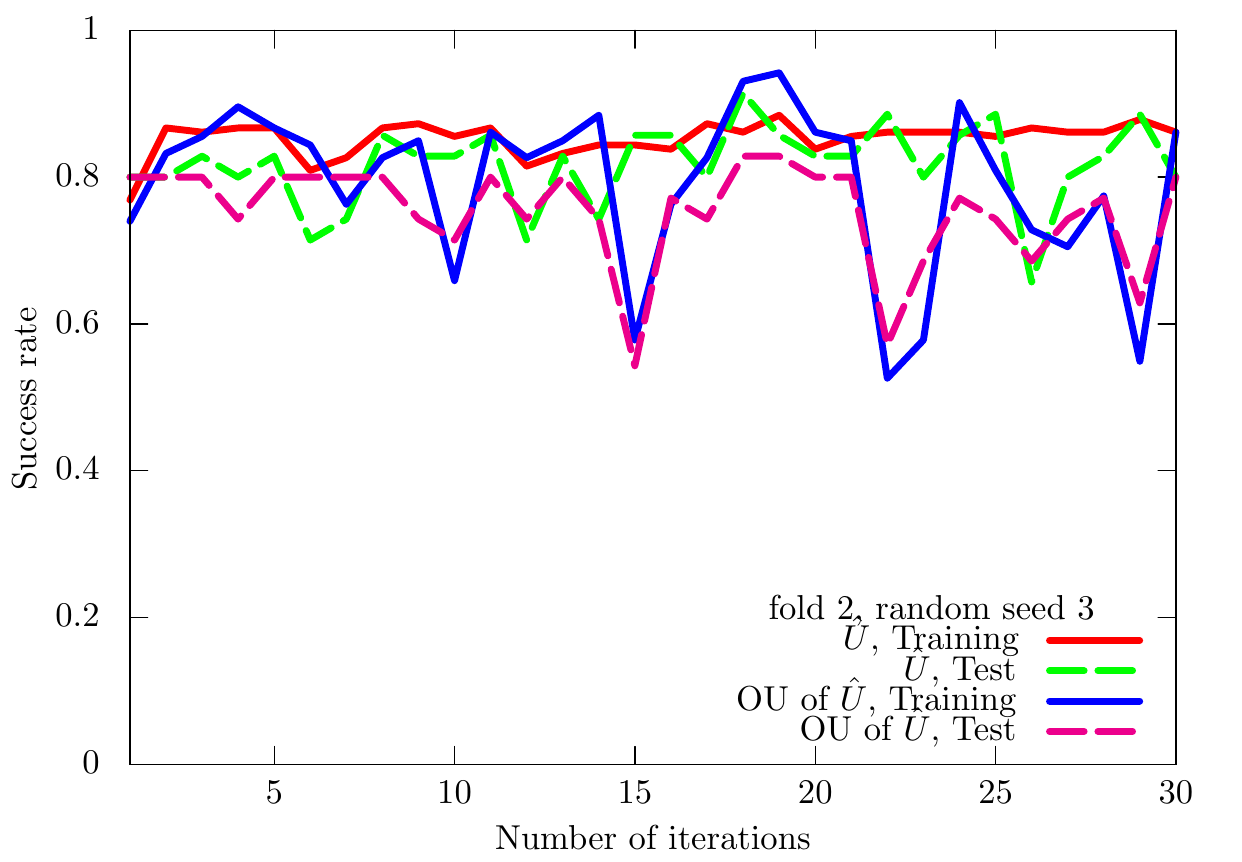}
\includegraphics[scale=0.25]{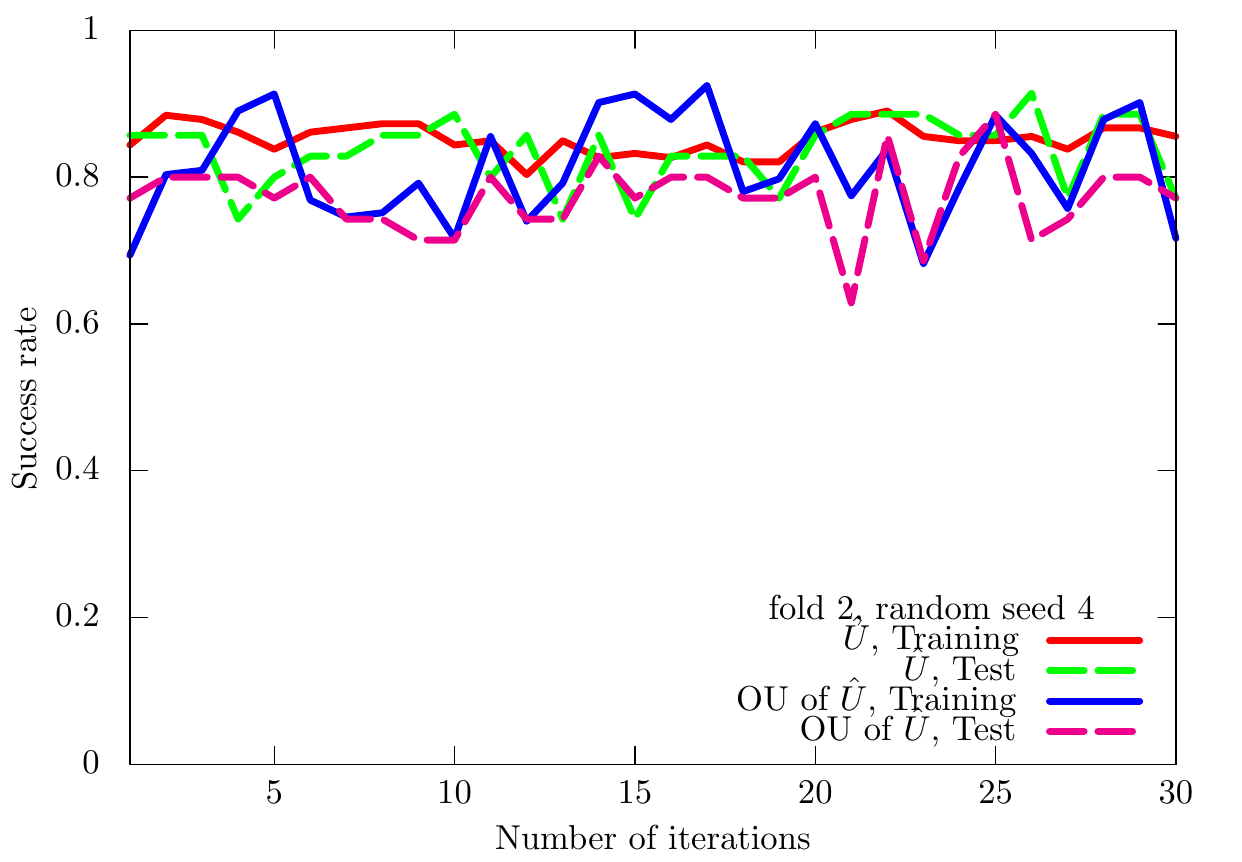}
\includegraphics[scale=0.25]{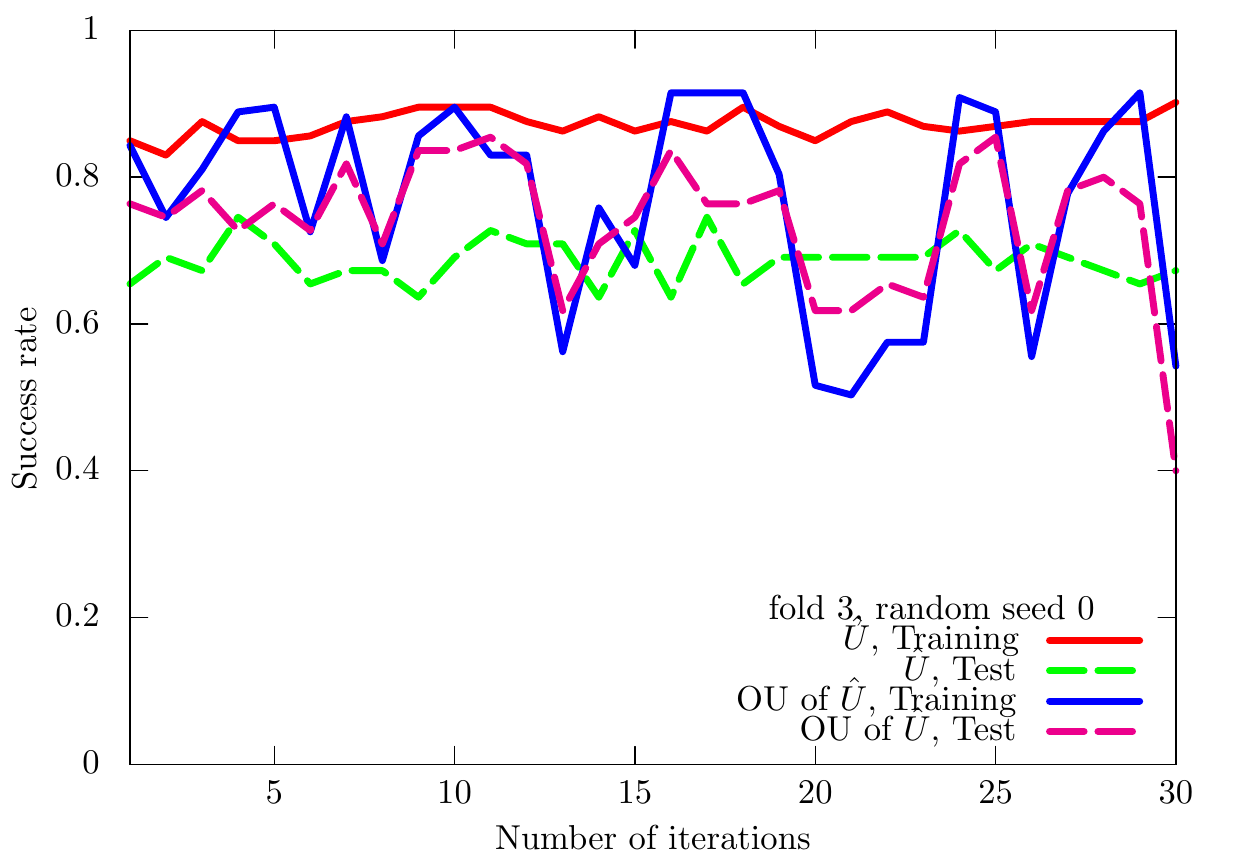}
\includegraphics[scale=0.25]{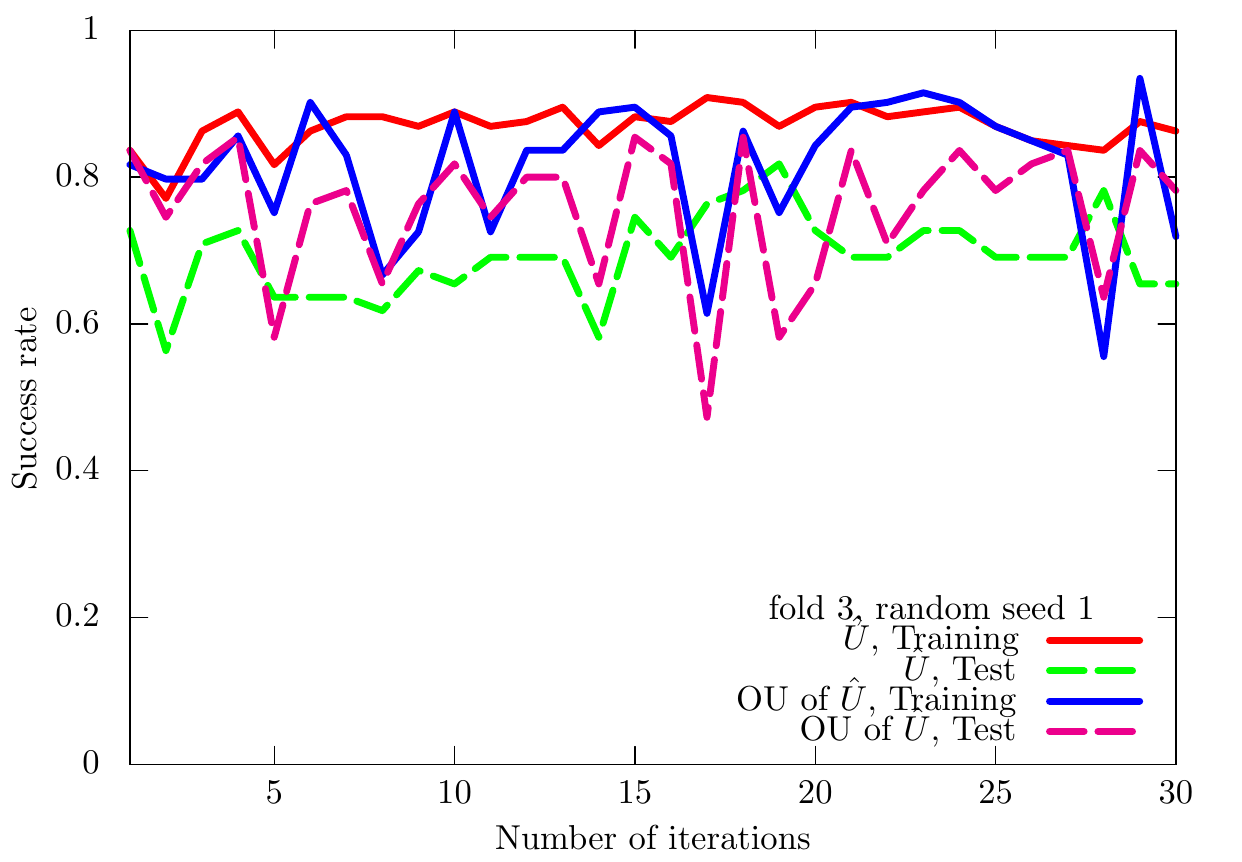}
\includegraphics[scale=0.25]{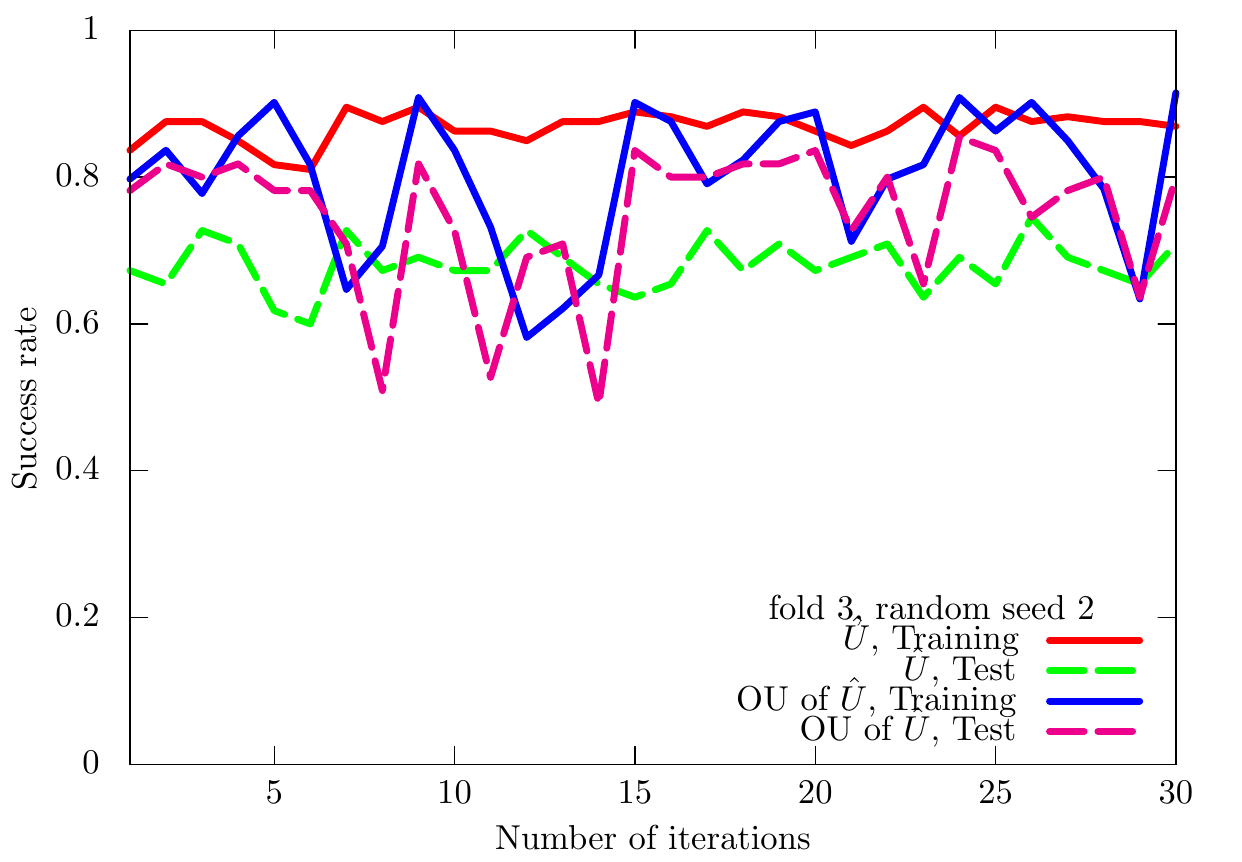}
\includegraphics[scale=0.25]{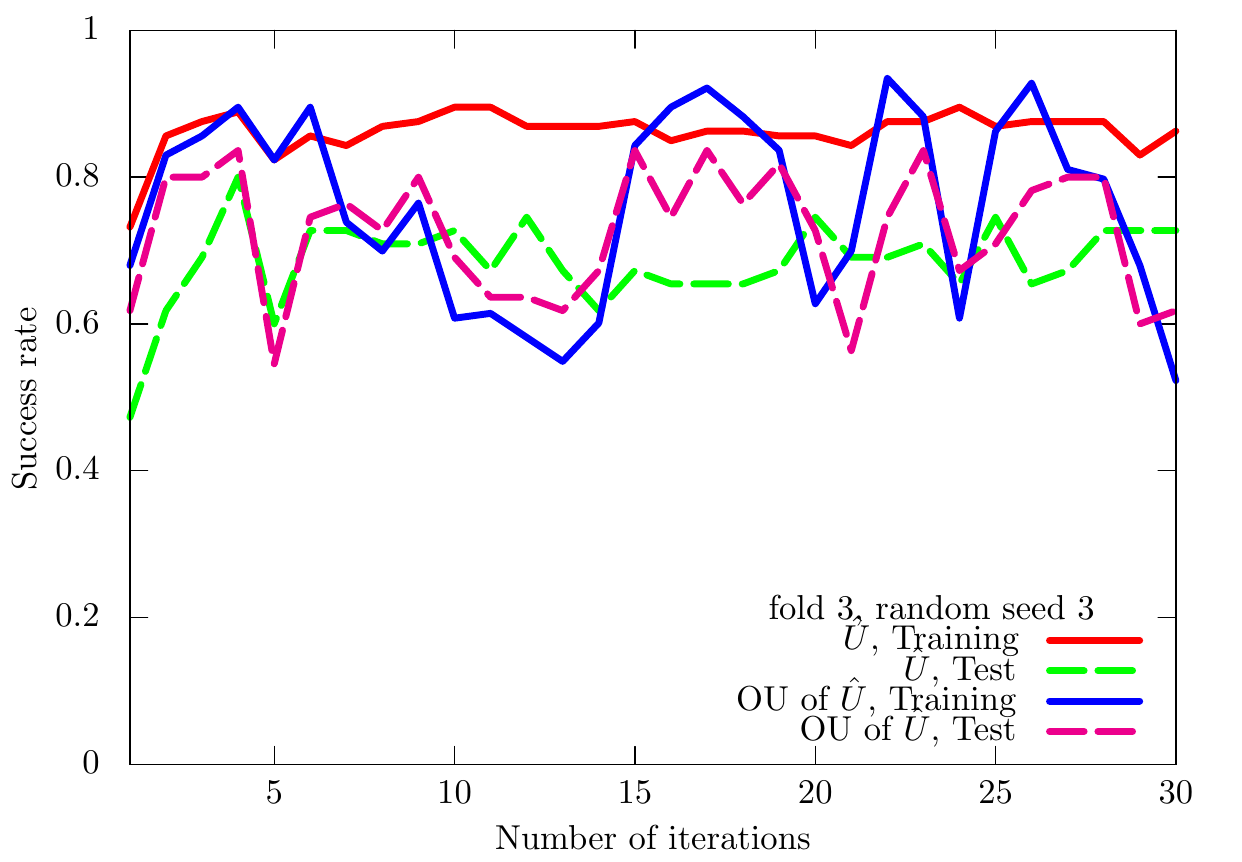}
\includegraphics[scale=0.25]{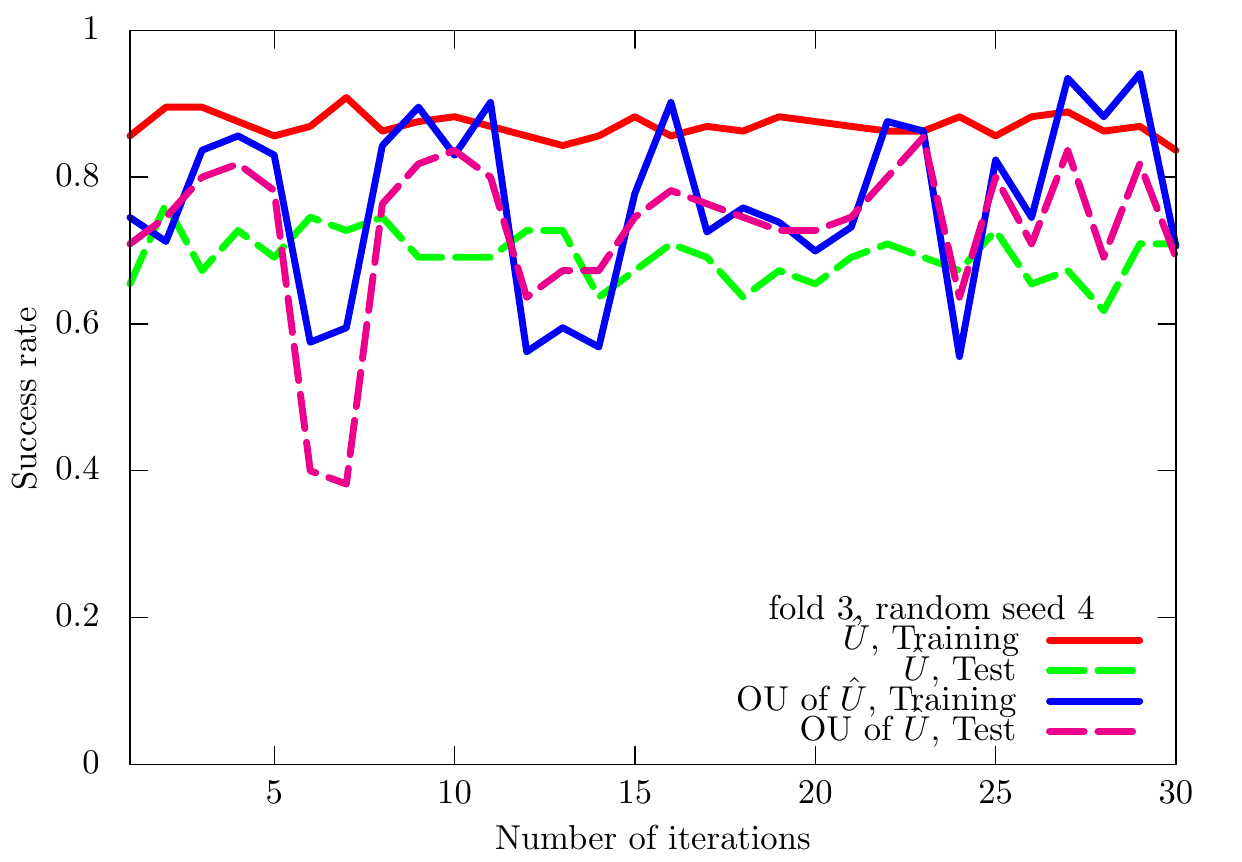}
\includegraphics[scale=0.25]{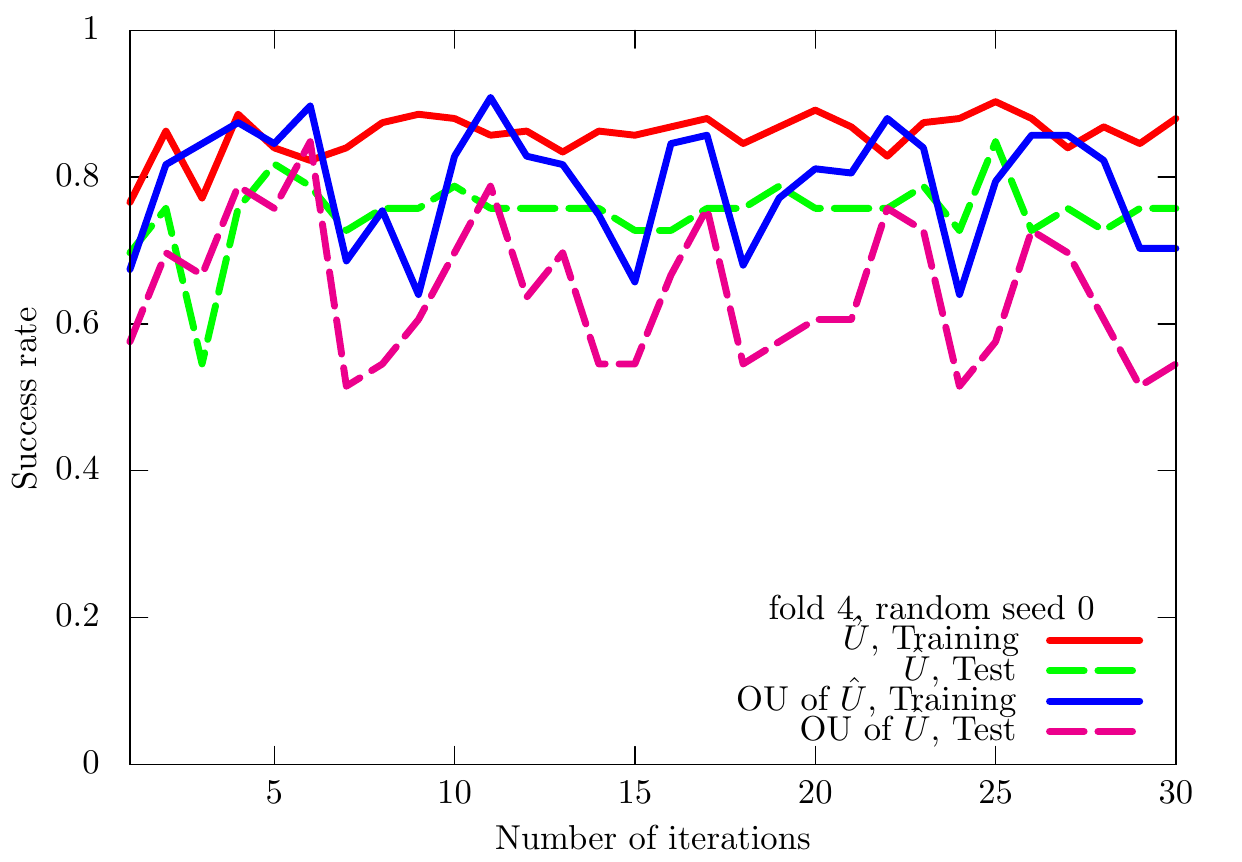}
\includegraphics[scale=0.25]{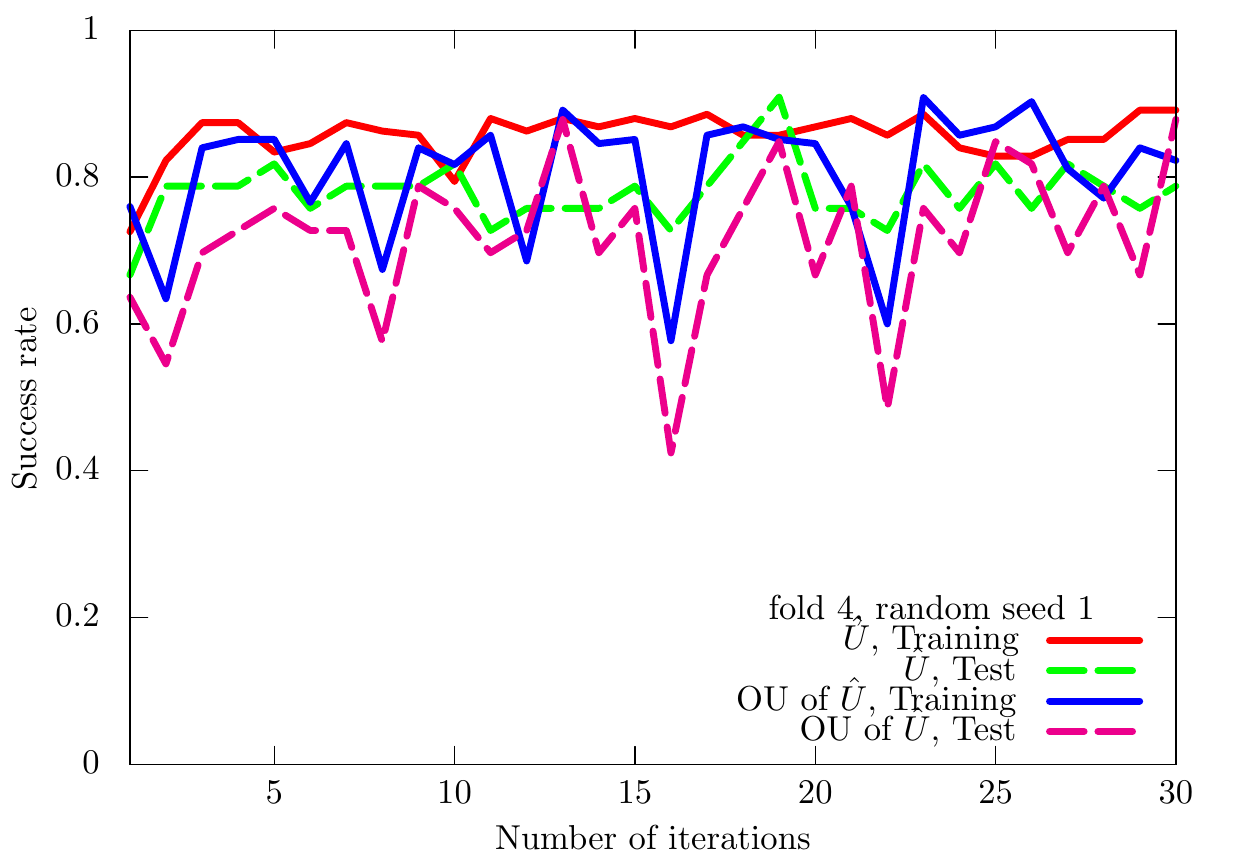}
\includegraphics[scale=0.25]{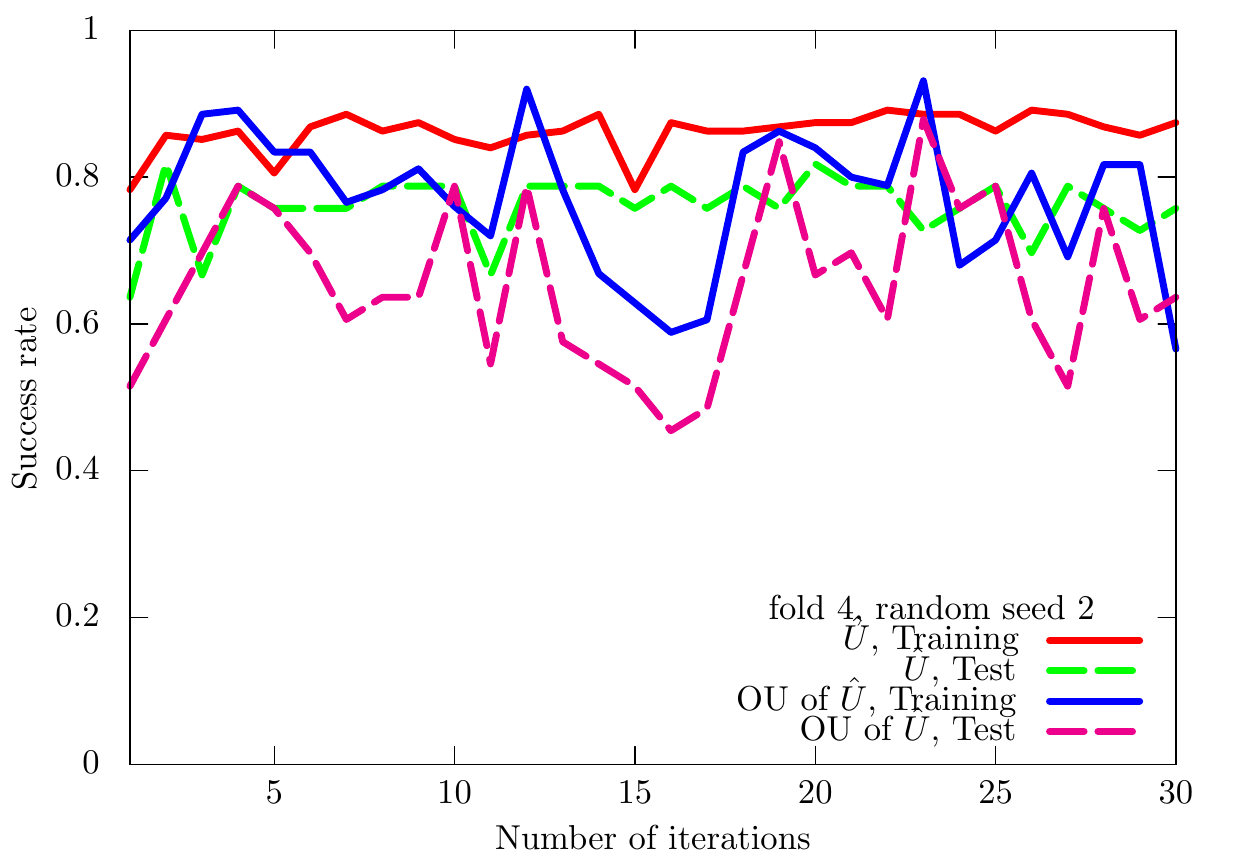}
\includegraphics[scale=0.25]{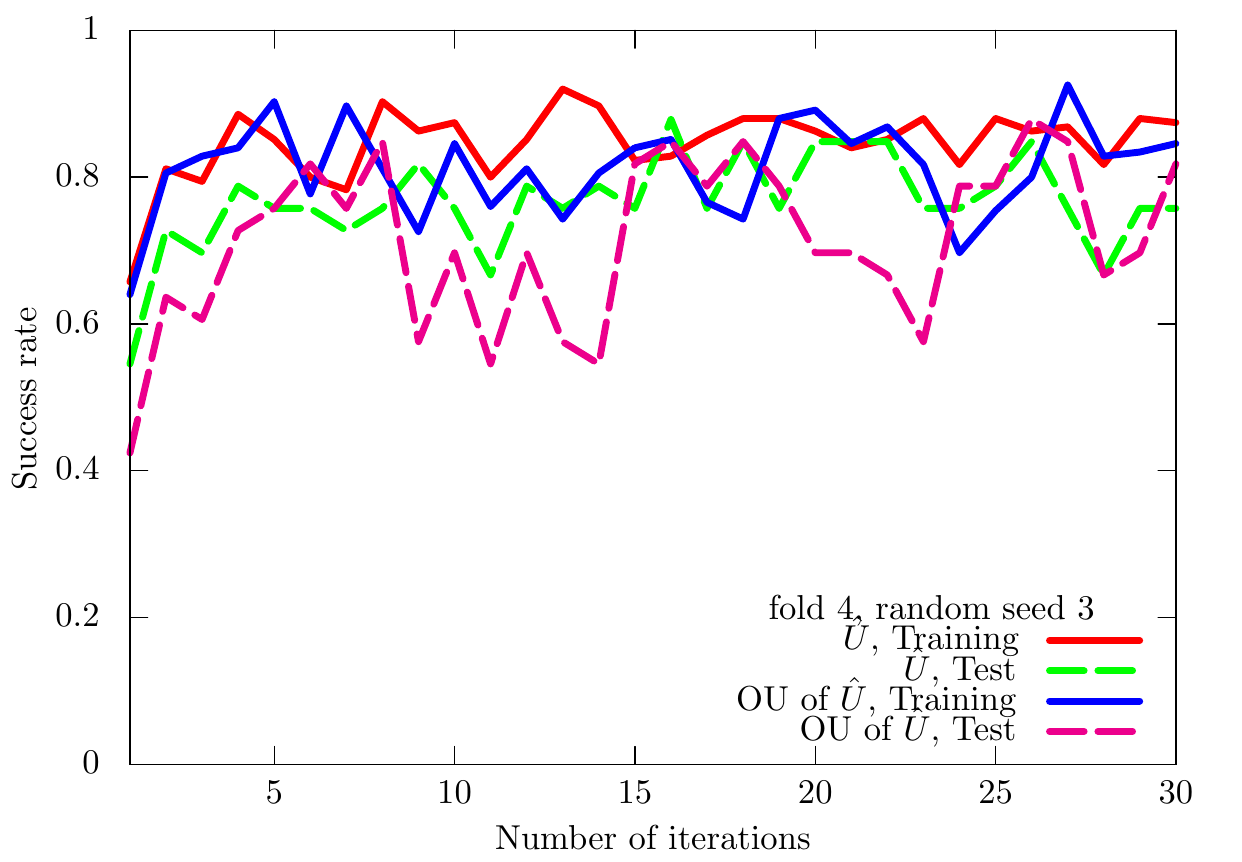}
\includegraphics[scale=0.25]{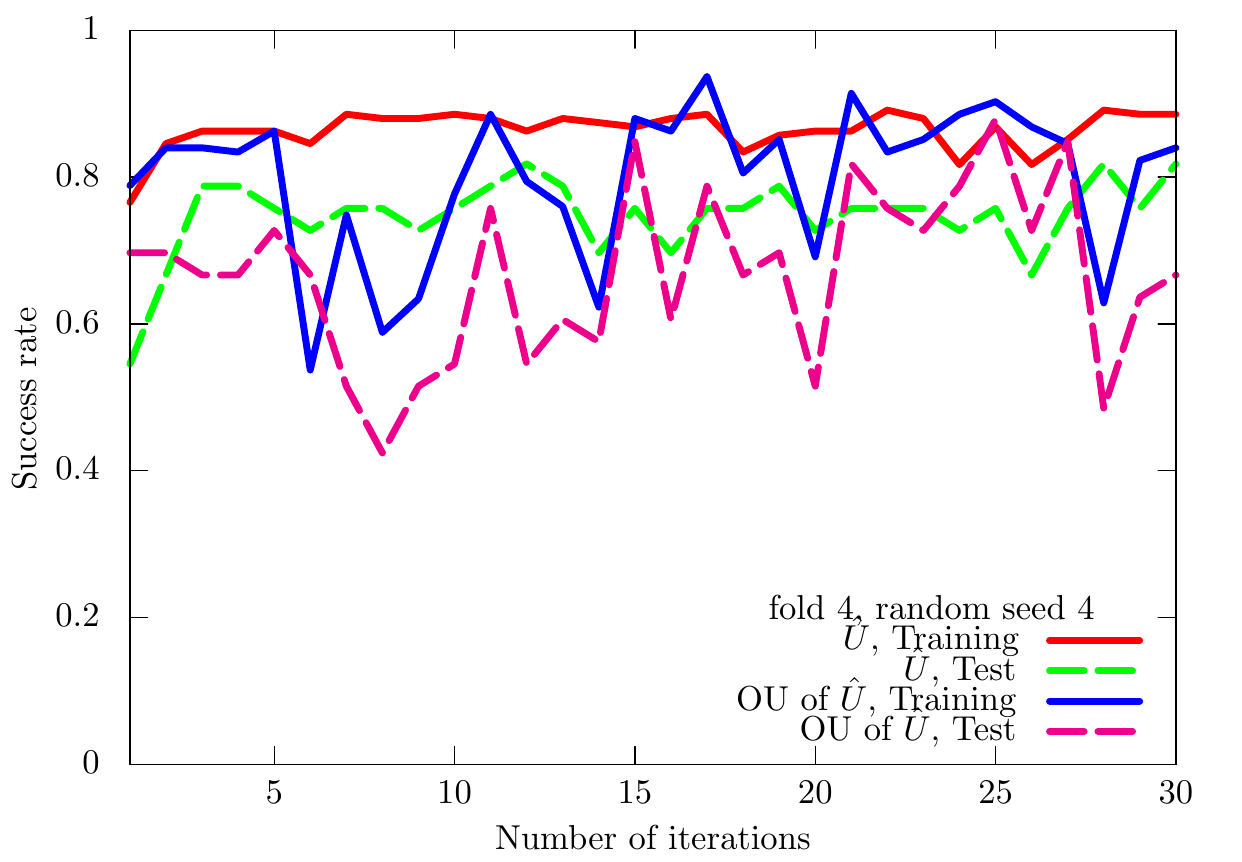}
\caption{Results of the UKM ($\hat{X}$ and OU of $\hat{X}$) on the $5$-fold datasets with $5$ different random seeds for the sonar dataset ($0$ or $1$). We use complex matrices and set $\theta_\mathrm{bias} = 0$. We set $r = 0.010$.}
\label{supp-arXiv-numerical-result-raw-data-fold-001-rand-001-UKM-OUU-UCI-sonar-0-1}
\end{figure*}

In Fig.~\ref{supp-arXiv-numerical-result-raw-data-fold-001-rand-001-QCL-UCI-sonar-0-1}, we show the numerical results of QCL for the $5$-fold datasets with $5$ different random seeds.
In Fig.~\ref{supp-arXiv-numerical-result-raw-data-fold-001-rand-001-UKM-P-UCI-sonar-0-1}, we show the numerical results of $\hat{P}$ of the UKM for the $5$-fold datasets with $5$ different random seeds.
In Fig.~\ref{supp-arXiv-numerical-result-raw-data-fold-001-rand-001-UKM-OUU-UCI-sonar-0-1}, we also show the numerical results of OU of $\hat{X}$ of the UKM for the $5$-fold datasets with $5$ different random seeds.

\begin{figure}[htb]
\centering
\includegraphics[scale=0.45]{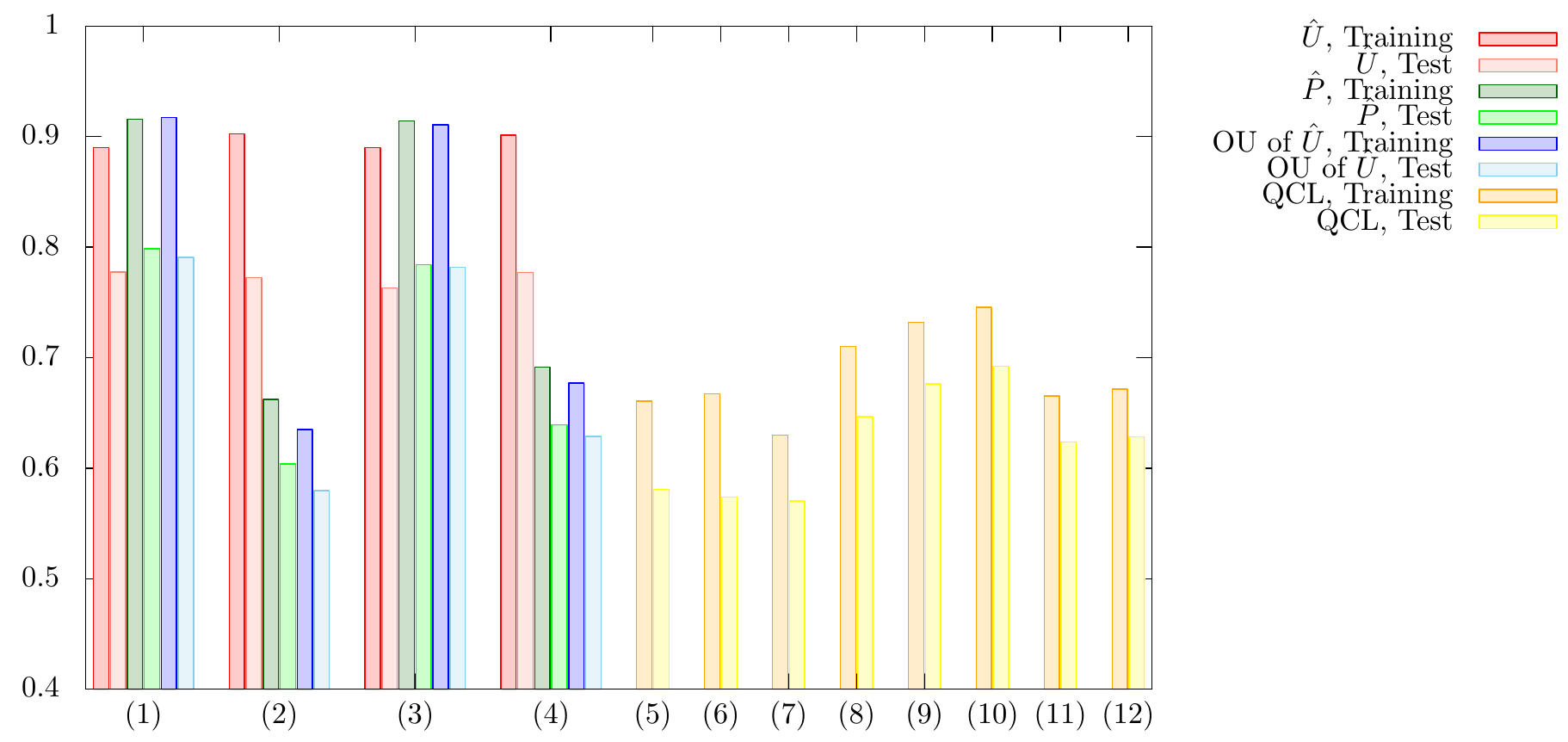}
\caption{Results of $5$-fold CV with $5$ different random seeds for the sonar dataset ($0$ or $1$). For the UKM, we put $r = 0.010$ and set $K = 30$ and $K' = 10$. For QCL, the number of layers $L$ is $5$ and the number of iterations is $300$. The numerical settings are as follows: (1) complex matrices without the bias term, (2) complex matrices with the bias term, (3) real matrices without the bias term, (4) real matrices with the bias term, (5) CNOT-based circuit without the bias term, (6) CNOT-based circuit with the bias term, (7) CRot-based circuit without the bias term, (8) CRot-based circuit with the bias term, (9) 1d Heisenberg circuit without the bias term, (10) 1d Heisenberg circuit with the bias term, (11) FC Heisenberg circuit without the bias term, and (12) FC Heisenberg circuit with the bias term.}
\label{supp-arXiv-numerical-result-performance-UKM-QCL-UCI-sonar-0-1}
\end{figure}

We summarize the results of 5-fold CV with 5 different random seeds of QCL and the UKM in Tables~\ref{supp-arXiv-table-UCI-sonar-0-1-002} and \ref{supp-arXiv-table-UCI-sonar-0-1-001}, respectively.
For QCL and the UKM, we select the best model for the training dataset over iterations to compute the performance.
\begin{table}[htb]
  \begin{tabular}{cc|cc}
    \hline \hline
    Algo. & Condition & Training & Test \\
    \hline
  QCL & CNOT-based, w/o bias & 0.6606 & 0.5810 \\
  QCL & CNOT-based, w/ bias & 0.6671 & 0.5739 \\
    \hline
  QCL & CRot-based, w/o bias & 0.6299 & 0.5703 \\
  QCL & CRot-based, w/ bias & 0.7102 & 0.6468 \\
    \hline
  QCL & 1d Heisenberg, w/o bias & 0.7320 & 0.6761 \\
  QCL & 1d Heisenberg, w/ bias & 0.7455 & 0.6924 \\
    \hline
  QCL & FC Heisenberg, w/o bias & 0.6653 & 0.6236 \\
  QCL & FC Heisenberg, w/ bias & 0.6715 & 0.6285 \\
    \hline \hline
  \end{tabular}
\caption{Results of $5$-fold CV with $5$ different random seeds of QCL for the sonar dataset ($0$ or $1$). The number of layers $L$ is set to $5$ and the number of iterations is set to $300$.}
\label{supp-arXiv-table-UCI-sonar-0-1-002}
\end{table}
\begin{table}[htb]
  \begin{tabular}{cc|cc}
    \hline \hline
    Algo. & Condition & Training & Test \\
    \hline
  UKM & $\hat{X}$, complex, w/o bias & 0.8903 & 0.7774 \\
  UKM & $\hat{P}$, complex, w/o bias & 0.9159 & 0.7985 \\
  UKM & OU of $\hat{X}$, complex, w/o bias & 0.9175 & 0.7909 \\
    \hline
  UKM & $\hat{X}$, complex, w/ bias & 0.9027 & 0.7723 \\
  UKM & $\hat{P}$, complex, w/ bias & 0.6621 & 0.6036 \\
  UKM & OU of $\hat{X}$, complex, w/ bias & 0.6351 & 0.5795 \\
    \hline
  UKM & $\hat{X}$, real, w/o bias & 0.8899 & 0.7630 \\
  UKM & $\hat{P}$, real, w/o bias & 0.9141 & 0.7842 \\
  UKM & OU of $\hat{X}$, real, w/o bias & 0.9108 & 0.7818 \\
    \hline
  UKM & $\hat{X}$, real, w/ bias & 0.9014 & 0.7772 \\
  UKM & $\hat{P}$, real, w/ bias & 0.6913 & 0.6391 \\
  UKM & OU of $\hat{X}$, real, w/ bias & 0.6769 & 0.6288 \\
    \hline \hline
  \end{tabular}
\caption{Results of $5$-fold CV with $5$ different random seeds of the UKM for the sonar dataset ($0$ or $1$). We put $r = 0.010$ and set $K = 30$ and $K' = 10$.}
\label{supp-arXiv-table-UCI-sonar-0-1-001}
\end{table}
In Fig.~\ref{supp-arXiv-numerical-result-performance-UKM-QCL-UCI-sonar-0-1}, we plot the data shown in Tables~\ref{supp-arXiv-table-UCI-sonar-0-1-002} and \ref{supp-arXiv-table-UCI-sonar-0-1-001}.
We also summarize the results of 5-fold CV with 5 different random seeds of the kernel method in Table~\ref{supp-arXiv-table-UCI-sonar-0-1-003}.
More specifically, we use Ridge classification in Sec.~\ref{supp-arXiv-sec-Ridge-001}.
We consider the linear functions and the second-order polynomial functions for $\phi (\cdot)$ in Eq.~\eqref{supp-arXiv-f-pred-kernel-method-001-002} with and without normalization.
We set $\lambda = 10^{-2}, 10^{-1}, 1$ where $\lambda$ is the coefficient of the regularization term.
\begin{table}[htb]
  \begin{tabular}{cc|cc}
    \hline \hline
    Algo. & Condition & Training & Test \\
    \hline
  Kernel method & Linear, w/o normalization, $\lambda = 10^{-2}$ & 0.9109 & 0.7574 \\
  Kernel method & Linear, w/o normalization, $\lambda = 10^{-1}$ & 0.8738 & 0.7717 \\
  Kernel method & Linear, w/o normalization, $\lambda = 1$ & 0.8667 & 0.7631 \\
    \hline
  Kernel method & Linear, w/ normalization, $\lambda = 10^{-2}$ & 0.8806 & 0.7694 \\
  Kernel method & Linear, w/ normalization, $\lambda = 10^{-1}$ & 0.8522 & 0.7500 \\
  Kernel method & Linear, w/ normalization, $\lambda = 1$ & 0.8093 & 0.7076 \\
    \hline
  Kernel method & Poly-2, w/o normalization, $\lambda = 10^{-2}$ & 1.0000 & 0.8198 \\
  Kernel method & Poly-2, w/o normalization, $\lambda = 10^{-1}$ & 1.0000 & 0.8084 \\
  Kernel method & Poly-2, w/o normalization, $\lambda = 1$ & 0.9865 & 0.8035 \\
    \hline
  Kernel method & Poly-2, w/ normalization, $\lambda = 10^{-2}$ & 0.9901 & 0.7973 \\
  Kernel method & Poly-2, w/ normalization, $\lambda = 10^{-1}$ & 0.9297 & 0.8009 \\
  Kernel method & Poly-2, w/ normalization, $\lambda = 1$ & 0.8348 & 0.7374 \\
    \hline \hline
  \end{tabular}
\caption{Results of 5-fold CV with 5 different random seeds of the kernel method for the sonar dataset ($0$ or $1$).}
\label{supp-arXiv-table-UCI-sonar-0-1-003}
\end{table}

\begin{figure}[htb]
\centering
\includegraphics[scale=0.45]{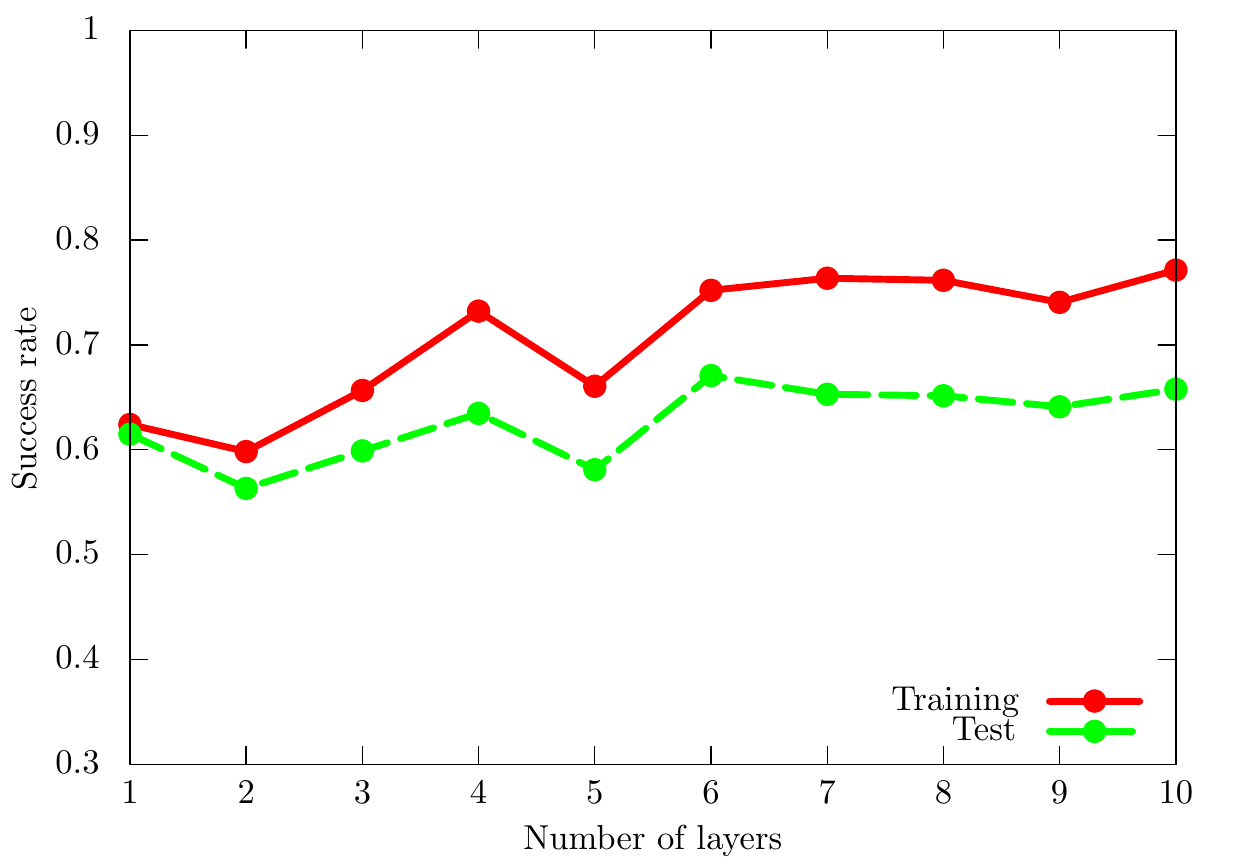}
\caption{Performance dependence of QCL on the number of layers $L$ for the sonar dataset ($0$ or $1$). We use the CNOT-based circuit geometry and set $\theta_\mathrm{bias} = 0$. We iterate the computation $300$ times.}
\label{supp-arXiv-numerical-result-layers-dependence-QCL-UCI-sonar-0-1}
\end{figure}

\begin{figure}[htb]
\centering
\includegraphics[scale=0.45]{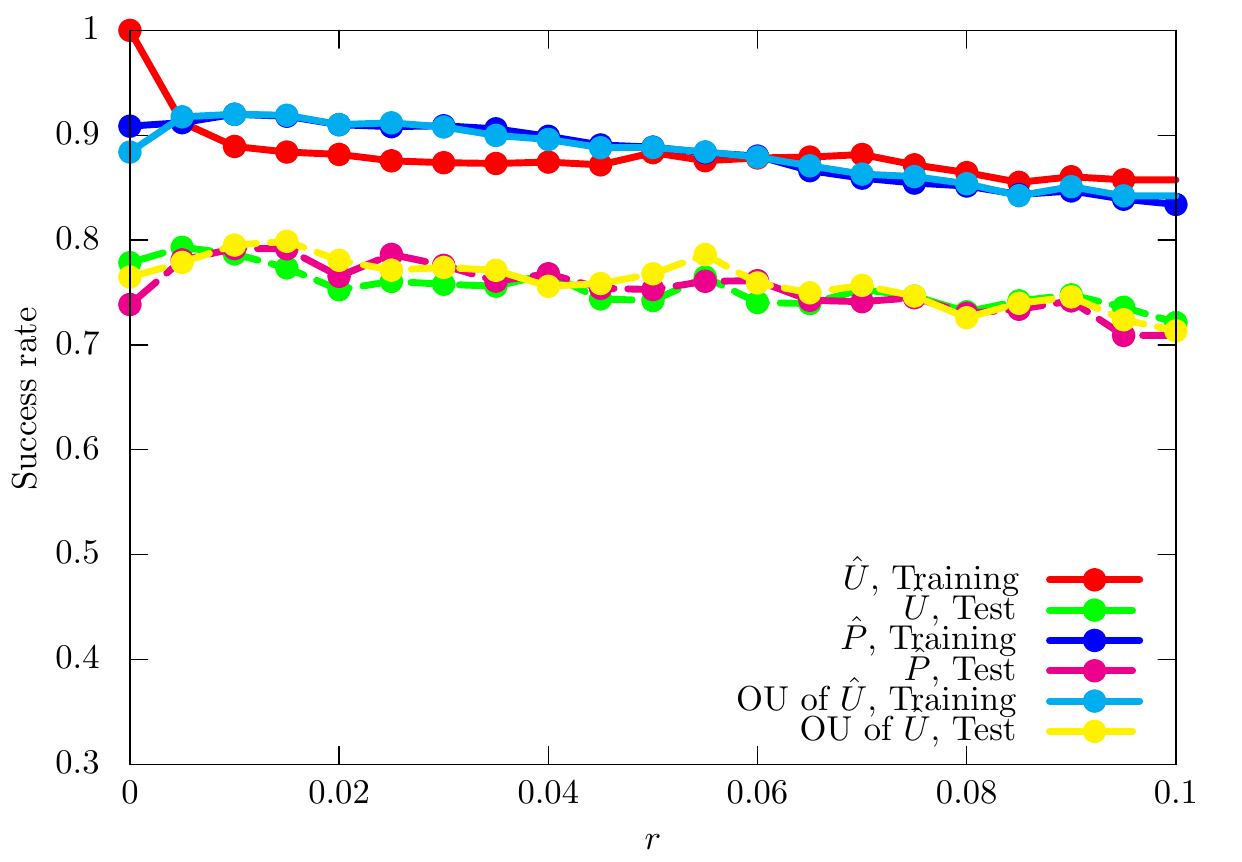}
\includegraphics[scale=0.45]{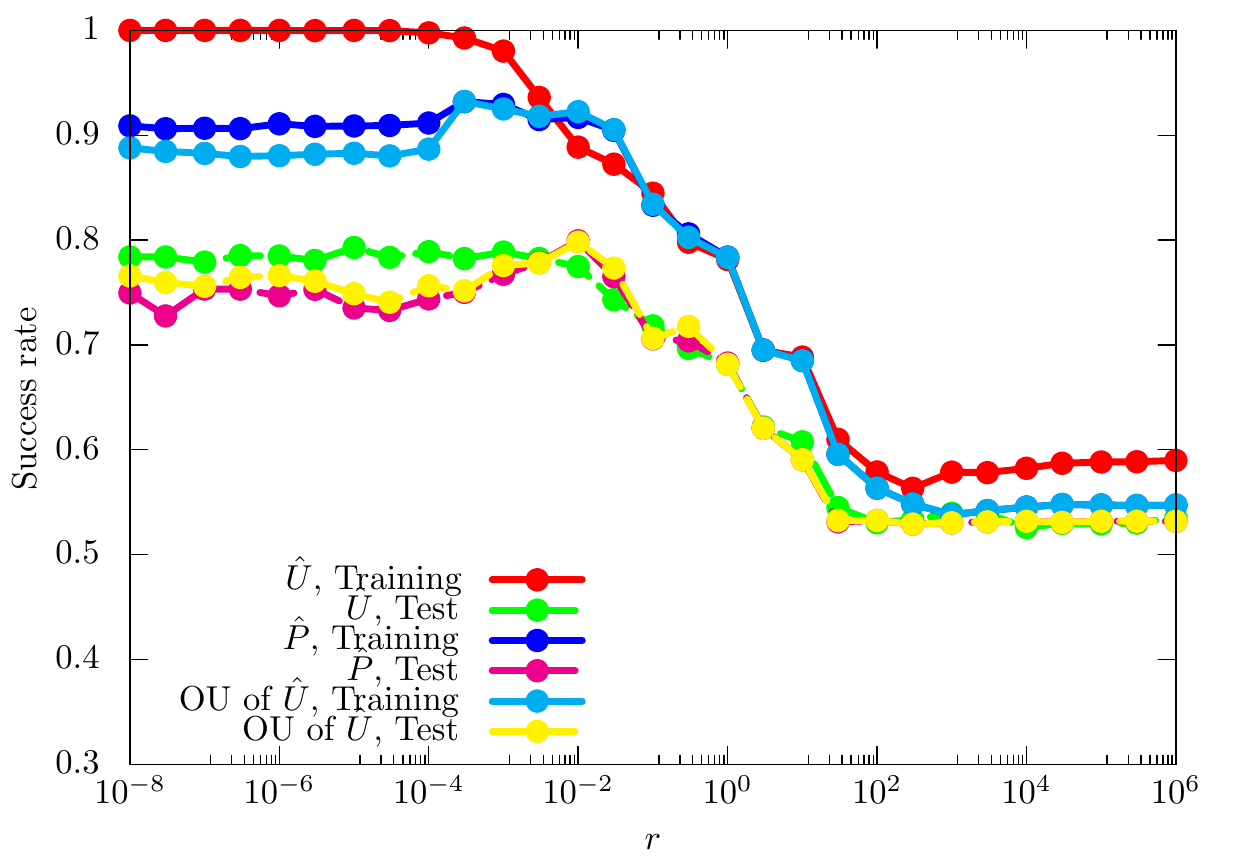}
\caption{Performance dependence of the UKM on $r$, which is the coefficient of the second term in the right-hand side of Eq.~\eqref{supp-arXiv-quantum-kernel-method-001-011} for the sonar dataset ($0$ or $1$). We use complex matrices and set $\theta_\mathrm{bias} = 0$. We set $K = 30$ and $K' = 10$.}
\label{supp-arXiv-numerical-result-r-dependence-UKM-UCI-sonar-0-1}
\end{figure}

\begin{figure}[htb]
\centering
\includegraphics[scale=0.45]{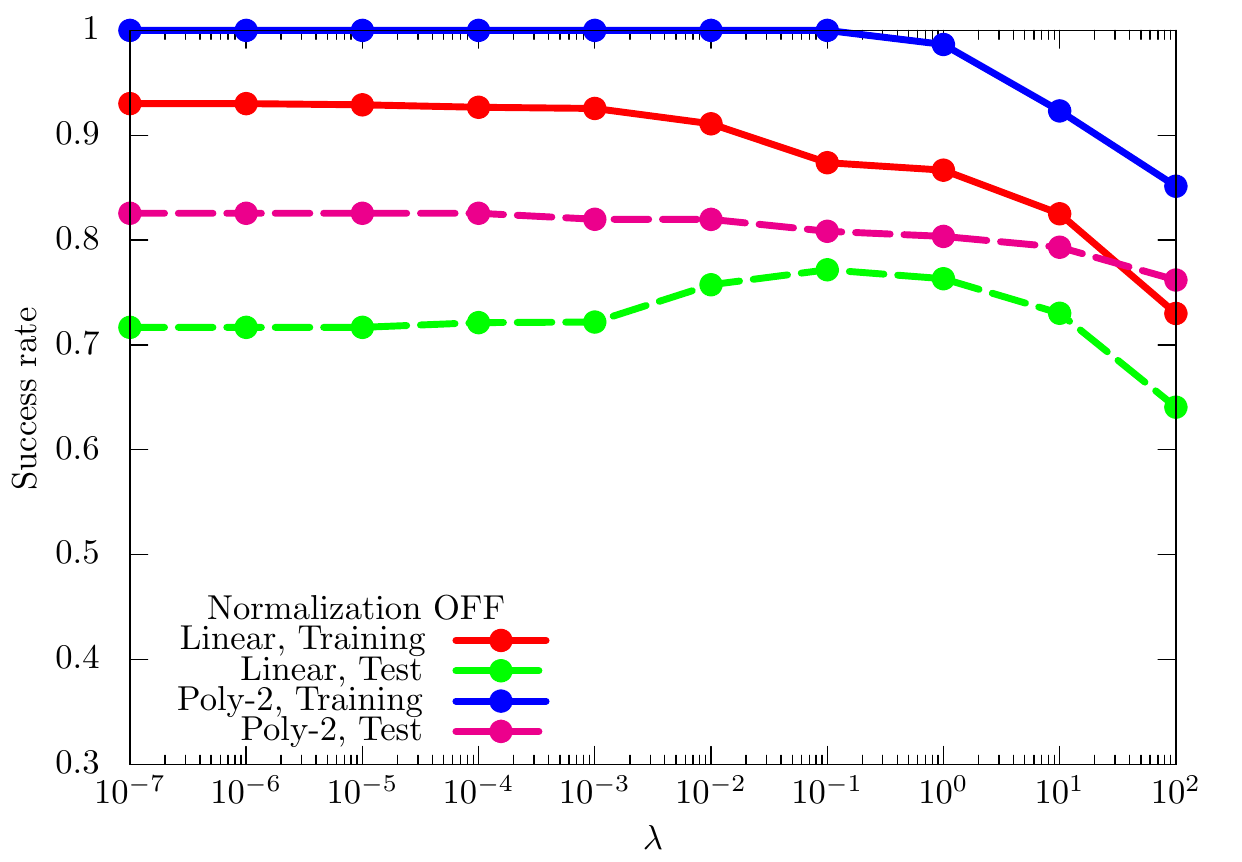}
\includegraphics[scale=0.45]{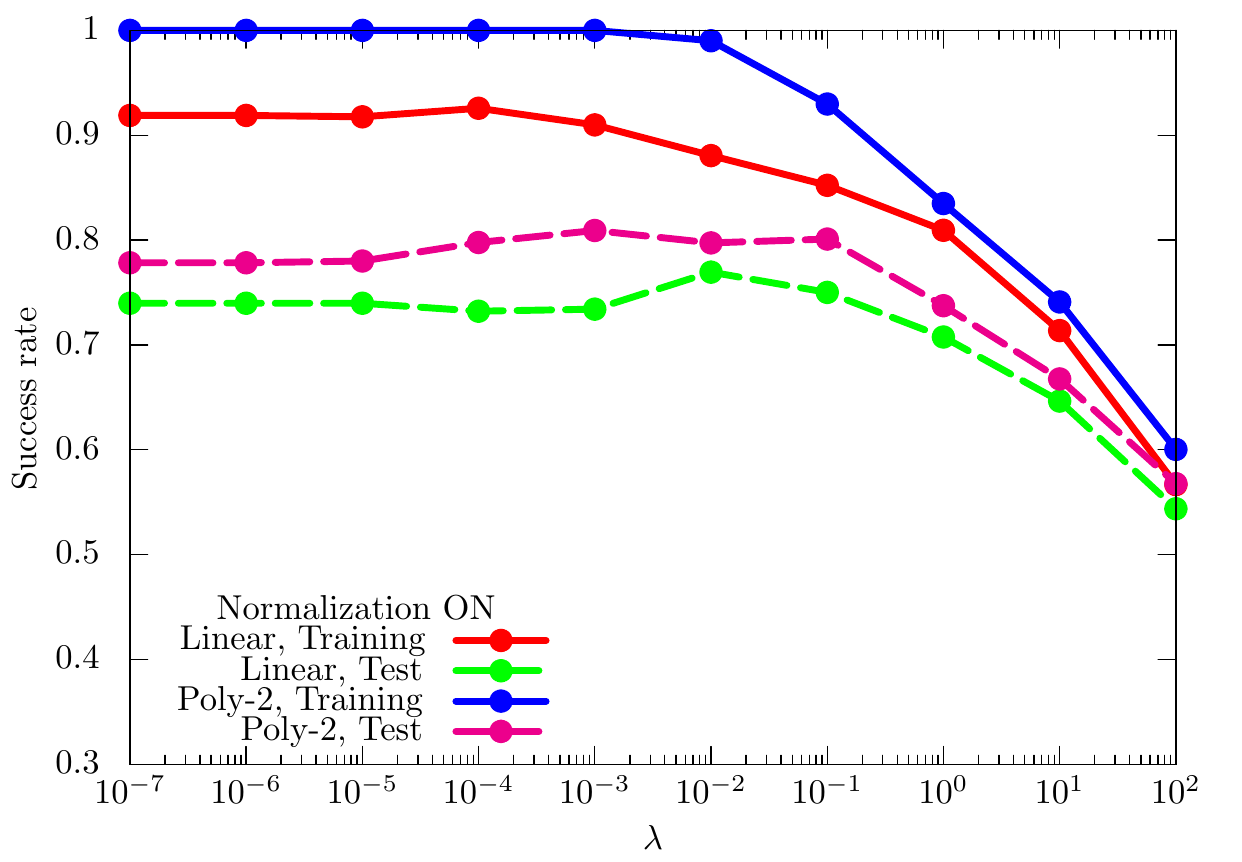}
\caption{Performance dependence of the kernel method on $\lambda$, which is the coefficient of the second term in the right-hand side of Eq.~\eqref{supp-arXiv-cost-function-kernel-method-001-002} for the sonar dataset ($0$ or $1$). For $\phi (\cdot)$ in Eq.~\eqref{supp-arXiv-f-pred-kernel-method-001-002}, we use the linear functions and the second-degree polynomial functions with and without normalization.}
\label{supp-arXiv-numerical-result-lambda-dependence-kernel-method-sonar-0-1}
\end{figure}

Next, we show the performance dependence of the three algorithms on their key parameters.
We see the performance dependence of QCL on the number of layers $L$.
The result is shown in Fig.~\ref{supp-arXiv-numerical-result-layers-dependence-QCL-UCI-sonar-0-1}.
We then see the performance dependence of the UKM on $r$, which is the coefficient of the second term in the right-hand side of Eq.~\eqref{supp-arXiv-quantum-kernel-method-001-011}.
The result is shown in Fig.~\ref{supp-arXiv-numerical-result-r-dependence-UKM-UCI-sonar-0-1}.
In Fig.~\ref{supp-arXiv-numerical-result-lambda-dependence-kernel-method-sonar-0-1}, we show the performance dependence of the kernel method on $\lambda$, which is the coefficient of the second term in the right-hand side of Eq.~\eqref{supp-arXiv-cost-function-kernel-method-001-002}.

\begin{figure}[htb]
\centering
\includegraphics[scale=0.45]{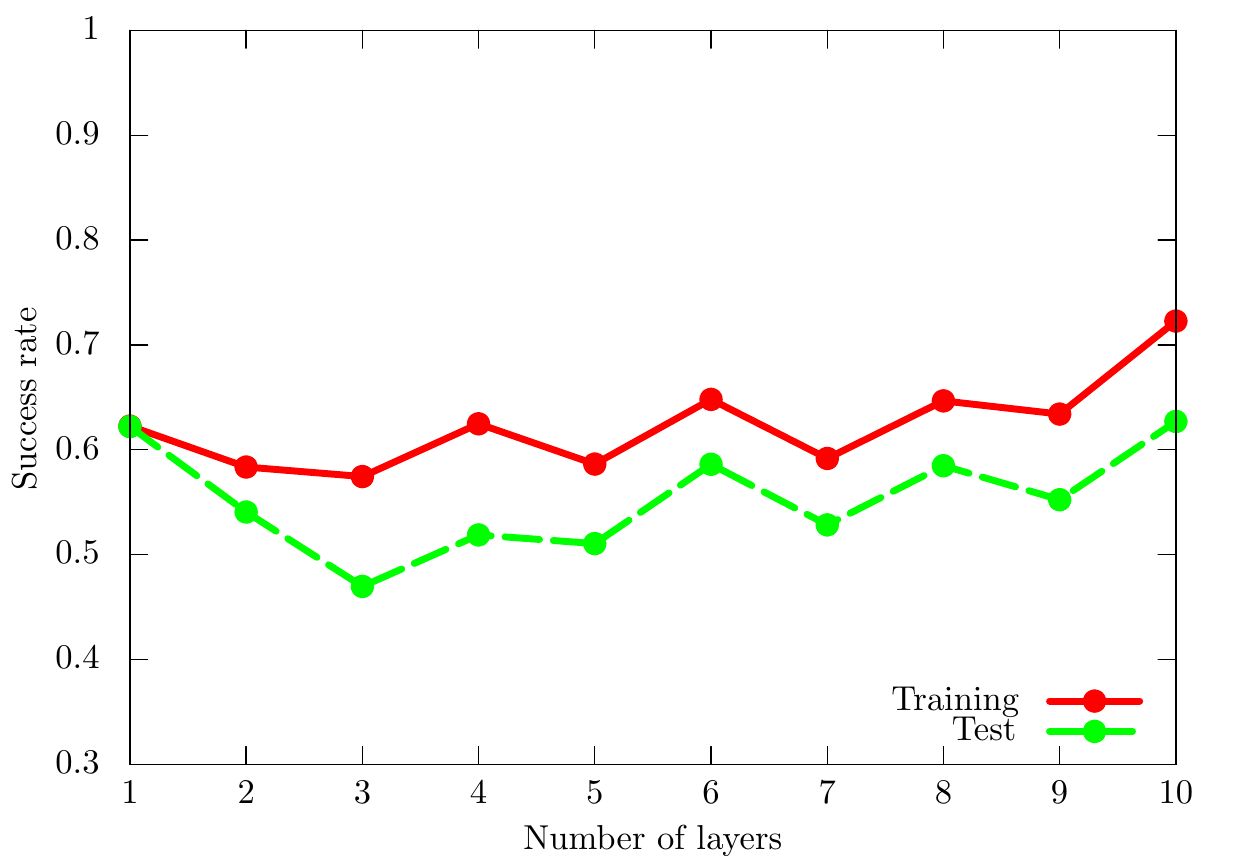}
\caption{Performance dependence of QCL on the number of layers $L$ for the sonar dataset ($0$ or $1$) in the case of the hinge function, Eq.~\eqref{supp-arXiv-hinge-function-001-001}. We use the CNOT-based circuit geometry and set $\theta_\mathrm{bias} = 0$. We iterate the computation $300$ times.}
\label{supp-arXiv-numerical-result-layers-dependence-QCL-UCI-sonar-0-1-hinge}
\end{figure}

\begin{figure}[htb]
\centering
\includegraphics[scale=0.45]{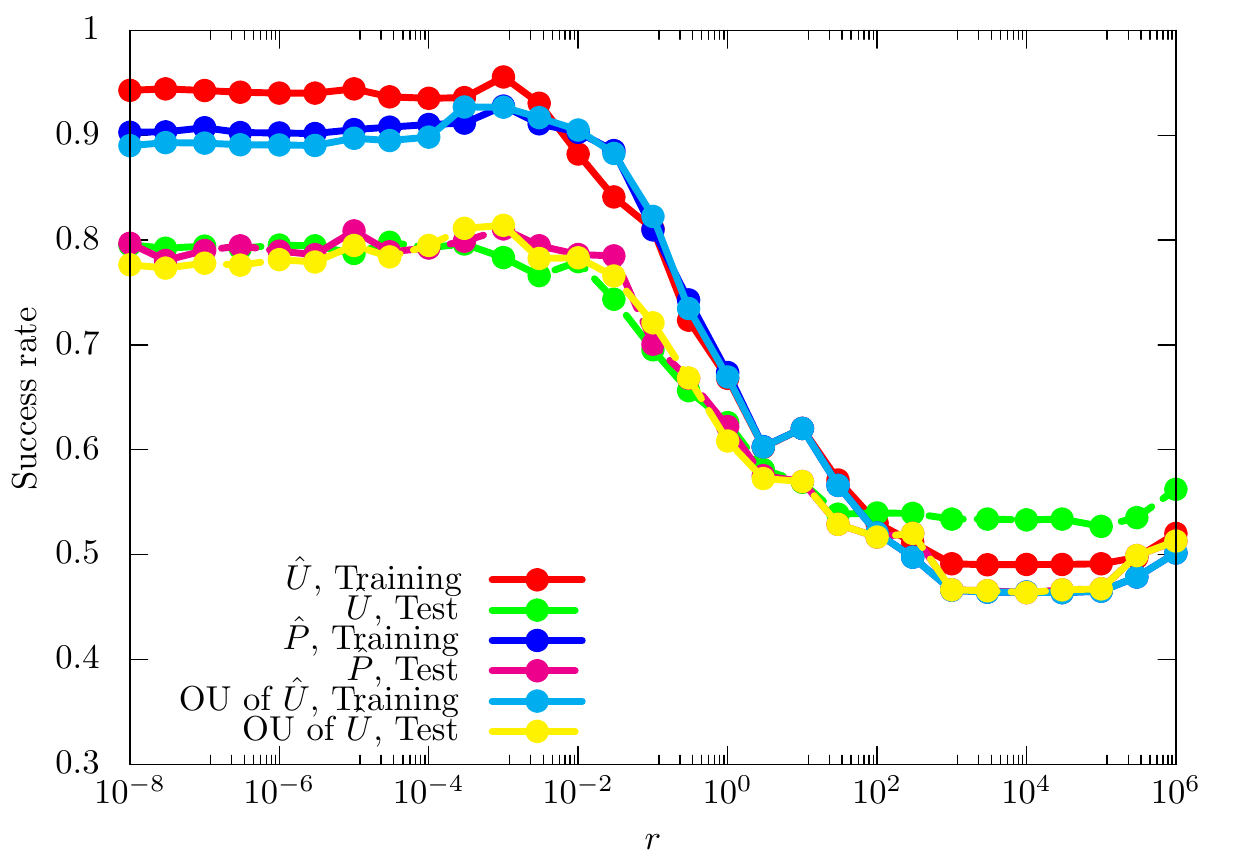}
\caption{Performance dependence of the UKM on $r$, which is the coefficient of the second term in the right-hand side of Eq.~\eqref{supp-arXiv-quantum-kernel-method-001-011} for the sonar dataset ($0$ or $1$) in the case of the hinge function, Eq.~\eqref{supp-arXiv-hinge-function-001-001}. We use complex matrices and set $\theta_\mathrm{bias} = 0$. We set $K = 30$ and $K' = 10$.}
\label{supp-arXiv-numerical-result-r-dependence-UKM-UCI-sonar-0-1-hinge}
\end{figure}

So far, we have used the squared error function, Eq.~\eqref{supp-arXiv-squared-error-function-001-001}.
In Fig.~\ref{supp-arXiv-numerical-result-layers-dependence-QCL-UCI-sonar-0-1-hinge}, we show the performance dependence of QCL on the number of layers $L$ in the case of the hinge function, Eq.~\eqref{supp-arXiv-hinge-function-001-001}.
In Fig.~\ref{supp-arXiv-numerical-result-r-dependence-UKM-UCI-sonar-0-1-hinge}, we show the performance dependence of the UKM on $r$, which is the coefficient of the second term in the right-hand side of Eq.~\eqref{supp-arXiv-quantum-kernel-method-001-011}, in the case of the hinge function, Eq.~\eqref{supp-arXiv-hinge-function-001-001}.

\clearpage

\subsection{Wine dataset ($0$ or non-$0$)}

We here show the numerical result for the wine dataset ($0$ or non-$0$).
For the UKM, we put $r = 0.010$ and set $K = 30$ and $K' = 10$ in Algo.~\ref{supp-arXiv-quantum-kernel-method-002-001}.
For QCL, we run iterations $300$ times.

\begin{figure*}[htb]
\centering
\includegraphics[scale=0.25]{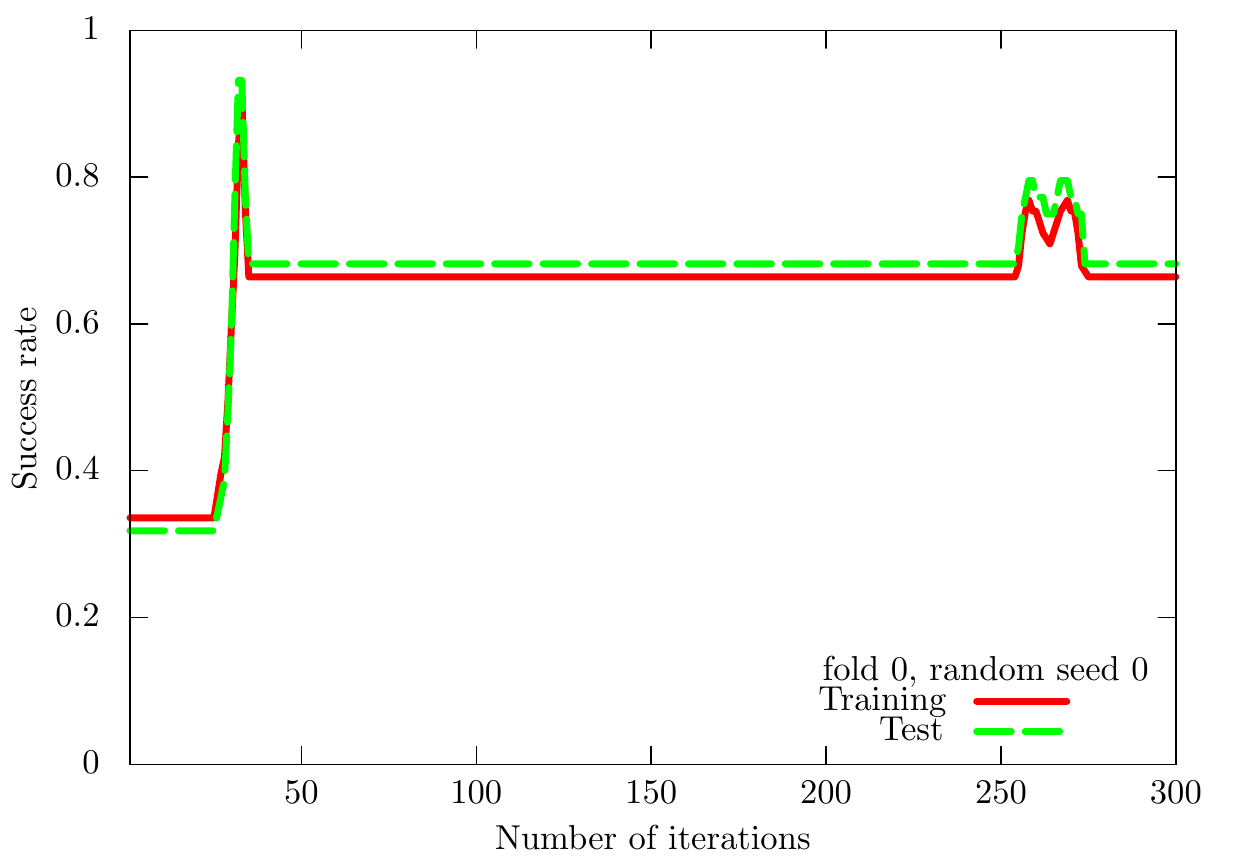}
\includegraphics[scale=0.25]{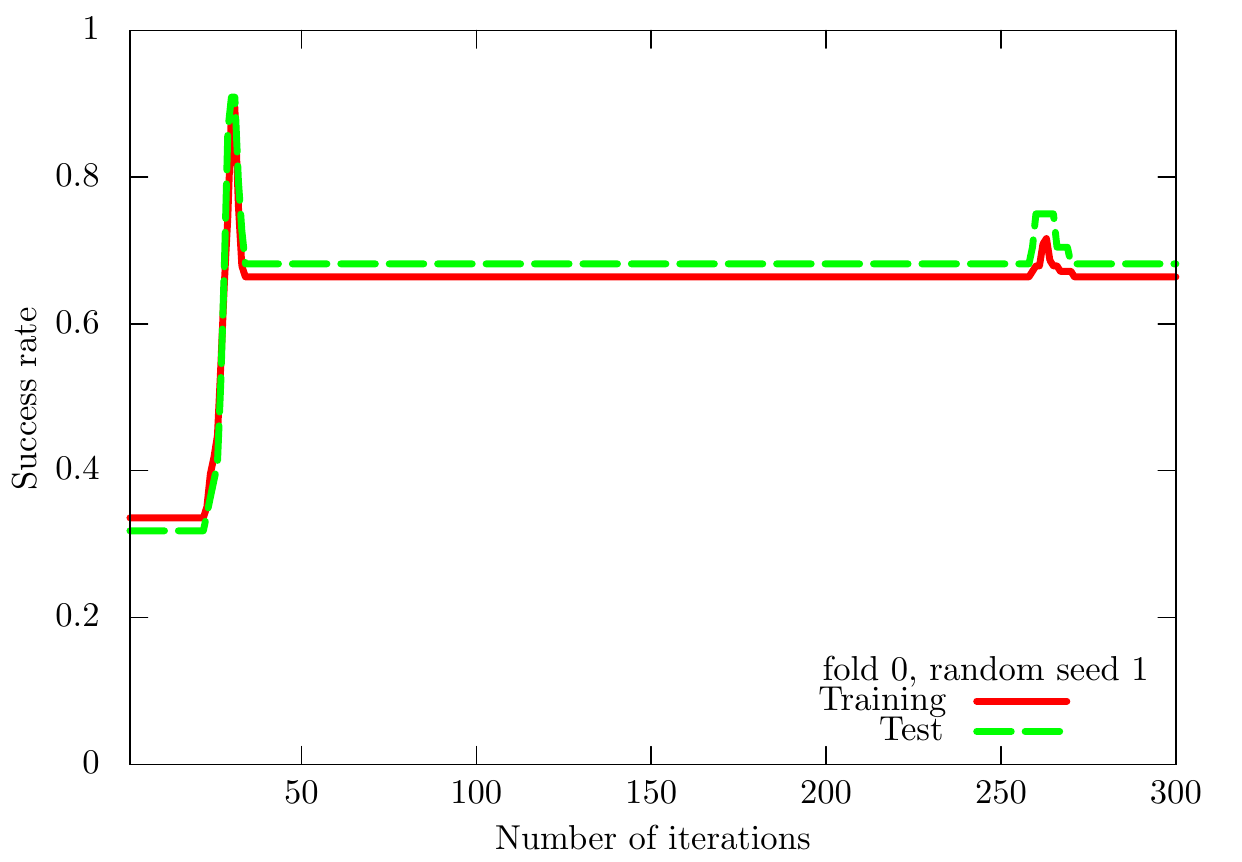}
\includegraphics[scale=0.25]{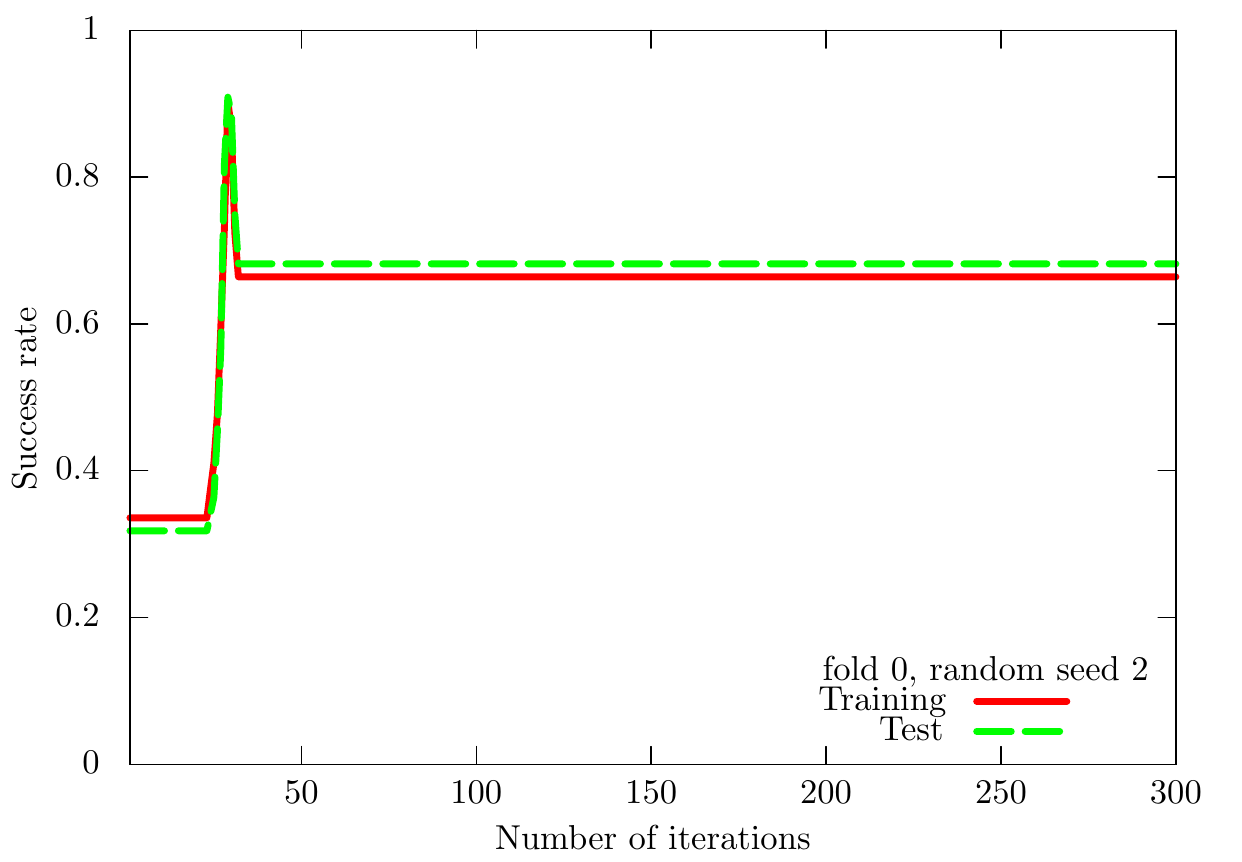}
\includegraphics[scale=0.25]{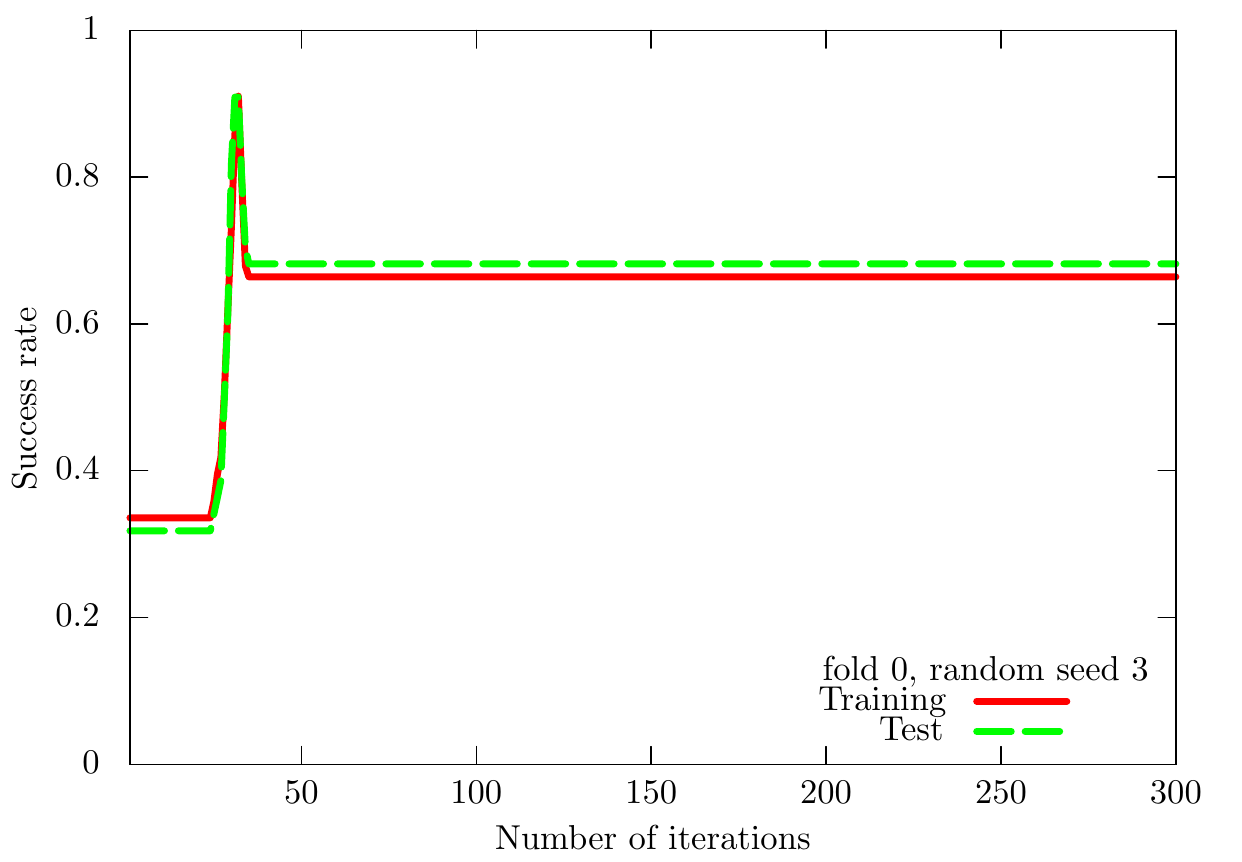}
\includegraphics[scale=0.25]{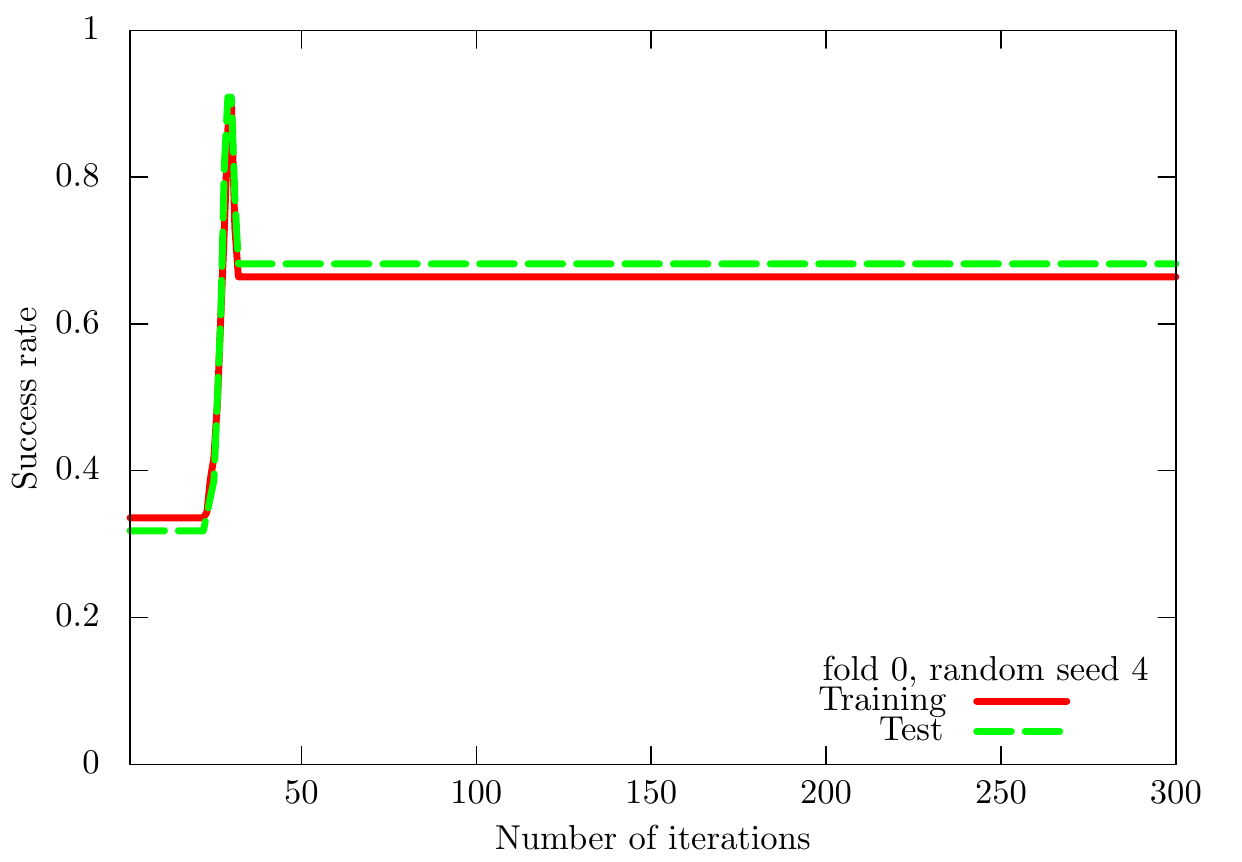}
\includegraphics[scale=0.25]{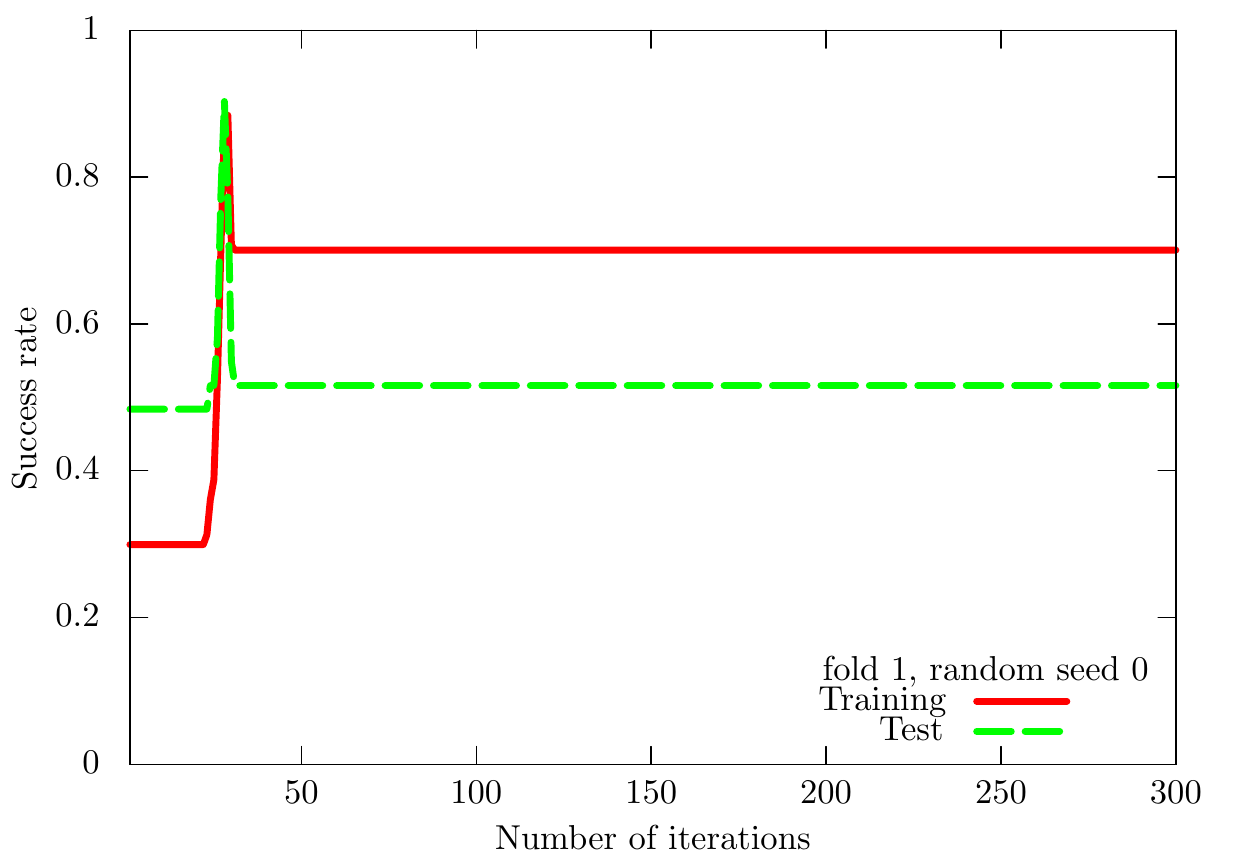}
\includegraphics[scale=0.25]{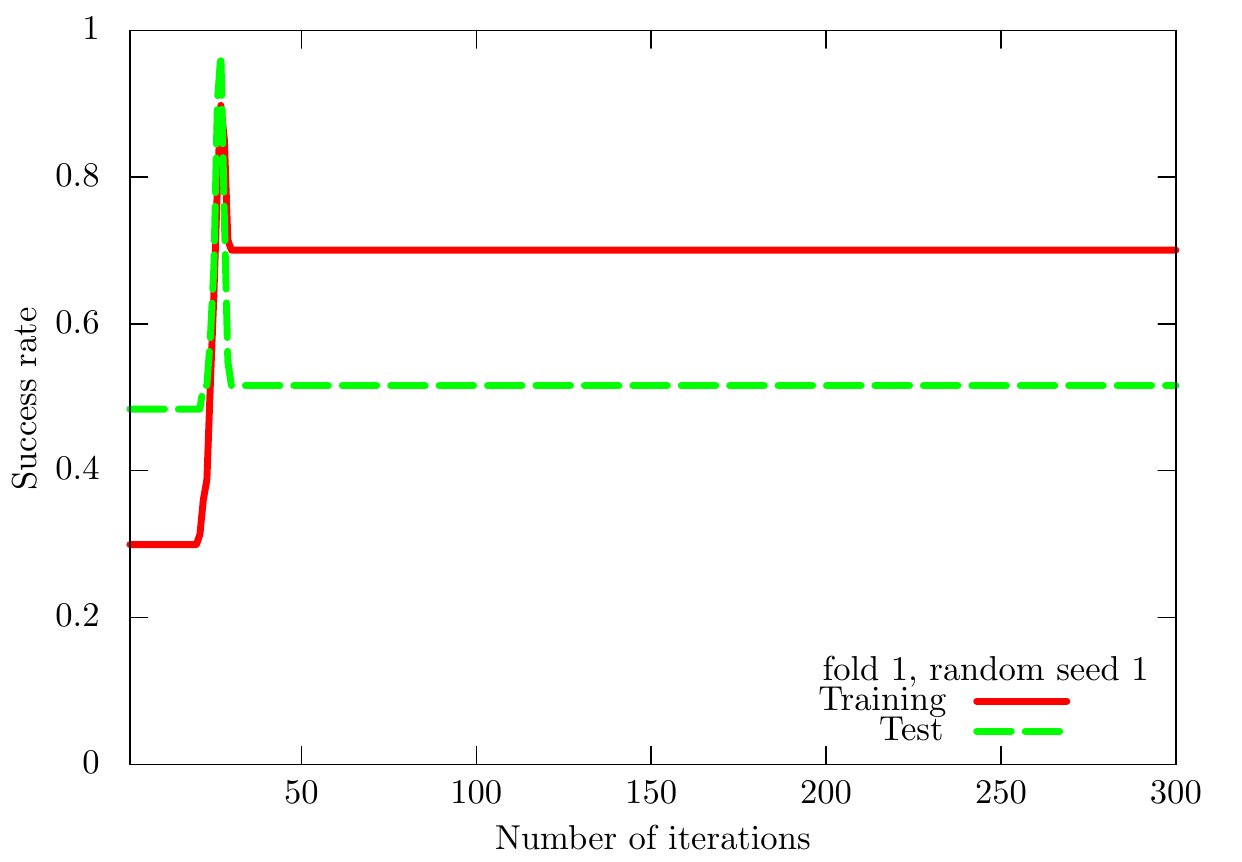}
\includegraphics[scale=0.25]{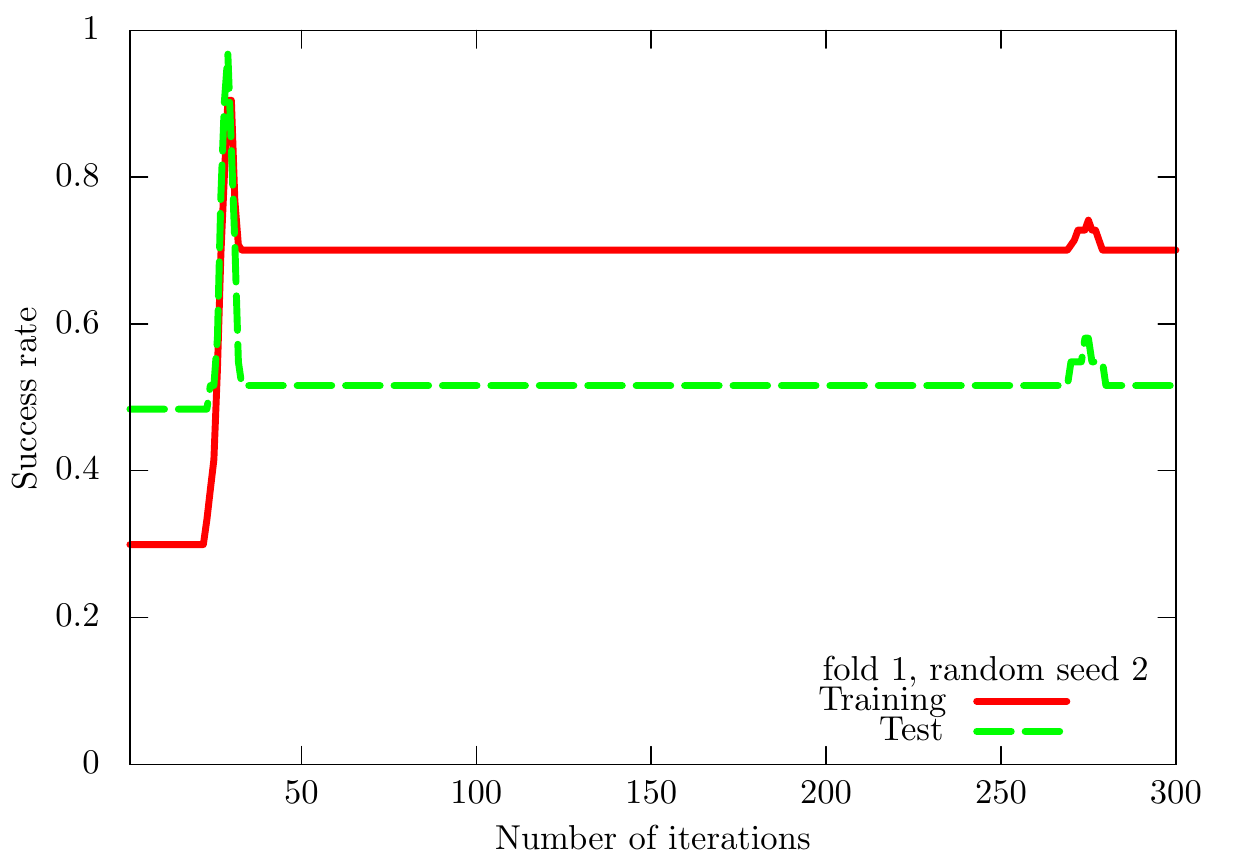}
\includegraphics[scale=0.25]{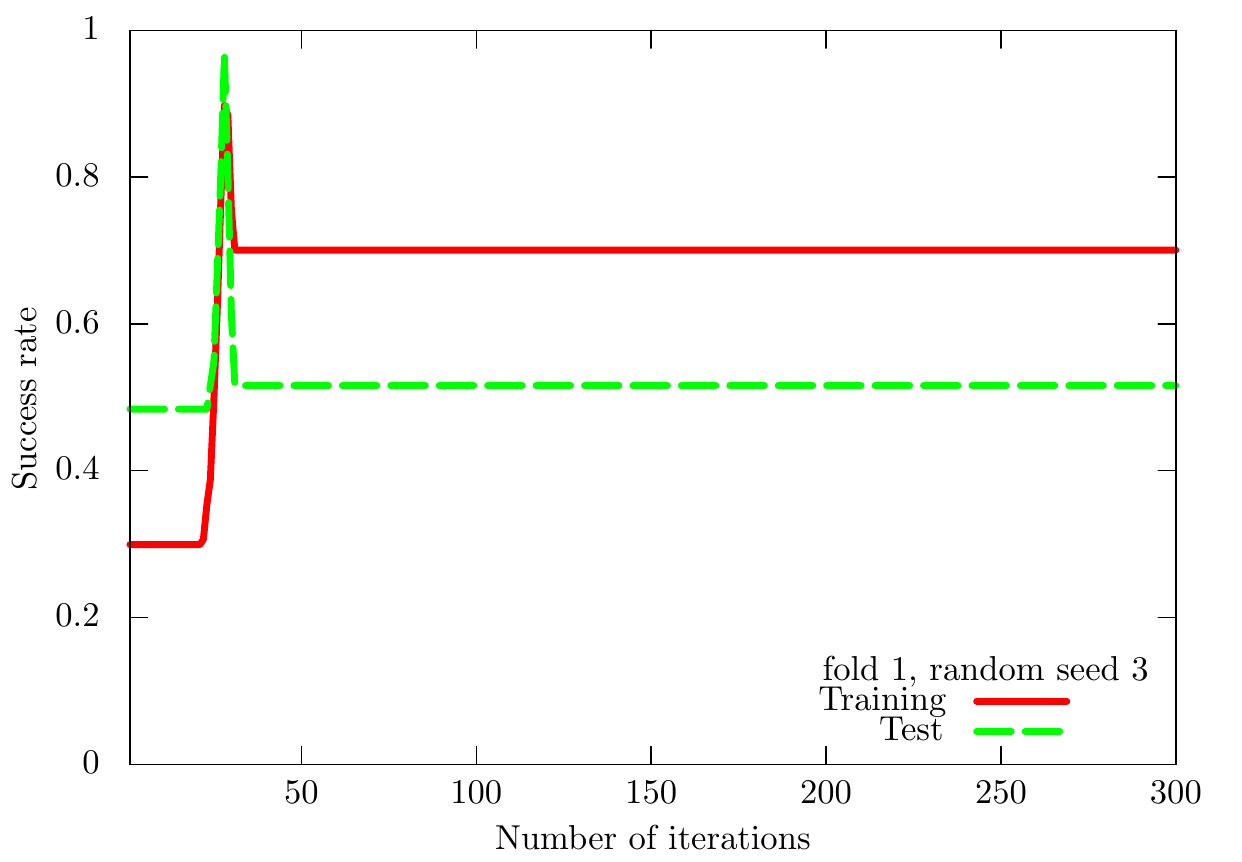}
\includegraphics[scale=0.25]{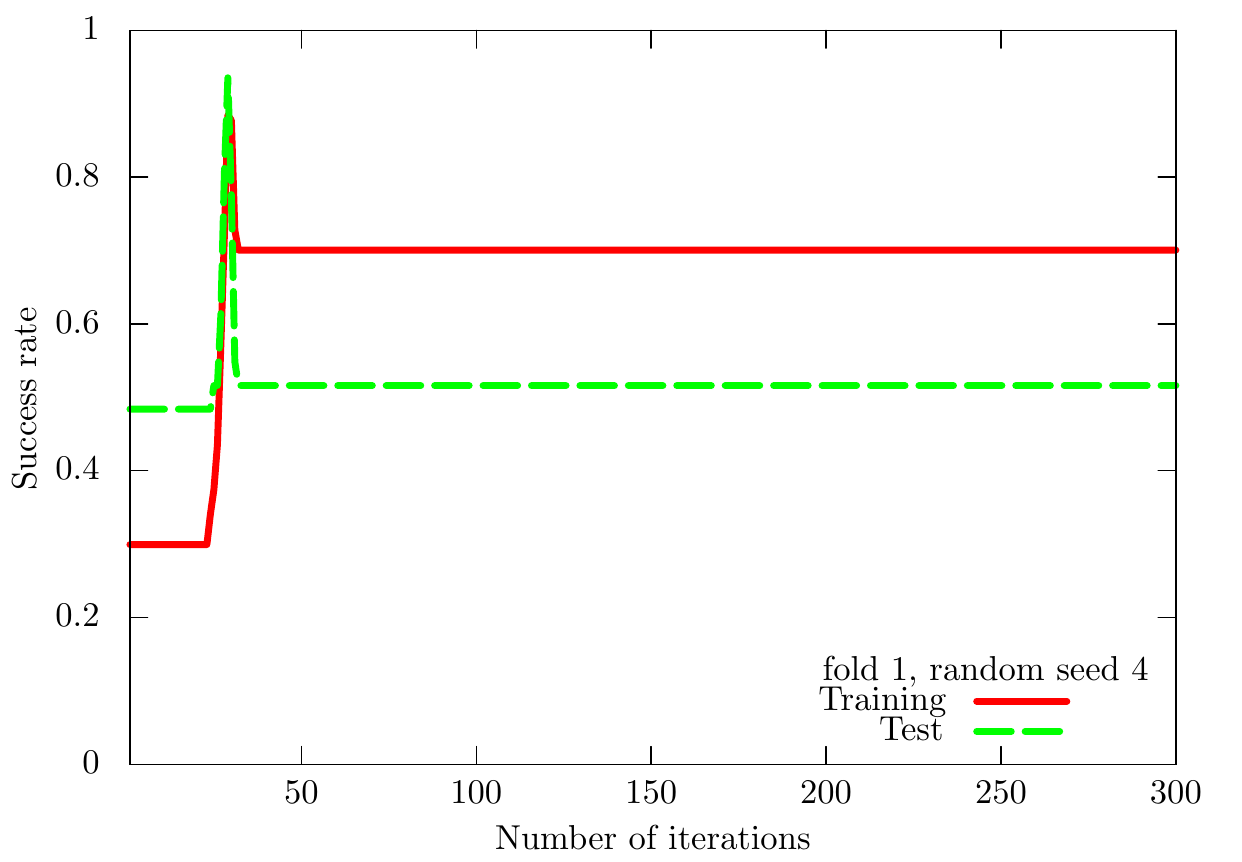}
\includegraphics[scale=0.25]{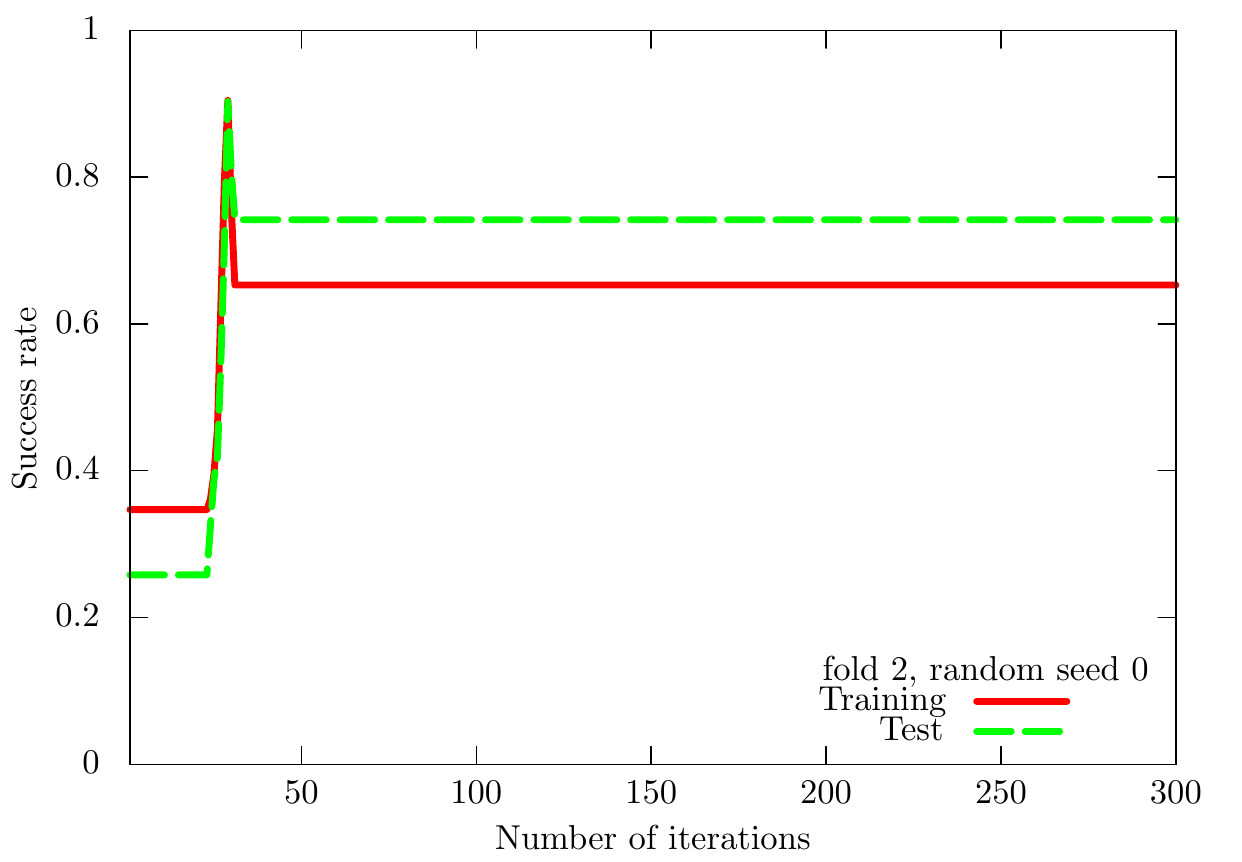}
\includegraphics[scale=0.25]{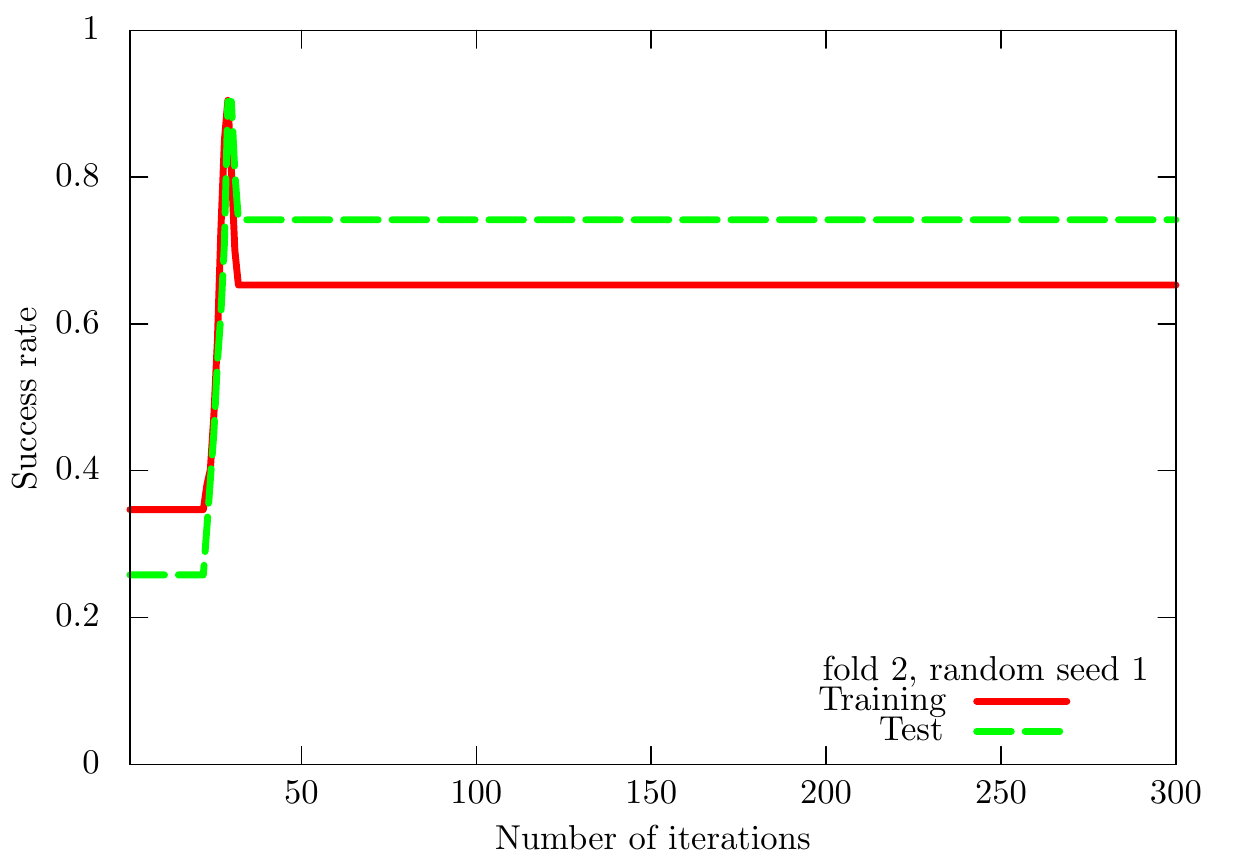}
\includegraphics[scale=0.25]{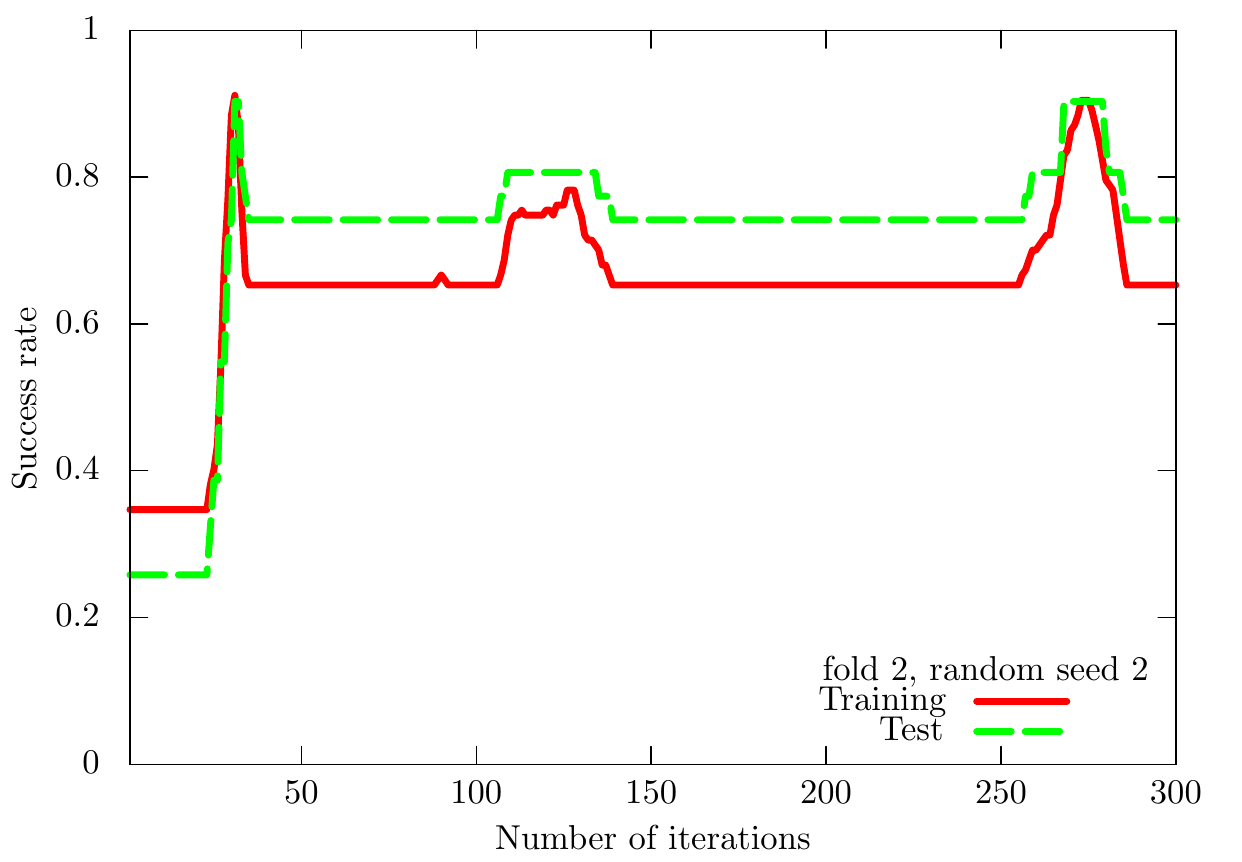}
\includegraphics[scale=0.25]{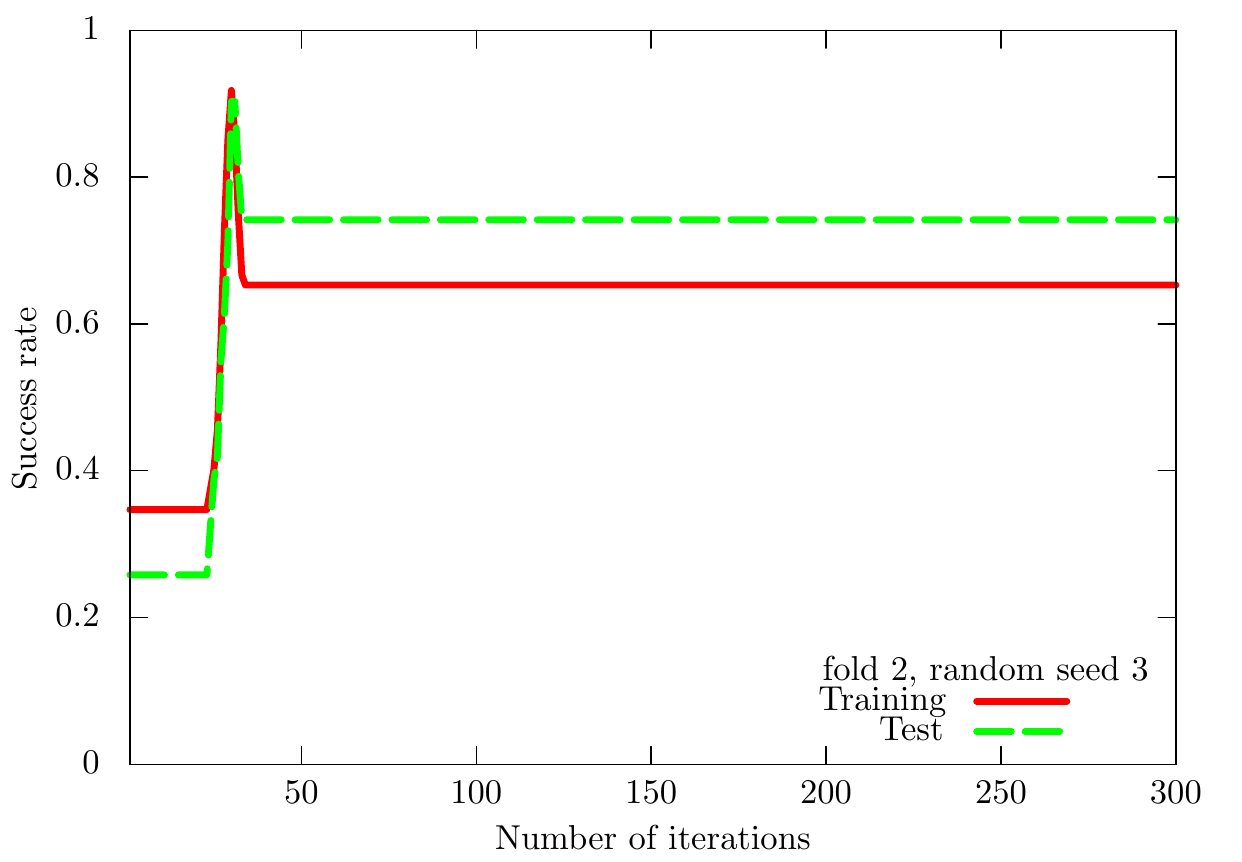}
\includegraphics[scale=0.25]{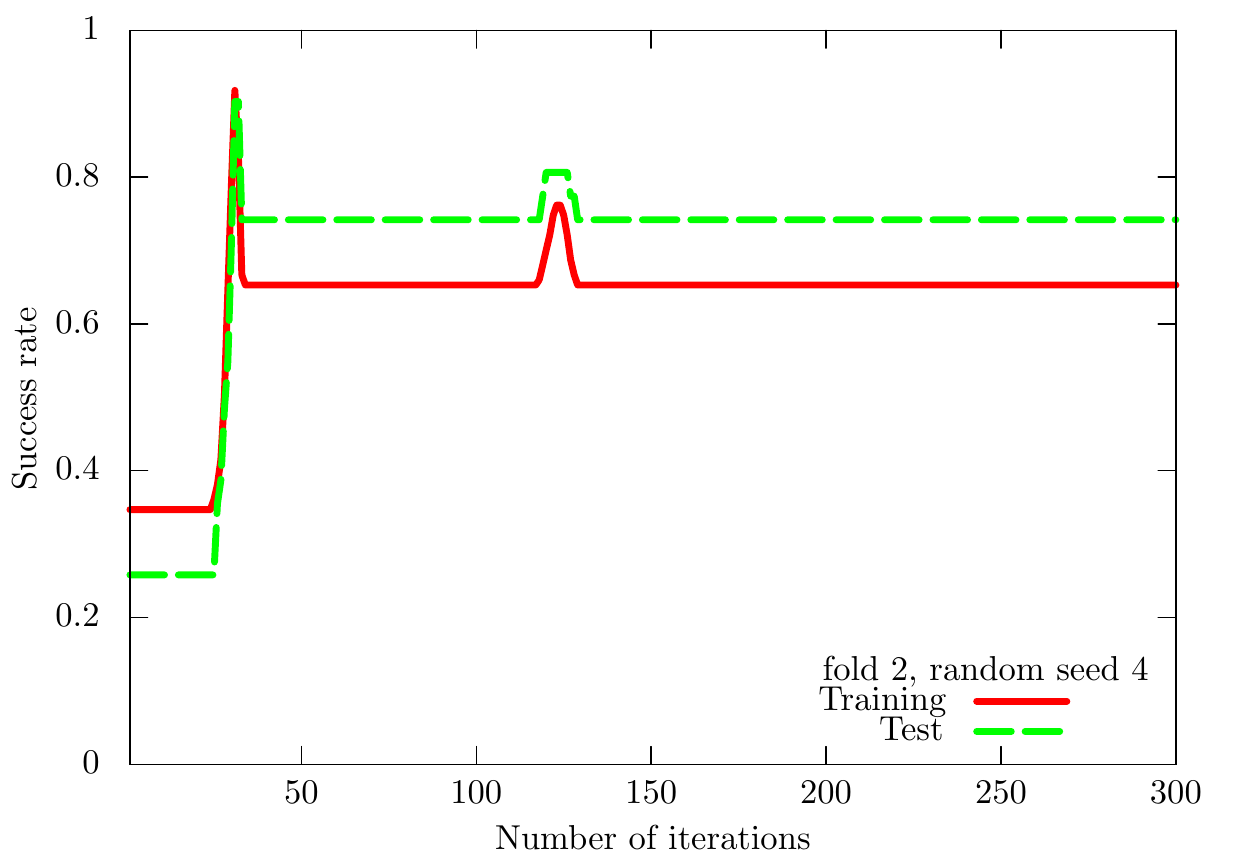}
\includegraphics[scale=0.25]{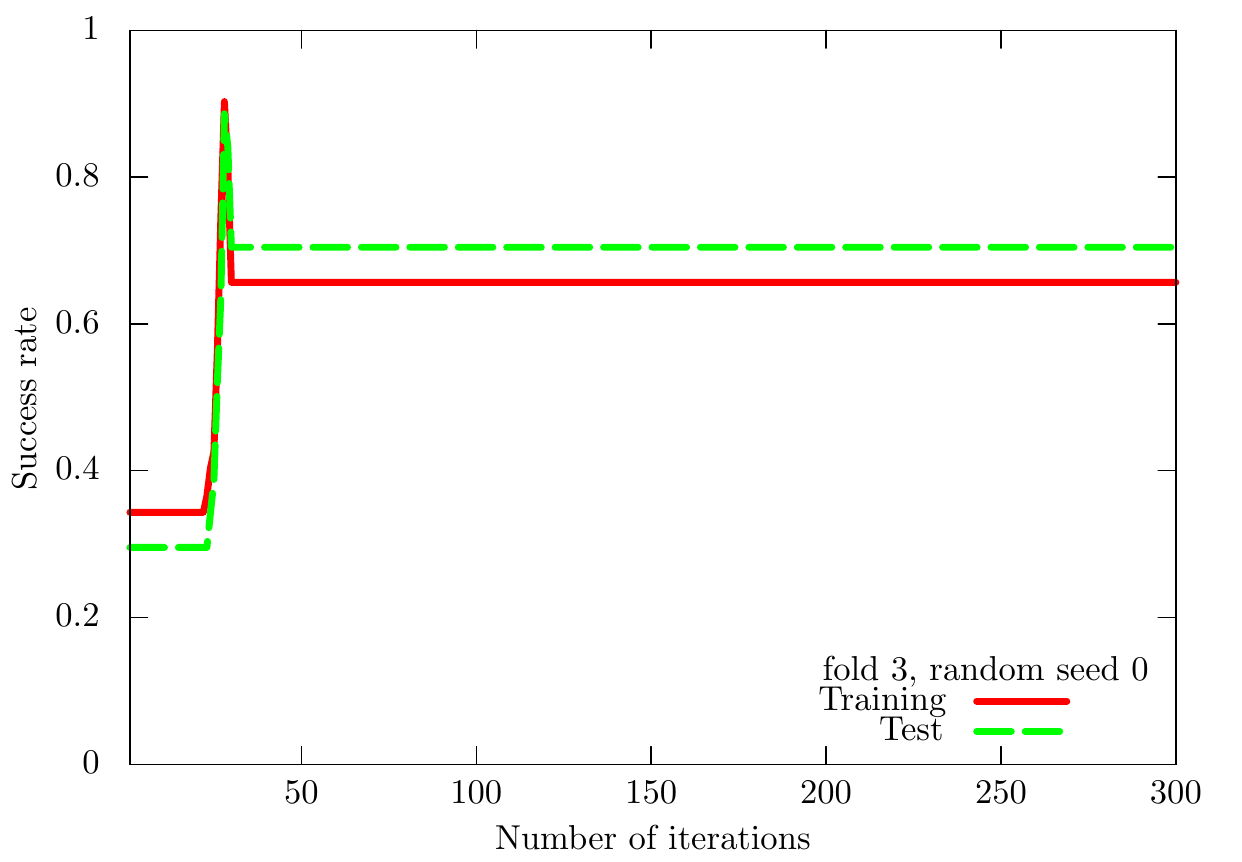}
\includegraphics[scale=0.25]{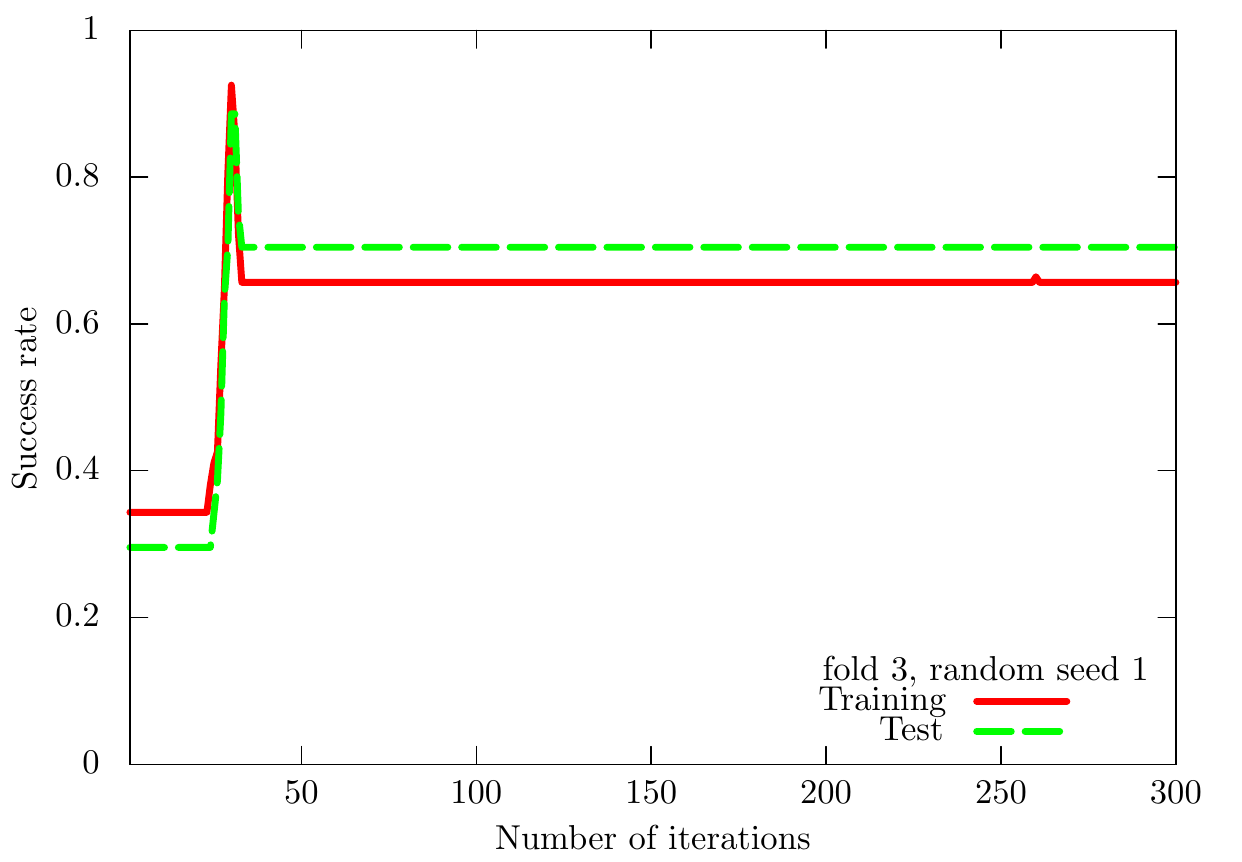}
\includegraphics[scale=0.25]{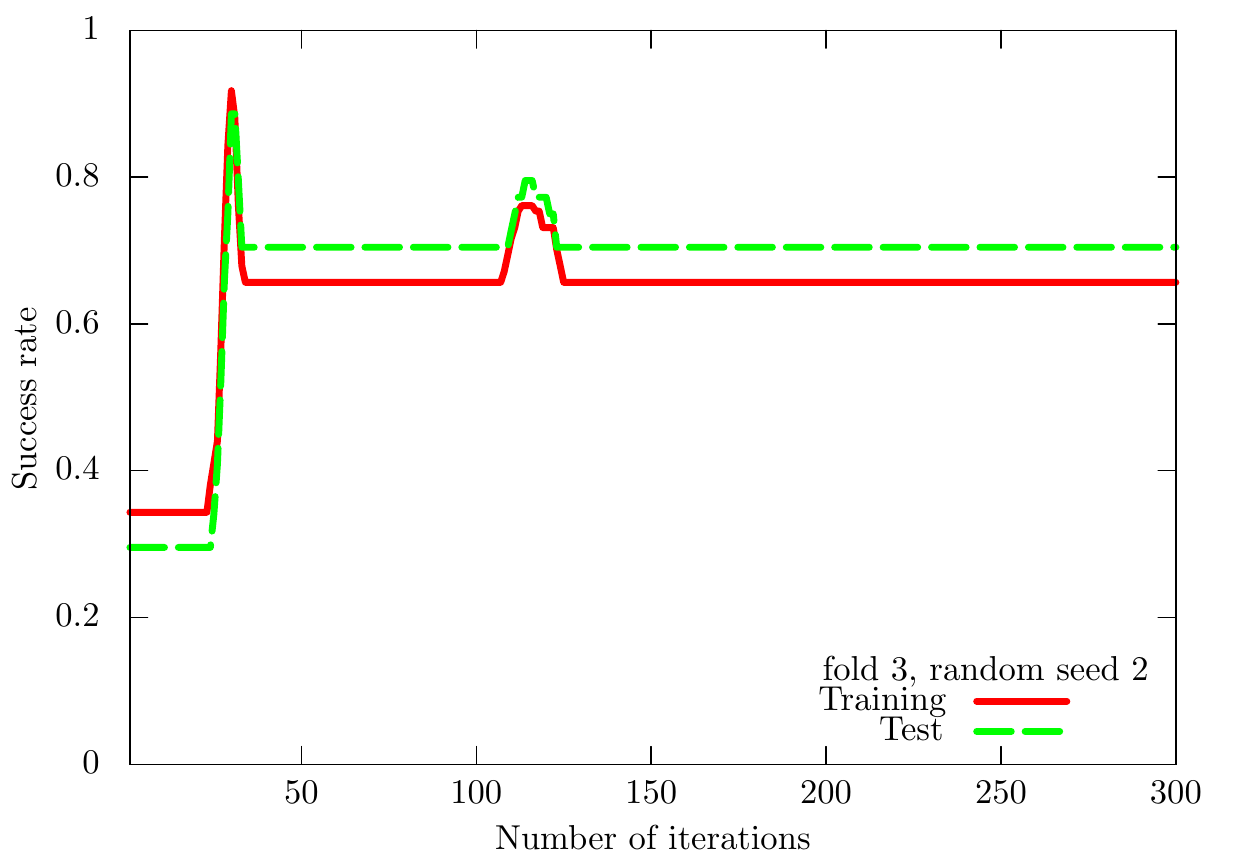}
\includegraphics[scale=0.25]{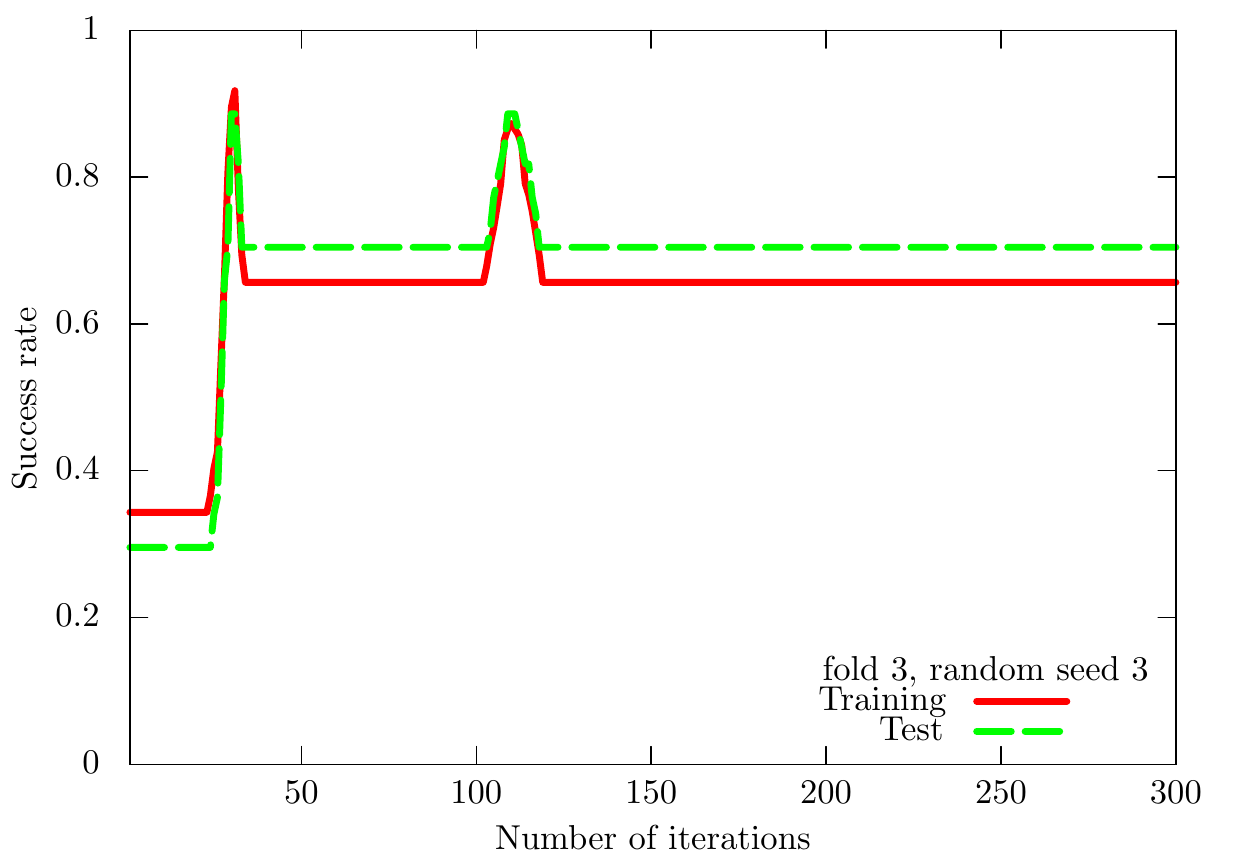}
\includegraphics[scale=0.25]{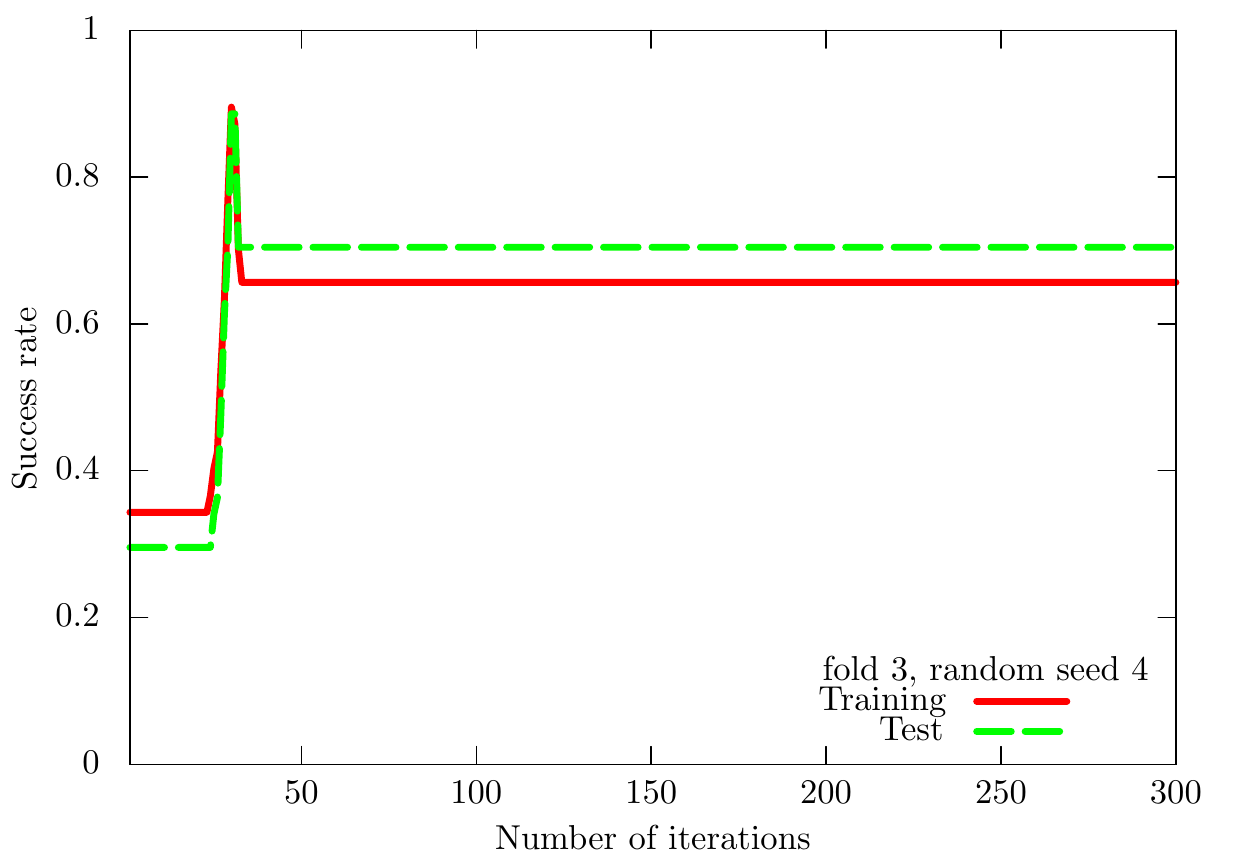}
\includegraphics[scale=0.25]{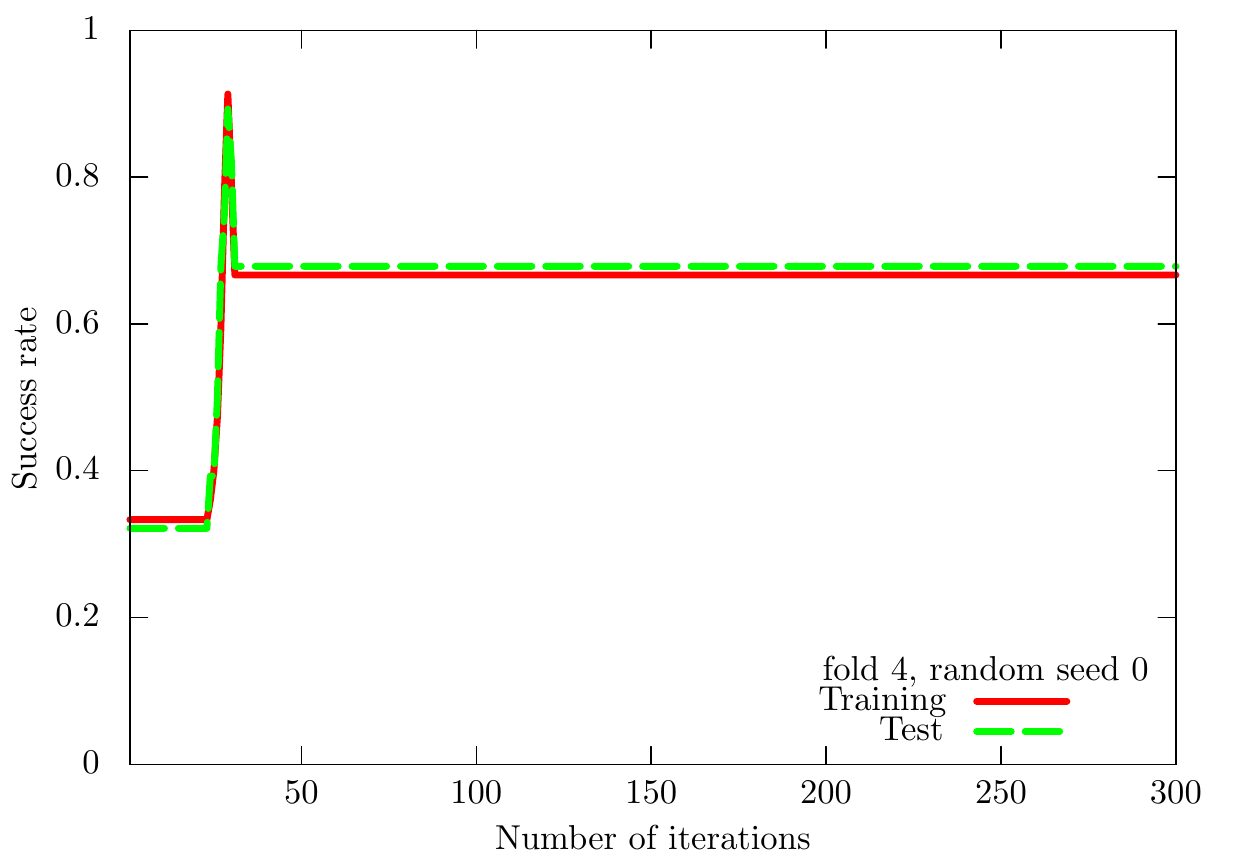}
\includegraphics[scale=0.25]{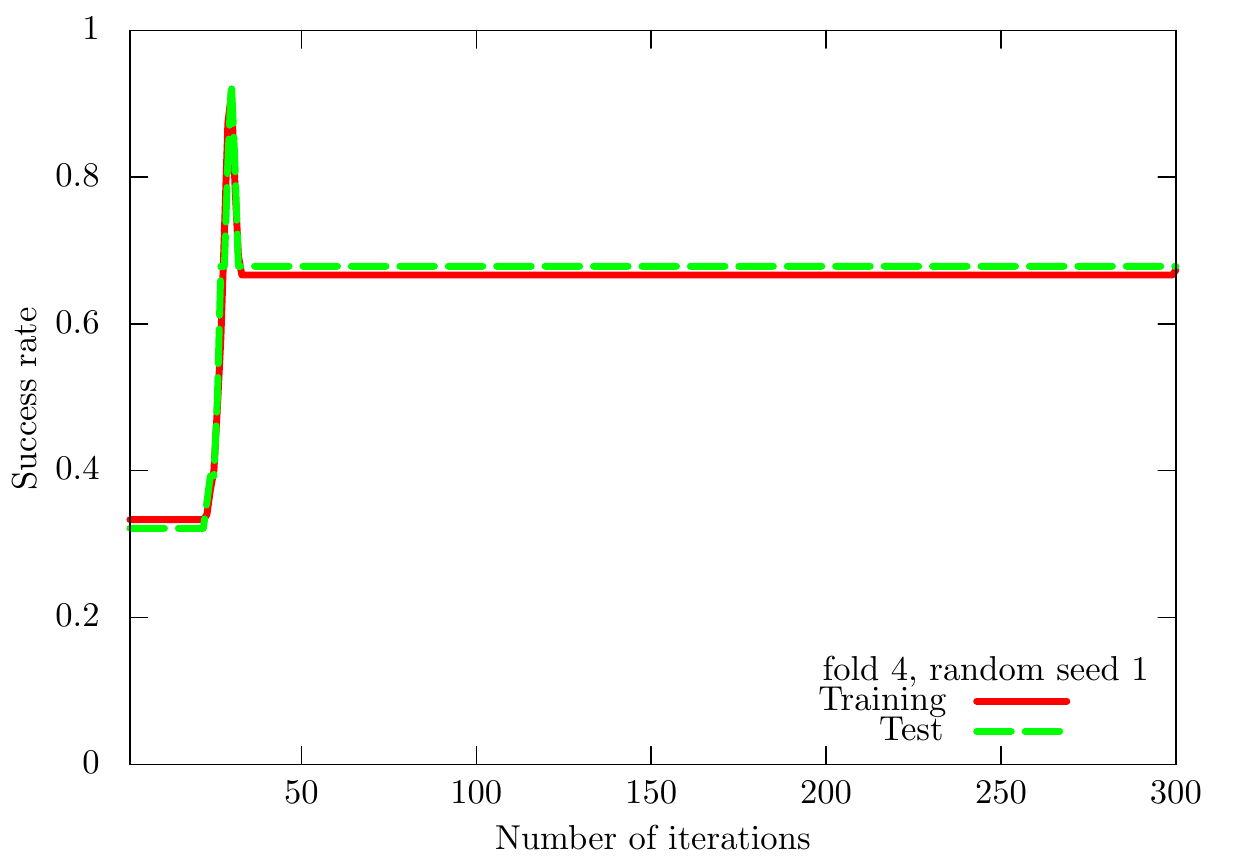}
\includegraphics[scale=0.25]{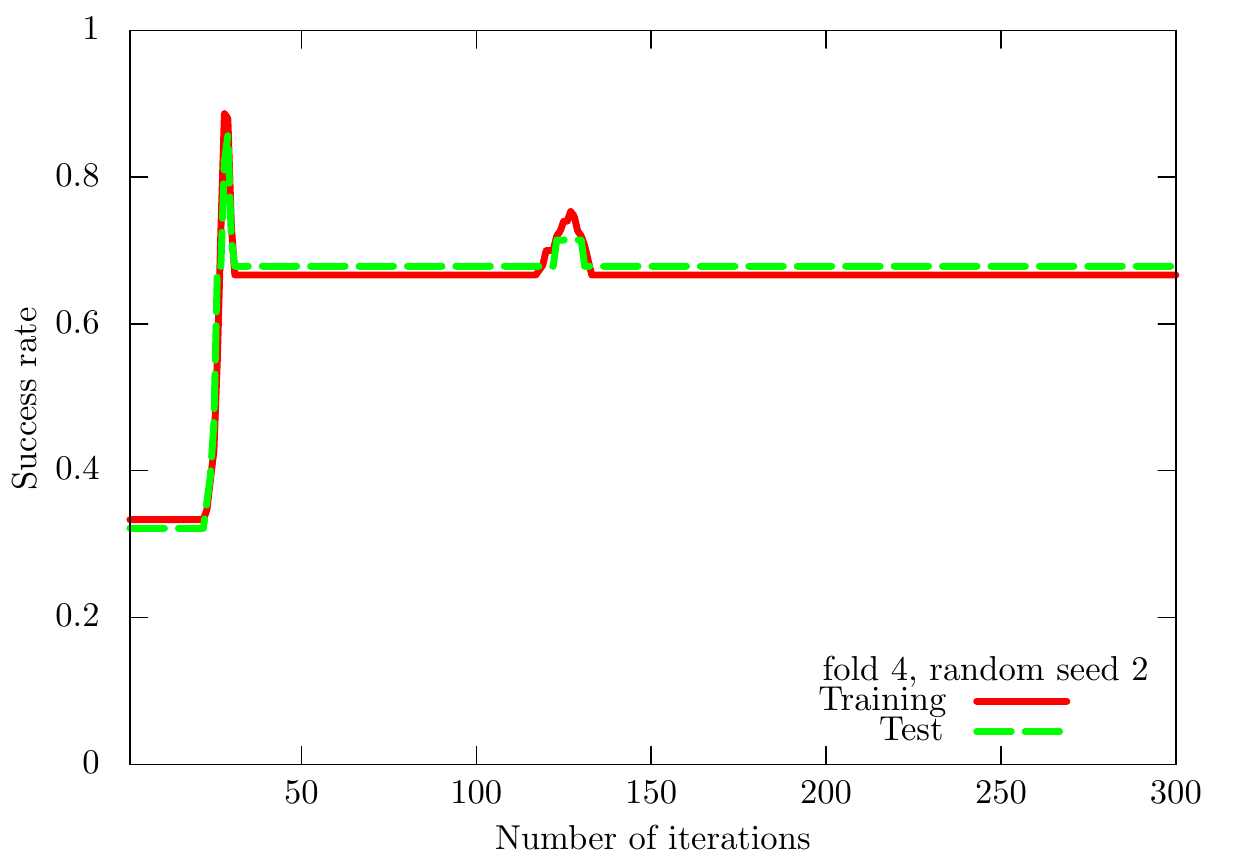}
\includegraphics[scale=0.25]{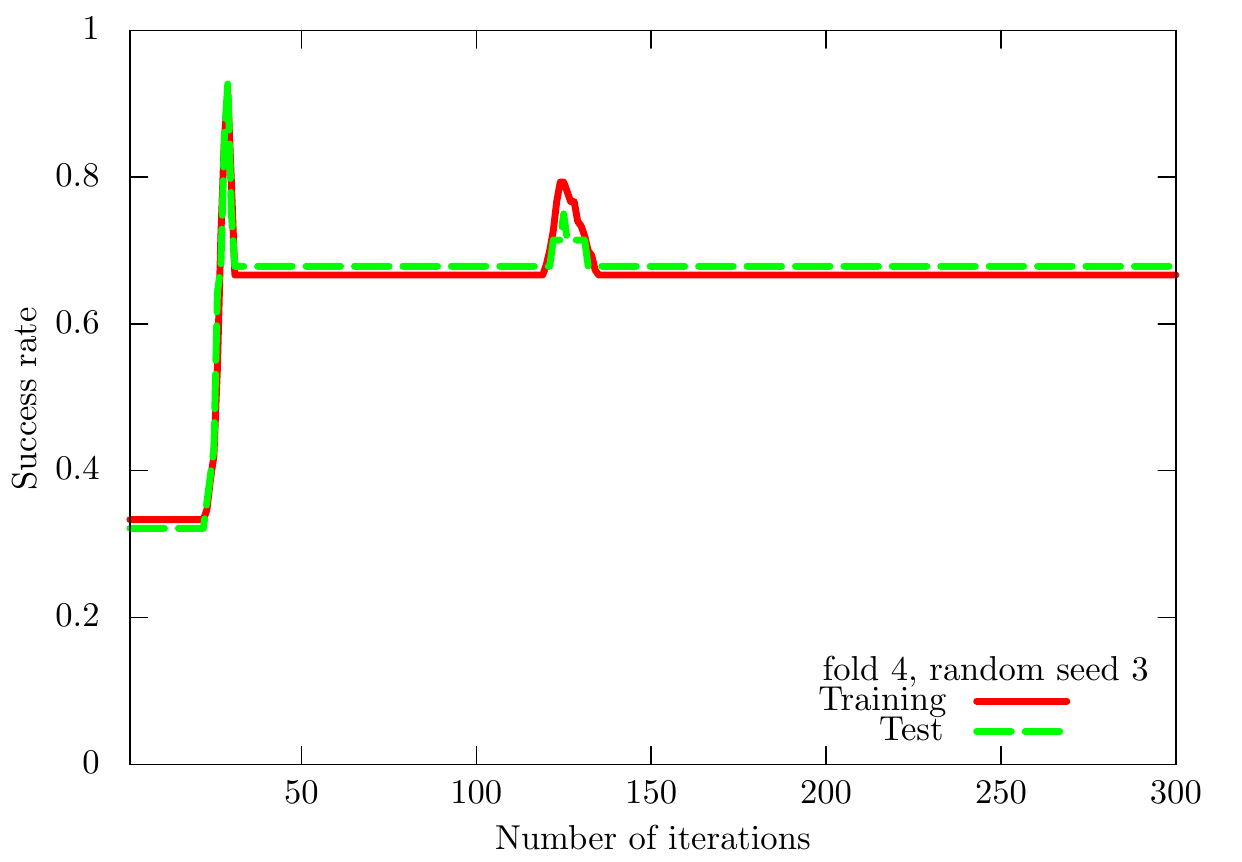}
\includegraphics[scale=0.25]{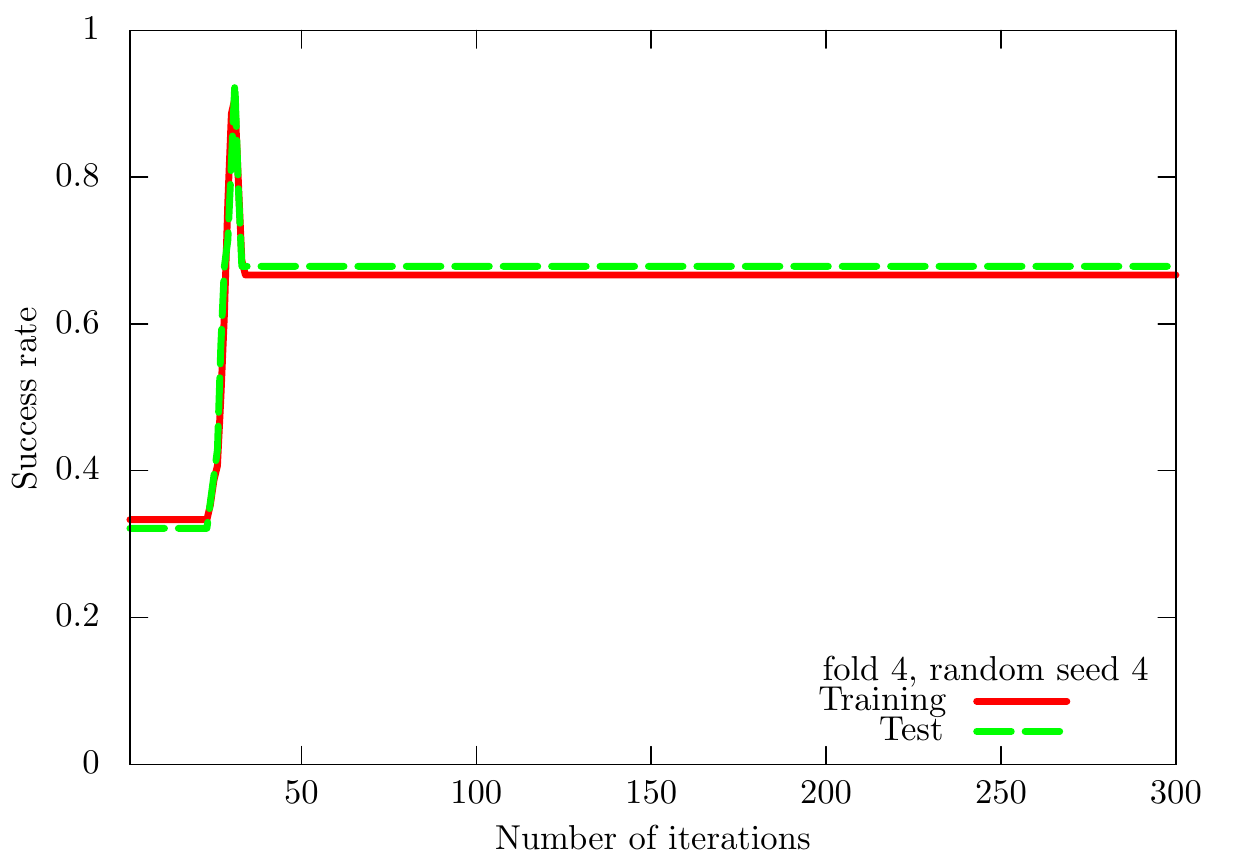}
\caption{Results of QCL on the $5$-fold datasets with $5$ different random seeds for the wine dataset ($0$ or non-$0$). We use the CNOT-based circuit and set $\theta_\mathrm{bias} = 0$. The number of layers $L$ is set to $5$.}
\label{supp-arXiv-numerical-result-raw-data-fold-001-rand-001-QCL-UCI-wine-0-non0}
\end{figure*}

\begin{figure*}[htb]
\centering
\includegraphics[scale=0.25]{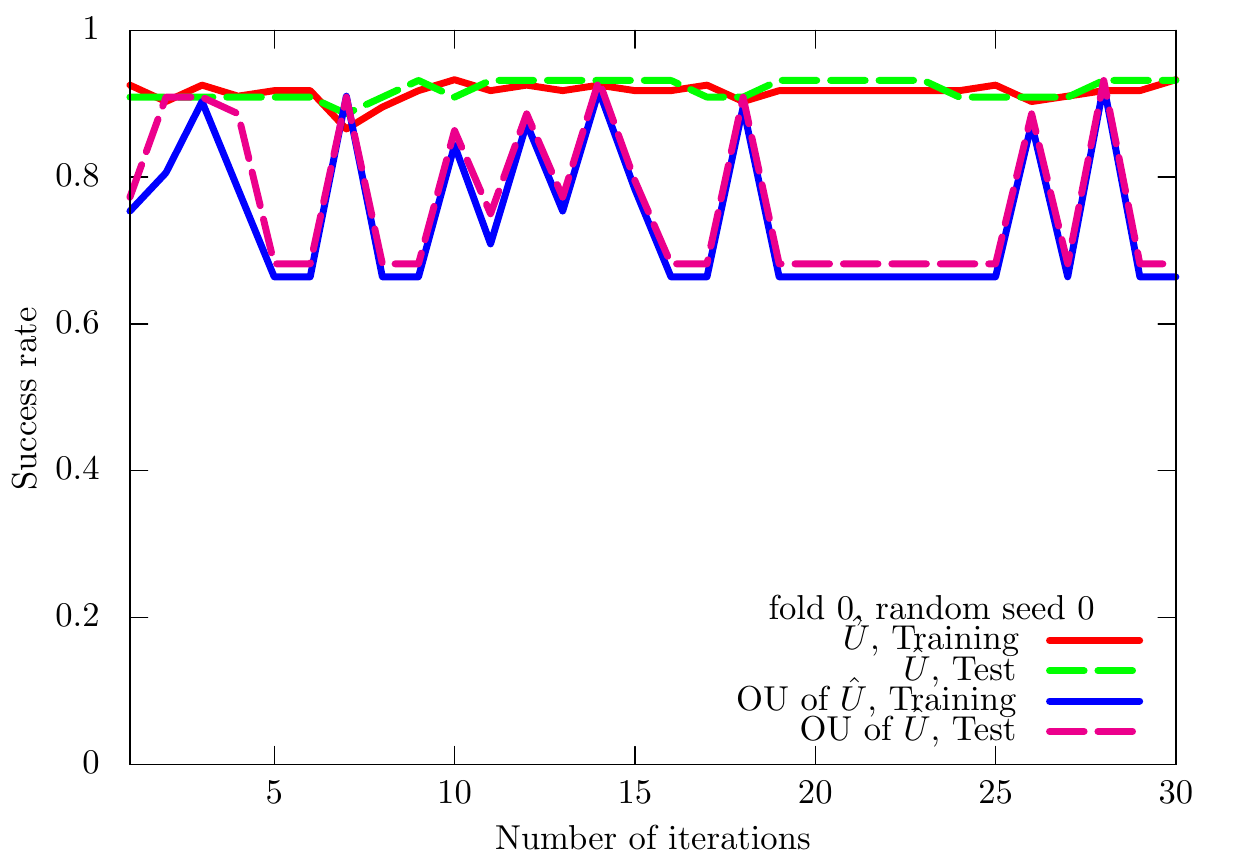}
\includegraphics[scale=0.25]{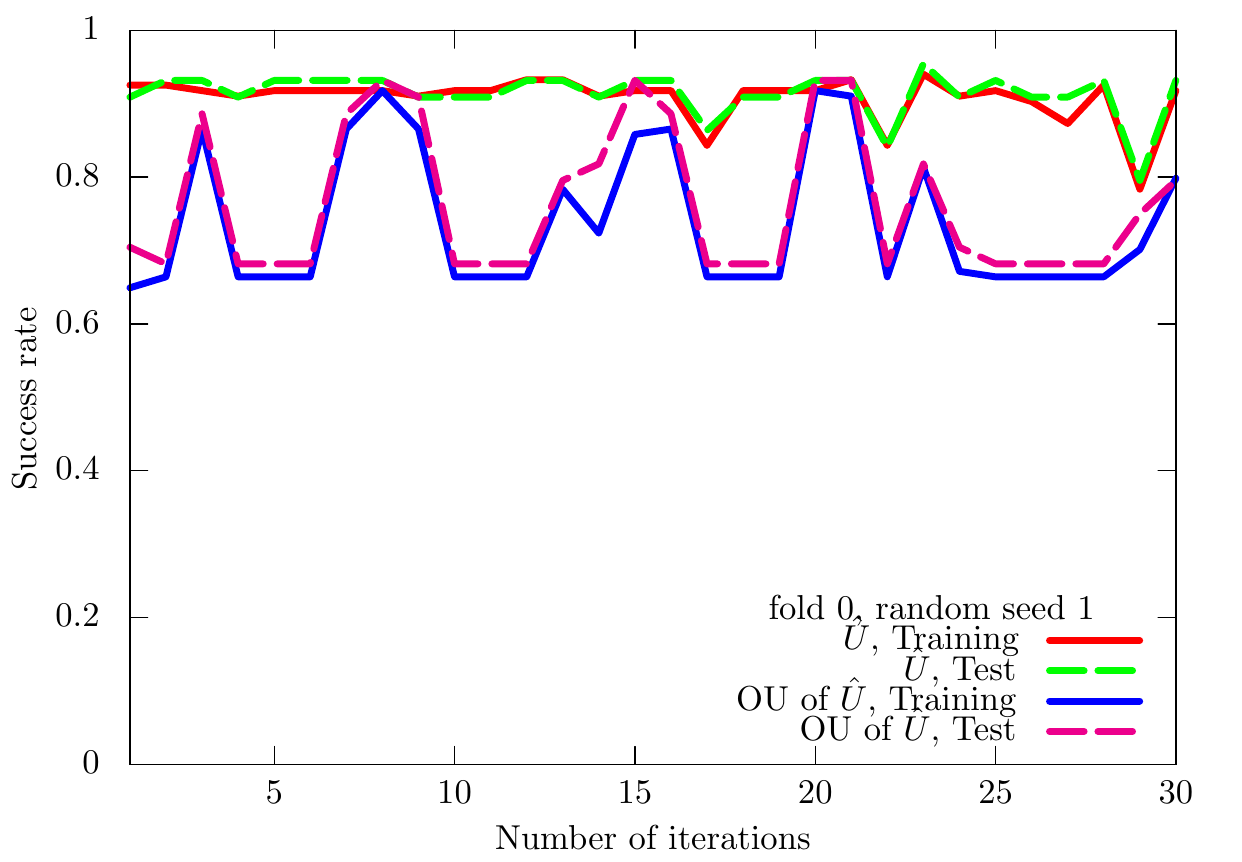}
\includegraphics[scale=0.25]{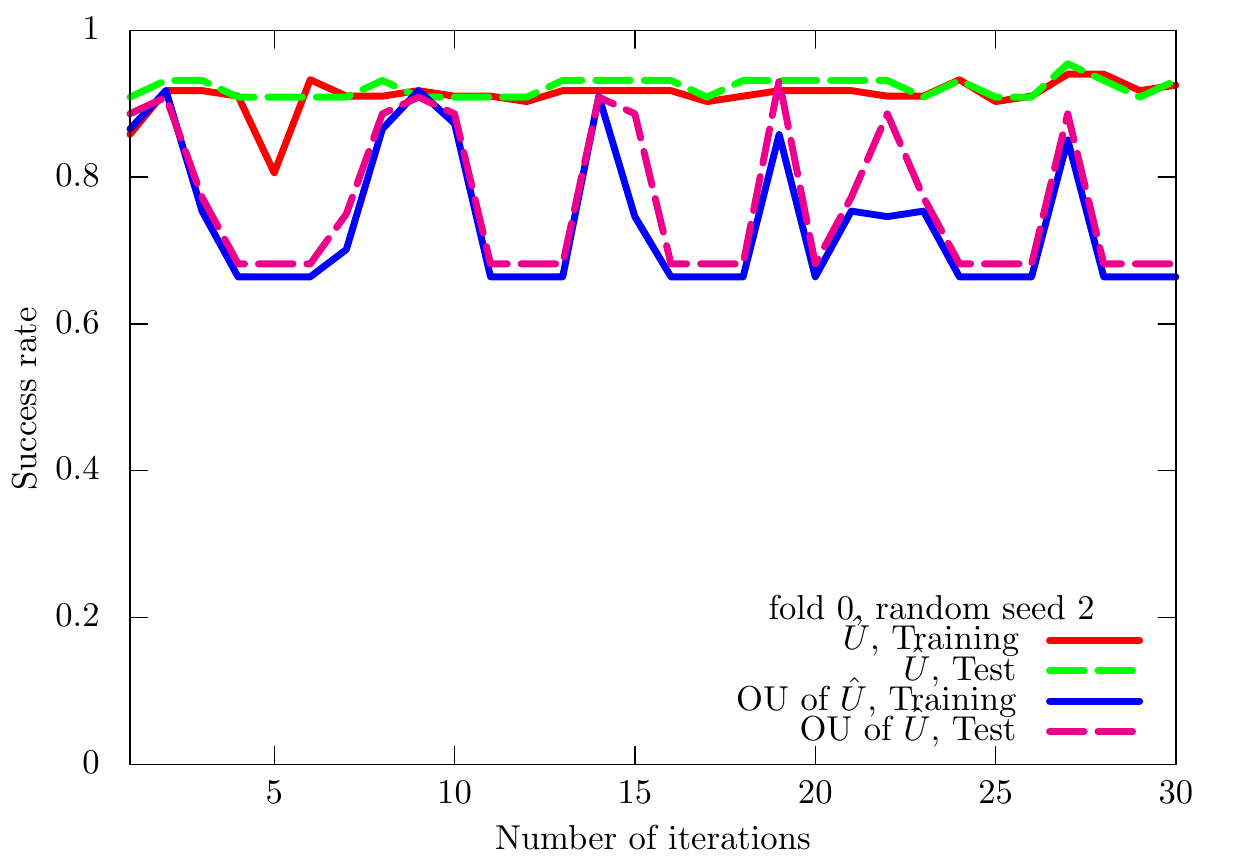}
\includegraphics[scale=0.25]{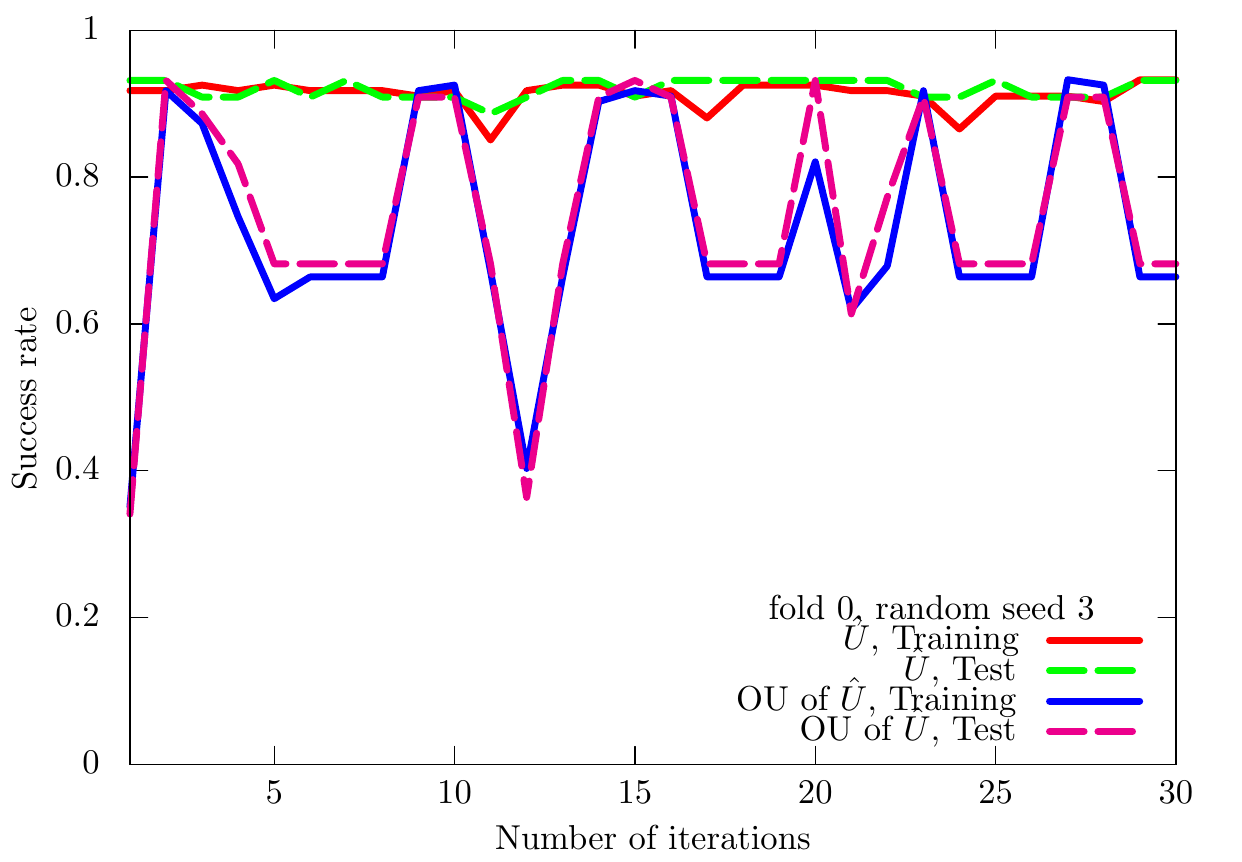}
\includegraphics[scale=0.25]{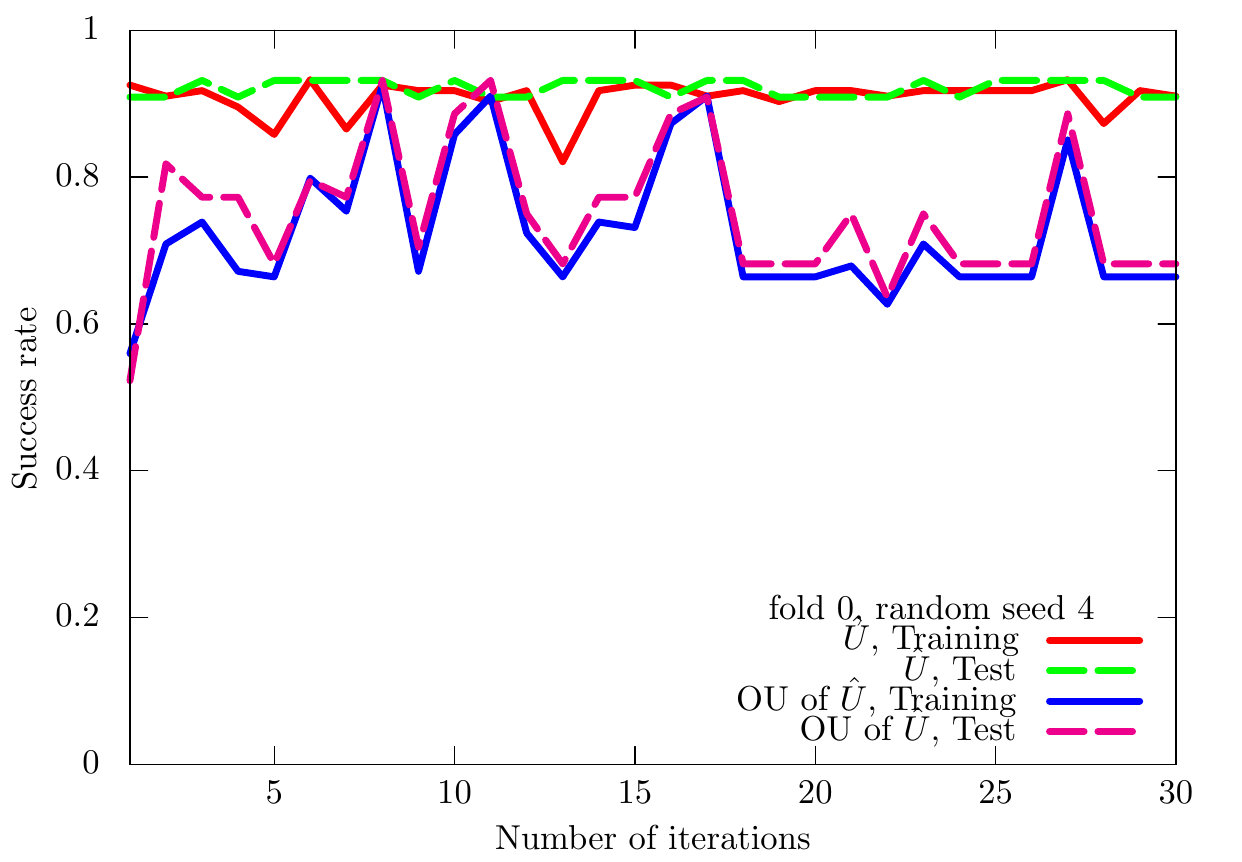}
\includegraphics[scale=0.25]{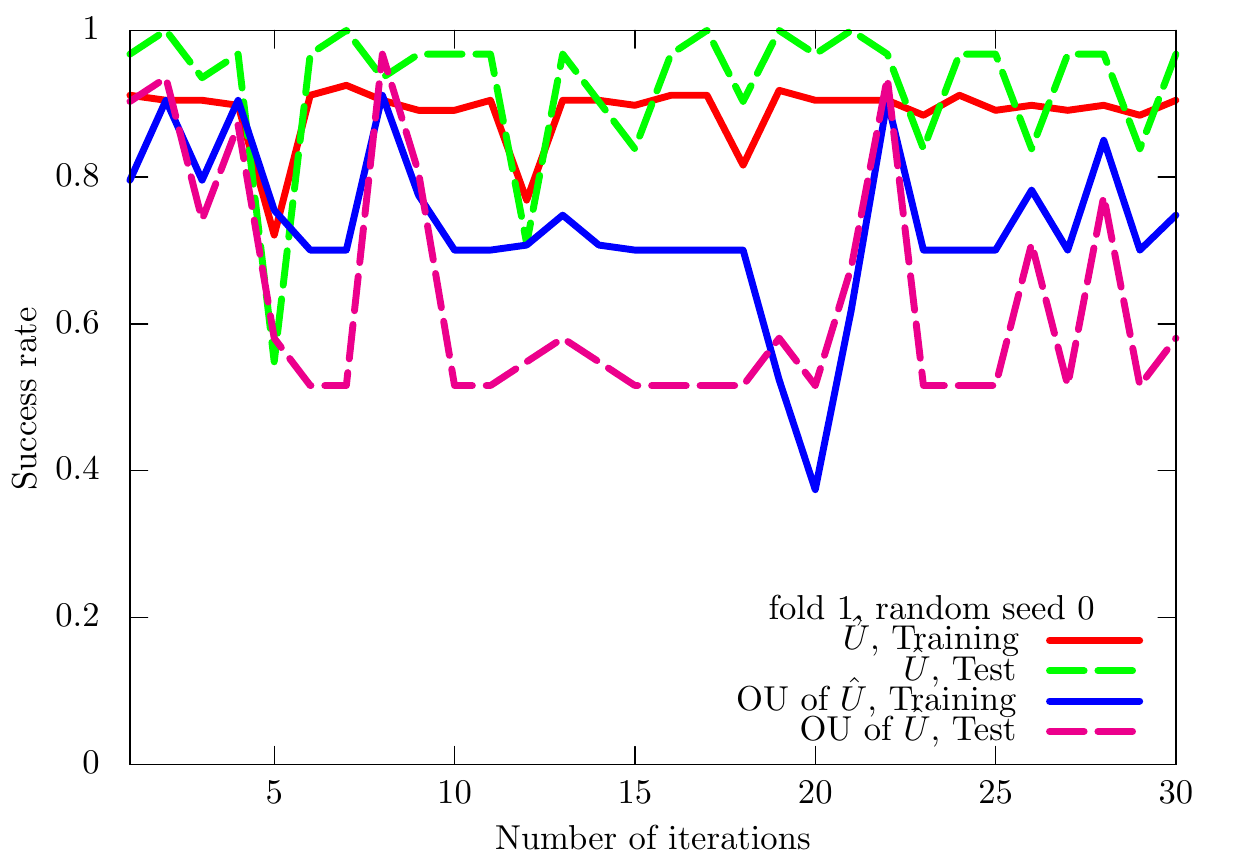}
\includegraphics[scale=0.25]{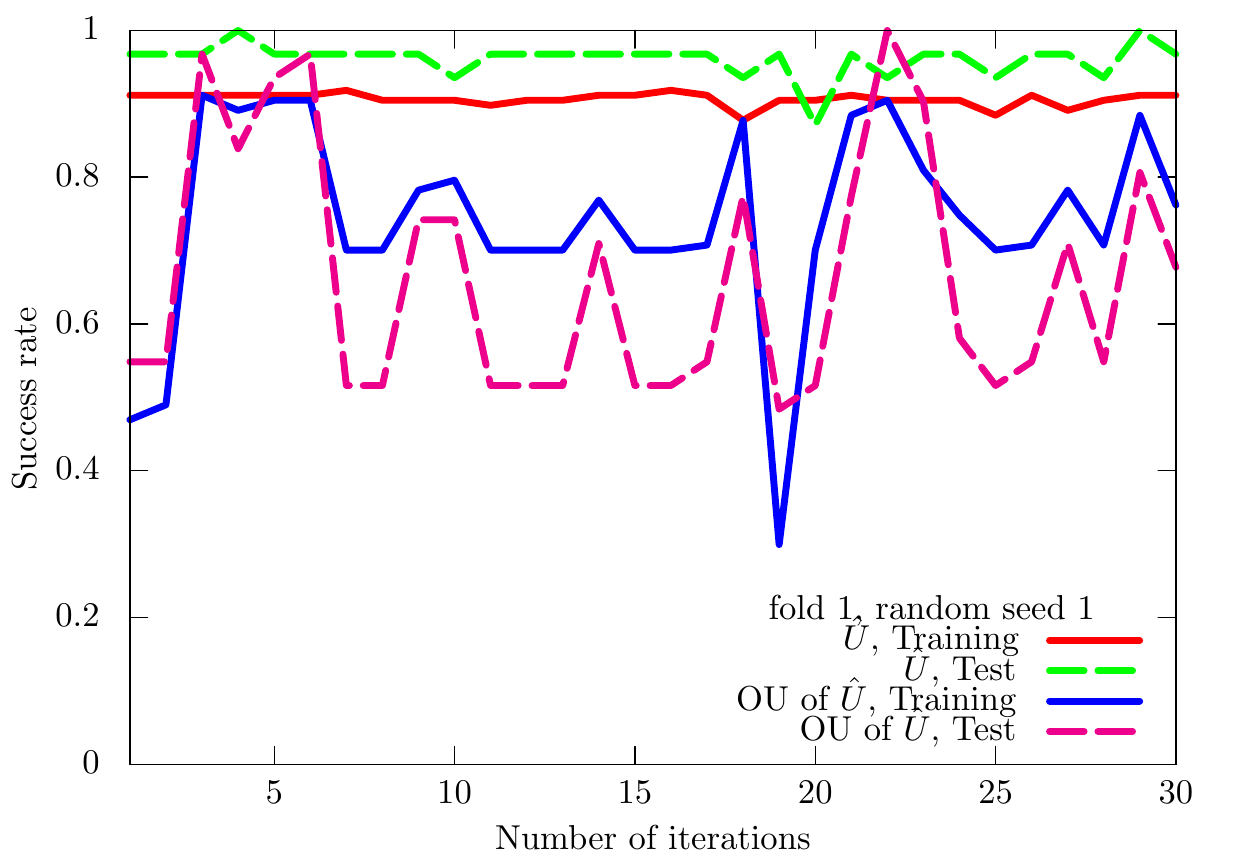}
\includegraphics[scale=0.25]{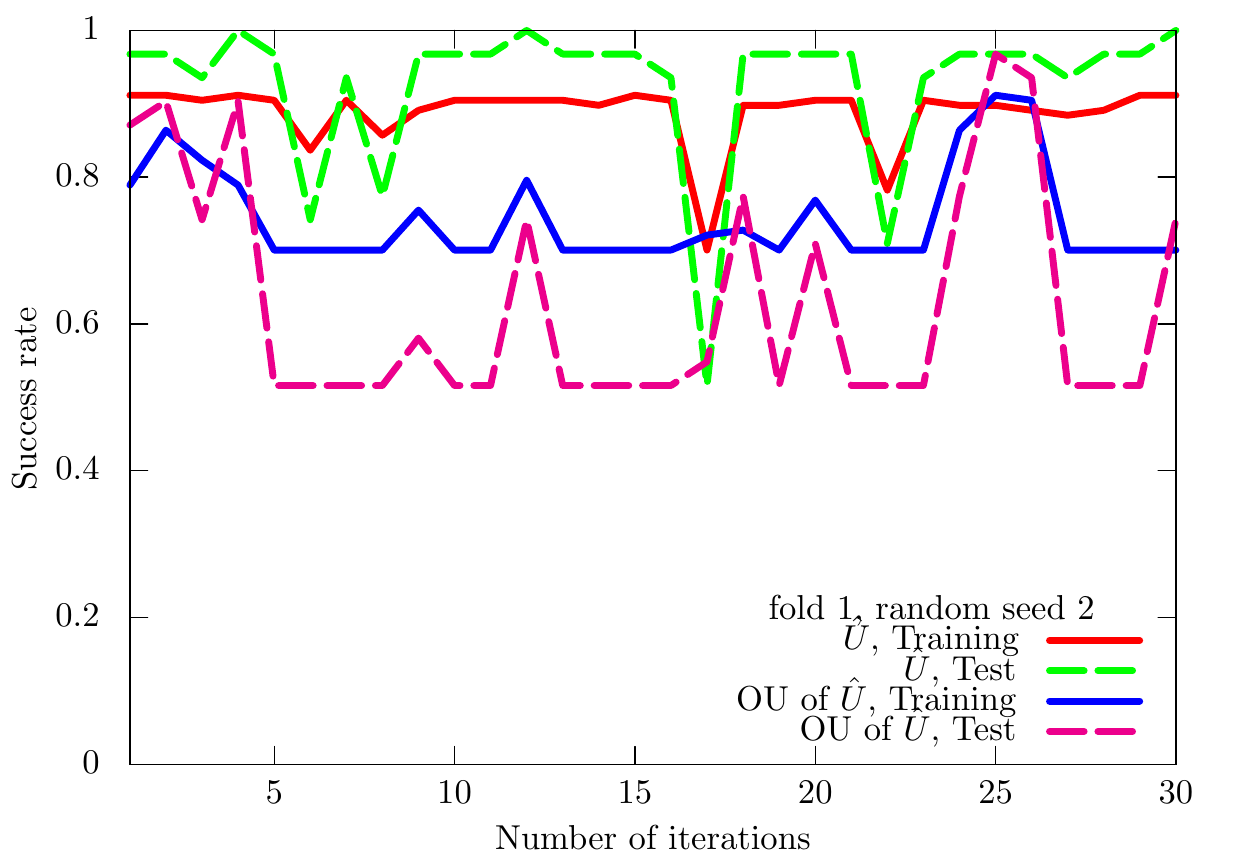}
\includegraphics[scale=0.25]{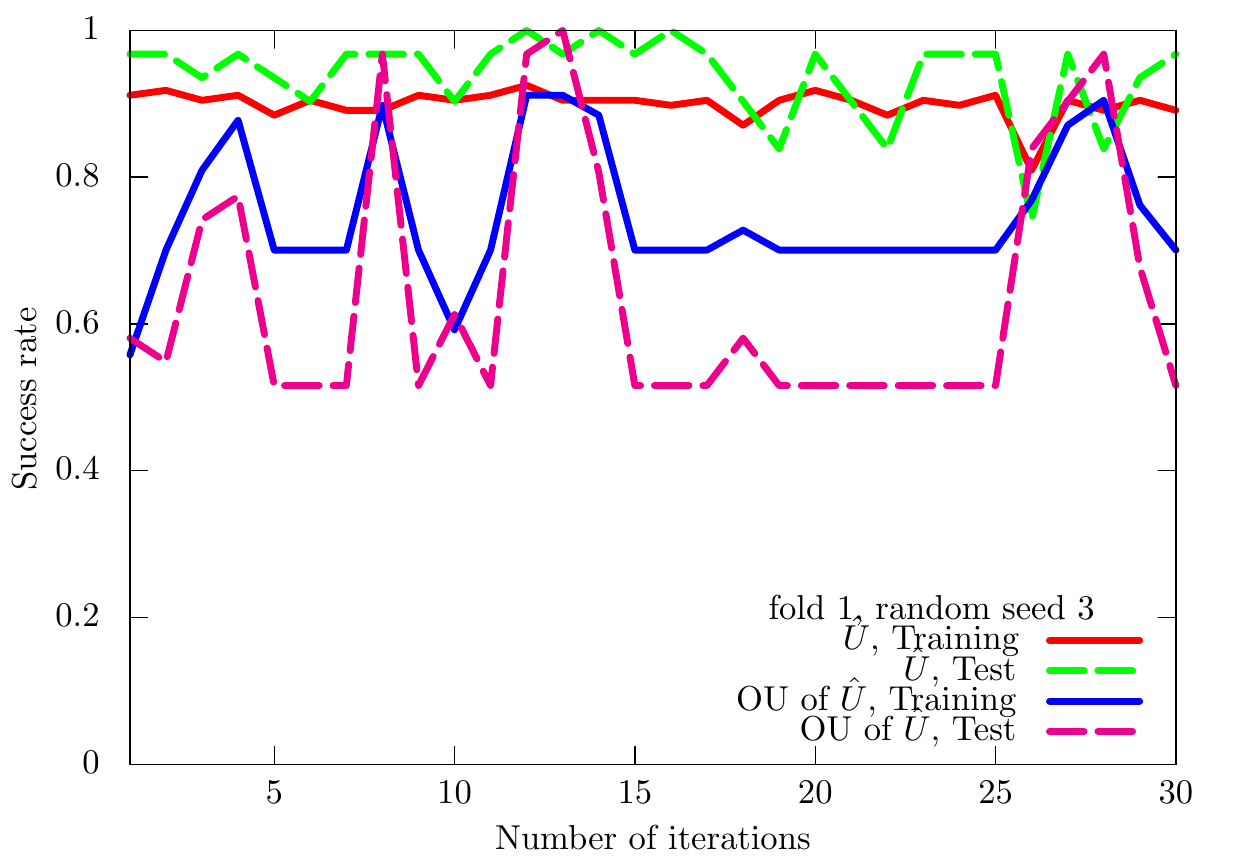}
\includegraphics[scale=0.25]{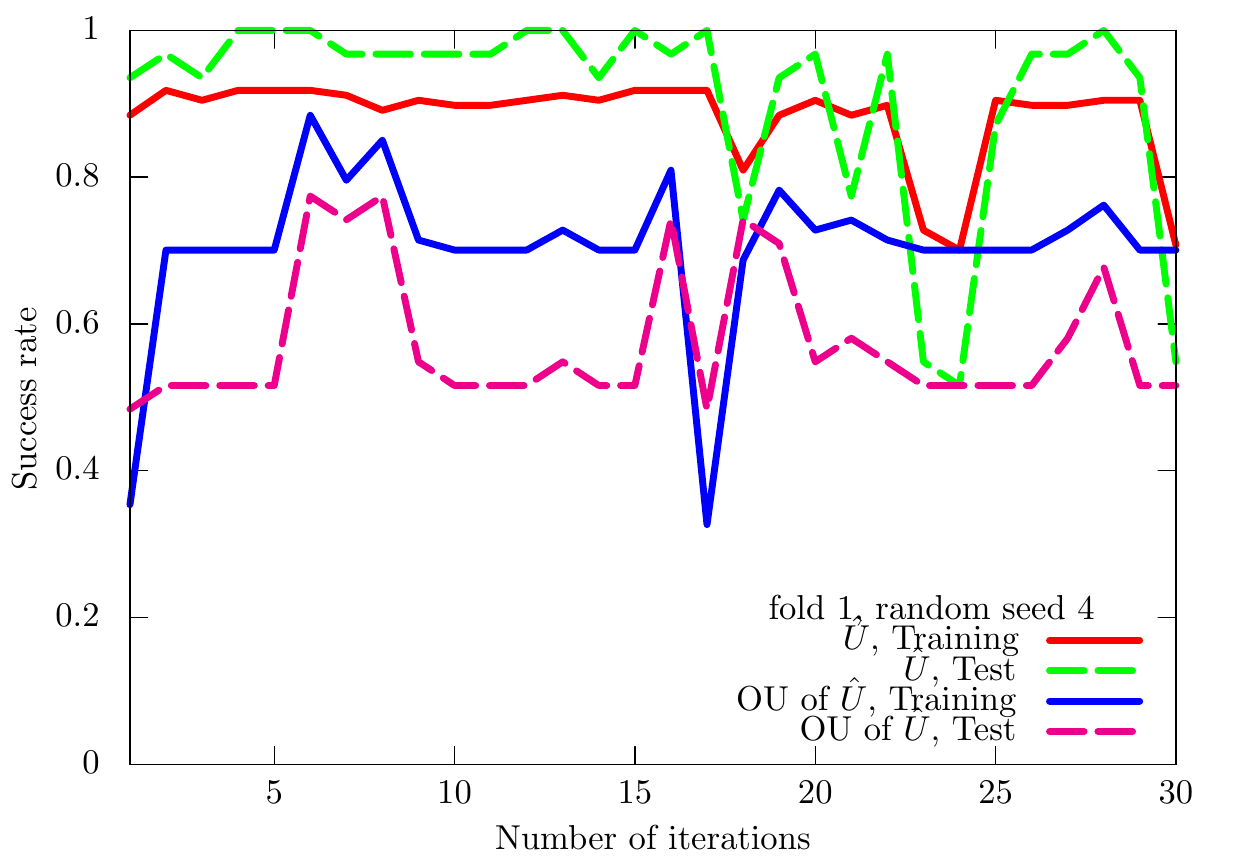}
\includegraphics[scale=0.25]{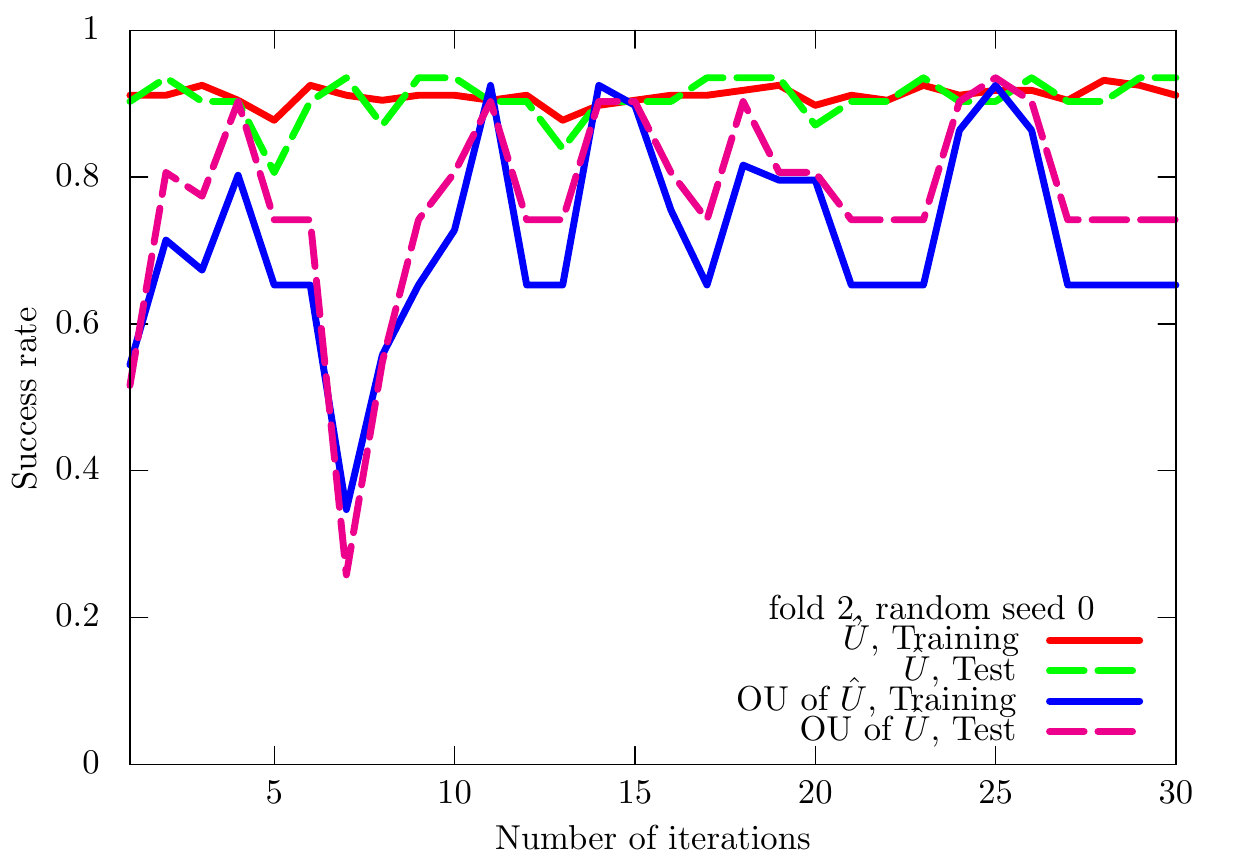}
\includegraphics[scale=0.25]{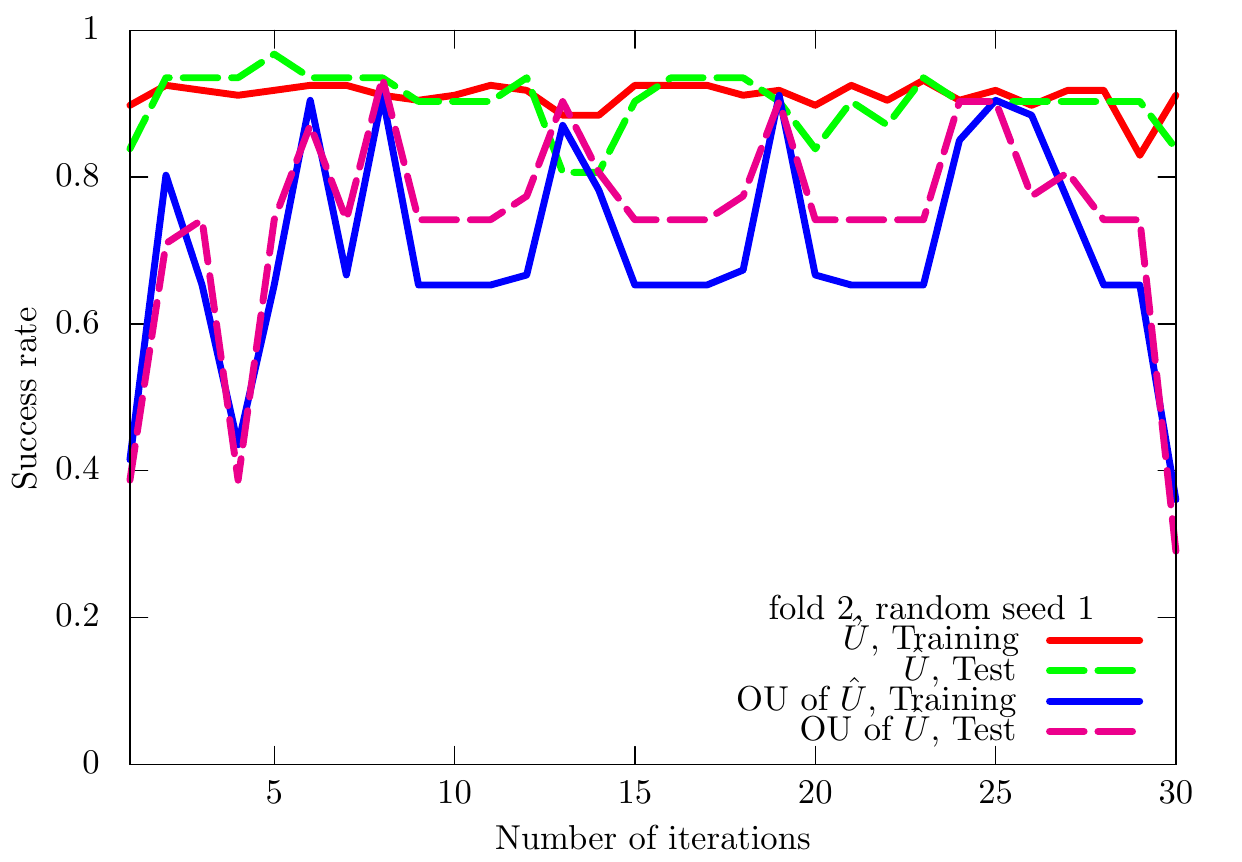}
\includegraphics[scale=0.25]{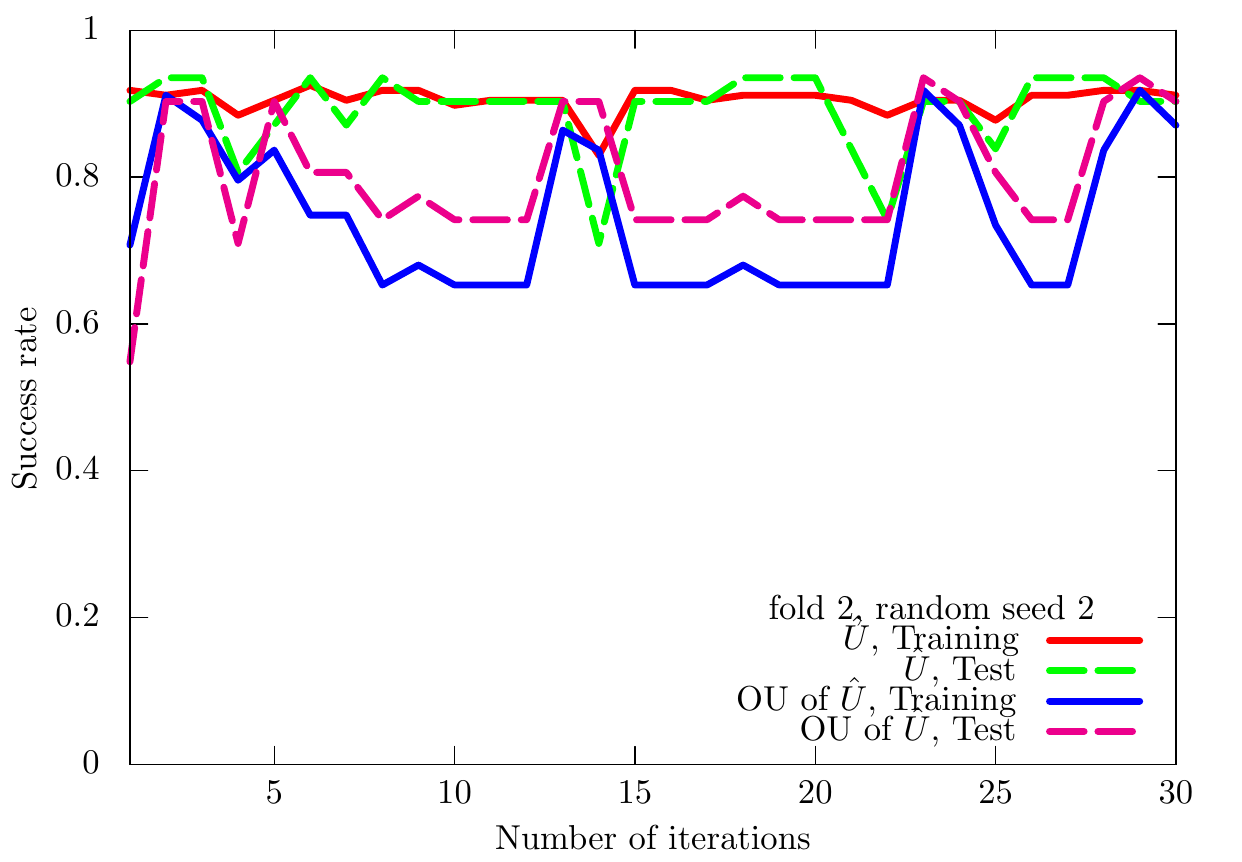}
\includegraphics[scale=0.25]{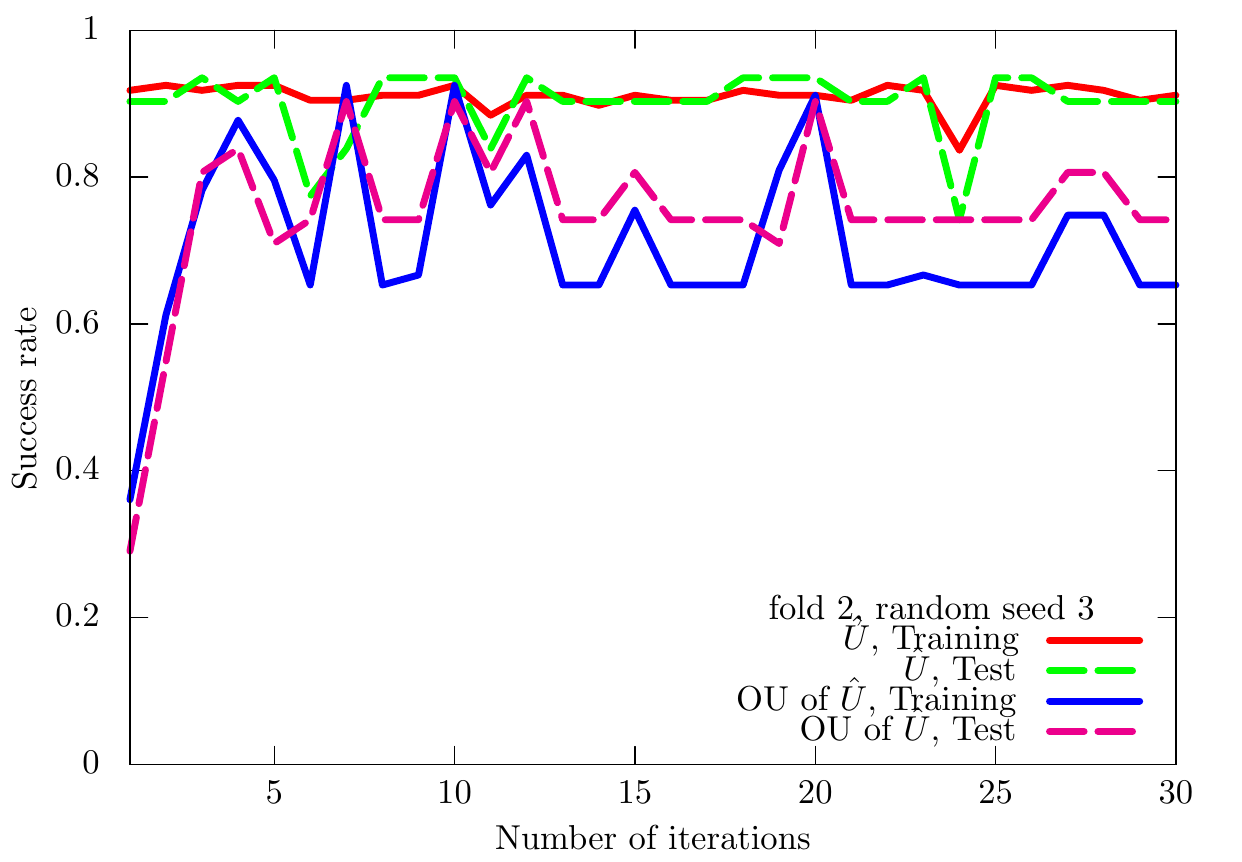}
\includegraphics[scale=0.25]{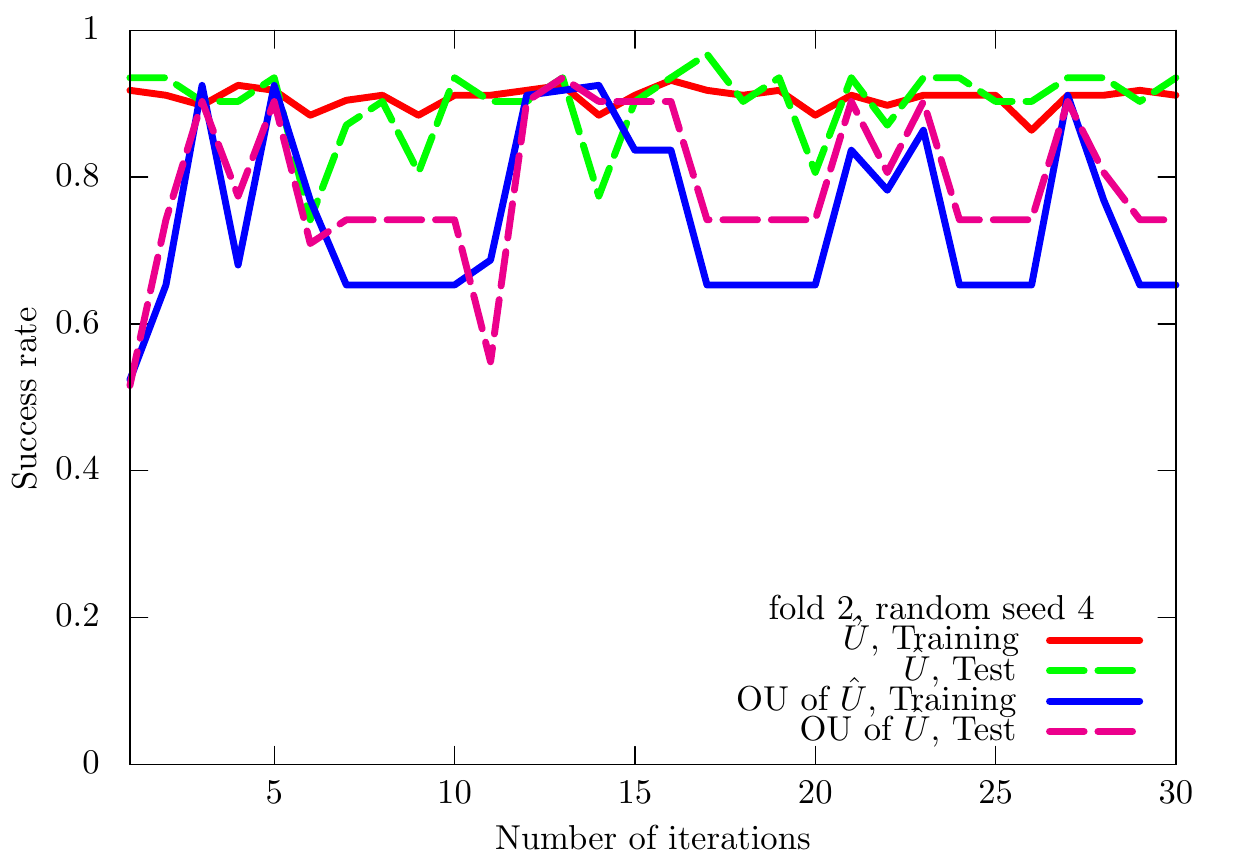}
\includegraphics[scale=0.25]{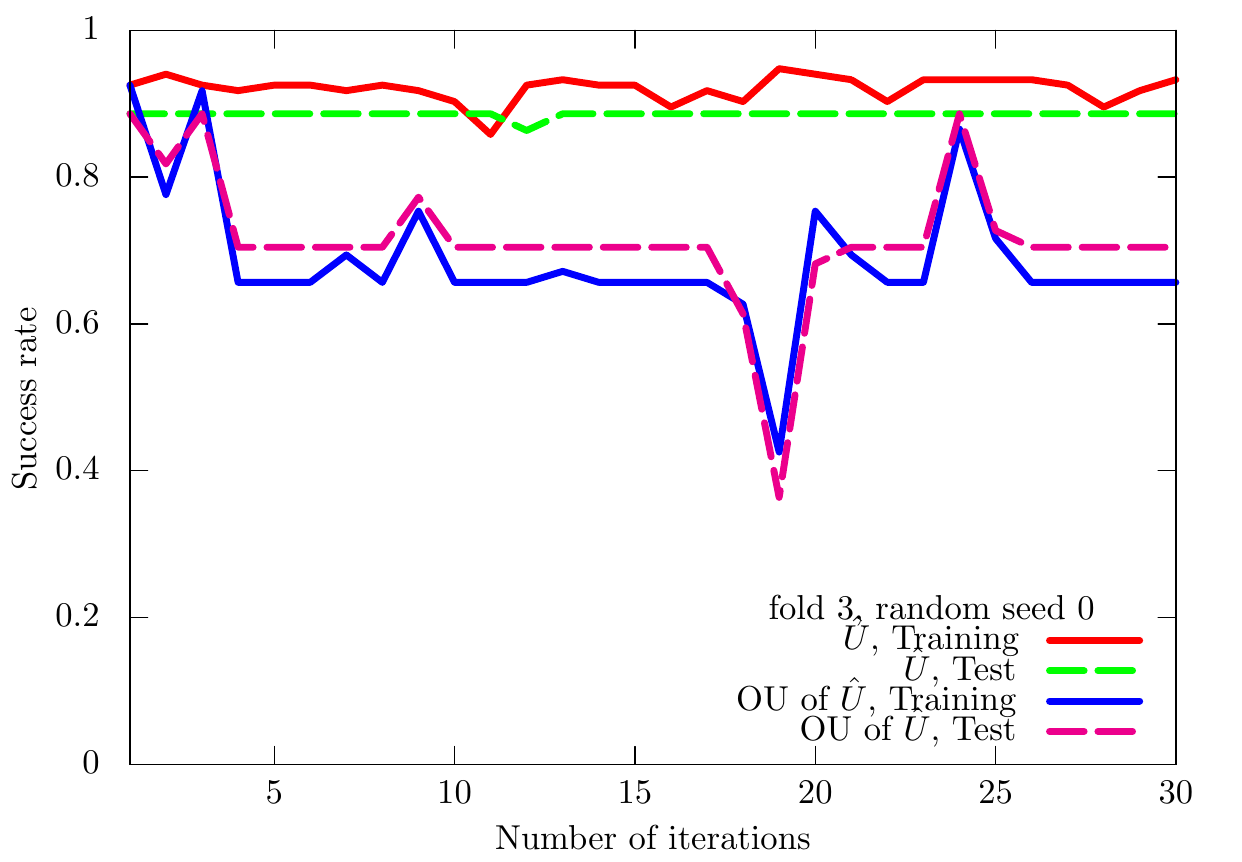}
\includegraphics[scale=0.25]{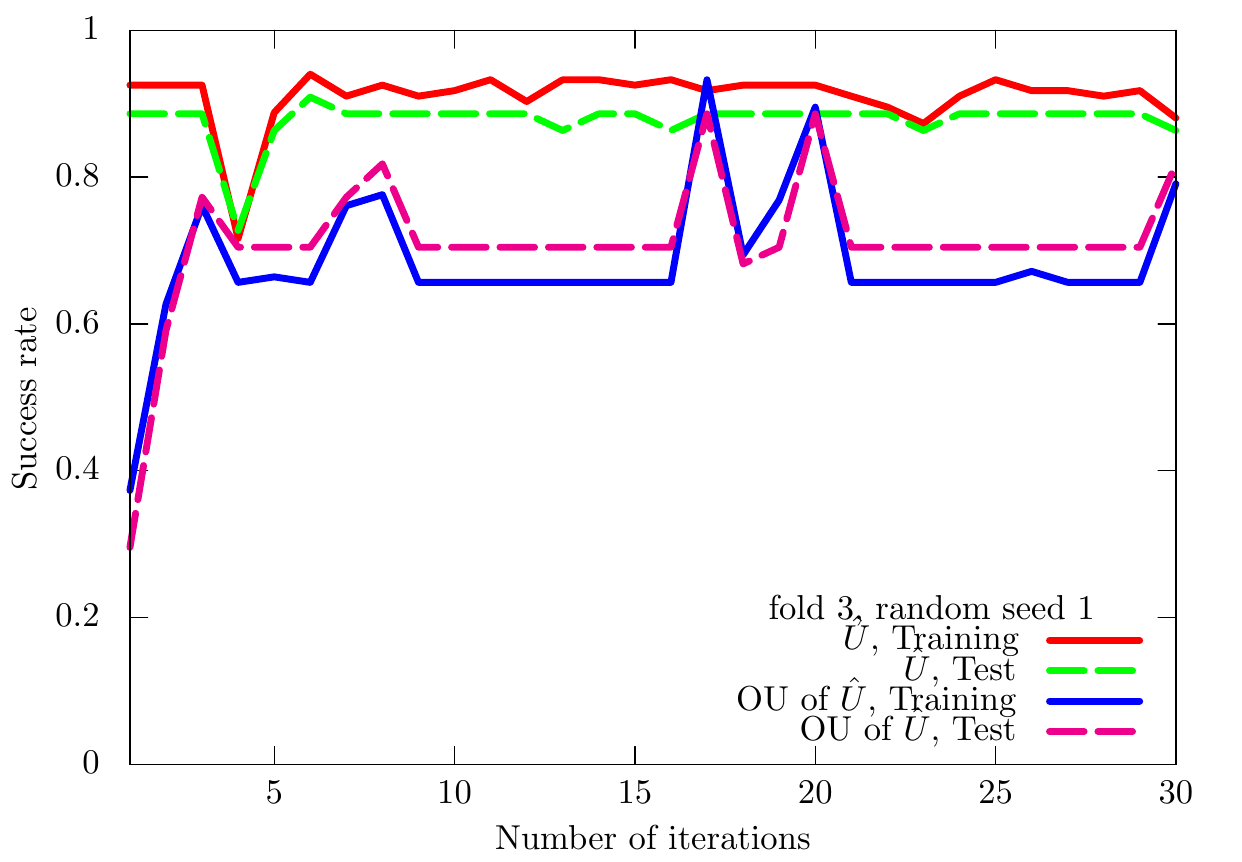}
\includegraphics[scale=0.25]{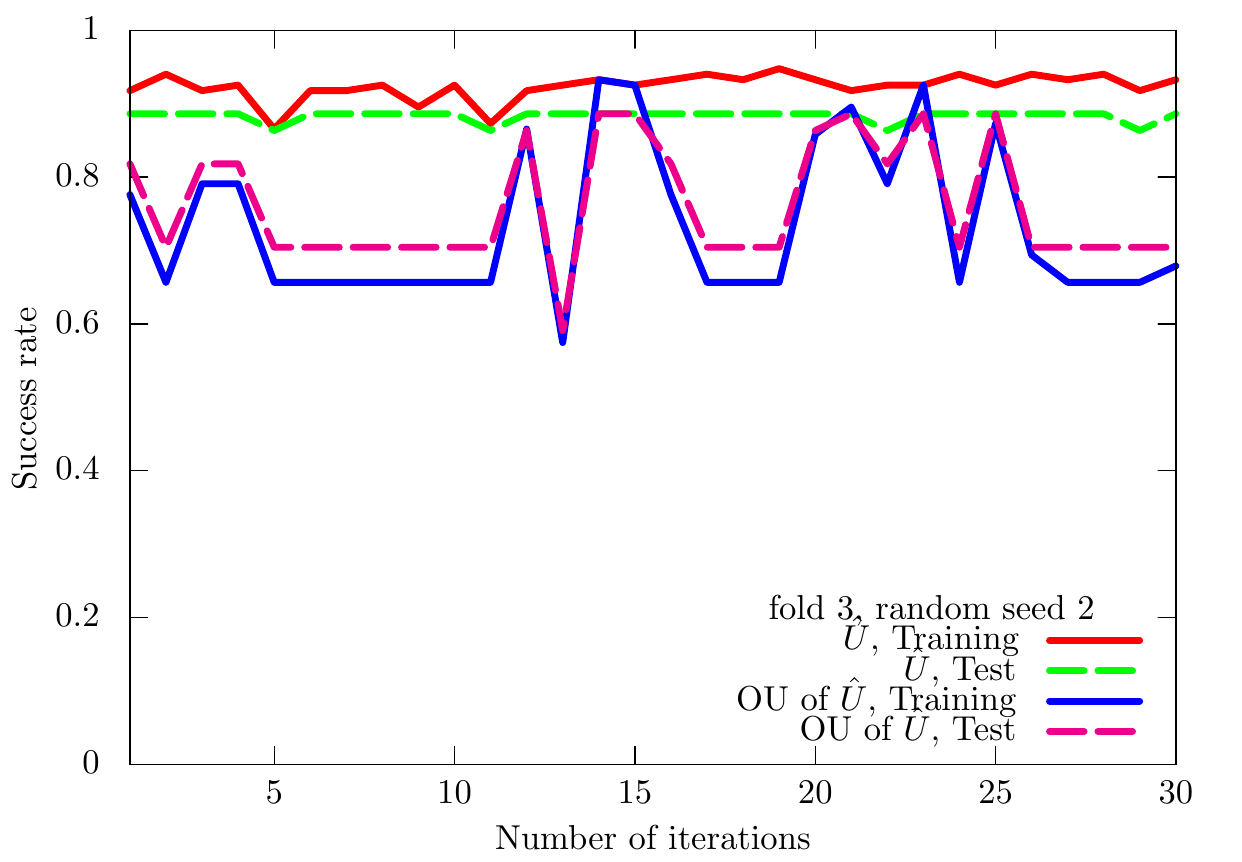}
\includegraphics[scale=0.25]{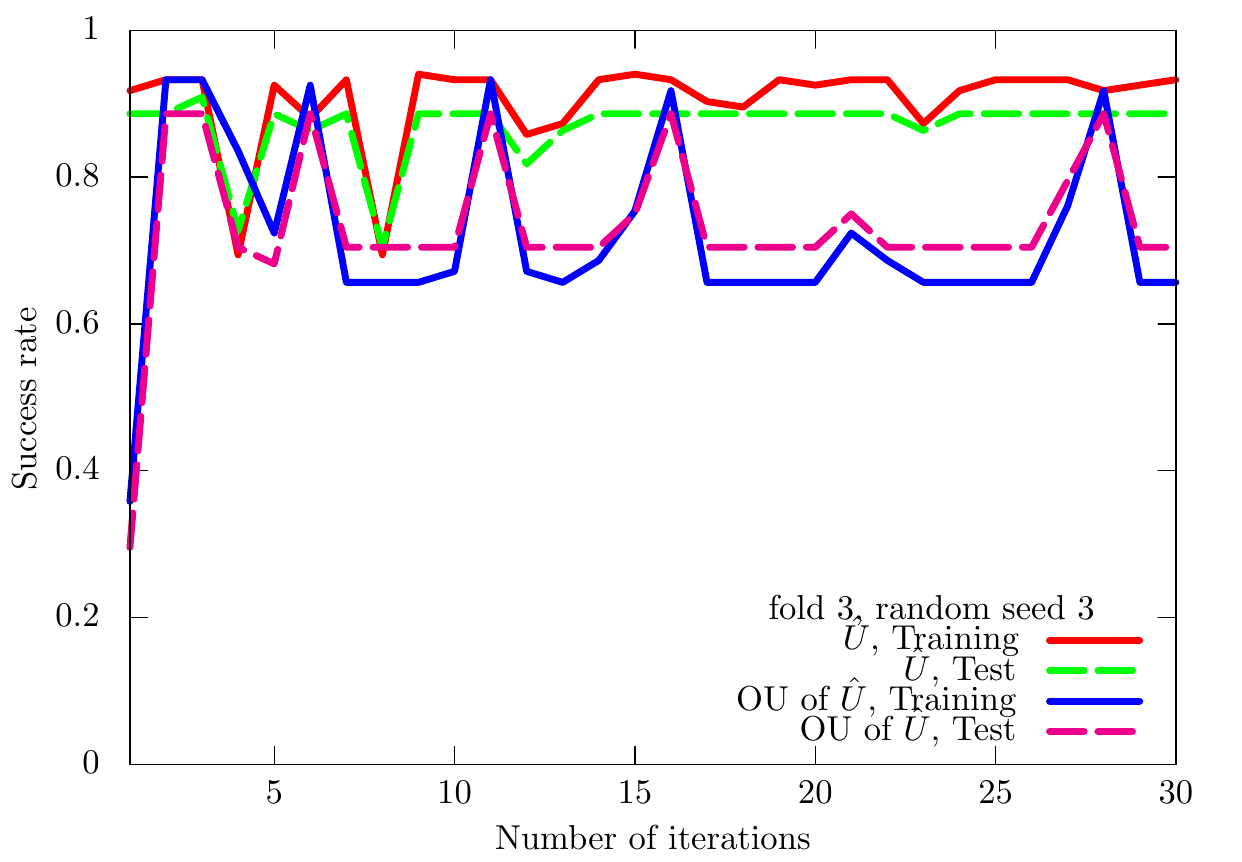}
\includegraphics[scale=0.25]{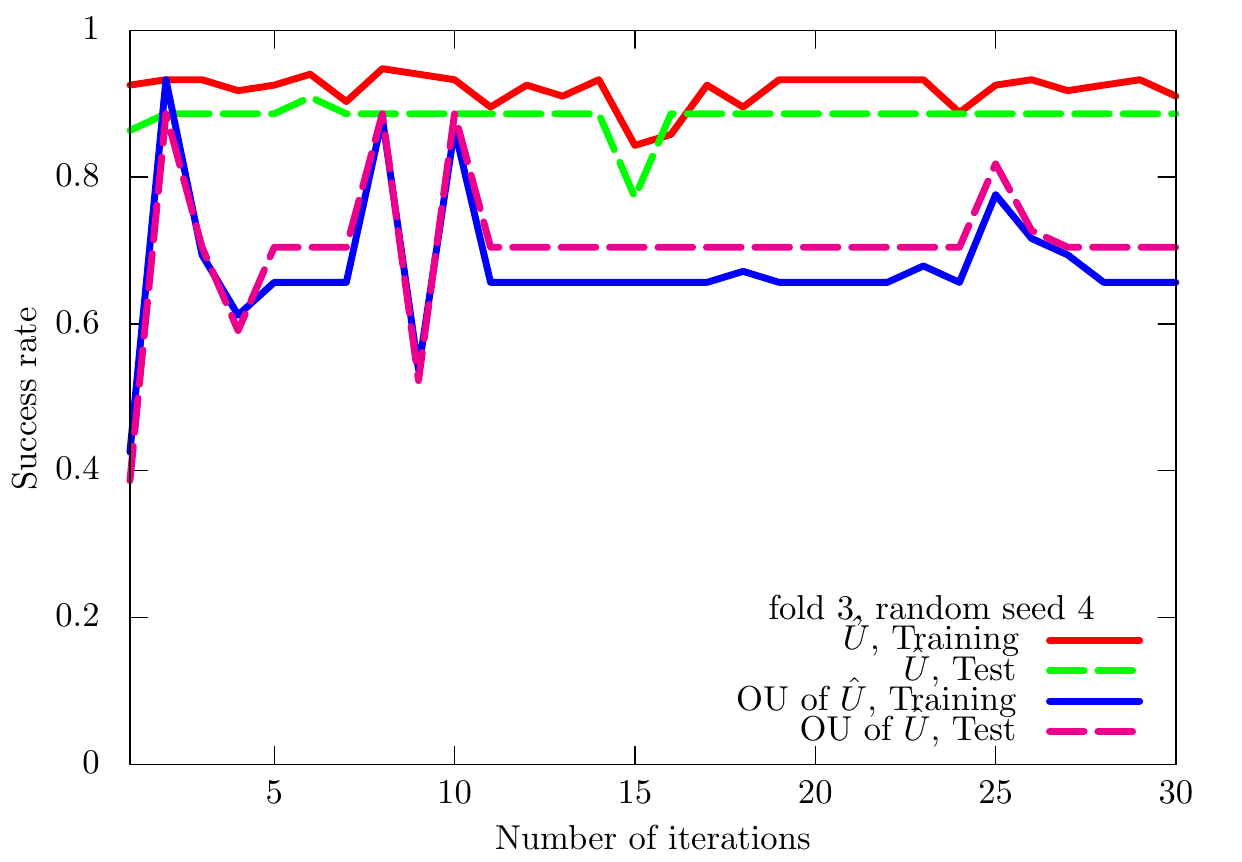}
\includegraphics[scale=0.25]{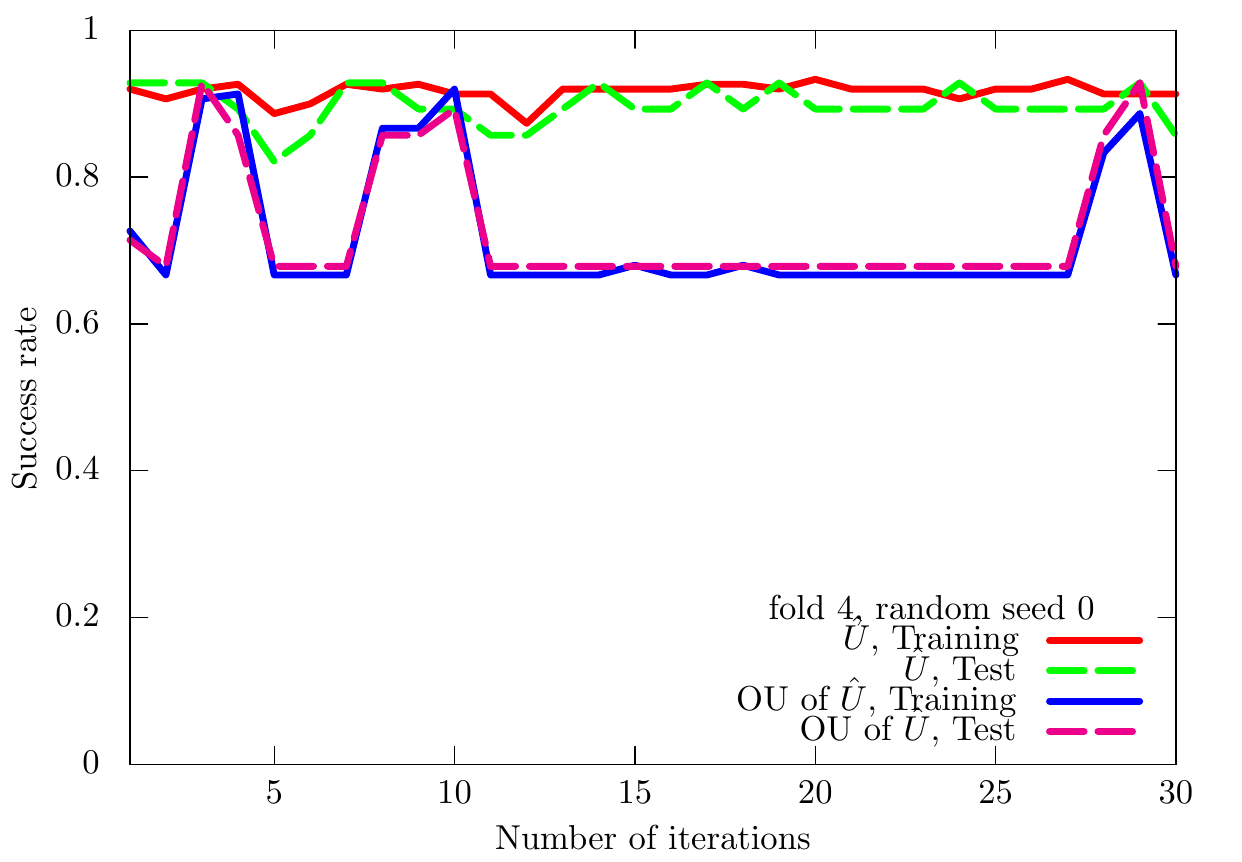}
\includegraphics[scale=0.25]{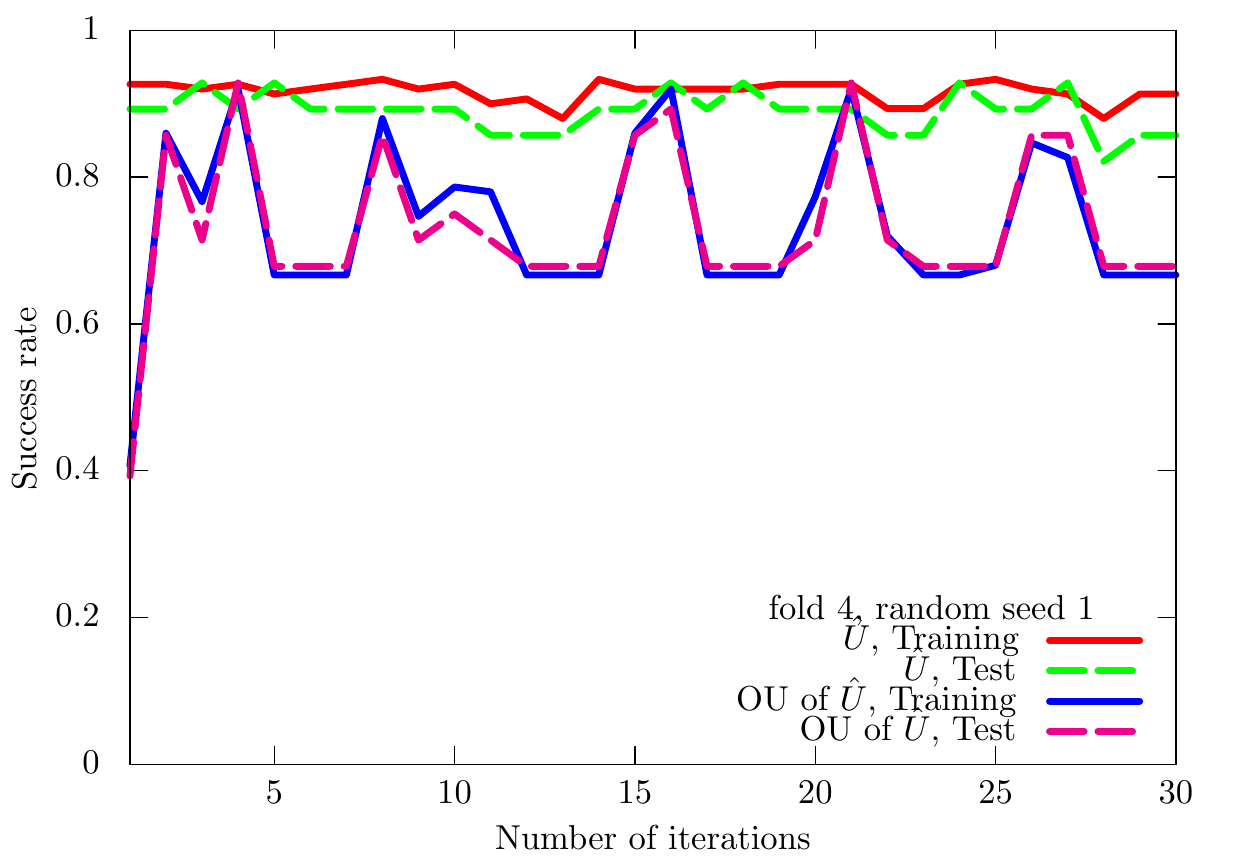}
\includegraphics[scale=0.25]{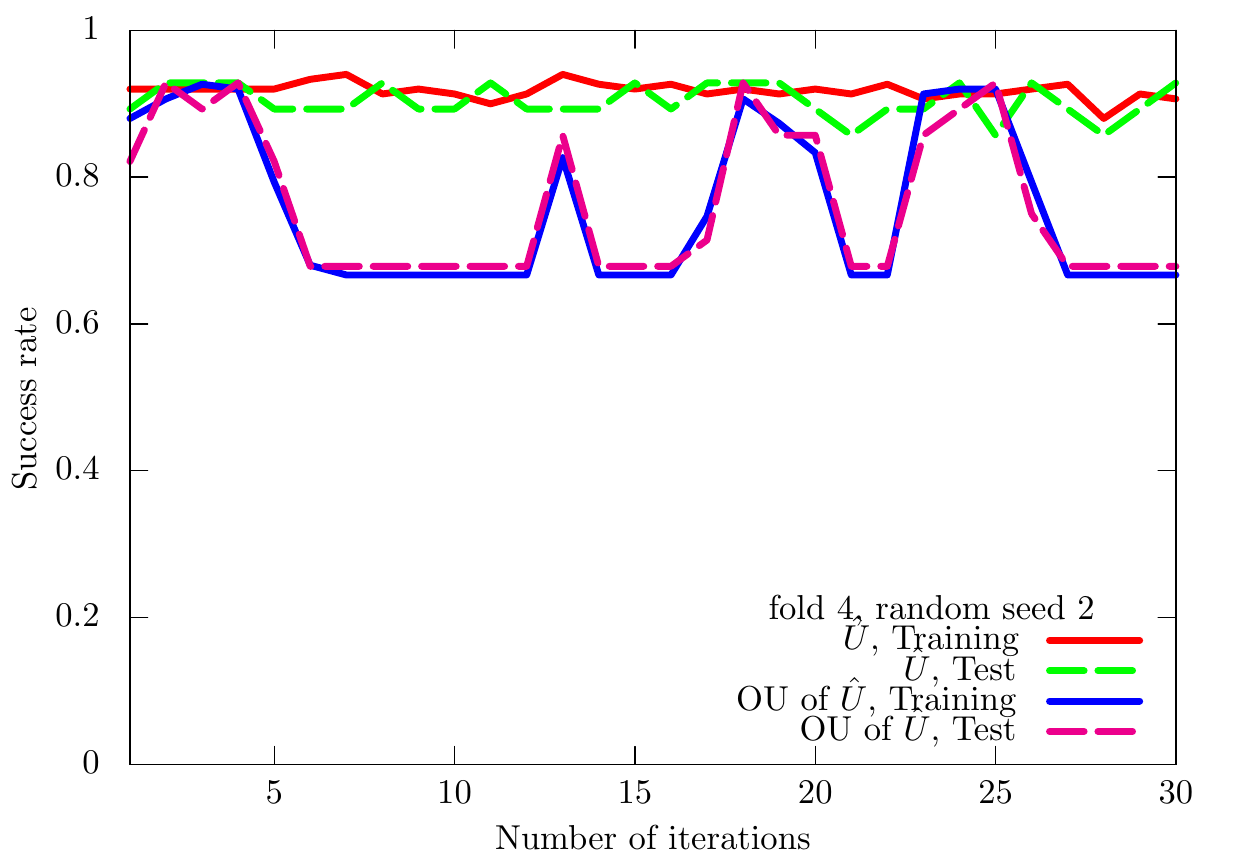}
\includegraphics[scale=0.25]{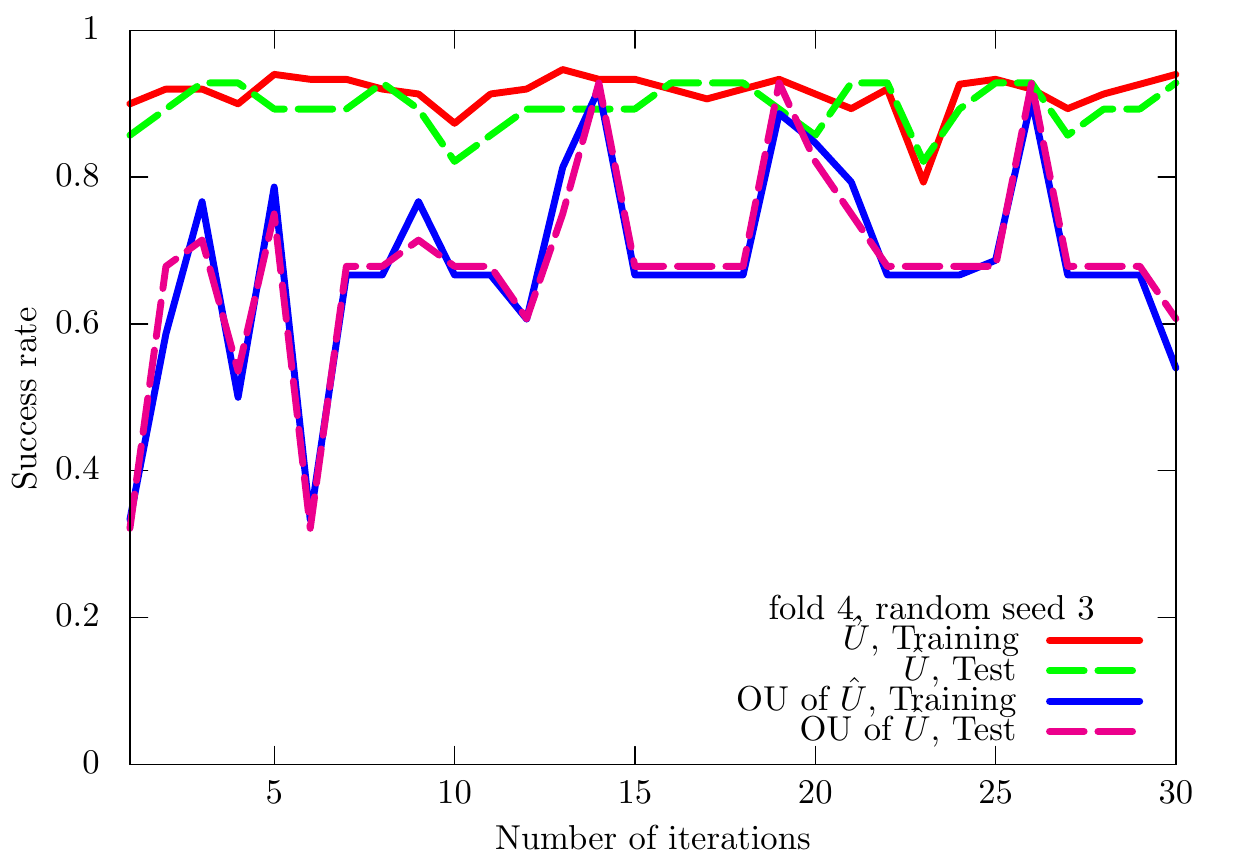}
\includegraphics[scale=0.25]{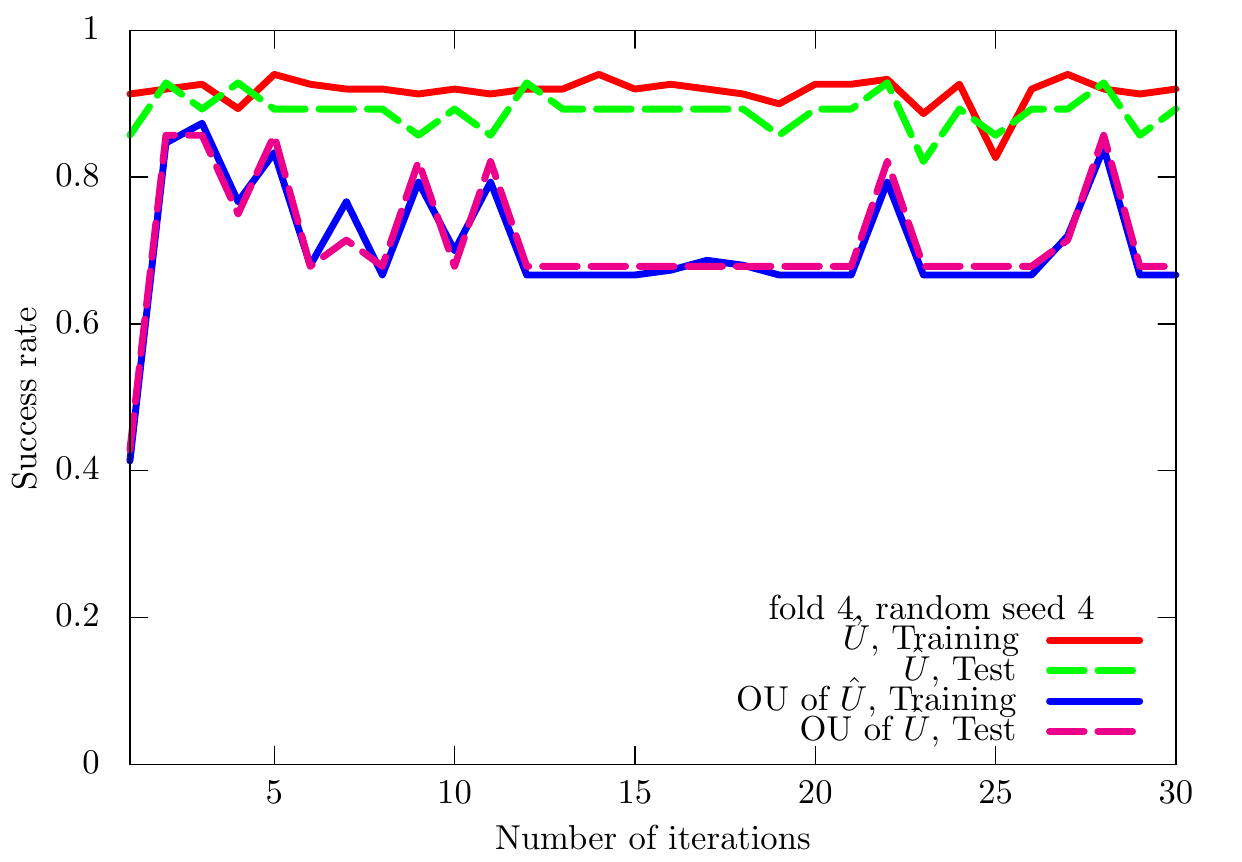}
\caption{Results of the UKM ($\hat{X}$ and $\hat{P}$) on the $5$-fold datasets with $5$ different random seeds for the wine dataset ($0$ or non-$0$). We use complex matrices and set $\theta_\mathrm{bias} = 0$. We set $r = 0.010$.}
\label{supp-arXiv-numerical-result-raw-data-fold-001-rand-001-UKM-P-UCI-wine-0-non0}
\end{figure*}

\begin{figure*}[htb]
\centering
\includegraphics[scale=0.25]{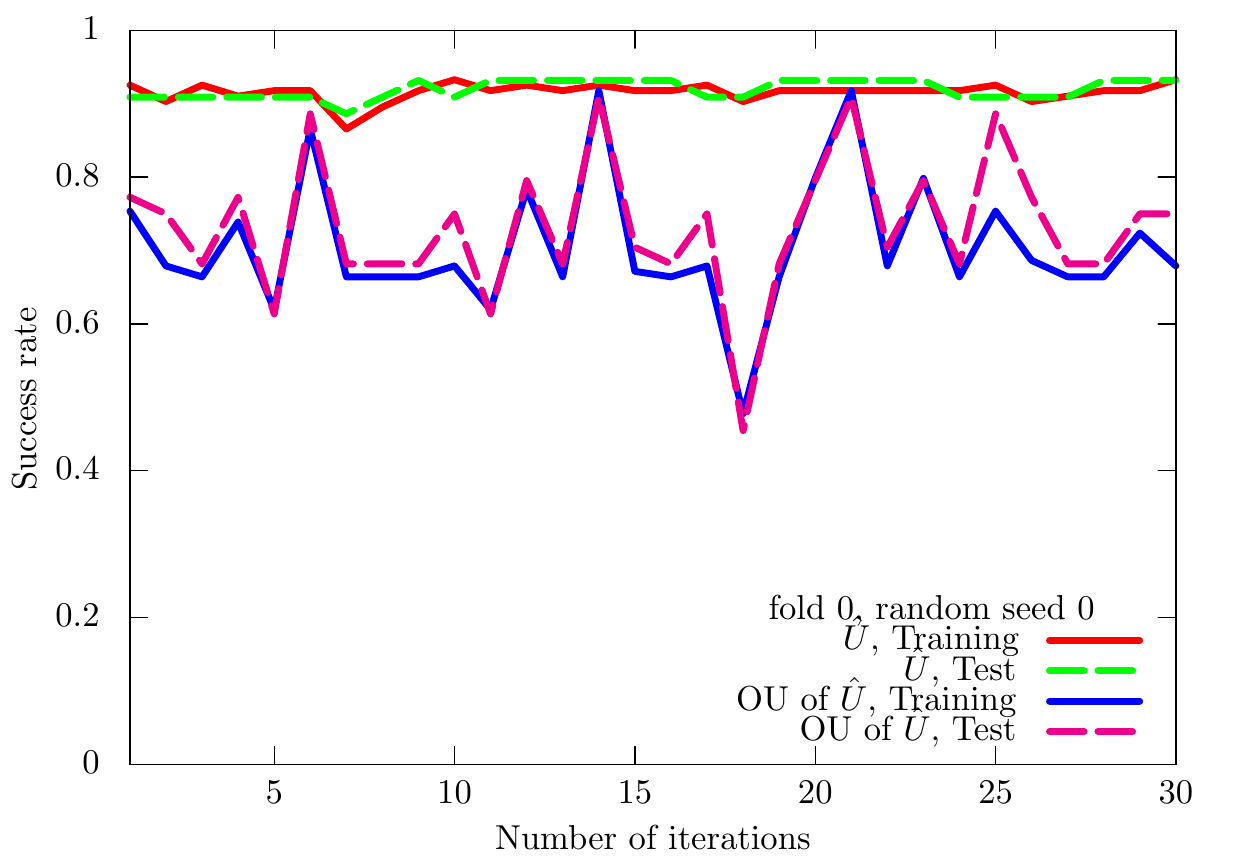}
\includegraphics[scale=0.25]{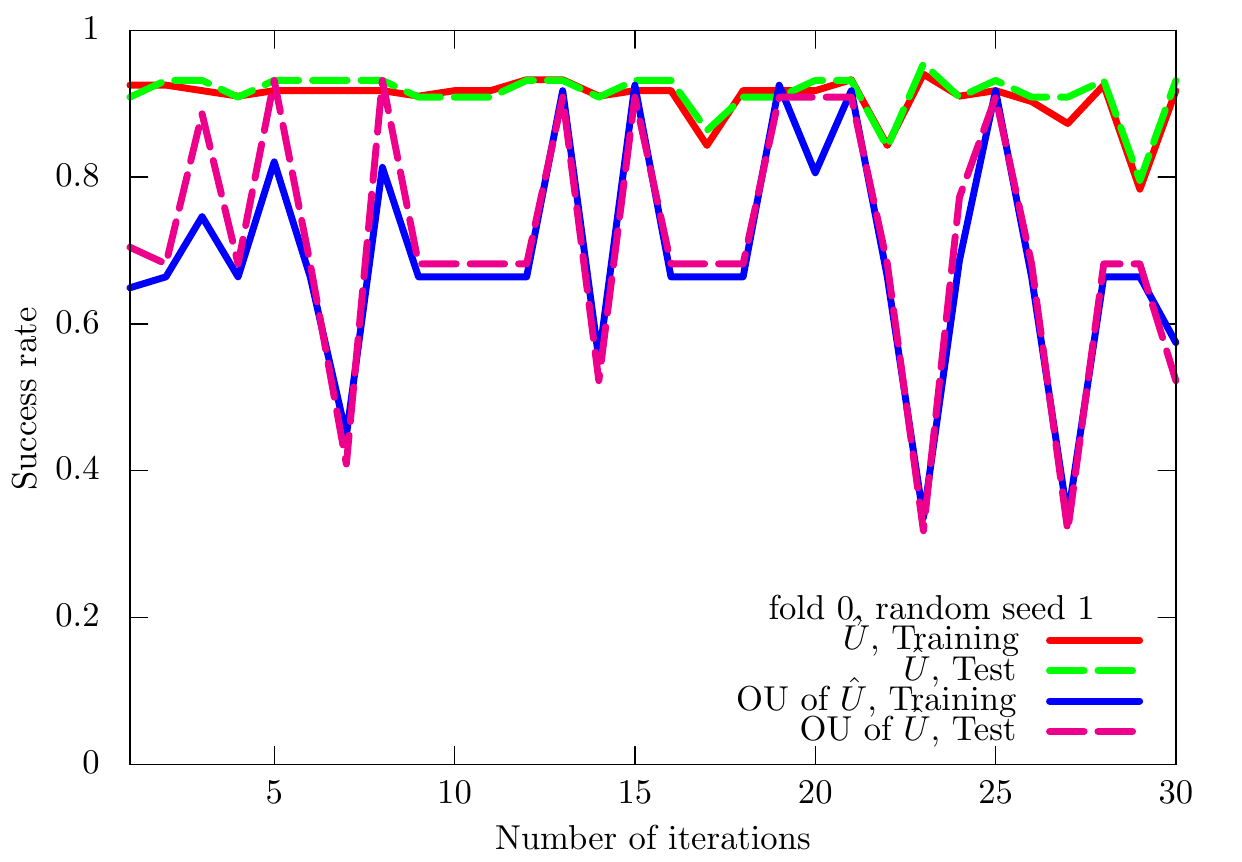}
\includegraphics[scale=0.25]{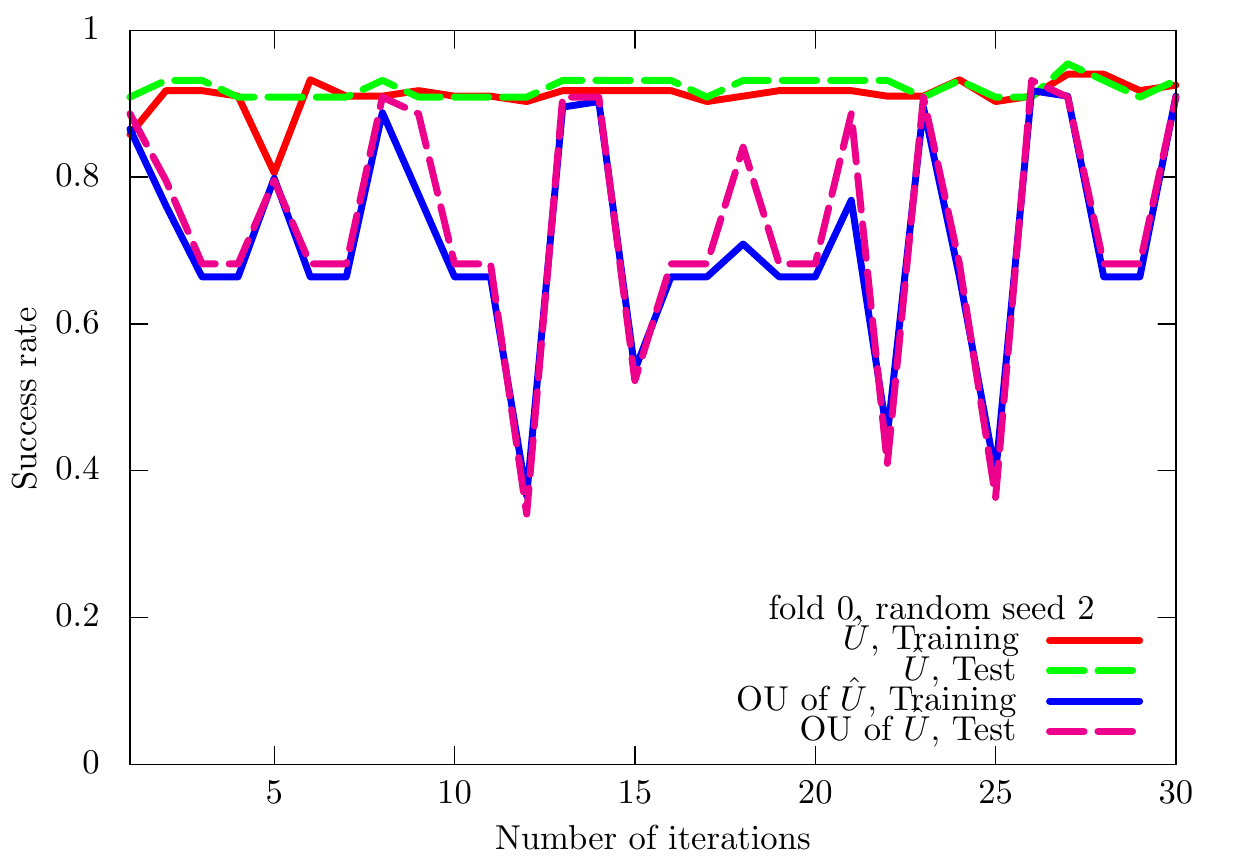}
\includegraphics[scale=0.25]{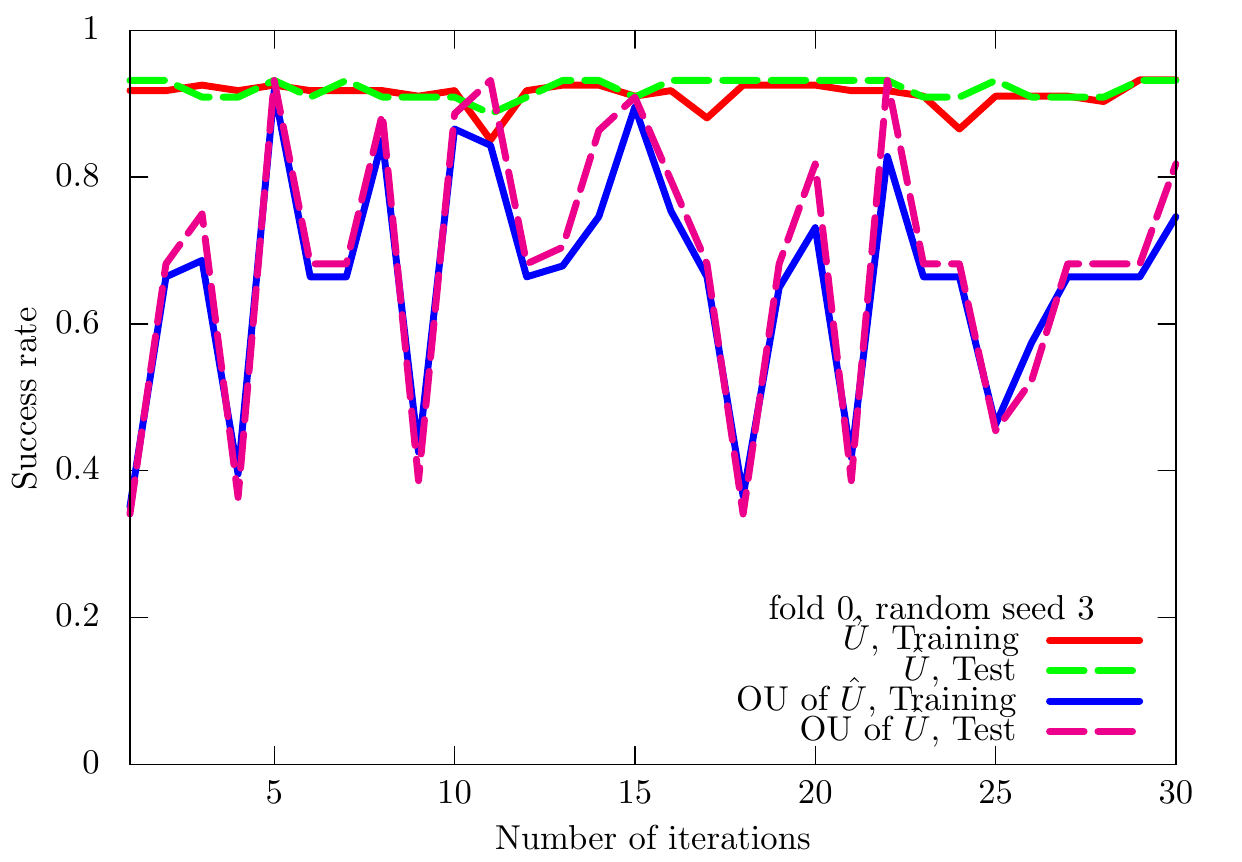}
\includegraphics[scale=0.25]{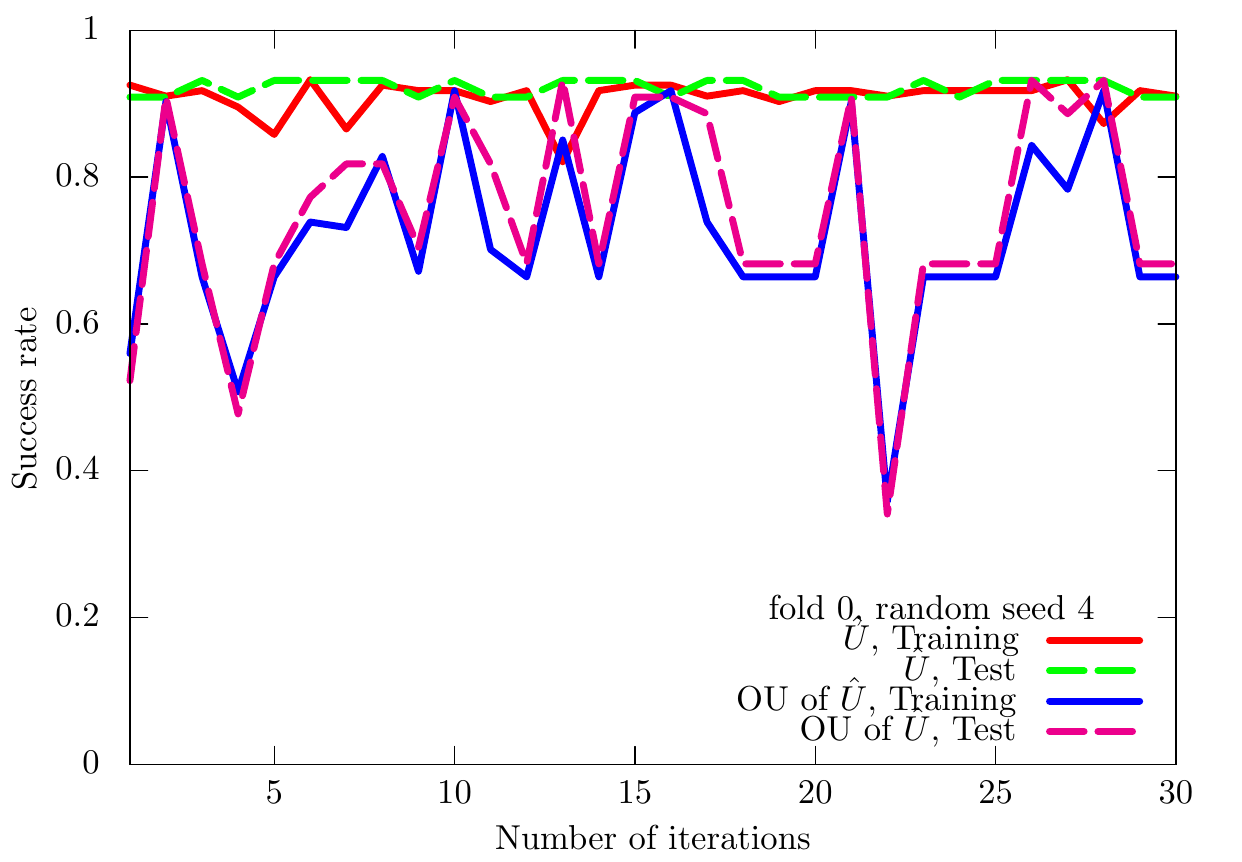}
\includegraphics[scale=0.25]{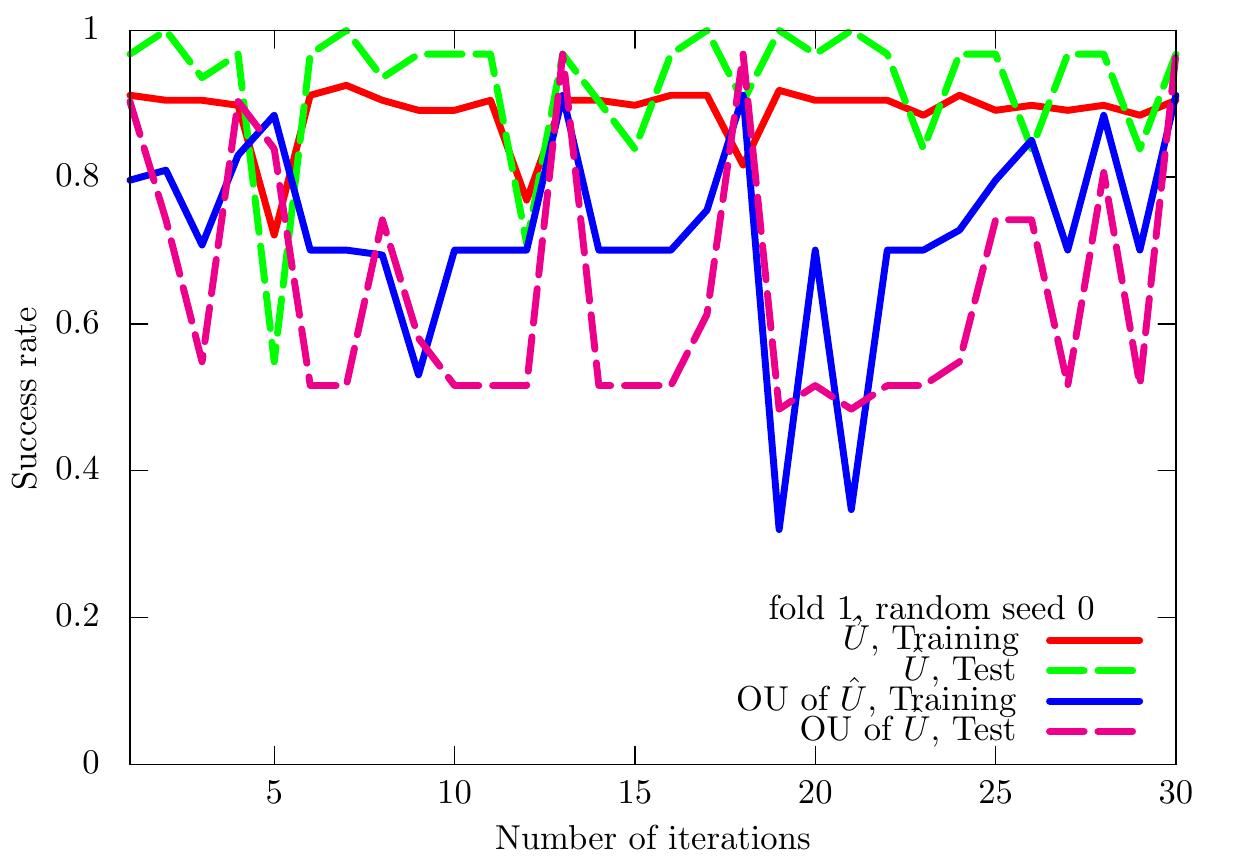}
\includegraphics[scale=0.25]{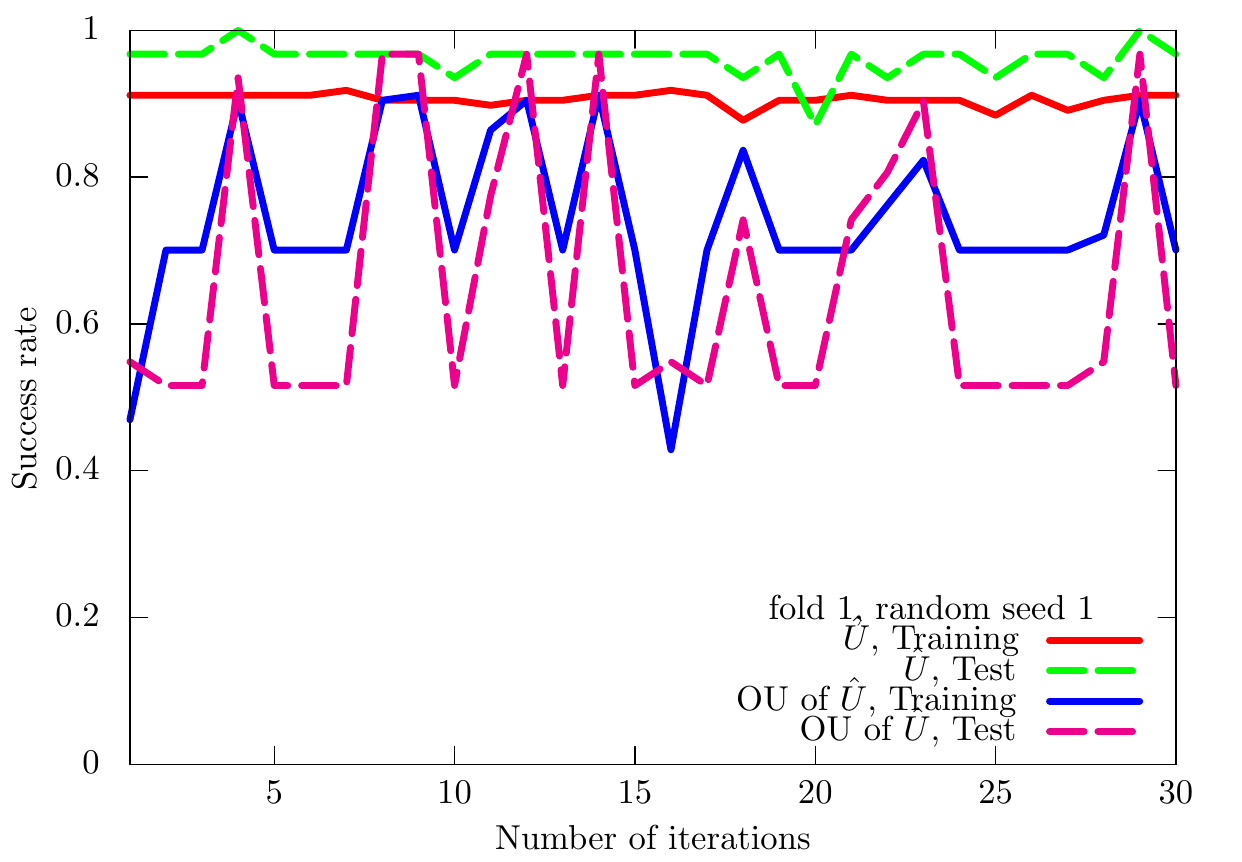}
\includegraphics[scale=0.25]{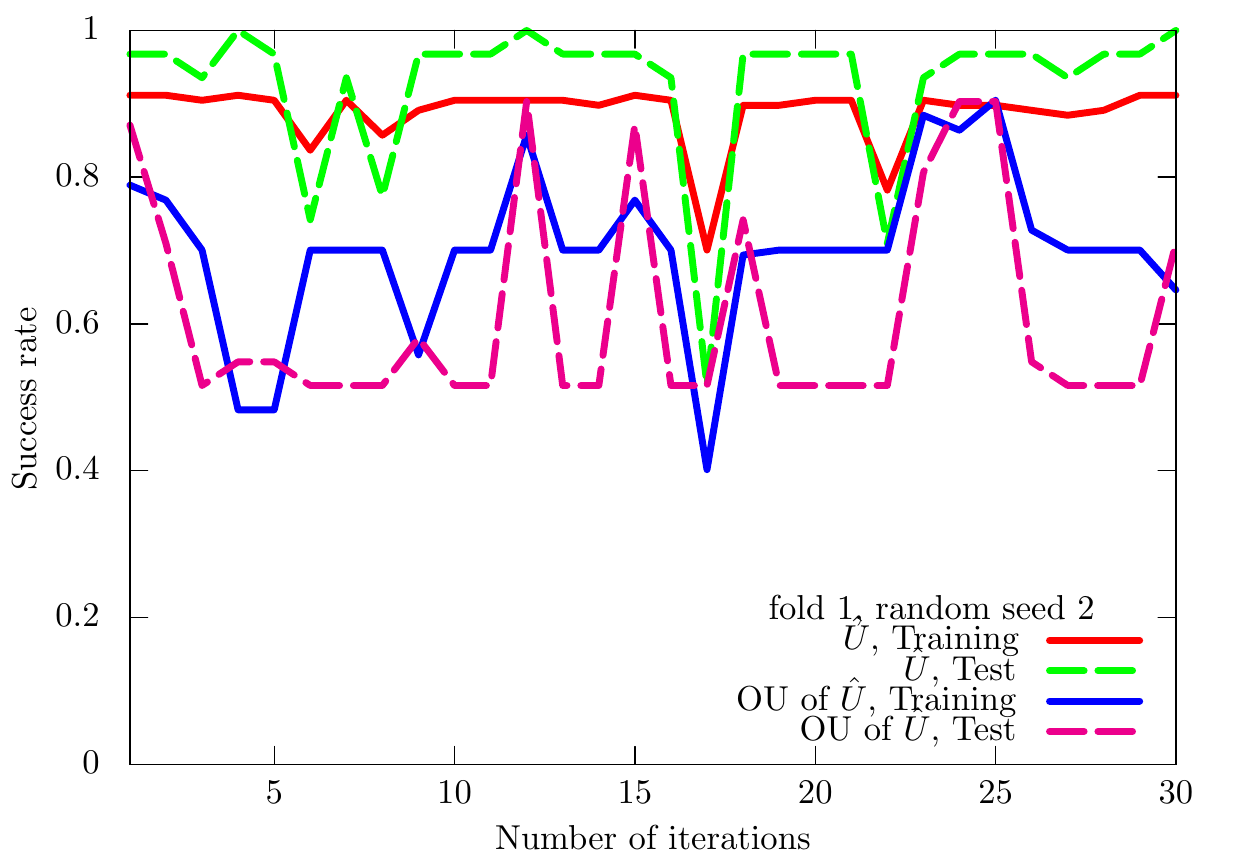}
\includegraphics[scale=0.25]{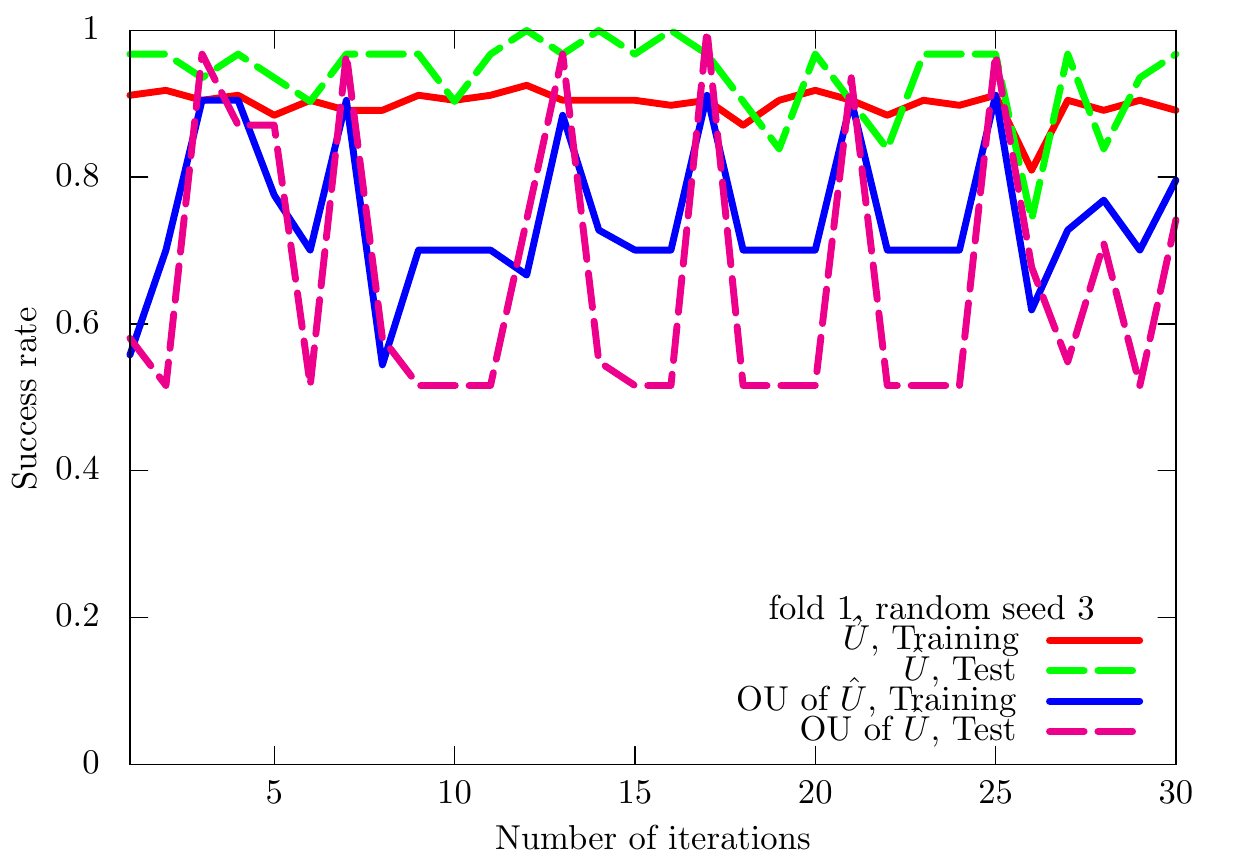}
\includegraphics[scale=0.25]{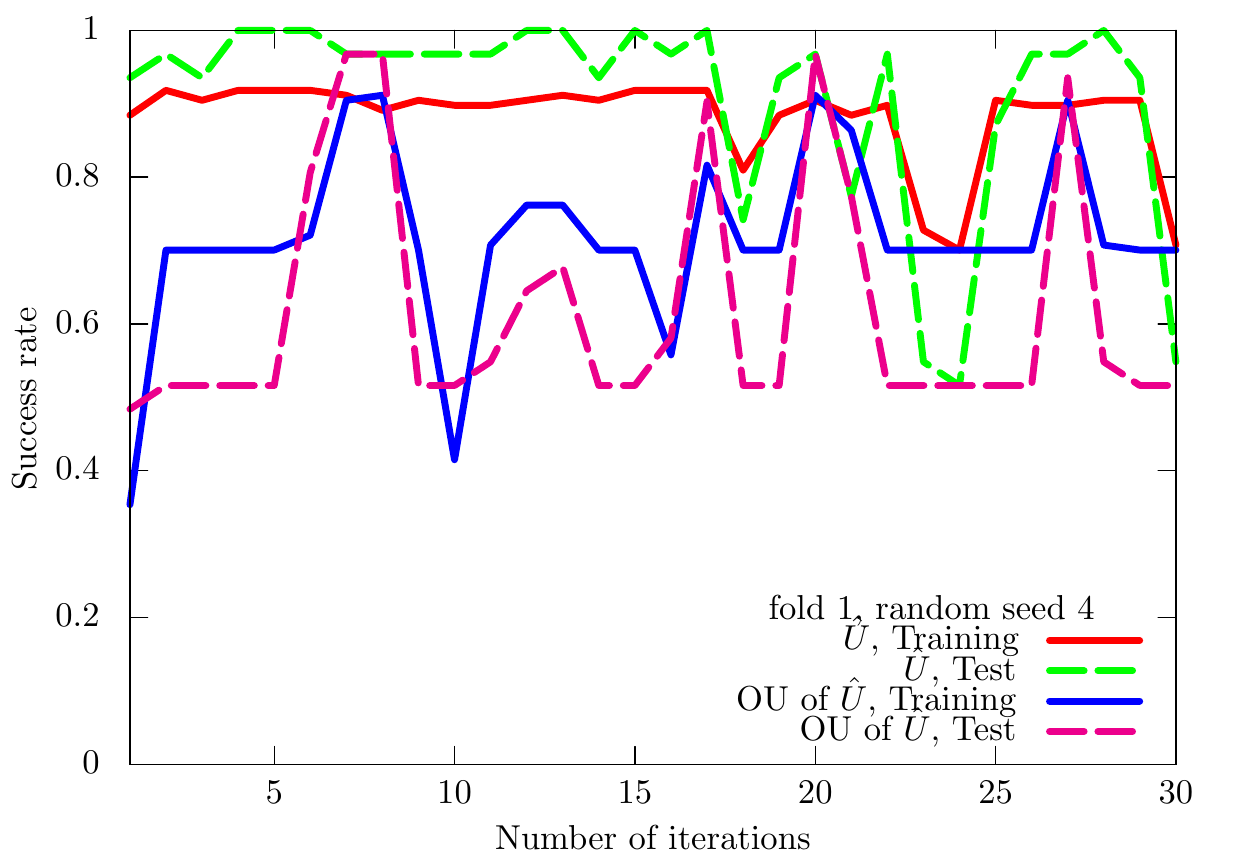}
\includegraphics[scale=0.25]{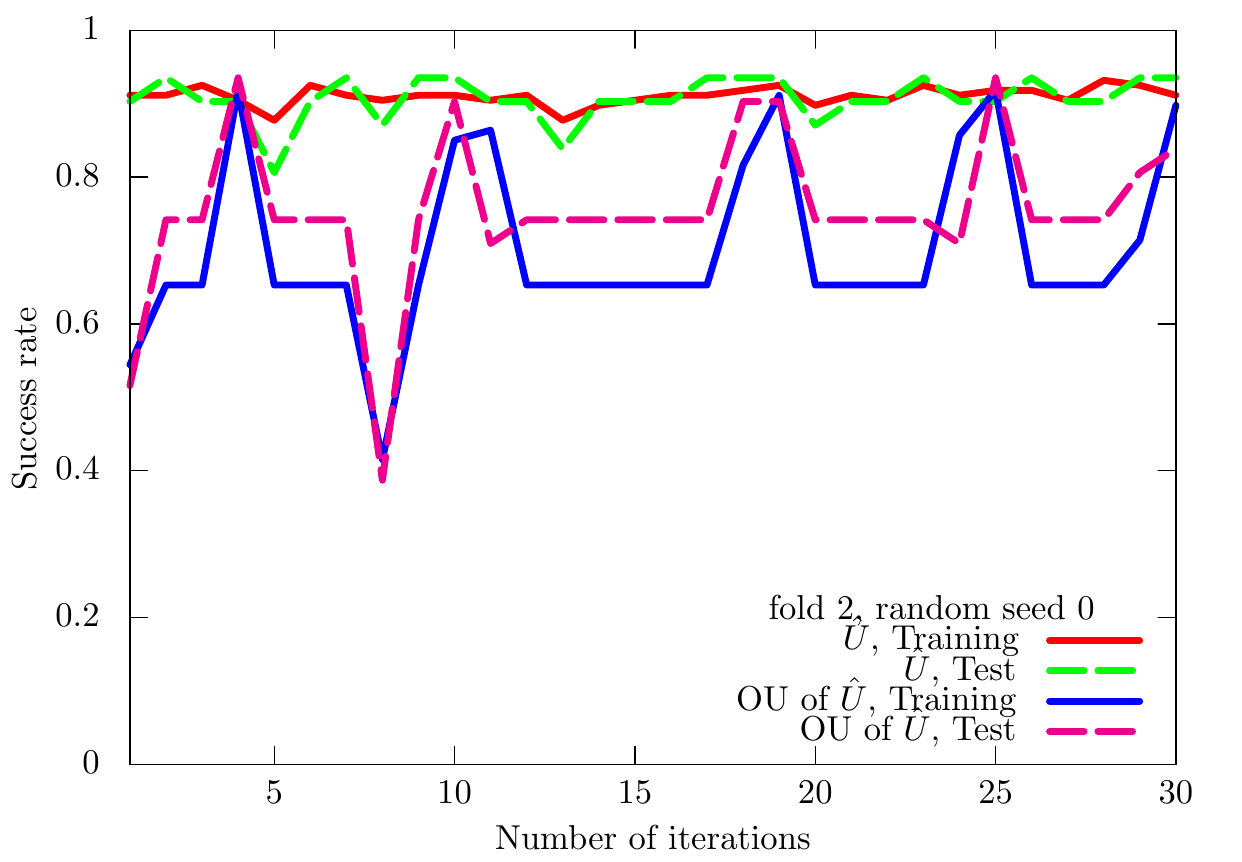}
\includegraphics[scale=0.25]{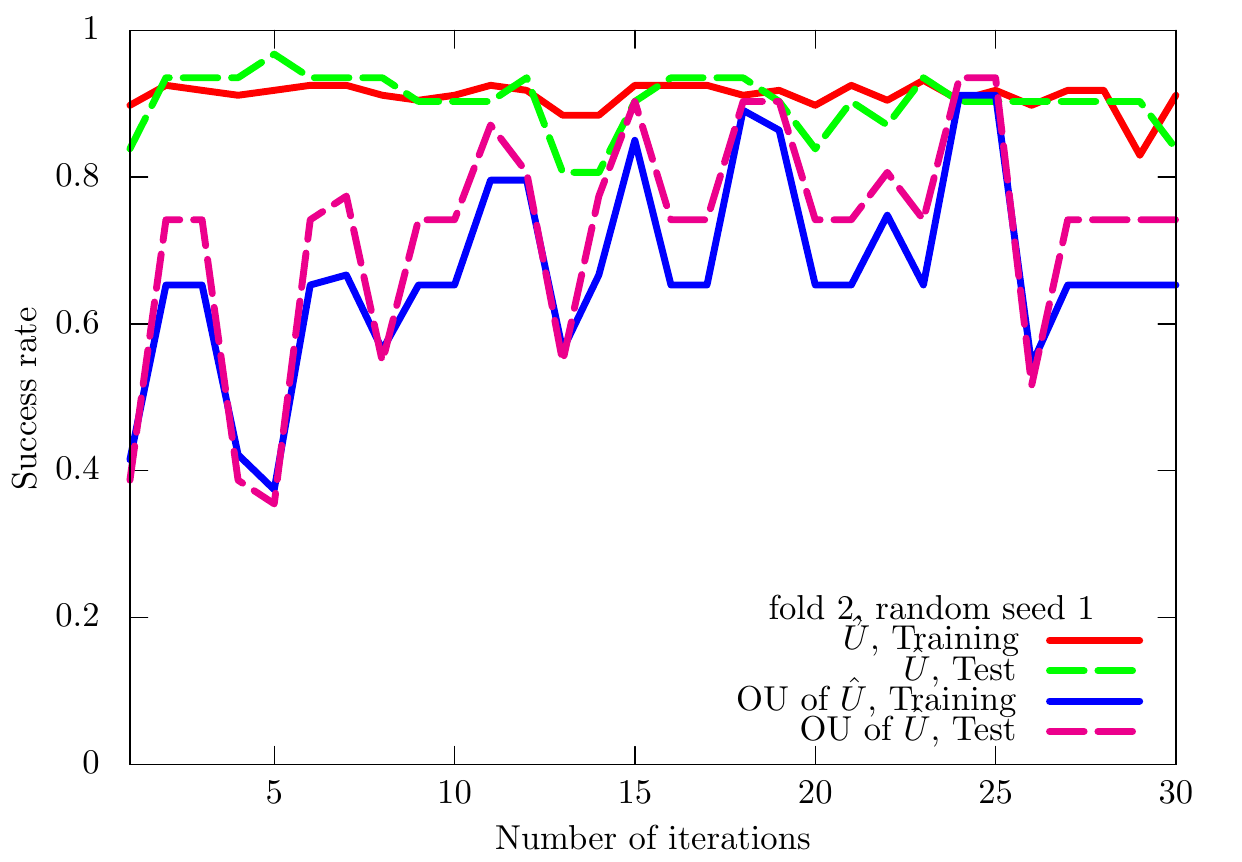}
\includegraphics[scale=0.25]{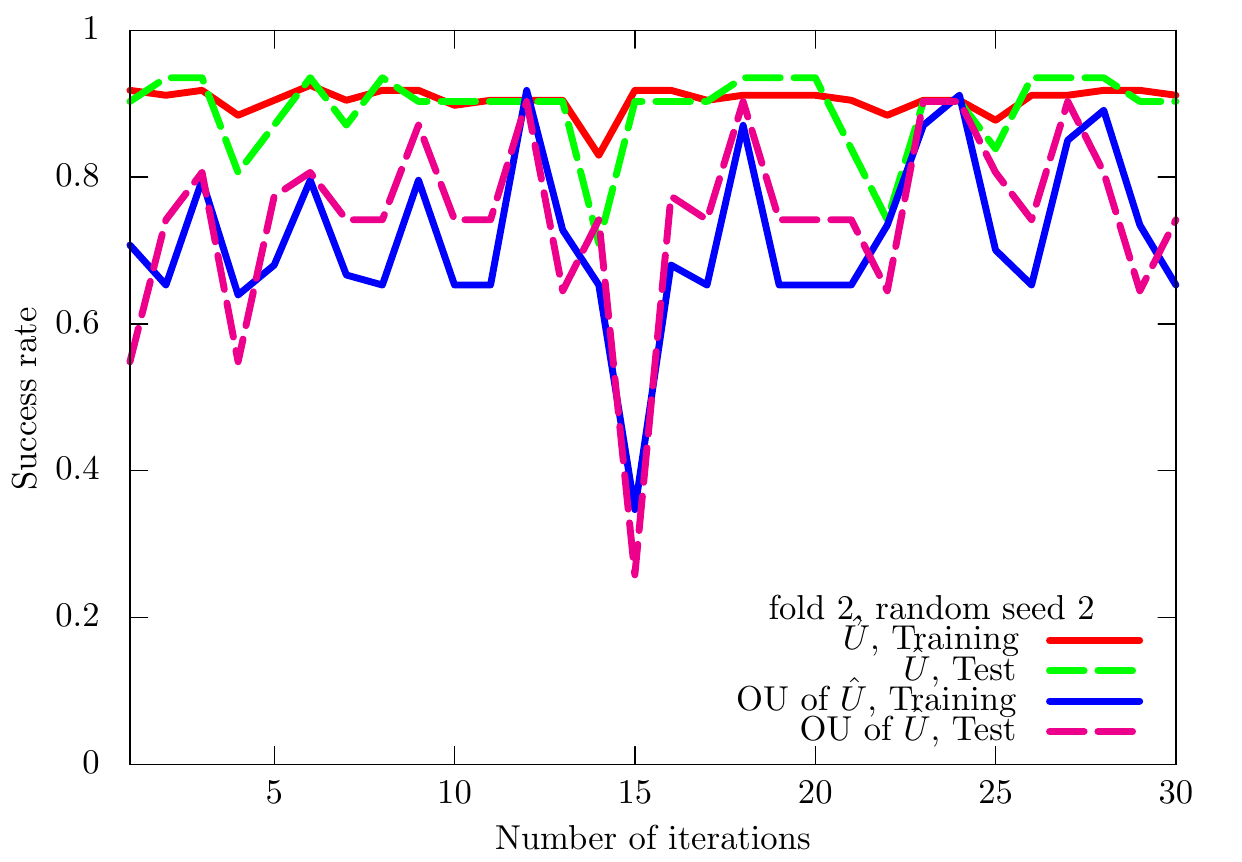}
\includegraphics[scale=0.25]{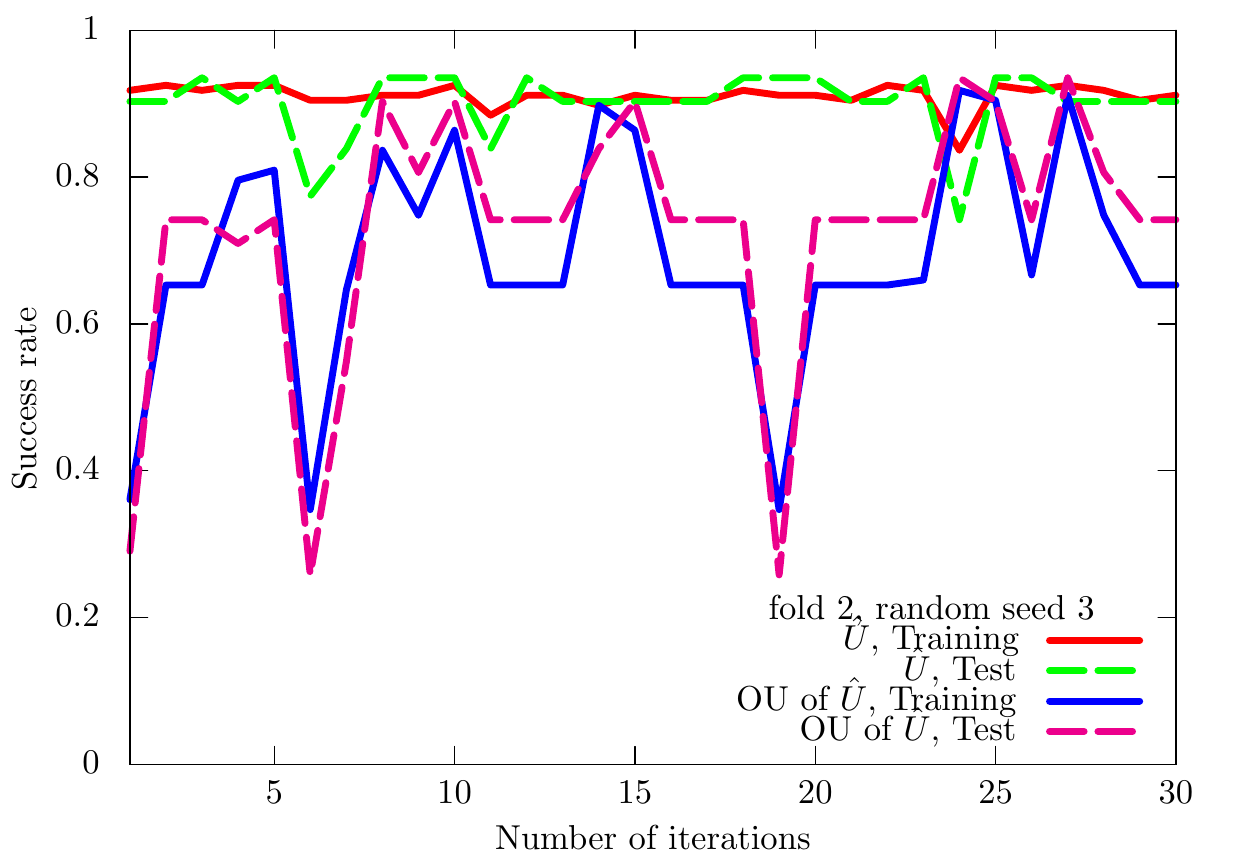}
\includegraphics[scale=0.25]{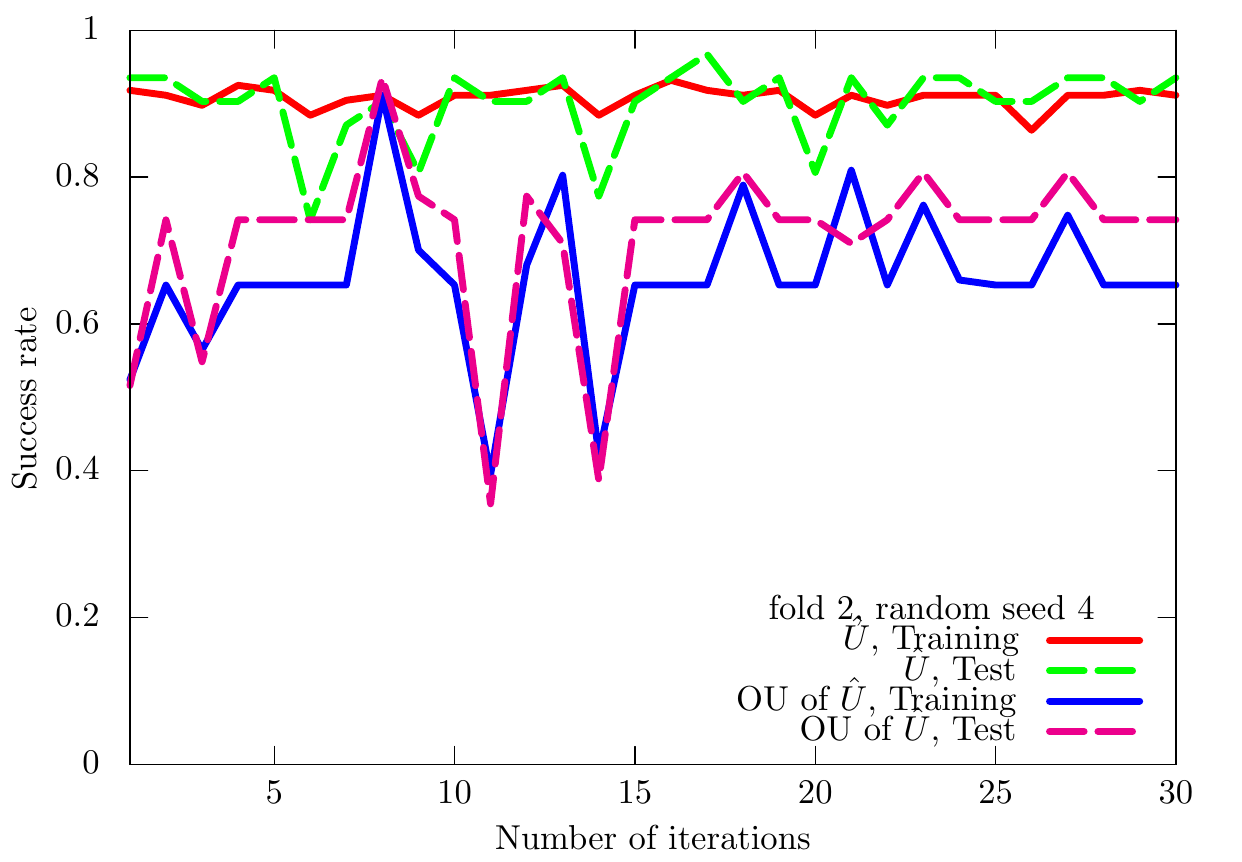}
\includegraphics[scale=0.25]{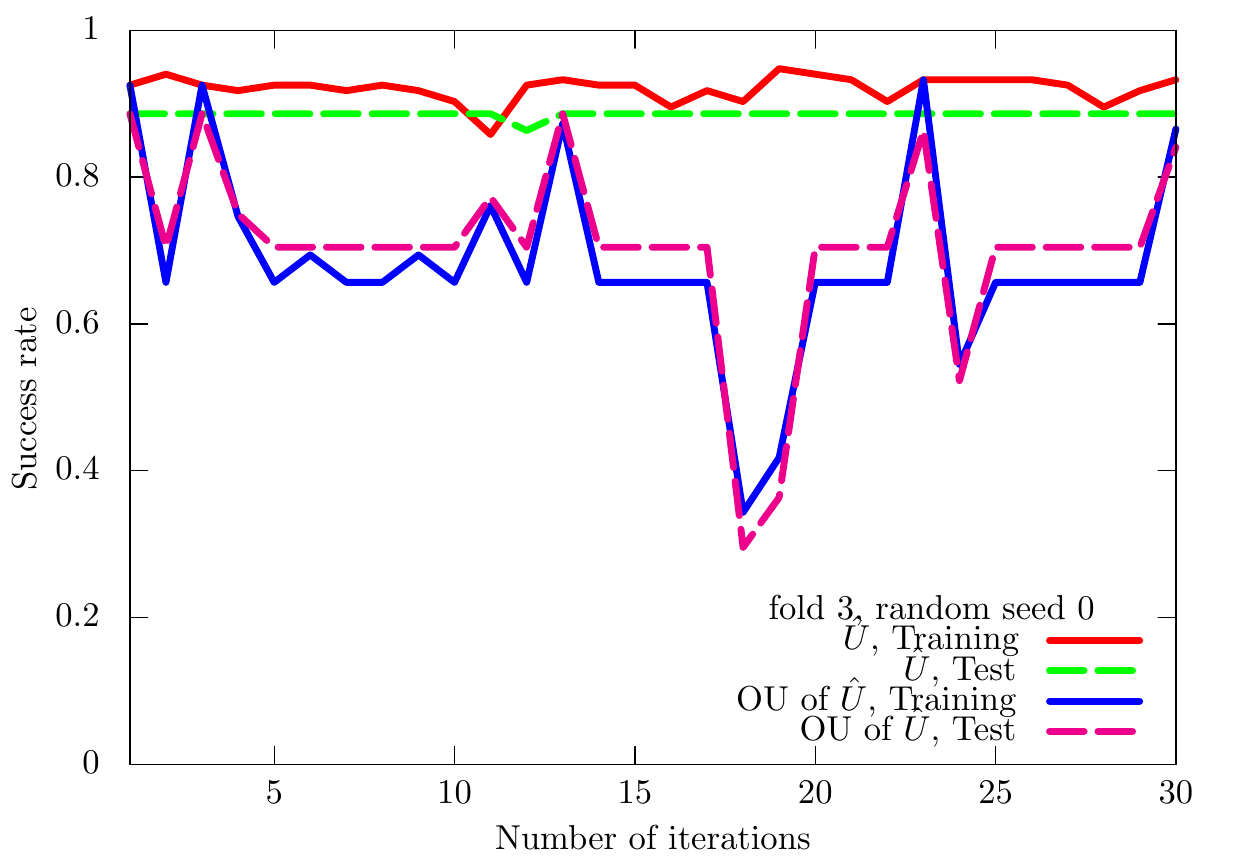}
\includegraphics[scale=0.25]{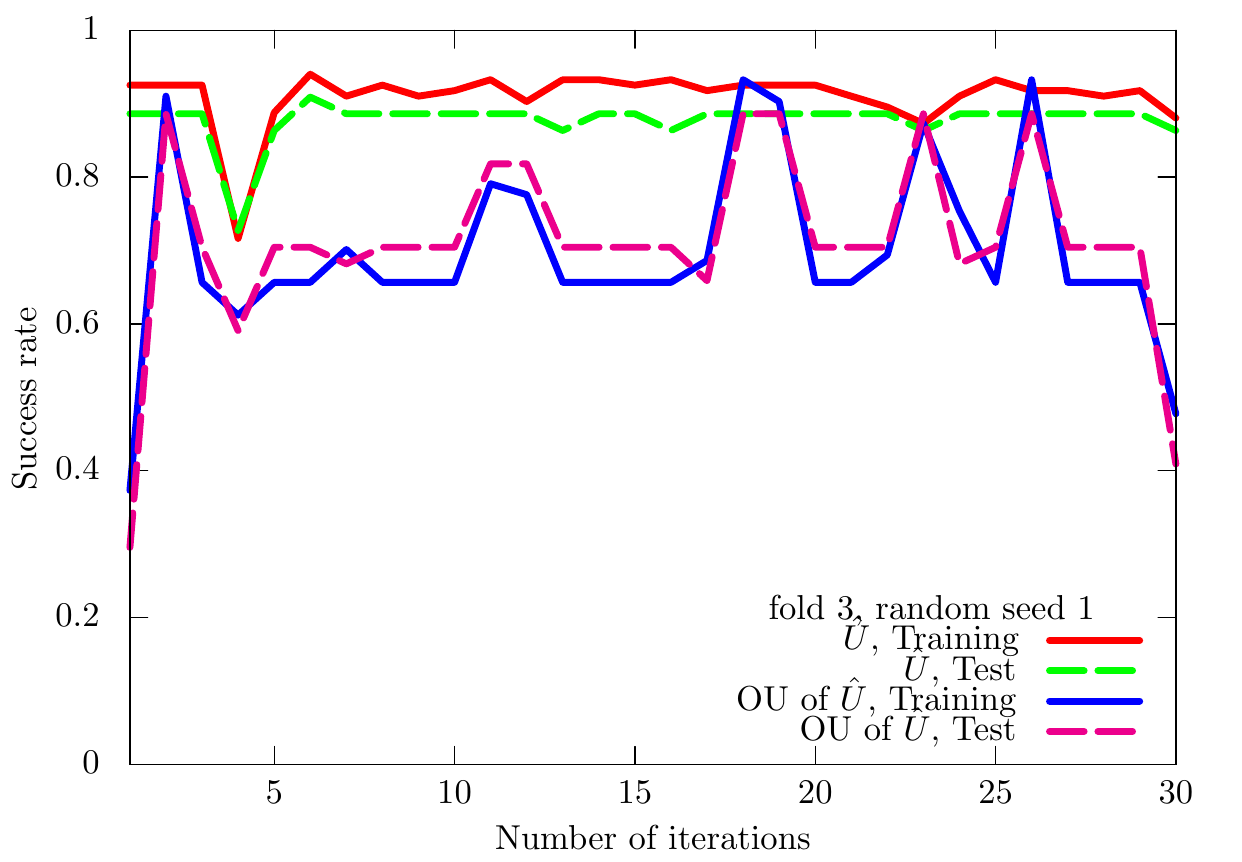}
\includegraphics[scale=0.25]{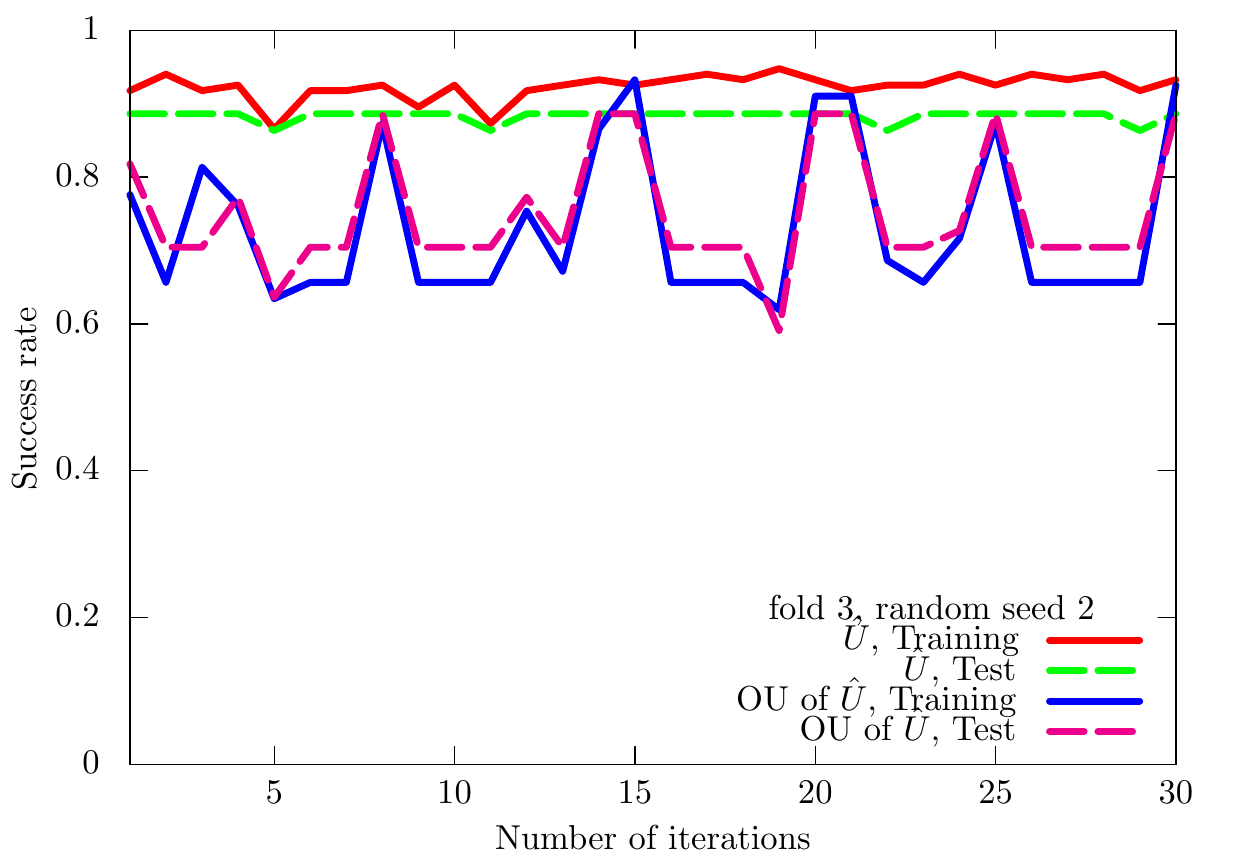}
\includegraphics[scale=0.25]{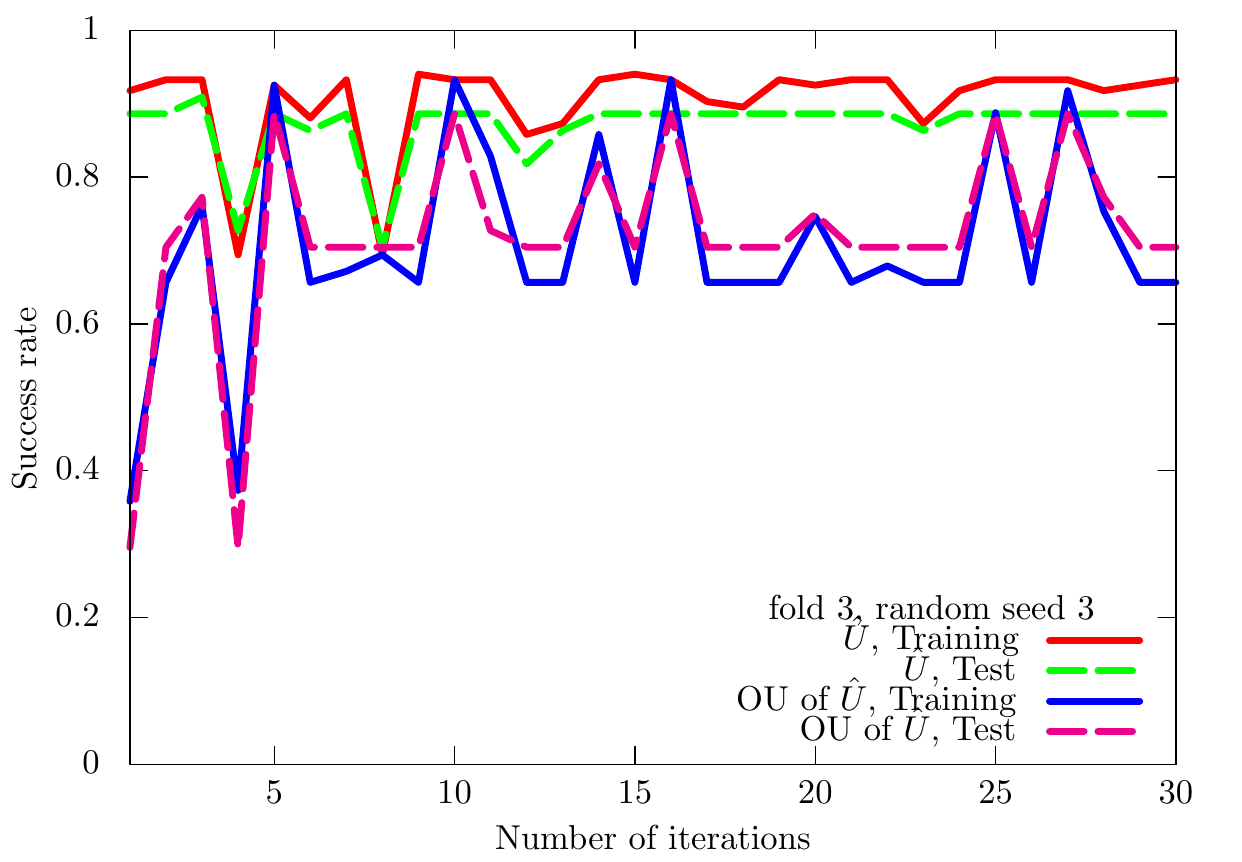}
\includegraphics[scale=0.25]{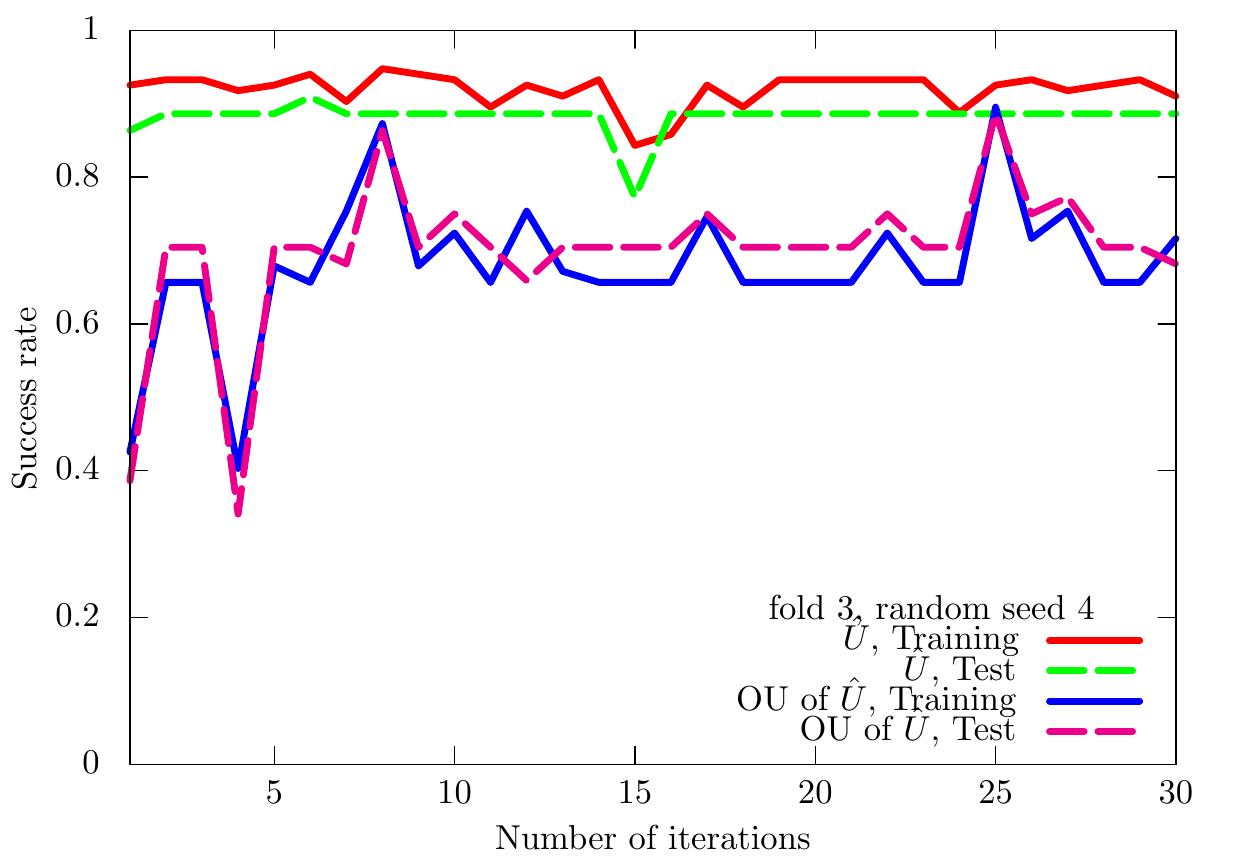}
\includegraphics[scale=0.25]{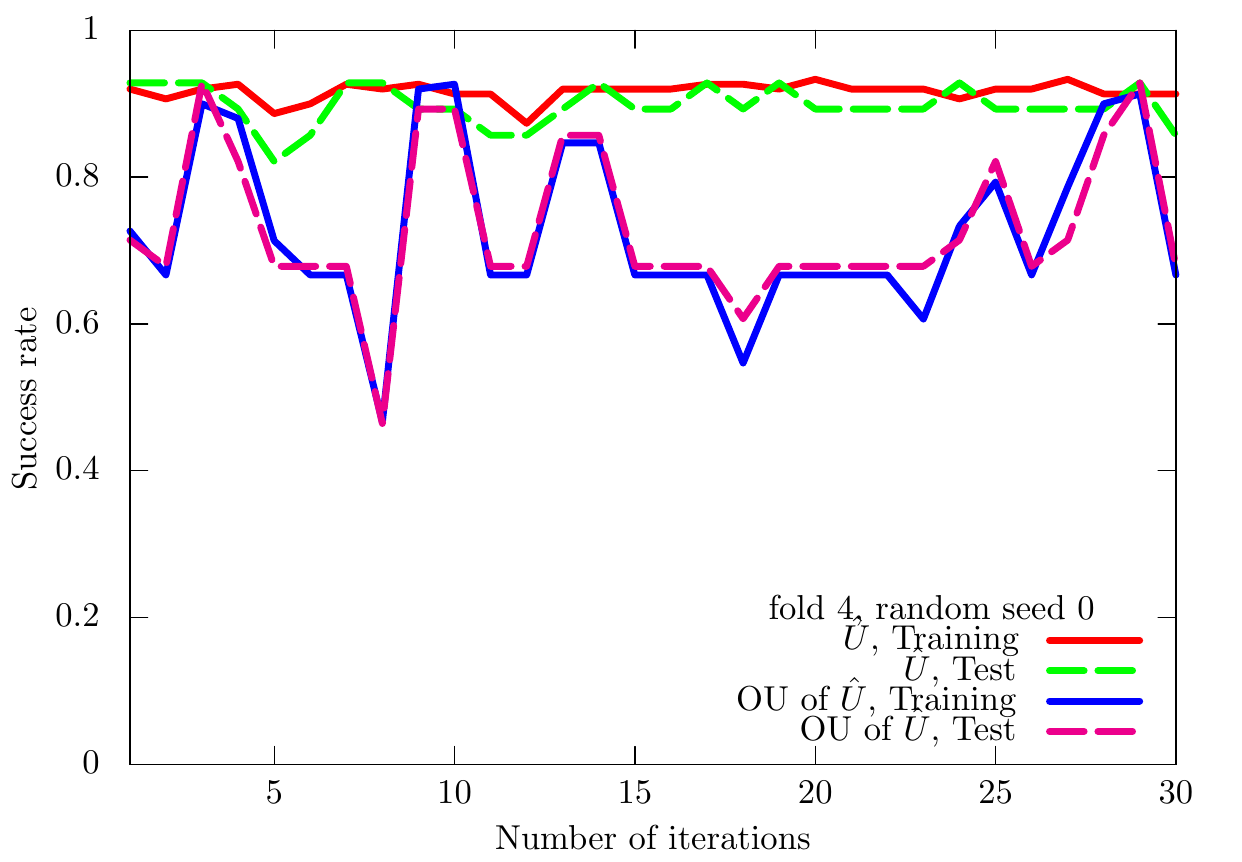}
\includegraphics[scale=0.25]{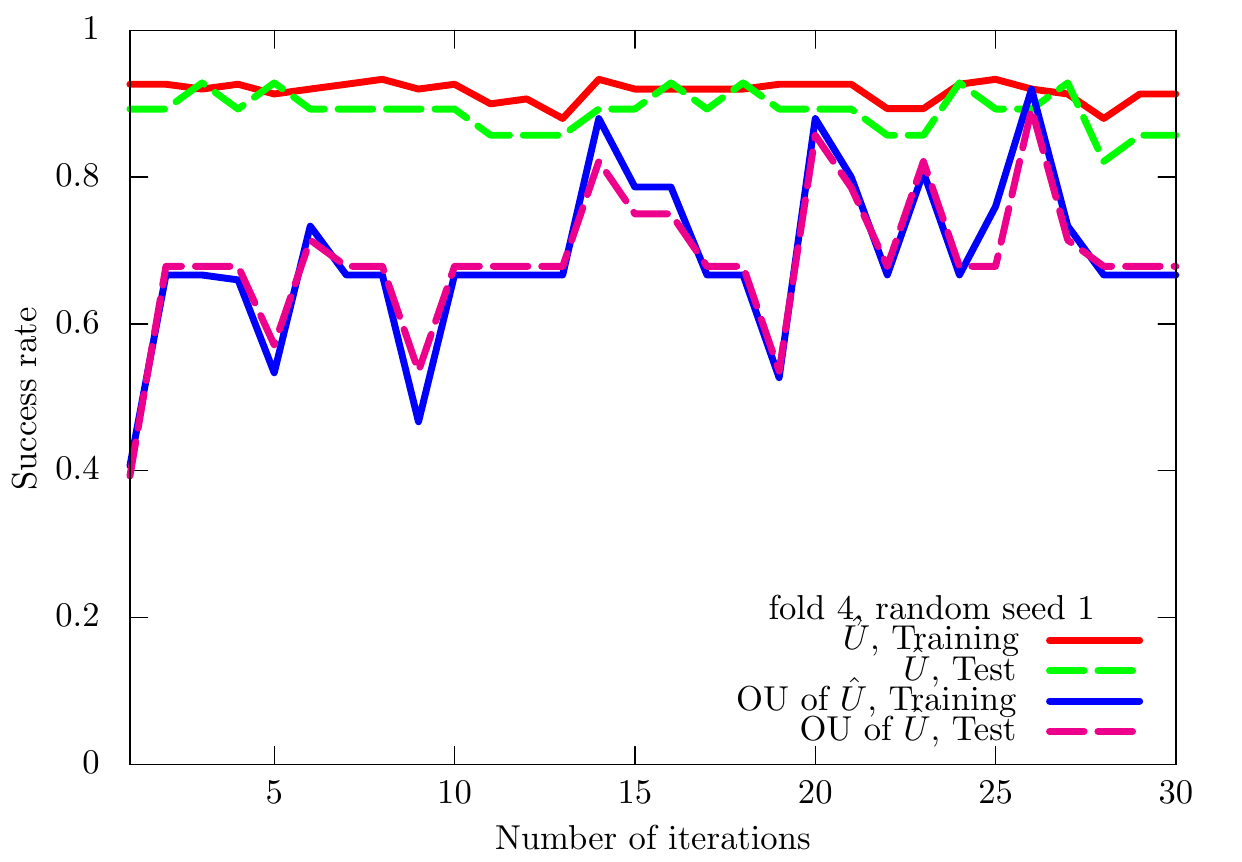}
\includegraphics[scale=0.25]{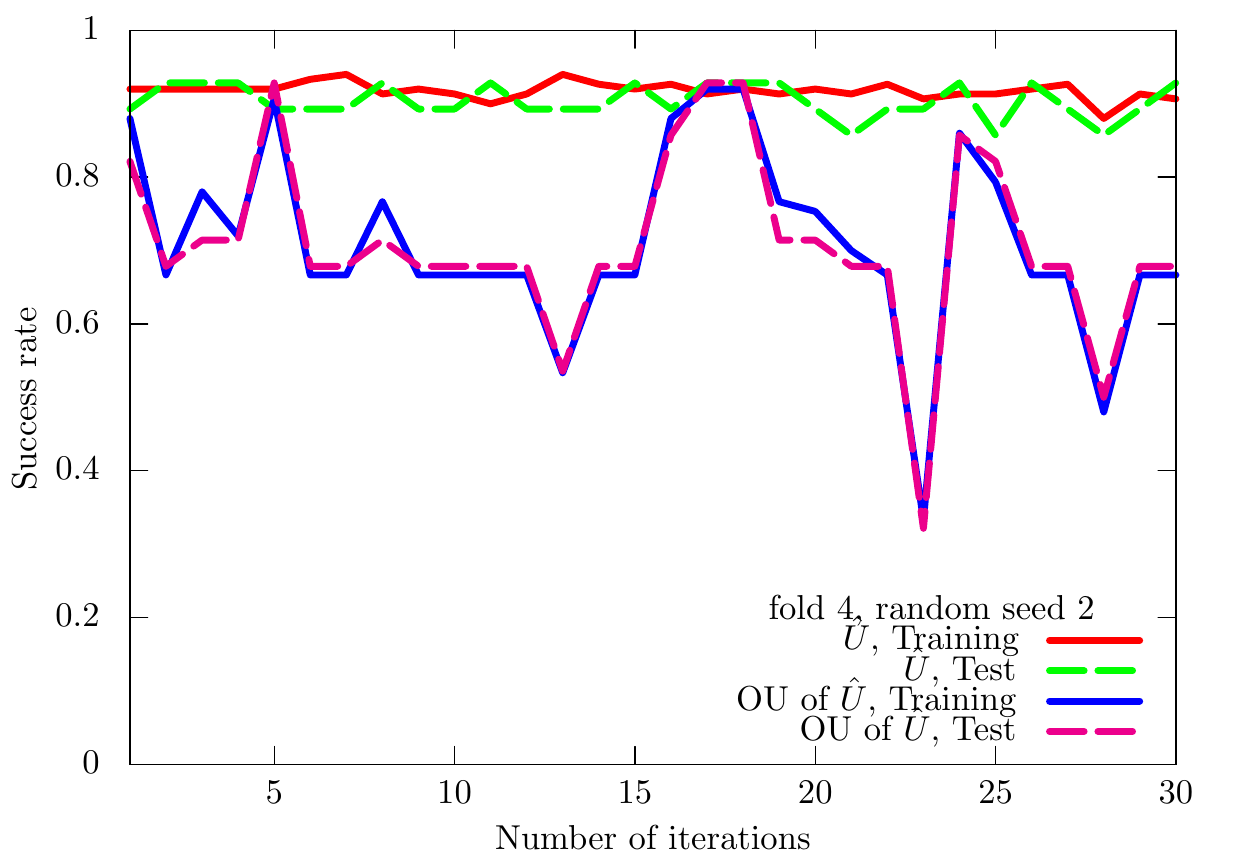}
\includegraphics[scale=0.25]{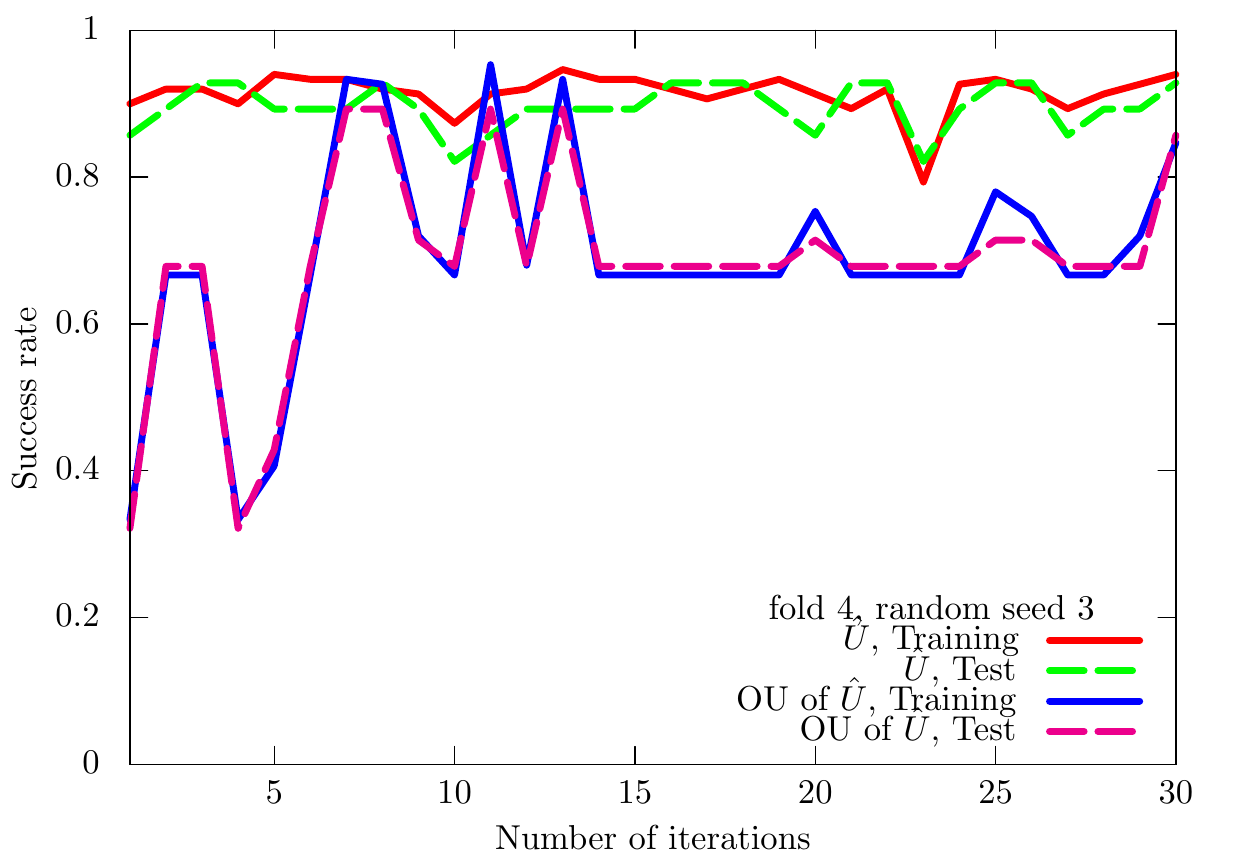}
\includegraphics[scale=0.25]{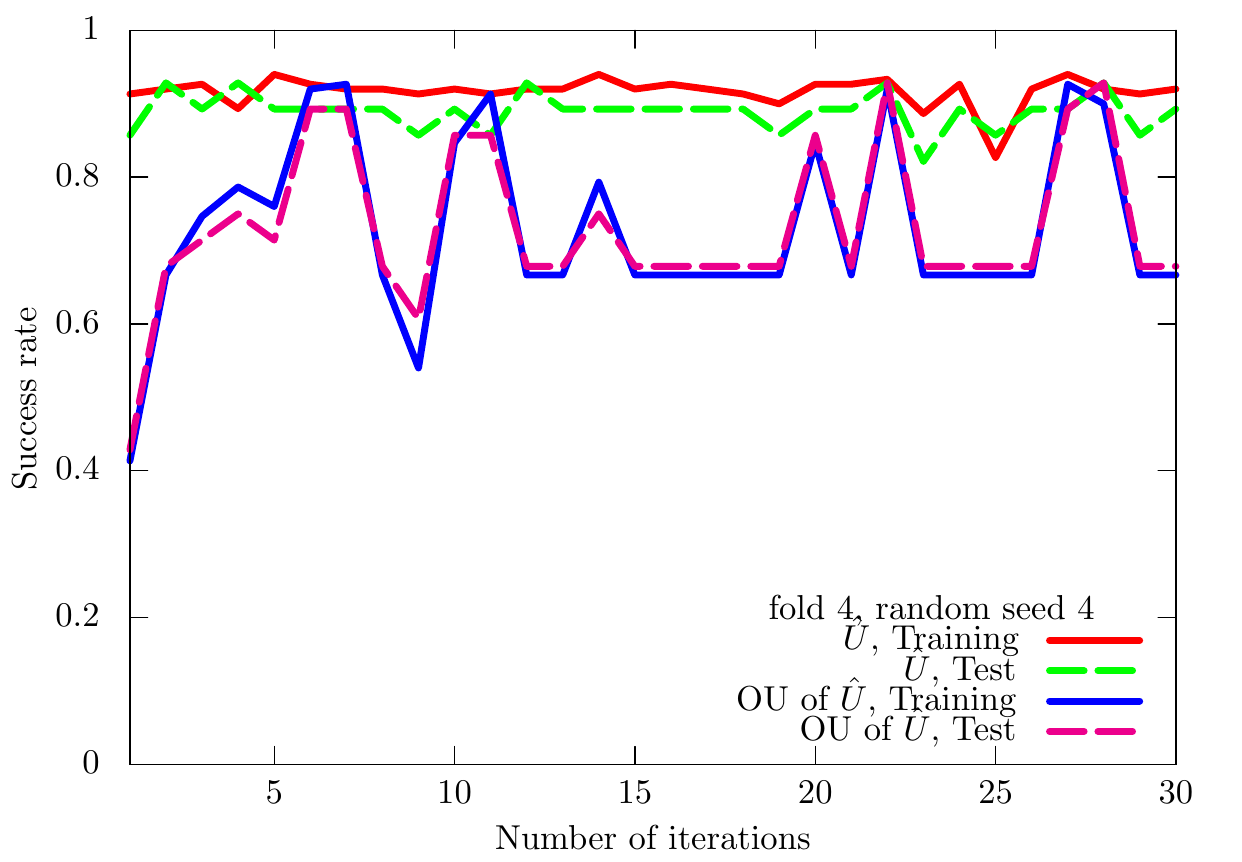}
\caption{Results of the UKM ($\hat{X}$ and OU of $\hat{X}$) on the $5$-fold datasets with $5$ different random seeds for the wine dataset ($0$ or non-$0$). We use complex matrices and set $\theta_\mathrm{bias} = 0$. We set $r = 0.010$.}
\label{supp-arXiv-numerical-result-raw-data-fold-001-rand-001-UKM-OUU-UCI-wine-0-non0}
\end{figure*}

In Fig.~\ref{supp-arXiv-numerical-result-raw-data-fold-001-rand-001-QCL-UCI-wine-0-non0}, we show the numerical results of QCL for the $5$-fold datasets with $5$ different random seeds.
In Fig.~\ref{supp-arXiv-numerical-result-raw-data-fold-001-rand-001-UKM-P-UCI-wine-0-non0}, we show the numerical results of $\hat{P}$ of the UKM for the $5$-fold datasets with $5$ different random seeds.
In Fig.~\ref{supp-arXiv-numerical-result-raw-data-fold-001-rand-001-UKM-OUU-UCI-wine-0-non0}, we also show the numerical results of OU of $\hat{X}$ of the UKM for the $5$-fold datasets with $5$ different random seeds.

\begin{figure}[htb]
\centering
\includegraphics[scale=0.45]{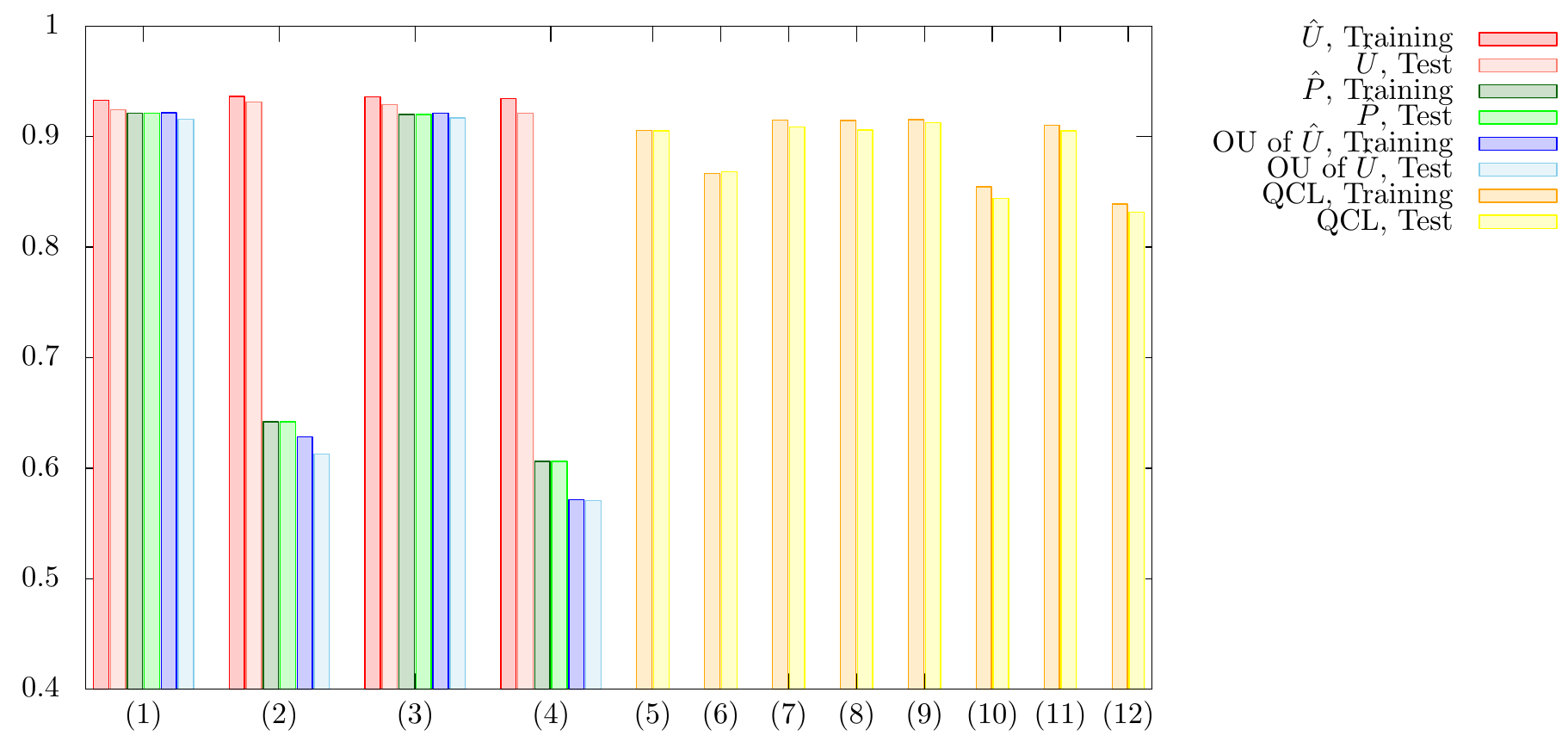}
\caption{Results of $5$-fold CV with $5$ different random seeds for the wine dataset ($0$ or non-$0$). For the UKM, we put $r = 0.010$ and set $K = 30$ and $K' = 10$. For QCL, the number of layers $L$ is $5$ and the number of iterations is $300$. The numerical settings are as follows: (1) complex matrices without the bias term, (2) complex matrices with the bias term, (3) real matrices without the bias term, (4) real matrices with the bias term, (5) CNOT-based circuit without the bias term, (6) CNOT-based circuit with the bias term, (7) CRot-based circuit without the bias term, (8) CRot-based circuit with the bias term, (9) 1d Heisenberg circuit without the bias term, (10) 1d Heisenberg circuit with the bias term, (11) FC Heisenberg circuit without the bias term, and (12) FC Heisenberg circuit with the bias term.}
\label{supp-arXiv-numerical-result-performance-UKM-QCL-UCI-wine-0-non0}
\end{figure}

We summarize the results of 5-fold CV with 5 different random seeds of QCL and the UKM in Tables~\ref{supp-arXiv-table-UCI-wine-0-non0-002} and \ref{supp-arXiv-table-UCI-wine-0-non0-001}, respectively.
For QCL and the UKM, we select the best model for the training dataset over iterations to compute the performance.
\begin{table}[htb]
  \begin{tabular}{cc|cc}
    \hline \hline
    Algo. & Condition & Training & Test \\
    \hline
  QCL & CNOT-based, w/o bias & 0.9057 & 0.9052 \\
  QCL & CNOT-based, w/ bias & 0.8665 & 0.8680 \\
    \hline
  QCL & CRot-based, w/o bias & 0.9151 & 0.9086 \\
  QCL & CRot-based, w/ bias & 0.9145 & 0.9059 \\
    \hline
  QCL & 1d Heisenberg, w/o bias & 0.9155 & 0.9126 \\
  QCL & 1d Heisenberg, w/ bias & 0.8547 & 0.8440 \\
    \hline
  QCL & FC Heisenberg, w/o bias & 0.9103 & 0.9052 \\
  QCL & FC Heisenberg, w/ bias & 0.8390 & 0.8318 \\
    \hline \hline
  \end{tabular}
\caption{Results of $5$-fold CV with $5$ different random seeds of QCL for the wine dataset ($0$ or non-$0$). The number of layers $L$ is set to $5$ and the number of iterations is set to $300$.}
\label{supp-arXiv-table-UCI-wine-0-non0-002}
\end{table}
\begin{table}[htb]
  \begin{tabular}{cc|cc}
    \hline \hline
    Algo. & Condition & Training & Test \\
    \hline
  UKM & $\hat{X}$, complex, w/o bias & 0.9328 & 0.9243 \\
  UKM & $\hat{P}$, complex, w/o bias & 0.9213 & 0.9116 \\
  UKM & OU of $\hat{X}$, complex, w/o bias & 0.9215 & 0.9159 \\
    \hline
  UKM & $\hat{X}$, complex, w/ bias & 0.9364 & 0.9313 \\
  UKM & $\hat{P}$, complex, w/ bias & 0.6420 & 0.6184 \\
  UKM & OU of $\hat{X}$, complex, w/ bias & 0.6282 & 0.6128 \\
    \hline
  UKM & $\hat{X}$, real, w/o bias & 0.9359 & 0.9292 \\
  UKM & $\hat{P}$, real, w/o bias & 0.9200 & 0.9185 \\
  UKM & OU of $\hat{X}$, real, w/o bias & 0.9212 & 0.9171 \\
    \hline
  UKM & $\hat{X}$, real, w/ bias & 0.9345 & 0.9212 \\
  UKM & $\hat{P}$, real, w/ bias & 0.6062 & 0.6055 \\
  UKM & OU of $\hat{X}$, real, w/ bias & 0.5716 & 0.5708 \\
    \hline \hline
  \end{tabular}
\caption{Results of $5$-fold CV with $5$ different random seeds of the UKM for the wine dataset ($0$ or non-$0$). We put $r = 0.010$ and set $K = 30$ and $K' = 10$.}
\label{supp-arXiv-table-UCI-wine-0-non0-001}
\end{table}
In Fig.~\ref{supp-arXiv-numerical-result-performance-UKM-QCL-UCI-wine-0-non0}, we plot the data shown in Tables~\ref{supp-arXiv-table-UCI-wine-0-non0-002} and \ref{supp-arXiv-table-UCI-wine-0-non0-001}.
We also summarize the results of 5-fold CV with 5 different random seeds of the kernel method in Table~\ref{supp-arXiv-table-UCI-wine-0-non0-003}.
More specifically, we use Ridge classification in Sec.~\ref{supp-arXiv-sec-Ridge-001}.
We consider the linear functions and the second-order polynomial functions for $\phi (\cdot)$ in Eq.~\eqref{supp-arXiv-f-pred-kernel-method-001-002} with and without normalization.
We set $\lambda = 10^{-2}, 10^{-1}, 1$ where $\lambda$ is the coefficient of the regularization term.
\begin{table}[htb]
  \begin{tabular}{cc|cc}
    \hline \hline
    Algo. & Condition & Training & Test \\
    \hline
  Kernel method & Linear, w/o normalization, $\lambda = 10^{-2}$ & 0.9987 & 0.9955 \\
  Kernel method & Linear, w/o normalization, $\lambda = 10^{-1}$ & 0.9987 & 0.9883 \\
  Kernel method & Linear, w/o normalization, $\lambda = 1$ & 0.9987 & 0.9883 \\
    \hline
  Kernel method & Linear, w/ normalization, $\lambda = 10^{-2}$ & 0.9357 & 0.9474 \\
  Kernel method & Linear, w/ normalization, $\lambda = 10^{-1}$ & 0.9230 & 0.9183 \\
  Kernel method & Linear, w/ normalization, $\lambda = 1$ & 0.6807 & 0.6711 \\
    \hline
  Kernel method & Poly-2, w/o normalization, $\lambda = 10^{-2}$ & 1.0000 & 0.9481 \\
  Kernel method & Poly-2, w/o normalization, $\lambda = 10^{-1}$ & 1.0000 & 0.9558 \\
  Kernel method & Poly-2, w/o normalization, $\lambda = 1$ & 1.0000 & 0.9683 \\
    \hline
  Kernel method & Poly-2, w/ normalization, $\lambda = 10^{-2}$ & 0.9455 & 0.9520 \\
  Kernel method & Poly-2, w/ normalization, $\lambda = 10^{-1}$ & 0.9259 & 0.9293 \\
  Kernel method & Poly-2, w/ normalization, $\lambda = 1$ & 0.8413 & 0.8258 \\
    \hline \hline
  \end{tabular}
\caption{Results of 5-fold CV with 5 different random seeds of the kernel method for the wine dataset ($0$ or non-$0$).}
\label{supp-arXiv-table-UCI-wine-0-non0-003}
\end{table}

\begin{figure}[htb]
\centering
\includegraphics[scale=0.45]{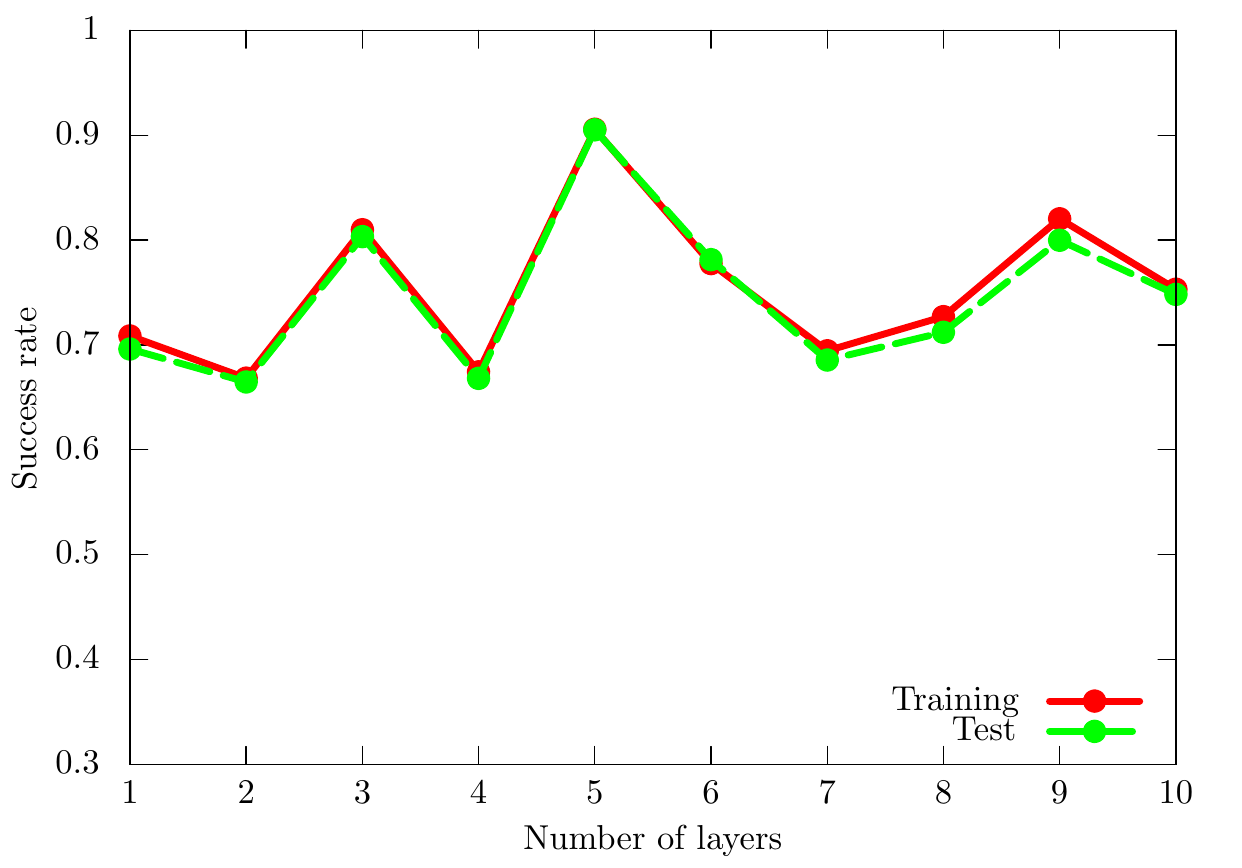}
\caption{Performance dependence of QCL on the number of layers $L$ for the wine dataset ($0$ or non-$0$). We use the CNOT-based circuit geometry and set $\theta_\mathrm{bias} = 0$. We iterate the computation $300$ times.}
\label{supp-arXiv-numerical-result-layers-dependence-QCL-UCI-wine-0-non0}
\end{figure}

\begin{figure}[htb]
\centering
\includegraphics[scale=0.45]{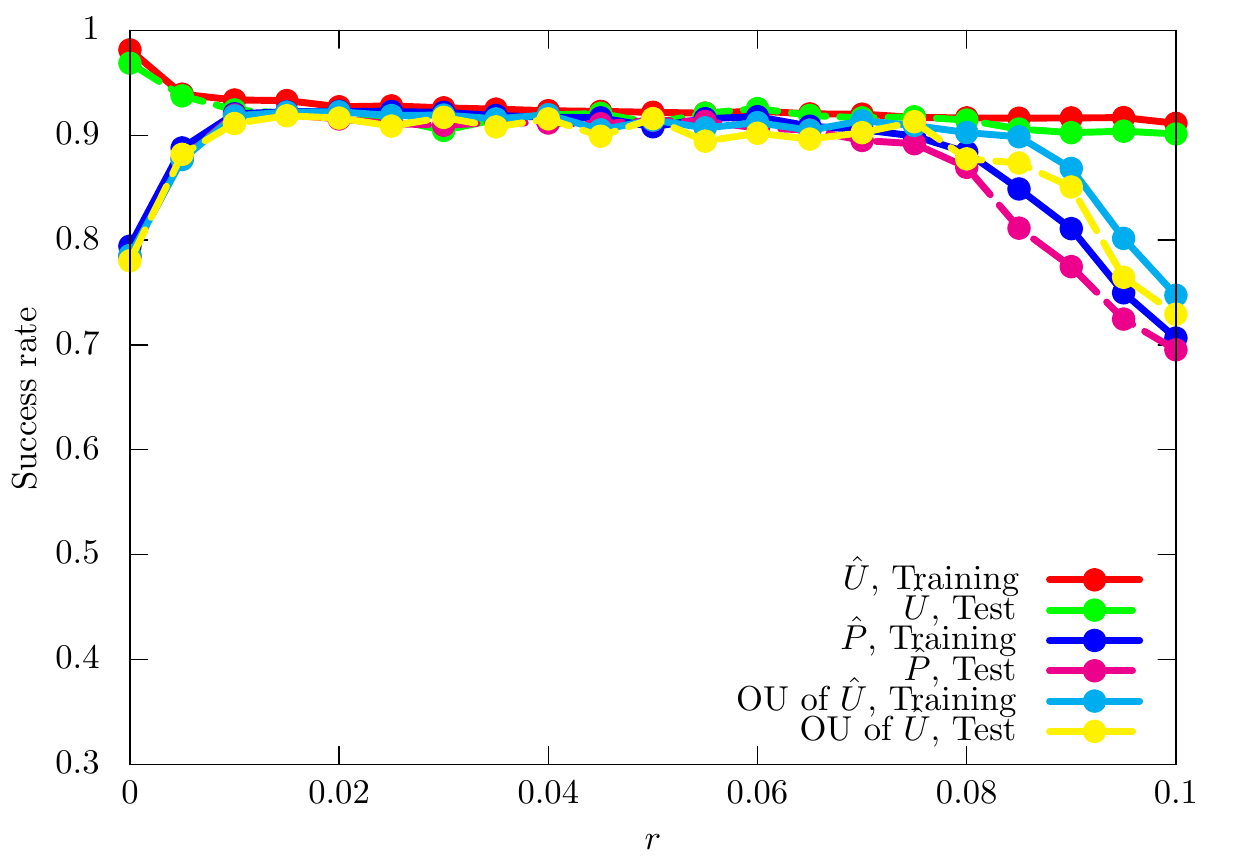}
\includegraphics[scale=0.45]{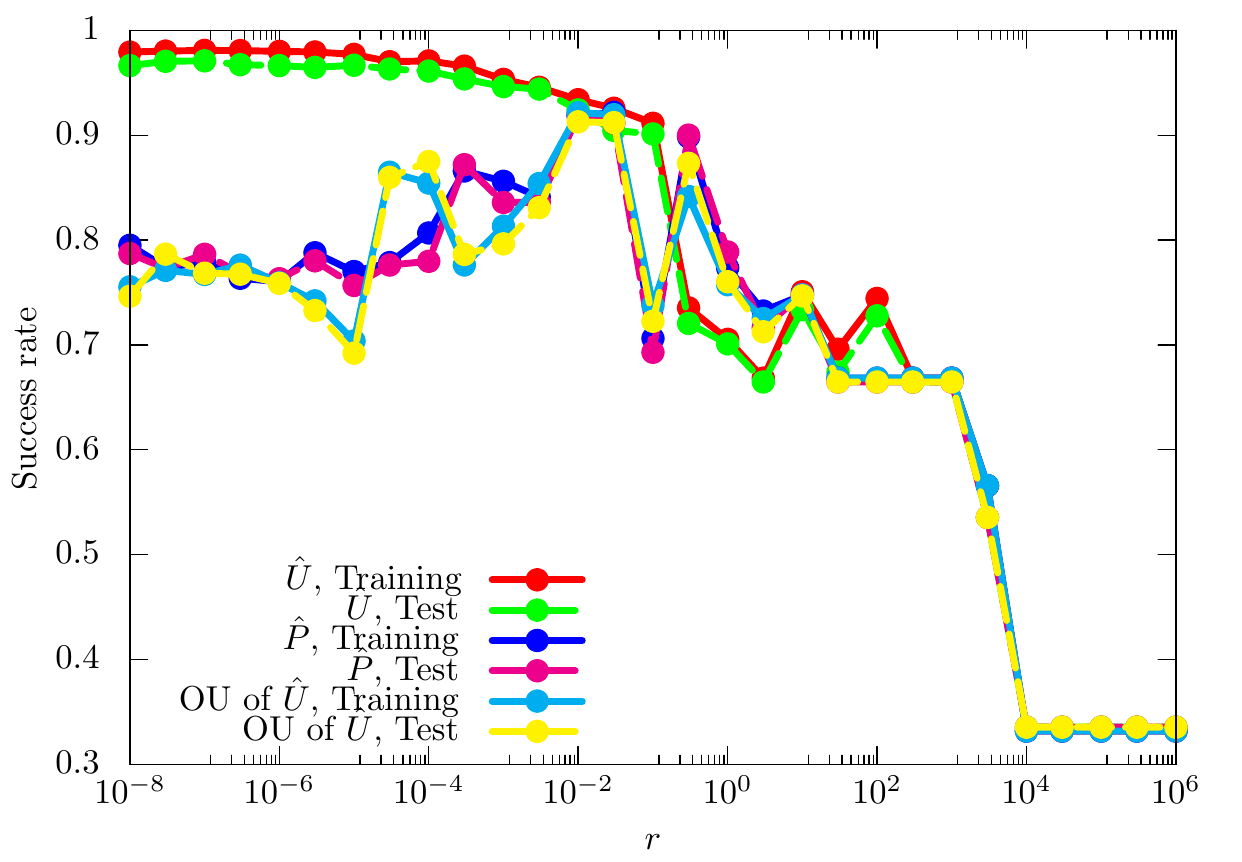}
\caption{Performance dependence of the UKM on $r$, which is the coefficient of the second term in the right-hand side of Eq.~\eqref{supp-arXiv-quantum-kernel-method-001-011} for the wine dataset ($0$ or non-$0$). We use complex matrices and set $\theta_\mathrm{bias} = 0$. We set $K = 30$ and $K' = 10$.}
\label{supp-arXiv-numerical-result-r-dependence-UKM-UCI-wine-0-non0}
\end{figure}

\begin{figure}[htb]
\centering
\includegraphics[scale=0.45]{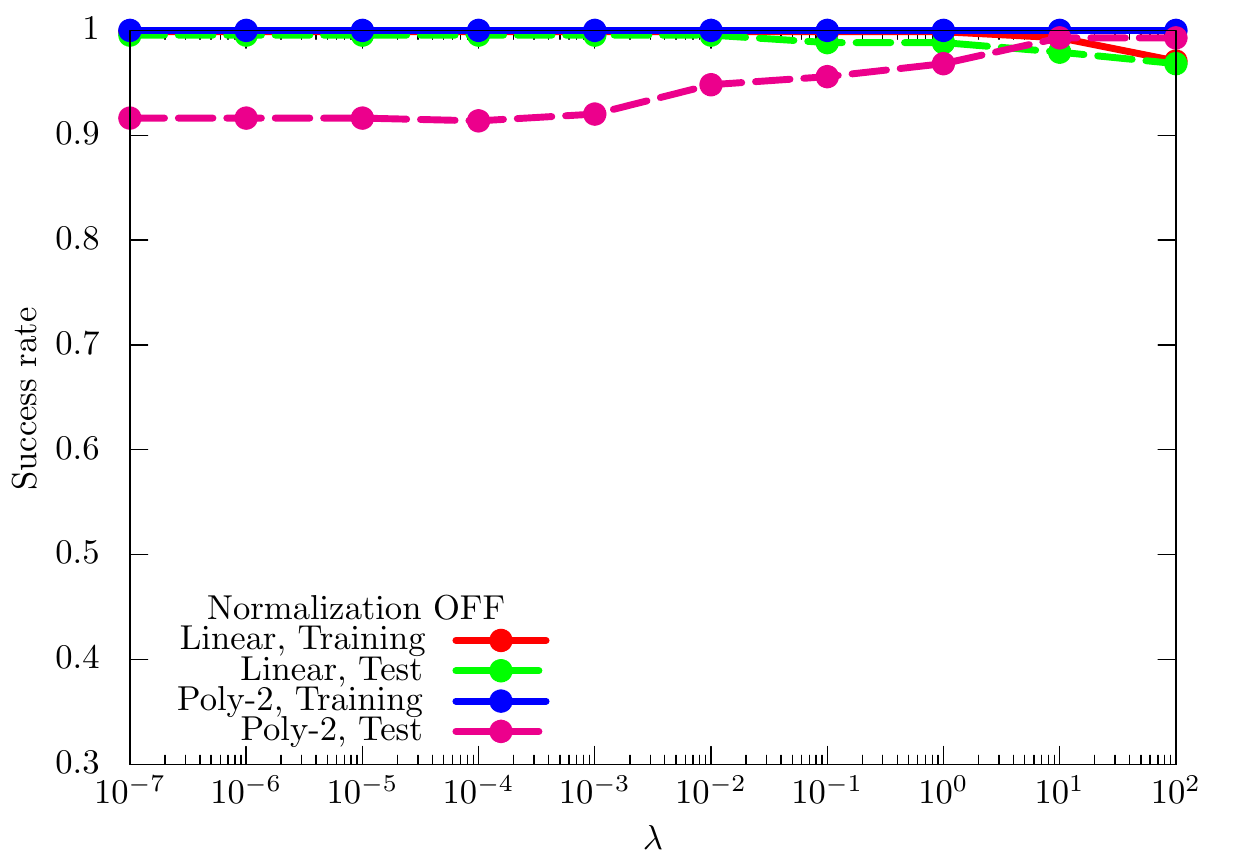}
\includegraphics[scale=0.45]{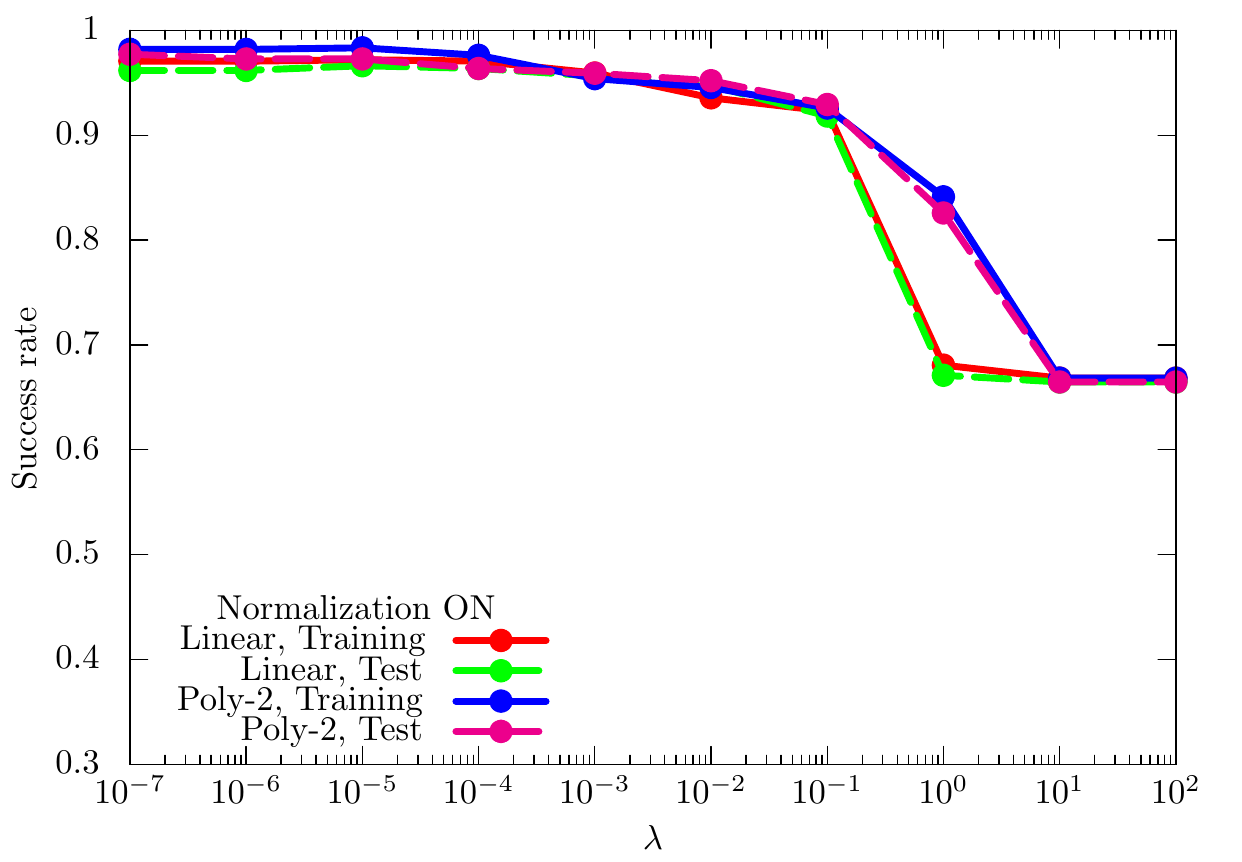}
\caption{Performance dependence of the kernel method on $\lambda$, which is the coefficient of the second term in the right-hand side of Eq.~\eqref{supp-arXiv-cost-function-kernel-method-001-002} for the wine dataset ($0$ or non-$0$). For $\phi (\cdot)$ in Eq.~\eqref{supp-arXiv-f-pred-kernel-method-001-002}, we use the linear functions and the second-degree polynomial functions with and without normalization.}
\label{supp-arXiv-numerical-result-lambda-dependence-kernel-method-wine-0-non0}
\end{figure}

Next, we show the performance dependence of the three algorithms on their key parameters.
We see the performance dependence of QCL on the number of layers $L$.
The result is shown in Fig.~\ref{supp-arXiv-numerical-result-layers-dependence-QCL-UCI-wine-0-non0}.
We then see the performance dependence of the UKM on $r$, which is the coefficient of the second term in the right-hand side of Eq.~\eqref{supp-arXiv-quantum-kernel-method-001-011}.
The result is shown in Fig.~\ref{supp-arXiv-numerical-result-r-dependence-UKM-UCI-wine-0-non0}.
In Fig.~\ref{supp-arXiv-numerical-result-lambda-dependence-kernel-method-wine-0-non0}, we show the performance dependence of the kernel method on $\lambda$, which is the coefficient of the second term in the right-hand side of Eq.~\eqref{supp-arXiv-cost-function-kernel-method-001-002}.

\begin{figure}[htb]
\centering
\includegraphics[scale=0.45]{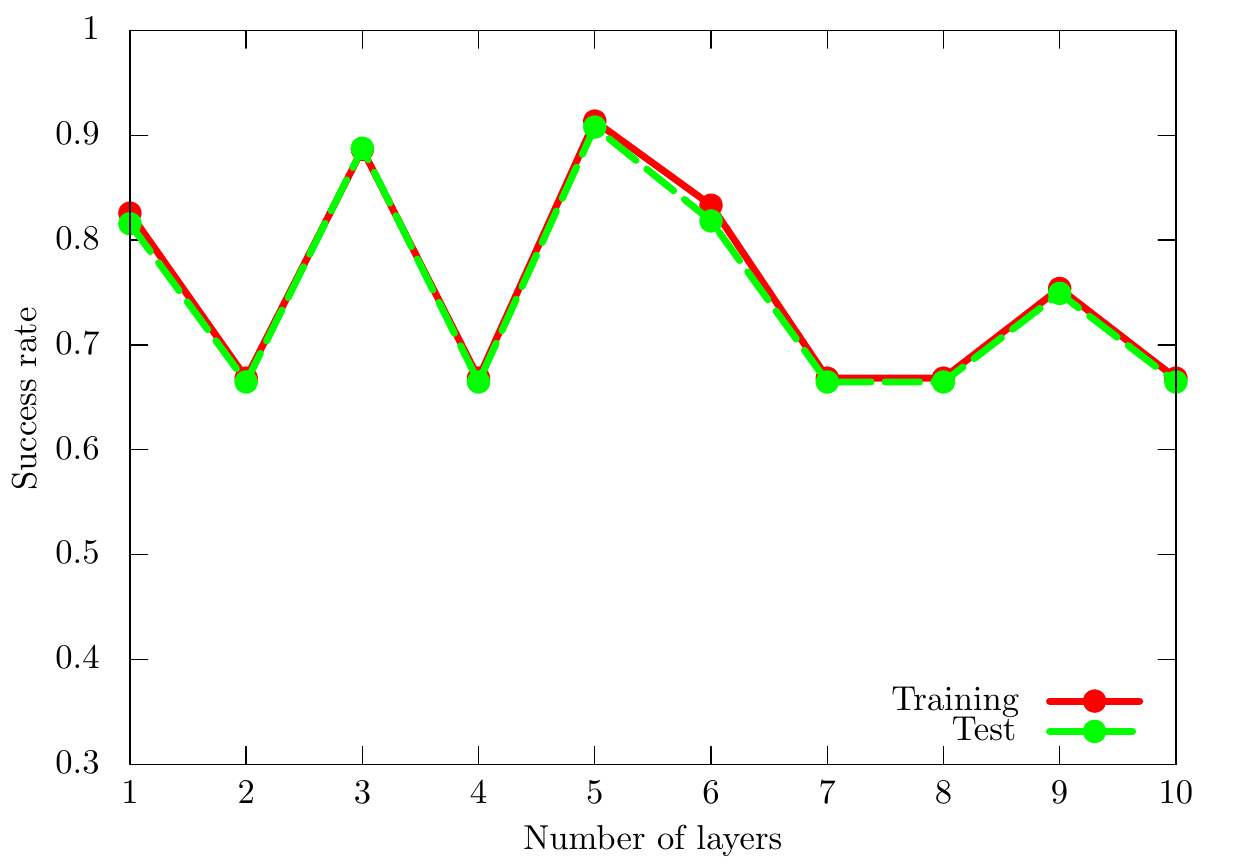}
\caption{Performance dependence of QCL on the number of layers $L$ for the wine dataset ($0$ or non-$0$) in the case of the hinge function, Eq.~\eqref{supp-arXiv-hinge-function-001-001}. We use the CNOT-based circuit geometry and set $\theta_\mathrm{bias} = 0$. We iterate the computation $300$ times.}
\label{supp-arXiv-numerical-result-layers-dependence-QCL-UCI-wine-0-non0-hinge}
\end{figure}

\begin{figure}[htb]
\centering
\includegraphics[scale=0.45]{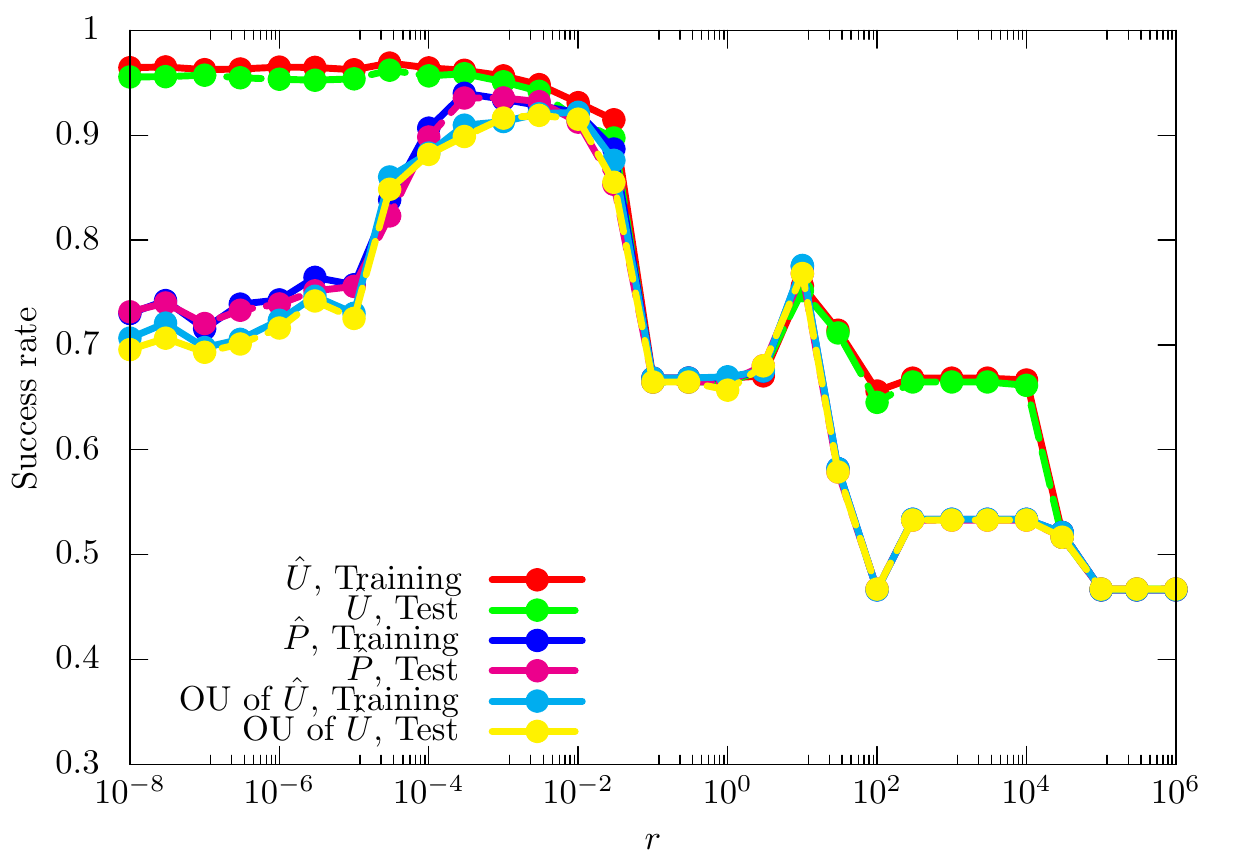}
\caption{Performance dependence of the UKM on $r$, which is the coefficient of the second term in the right-hand side of Eq.~\eqref{supp-arXiv-quantum-kernel-method-001-011} for the wine dataset ($0$ or non-$0$) in the case of the hinge function, Eq.~\eqref{supp-arXiv-hinge-function-001-001}. We use complex matrices and set $\theta_\mathrm{bias} = 0$. We set $K = 30$ and $K' = 10$.}
\label{supp-arXiv-numerical-result-r-dependence-UKM-UCI-wine-0-non0-hinge}
\end{figure}

So far, we have used the squared error function, Eq.~\eqref{supp-arXiv-squared-error-function-001-001}.
In Fig.~\ref{supp-arXiv-numerical-result-layers-dependence-QCL-UCI-wine-0-non0-hinge}, we show the performance dependence of QCL on the number of layers $L$ in the case of the hinge function, Eq.~\eqref{supp-arXiv-hinge-function-001-001}.
In Fig.~\ref{supp-arXiv-numerical-result-r-dependence-UKM-UCI-wine-0-non0-hinge}, we show the performance dependence of the UKM on $r$, which is the coefficient of the second term in the right-hand side of Eq.~\eqref{supp-arXiv-quantum-kernel-method-001-011}, in the case of the hinge function, Eq.~\eqref{supp-arXiv-hinge-function-001-001}.

\clearpage

\subsection{Semeion dataset ($0$ or $1$)}

We here show the numerical result for the semeion dataset ($0$ or $1$).
For the UKM, we put $r = 0.010$ and set $K = 20$ and $K' = 10$ in Algo.~\ref{supp-arXiv-quantum-kernel-method-002-001}.
For QCL, we run iterations $100$ times.

\begin{figure*}[htb]
\centering
\includegraphics[scale=0.25]{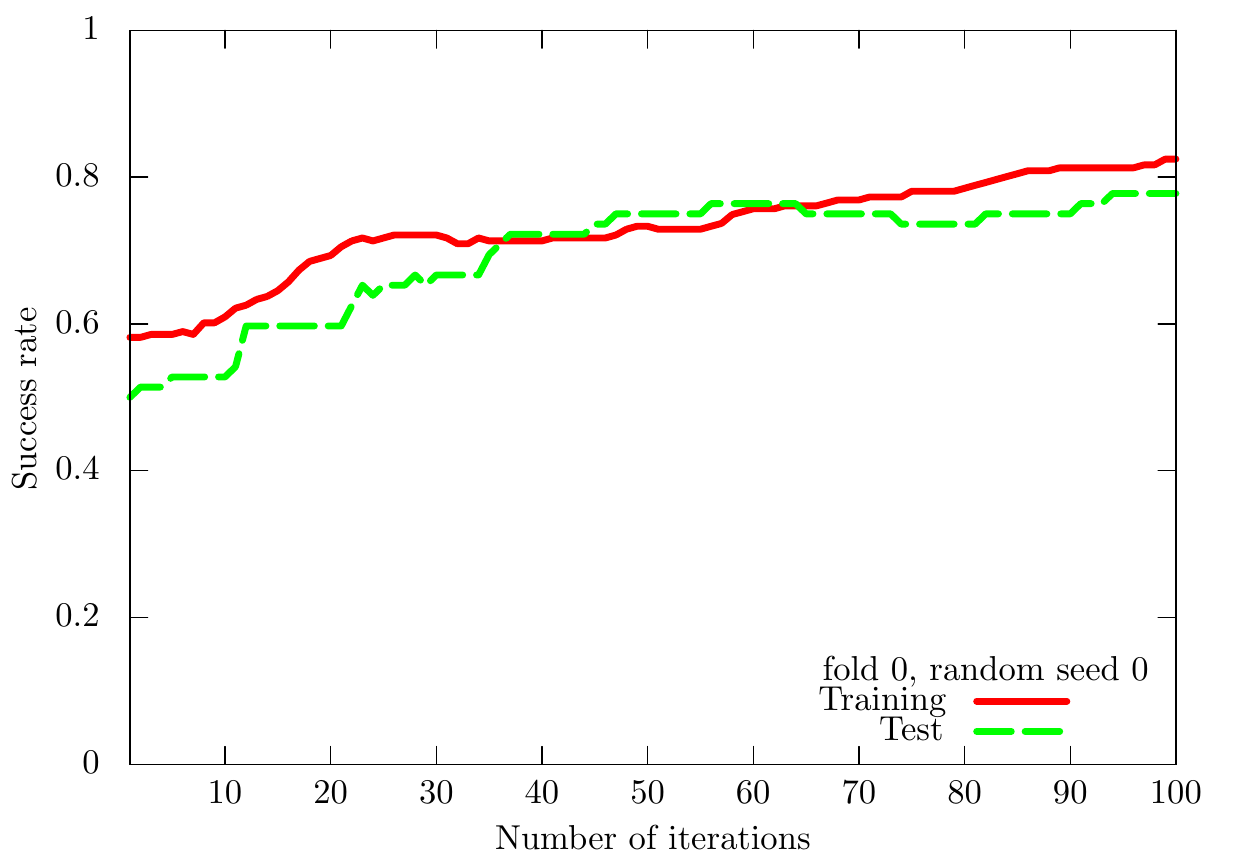}
\includegraphics[scale=0.25]{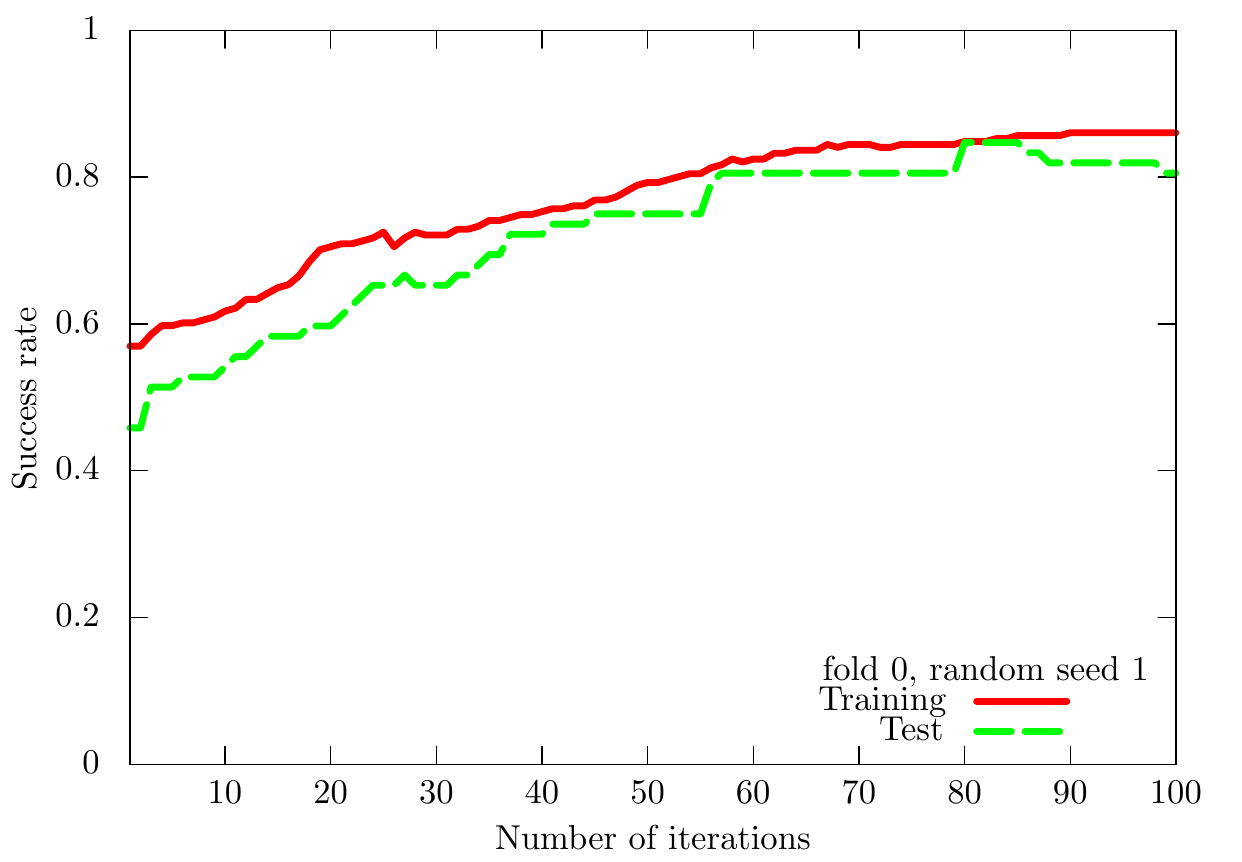}
\includegraphics[scale=0.25]{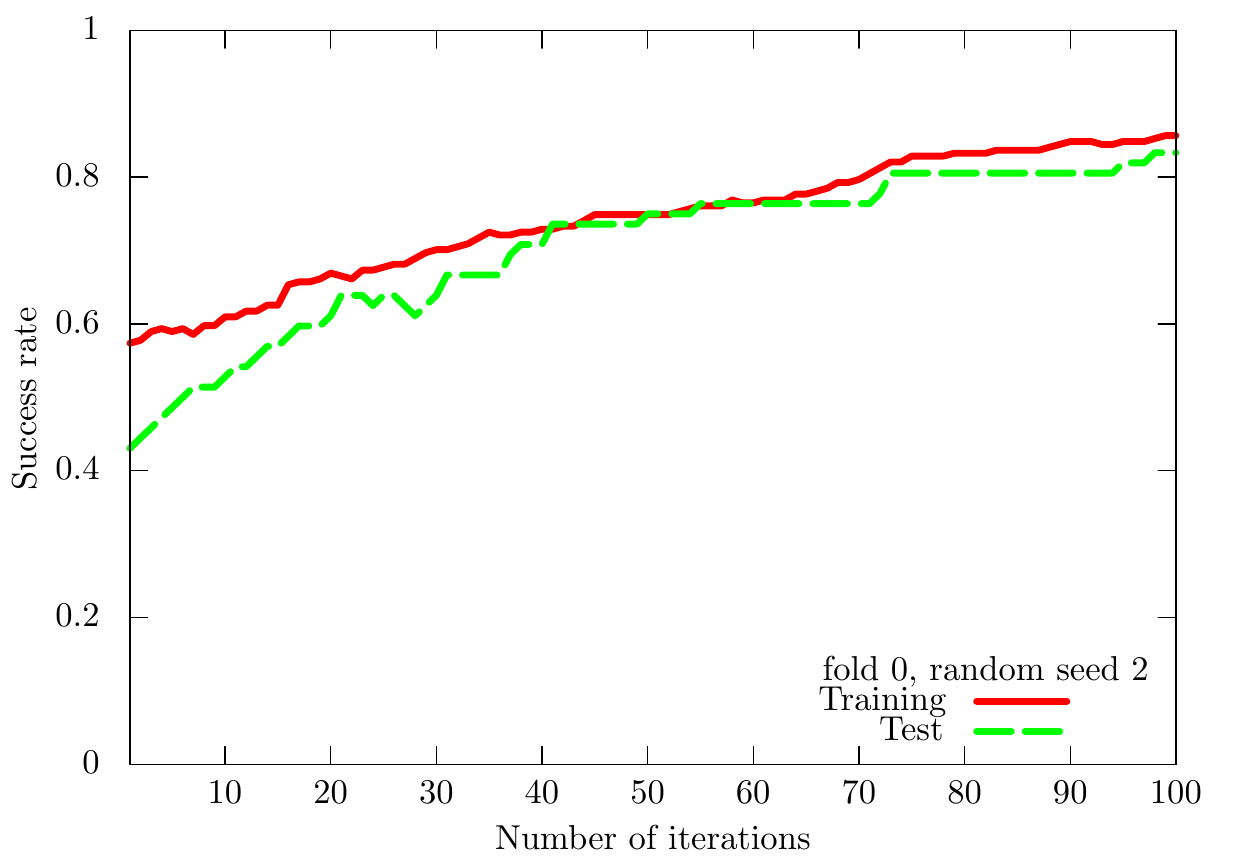}
\includegraphics[scale=0.25]{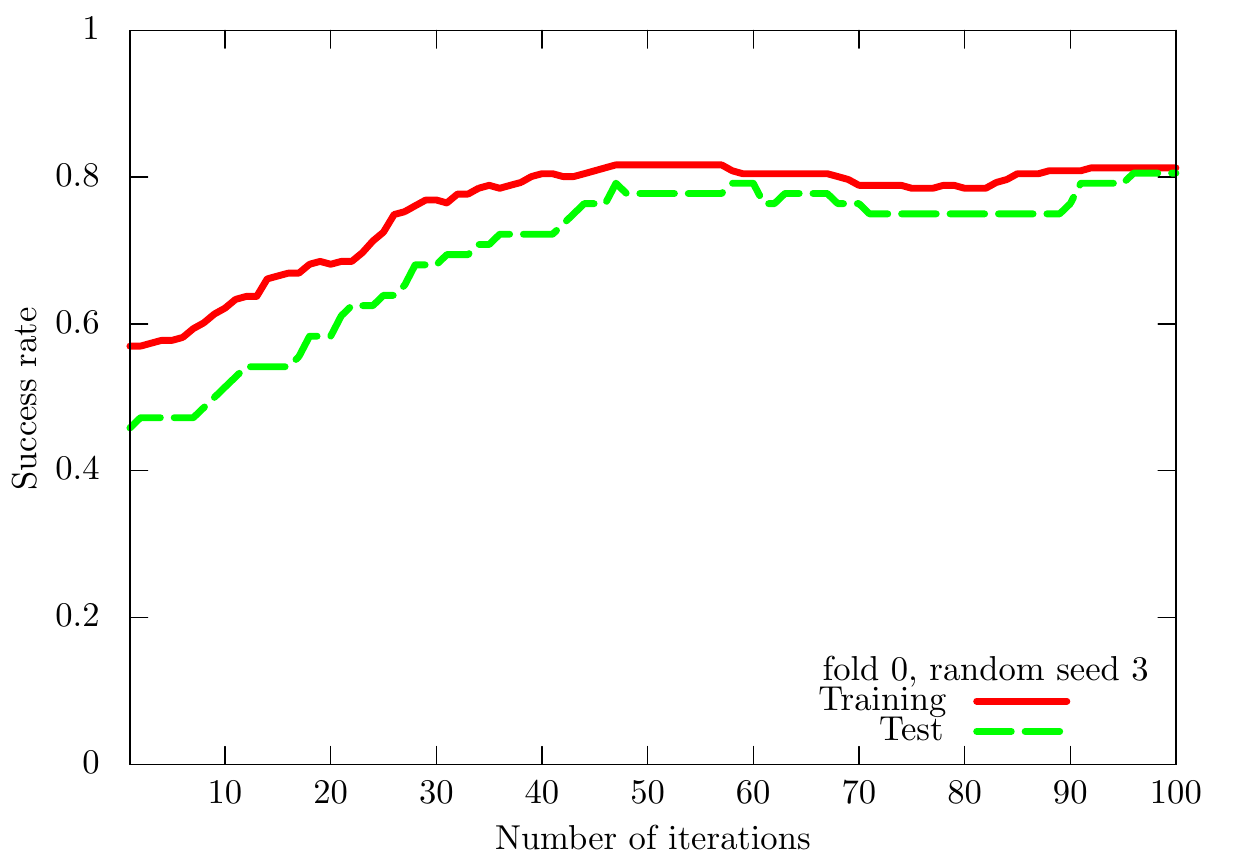}
\includegraphics[scale=0.25]{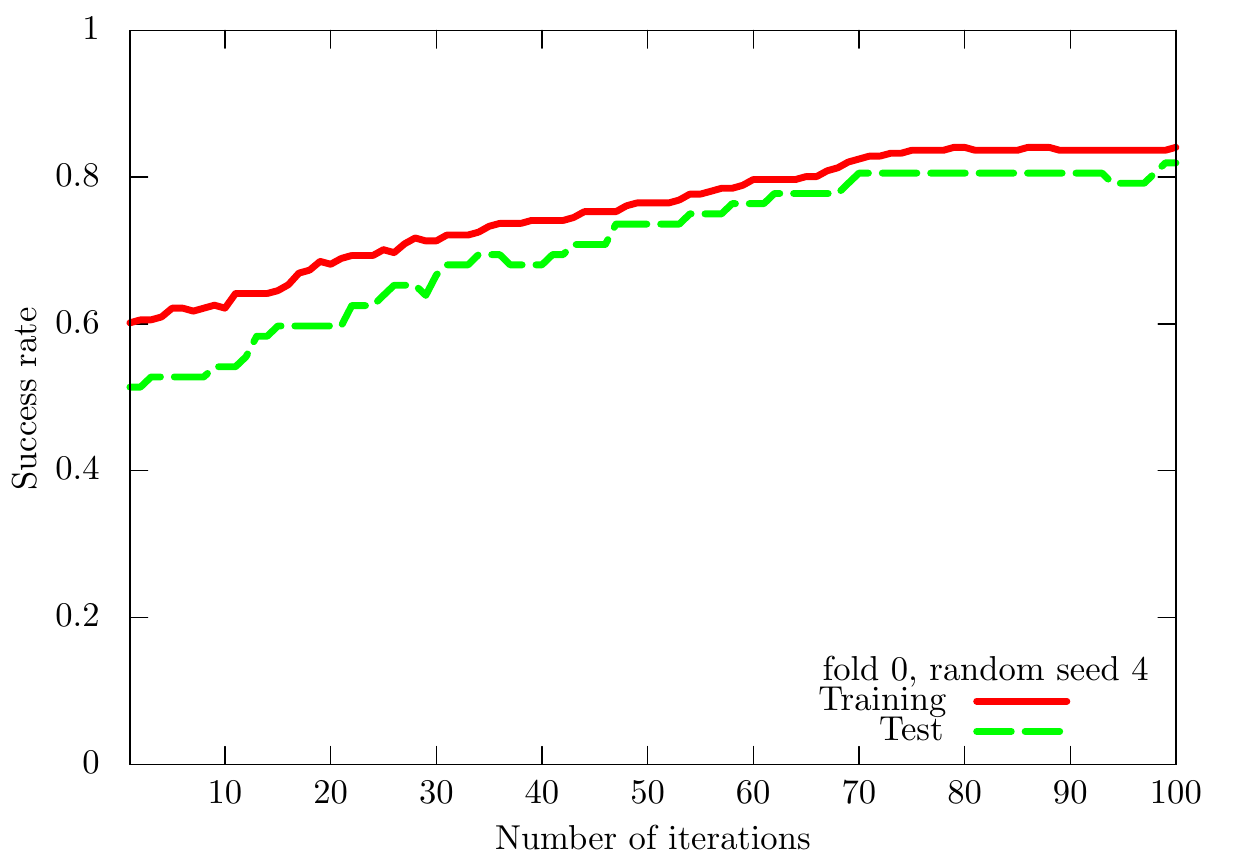}
\includegraphics[scale=0.25]{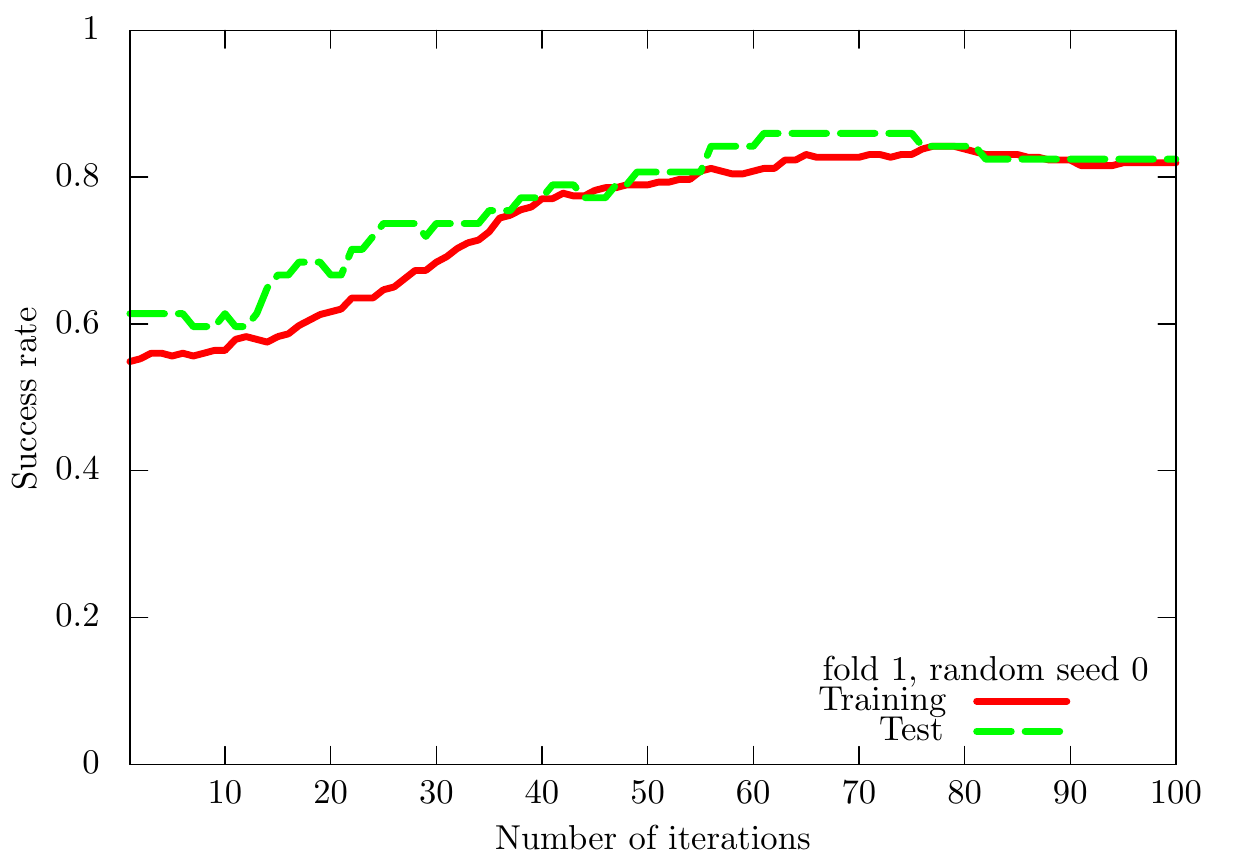}
\includegraphics[scale=0.25]{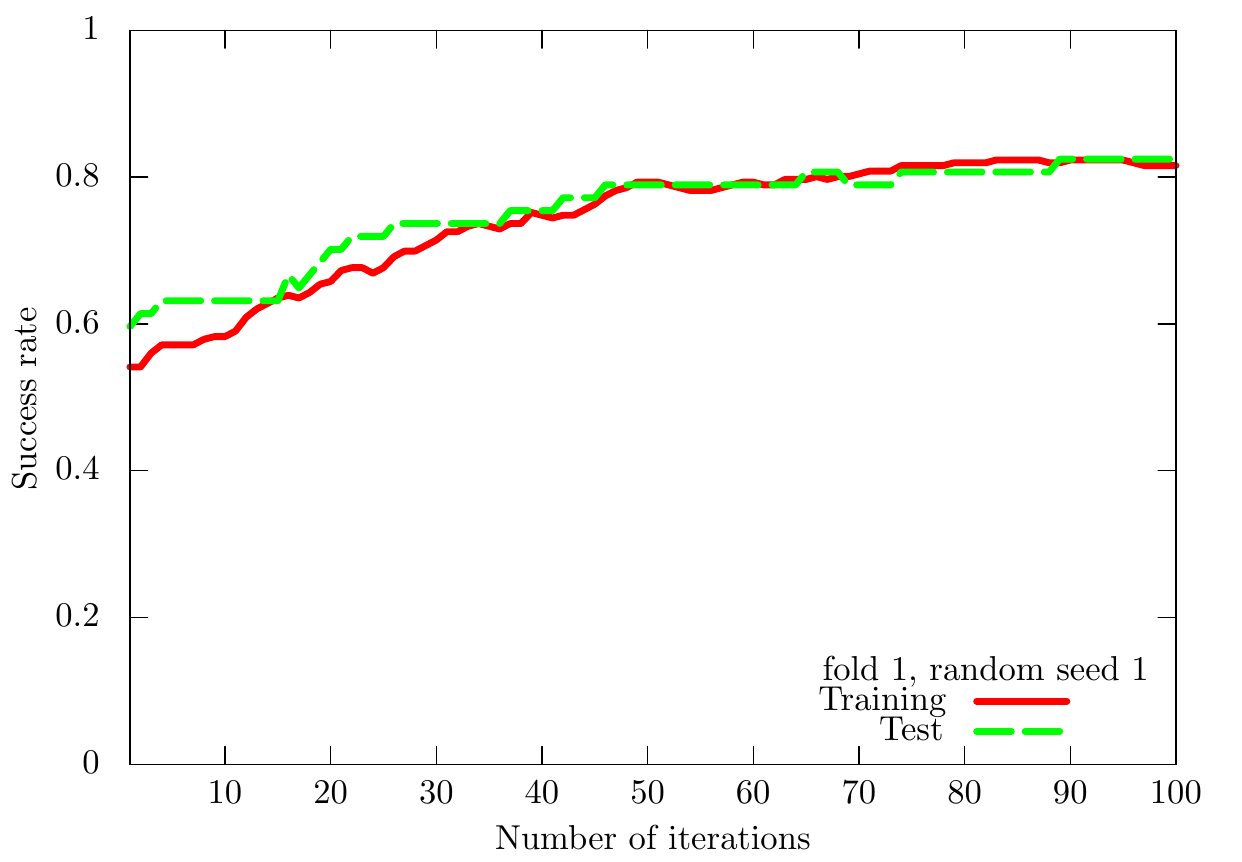}
\includegraphics[scale=0.25]{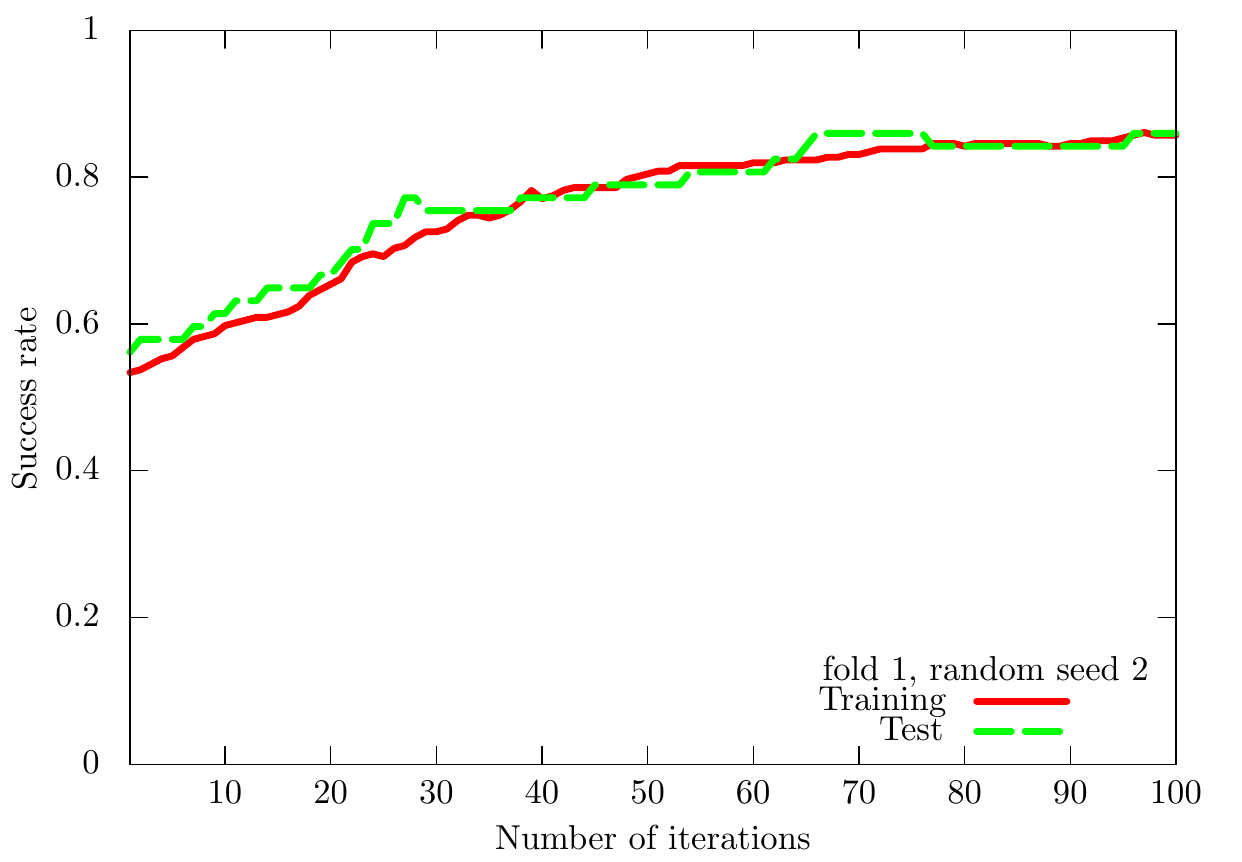}
\includegraphics[scale=0.25]{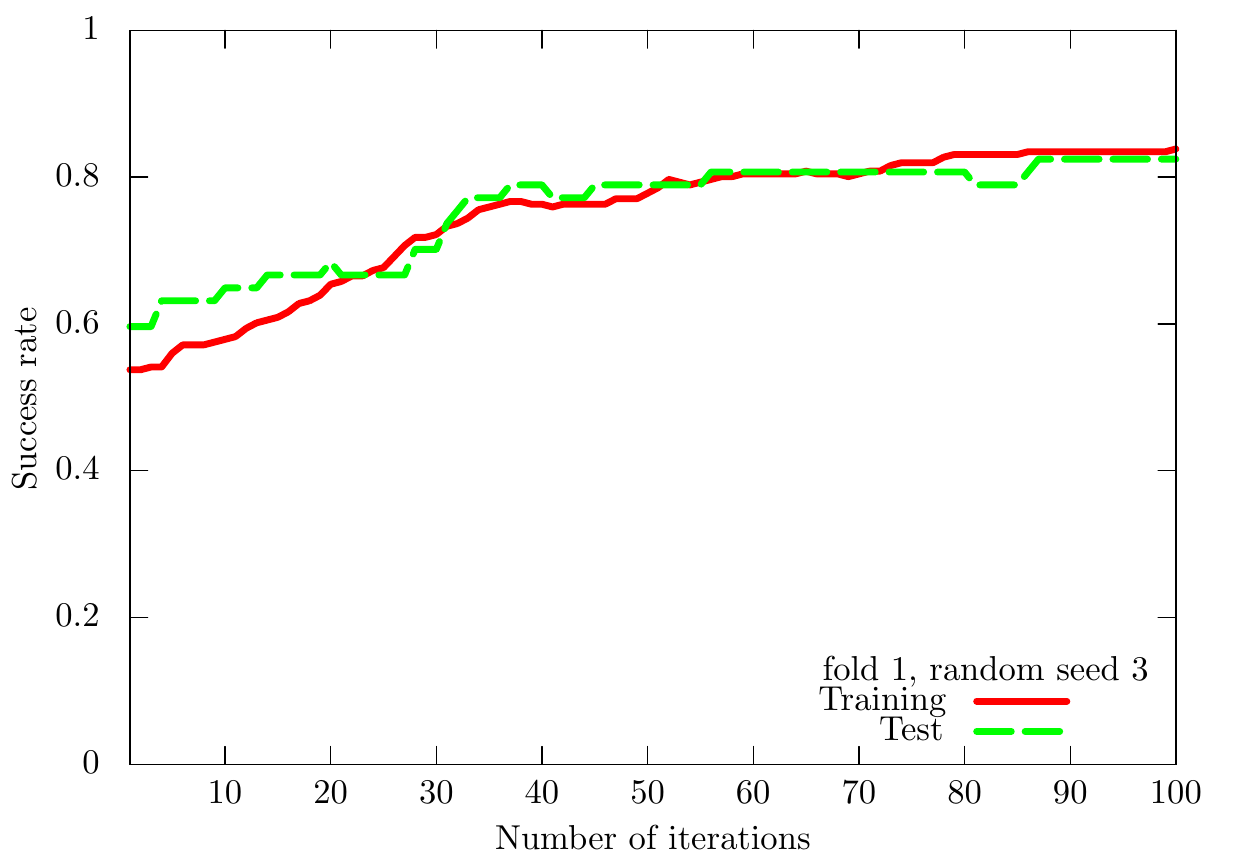}
\includegraphics[scale=0.25]{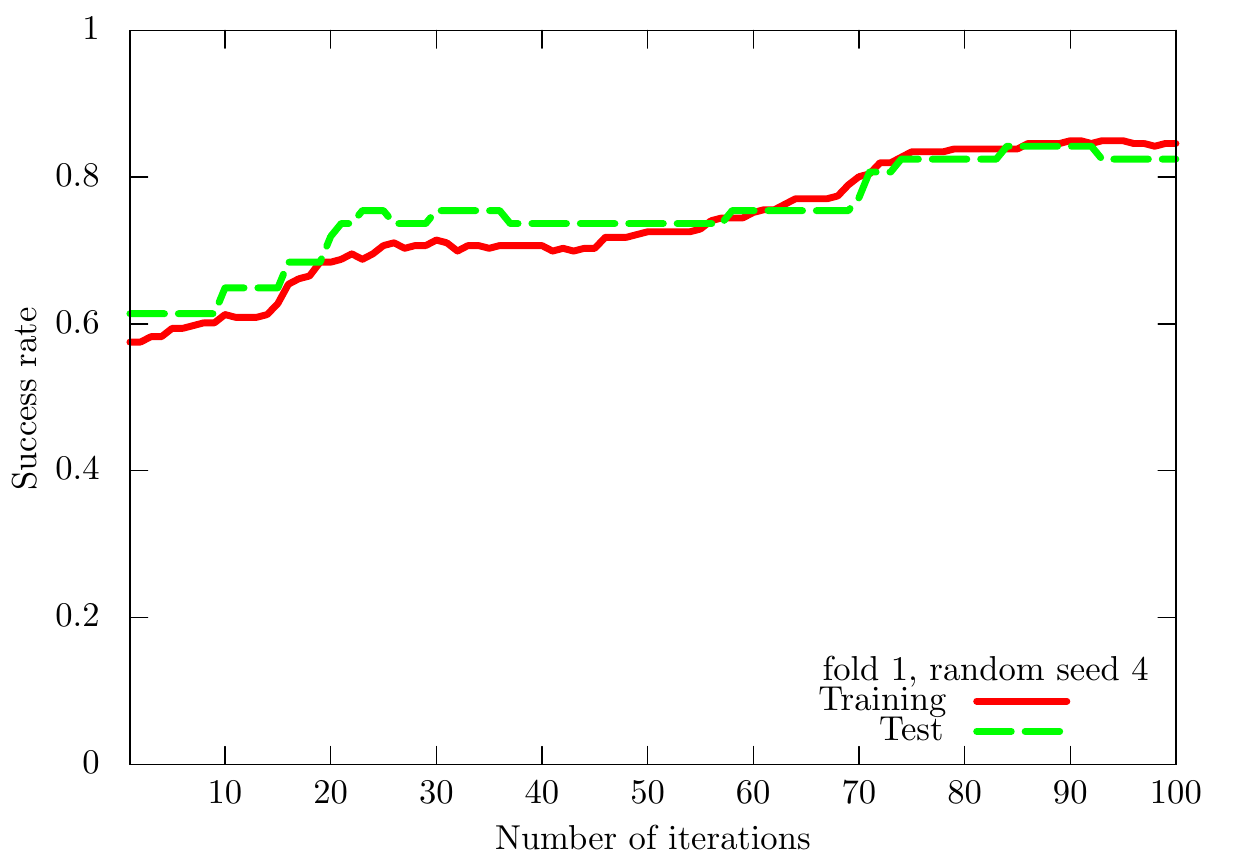}
\includegraphics[scale=0.25]{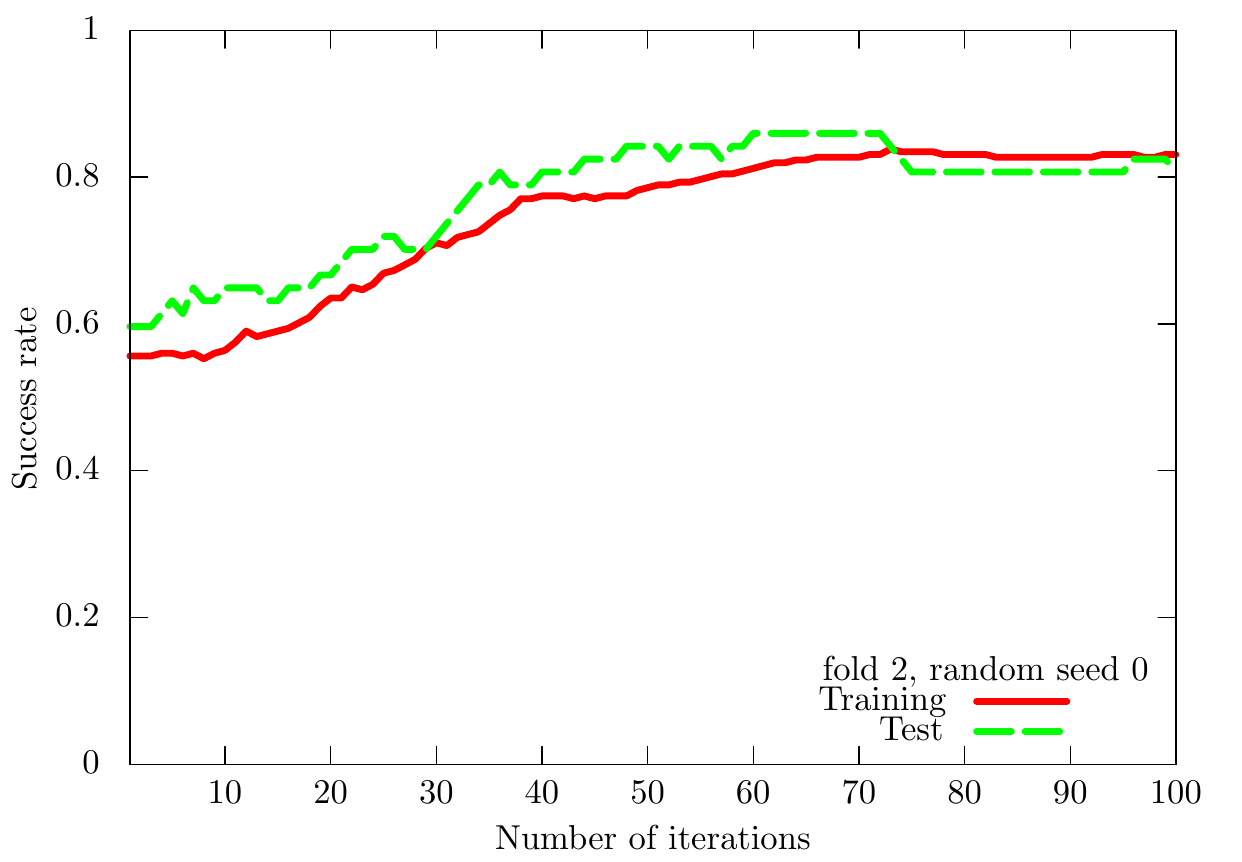}
\includegraphics[scale=0.25]{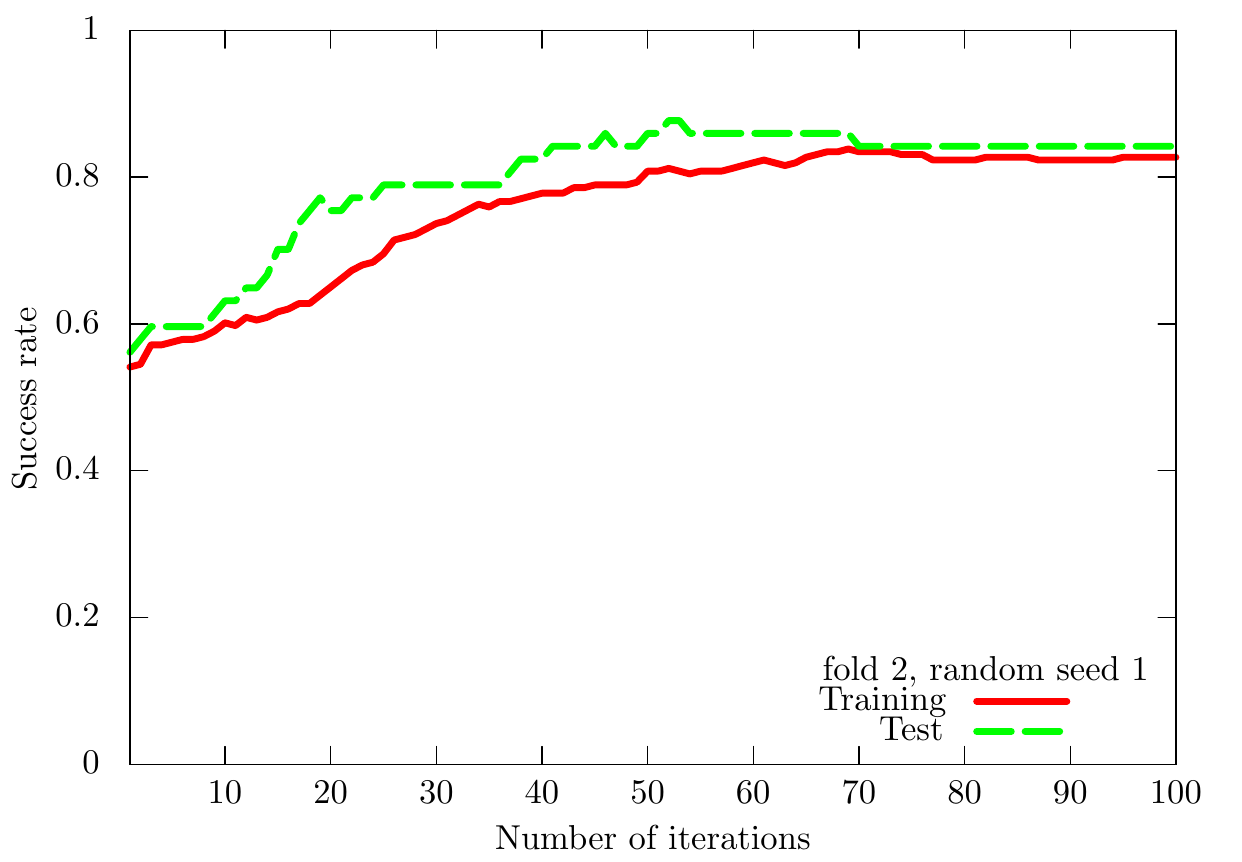}
\includegraphics[scale=0.25]{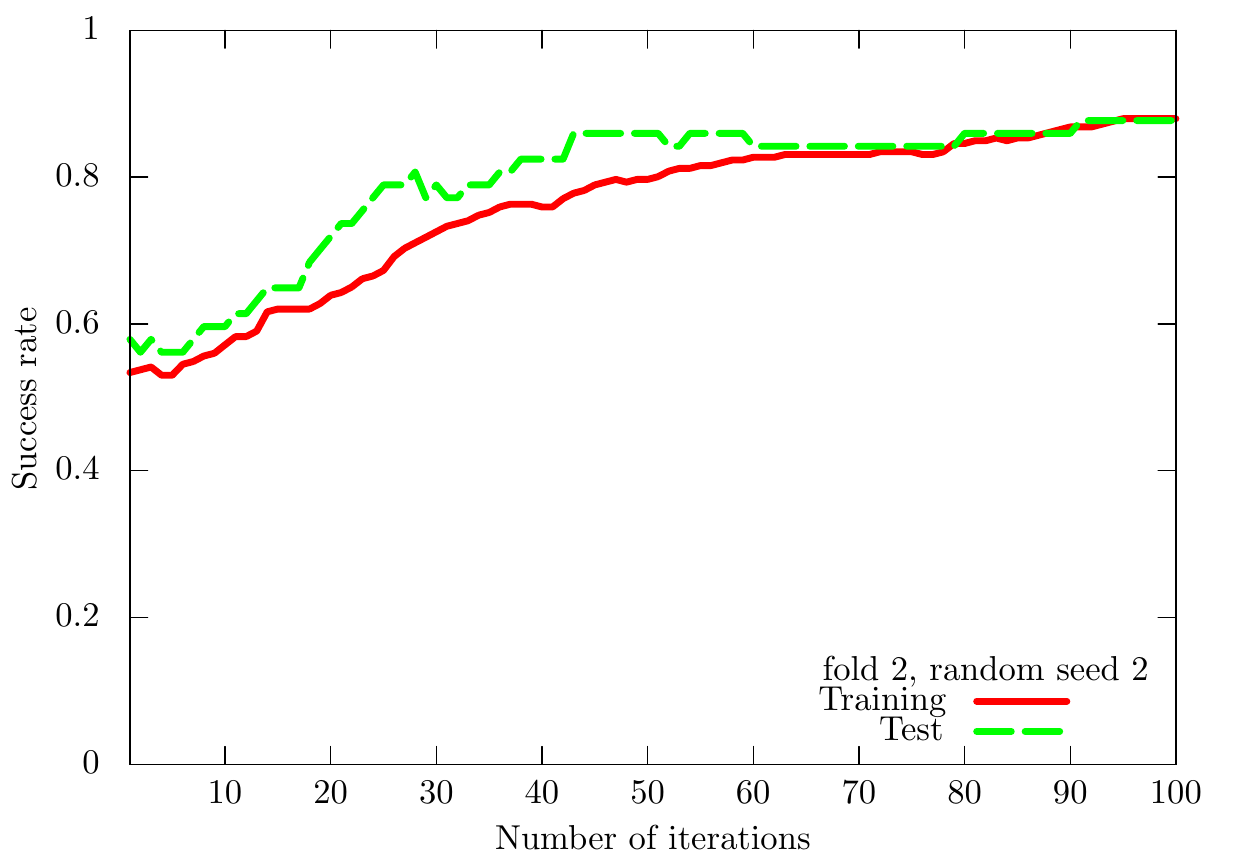}
\includegraphics[scale=0.25]{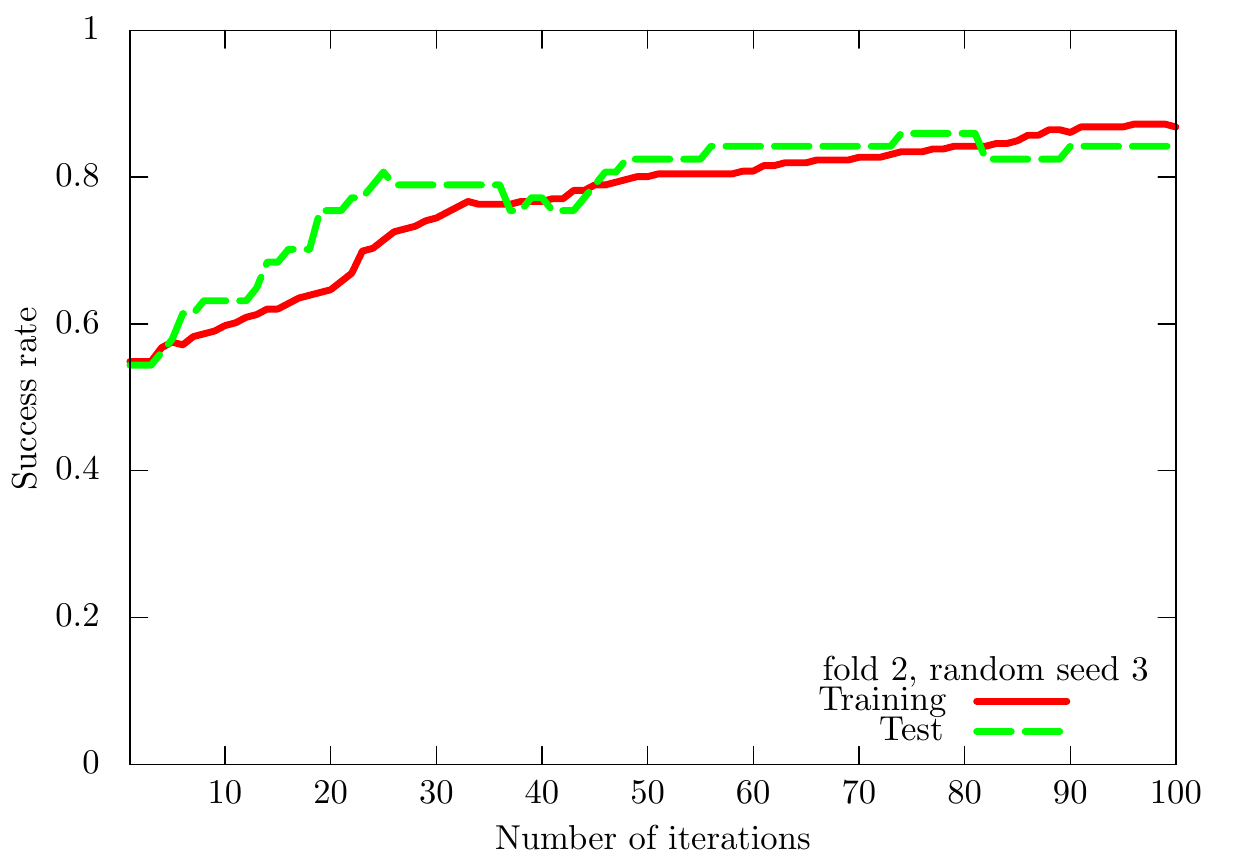}
\includegraphics[scale=0.25]{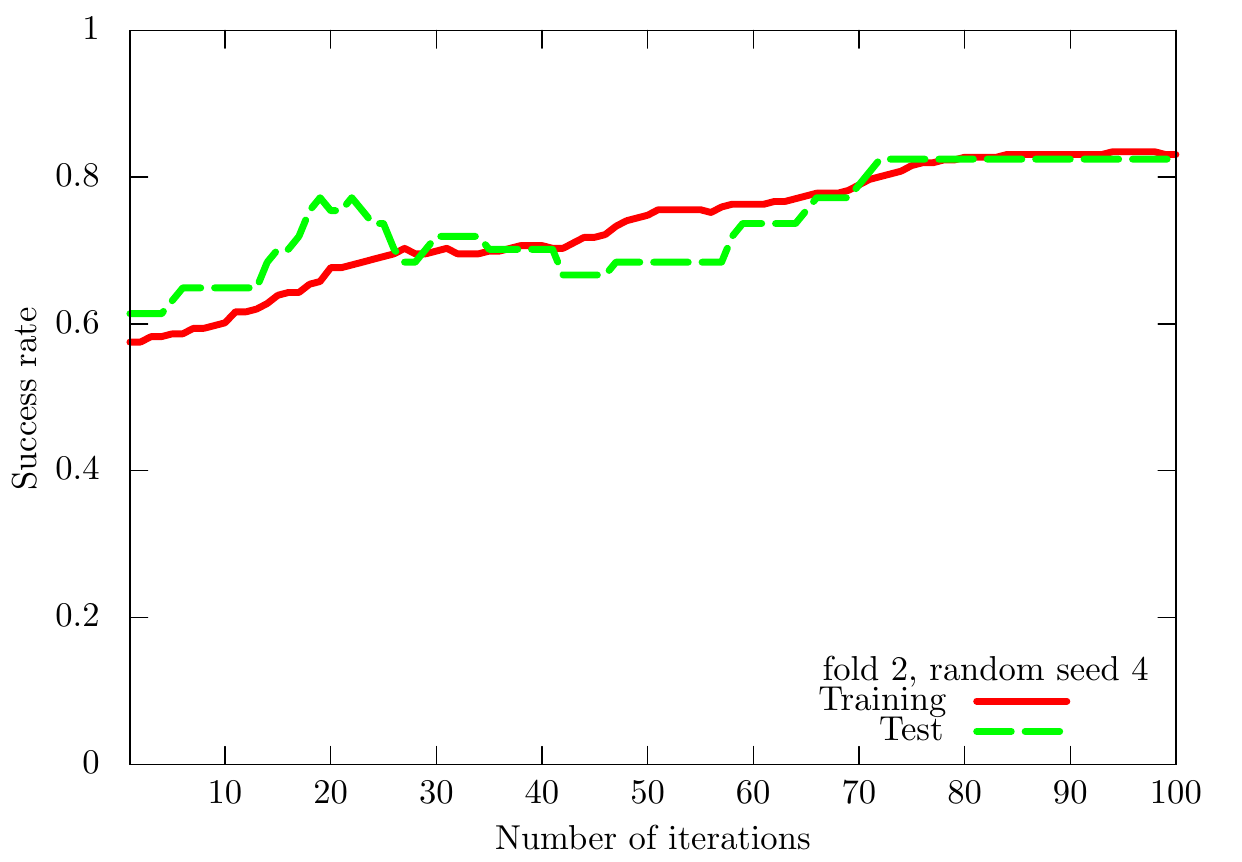}
\includegraphics[scale=0.25]{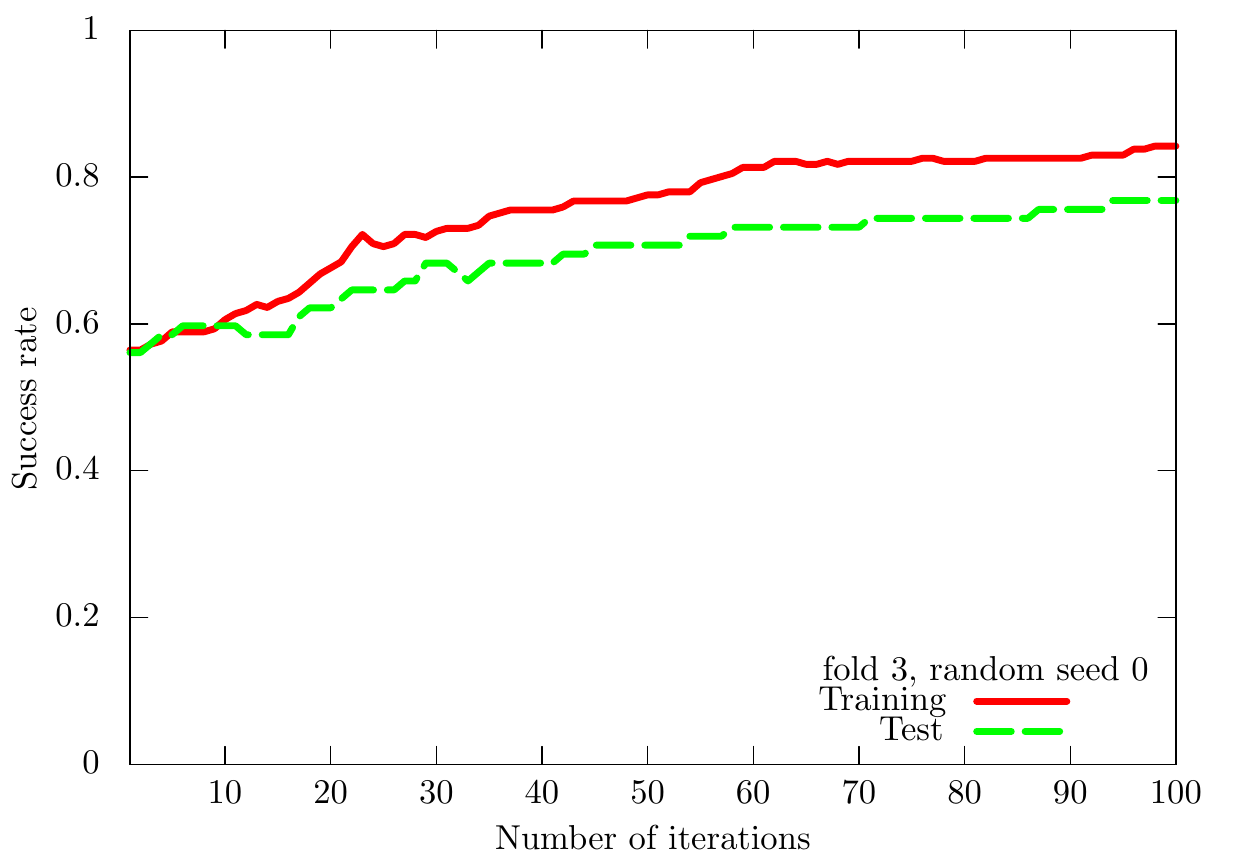}
\includegraphics[scale=0.25]{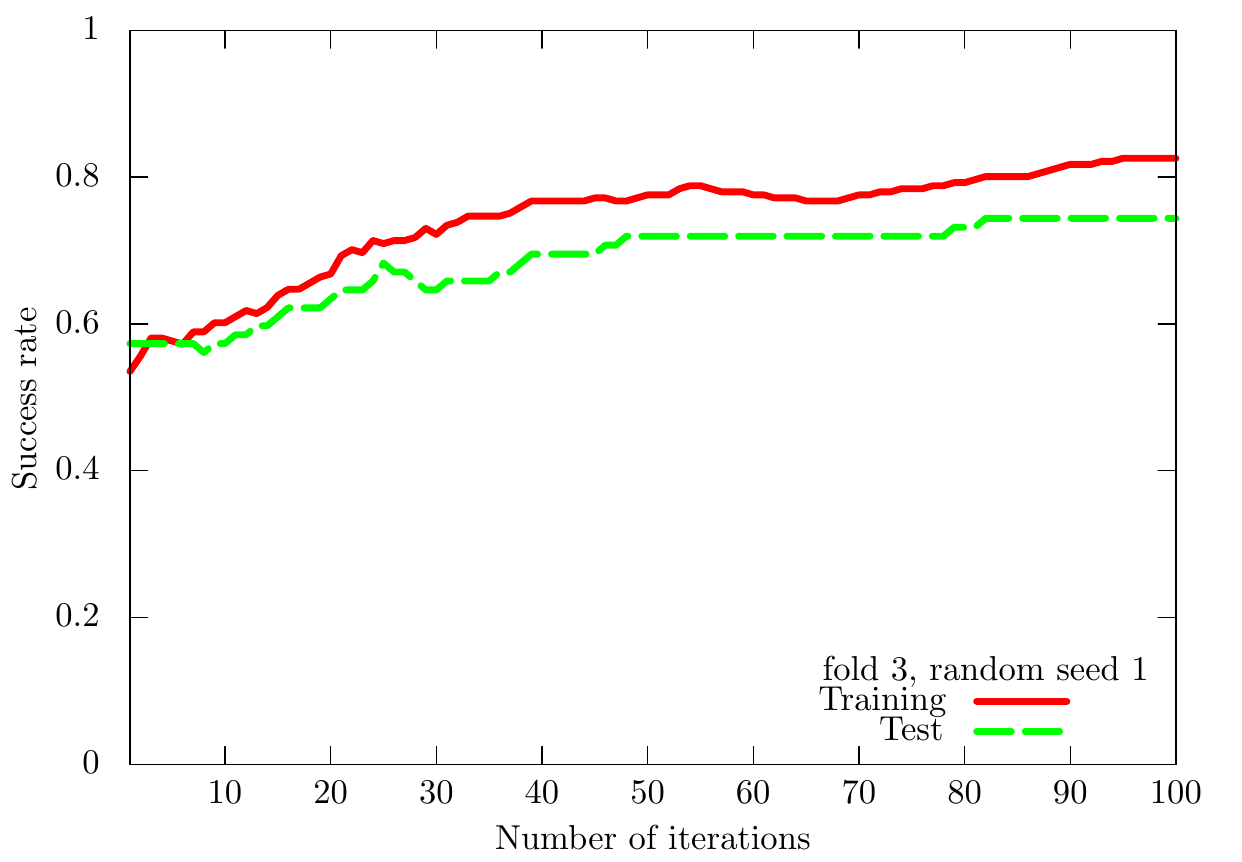}
\includegraphics[scale=0.25]{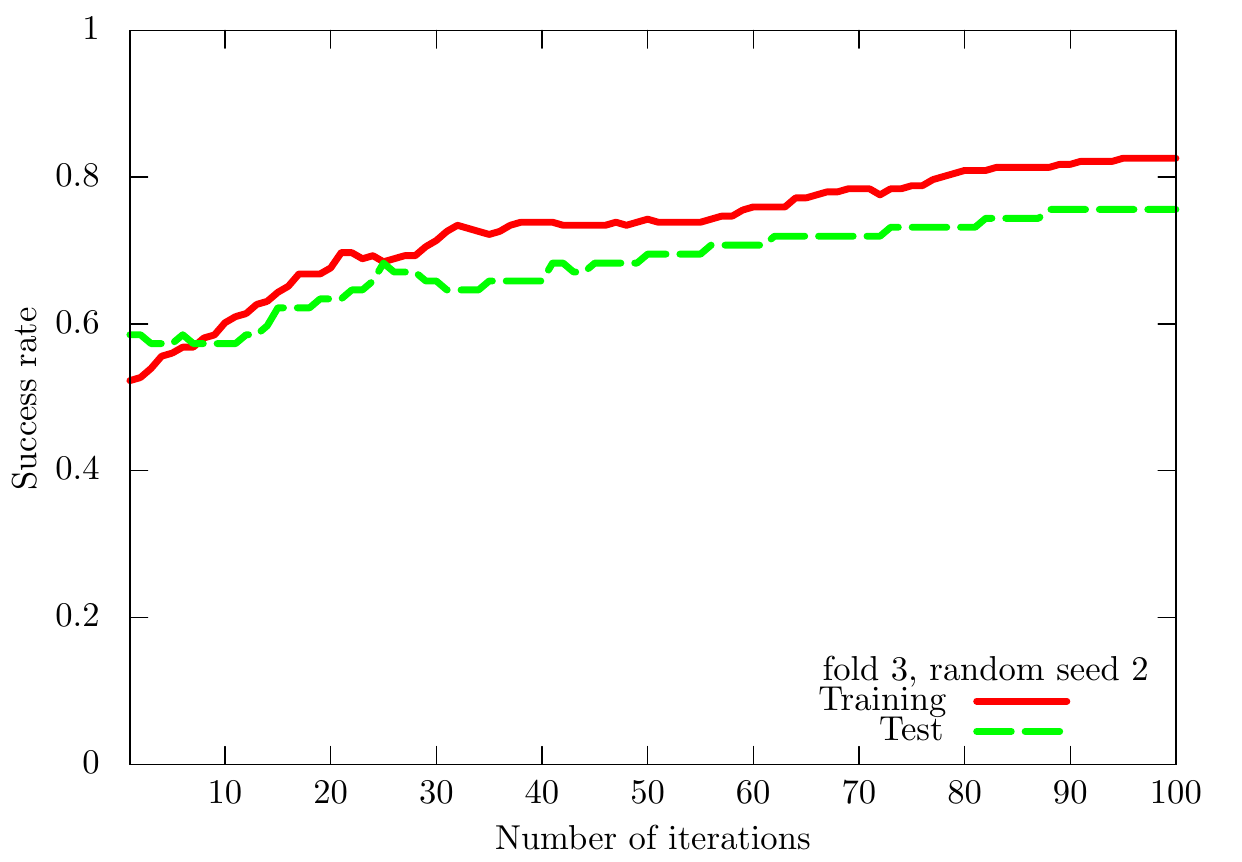}
\includegraphics[scale=0.25]{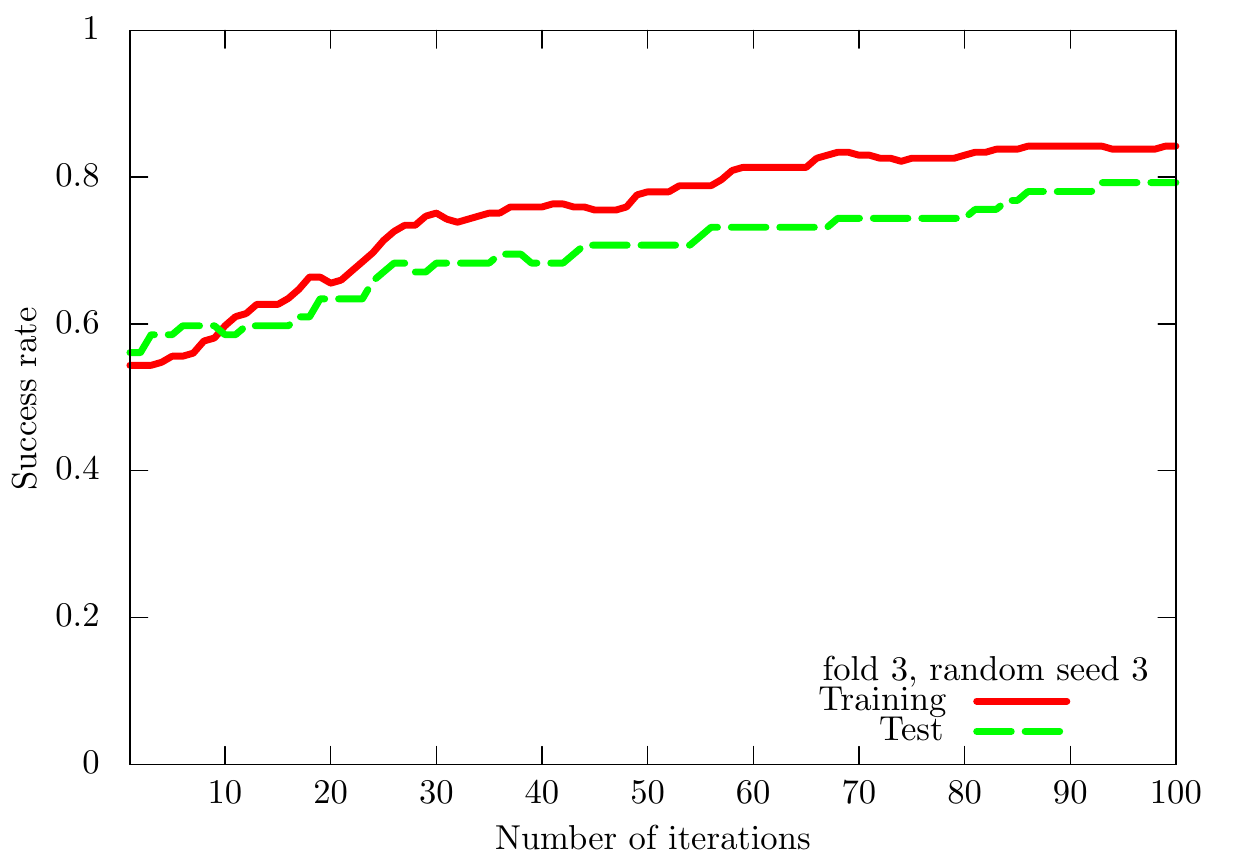}
\includegraphics[scale=0.25]{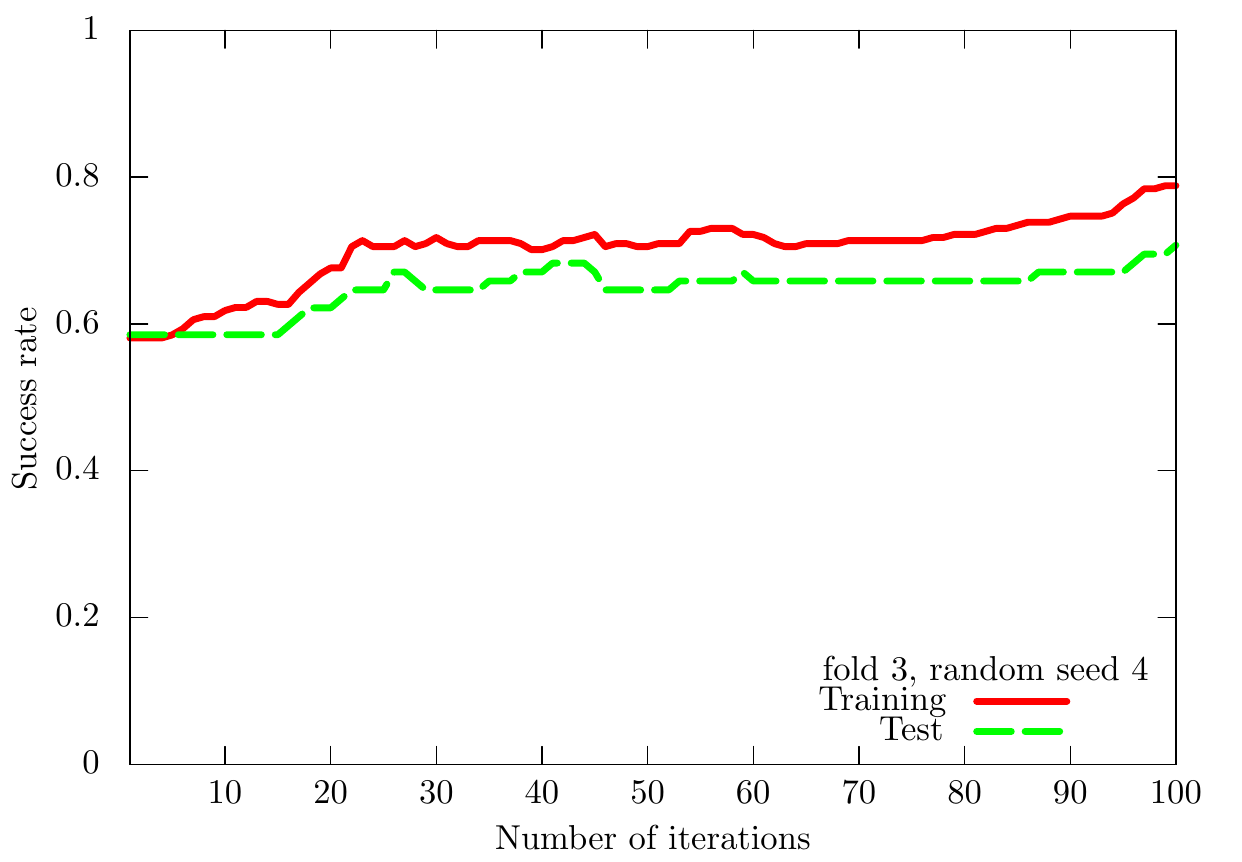}
\includegraphics[scale=0.25]{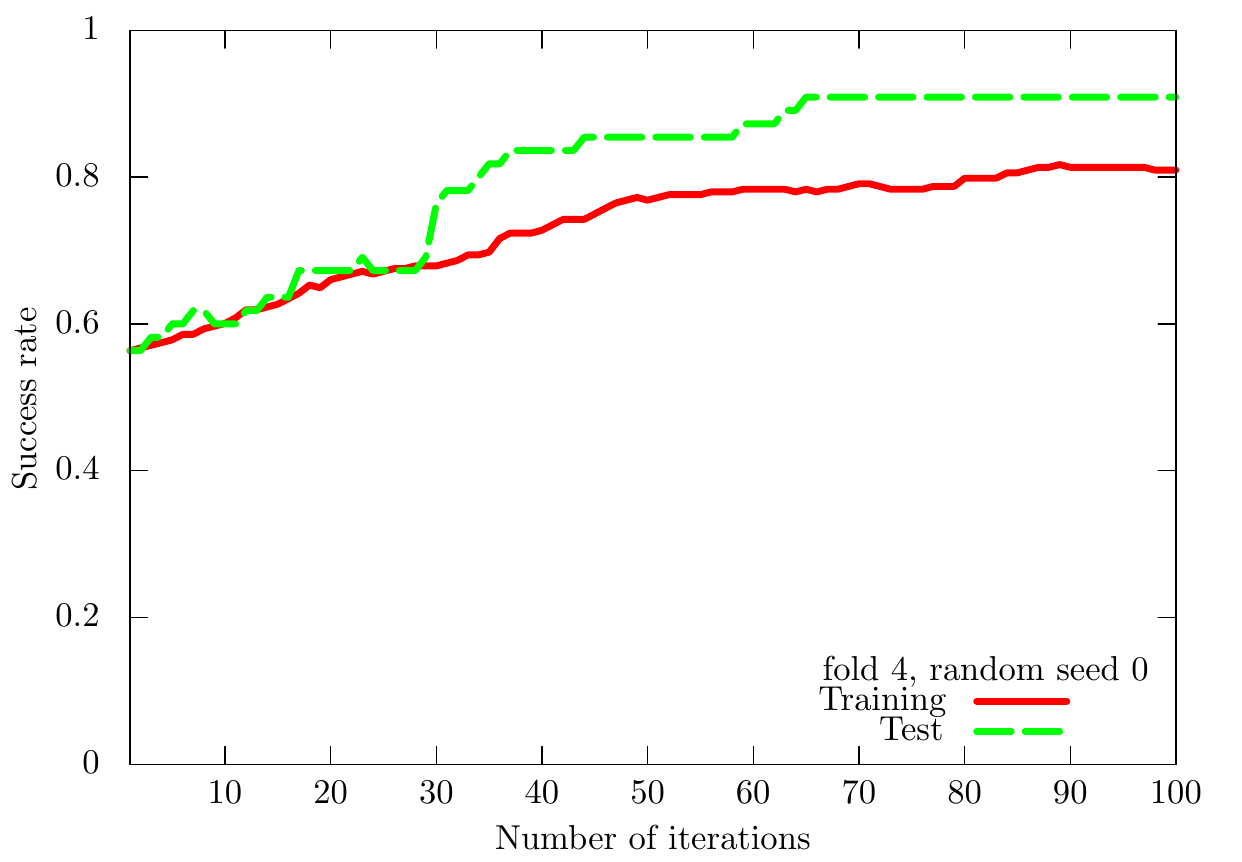}
\includegraphics[scale=0.25]{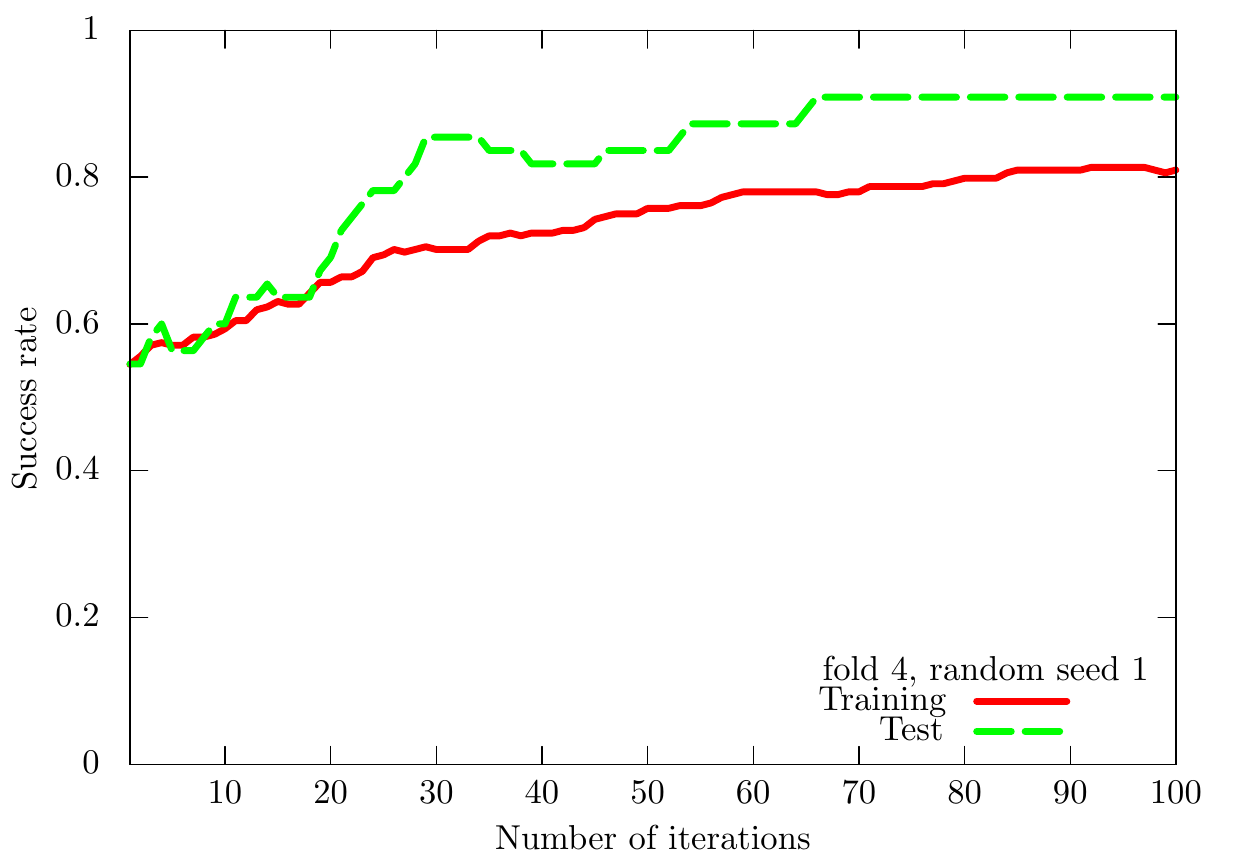}
\includegraphics[scale=0.25]{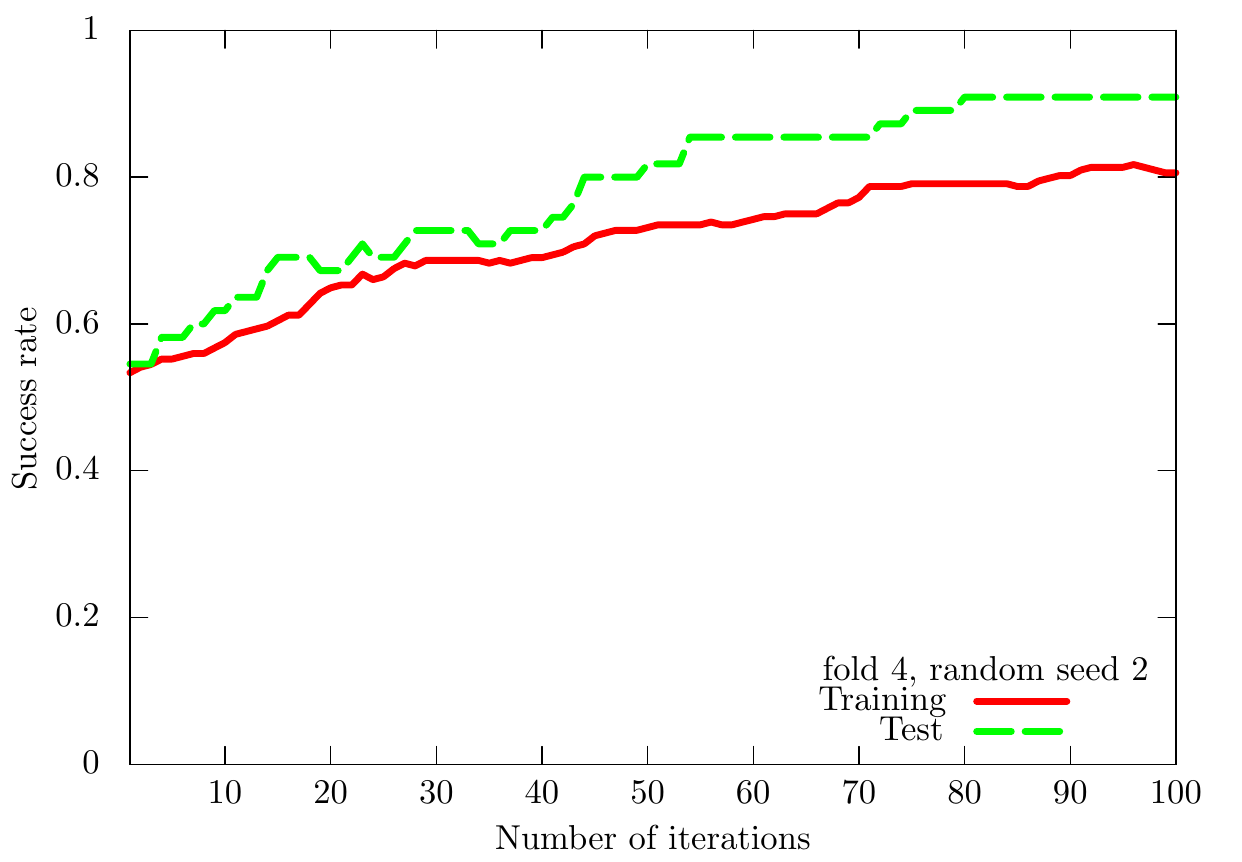}
\includegraphics[scale=0.25]{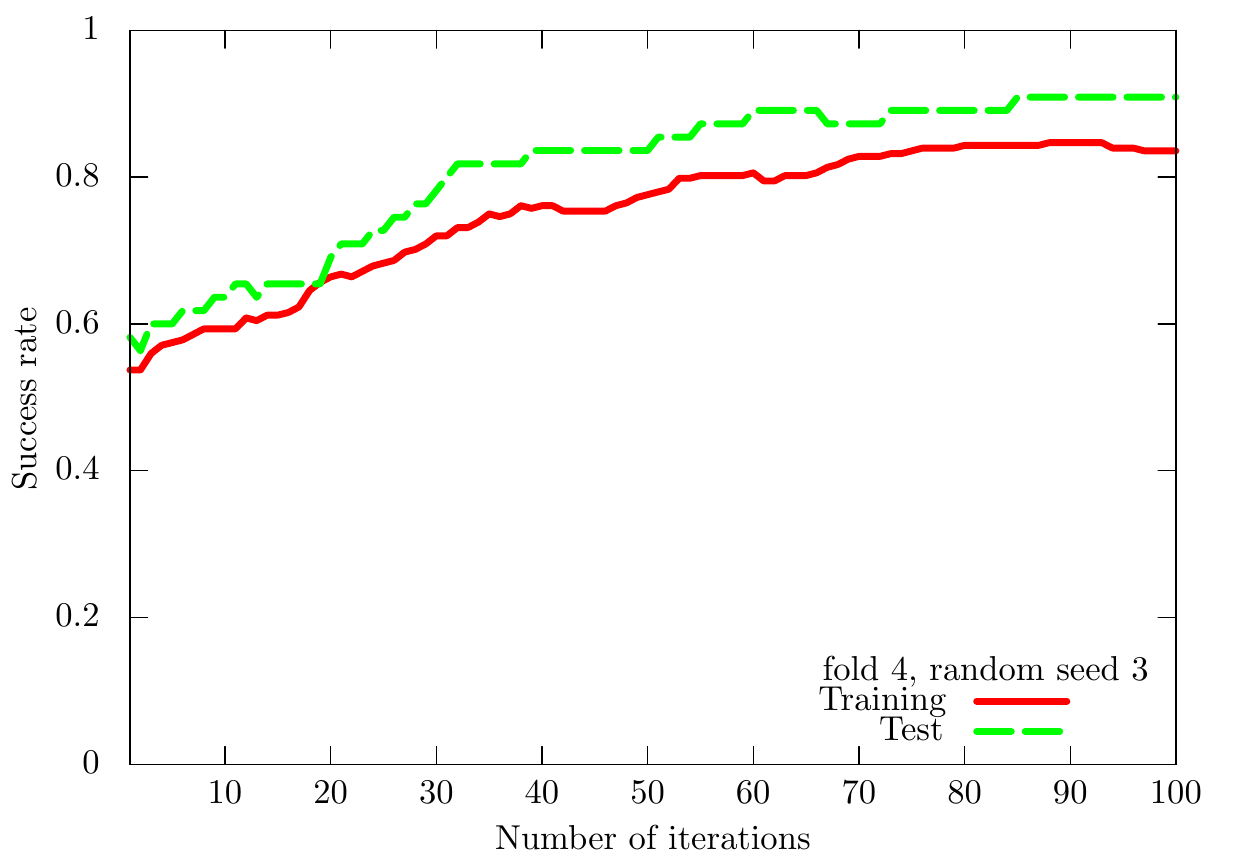}
\includegraphics[scale=0.25]{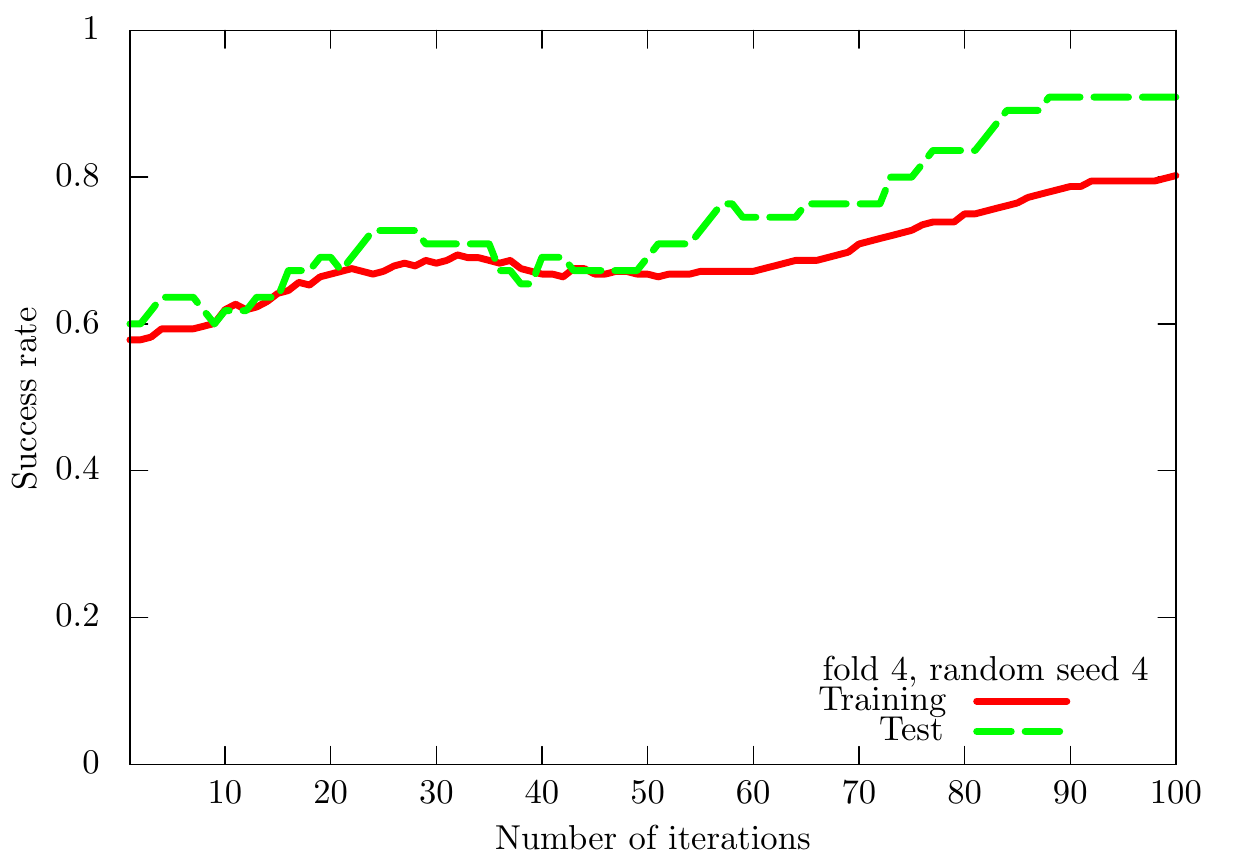}
\caption{Results of QCL on the $5$-fold datasets with $5$ different random seeds for the semeion dataset ($0$ or $1$). We use the CNOT-based circuit and set $\theta_\mathrm{bias} = 0$. The number of layers $L$ is set to $5$.}
\label{supp-arXiv-numerical-result-raw-data-fold-001-rand-001-QCL-UCI-semeion-0-1}
\end{figure*}

\begin{figure*}[htb]
\centering
\includegraphics[scale=0.25]{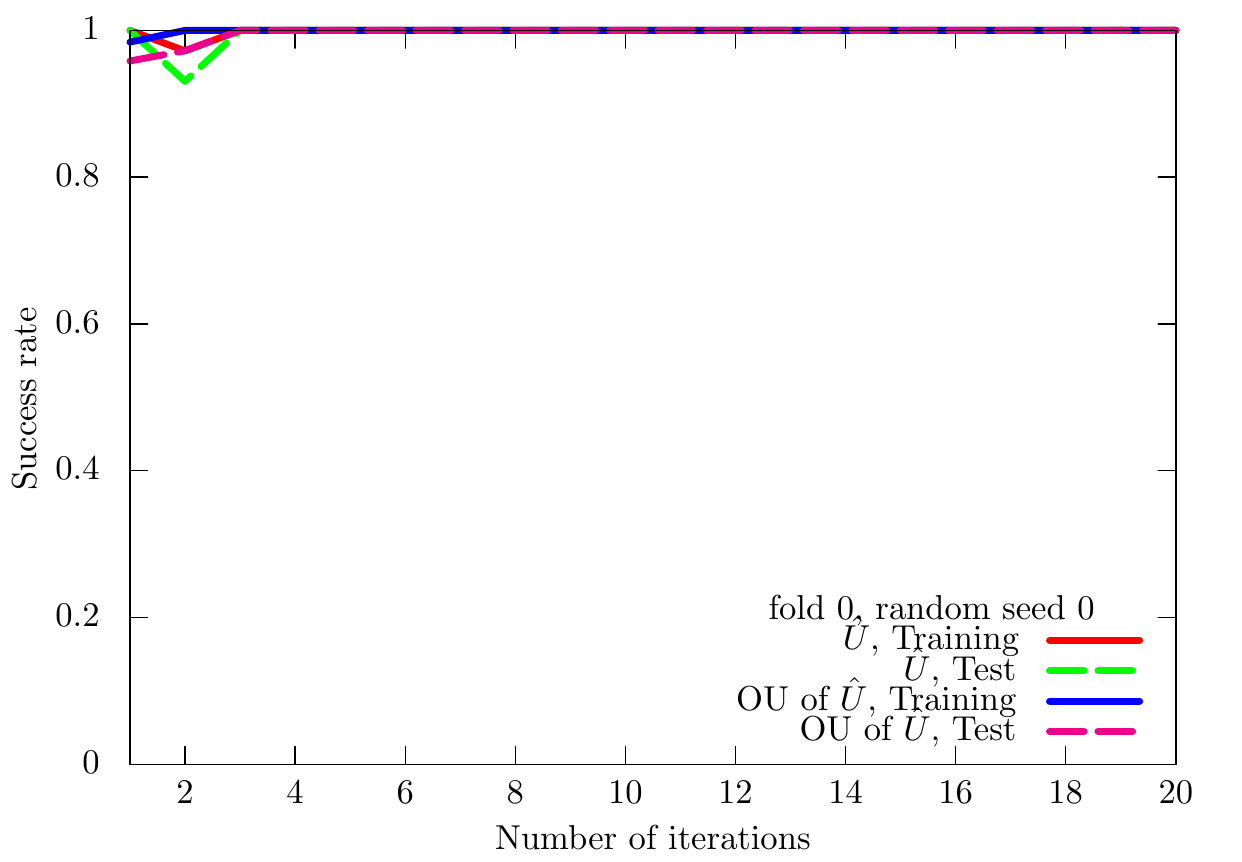}
\includegraphics[scale=0.25]{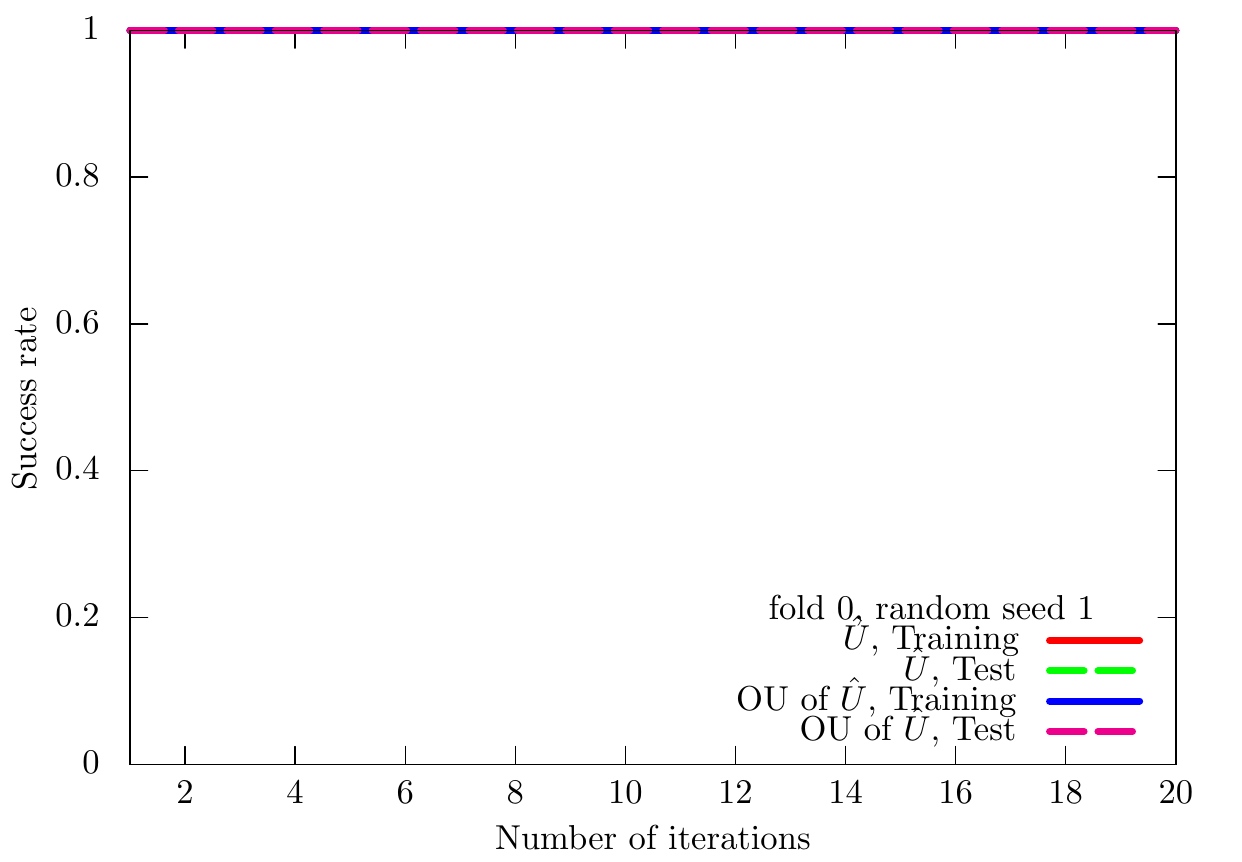}
\includegraphics[scale=0.25]{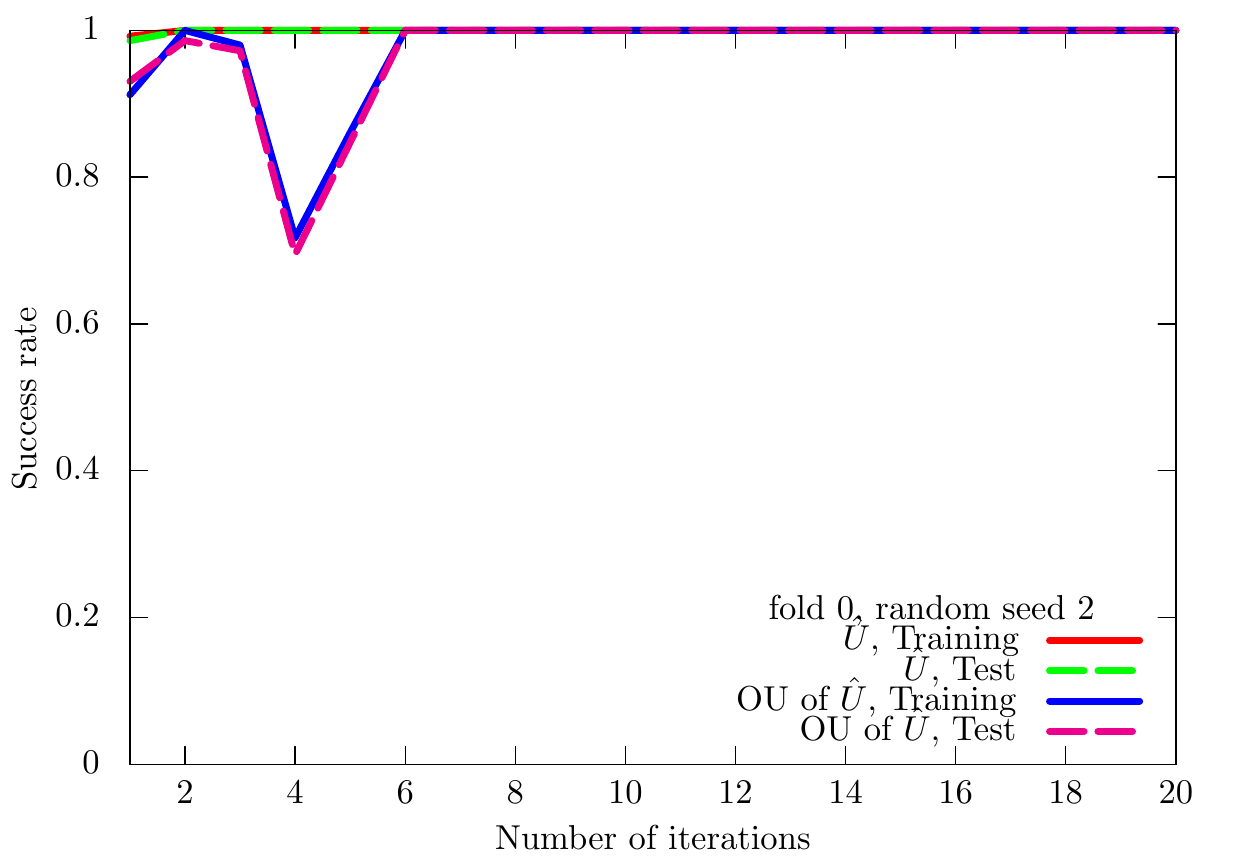}
\includegraphics[scale=0.25]{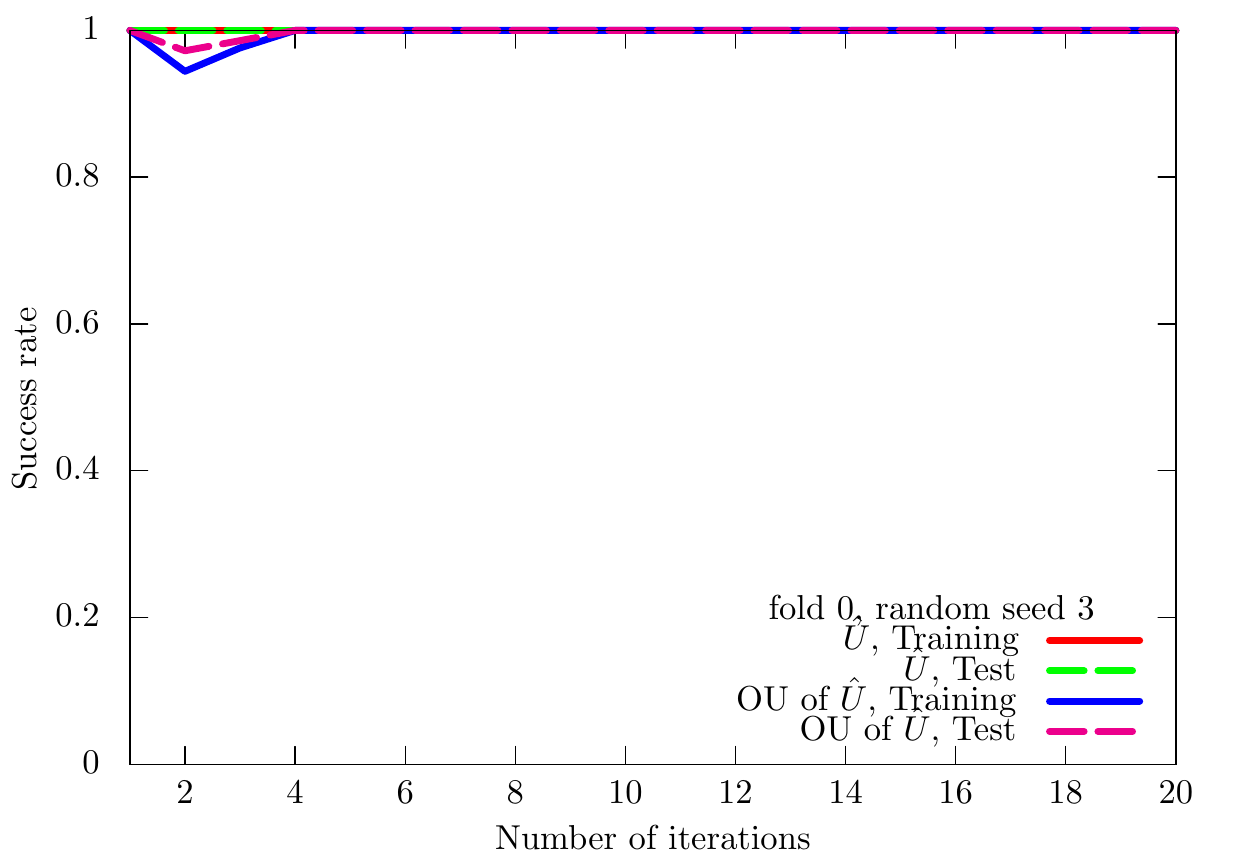}
\includegraphics[scale=0.25]{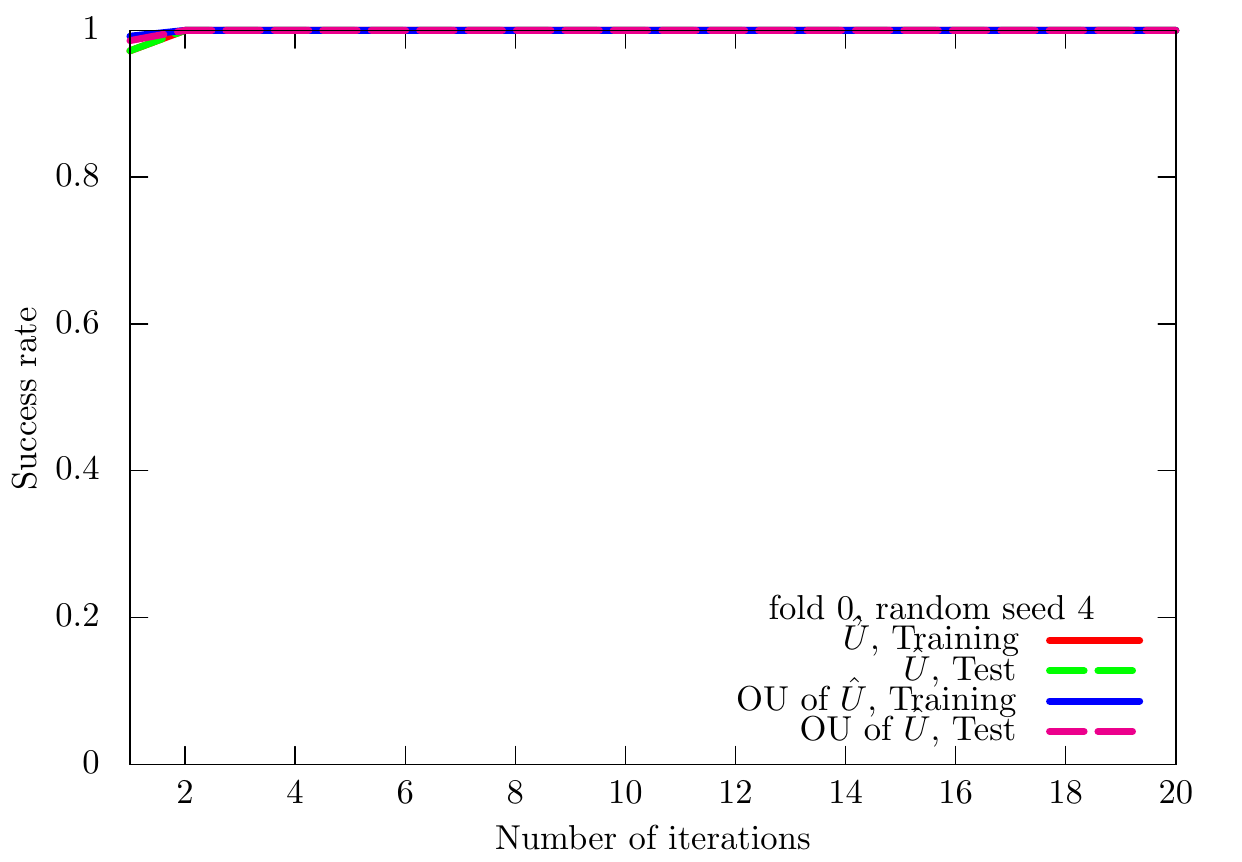}
\includegraphics[scale=0.25]{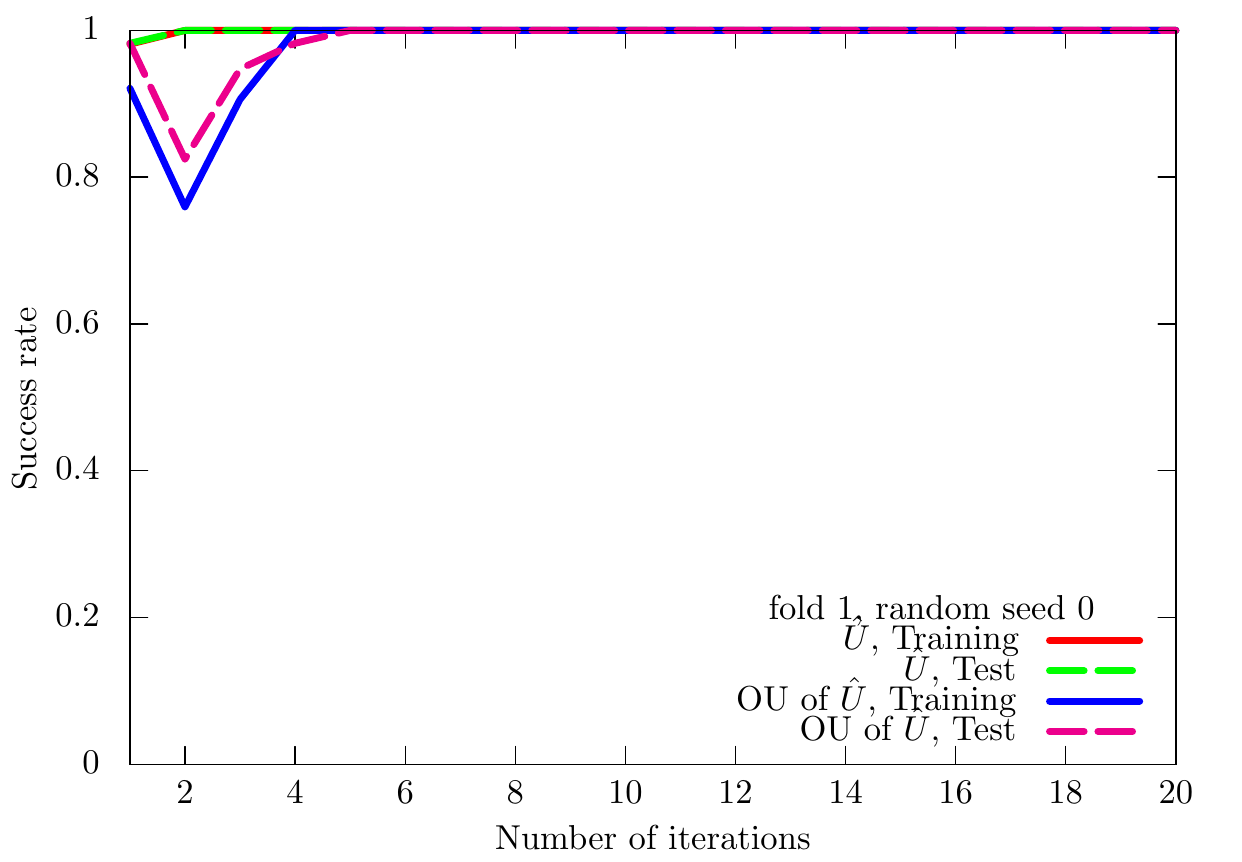}
\includegraphics[scale=0.25]{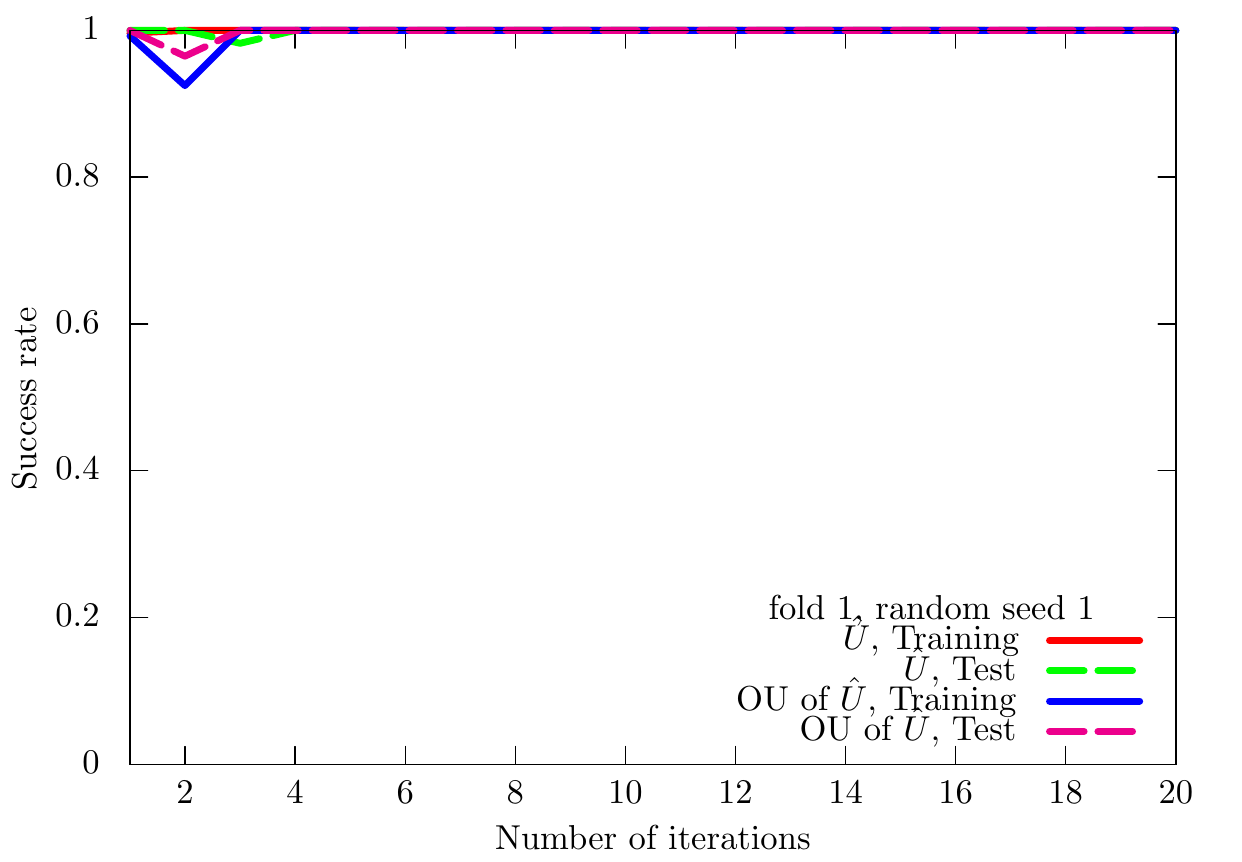}
\includegraphics[scale=0.25]{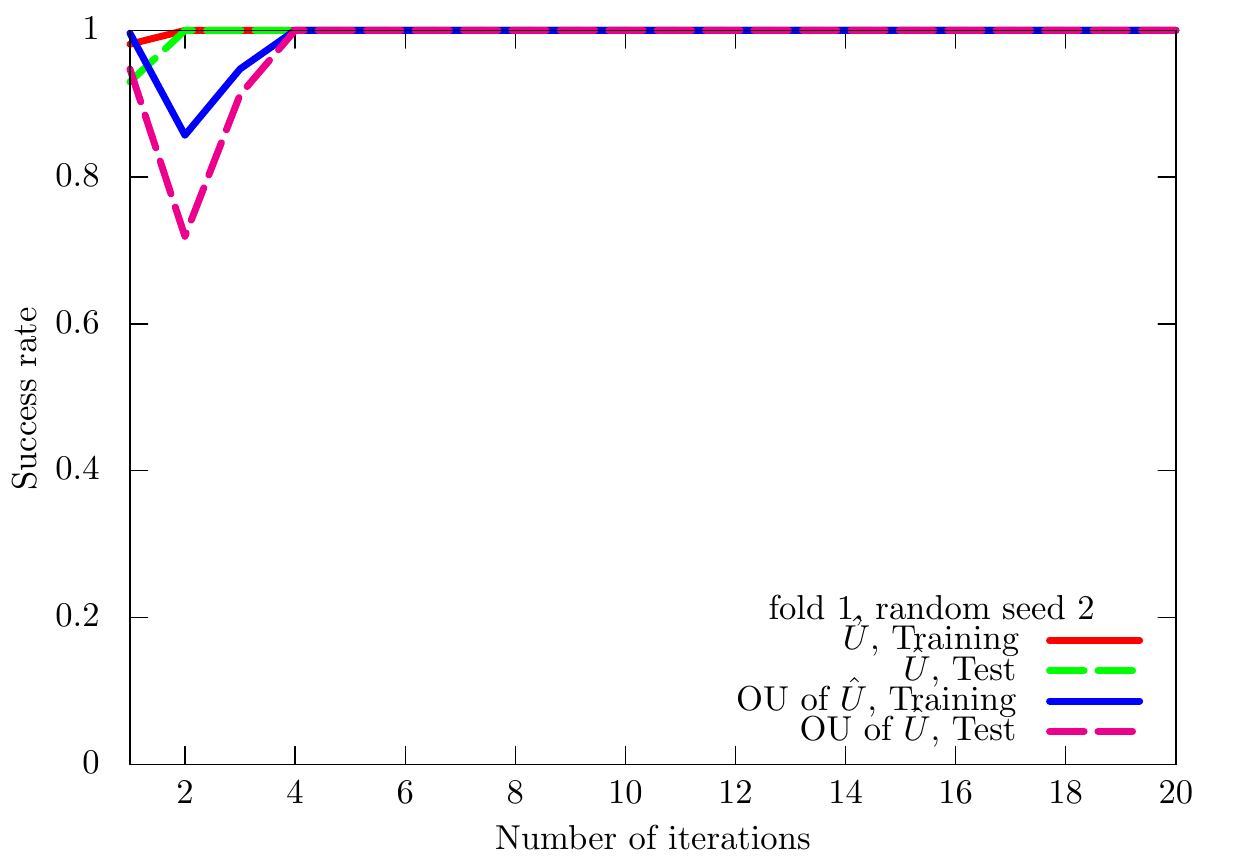}
\includegraphics[scale=0.25]{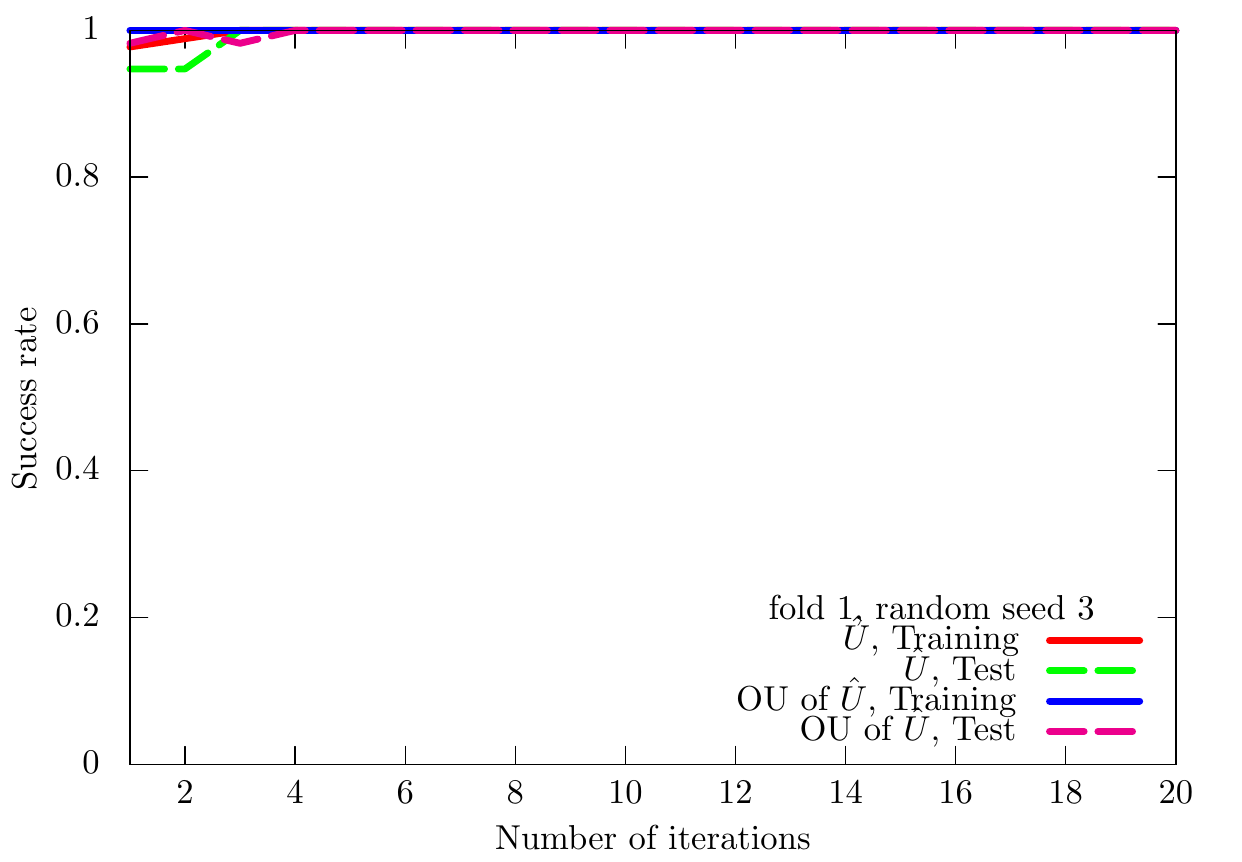}
\includegraphics[scale=0.25]{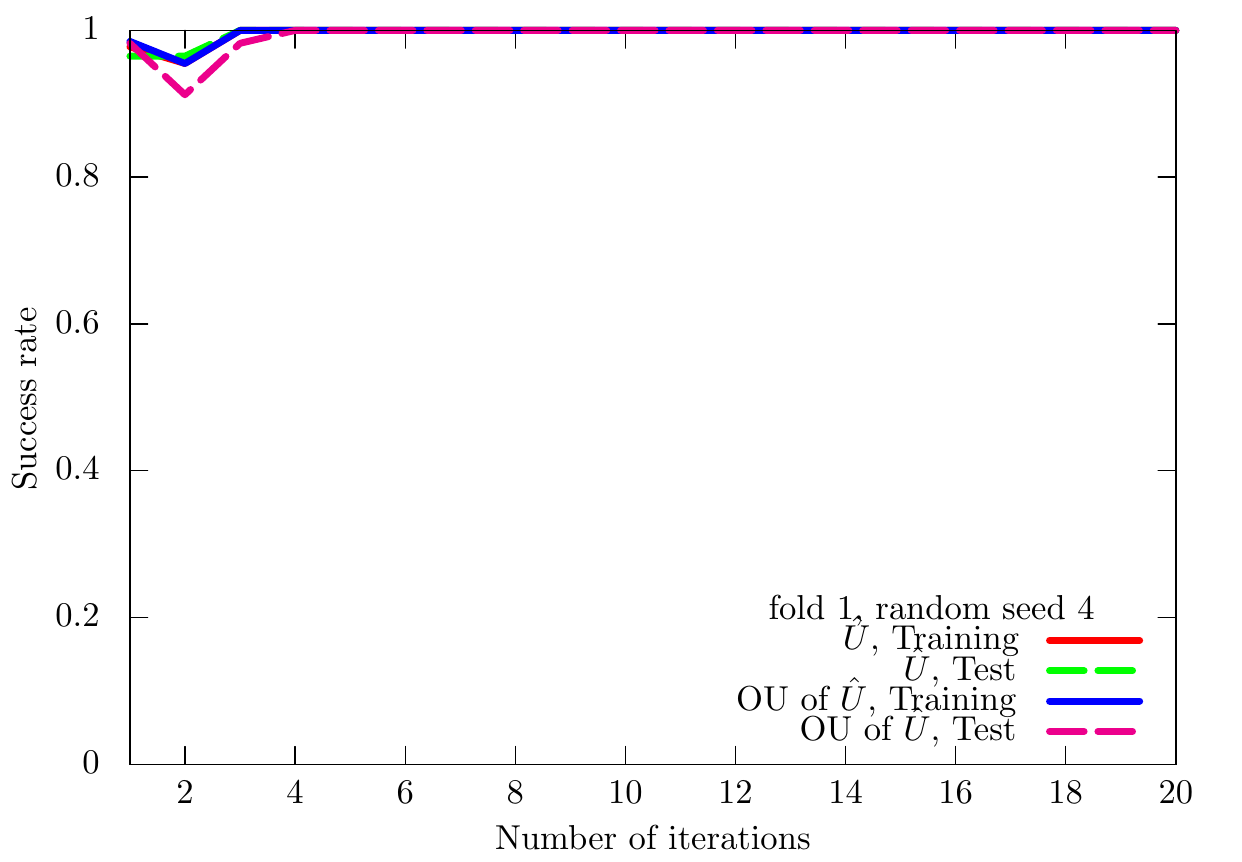}
\includegraphics[scale=0.25]{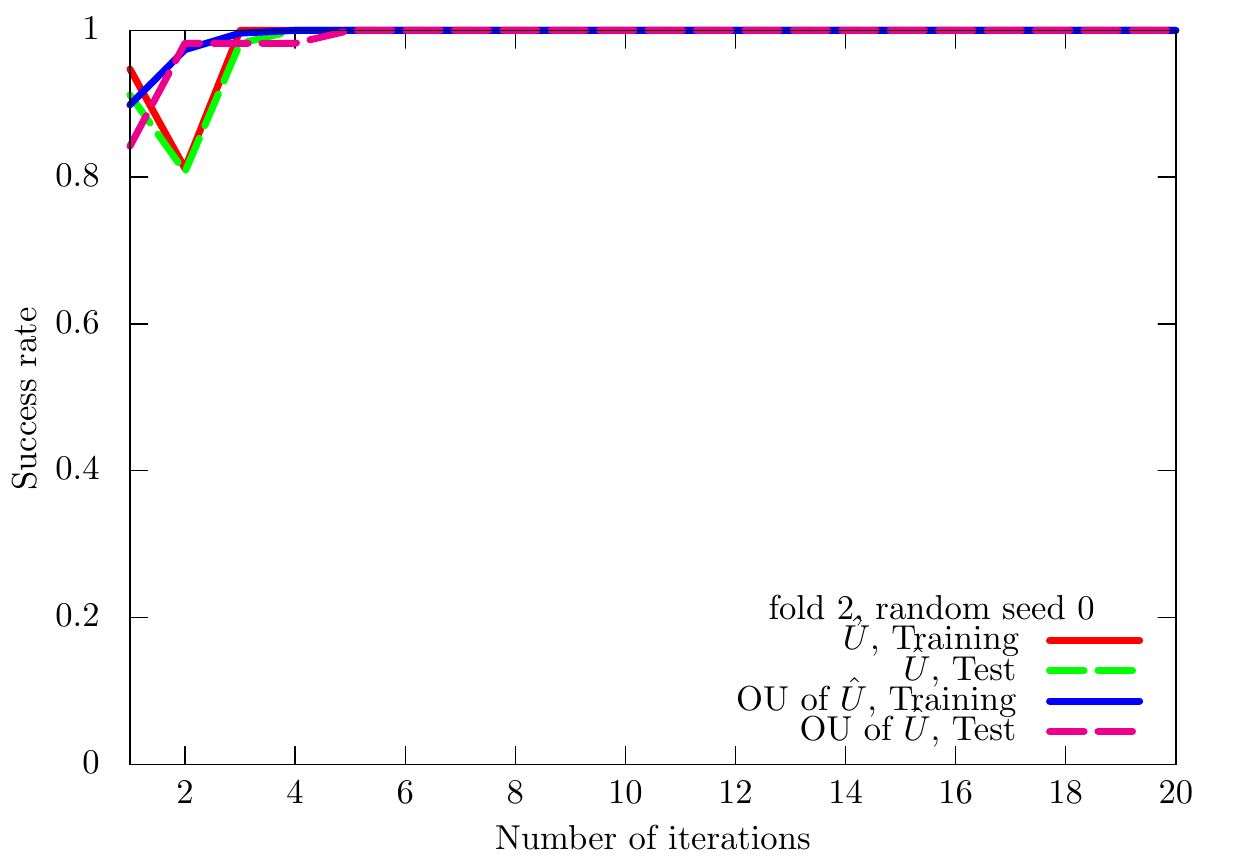}
\includegraphics[scale=0.25]{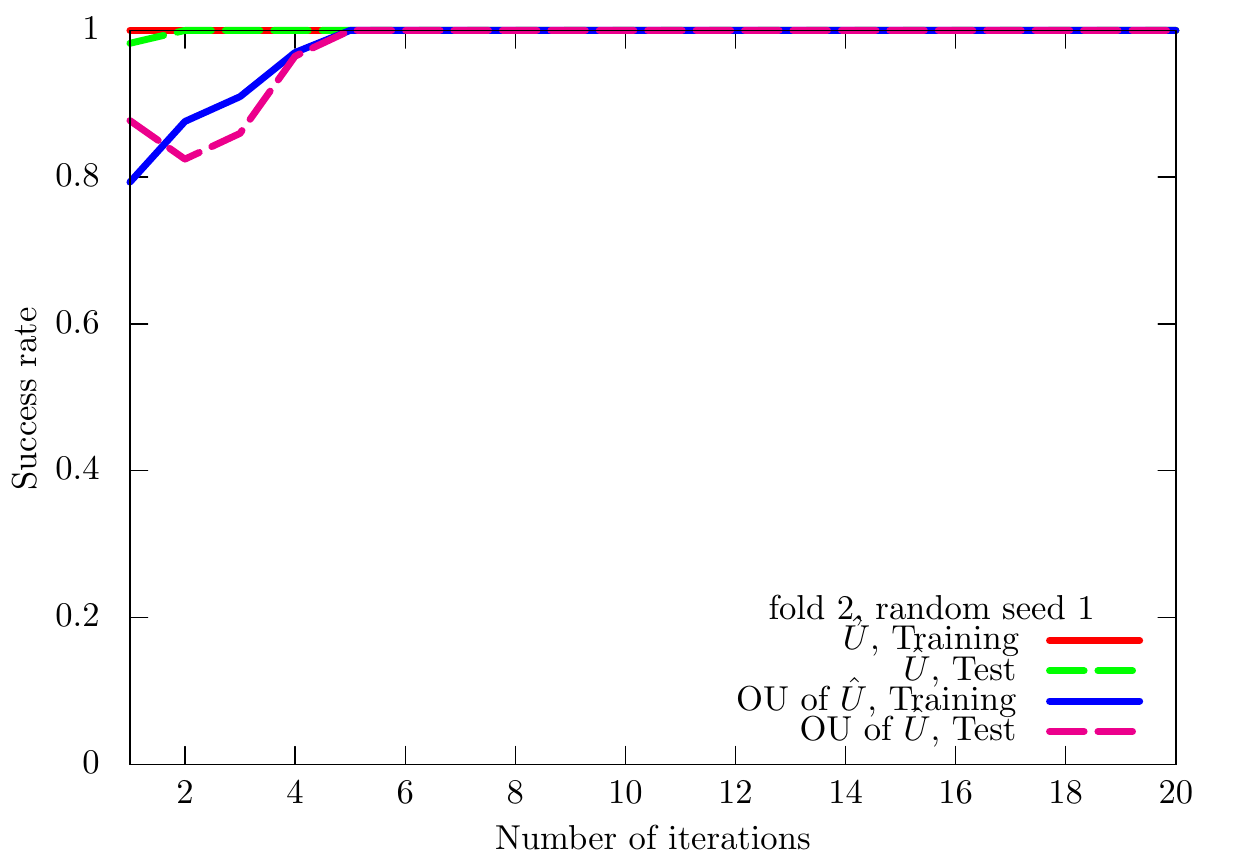}
\includegraphics[scale=0.25]{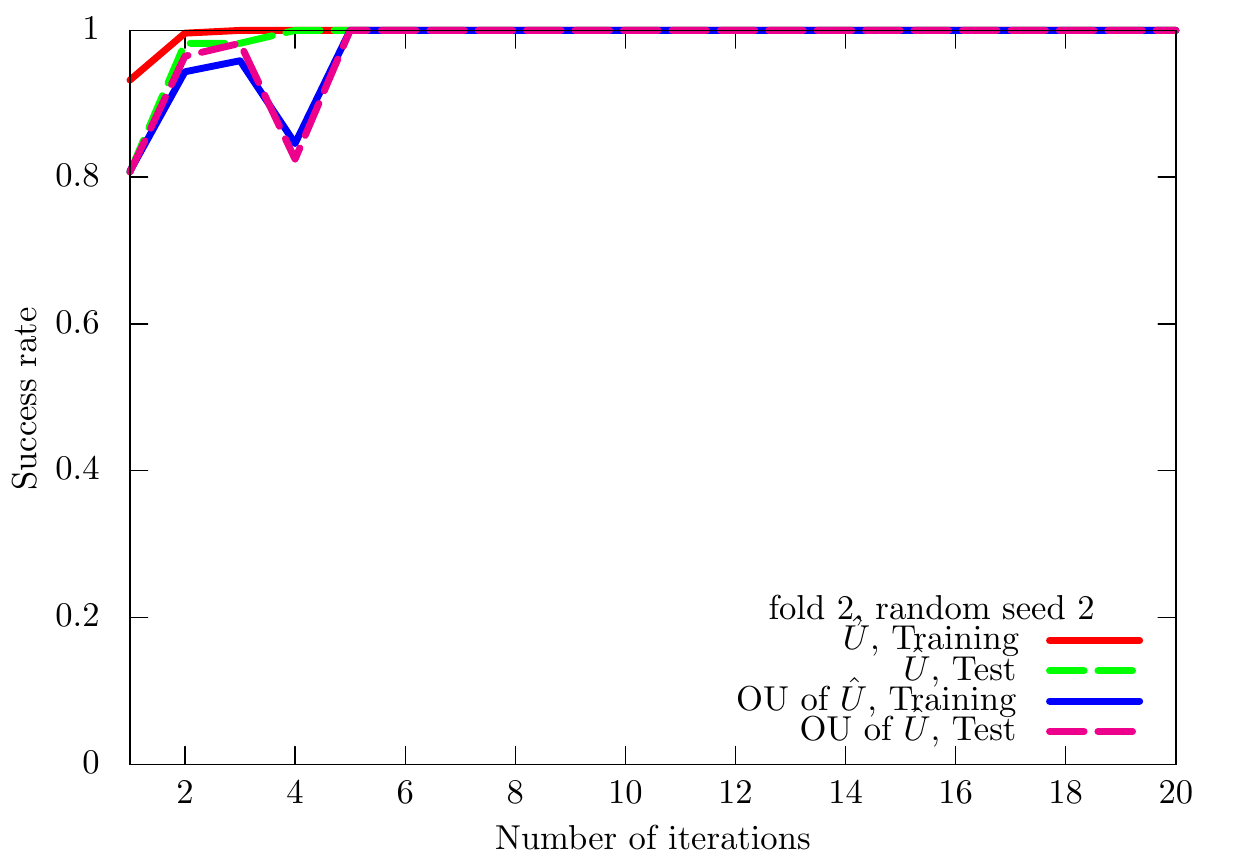}
\includegraphics[scale=0.25]{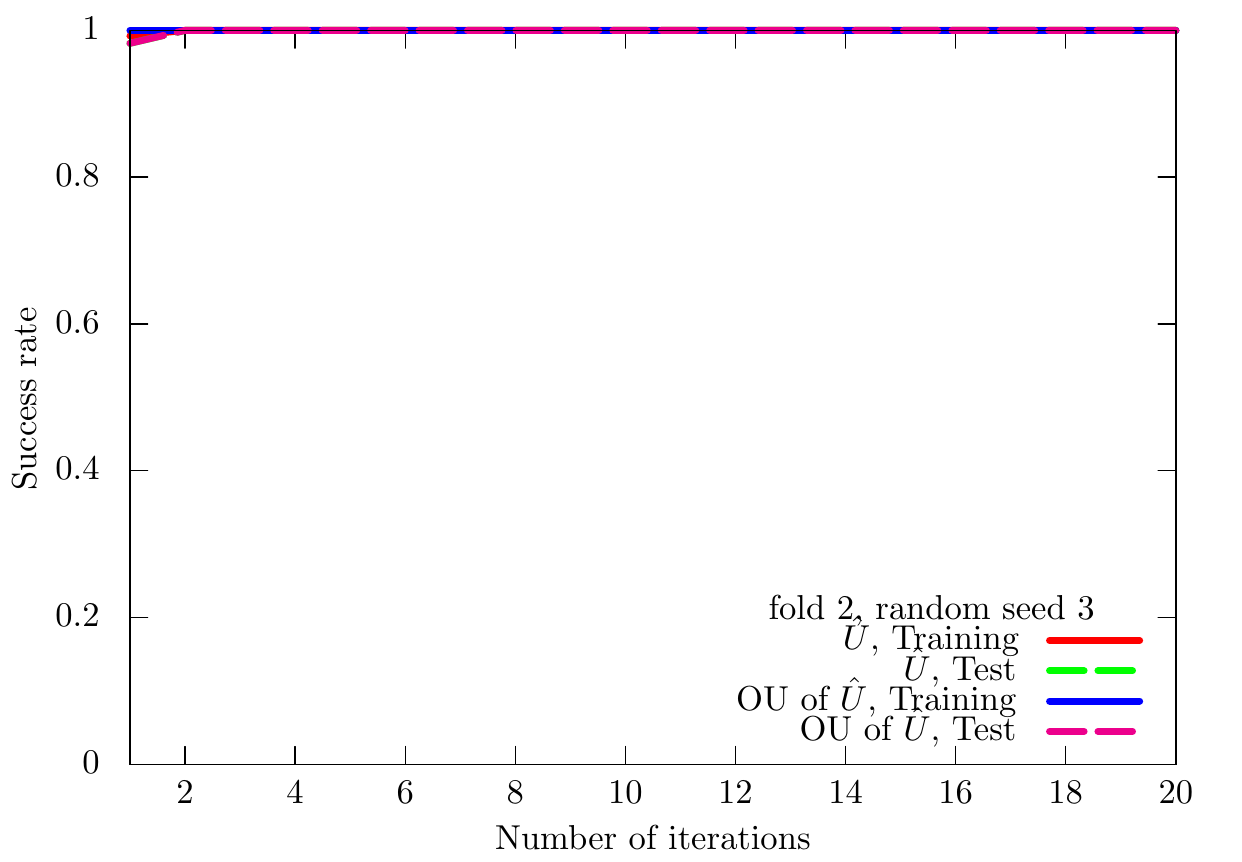}
\includegraphics[scale=0.25]{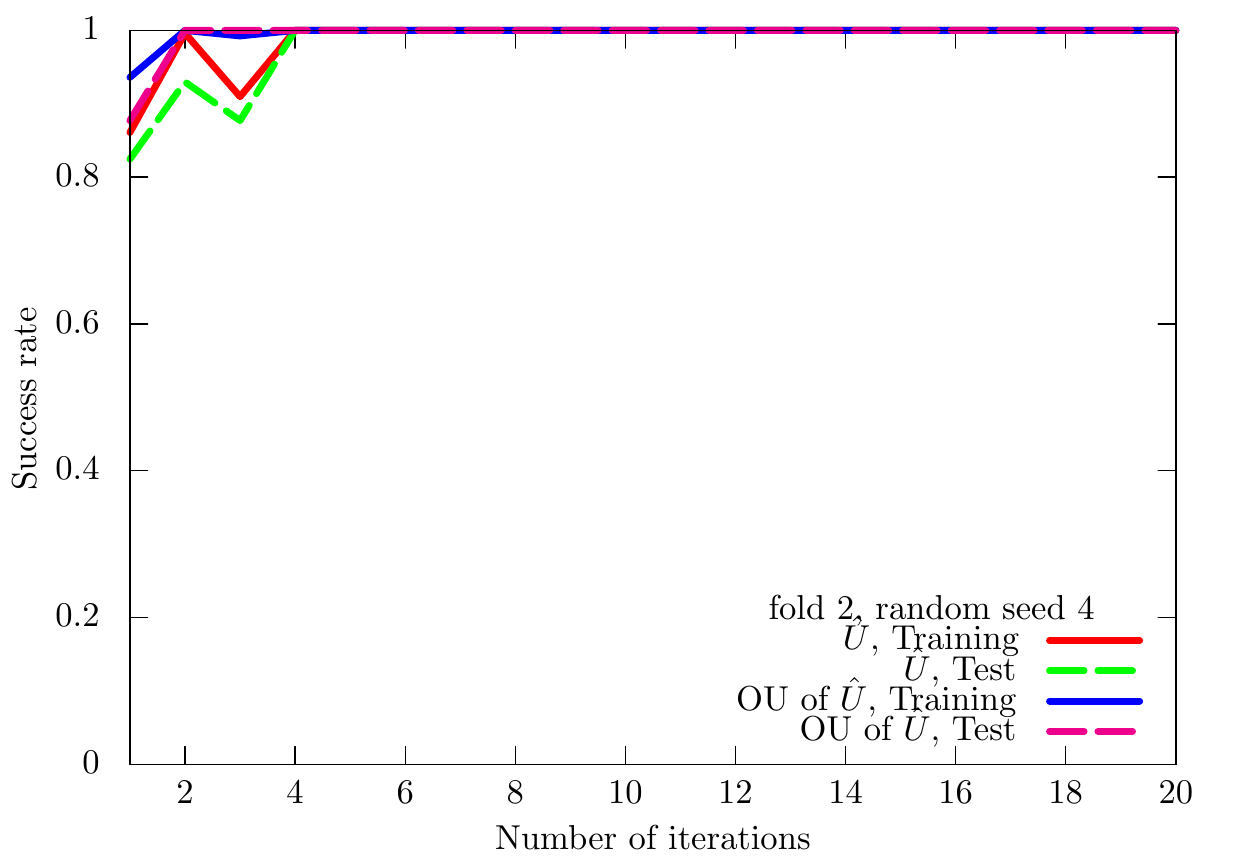}
\includegraphics[scale=0.25]{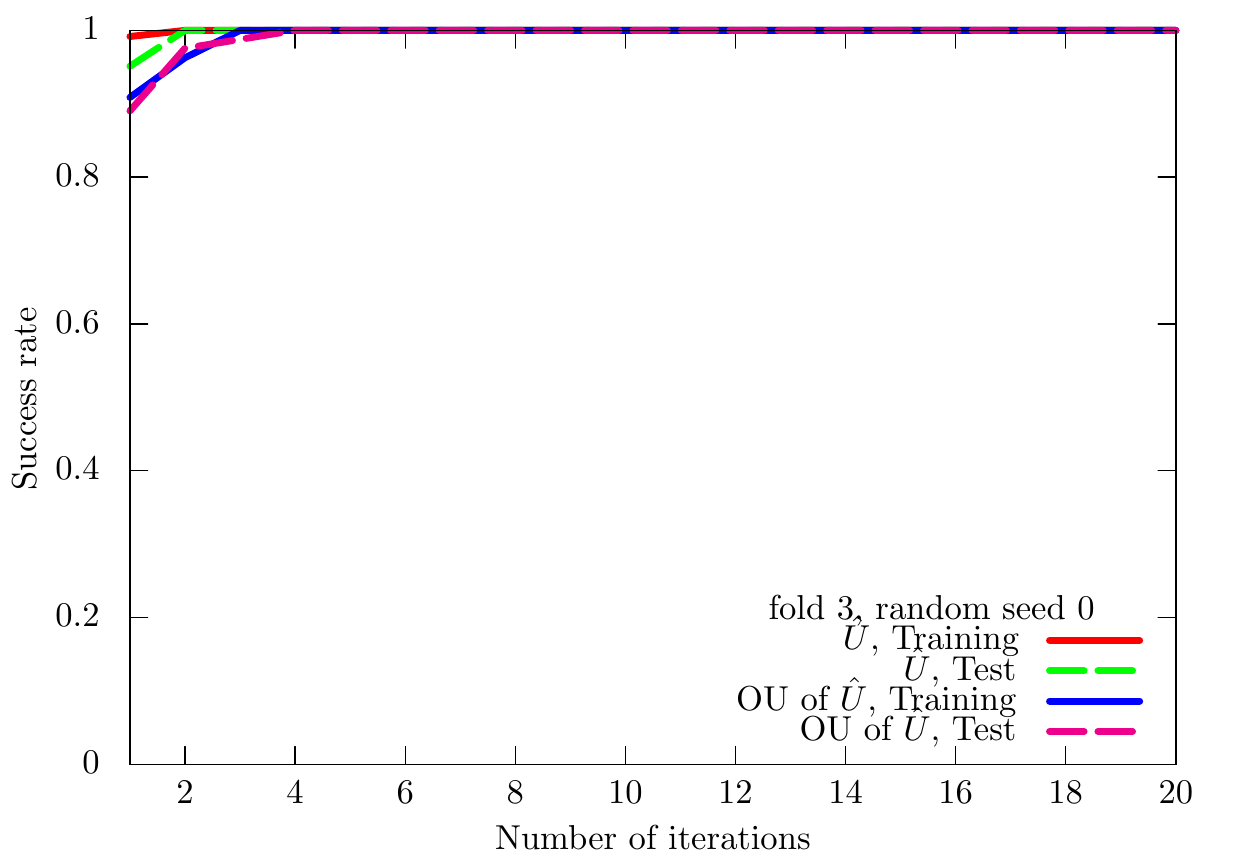}
\includegraphics[scale=0.25]{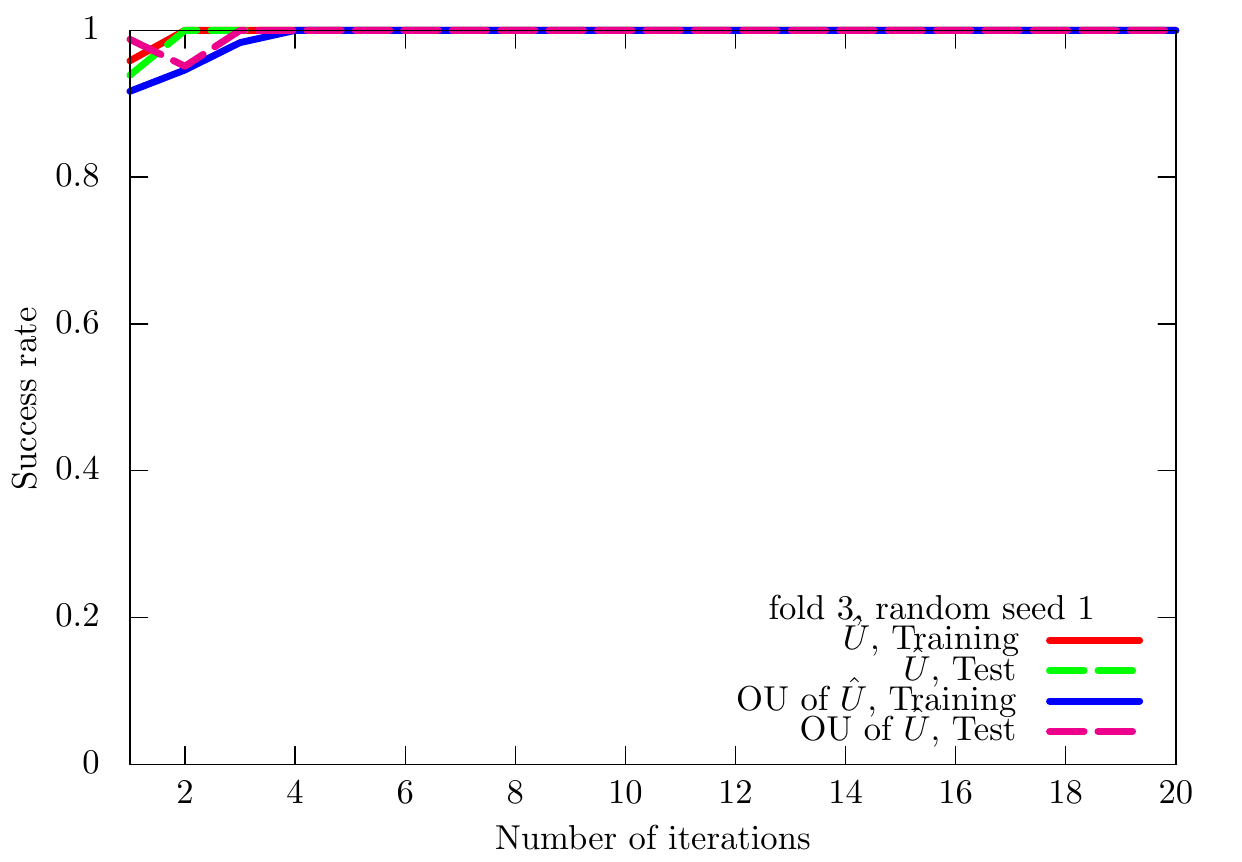}
\includegraphics[scale=0.25]{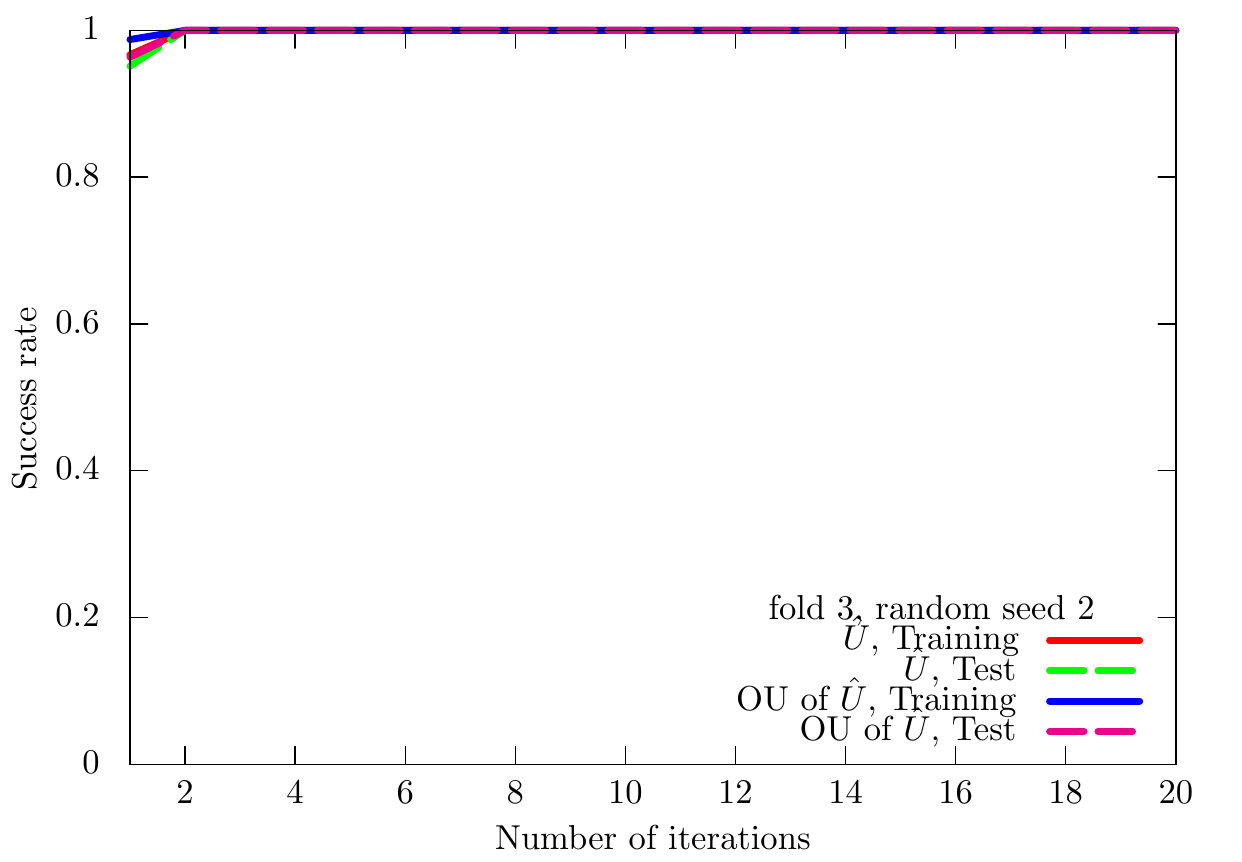}
\includegraphics[scale=0.25]{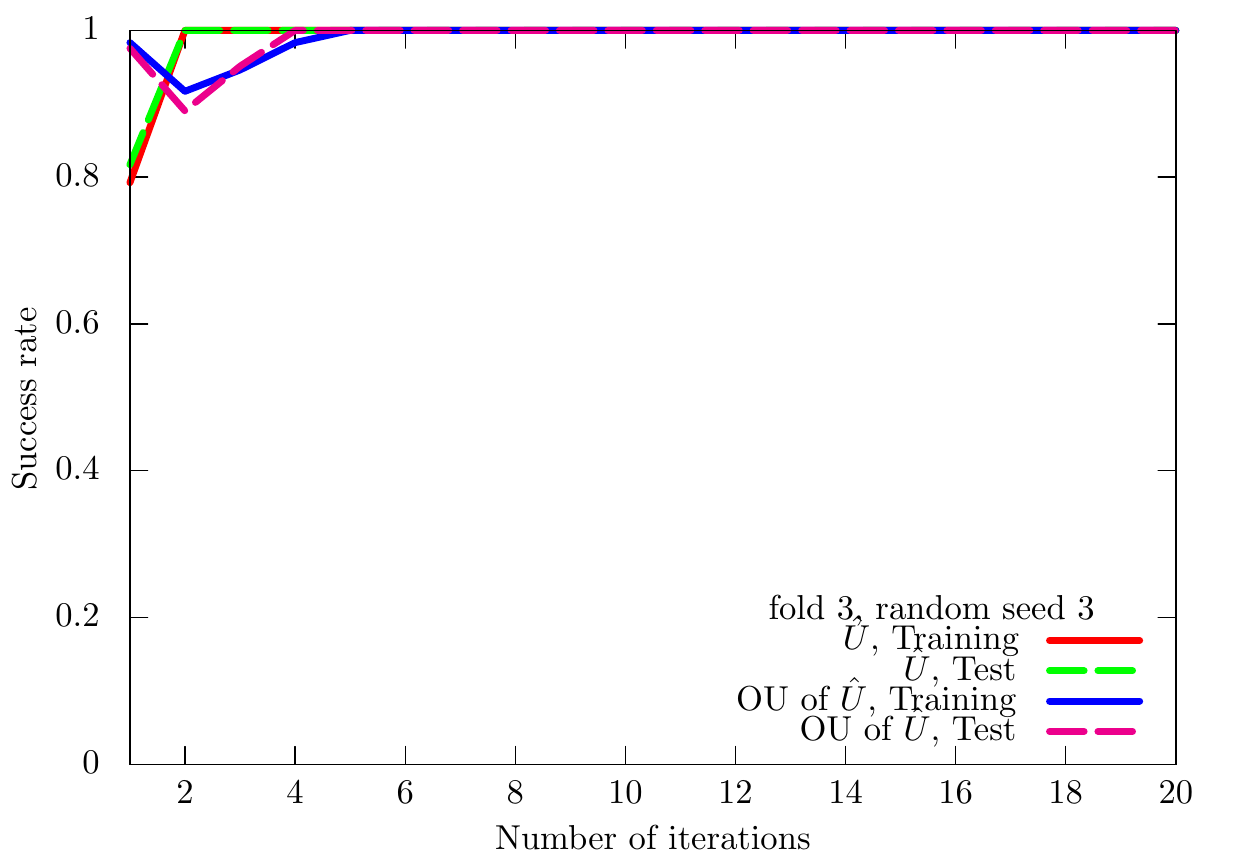}
\includegraphics[scale=0.25]{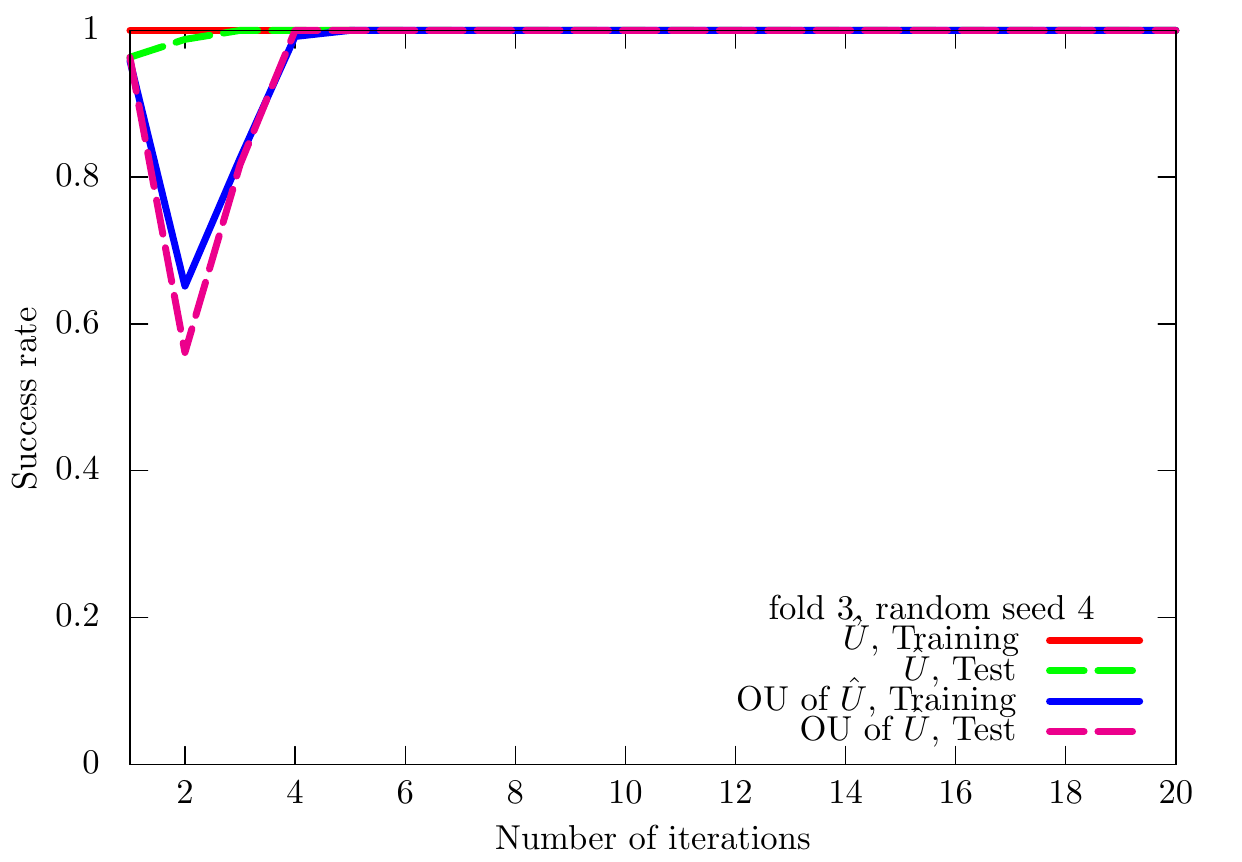}
\includegraphics[scale=0.25]{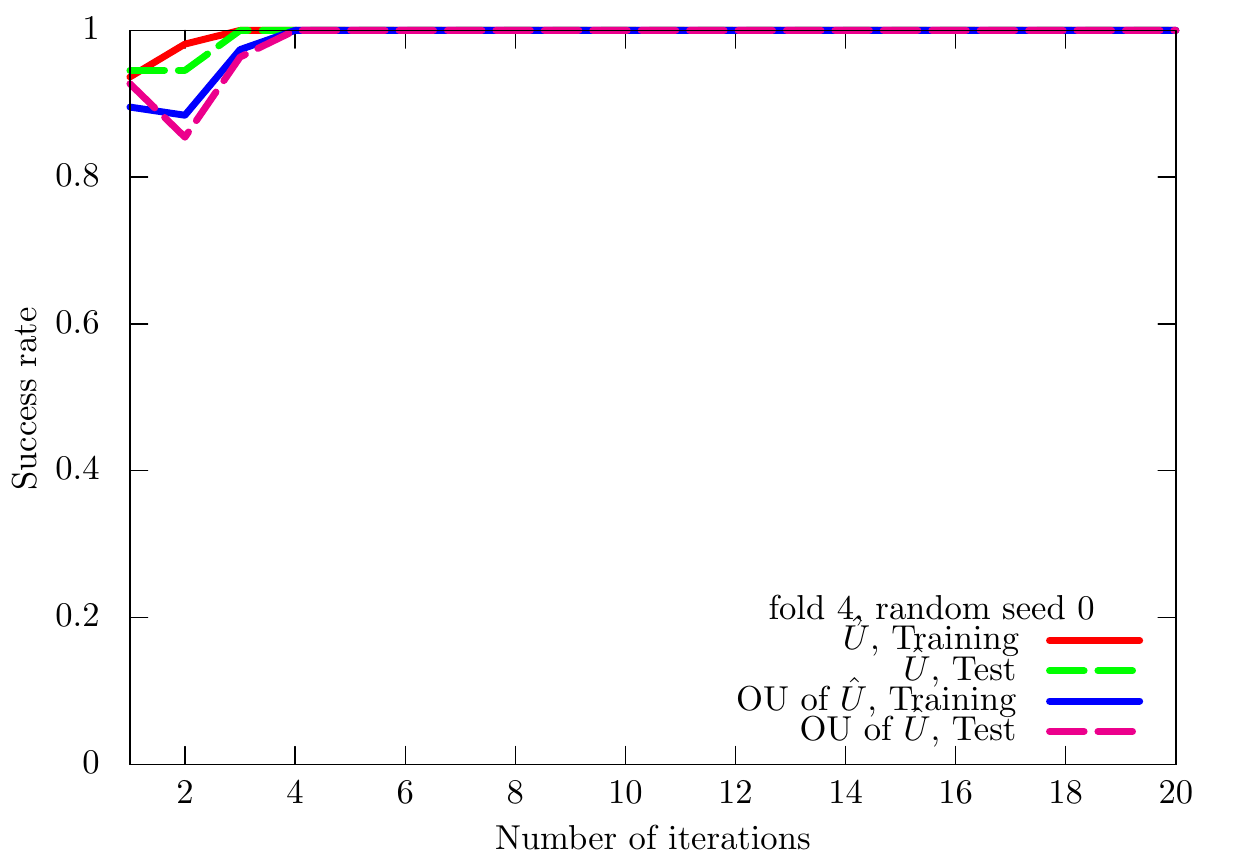}
\includegraphics[scale=0.25]{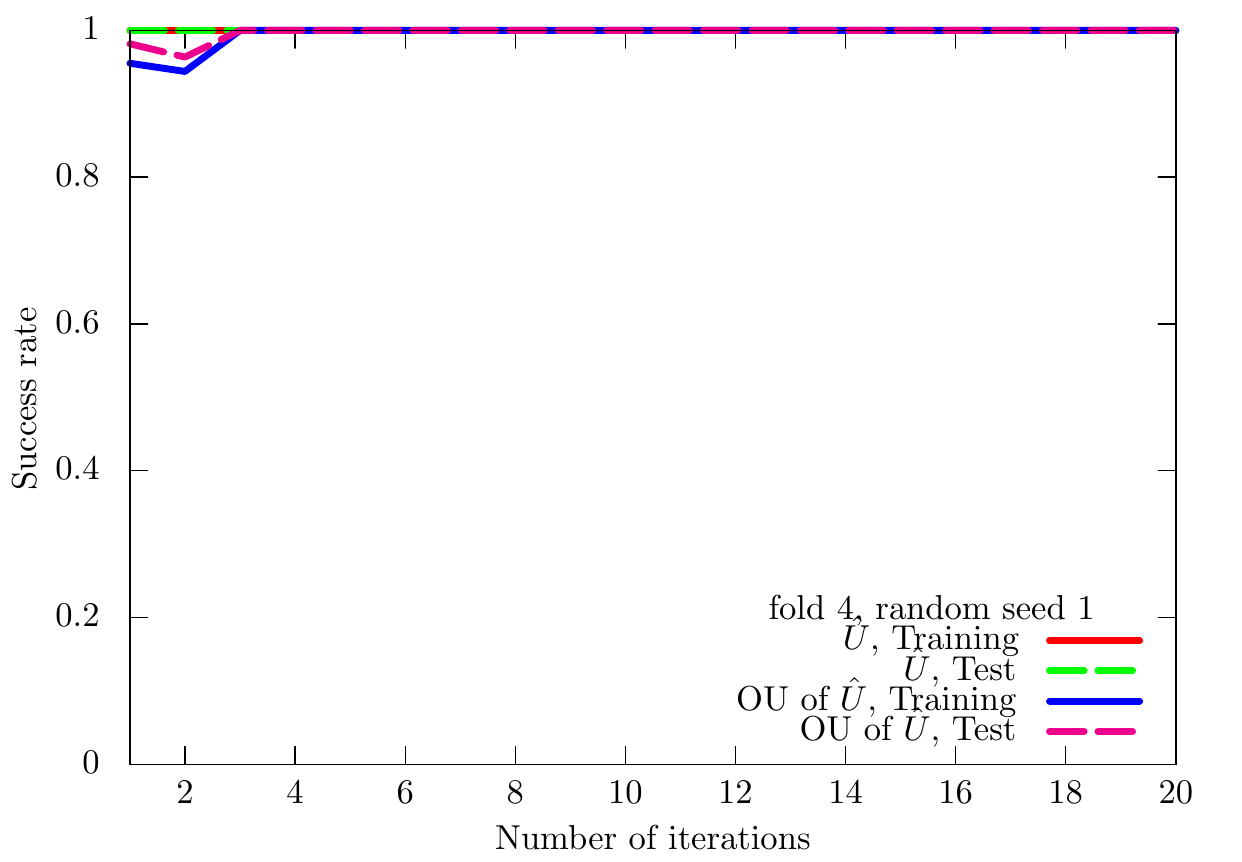}
\includegraphics[scale=0.25]{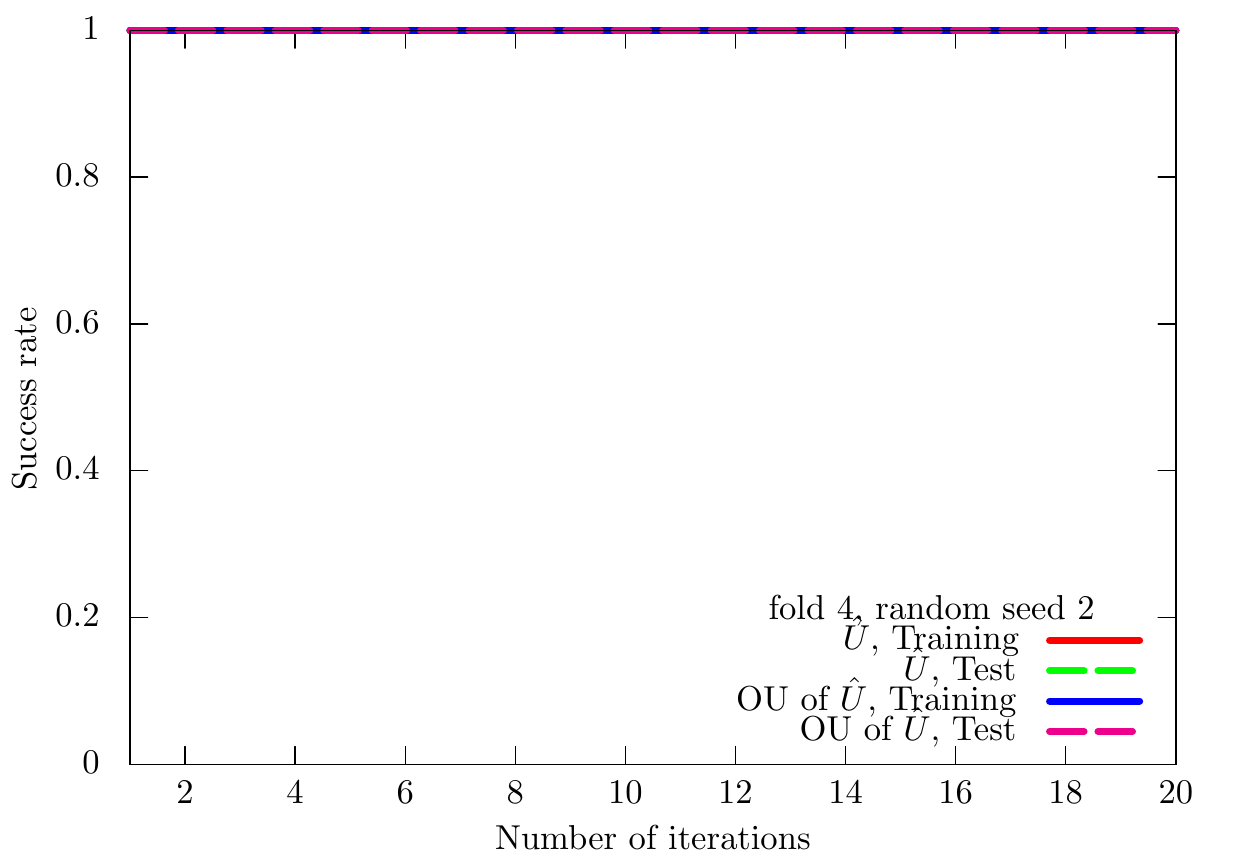}
\includegraphics[scale=0.25]{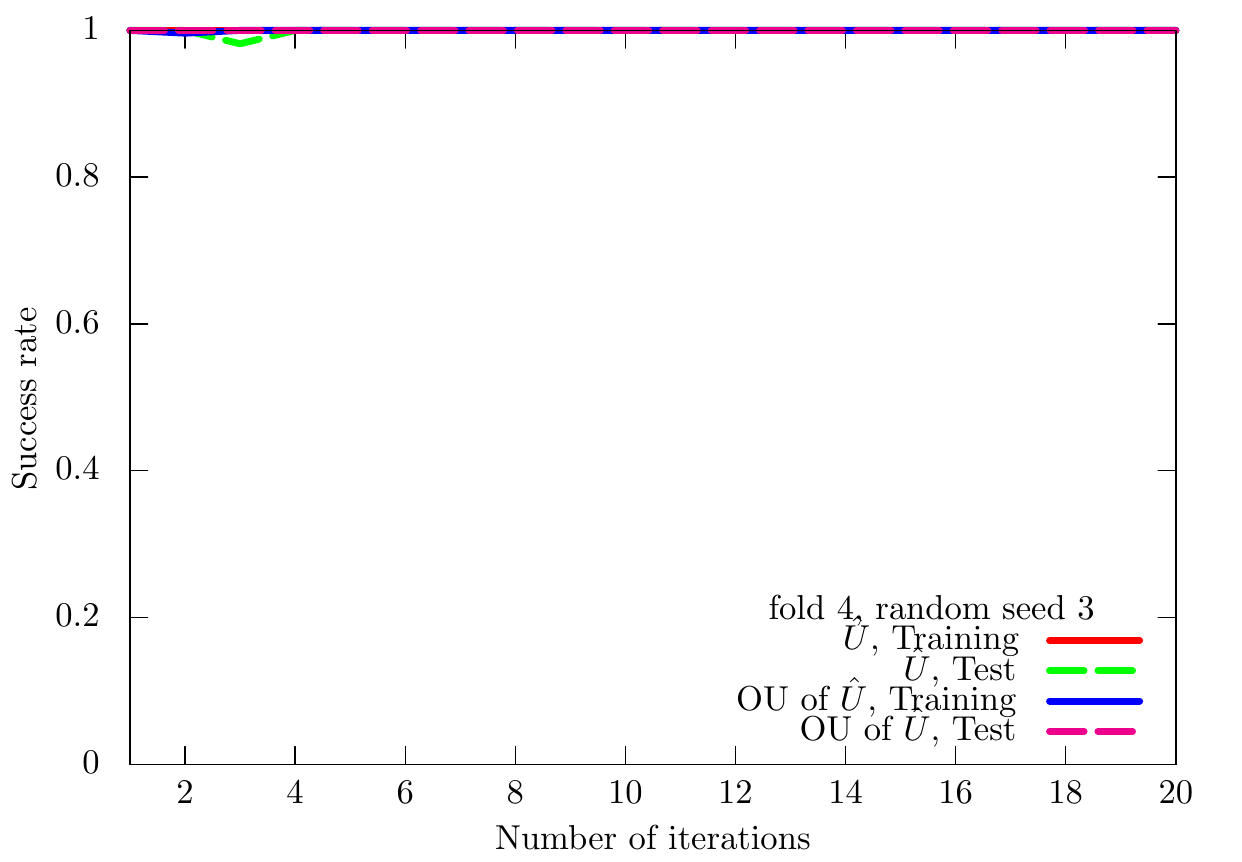}
\includegraphics[scale=0.25]{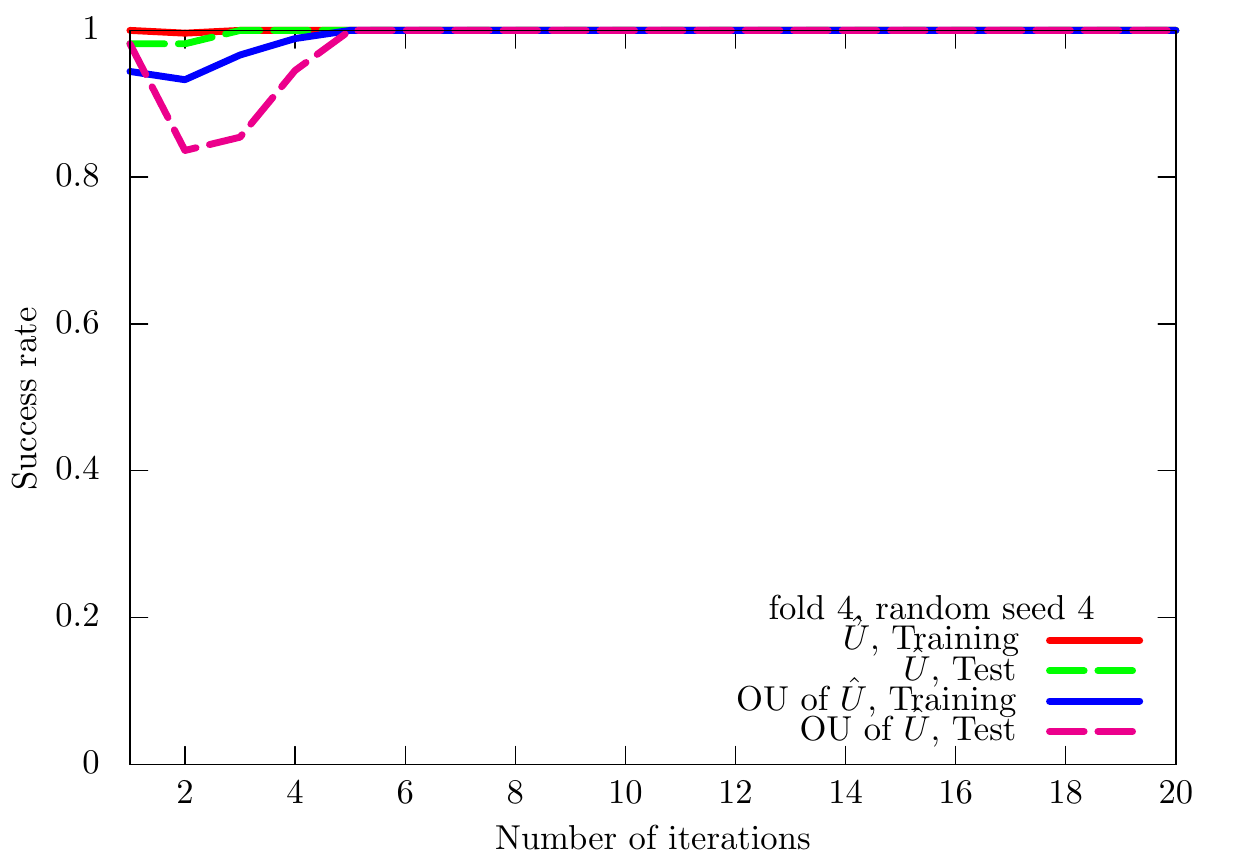}
\caption{Results of the UKM ($\hat{X}$ and $\hat{P}$) on the $5$-fold datasets with $5$ different random seeds for the semeion dataset ($0$ or $1$). We use complex matrices and set $\theta_\mathrm{bias} = 0$. We set $r = 0.010$.}
\label{supp-arXiv-numerical-result-raw-data-fold-001-rand-001-UKM-P-UCI-semeion-0-1}
\end{figure*}

\begin{figure*}[htb]
\centering
\includegraphics[scale=0.25]{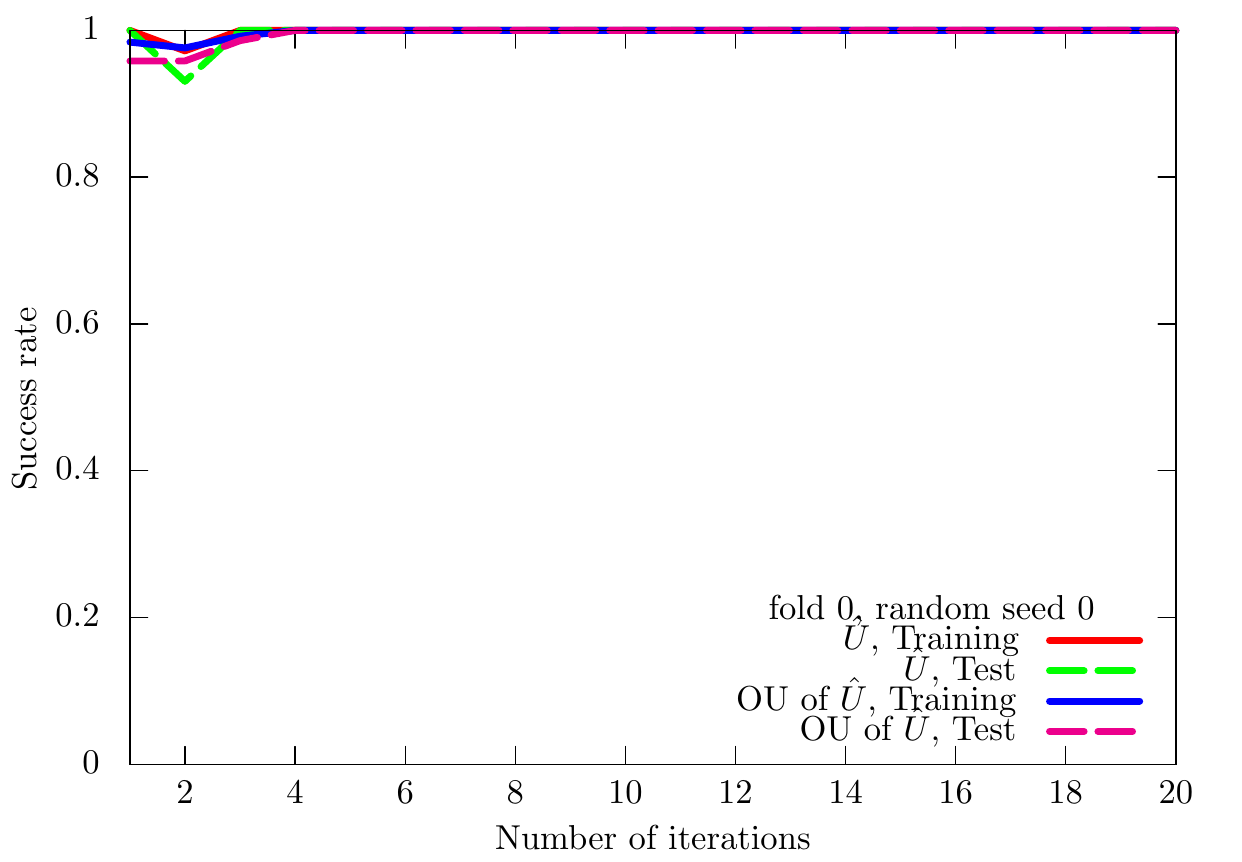}
\includegraphics[scale=0.25]{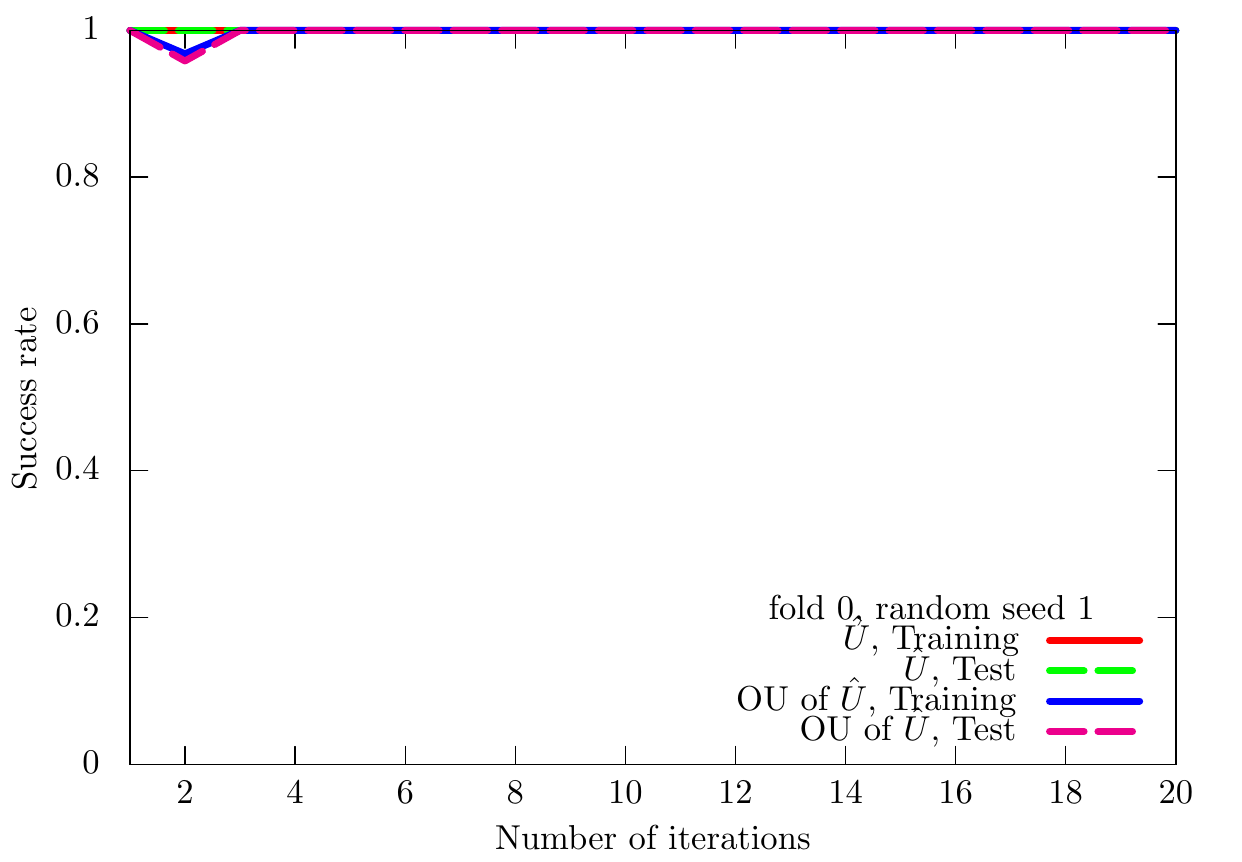}
\includegraphics[scale=0.25]{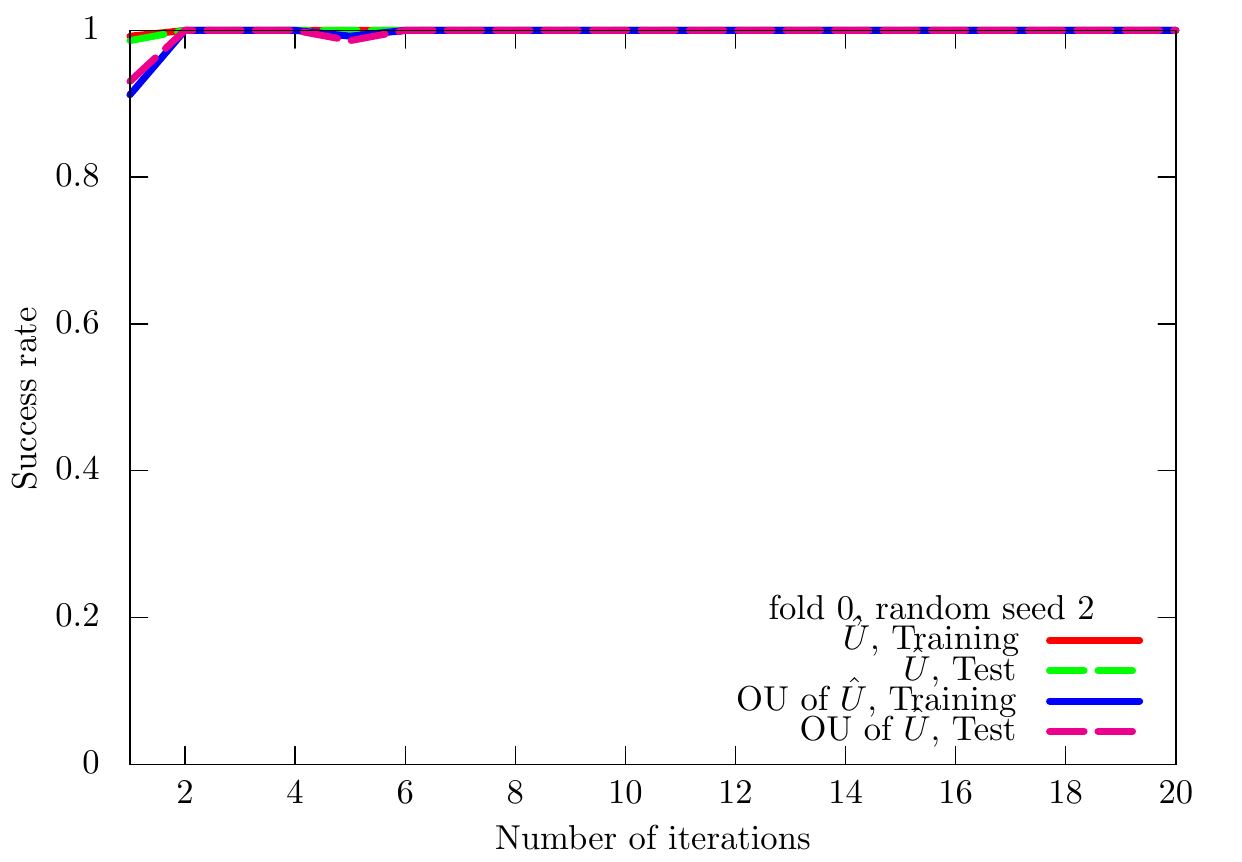}
\includegraphics[scale=0.25]{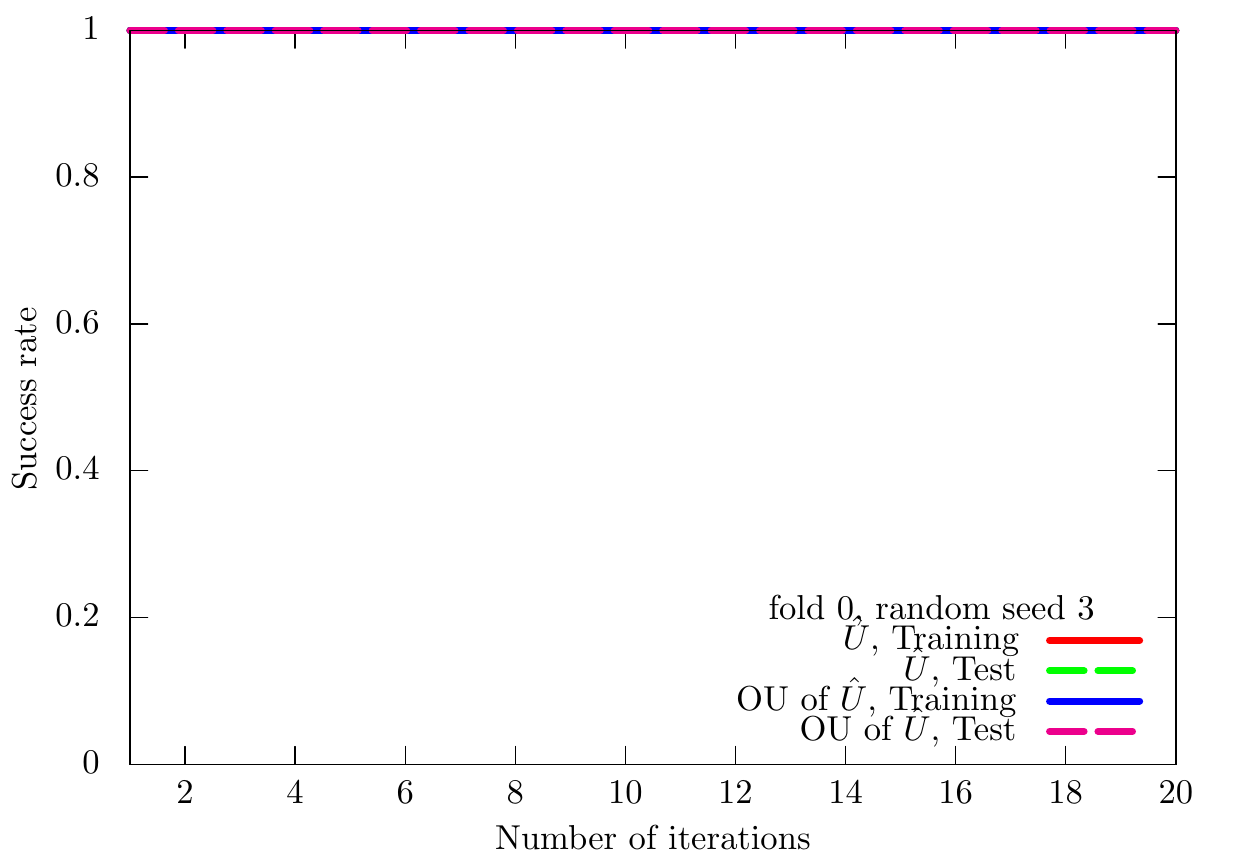}
\includegraphics[scale=0.25]{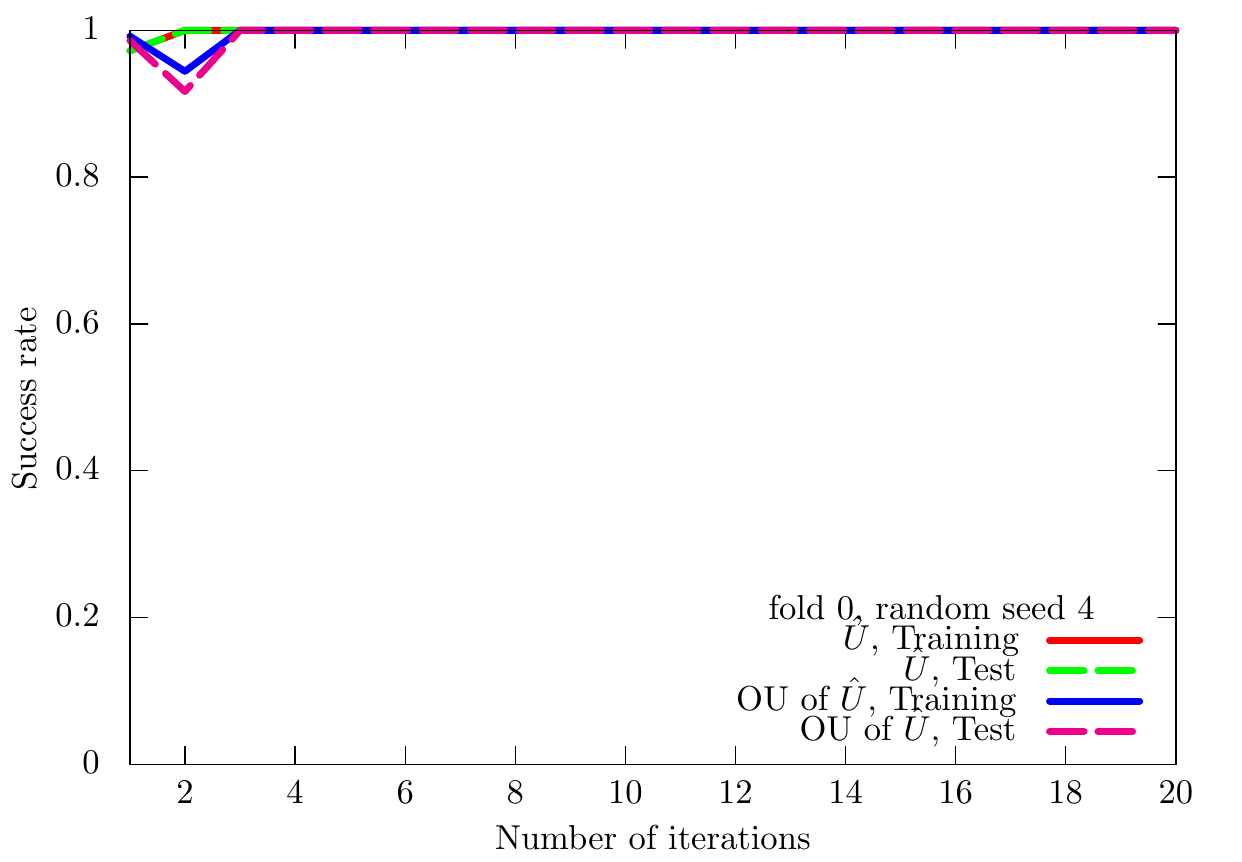}
\includegraphics[scale=0.25]{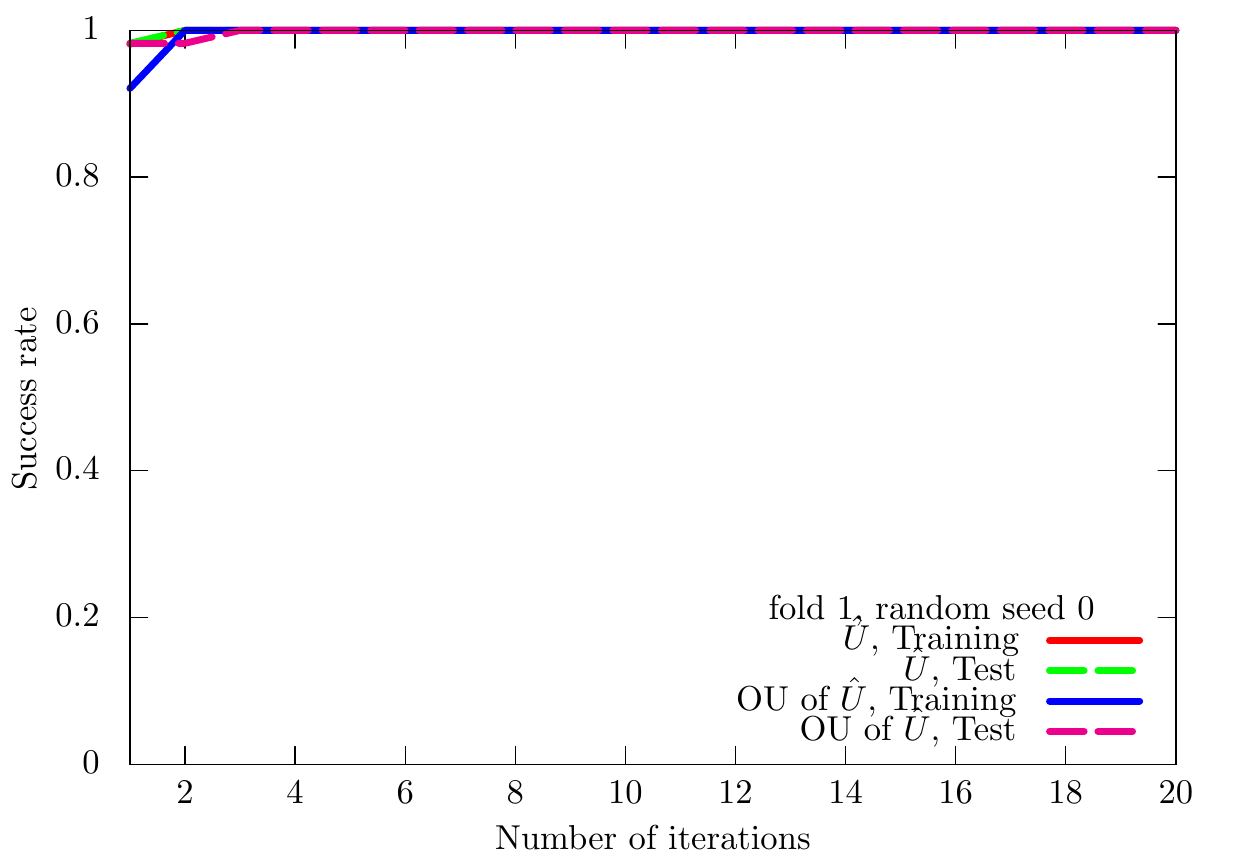}
\includegraphics[scale=0.25]{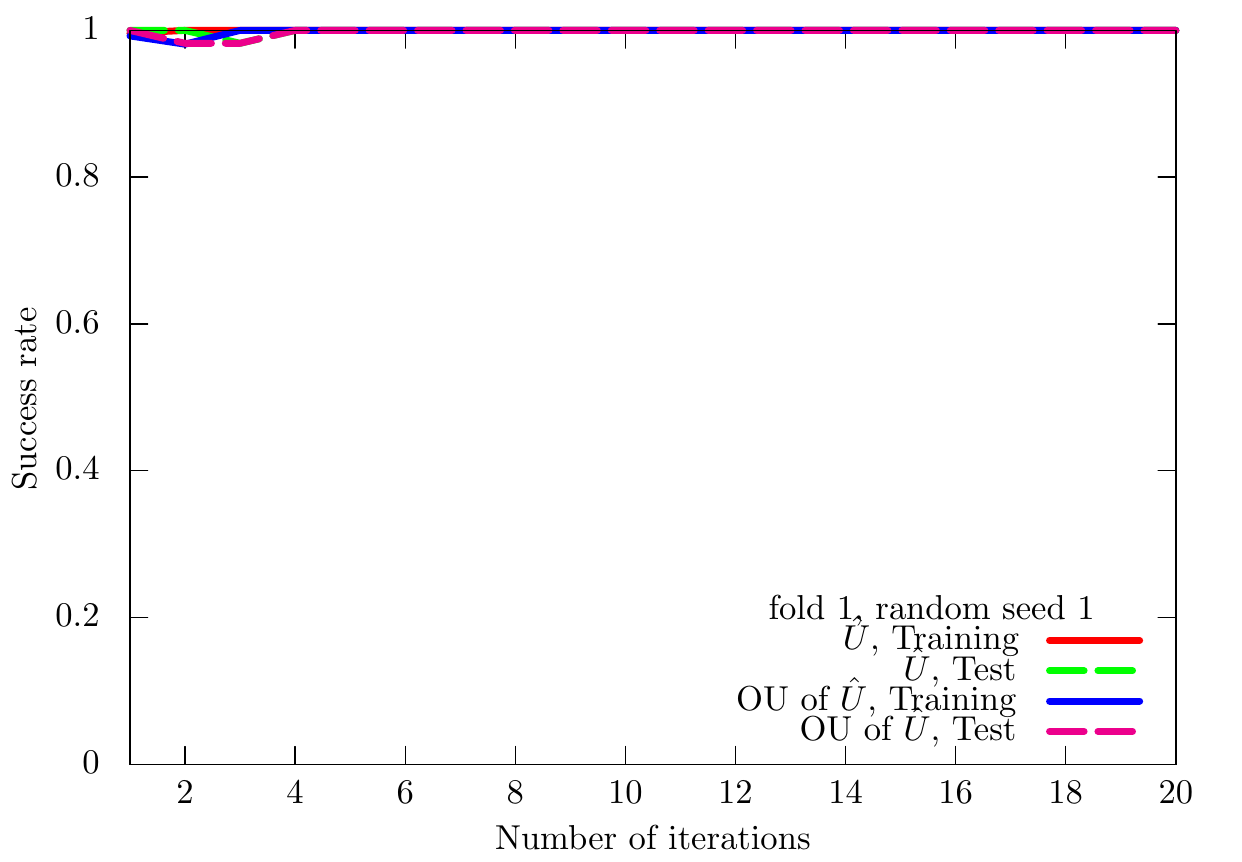}
\includegraphics[scale=0.25]{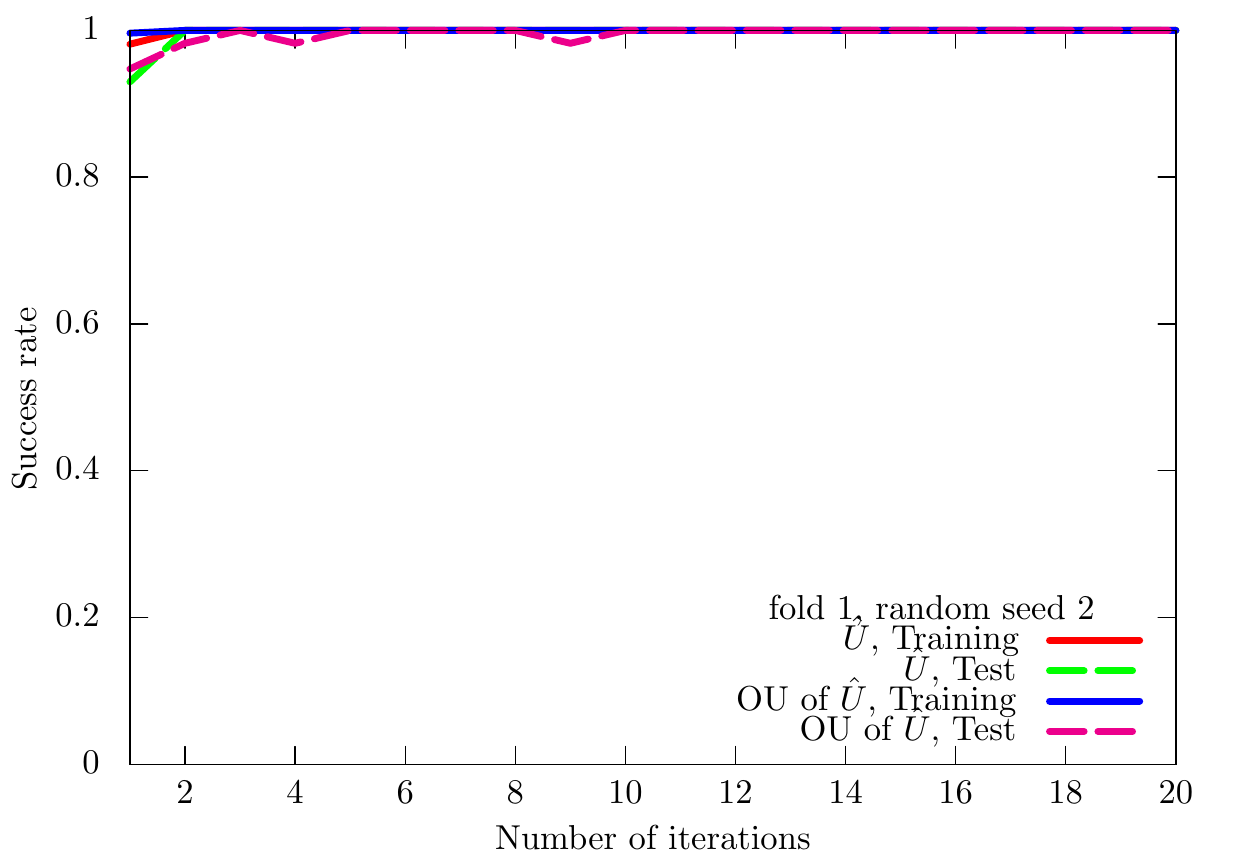}
\includegraphics[scale=0.25]{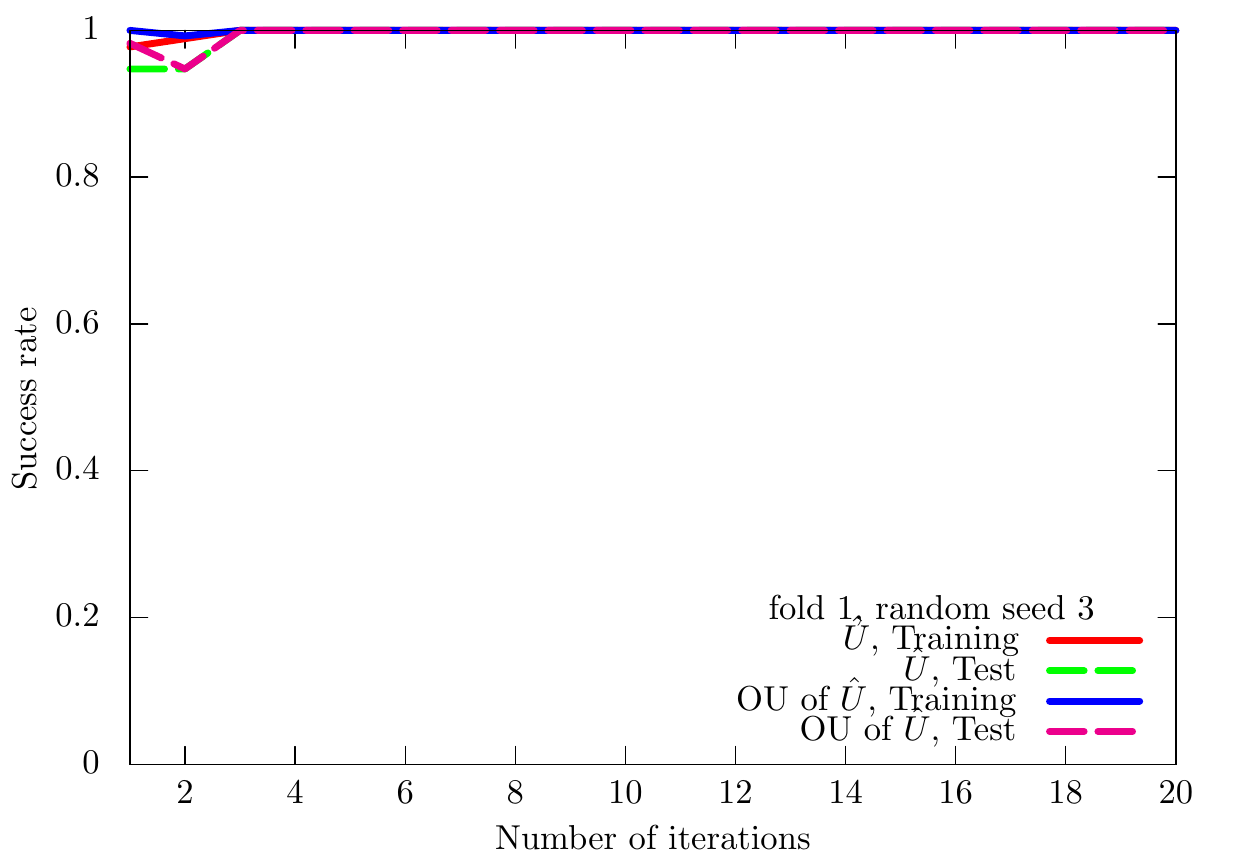}
\includegraphics[scale=0.25]{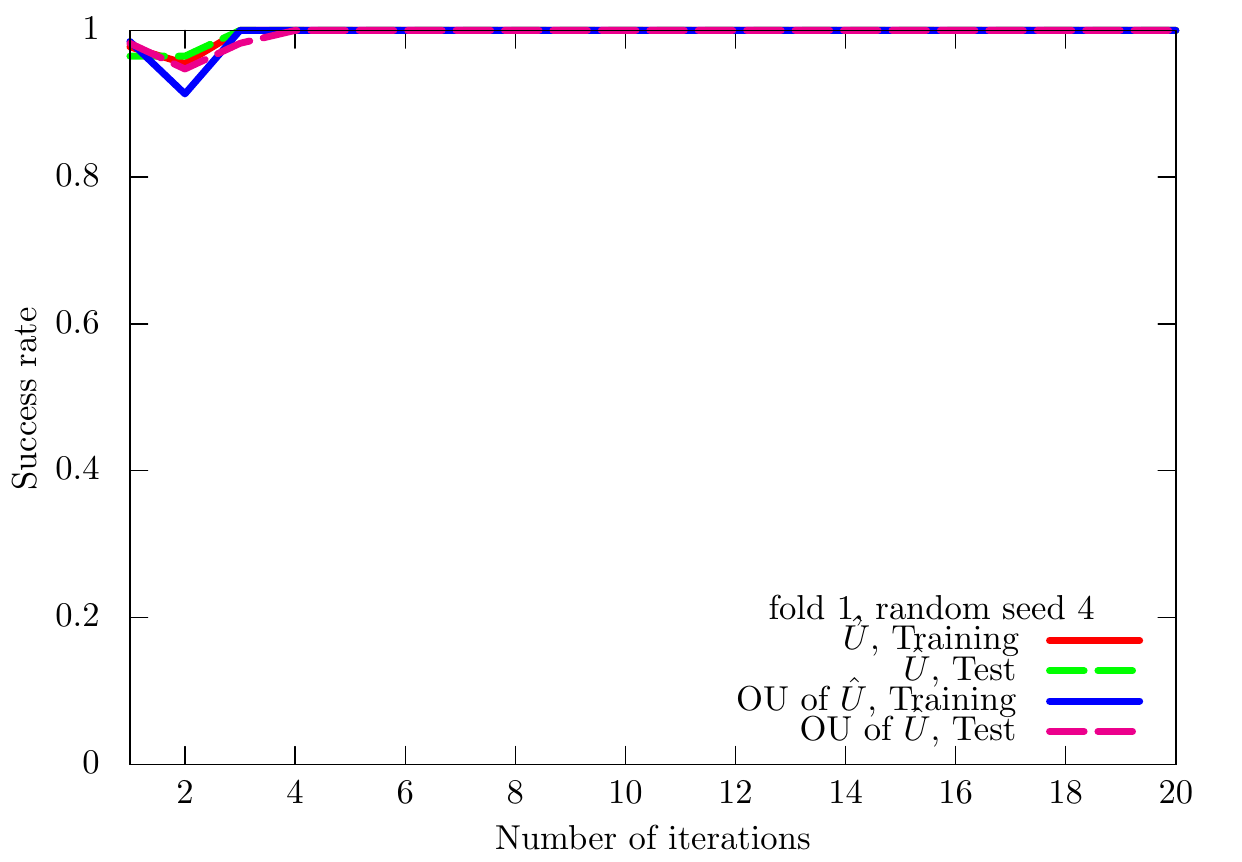}
\includegraphics[scale=0.25]{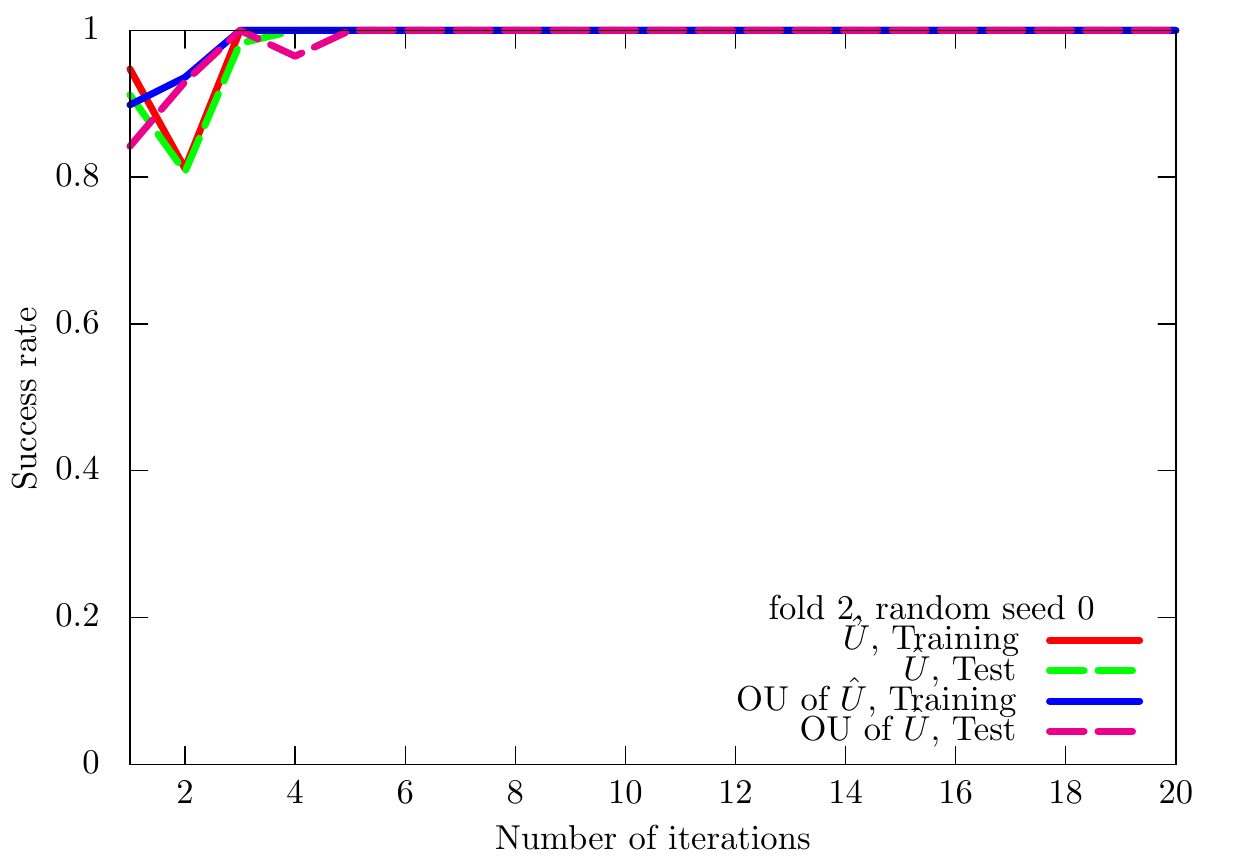}
\includegraphics[scale=0.25]{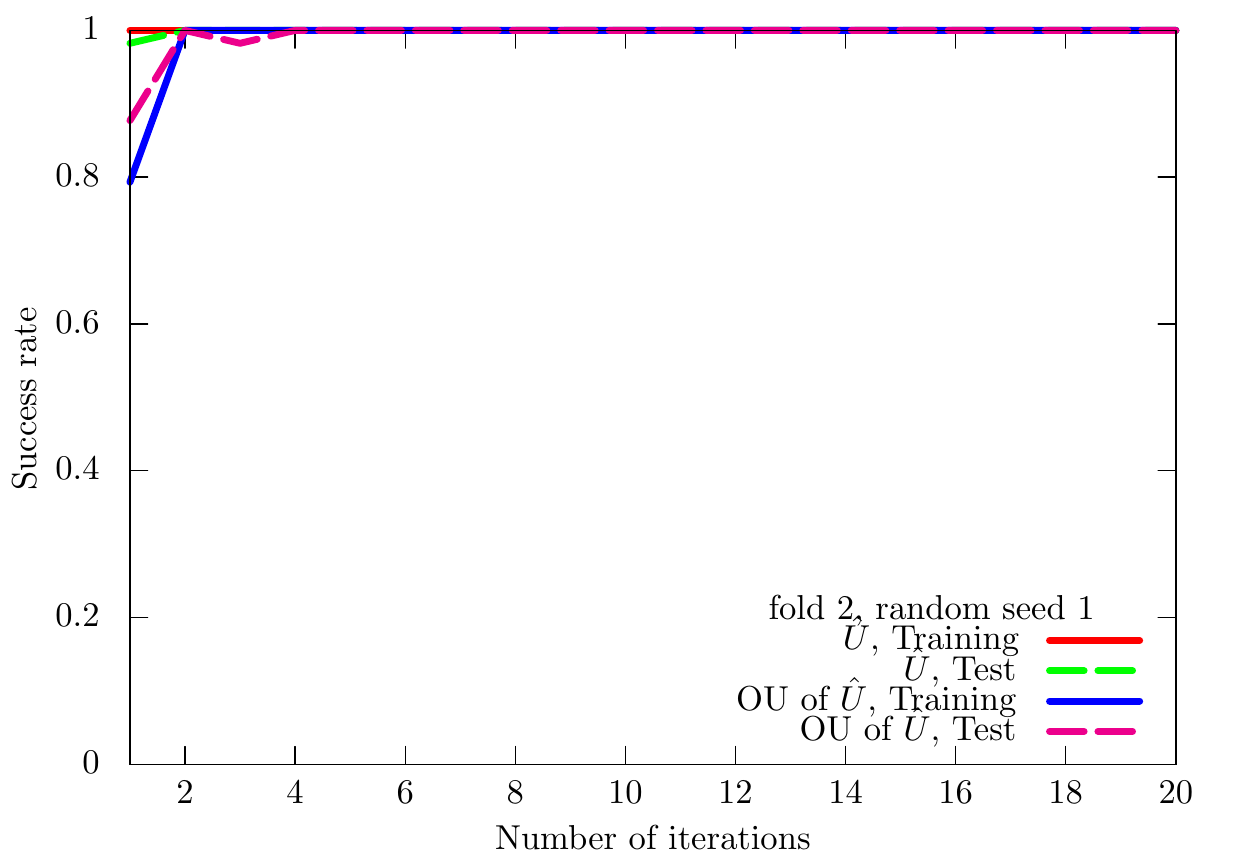}
\includegraphics[scale=0.25]{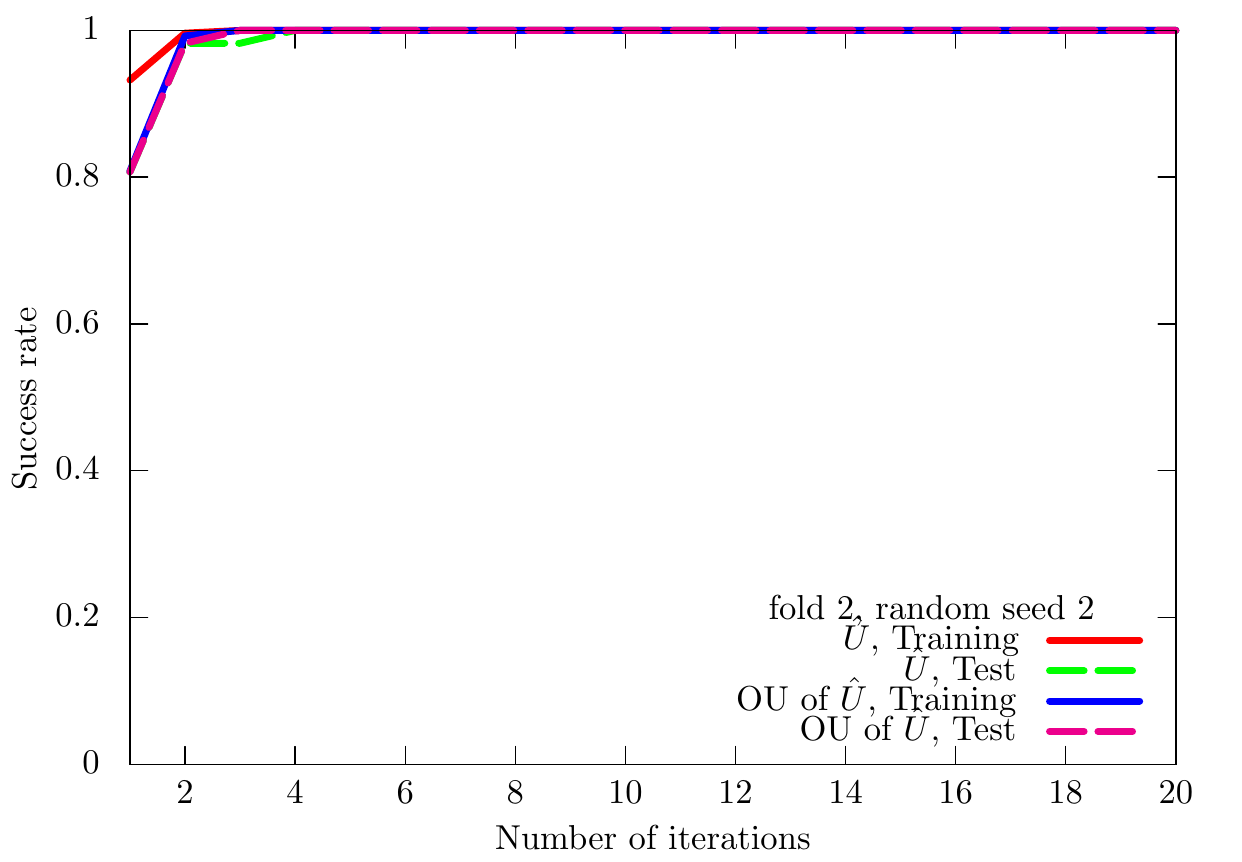}
\includegraphics[scale=0.25]{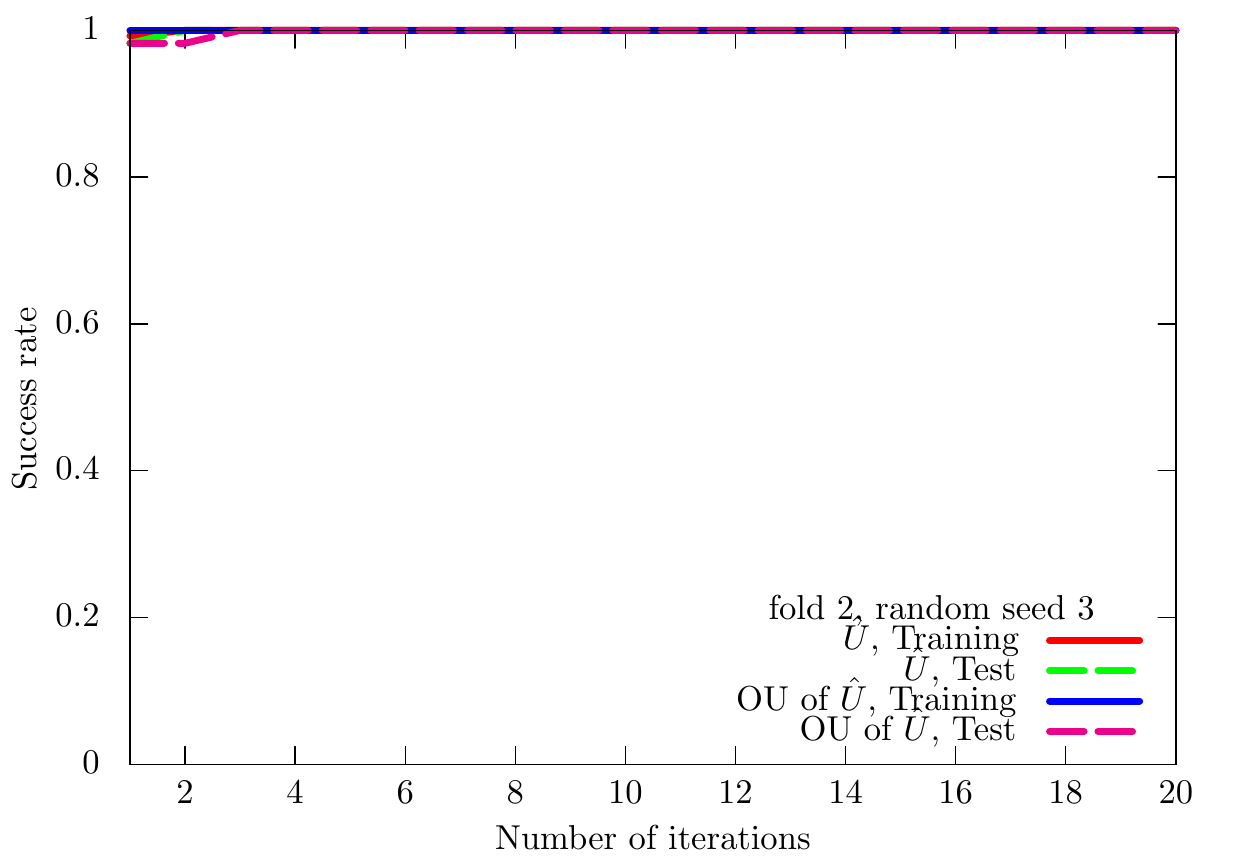}
\includegraphics[scale=0.25]{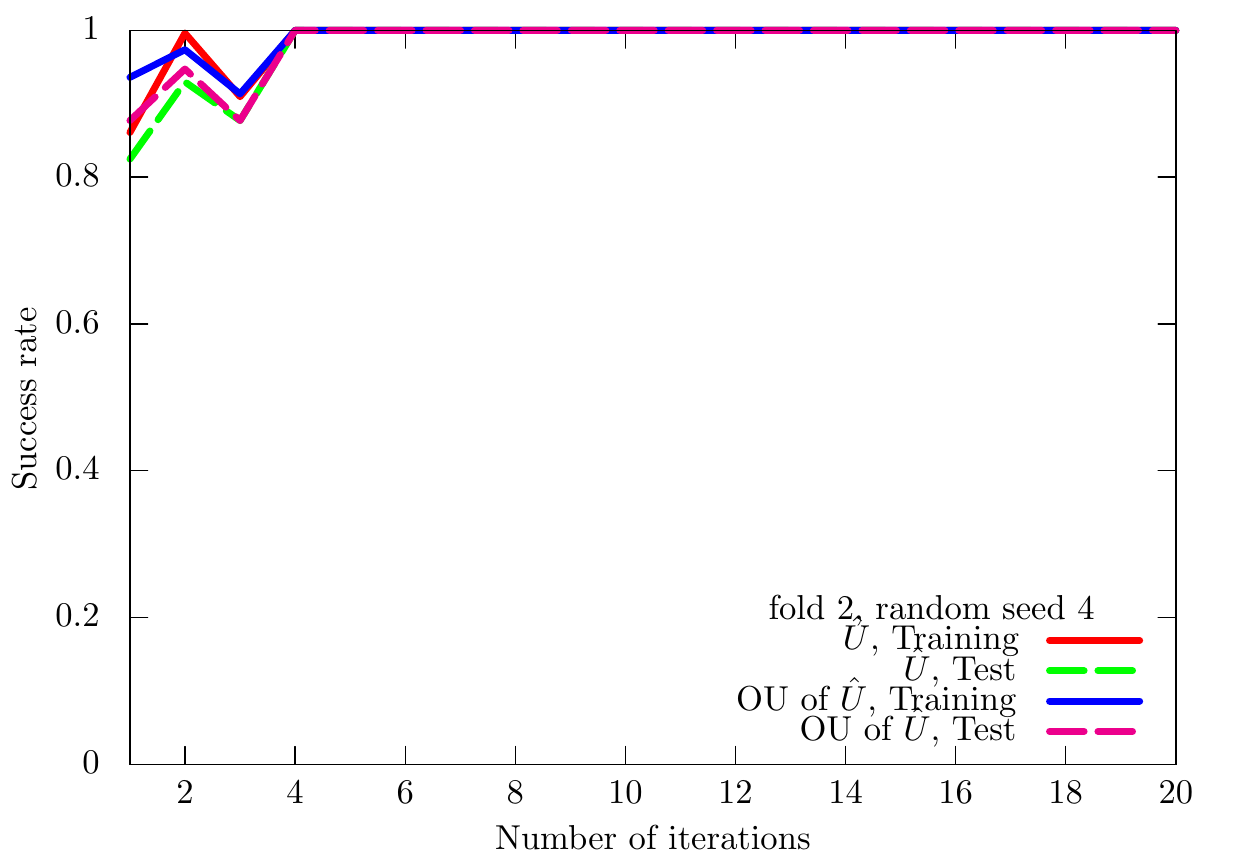}
\includegraphics[scale=0.25]{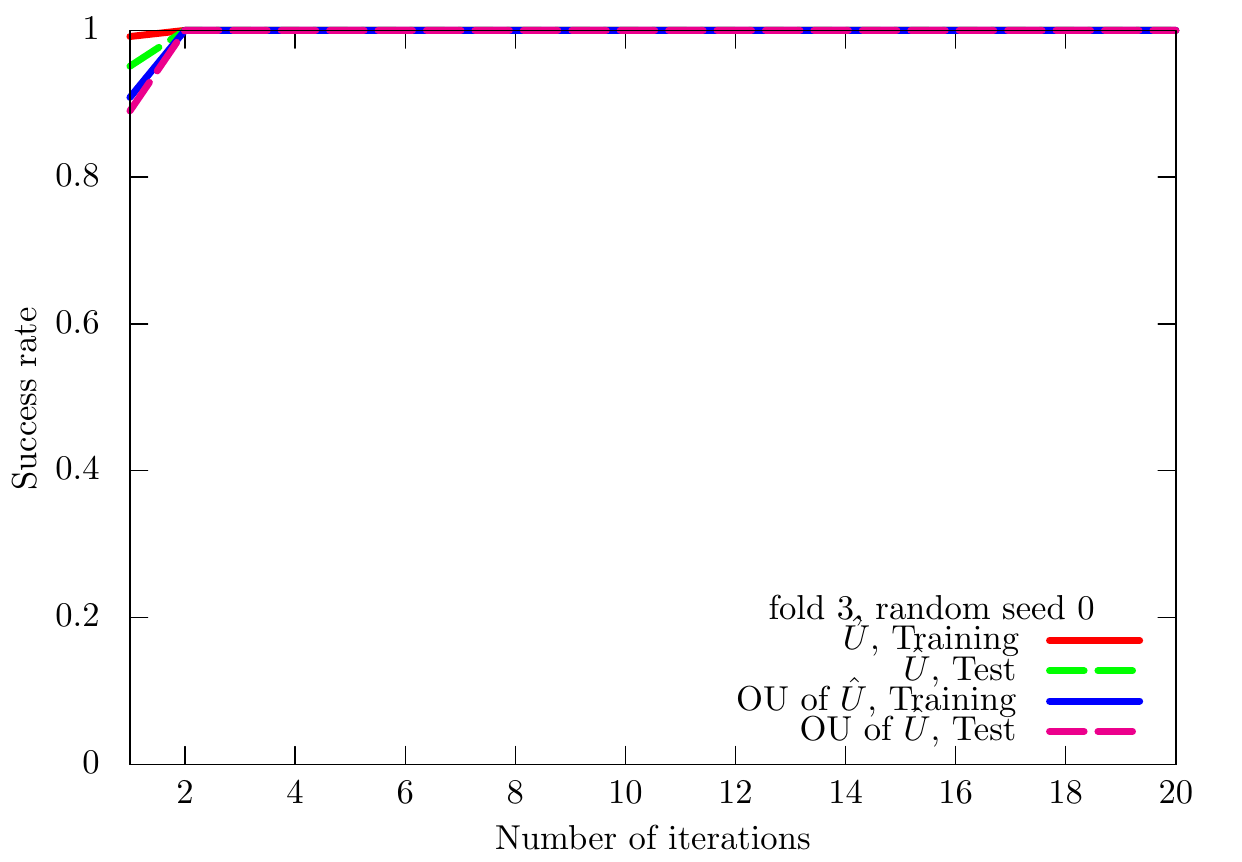}
\includegraphics[scale=0.25]{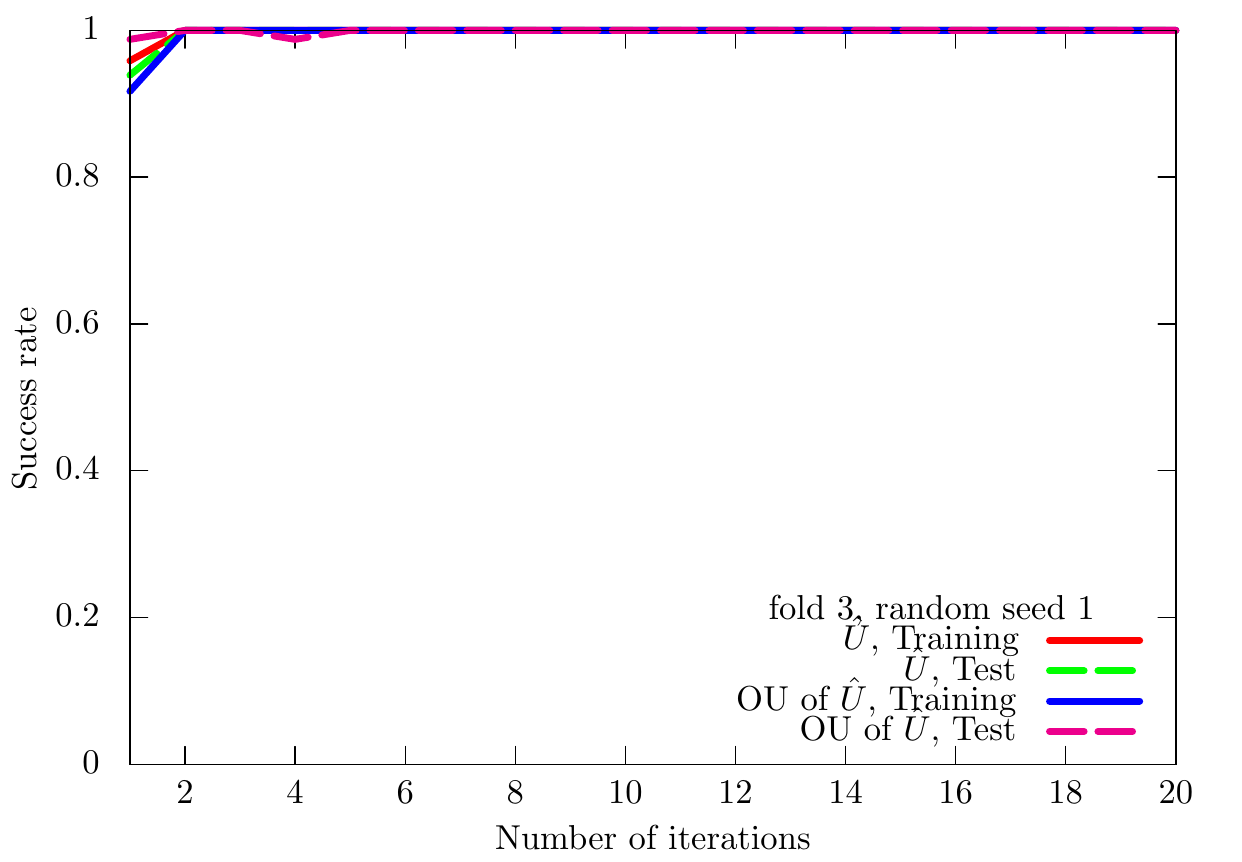}
\includegraphics[scale=0.25]{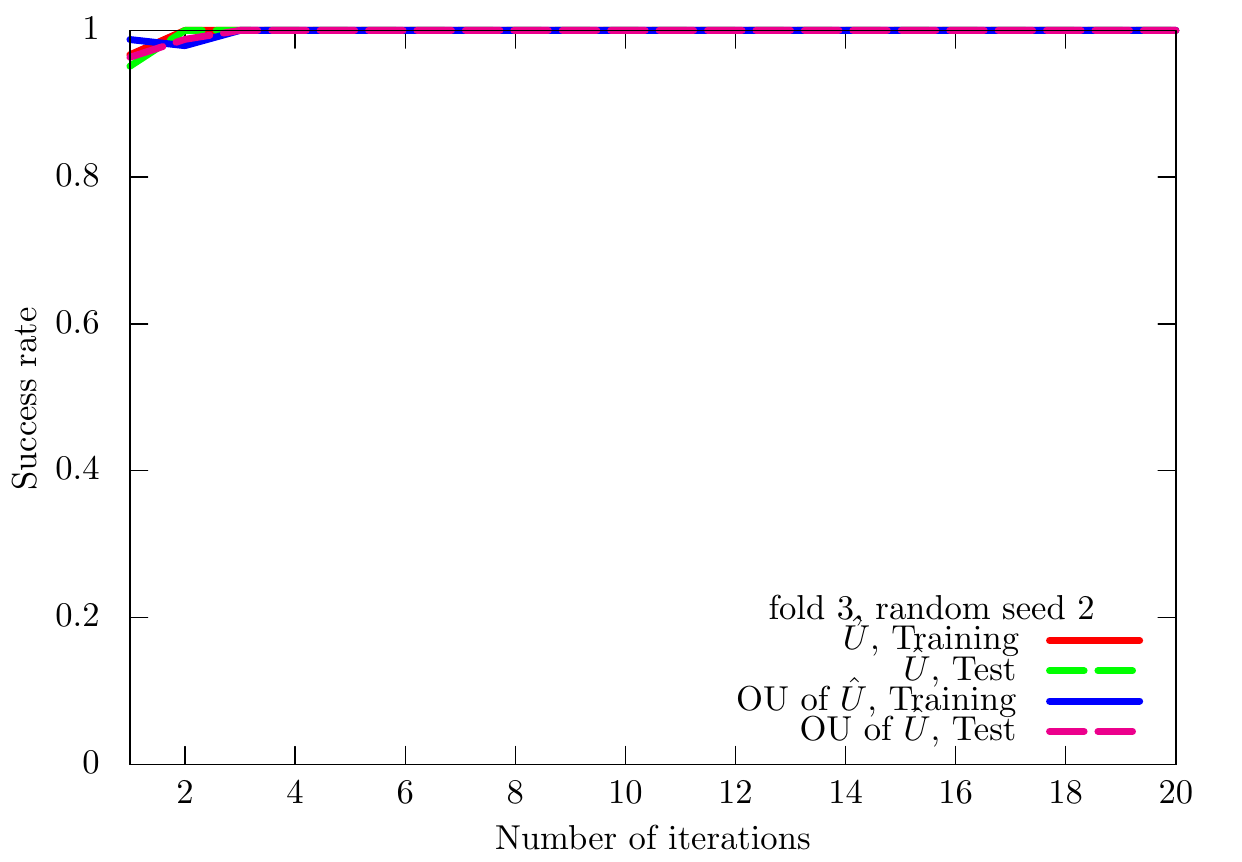}
\includegraphics[scale=0.25]{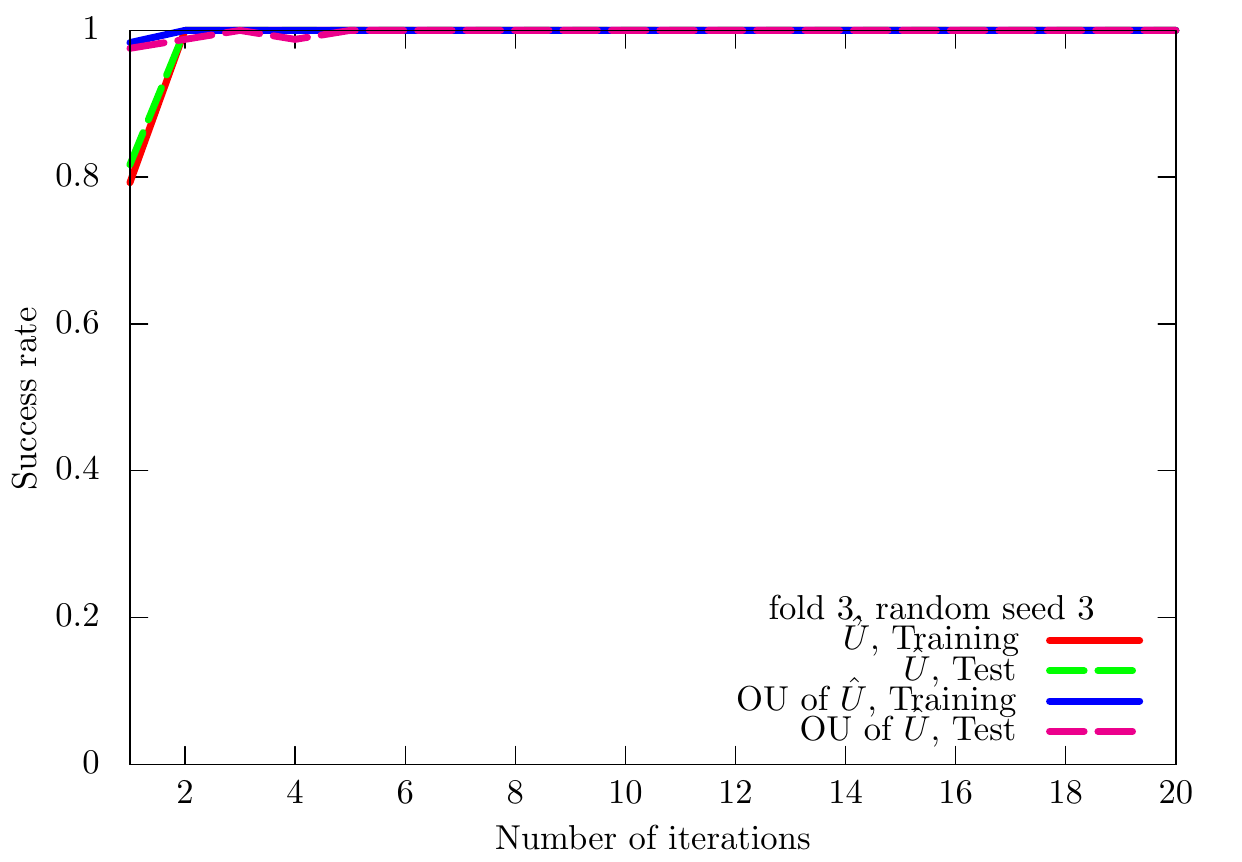}
\includegraphics[scale=0.25]{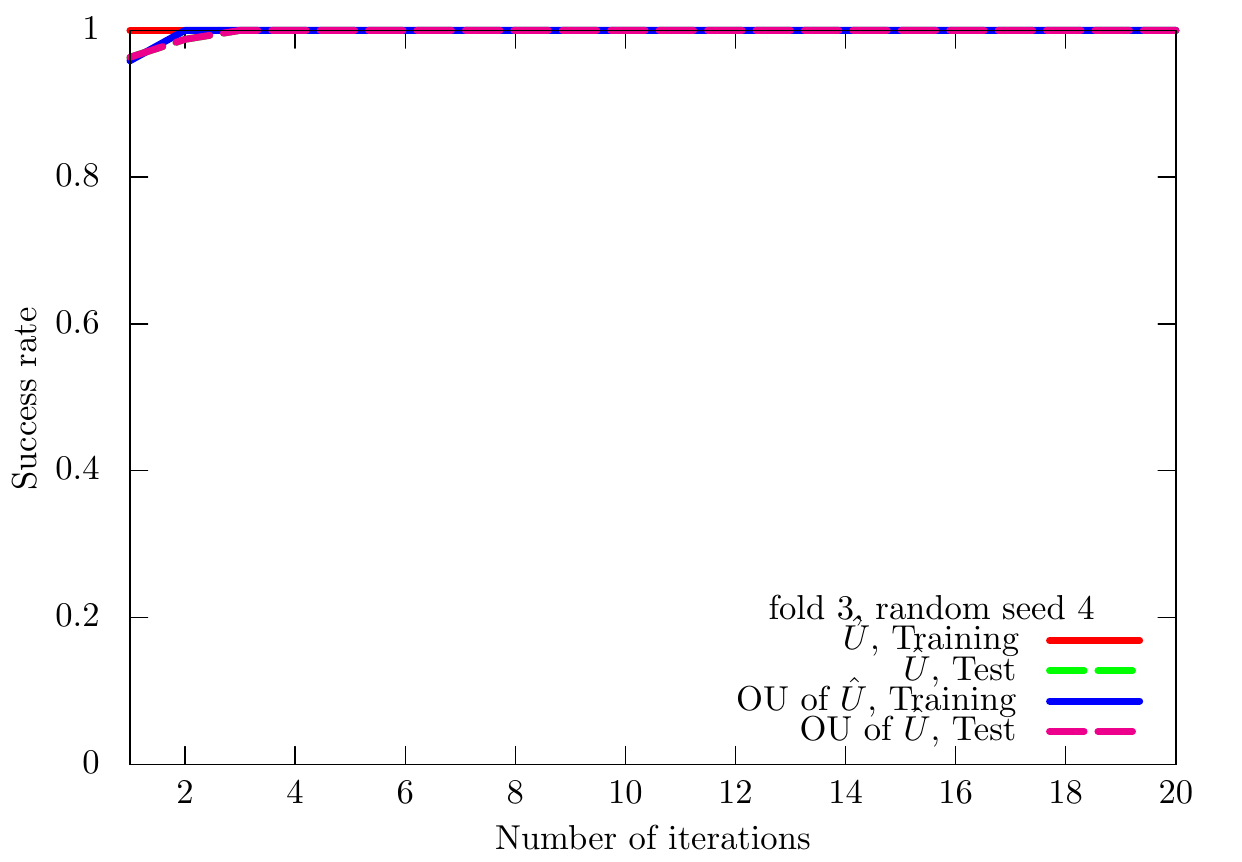}
\includegraphics[scale=0.25]{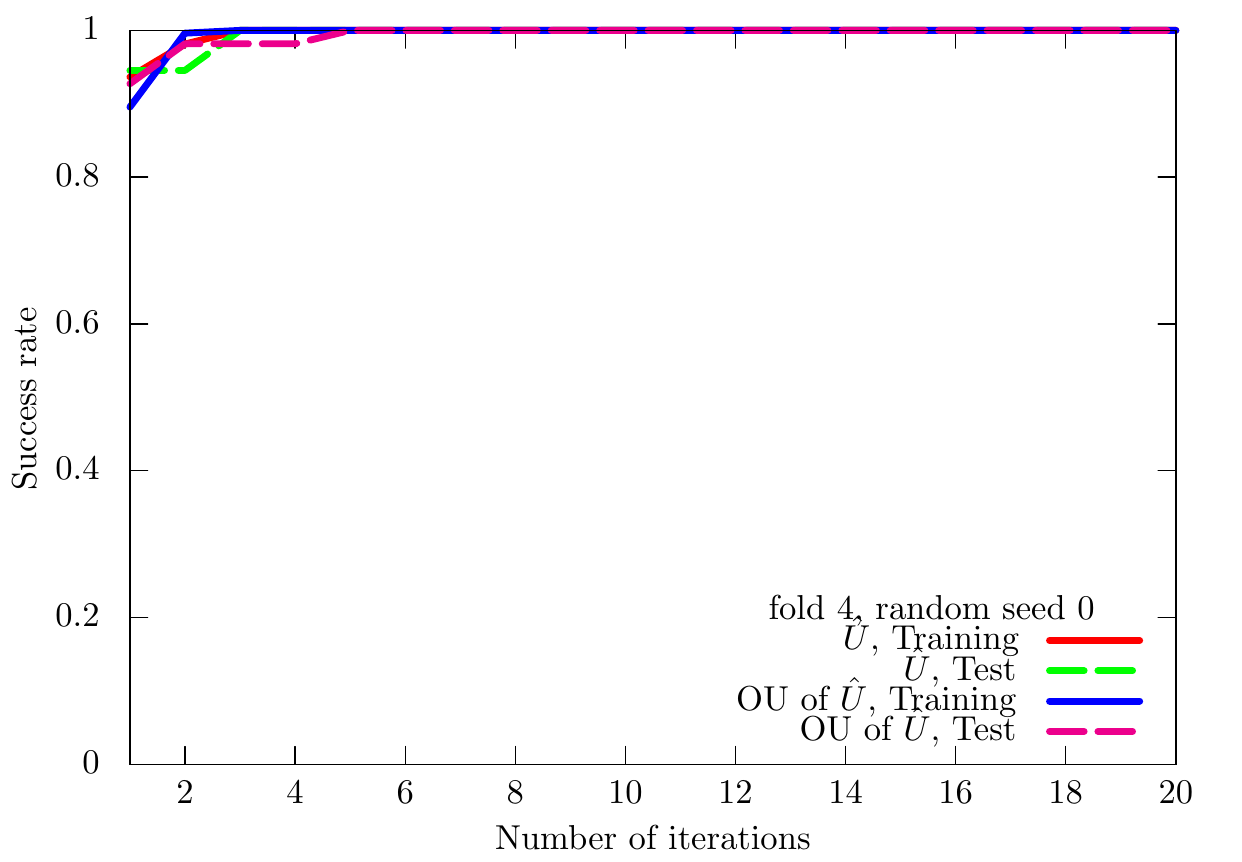}
\includegraphics[scale=0.25]{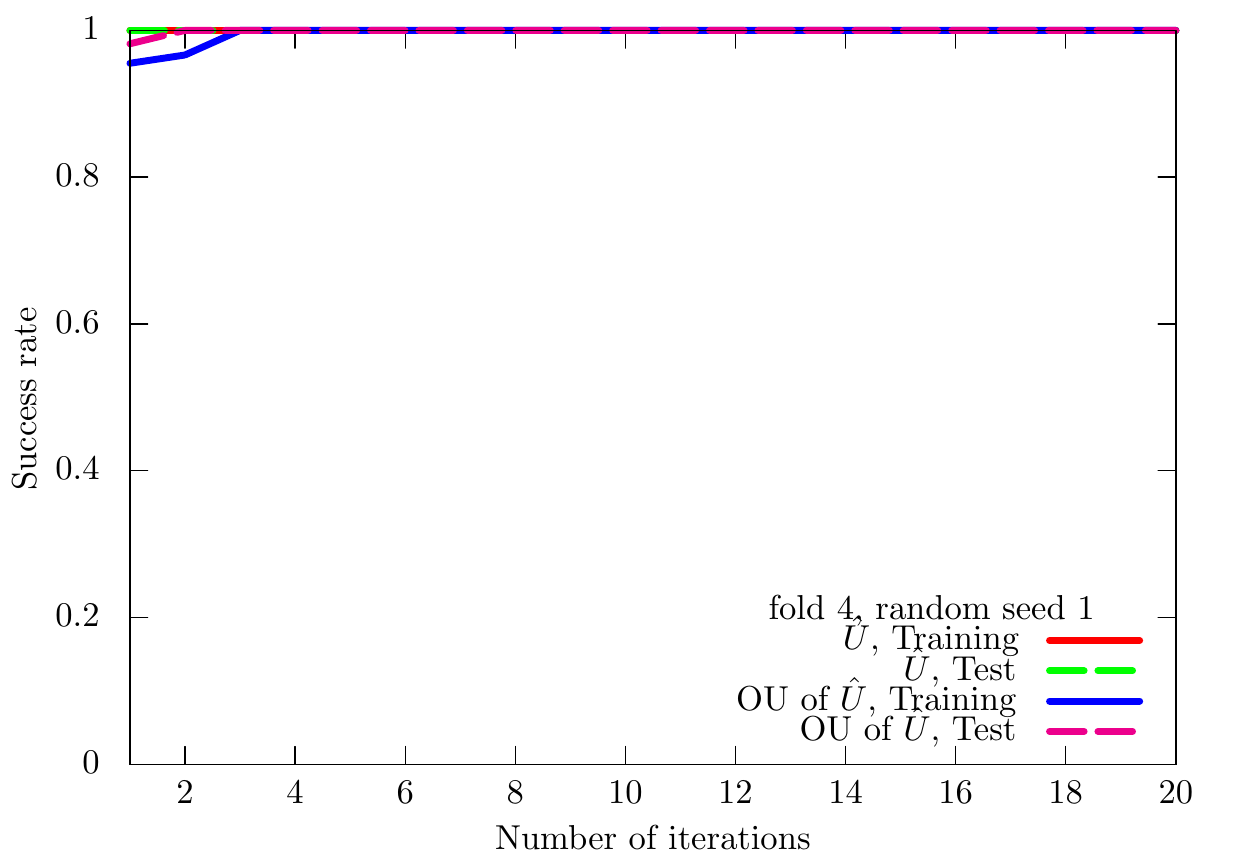}
\includegraphics[scale=0.25]{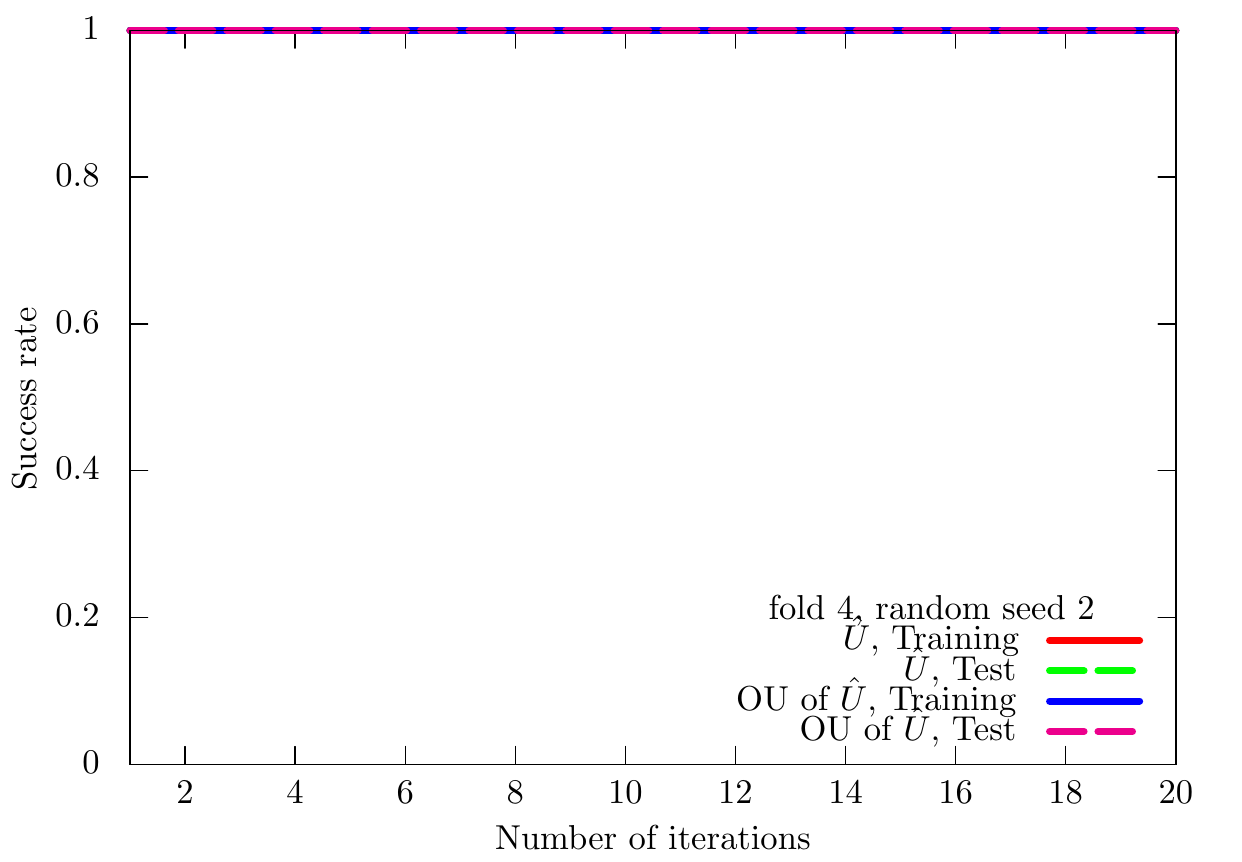}
\includegraphics[scale=0.25]{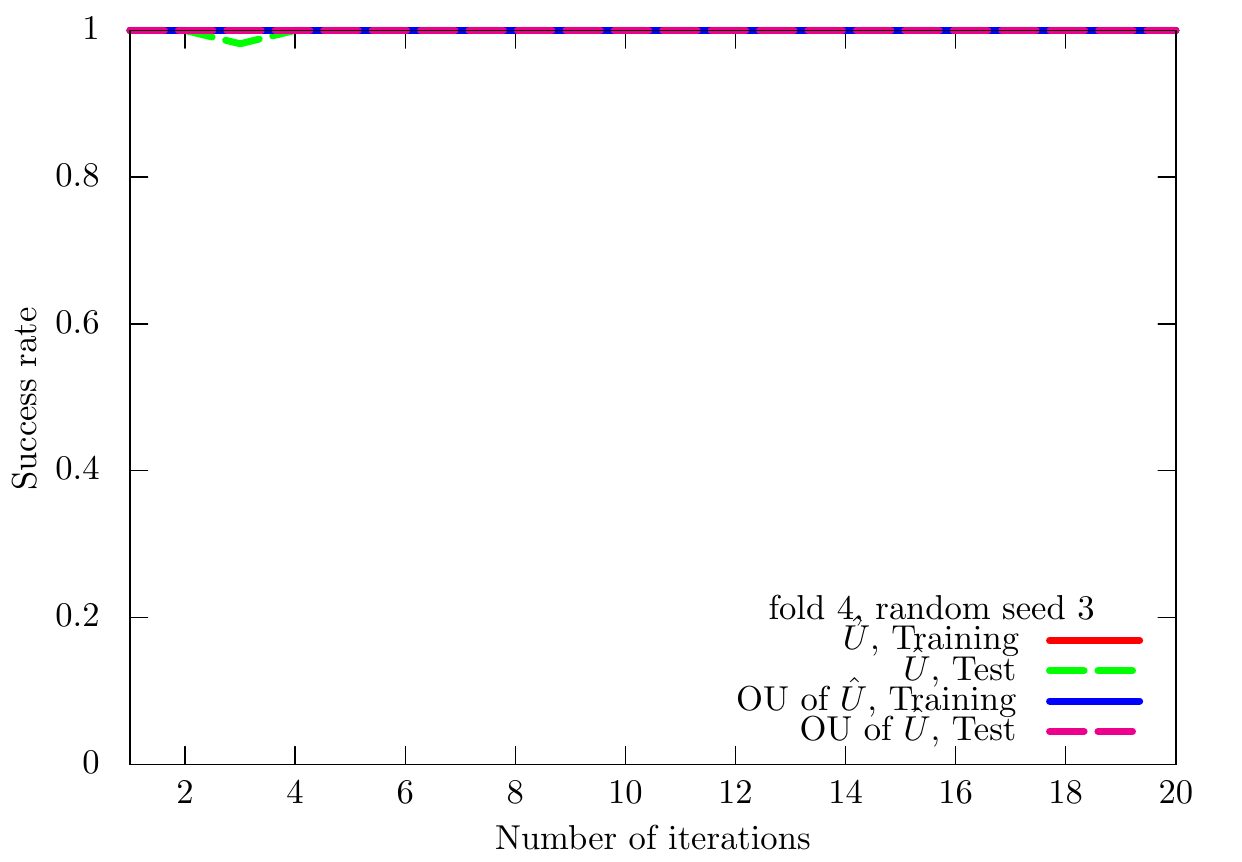}
\includegraphics[scale=0.25]{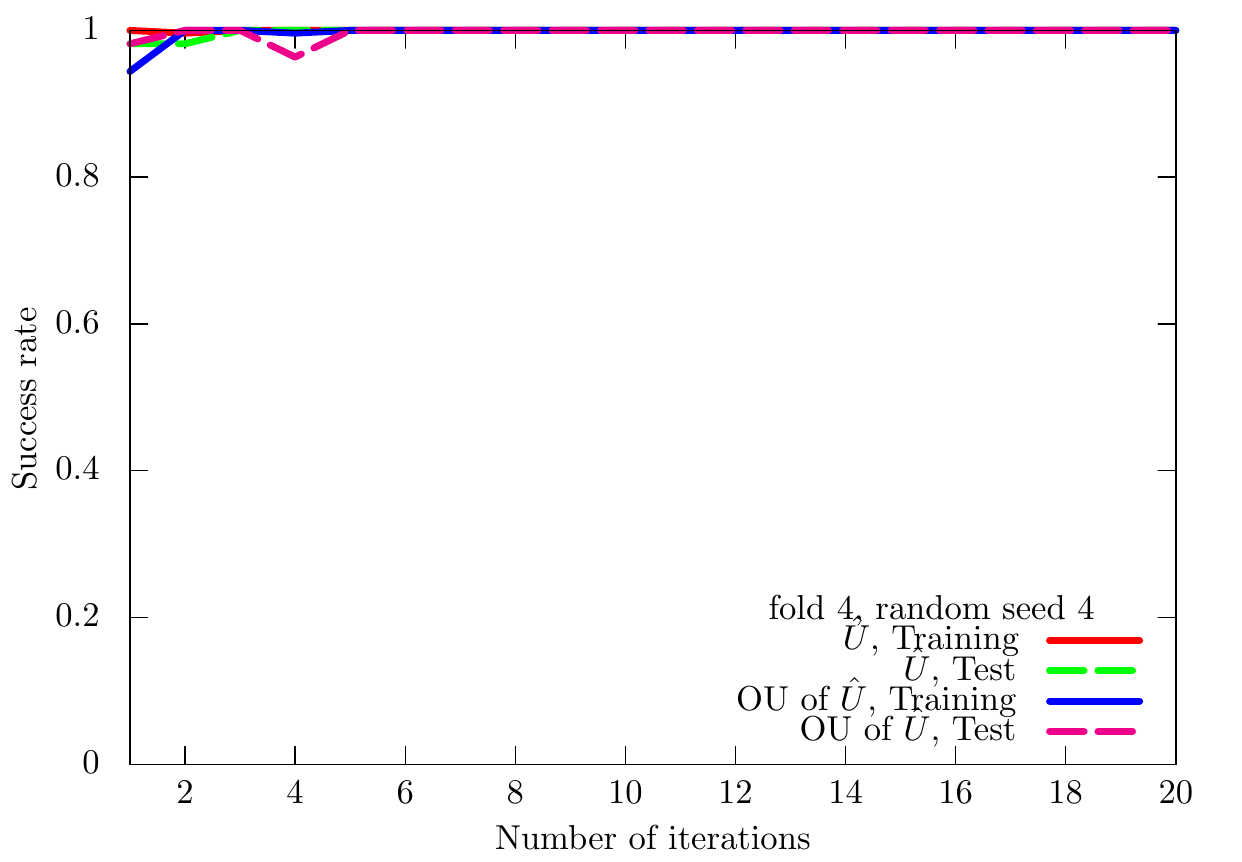}
\caption{Results of the UKM ($\hat{X}$ and OU of $\hat{X}$) on the $5$-fold datasets with $5$ different random seeds for the semeion dataset ($0$ or $1$). We use complex matrices and set $\theta_\mathrm{bias} = 0$. We set $r = 0.010$.}
\label{supp-arXiv-numerical-result-raw-data-fold-001-rand-001-UKM-OUU-UCI-semeion-0-1}
\end{figure*}

In Fig.~\ref{supp-arXiv-numerical-result-raw-data-fold-001-rand-001-QCL-UCI-semeion-0-1}, we show the numerical results of QCL for the $5$-fold datasets with $5$ different random seeds.
In Fig.~\ref{supp-arXiv-numerical-result-raw-data-fold-001-rand-001-UKM-P-UCI-semeion-0-1}, we show the numerical results of $\hat{P}$ of the UKM for the $5$-fold datasets with $5$ different random seeds.
In Fig.~\ref{supp-arXiv-numerical-result-raw-data-fold-001-rand-001-UKM-OUU-UCI-semeion-0-1}, we also show the numerical results of OU of $\hat{X}$ of the UKM for the $5$-fold datasets with $5$ different random seeds.

\begin{figure}[htb]
\centering
\includegraphics[scale=0.45]{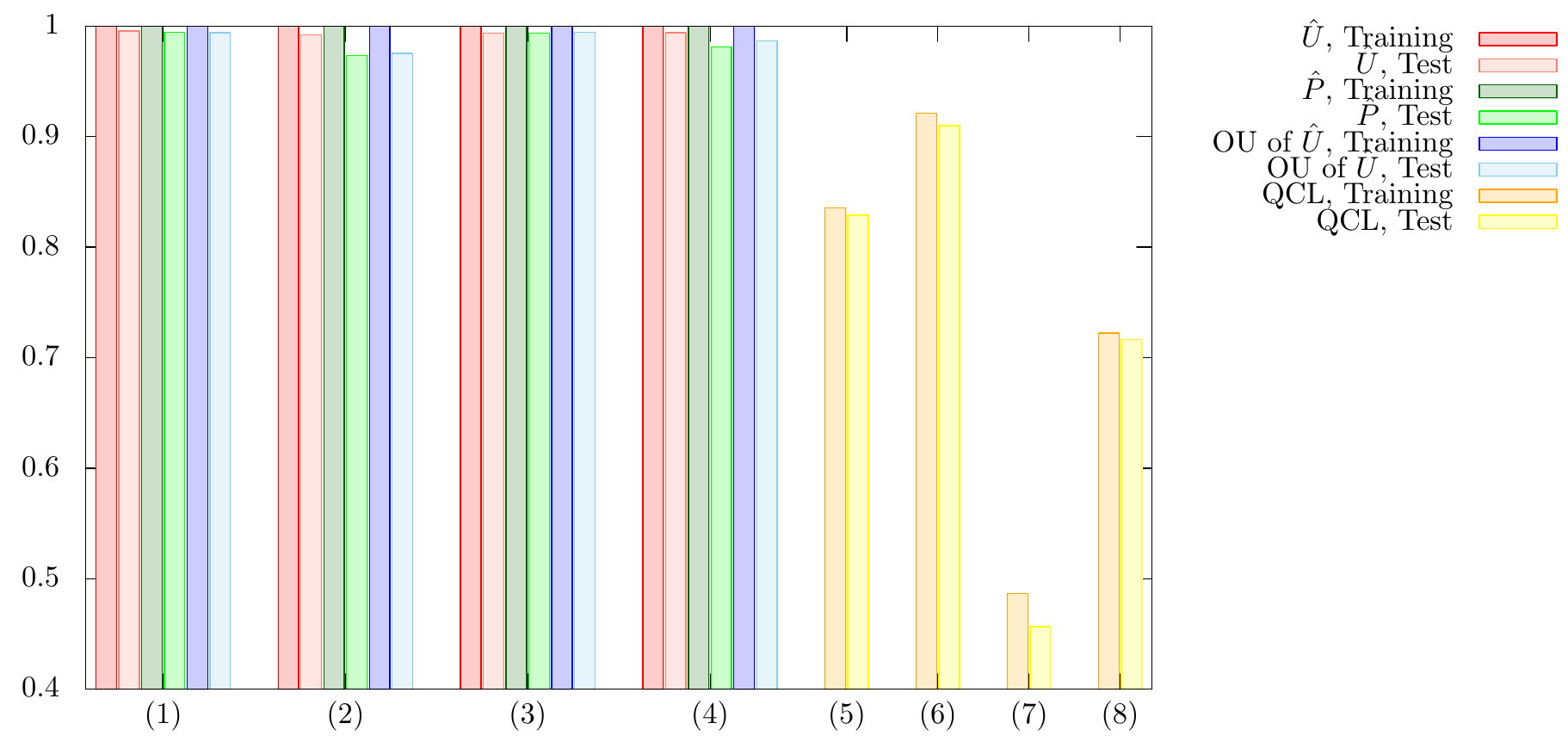}
\caption{Results of $5$-fold CV with $5$ different random seeds for the semeion dataset ($0$ or $1$). For the UKM, we put $r = 0.010$ and set $K = 20$ and $K' = 10$. For QCL, the number of layers $L$ is $5$ and the number of iterations is $100$. The numerical settings are as follows: (1) complex matrices without the bias term, (2) complex matrices with the bias term, (3) real matrices without the bias term, (4) real matrices with the bias term, (5) CNOT-based circuit without the bias term, (6) CNOT-based circuit with the bias term, (7) CRot-based circuit without the bias term, (8) CRot-based circuit with the bias term, (9) 1d Heisenberg circuit without the bias term, (10) 1d Heisenberg circuit with the bias term, (11) FC Heisenberg circuit without the bias term, and (12) FC Heisenberg circuit with the bias term.}
\label{supp-arXiv-numerical-result-performance-UKM-QCL-UCI-semeion-0-1}
\end{figure}

We summarize the results of 5-fold CV with 5 different random seeds of QCL and the UKM in Tables~\ref{supp-arXiv-table-UCI-semeion-0-1-002} and \ref{supp-arXiv-table-UCI-semeion-0-1-001}, respectively.
For QCL and the UKM, we select the best model for the training dataset over iterations to compute the performance.
\begin{table}[htb]
  \begin{tabular}{cc|cc}
    \hline \hline
    Algo. & Condition & Training & Test \\
    \hline
  QCL & CNOT-based, w/o bias & 0.8356 & 0.8288 \\
  QCL & CNOT-based, w/ bias & 0.9210 & 0.9099 \\
    \hline
  QCL & CRot-based, w/o bias & 0.4867 & 0.4566 \\
  QCL & CRot-based, w/ bias & 0.7222 & 0.7162 \\
    \hline \hline
  \end{tabular}
\caption{Results of $5$-fold CV with $5$ different random seeds of QCL for the semeion dataset ($0$ or $1$). The number of layers $L$ is $5$ and the number of iterations is $100$.}
\label{supp-arXiv-table-UCI-semeion-0-1-002}
\end{table}
\begin{table}[htb]
  \begin{tabular}{cc|cc}
    \hline \hline
    Algo. & Condition & Training & Test \\
    \hline
  UKM & $\hat{X}$, complex, w/o bias & 1.0 & 0.9957 \\
  UKM & $\hat{P}$, complex, w/o bias & 1.0 & 0.9943 \\
  UKM & OU of $\hat{X}$, complex, w/o bias & 1.0 & 0.9941 \\
    \hline
  UKM & $\hat{X}$, complex, w/ bias & 1.0 & 0.9922 \\
  UKM & $\hat{P}$, complex, w/ bias & 1.0 & 0.9735 \\
  UKM & OU of $\hat{X}$, complex, w/ bias & 1.0 & 0.9755 \\
    \hline
  UKM & $\hat{X}$, real, w/o bias & 1.0 & 0.9939 \\
  UKM & $\hat{P}$, real, w/o bias & 1.0 & 0.9939 \\
  UKM & OU of $\hat{X}$, real, w/o bias & 1.0 & 0.9945 \\
    \hline
  UKM & $\hat{X}$, real, w/ bias & 1.0 & 0.9941 \\
  UKM & $\hat{P}$, real, w/ bias & 1.0 & 0.9814 \\
  UKM & OU of $\hat{X}$, real, w/ bias & 1.0 & 0.9865 \\
    \hline \hline
  \end{tabular}
\caption{Results of $5$-fold CV with $5$ different random seeds of the UKM for the semeion dataset ($0$ or $1$). We put $r = 0.010$ and set $K = 20$ and $K = 10$.}
\label{supp-arXiv-table-UCI-semeion-0-1-001}
\end{table}
In Fig.~\ref{supp-arXiv-numerical-result-performance-UKM-QCL-UCI-semeion-0-1}, we plot the data shown in Tables~\ref{supp-arXiv-table-UCI-semeion-0-1-002} and \ref{supp-arXiv-table-UCI-semeion-0-1-001}.
We also summarize the results of 5-fold CV with 5 different random seeds of the kernel method in Table~\ref{supp-arXiv-table-UCI-semeion-0-1-003}.
More specifically, we use Ridge classification in Sec.~\ref{supp-arXiv-sec-Ridge-001}.
We consider the linear functions and the second-order polynomial functions for $\phi (\cdot)$ in Eq.~\eqref{supp-arXiv-f-pred-kernel-method-001-002} with and without normalization.
We set $\lambda = 10^{-2}, 10^{-1}, 1$ where $\lambda$ is the coefficient of the regularization term.
\begin{table}[htb]
  \begin{tabular}{cc|cc}
    \hline \hline
    Algo. & Condition & Training & Test \\
    \hline
  Kernel method & Linear, w/o normalization, $\lambda = 10^{-2}$ & 1.0000 & 0.8688 \\
  Kernel method & Linear, w/o normalization, $\lambda = 10^{-1}$ & 1.0000 & 0.9413 \\
  Kernel method & Linear, w/o normalization, $\lambda = 1$ & 1.0000 & 0.9822 \\
    \hline
  Kernel method & Linear, w/ normalization, $\lambda = 10^{-2}$ & 1.0000 & 0.9858 \\
  Kernel method & Linear, w/ normalization, $\lambda = 10^{-1}$ & 1.0000 & 0.9965 \\
  Kernel method & Linear, w/ normalization, $\lambda = 1$ & 1.0000 & 1.0000 \\
    \hline
  Kernel method & Poly-2, w/o normalization, $\lambda = 10^{-2}$ & 1.0000 & 0.9895 \\
  Kernel method & Poly-2, w/o normalization, $\lambda = 10^{-1}$ & 1.0000 & 0.9895 \\
  Kernel method & Poly-2, w/o normalization, $\lambda = 1$ & 1.0000 & 0.9895 \\
    \hline
  Kernel method & Poly-2, w/ normalization, $\lambda = 10^{-2}$ & 1.0000 & 0.9965 \\
  Kernel method & Poly-2, w/ normalization, $\lambda = 10^{-1}$ & 1.0000 & 0.9965 \\
  Kernel method & Poly-2, w/ normalization, $\lambda = 1$ & 1.0000 & 1.0000 \\
    \hline \hline
  \end{tabular}
\caption{Results of 5-fold CV with 5 different random seeds of the kernel method for the semeion dataset ($0$ or $1$).}
\label{supp-arXiv-table-UCI-semeion-0-1-003}
\end{table}

\begin{figure}[htb]
\centering
\includegraphics[scale=0.45]{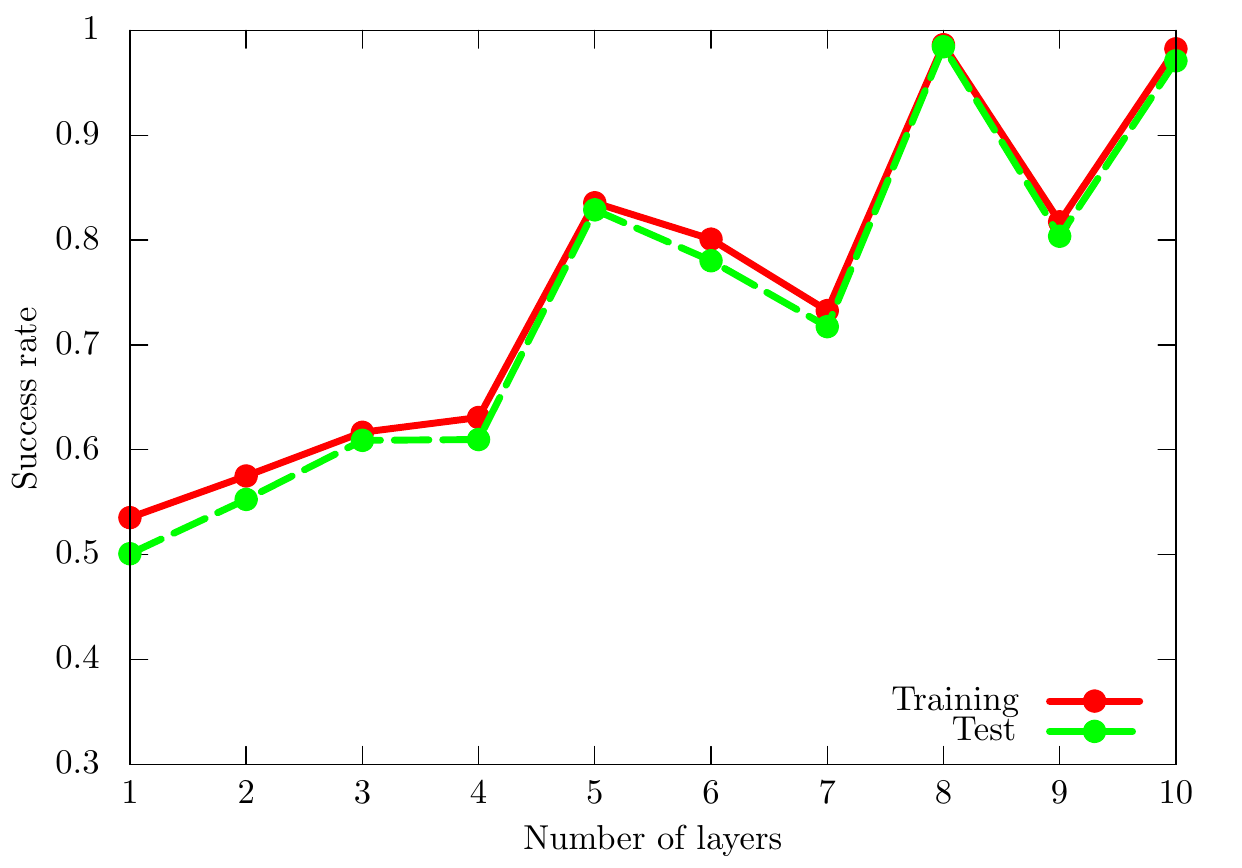}
\caption{Performance dependence of QCL on the number of layers $L$ for the semeion dataset ($0$ or $1$). We use the CNOT-based circuit geometry and set $\theta_\mathrm{bias} = 0$. We iterate the computation $100$ times.}
\label{supp-arXiv-numerical-result-layers-dependence-QCL-UCI-semeion-0-1}
\end{figure}

\begin{figure}[htb]
\centering
\includegraphics[scale=0.45]{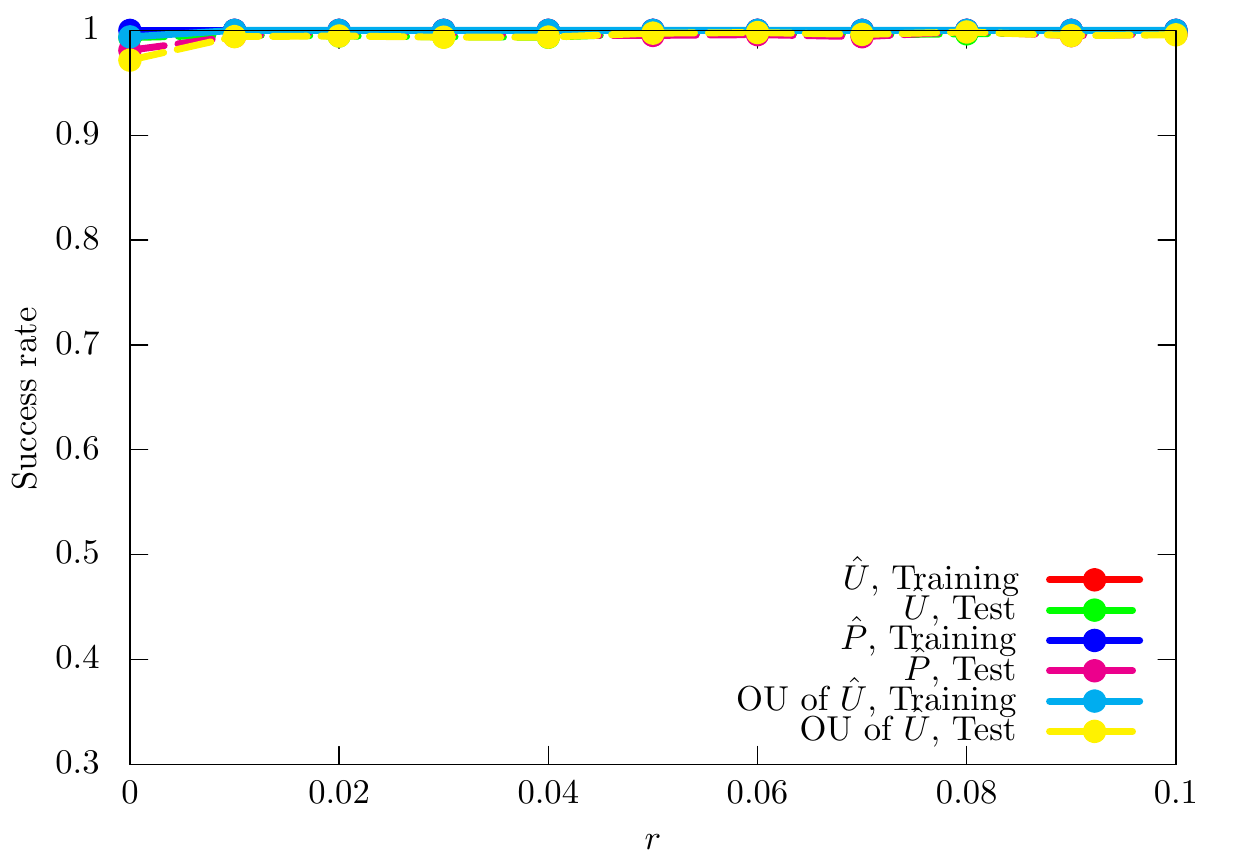}
\includegraphics[scale=0.45]{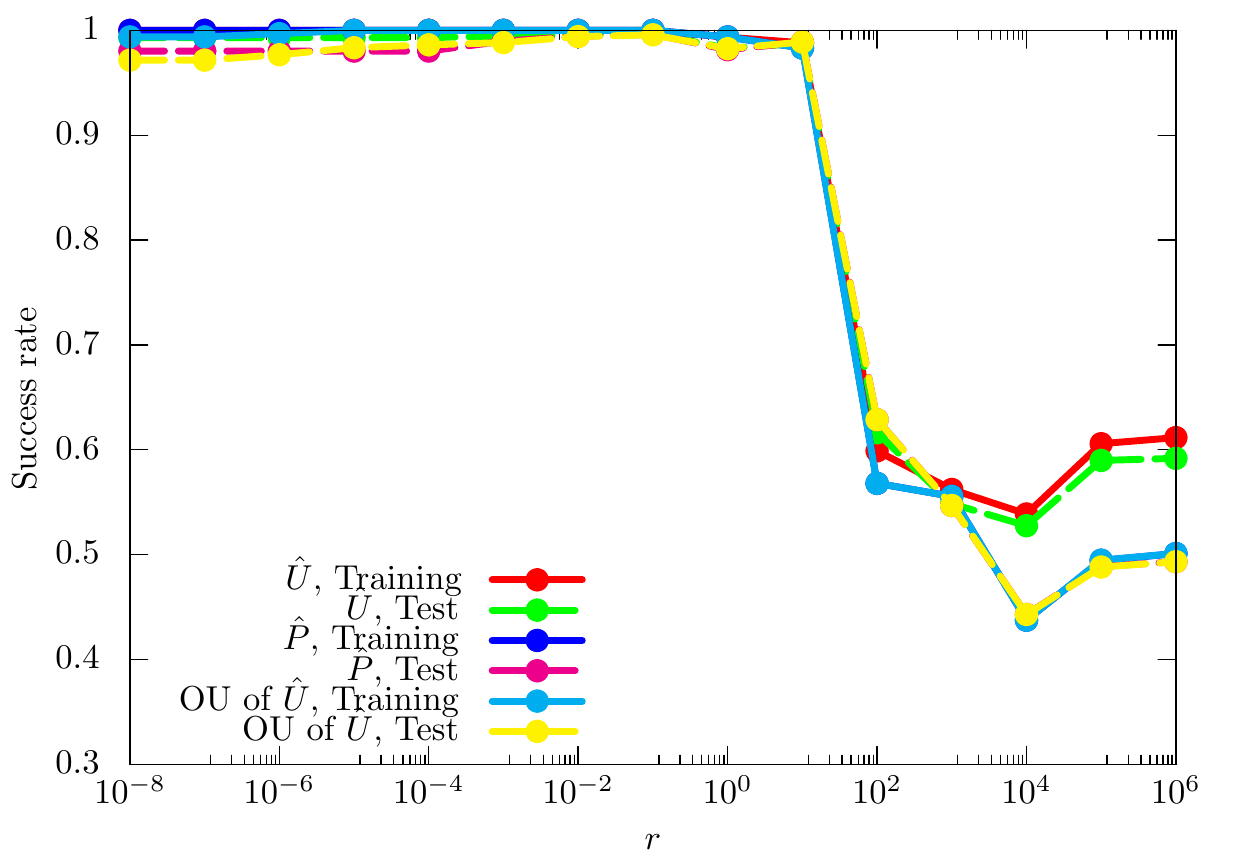}
\caption{Performance dependence of the UKM on $r$, which is the coefficient of the second term in the right-hand side of Eq.~\eqref{supp-arXiv-quantum-kernel-method-001-011} for the semeion dataset ($0$ or $1$). We use complex matrices and set $\theta_\mathrm{bias} = 0$. We set $K = 20$ and $K' = 10$.}
\label{supp-arXiv-numerical-result-r-dependence-UKM-UCI-semeion-0-1}
\end{figure}

\begin{figure}[htb]
\centering
\includegraphics[scale=0.45]{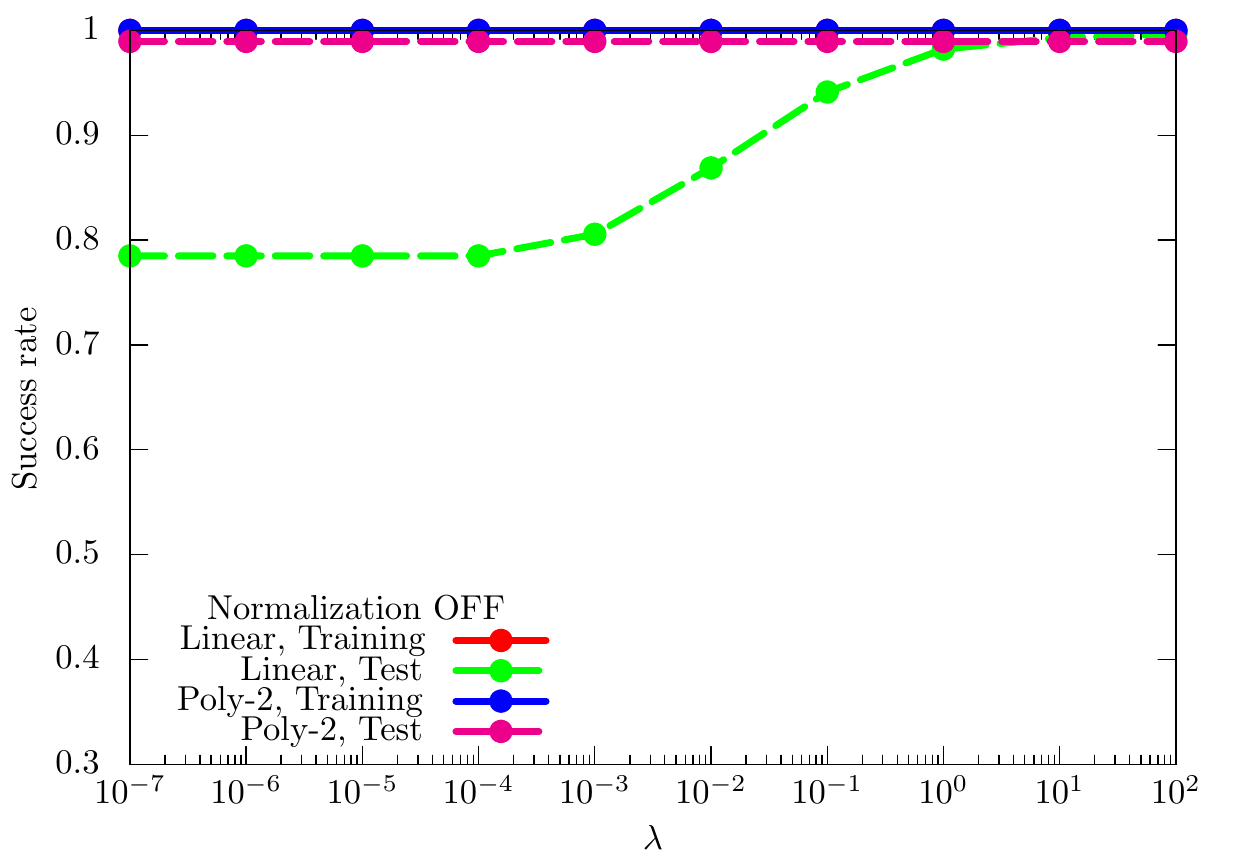}
\includegraphics[scale=0.45]{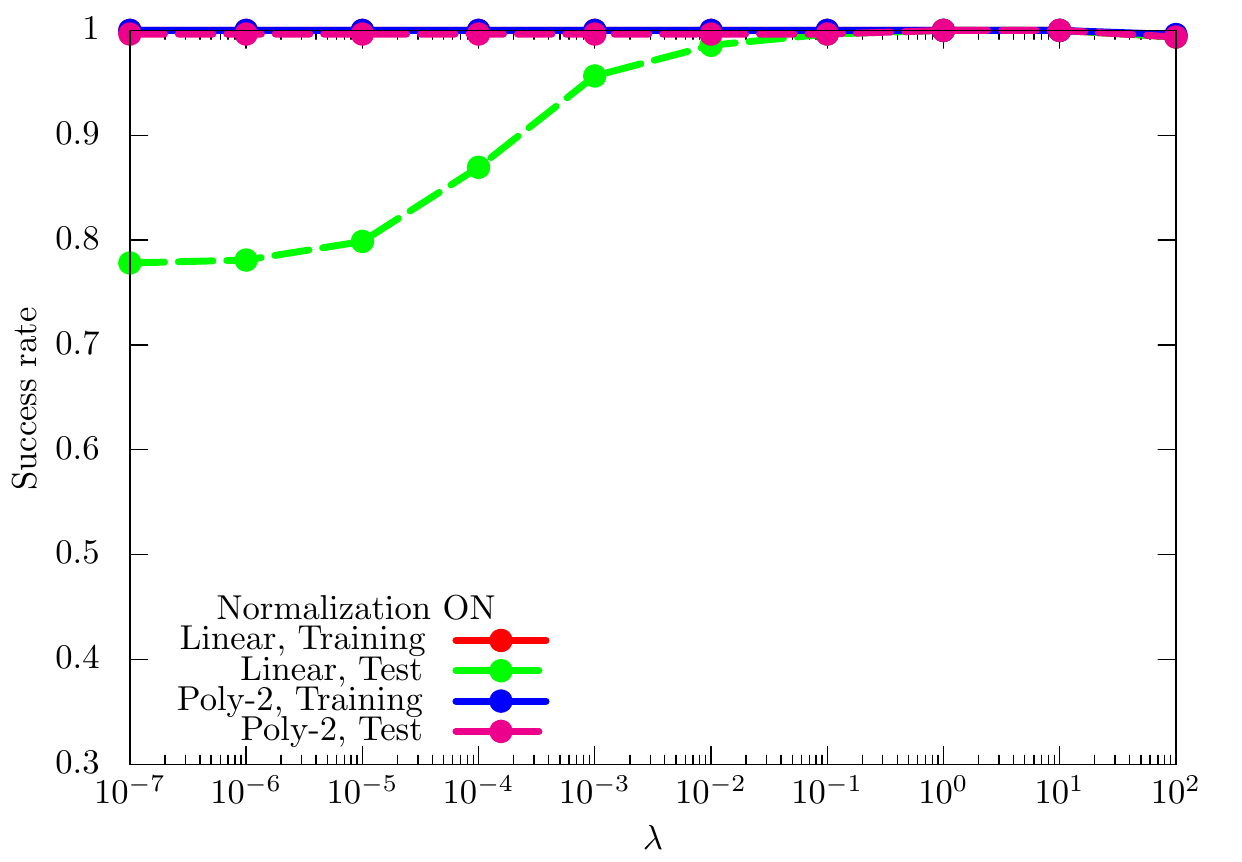}
\caption{Performance dependence of the kernel method on $\lambda$, which is the coefficient of the second term in the right-hand side of Eq.~\eqref{supp-arXiv-cost-function-kernel-method-001-002} for the semeion dataset ($0$ or $1$). For $\phi (\cdot)$ in Eq.~\eqref{supp-arXiv-f-pred-kernel-method-001-002}, we use the linear functions and the second-degree polynomial functions with and without normalization.}
\label{supp-arXiv-numerical-result-lambda-dependence-kernel-method-semeion-0-1}
\end{figure}

Next, we show the performance dependence of the three algorithms on their key parameters.
We see the performance dependence of QCL on the number of layers $L$.
The result is shown in Fig.~\ref{supp-arXiv-numerical-result-layers-dependence-QCL-UCI-semeion-0-1}.
We then see the performance dependence of the UKM on $r$, which is the coefficient of the second term in the right-hand side of Eq.~\eqref{supp-arXiv-quantum-kernel-method-001-011}.
The result is shown in Fig.~\ref{supp-arXiv-numerical-result-r-dependence-UKM-UCI-semeion-0-1}.
In Fig.~\ref{supp-arXiv-numerical-result-lambda-dependence-kernel-method-semeion-0-1}, we show the performance dependence of the kernel method on $\lambda$, which is the coefficient of the second term in the right-hand side of Eq.~\eqref{supp-arXiv-cost-function-kernel-method-001-002}.

\begin{figure}[htb]
\centering
\includegraphics[scale=0.45]{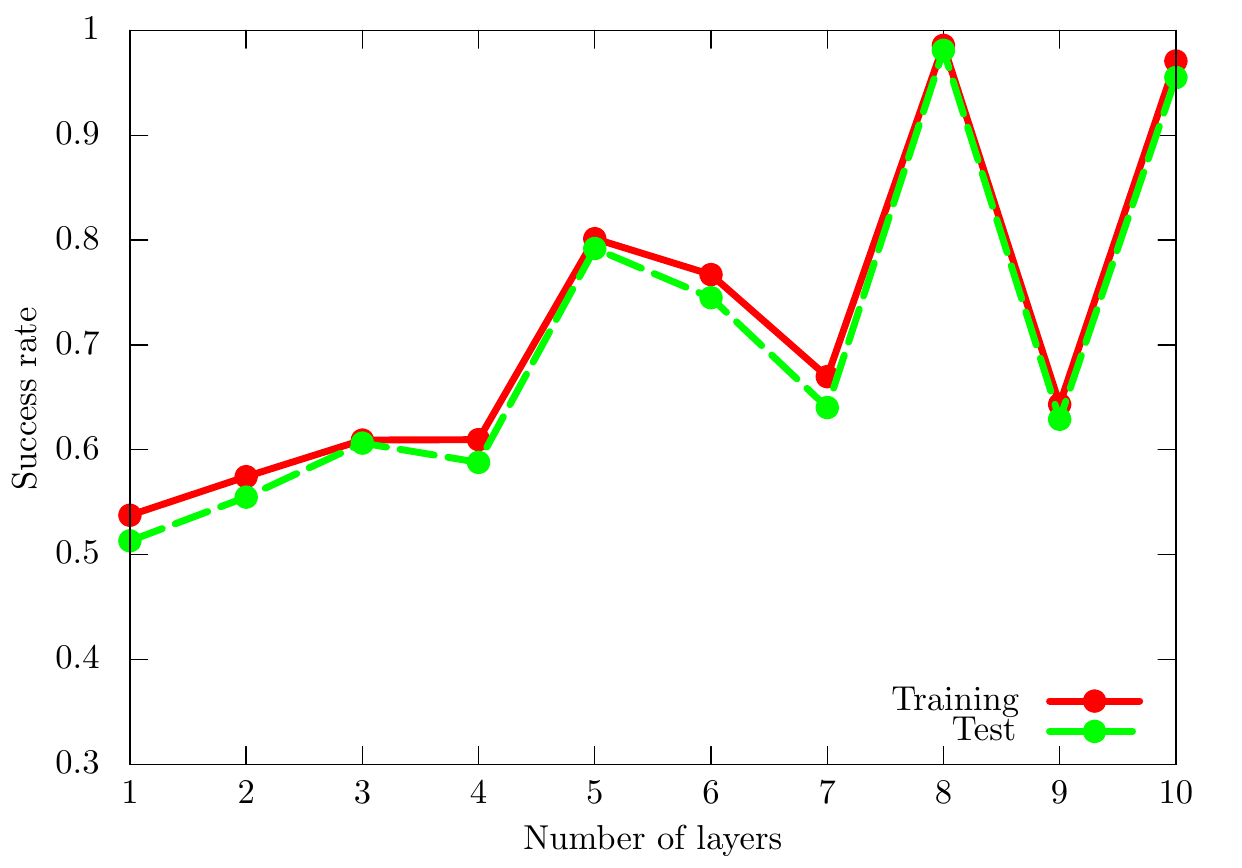}
\caption{Performance dependence of QCL on the number of layers $L$ for the semeion dataset ($0$ or $1$) in the case of the hinge function, Eq.~\eqref{supp-arXiv-hinge-function-001-001}. We use the CNOT-based circuit geometry and set $\theta_\mathrm{bias} = 0$. We iterate the computation $300$ times.}
\label{supp-arXiv-numerical-result-layers-dependence-QCL-UCI-semeion-0-1-hinge}
\end{figure}

\begin{figure}[htb]
\centering
\includegraphics[scale=0.45]{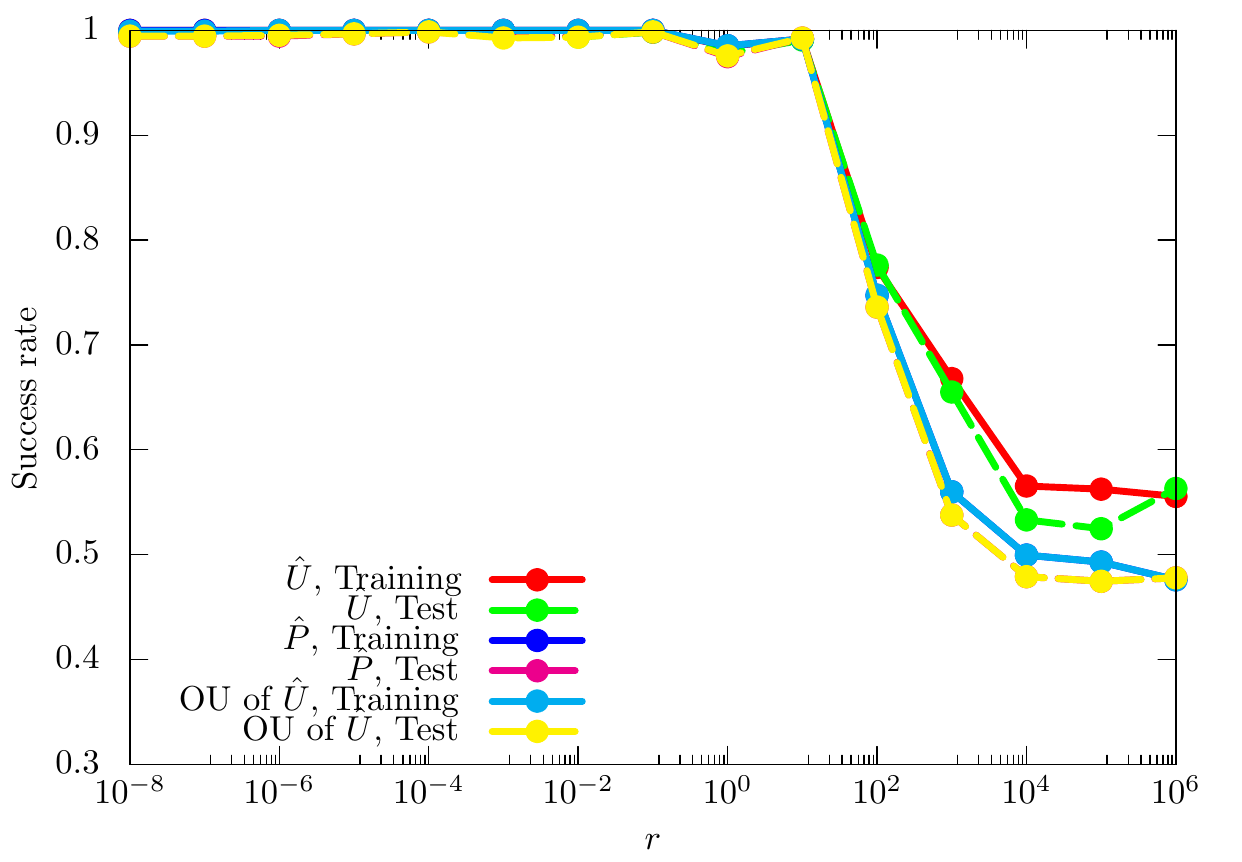}
\caption{Performance dependence of the UKM on $r$, which is the coefficient of the second term in the right-hand side of Eq.~\eqref{supp-arXiv-quantum-kernel-method-001-011} for the semeion dataset ($0$ or $1$) in the case of the hinge function, Eq.~\eqref{supp-arXiv-hinge-function-001-001}. We use complex matrices and set $\theta_\mathrm{bias} = 0$. We set $K = 30$ and $K' = 10$.}
\label{supp-arXiv-numerical-result-r-dependence-UKM-UCI-semeion-0-1-hinge}
\end{figure}

So far, we have used the squared error function, Eq.~\eqref{supp-arXiv-squared-error-function-001-001}.
In Fig.~\ref{supp-arXiv-numerical-result-layers-dependence-QCL-UCI-semeion-0-1-hinge}, we show the performance dependence of QCL on the number of layers $L$ in the case of the hinge function, Eq.~\eqref{supp-arXiv-hinge-function-001-001}.
In Fig.~\ref{supp-arXiv-numerical-result-r-dependence-UKM-UCI-semeion-0-1-hinge}, we show the performance dependence of the UKM on $r$, which is the coefficient of the second term in the right-hand side of Eq.~\eqref{supp-arXiv-quantum-kernel-method-001-011}, in the case of the hinge function, Eq.~\eqref{supp-arXiv-hinge-function-001-001}.

\clearpage

\subsection{Semeion dataset ($0$ or non-$0$)}

We here show the numerical result for the semeion dataset ($0$ or non-$0$).
For the UKM, we put $r = 0.010$ and set $K = 10$ and $K' = 5$ in Algo.~\ref{supp-arXiv-quantum-kernel-method-002-001}.
For QCL, we run iterations $50$ times.

\begin{figure*}[htb]
\centering
\includegraphics[scale=0.25]{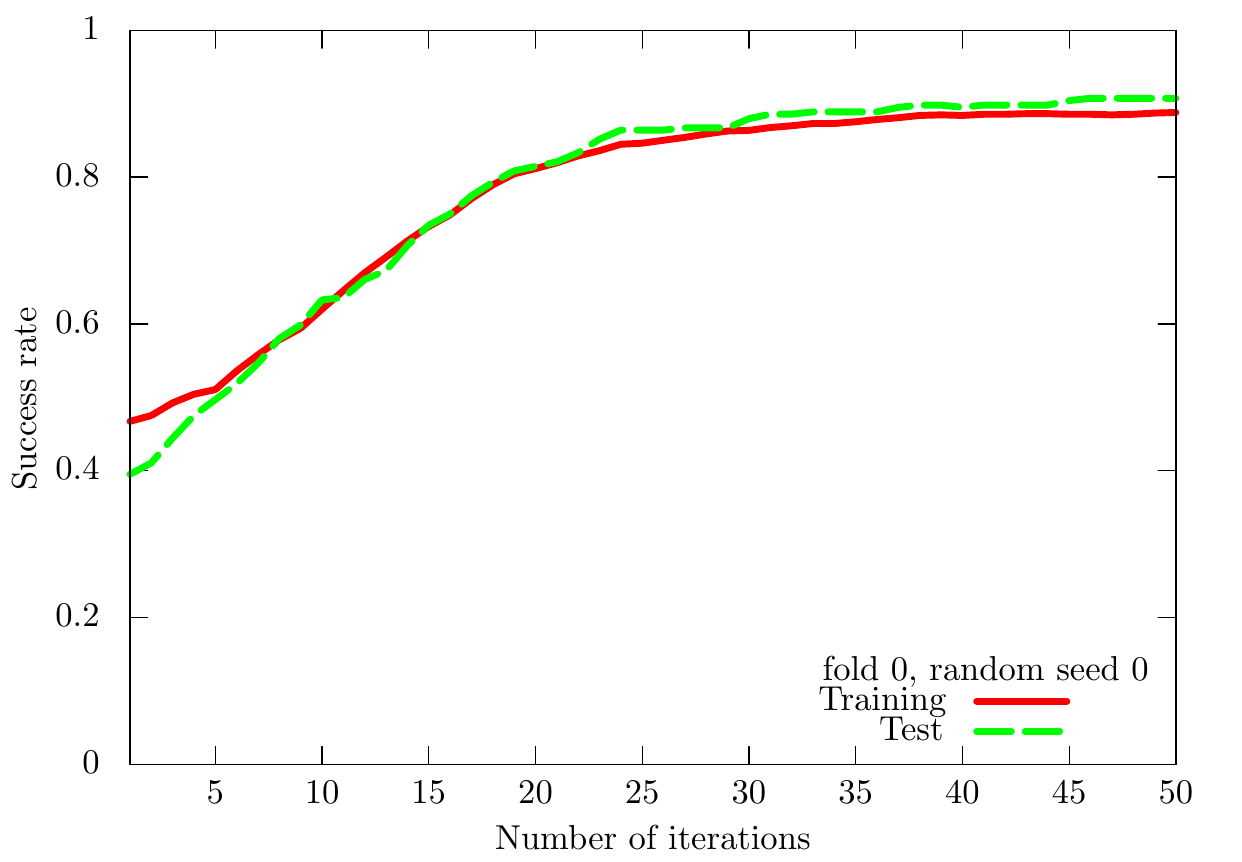}
\includegraphics[scale=0.25]{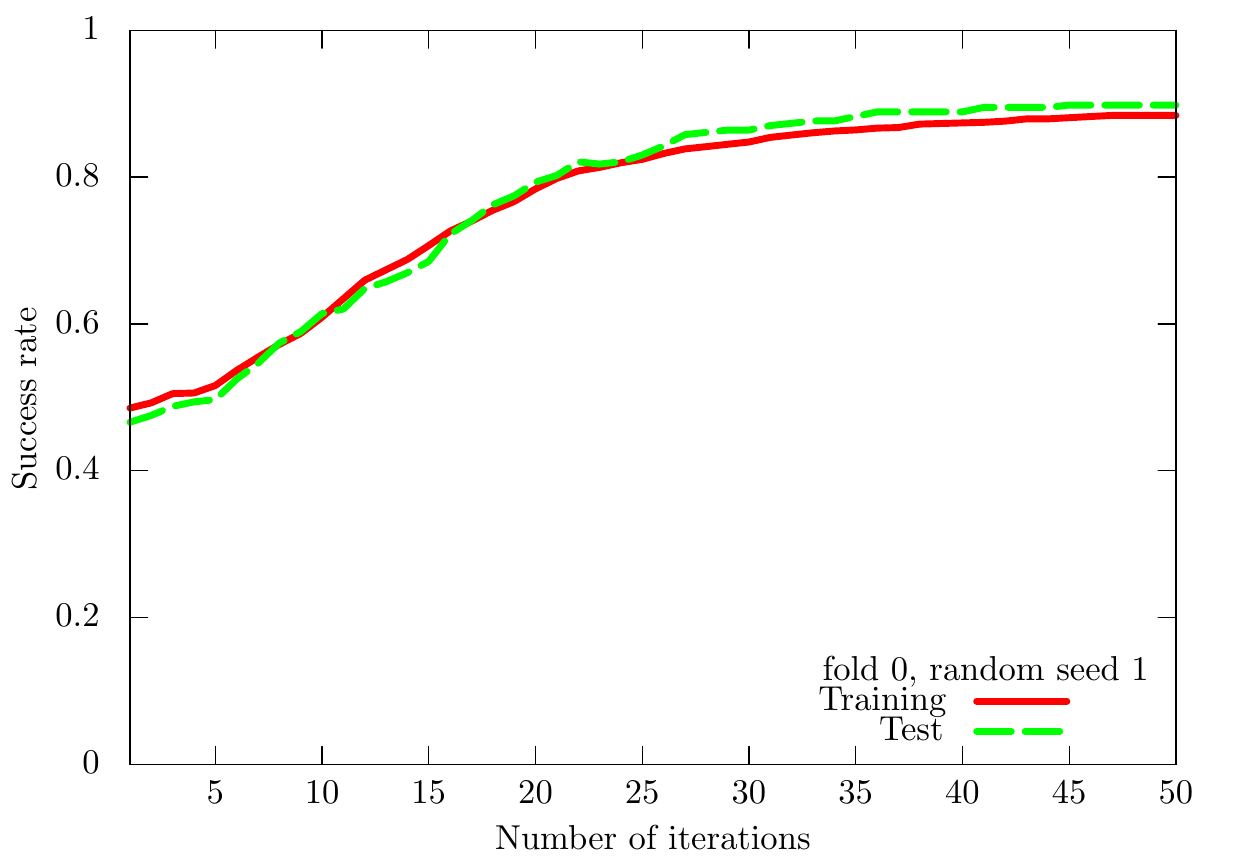}
\includegraphics[scale=0.25]{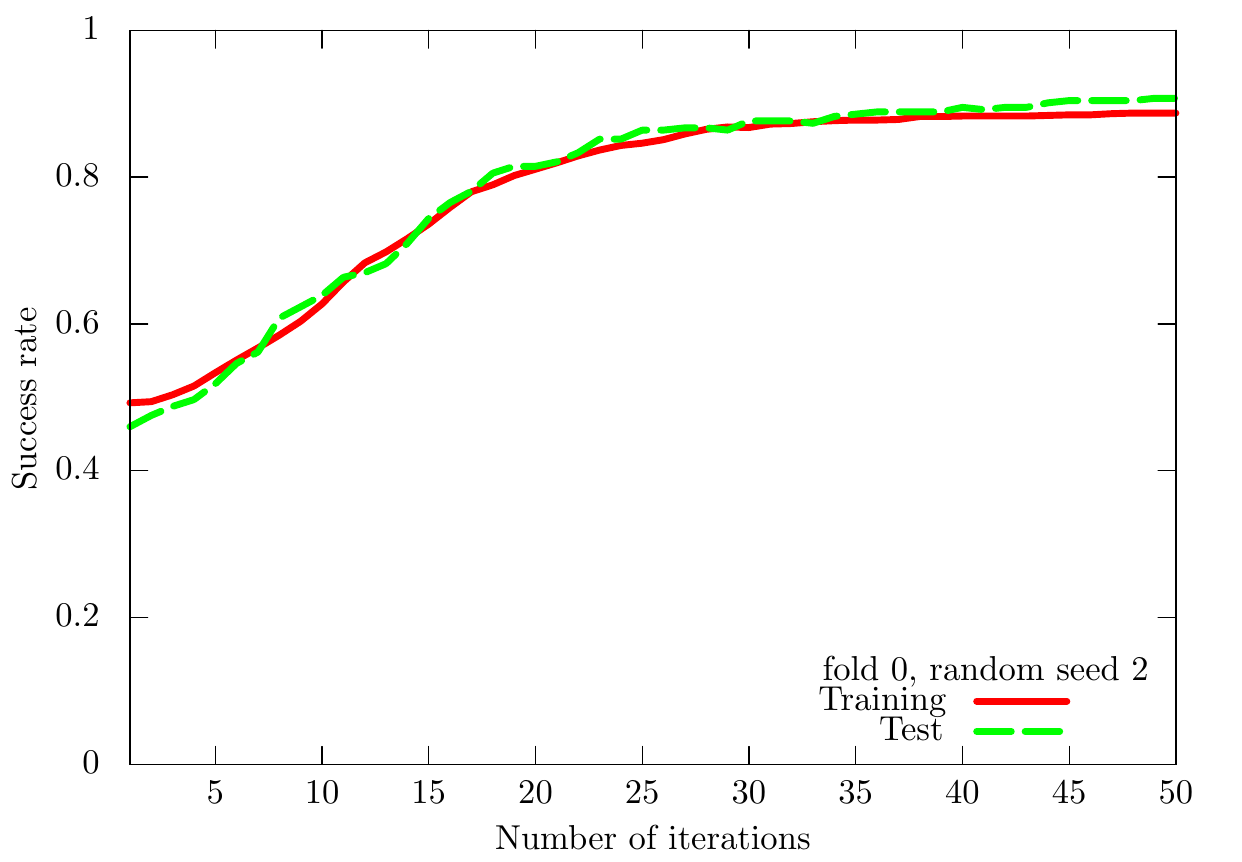}
\includegraphics[scale=0.25]{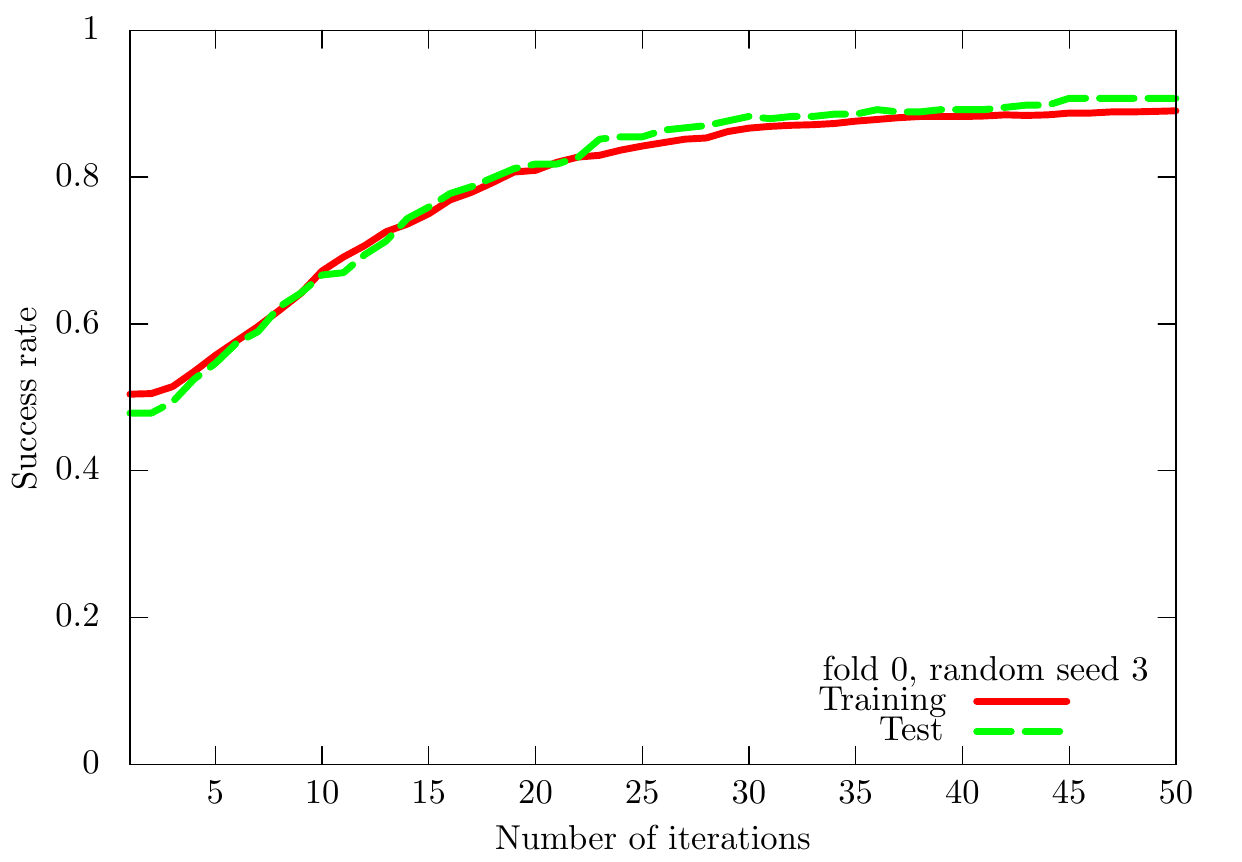}
\includegraphics[scale=0.25]{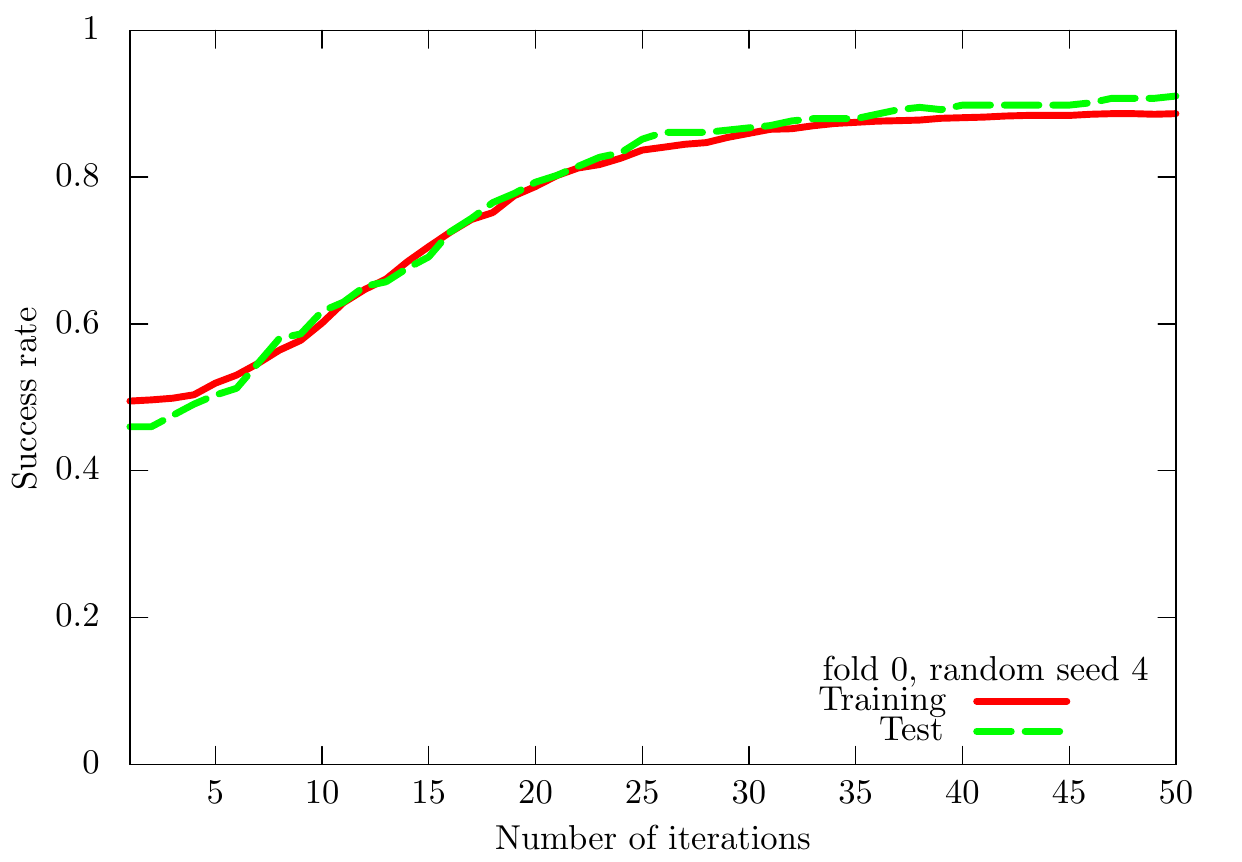}
\includegraphics[scale=0.25]{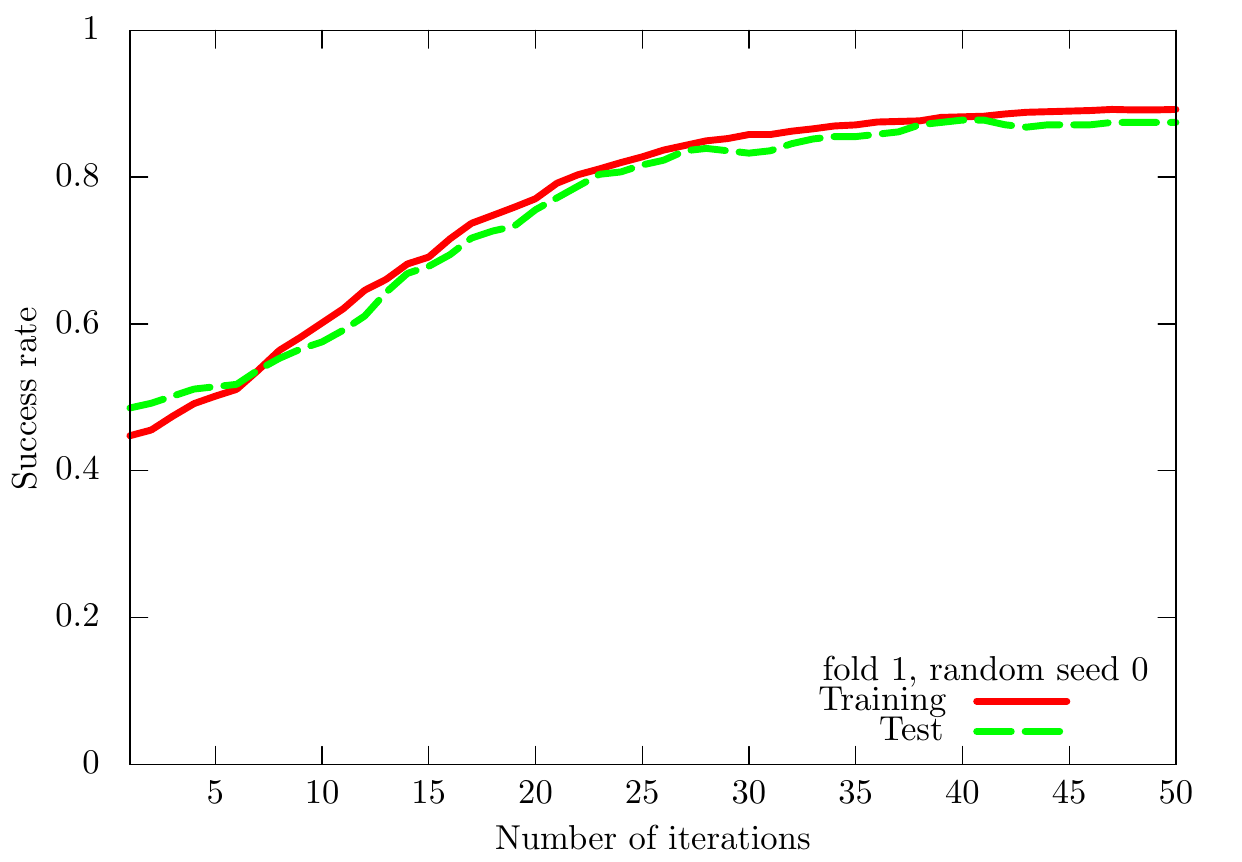}
\includegraphics[scale=0.25]{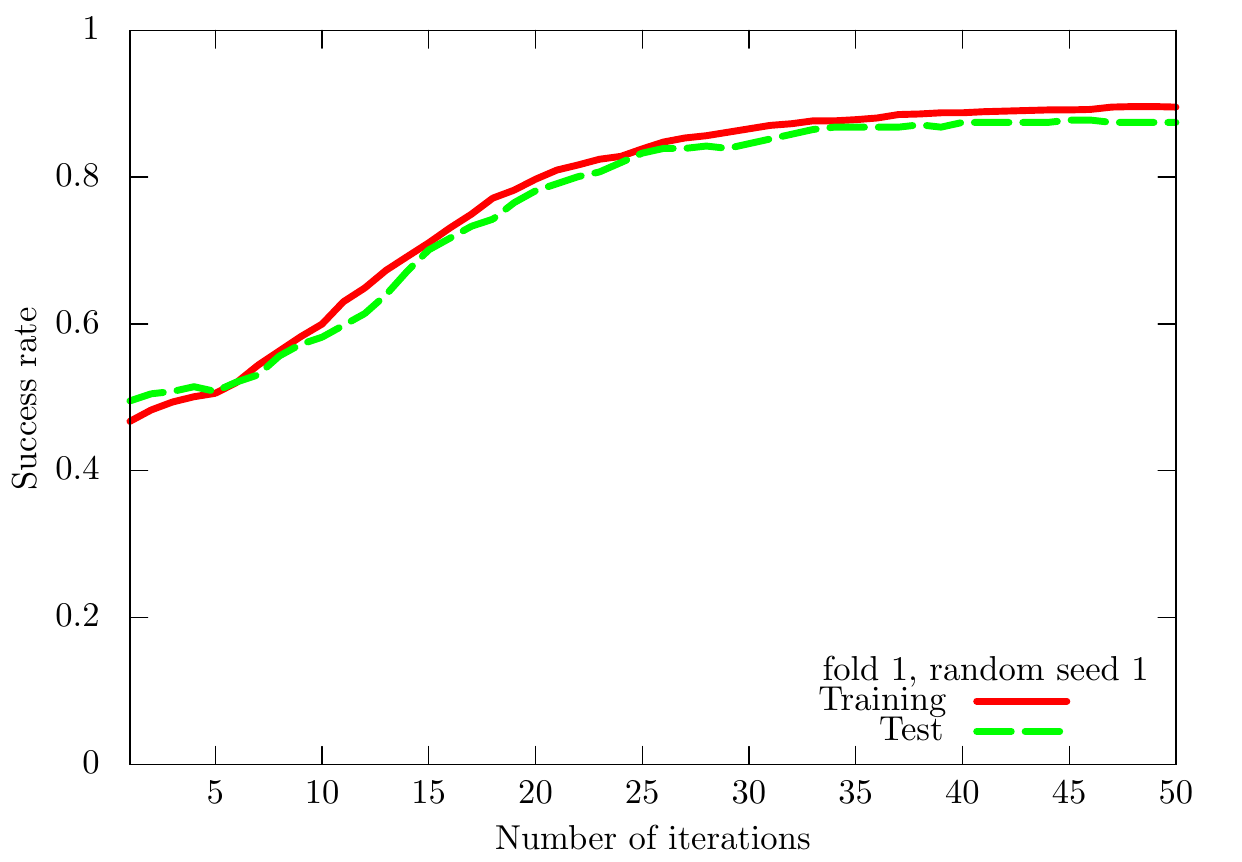}
\includegraphics[scale=0.25]{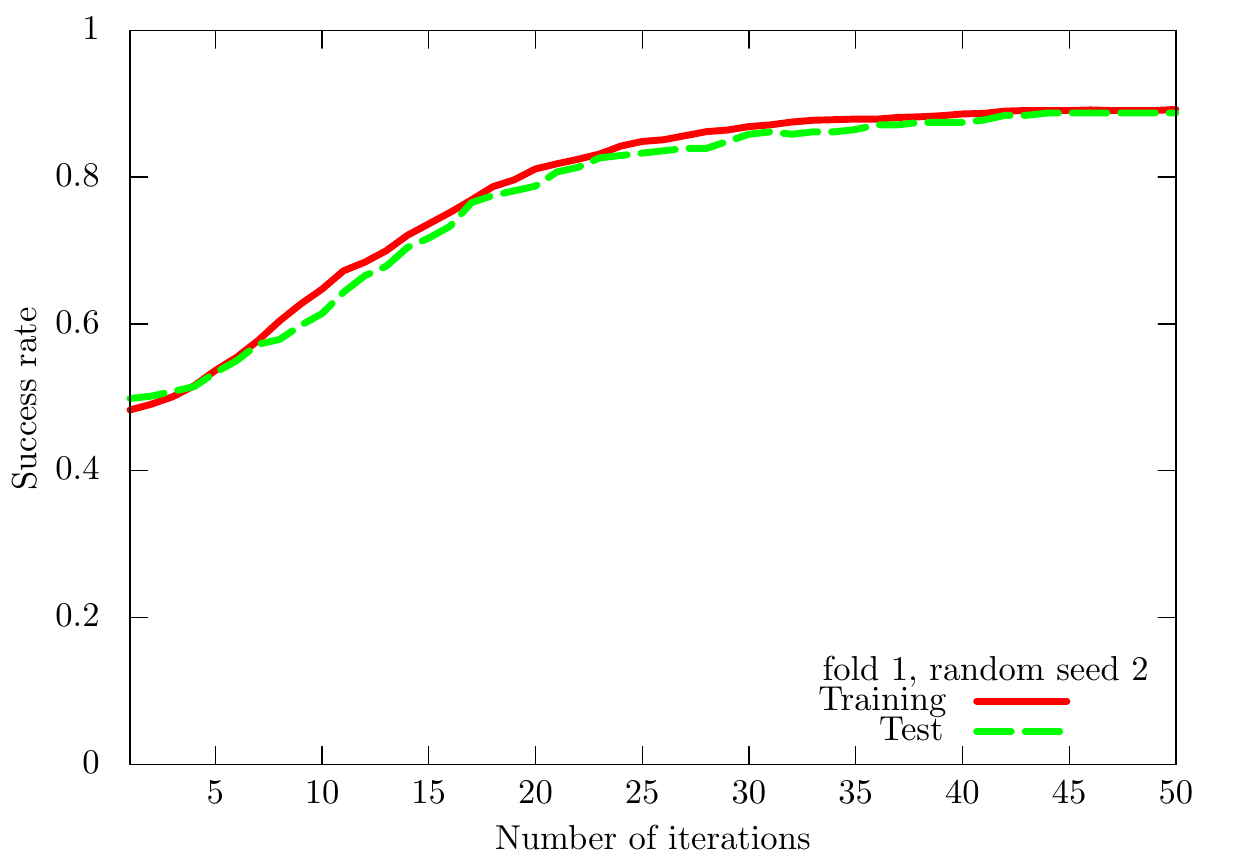}
\includegraphics[scale=0.25]{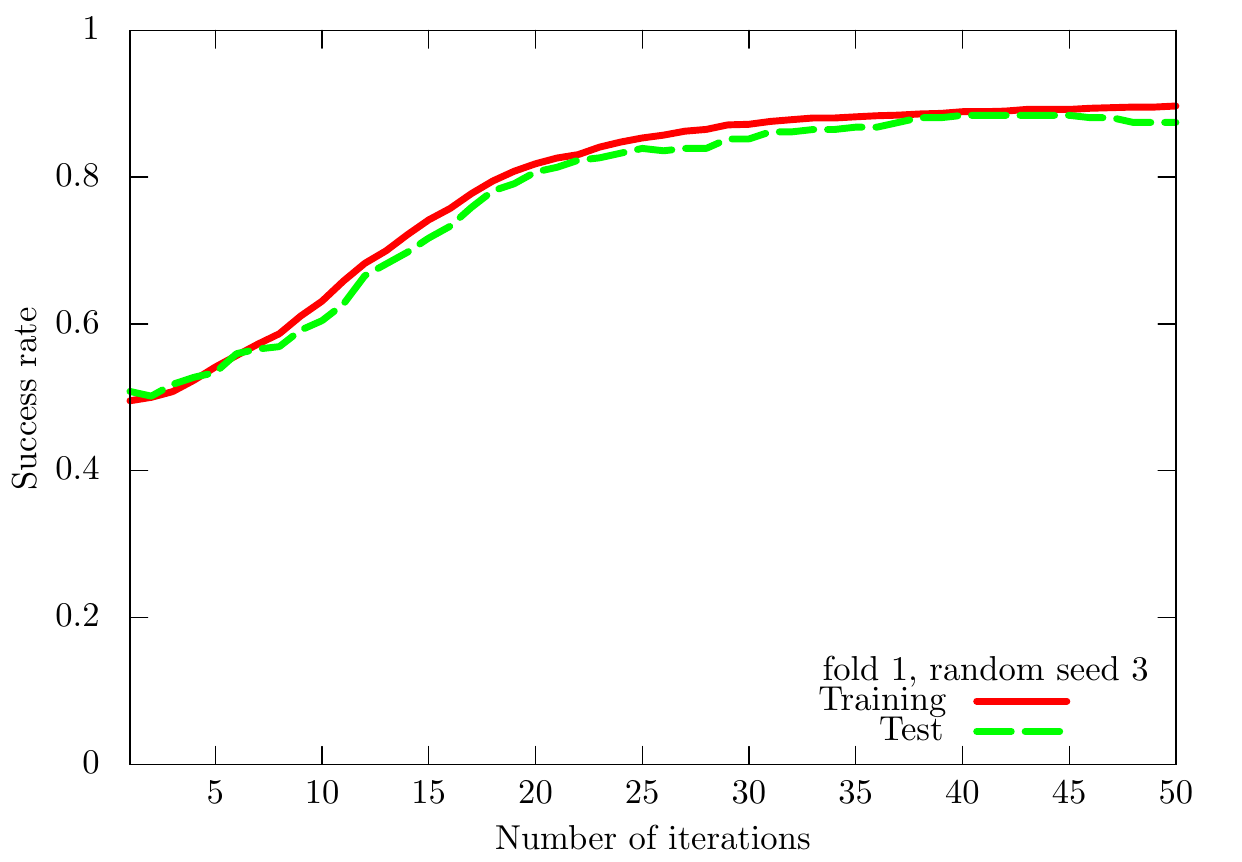}
\includegraphics[scale=0.25]{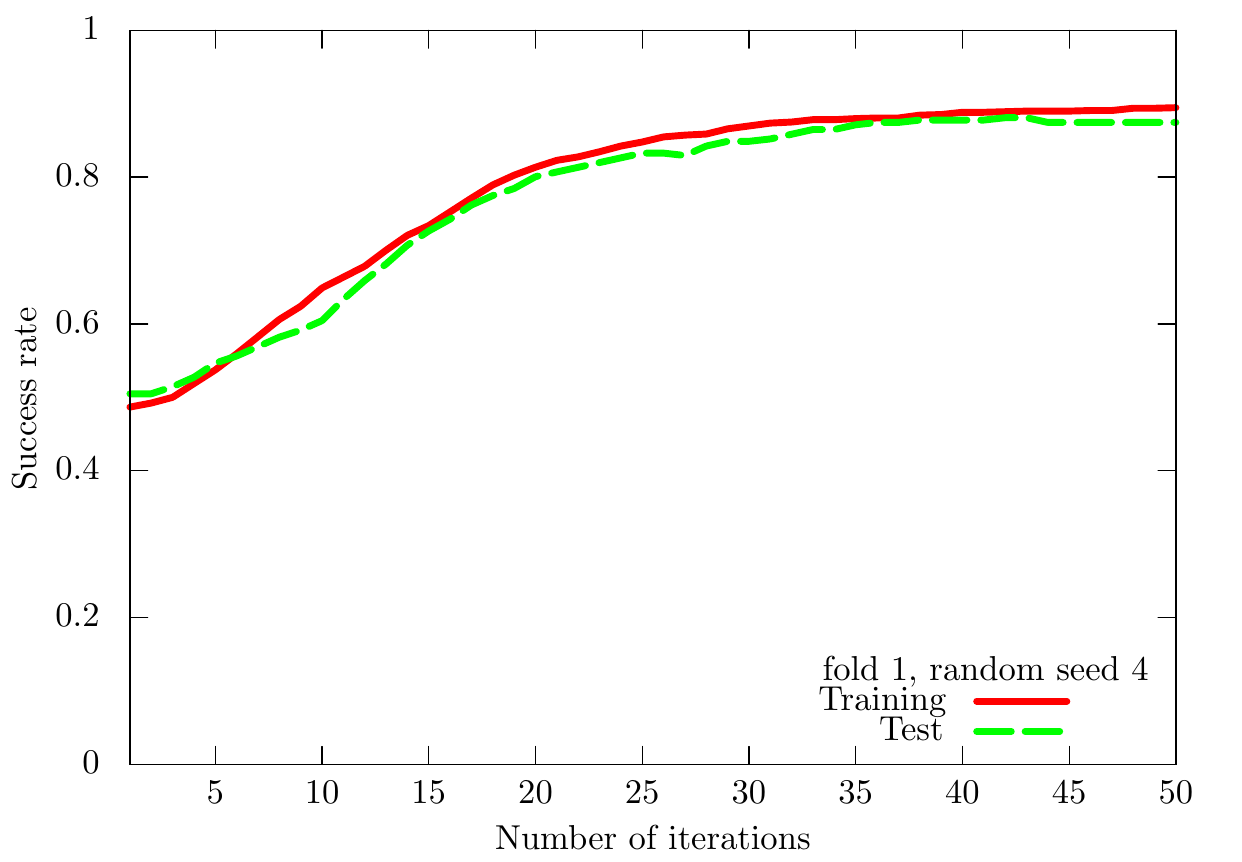}
\includegraphics[scale=0.25]{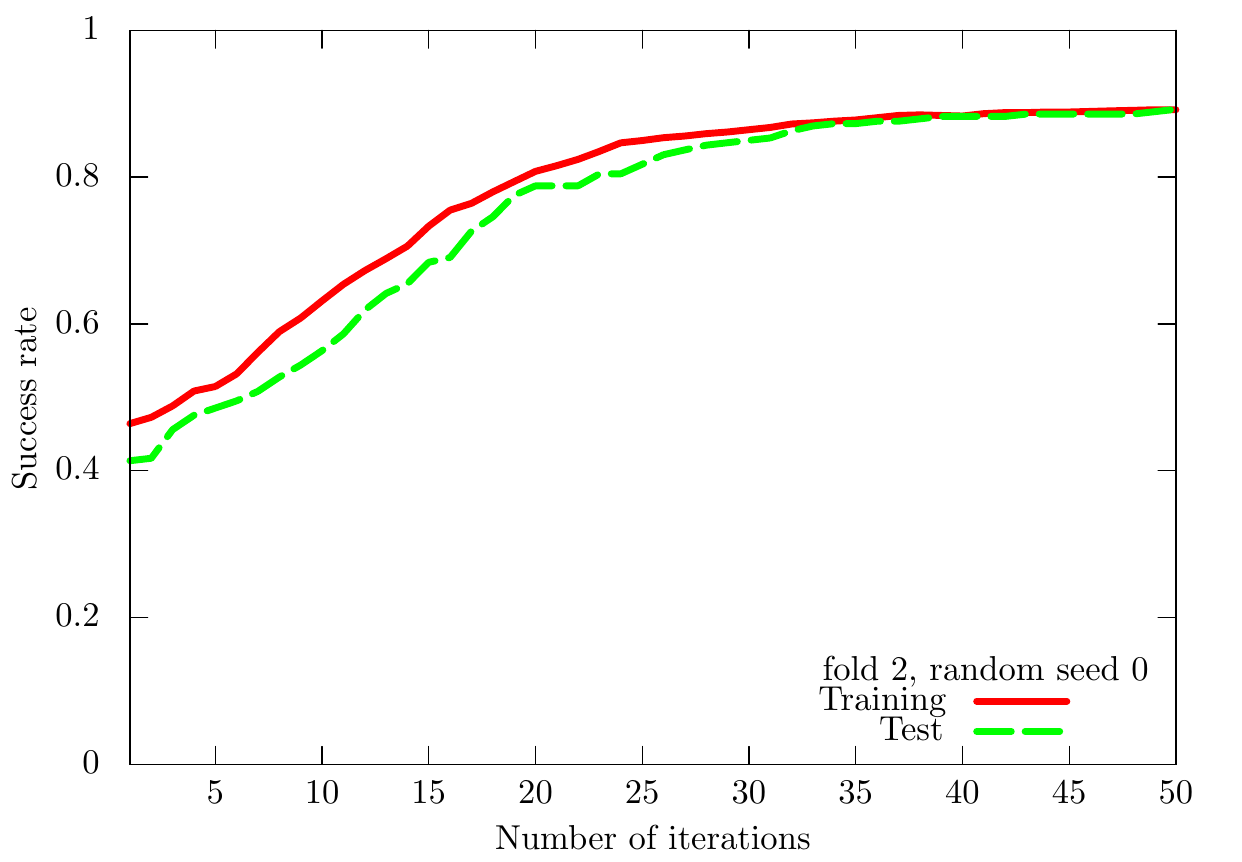}
\includegraphics[scale=0.25]{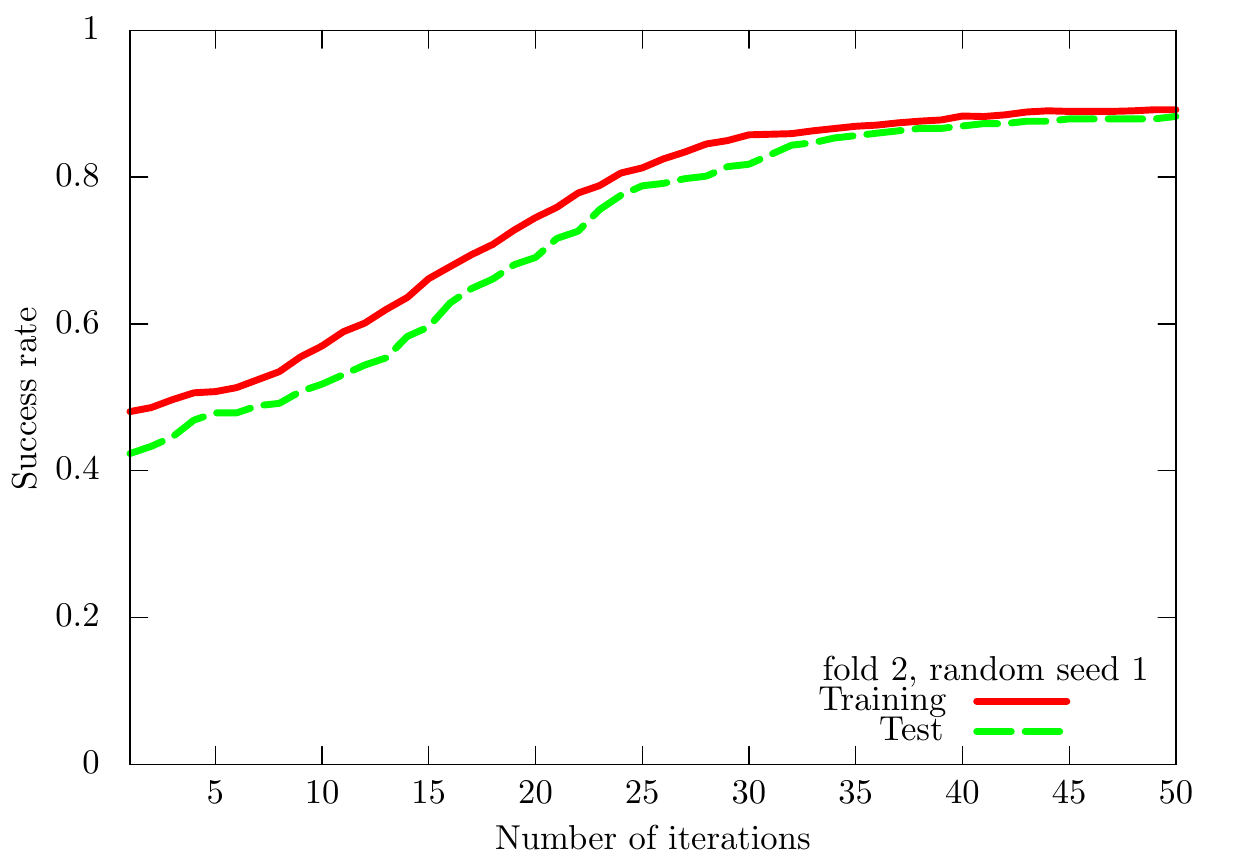}
\includegraphics[scale=0.25]{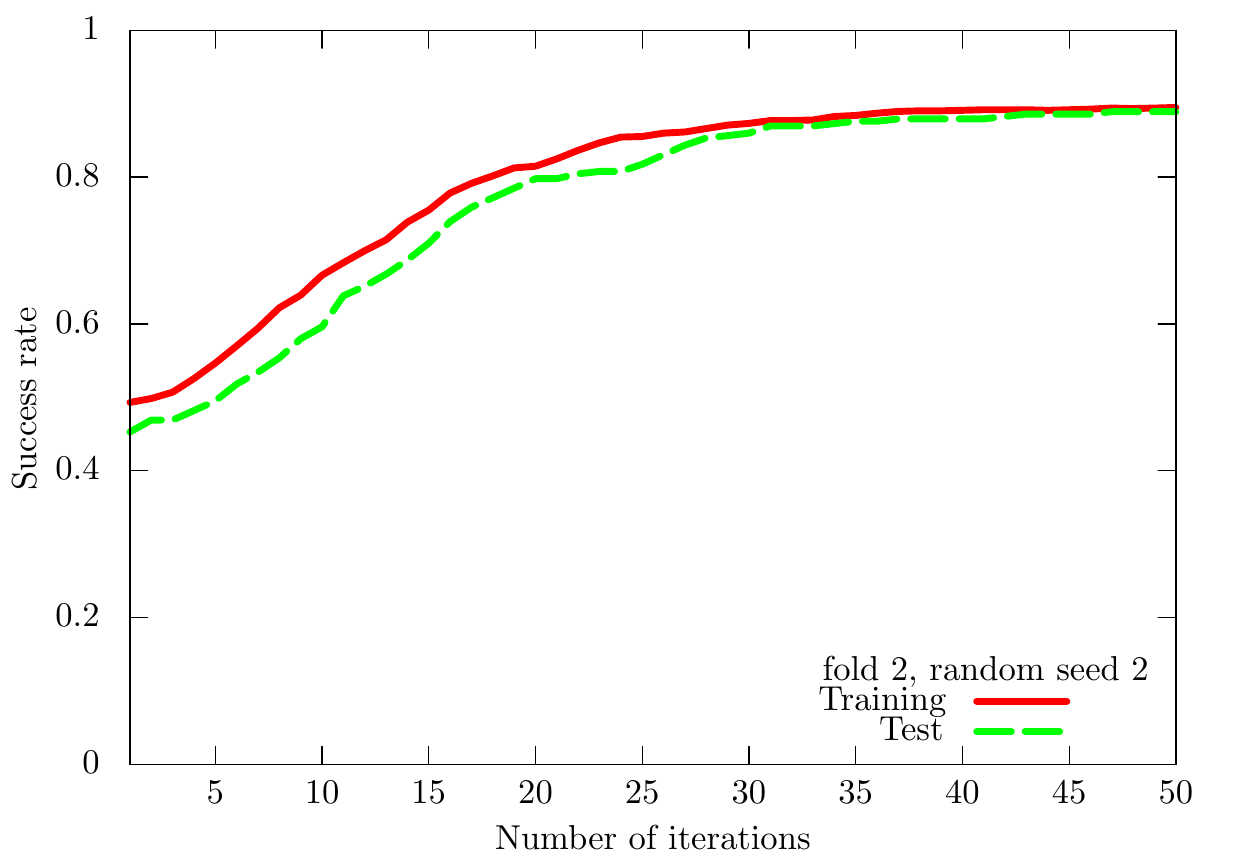}
\includegraphics[scale=0.25]{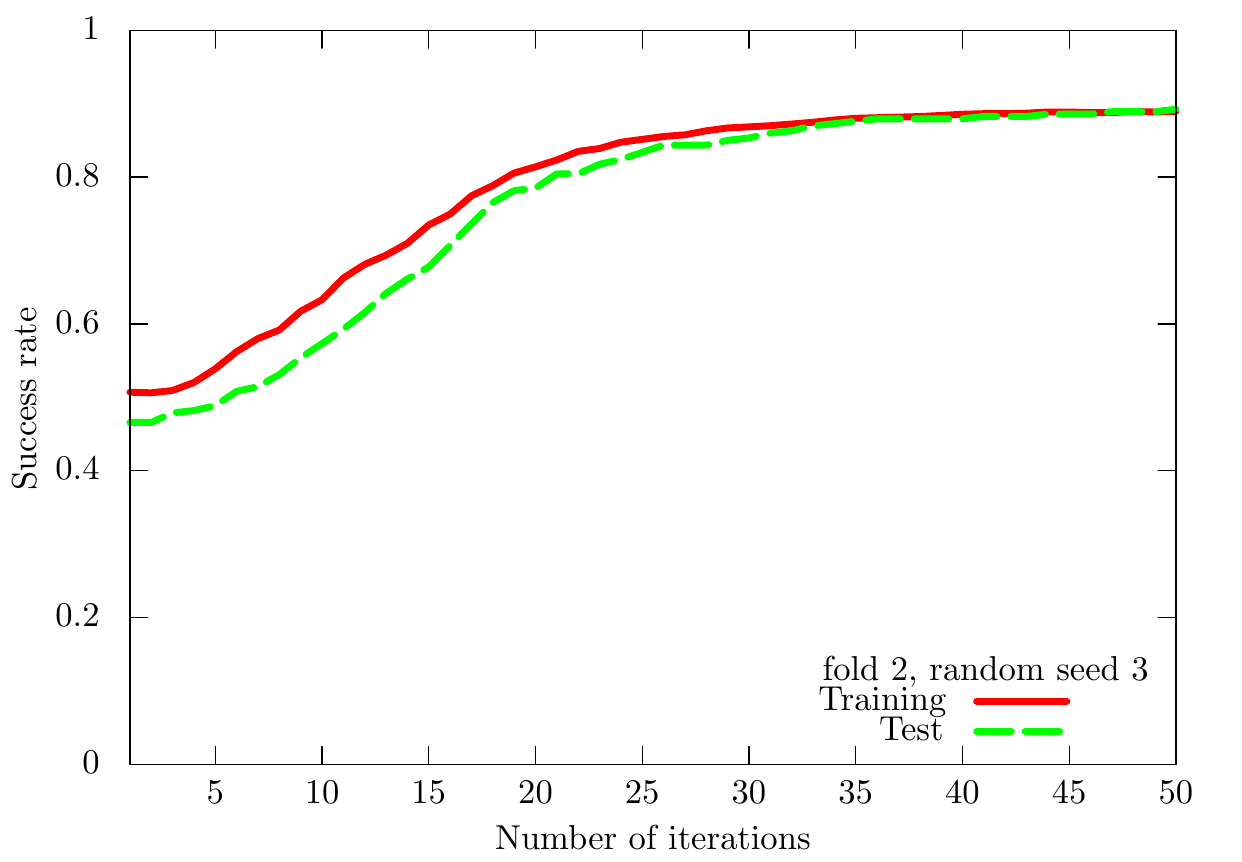}
\includegraphics[scale=0.25]{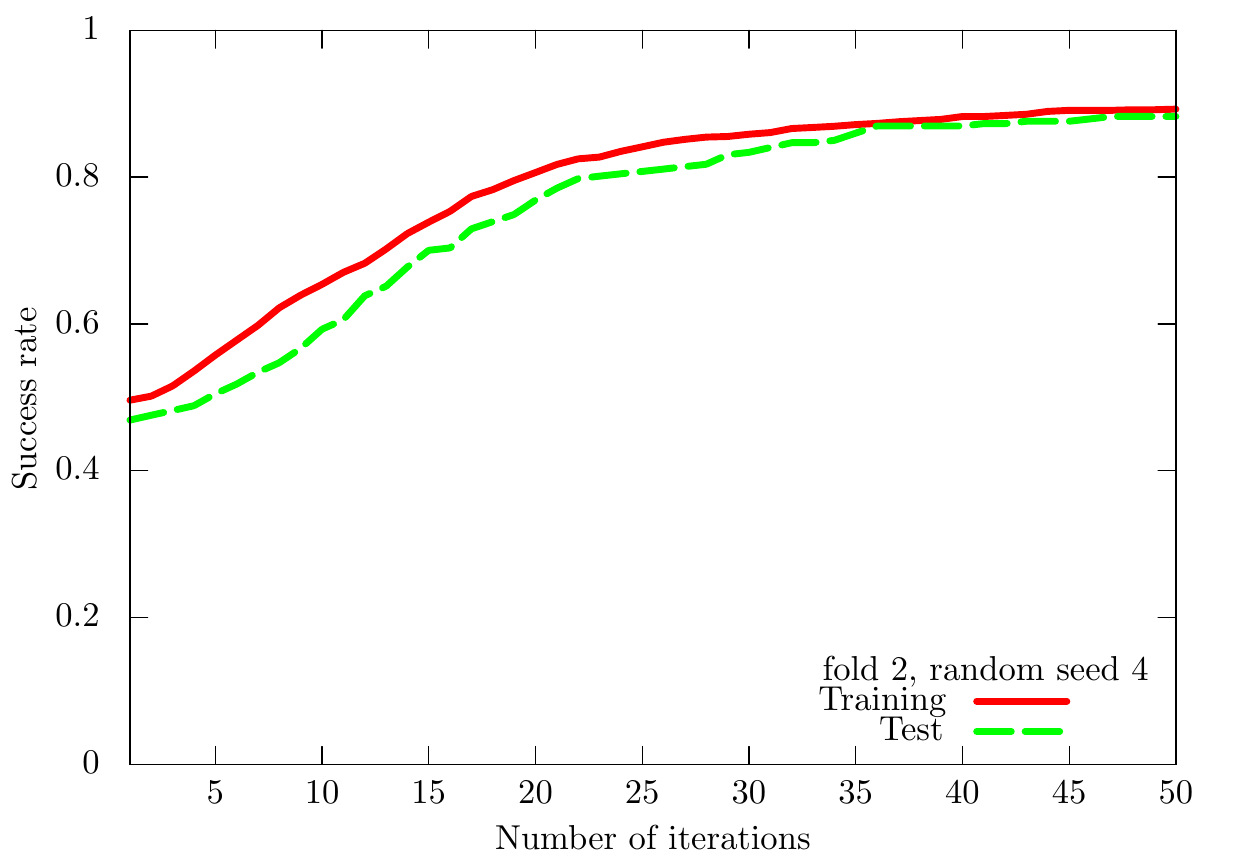}
\includegraphics[scale=0.25]{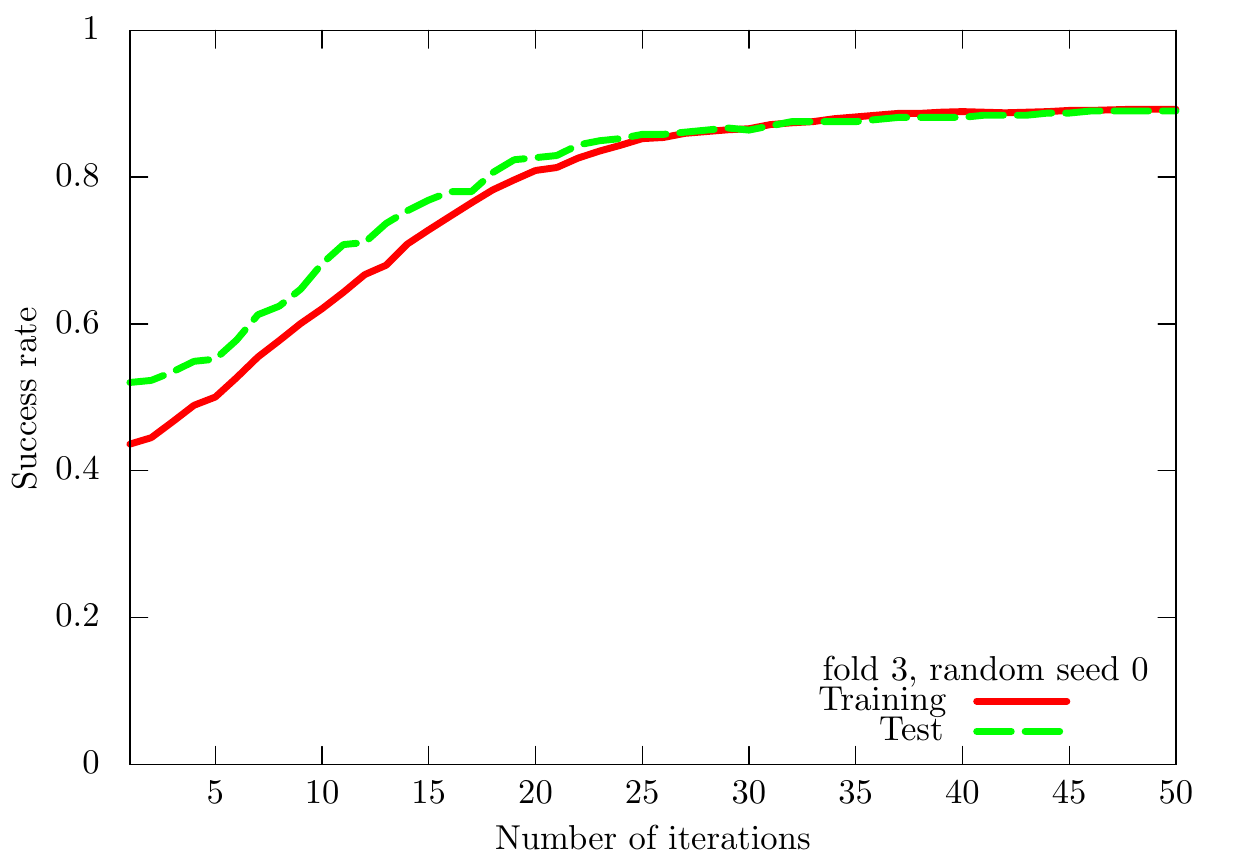}
\includegraphics[scale=0.25]{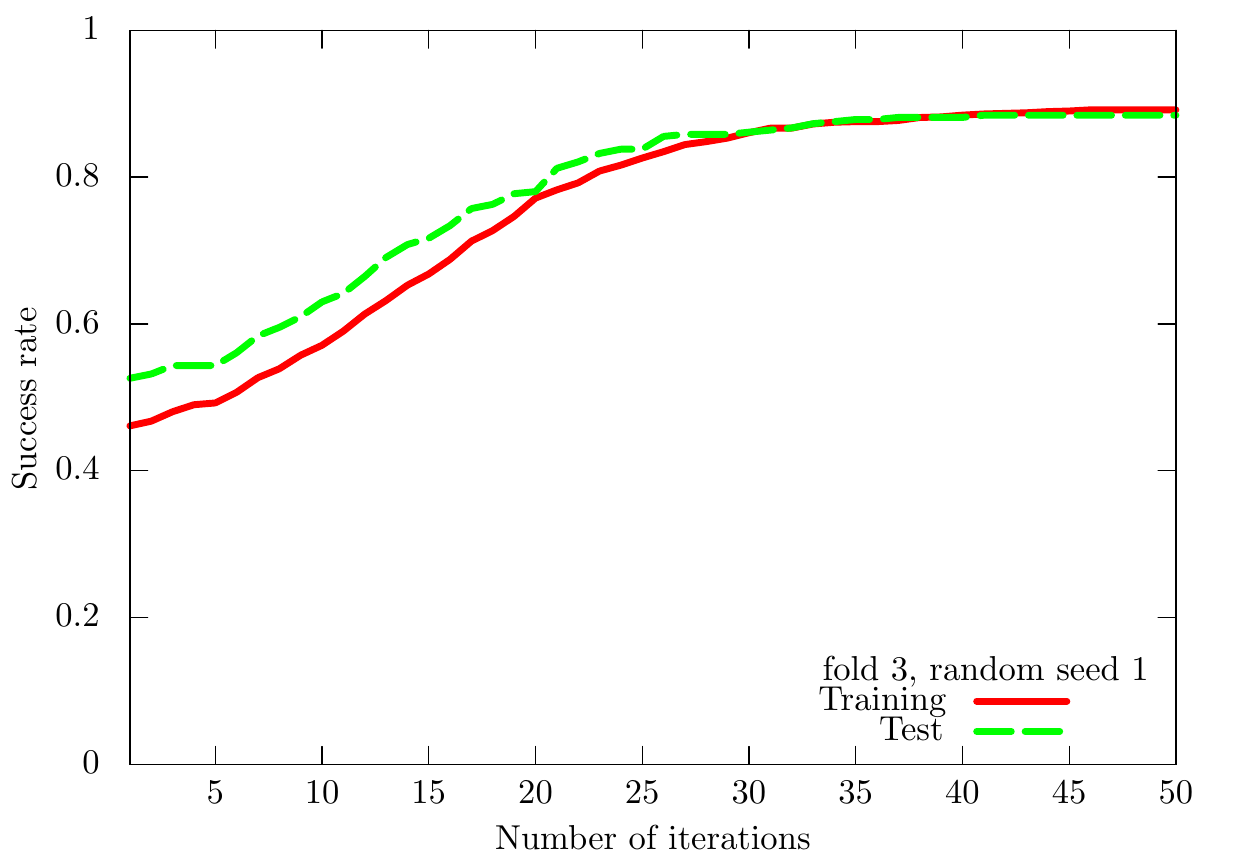}
\includegraphics[scale=0.25]{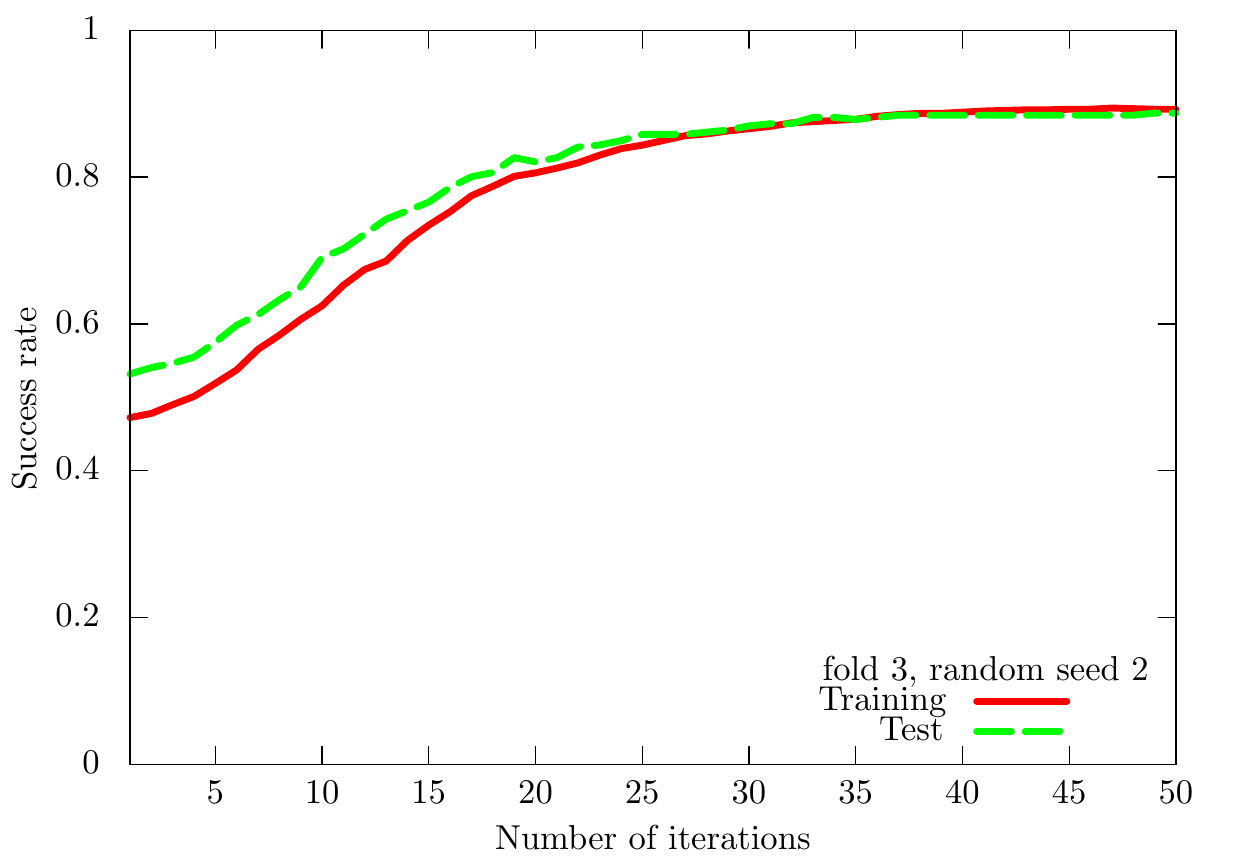}
\includegraphics[scale=0.25]{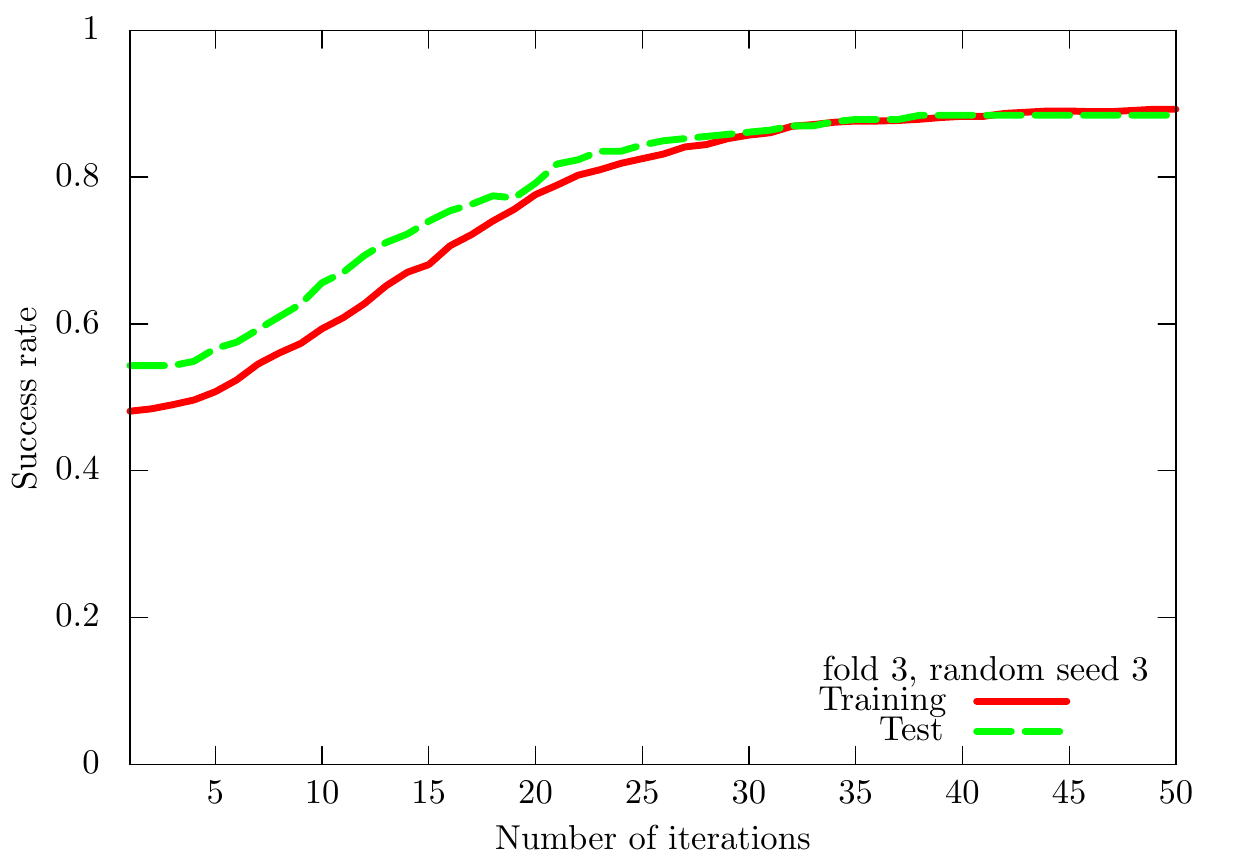}
\includegraphics[scale=0.25]{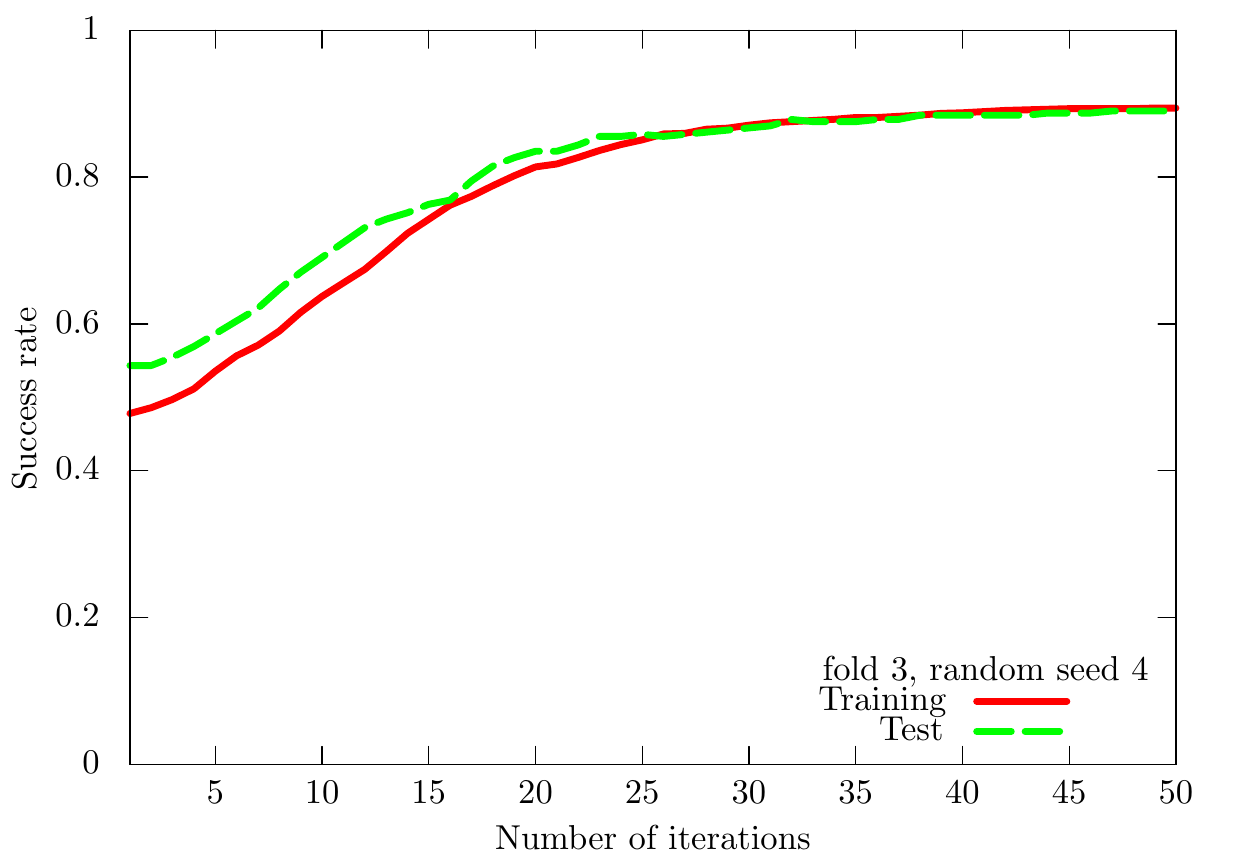}
\includegraphics[scale=0.25]{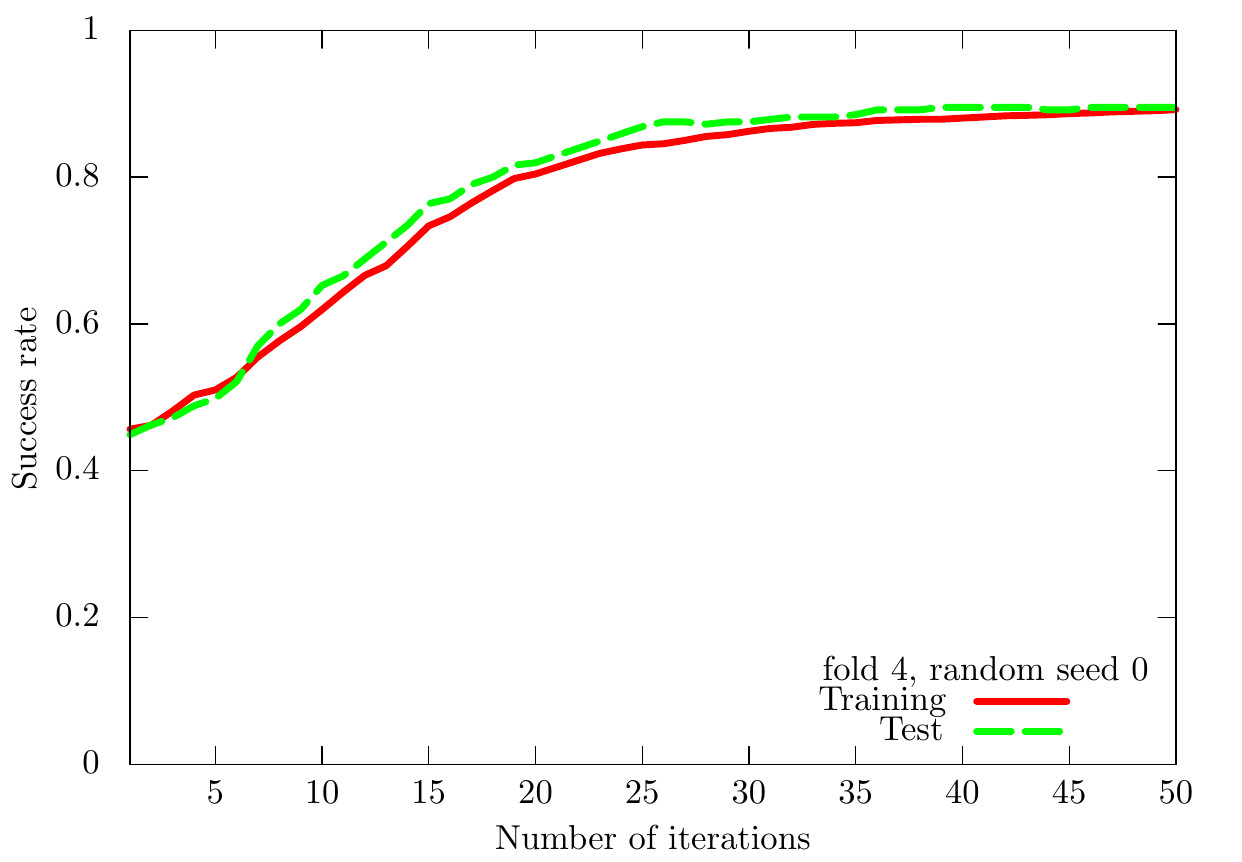}
\includegraphics[scale=0.25]{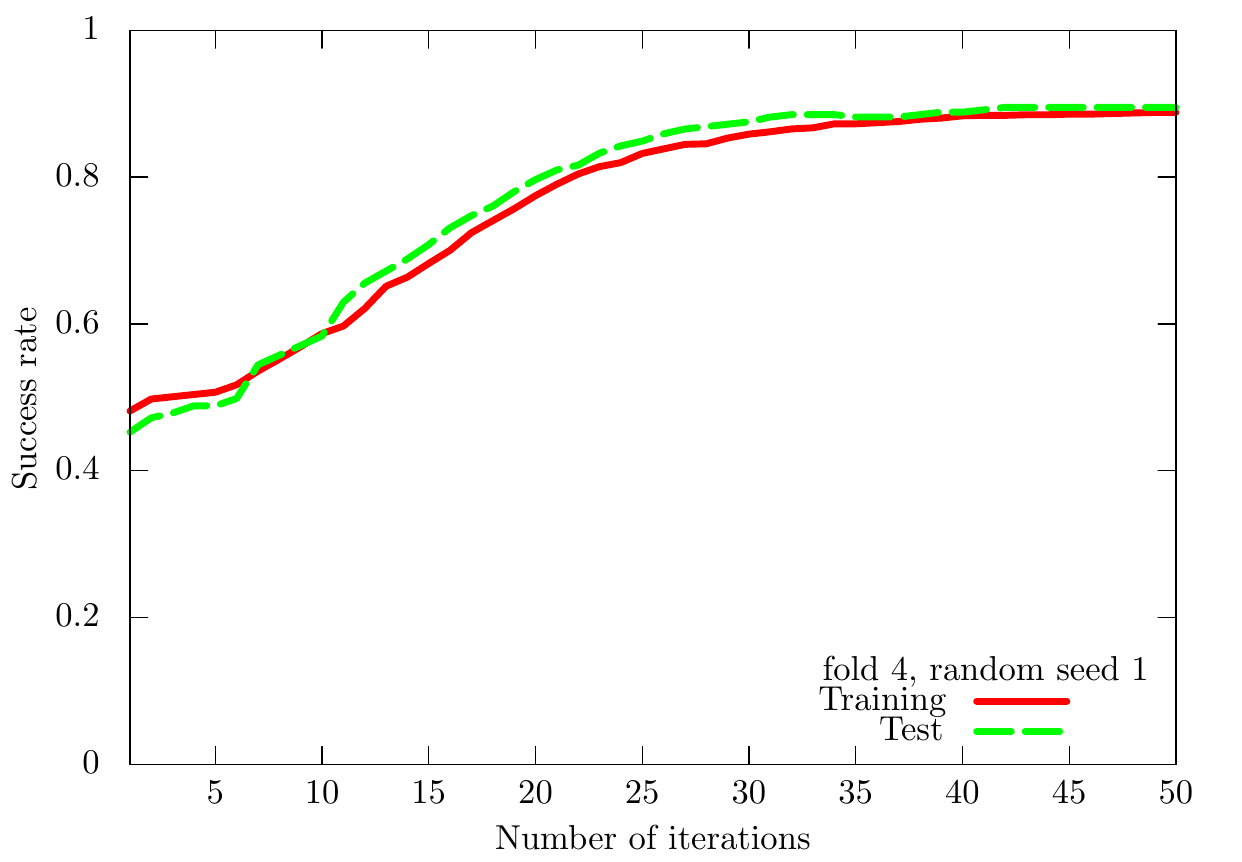}
\includegraphics[scale=0.25]{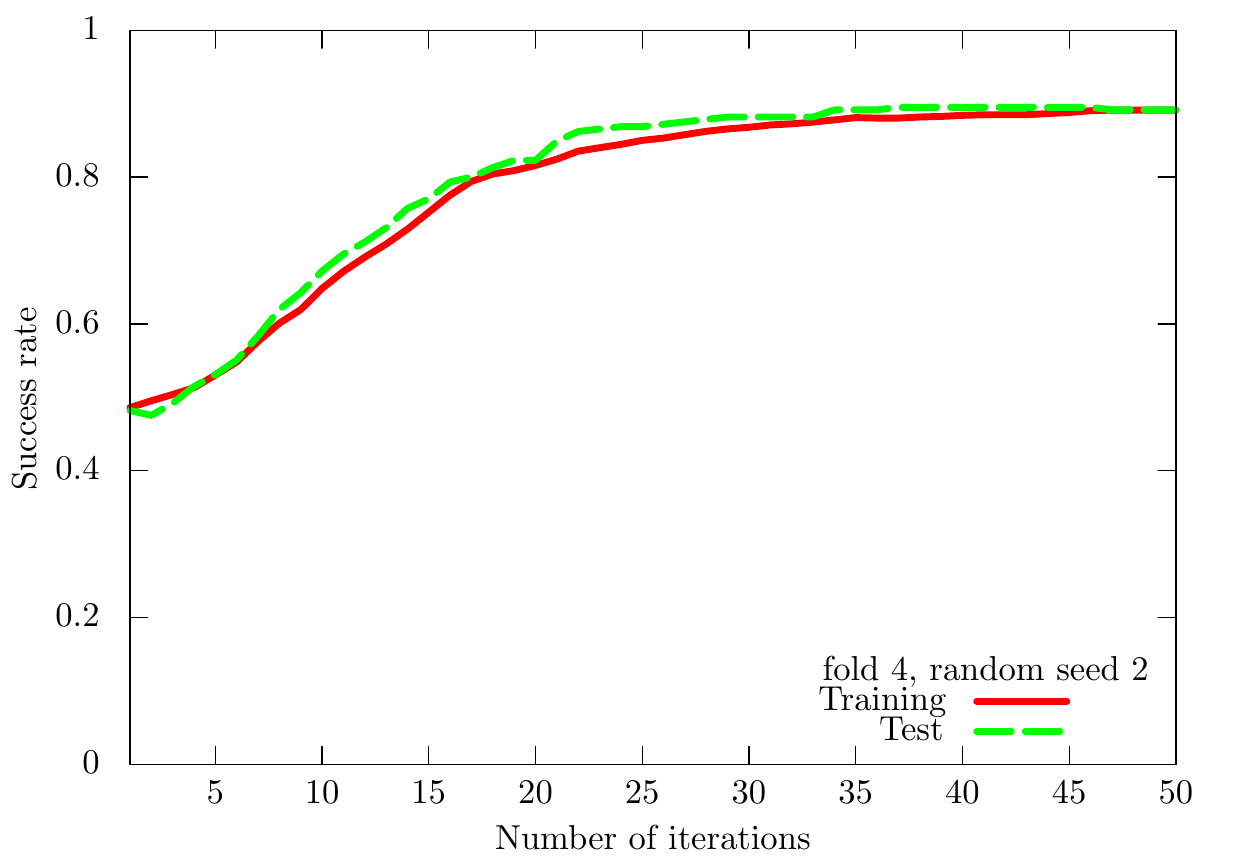}
\includegraphics[scale=0.25]{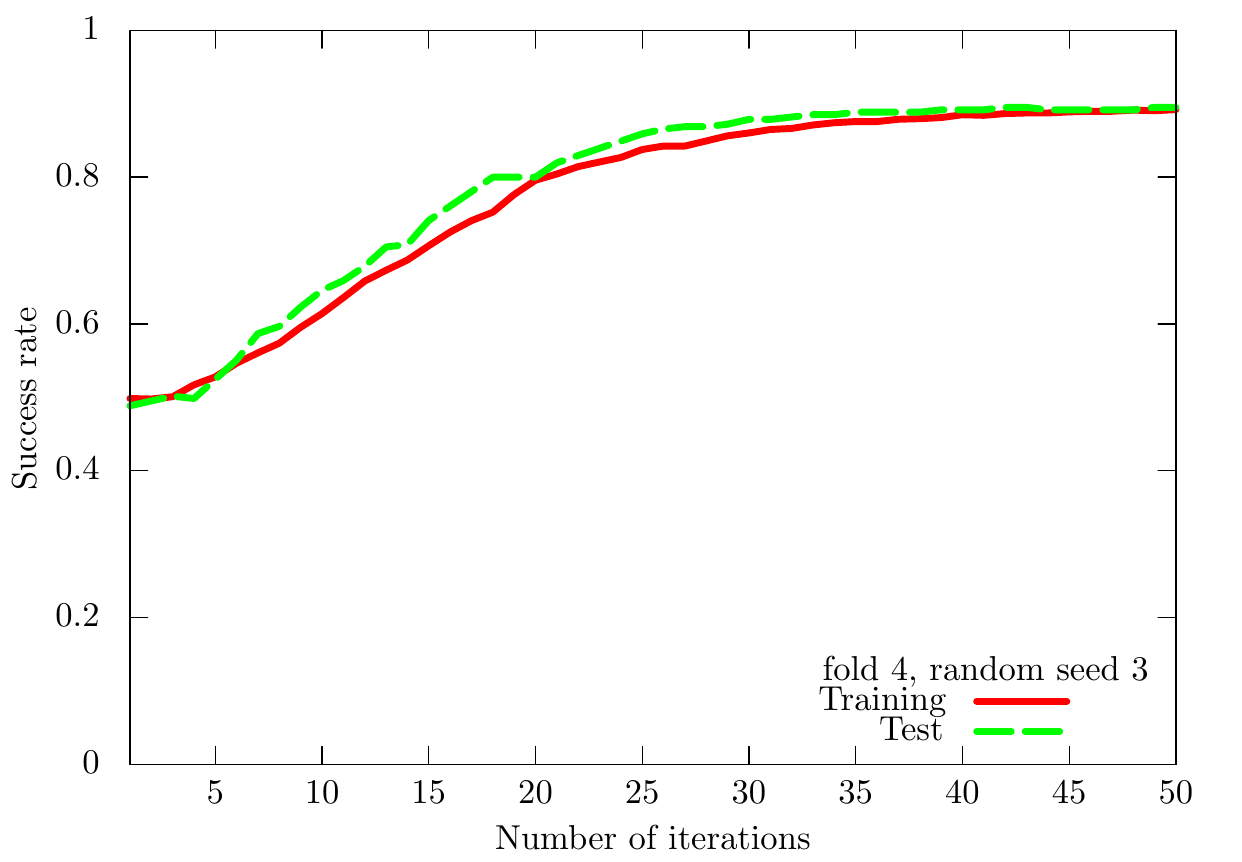}
\includegraphics[scale=0.25]{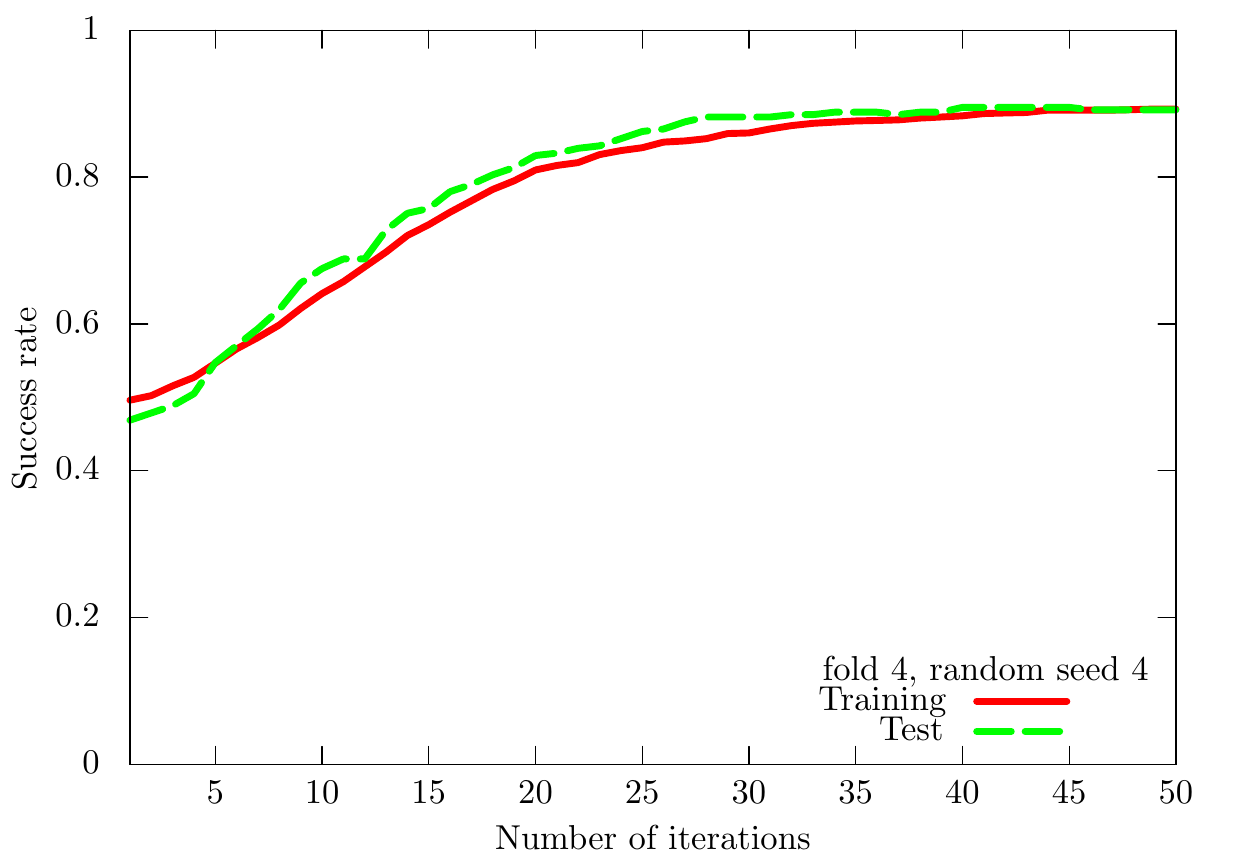}
\caption{Results of QCL on the $5$-fold datasets with $5$ different random seeds for the semeion dataset ($0$ or non-$0$). We use the CNOT-based circuit and set $\theta_\mathrm{bias} = 0$. The number of layers $L$ is set to $5$.}
\label{supp-arXiv-numerical-result-raw-data-fold-001-rand-001-QCL-UCI-semeion-0-non0}
\end{figure*}

\begin{figure*}[htb]
\centering
\includegraphics[scale=0.25]{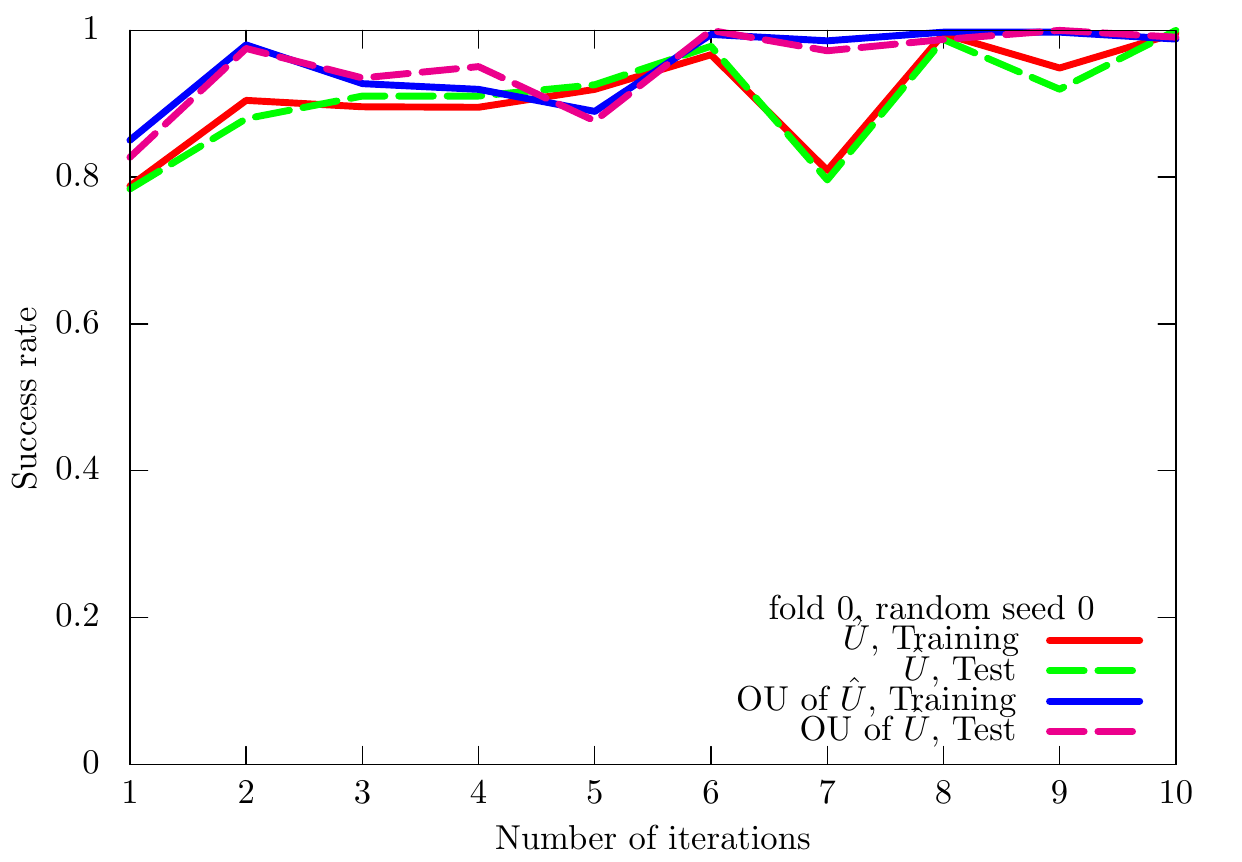}
\includegraphics[scale=0.25]{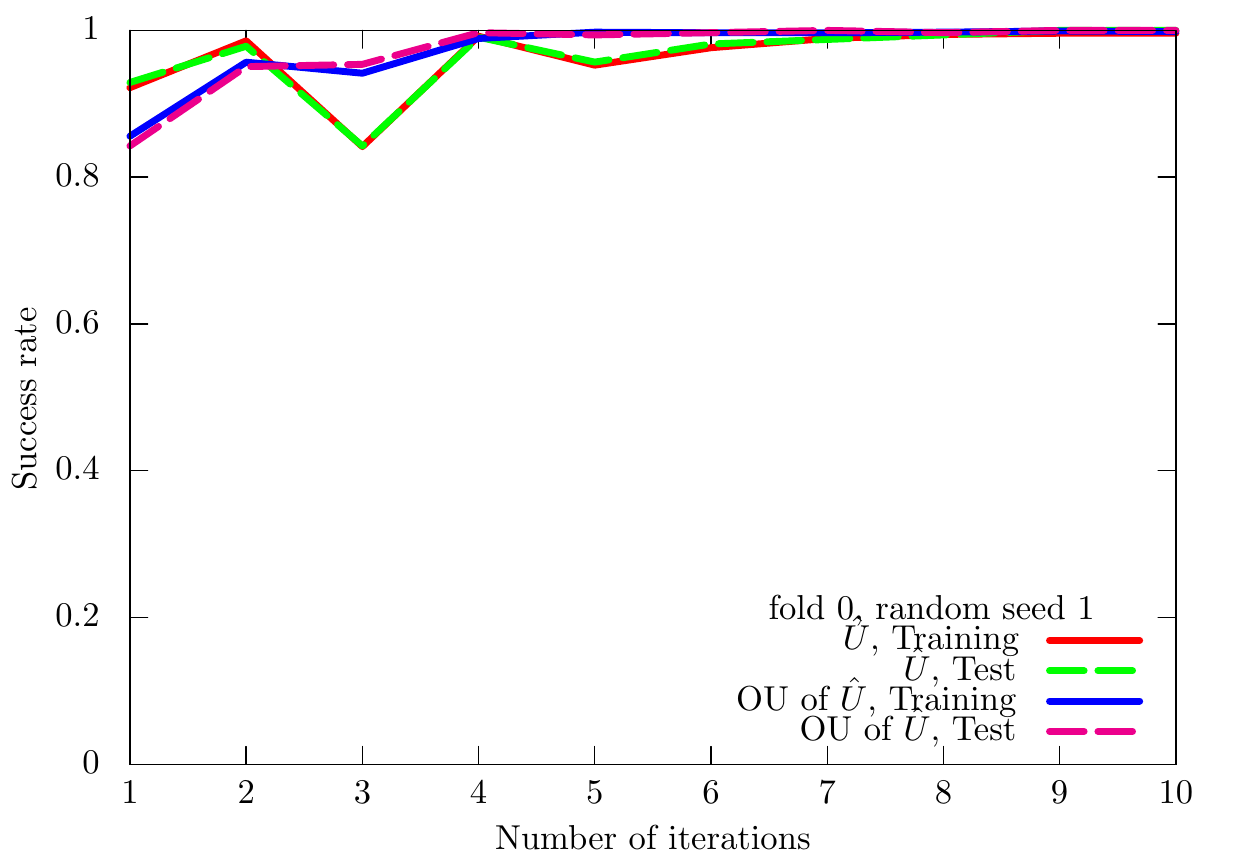}
\includegraphics[scale=0.25]{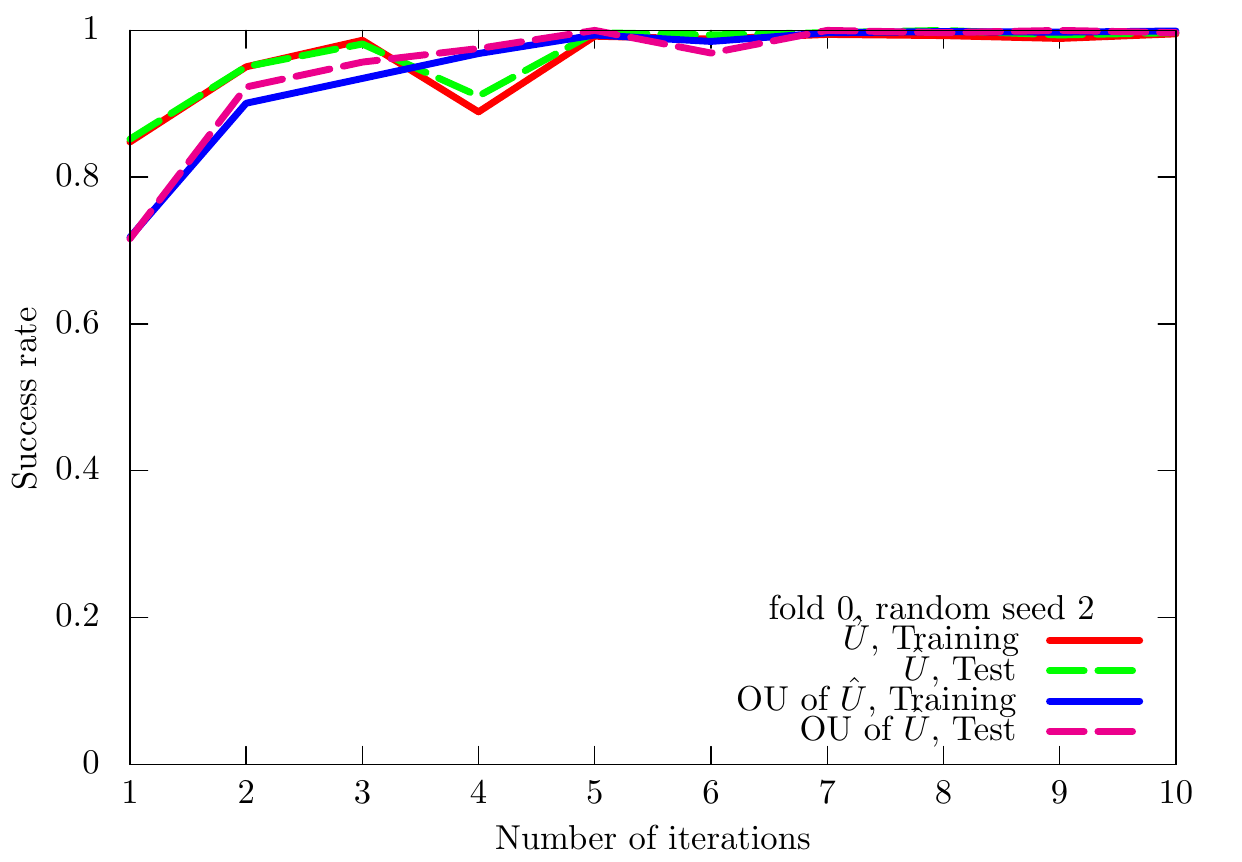}
\includegraphics[scale=0.25]{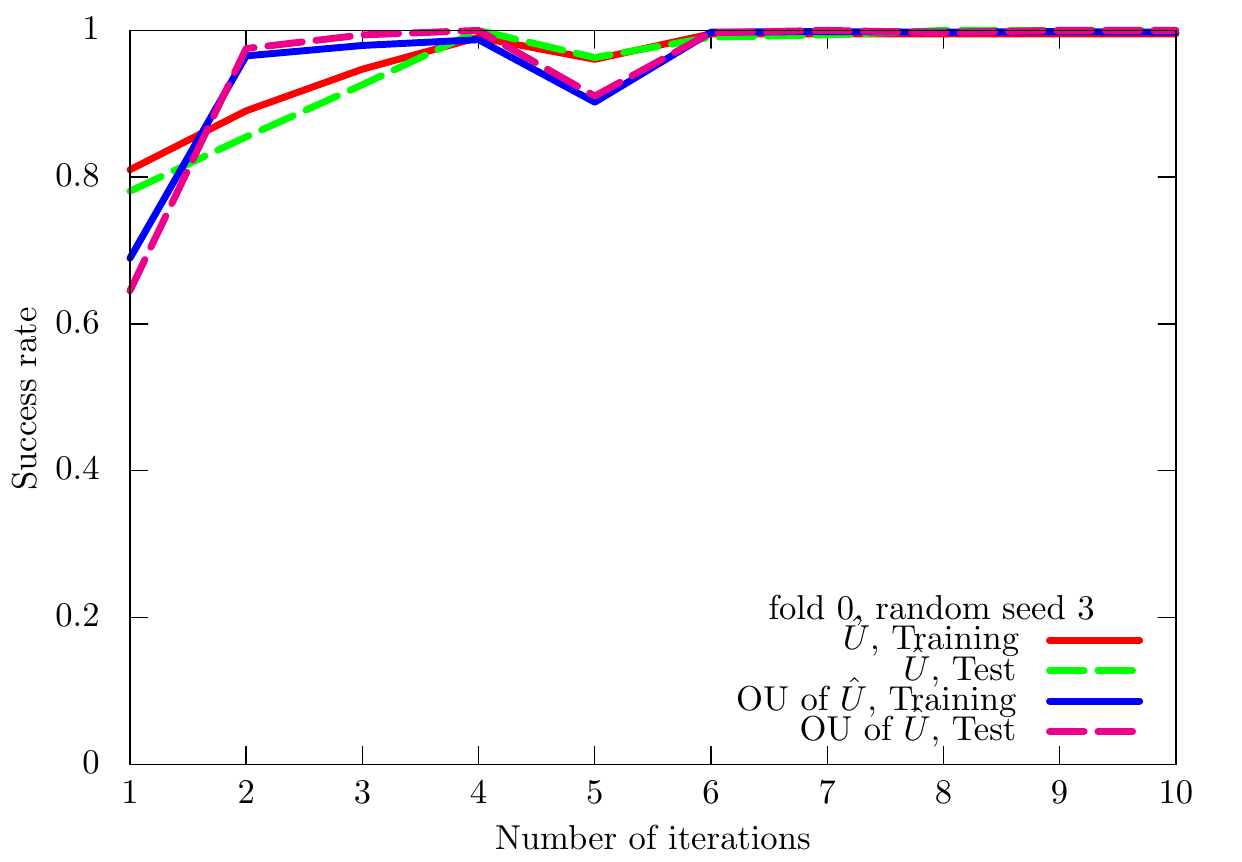}
\includegraphics[scale=0.25]{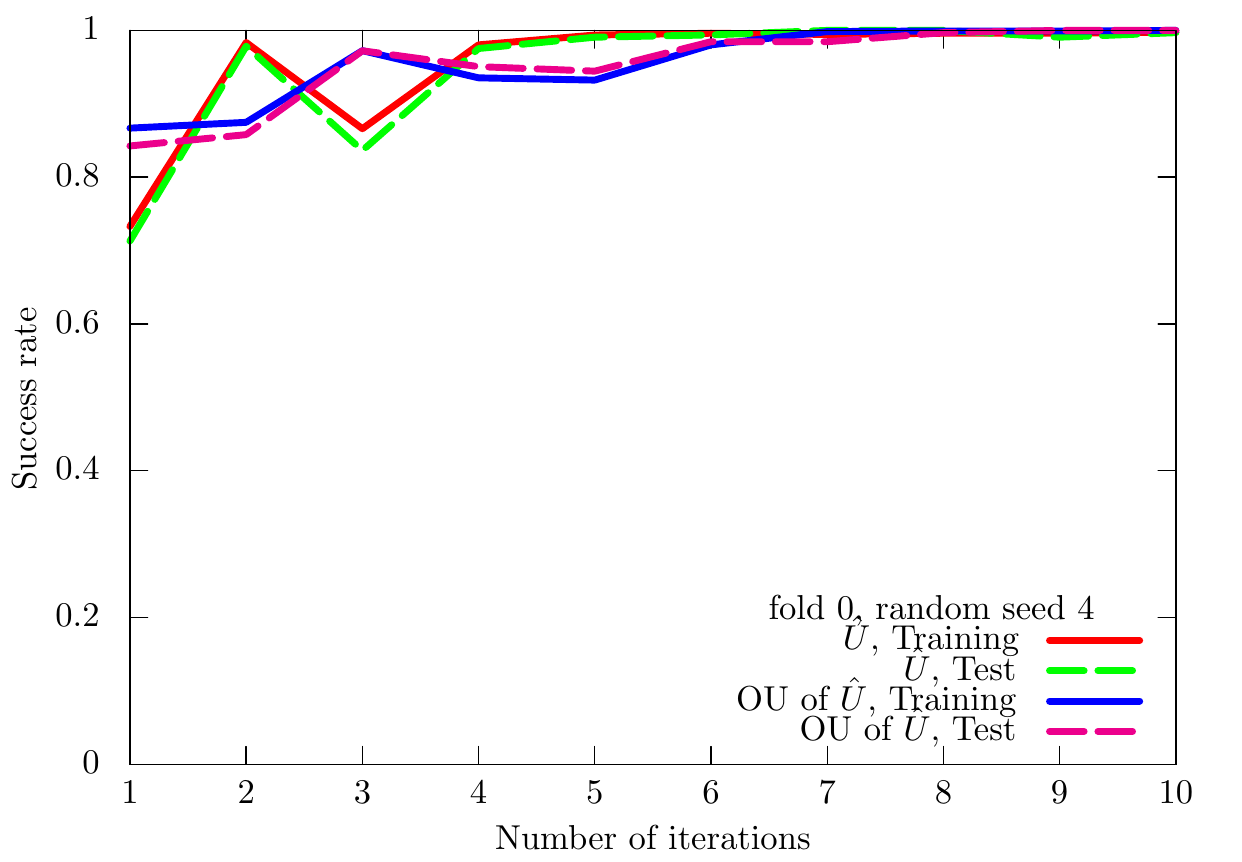}
\includegraphics[scale=0.25]{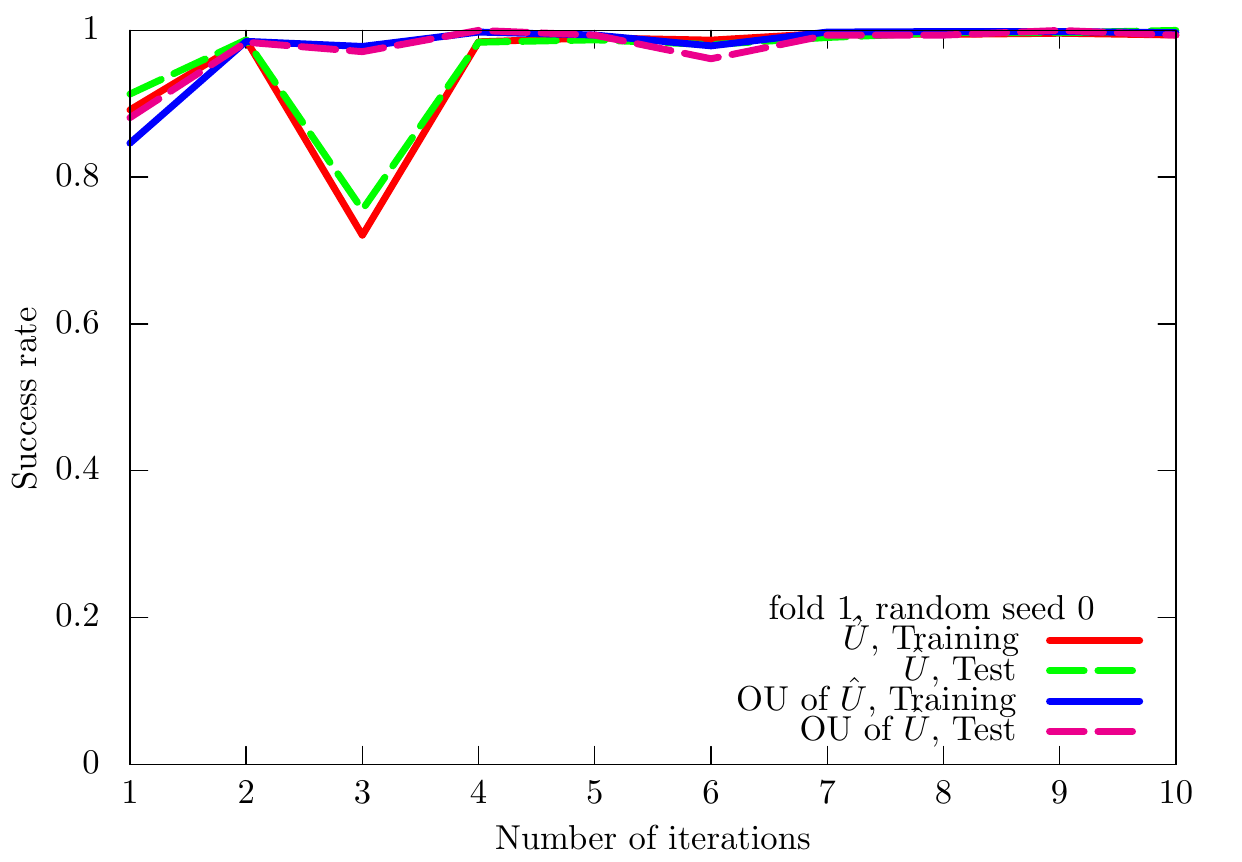}
\includegraphics[scale=0.25]{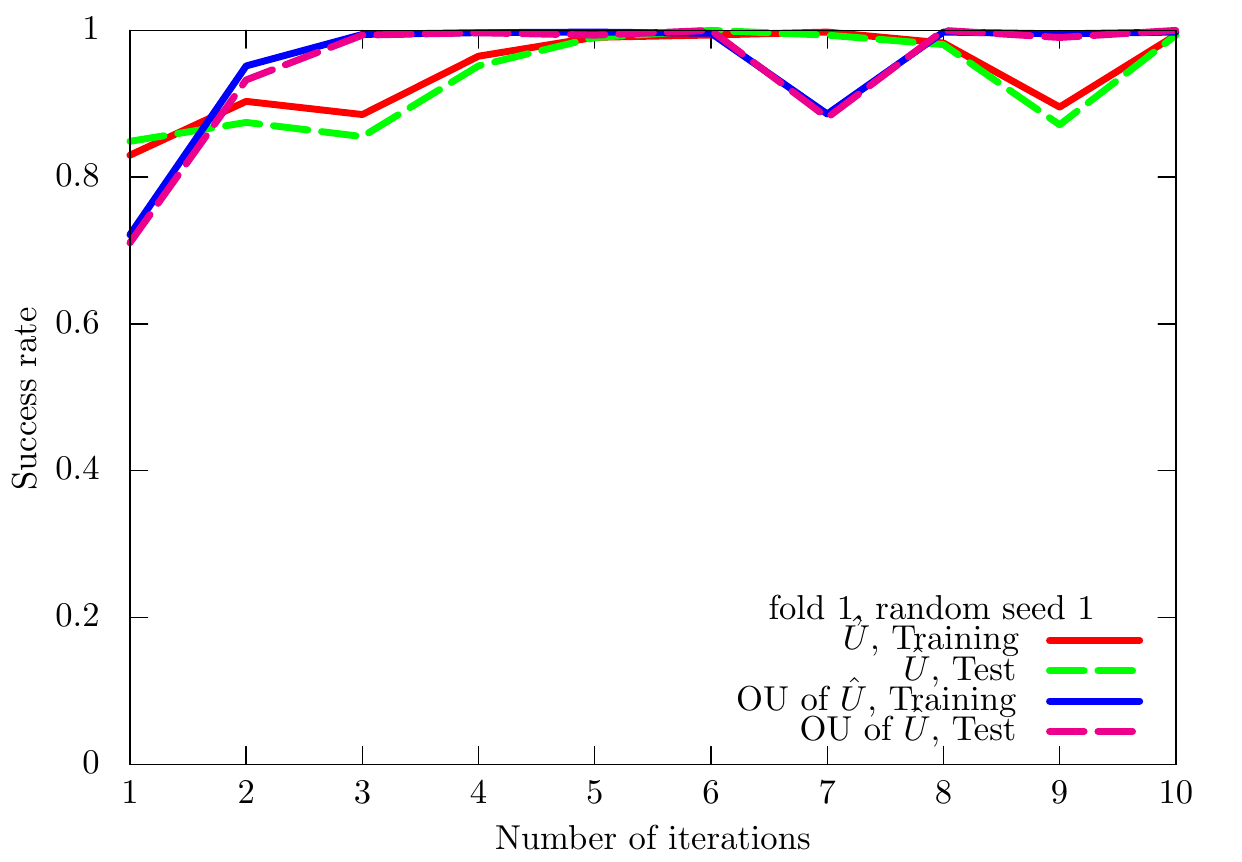}
\includegraphics[scale=0.25]{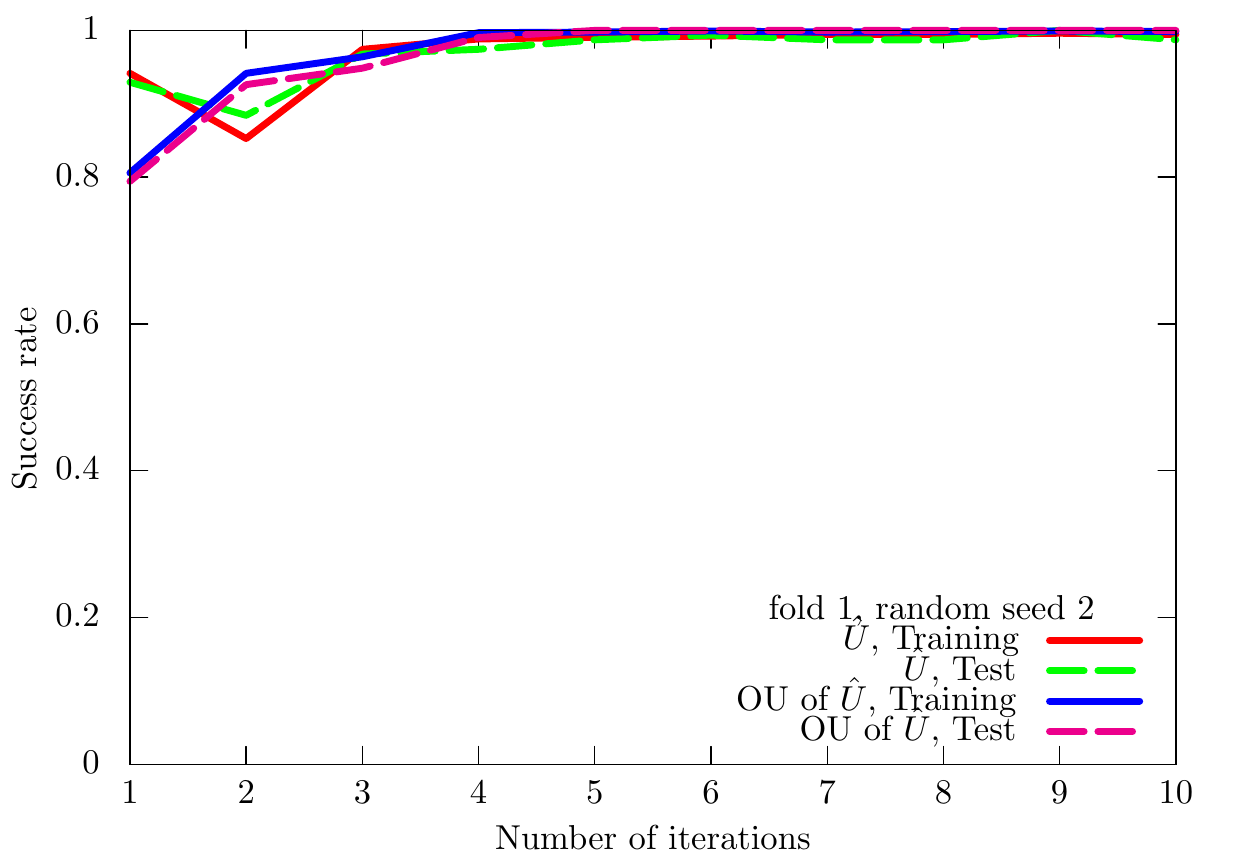}
\includegraphics[scale=0.25]{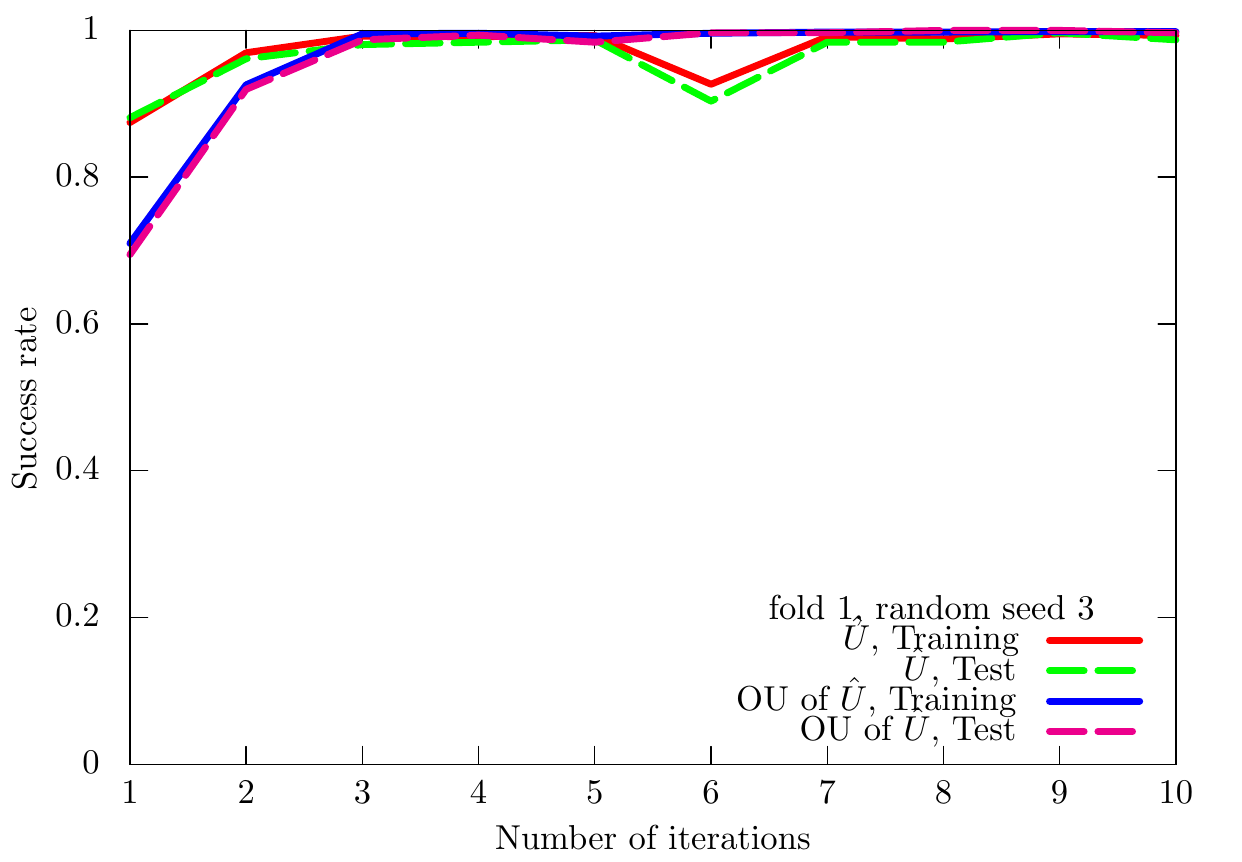}
\includegraphics[scale=0.25]{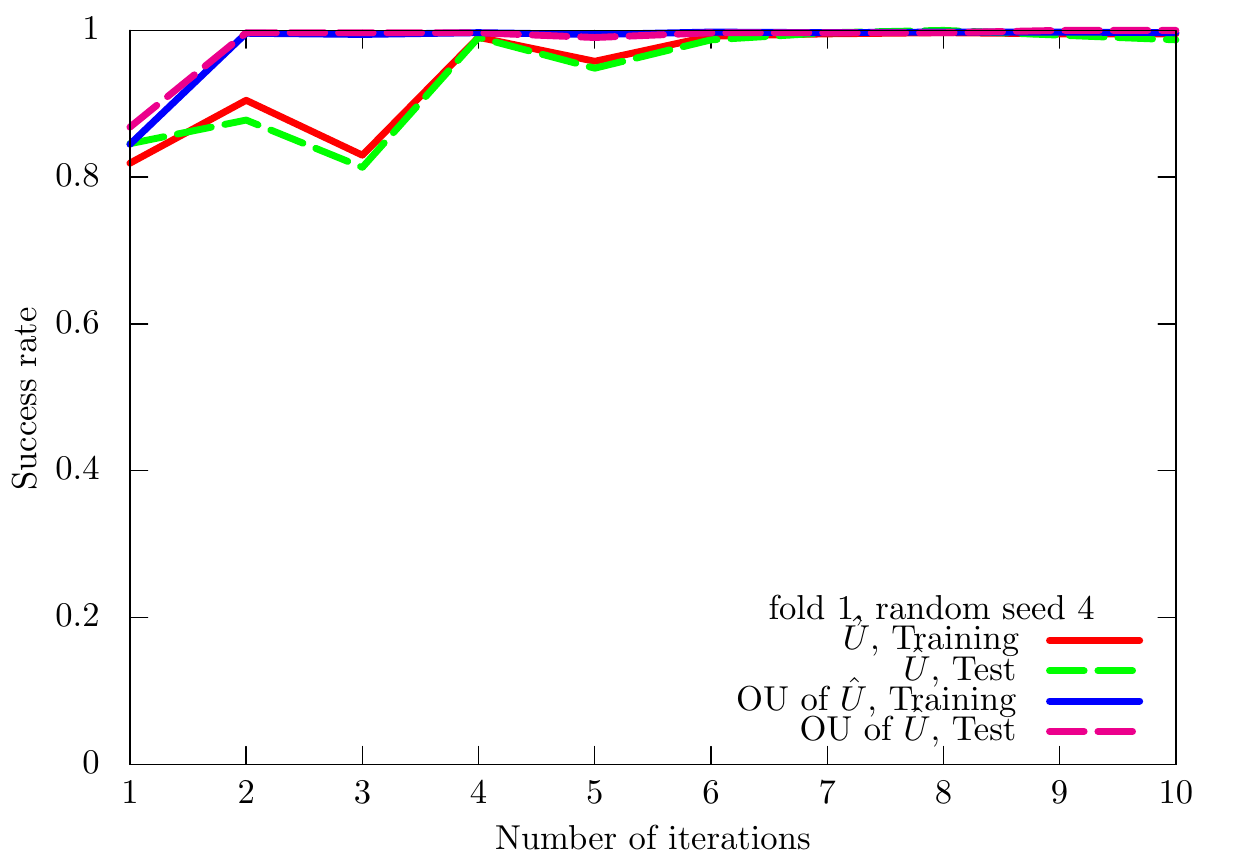}
\includegraphics[scale=0.25]{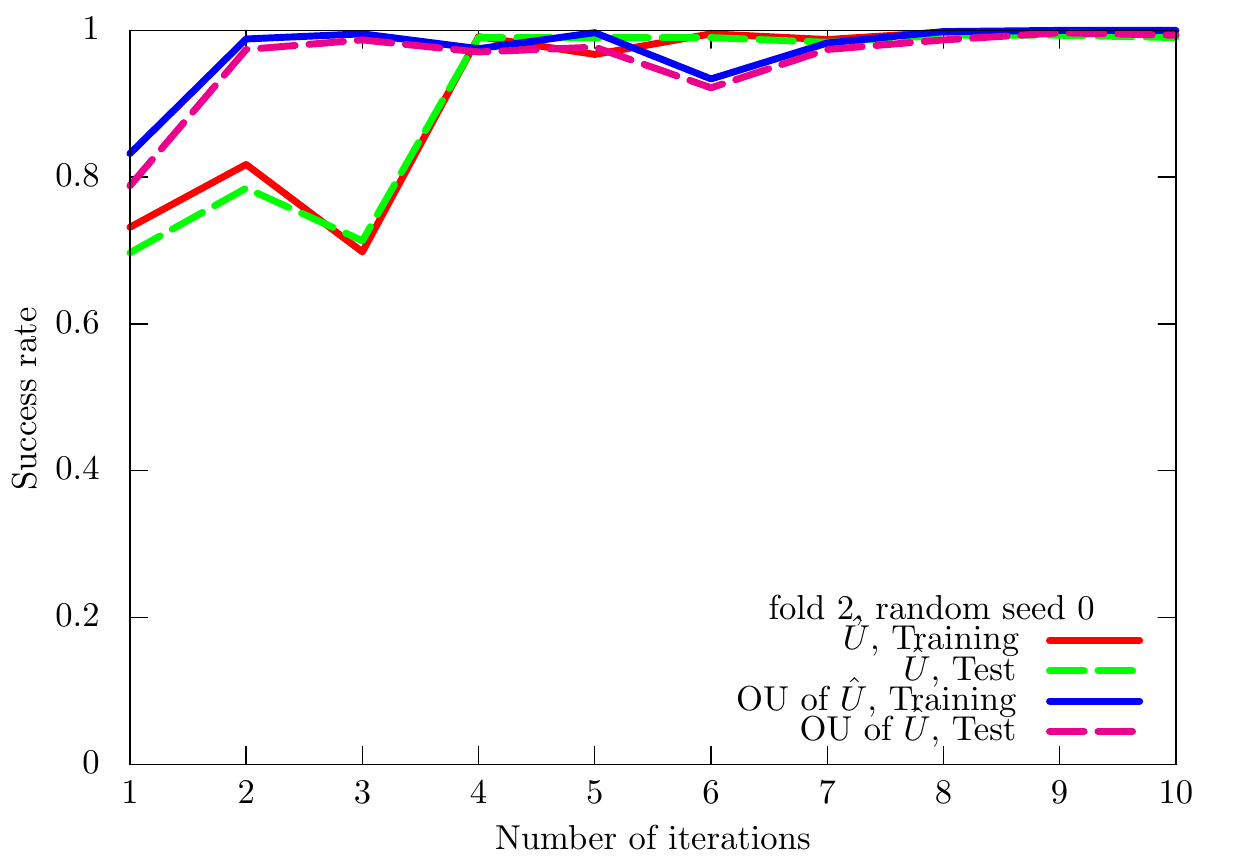}
\includegraphics[scale=0.25]{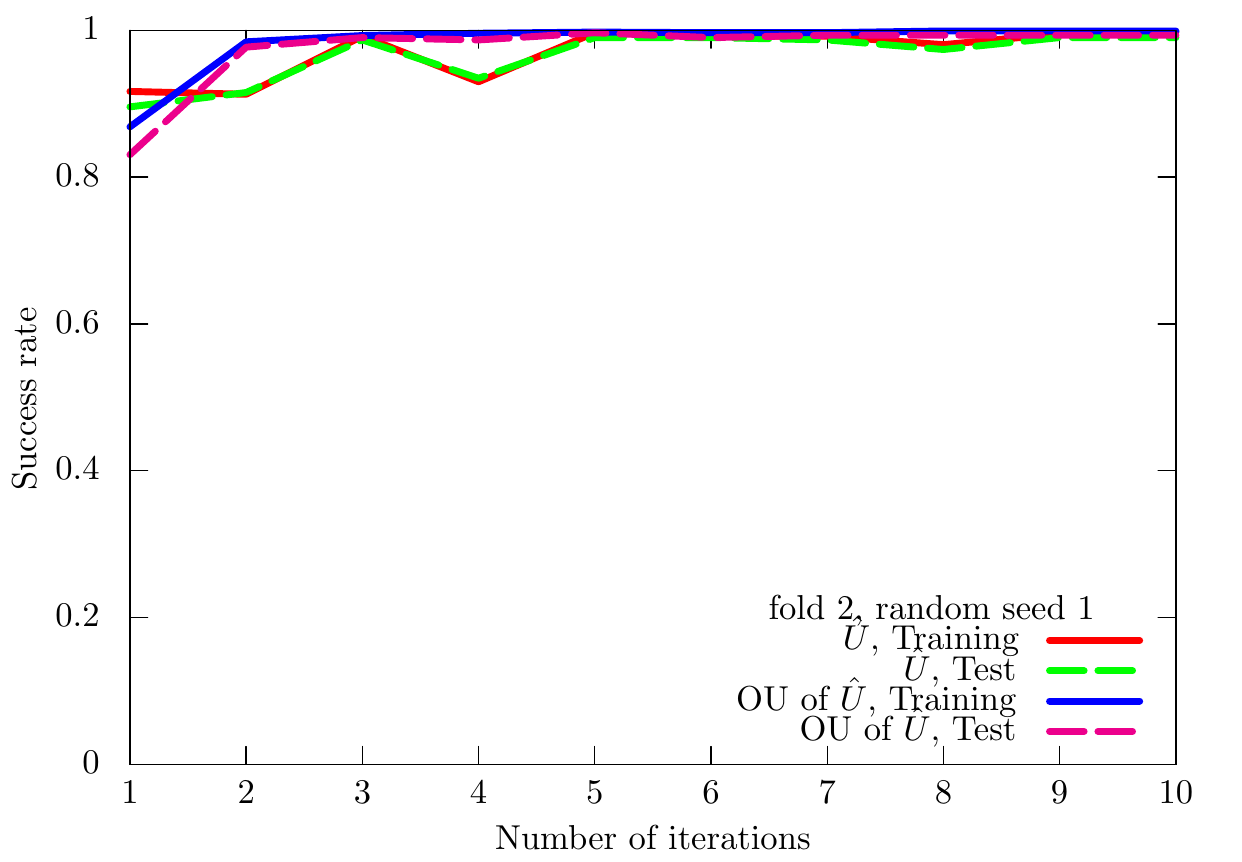}
\includegraphics[scale=0.25]{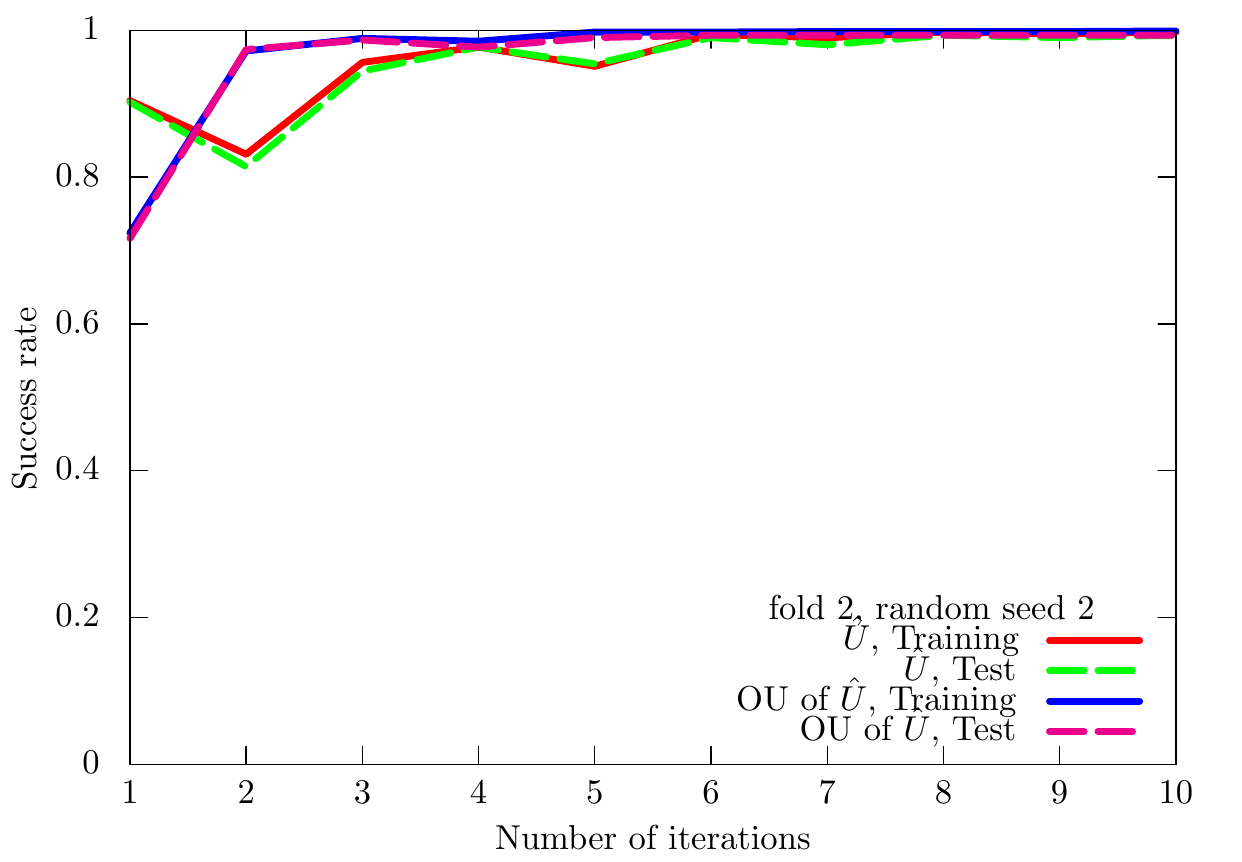}
\includegraphics[scale=0.25]{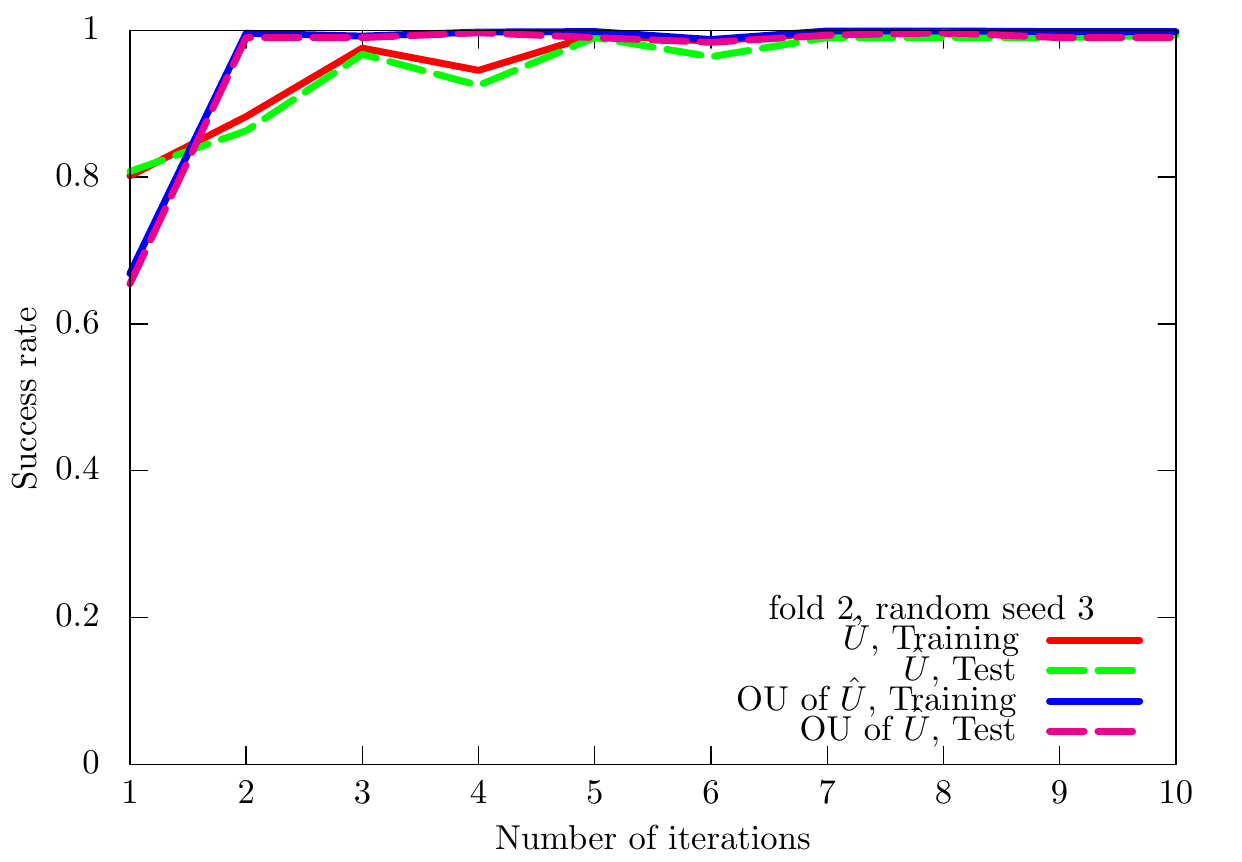}
\includegraphics[scale=0.25]{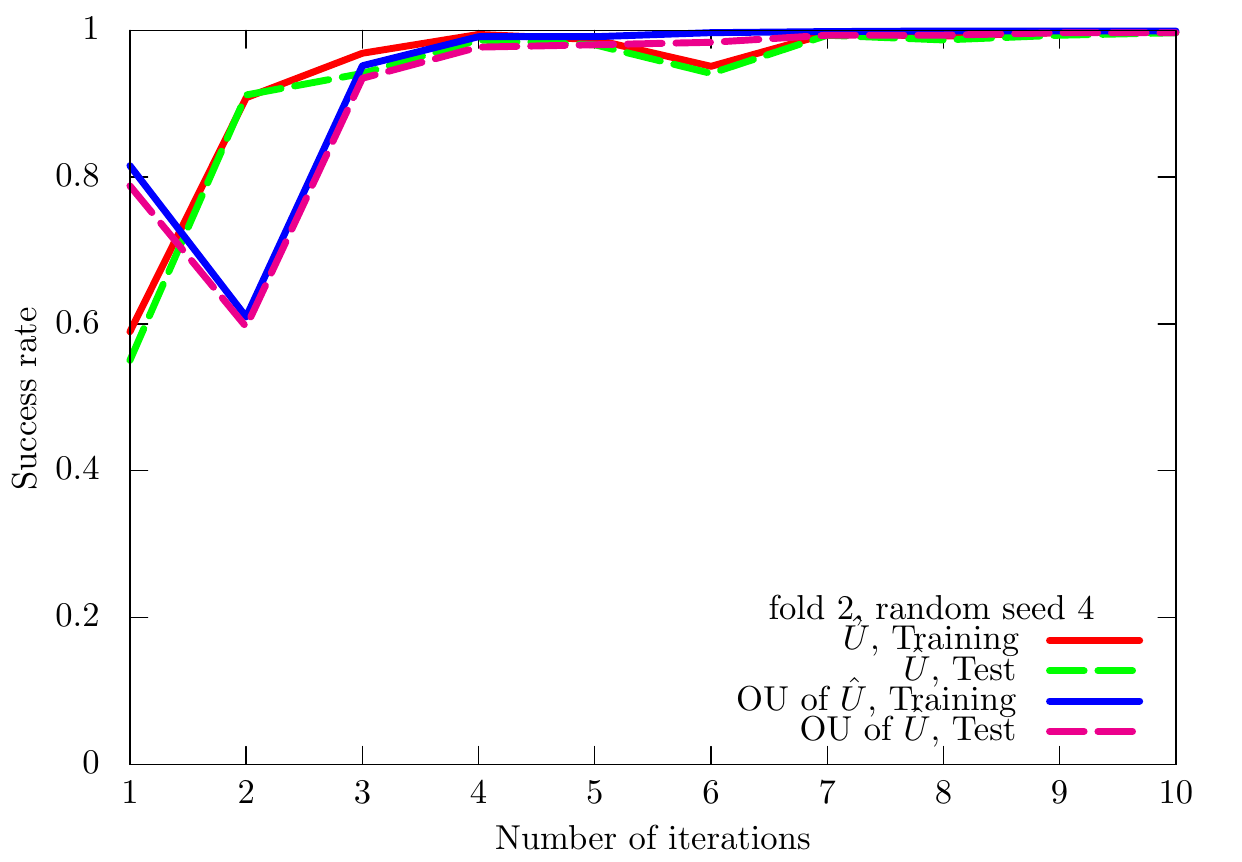}
\includegraphics[scale=0.25]{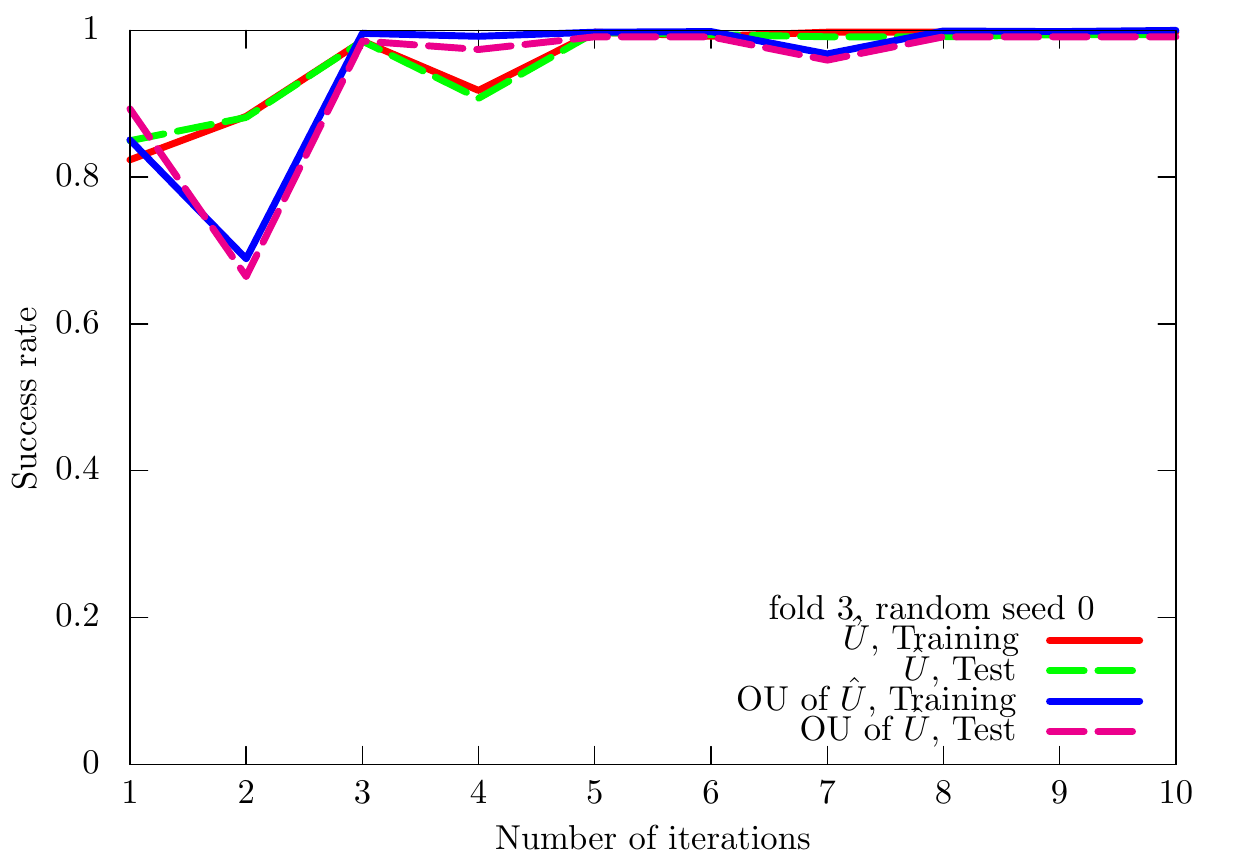}
\includegraphics[scale=0.25]{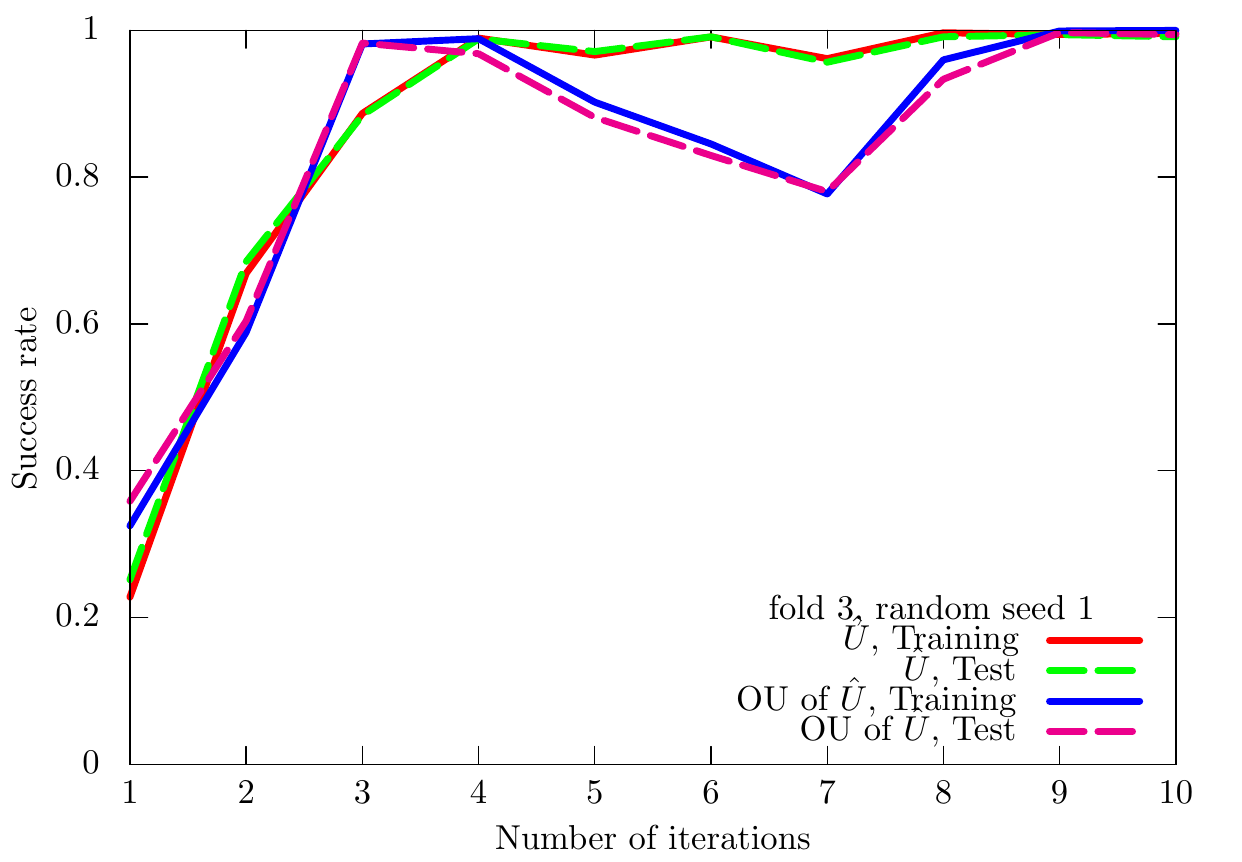}
\includegraphics[scale=0.25]{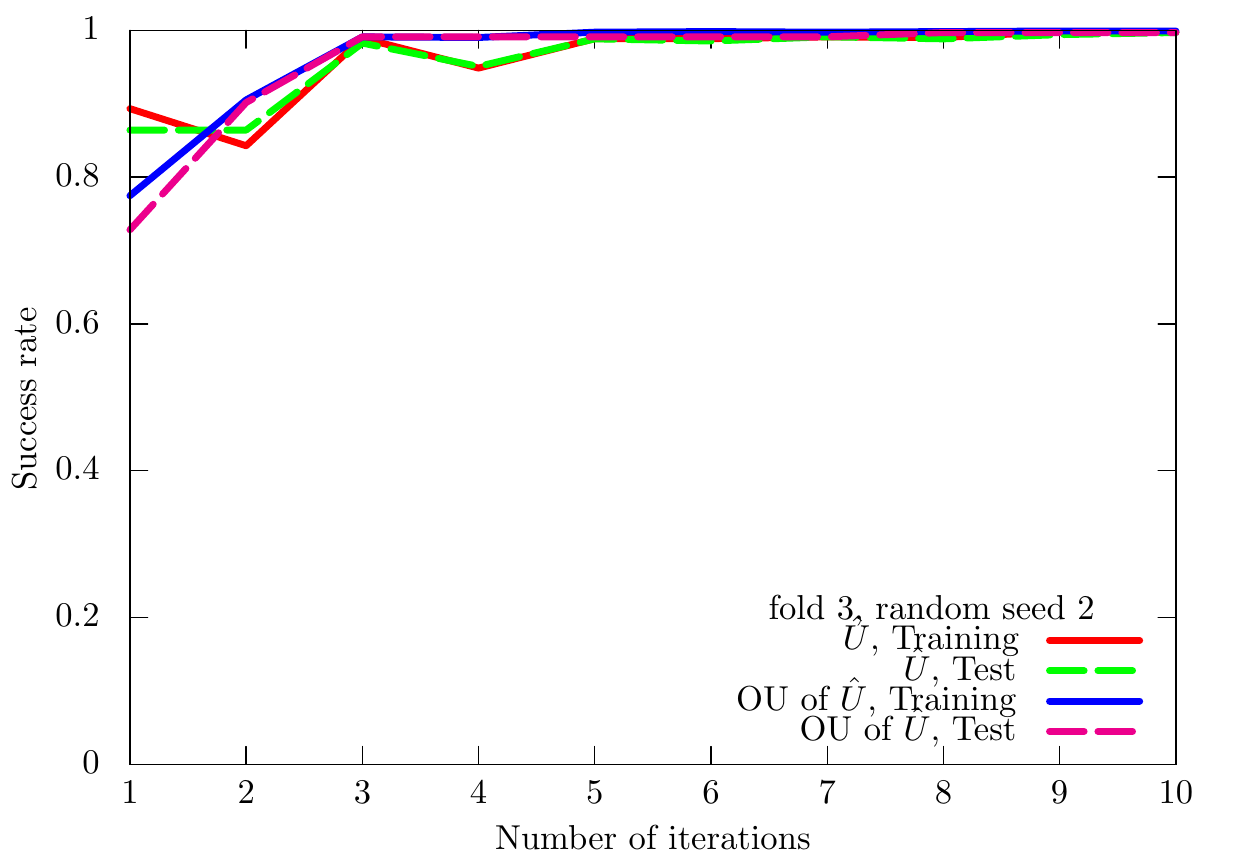}
\includegraphics[scale=0.25]{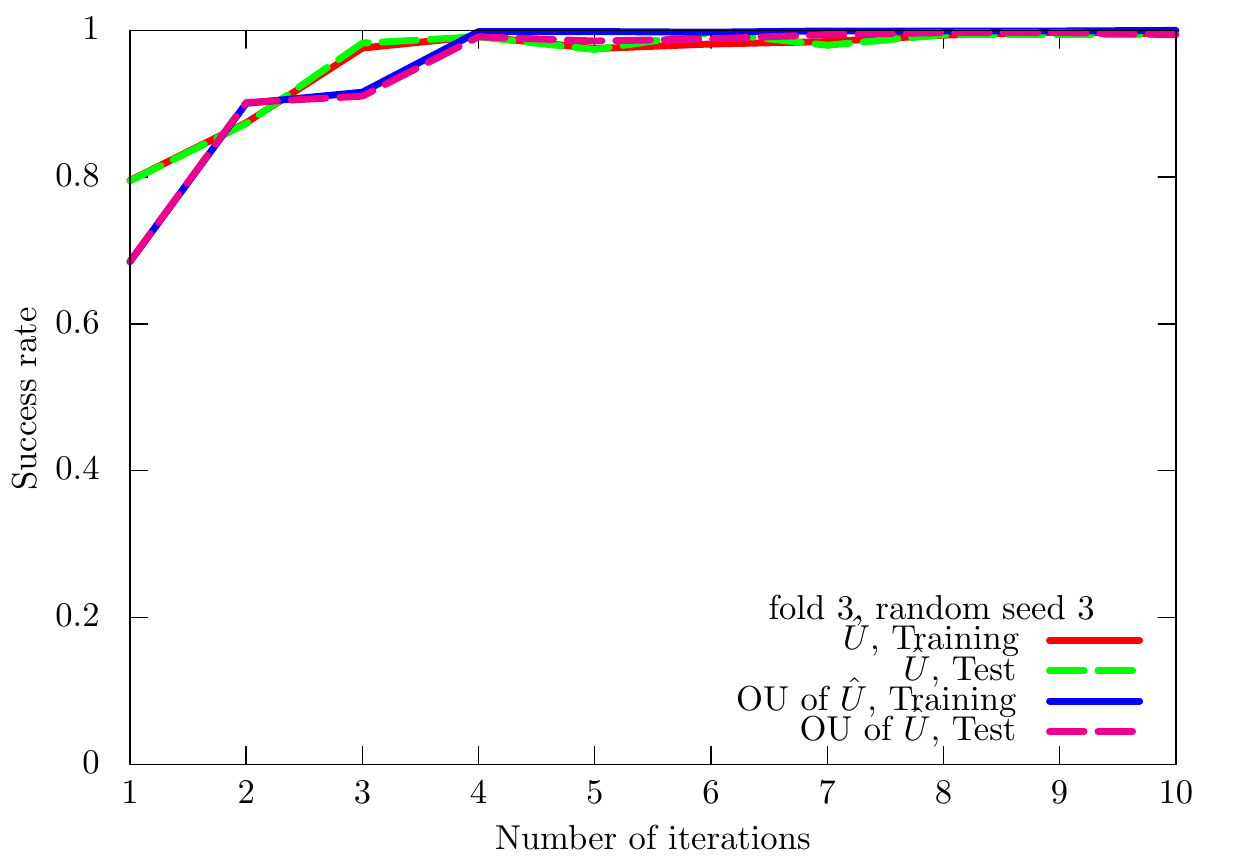}
\includegraphics[scale=0.25]{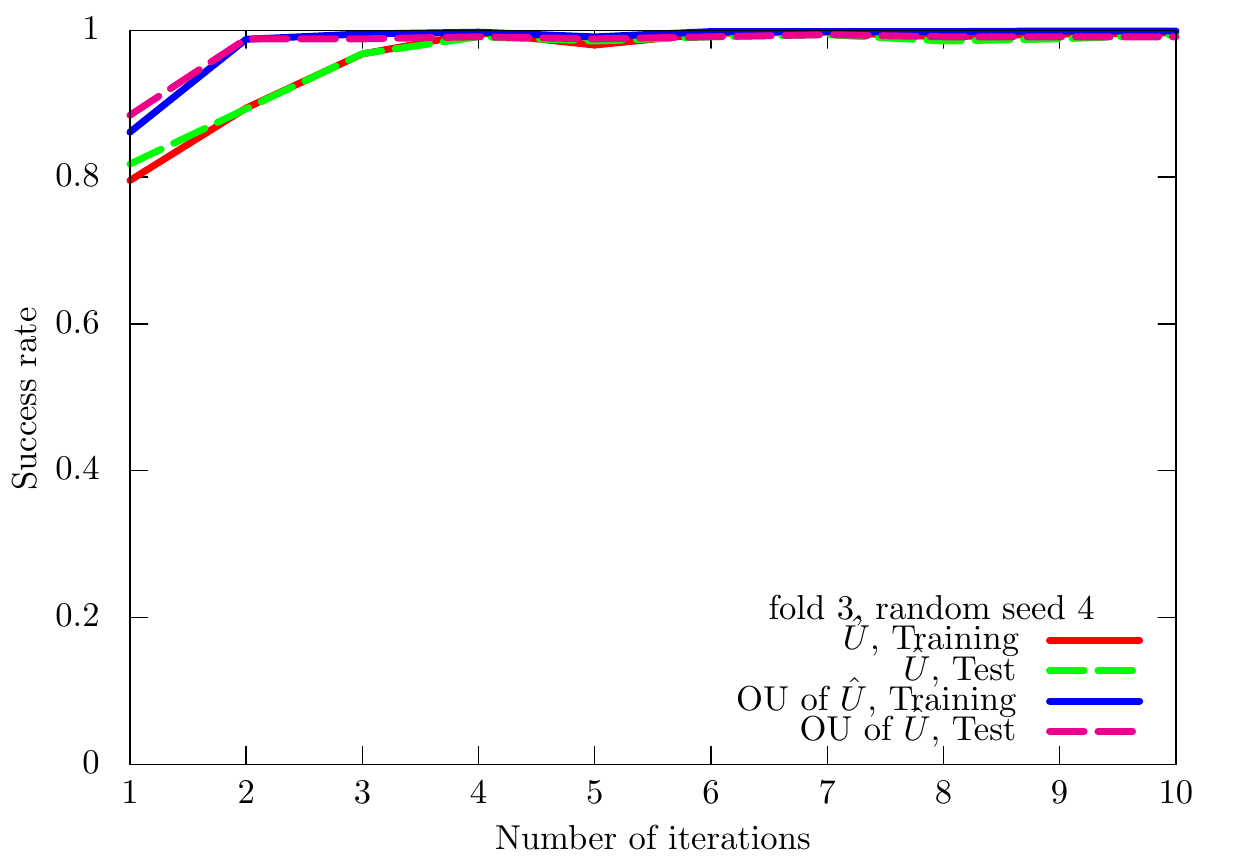}
\includegraphics[scale=0.25]{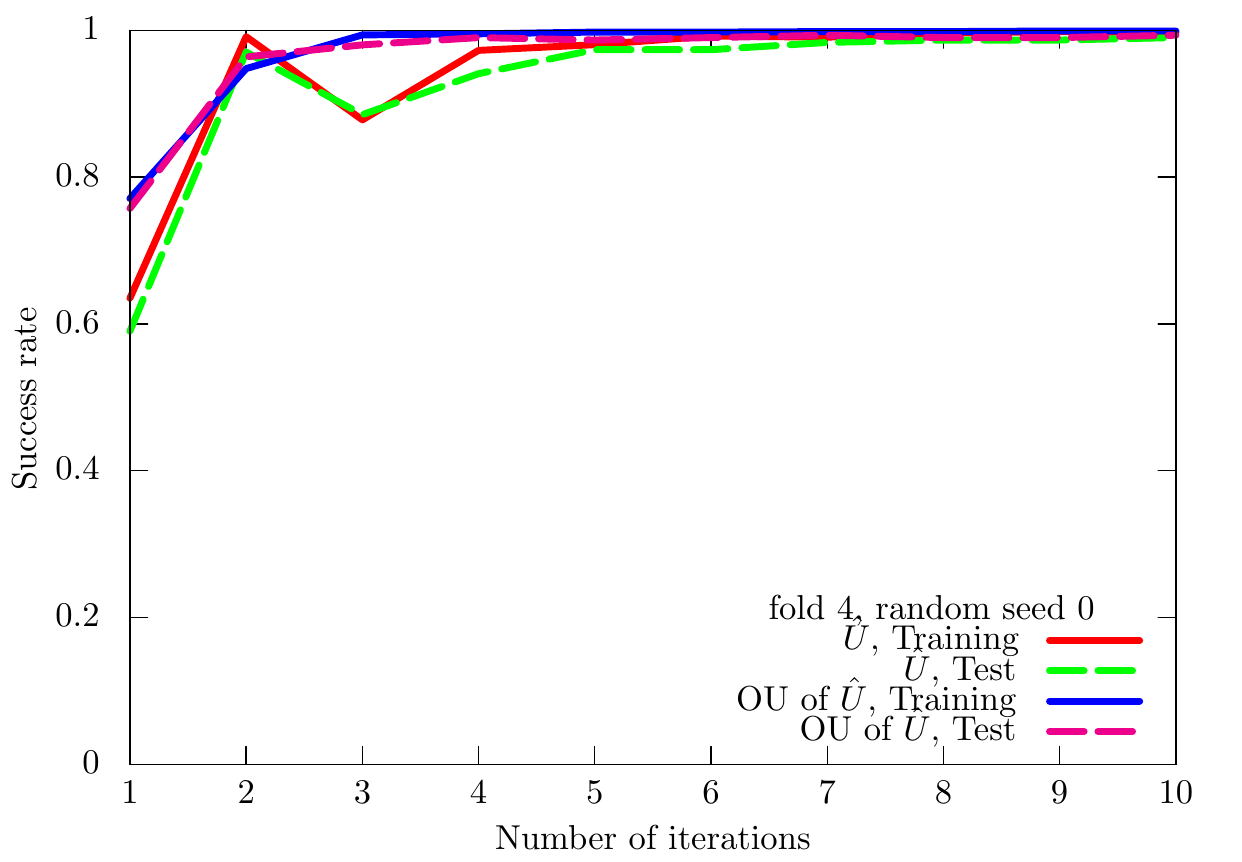}
\includegraphics[scale=0.25]{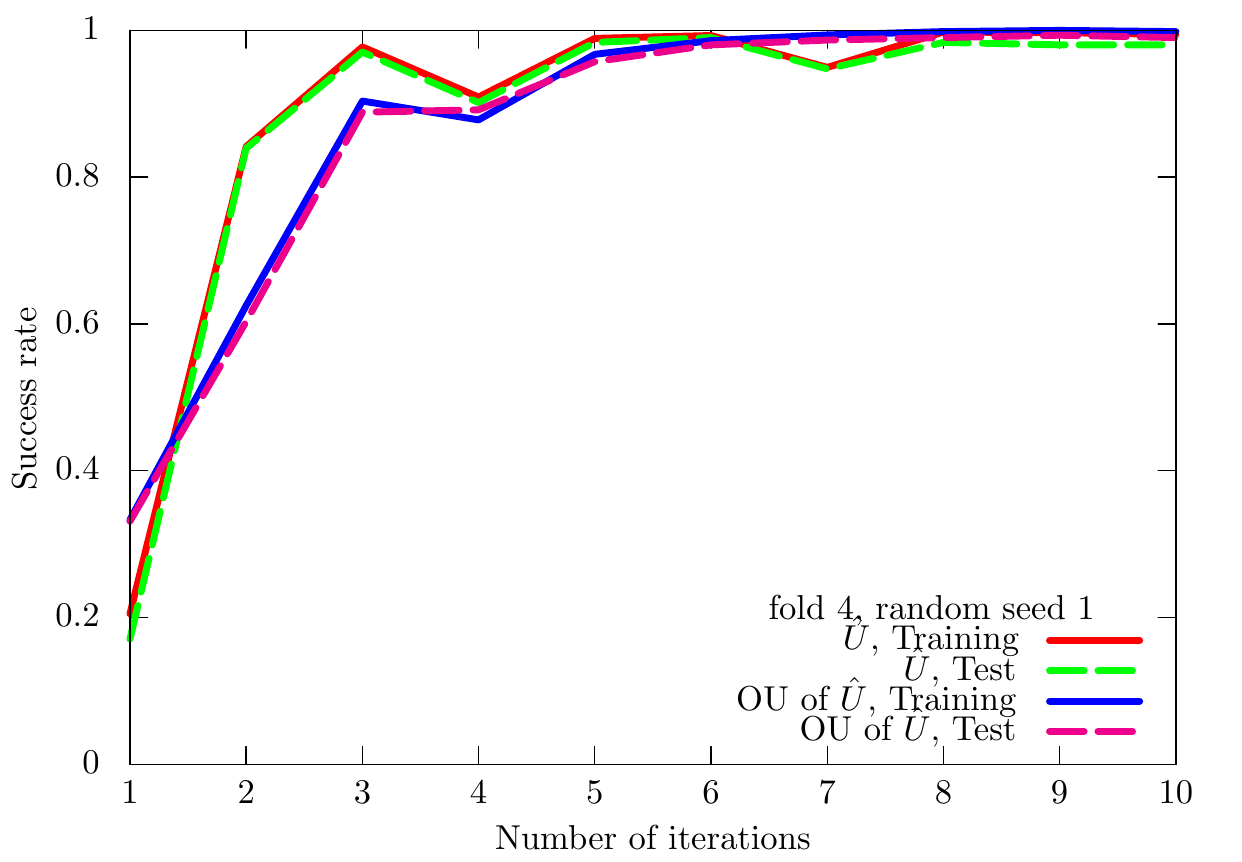}
\includegraphics[scale=0.25]{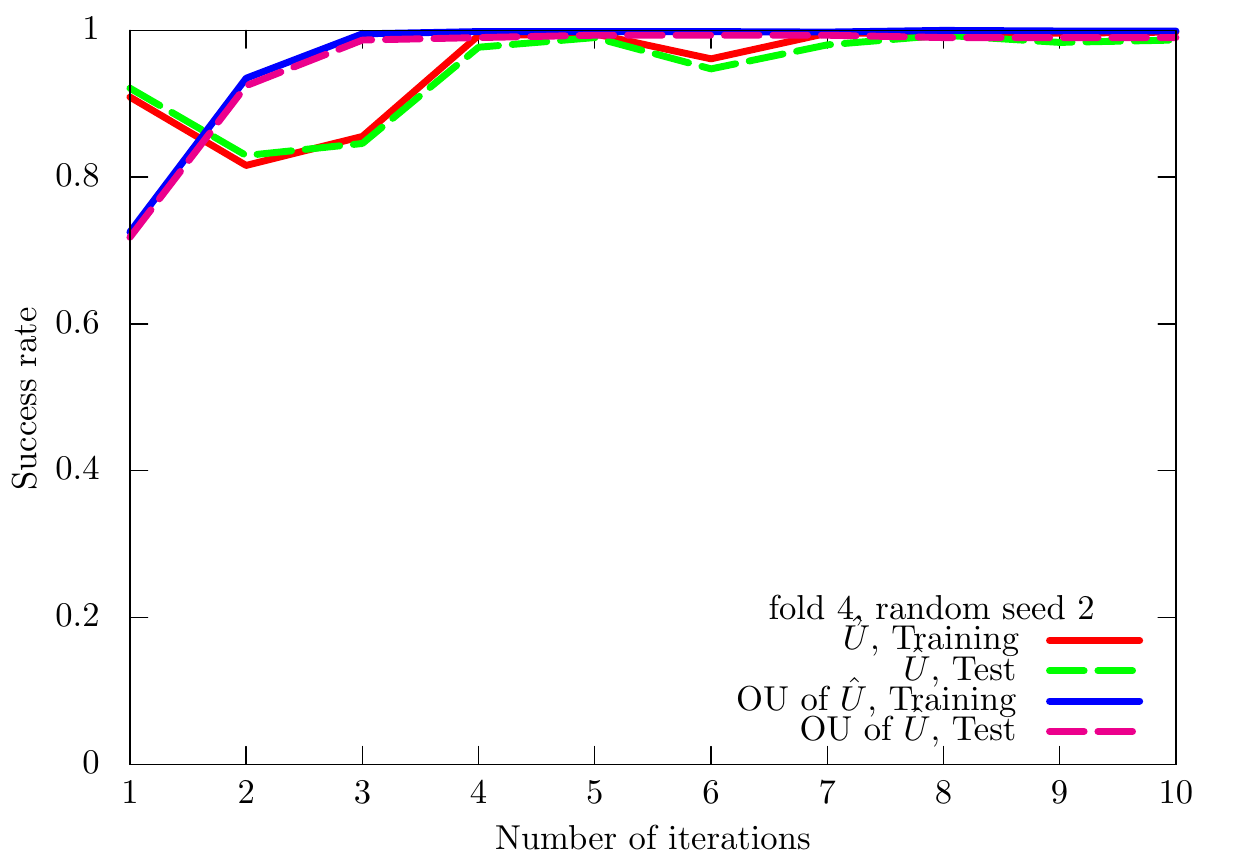}
\includegraphics[scale=0.25]{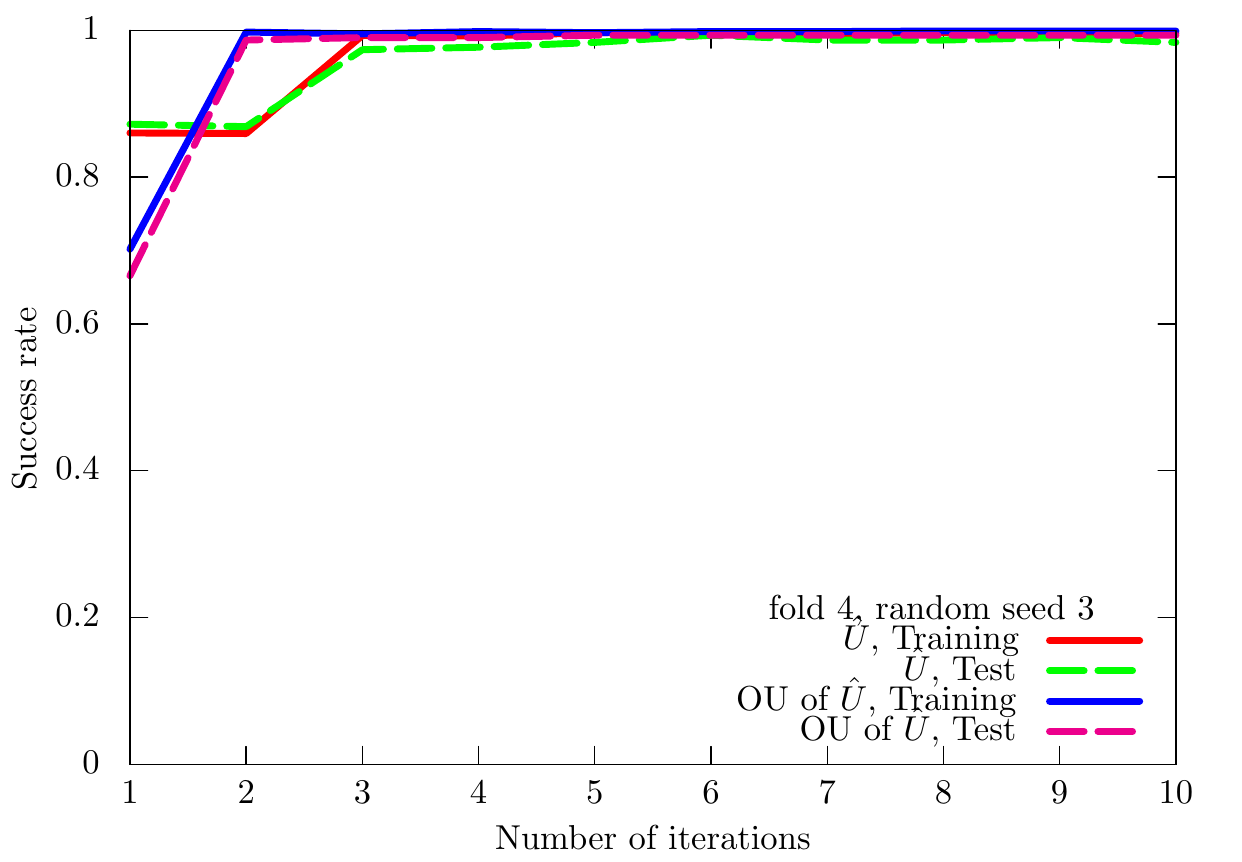}
\includegraphics[scale=0.25]{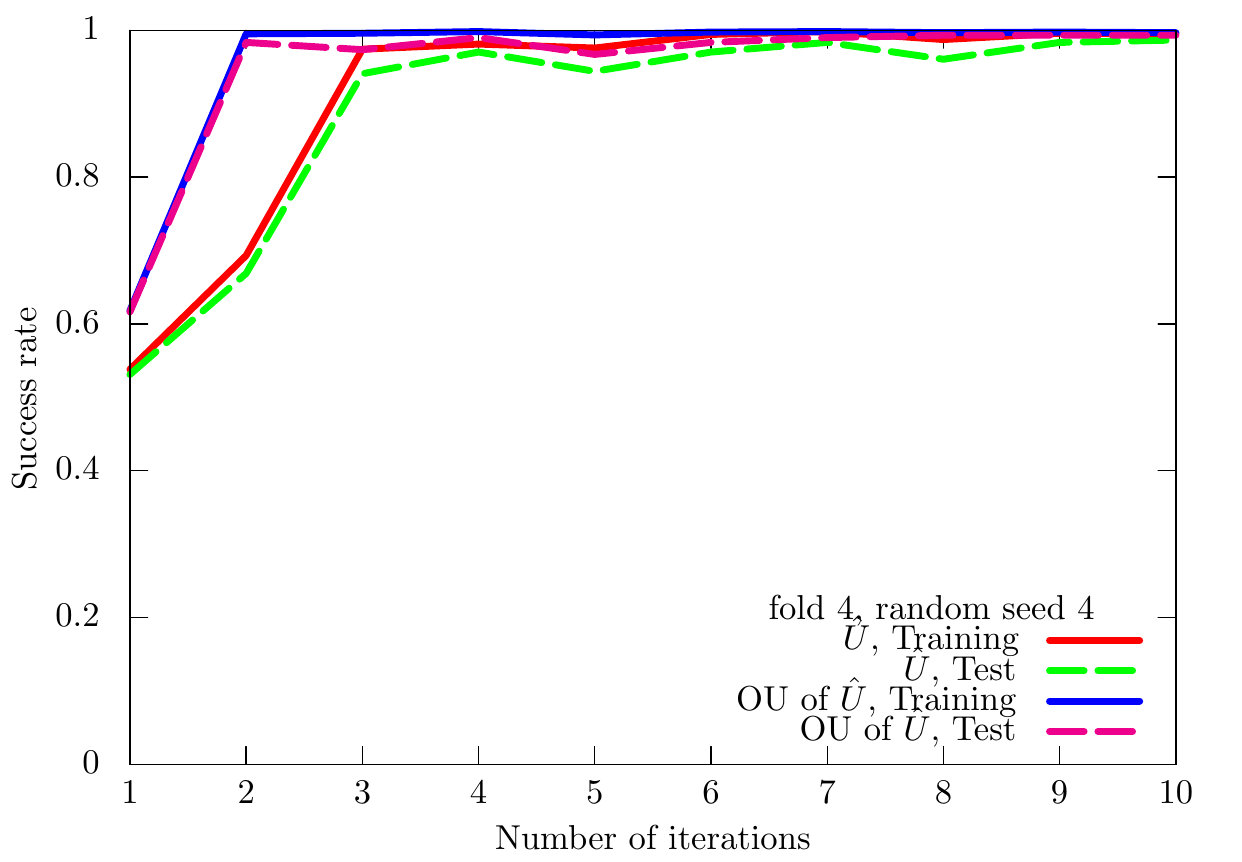}
\caption{Results of the UKM ($\hat{X}$ and $\hat{P}$) on the $5$-fold datasets with $5$ different random seeds for the semeion dataset ($0$ or non-$0$). We use complex matrices and set $\theta_\mathrm{bias} = 0$. We set $r = 0.010$.}
\label{supp-arXiv-numerical-result-raw-data-fold-001-rand-001-UKM-P-UCI-semeion-0-non0}
\end{figure*}

\begin{figure*}[htb]
\centering
\includegraphics[scale=0.25]{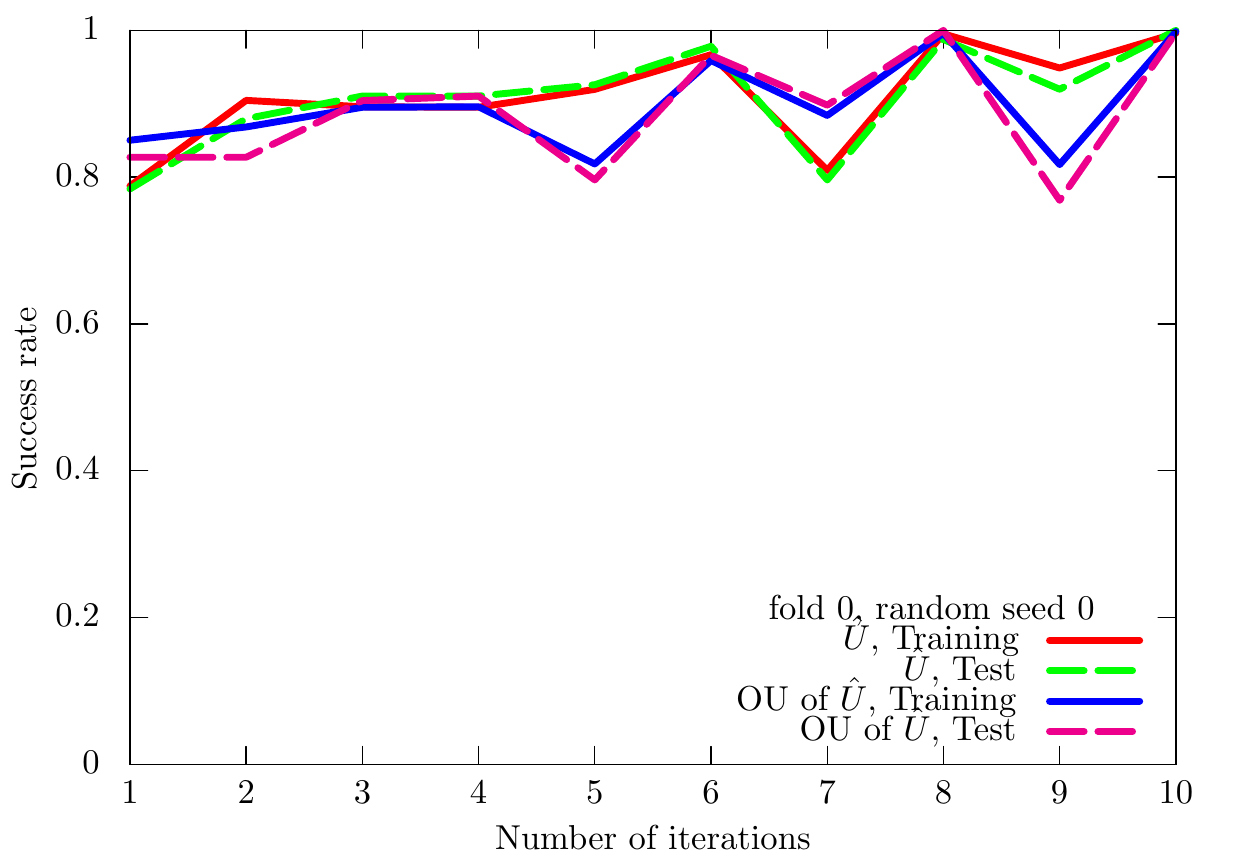}
\includegraphics[scale=0.25]{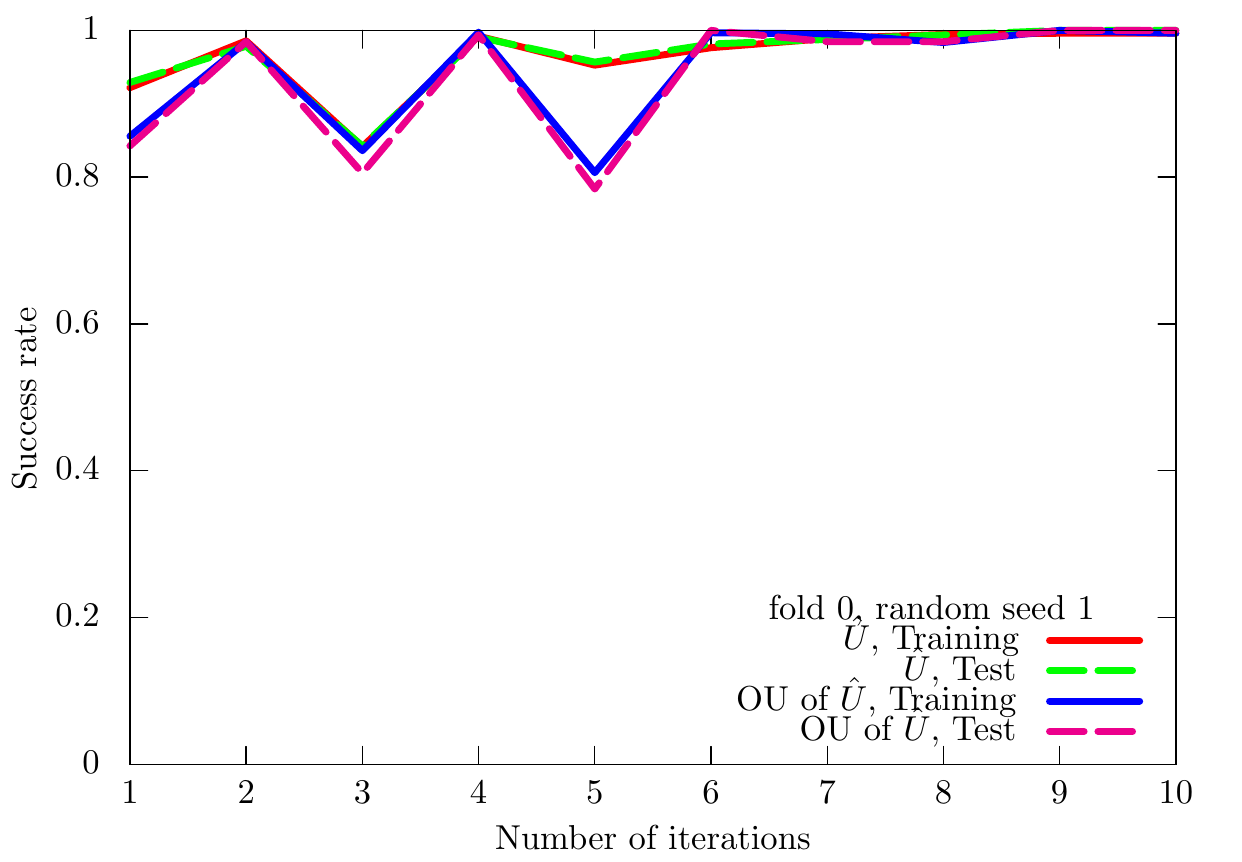}
\includegraphics[scale=0.25]{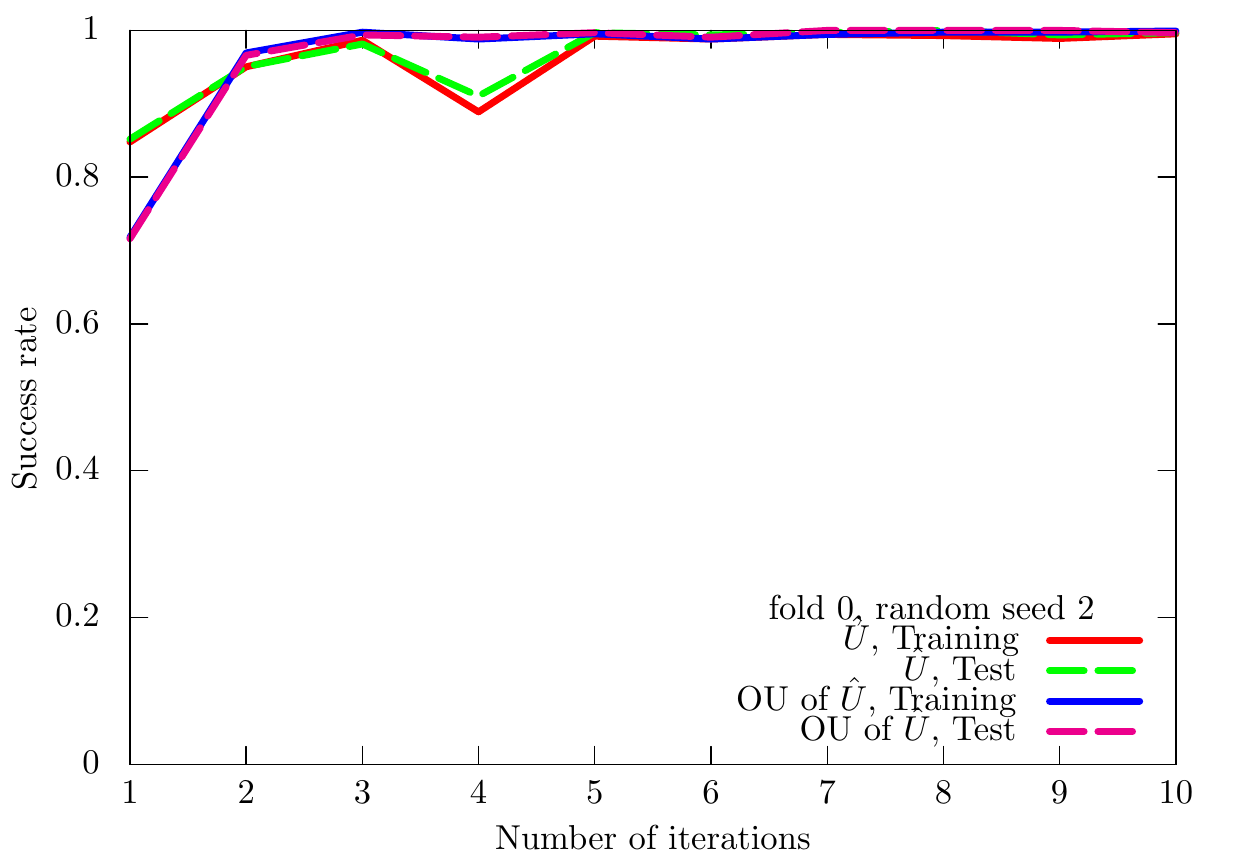}
\includegraphics[scale=0.25]{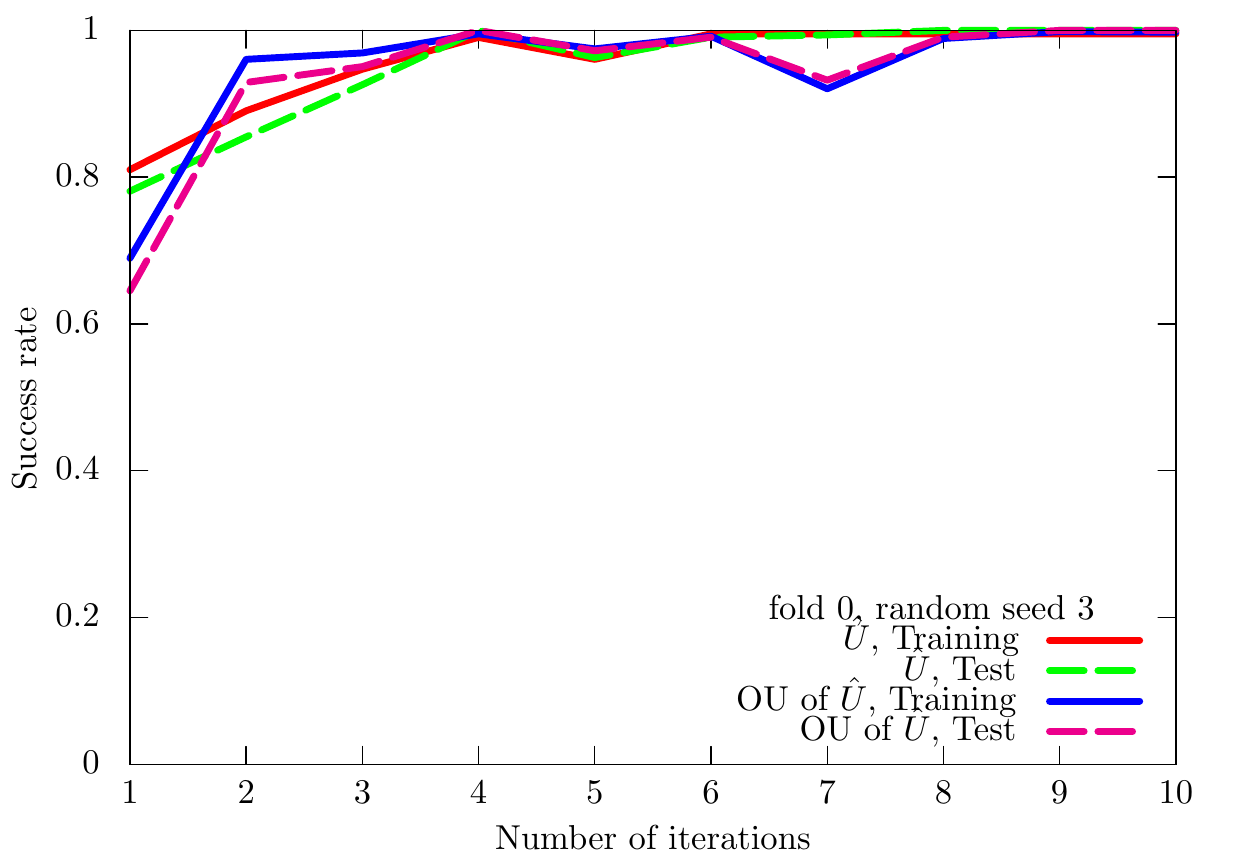}
\includegraphics[scale=0.25]{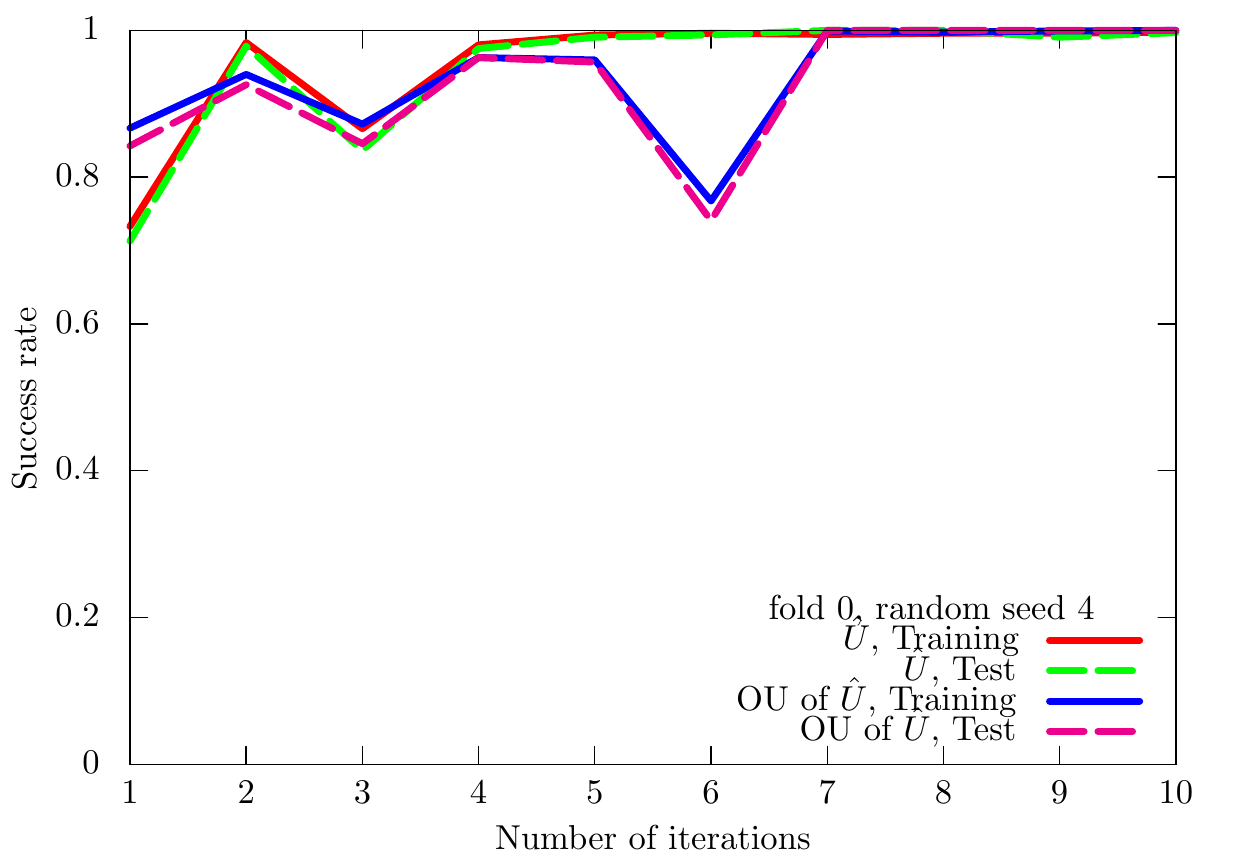}
\includegraphics[scale=0.25]{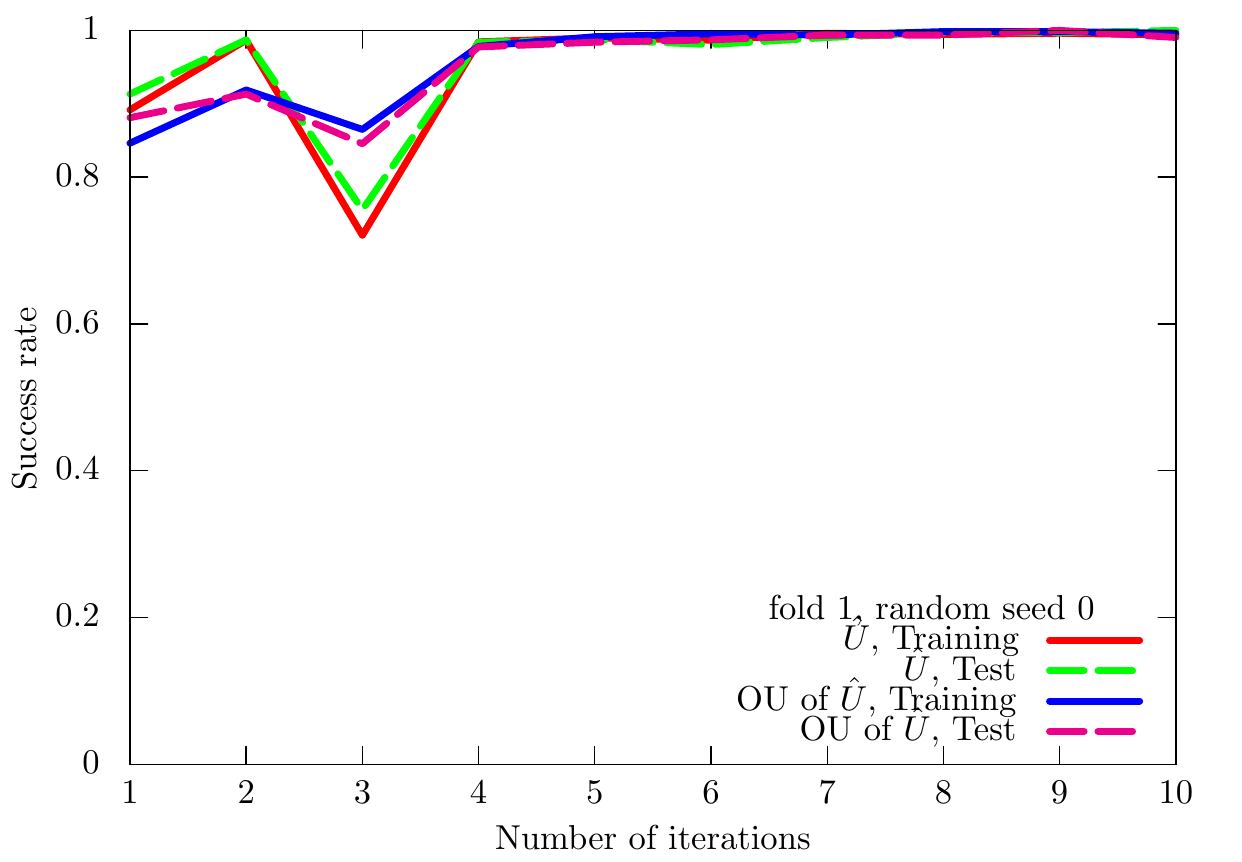}
\includegraphics[scale=0.25]{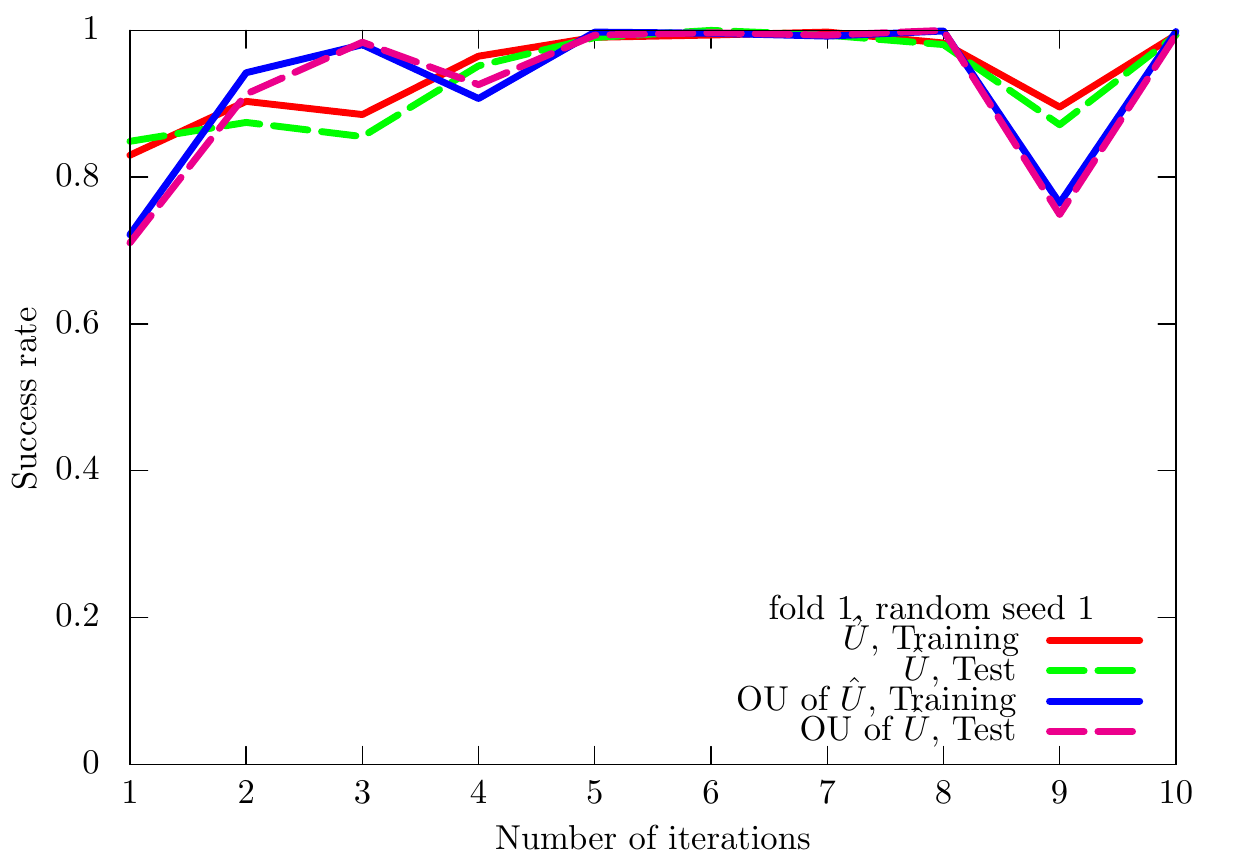}
\includegraphics[scale=0.25]{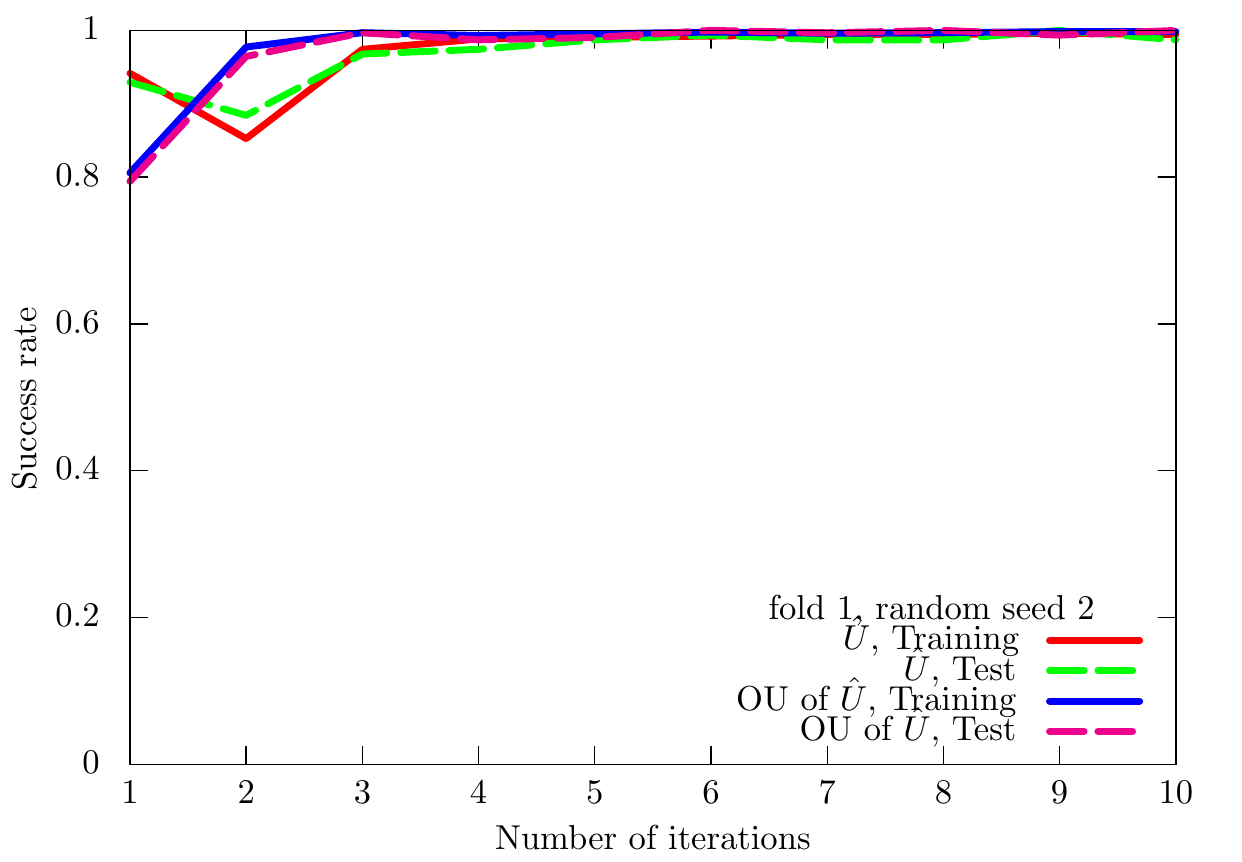}
\includegraphics[scale=0.25]{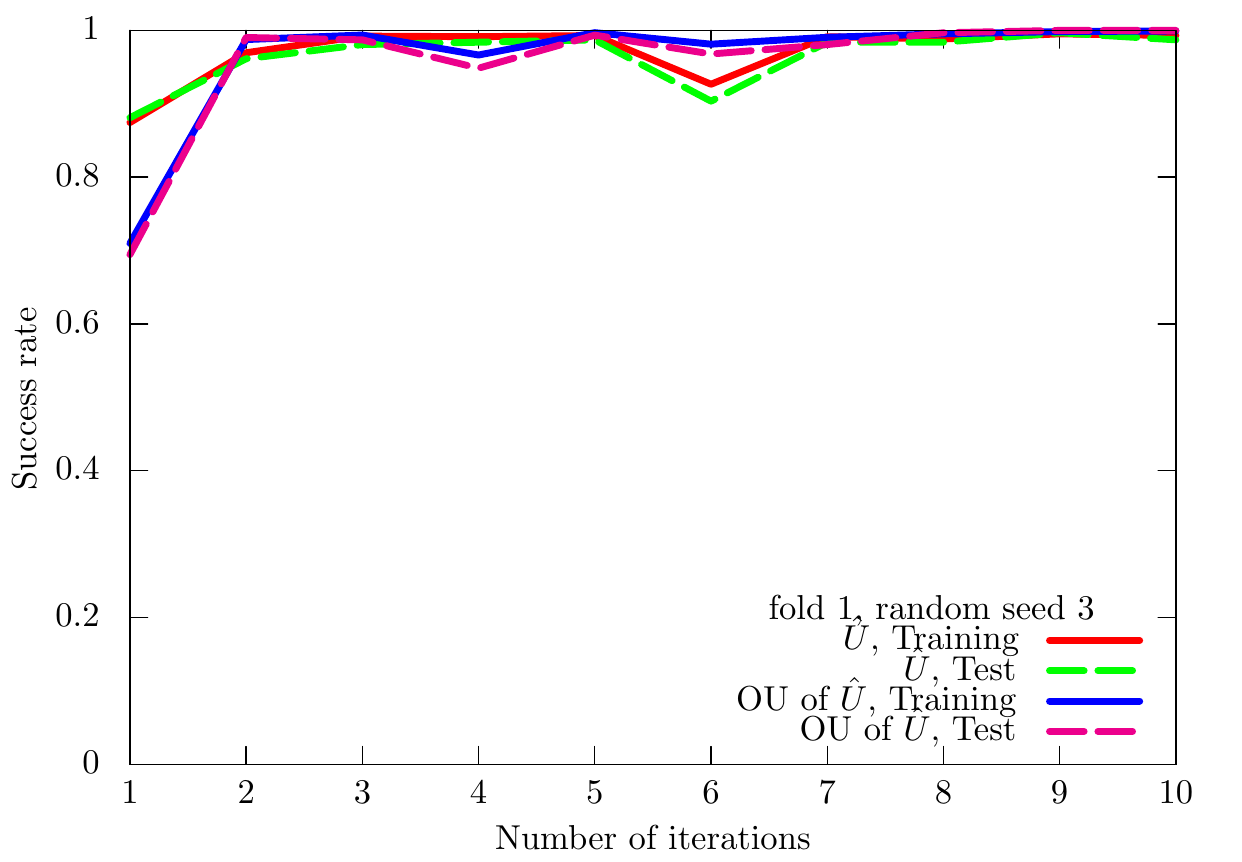}
\includegraphics[scale=0.25]{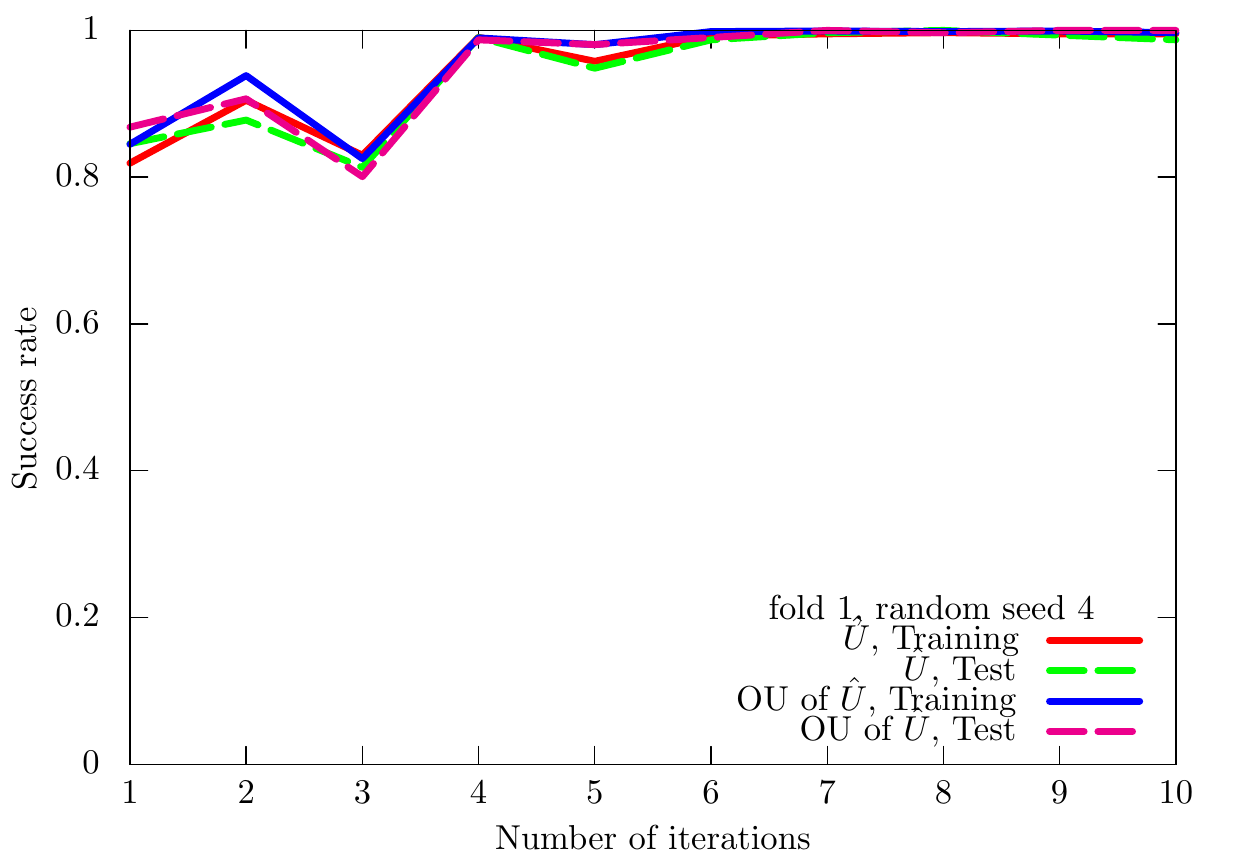}
\includegraphics[scale=0.25]{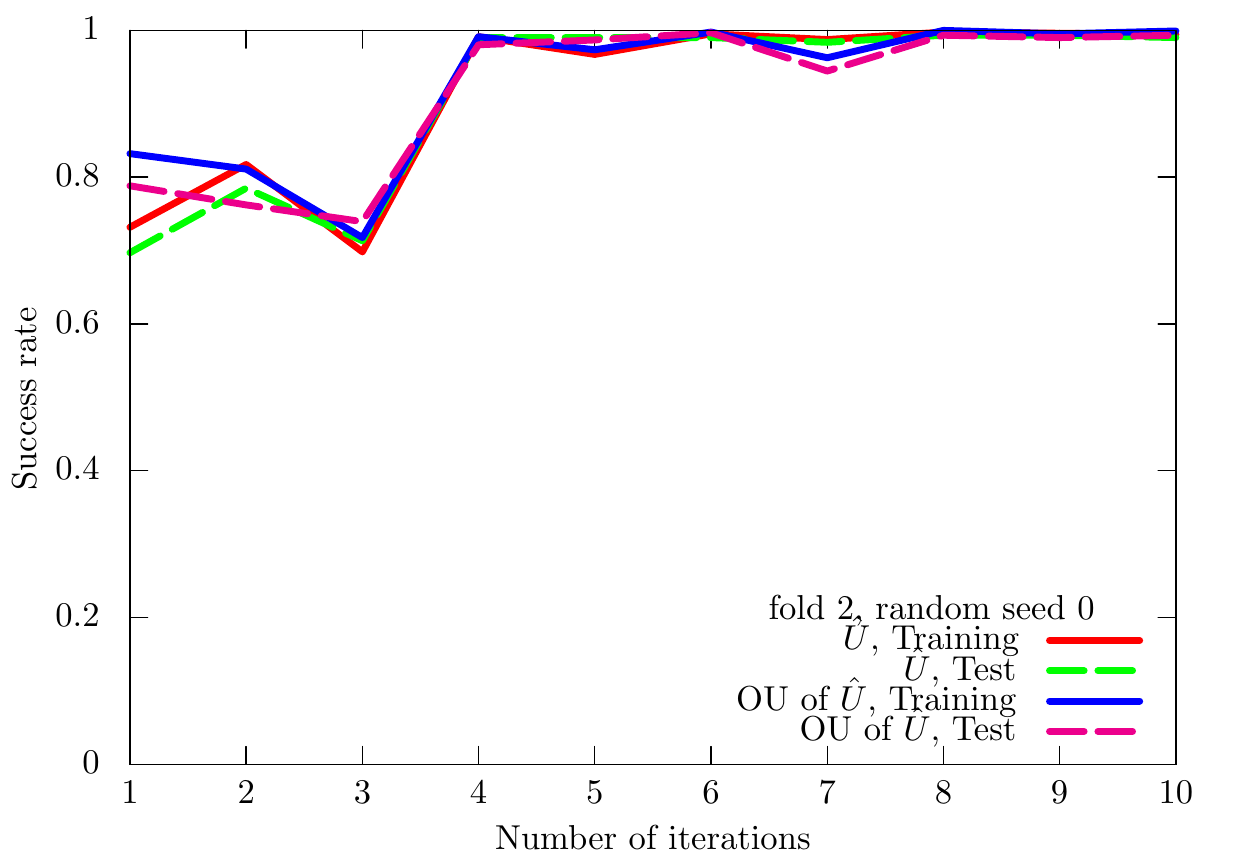}
\includegraphics[scale=0.25]{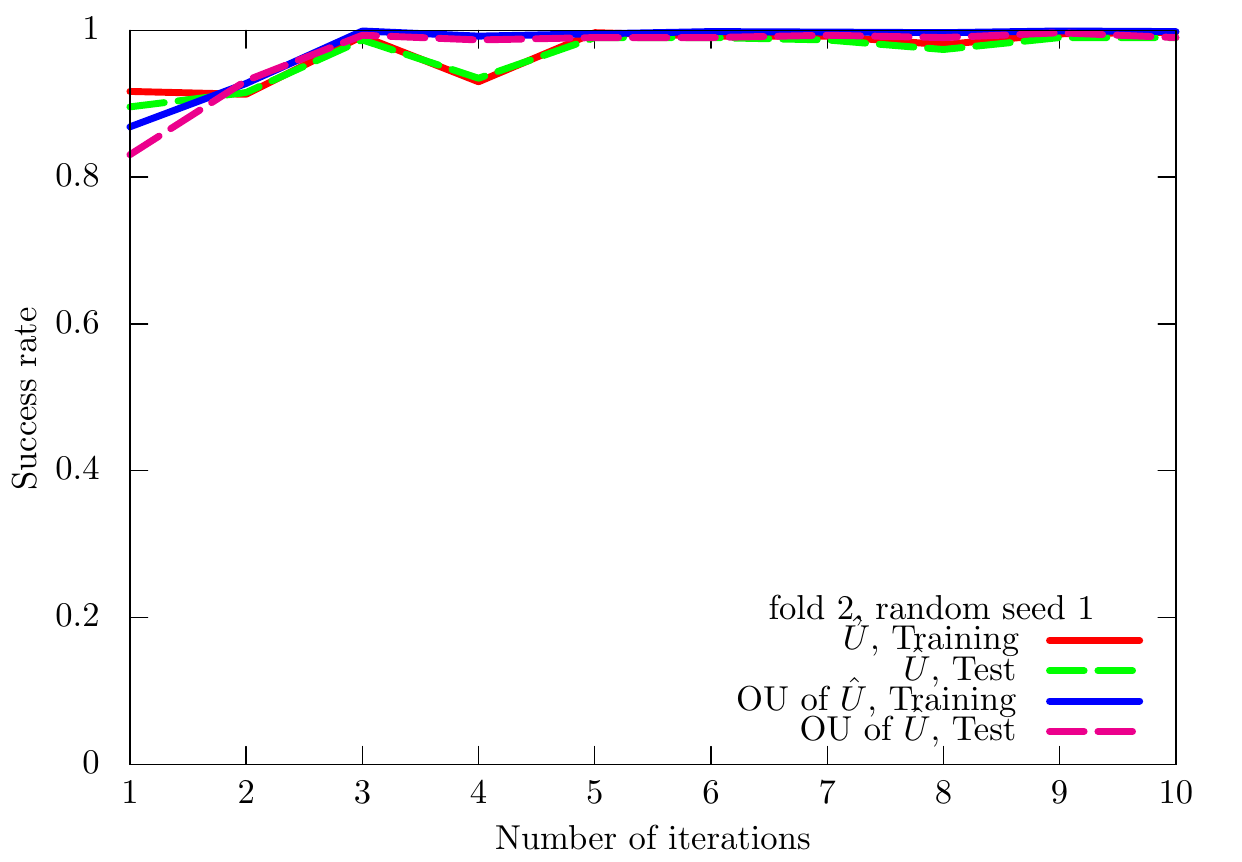}
\includegraphics[scale=0.25]{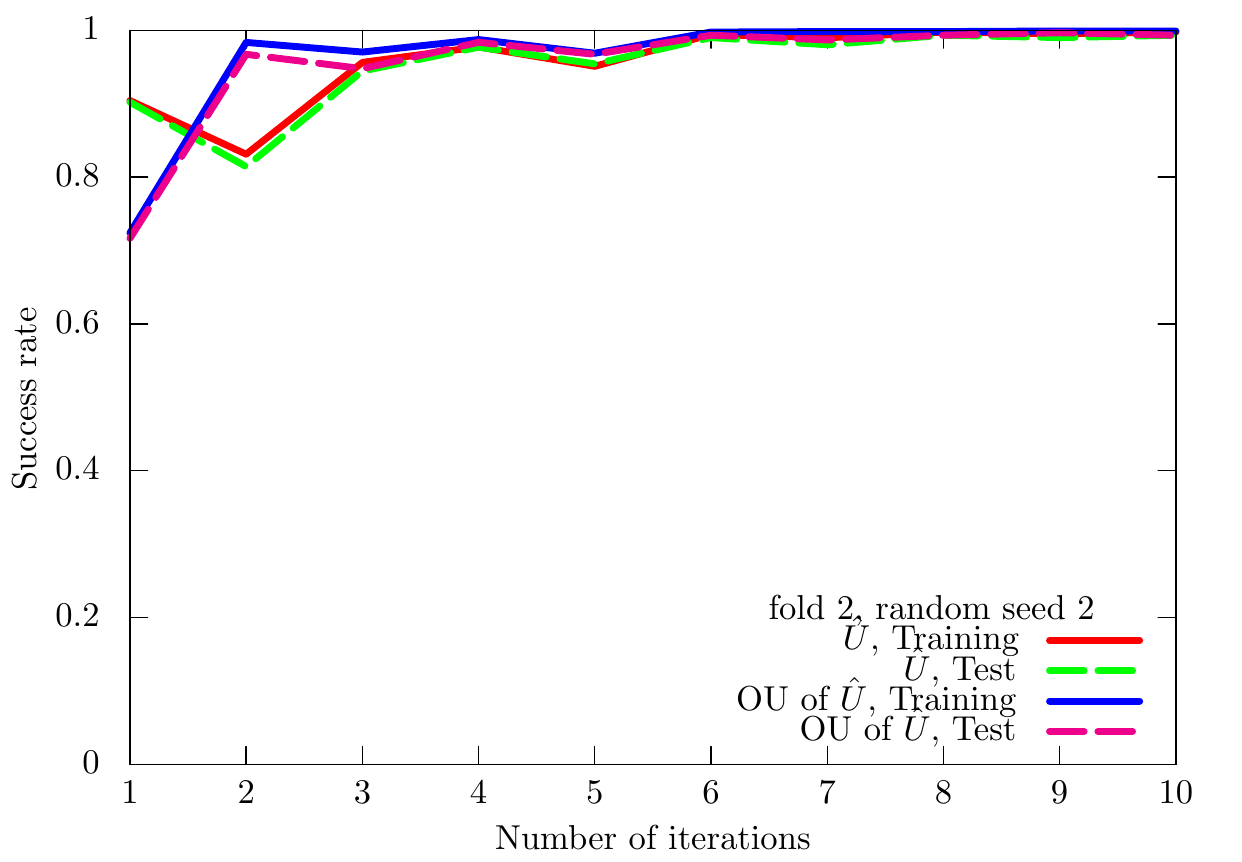}
\includegraphics[scale=0.25]{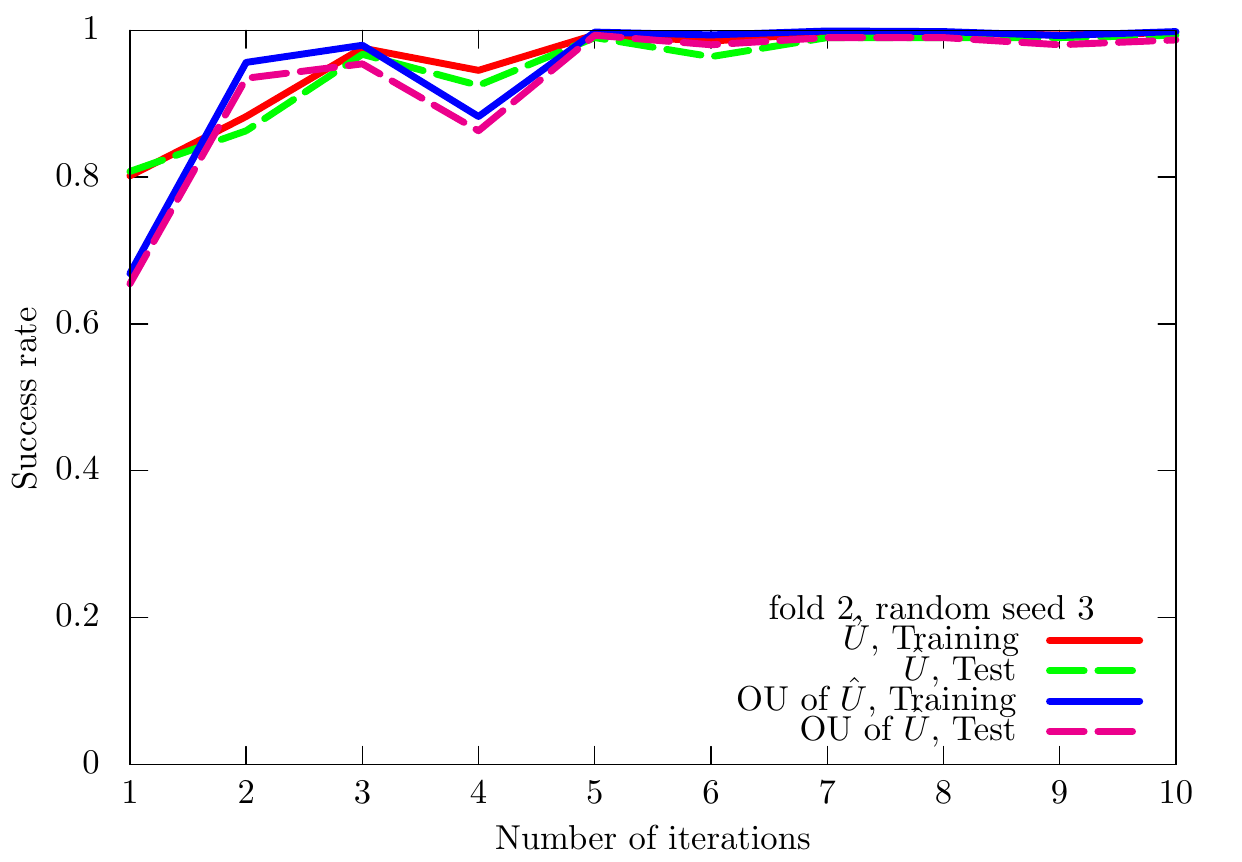}
\includegraphics[scale=0.25]{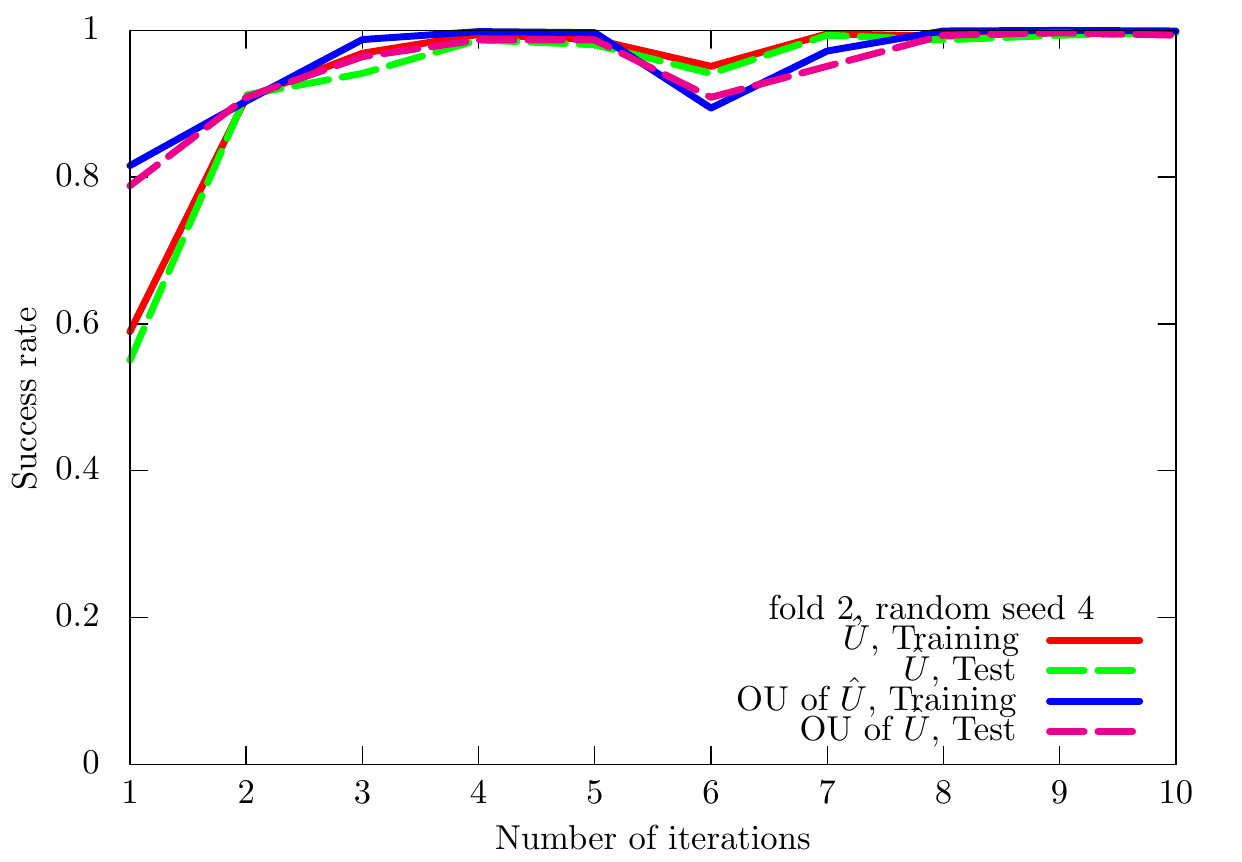}
\includegraphics[scale=0.25]{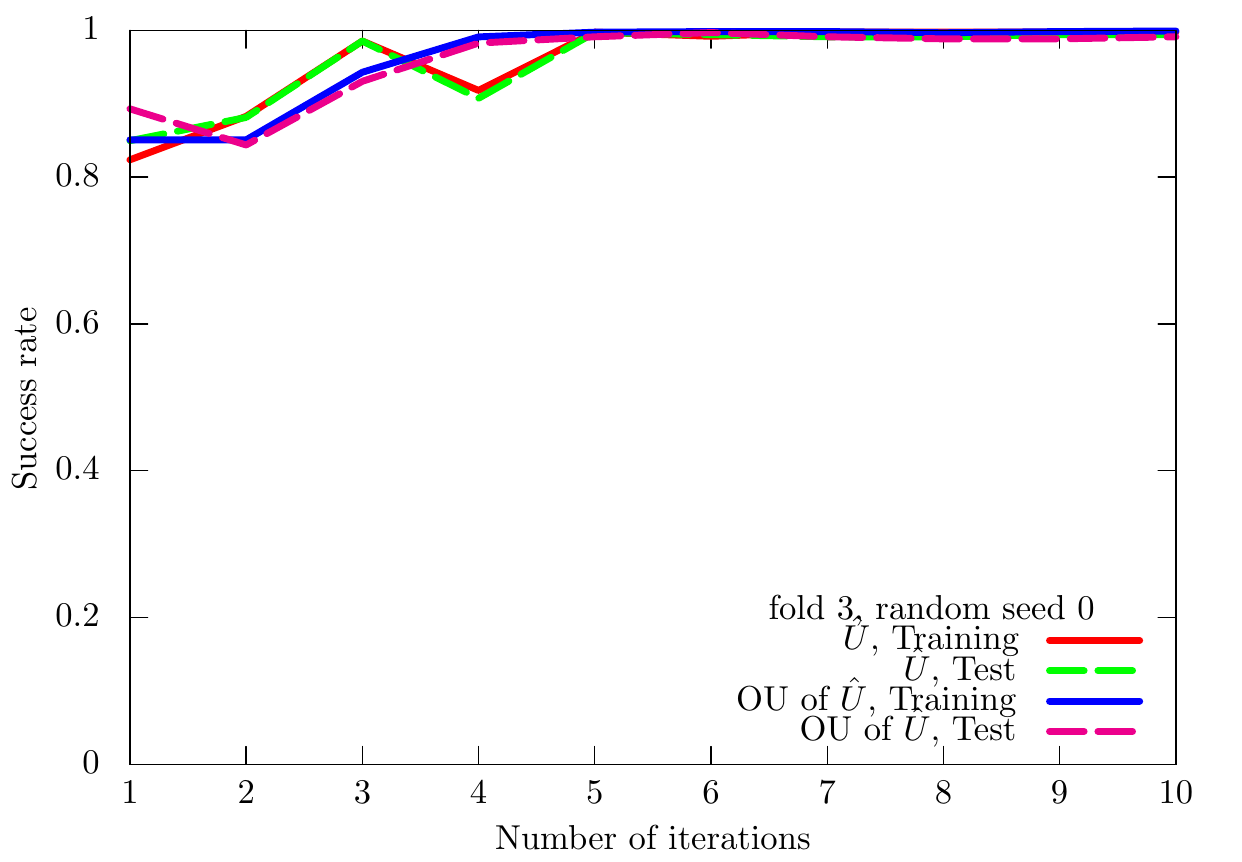}
\includegraphics[scale=0.25]{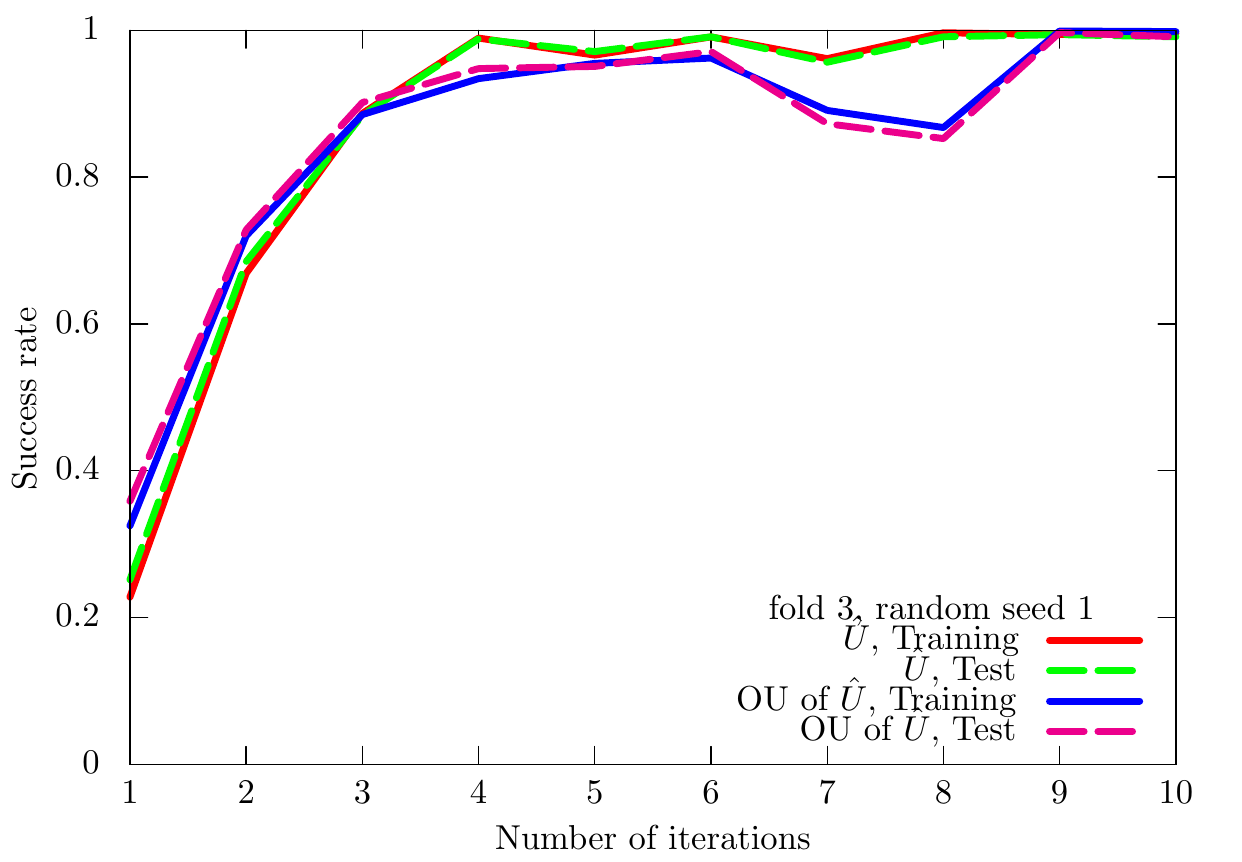}
\includegraphics[scale=0.25]{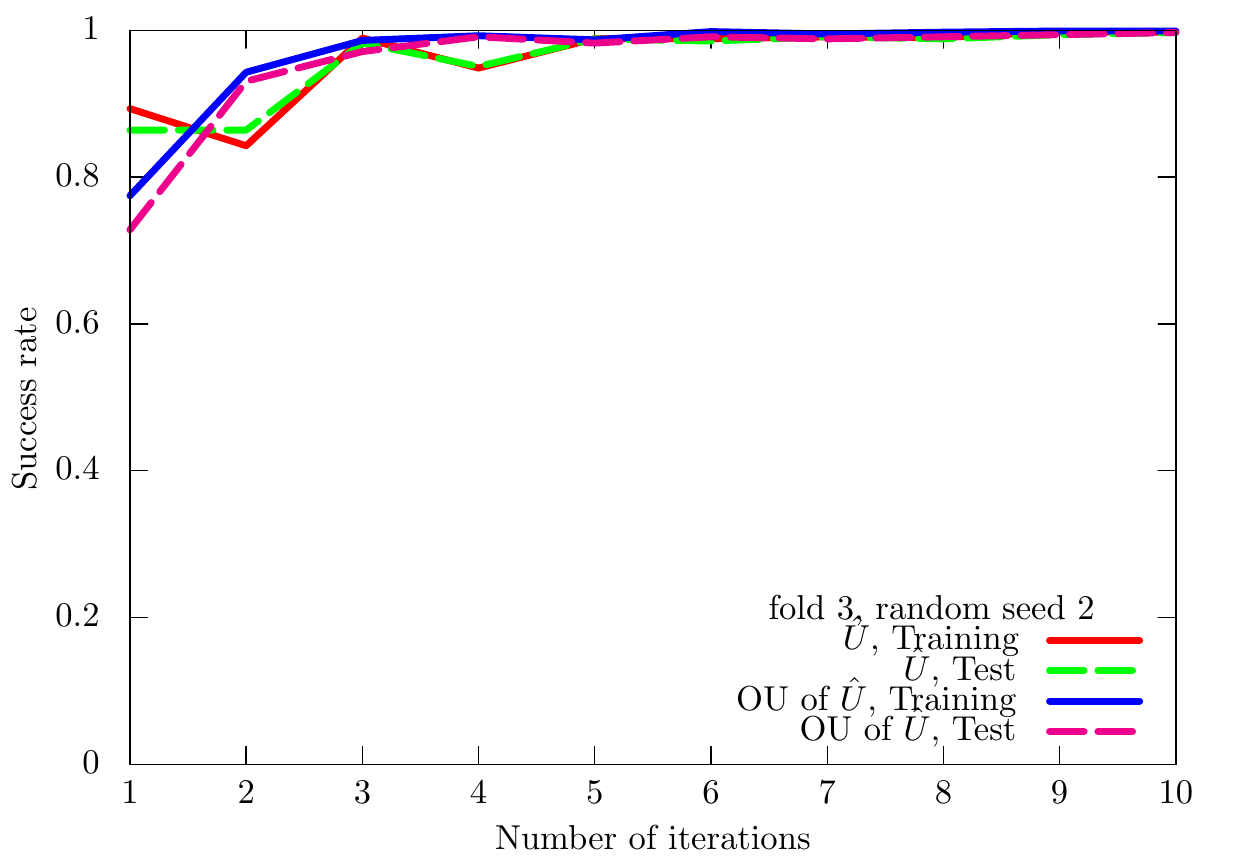}
\includegraphics[scale=0.25]{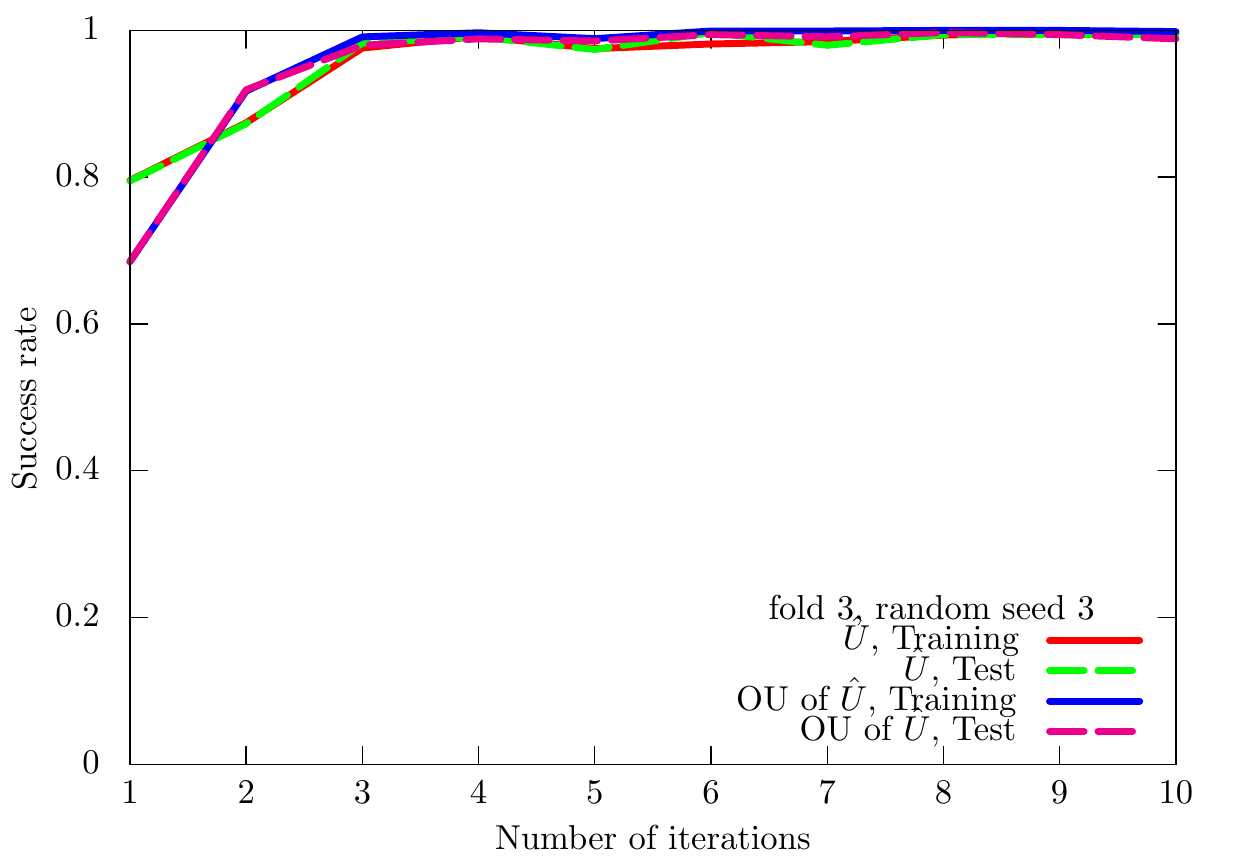}
\includegraphics[scale=0.25]{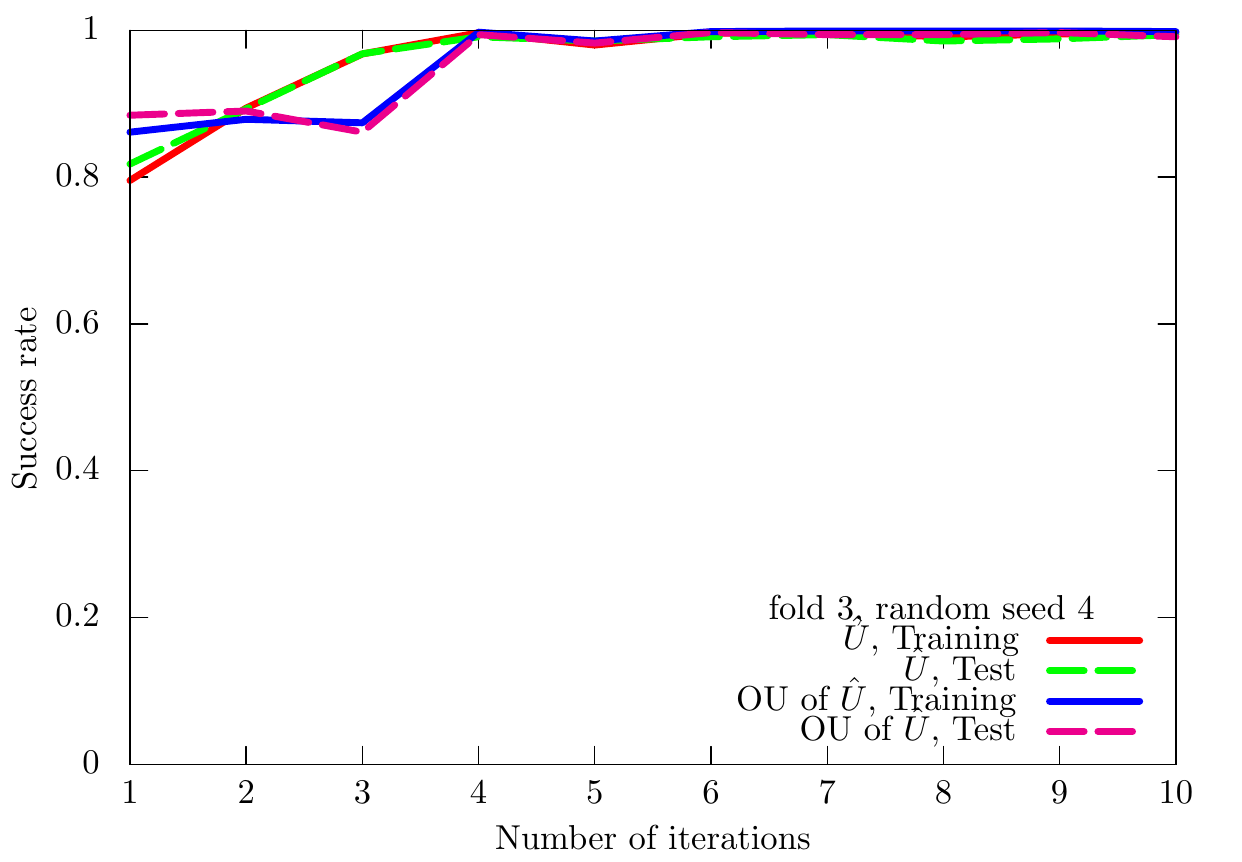}
\includegraphics[scale=0.25]{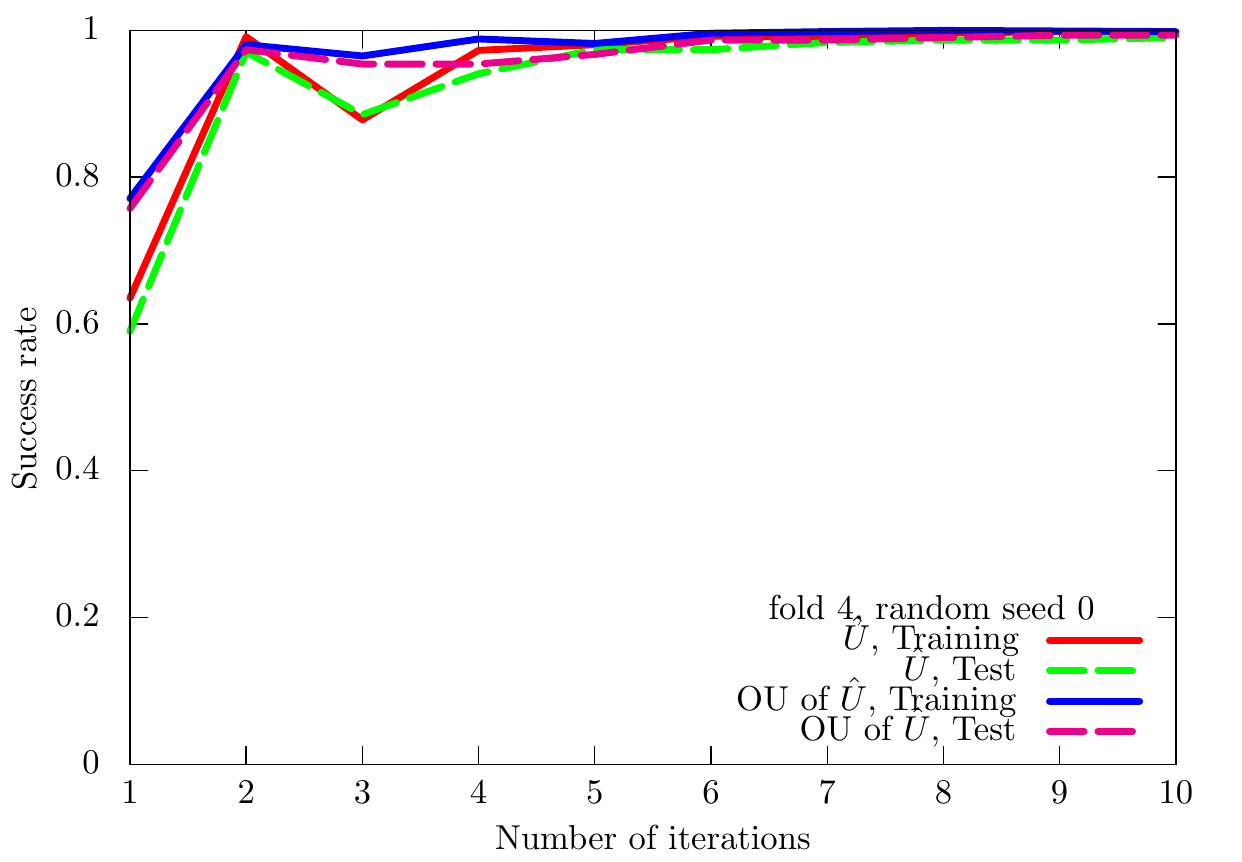}
\includegraphics[scale=0.25]{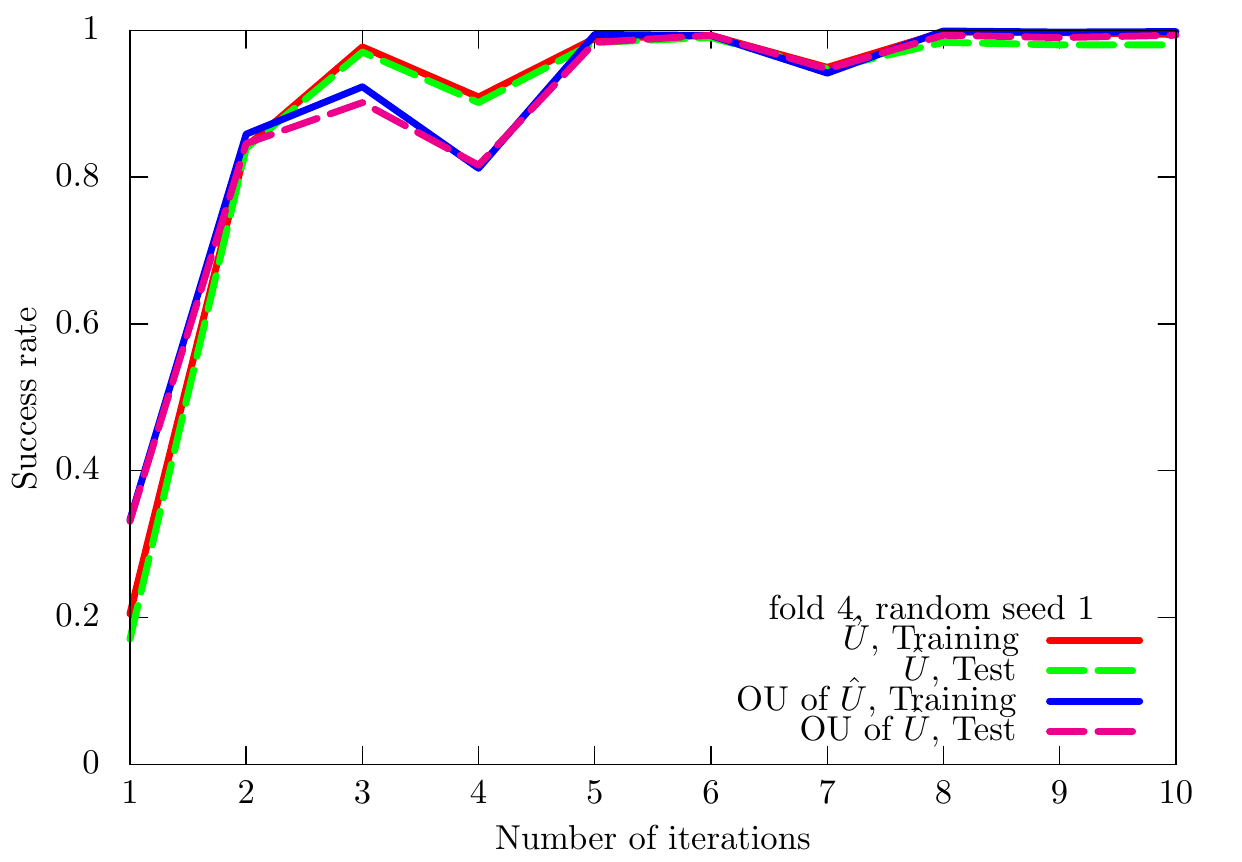}
\includegraphics[scale=0.25]{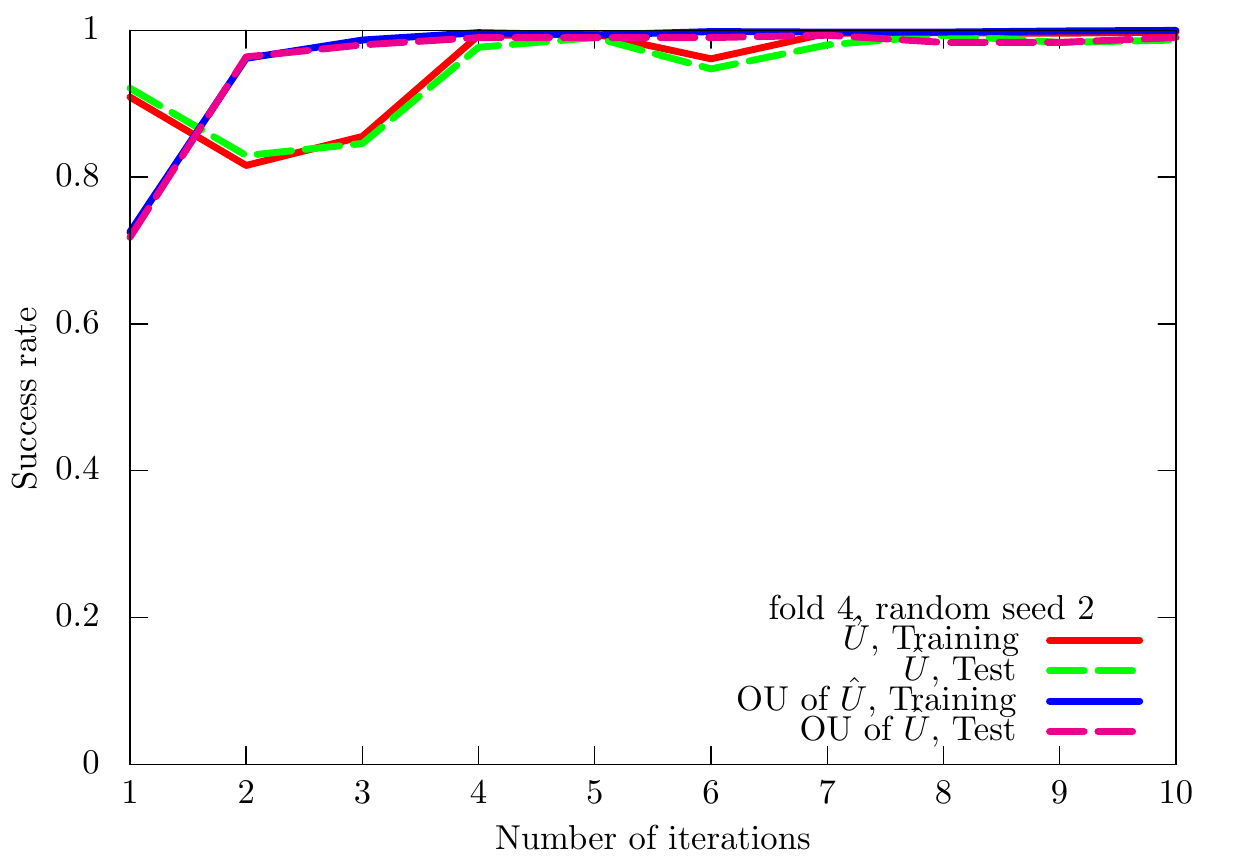}
\includegraphics[scale=0.25]{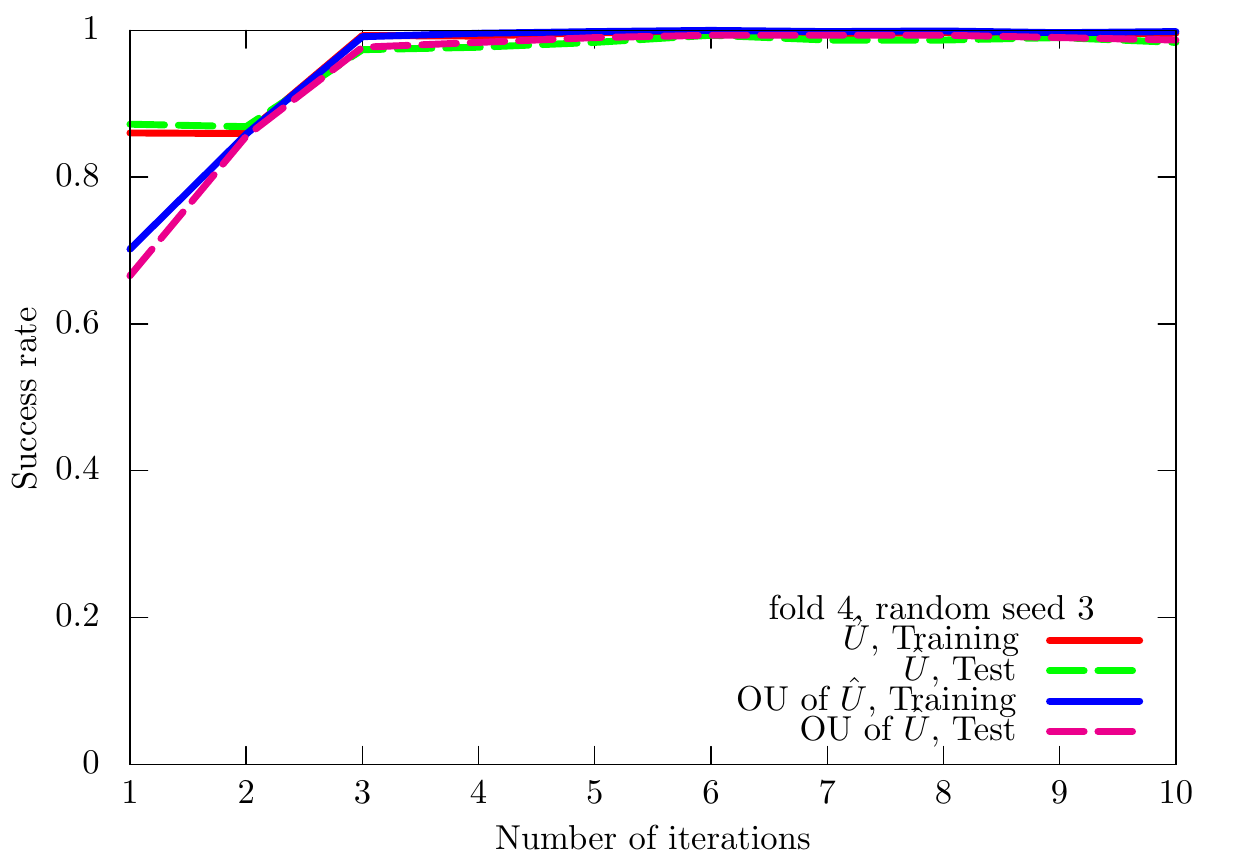}
\includegraphics[scale=0.25]{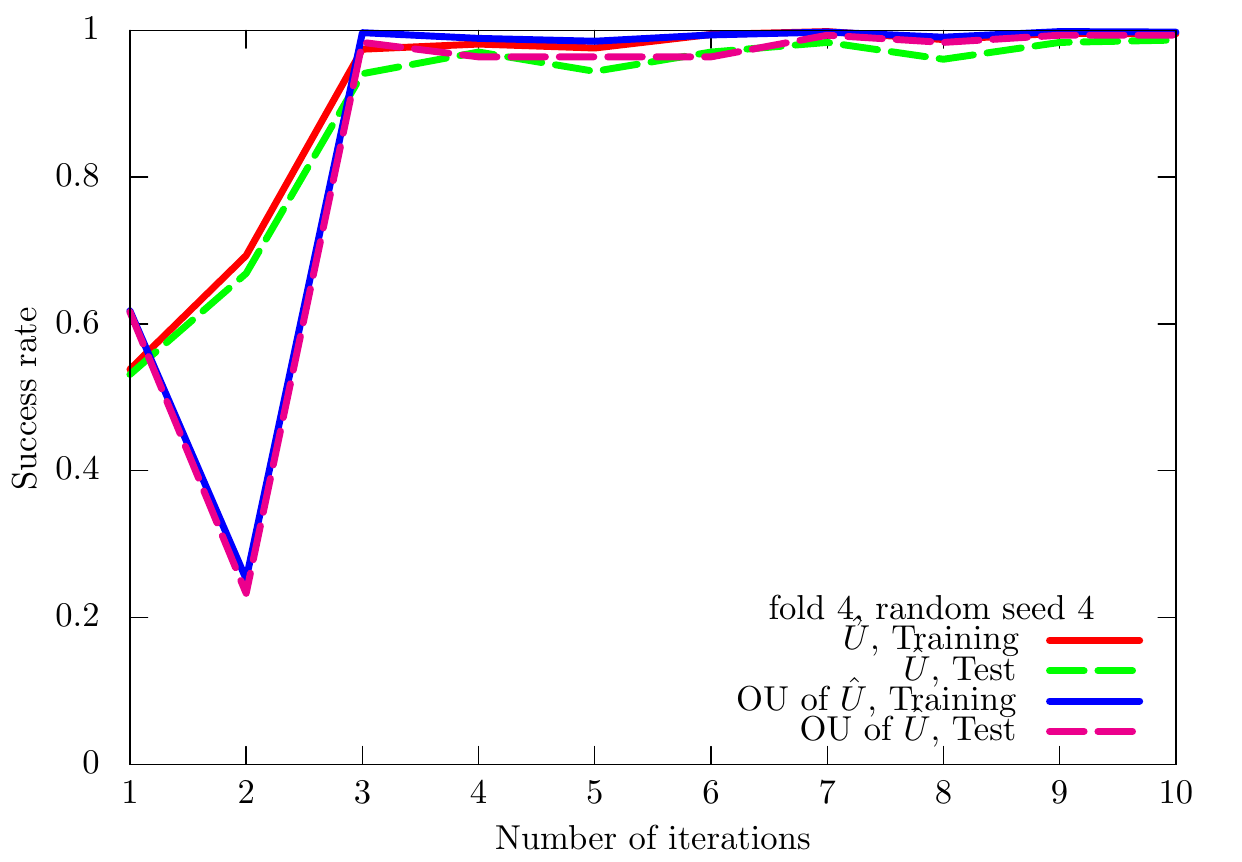}
\caption{Results of the UKM ($\hat{X}$ and OU of $\hat{X}$) on the $5$-fold datasets with $5$ different random seeds for the semeion dataset ($0$ or non-$0$). We use complex matrices and set $\theta_\mathrm{bias} = 0$. We set $r = 0.010$.}
\label{supp-arXiv-numerical-result-raw-data-fold-001-rand-001-UKM-OUU-UCI-semeion-0-non0}
\end{figure*}

In Fig.~\ref{supp-arXiv-numerical-result-raw-data-fold-001-rand-001-QCL-UCI-semeion-0-non0}, we show the numerical results of QCL for the $5$-fold datasets with $5$ different random seeds.
In Fig.~\ref{supp-arXiv-numerical-result-raw-data-fold-001-rand-001-UKM-P-UCI-semeion-0-non0}, we show the numerical results of $\hat{P}$ of the UKM for the $5$-fold datasets with $5$ different random seeds.
In Fig.~\ref{supp-arXiv-numerical-result-raw-data-fold-001-rand-001-UKM-OUU-UCI-semeion-0-non0}, we also show the numerical results of OU of $\hat{X}$ of the UKM for the $5$-fold datasets with $5$ different random seeds.

\begin{figure}[htb]
\centering
\includegraphics[scale=0.45]{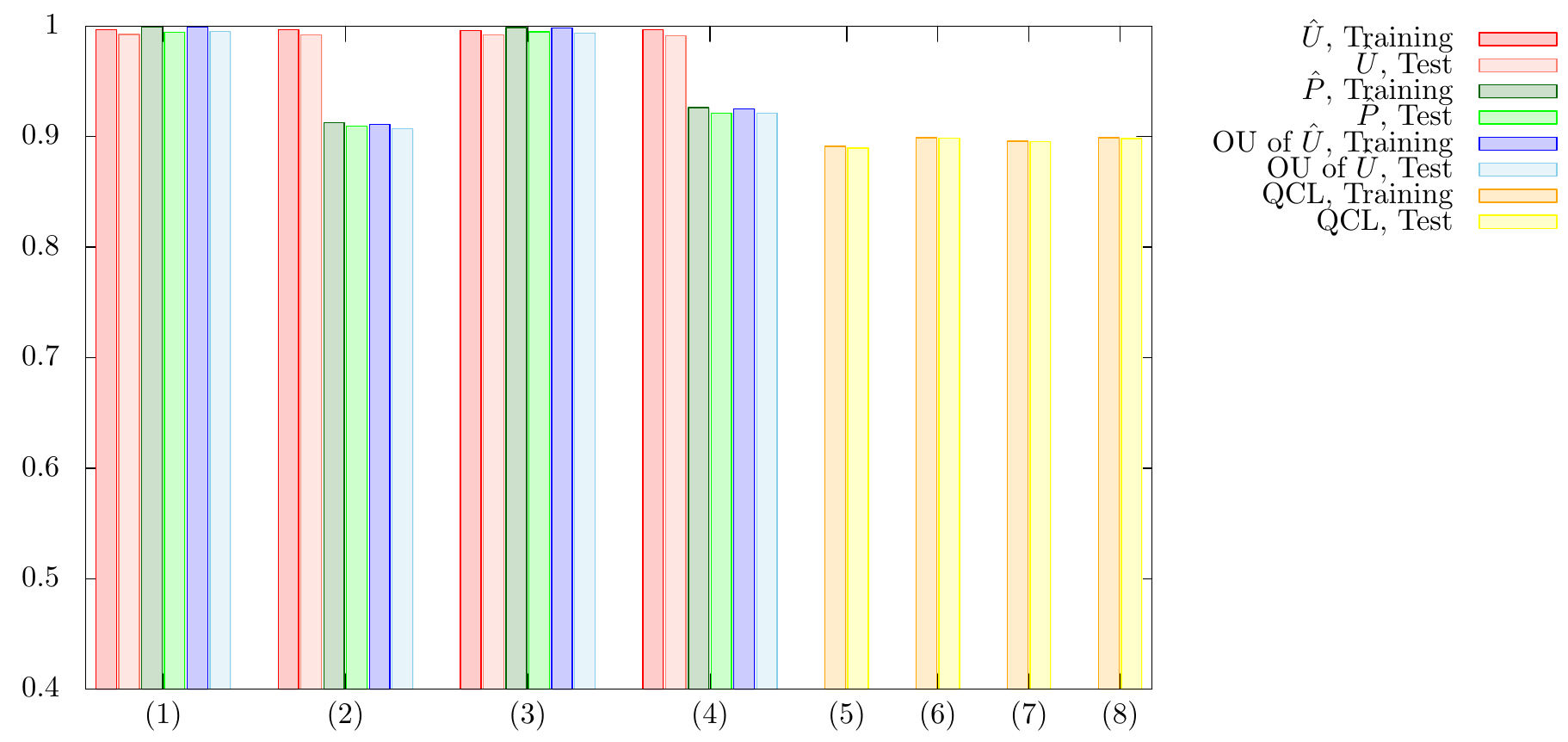}
\caption{Results of $5$-fold CV with $5$ different random seeds for the semeion dataset ($0$ or non-$0$). For the UKM, we put $r = 0.010$ and set $K = 10$ and $K' = 5$. For QCL, the number of layers $L$ is $5$ and the number of iterations is $50$. The numerical settings are as follows: (1) complex matrices without the bias term, (2) complex matrices with the bias term, (3) real matrices without the bias term, (4) real matrices with the bias term, (5) CNOT-based circuit without the bias term, (6) CNOT-based circuit with the bias term, (7) CRot-based circuit without the bias term, (8) CRot-based circuit with the bias term, (9) 1d Heisenberg circuit without the bias term, (10) 1d Heisenberg circuit with the bias term, (11) FC Heisenberg circuit without the bias term, and (12) FC Heisenberg circuit with the bias term.}
\label{supp-arXiv-numerical-result-performance-UKM-QCL-UCI-semeion-0-non0}
\end{figure}

We summarize the results of 5-fold CV with 5 different random seeds of QCL and the UKM in Tables~\ref{supp-arXiv-table-UCI-semeion-0-non0-002} and \ref{supp-arXiv-table-UCI-semeion-0-non0-001}, respectively.
For QCL and the UKM, we select the best model for the training dataset over iterations to compute the performance.
\begin{table}[htb]
  \begin{tabular}{cc|cc}
    \hline \hline
    Algo. & Condition & Training & Test \\
    \hline
    QCL & CNOT-based, w/o bias & 0.8913 & 0.8896 \\
    QCL & CNOT-based, w/ bias & 0.8989 & 0.8987 \\
    \hline
    QCL & CRot-based, w/o bias & 0.8959 & 0.8957 \\
    QCL & CRot-based, w/ bias & 0.8989 & 0.8982 \\
    \hline \hline
  \end{tabular}
\caption{Results of $5$-fold CV with $5$ different random seeds of QCL for the semeion dataset ($0$ or non-$0$). The number of layers $L$ is $5$ and the number of iterations is $50$.}
\label{supp-arXiv-table-UCI-semeion-0-non0-002}
\end{table}
\begin{table}[htb]
  \begin{tabular}{cc|cc}
    \hline \hline
    Algo. & Condition & Training & Test \\
    \hline
    UKM & $\hat{X}$, complex, w/o bias & 0.9969 & 0.9925 \\
    UKM & $\hat{P}$, complex, w/o bias & 0.9989 & 0.9945 \\
    UKM & OU of $\hat{X}$, complex, w/o bias & 0.9990 & 0.9953 \\
    \hline
    UKM & $\hat{X}$, complex, w/ bias & 0.9968 & 0.9921 \\
    UKM & $\hat{P}$, complex, w/ bias & 0.9127 & 0.9094 \\
    UKM & OU of $\hat{X}$, complex, w/ bias & 0.9113 & 0.9071 \\
    \hline
    UKM & $\hat{X}$, real, w/o bias & 0.9961 & 0.9920 \\
    UKM & $\hat{P}$, real, w/o bias & 0.9988 & 0.9949 \\
    UKM & OU of $\hat{X}$, real, w/o bias & 0.9985 & 0.9938 \\
    \hline
    UKM & $\hat{X}$, real, w/ bias & 0.9969 & 0.9912 \\
    UKM & $\hat{P}$, real, w/ bias & 0.9263 & 0.9214 \\
    UKM & OU of $\hat{X}$, real, w/ bias & 0.9251 & 0.9213 \\
    \hline \hline
  \end{tabular}
\caption{Results of $5$-fold CV with $5$ different random seeds of the UKM for the semeion dataset ($0$ or non-$0$). We put $r = 0.010$ and set $K = 10$ and $K' = 5$.}
\label{supp-arXiv-table-UCI-semeion-0-non0-001}
\end{table}
In Fig.~\ref{supp-arXiv-numerical-result-performance-UKM-QCL-UCI-semeion-0-non0}, we plot the data shown in Tables~\ref{supp-arXiv-table-UCI-semeion-0-non0-002} and \ref{supp-arXiv-table-UCI-semeion-0-non0-001}.
We also summarize the results of 5-fold CV with 5 different random seeds of the kernel method in Table~\ref{supp-arXiv-table-UCI-semeion-0-non0-003}.
More specifically, we use Ridge classification in Sec.~\ref{supp-arXiv-sec-Ridge-001}.
We consider the linear functions and the second-order polynomial functions for $\phi (\cdot)$ in Eq.~\eqref{supp-arXiv-f-pred-kernel-method-001-002} with and without normalization.
We set $\lambda = 10^{-2}, 10^{-1}, 1$ where $\lambda$ is the coefficient of the regularization term.
\begin{table}[htb]
  \begin{tabular}{cc|cc}
    \hline \hline
    Algo. & Condition & Training & Test \\
    \hline
  Kernel method & Linear, w/o normalization, $\lambda = 10^{-2}$ & 0.9937 & 0.9887 \\
  Kernel method & Linear, w/o normalization, $\lambda = 10^{-1}$ & 0.9937 & 0.9887 \\
  Kernel method & Linear, w/o normalization, $\lambda = 1$ & 0.9937 & 0.9887 \\
    \hline
  Kernel method & Linear, w/ normalization, $\lambda = 10^{-2}$ & 0.9936 & 0.9873 \\
  Kernel method & Linear, w/ normalization, $\lambda = 10^{-1}$ & 0.9934 & 0.9892 \\
  Kernel method & Linear, w/ normalization, $\lambda = 1$ & 0.9931 & 0.9898 \\
    \hline
  Kernel method & Poly-2, w/o normalization, $\lambda = 10^{-2}$ & 1.0000 & 0.9956 \\
  Kernel method & Poly-2, w/o normalization, $\lambda = 10^{-1}$ & 1.0000 & 0.9956 \\
  Kernel method & Poly-2, w/o normalization, $\lambda = 1$ & 1.0000 & 0.9956 \\
    \hline
  Kernel method & Poly-2, w/ normalization, $\lambda = 10^{-2}$ & 1.0000 & 0.9955 \\
  Kernel method & Poly-2, w/ normalization, $\lambda = 10^{-1}$ & 1.0000 & 0.9955 \\
  Kernel method & Poly-2, w/ normalization, $\lambda = 1$ & 0.9980 & 0.9949 \\
    \hline \hline
  \end{tabular}
\caption{Results of 5-fold CV with 5 different random seeds of the kernel method for the semeion dataset ($0$ or non-$0$).}
\label{supp-arXiv-table-UCI-semeion-0-non0-003}
\end{table}

\begin{figure}[htb]
\centering
\includegraphics[scale=0.45]{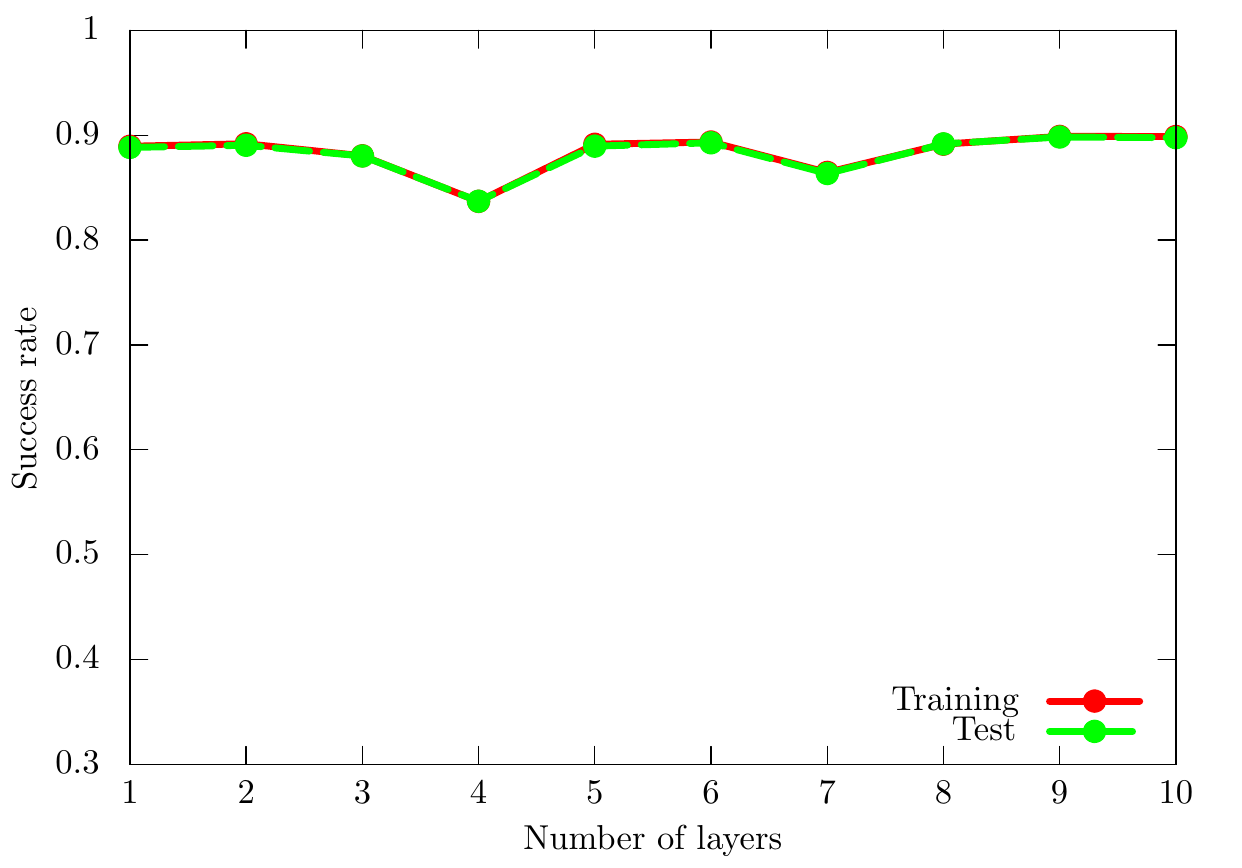}
\caption{Performance dependence of QCL on the number of layers $L$ for the semeion dataset ($0$ or non-$0$). We use the CNOT-based circuit geometry and set $\theta_\mathrm{bias} = 0$. We iterate the computation $50$ times.}
\label{supp-arXiv-numerical-result-layers-dependence-QCL-UCI-semeion-0-non0}
\end{figure}

\begin{figure}[htb]
\centering
\includegraphics[scale=0.45]{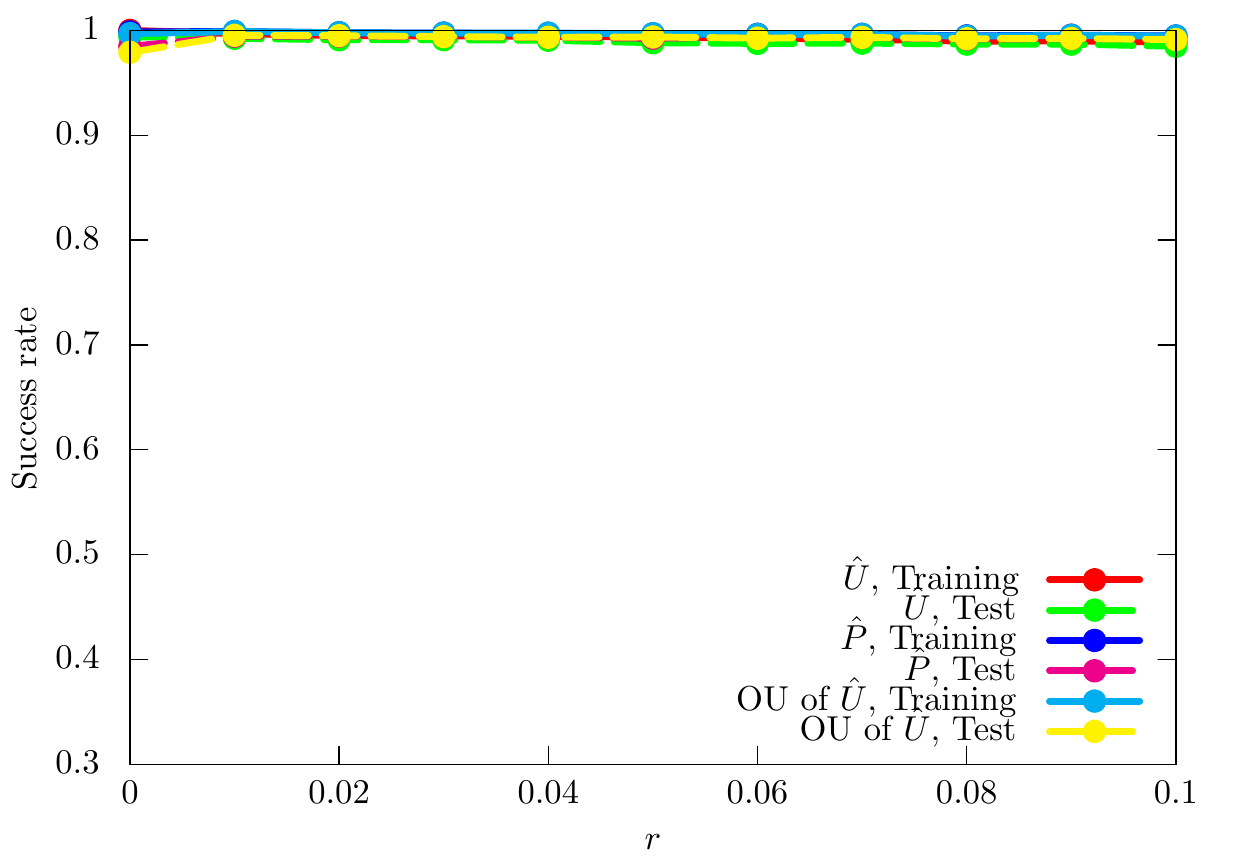}
\includegraphics[scale=0.45]{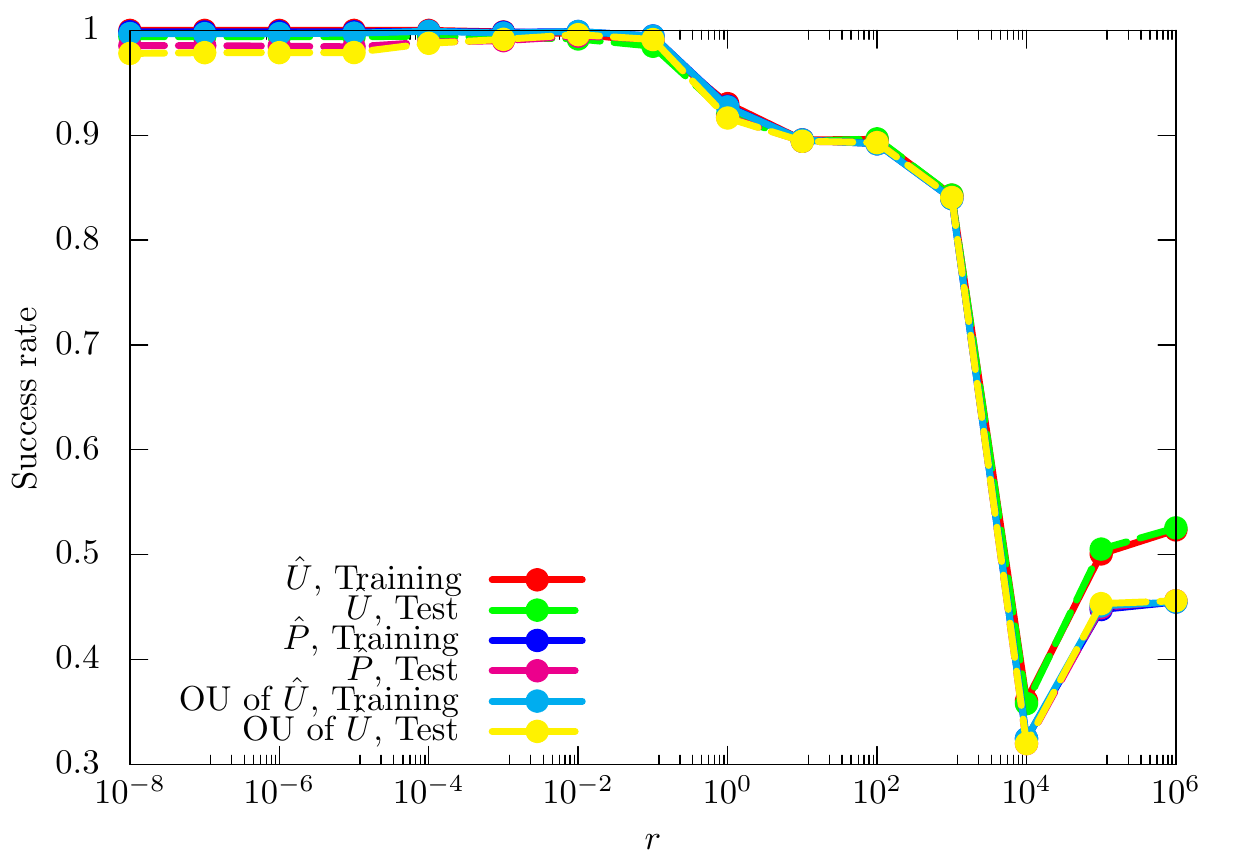}
\caption{Performance dependence of the UKM on $r$, which is the coefficient of the second term in the right-hand side of Eq.~\eqref{supp-arXiv-quantum-kernel-method-001-011} for the semeion dataset ($0$ or non-$0$). We use complex matrices and set $\theta_\mathrm{bias} = 0$. We set $K = 10$ and $K' = 5$.}
\label{supp-arXiv-numerical-result-r-dependence-UKM-UCI-semeion-0-non0}
\end{figure}

\begin{figure}[htb]
\centering
\includegraphics[scale=0.45]{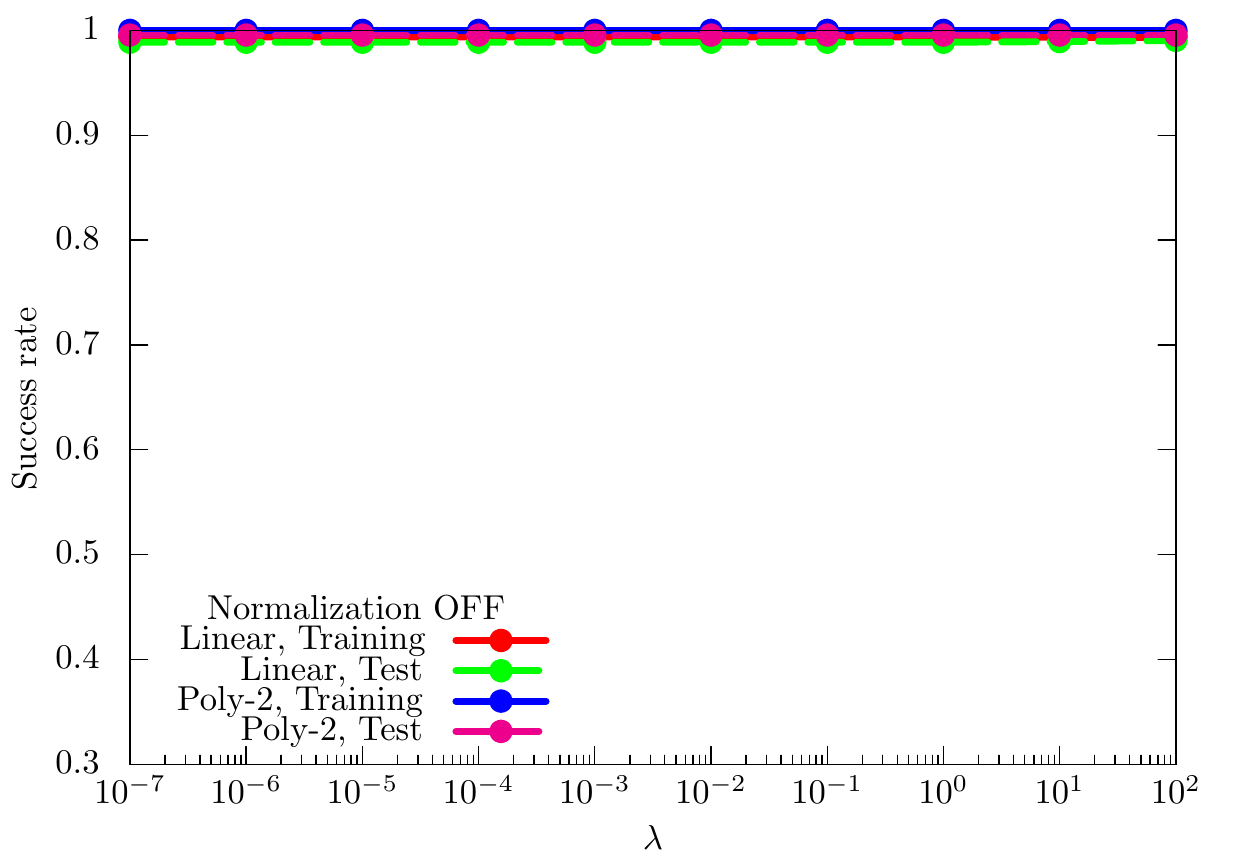}
\includegraphics[scale=0.45]{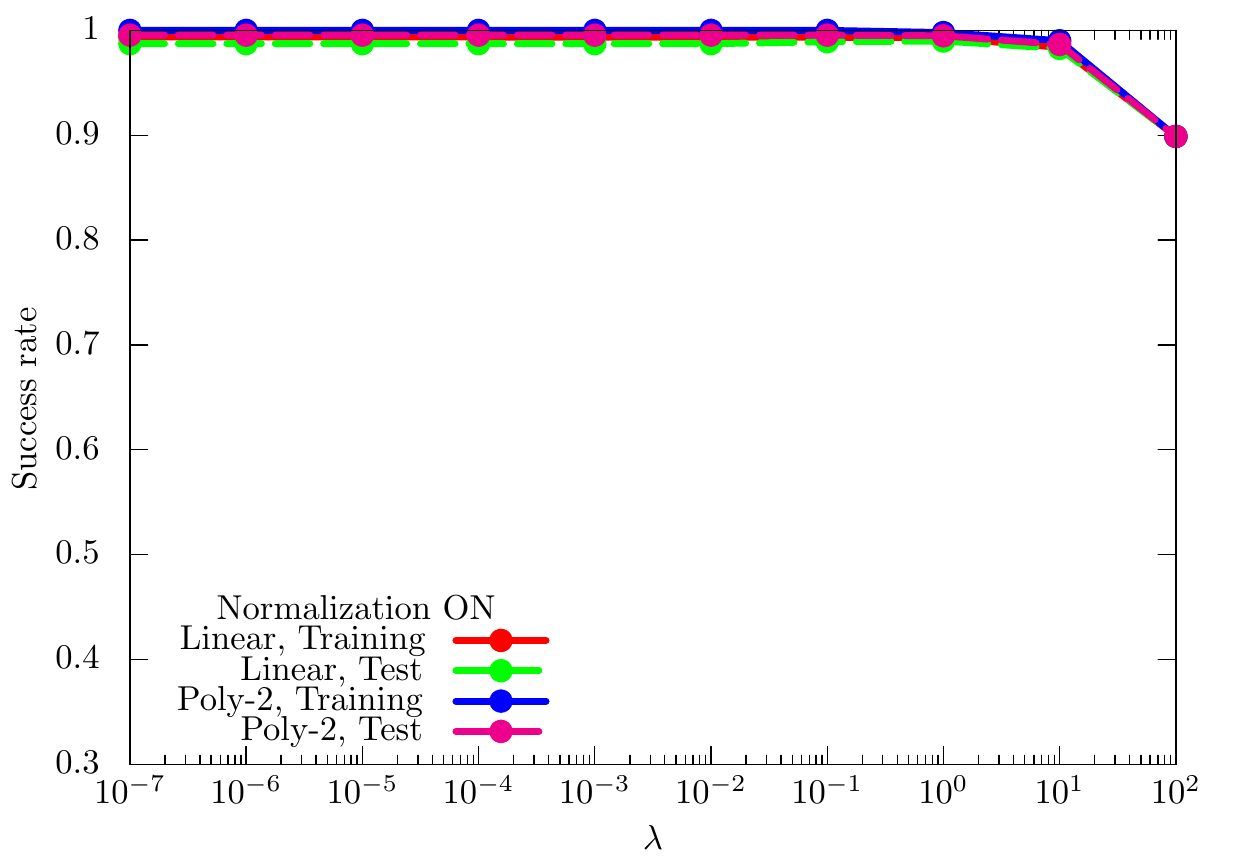}
\caption{Performance dependence of the kernel method on $\lambda$, which is the coefficient of the second term in the right-hand side of Eq.~\eqref{supp-arXiv-cost-function-kernel-method-001-002} for the semeion dataset ($0$ or non-$0$). For $\phi (\cdot)$ in Eq.~\eqref{supp-arXiv-f-pred-kernel-method-001-002}, we use the linear functions and the second-degree polynomial functions with and without normalization.}
\label{supp-arXiv-numerical-result-lambda-dependence-kernel-method-semeion-0-non0}
\end{figure}

Next, we show the performance dependence of the three algorithms on their key parameters.
We see the performance dependence of QCL on the number of layers $L$.
The result is shown in Fig.~\ref{supp-arXiv-numerical-result-layers-dependence-QCL-UCI-semeion-0-non0}.
We then see the performance dependence of the UKM on $r$, which is the coefficient of the second term in the right-hand side of Eq.~\eqref{supp-arXiv-quantum-kernel-method-001-011}.
The result is shown in Fig.~\ref{supp-arXiv-numerical-result-r-dependence-UKM-UCI-semeion-0-non0}.
In Fig.~\ref{supp-arXiv-numerical-result-lambda-dependence-kernel-method-semeion-0-non0}, we show the performance dependence of the kernel method on $\lambda$, which is the coefficient of the second term in the right-hand side of Eq.~\eqref{supp-arXiv-cost-function-kernel-method-001-002}.

\begin{figure}[htb]
\centering
\includegraphics[scale=0.45]{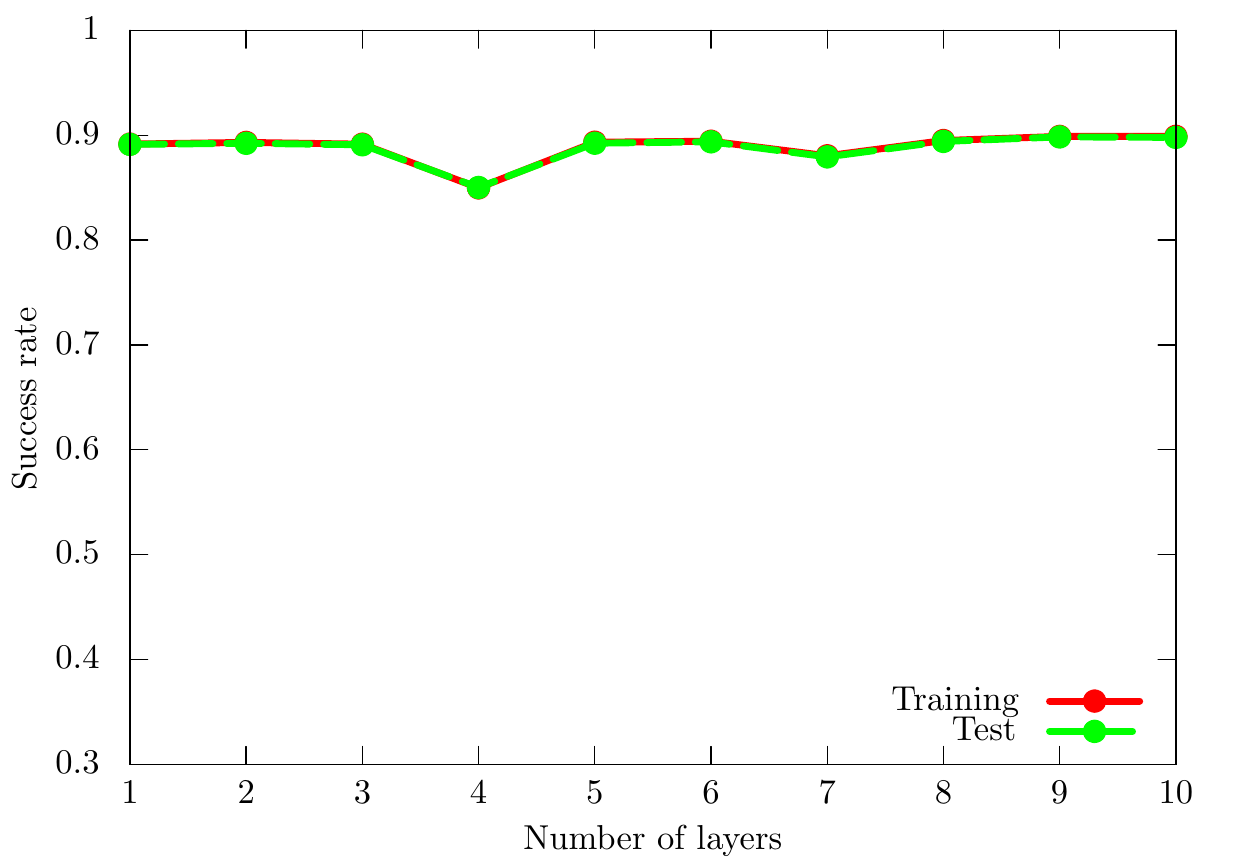}
\caption{Performance dependence of QCL on the number of layers $L$ for the semeion dataset ($0$ or non-$0$) in the case of the hinge function, Eq.~\eqref{supp-arXiv-hinge-function-001-001}. We use the CNOT-based circuit geometry and set $\theta_\mathrm{bias} = 0$. We iterate the computation $300$ times.}
\label{supp-arXiv-numerical-result-layers-dependence-QCL-UCI-semeion-0-non0-hinge}
\end{figure}

\begin{figure}[htb]
\centering
\includegraphics[scale=0.45]{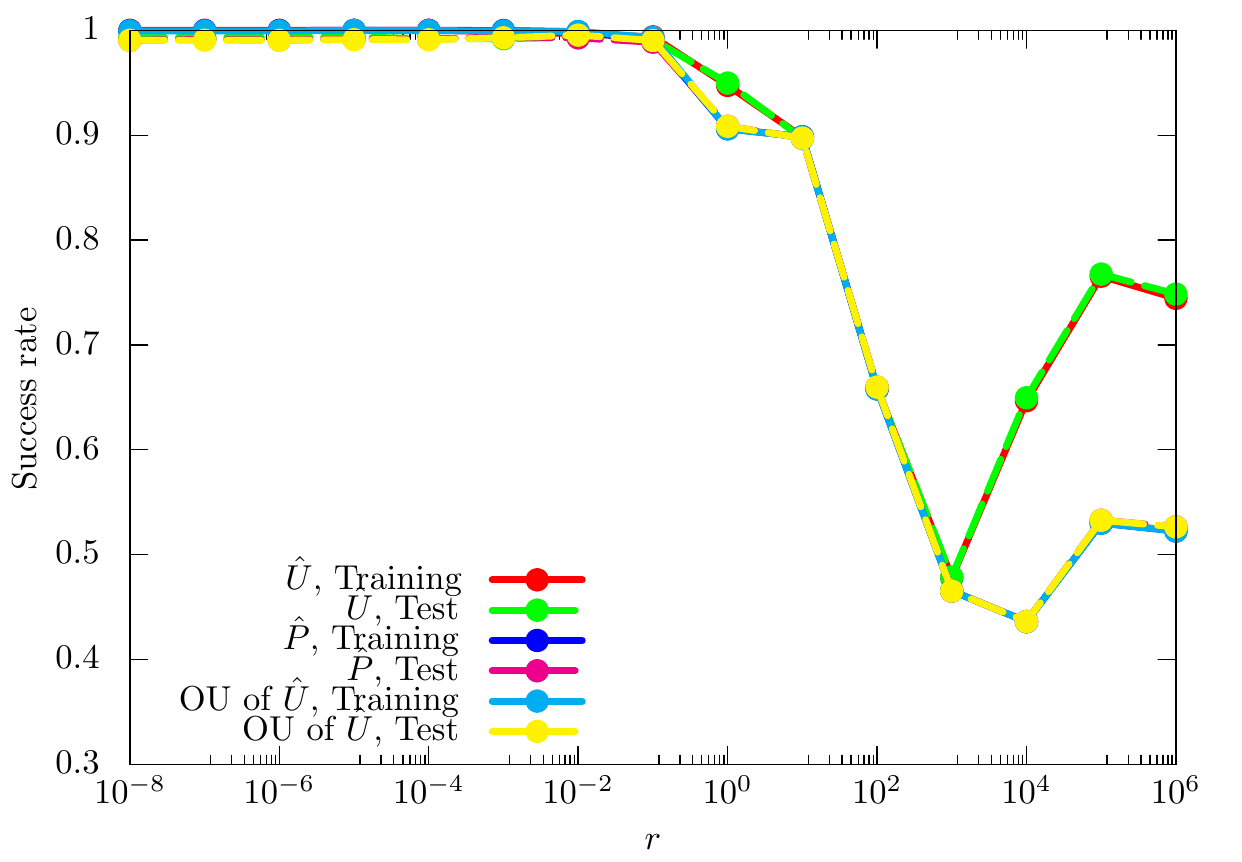}
\caption{Performance dependence of the UKM on $r$, which is the coefficient of the second term in the right-hand side of Eq.~\eqref{supp-arXiv-quantum-kernel-method-001-011} for the semeion dataset ($0$ or non-$0$) in the case of the hinge function, Eq.~\eqref{supp-arXiv-hinge-function-001-001}. We use complex matrices and set $\theta_\mathrm{bias} = 0$. We set $K = 30$ and $K' = 10$.}
\label{supp-arXiv-numerical-result-r-dependence-UKM-UCI-semeion-0-non0-hinge}
\end{figure}

So far, we have used the squared error function, Eq.~\eqref{supp-arXiv-squared-error-function-001-001}.
In Fig.~\ref{supp-arXiv-numerical-result-layers-dependence-QCL-UCI-semeion-0-non0-hinge}, we show the performance dependence of QCL on the number of layers $L$ in the case of the hinge function, Eq.~\eqref{supp-arXiv-hinge-function-001-001}.
In Fig.~\ref{supp-arXiv-numerical-result-r-dependence-UKM-UCI-semeion-0-non0-hinge}, we show the performance dependence of the UKM on $r$, which is the coefficient of the second term in the right-hand side of Eq.~\eqref{supp-arXiv-quantum-kernel-method-001-011}, in the case of the hinge function, Eq.~\eqref{supp-arXiv-hinge-function-001-001}.

\clearpage

\subsection{MNIST256 dataset ($0$ or $1$)}

We here show the numerical result for the MNIST256 dataset ($0$ or $1$).
For the UKM, we put $r = 0.010$ and set $K = 20$ and $K' = 10$ in Algo.~\ref{supp-arXiv-quantum-kernel-method-002-001}.
For QCL, we run iterations $100$ times.

\begin{figure*}[htb]
\centering
\includegraphics[scale=0.25]{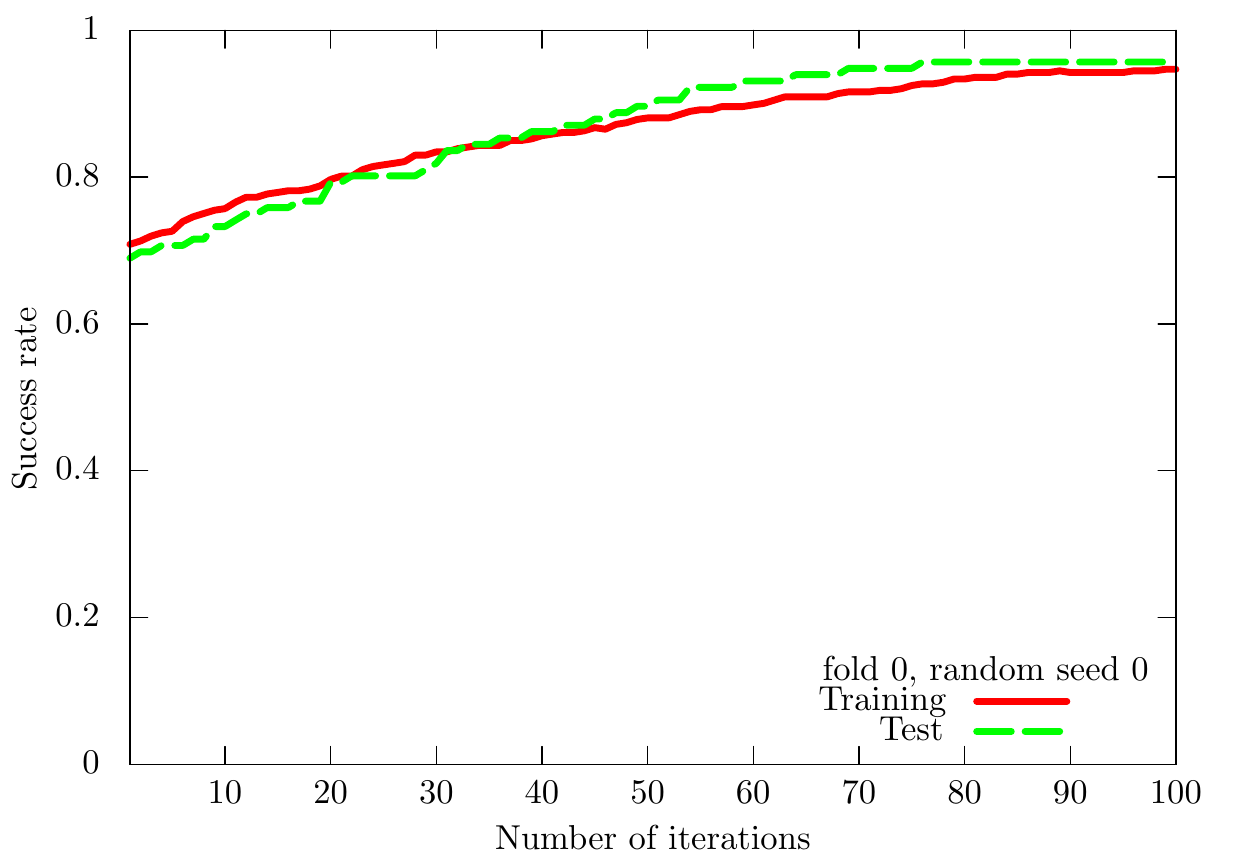}
\includegraphics[scale=0.25]{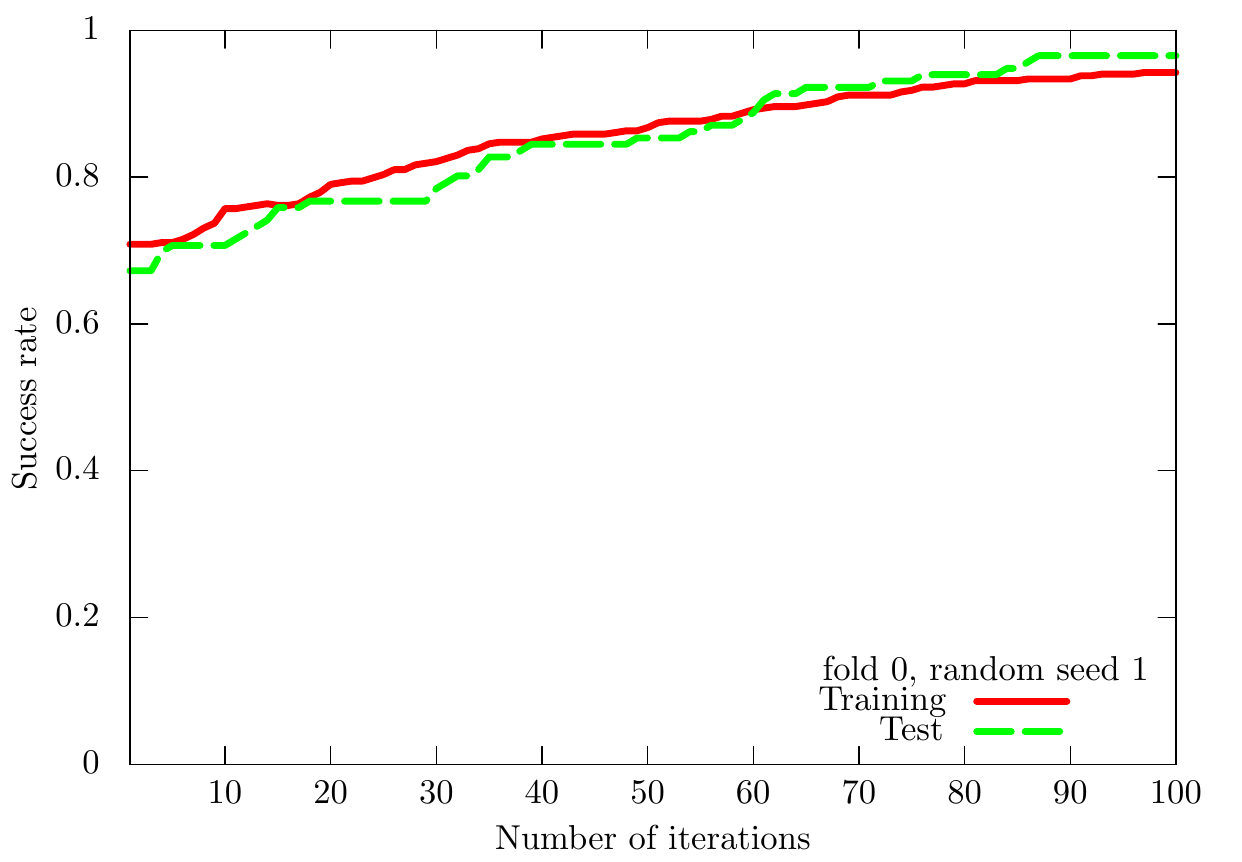}
\includegraphics[scale=0.25]{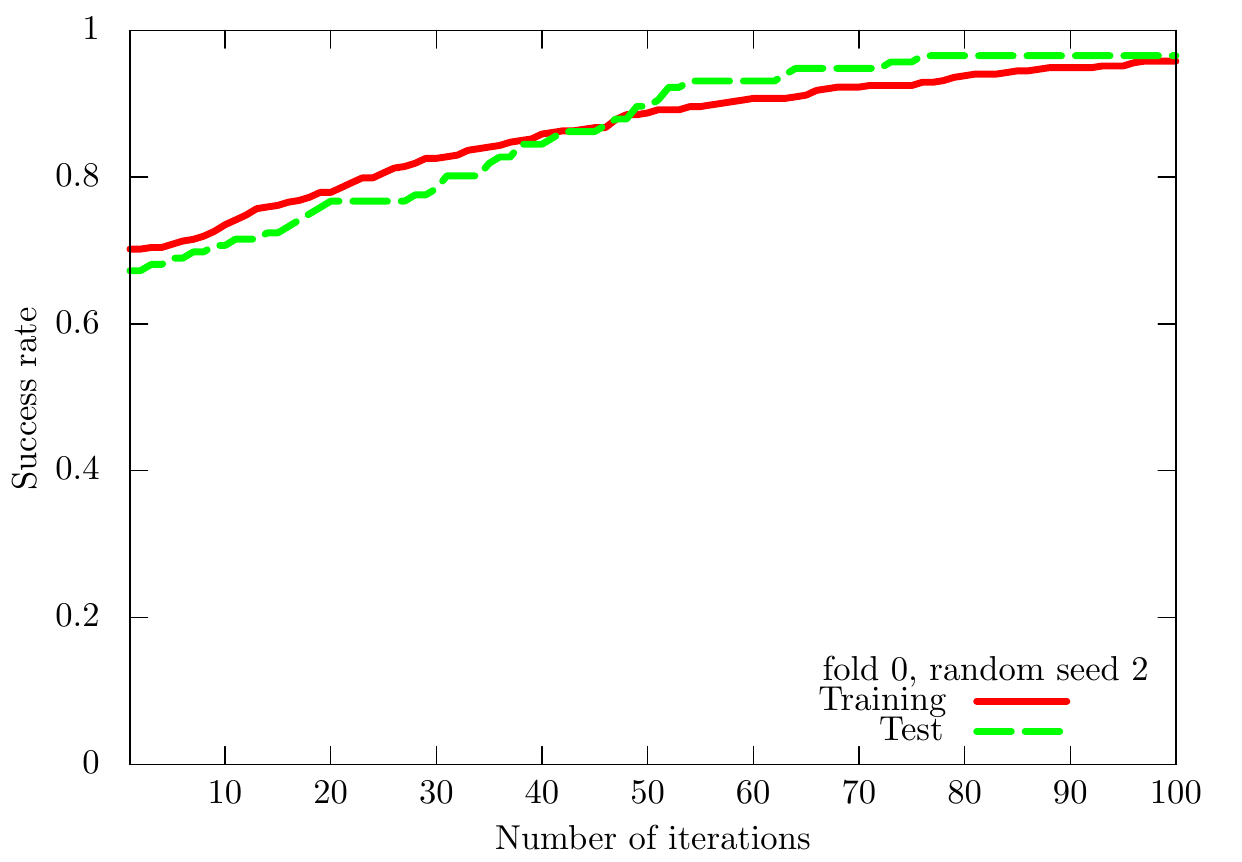}
\includegraphics[scale=0.25]{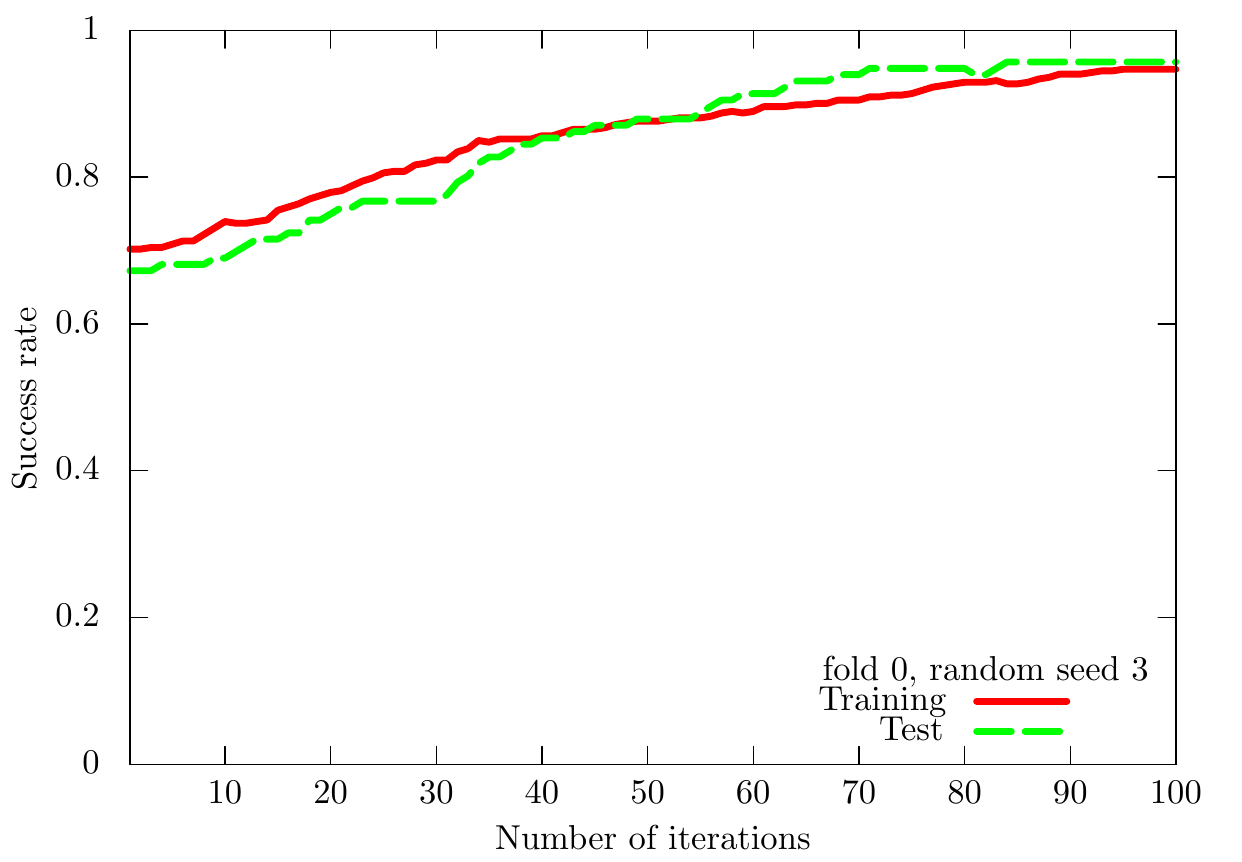}
\includegraphics[scale=0.25]{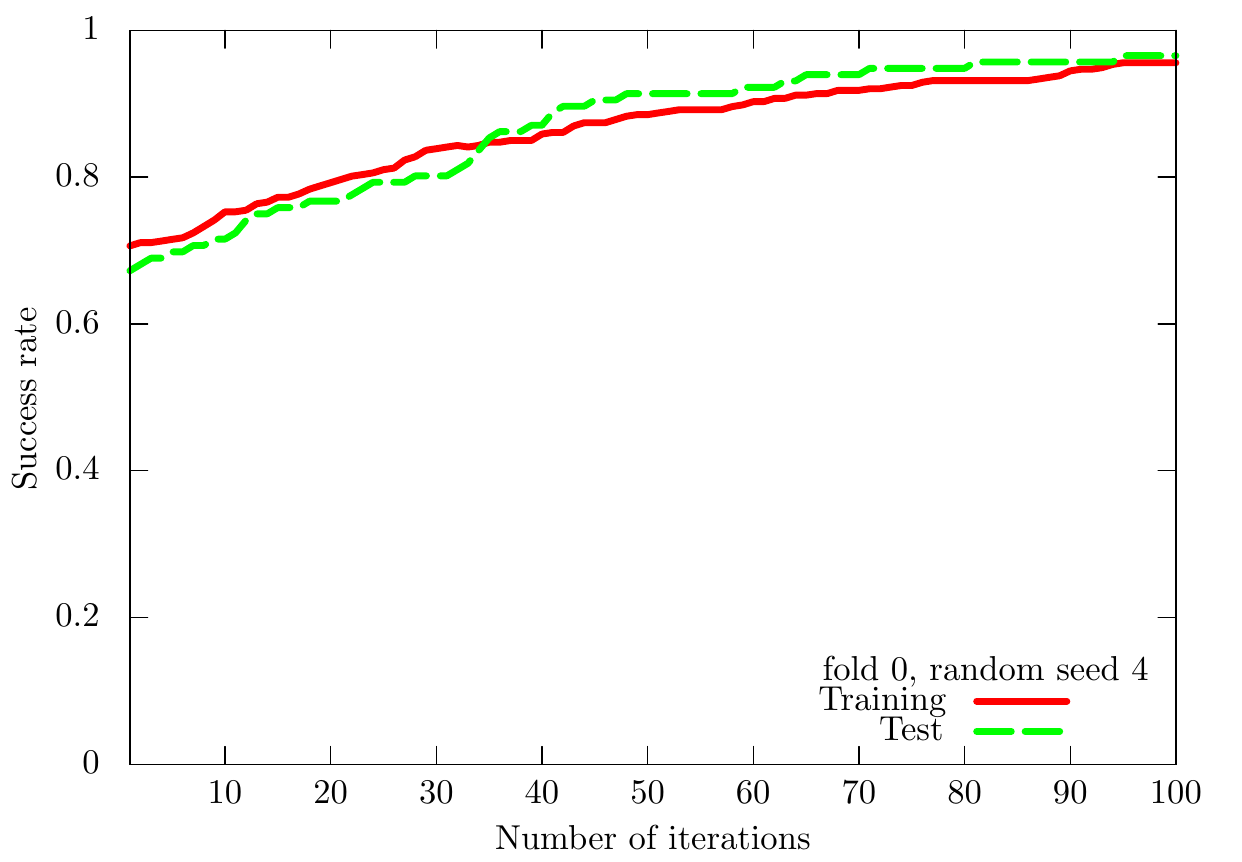}
\includegraphics[scale=0.25]{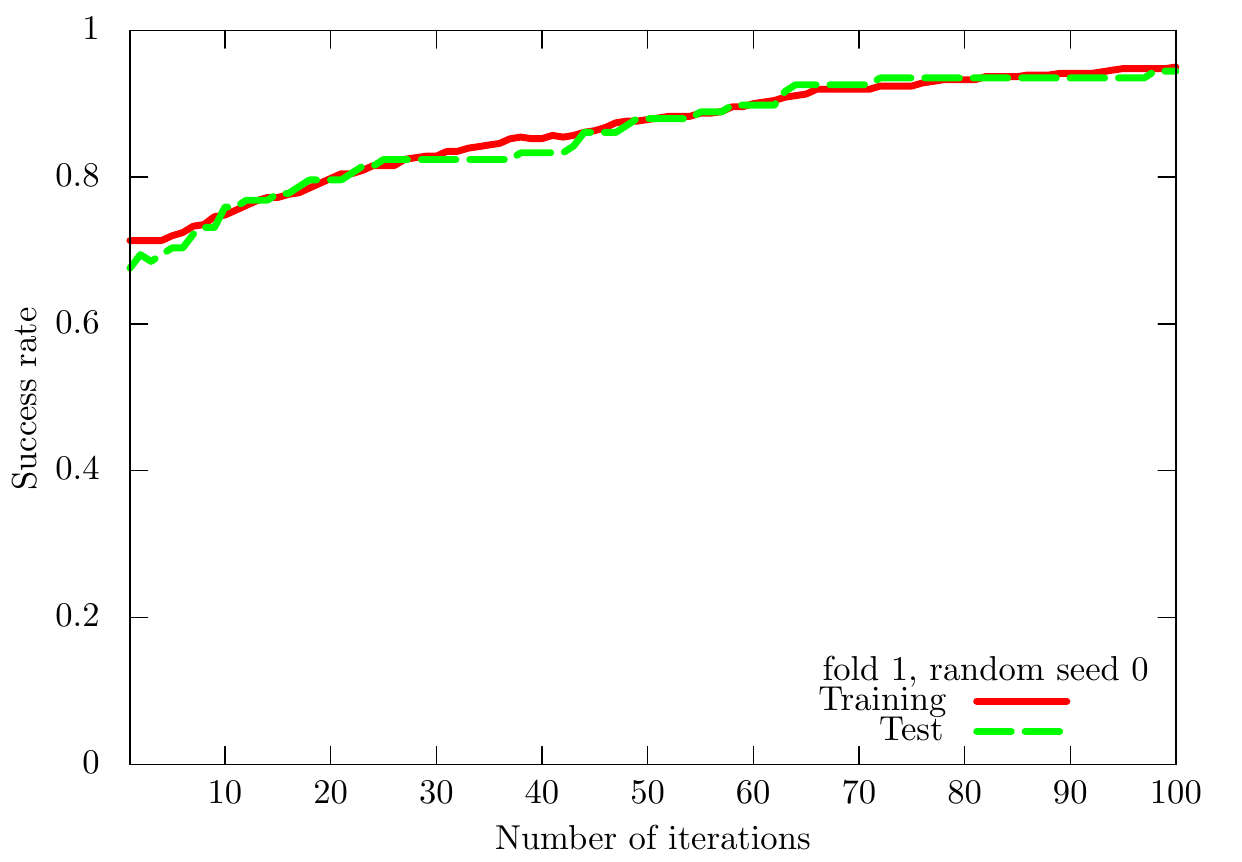}
\includegraphics[scale=0.25]{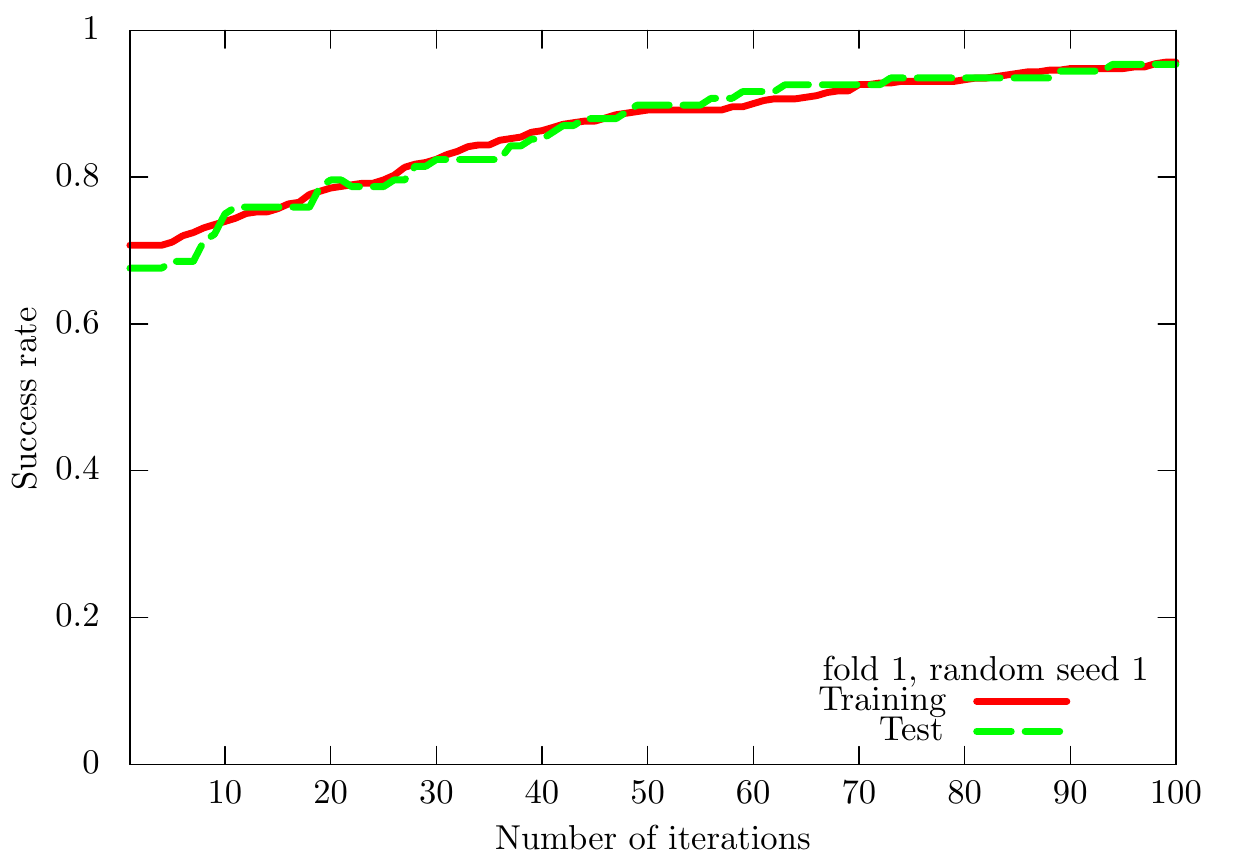}
\includegraphics[scale=0.25]{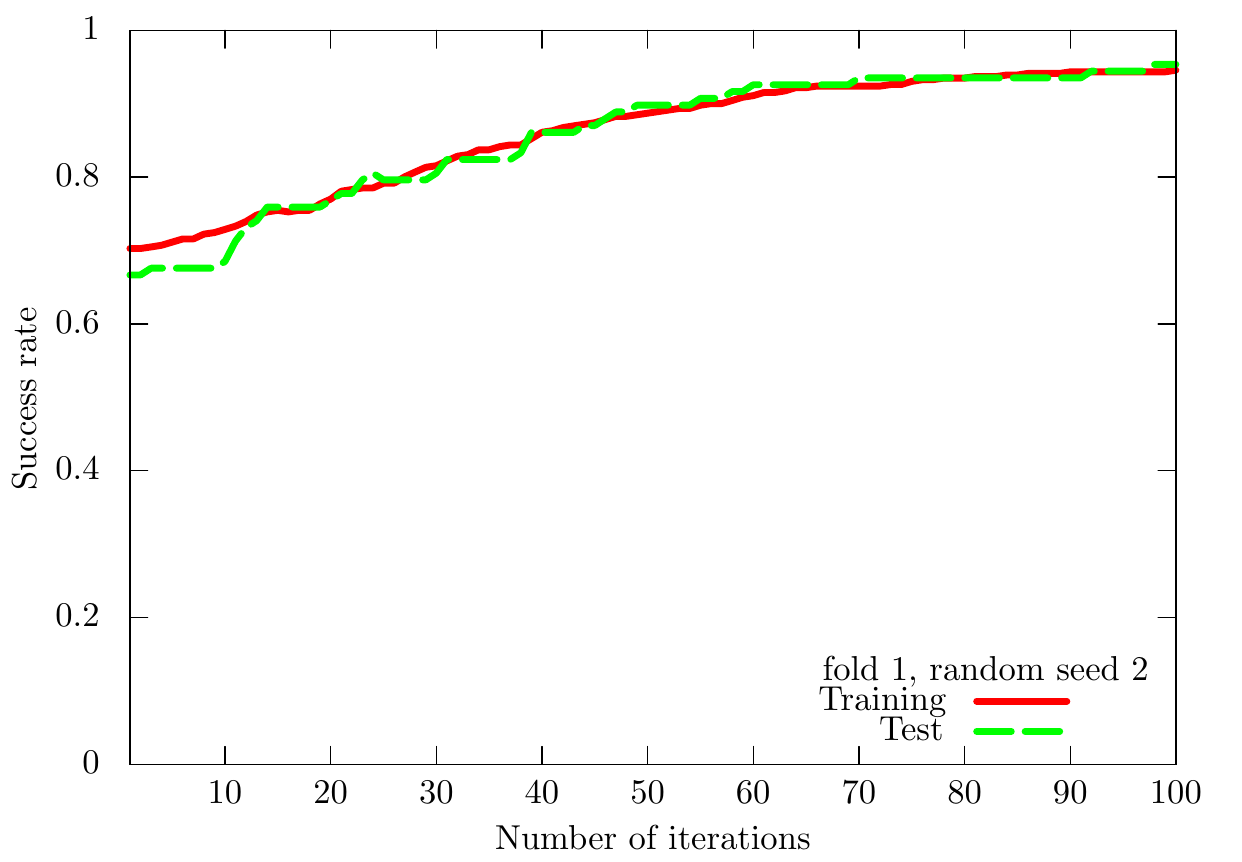}
\includegraphics[scale=0.25]{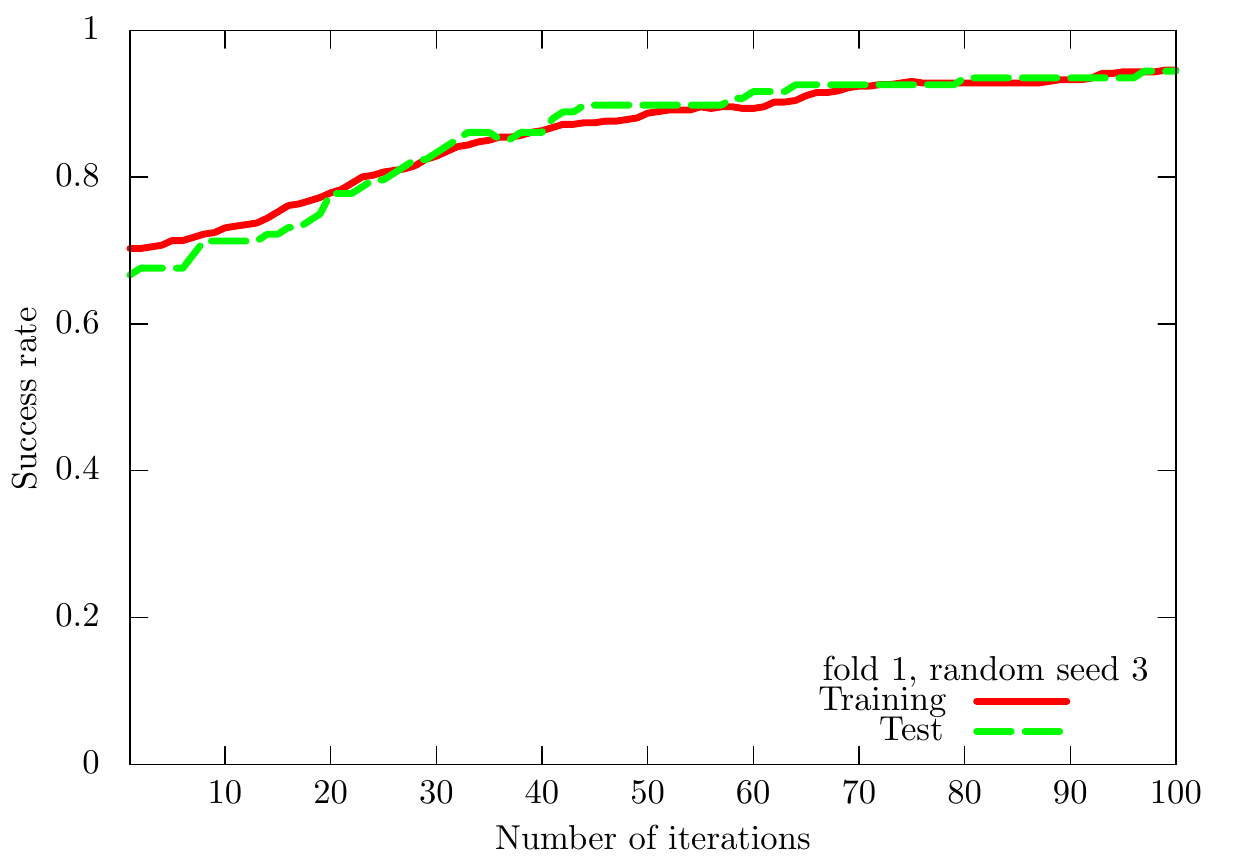}
\includegraphics[scale=0.25]{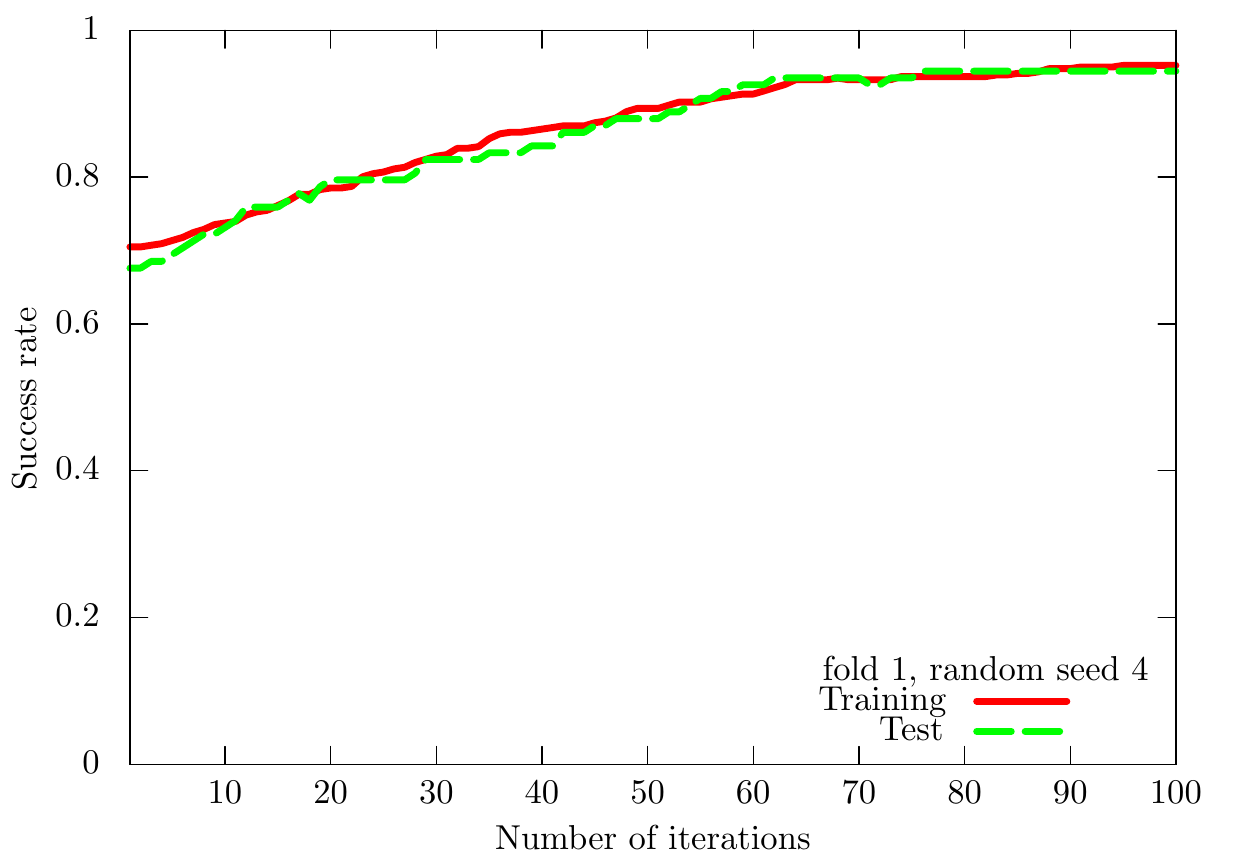}
\includegraphics[scale=0.25]{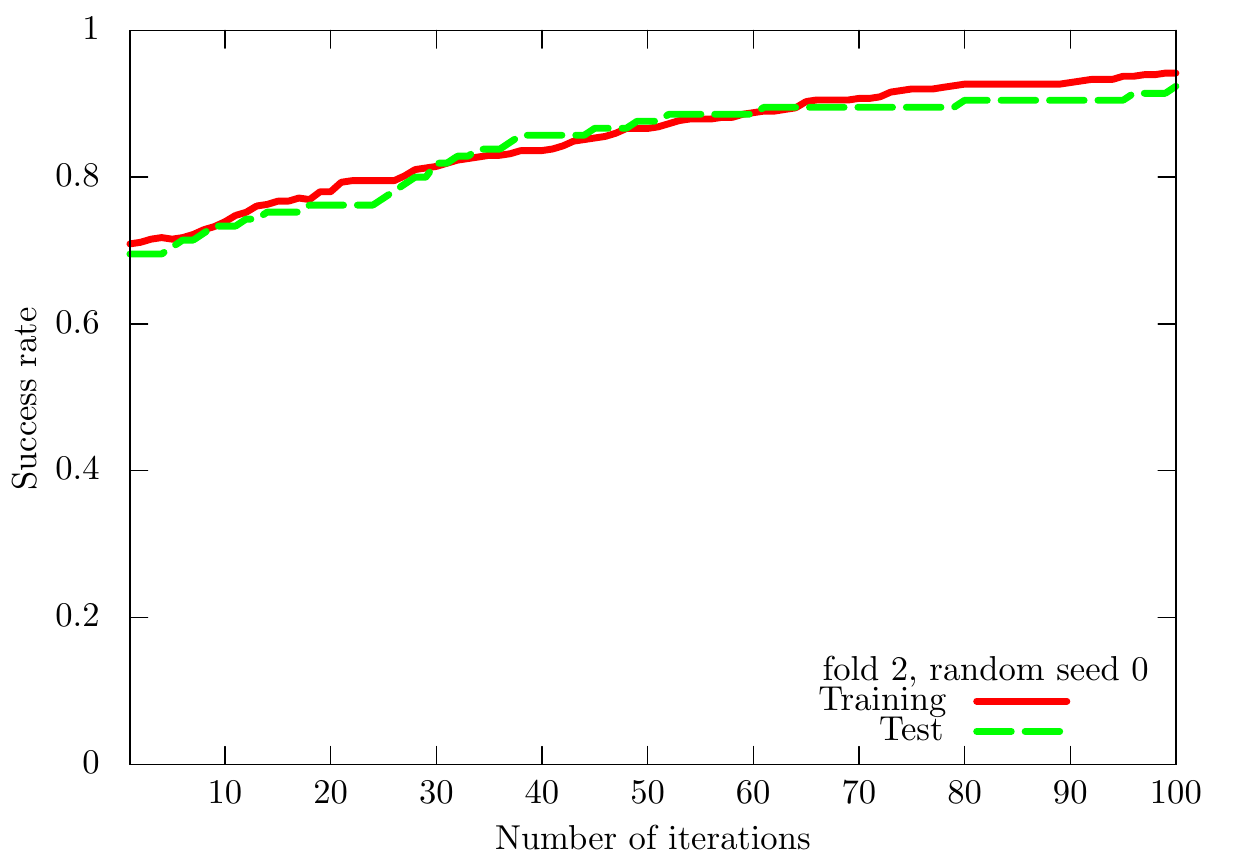}
\includegraphics[scale=0.25]{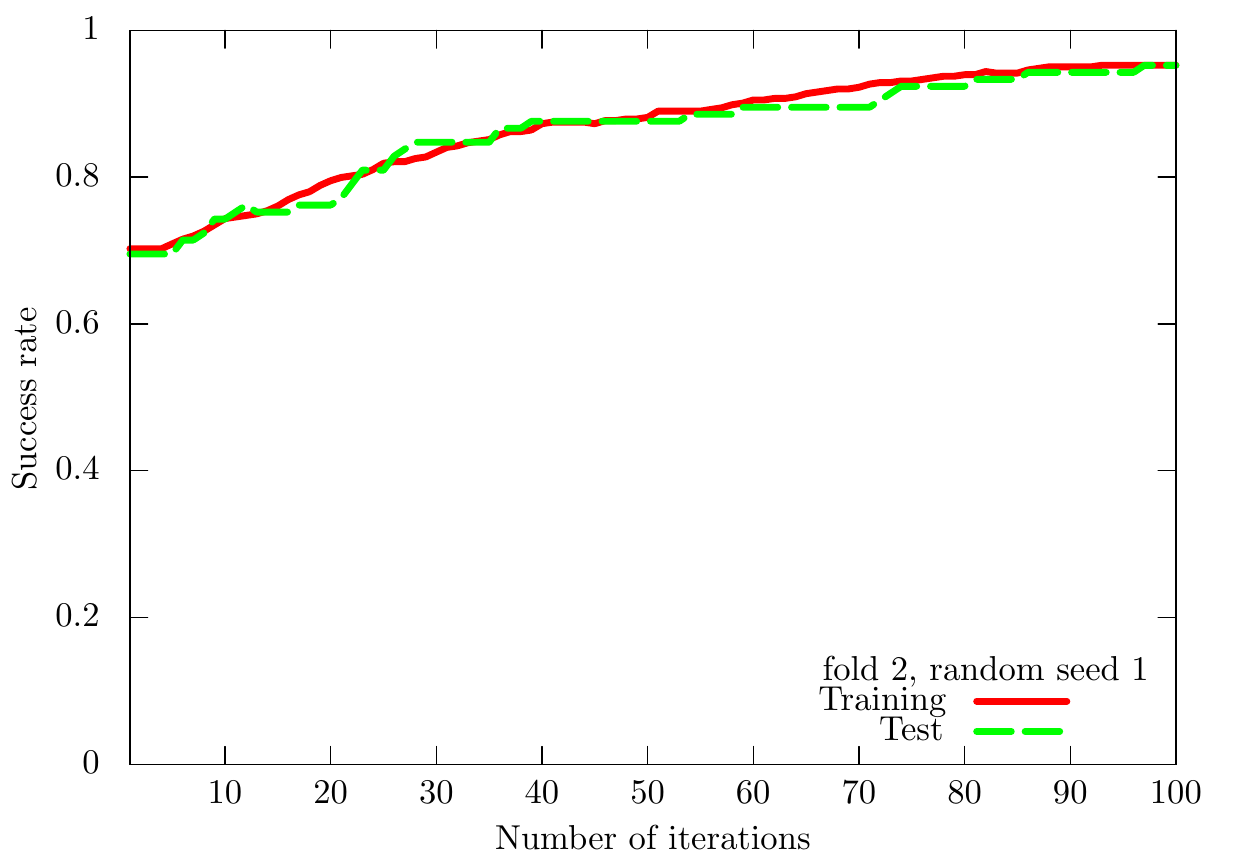}
\includegraphics[scale=0.25]{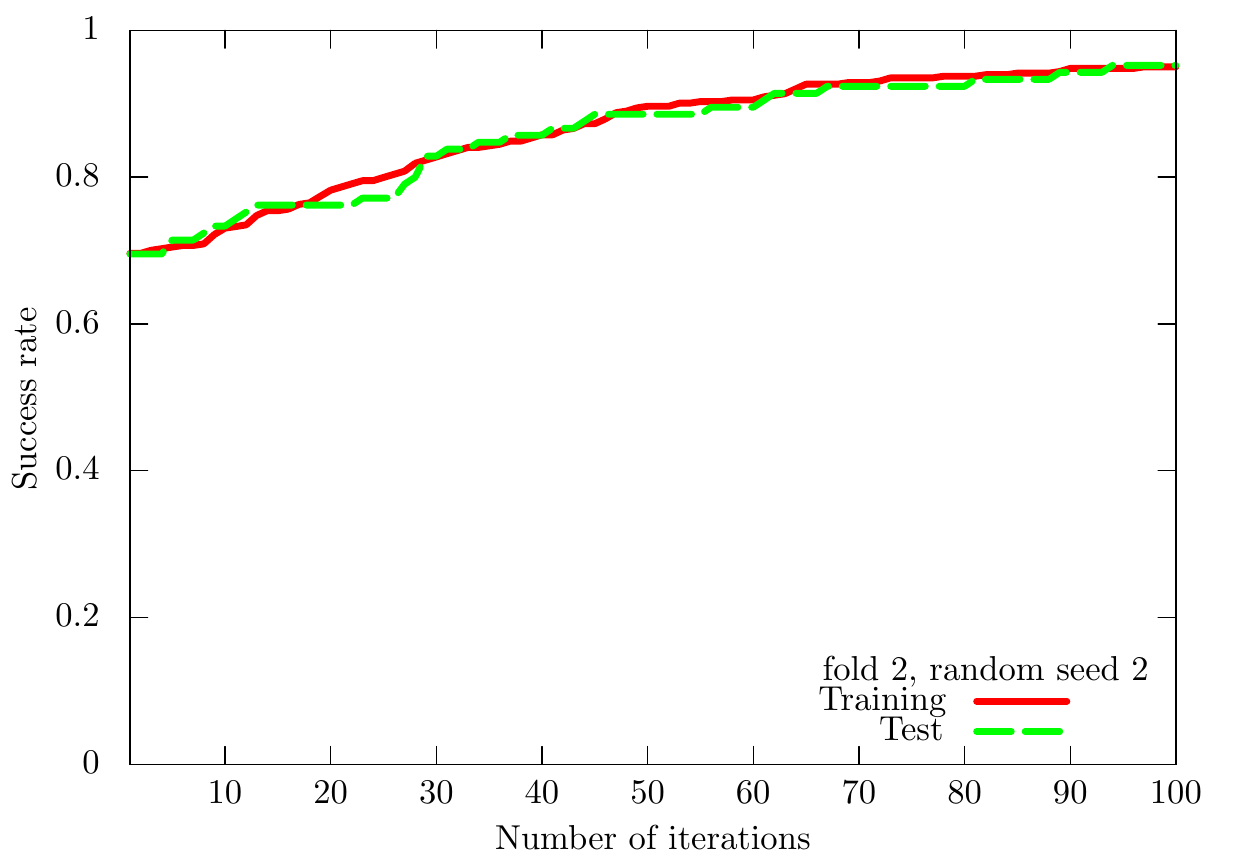}
\includegraphics[scale=0.25]{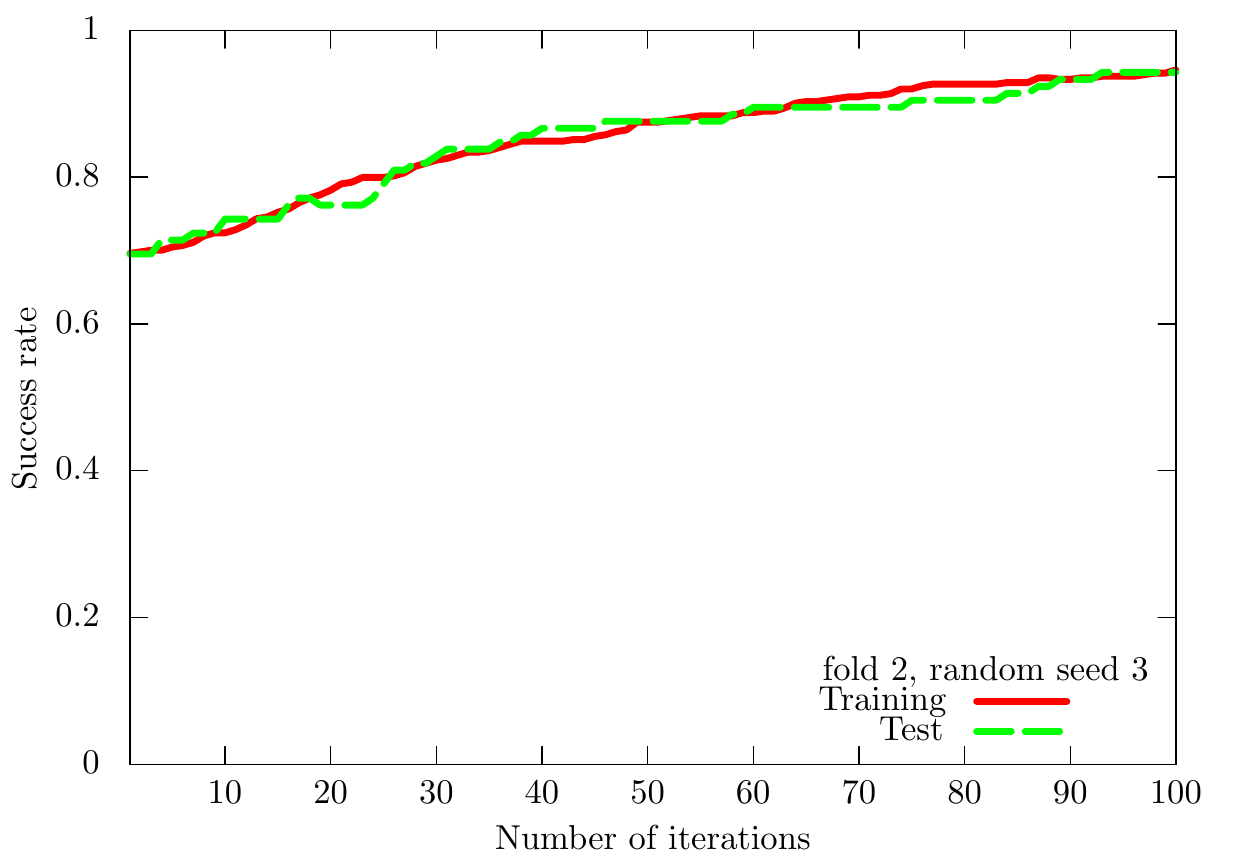}
\includegraphics[scale=0.25]{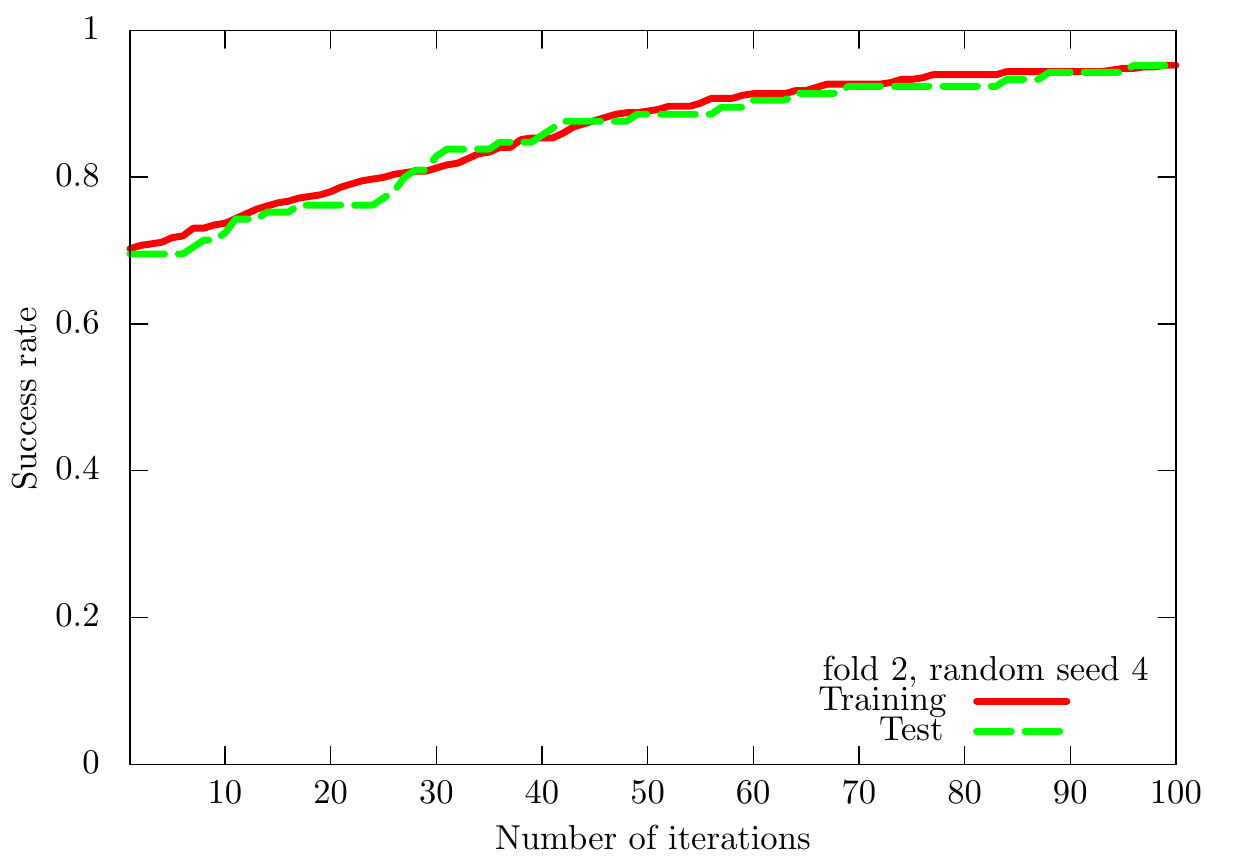}
\includegraphics[scale=0.25]{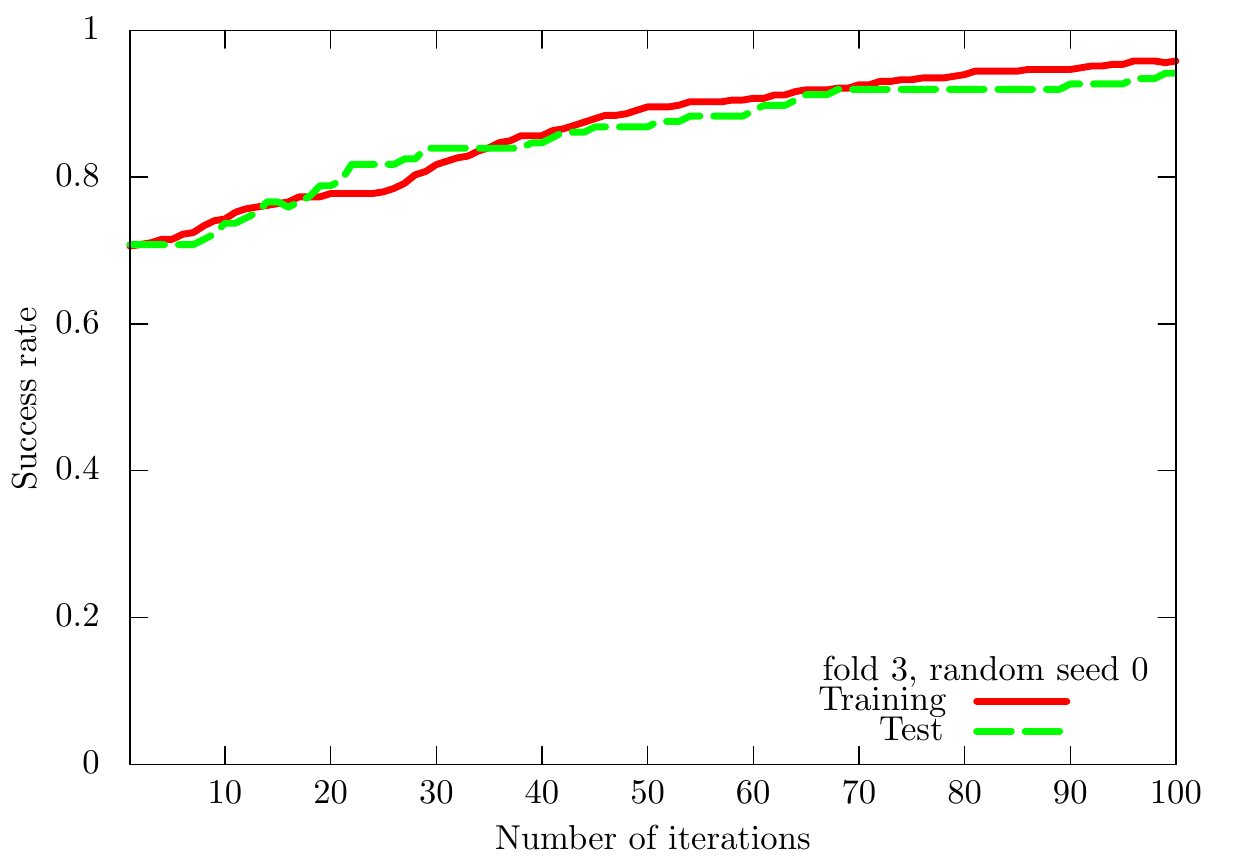}
\includegraphics[scale=0.25]{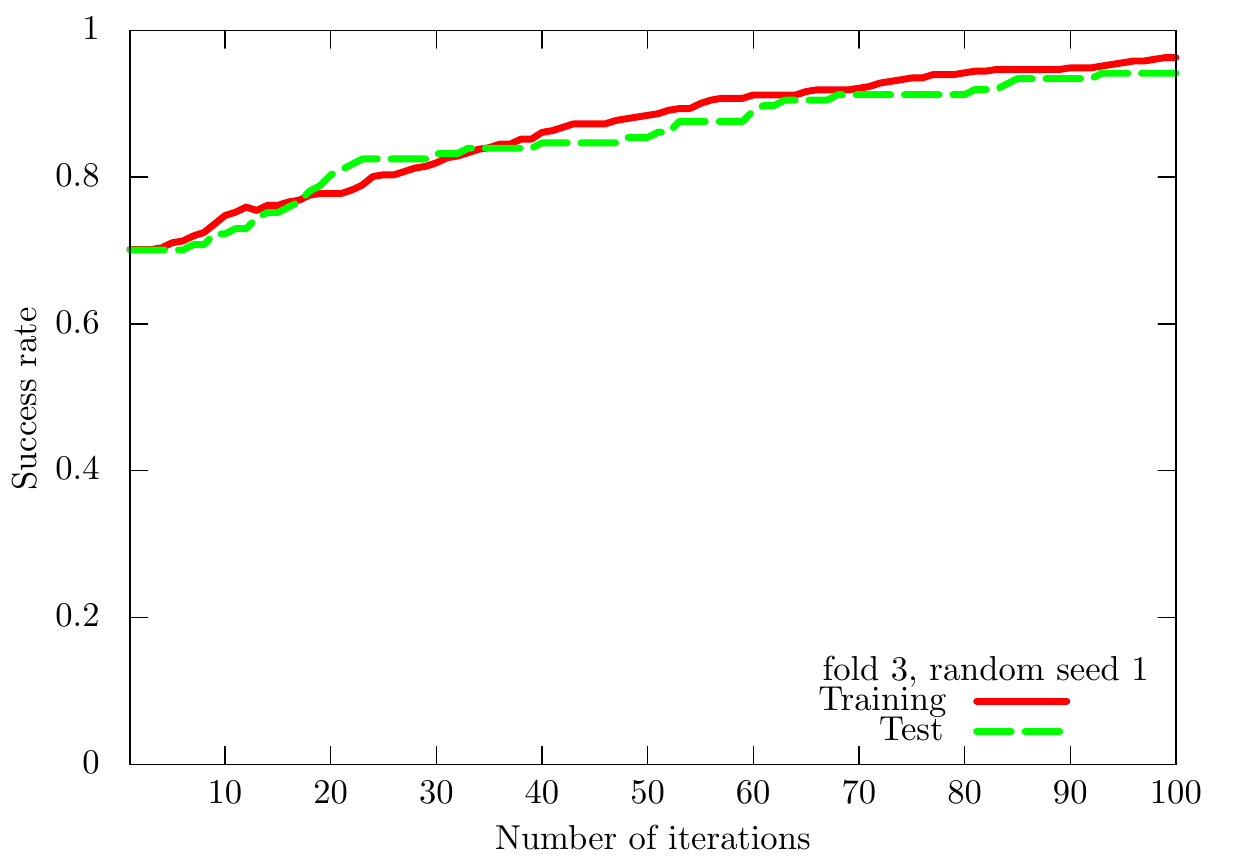}
\includegraphics[scale=0.25]{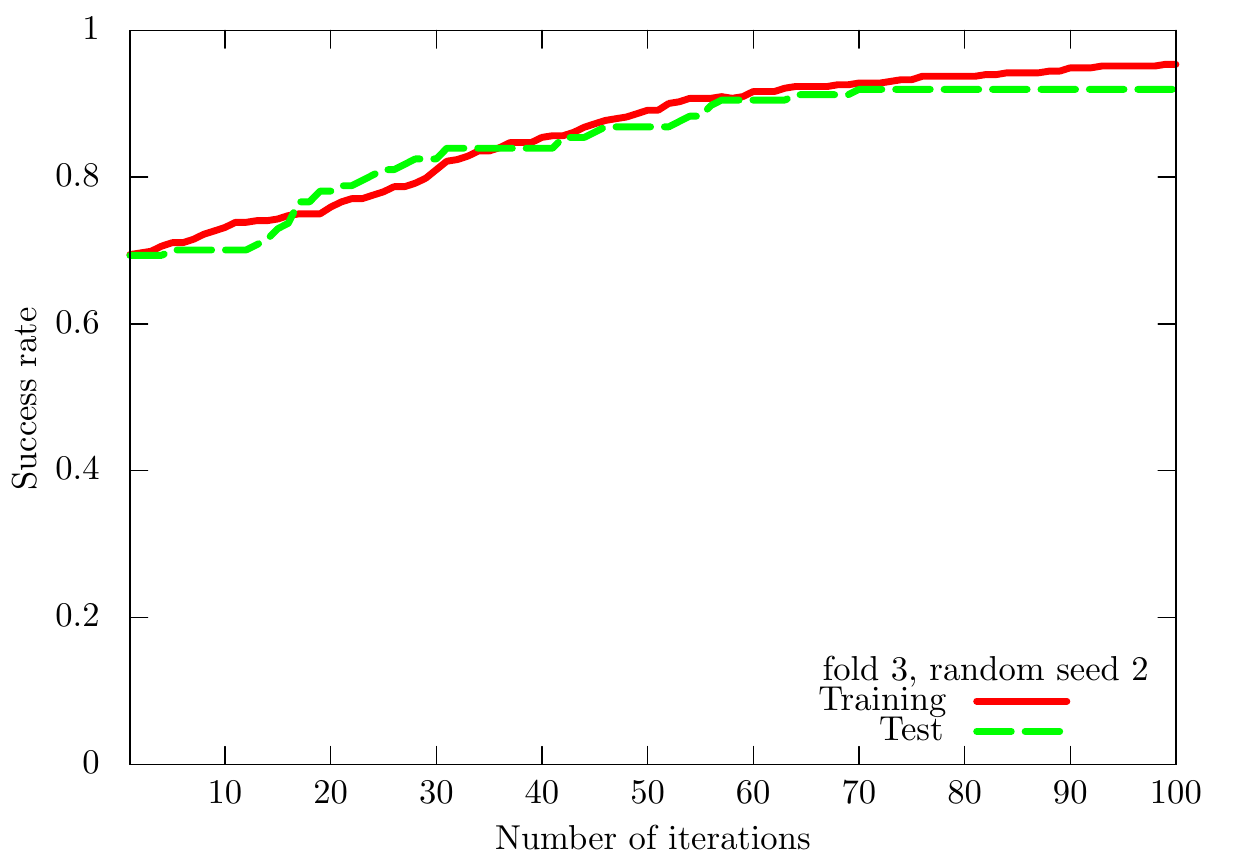}
\includegraphics[scale=0.25]{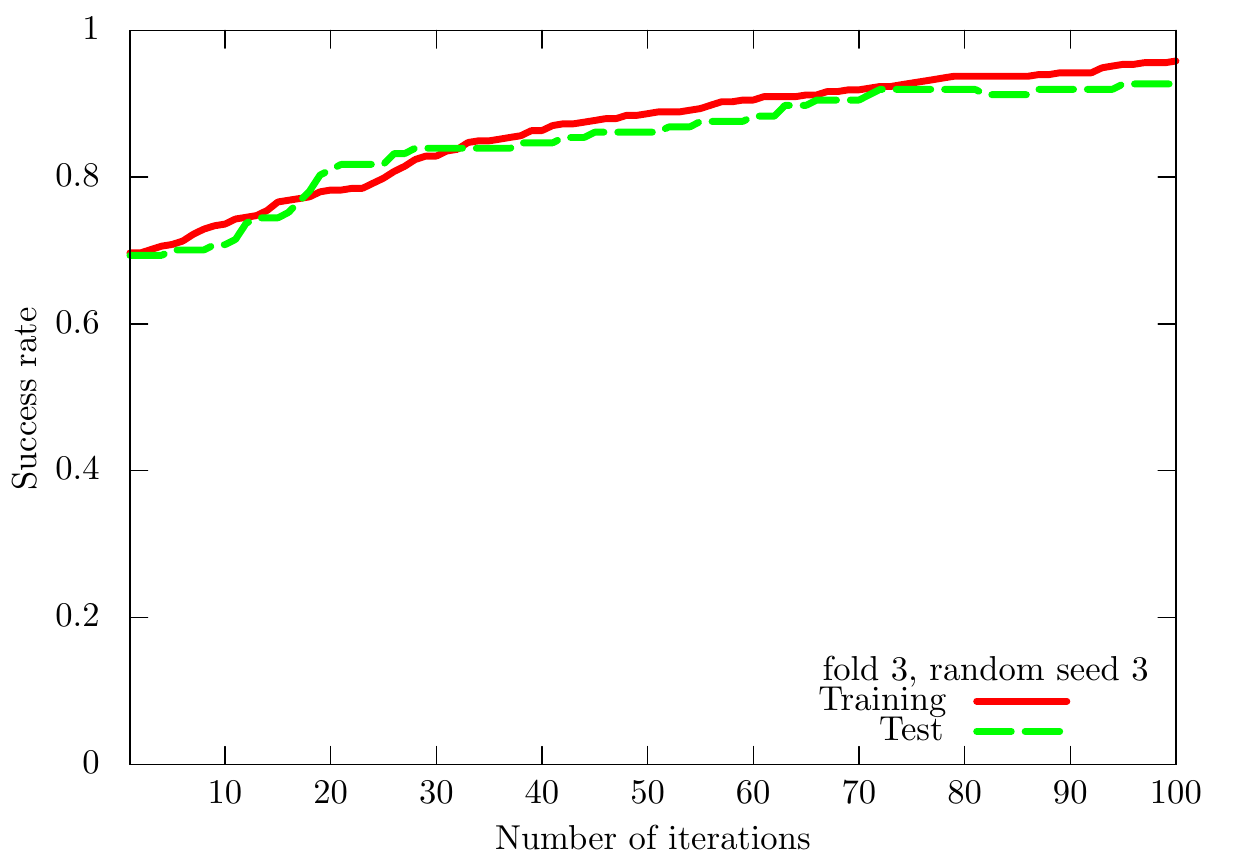}
\includegraphics[scale=0.25]{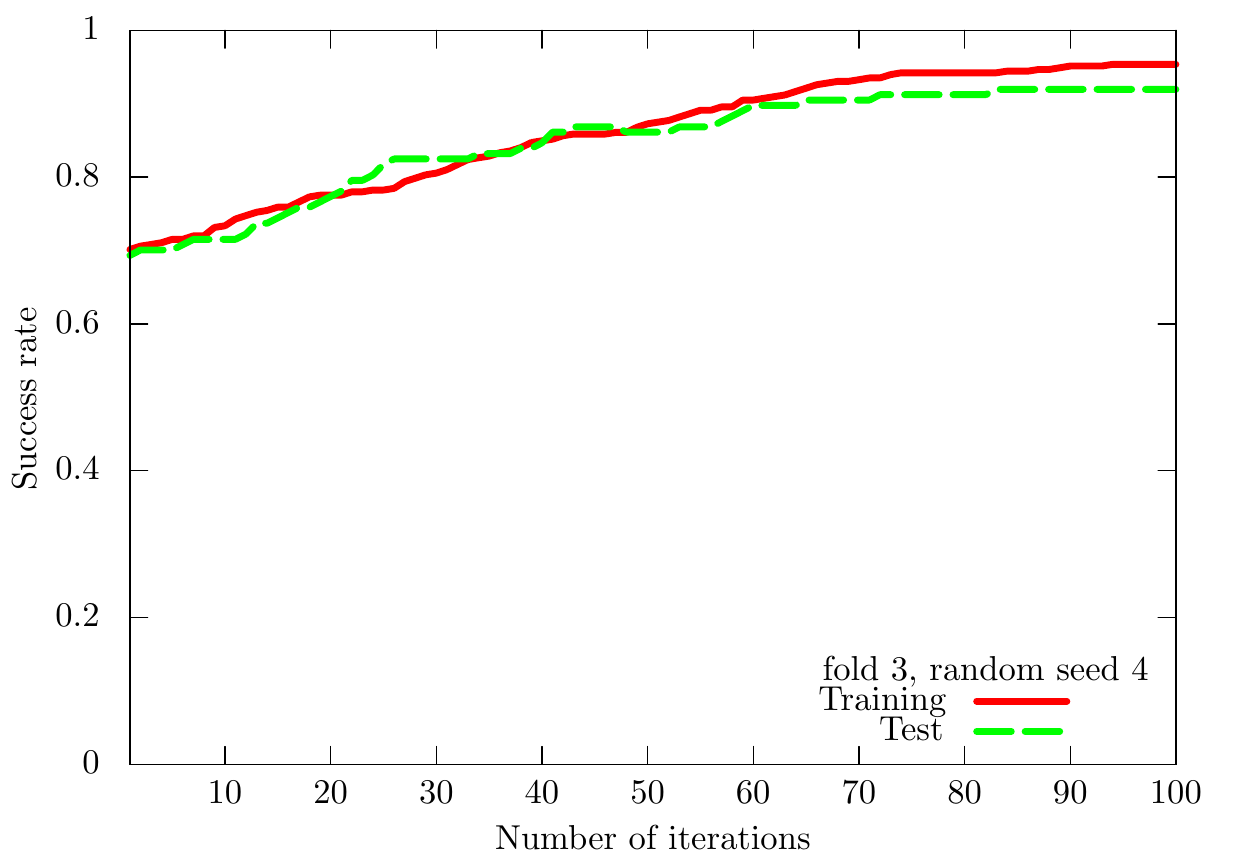}
\includegraphics[scale=0.25]{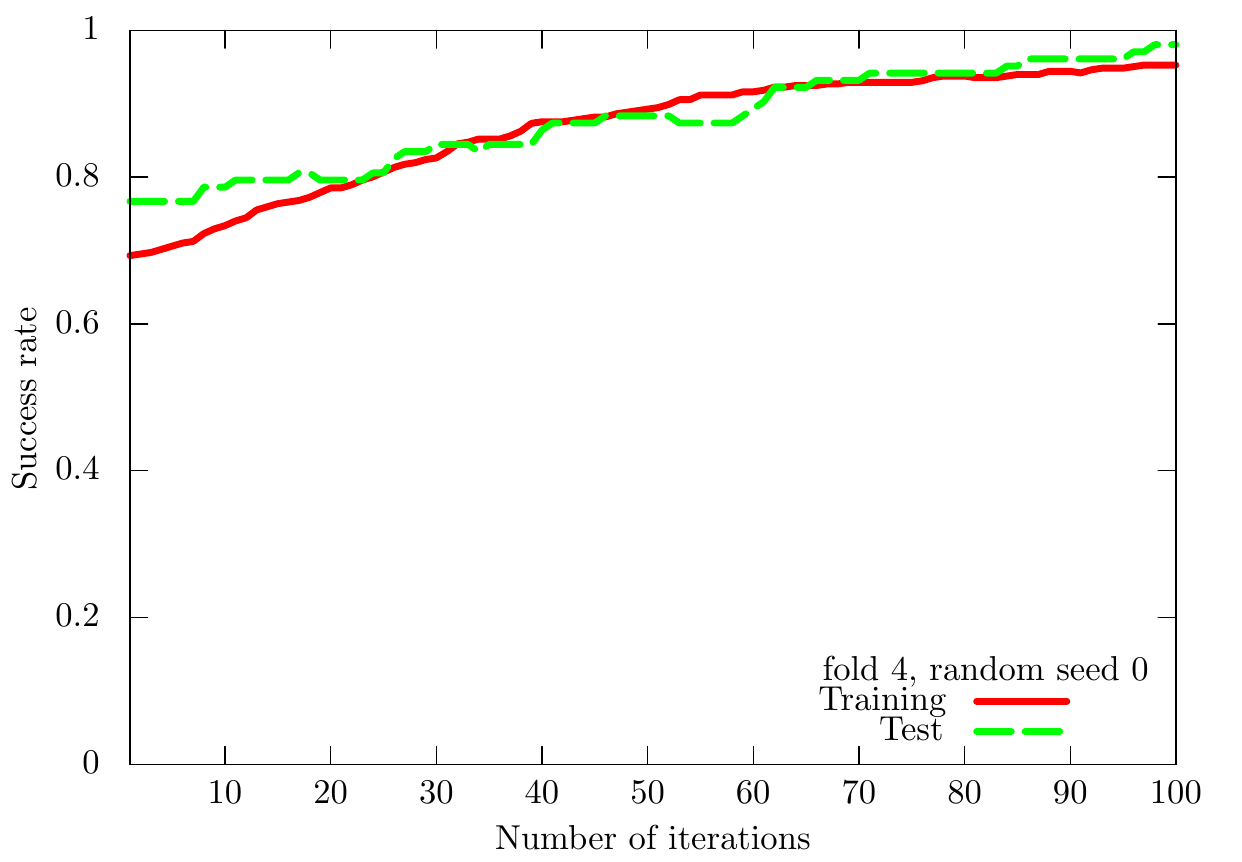}
\includegraphics[scale=0.25]{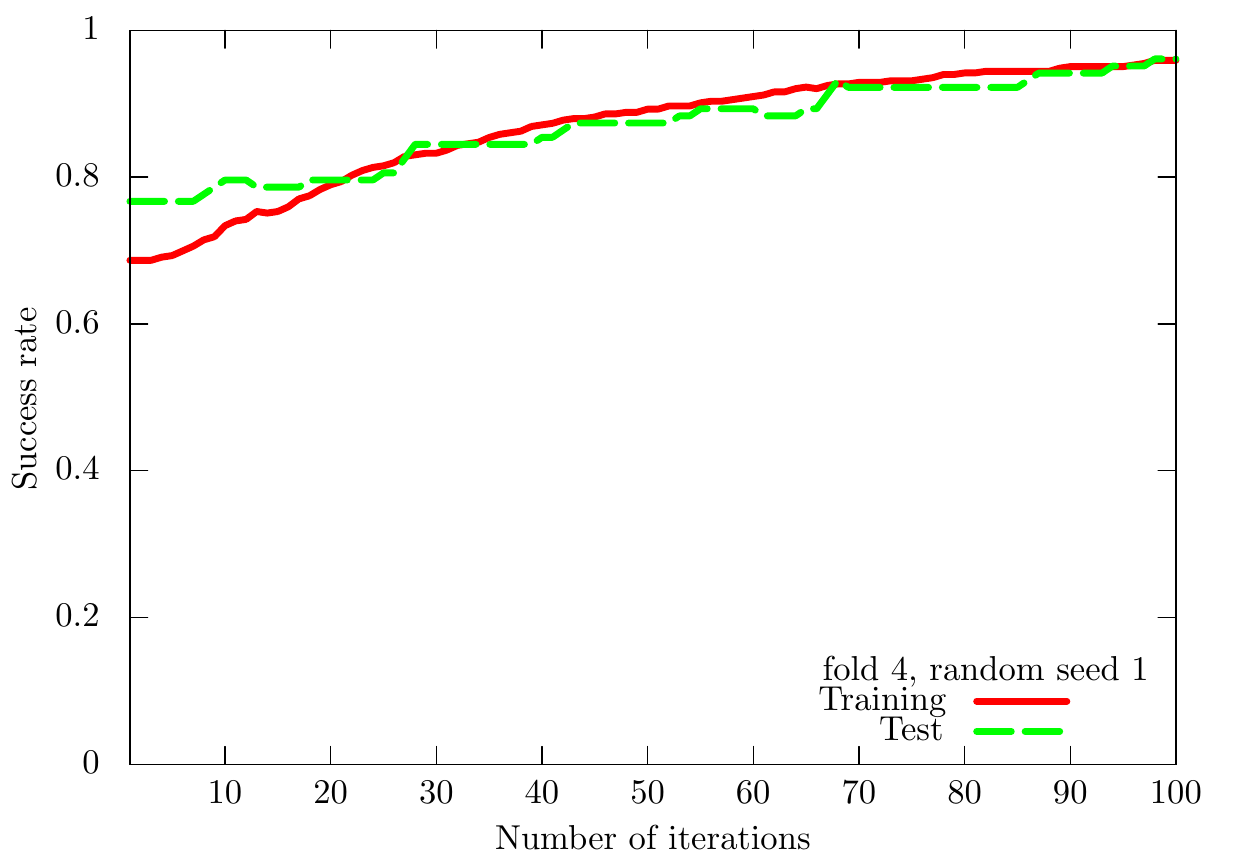}
\includegraphics[scale=0.25]{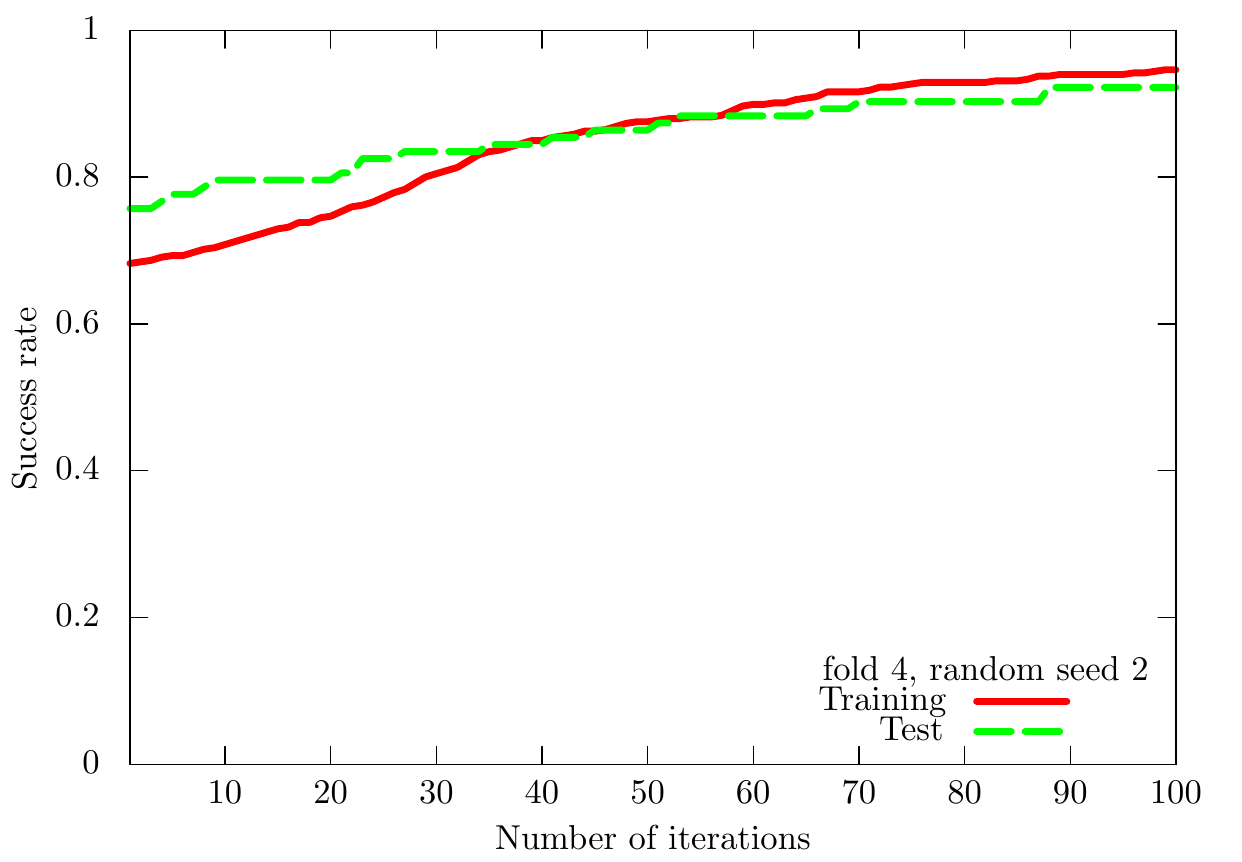}
\includegraphics[scale=0.25]{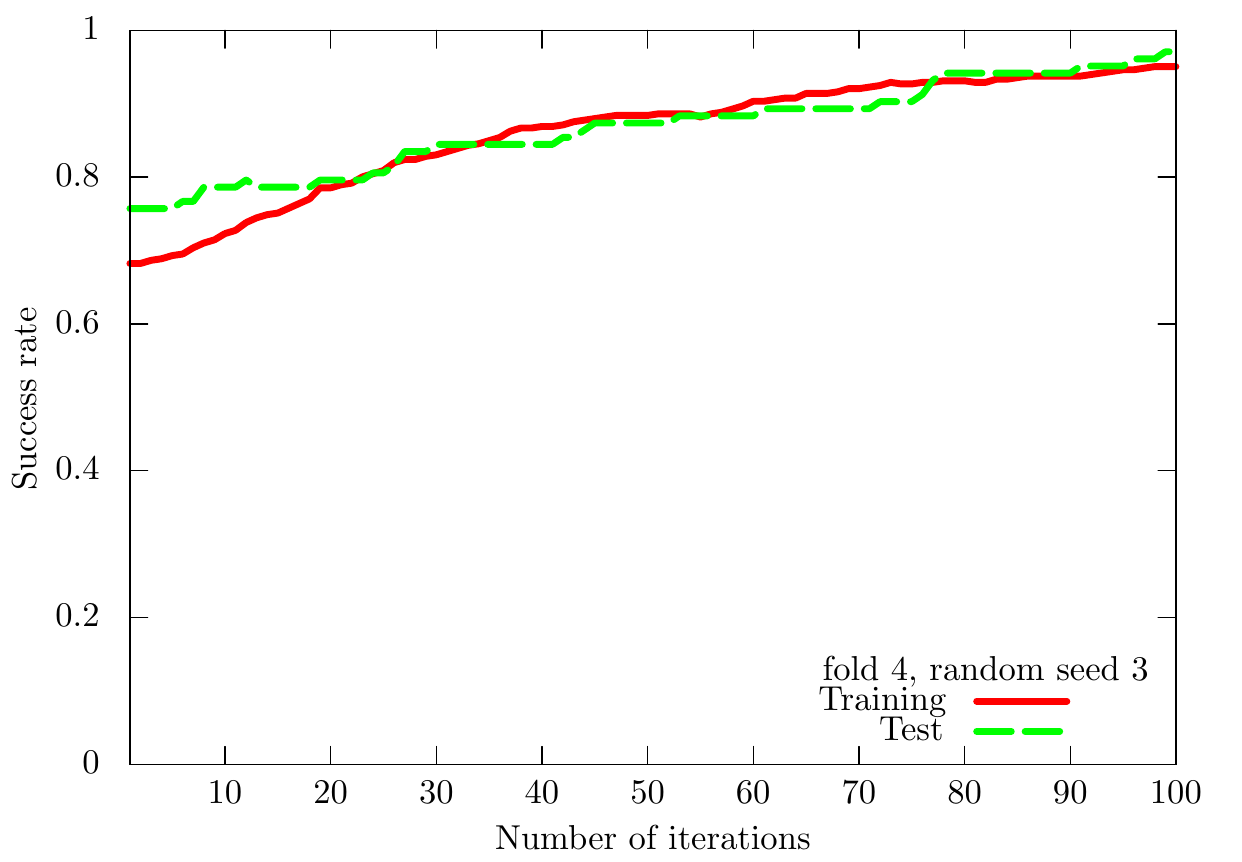}
\includegraphics[scale=0.25]{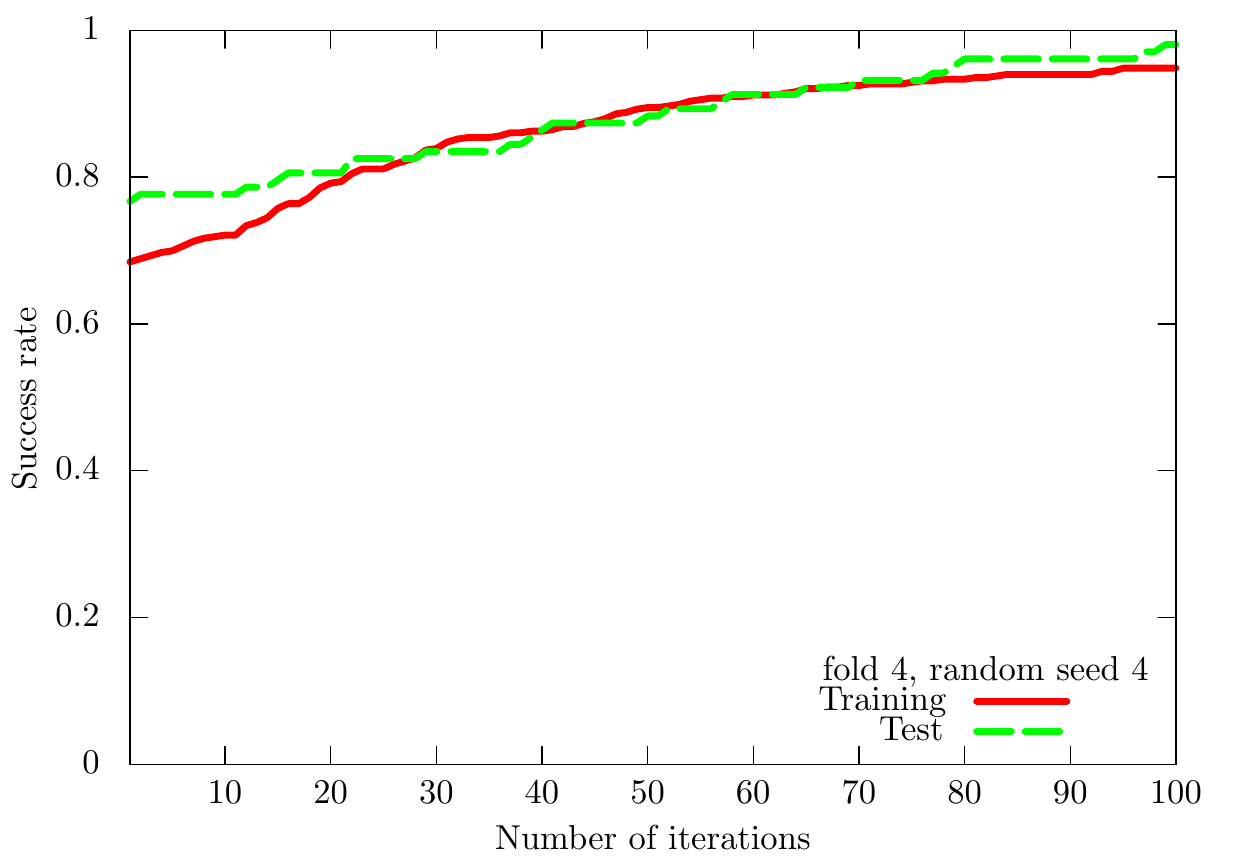}
\caption{Results of QCL on the $5$-fold datasets with $5$ different random seeds for the MNIST256 dataset ($0$ or $1$). We use the CNOT-based circuit and set $\theta_\mathrm{bias} = 0$. The number of layers $L$ is set to $5$.}
\label{supp-arXiv-numerical-result-raw-data-fold-001-rand-001-QCL-MNIST256-0-1}
\end{figure*}

\begin{figure*}[htb]
\centering
\includegraphics[scale=0.25]{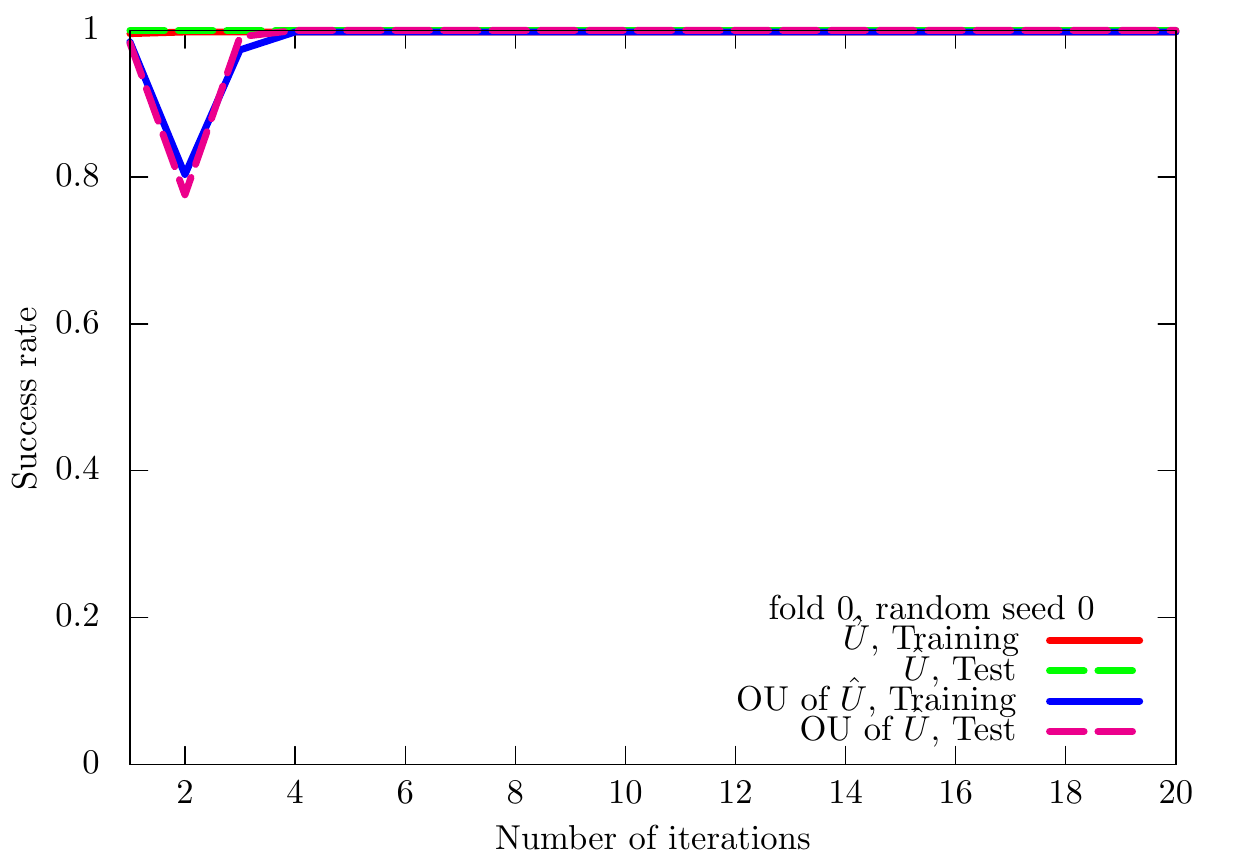}
\includegraphics[scale=0.25]{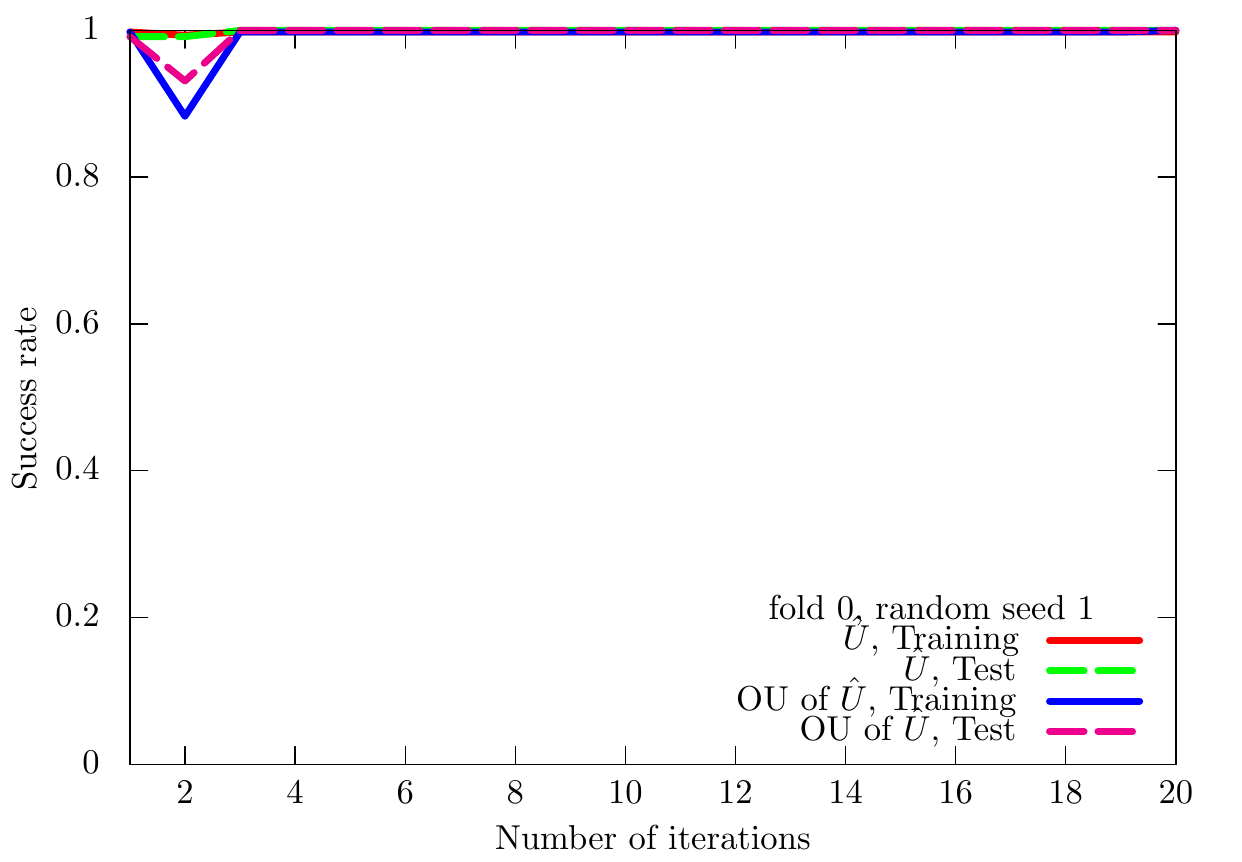}
\includegraphics[scale=0.25]{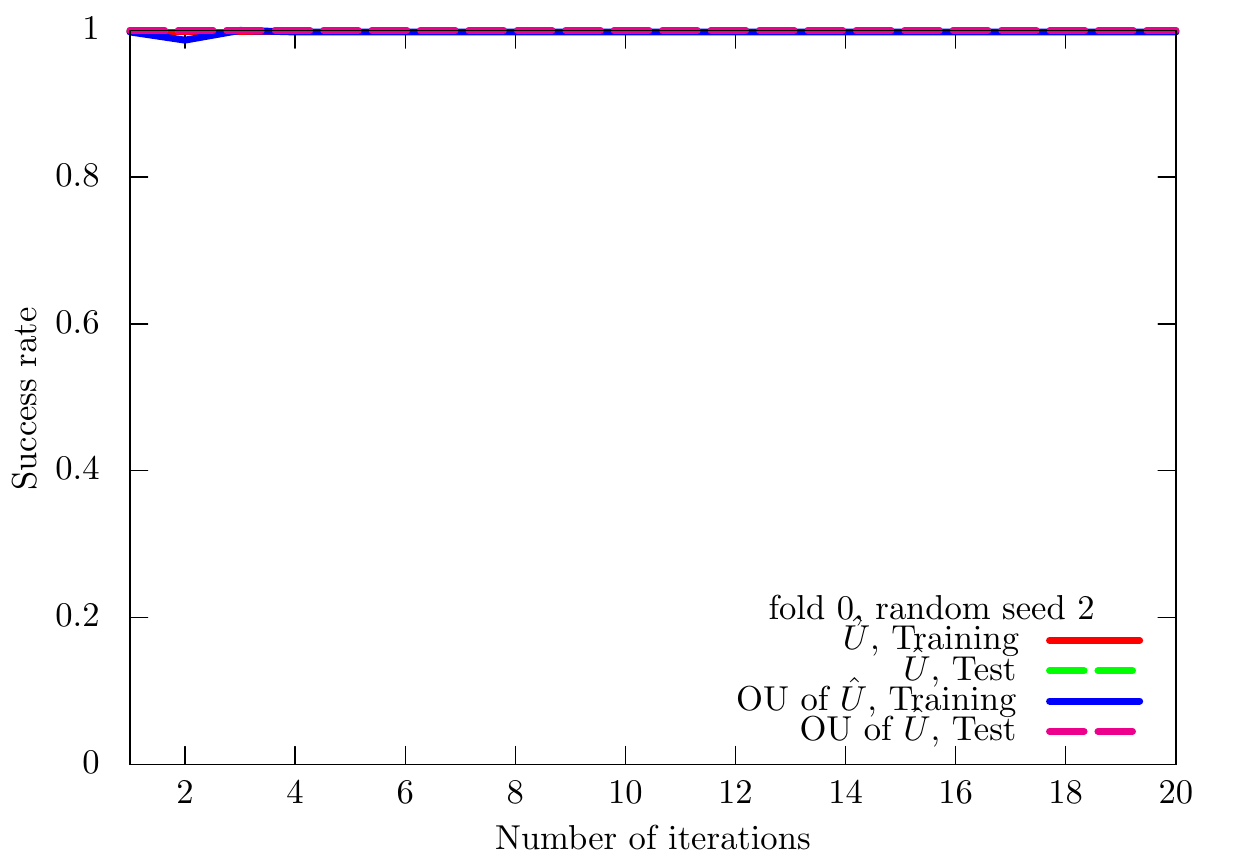}
\includegraphics[scale=0.25]{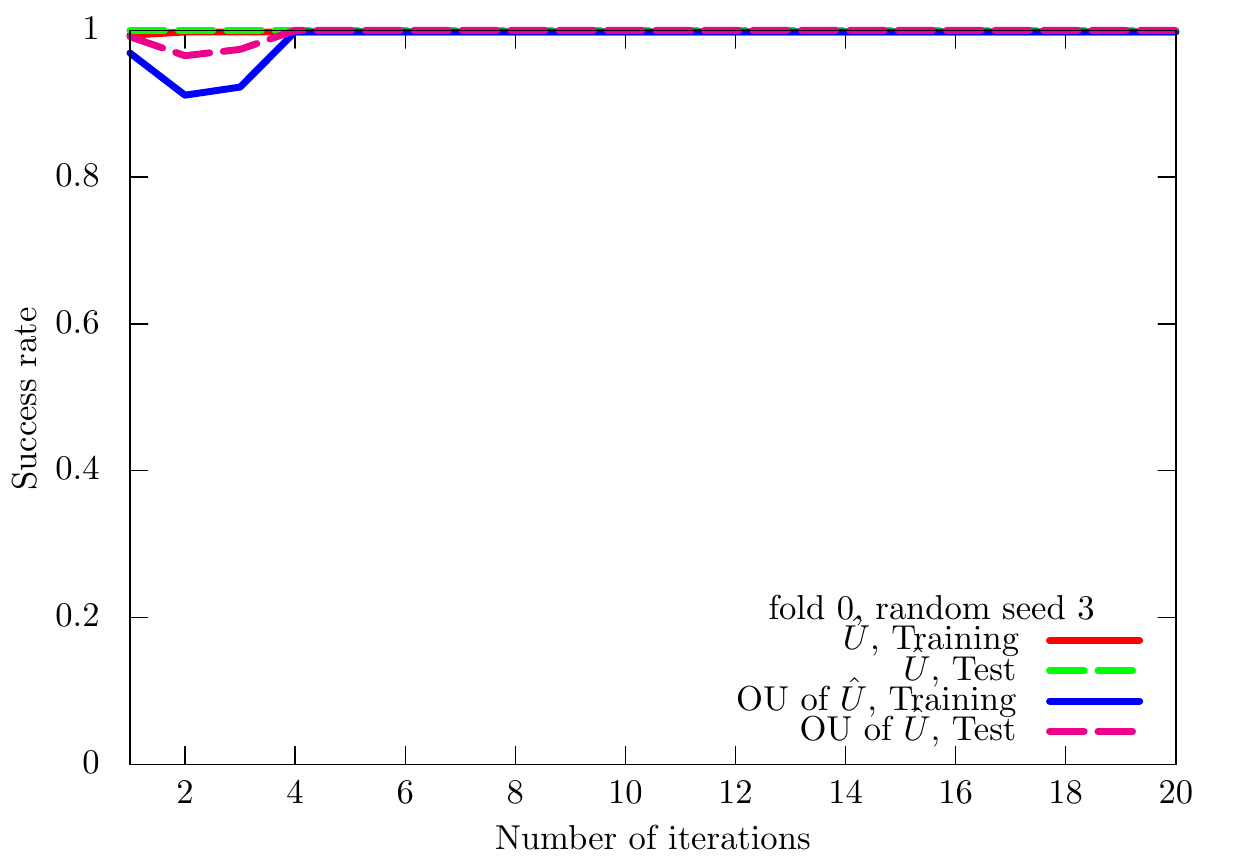}
\includegraphics[scale=0.25]{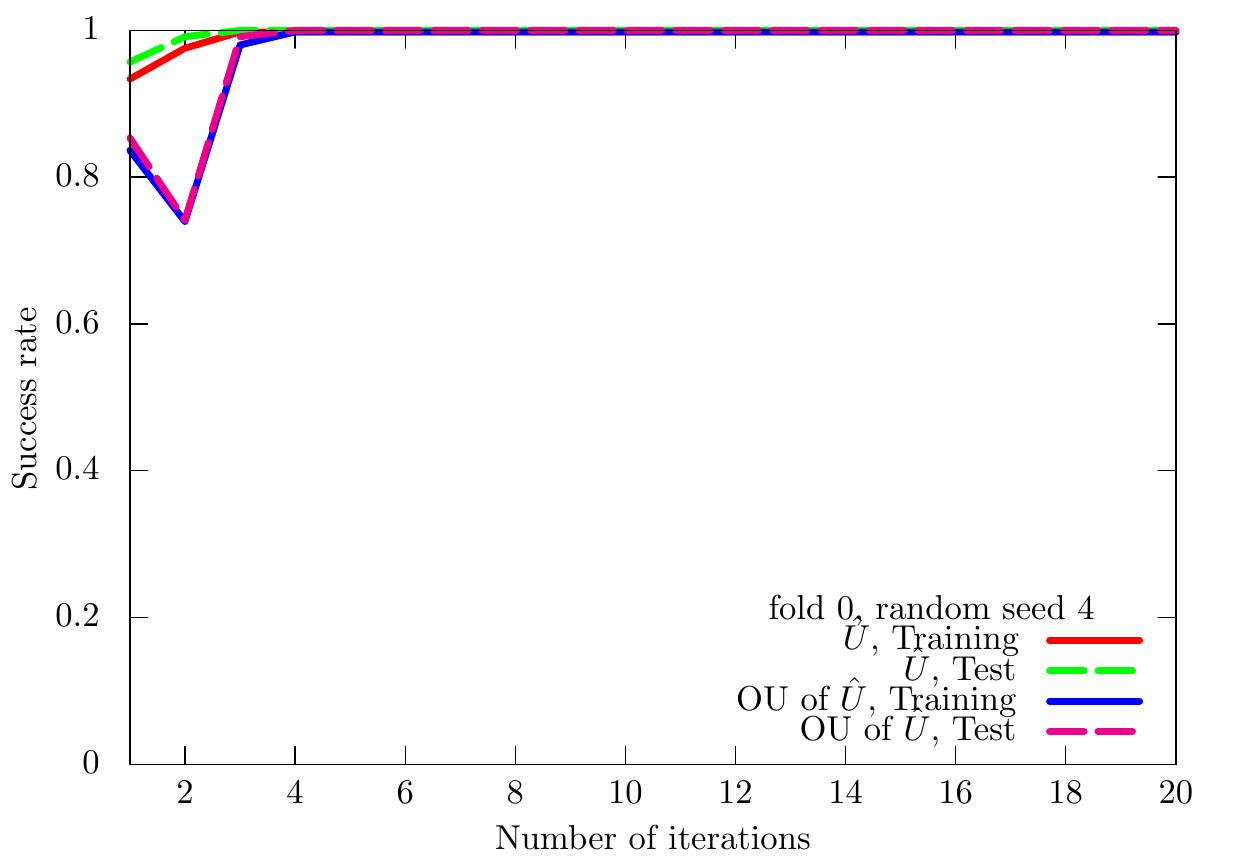}
\includegraphics[scale=0.25]{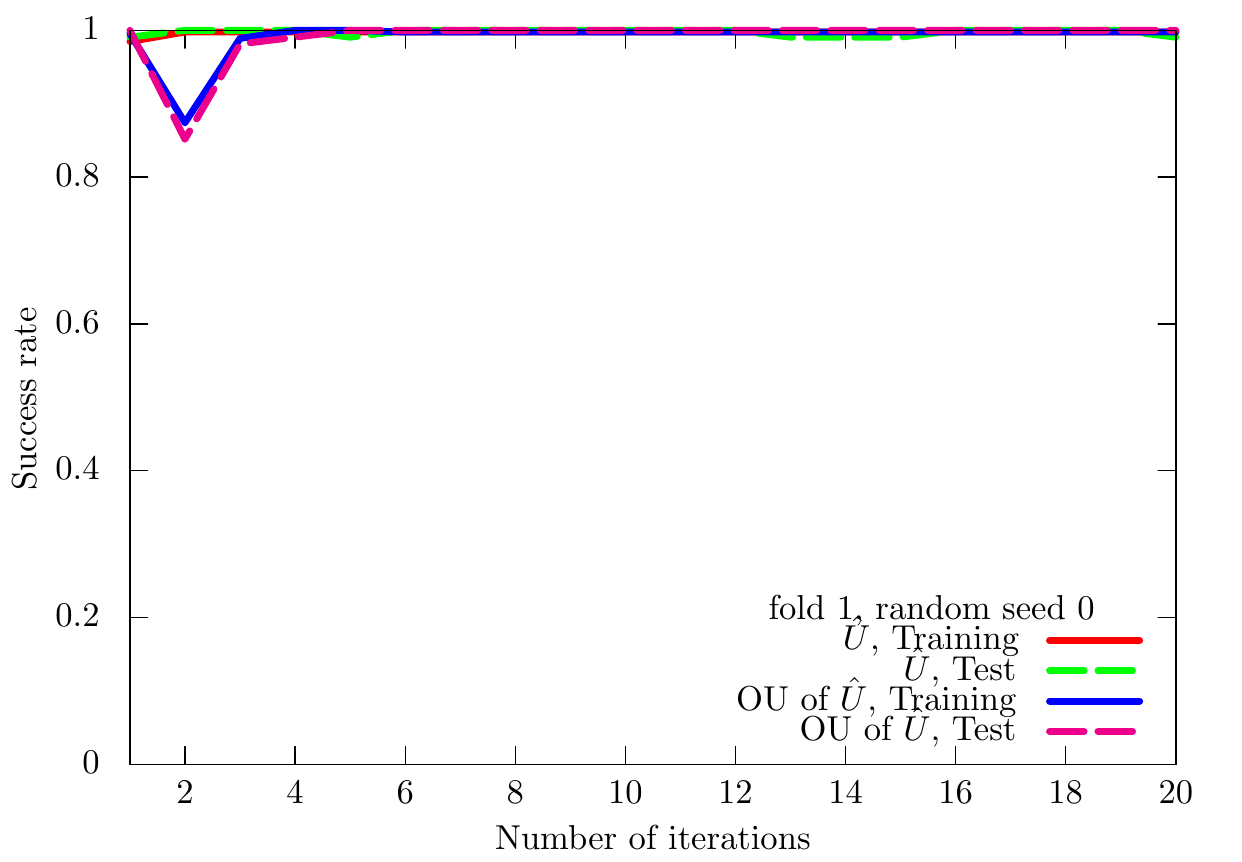}
\includegraphics[scale=0.25]{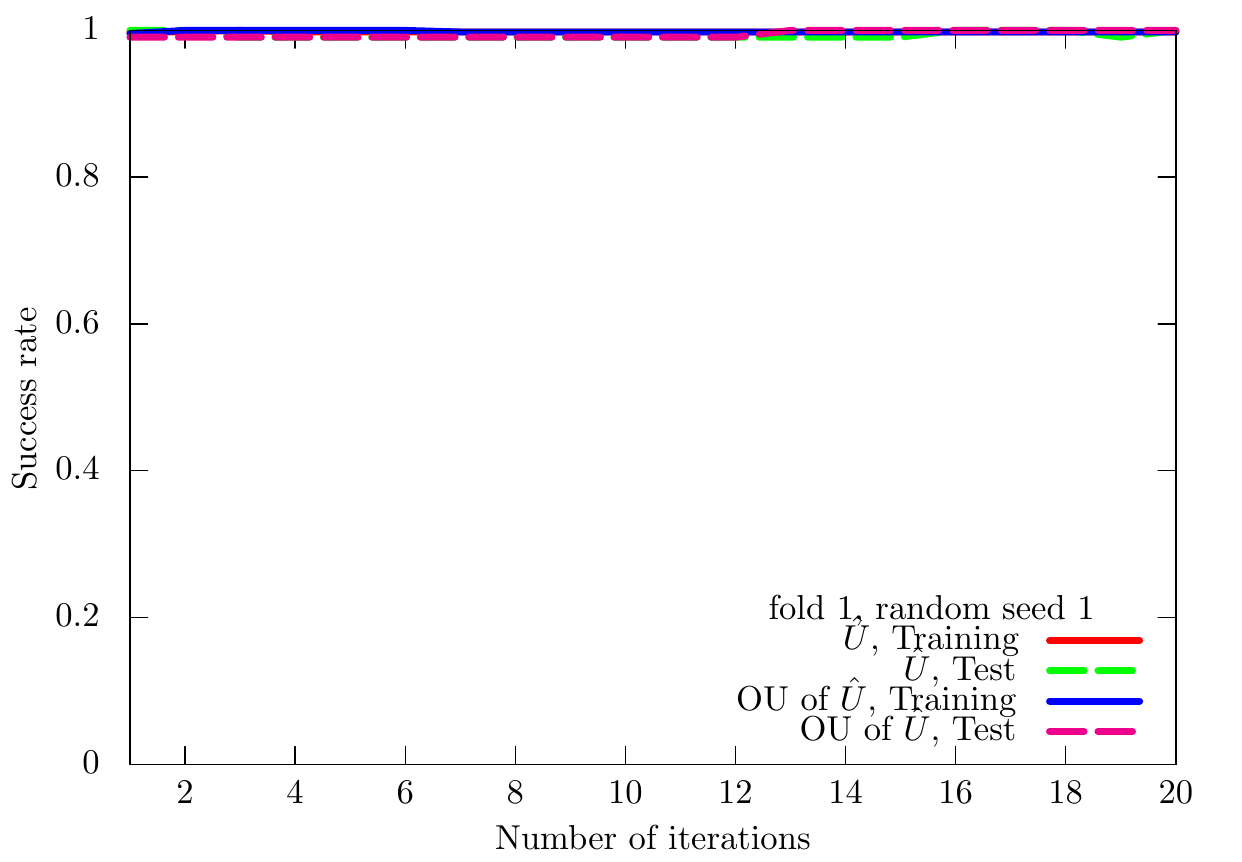}
\includegraphics[scale=0.25]{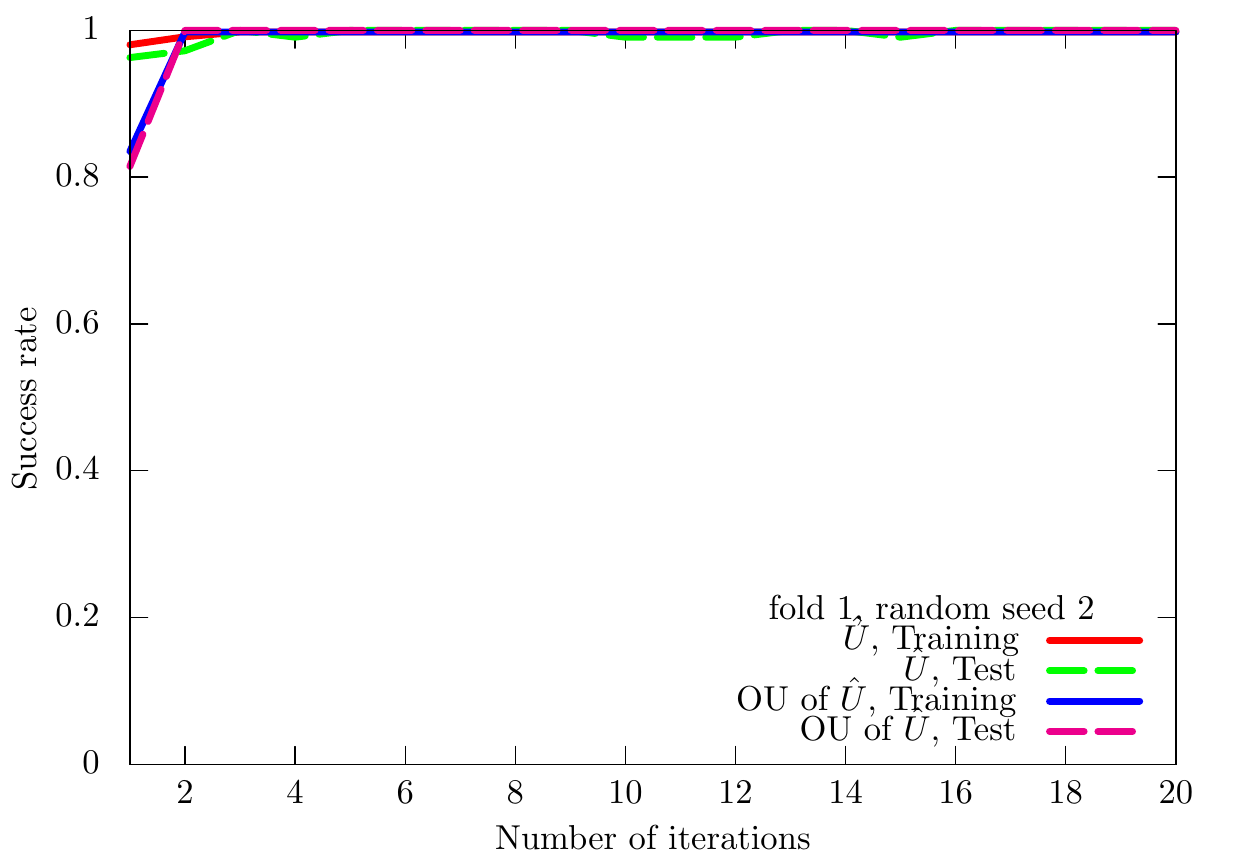}
\includegraphics[scale=0.25]{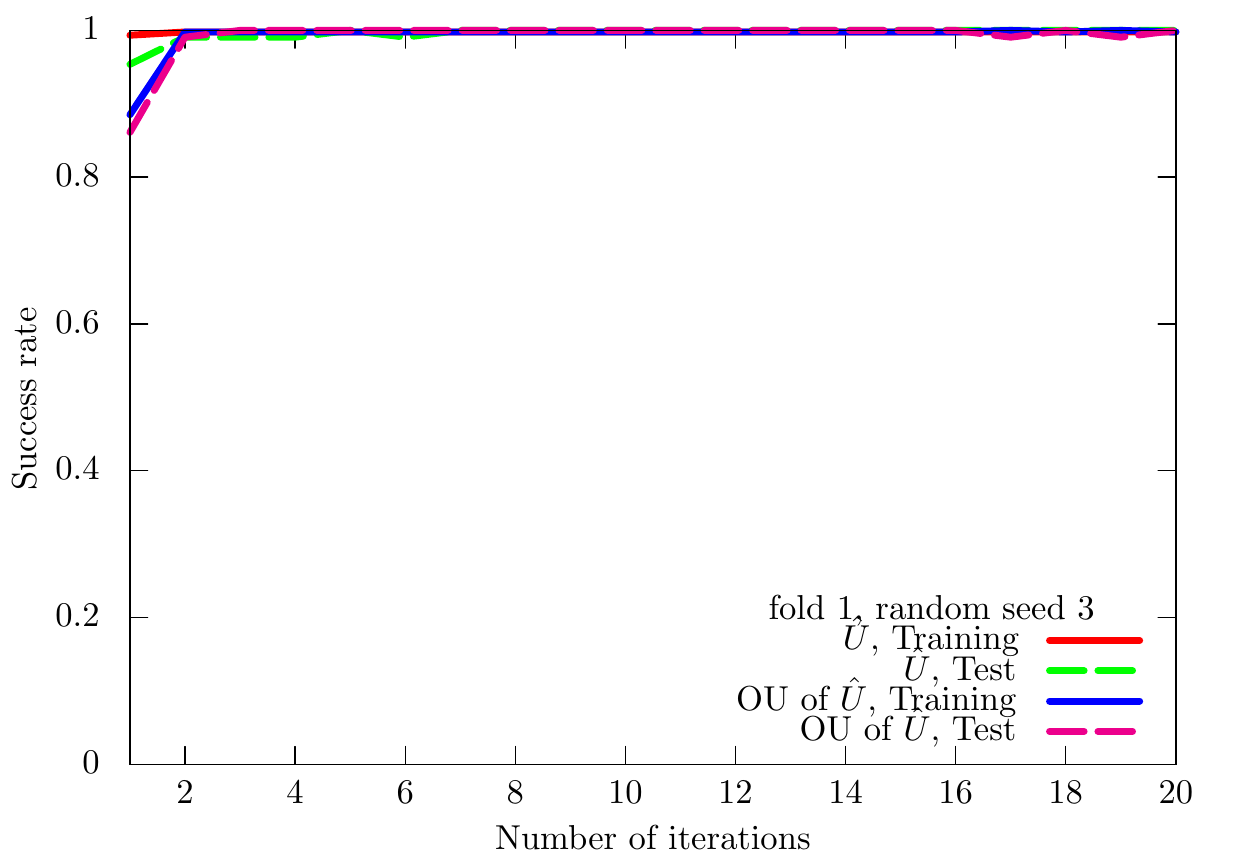}
\includegraphics[scale=0.25]{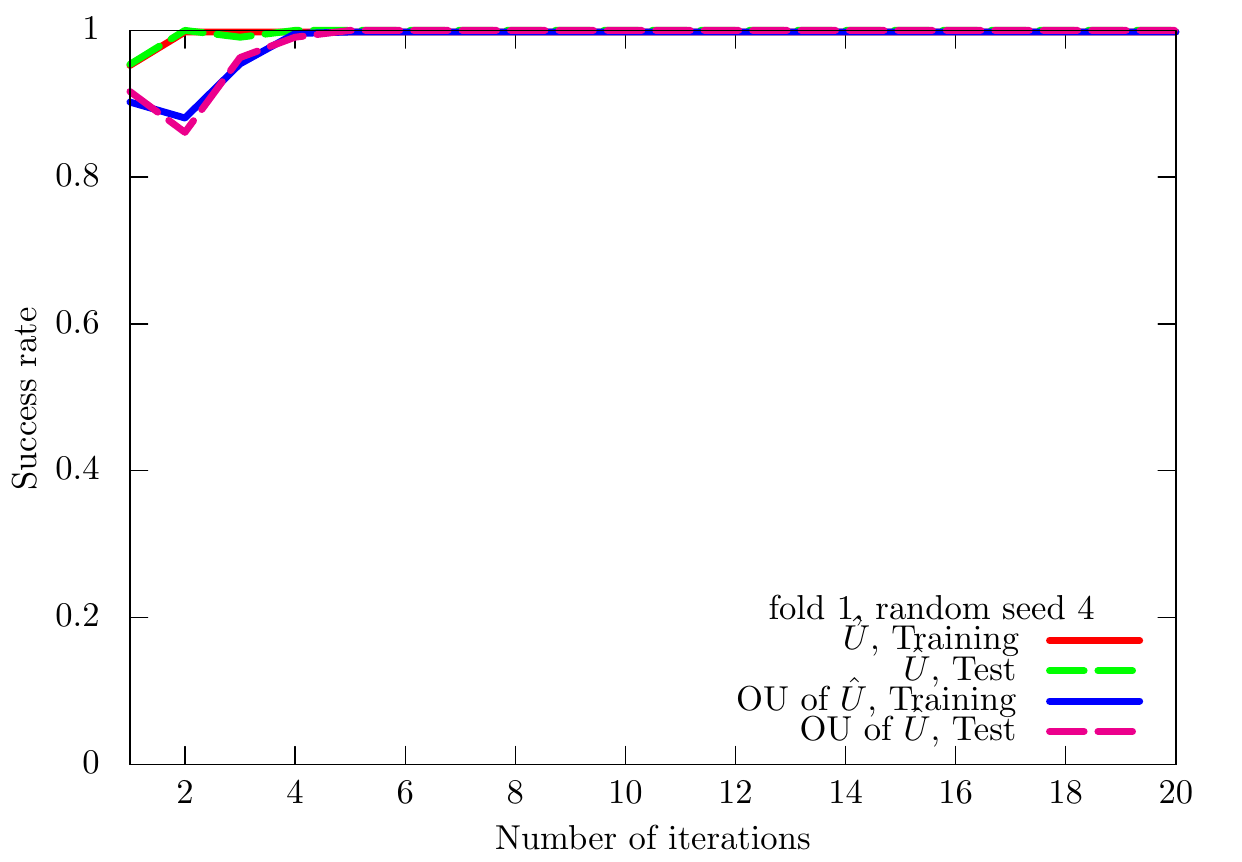}
\includegraphics[scale=0.25]{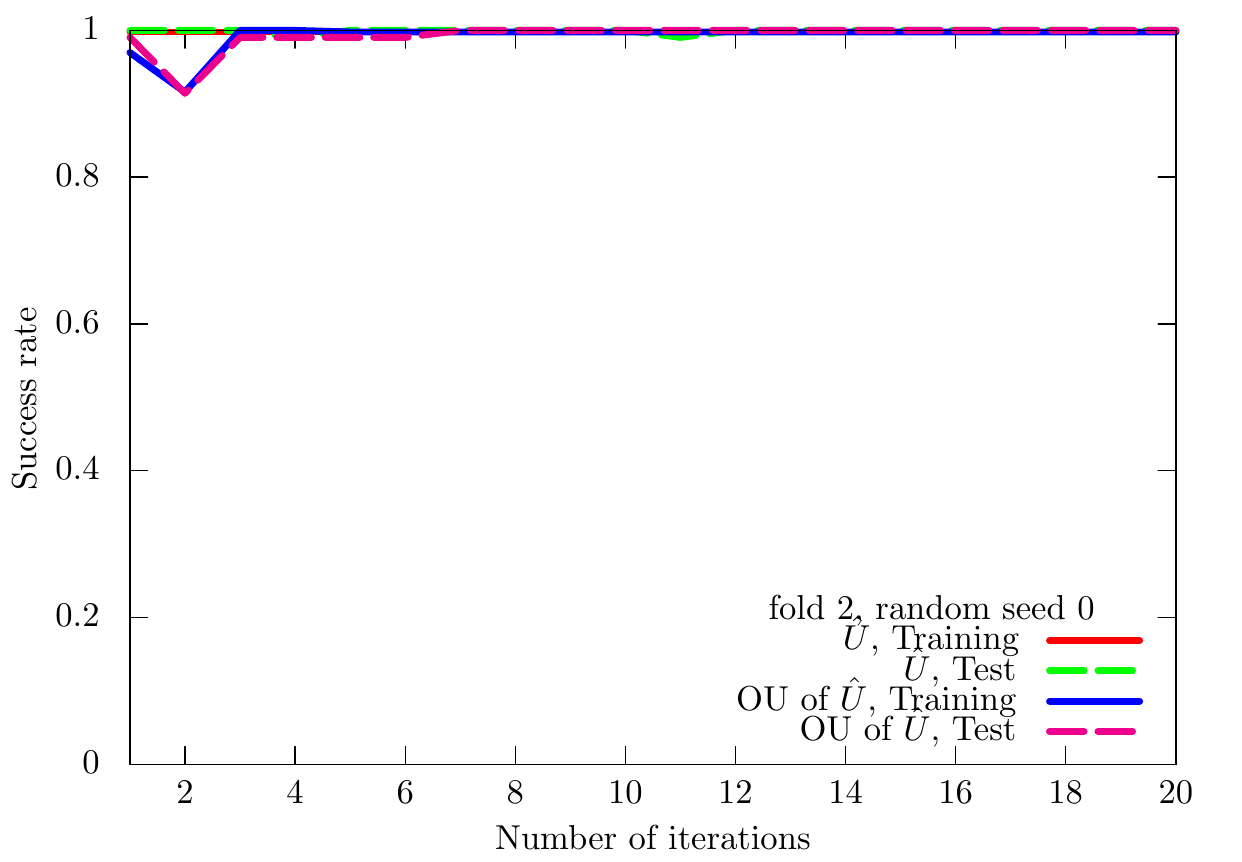}
\includegraphics[scale=0.25]{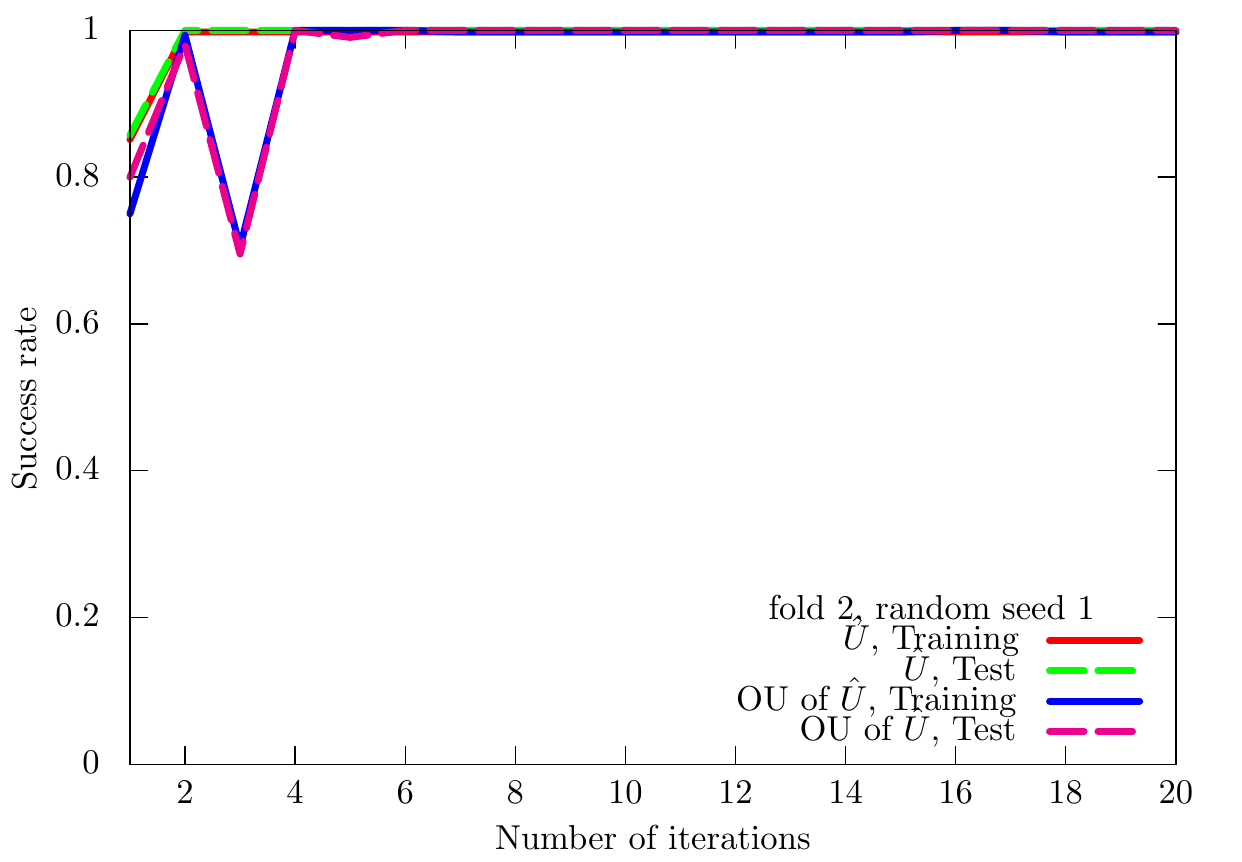}
\includegraphics[scale=0.25]{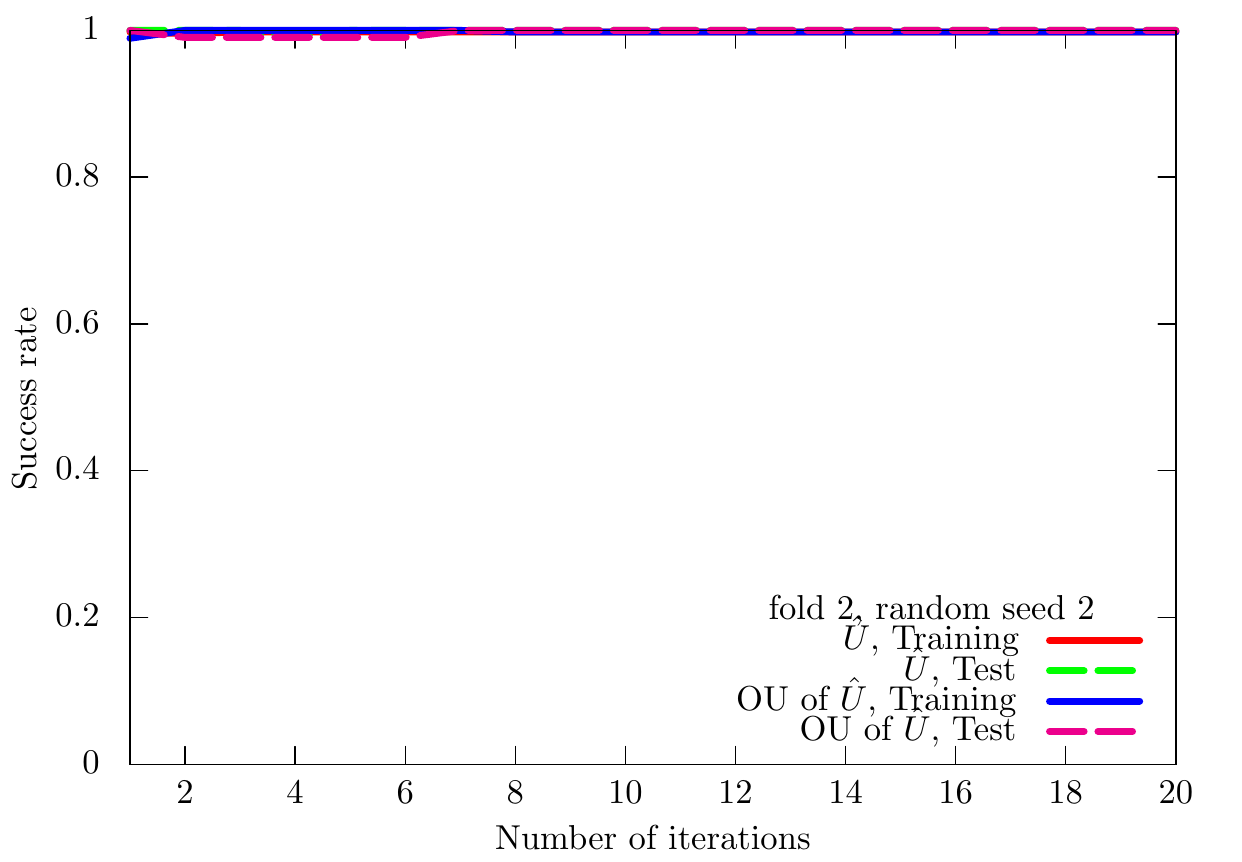}
\includegraphics[scale=0.25]{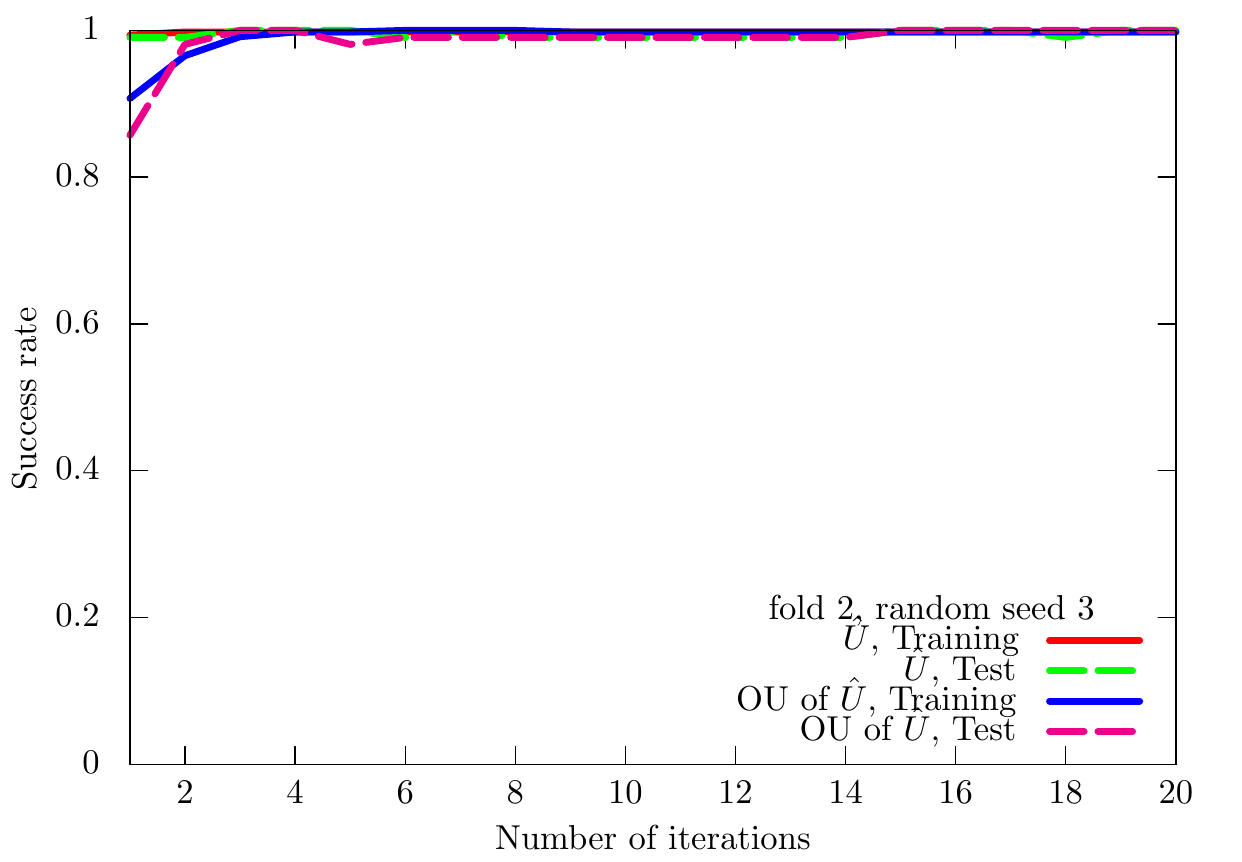}
\includegraphics[scale=0.25]{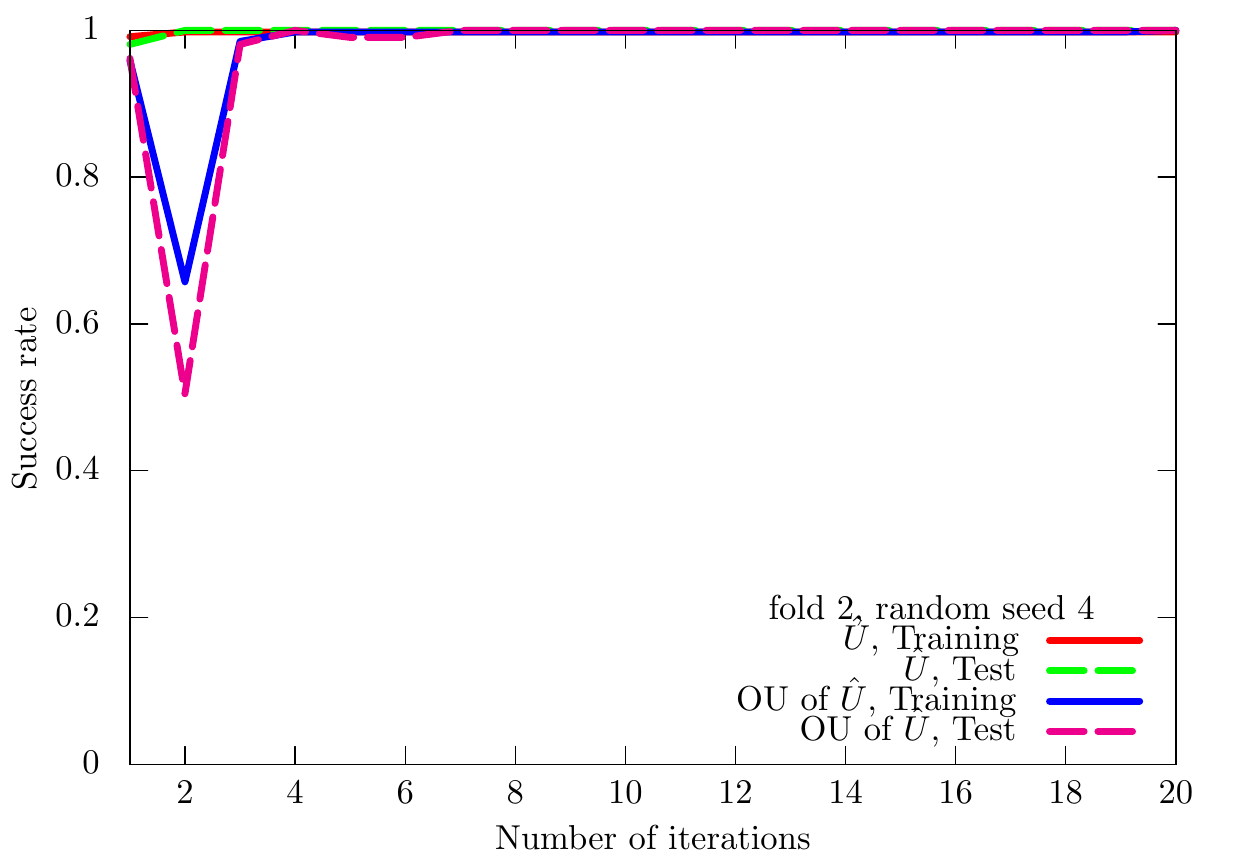}
\includegraphics[scale=0.25]{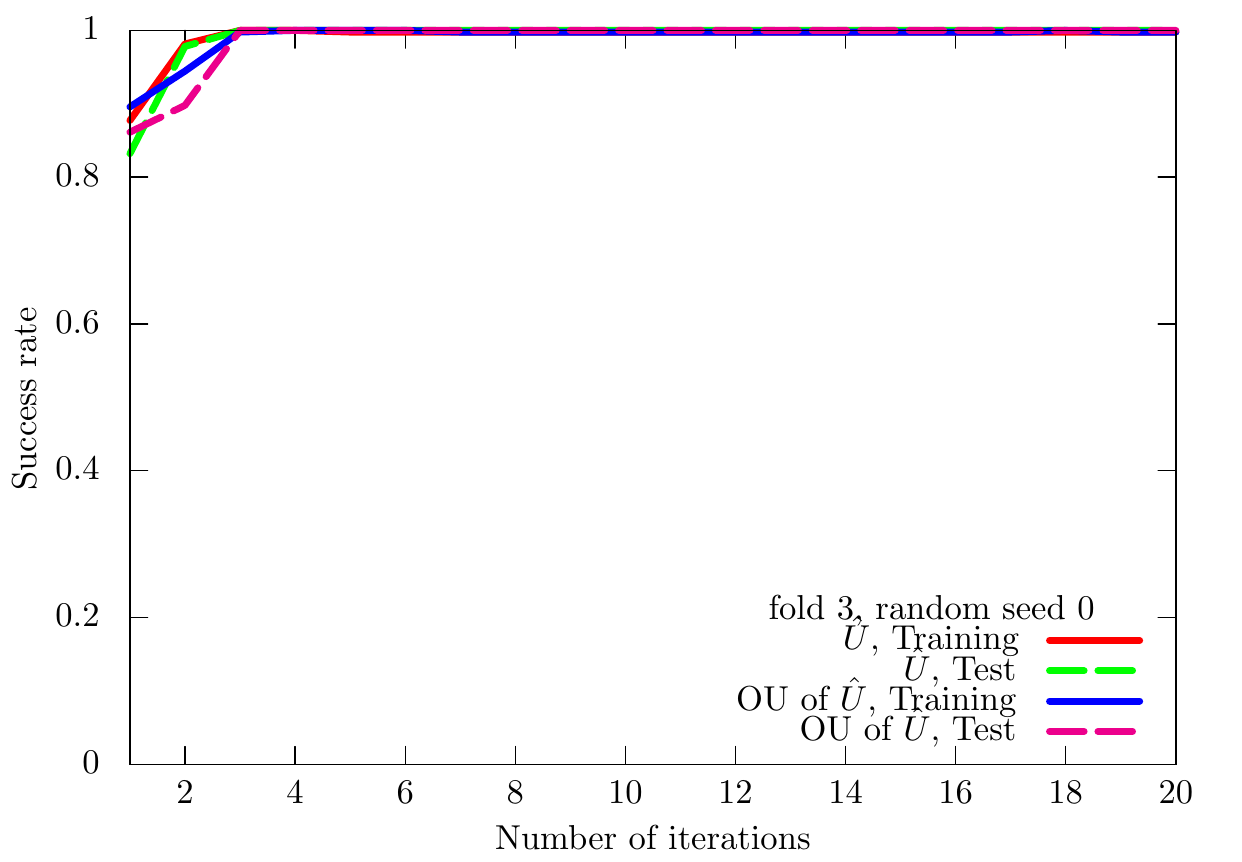}
\includegraphics[scale=0.25]{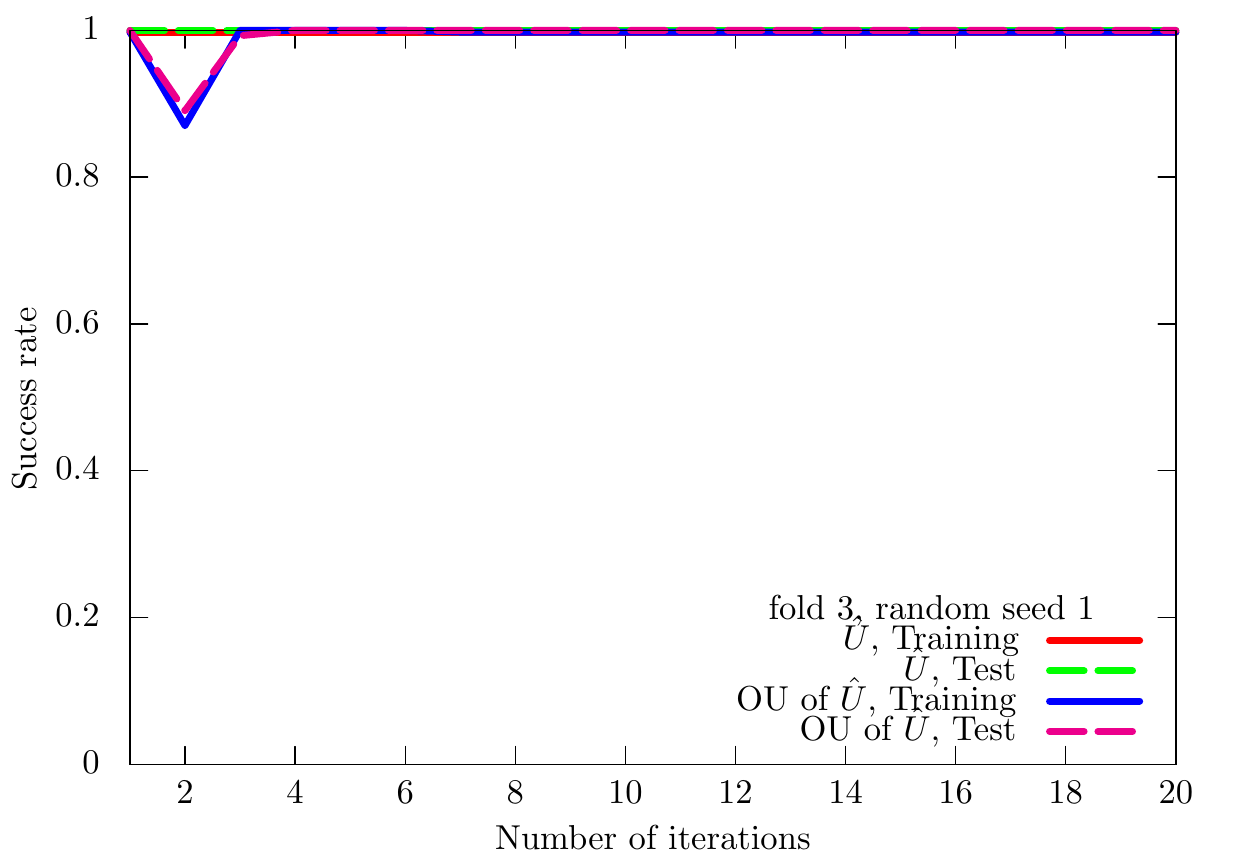}
\includegraphics[scale=0.25]{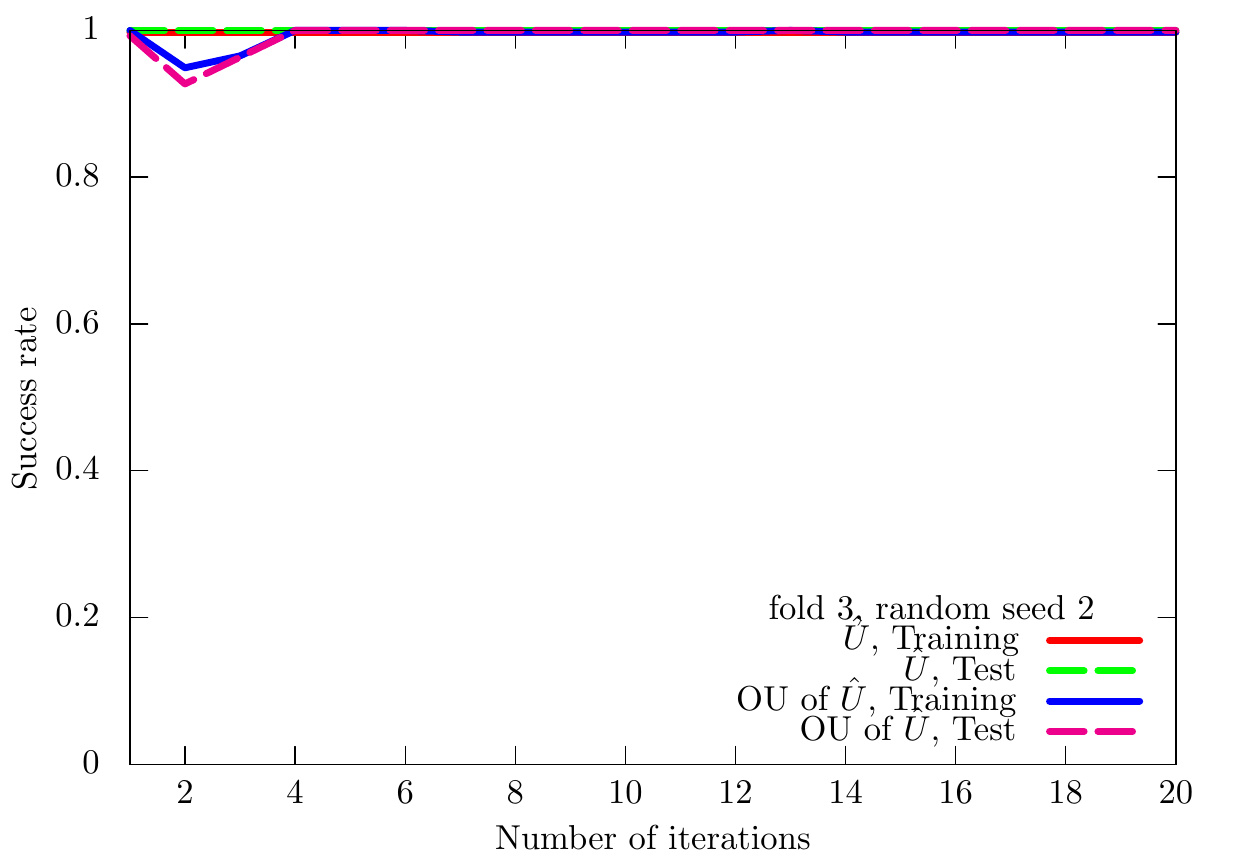}
\includegraphics[scale=0.25]{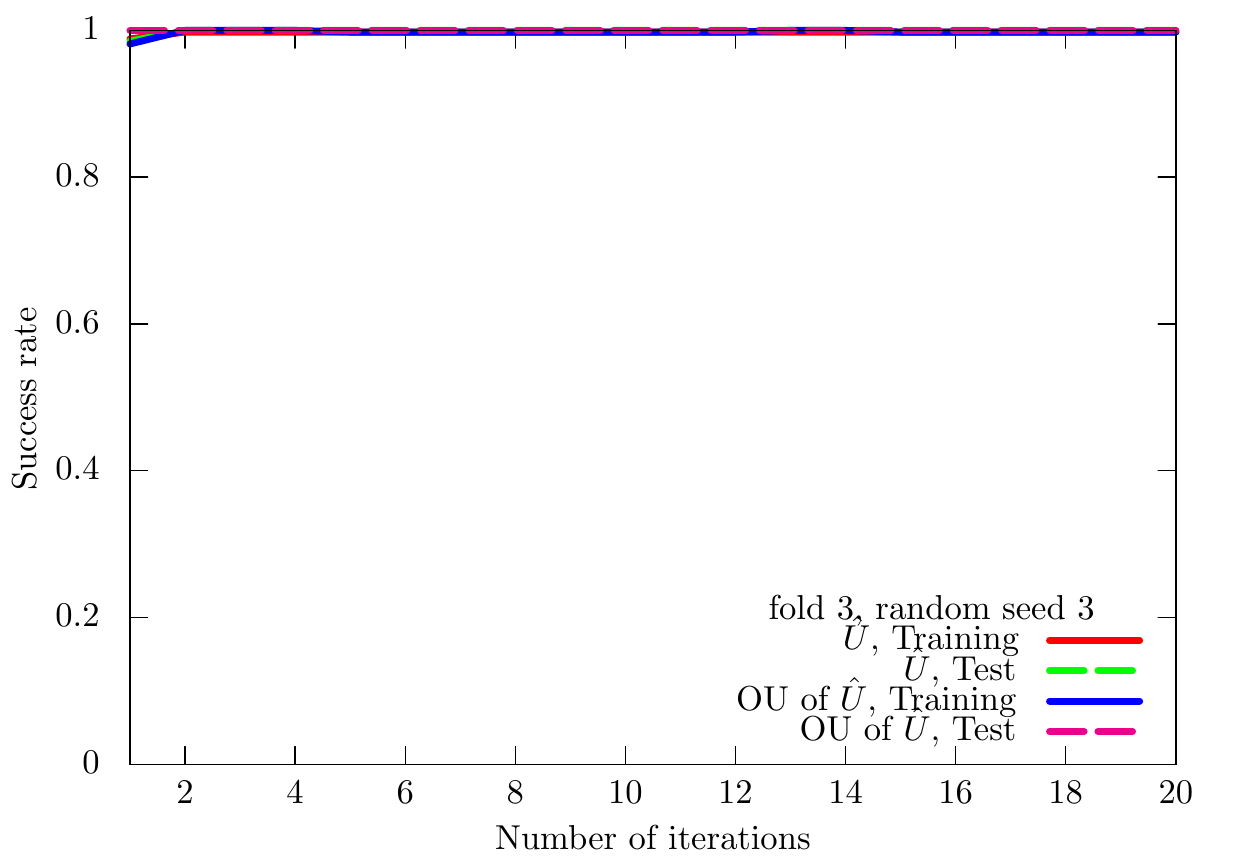}
\includegraphics[scale=0.25]{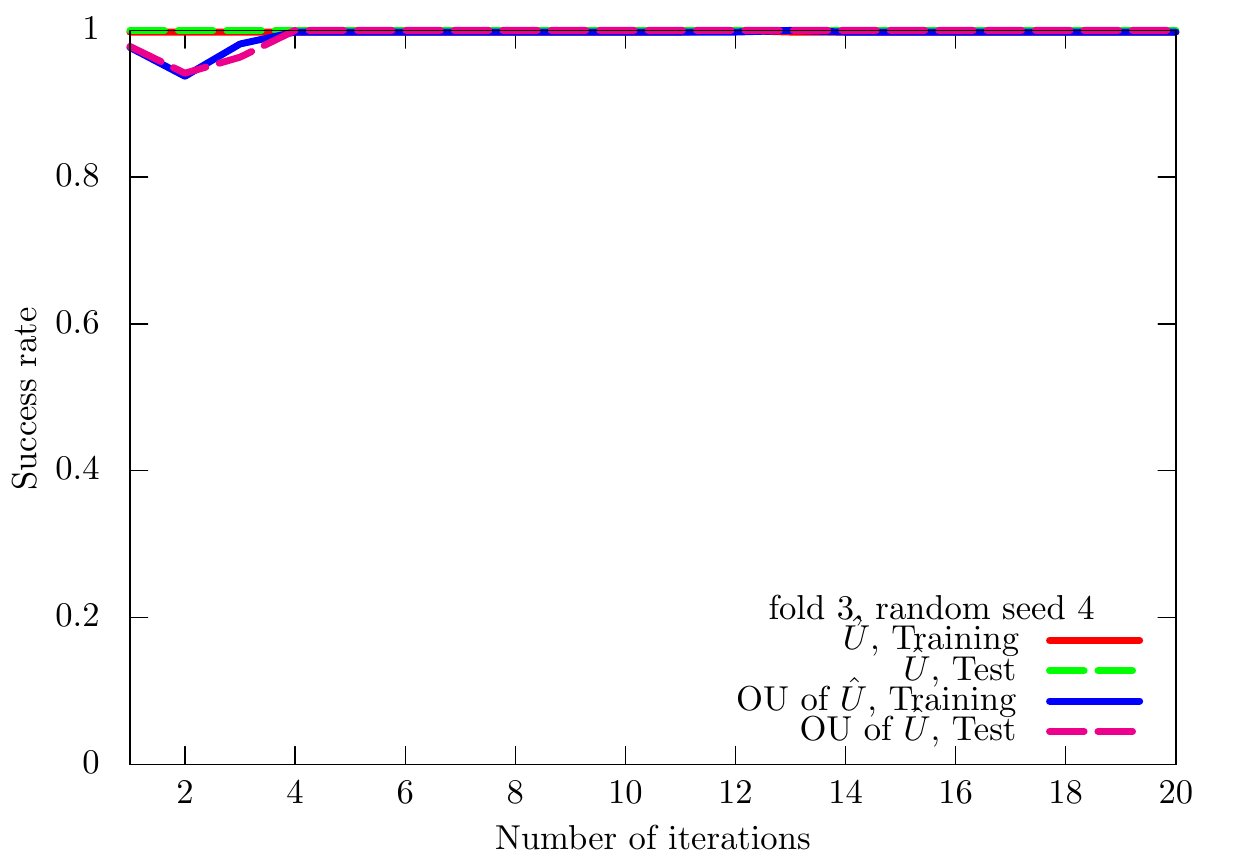}
\includegraphics[scale=0.25]{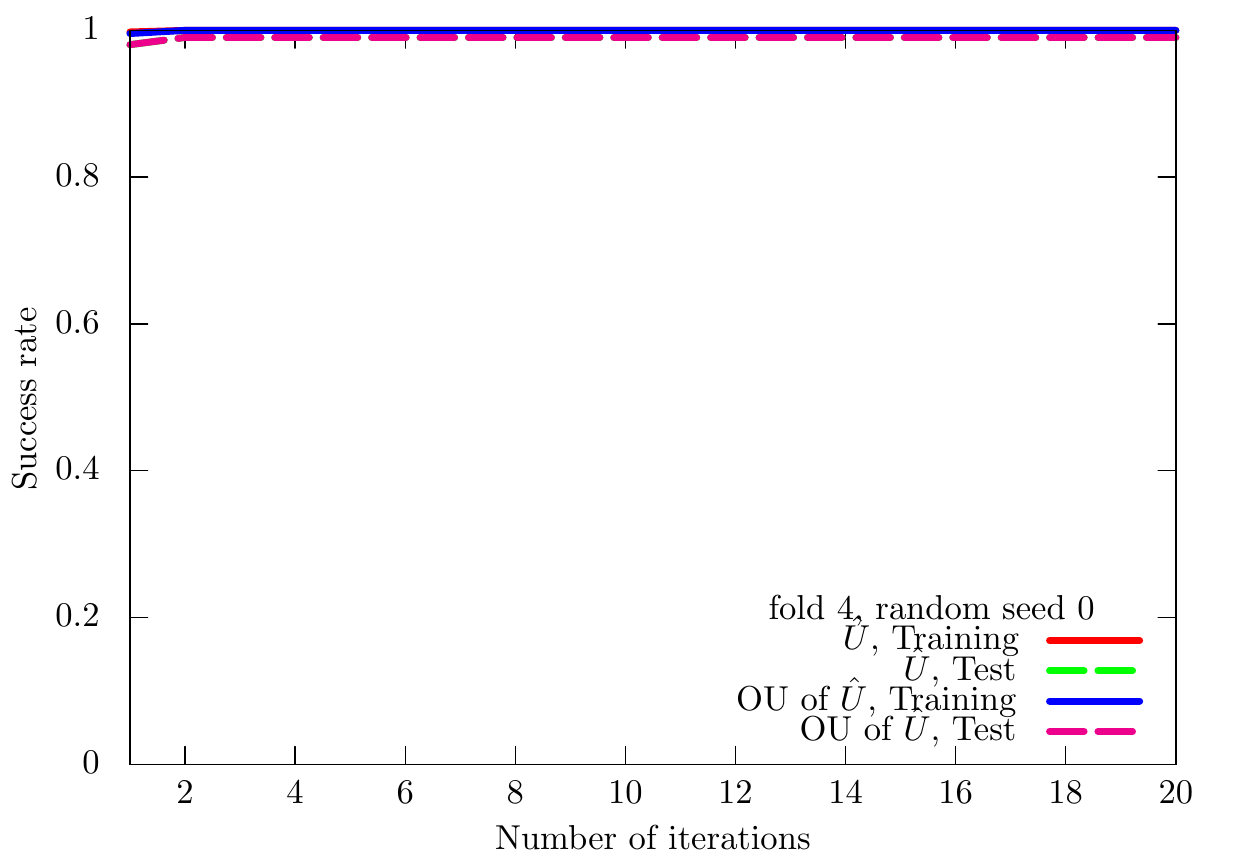}
\includegraphics[scale=0.25]{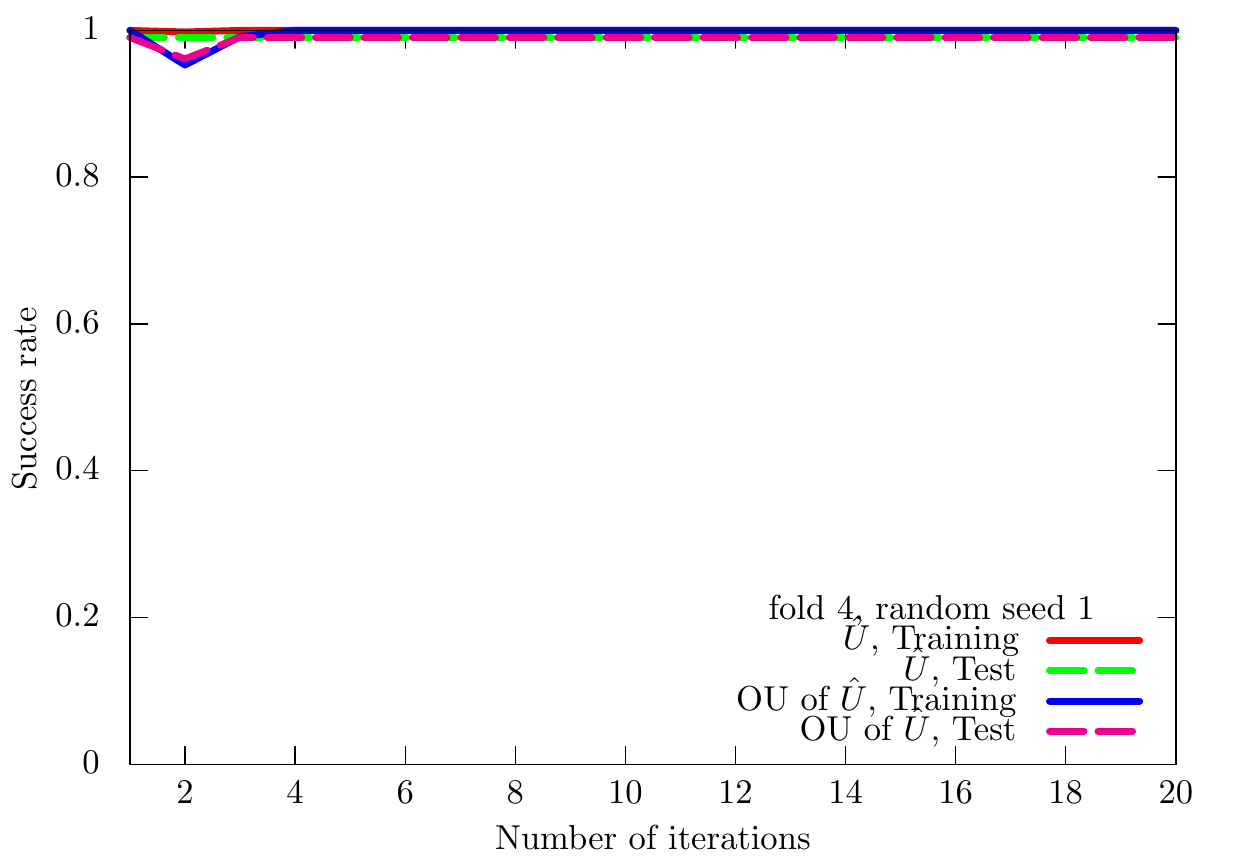}
\includegraphics[scale=0.25]{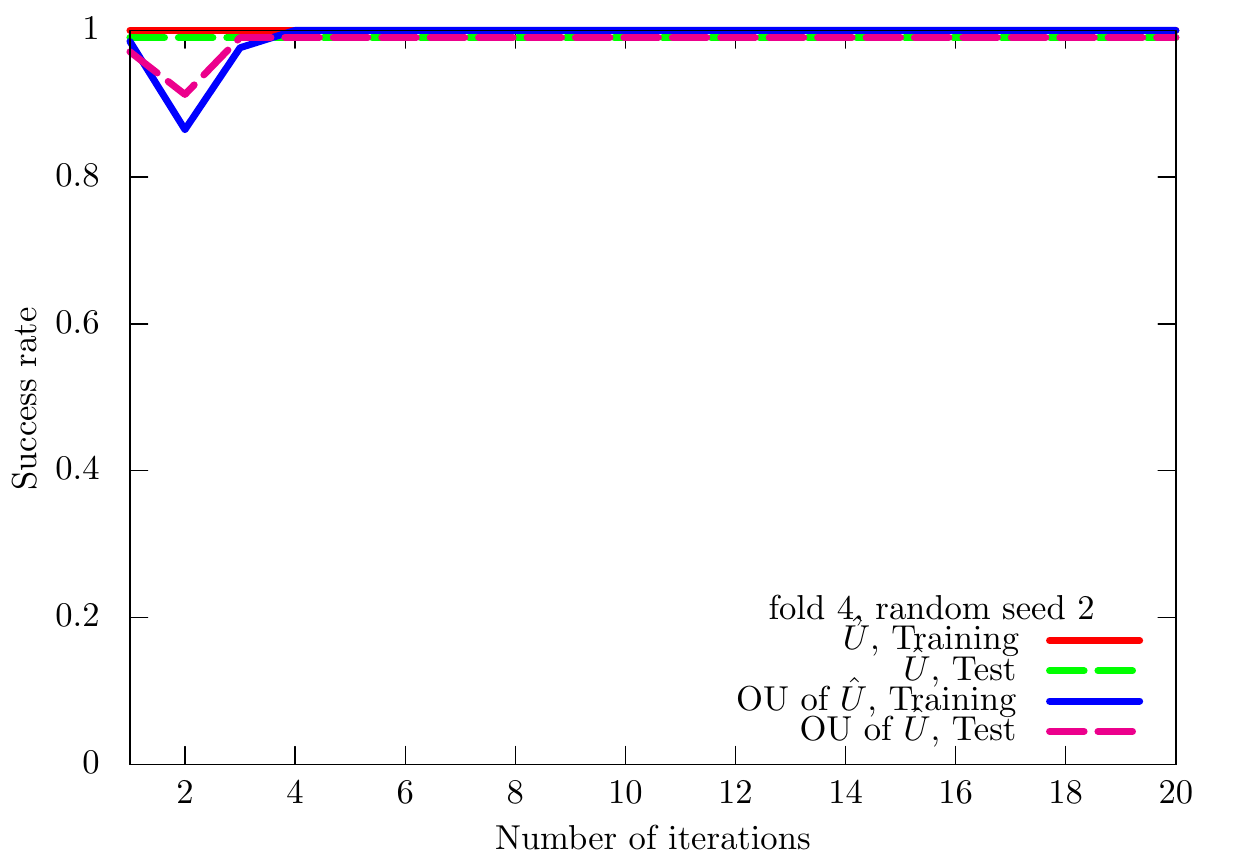}
\includegraphics[scale=0.25]{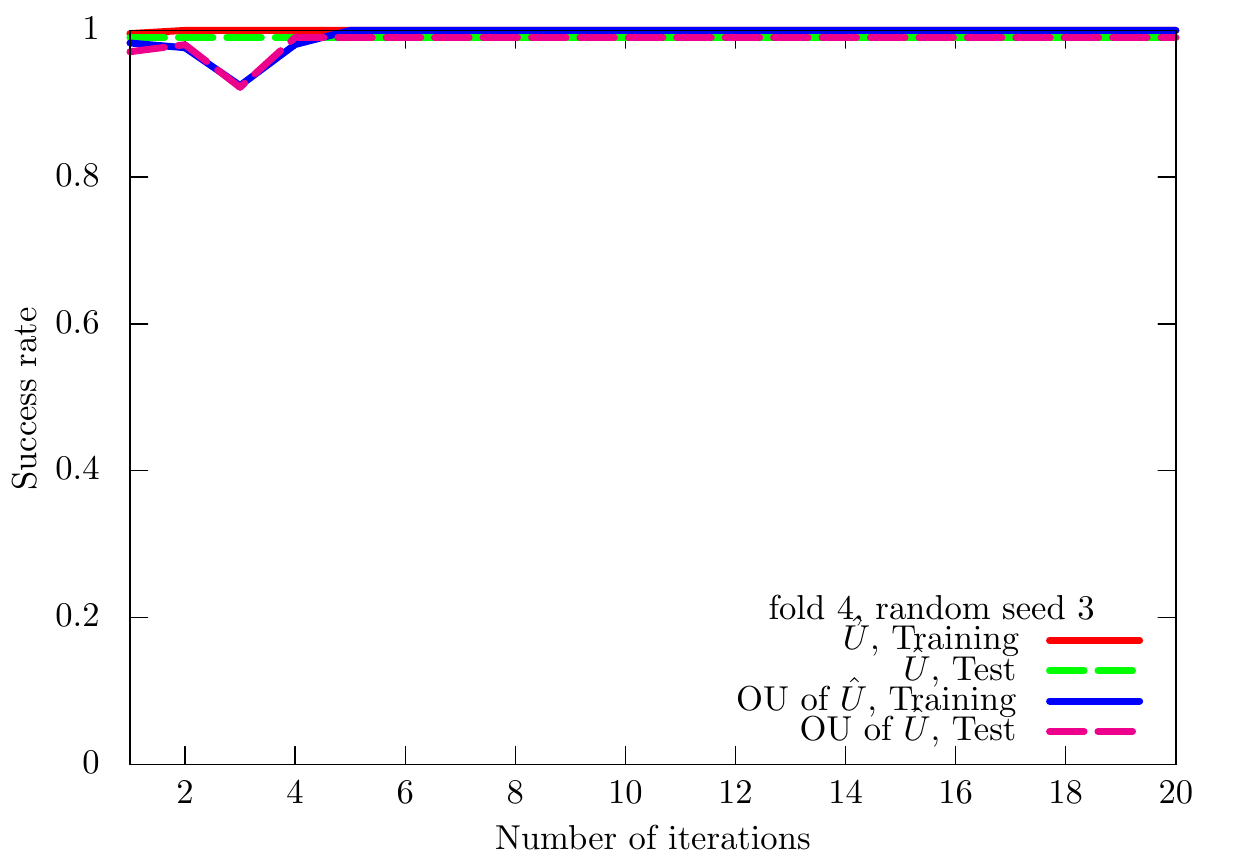}
\includegraphics[scale=0.25]{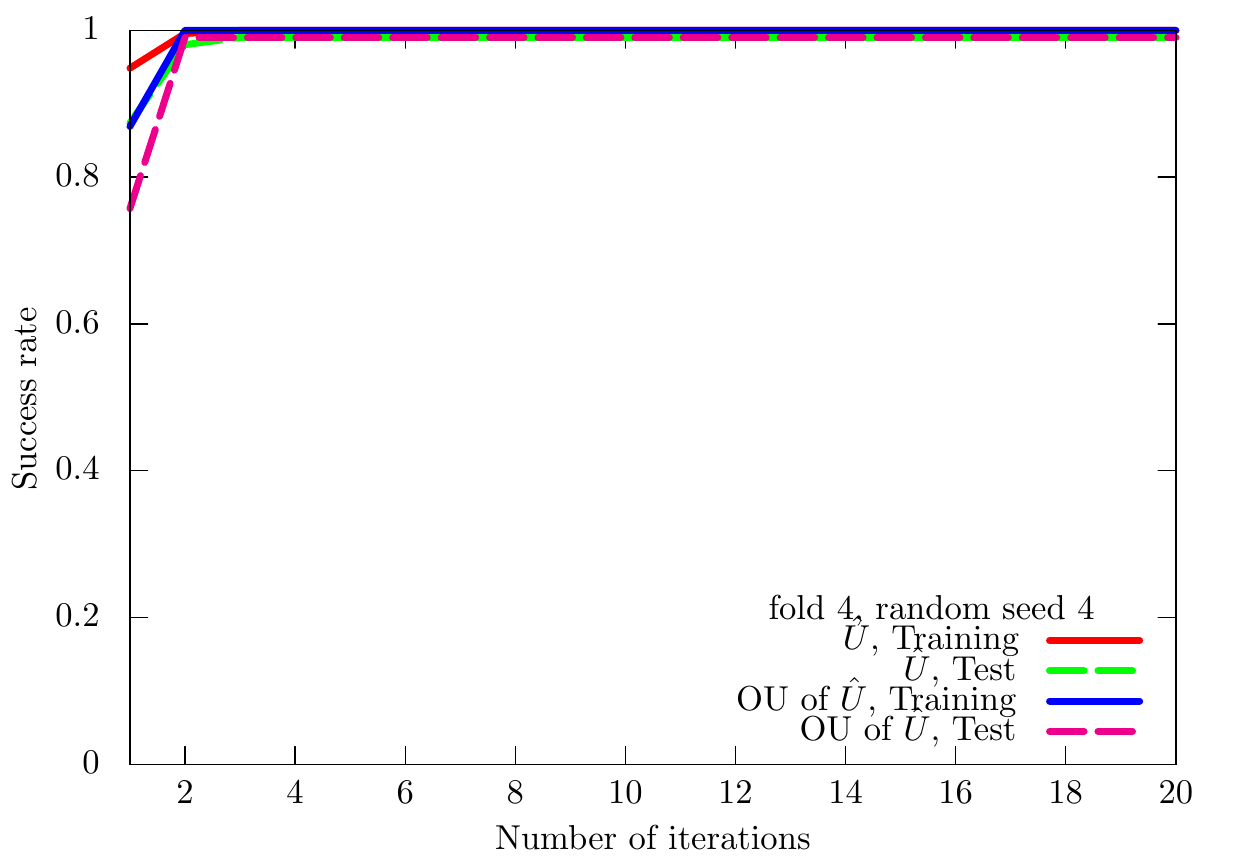}
\caption{Results of the UKM ($\hat{X}$ and $\hat{P}$) on the $5$-fold datasets with $5$ different random seeds for the MNIST256 dataset ($0$ or $1$). We use complex matrices and set $\theta_\mathrm{bias} = 0$. We set $r = 0.010$.}
\label{supp-arXiv-numerical-result-raw-data-fold-001-rand-001-UKM-P-MNIST256-0-1}
\end{figure*}

\begin{figure*}[htb]
\centering
\includegraphics[scale=0.25]{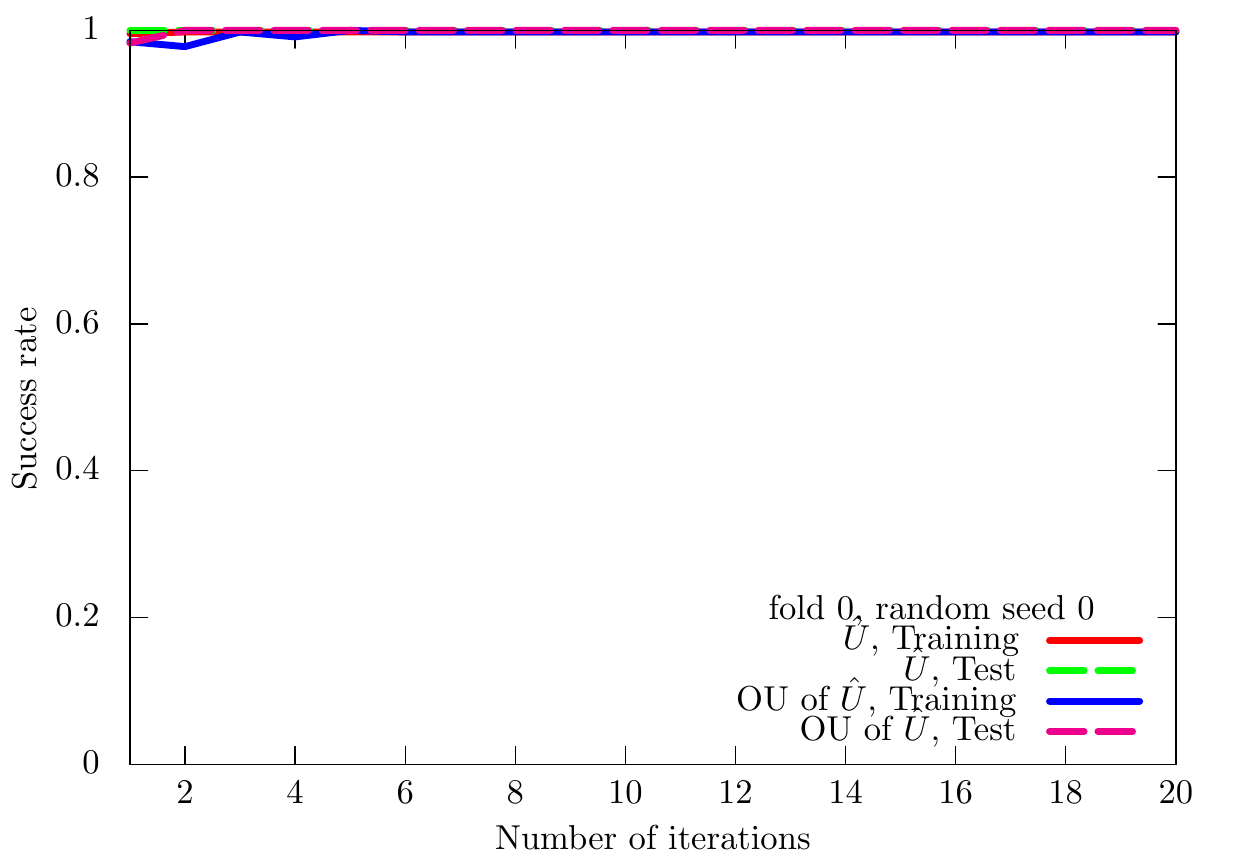}
\includegraphics[scale=0.25]{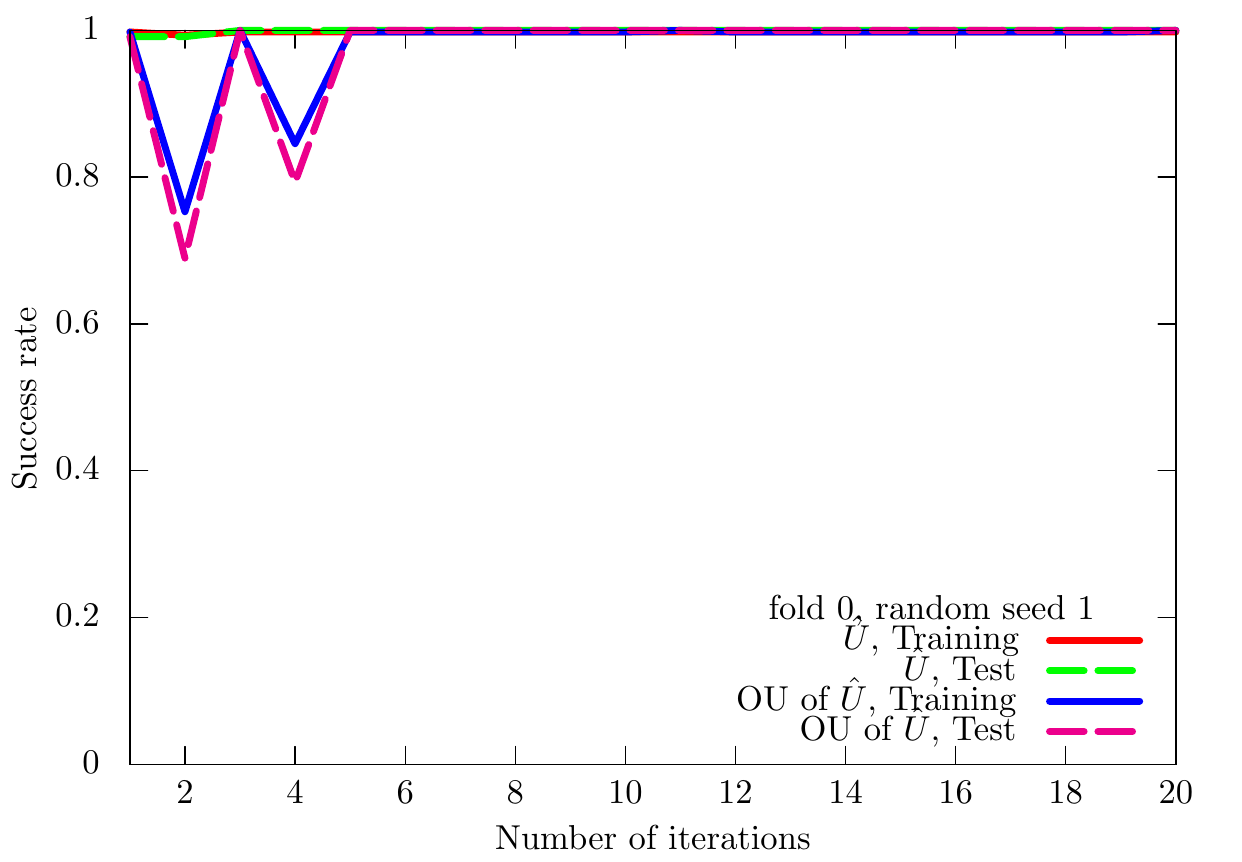}
\includegraphics[scale=0.25]{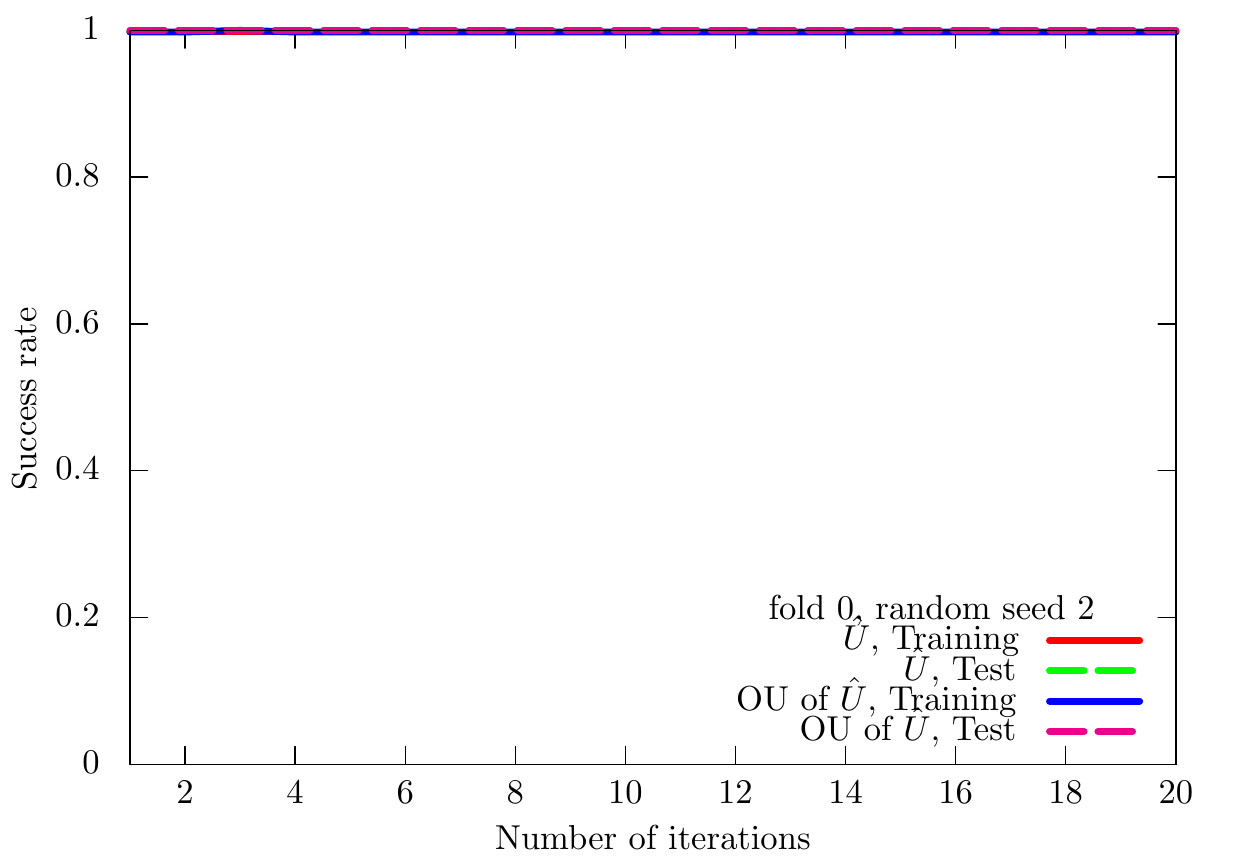}
\includegraphics[scale=0.25]{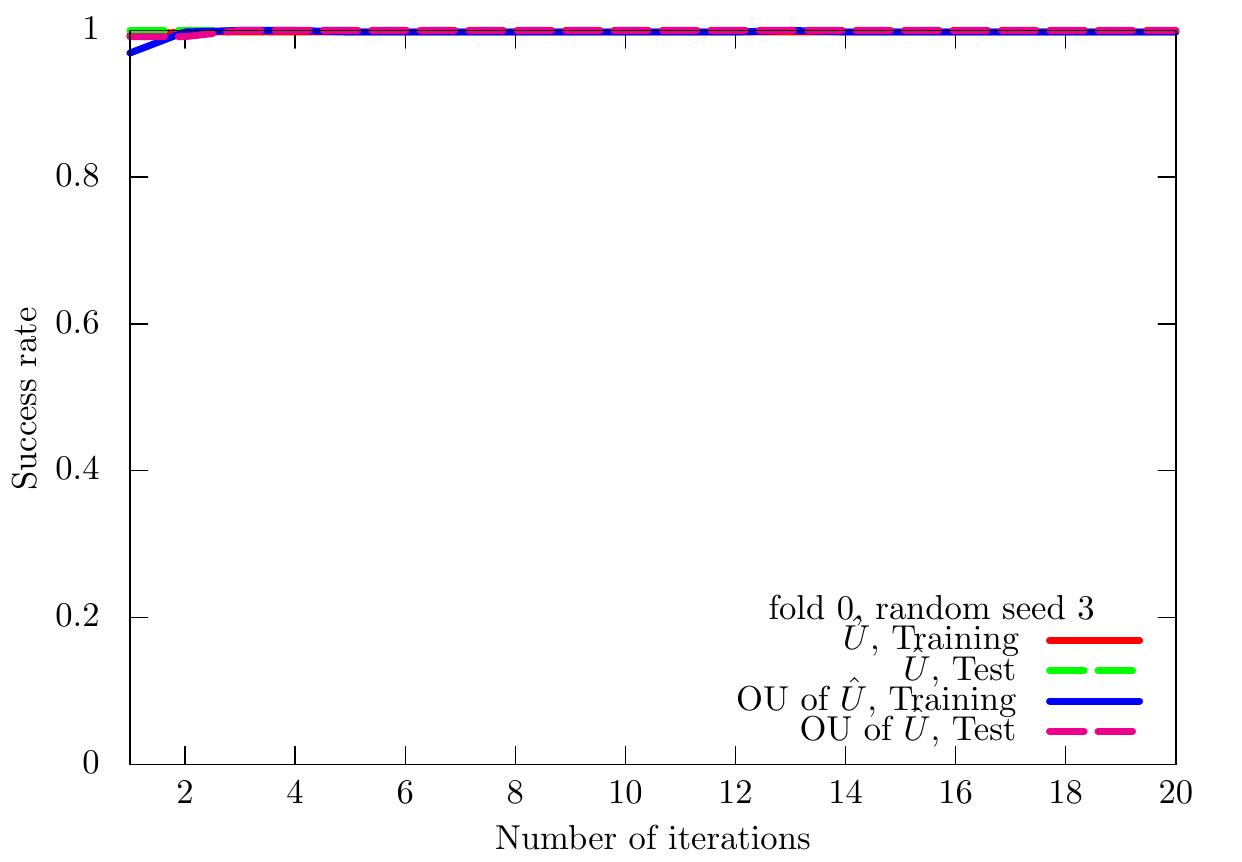}
\includegraphics[scale=0.25]{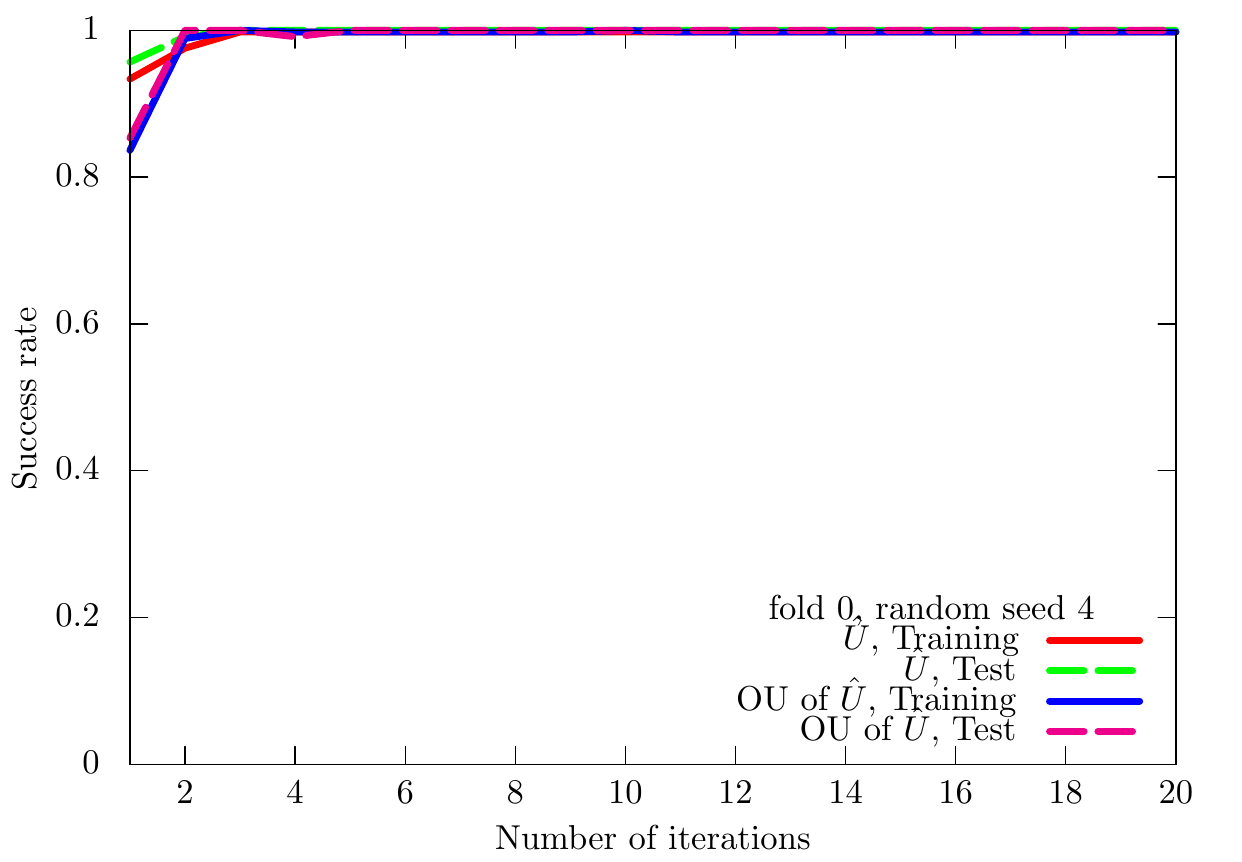}
\includegraphics[scale=0.25]{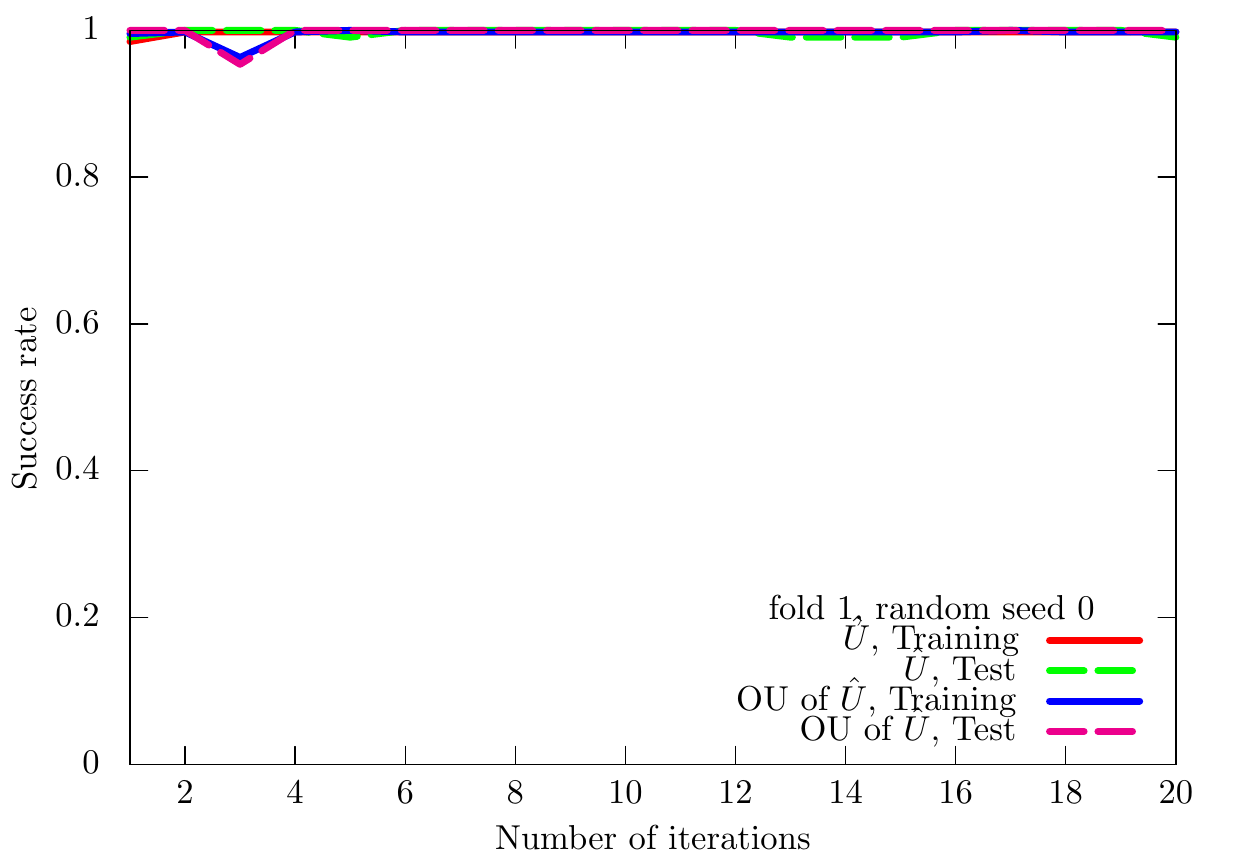}
\includegraphics[scale=0.25]{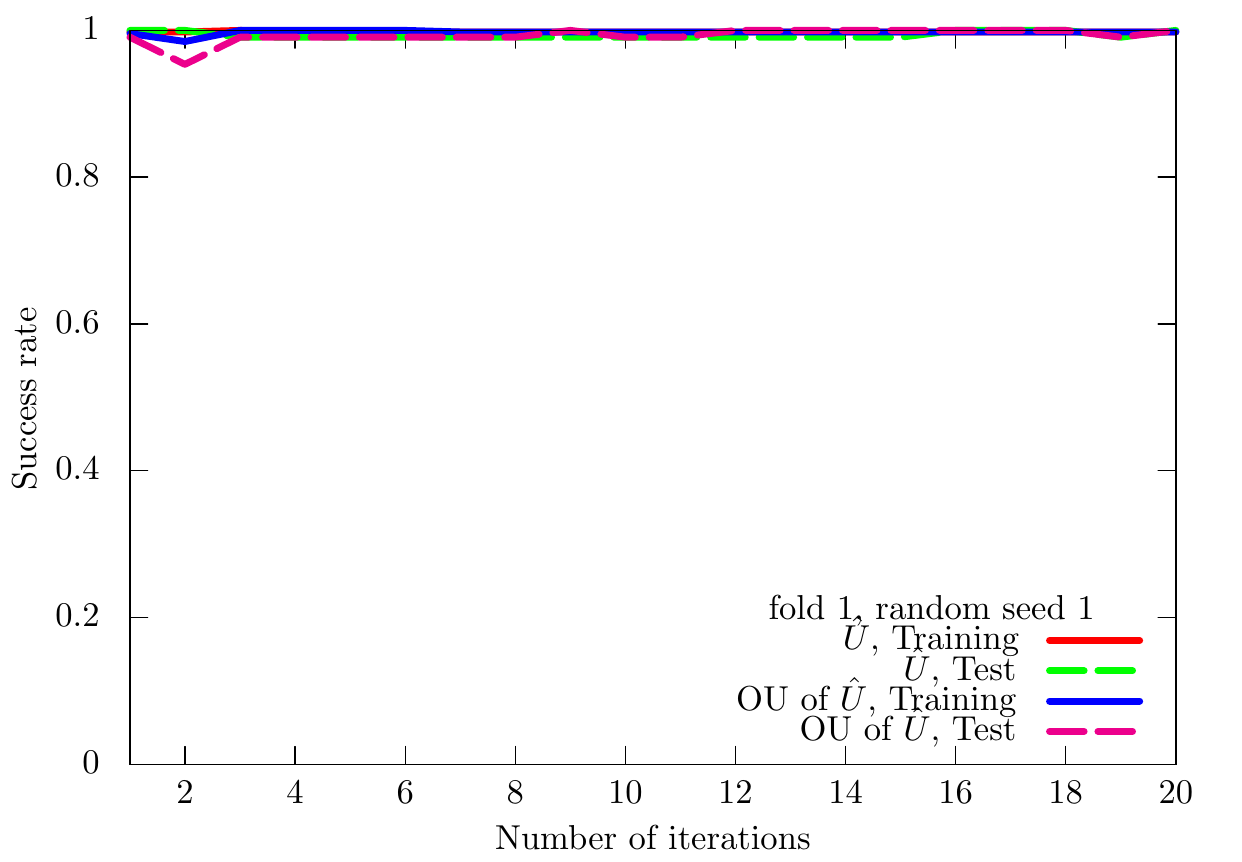}
\includegraphics[scale=0.25]{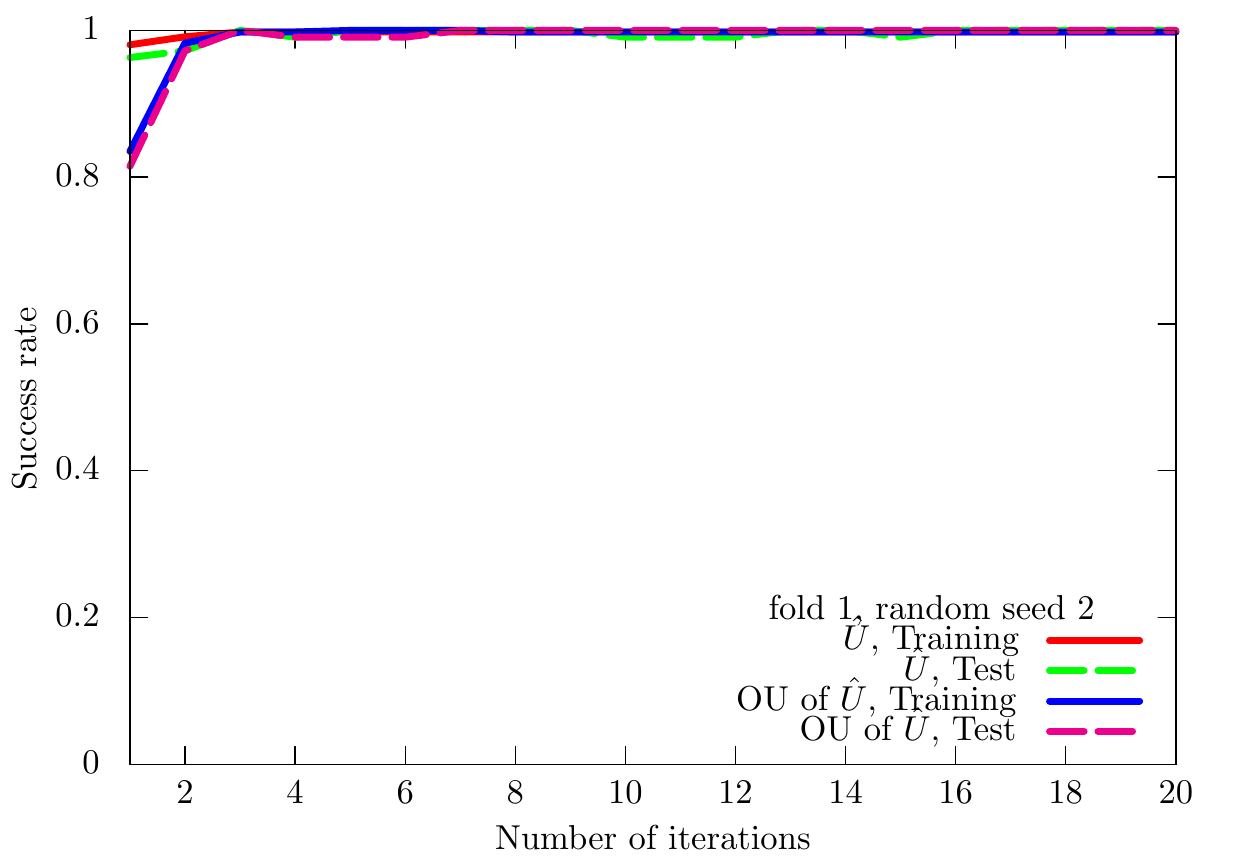}
\includegraphics[scale=0.25]{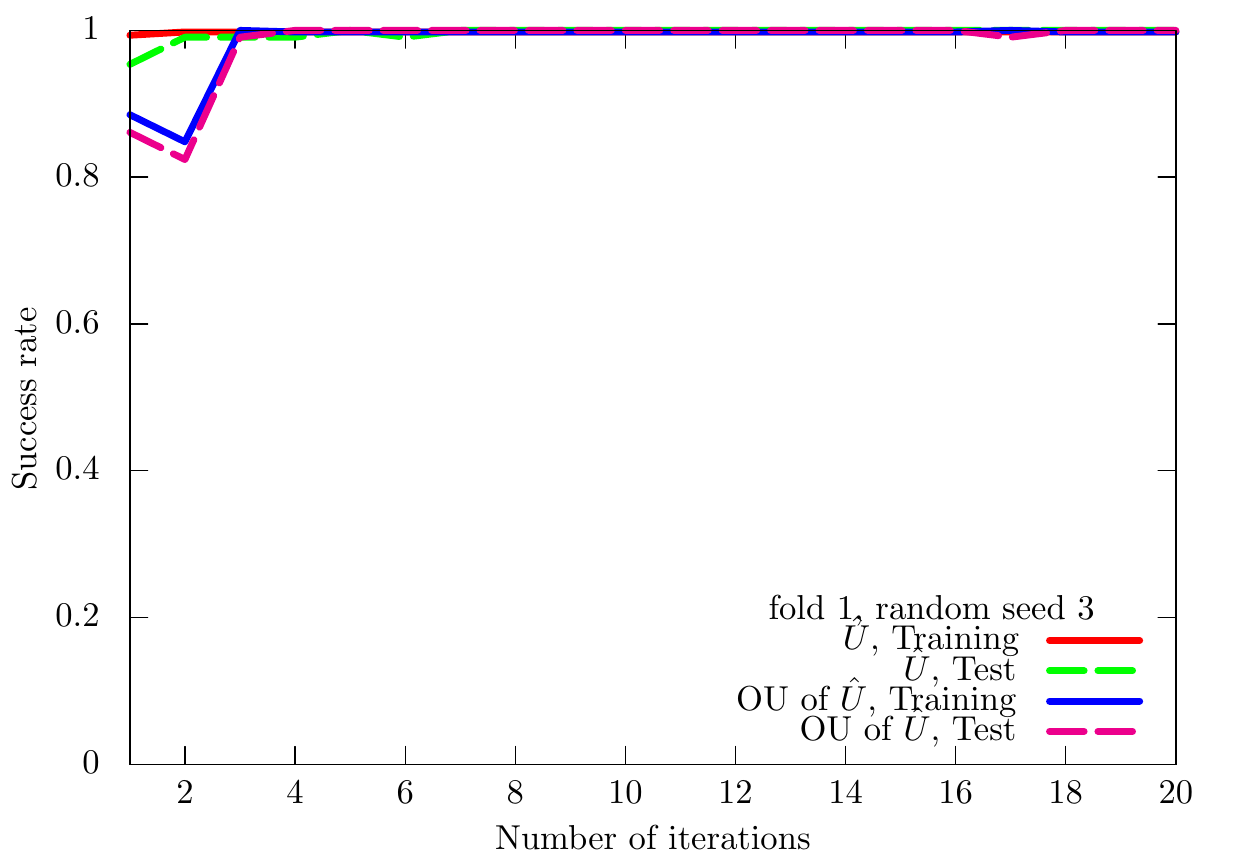}
\includegraphics[scale=0.25]{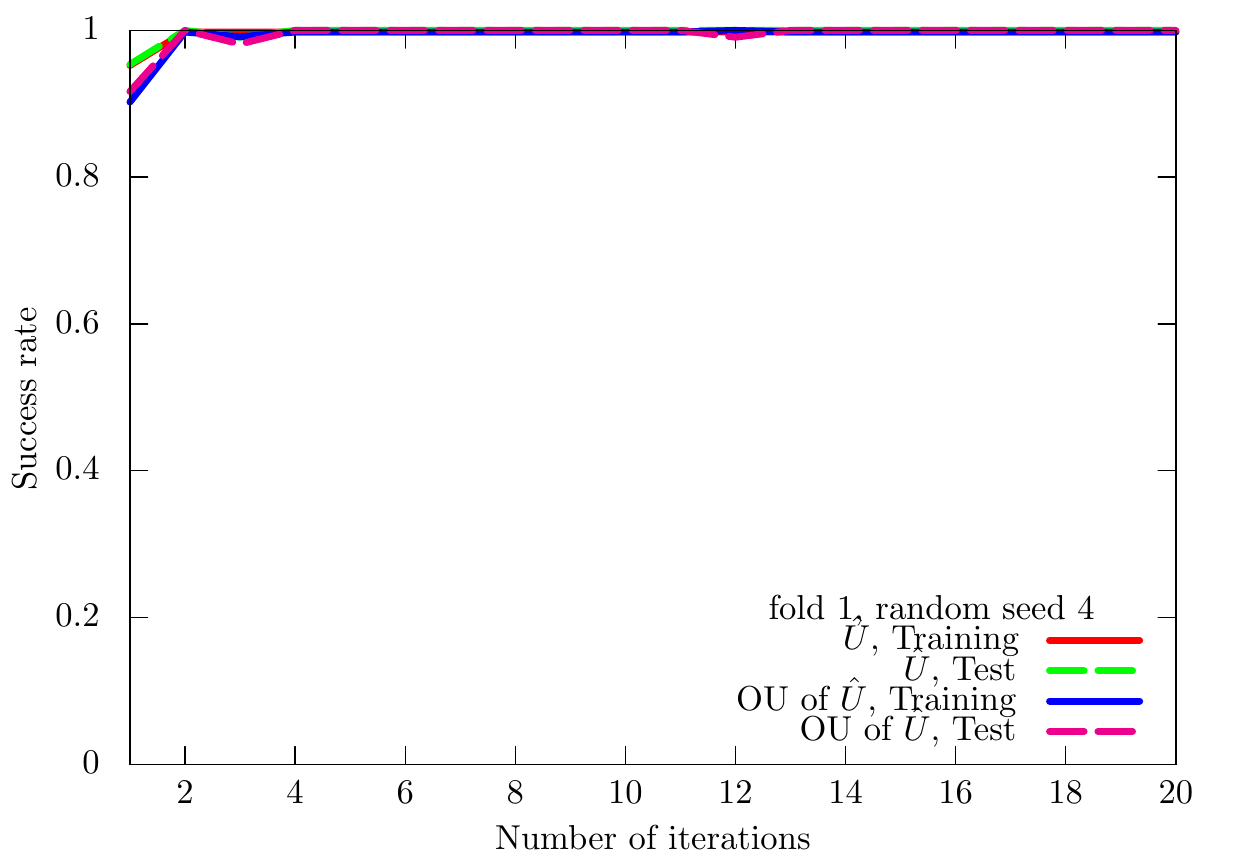}
\includegraphics[scale=0.25]{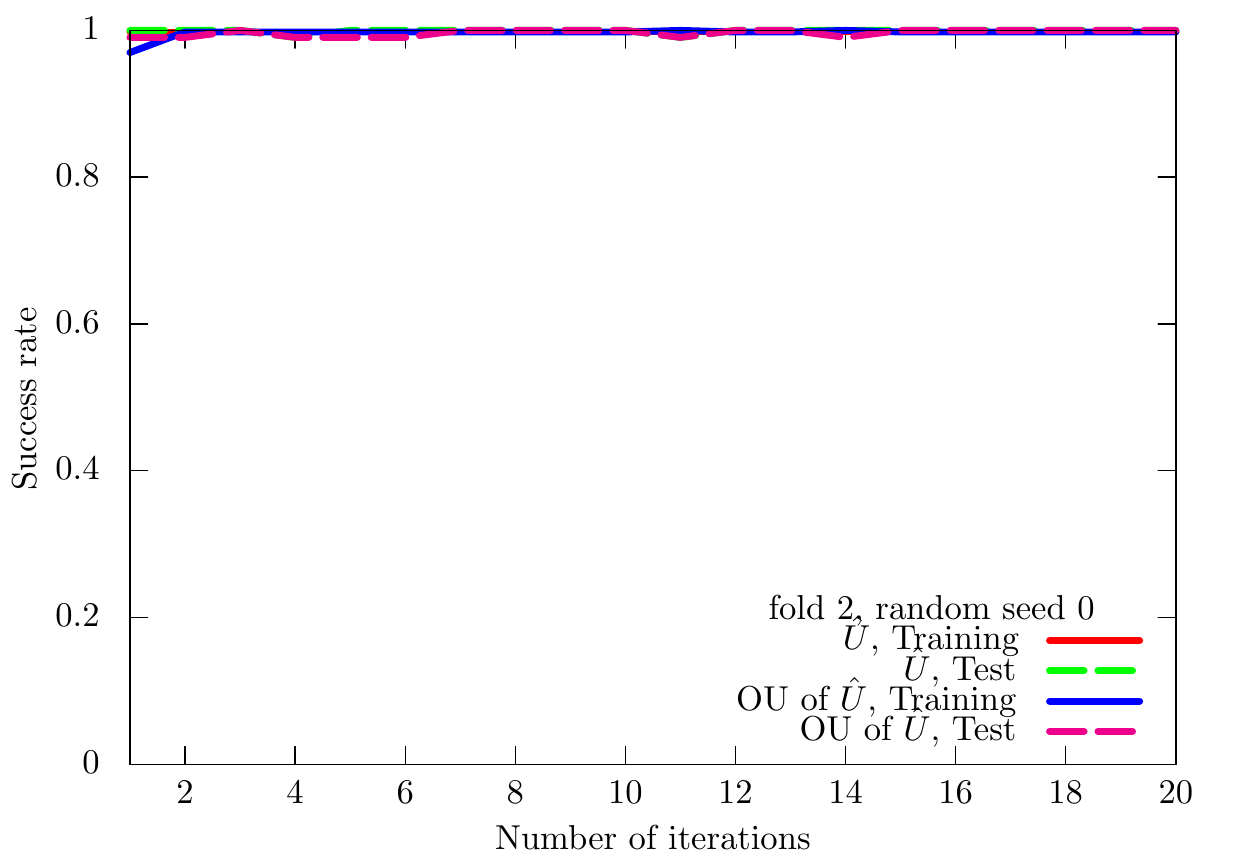}
\includegraphics[scale=0.25]{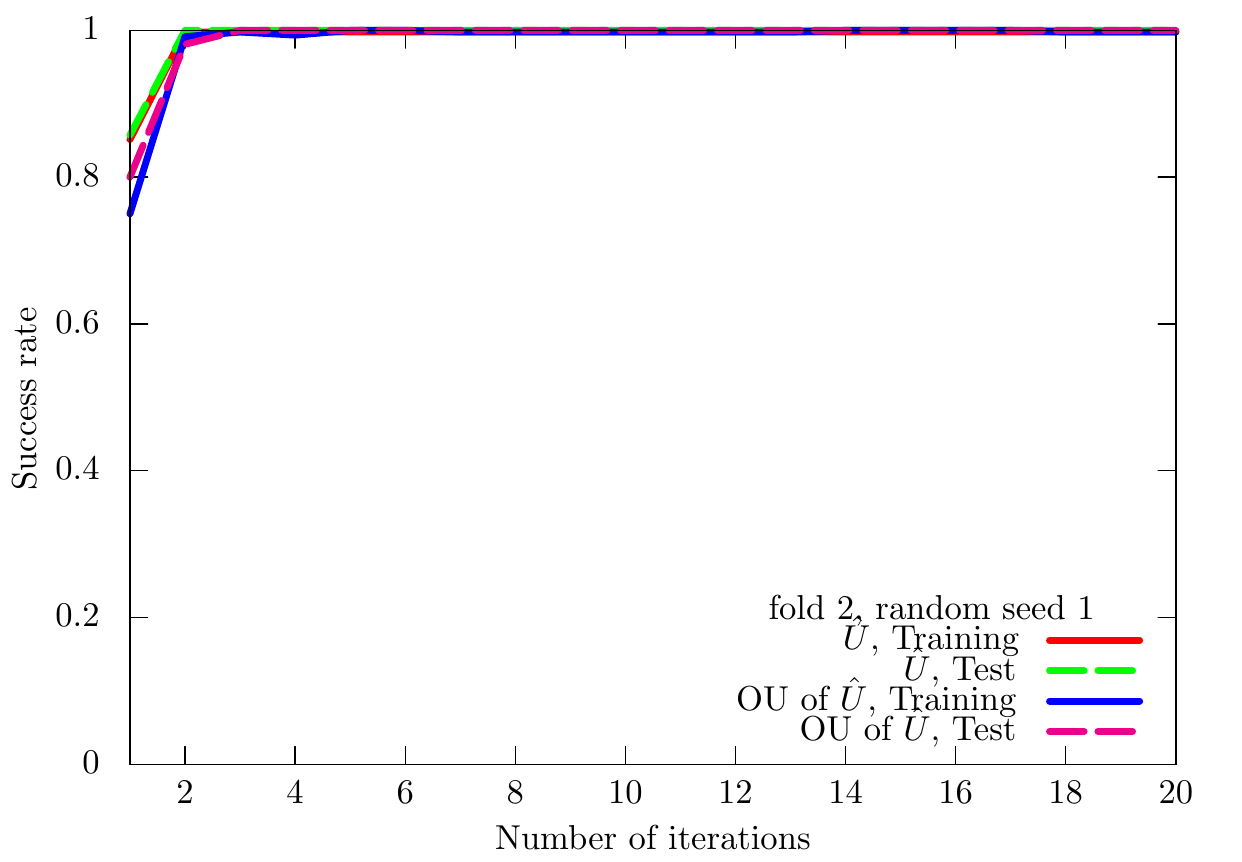}
\includegraphics[scale=0.25]{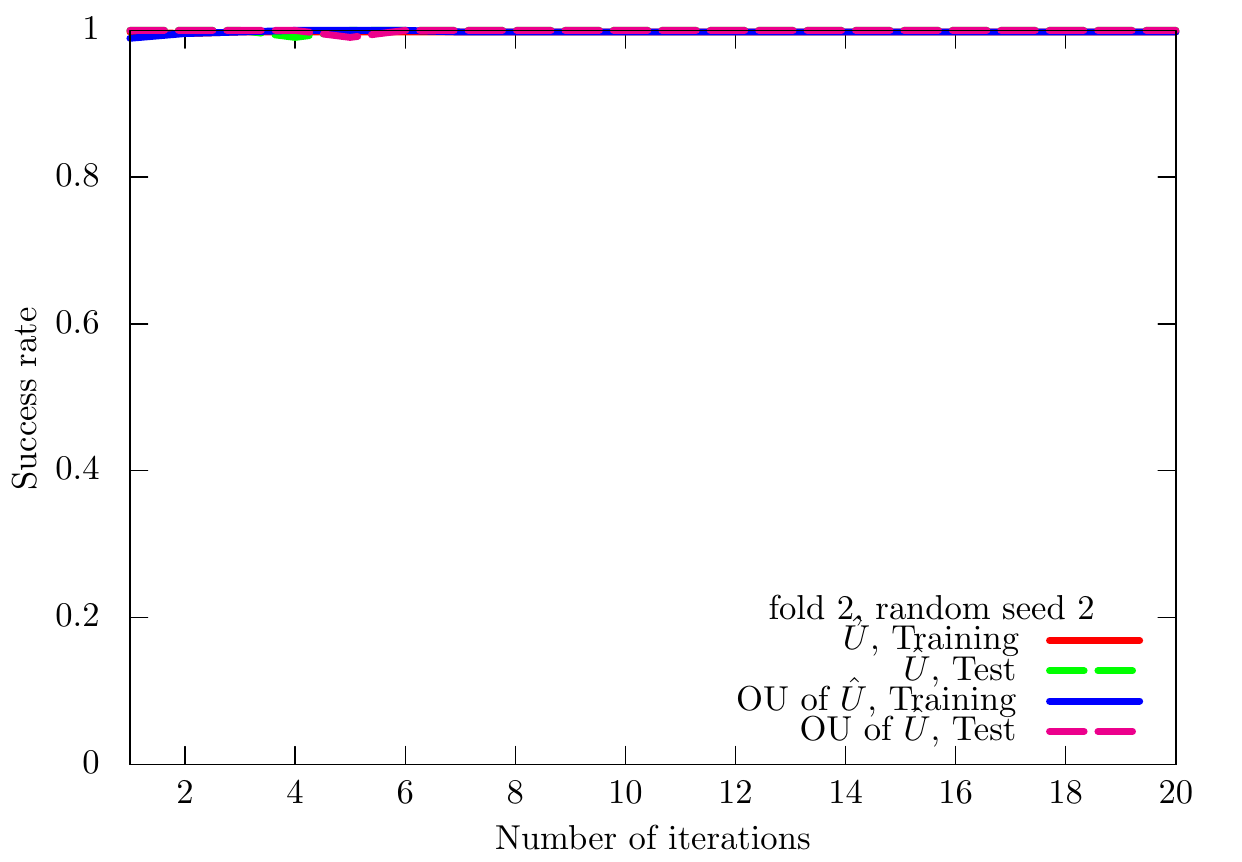}
\includegraphics[scale=0.25]{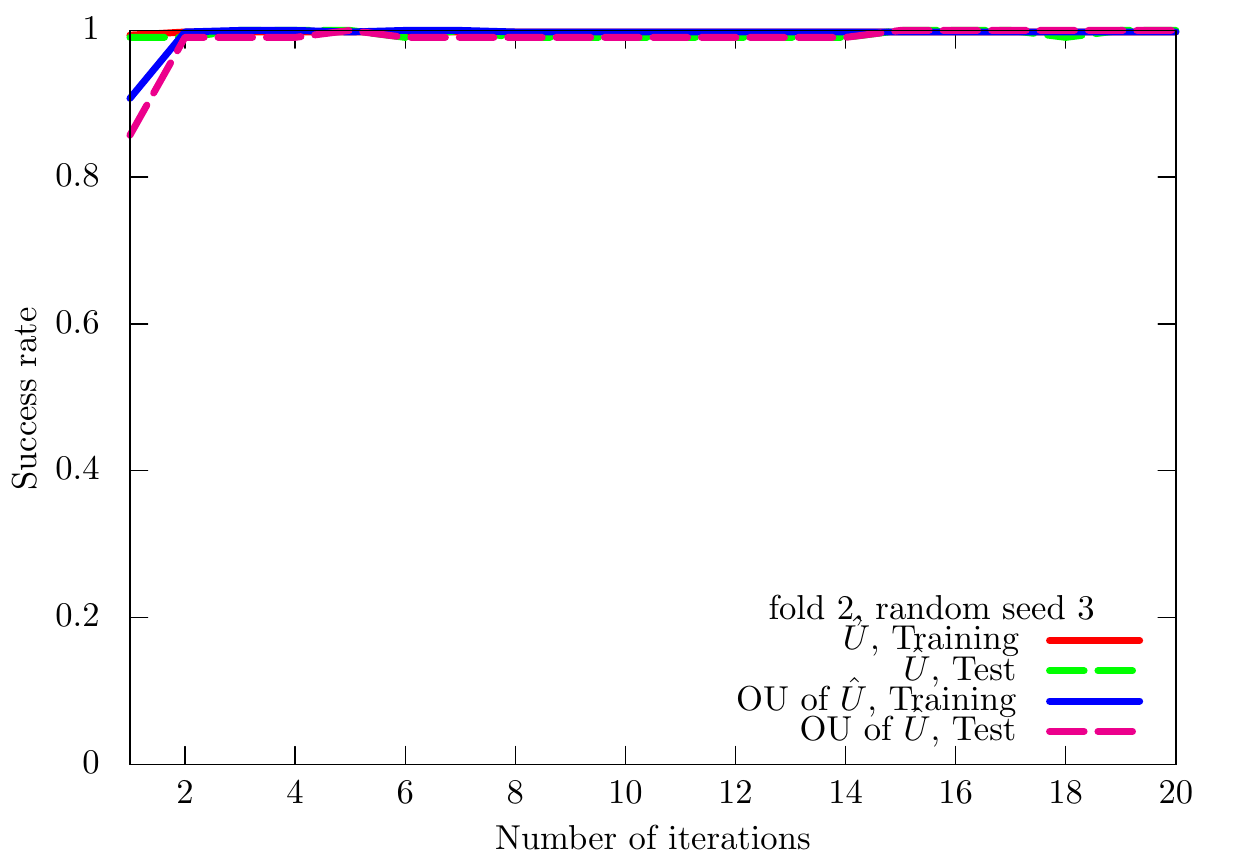}
\includegraphics[scale=0.25]{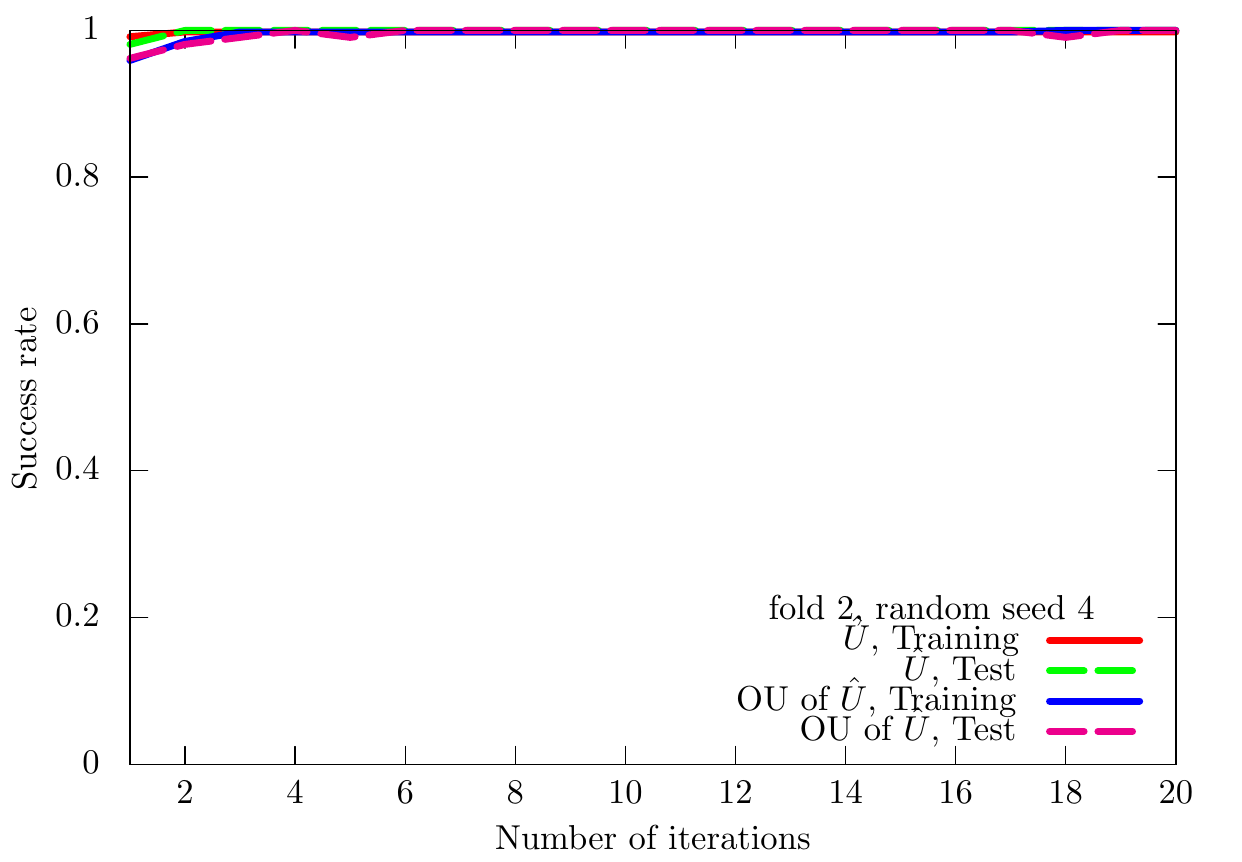}
\includegraphics[scale=0.25]{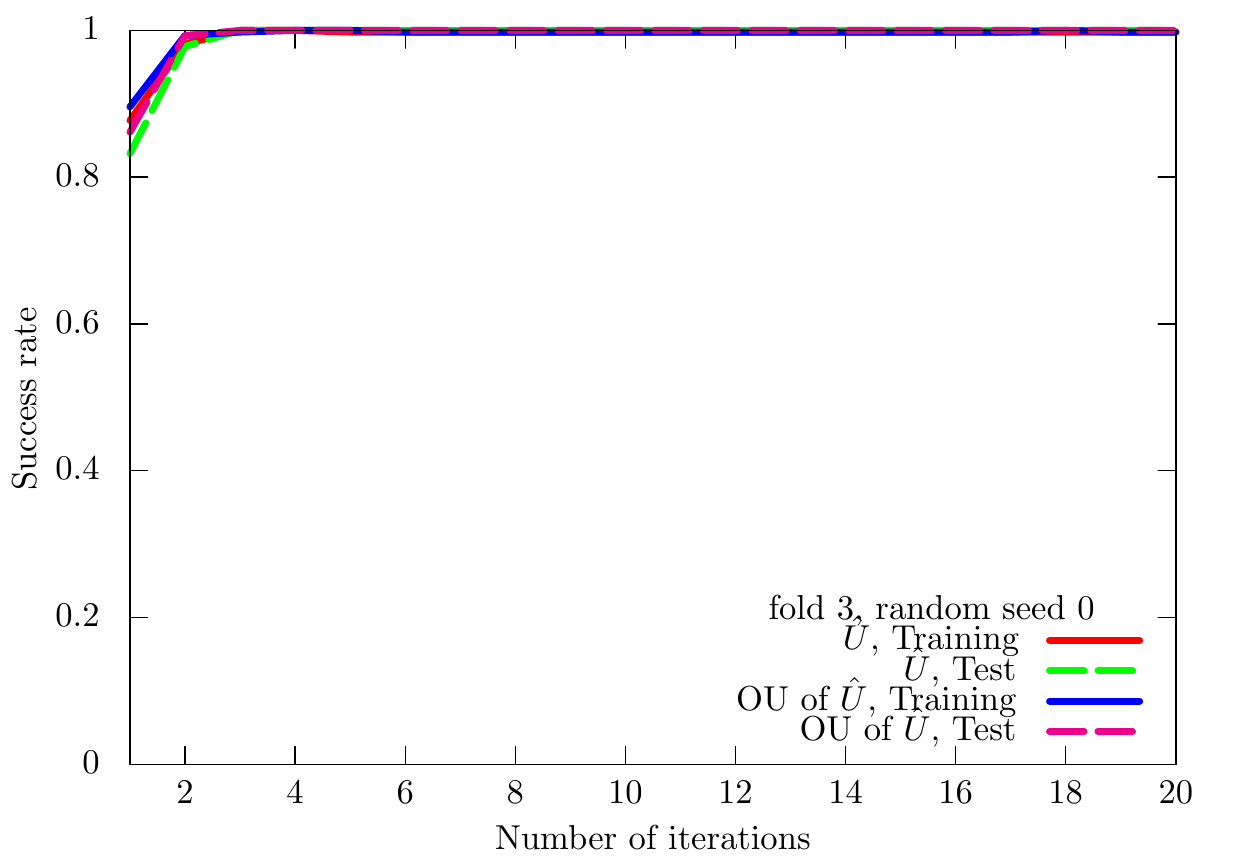}
\includegraphics[scale=0.25]{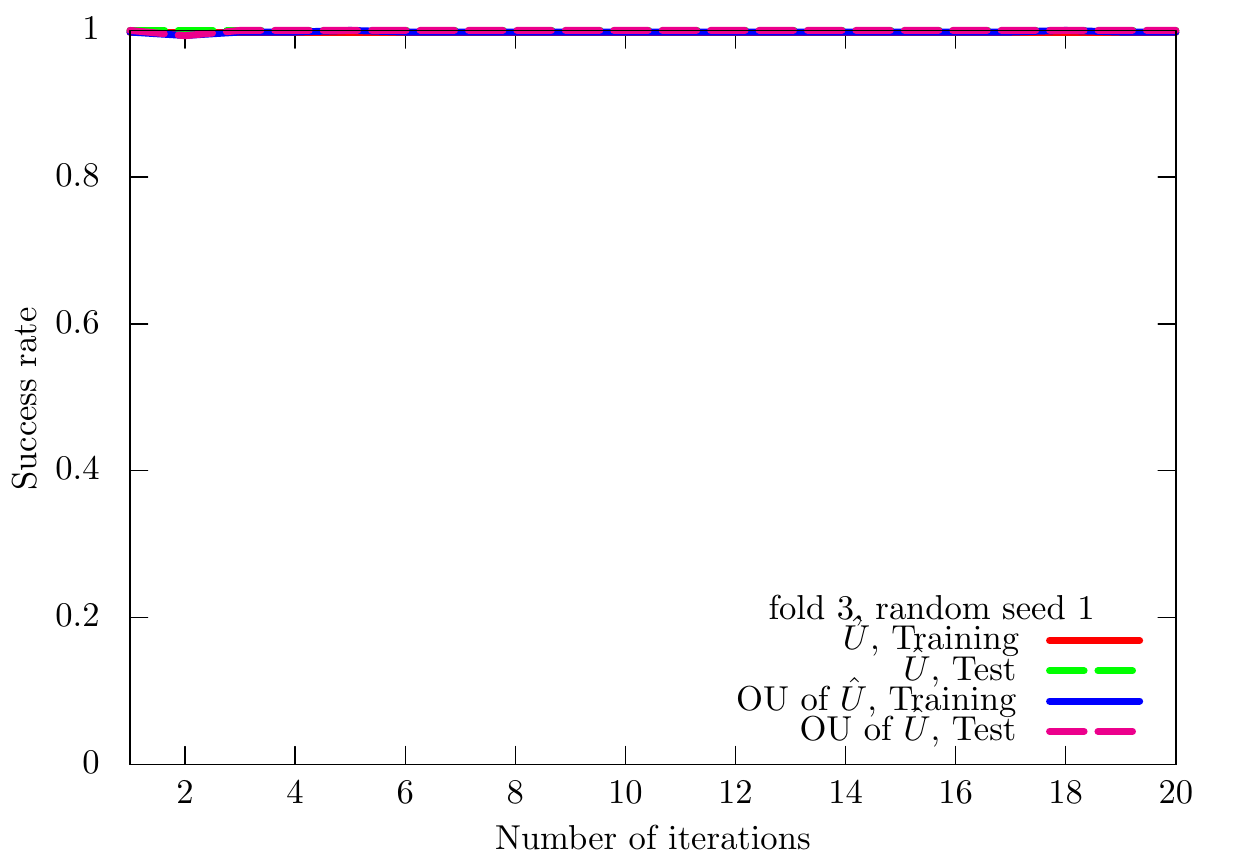}
\includegraphics[scale=0.25]{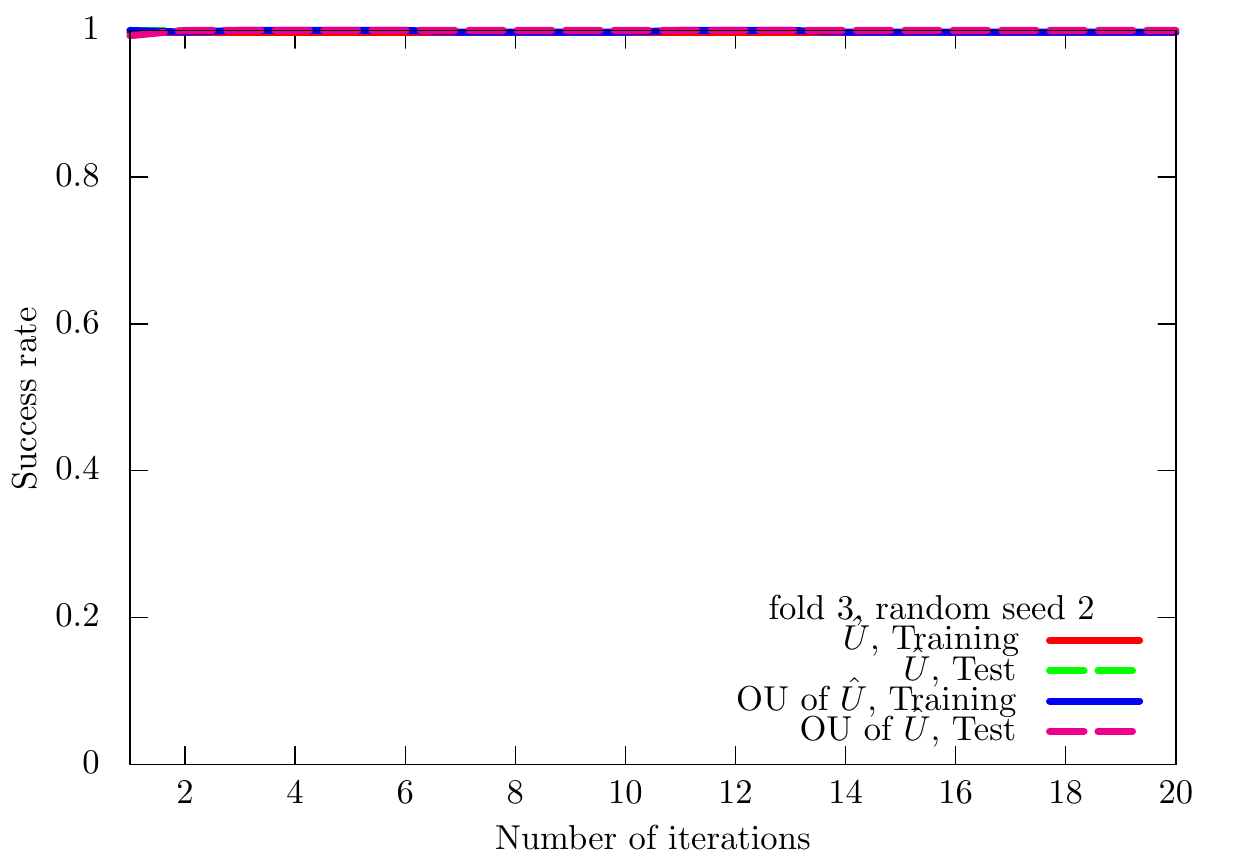}
\includegraphics[scale=0.25]{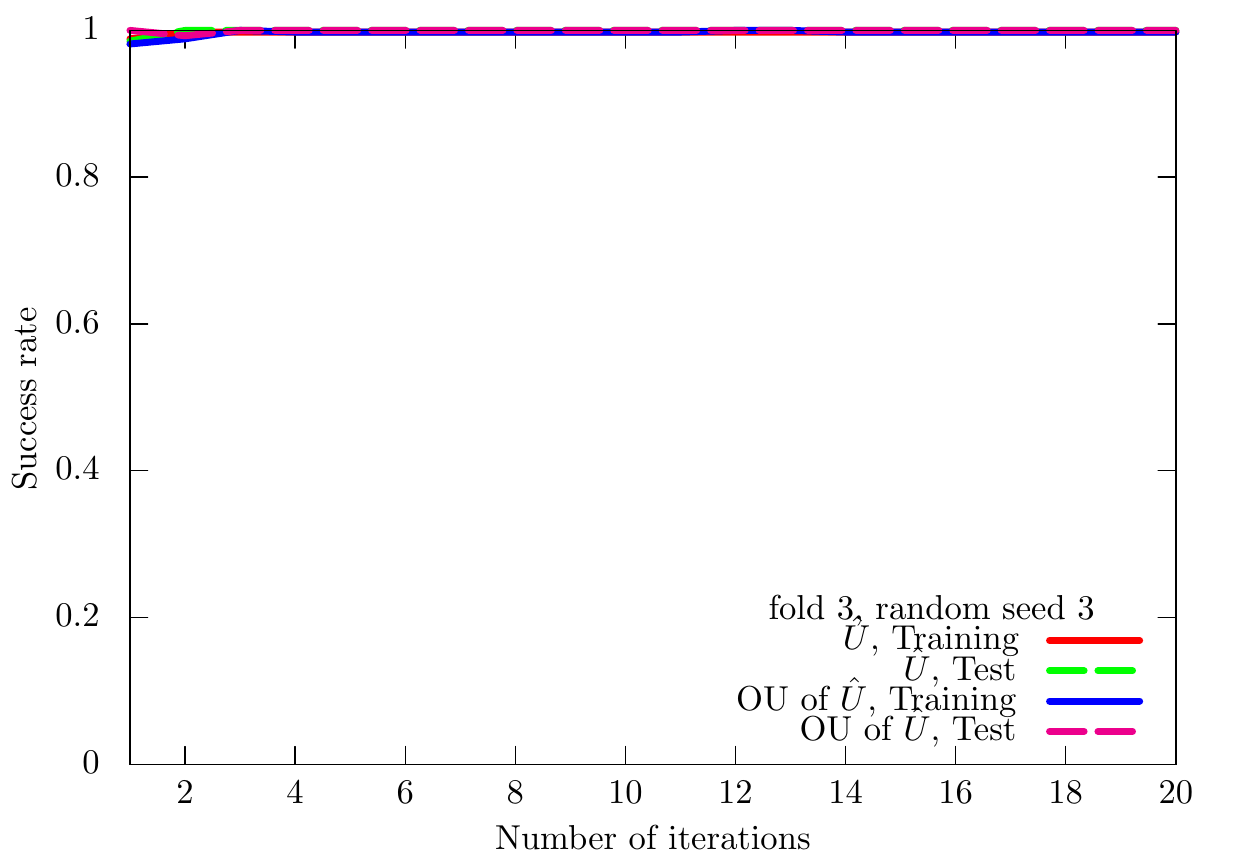}
\includegraphics[scale=0.25]{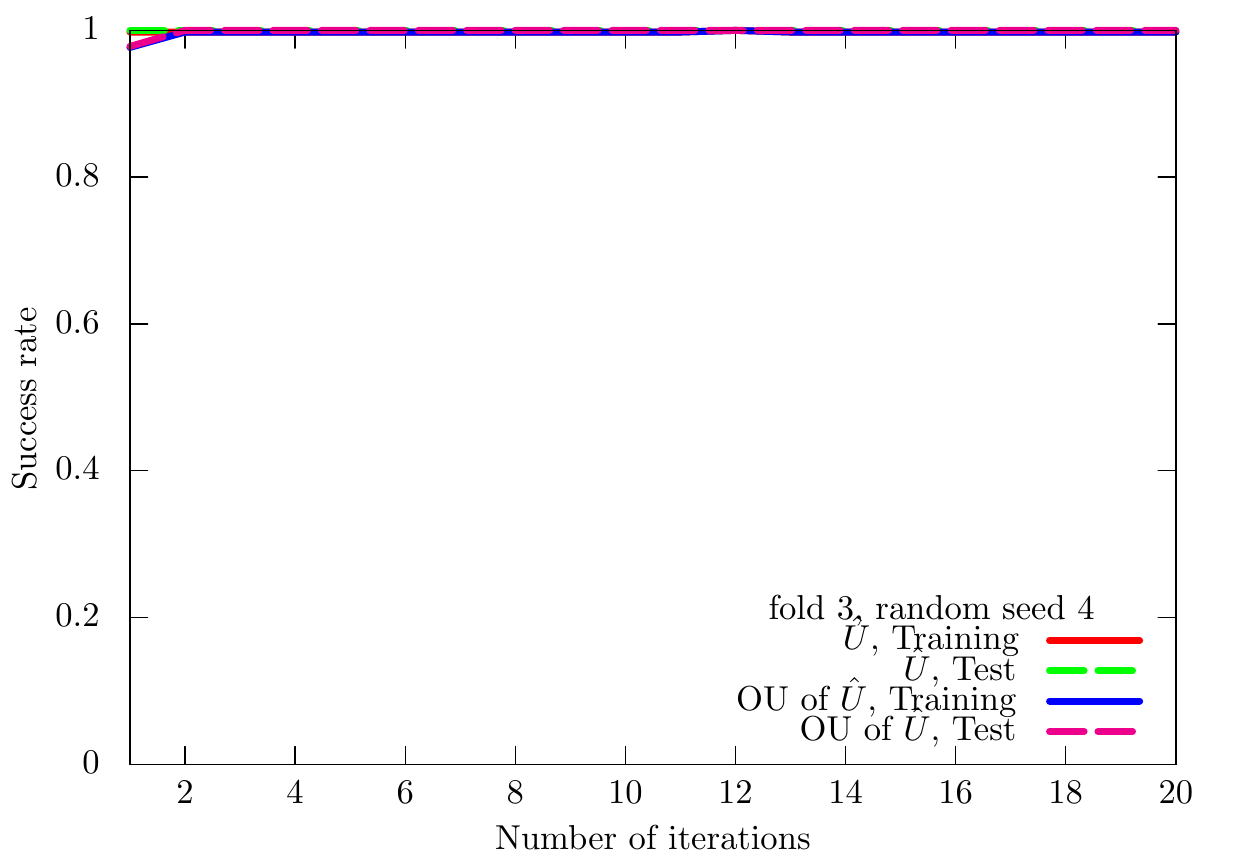}
\includegraphics[scale=0.25]{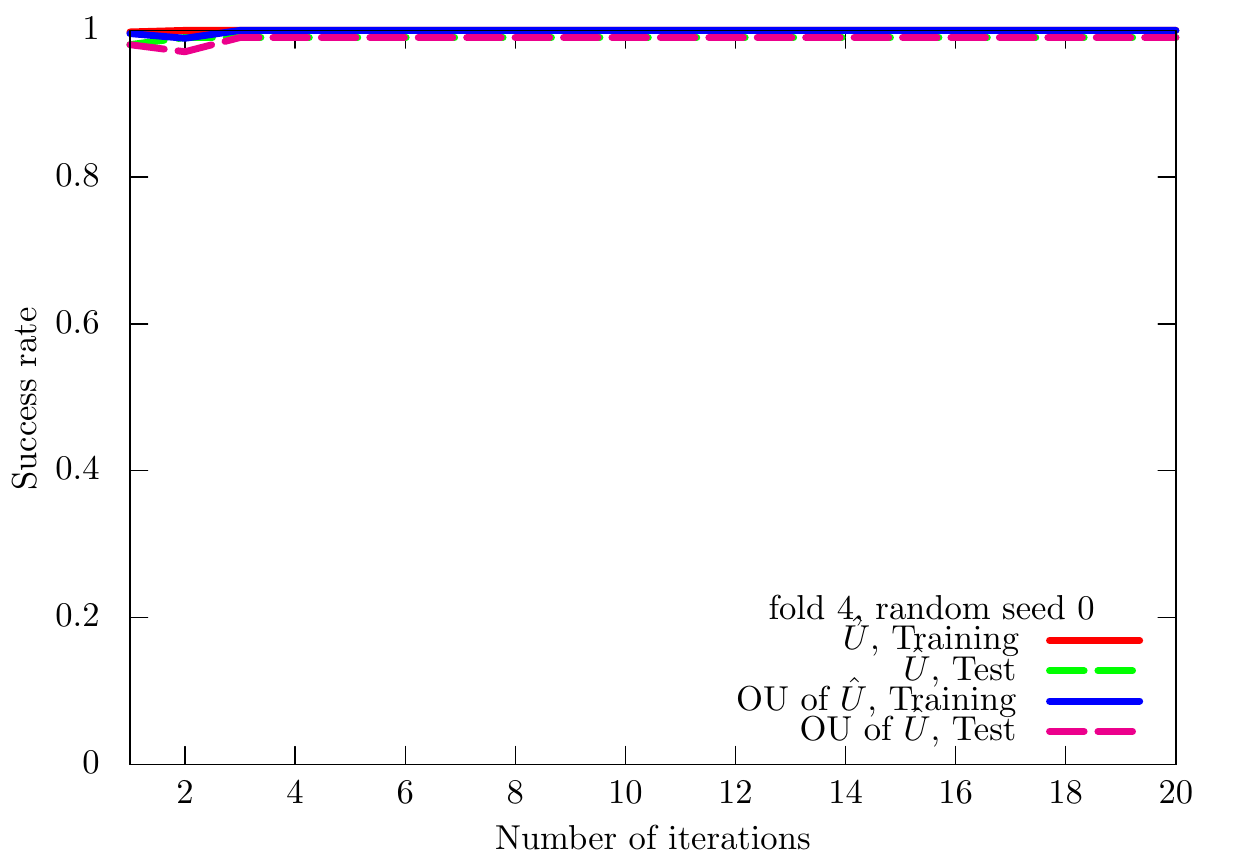}
\includegraphics[scale=0.25]{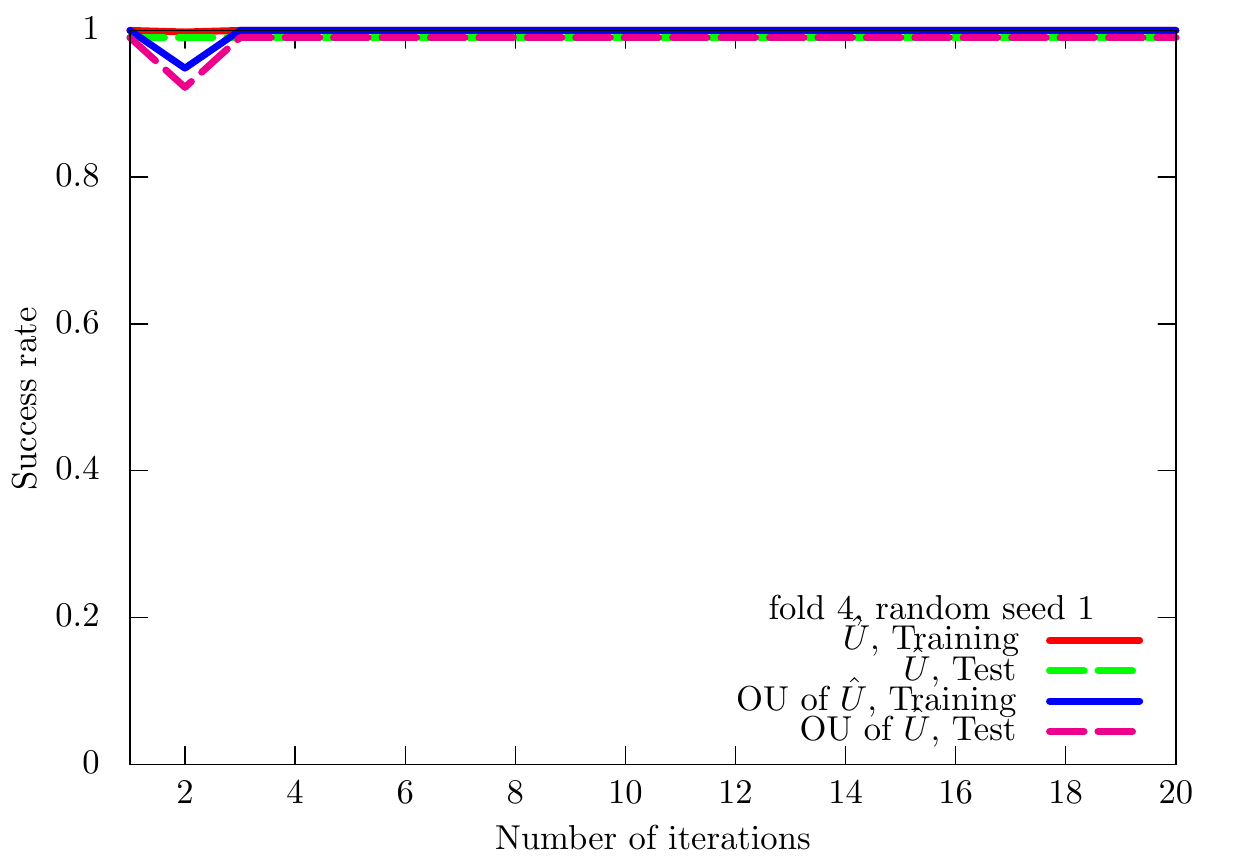}
\includegraphics[scale=0.25]{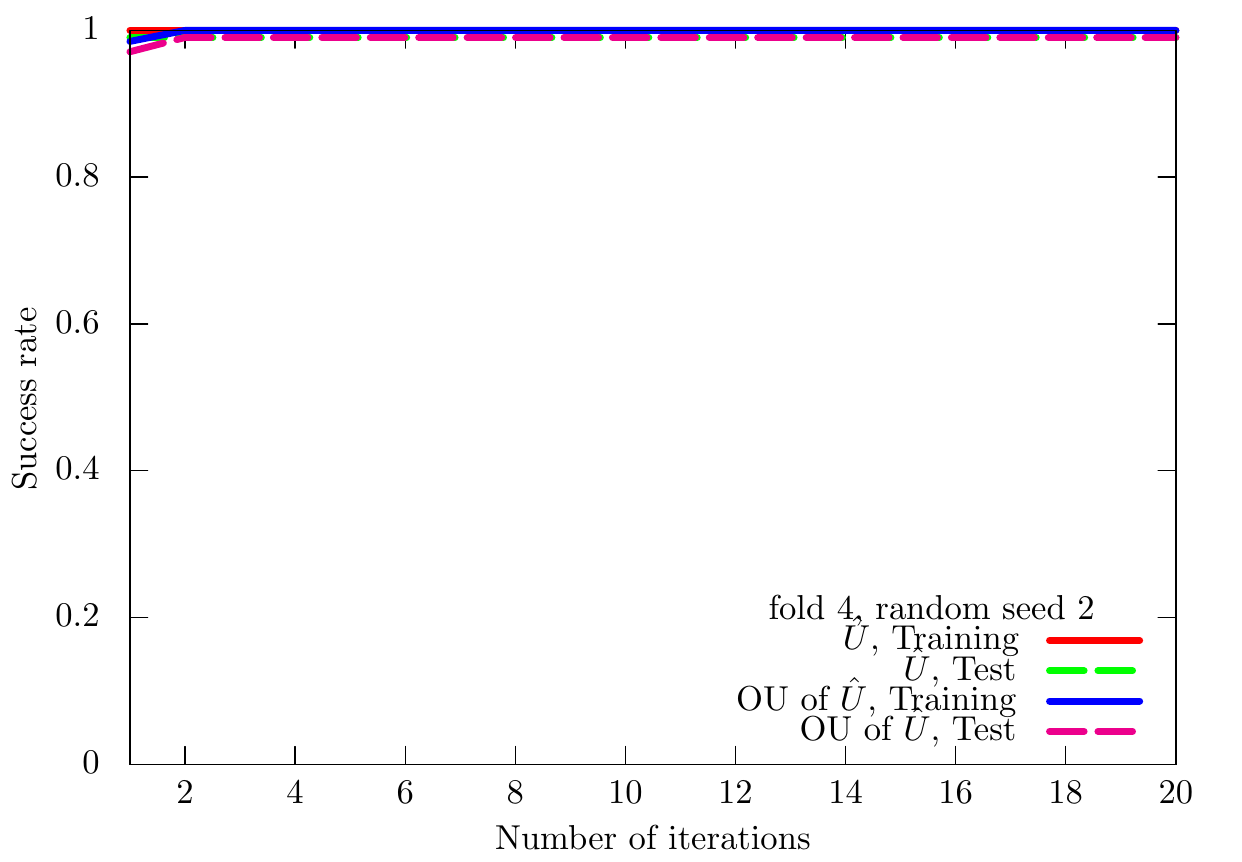}
\includegraphics[scale=0.25]{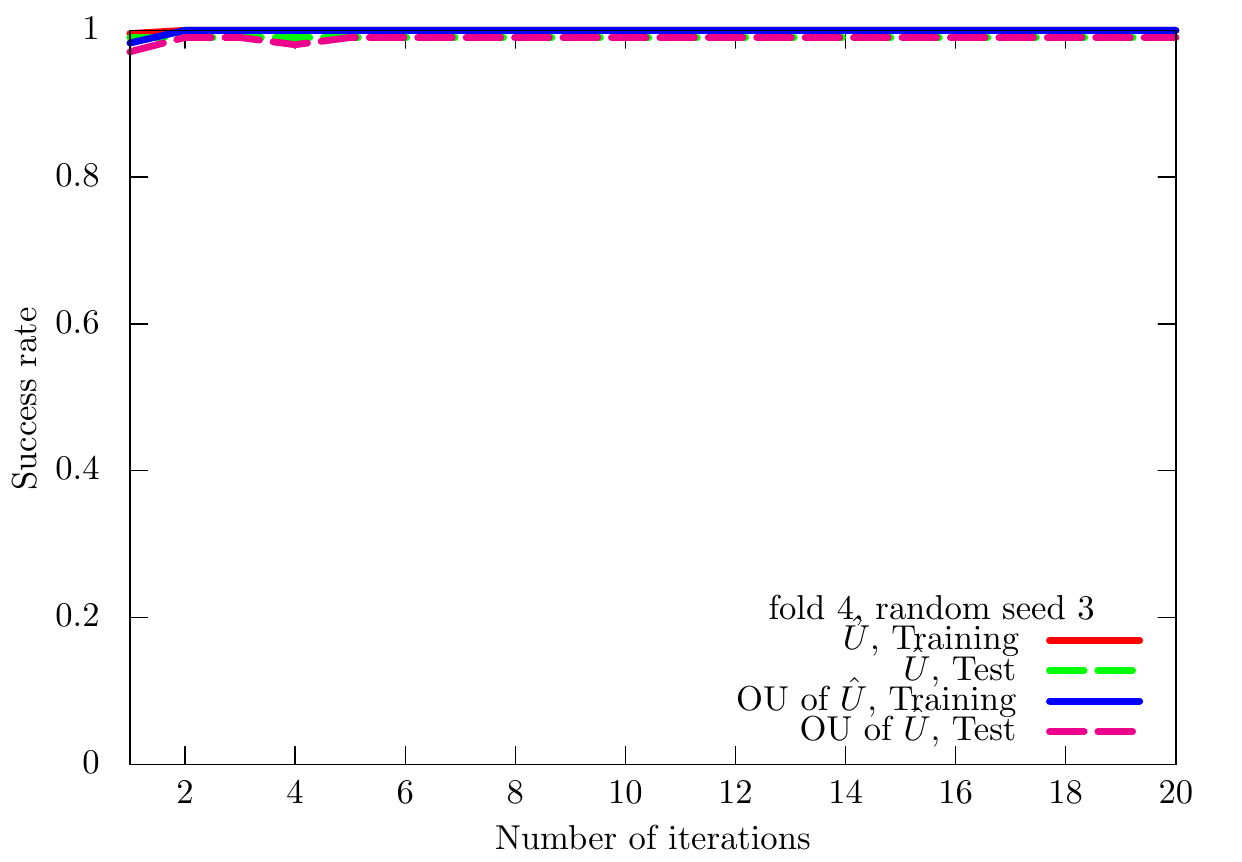}
\includegraphics[scale=0.25]{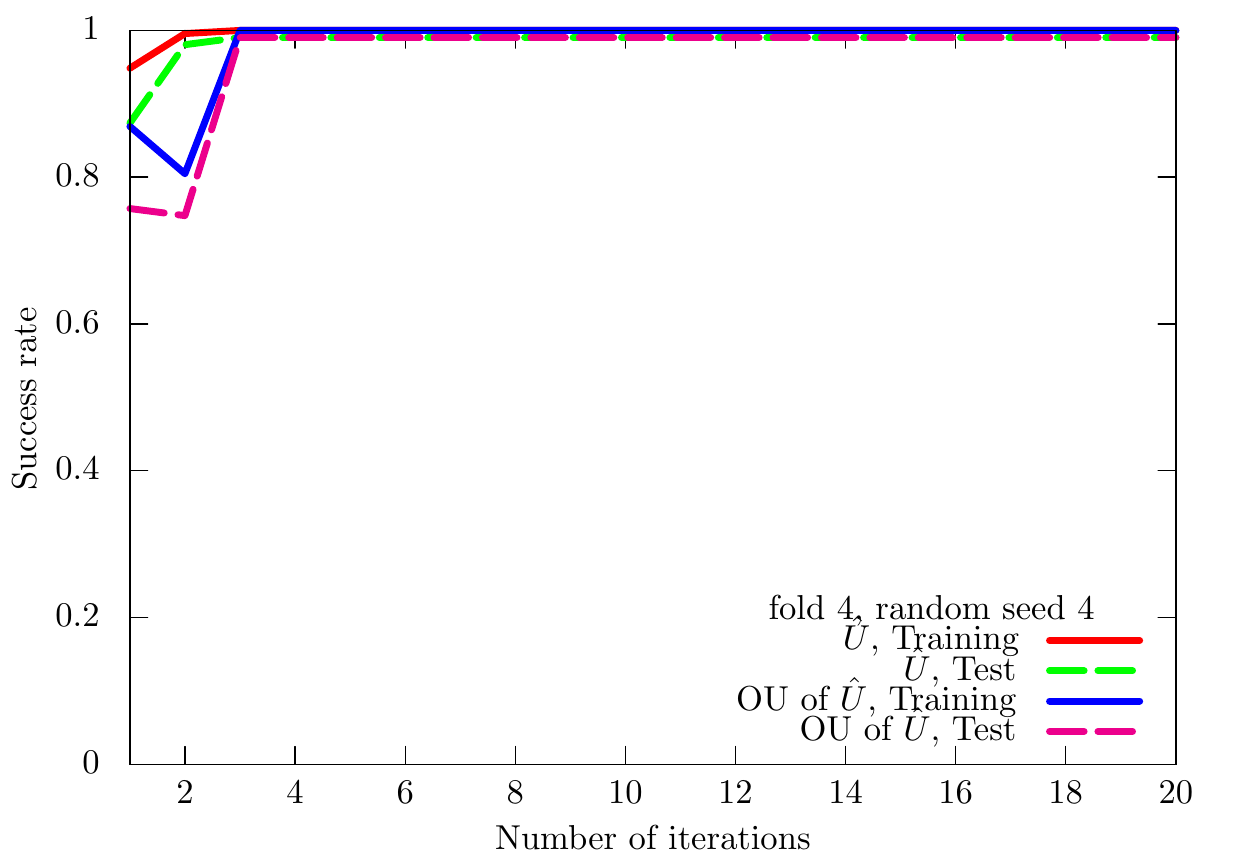}
\caption{Results of the UKM ($\hat{X}$ and OU of $\hat{X}$) on the $5$-fold datasets with $5$ different random seeds for the MNIST256 dataset ($0$ or $1$). We use complex matrices and set $\theta_\mathrm{bias} = 0$. We set $r = 0.010$.}
\label{supp-arXiv-numerical-result-raw-data-fold-001-rand-001-UKM-OUU-MNIST256-0-1}
\end{figure*}

In Fig.~\ref{supp-arXiv-numerical-result-raw-data-fold-001-rand-001-QCL-MNIST256-0-1}, we show the numerical results of QCL for the $5$-fold datasets with $5$ different random seeds.
In Fig.~\ref{supp-arXiv-numerical-result-raw-data-fold-001-rand-001-UKM-P-MNIST256-0-1}, we show the numerical results of $\hat{P}$ of the UKM for the $5$-fold datasets with $5$ different random seeds.
In Fig.~\ref{supp-arXiv-numerical-result-raw-data-fold-001-rand-001-UKM-OUU-MNIST256-0-1}, we also show the numerical results of OU of $\hat{X}$ of the UKM for the $5$-fold datasets with $5$ different random seeds.

\begin{figure}[htb]
\centering
\includegraphics[scale=0.45]{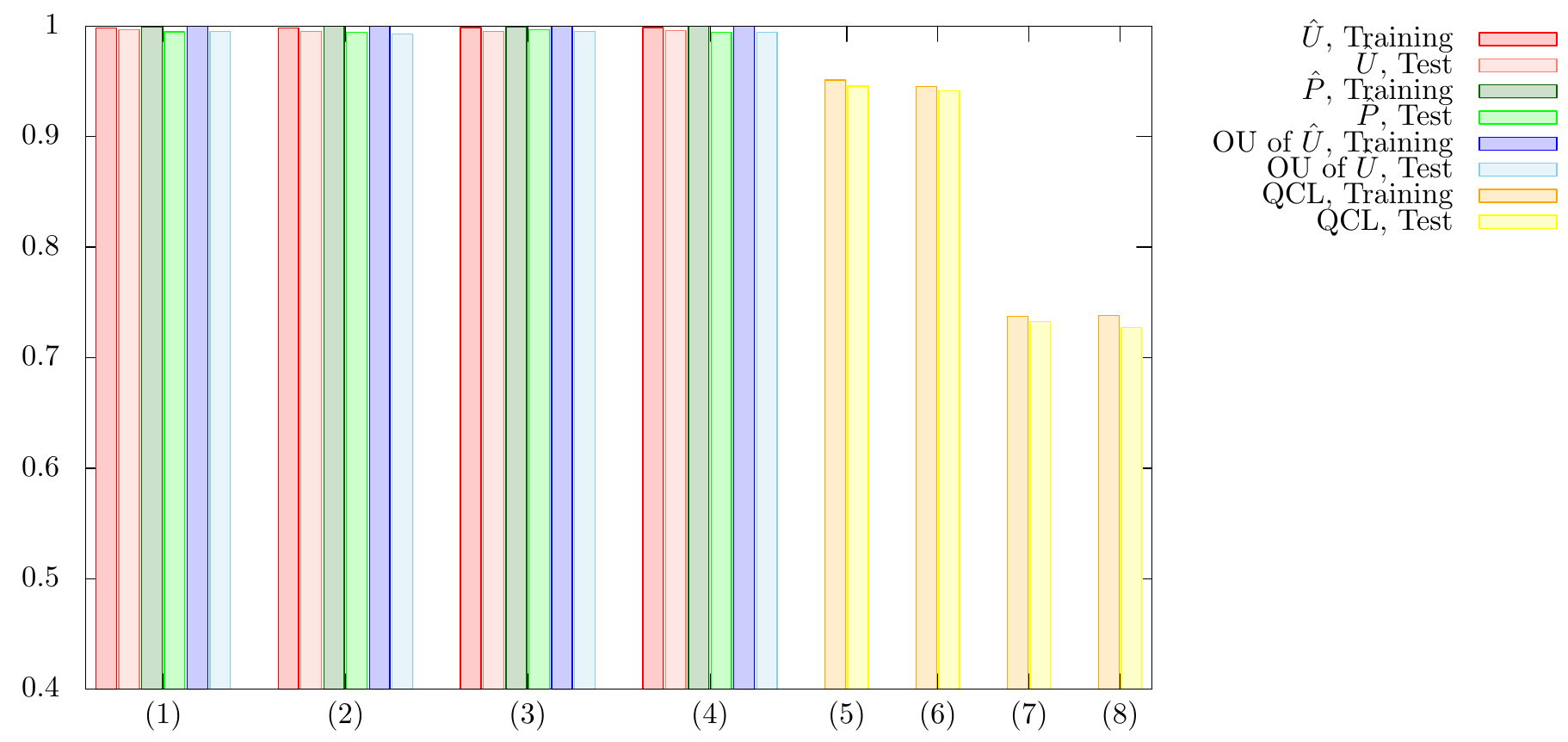}
\caption{Results of $5$-fold CV with $5$ different random seeds for the MNIST256 dataset ($0$ or $1$). For the UKM, we put $r = 0.010$ and set $K = 20$ and $K' = 10$. For QCL, the number of layers $L$ is $5$ and the number of iterations is $100$. The numerical settings are as follows: (1) complex matrices without the bias term, (2) complex matrices with the bias term, (3) real matrices without the bias term, (4) real matrices with the bias term, (5) CNOT-based circuit without the bias term, (6) CNOT-based circuit with the bias term, (7) CRot-based circuit without the bias term, (8) CRot-based circuit with the bias term, (9) 1d Heisenberg circuit without the bias term, (10) 1d Heisenberg circuit with the bias term, (11) FC Heisenberg circuit without the bias term, and (12) FC Heisenberg circuit with the bias term.}
\label{supp-arXiv-numerical-result-performance-UKM-QCL-MNIST256-0-1}
\end{figure}

We summarize the results of 5-fold CV with 5 different random seeds of QCL and the UKM in Tables~\ref{supp-arXiv-table-MNIST256-0-1-002} and \ref{supp-arXiv-table-MNIST256-0-1-001}, respectively.
For QCL and the UKM, we select the best model for the training dataset over iterations to compute the performance.
\begin{table}[htb]
  \begin{tabular}{cc|cc}
    \hline \hline
    Algo. & Condition & Training & Test \\
    \hline
  QCL & CNOT-based, w/o bias & 0.9511 & 0.9459 \\
  QCL & CNOT-based, w/ bias & 0.9452 & 0.9413 \\
    \hline
  QCL & CRot-based, w/o bias & 0.7372 & 0.7326 \\
  QCL & CRot-based, w/ bias & 0.7383 & 0.7273 \\
    \hline \hline
  \end{tabular}
\caption{Results of $5$-fold CV with $5$ different random seeds of QCL for the MNIST256 dataset ($0$ or $1$). The number of layers $L$ is $5$ and the number of iterations is $100$.}
\label{supp-arXiv-table-MNIST256-0-1-002}
\end{table}
\begin{table}[htb]
  \begin{tabular}{cc|cc}
    \hline \hline
    Algo. & Condition & Training & Test \\
    \hline
  UKM & $\hat{X}$, complex, w/o bias & 0.9985 & 0.9966 \\
  UKM & $\hat{P}$, complex, w/o bias & 0.9992 & 0.9949 \\
  UKM & OU of $\hat{X}$, complex, w/o bias & 1.0 & 0.9951 \\
    \hline
  UKM & $\hat{X}$, complex, w/ bias & 0.9984 & 0.9951 \\
  UKM & $\hat{P}$, complex, w/ bias & 1.0 & 0.9945 \\
  UKM & OU of $\hat{X}$, complex, w/ bias & 1.0 & 0.9927 \\
    \hline
  UKM & $\hat{X}$, real, w/o bias & 0.9987 & 0.9954 \\
  UKM & $\hat{P}$, real, w/o bias & 0.9991 & 0.9969 \\
  UKM & OU of $\hat{X}$, real, w/o bias & 0.9997 & 0.9950 \\
    \hline
  UKM & $\hat{X}$, real, w/ bias & 0.9987 & 0.9962 \\
  UKM & $\hat{P}$, real, w/ bias & 0.9997 & 0.9944 \\
  UKM & OU of $\hat{X}$, real, w/ bias & 0.9997 & 0.9947 \\
    \hline \hline
  \end{tabular}
\caption{Results of $5$-fold CV with $5$ different random seeds of the UKM for the MNIST256 dataset ($0$ or $1$). We put $r = 0.010$ and set $K = 20$ and $K' = 10$.}
\label{supp-arXiv-table-MNIST256-0-1-001}
\end{table}
In Fig.~\ref{supp-arXiv-numerical-result-performance-UKM-QCL-MNIST256-0-1}, we plot the data shown in Tables~\ref{supp-arXiv-table-MNIST256-0-1-002} and \ref{supp-arXiv-table-MNIST256-0-1-001}.
We also summarize the results of 5-fold CV with 5 different random seeds of the kernel method in Table~\ref{supp-arXiv-table-UCI-MNIST256-0-1-003}.
More specifically, we use Ridge classification in Sec.~\ref{supp-arXiv-sec-Ridge-001}.
We consider the linear functions and the second-order polynomial functions for $\phi (\cdot)$ in Eq.~\eqref{supp-arXiv-f-pred-kernel-method-001-002} with and without normalization.
We set $\lambda = 10^{-2}, 10^{-1}, 1$ where $\lambda$ is the coefficient of the regularization term.
\begin{table}[htb]
  \begin{tabular}{cc|cc}
    \hline \hline
    Algo. & Condition & Training & Test \\
    \hline
  Kernel method & Linear, w/o normalization, $\lambda = 10^{-2}$ & 1.0000 & 1.0000 \\
  Kernel method & Linear, w/o normalization, $\lambda = 10^{-1}$ & 1.0000 & 1.0000 \\
  Kernel method & Linear, w/o normalization, $\lambda = 1$ & 1.0000 & 1.0000 \\
    \hline
  Kernel method & Linear, w/ normalization, $\lambda = 10^{-2}$ & 0.9991 & 0.9981 \\
  Kernel method & Linear, w/ normalization, $\lambda = 10^{-1}$ & 0.9982 & 0.9981 \\
  Kernel method & Linear, w/ normalization, $\lambda = 1$ & 0.9982 & 0.9981 \\
    \hline
  Kernel method & Poly-2, w/o normalization, $\lambda = 10^{-2}$ & 1.0000 & 0.9897 \\
  Kernel method & Poly-2, w/o normalization, $\lambda = 10^{-1}$ & 1.0000 & 0.9897 \\
  Kernel method & Poly-2, w/o normalization, $\lambda = 1$ & 1.0000 & 0.9897 \\
    \hline
  Kernel method & Poly-2, w/ normalization, $\lambda = 10^{-2}$ & 1.0000 & 0.9962 \\
  Kernel method & Poly-2, w/ normalization, $\lambda = 10^{-1}$ & 1.0000 & 0.9981 \\
  Kernel method & Poly-2, w/ normalization, $\lambda = 1$ & 0.9982 & 0.9981 \\
    \hline \hline
  \end{tabular}
\caption{Results of 5-fold CV with 5 different random seeds of the kernel method for the MNIST256 dataset ($0$ or $1$).}
\label{supp-arXiv-table-UCI-MNIST256-0-1-003}
\end{table}

\begin{figure}[htb]
\centering
\includegraphics[scale=0.45]{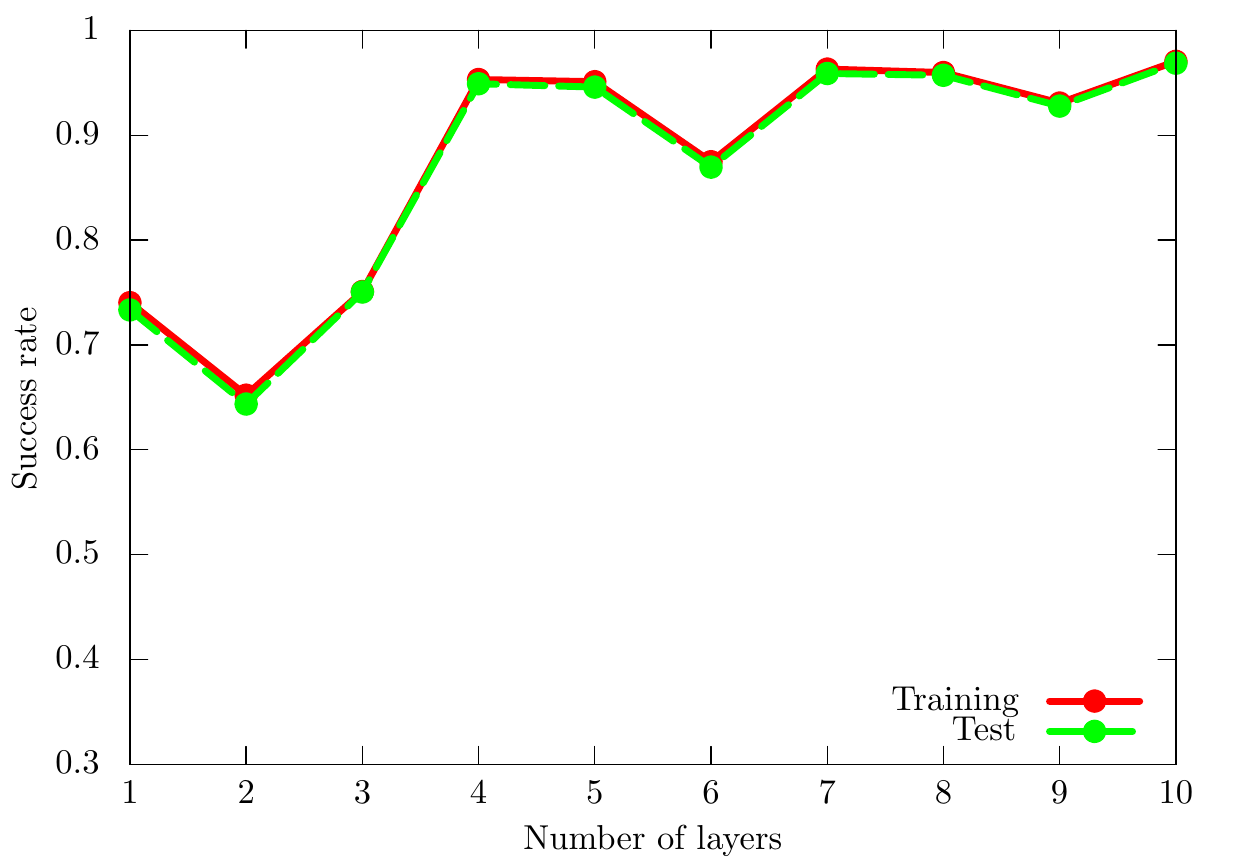}
\caption{Performance dependence of QCL on the number of layers $L$ for the MNIST256 dataset ($0$ or $1$). We use the CNOT-based circuit geometry and set $\theta_\mathrm{bias} = 0$. We iterate the computation $100$ times.}
\label{supp-arXiv-numerical-result-layers-dependence-QCL-MNIST256-0-1}
\end{figure}

\begin{figure}[htb]
\centering
\includegraphics[scale=0.45]{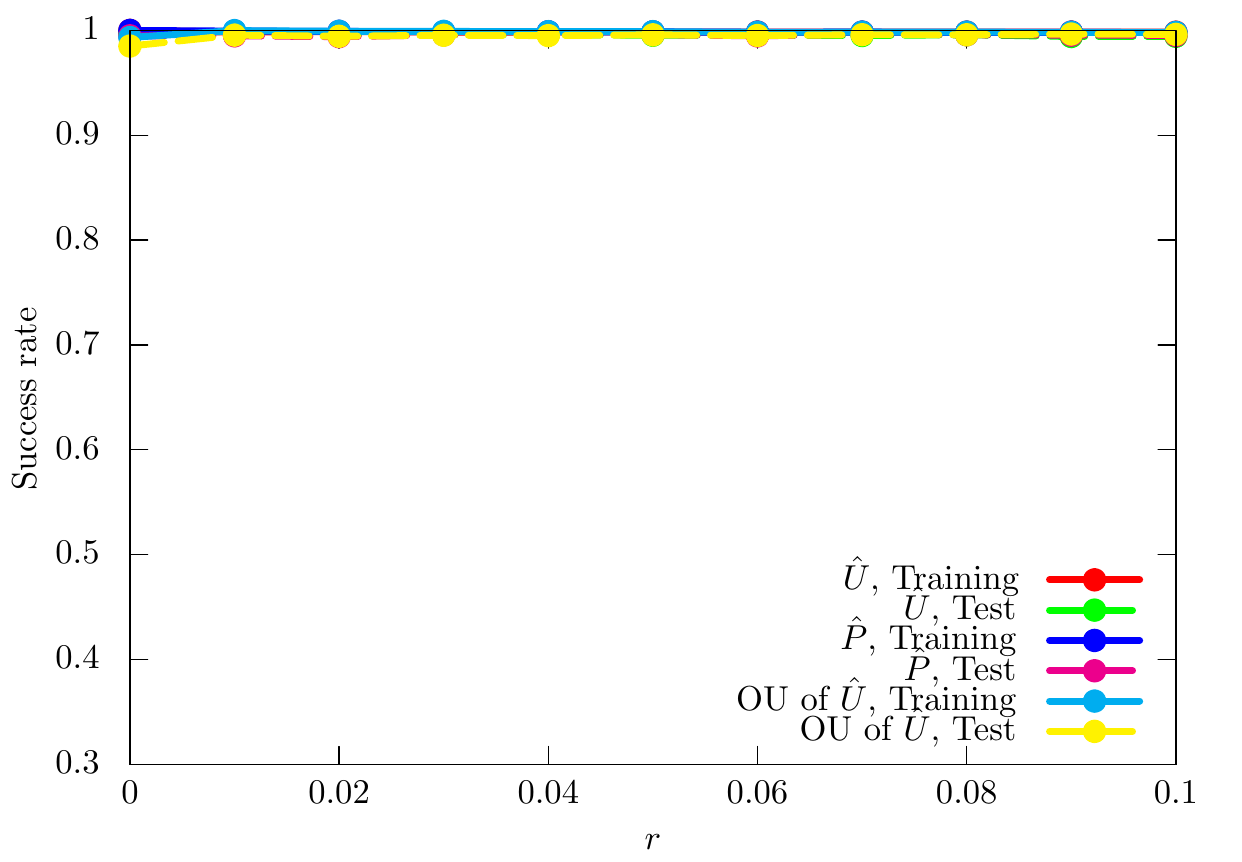}
\includegraphics[scale=0.45]{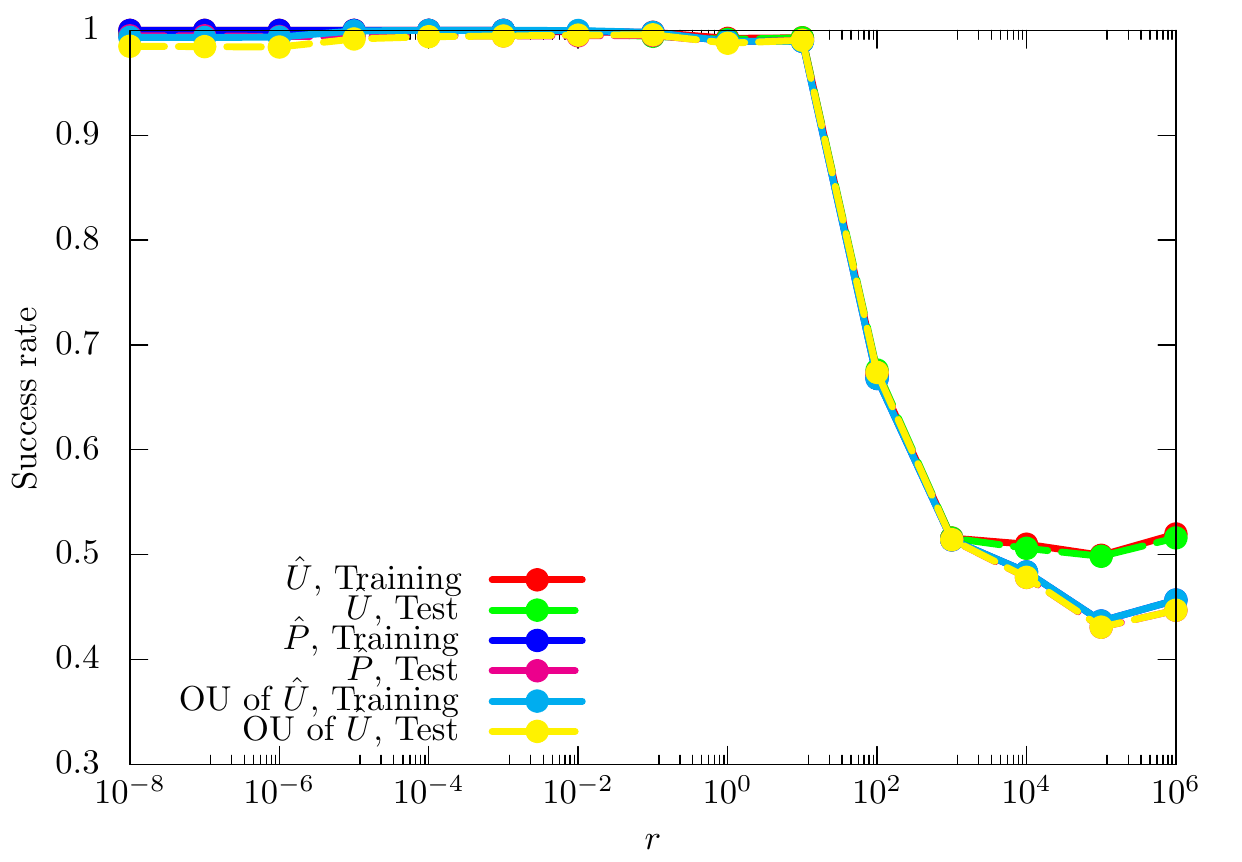}
\caption{Performance dependence of the UKM on $r$, which is the coefficient of the second term in the right-hand side of Eq.~\eqref{supp-arXiv-quantum-kernel-method-001-011} for the MNIST256 dataset ($0$ or $1$). We use complex matrices and set $\theta_\mathrm{bias} = 0$. We set $K = 20$ and $K' = 10$.}
\label{supp-arXiv-numerical-result-r-dependence-UKM-MNIST256-0-1}
\end{figure}

\begin{figure}[htb]
\centering
\includegraphics[scale=0.45]{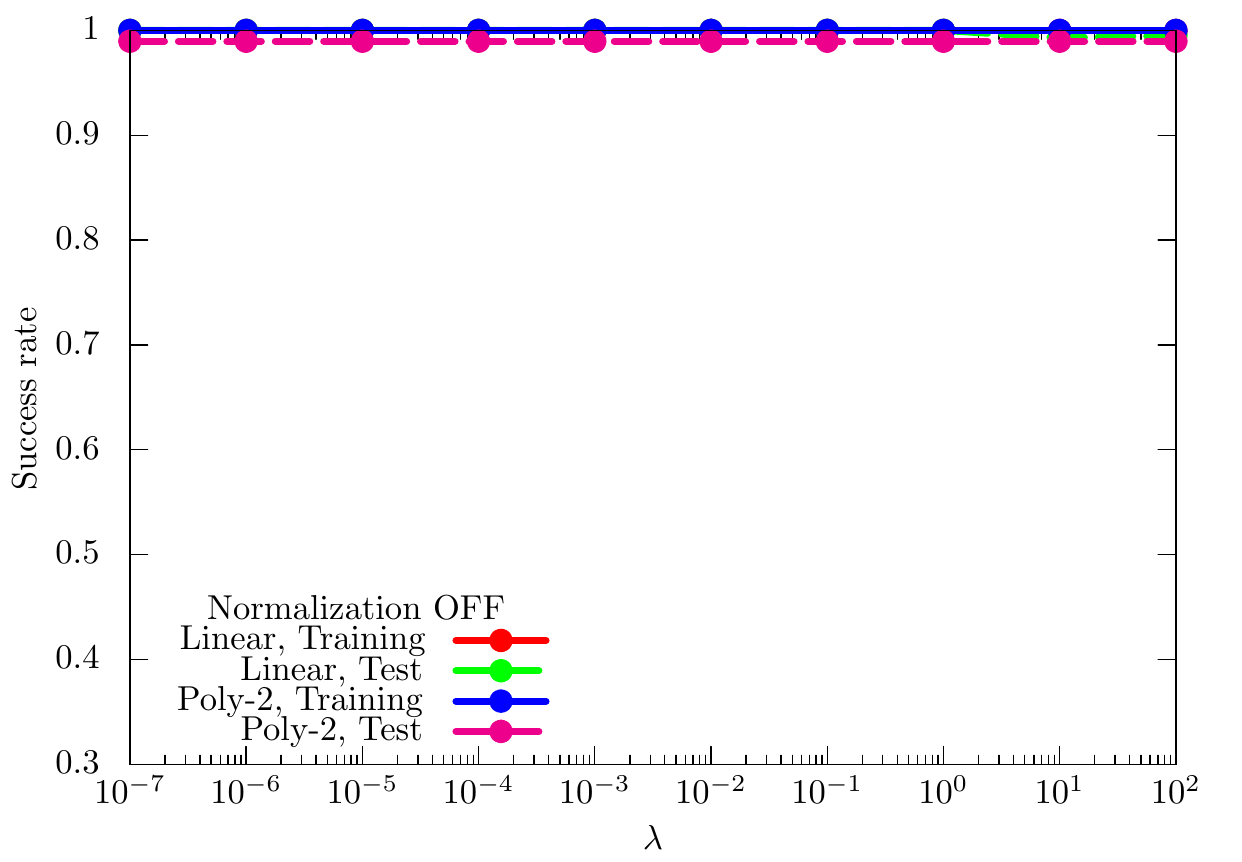}
\includegraphics[scale=0.45]{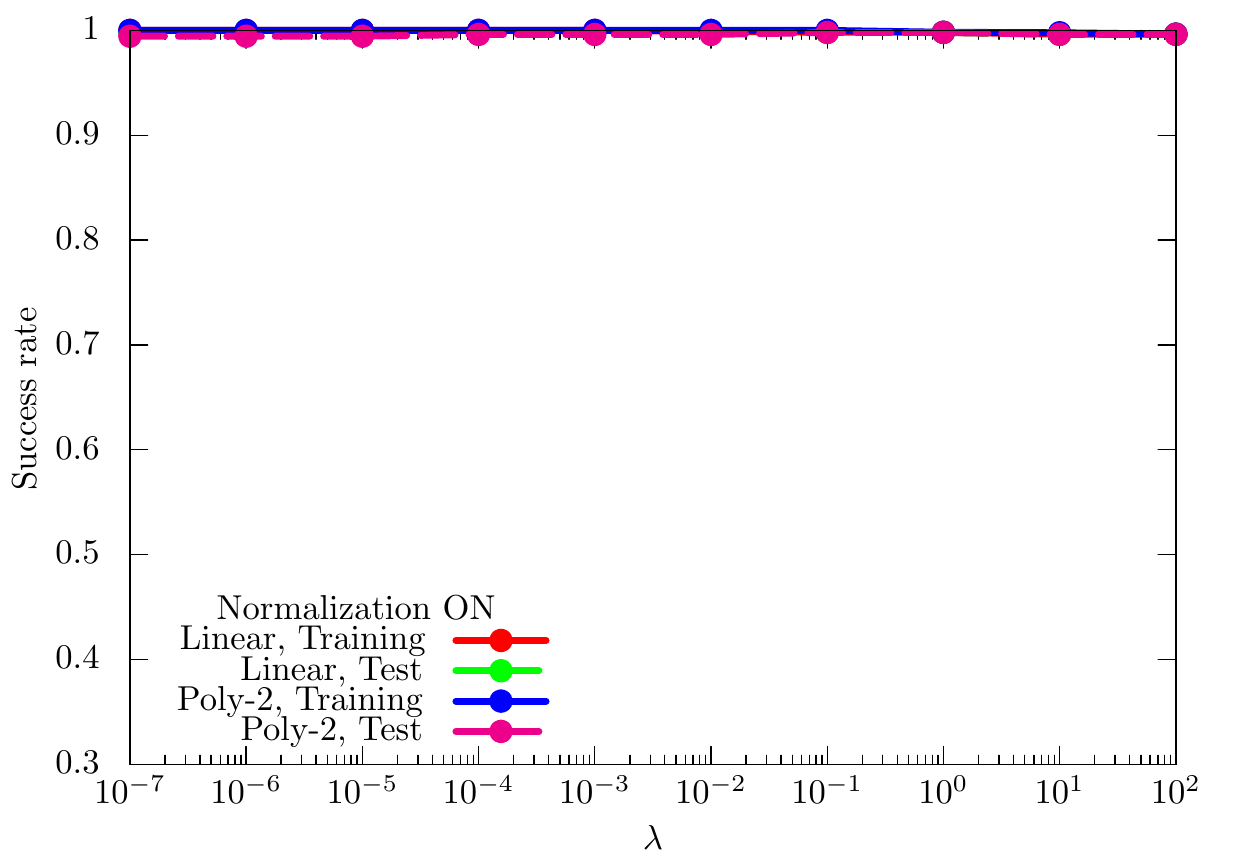}
\caption{Performance dependence of the kernel method on $\lambda$, which is the coefficient of the second term in the right-hand side of Eq.~\eqref{supp-arXiv-cost-function-kernel-method-001-002} for the MNIST256 dataset ($0$ or $1$). For $\phi (\cdot)$ in Eq.~\eqref{supp-arXiv-f-pred-kernel-method-001-002}, we use the linear functions and the second-degree polynomial functions with and without normalization.}
\label{supp-arXiv-numerical-result-lambda-dependence-kernel-method-MNIST256-0-1}
\end{figure}

Next, we show the performance dependence of the three algorithms on their key parameters.
We see the performance dependence of QCL on the number of layers $L$.
The result is shown in Fig.~\ref{supp-arXiv-numerical-result-layers-dependence-QCL-MNIST256-0-1}.
We then see the performance dependence of the UKM on $r$, which is the coefficient of the second term in the right-hand side of Eq.~\eqref{supp-arXiv-quantum-kernel-method-001-011}.
The result is shown in Fig.~\ref{supp-arXiv-numerical-result-r-dependence-UKM-MNIST256-0-1}.
In Fig.~\ref{supp-arXiv-numerical-result-lambda-dependence-kernel-method-MNIST256-0-1}, we show the performance dependence of the kernel method on $\lambda$, which is the coefficient of the second term in the right-hand side of Eq.~\eqref{supp-arXiv-cost-function-kernel-method-001-002}.

\begin{figure}[htb]
\centering
\includegraphics[scale=0.45]{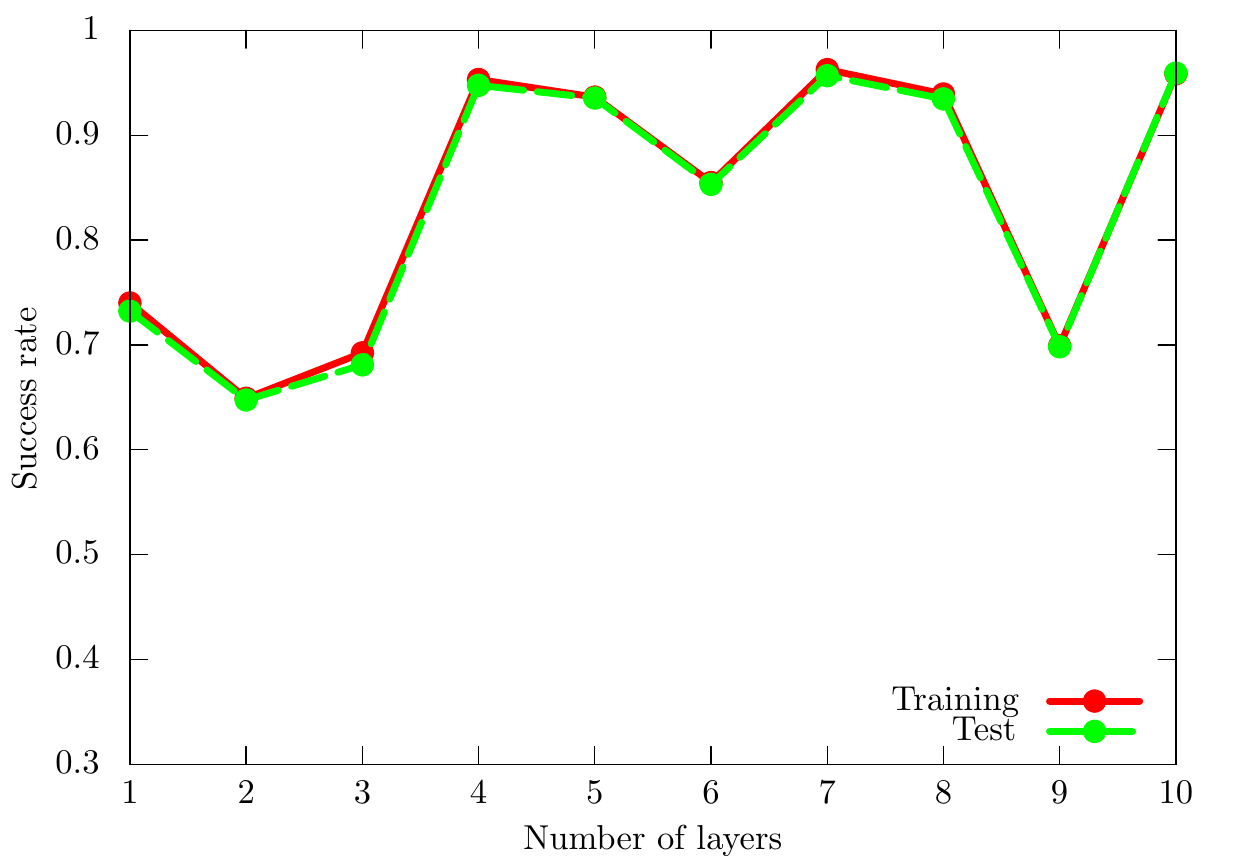}
\caption{Performance dependence of QCL on the number of layers $L$ for the MNIST256 dataset ($0$ or $1$) in the case of the hinge function, Eq.~\eqref{supp-arXiv-hinge-function-001-001}. We use the CNOT-based circuit geometry and set $\theta_\mathrm{bias} = 0$. We iterate the computation $300$ times.}
\label{supp-arXiv-numerical-result-layers-dependence-QCL-MNIST256-0-1-hinge}
\end{figure}

\begin{figure}[htb]
\centering
\includegraphics[scale=0.45]{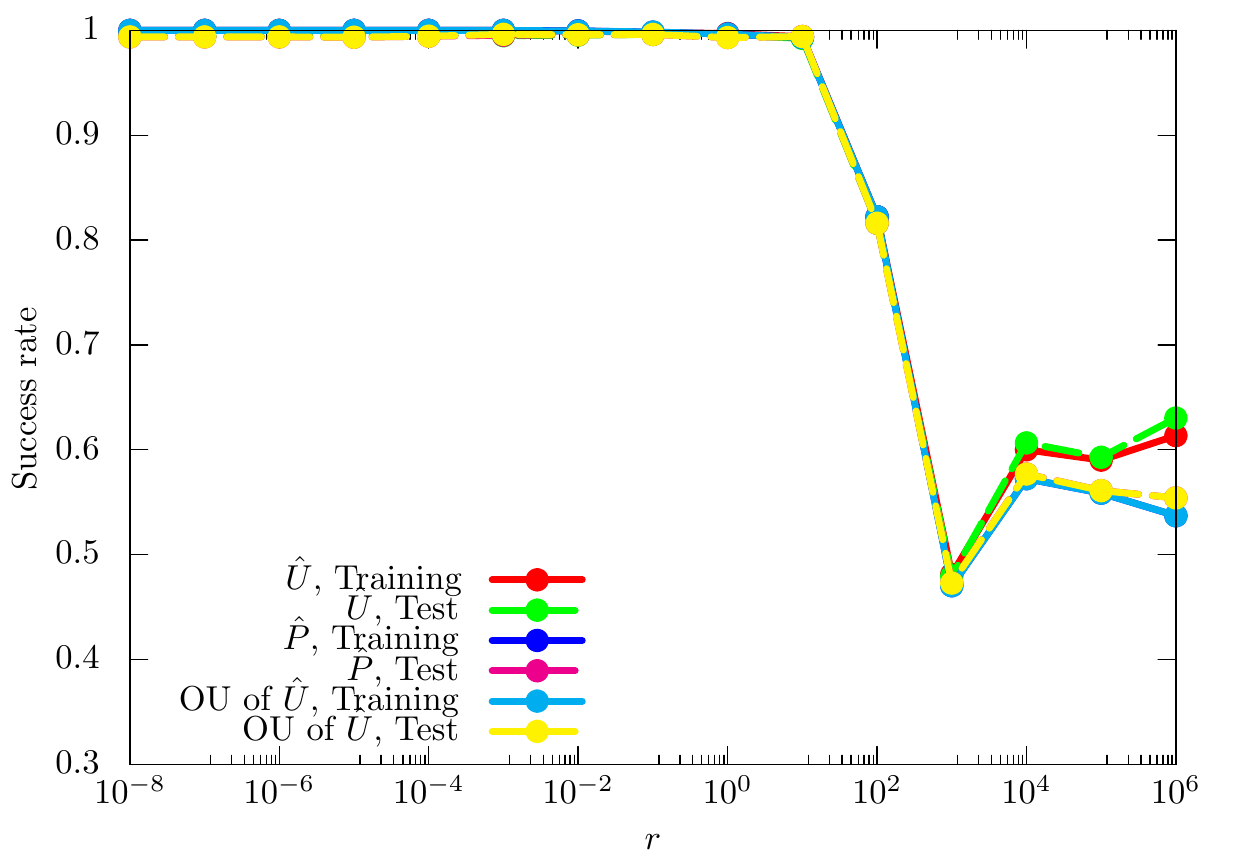}
\caption{Performance dependence of the UKM on $r$, which is the coefficient of the second term in the right-hand side of Eq.~\eqref{supp-arXiv-quantum-kernel-method-001-011} for the MNIST256 dataset ($0$ or $1$) in the case of the hinge function, Eq.~\eqref{supp-arXiv-hinge-function-001-001}. We use complex matrices and set $\theta_\mathrm{bias} = 0$. We set $K = 30$ and $K' = 10$.}
\label{supp-arXiv-numerical-result-r-dependence-UKM-MNIST256-0-1-hinge}
\end{figure}

So far, we have used the squared error function, Eq.~\eqref{supp-arXiv-squared-error-function-001-001}.
In Fig.~\ref{supp-arXiv-numerical-result-layers-dependence-QCL-MNIST256-0-1-hinge}, we show the performance dependence of QCL on the number of layers $L$ in the case of the hinge function, Eq.~\eqref{supp-arXiv-hinge-function-001-001}.
In Fig.~\ref{supp-arXiv-numerical-result-r-dependence-UKM-MNIST256-0-1-hinge}, we show the performance dependence of the UKM on $r$, which is the coefficient of the second term in the right-hand side of Eq.~\eqref{supp-arXiv-quantum-kernel-method-001-011}, in the case of the hinge function, Eq.~\eqref{supp-arXiv-hinge-function-001-001}.

\clearpage

\subsection{MNIST256 dataset ($0$ or non-$0$)}

We here show the numerical result for the MNIST256 dataset ($0$ or non-$0$).
For the UKM, we put $r = 0.010$ and set $K = 10$ and $K' = 5$ in Algo.~\ref{supp-arXiv-quantum-kernel-method-002-001}.
For QCL, we run iterations $50$ times.

\begin{figure*}[htb]
\centering
\includegraphics[scale=0.25]{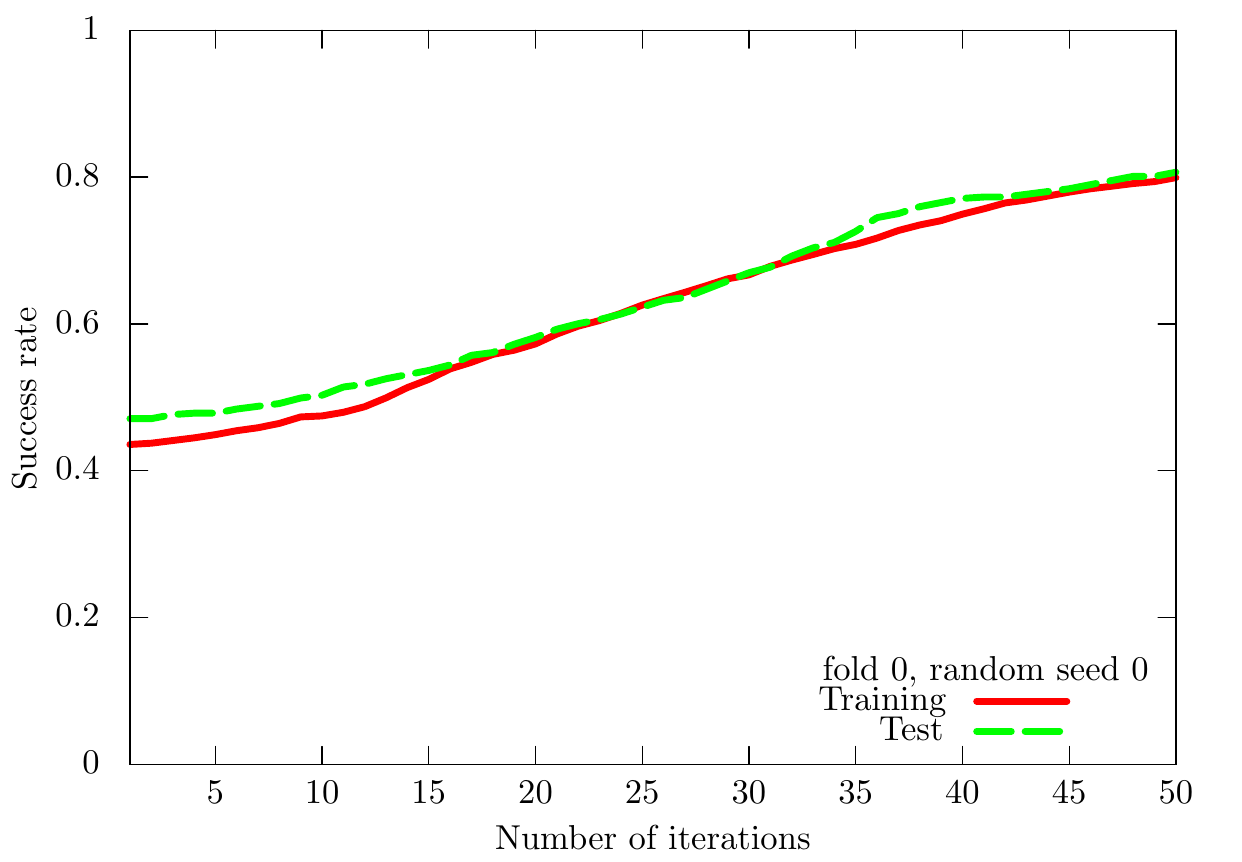}
\includegraphics[scale=0.25]{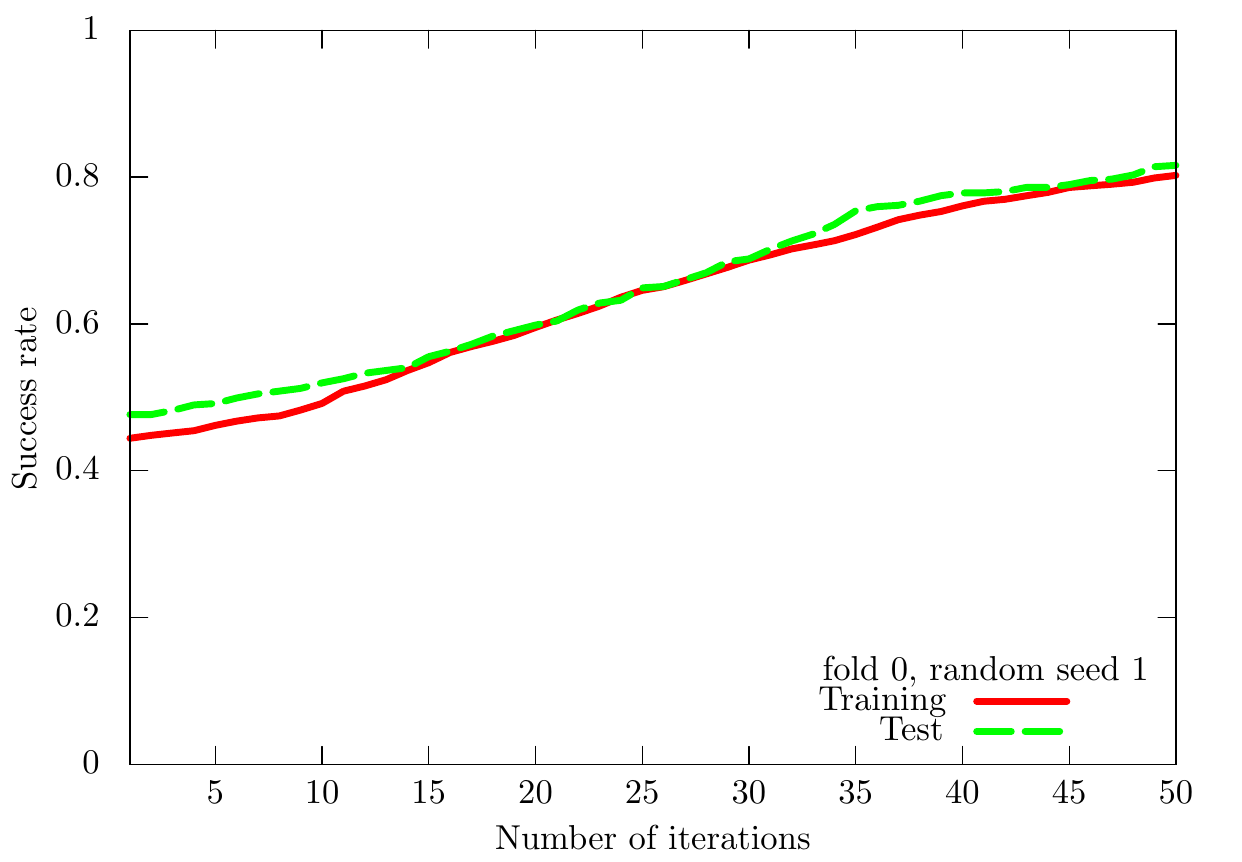}
\includegraphics[scale=0.25]{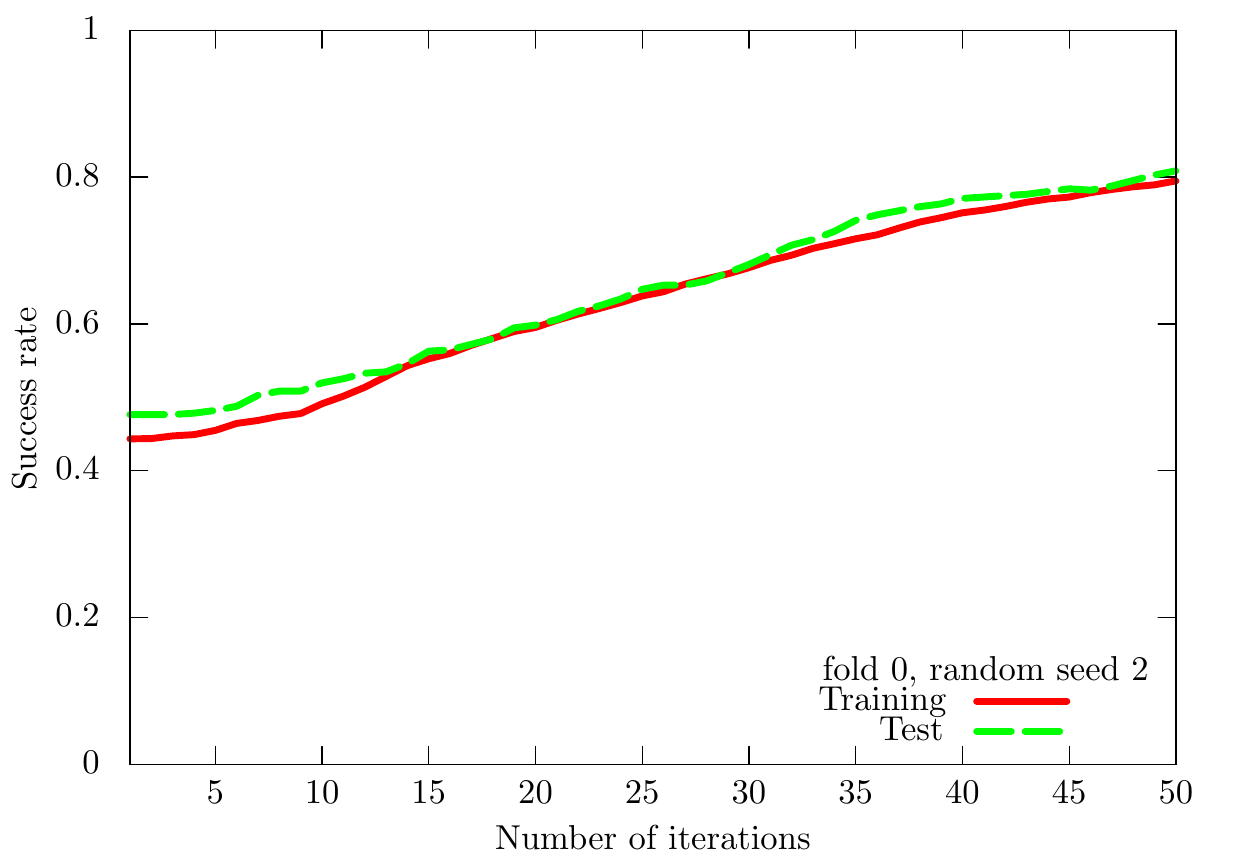}
\includegraphics[scale=0.25]{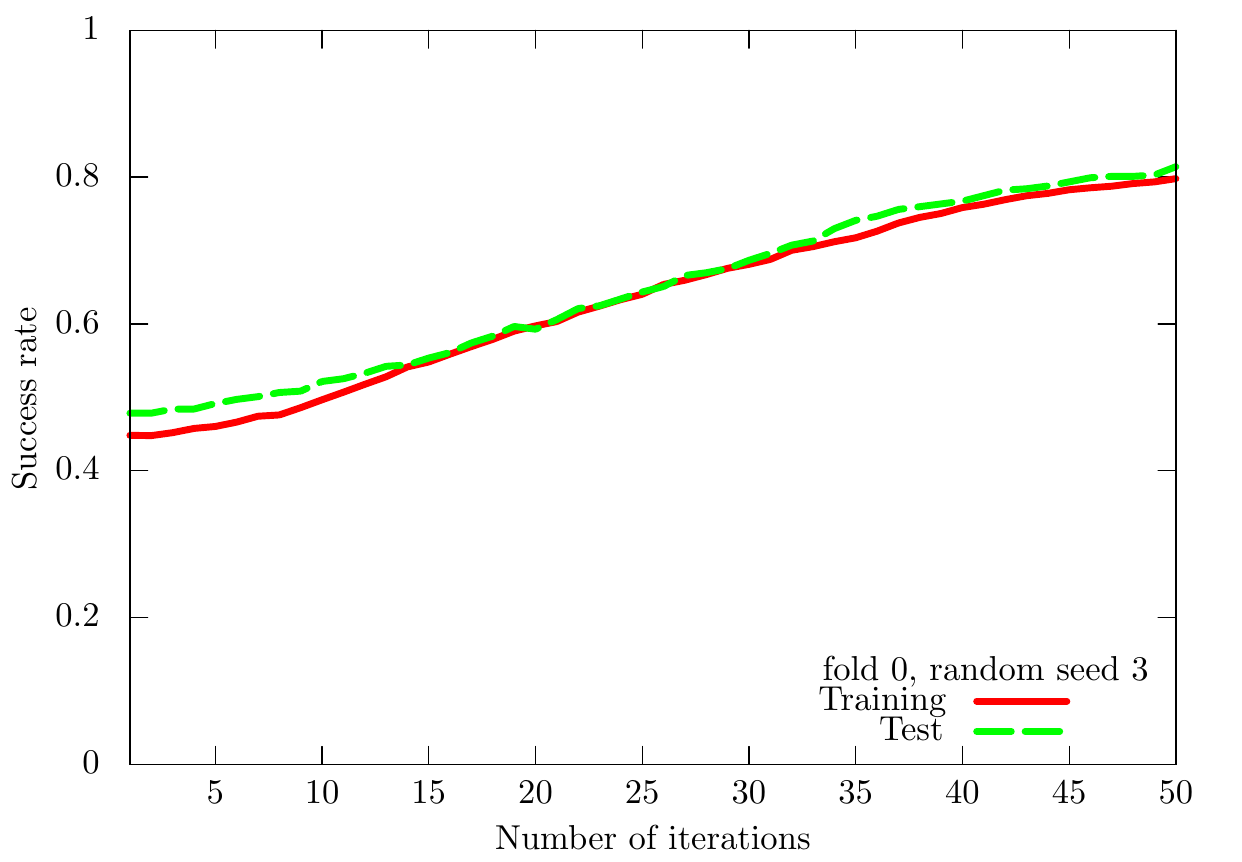}
\includegraphics[scale=0.25]{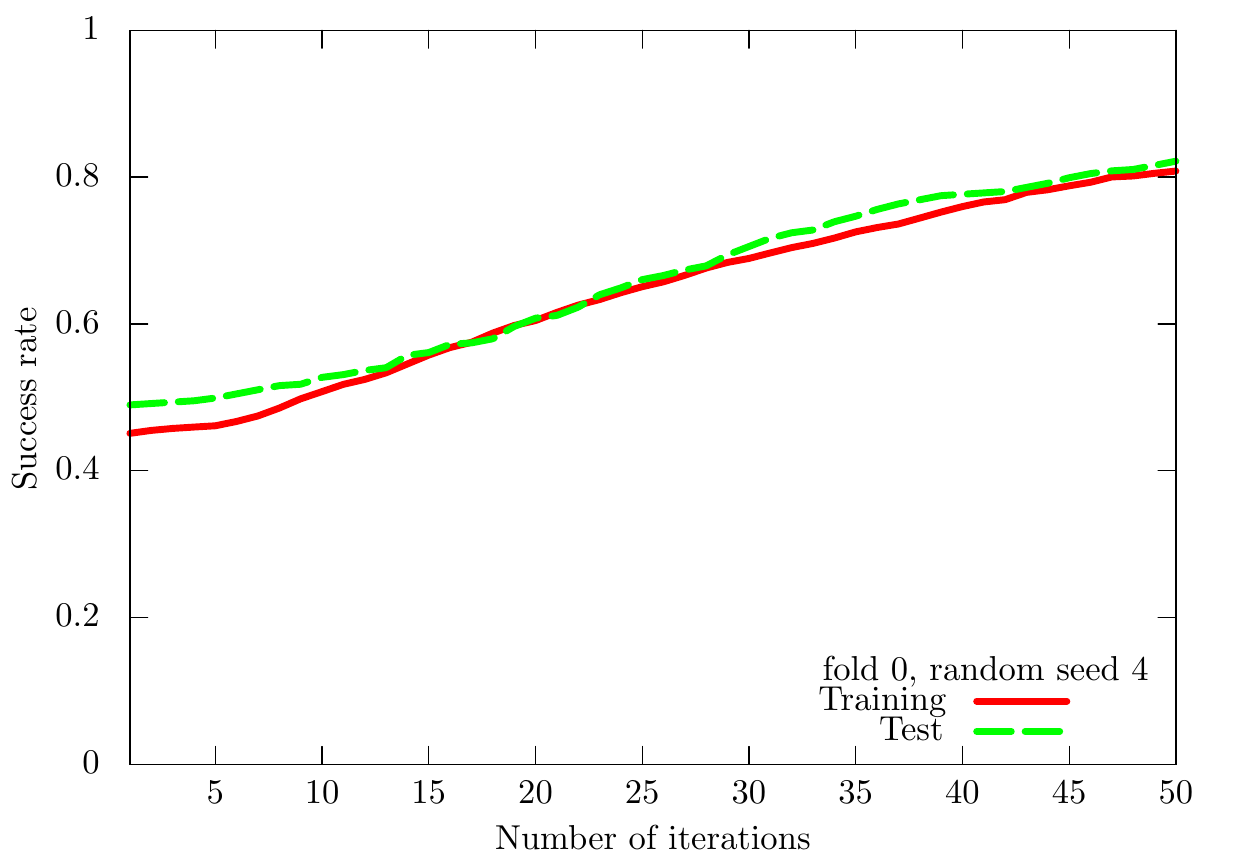}
\includegraphics[scale=0.25]{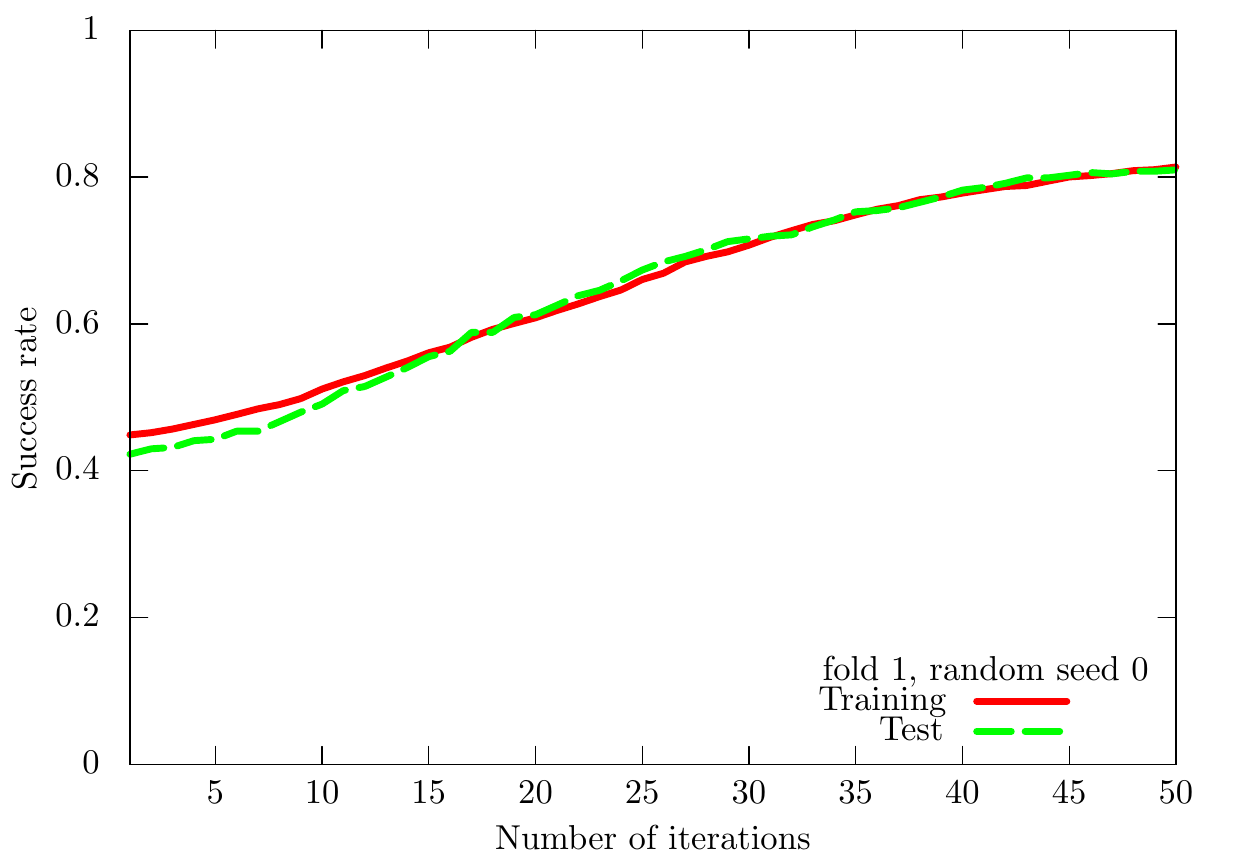}
\includegraphics[scale=0.25]{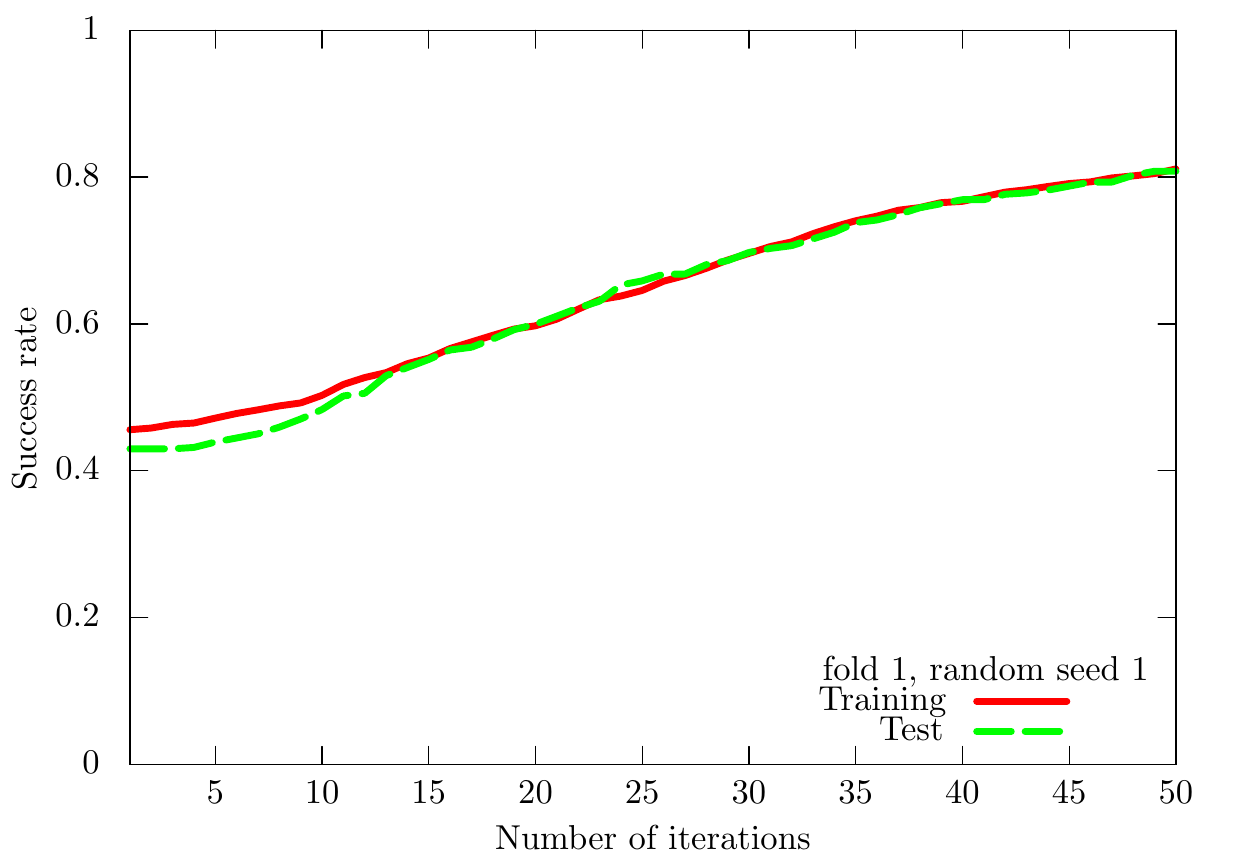}
\includegraphics[scale=0.25]{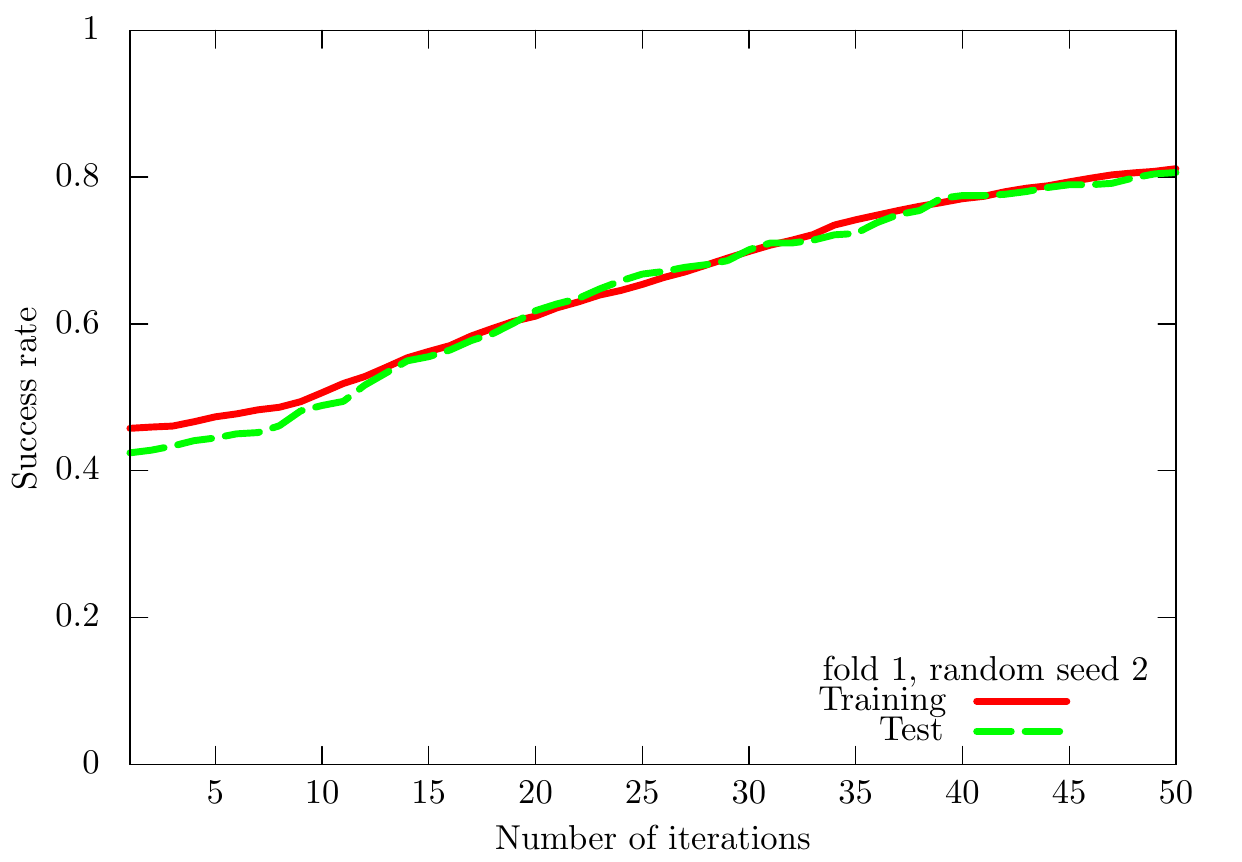}
\includegraphics[scale=0.25]{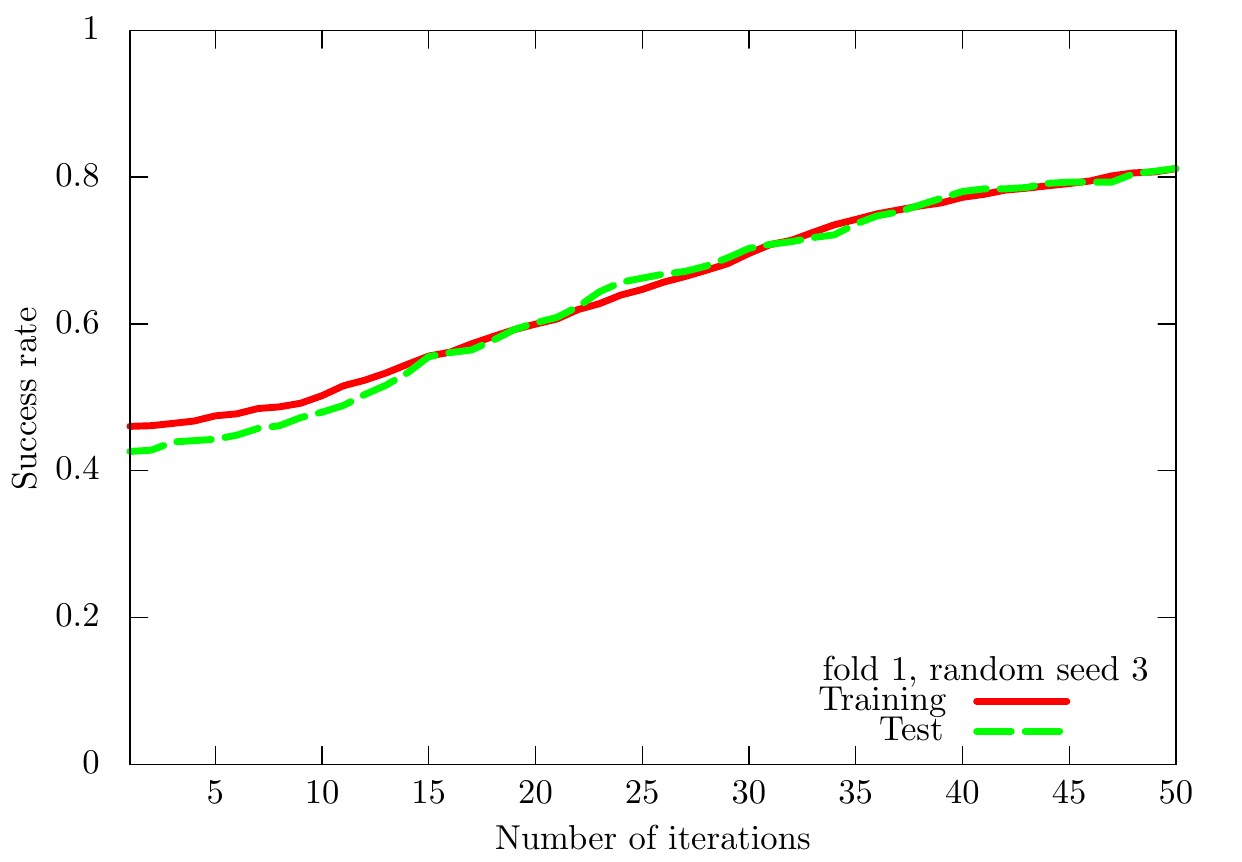}
\includegraphics[scale=0.25]{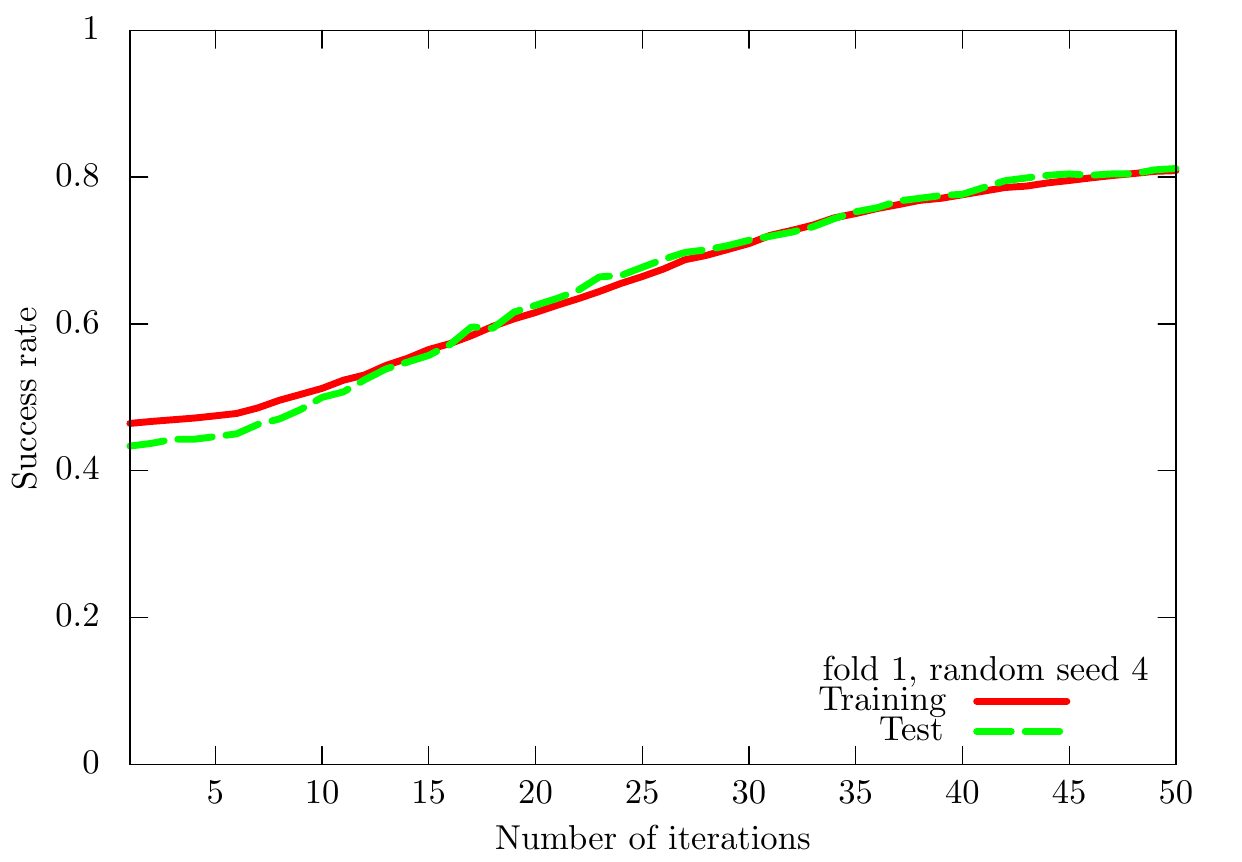}
\includegraphics[scale=0.25]{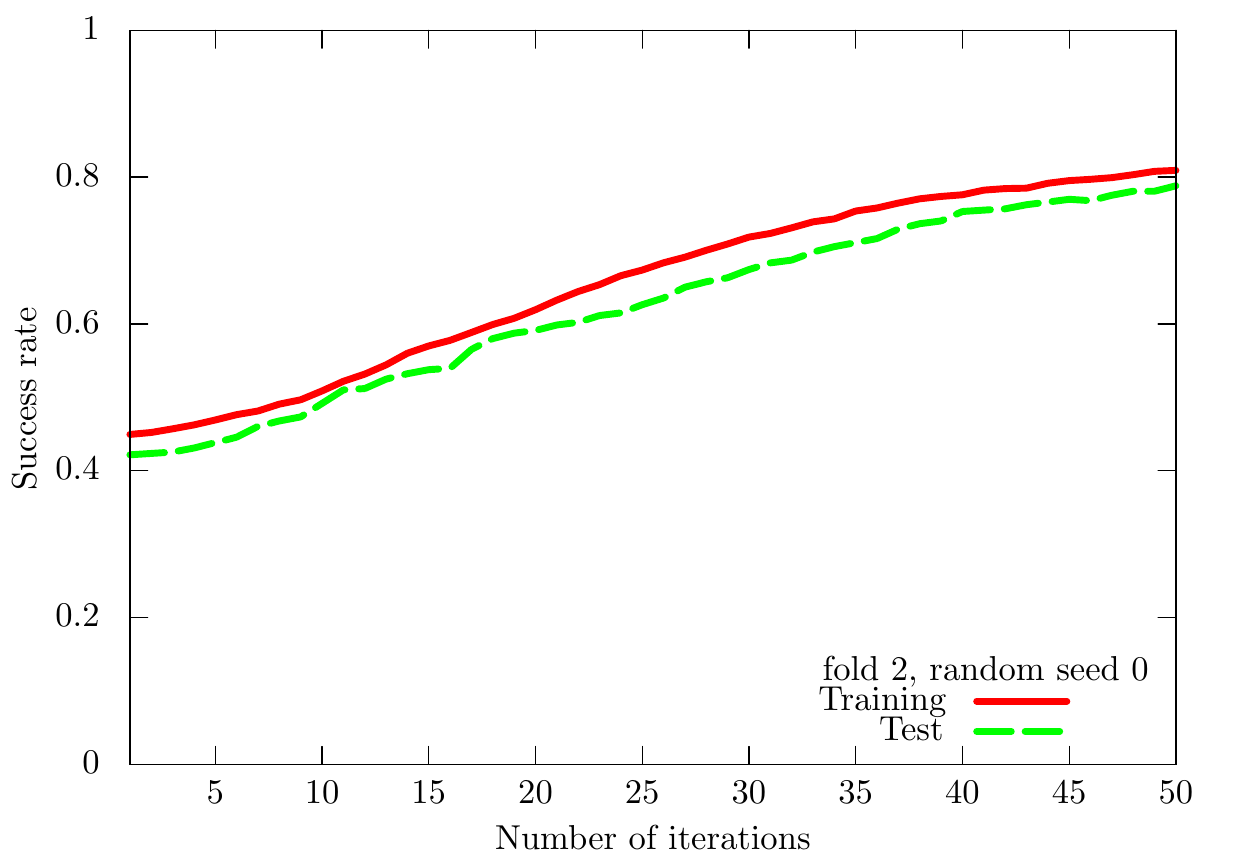}
\includegraphics[scale=0.25]{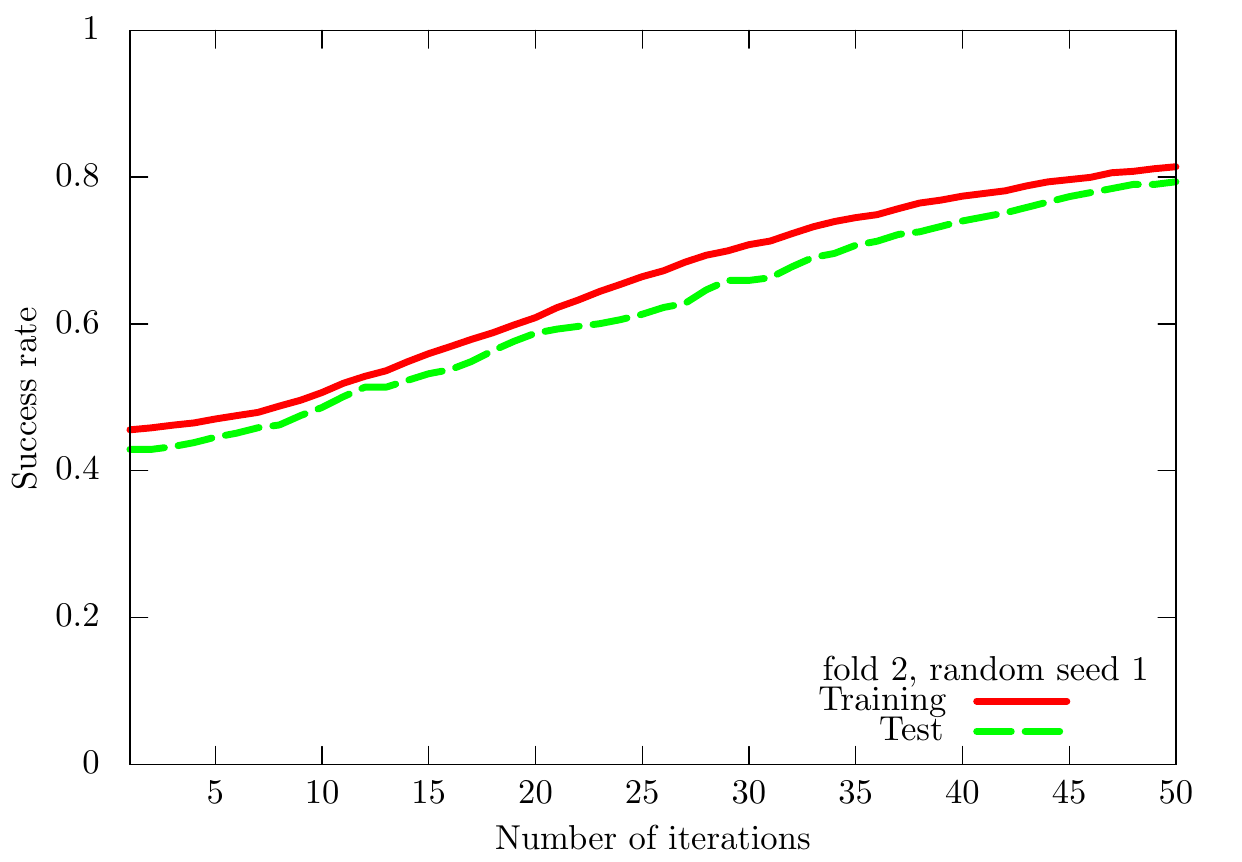}
\includegraphics[scale=0.25]{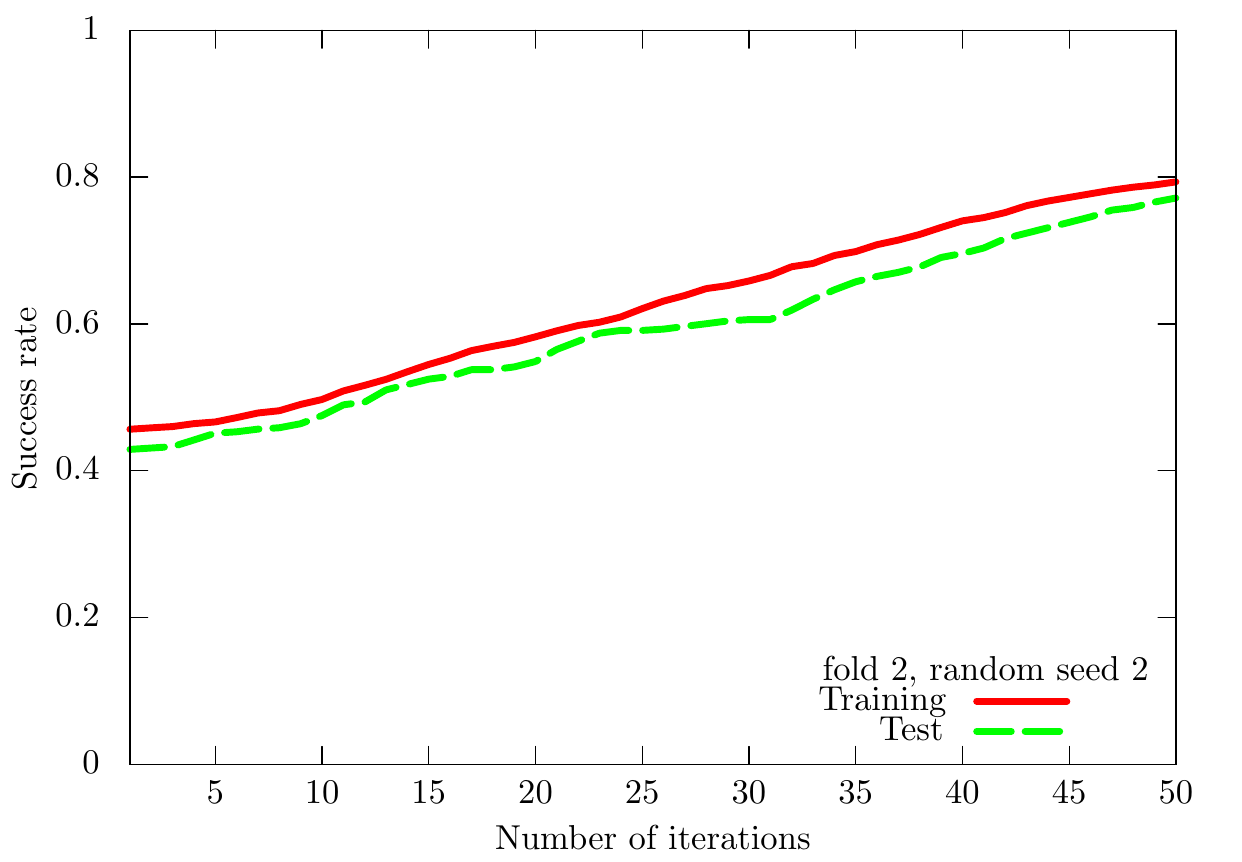}
\includegraphics[scale=0.25]{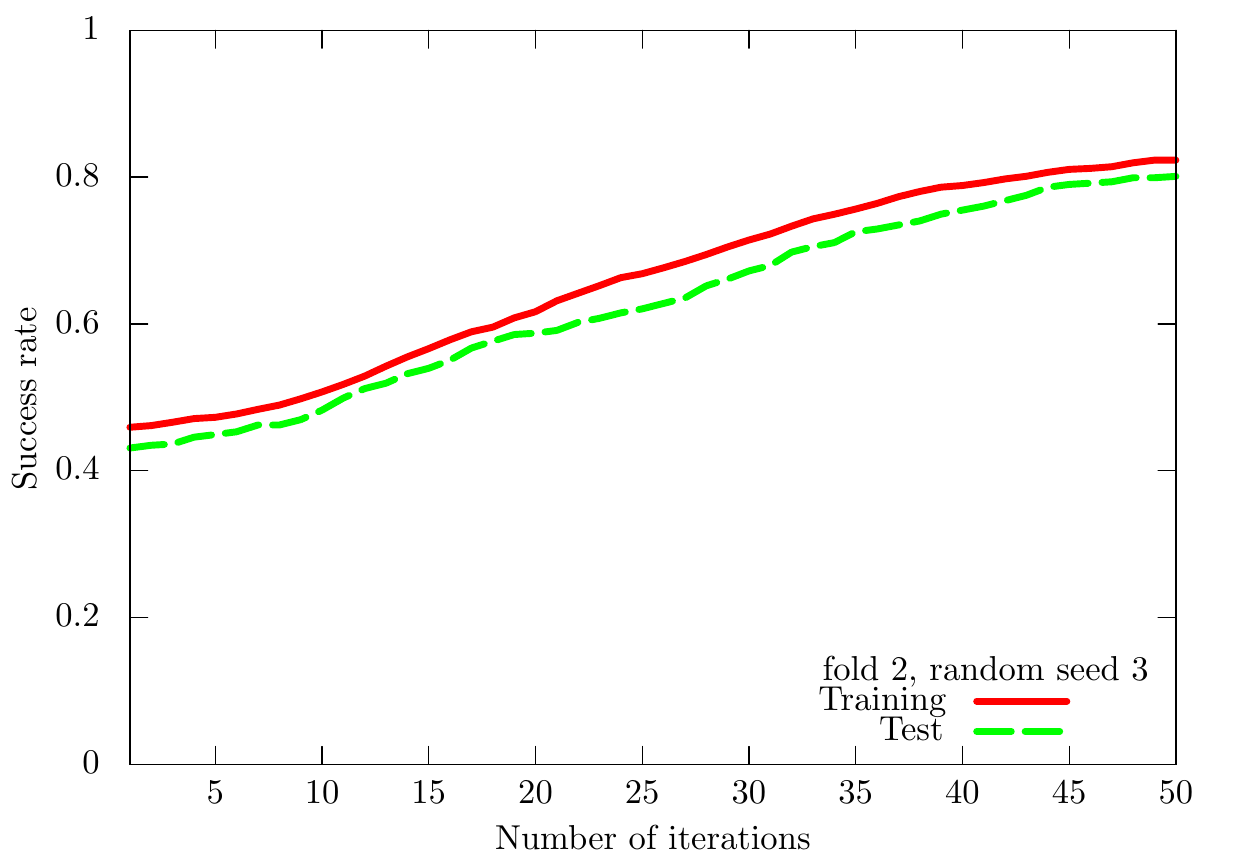}
\includegraphics[scale=0.25]{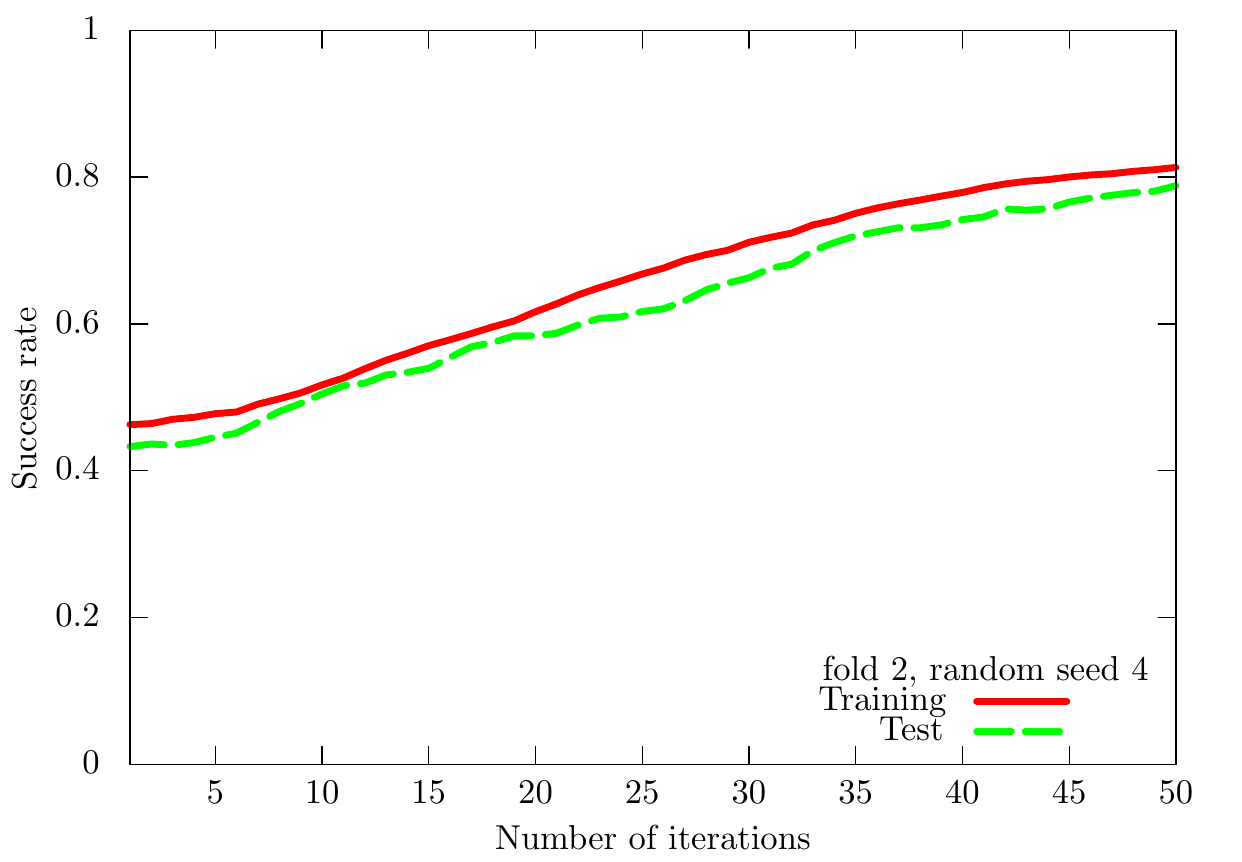}
\includegraphics[scale=0.25]{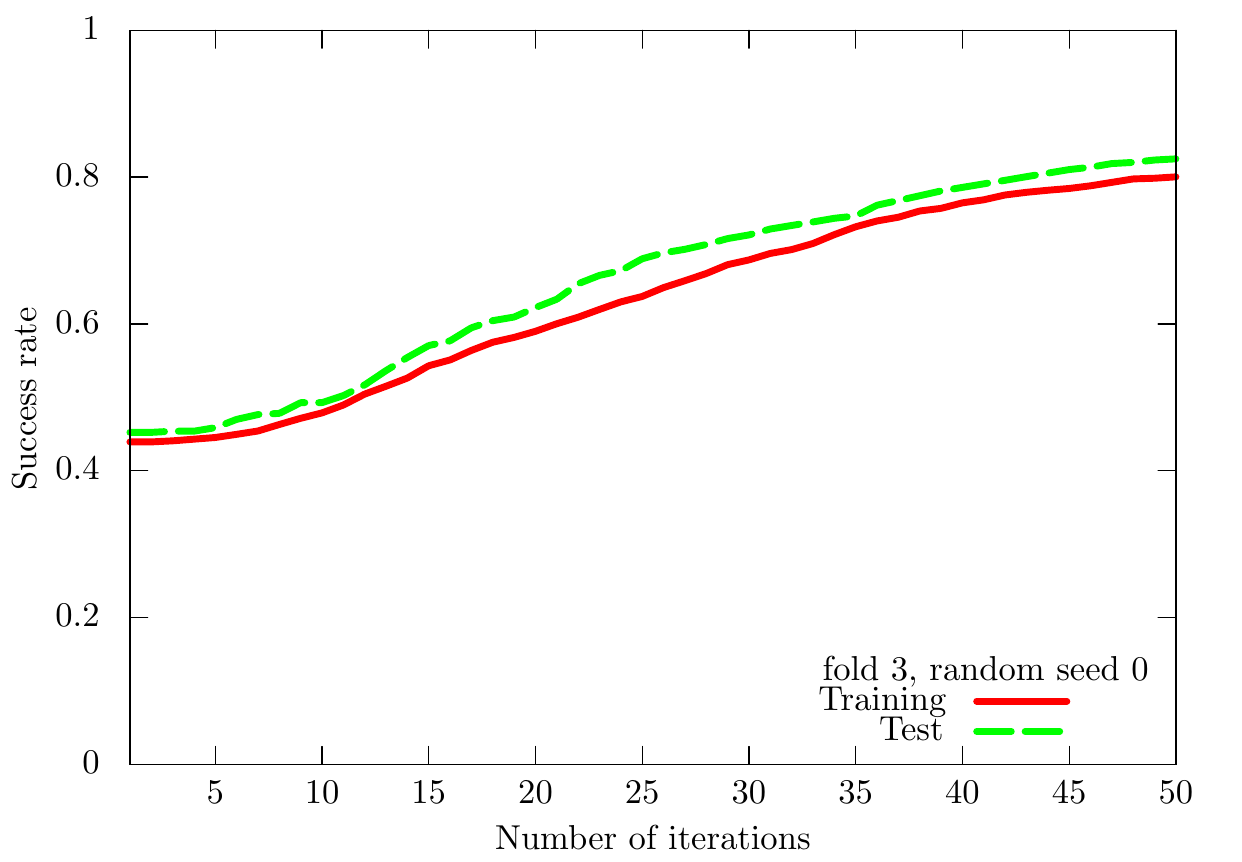}
\includegraphics[scale=0.25]{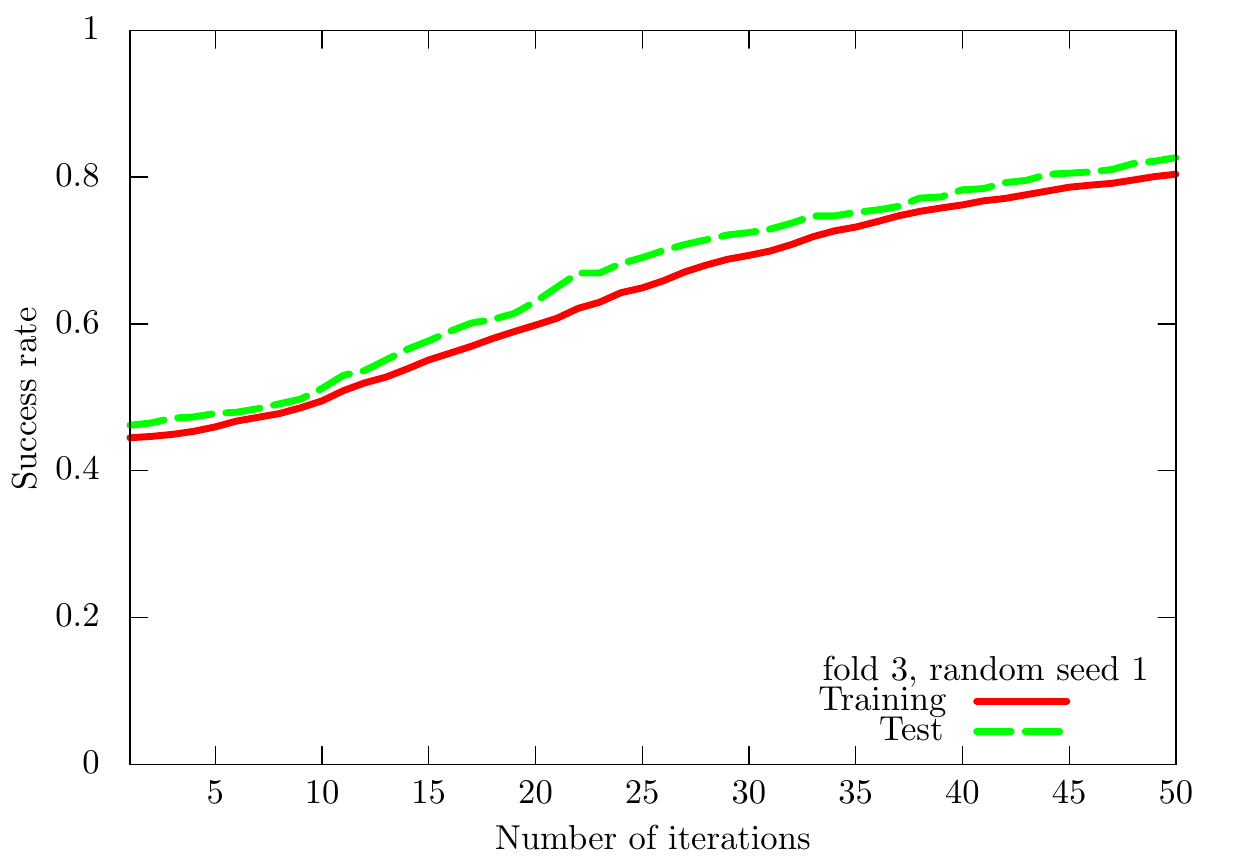}
\includegraphics[scale=0.25]{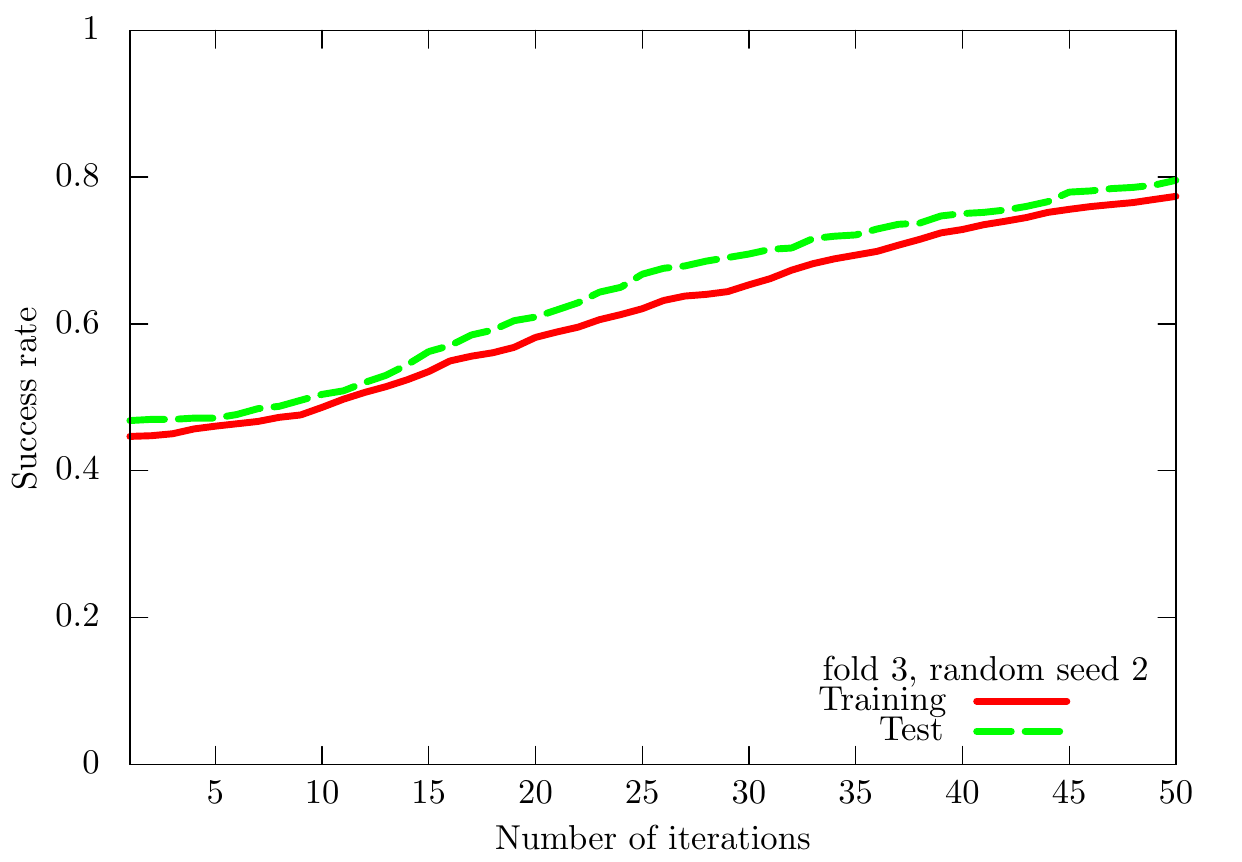}
\includegraphics[scale=0.25]{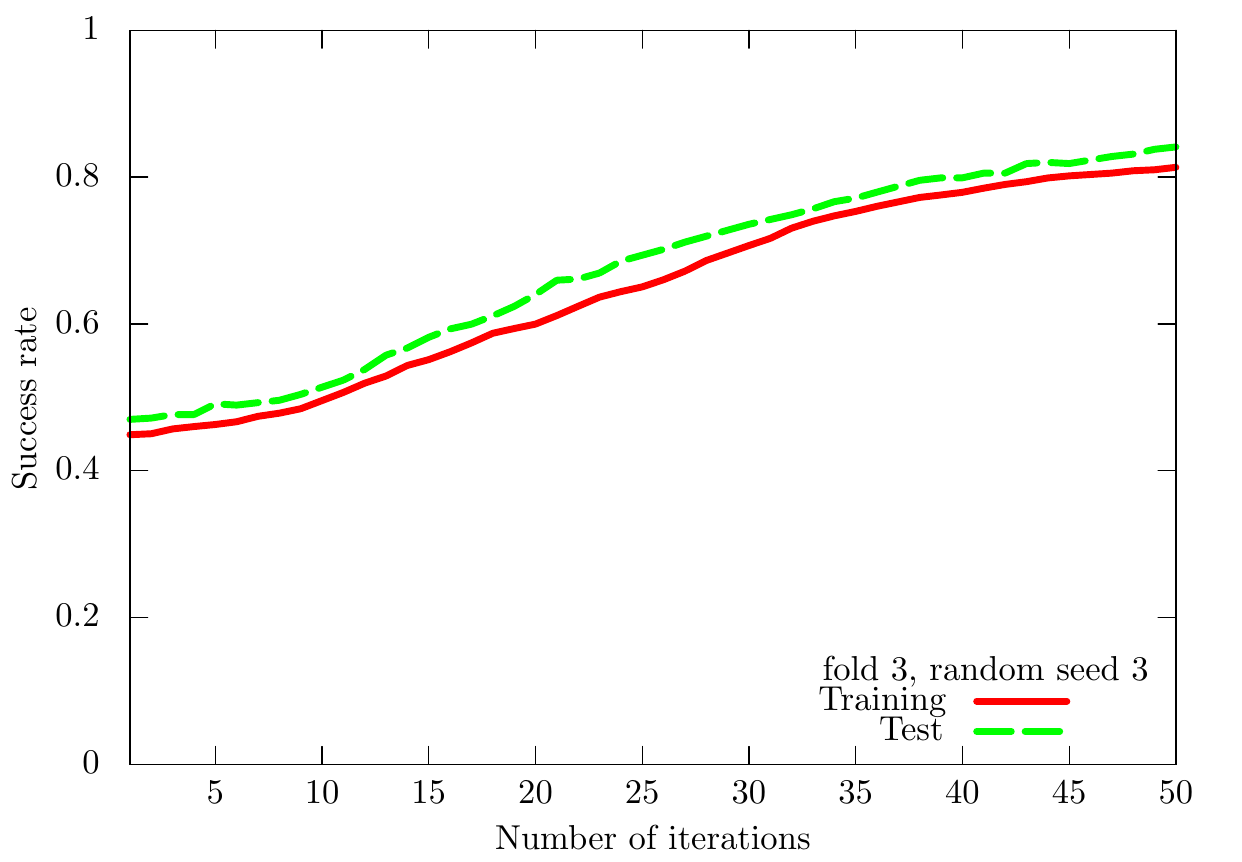}
\includegraphics[scale=0.25]{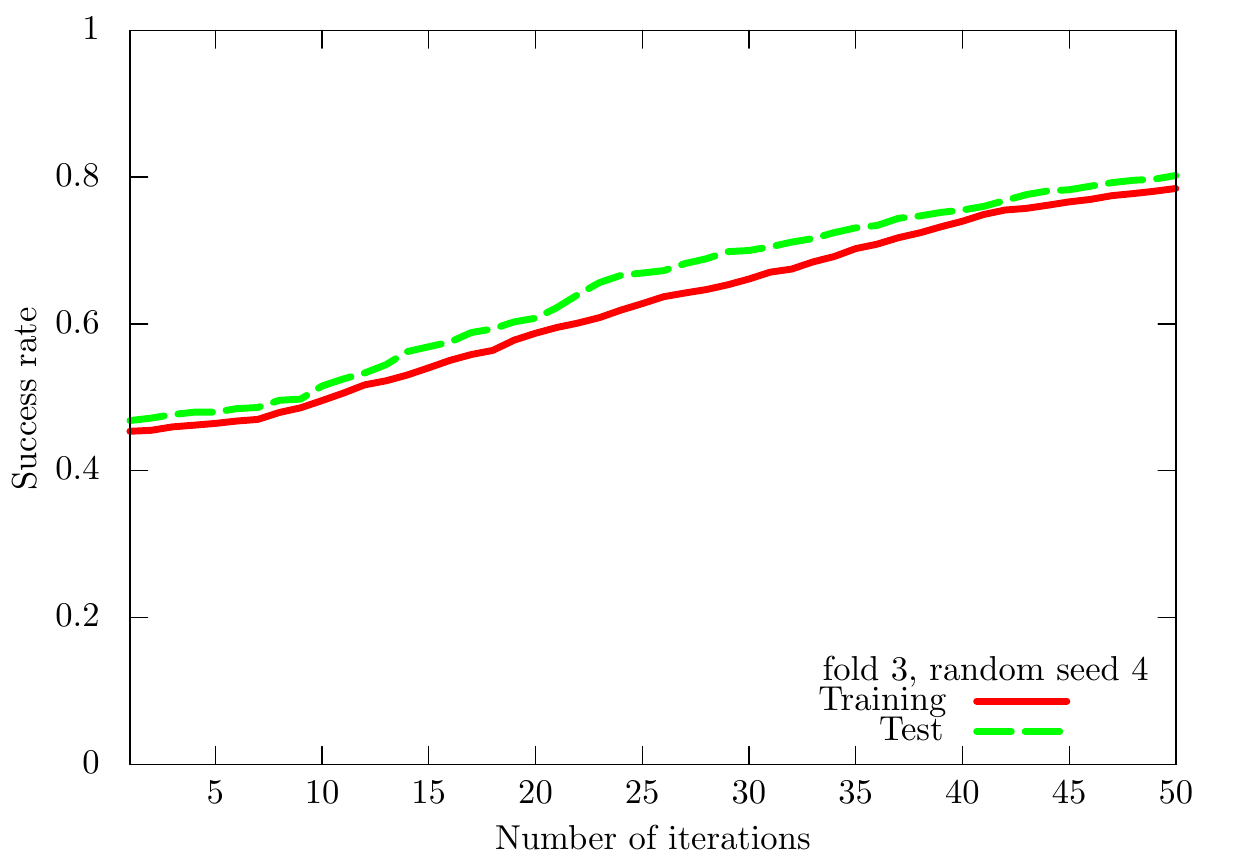}
\includegraphics[scale=0.25]{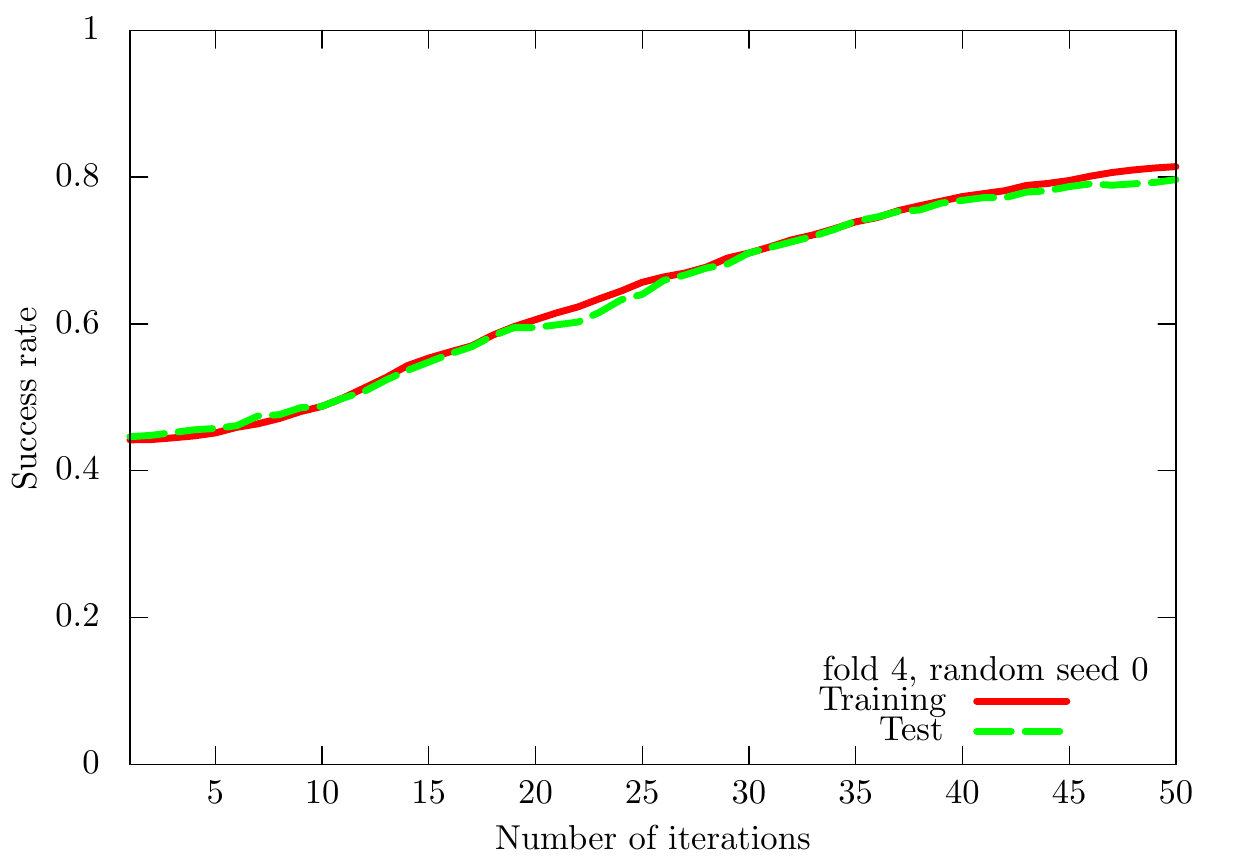}
\includegraphics[scale=0.25]{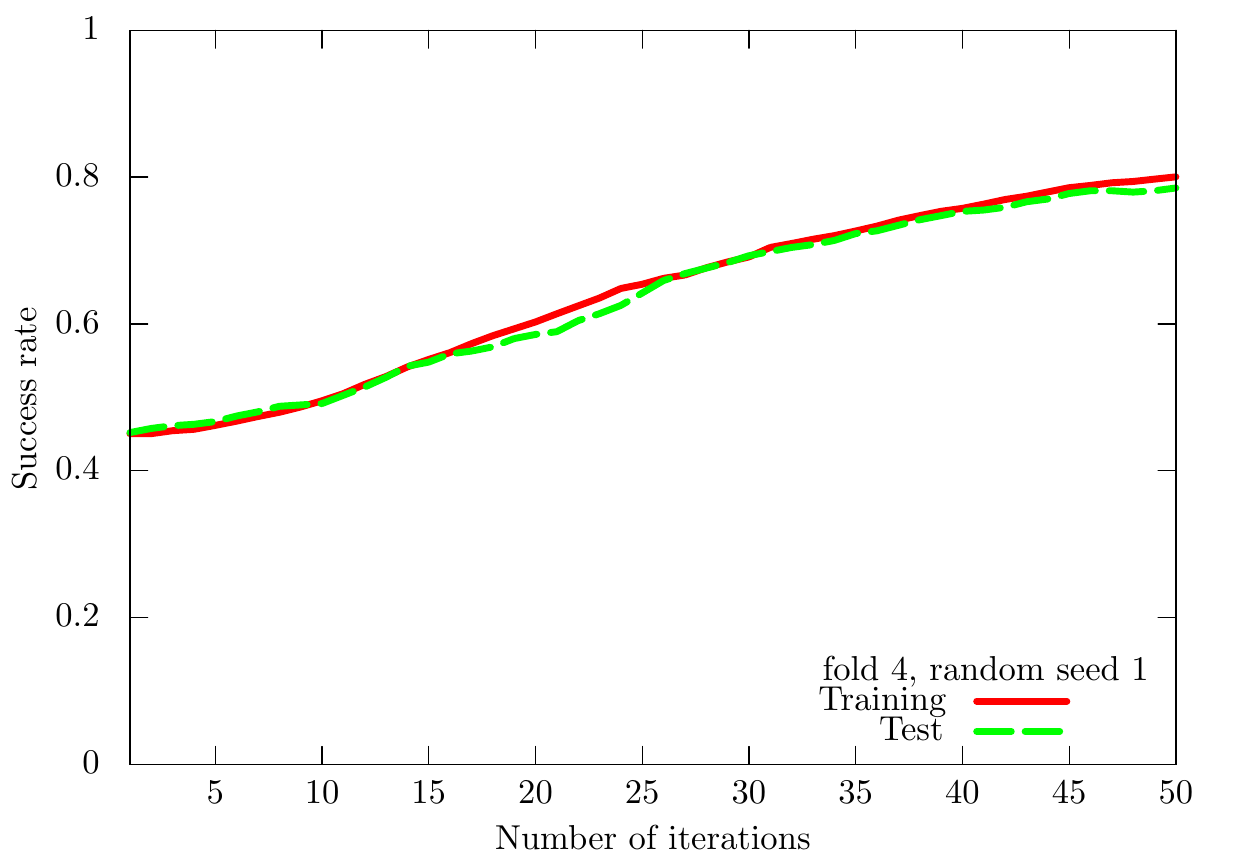}
\includegraphics[scale=0.25]{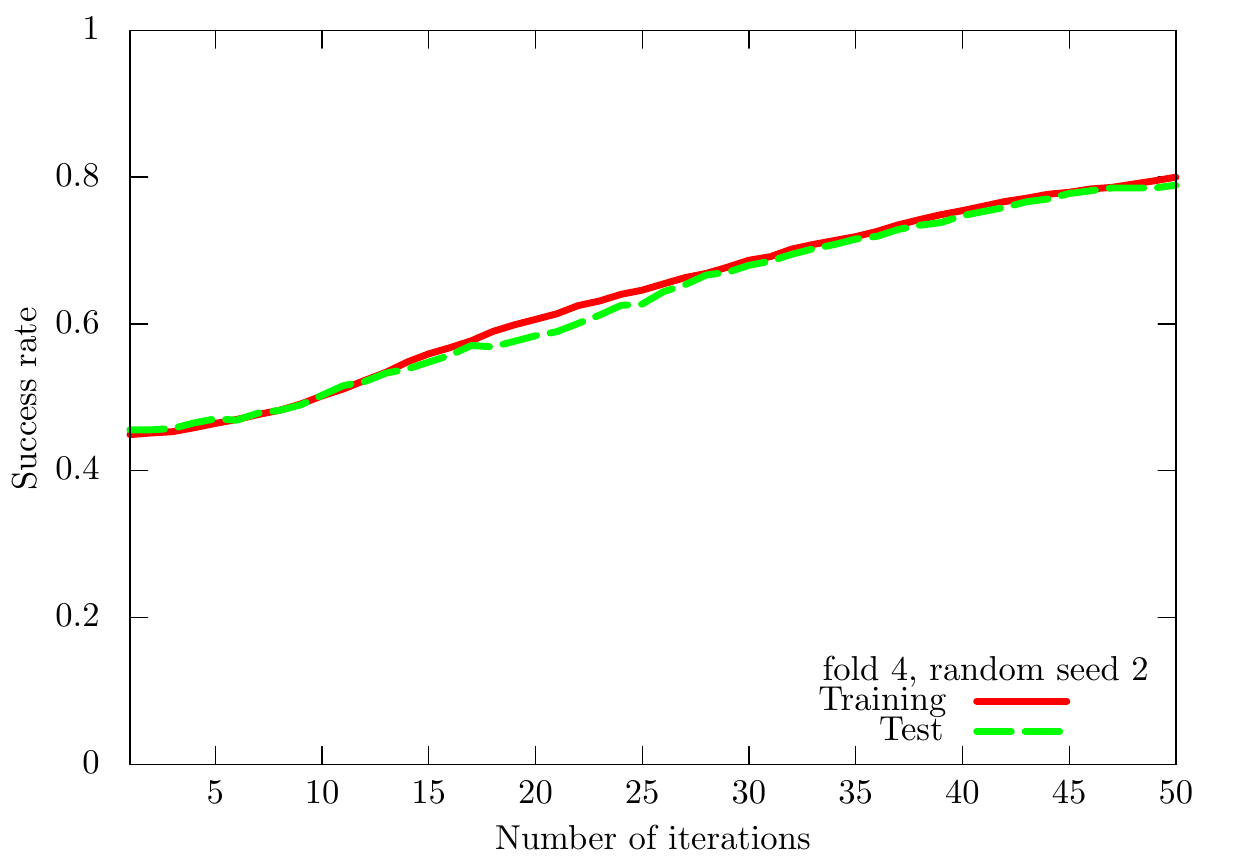}
\includegraphics[scale=0.25]{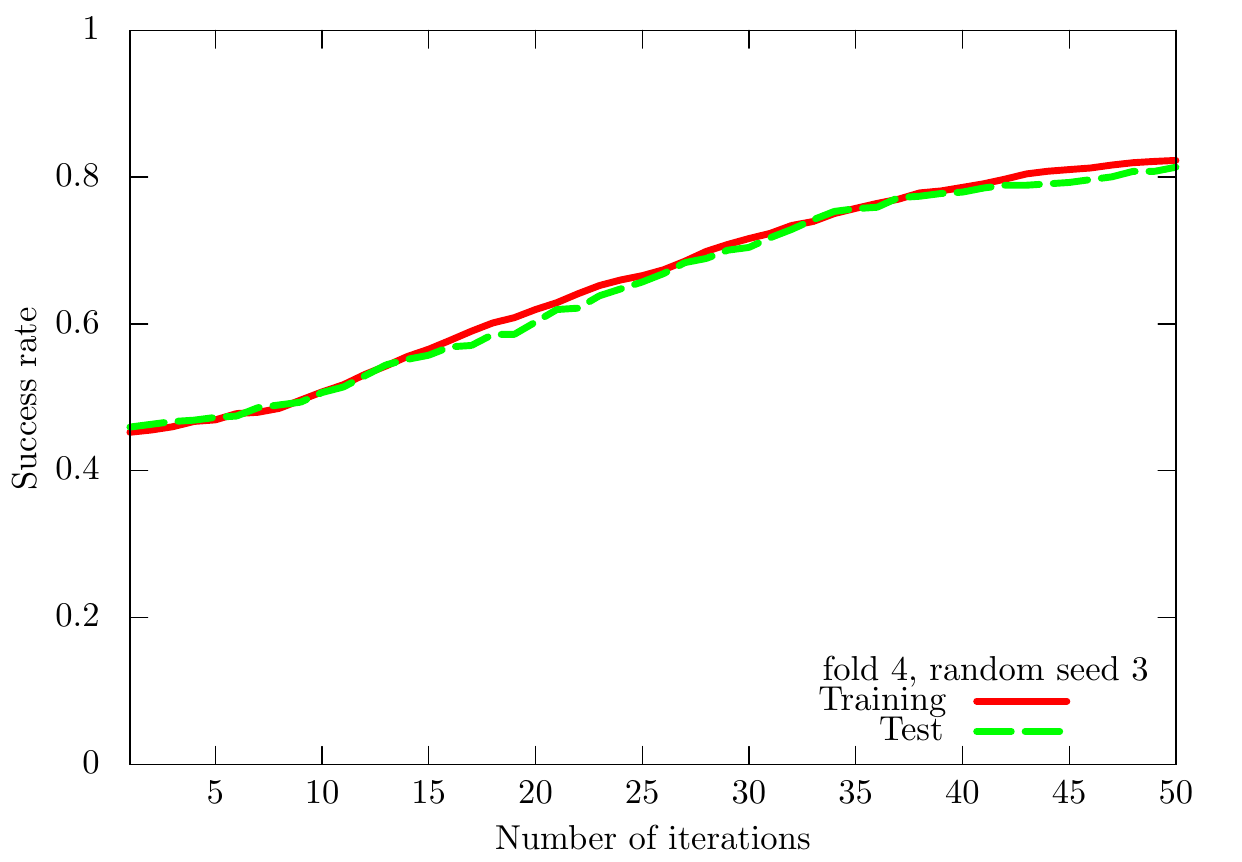}
\includegraphics[scale=0.25]{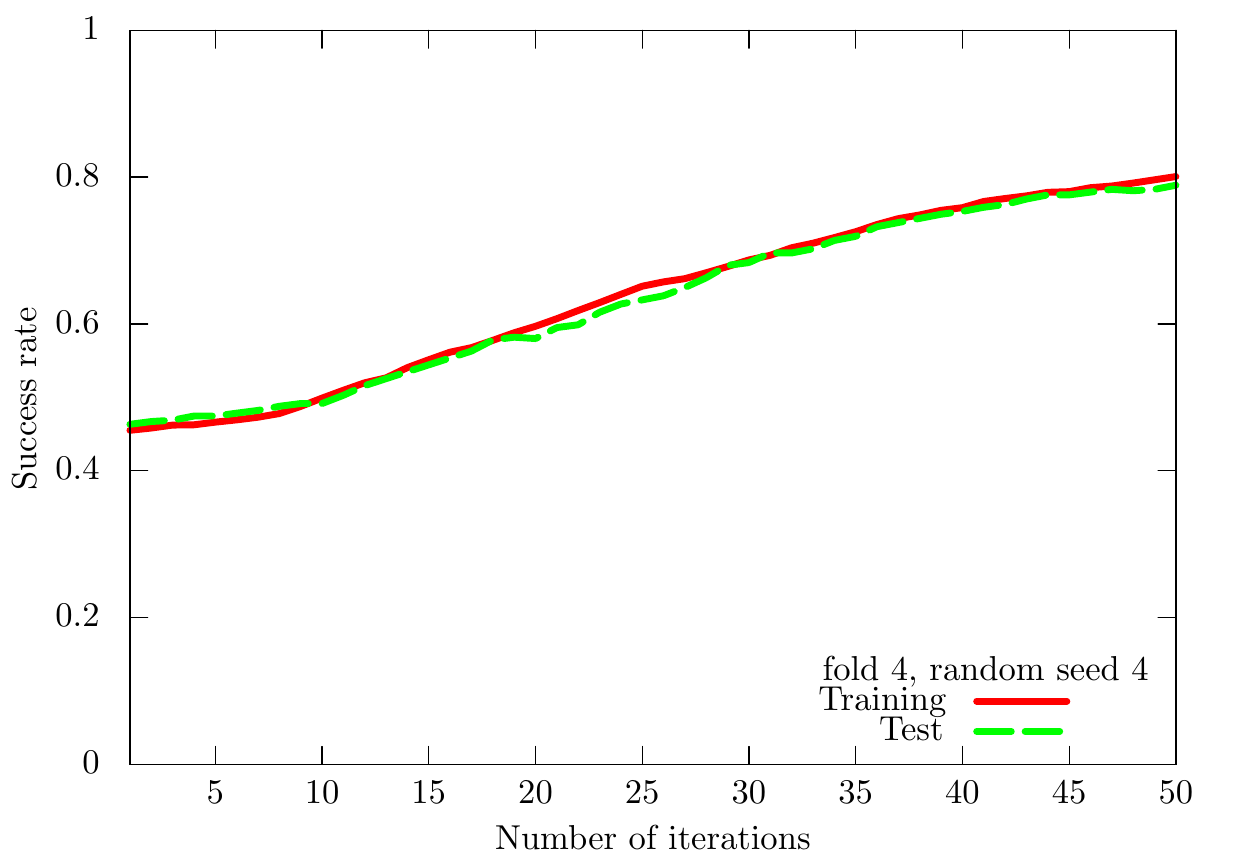}
\caption{Results of QCL on the $5$-fold datasets with $5$ different random seeds for the MNIST256 dataset ($0$ or non-$0$). We use the CNOT-based circuit and set $\theta_\mathrm{bias} = 0$. The number of layers $L$ is set to $5$.}
\label{supp-arXiv-numerical-result-raw-data-fold-001-rand-001-QCL-MNIST256-0-non0}
\end{figure*}

\begin{figure*}[htb]
\centering
\includegraphics[scale=0.25]{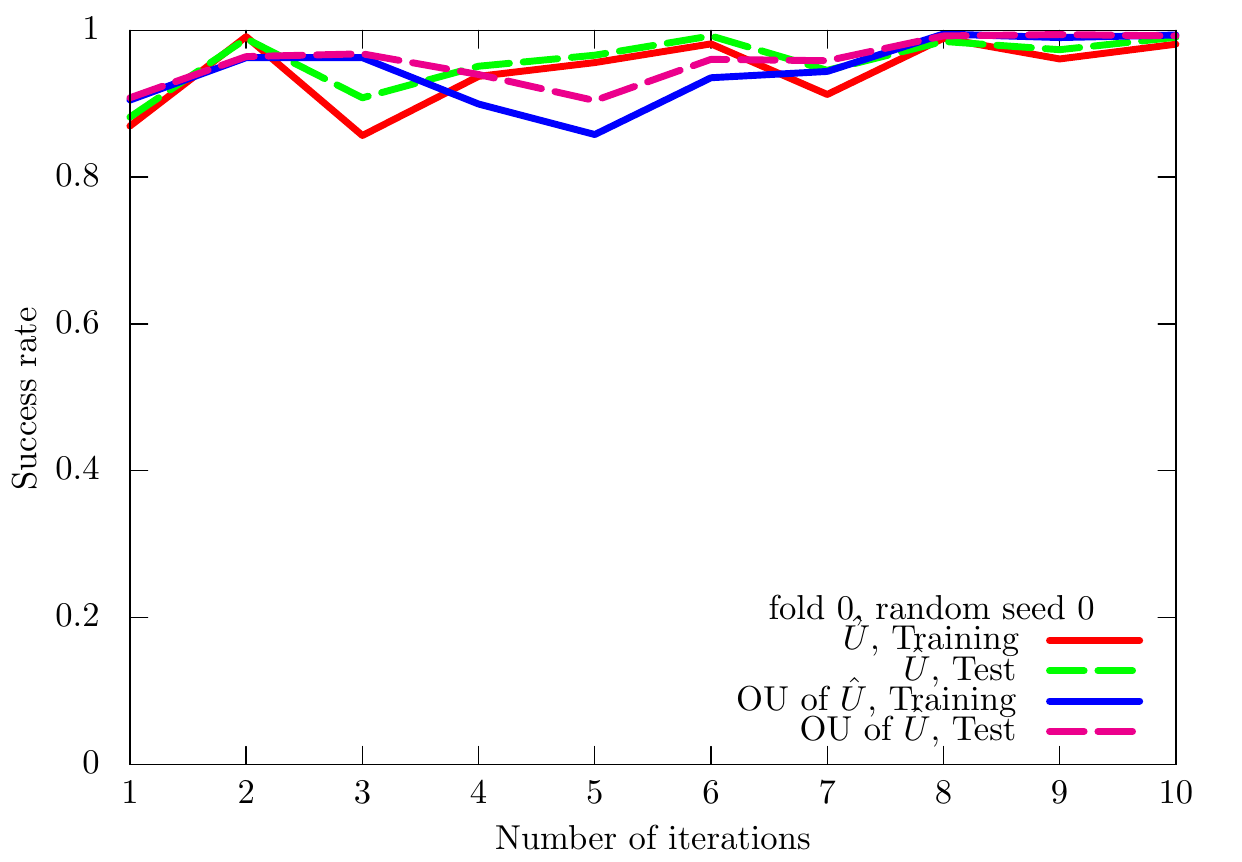}
\includegraphics[scale=0.25]{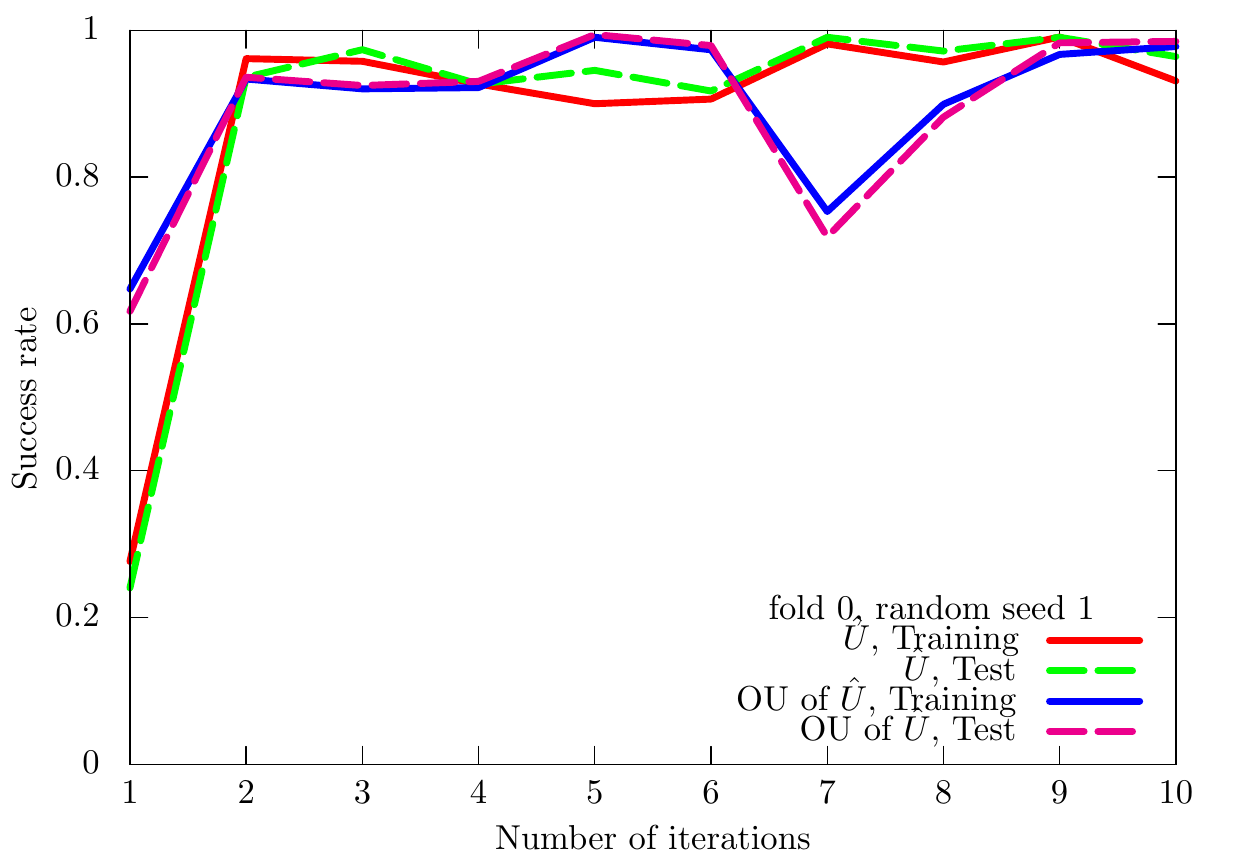}
\includegraphics[scale=0.25]{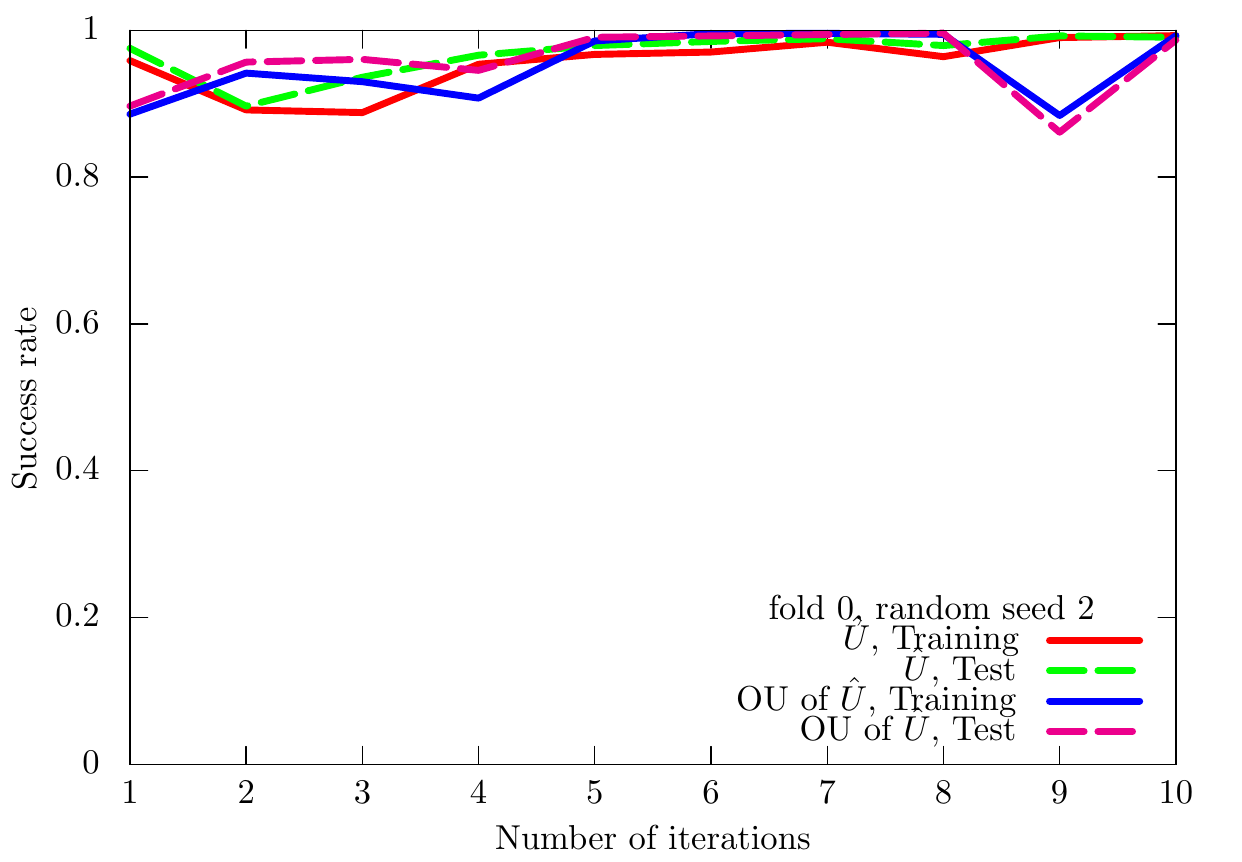}
\includegraphics[scale=0.25]{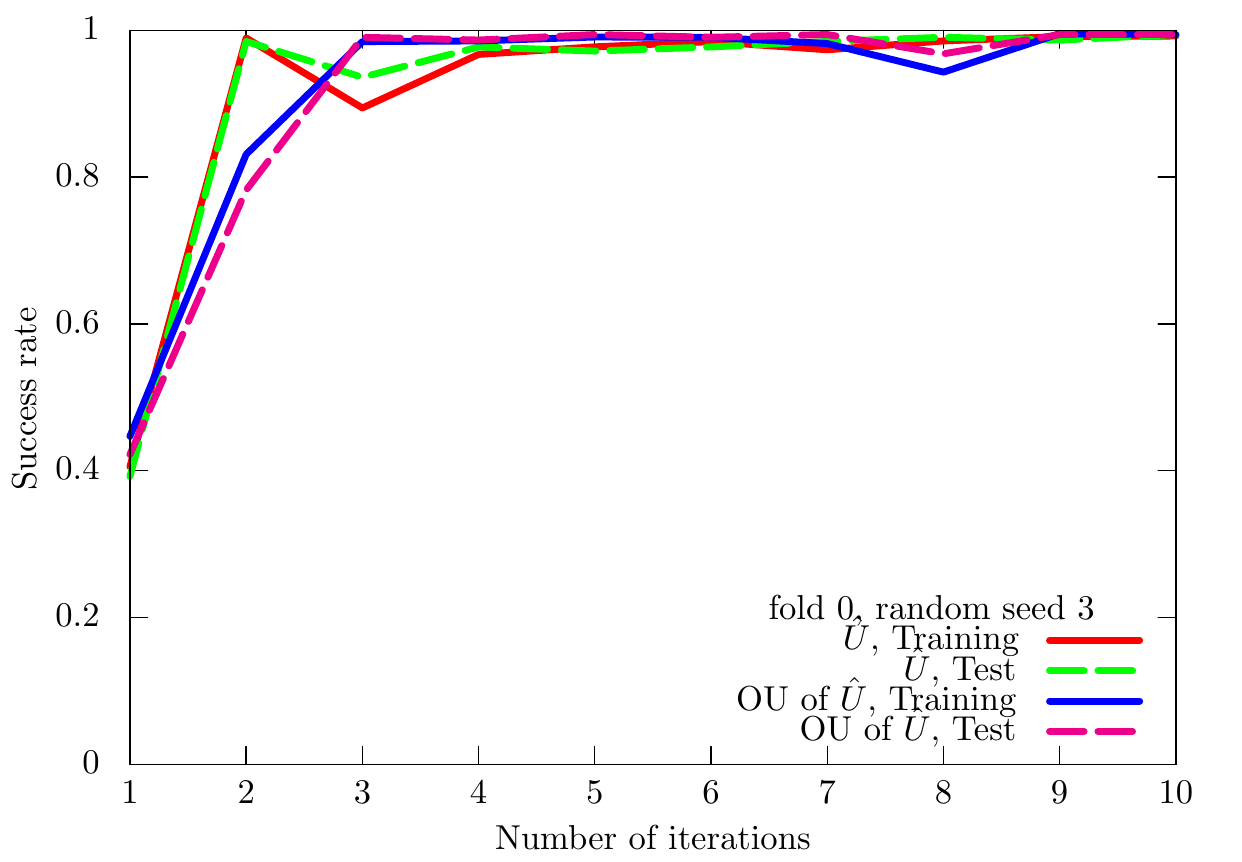}
\includegraphics[scale=0.25]{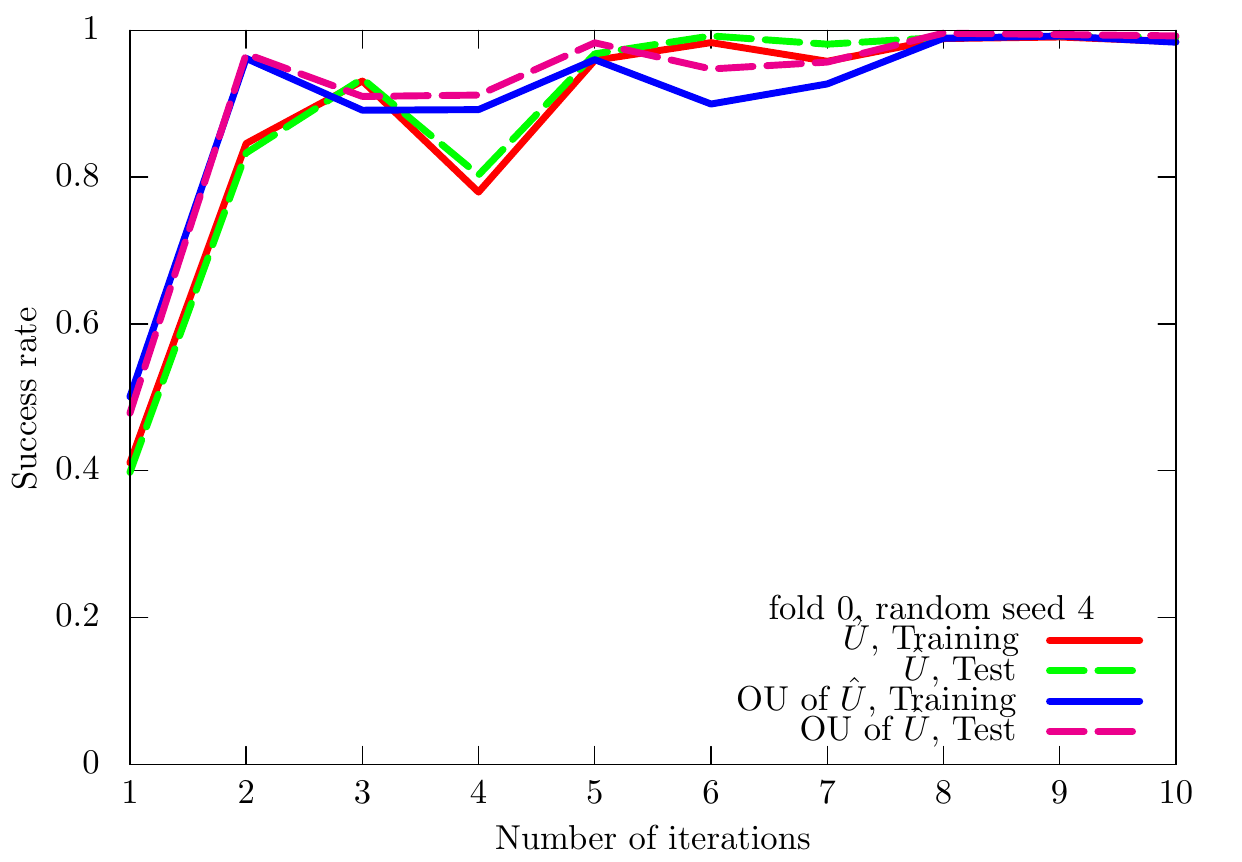}
\includegraphics[scale=0.25]{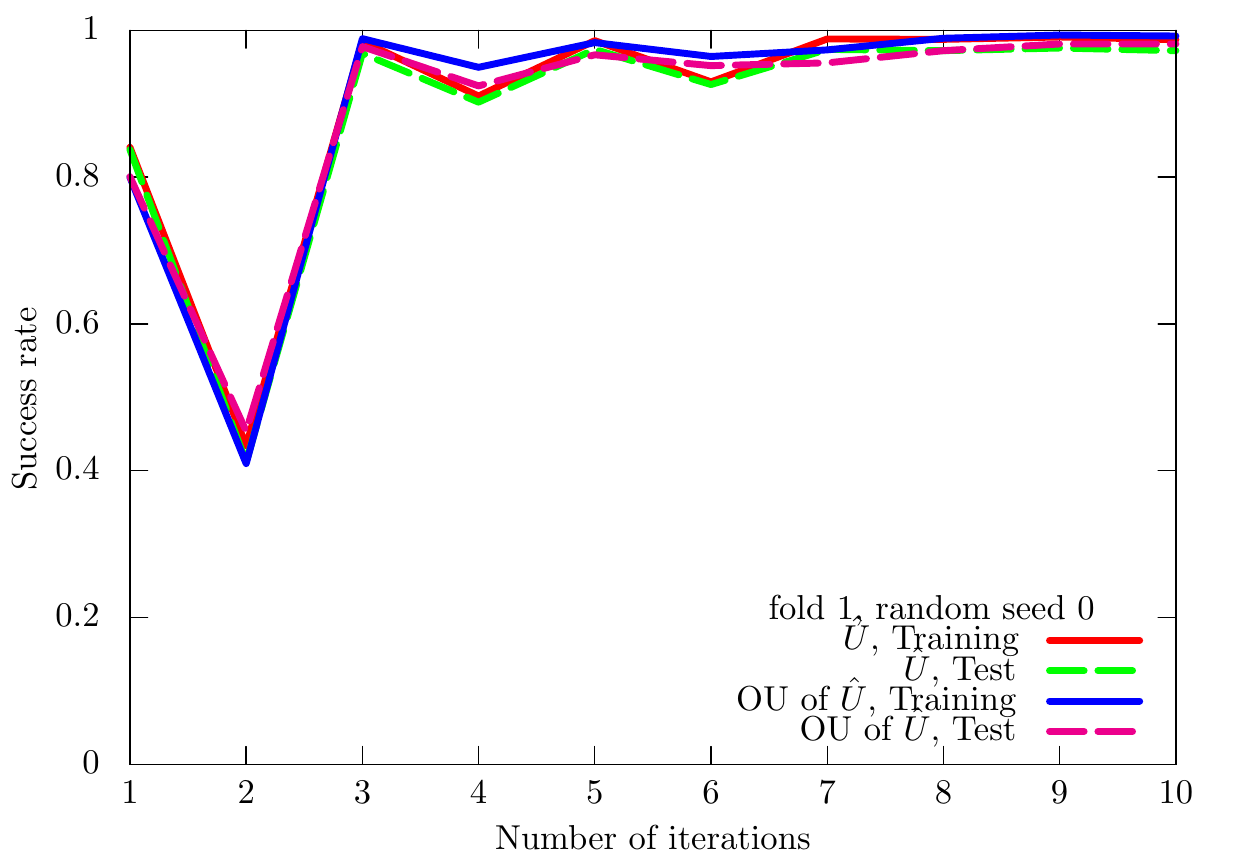}
\includegraphics[scale=0.25]{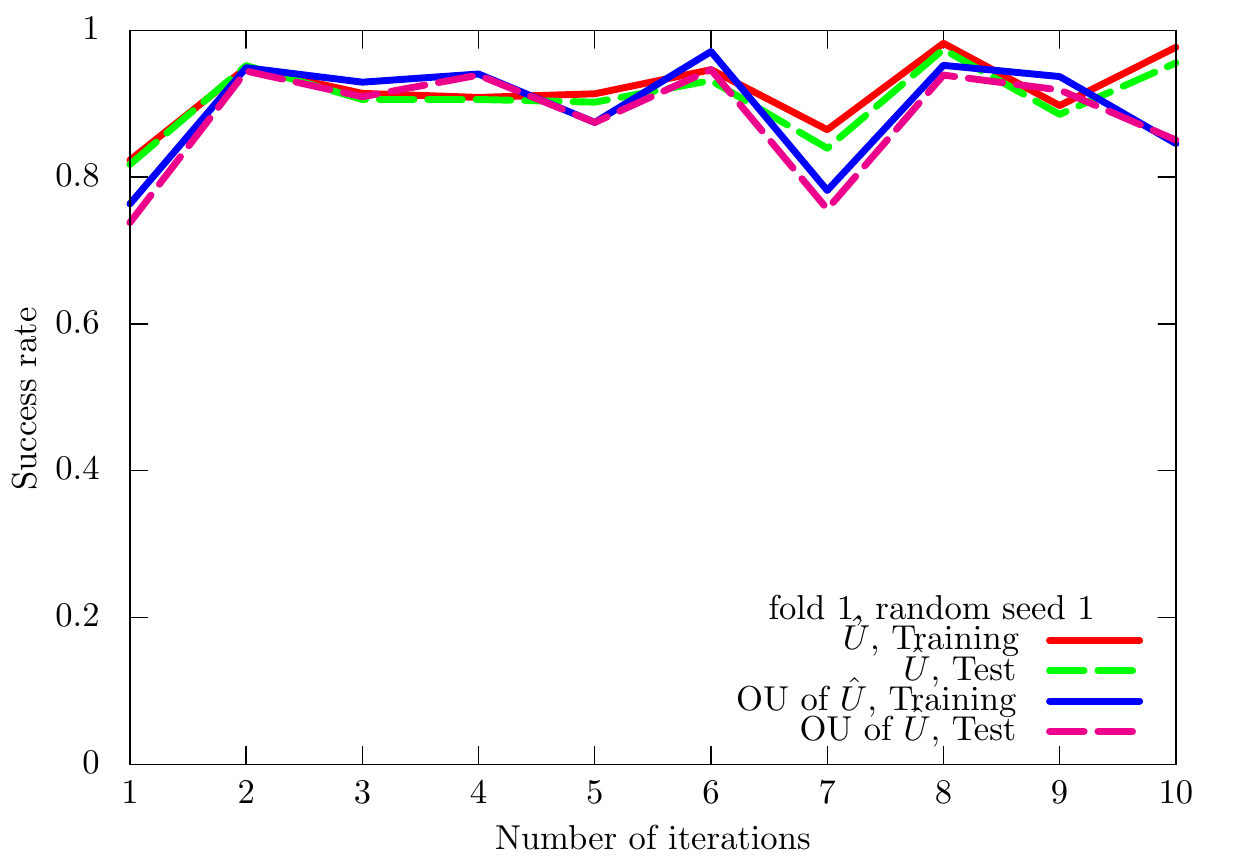}
\includegraphics[scale=0.25]{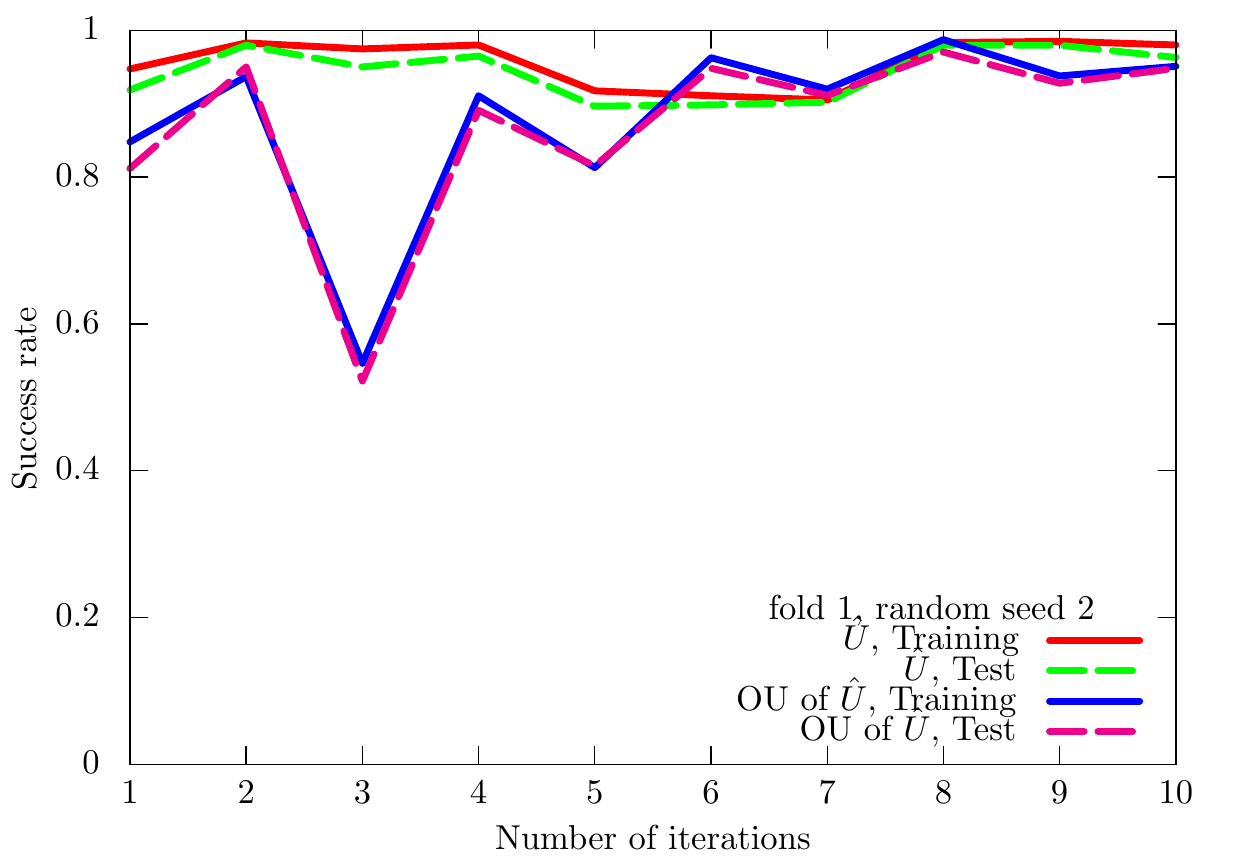}
\includegraphics[scale=0.25]{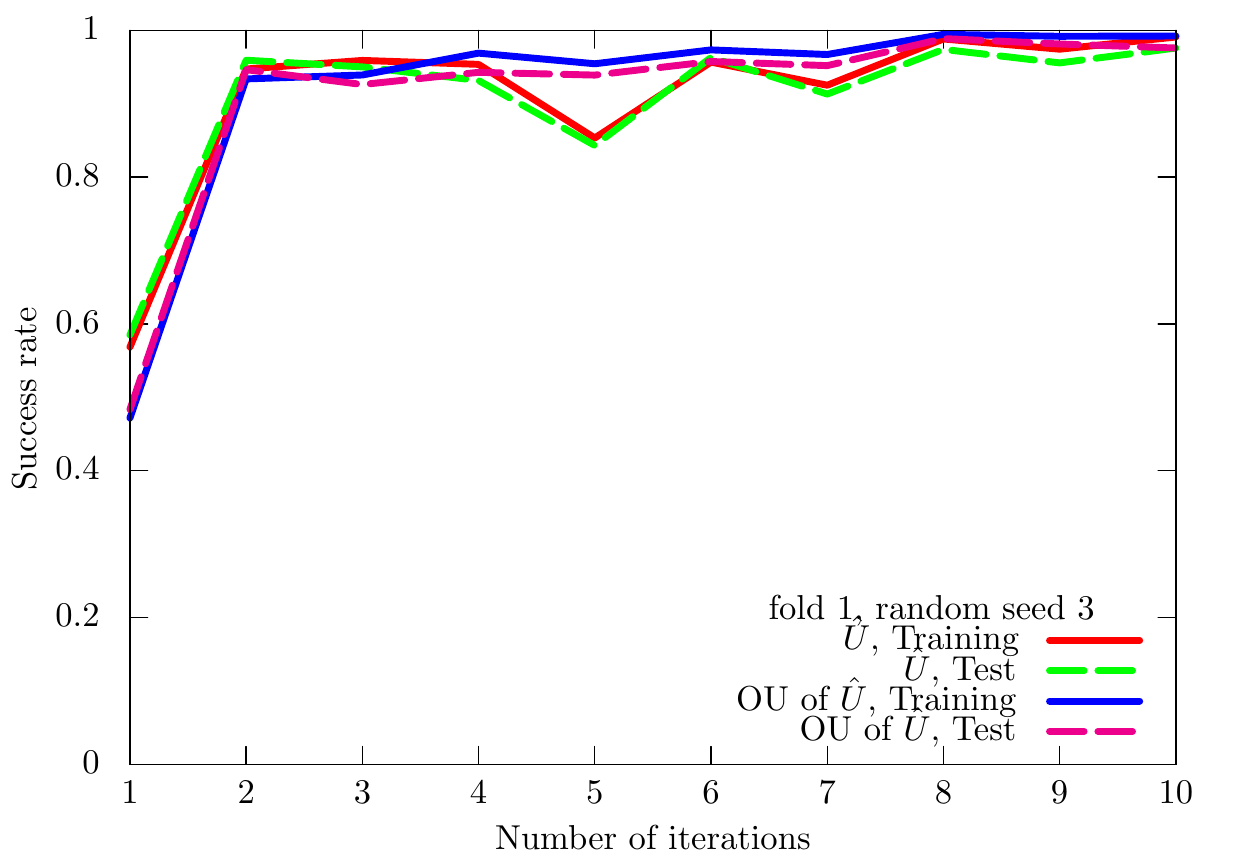}
\includegraphics[scale=0.25]{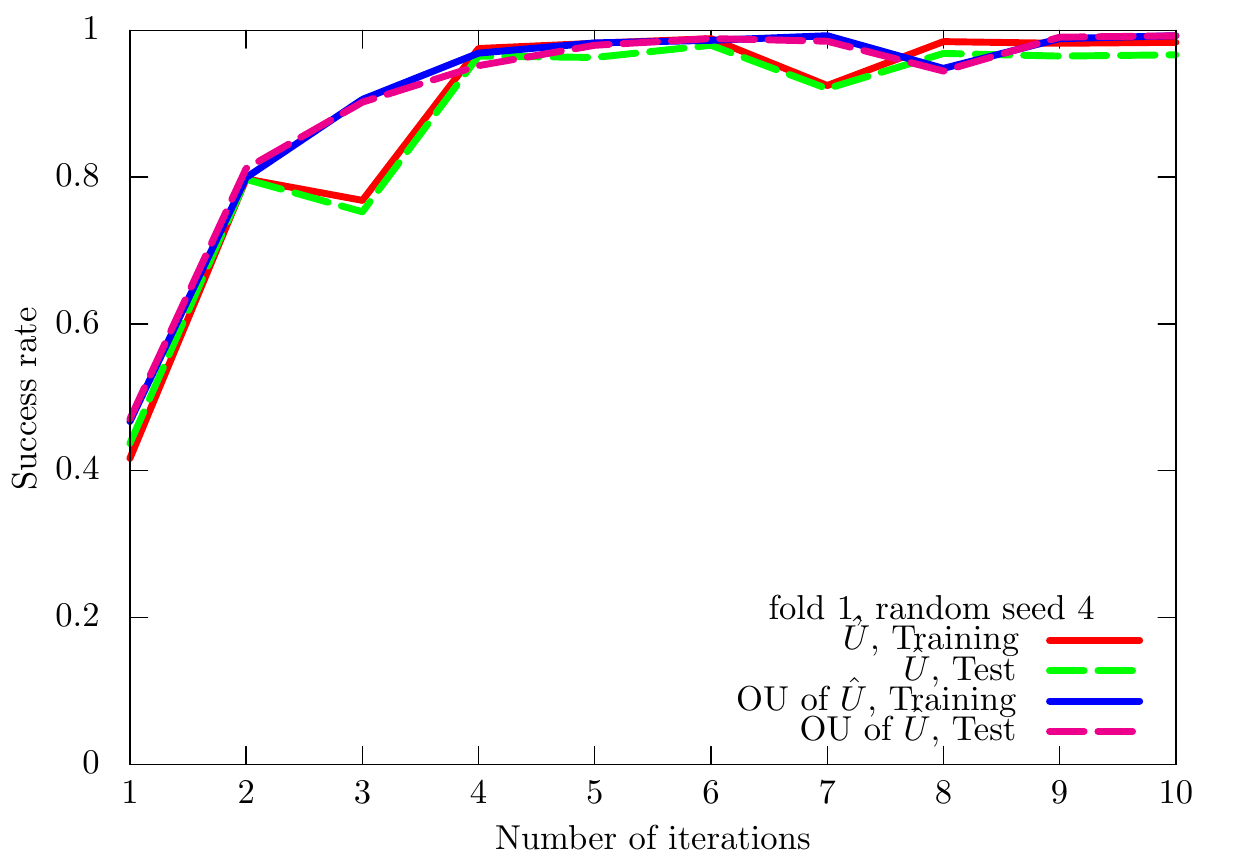}
\includegraphics[scale=0.25]{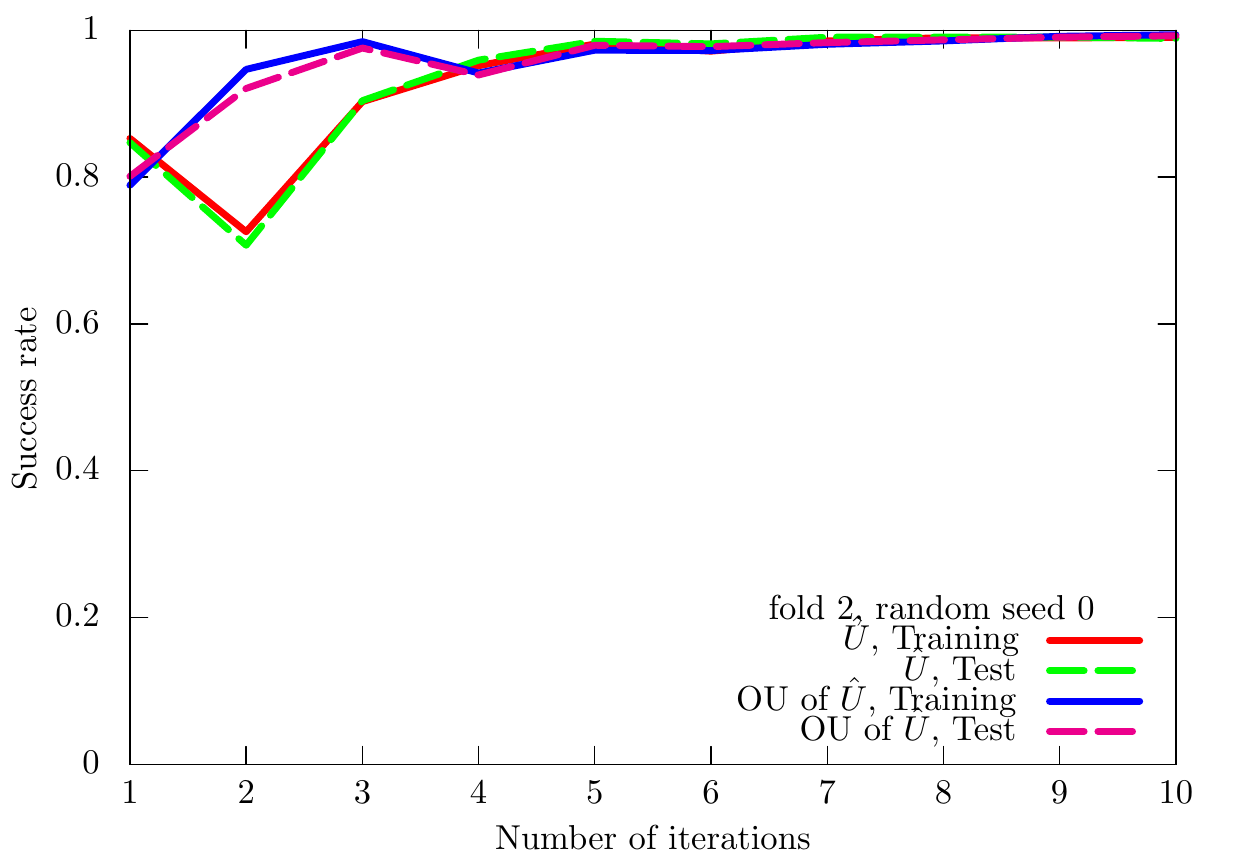}
\includegraphics[scale=0.25]{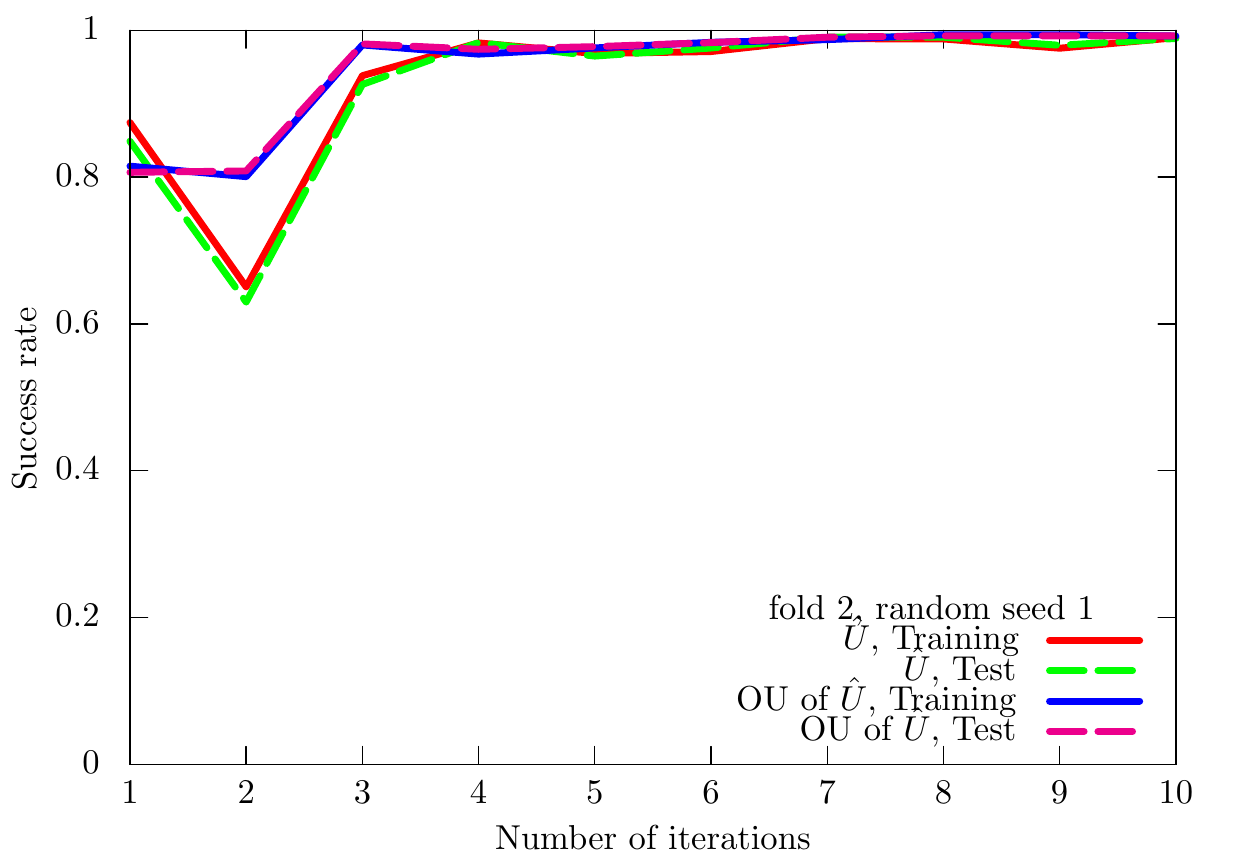}
\includegraphics[scale=0.25]{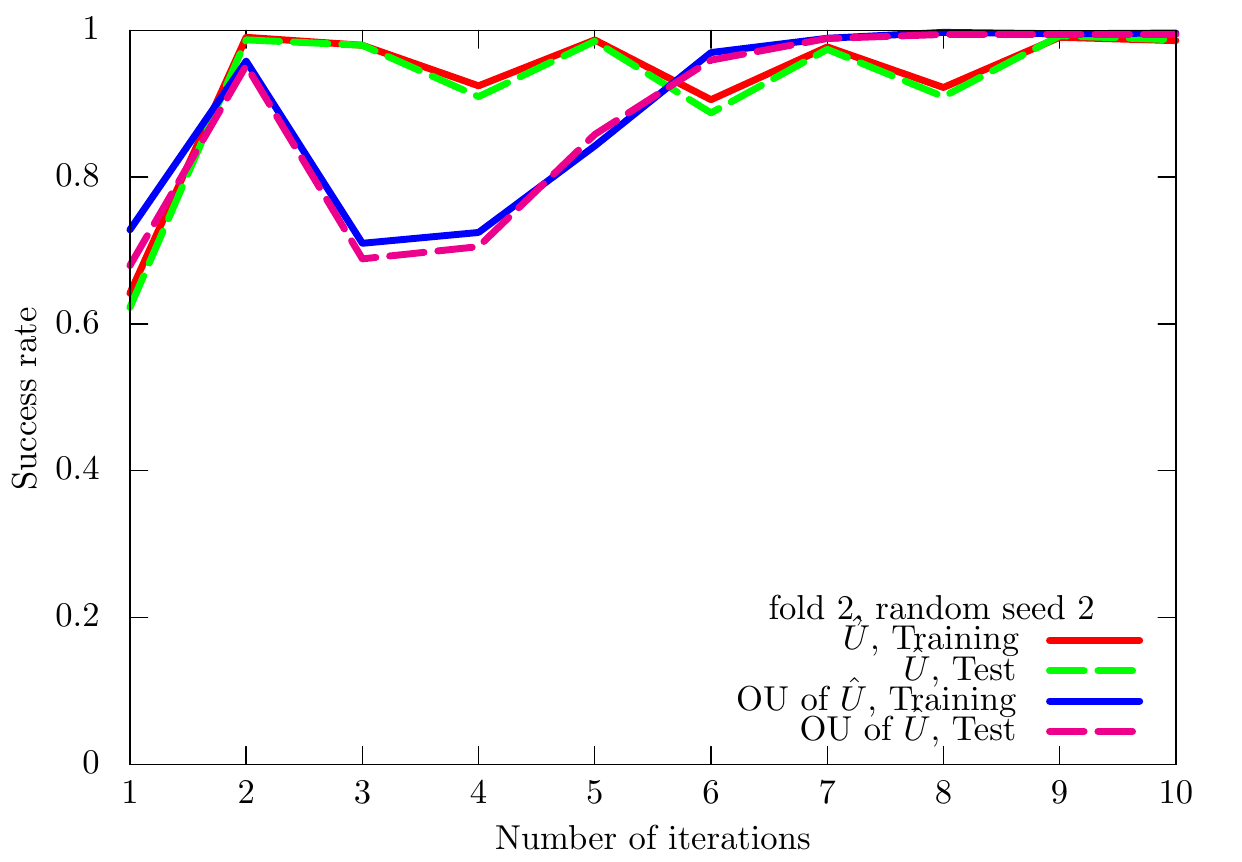}
\includegraphics[scale=0.25]{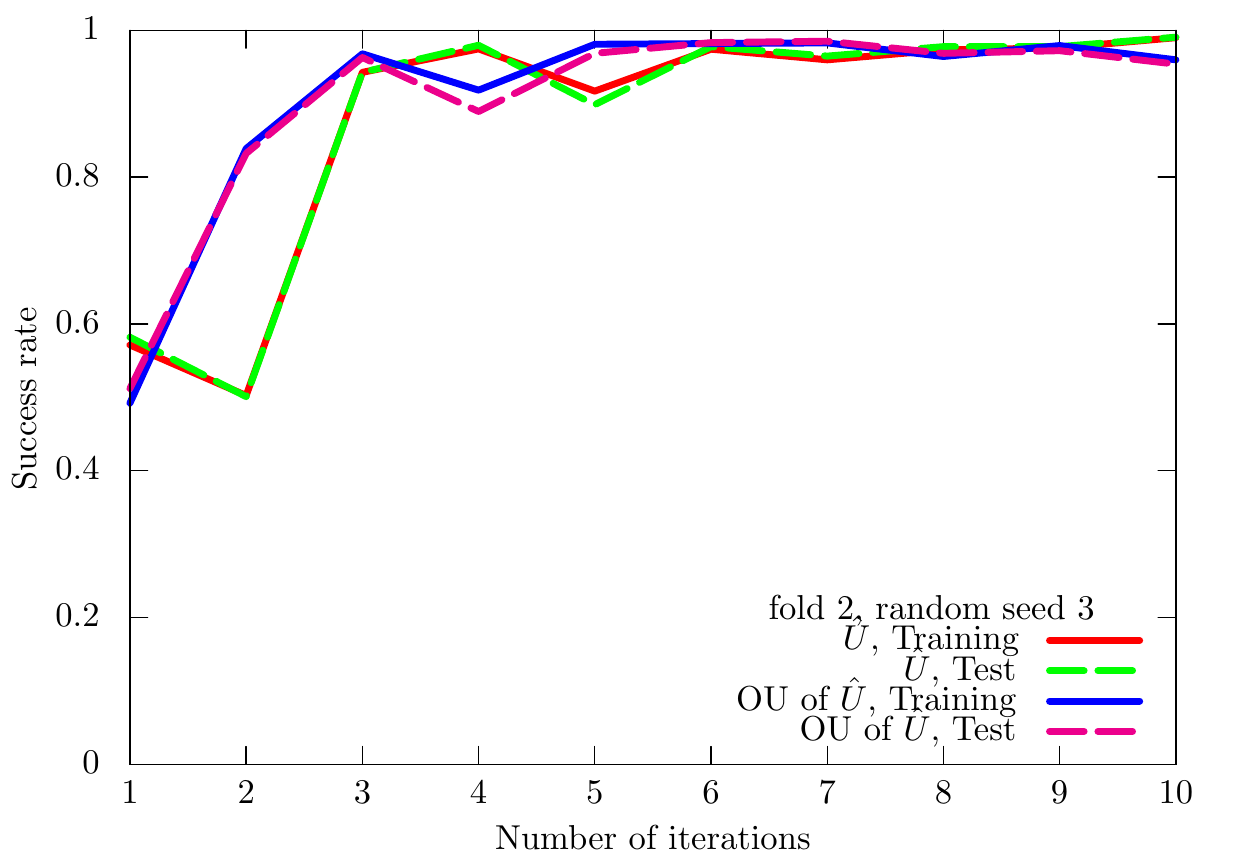}
\includegraphics[scale=0.25]{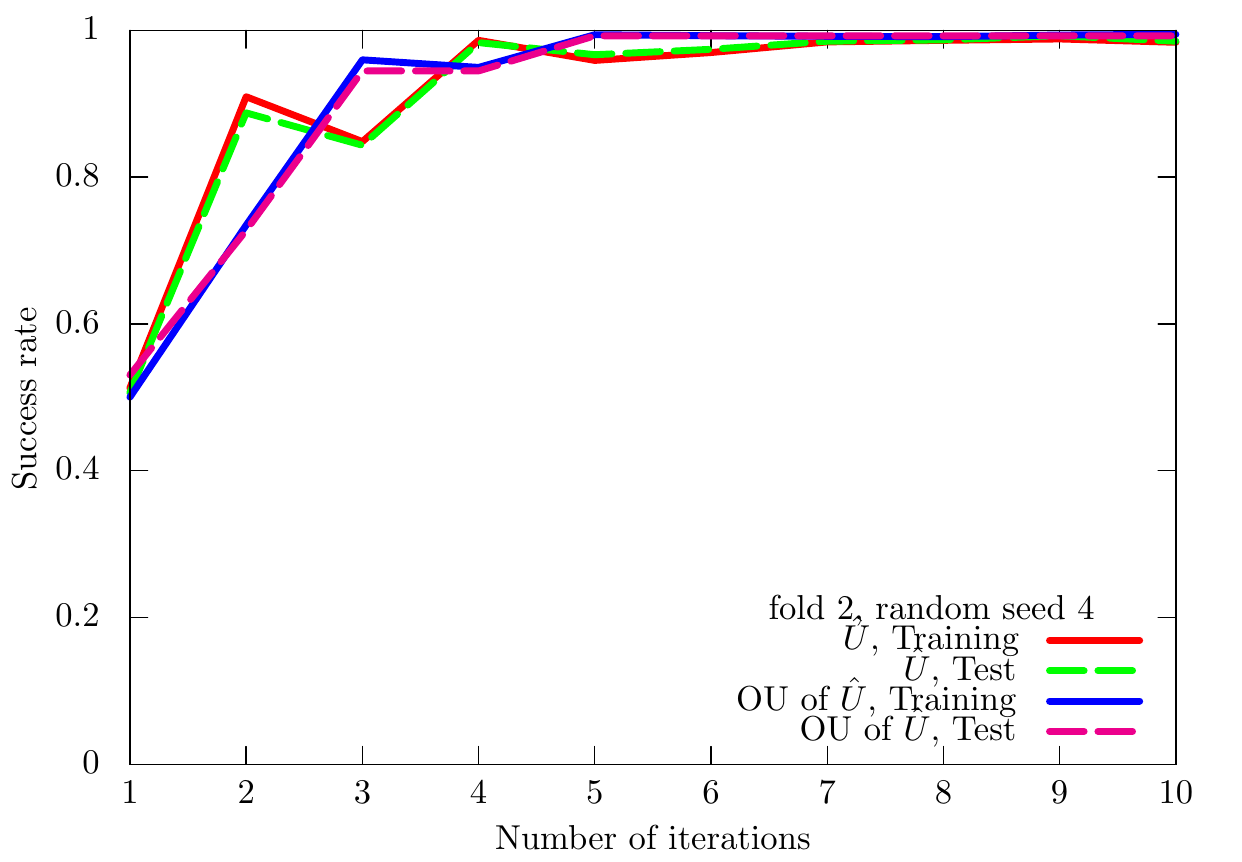}
\includegraphics[scale=0.25]{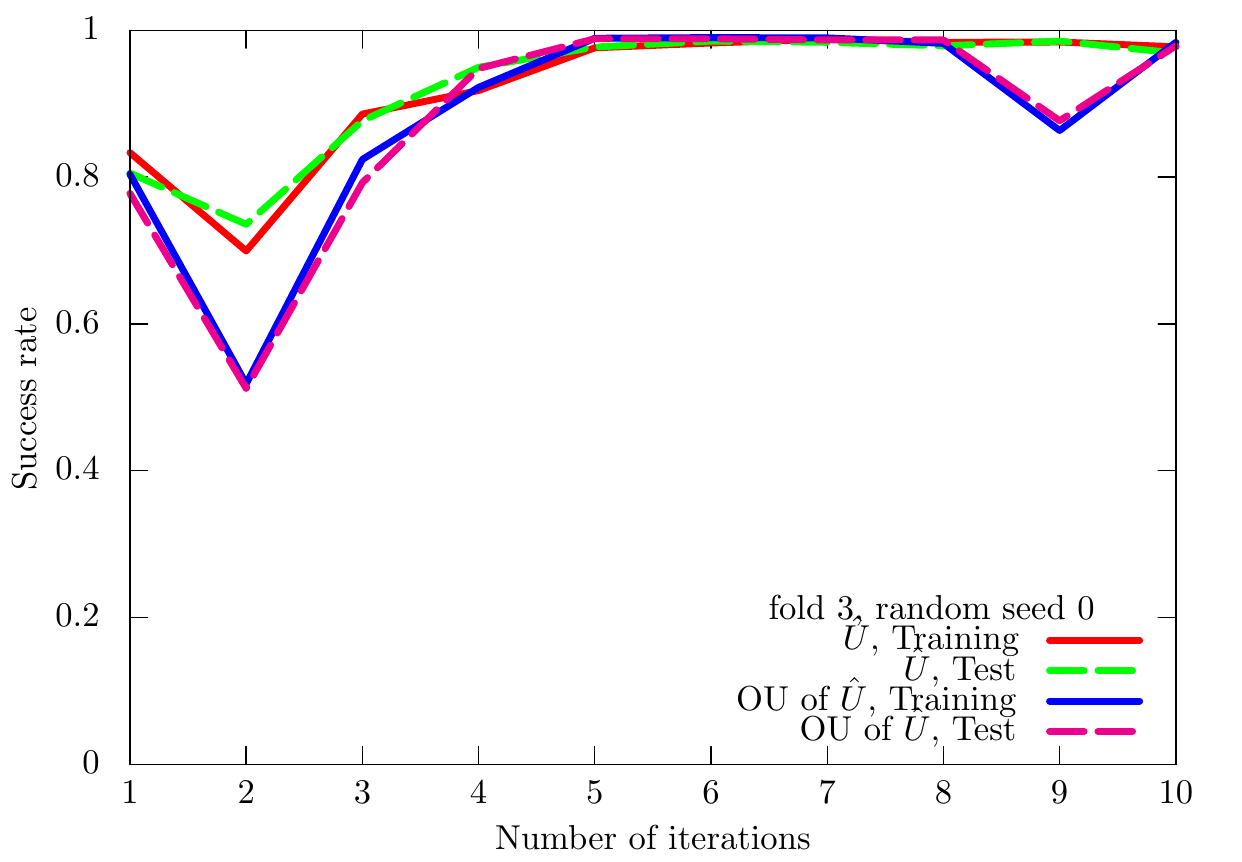}
\includegraphics[scale=0.25]{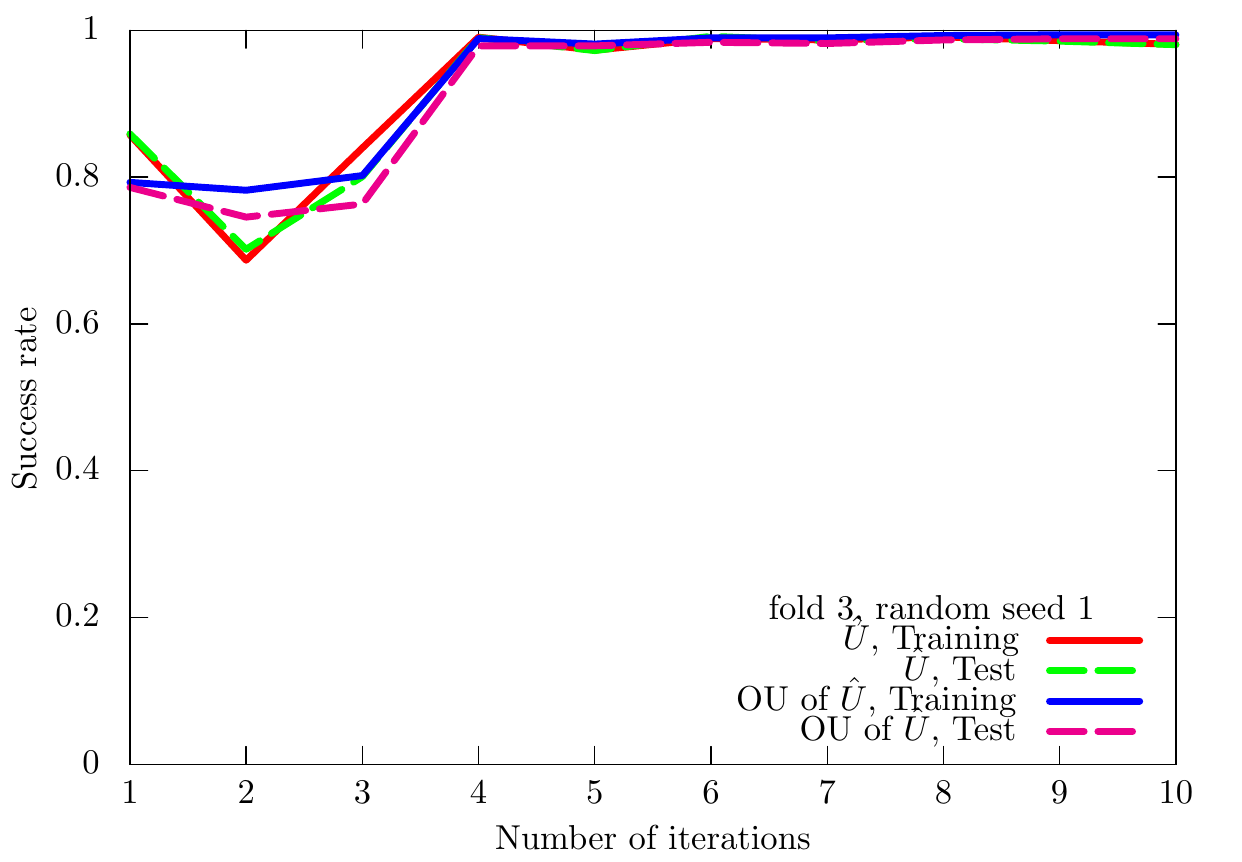}
\includegraphics[scale=0.25]{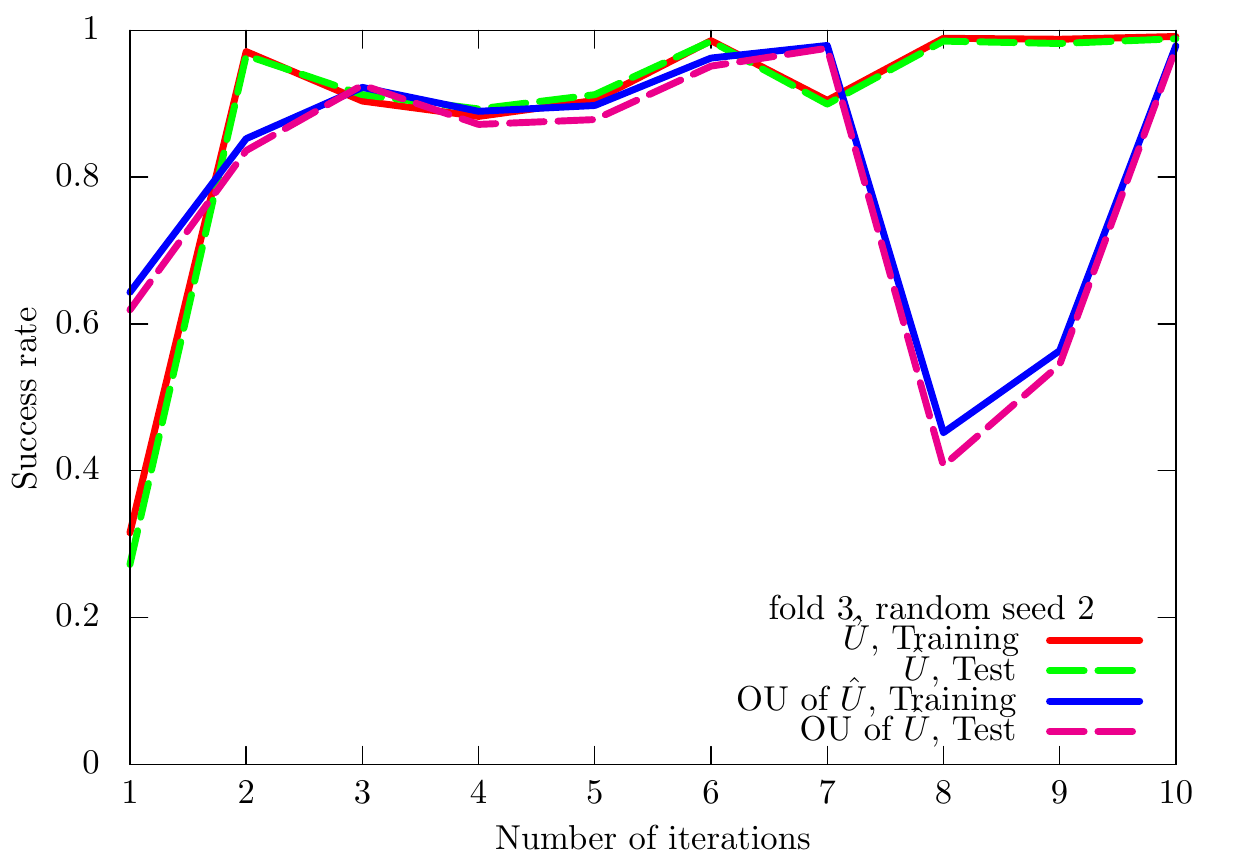}
\includegraphics[scale=0.25]{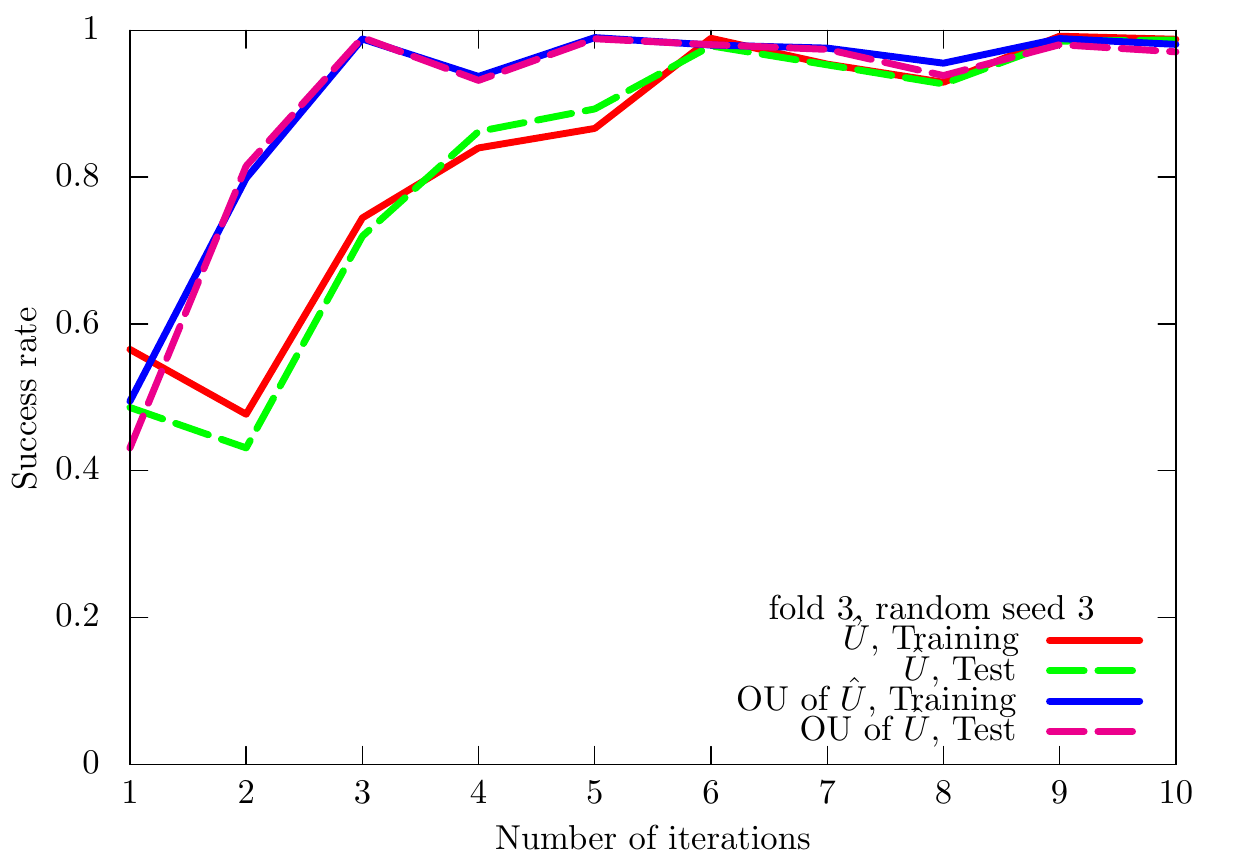}
\includegraphics[scale=0.25]{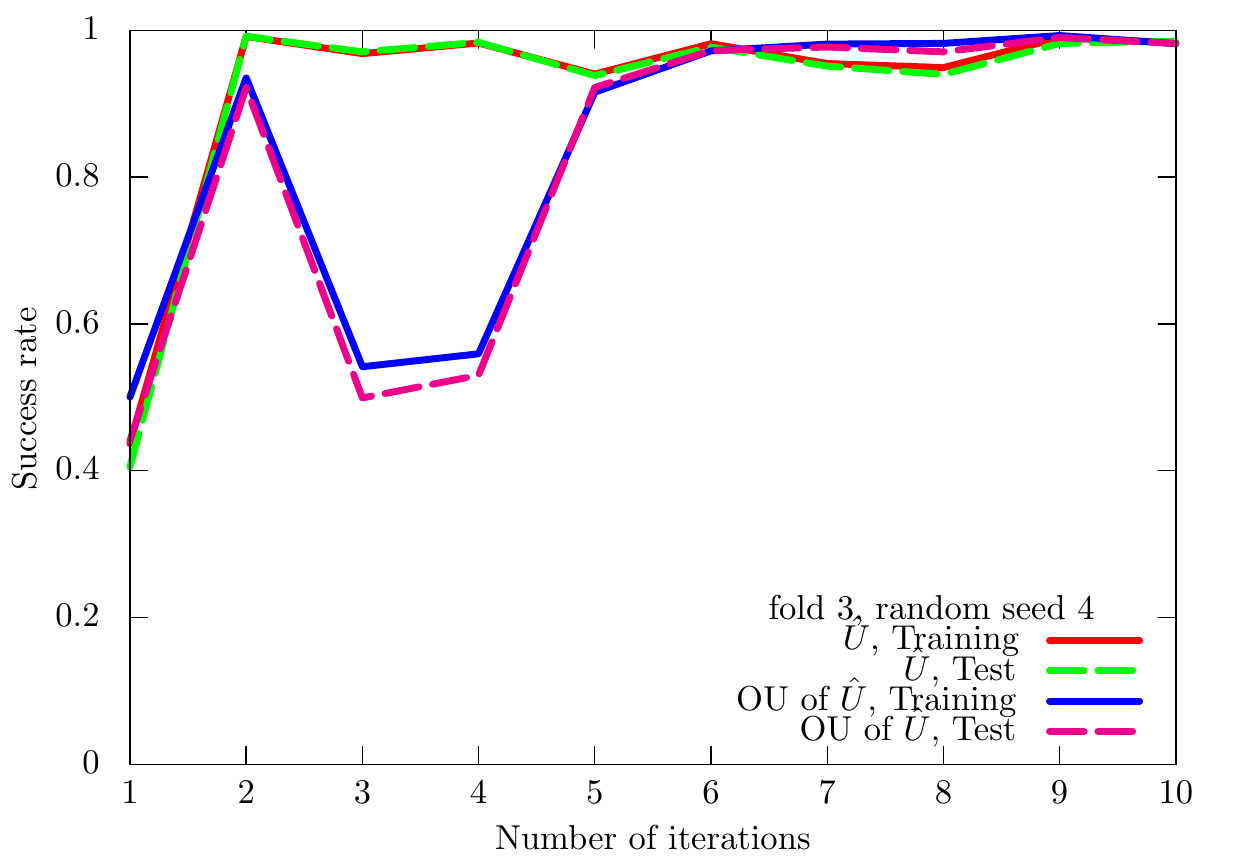}
\includegraphics[scale=0.25]{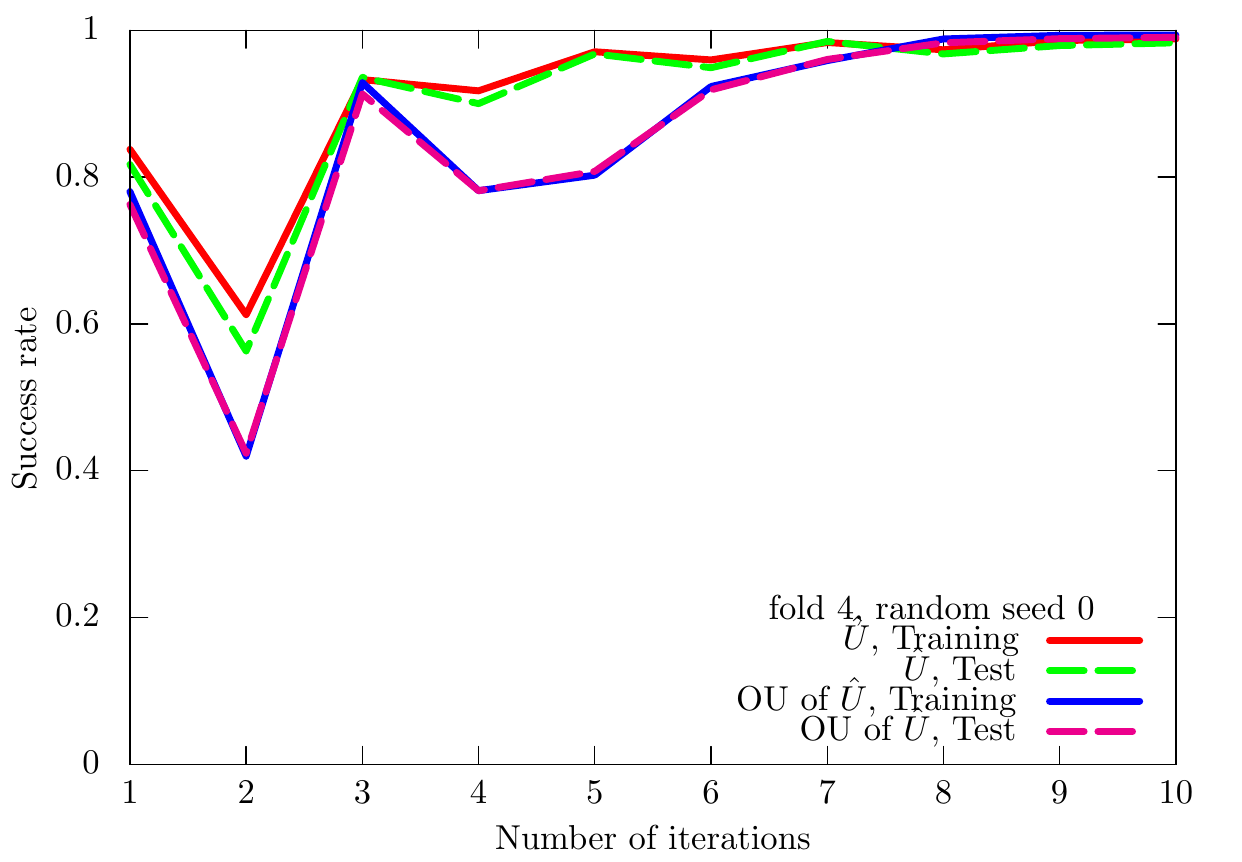}
\includegraphics[scale=0.25]{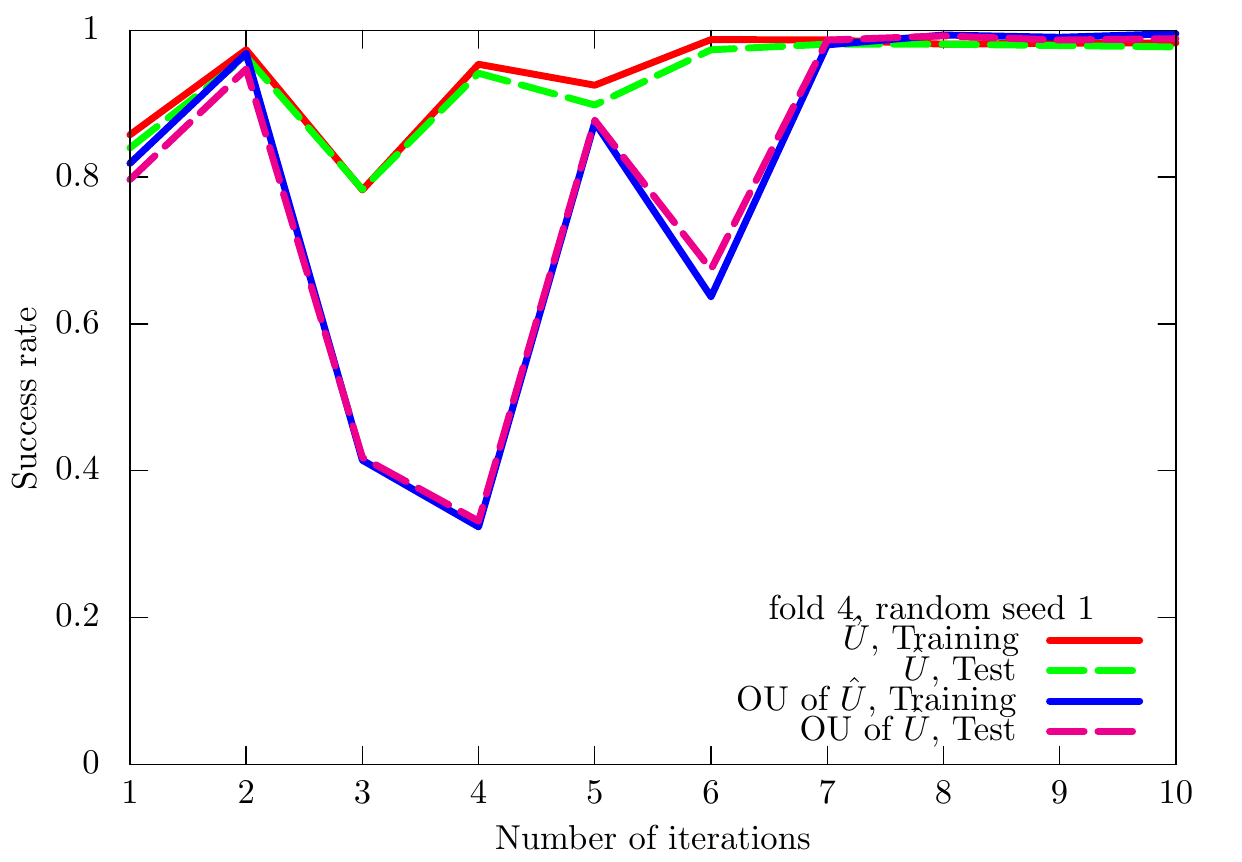}
\includegraphics[scale=0.25]{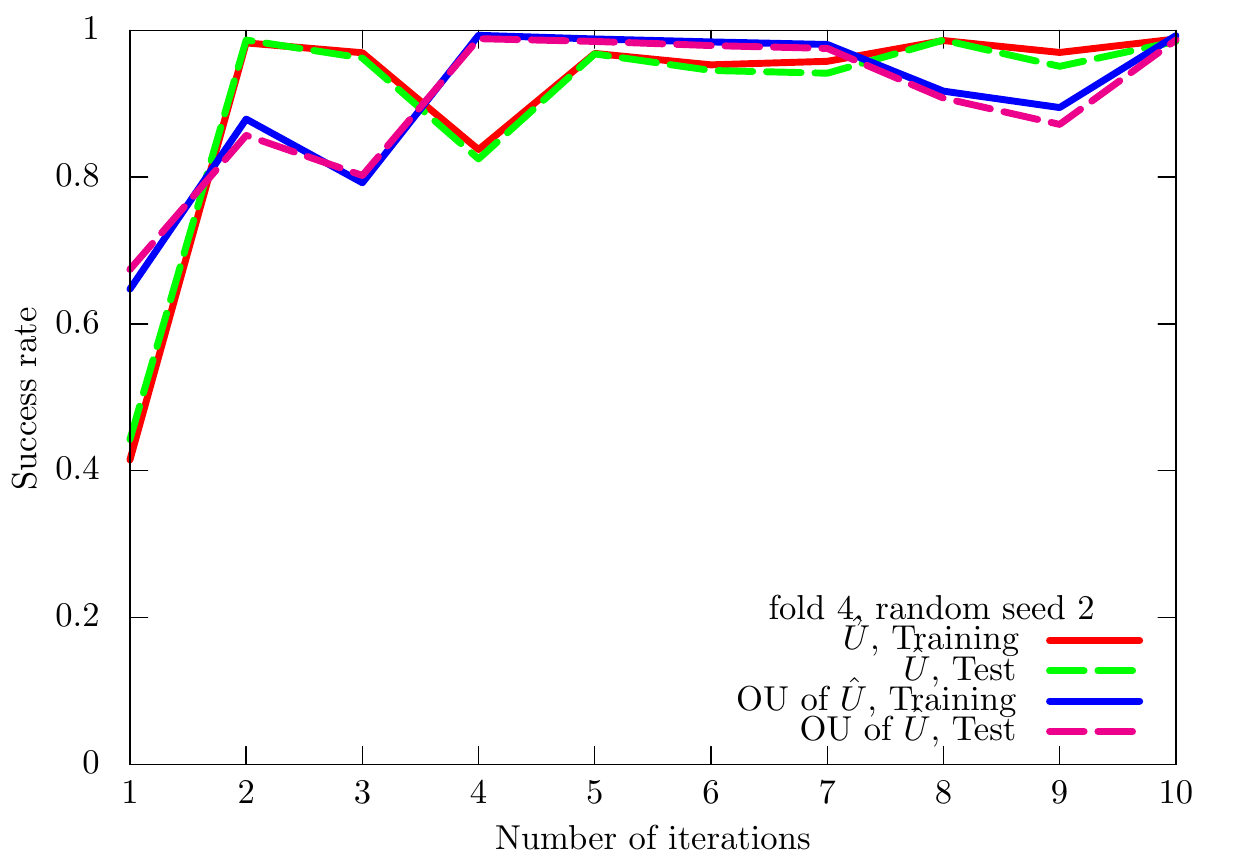}
\includegraphics[scale=0.25]{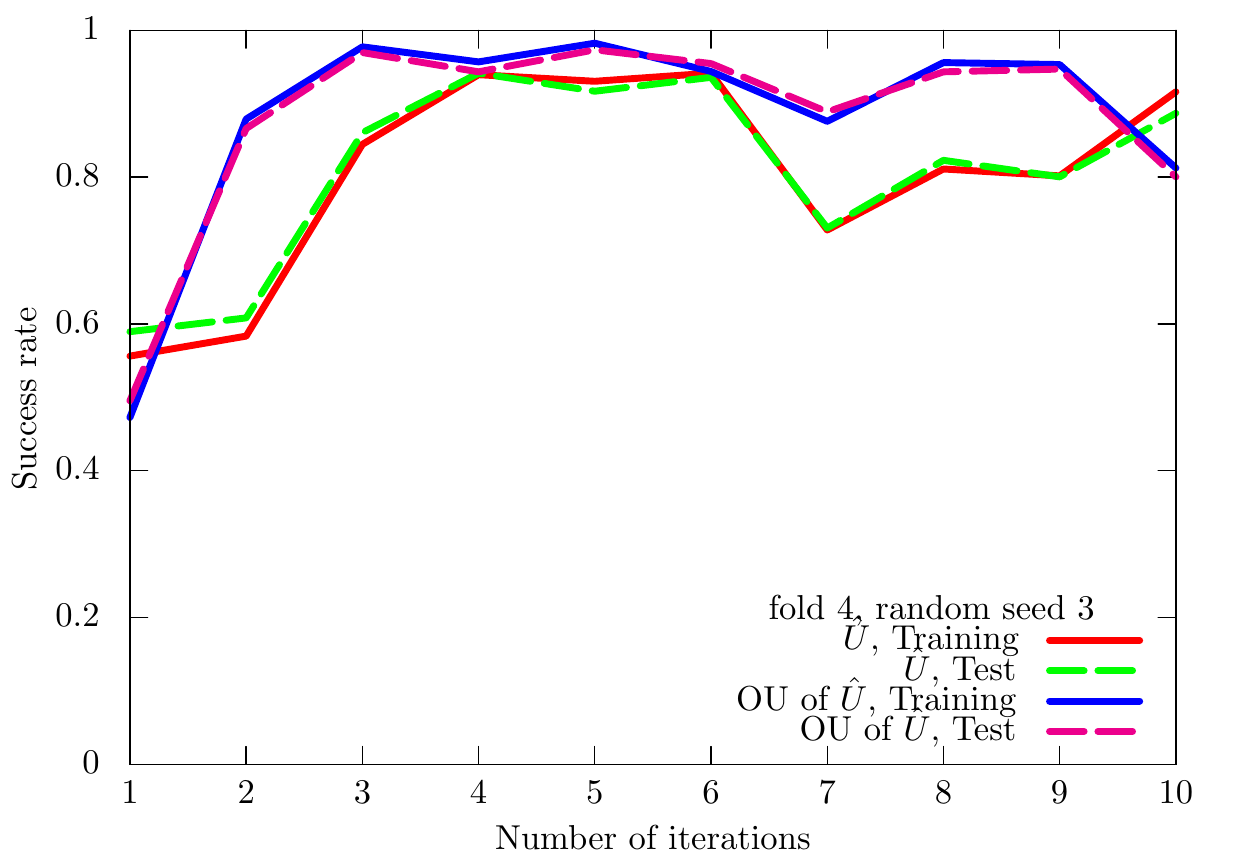}
\includegraphics[scale=0.25]{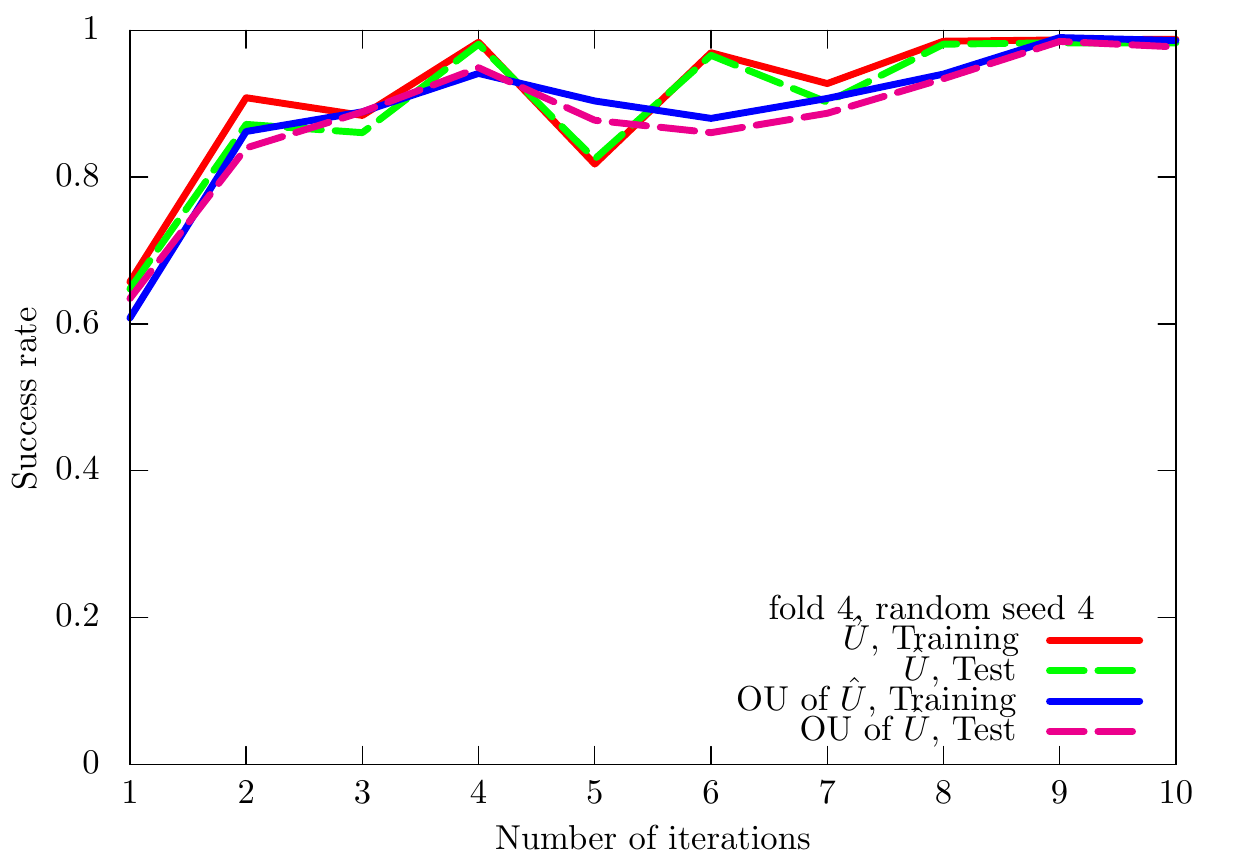}
\caption{Results of the UKM ($\hat{X}$ and $\hat{P}$) on the $5$-fold datasets with $5$ different random seeds for the MNIST256 dataset ($0$ or non-$0$). We use complex matrices and set $\theta_\mathrm{bias} = 0$. We set $r = 0.010$.}
\label{supp-arXiv-numerical-result-raw-data-fold-001-rand-001-UKM-P-MNIST256-0-non0}
\end{figure*}

\begin{figure*}[htb]
\centering
\includegraphics[scale=0.25]{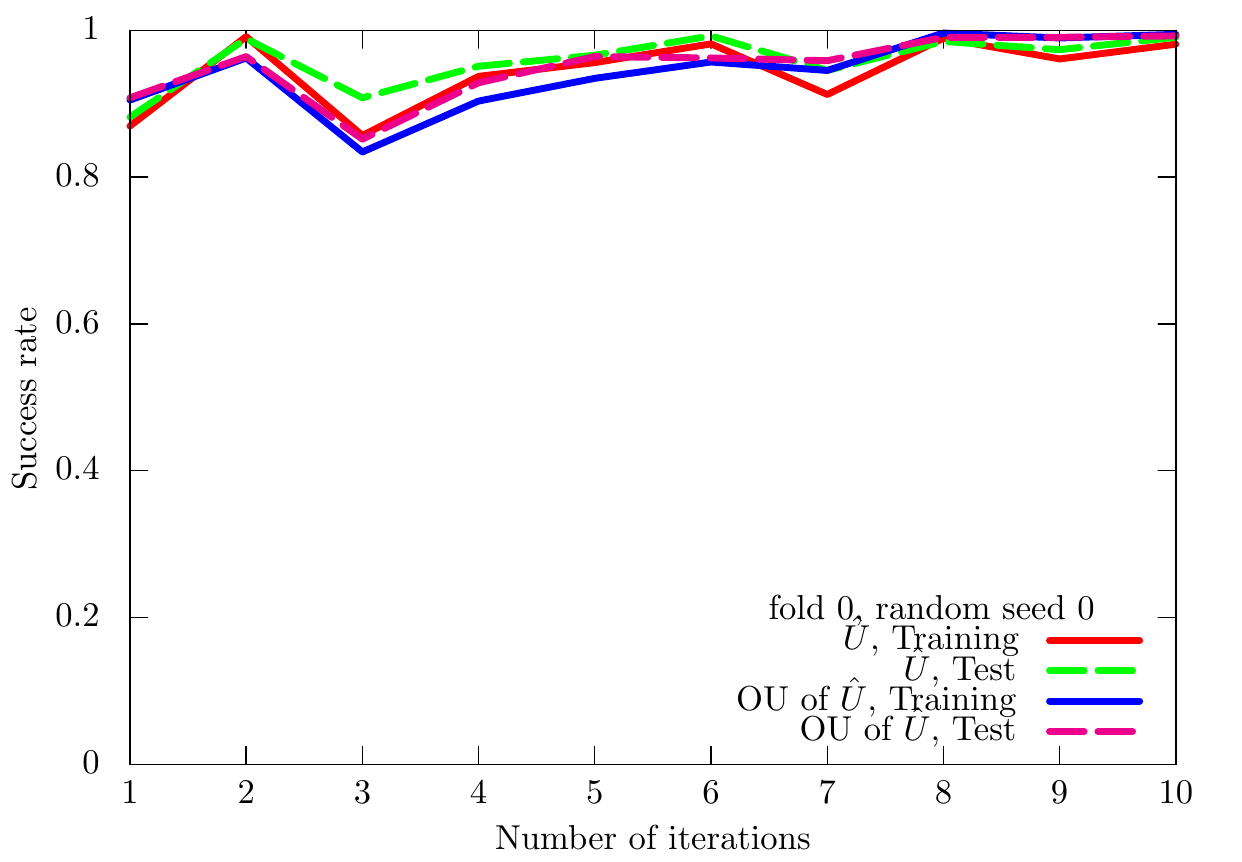}
\includegraphics[scale=0.25]{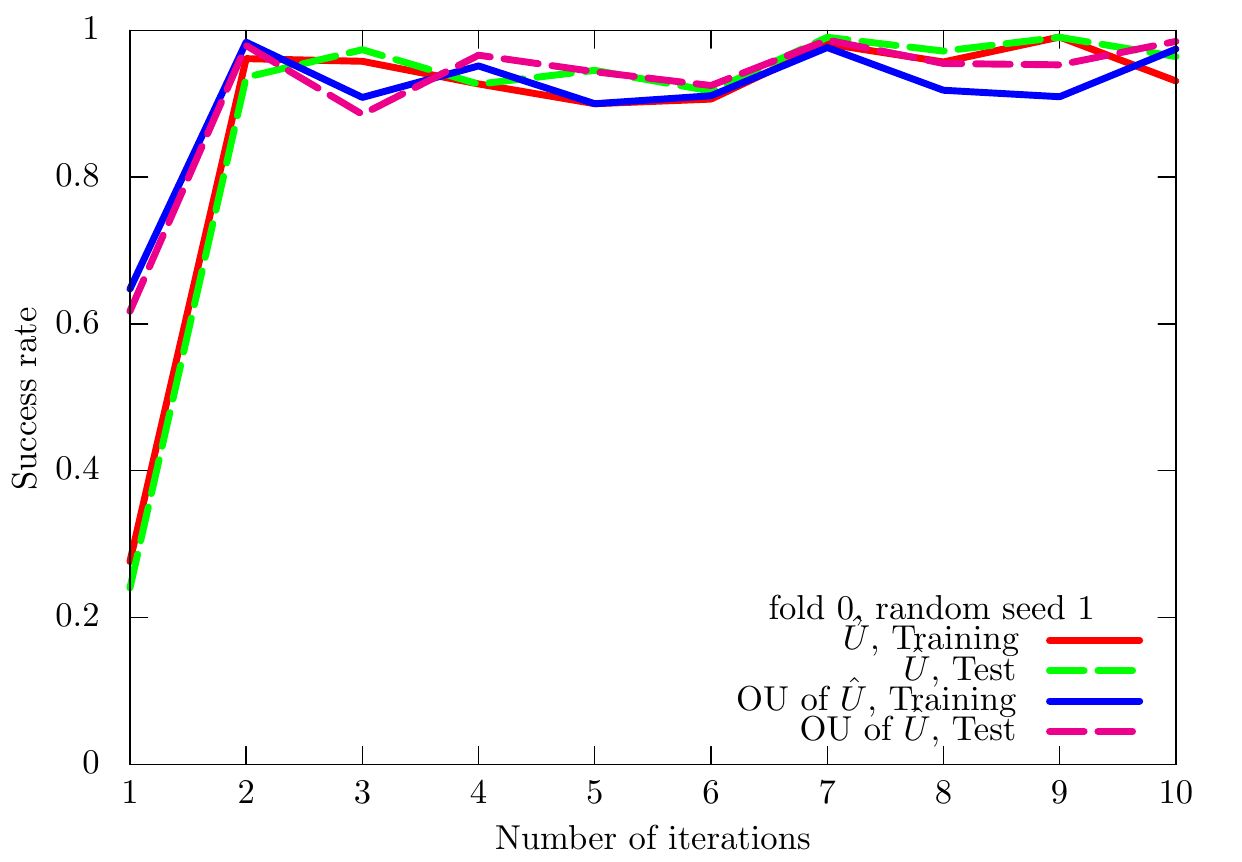}
\includegraphics[scale=0.25]{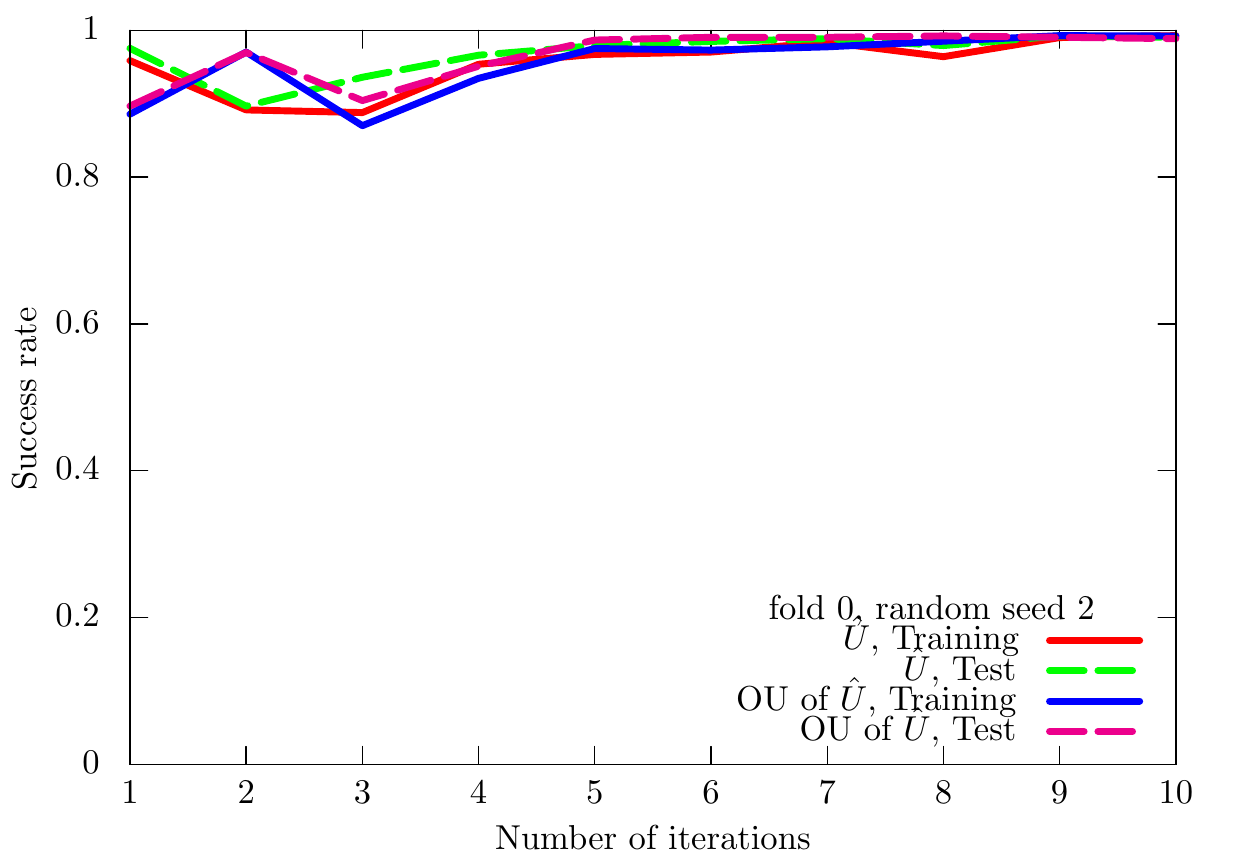}
\includegraphics[scale=0.25]{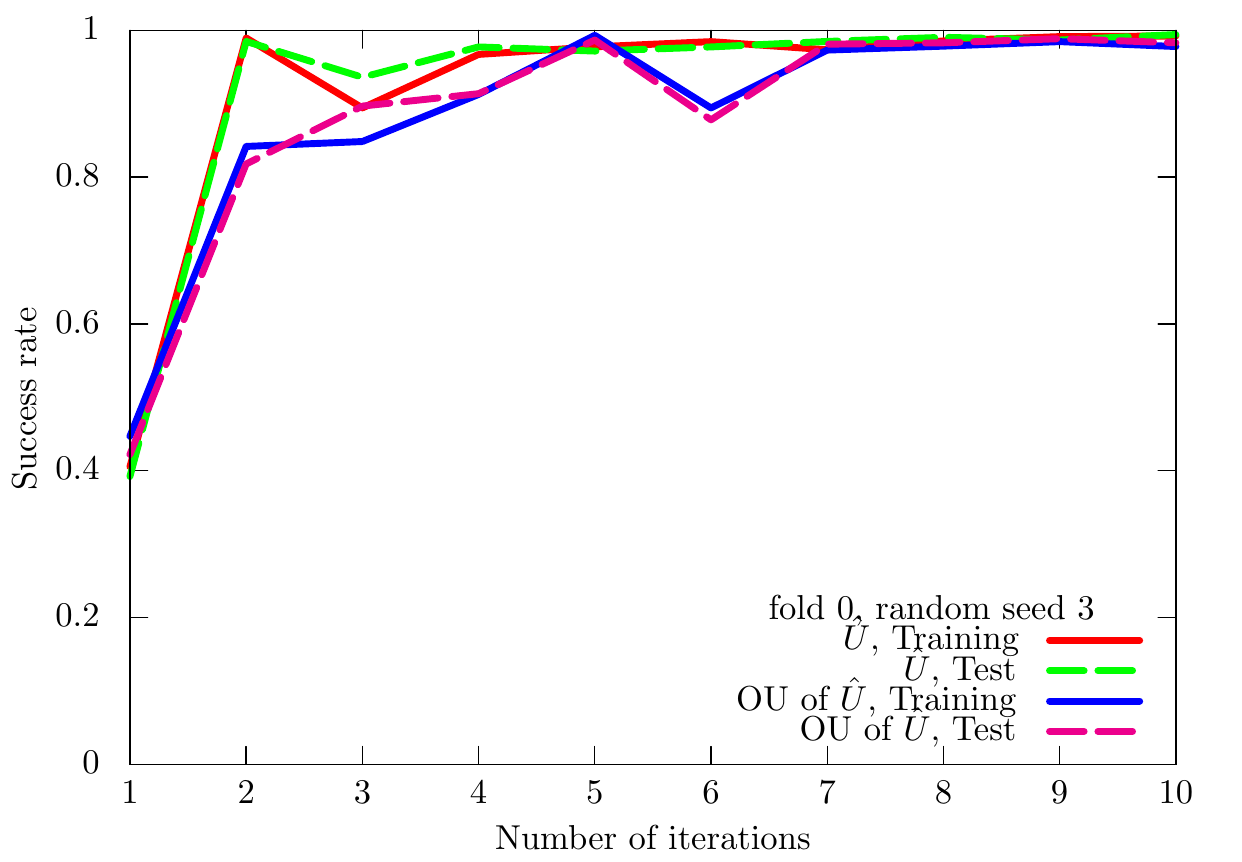}
\includegraphics[scale=0.25]{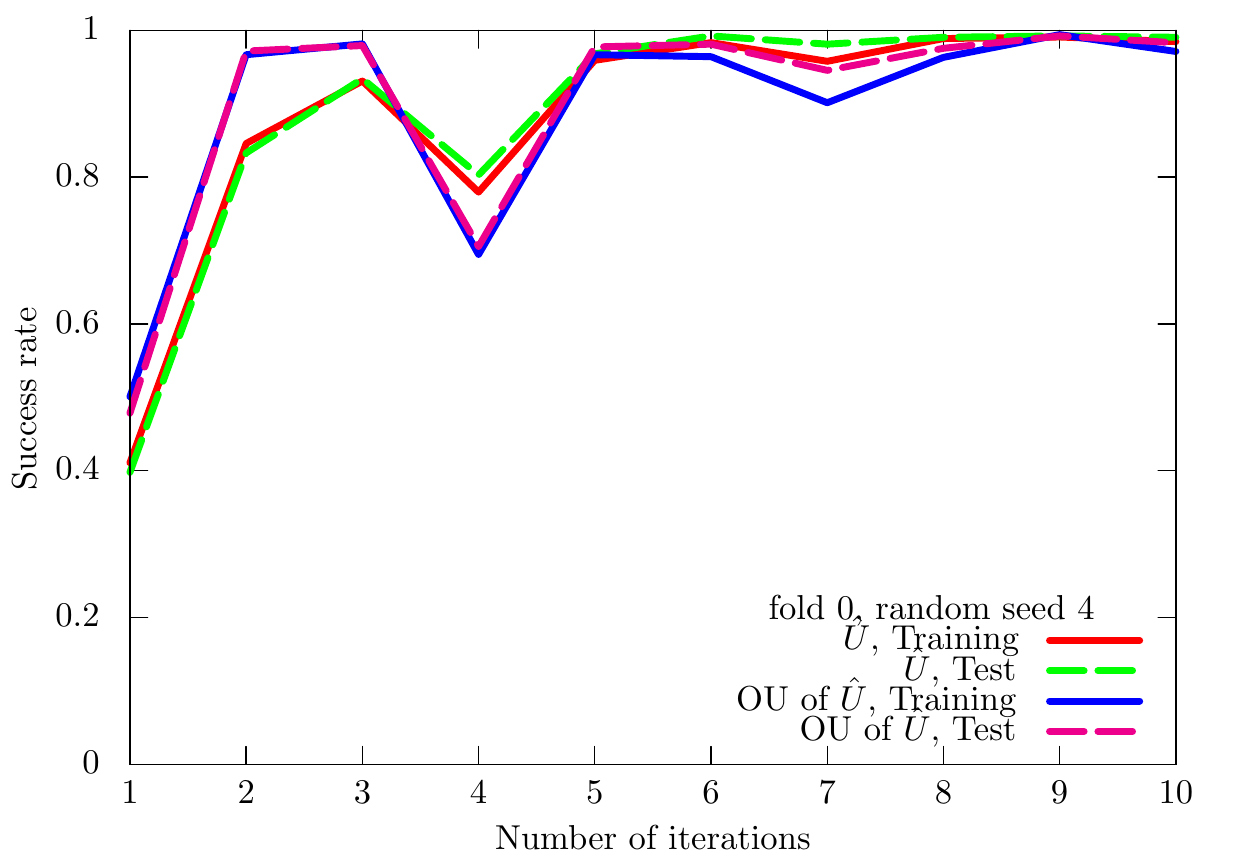}
\includegraphics[scale=0.25]{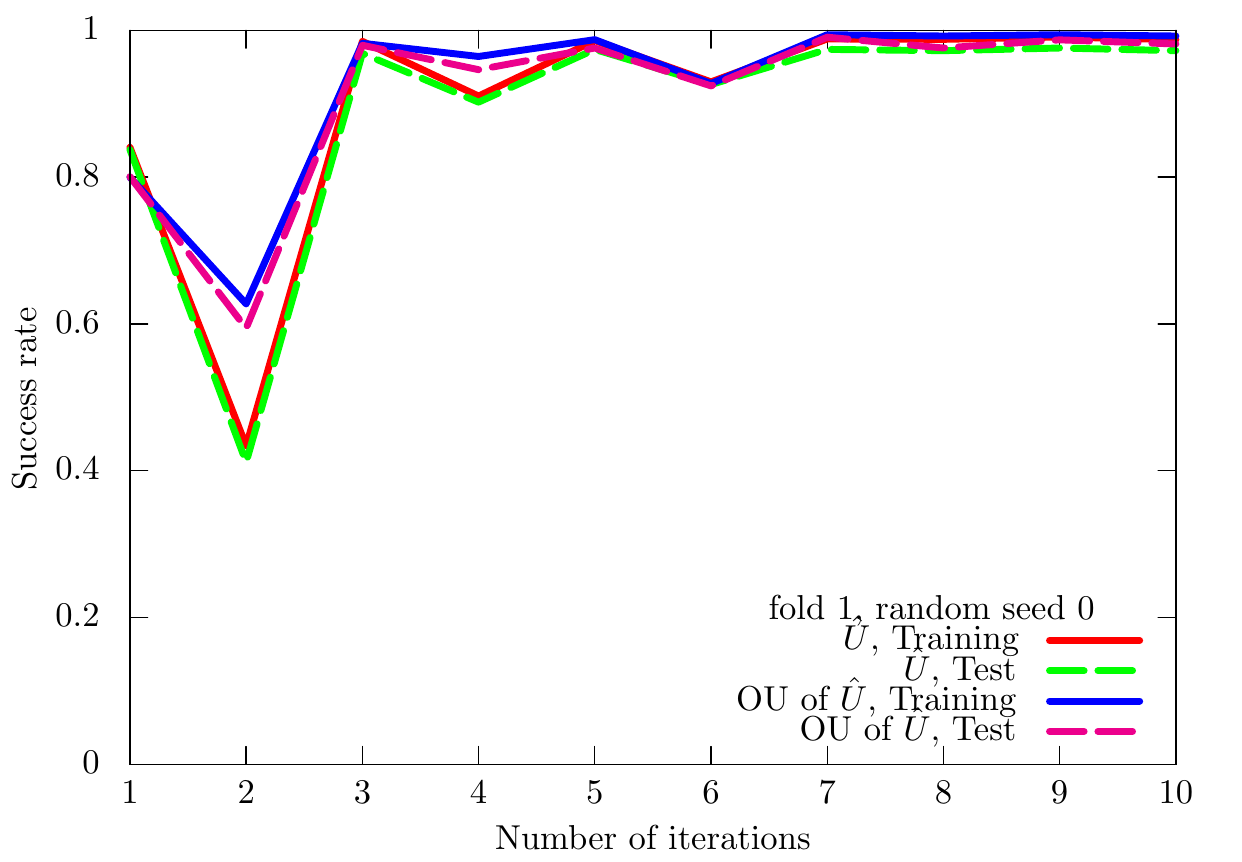}
\includegraphics[scale=0.25]{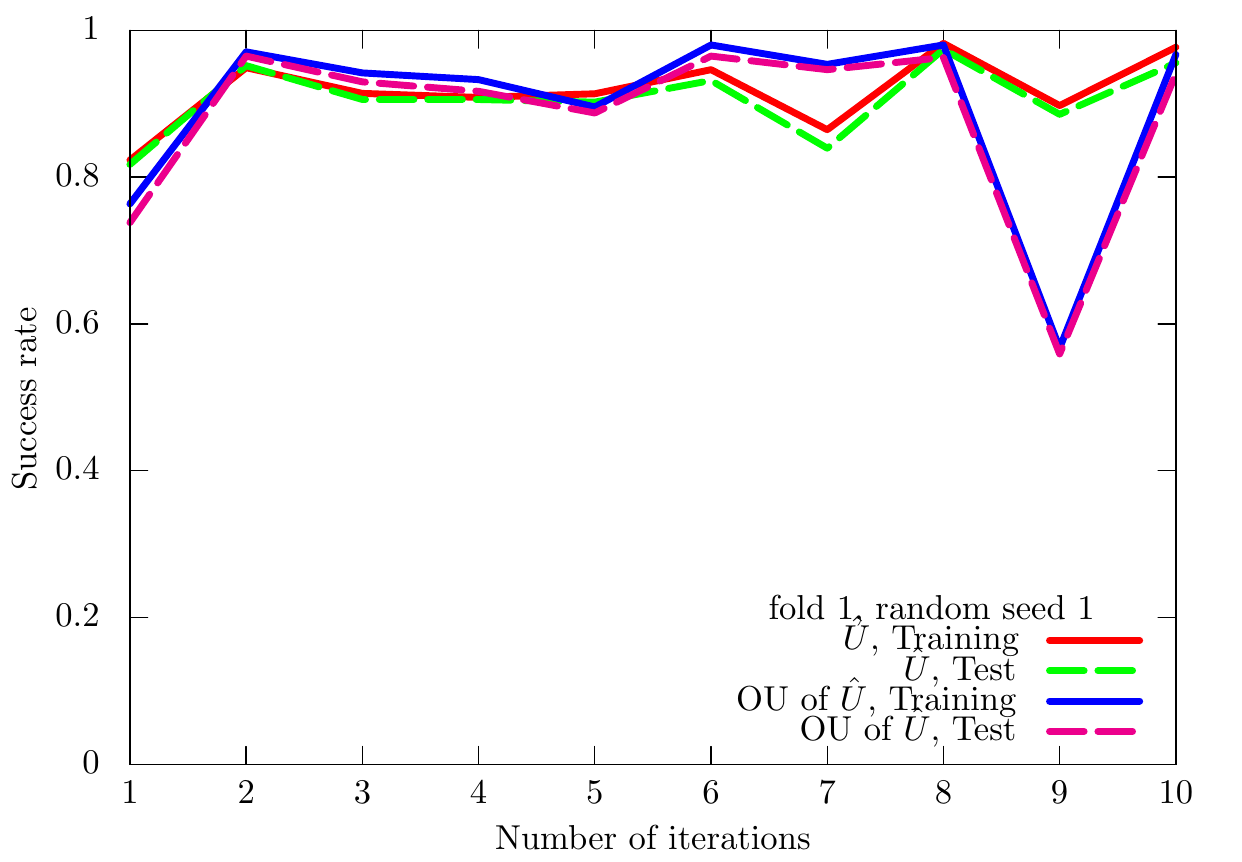}
\includegraphics[scale=0.25]{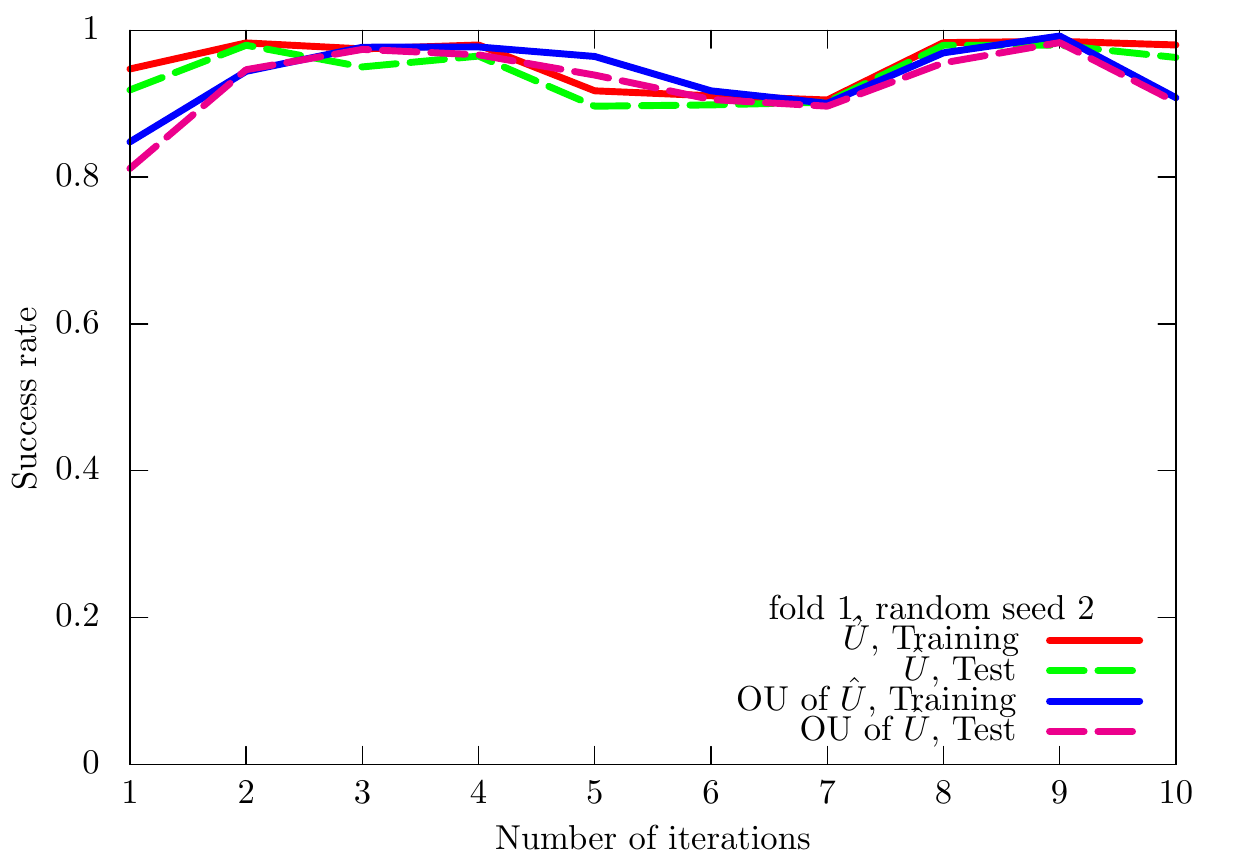}
\includegraphics[scale=0.25]{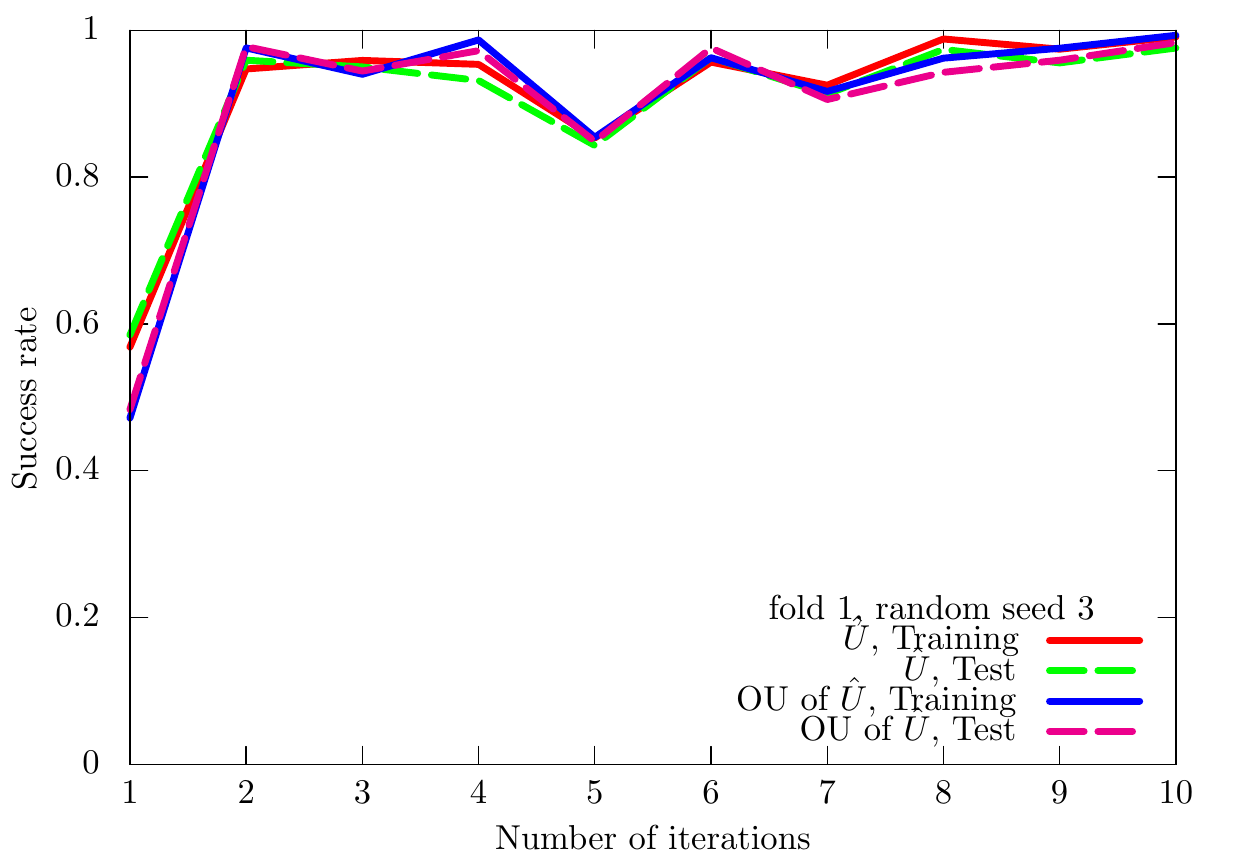}
\includegraphics[scale=0.25]{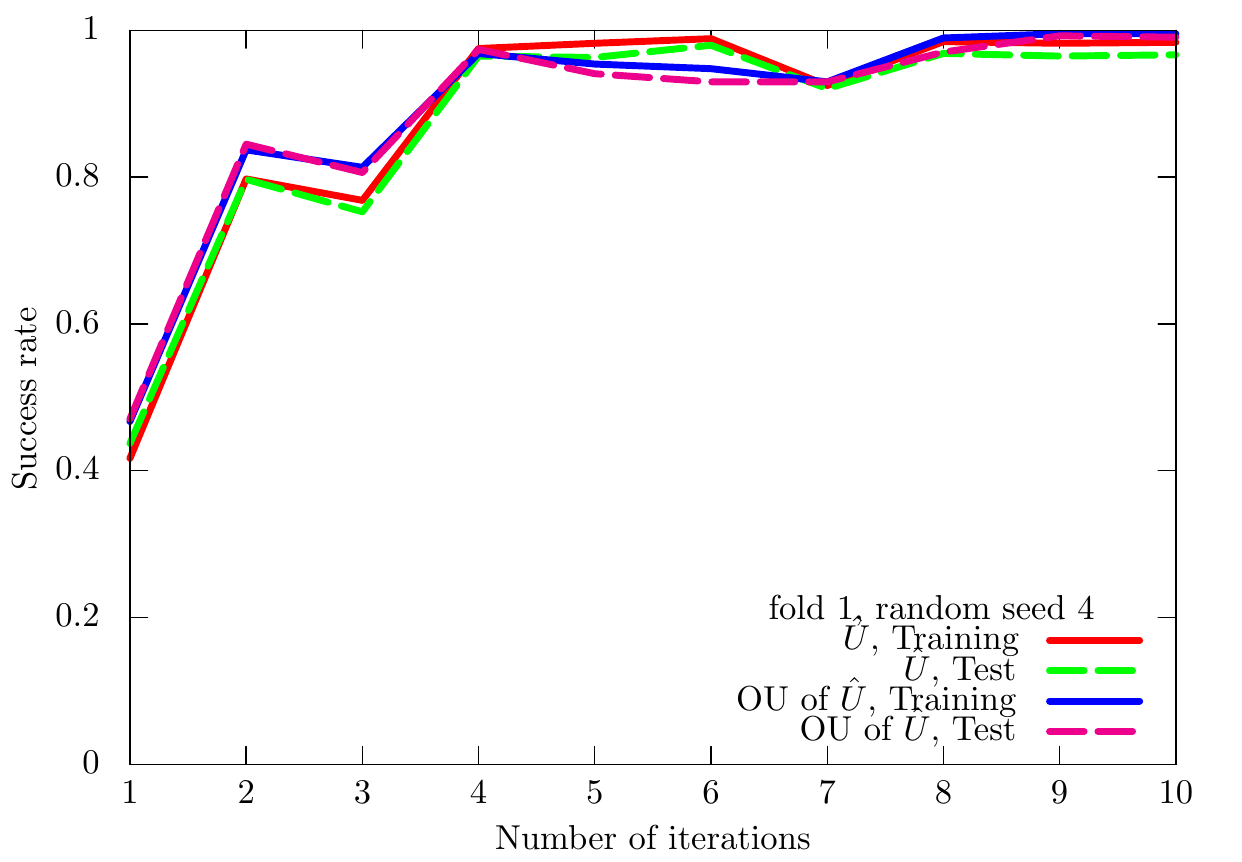}
\includegraphics[scale=0.25]{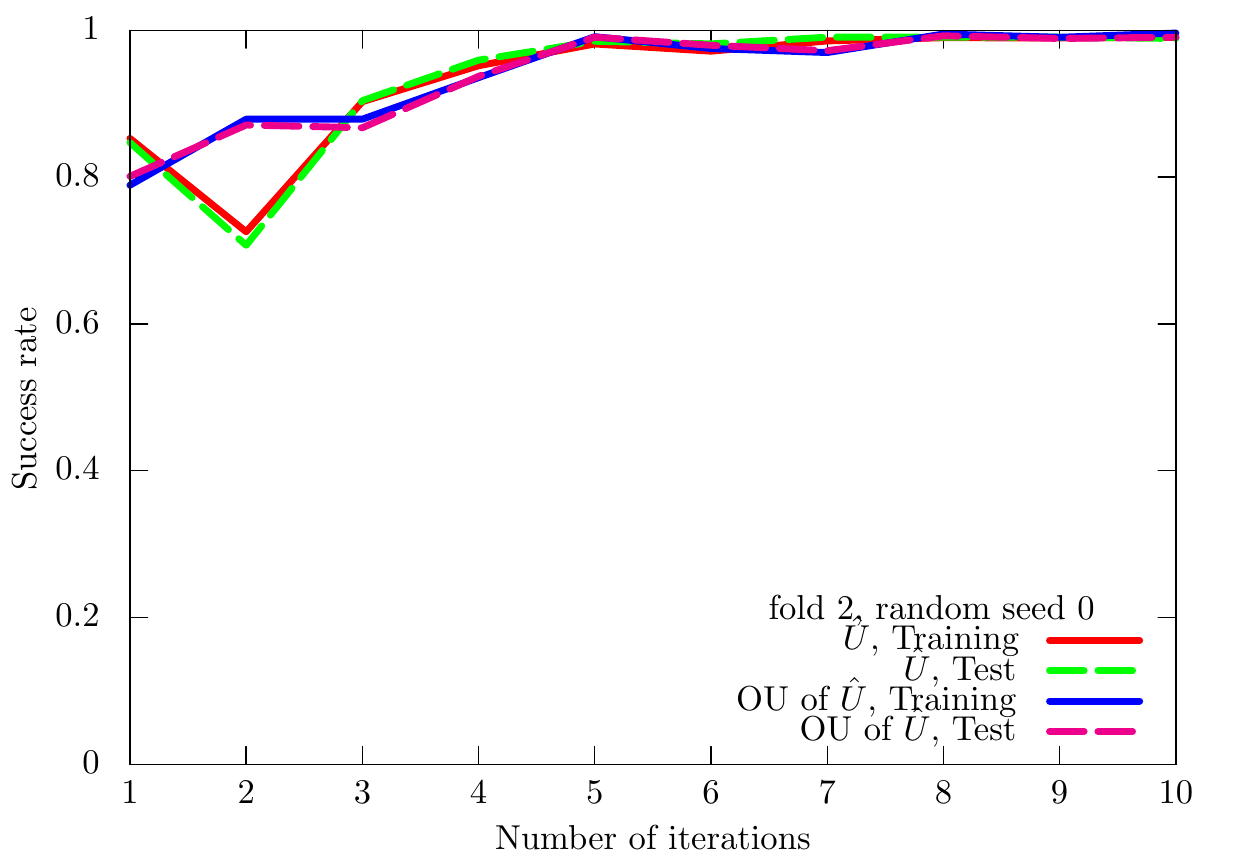}
\includegraphics[scale=0.25]{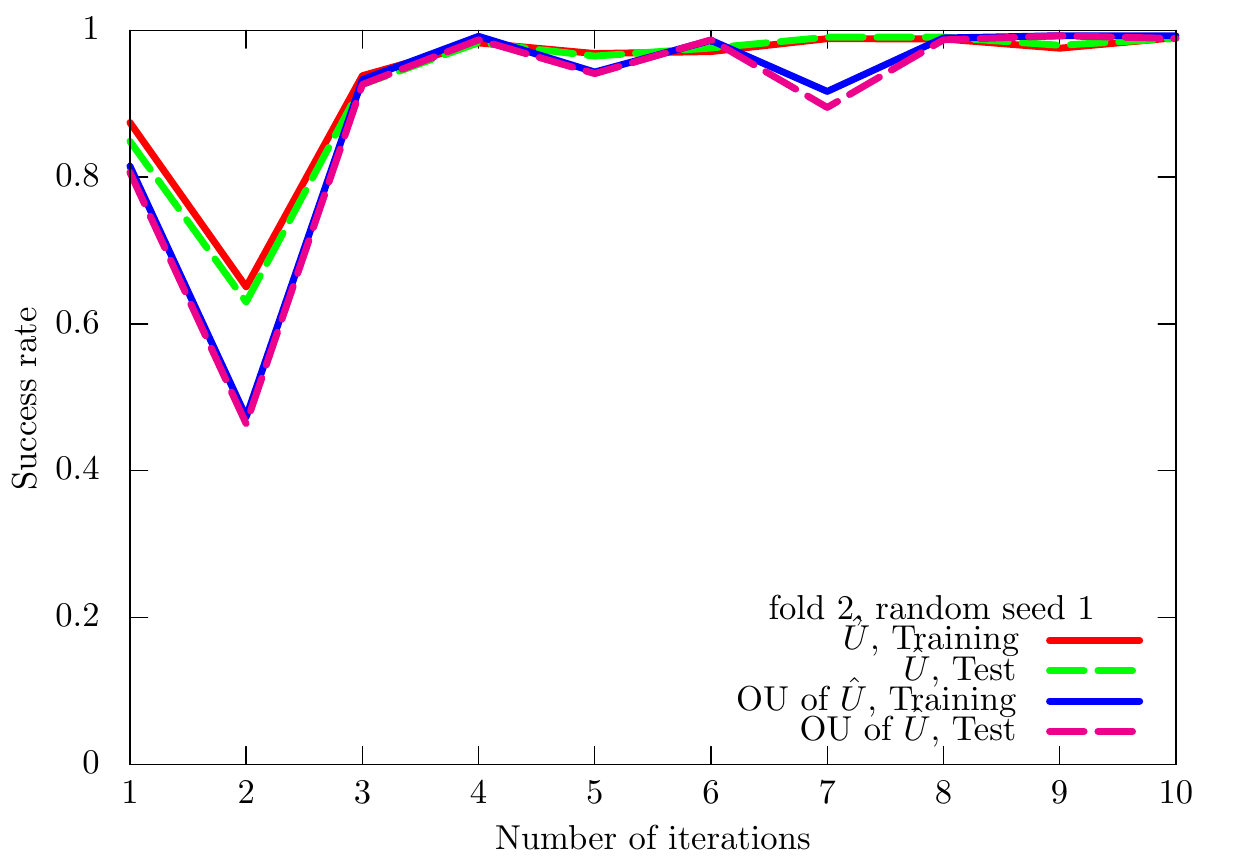}
\includegraphics[scale=0.25]{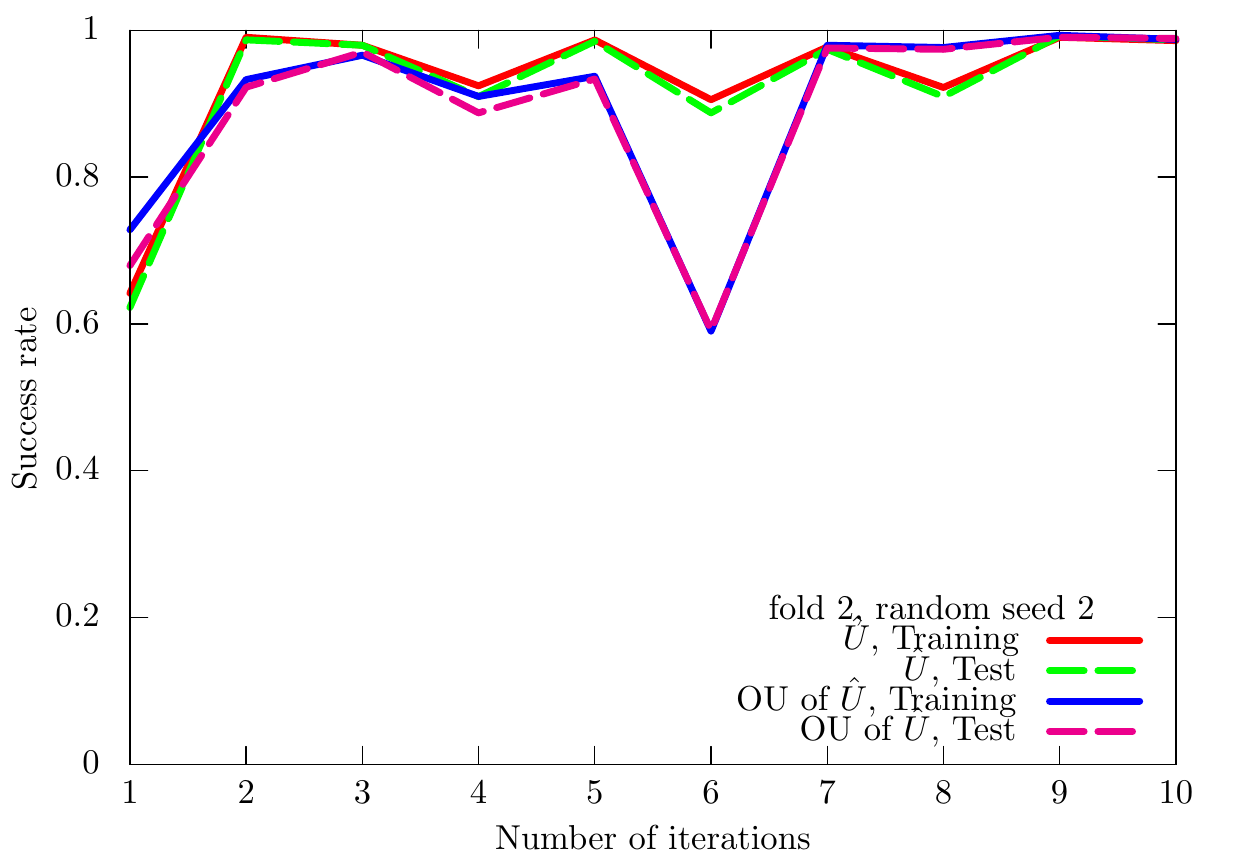}
\includegraphics[scale=0.25]{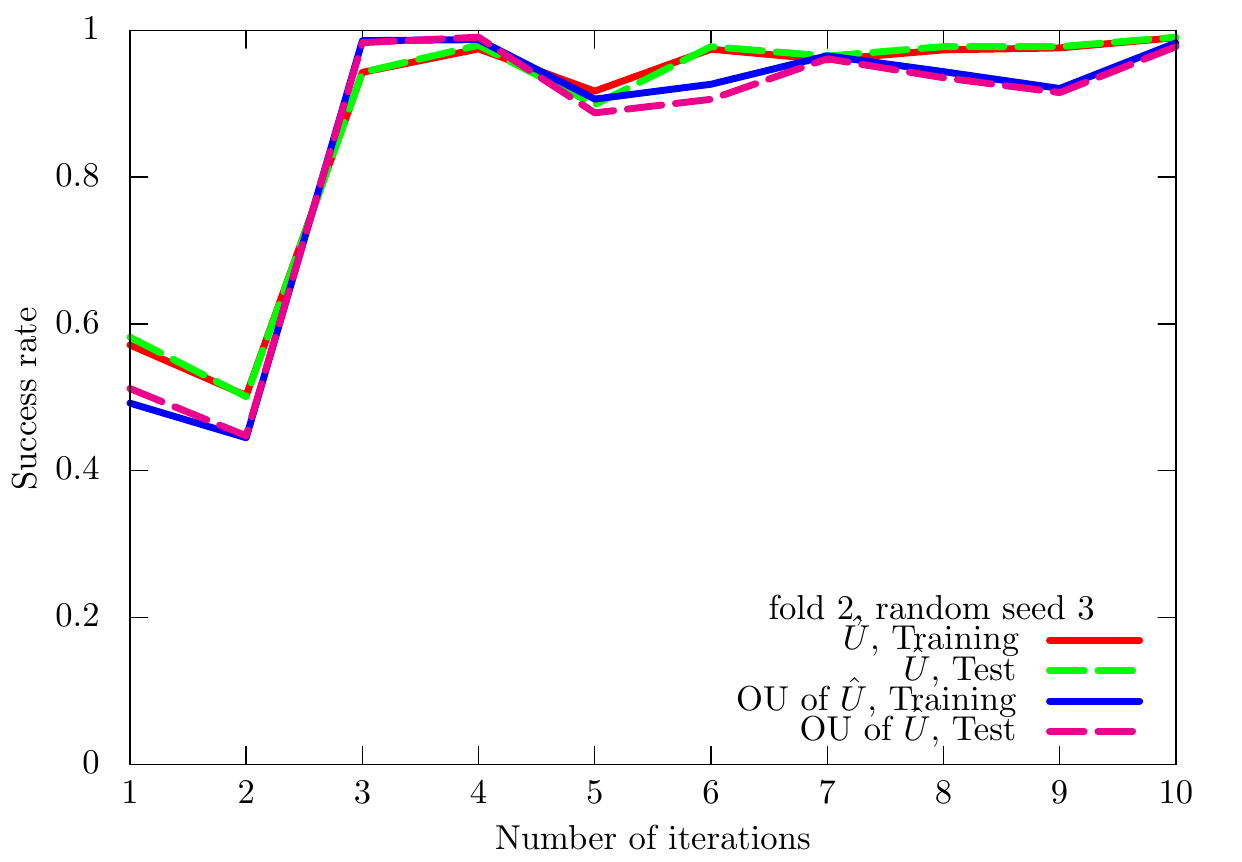}
\includegraphics[scale=0.25]{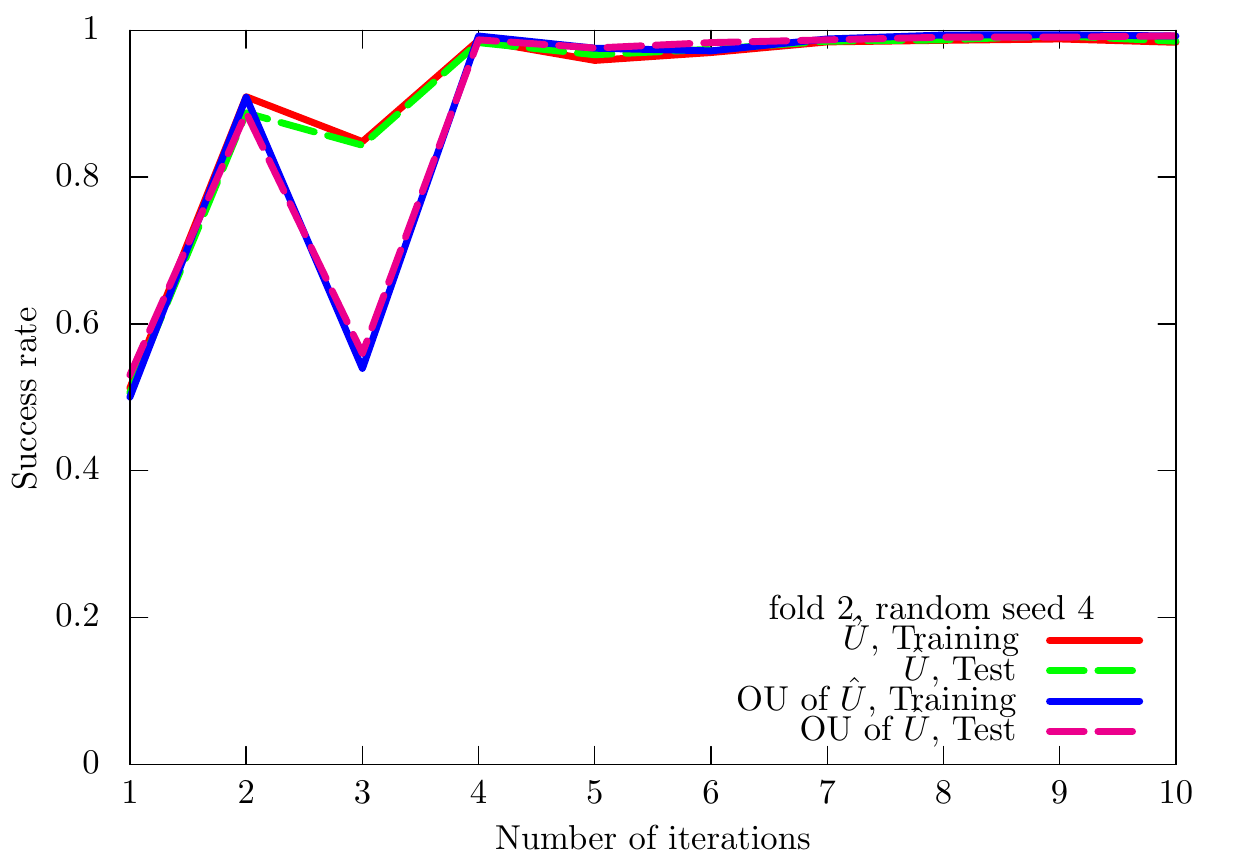}
\includegraphics[scale=0.25]{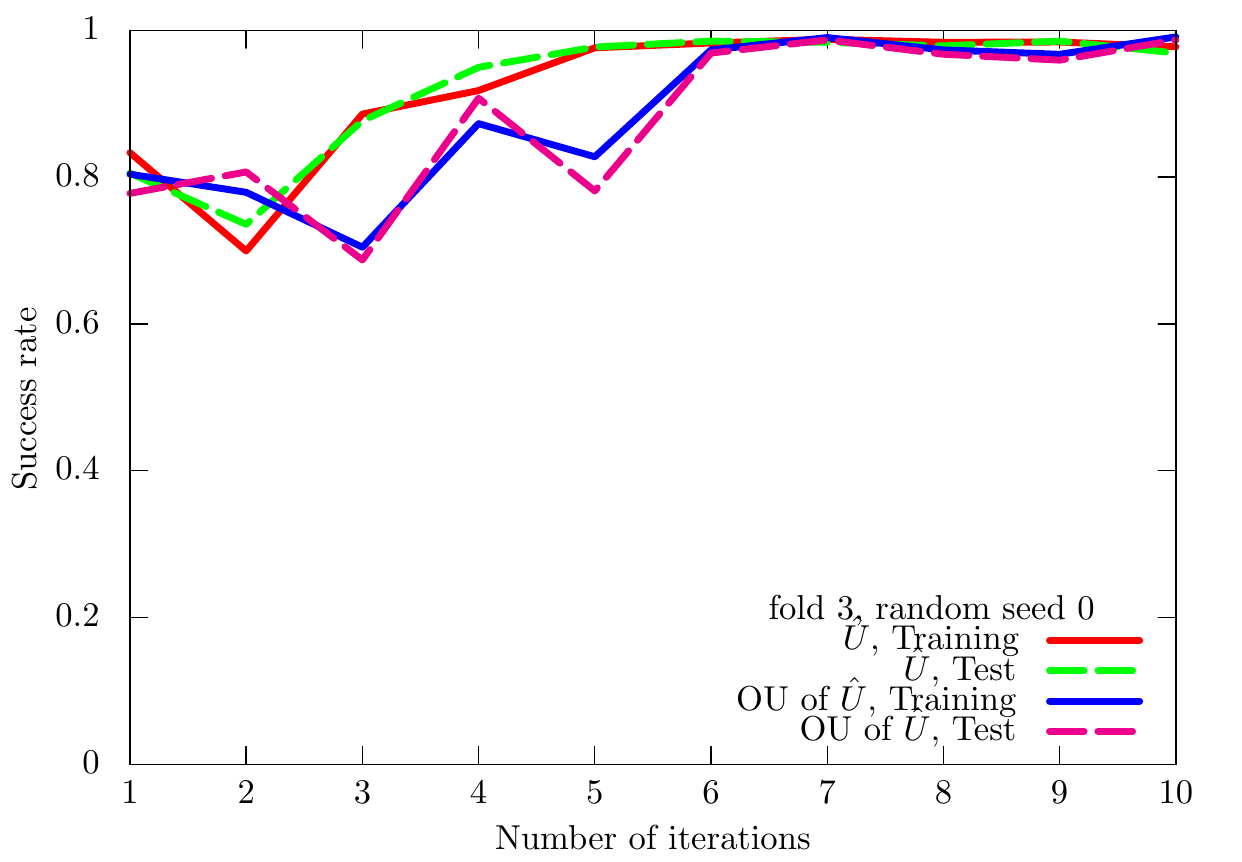}
\includegraphics[scale=0.25]{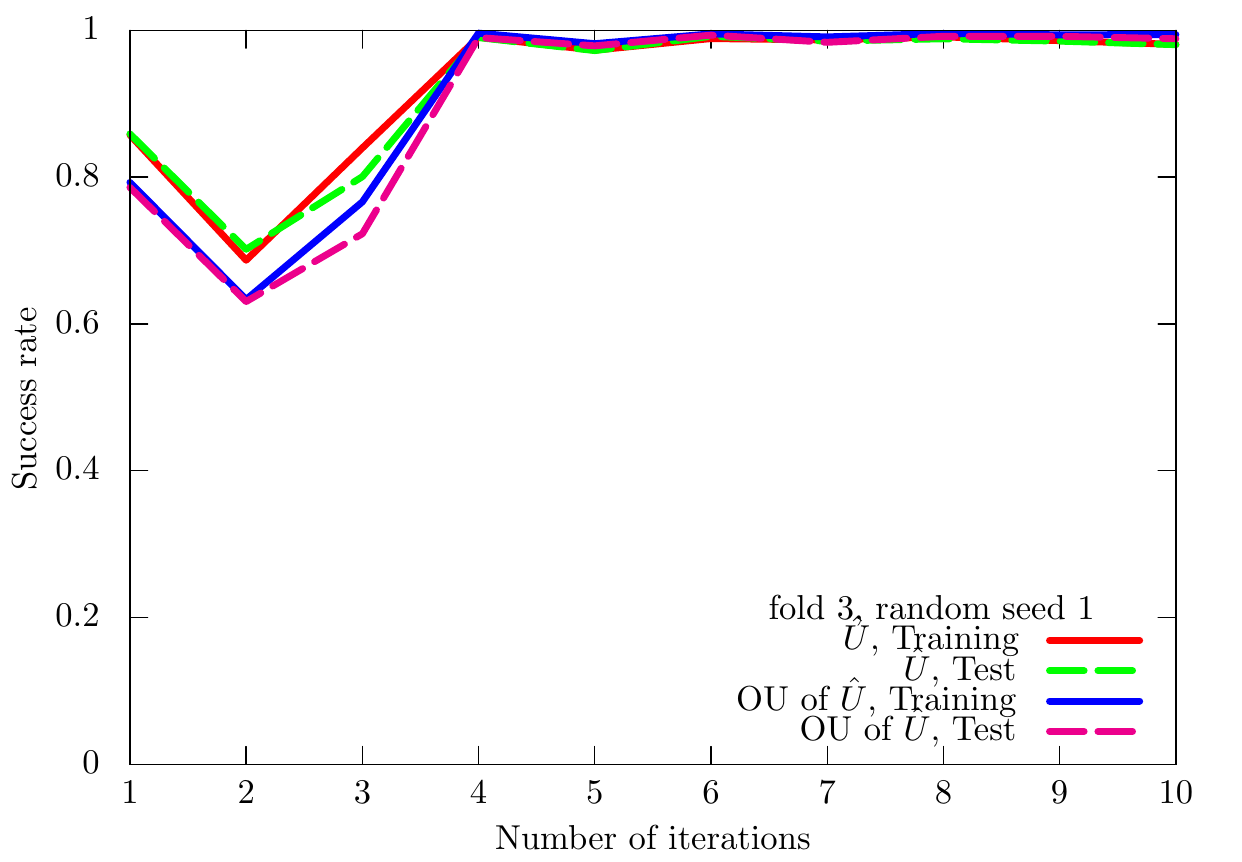}
\includegraphics[scale=0.25]{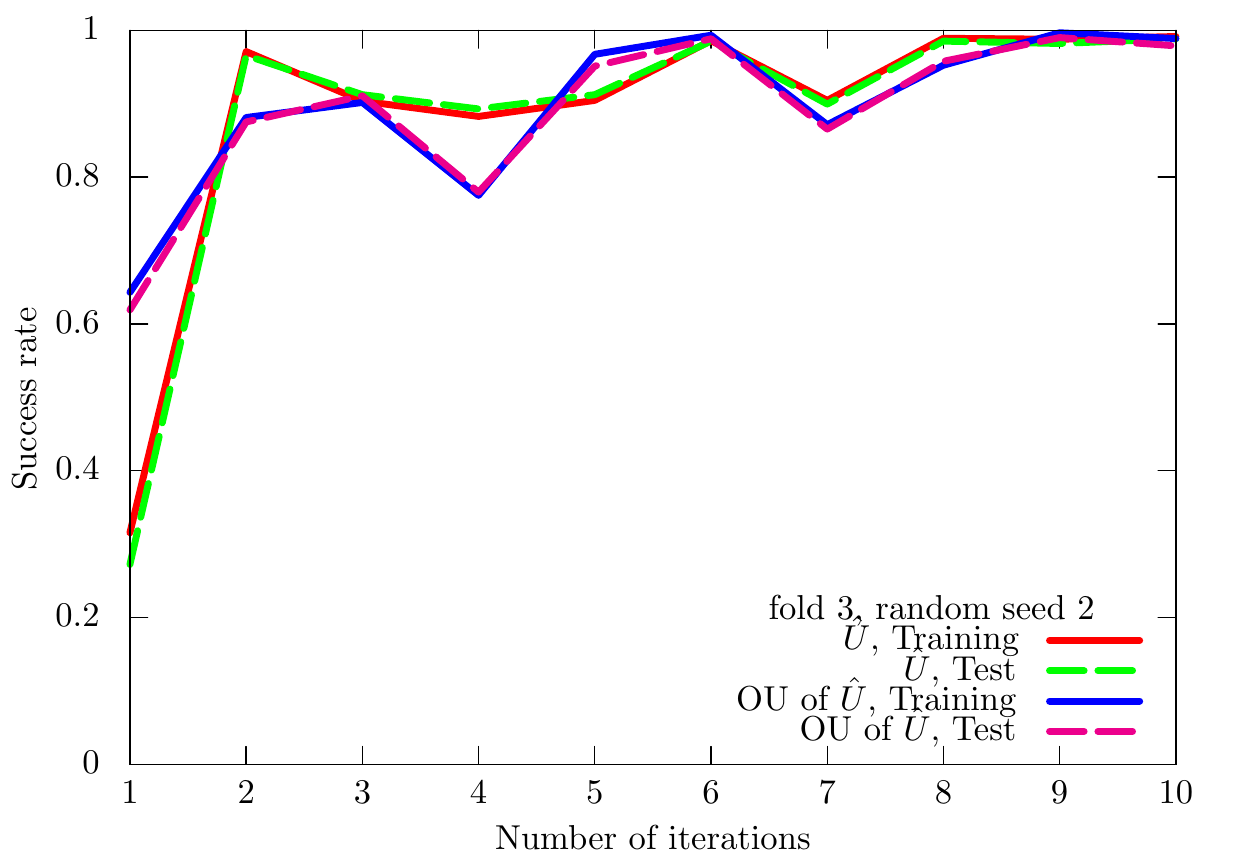}
\includegraphics[scale=0.25]{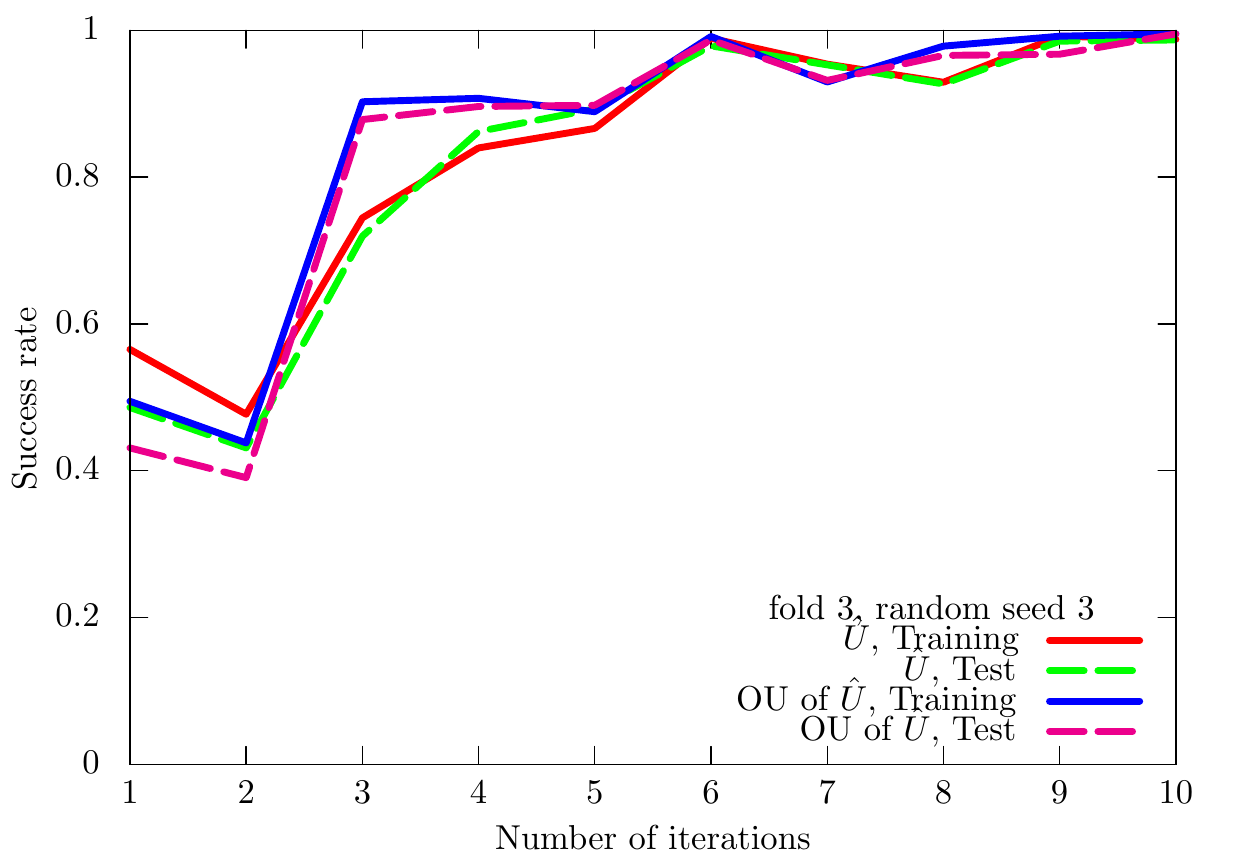}
\includegraphics[scale=0.25]{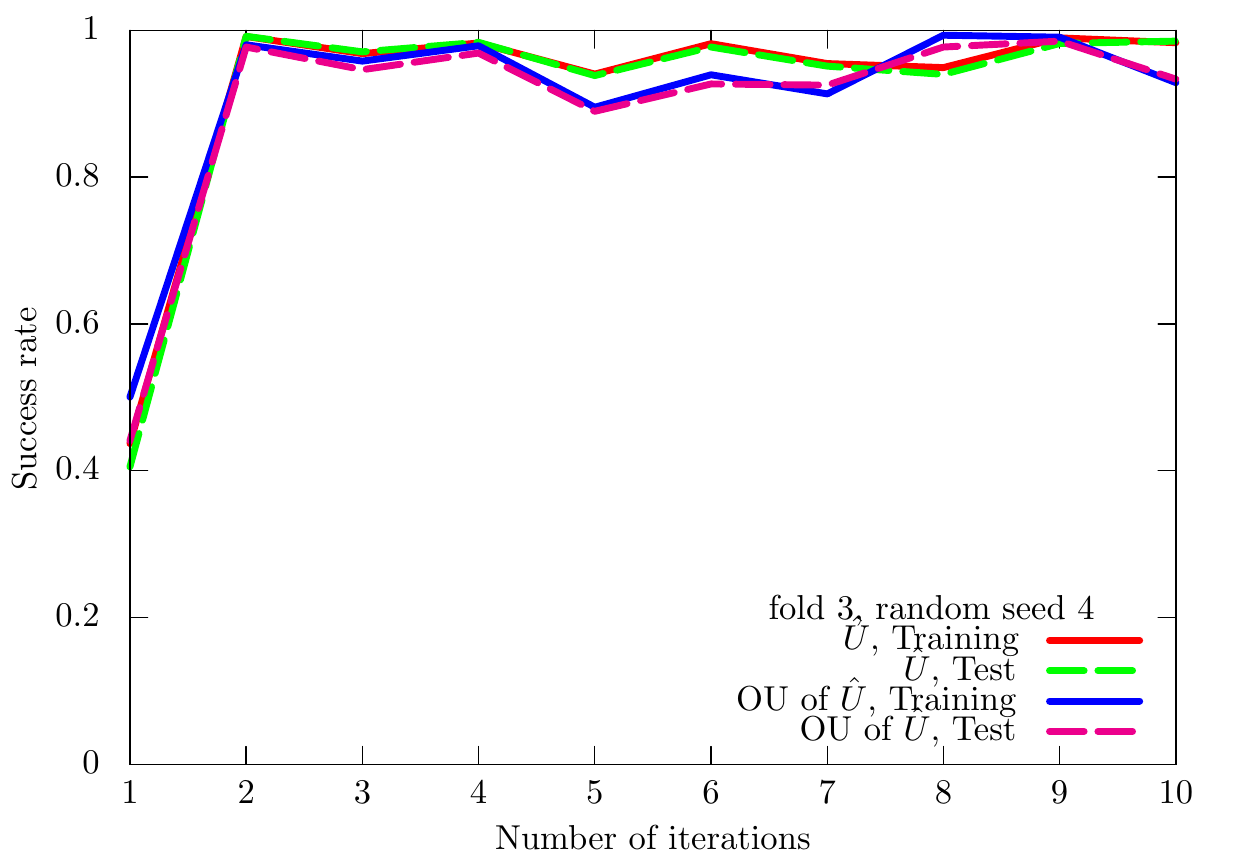}
\includegraphics[scale=0.25]{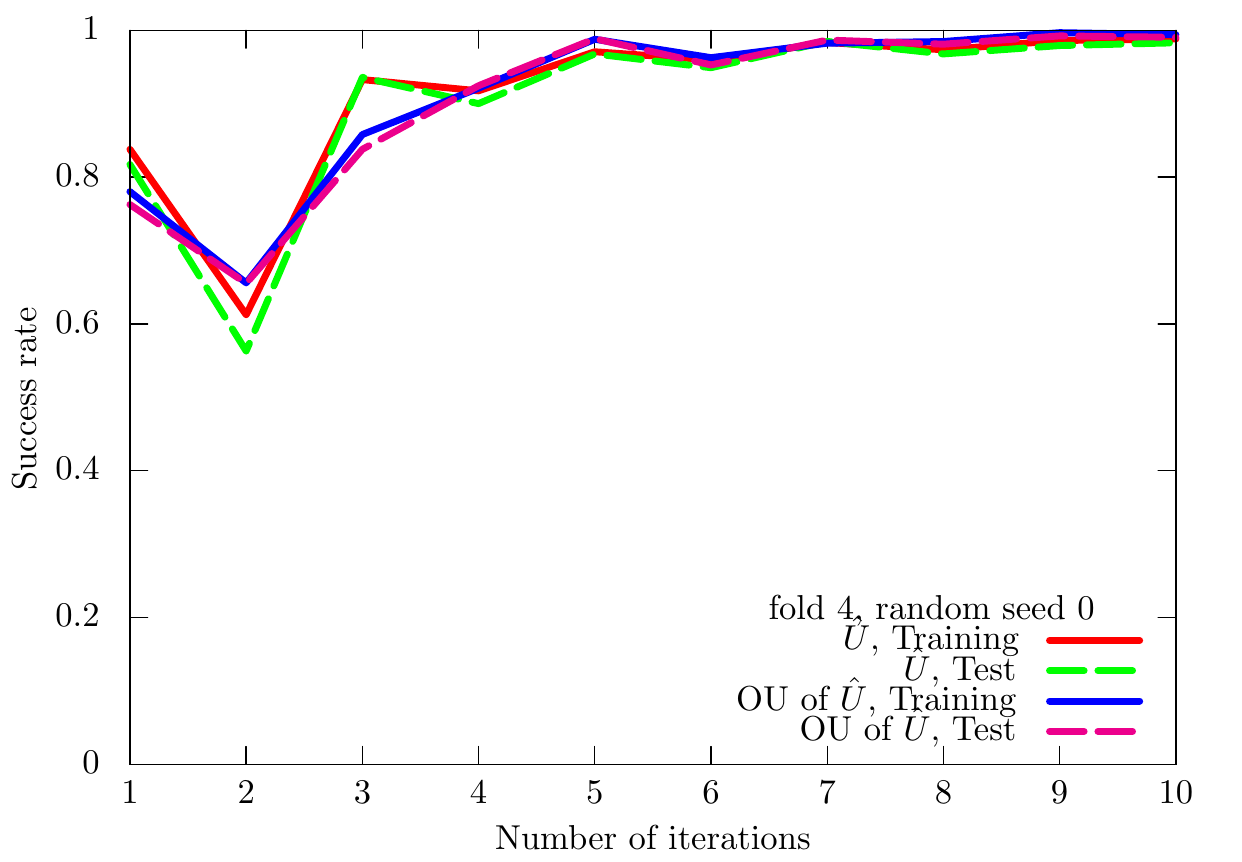}
\includegraphics[scale=0.25]{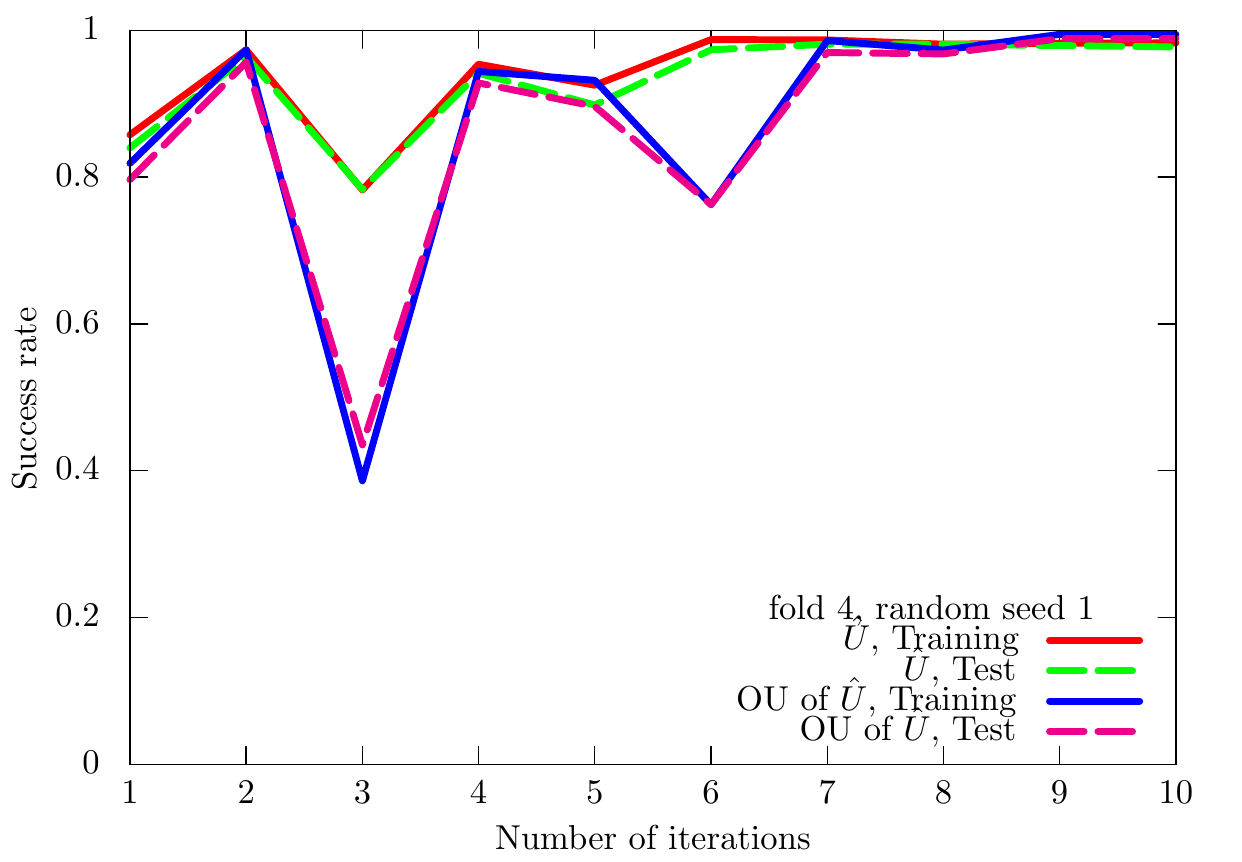}
\includegraphics[scale=0.25]{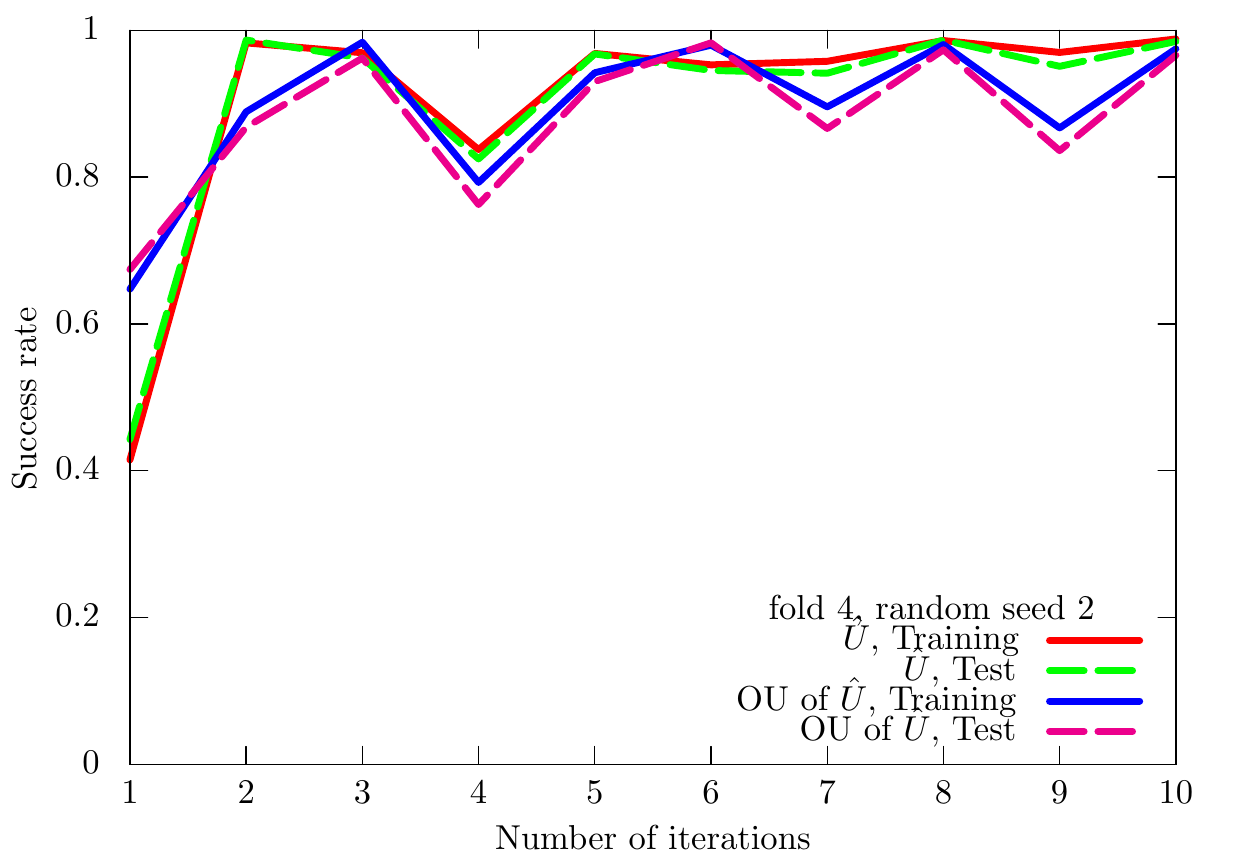}
\includegraphics[scale=0.25]{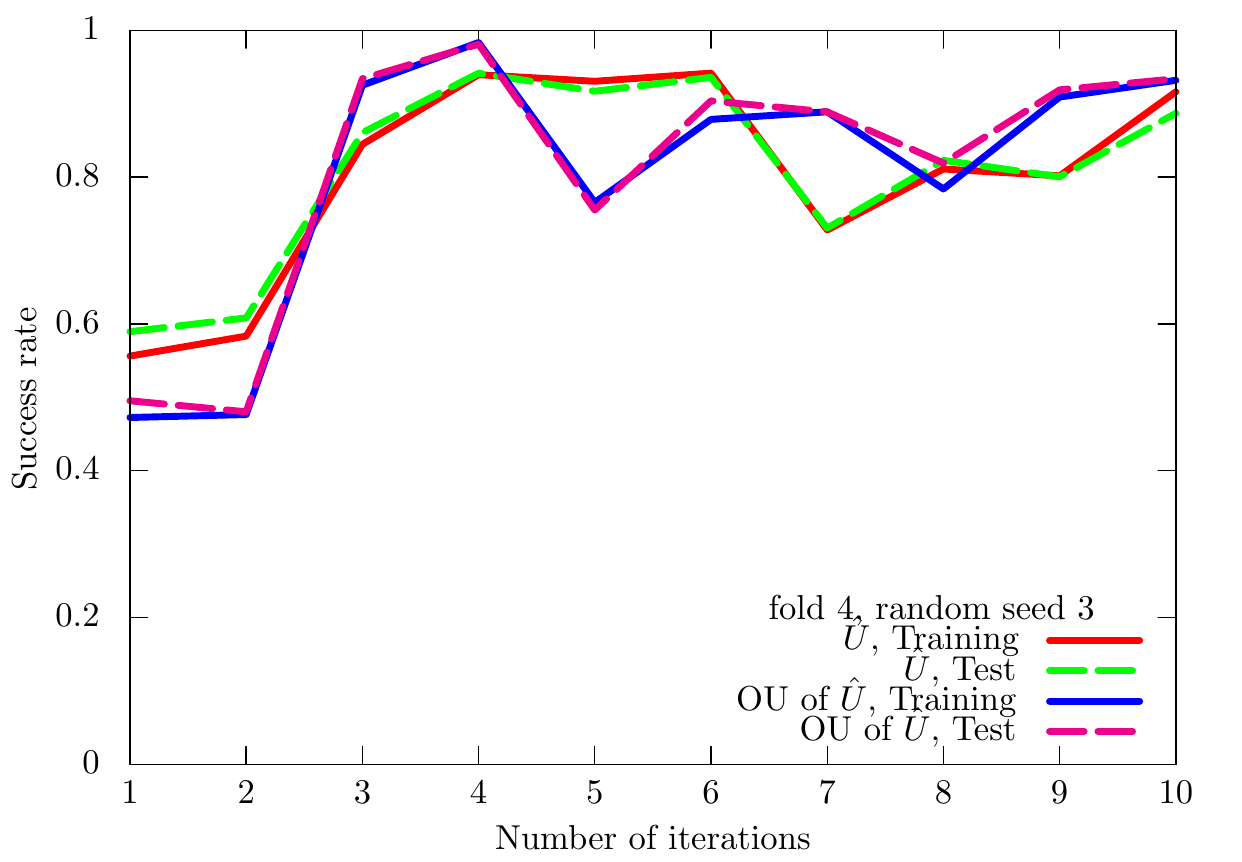}
\includegraphics[scale=0.25]{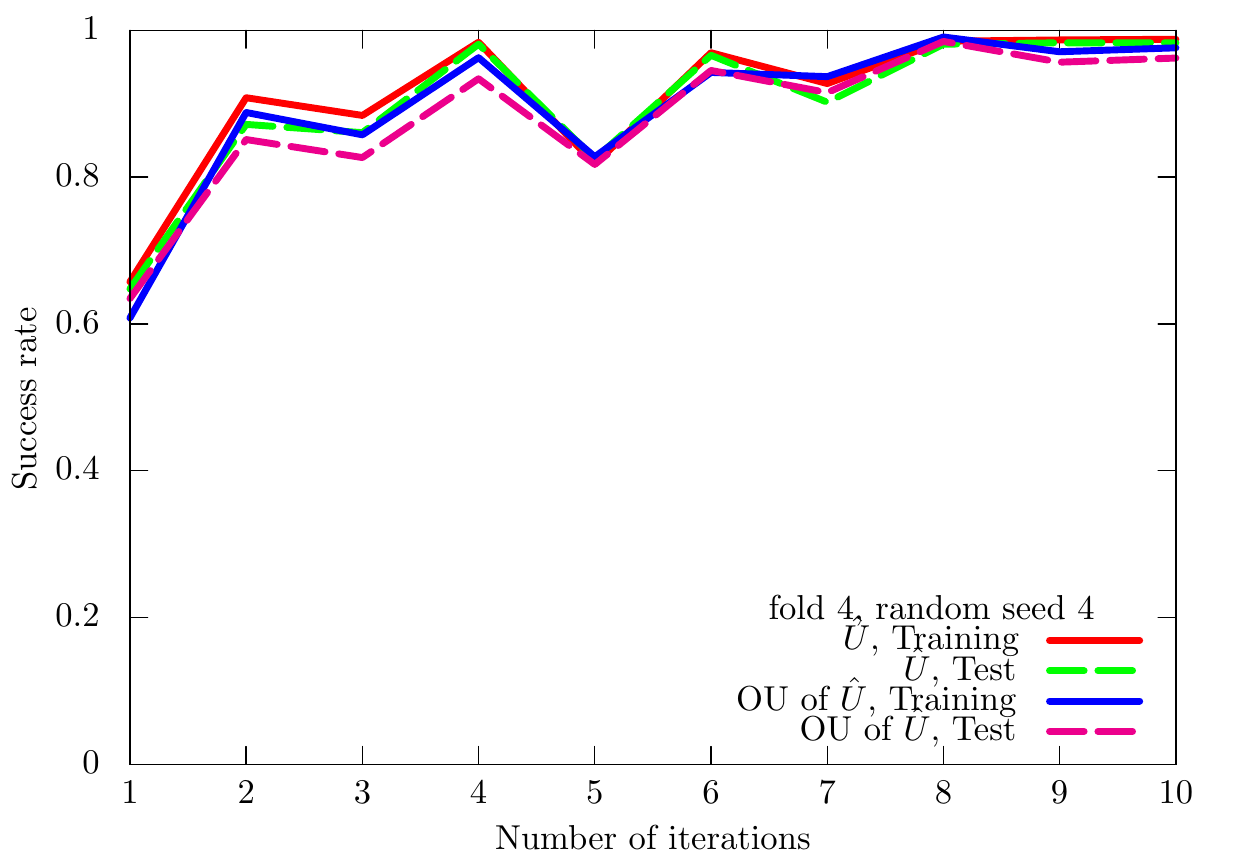}
\caption{Results of the UKM ($\hat{X}$ and OU of $\hat{X}$) on the $5$-fold datasets with $5$ different random seeds for the MNIST256 dataset ($0$ or non-$0$). We use complex matrices and set $\theta_\mathrm{bias} = 0$. We set $r = 0.010$.}
\label{supp-arXiv-numerical-result-raw-data-fold-001-rand-001-UKM-OUU-MNIST256-0-non0}
\end{figure*}

In Fig.~\ref{supp-arXiv-numerical-result-raw-data-fold-001-rand-001-QCL-MNIST256-0-non0}, we show the numerical results of QCL for the $5$-fold datasets with $5$ different random seeds.
In Fig.~\ref{supp-arXiv-numerical-result-raw-data-fold-001-rand-001-UKM-P-MNIST256-0-non0}, we show the numerical results of $\hat{P}$ of the UKM for the $5$-fold datasets with $5$ different random seeds.
In Fig.~\ref{supp-arXiv-numerical-result-raw-data-fold-001-rand-001-UKM-OUU-MNIST256-0-non0}, we also show the numerical results of OU of $\hat{X}$ of the UKM for the $5$-fold datasets with $5$ different random seeds.

\begin{figure}[htb]
\centering
\includegraphics[scale=0.45]{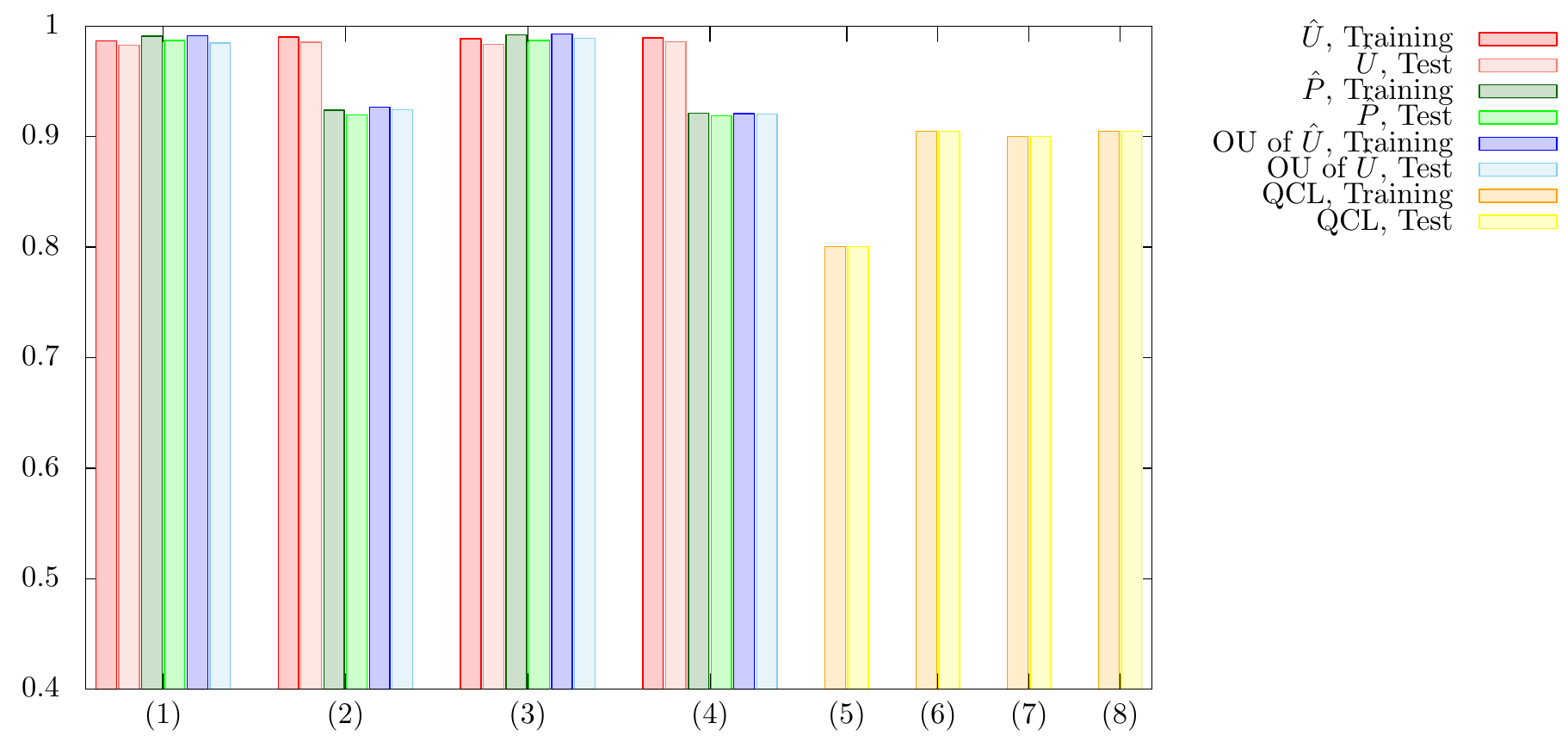}
\caption{Results of $5$-fold CV with $5$ different random seeds for the MNIST256 dataset ($0$ or non-$0$). For the UKM, we put $r = 0.010$ and set $K = 10$ and $K' = 5$. For QCL, the number of layers $L$ is $5$ and the number of iterations is $50$. The numerical settings are as follows: (1) complex matrices without the bias term, (2) complex matrices with the bias term, (3) real matrices without the bias term, (4) real matrices with the bias term, (5) CNOT-based circuit without the bias term, (6) CNOT-based circuit with the bias term, (7) CRot-based circuit without the bias term, (8) CRot-based circuit with the bias term, (9) 1d Heisenberg circuit without the bias term, (10) 1d Heisenberg circuit with the bias term, (11) FC Heisenberg circuit without the bias term, and (12) FC Heisenberg circuit with the bias term.}
\label{supp-arXiv-numerical-result-performance-UKM-QCL-MNIST256-0-non0}
\end{figure}

We summarize the results of 5-fold CV with 5 different random seeds of QCL and the UKM in Tables~\ref{supp-arXiv-table-MNIST256-0-non0-002} and \ref{supp-arXiv-table-MNIST256-0-non0-001}, respectively.
For QCL and the UKM, we select the best model for the training dataset over iterations to compute the performance.
\begin{table}[htb]
  \begin{tabular}{cc|cc}
    \hline \hline
    Algo. & Condition & Training & Test \\
    \hline
  QCL & CNOT-based, w/o bias & 0.8018 & 0.8006 \\
  QCL & CNOT-based, w/ bias & 0.9053 & 0.9050 \\
    \hline
  QCL & CRot-based, w/o bias & 0.9010 & 0.9002 \\
  QCL & CRot-based, w/ bias & 0.9053 & 0.9048 \\
    \hline \hline
  \end{tabular}
\caption{Results of $5$-fold CV with $5$ different random seeds of QCL for the MNIST256 dataset ($0$ or non-$0$). The number of layers $L$ is $5$ and the number of iterations is $50$.}
\label{supp-arXiv-table-MNIST256-0-non0-002}
\end{table}
\begin{table}[htb]
  \begin{tabular}{cc|cc}
    \hline \hline
    Algo. & Condition & Training & Test \\
    \hline
  UKM & $\hat{X}$, complex, w/o bias & 0.9868 & 0.9829 \\
  UKM & $\hat{P}$, complex, w/o bias & 0.9911 & 0.9870 \\
  UKM & OU of $\hat{X}$, complex, w/o bias & 0.9915 & 0.9848 \\
    \hline
  UKM & $\hat{X}$, complex, w/ bias & 0.9902 & 0.9856 \\
  UKM & $\hat{P}$, complex, w/ bias & 0.9238 & 0.9199 \\
  UKM & OU of $\hat{X}$, complex, w/ bias & 0.9264 & 0.9244 \\
    \hline
  UKM & $\hat{X}$, real, w/o bias & 0.9886 & 0.9836 \\
  UKM & $\hat{P}$, real, w/o bias & 0.9922 & 0.9871 \\
  UKM & OU of $\hat{X}$, real, w/o bias & 0.9927 & 0.9889 \\
    \hline
  UKM & $\hat{X}$, real, w/ bias & 0.9894 & 0.9859 \\
  UKM & $\hat{P}$, real, w/ bias & 0.9213 & 0.9188 \\
  UKM & OU of $\hat{X}$, real, w/ bias & 0.9208 & 0.9205 \\
    \hline \hline
  \end{tabular}
\caption{Results of $5$-fold CV with $5$ different random seeds of the UKM for the MNIST256 dataset ($0$ or non-$0$). We put $r = 0.010$ and set $K = 10$ and $K' = 5$.}
\label{supp-arXiv-table-MNIST256-0-non0-001}
\end{table}
In Fig.~\ref{supp-arXiv-numerical-result-performance-UKM-QCL-MNIST256-0-non0}, we plot the data shown in Tables~\ref{supp-arXiv-table-MNIST256-0-non0-002} and \ref{supp-arXiv-table-MNIST256-0-non0-001}.
We also summarize the results of 5-fold CV with 5 different random seeds of the kernel method in Table~\ref{supp-arXiv-table-UCI-MNIST256-0-non0-003}.
More specifically, we use Ridge classification in Sec.~\ref{supp-arXiv-sec-Ridge-001}.
We consider the linear functions and the second-order polynomial functions for $\phi (\cdot)$ in Eq.~\eqref{supp-arXiv-f-pred-kernel-method-001-002} with and without normalization.
We set $\lambda = 10^{-2}, 10^{-1}, 1$ where $\lambda$ is the coefficient of the regularization term.
\begin{table}[htb]
  \begin{tabular}{cc|cc}
    \hline \hline
    Algo. & Condition & Training & Test \\
    \hline
  Kernel method & Linear, w/o normalization, $\lambda = 10^{-2}$ & 0.9963 & 0.9921 \\
  Kernel method & Linear, w/o normalization, $\lambda = 10^{-1}$ & 0.9963 & 0.9921 \\
  Kernel method & Linear, w/o normalization, $\lambda = 1$ & 0.9963 & 0.9921 \\
    \hline
  Kernel method & Linear, w/ normalization, $\lambda = 10^{-2}$ & 0.9893 & 0.98189 \\
  Kernel method & Linear, w/ normalization, $\lambda = 10^{-1}$ & 0.9856 & 0.9792 \\
  Kernel method & Linear, w/ normalization, $\lambda = 1$ & 0.9843 & 0.9787 \\
    \hline
  Kernel method & Poly-2, w/o normalization, $\lambda = 10^{-2}$ & 1.0000 & 0.9906 \\
  Kernel method & Poly-2, w/o normalization, $\lambda = 10^{-1}$ & 1.0000 & 0.9906 \\
  Kernel method & Poly-2, w/o normalization, $\lambda = 1$ & 1.0000 & 0.9906 \\
    \hline
  Kernel method & Poly-2, w/ normalization, $\lambda = 10^{-2}$ & 0.9997 & 0.9949 \\
  Kernel method & Poly-2, w/ normalization, $\lambda = 10^{-1}$ & 0.9992 & 0.9953 \\
  Kernel method & Poly-2, w/ normalization, $\lambda = 1$ & 0.9946 & 0.9916 \\
    \hline \hline
  \end{tabular}
\caption{Results of 5-fold CV with 5 different random seeds of the kernel method for the MNIST256 dataset ($0$ or non-$0$).}
\label{supp-arXiv-table-UCI-MNIST256-0-non0-003}
\end{table}

\begin{figure}[htb]
\centering
\includegraphics[scale=0.45]{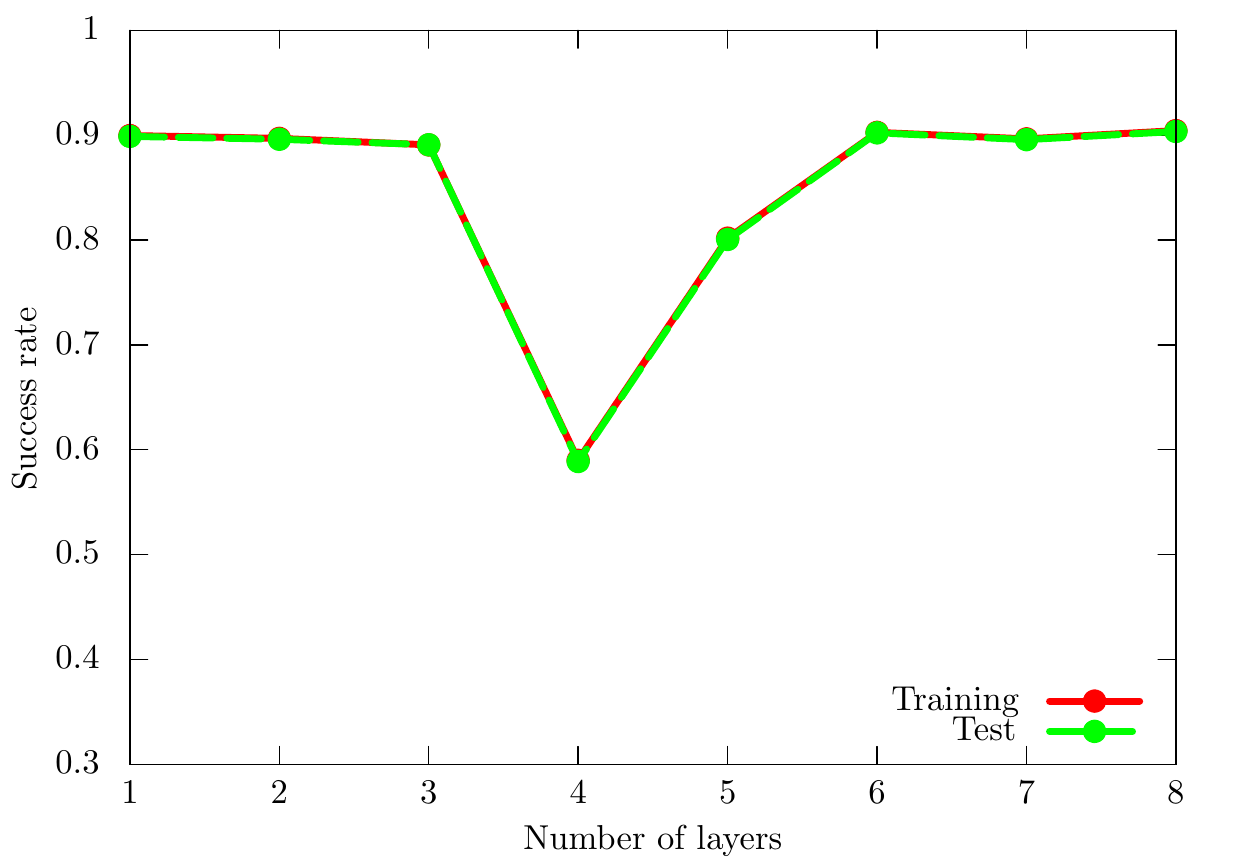}
\caption{Performance dependence of QCL on the number of layers $L$ for the MNIST256 dataset ($0$ or non-$0$). We use the CNOT-based circuit geometry and set $\theta_\mathrm{bias} = 0$. We iterate the computation $50$ times.}
\label{supp-arXiv-numerical-result-layers-dependence-QCL-MNIST256-0-non0}
\end{figure}

\begin{figure}[htb]
\centering
\includegraphics[scale=0.45]{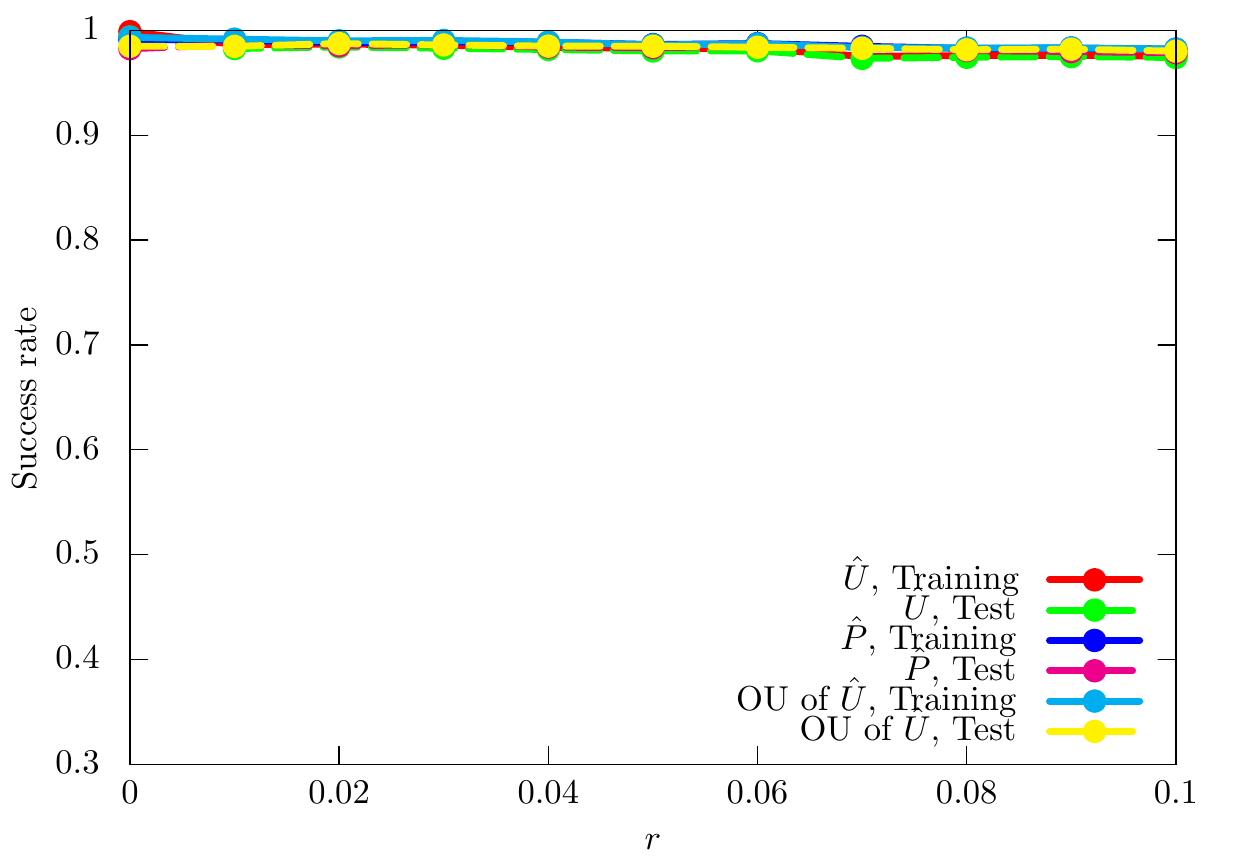}
\includegraphics[scale=0.45]{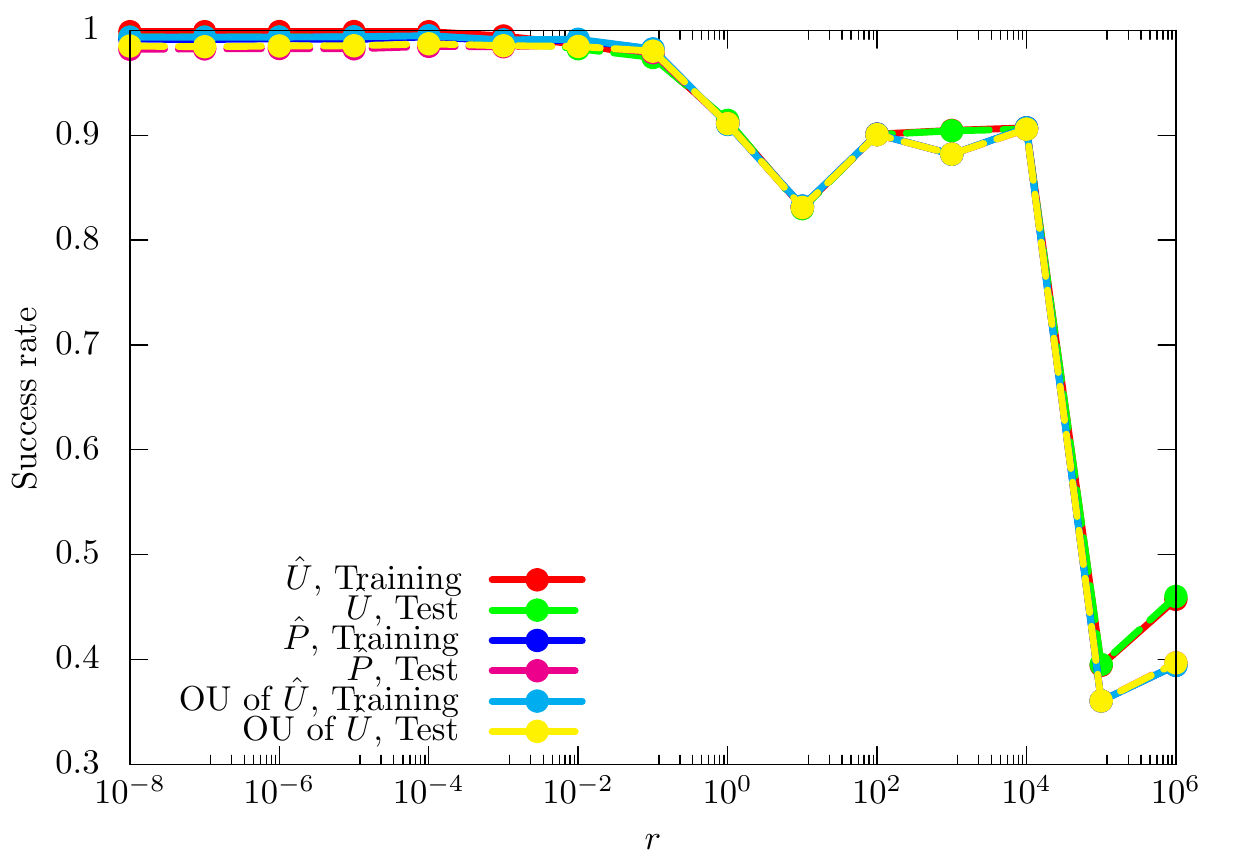}
\caption{Performance dependence of the UKM on $r$, which is the coefficient of the second term in the right-hand side of Eq.~\eqref{supp-arXiv-quantum-kernel-method-001-011} for the MNIST256 dataset ($0$ or non-$0$). We use complex matrices and set $\theta_\mathrm{bias} = 0$. We set $K = 10$ and $K' = 5$.}
\label{supp-arXiv-numerical-result-r-dependence-UKM-MNIST256-0-non0}
\end{figure}

\begin{figure}[htb]
\centering
\includegraphics[scale=0.45]{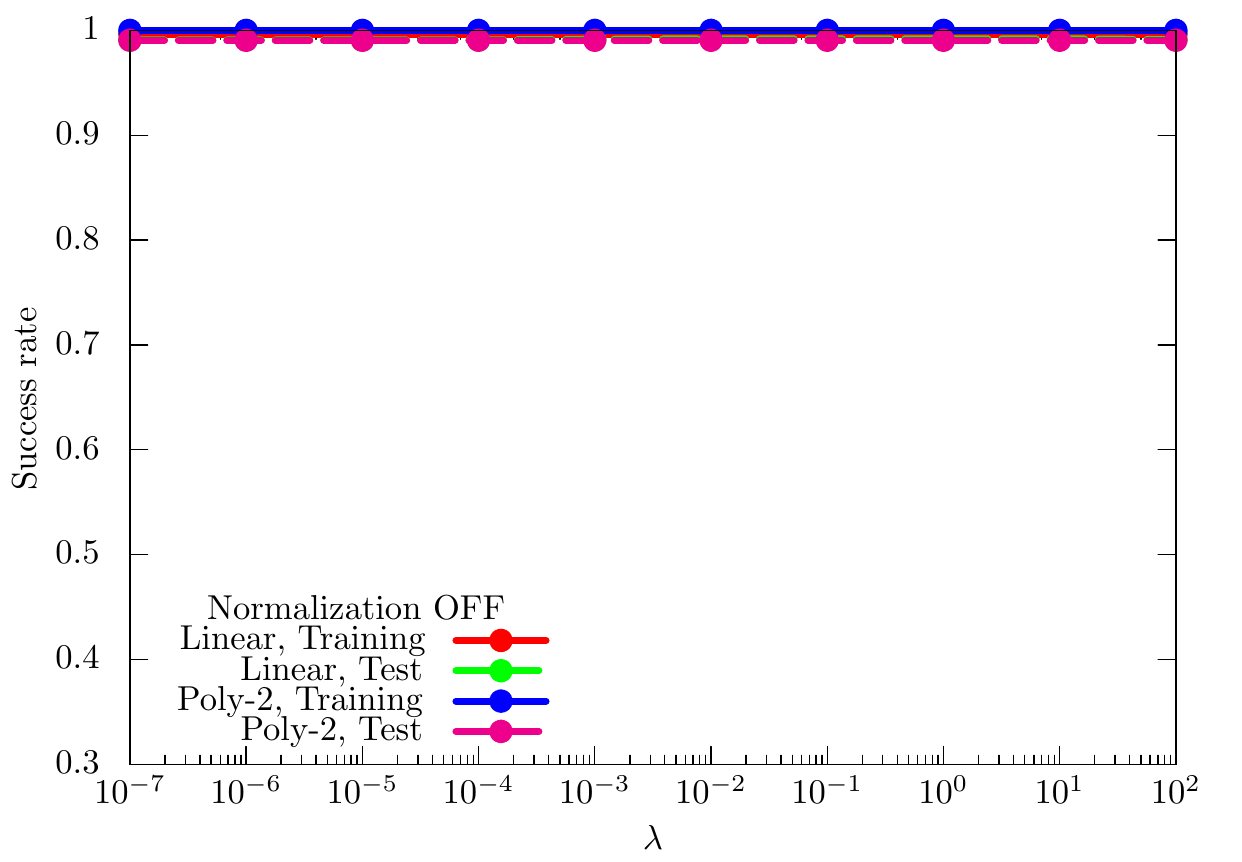}
\includegraphics[scale=0.45]{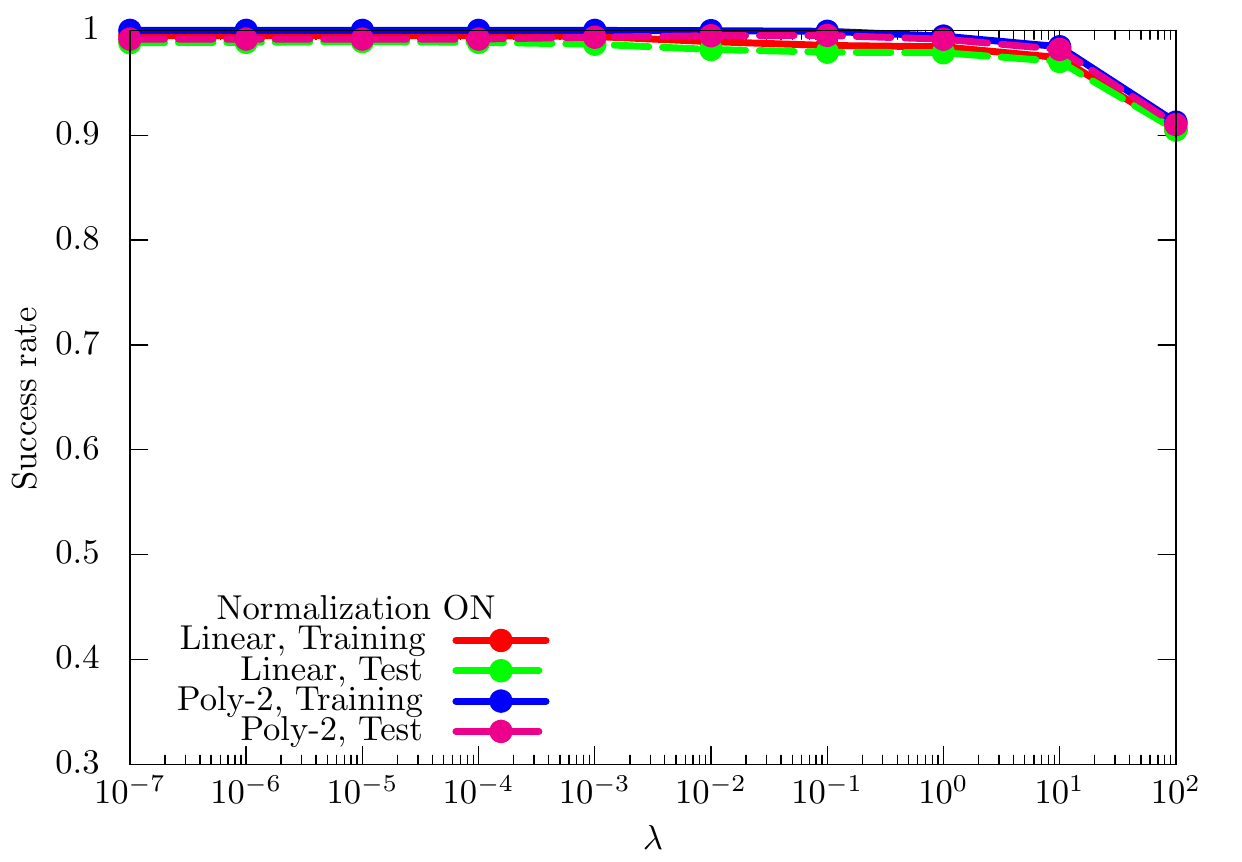}
\caption{Performance dependence of the kernel method on $\lambda$, which is the coefficient of the second term in the right-hand side of Eq.~\eqref{supp-arXiv-cost-function-kernel-method-001-002} for the MNIST256 dataset ($0$ or non-$0$). For $\phi (\cdot)$ in Eq.~\eqref{supp-arXiv-f-pred-kernel-method-001-002}, we use the linear functions and the second-degree polynomial functions with and without normalization.}
\label{supp-arXiv-numerical-result-lambda-dependence-kernel-method-MNIST256-0-non0}
\end{figure}

Next, we show the performance dependence of the three algorithms on their key parameters.
We see the performance dependence of QCL on the number of layers $L$.
The result is shown in Fig.~\ref{supp-arXiv-numerical-result-layers-dependence-QCL-MNIST256-0-non0}.
We then see the performance dependence of the UKM on $r$, which is the coefficient of the second term in the right-hand side of Eq.~\eqref{supp-arXiv-quantum-kernel-method-001-011}.
The result is shown in Fig.~\ref{supp-arXiv-numerical-result-r-dependence-UKM-MNIST256-0-non0}.
In Fig.~\ref{supp-arXiv-numerical-result-lambda-dependence-kernel-method-MNIST256-0-non0}, we show the performance dependence of the kernel method on $\lambda$, which is the coefficient of the second term in the right-hand side of Eq.~\eqref{supp-arXiv-cost-function-kernel-method-001-002}.

\begin{figure}[htb]
\centering
\includegraphics[scale=0.45]{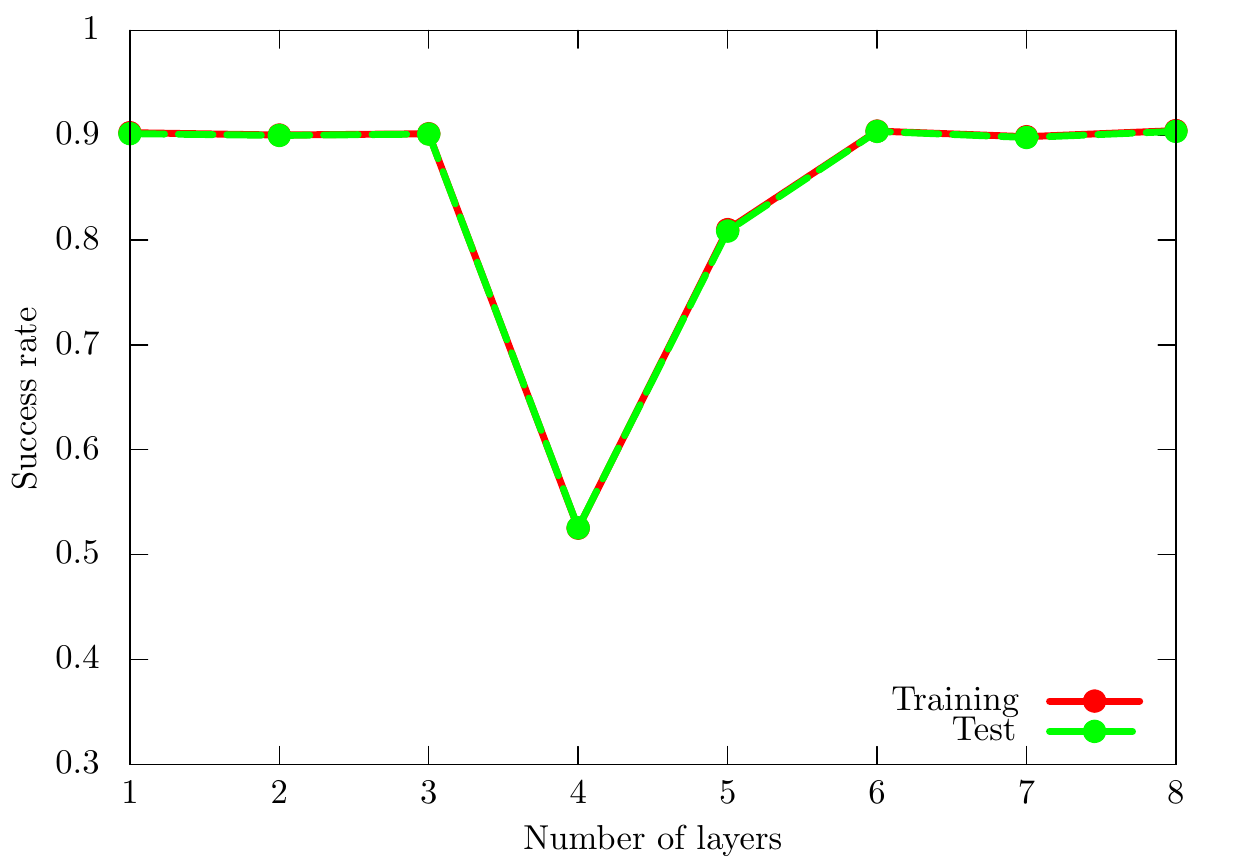}
\caption{Performance dependence of QCL on the number of layers $L$ for the MNIST256 dataset ($0$ or non-$0$) in the case of the hinge function, Eq.~\eqref{supp-arXiv-hinge-function-001-001}. We use the CNOT-based circuit geometry and set $\theta_\mathrm{bias} = 0$. We iterate the computation $300$ times.}
\label{supp-arXiv-numerical-result-layers-dependence-QCL-MNIST256-0-non0-hinge}
\end{figure}

\begin{figure}[htb]
\centering
\includegraphics[scale=0.45]{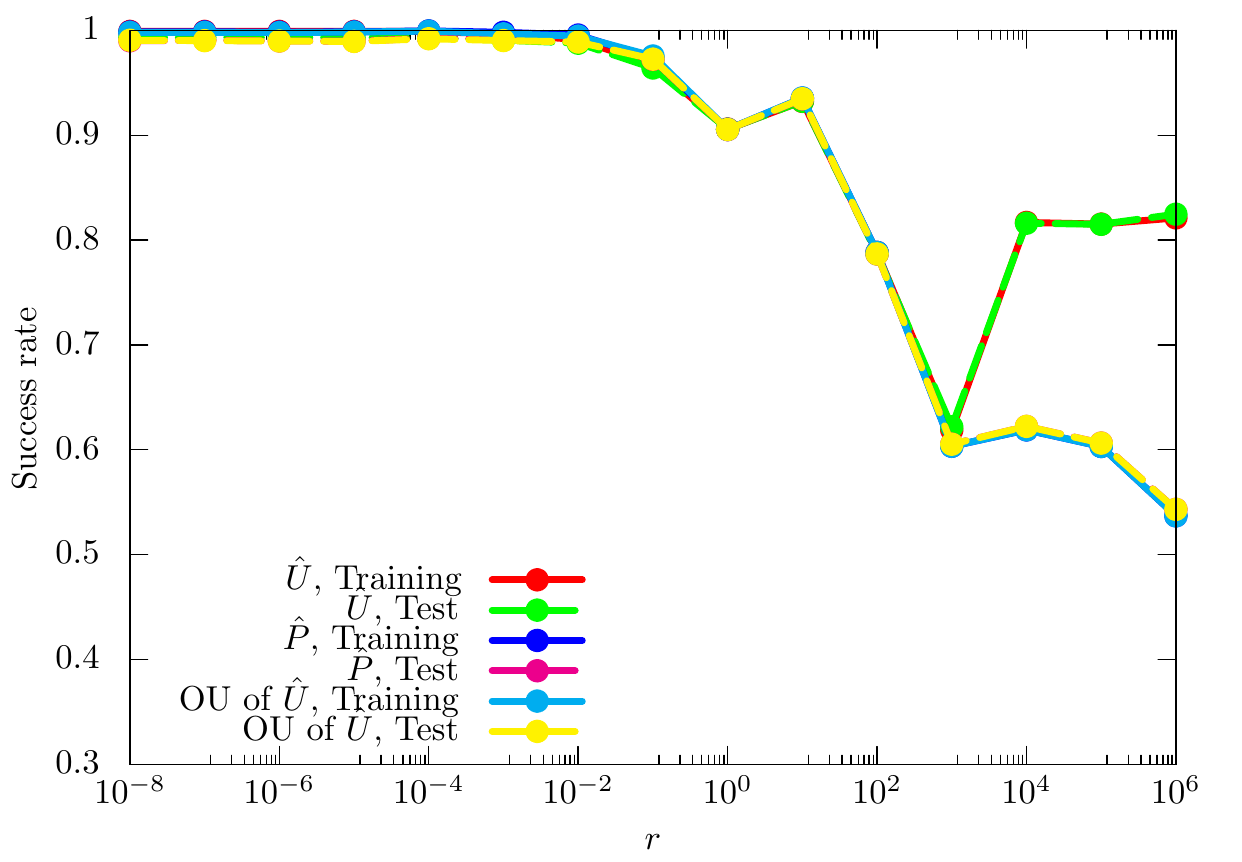}
\caption{Performance dependence of the UKM on $r$, which is the coefficient of the second term in the right-hand side of Eq.~\eqref{supp-arXiv-quantum-kernel-method-001-011} for the MNIST256 dataset ($0$ or non-$0$) in the case of the hinge function, Eq.~\eqref{supp-arXiv-hinge-function-001-001}. We use complex matrices and set $\theta_\mathrm{bias} = 0$. We set $K = 30$ and $K' = 10$.}
\label{supp-arXiv-numerical-result-r-dependence-UKM-MNIST256-0-non0-hinge}
\end{figure}

So far, we have used the squared error function, Eq.~\eqref{supp-arXiv-squared-error-function-001-001}.
In Fig.~\ref{supp-arXiv-numerical-result-layers-dependence-QCL-MNIST256-0-non0-hinge}, we show the performance dependence of QCL on the number of layers $L$ in the case of the hinge function, Eq.~\eqref{supp-arXiv-hinge-function-001-001}.
In Fig.~\ref{supp-arXiv-numerical-result-r-dependence-UKM-MNIST256-0-non0-hinge}, we show the performance dependence of the UKM on $r$, which is the coefficient of the second term in the right-hand side of Eq.~\eqref{supp-arXiv-quantum-kernel-method-001-011}, in the case of the hinge function, Eq.~\eqref{supp-arXiv-hinge-function-001-001}.

\clearpage

\subsection{Discussions}

In the main text, we give the disccuions on the numerical results.
The results shown in this section are basically consistent with the results shown in the main text and support the statement.

\section{Numerical simulation of the VCR} \label{supp-arXiv-sec-numerical-result-VCR-001}

In this section, we show numerical simulation on the VCR.
For $\hat{U}_\mathrm{c} (\theta)$, we use the CNOT-based circuit geometry~\eqref{supp-arXiv-gate-CNOT-001-001}.

\subsection{Numerical setting} \label{supp-arXiv-sec-numerical-settings-VCR-001}

In this section, we use Eq.~\eqref{supp-arXiv-def-g2-001} with $p = 2$ as the optimization function of the VCR.
For $\hat{U}_\mathrm{c} (\theta)$, we use the CNOT-based circuit.

\subsection{Datasets} \label{supp-arXiv-sec-datasets-VCR-001-001}

In this section, we consider the iris dataset ($1$ or non-$1$), the cancer dataset ($0$ or $1$), and the wine dataset ($0$ or non-$0$) described in Sec.~\ref{supp-arXiv-sec-datasets-UKM-001-001}.
Refer to Sec.~\ref{supp-arXiv-sec-datasets-UKM-001-001} for the details.

\subsection{Iris dataset ($1$ or non-$1$)}

\begin{figure}[t]
\centering
\includegraphics[scale=0.45]{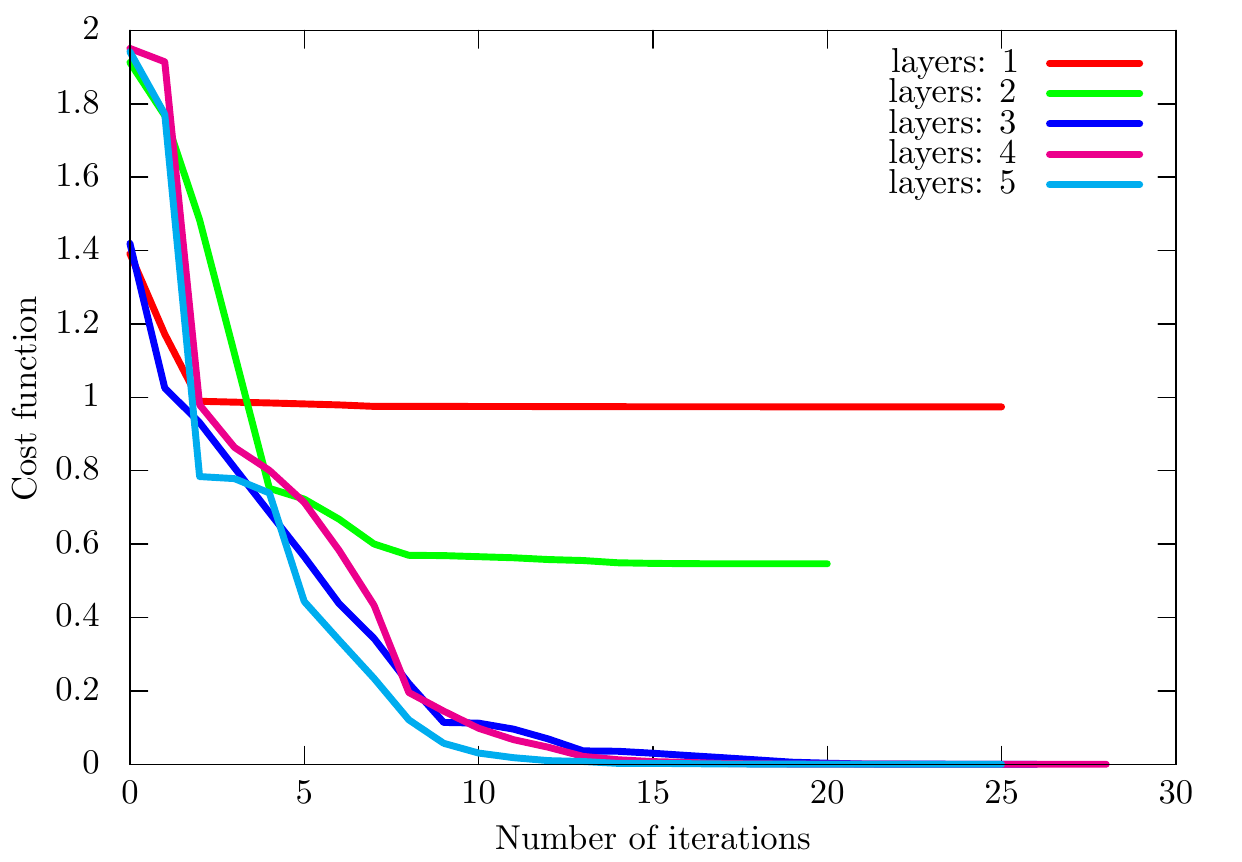}
\caption{Values of the cost function $\mathcal{J}_\mathrm{cost} (\theta, \lambda; L, 2, \hat{U})$, Eq.~\eqref{supp-arXiv-def-g2-001} with $p = 2$, for the iris dataset ($1$ or non-$1$).}
\label{supp-arXiv-numerical-result-VCR-UCI-iris-1-non1-001}
\end{figure}

Let us consider the iris dataset ($1$ or non-$1$) and minimizing Eq.~\eqref{supp-arXiv-def-g2-001} with $p = 2$.
As $\hat{U}$, we use $\hat{P}$ computed by the UKM under the condition of real matrices, $r = 0.010$, and $\theta_\mathrm{b} = 0$ that gives the success rate for the training dataset $0.7565$ and that for the test dataset $0.8286$.
In Fig.~\ref{supp-arXiv-numerical-result-VCR-UCI-iris-1-non1-001}, we show the values of the cost function $\mathcal{J}_\mathrm{cost} (\theta, \lambda; L, 2, \hat{U})$, Eq.~\eqref{supp-arXiv-def-g2-001} with $p = 2$.
We vary the number of layers $L$.
In Table~\ref{supp-arXiv-table-VCR-UCI-iris-1-non1-001}, we summarize the performance of the input unitary operator, QCL, and the circuit geometry computed by the VCR.
We have $L_{0.001} = 3$.
\begin{table}[htb]
  \begin{tabular}{cc|ccc}
    \hline \hline
    Algo. & Condition & Cost & Training & Test \\
    \hline
    Input & UKM, $\hat{P}$, real, w/o bias & --- & 0.7565 & 0.8286 \\
    \hline
    VCR & \# of layers: 1 & 0.9740 & 0.3565 & 0.2571 \\
    VCR & \# of layers: 2 & 0.5465 & 0.3565 & 0.2571 \\
    VCR & \# of layers: 3 & 0.0000 & 0.7565 & 0.8286 \\
    VCR & \# of layers: 4 & 0.0000 & 0.7565 & 0.8286 \\
    VCR & \# of layers: 5 & 0.0000 & 0.7565 & 0.8286 \\
    \hline
    UKM & $\hat{P}$, real, w/o bias & --- & 0.7880 & 0.7789 \\
    QCL & \# of layers: 5 & --- & 0.6851 & 0.5845 \\
    \hline \hline
  \end{tabular}
\caption{Performance of the VCR for the iris dataset ($1$ or non-$1$). We show the success rates for the training and test datasets and the values of the cost function $\mathcal{J}_\mathrm{cost} (\theta, \lambda; L, 2, \hat{U})$, Eq.~\eqref{supp-arXiv-def-g2-001} with $p = 2$. The input for the VCR is $\hat{P}$ created by the UKM under the condition of real matrices, $r = 0.010$, and $\theta_\mathrm{b} = 0$. For reference, we add the last two rows that show the results of 5-fold CV.}
\label{supp-arXiv-table-VCR-UCI-iris-1-non1-001}
\end{table}

\subsection{Cancer dataset ($0$ or $1$)}

\begin{figure}[t]
\centering
\includegraphics[scale=0.45]{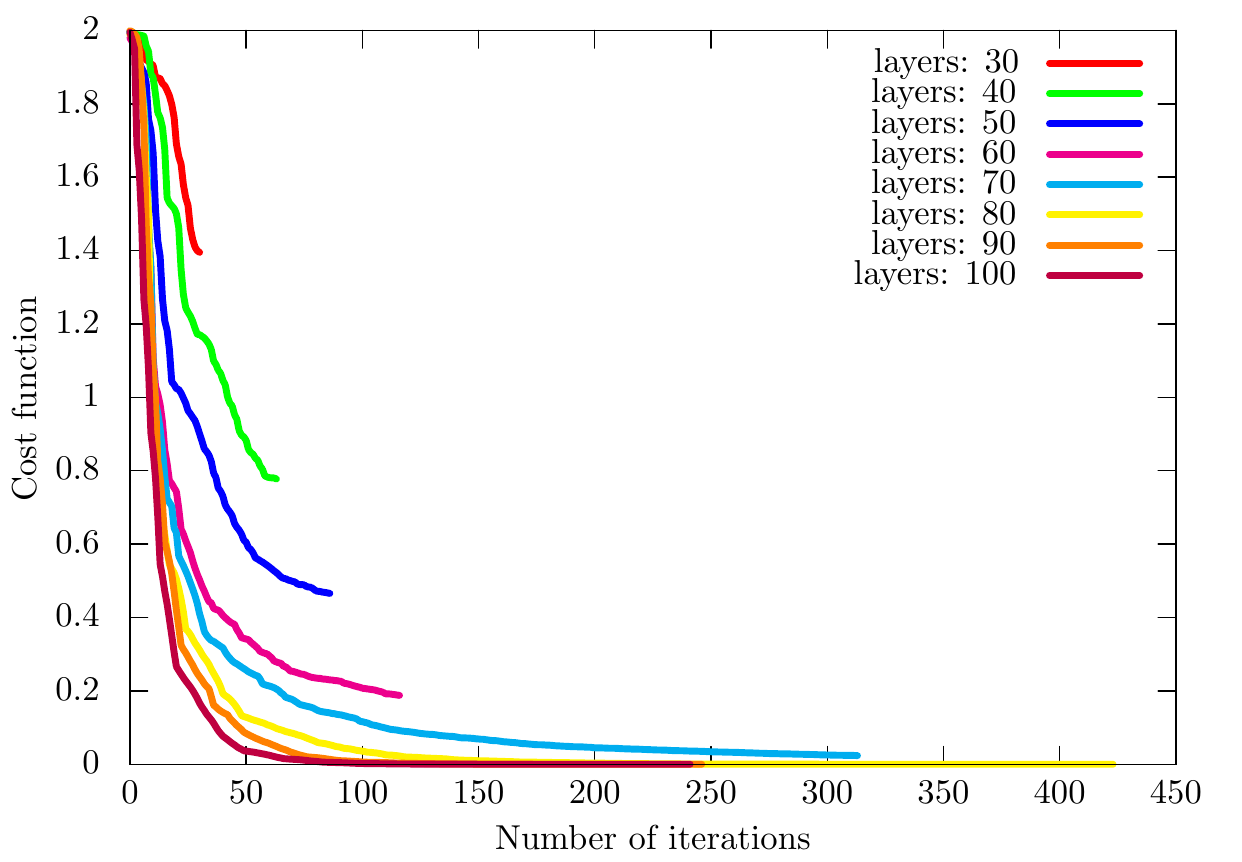}
\caption{Values of the cost function $\mathcal{J}_\mathrm{cost} (\theta, \lambda; L, 2, \hat{U})$, Eq.~\eqref{supp-arXiv-def-g2-001} with $p = 2$, for the cancer dataset ($0$ or $1$).}
\label{supp-arXiv-numerical-result-VCR-UCI-cancer-0-1-001}
\end{figure}

Let us consider the cancer dataset ($0$ or $1$) and minimizing Eq.~\eqref{supp-arXiv-def-g2-001} with $p = 2$.
As $\hat{U}$, we use $\hat{P}$ computed by the UKM under the condition of real matrices, $r = 0.010$, and $\theta_\mathrm{b} = 0$ that gives the success rate for the training dataset $0.9139$ and that for the test dataset $0.9483$.
In Fig.~\ref{supp-arXiv-numerical-result-VCR-UCI-cancer-0-1-001}, we show the values of the cost function $\mathcal{J}_\mathrm{cost} (\theta, \lambda; L, 2, \hat{U})$, Eq.~\eqref{supp-arXiv-def-g2-001} with $p = 2$.
We vary the number of layers $L$.
In Table~\ref{supp-arXiv-table-VCR-UCI-cancer-0-1-001}, we summarize the performance of the input unitary operator, QCL, and the circuit geometry computed by the VCR.
We have $L_{0.001} = 80$.
\begin{table}[htb]
  \begin{tabular}{cc|ccc}
    \hline \hline
    Algo. & Condition & Cost & Training & Test \\
    \hline
    Input & UKM, $\hat{P}$, real, w/o bias & --- & 0.9139 & 0.9483 \\
    \hline
    VCR & \# of layers: 10 & 1.9694 & 0.3929 & 0.2931 \\
    VCR & \# of layers: 20 & 1.9734 & 0.6071 & 0.7069 \\
    VCR & \# of layers: 30 & 1.3950 & 0.6071 & 0.7069 \\
    VCR & \# of layers: 40 & 0.7777 & 0.6909 & 0.7586 \\
    VCR & \# of layers: 50 & 0.4657 & 0.8499 & 0.9224 \\
    VCR & \# of layers: 60 & 0.1877 & 0.9073 & 0.9483 \\
    VCR & \# of layers: 70 & 0.0236 & 0.9073 & 0.9483 \\
    VCR & \# of layers: 80 & 0.0000 & 0.9139 & 0.9483 \\
    VCR & \# of layers: 90 & 0.0000 & 0.9139 & 0.9483 \\
    VCR & \# of layers: 100 & 0.0000 & 0.9139 & 0.9483 \\
    \hline
    UKM & $\hat{P}$, real, w/o bias & --- & 0.9194 & 0.9131 \\
    QCL & \# of layers: 5 & --- & 0.8798 & 0.8768 \\
    QCL & \# of layers: 10 & --- & 0.7814 & 0.7767 \\
    \hline \hline
  \end{tabular}
\caption{Performance of the VCR for the cancer dataset ($0$ or $1$). We show the success rates for the training and test datasets and the values of the cost function $\mathcal{J}_\mathrm{cost} (\theta, \lambda; L, 2, \hat{U})$, Eq.~\eqref{supp-arXiv-def-g2-001} with $p = 2$. The input for the VCR is $\hat{P}$ created by the UKM under the condition of real matrices, $r = 0.010$, and $\theta_\mathrm{b} = 0$. For reference, we add the last three rows that show the results of 5-fold CV.}
\label{supp-arXiv-table-VCR-UCI-cancer-0-1-001}
\end{table}

\subsection{Wine dataset ($0$ or non-$0$)}

\begin{figure}[t]
\centering
\includegraphics[scale=0.45]{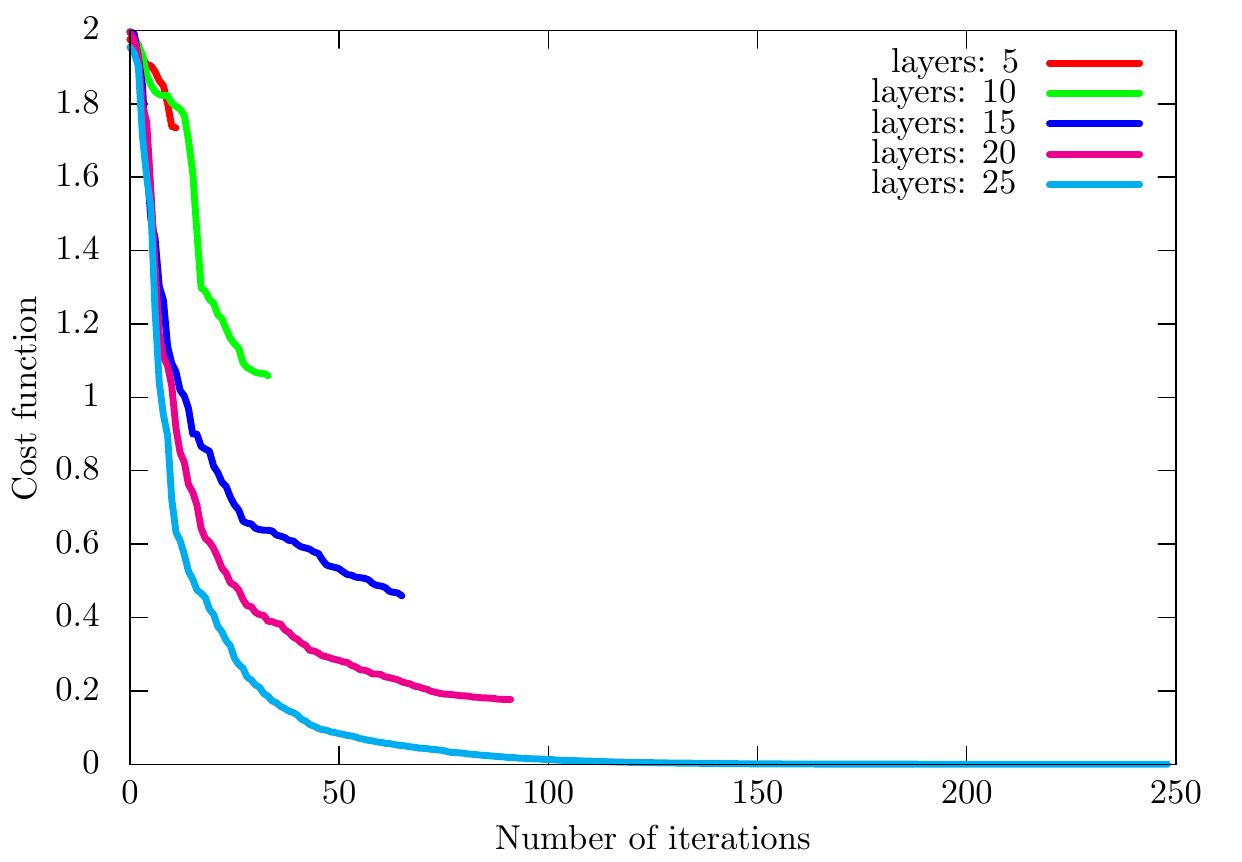}
\caption{Values of the cost function $\mathcal{J}_\mathrm{cost} (\theta, \lambda; L, 2, \hat{U})$, Eq.~\eqref{supp-arXiv-def-g2-001} with $p = 2$, for the wine dataset ($0$ or non-$0$).}
\label{supp-arXiv-numerical-result-VCR-UCI-wine-0-non0-001}
\end{figure}

Let us consider the wine dataset ($0$ or non-$0$) and minimizing Eq.~\eqref{supp-arXiv-def-g2-001} with $p = 2$.
As $\hat{U}$, we use $\hat{P}$ computed by the UKM under the condition of real matrices, $r = 0.010$, and $\theta_\mathrm{b} = 0$ that gives the success rate for the training dataset $0.9179$ and that for the test dataset $0.9091$.
In Fig.~\ref{supp-arXiv-numerical-result-VCR-UCI-wine-0-non0-001}, we show the values of the cost function $\mathcal{J}_\mathrm{cost} (\theta, \lambda; L, 2, \hat{U})$, Eq.~\eqref{supp-arXiv-def-g2-001} with $p = 2$.
We vary the number of layers $L$.
In Table~\ref{supp-arXiv-table-VCR-UCI-wine-0-non0-001}, we summarize the performance of the input unitary operator, QCL, and the circuit geometry computed by the VCR.
We have $L_{0.001} = 25$.
\begin{table}[htb]
  \begin{tabular}{cc|ccc}
    \hline \hline
    Algo. & Condition & Cost & Training & Test \\
    \hline
    Input & UKM, $\hat{P}$, real, w/o bias & --- & 0.9179 & 0.9091 \\
    \hline
    VCR & \# of layers: 5 & 1.7343 & 0.6642 & 0.6818 \\
    VCR & \# of layers: 10 & 1.0591 & 0.4254 & 0.4091 \\
    VCR & \# of layers: 15 & 0.4588 & 0.6791 & 0.7045 \\
    VCR & \# of layers: 20 & 0.1765 & 0.7836 & 0.8864 \\
    VCR & \# of layers: 25 & 0.0000 & 0.9179 & 0.9091 \\
    \hline
    UKM & $\hat{P}$, real, w/o bias & --- & 0.9200 & 0.9185 \\
    QCL & \# of layers: 5 & --- & 0.9057 & 0.9052 \\
    \hline \hline
  \end{tabular}
\caption{Performance of the VCR for the wine dataset ($0$ or non-$0$). We show the success rates for the training and test datasets and the values of the cost function $\mathcal{J}_\mathrm{cost} (\theta, \lambda; L, 2, \hat{U})$, Eq.~\eqref{supp-arXiv-def-g2-001} with $p = 2$. The input for the VCR is $\hat{P}$ created by the UKM under the condition of real matrices, $r = 0.010$, and $\theta_\mathrm{b} = 0$. For reference, we add the last two rows that show the results of 5-fold CV.}
\label{supp-arXiv-table-VCR-UCI-wine-0-non0-001}
\end{table}

\subsection{Discussions}

The discussions on the VCR is given in the main text.
In this section, we give the results of the VCR on three different datasets: the iris dataset ($1$ or non-$1$), the cancer dataset ($0$ or $1$), and the wine dataset ($0$ or non-$0$).
The results shown in this section support the statement in the main text.

\section{Conclusion} \label{supp-arXiv-conc-01}

In the SI, we have provided supplemental information on "Ansatz-Independent Variational Quantum Classifier."

\end{document}